\documentclass[12pt]{article}
\makeatletter \@addtoreset{equation}{section} \makeatother
\renewcommand{\theequation}{\thesection.\arabic{equation}}
\newcommand{\ba}{\begin{array}}
\newcommand{\ea}{\end{array}}
\newcommand{\beq}{\begin{equation}}
\newcommand{\eeq}{\end{equation}}
\newcommand{\bea}{\begin{eqnarray}}
\newcommand{\eea}{\end{eqnarray}}

\def\bce{\begin{center}}
\def\ece{\end{center}}

\def\nonu{\nonumber}

\def\pa{\partial}

\def\be{\beta}

\def\ep{\epsilon}

\def\eps6{{\displaystyle \mathop{\epsilon}^{6}}{}}
\def\g6{{\displaystyle \mathop{g}^{6}}{}}

\def\nab6{{\displaystyle \mathop{\nabla}^{6}}{}}

\def\0{{\sst{(0)}}}
\def\1{{\sst{(1)}}}
\def\2{{\sst{(2)}}}
\def\3{{\sst{(3)}}}
\def\4{{\sst{(4)}}}
\def\5{{\sst{(5)}}}
\def\6{{\sst{(6)}}}
\def\7{{\sst{(7)}}}
\def\8{{\sst{(8)}}}

\def\ba{\begin{array}}
\def\ea{\end{array}}
\def\beq{\begin{equation}}
\def\eeq{\end{equation}}
\def\be{\begin{equation}}
\def\ee{\end{equation}}

\def\eps{\epsilon}

\def\c{{\gamma}}

\def\ba{\begin{array}}
\def\ea{\end{array}}
\def\beq{\begin{equation}}
\def\eeq{\end{equation}}
\def\be{\begin{equation}}
\def\ee{\end{equation}}

\def\eps{\epsilon}

\def\c{{\gamma}}

\def\eps6{{\displaystyle \mathop{\epsilon}^{6}}{}}

\def\nab6{{\displaystyle \mathop{\nabla}^{6}}{}}

\newcommand{\bean}{\begin{eqnarray*}}
\newcommand{\eean}{\end{eqnarray*}}

\begin{document}
\thispagestyle{empty} \addtocounter{page}{-1}
   \begin{flushright}
\end{flushright}

\vspace*{1.3cm}
  
\centerline{ \Large \bf   
The Operator  Product  Expansion  between}
\vspace*{0.3cm}
\centerline{ \Large \bf the $16$ Lowest Higher Spin Currents
in the ${\cal N}=4$ Superspace } 
\vspace*{1.5cm}
\centerline{{\bf Changhyun Ahn } and {\bf Man Hea Kim}
} 
\vspace*{1.0cm} 
\centerline{\it 
Department of Physics, Kyungpook National University, Taegu
41566, Korea} 
\vspace*{0.8cm} 
\centerline{\tt ahn@knu.ac.kr, \qquad manhea@knu.ac.kr 
} 
\vskip1cm

\centerline{\bf Abstract}
\vspace*{0.5cm}

Some of the operator product expansions (OPEs) between 
the lowest $16$ higher spin currents of spins 
$(1, \frac{3}{2}, \frac{3}{2}, \frac{3}{2}, \frac{3}{2}, 2, 2, 2, 2, 2, 2,
\frac{5}{2}, \frac{5}{2}, \frac{5}{2}, \frac{5}{2}, 3)$
in an extension of the large ${\cal N}=4$ linear superconformal algebra
 were constructed in the ${\cal N}=4$ superconformal coset 
$\frac{SU(5)}{SU(3)}$ theory previously.
In this paper, by rewriting the above OPEs in the 
${\cal N}=4$ superspace developed by Schoutens (and other groups),  
the remaining undetermined OPEs where the corresponding 
singular terms possess the composite fields 
with spins $s =\frac{7}{2}, 4, \frac{9}{2}, 5$ 
are completely determined.
Furthermore, by introducing the arbitrary coefficients 
in front of the composite fields
in the right hand sides of the above complete $136$ OPEs,
reexpressing them in the ${\cal N}=2$ superspace 
and using the ${\cal N}=2$ OPEs 
mathematica package by Krivonos and Thielemans,
the complete structures of the above OPEs with fixed coefficient functions
are obtained with the help of 
various Jacobi identities.
Then one obtains ten ${\cal N }=2$ super OPEs between 
the four ${\cal N}=2$ 
higher spin currents denoted by 
$(1, \frac{3}{2}, \frac{3}{2}, 2), (\frac{3}{2}, 2, 2, \frac{5}{2}), 
(\frac{3}{2}, 2, 2, \frac{5}{2})$ and $(2, \frac{5}{2}, \frac{5}{2}, 3)$ 
(corresponding $136$ OPEs in the component 
approach) in 
 the ${\cal N}=4$ superconformal coset 
$\frac{SU(N+2)}{SU(N)}$ theory.
Finally, one describes them as one single ${\cal N}=4$ super OPE 
between the above sixteen 
higher spin currents in the ${\cal N}=4$ superspace. 
The fusion rule for this OPE contains 
the next $16$ higher spin currents of spins 
of
$(2, \frac{5}{2}, \frac{5}{2}, \frac{5}{2}, \frac{5}{2}, 3, 3, 3, 3, 3, 3,
\frac{7}{2}, \frac{7}{2}, \frac{7}{2}, \frac{7}{2}, 4)$
in addition to 
the quadratic ${\cal N}=4$ lowest higher spin multiplet and 
the large ${\cal N}=4$ linear superconformal family of 
the identity operator.
The various structure constants (fixed coefficient functions) 
appearing in the 
right hand side of this OPE depend on $N$ and the level $k$
of the bosonic spin-$1$ affine Kac-Moody current.
For convenience, the above $136$ OPEs in the component 
approach for generic $(N,k)$ with simplified notations are given.

\vskip0.5cm
\centerline{\it 
Dedicated to the Dept. of Physics, Yonsei Univ. on the 
occasion of its $100^{th}$ anniversary}

\baselineskip=18pt
\newpage
\renewcommand{\theequation}
{\arabic{section}\mbox{.}\arabic{equation}}

\section{Introduction}

In the large ${\cal N}=4$ holography observed in \cite{GG1305},
the duality between matrix extended higher spin theories on 
$AdS_3$ space with large ${\cal N}=4$ supersymmetry and 
the large ${\cal N}=4$ coset theory in two dimensional conformal 
field theory (CFT) was proposed \footnote{In the large level limit
of the type IIB string theory 
on $AdS_3 \times {\bf S}^3 \times {\bf S}^3 \times {\bf S}^1$, 
one of the two three spheres becomes flat and decompactifies to $R^3$
(and to three torus ${\bf T}^3$)
and then one obtains the type IIB string theory 
on $AdS_3 \times {\bf S}^3 \times {\bf T}^4$. In \cite{GG1406}, 
the better understanding for the string theory 
in the tensionless limit was described from the Vasiliev higher spin theory. 
In particular, they found that 
the perturbative Vasiliev theory is a subsector of the tensionless string 
theory. 
In \cite{GG1501}, the unbroken stringy symmetry algebra
is studied further. The corresponding 
$AdS_3 \times {\bf S}^3 
\times {\bf K}^3$ theory on the tensionless point is described 
in \cite{BGP}. 
Very recently, in \cite{GPZ},   
using the CFT perturbation theory that gives the nonzero string tension, 
it is shown that  
the symmetry generators of the symmetric orbifold theory  
of ${\bf T}^4$ describe the Regge trajectories. 
The leading Regge trajectory have lowest mass  
for a given spin and the higher spin generators are written in terms of
quadratic free fields.
The subleading Regge trajectories where the higher spin generators 
are written in terms of  
cubic and higher order free fields are also analyzed.}. 
One of the consistency checks for this duality 
is based on the matching of the correlation functions. The simplest 
three-point functions consist of two scalar primaries and 
one higher spin current. Then the zero mode eigenvalue equations 
of the higher spin Casimir current in the two dimensional 
${\cal N}=4$ coset 
model should coincide with 
   the zero mode eigenvalue equations 
of the higher spin field in an asymptotic (quantum) symmetry algebra
 of the higher spin theory on the 
$AdS_3$ space.     
Recently, in \cite{AK1506}, 
the zero mode eigenvalue equations (and corresponding three-point 
functions) of the bosonic (higher spin) currents 
of spins $s=1, 2, 3$
with two scalars  for any finite $N$ and $k$ (and for large 
$N$ 't Hooft limit) were obtained (Recall 
that the group $G$ of the ${\cal N}=4$ coset theory
is given by $G=SU(N+2)$ and the level of spin-$1$ Kac-Moody current 
is given by the positive integer $k$). 

What happens for the $AdS_3$ bulk theory?
How one can observe the findings \cite{AK1506} on the eigenvalue equations and 
three-point functions in two dimensional CFT computations in the context
of  the higher spin theory on the 
$AdS_3$ space? 
It is known, in the similar bosonic duality studied in \cite{GG1011,GG1205}, 
that  
the three-point functions for the spin-$s$ ($s=2, 3, 4, 5$)
Casimir current with two scalars  
in two dimensional CFT  for any finite 
$N$ and $k$ (and for the large $N$ 't Hooft limit) 
matched with those for the spin-$s$  field in 
the asymptotic symmetry algebra of the higher spin theory on the $AdS_3$
space \cite{AK1308}. 
See also \cite{Ahn1111}.
It was crucial for this consistency check 
to have the asymptotic symmetry algebra 
for any $N$ and $k$ explicitly in order to calculate the eigenvalue equations 
of the higher spin-$s$ field in an asymptotic quantum symmetry algebra
of the higher spin theory on the 
$AdS_3$. In the large ${\cal N}=4$ holography, the corresponding 
  asymptotic symmetry algebra
of the higher spin theory on the 
$AdS_3$ is not known so far.
Therefore, one should determine the asymptotic symmetry algebra 
 in order to calculate the three-point functions of the higher spin theory
on $AdS_3$ space.
Along the line of \cite{GG1205,AK1308}, 
one would like to construct the OPE between the lowest 
$16$ higher spin currents for generic $N$ and $k$.

One of the main results in \cite{Ahn1504} was 
that there exist the complete OPEs, for $N=3$ and arbitrary $k$, 
between the 
$16$ higher spin currents (one spin-$1$ current,
four spin-$\frac{3}{2}$ currents, six spin-$2$ currents, four 
spin-$\frac{5}{2}$ currents and one spin-$2$ current) except 
some singular terms possessing the composite fields 
with spins $s =\frac{7}{2}, 4, \frac{9}{2}, 5$.
We would like to construct the complete OPEs between the 
$16$ higher spin currents for any $N$ and $k$.
The ultimate goal for this direction (which will appear in near future)
is 
to calculate the eigenvalue equations and corresponding three-point functions
of the higher spin fields   
in an asymptotic symmetry algebra
of the higher spin theory on the $AdS_3$.

How one can obtain the general $N$-behavior form the $N=3$ results
in \cite{Ahn1504}? 
One way to obtain the $N$-behavior is to calculate the OPEs from the 
$N$-dependent higher spin currents (written in terms of various 
multiple product of WZW currents) by hand as done in \cite{AK1308}.
In principle, this is possible to do, but 
there exist too many (higher spin) currents: 
$16$ higher spin currents as well as other $16$ currents 
from the large ${\cal N}=4$ superconformal algebra.
One should calculate the $136(=16+ 15+ \cdots +2+1)$ 
OPEs and rearrange them in terms 
of the composite fields from the above $(16+16)$ currents and their 
derivatives.  
Then one should use other approach to obtain the $N$-dependence explicitly
in the OPEs between the higher spin currents. 
One can also proceed what has been done in \cite{Ahn1504}
for different values of $N$. For example, one can increase 
the value of $N$ like as $N =5,7,9,11,13, \cdots$.
With the help of Thielemans' package \cite{Thielemans}, 
the several $N$-cases can give us the consistency check 
in order to make sure the particular 
singular term. 
From the explicit OPEs in \cite{Ahn1504}, some of the 
OPEs have very complicated 
$k$-dependence in the fractional structure constants and the numerical 
values appearing in the power of $k$ will be generalized to
$N$-dependent terms (some power of $N$). For the fractional $k$-dependent
coefficient functions, where the power of $k$ is large, one need 
to consider more finite $N$-values (for example, beyond $N=13$). Therefore,
one cannot determine the complete OPEs from these several $N$-cases
because there are still undetermined OPEs for $N=3$ and one should 
repeat the above procedures without knowing the maximum value for the $N$. 

On the other hands, one uses the power of the ${\cal N}=4$ supersymmetry.
One expects that if one uses this ${\cal N}=4$ supersymmetry, then 
some of the unknown structures in the (higher spin) currents or 
some OPEs between them can be fixed. In other words, 
the ${\cal N}=4$ supersymmetry is  the kind of constraints which preserve
behind an extension of large ${\cal N}=4$ linear superconformal algebra
one is looking for. One also uses the Jacobi identity 
between the (higher spin) currents.
It is known that the above WZW construction (that is, the construction of 
higher spin currents using the WZW currents) satisfies the Jacobi identities
automatically \cite{BS}. One can easily see that the previous results 
for $N=3$ \cite{Ahn1504} satisfy the Jacobi identities.
In the Thielemans' package \cite{Thielemans}, one can check 
the Jacobi identities for any given OPEs.    
It is natural to ask whether one can find the complete OPEs for generic 
$N$ and $k$ by taking the same OPEs for $N=3$ case with the replacement of 
the structure constants as undetermined coefficients. Of course, 
these coefficients reduce to the $N=3$ results if one restricts to have 
the particular $N=3$ case as done in \cite{Ahn1504}.   
It will turn out that there exist consistent solutions for the above
coefficients by solving the various Jacobi identities.  

The Thielemans package \cite{Thielemans} 
is based on the OPEs in the component approach.
If one uses the Jacobi identities inside of this package, then
one should introduce too many unknown coefficients (which 
should be determined 
later). By some experience in dealing with the OPE package, 
it is better to solve the Jacobi identities with small number of
unknown coefficients. Fortunately, there exists the other package by 
Krivonos and Thielemans \cite{KT}, 
which is based on the OPEs in ${\cal N}=2$ superspace.       
Because any current in ${\cal N}=2$ superspace contains four component 
currents, one has the reduced number of unknown coefficients. Therefore, 
one uses the Jacobi identities in ${\cal N}=2$ superspace while 
in order to check the consistency check in the component approach
one can use the Jacobi identities in the component approach (for fixed 
structure constants it does not take too much time to check the Jacobi 
identities).  
One can easily go from the component results to the ${\cal N}=2$ superspace
results inside of \cite{KT} and vice versa.

How one can determine the complete structures (that is, the possible 
composite fields up to the spin-$5$ where the OPE between 
the higher spin-$3$ currents can have the composite fields with spin-$5$
at the first-order
pole) appearing in the 
right hand side of above $136$ OPEs? 
As mentioned before, one can go to the ${\cal N}=4$ superspace (or 
to the ${\cal N}=2$ superspace) from the component results 
in \cite{Ahn1504}.
 The 
${\cal N}=4$ single OPE between the ${\cal N}=4$ multiplet and itself   
can be expanded in terms of Grassmann (or fermionic) 
coordinates. Then the  coordinate 
difference for the singular terms is given by the difference for the 
ordinary coordinate.
In order to write the OPE in ${\cal N}=4$ superspace, one should reexpress
this difference for the ordinary coordinate in terms of 
the difference for the ${\cal N}=4$ superspace coordinate   
by subtracting  the product of Grassmann 
coordinates (and adding the same quantity) 
from the above difference for the 
ordinary coordinate.
Then any fractional power of the difference for the ordinary coordinate
can be expanded in terms of the sum of
fractional power of the difference for 
the ${\cal N}=4$ superspace coordinate    
by using the Taylor expansion with the help of 
property of Grassmann coordinates.
Then any singular term in the single ${\cal N}=4$ OPE
contains the product of Grassmann coordinates, the fractional power of
 the difference for 
the ${\cal N}=4$ superspace coordinate and the ordinary coordinate-dependent 
pole terms from the OPEs between  
the various $16$ higher spin currents.  
After simplifying these complicated singular terms in ${\cal N}=4$
supersymmetric way, one can obtain  the results in \cite{Ahn1504}
in terms of a single ${\cal N}=4$ OPE.
In other words, the undetermined parts of the $136$ OPEs in \cite{Ahn1504}
are completely fixed. 

Now one can go to the ${\cal N}=2$ superspace from 
these component results and all the expressions are given for $N=3$.
Let us replace the structure constants, which have $k$-dependence explicitly
with the arbitrary coefficients. Then one has the complete OPEs in 
${\cal N}=2$ superspace with undetermined structure constants
\footnote{So far, the field contents in the right hand side of the OPEs 
are the same as the ones in \cite{Ahn1504} written in ${\cal N}=2$ 
superspace. For the lowest ${\cal N}=4$ (higher spin) 
multiplet, the exact relations
between the $16$ higher spin currents and the ones in this paper (or 
the ones in \cite{BCG}) are known explicitly. For the next lowest 
${\cal N}=4$ higher spin 
multiplet, they will depend on $(N, k)$ in complicated 
way. For example, the higher spin-$2$ current living in 
the next lowest higher spin current denoted by ${\bf P^{(2)}}(z)$
in \cite{Ahn1504} contains $\Phi_0^{(2)}(z)$ (or equivalently $V_0^{(2)}(z)$) 
as well as many extra terms
with $(N, k)$-dependent fractional coefficient functions.
The merit of Jacobi identity is that one does not have to 
determine the exact relations between the next lowest higher spin currents
and the ones in this paper. We simply introduce the unknown 
coefficient functions in front of the possible composite fields.
They will be fixed by using the Jacobi identity. On the other hands,
it will turn out that if one 
obtains the complete OPEs between the 
higher spin-$1$ current and the $16$ higher spin current in the 
$SO(4)$ manifest basis for generic $(N,k)$, then all the remaining 
structure constants are automatically determined via ${\cal N}=4$ 
supersymmetry.        }. One 
uses the Jacobi identities in order to fix the structure constants.
In general, the new ${\cal N}=2$ primary current (which transforms as 
a primary current under the ${\cal N}=2$ stress energy tensor)  can occur 
in the right hand side of the OPEs as the spins of the currents 
increase. The above $16$ higher spin currents can be represented by 
four ${\cal N}=2$ multiplets. Similarly the $16$ currents of 
large ${\cal N}=4$ linear superconformal algebra can be combined into 
four ${\cal N}=2$ multiplets (where one them is given by chiral current 
and antichiral current). Then on can use the Jacobi identities 
by choosing 
one ${\cal N}=2$ current and two ${\cal N}=2$ higher spin currents.        
One cannot use the Jacobi identities by taking  
three ${\cal N}=2$ higher spin 
currents because if one considers the OPE between 
any ${\cal N}=2$ higher spin current and other new ${\cal N}=2$ higher spin 
current, one does not know this OPE at this level.
Therefore, the three quantities in using the Jacobi identities
are given by one ${\cal N}=2$ current and two ${\cal N}=2$ higher spin
currents. One can also consider the combination of one ${\cal N}=2$ higher
spin current and two ${\cal N}=2$ currents  or the case with three 
${\cal N}=2$ currents but these will do not produce any nontrivial equations
for the unknown coefficients. They are satisfied trivially.

One worries about whether one can fix all the unknown coefficients 
completely even though one does not exhaust all the Jacobi identities.       
It will 
turn out that all the structure constants are determined except one arbitrary 
unknown coefficient. In other words, all the known structure constants 
are expressed in terms of this unknown coefficient. How one can fix this 
unknown quantity?
One way to determine this quantity is that one should consider 
the other four ${\cal N}=2$ higher spin currents 
in principle, but this is not 
so useful.  This leads to another new problem because one should consider 
the OPE between the previous ${\cal N}=4$ higher spin 
multiplet of super spin $1$ 
and the new 
${\cal N}=4$ higher spin multiplet of super spin $2$.
Or one can find the above unknown coefficient by looking at 
the particular OPE as one varies the $N$-values.
Eventually one can determine the final unknown coefficient 
by using this analysis. In the component approach, 
the OPE between the higher spin-$1$ current and the 
higher spin-$3$ current for several $N$-values provides the $(N,k)$ 
dependence of the above unknown coefficient.   

In section $2$, the review for the large ${\cal N}=4$ linear superconformal 
algebra is described. The $16$ currents are expanded in an expansion of 
the Grassmann coordinates in ${\cal N}=4$ superspace.
The single OPE between the $16$ currents in ${\cal N}=4$ superspace
is given. The various OPEs between the $16$ currents in the component
approach are described. The different basis for the same 
large ${\cal N}=4$ linear superconformal algebra is provided.
The coset realization for   the large ${\cal N}=4$ linear superconformal 
algebra is given
by identifying the two levels of the  
the large ${\cal N}=4$ linear superconformal 
algebra with two parameters appearing in the ${\cal N}=4$ coset theory.

In section $3$, the $16$ higher spin-$s$ currents are introduced
in ${\cal N}=4$ superspace. The definition for the ${\cal N}=4$ primary 
current in ${\cal N}=4$ superspace
is described. The component results by projection into this definition
can be obtained. One can also present the above 
$16$ higher spin-$s$ currents in the primary basis. In other words, 
they are primary currents under the stress energy tensor.
The other component results, where the $SU(2) \times SU(2)$ symmetry 
is manifest, are described. In this basis, the higher spin-$(s+2)$ current,
which is the last component of this ${\cal N}=4$ multiplet (and its spin is 
maximum inside of  this multiplet), 
is not a primary current under the stress energy tensor.
The precise simple relations between the above 
lowest $16$ higher spin currents with the ones in \cite{Ahn1504} 
are given explicitly. 

In section $4$, the description given in section $2$ 
in ${\cal N}=2$ superspace is reviewed \cite{RASS,BO} after introducing the 
${\cal N}=2$ superspace.
One can represent the above $16$ currents in terms of 
four ${\cal N}=2$ multiplets (where one of them is given by 
one chiral current and one antichiral current).
Also one can rewrite the single ${\cal N}=4$ stress energy tensor 
introduced in section $2$ in terms of four ${\cal N}=2$ multiplets 
by expanding two Grassmann coordinates.

In section $5$,
the description of section $3$ is described in ${\cal N}=2$ superspace.
The ${\cal N}=4$ primary current condition introduced in section 
$3$ is rewritten in ${\cal N}=2$ superspace. 
The four ${\cal N}=2$ multiplets of spins $s, (s+\frac{1}{2}),(s+\frac{1}{2}),
(s+1)$ are located at the particular components of 
above ${\cal N}=4$ multiplet 
in an expansion of two Grassmann coordinates.
The precise relations between the four ${\cal N}=2$ multiplets
and its $16$ component currents are described.

In section $6$,
based on the previous results for $N=3$ in \cite{Ahn1504}, 
one can go on to the ${\cal N}=4$ superspace.
By introducing the arbitrary coefficients in the ten ${\cal N}=2$
OPEs, one would like to solve the various Jacobi identities in order
to determine the above unknown coefficients which depend on $(N,k)$.
The final ten OPEs in ${\cal N}=2$ superspace are given.      
The fusion rules are described.

In section $7$,
one can go to the component approach from the ${\cal N}=2$ results 
in section $6$. From the precise relations found in section $6$,
one obtains the final ${\cal N}=4$ OPE for generic $N$ and $k$.
The fusion rule is given. 

In section $8$,
the summary of this paper is given and the possible 
future works are presented. 

In Appendices $A$-$J$, 
the detailed computations in sections $2$-$7$
are presented.

If you are interested in the construction of ${\cal N}=2$ superspace 
description, the main text and Appendix $G$ are useful.
For the component approach, the main text and Appendix $H$
are crucial. If you are interested in the 
${\cal N}=4$ superspace description, go to the main text and Appendix $I$.  

The packages in \cite{Thielemans,KT} are used in this paper
\footnote{One describes the $16$ currents of the large ${\cal N}=4$
linear superconformal algebra as ``the $16$ currents'' simply 
and the $16$ lowest higher spin currents of the ${\cal N}=4$ multiplet
as ``$16$ higher spin currents''.  Sometimes one ignores the ``super'' 
appearing in  the super current, super field  or super OPE because
it occurs many times in this paper.    }.  

\section{The OPEs between the $16$ currents in 
${\cal N}=4$ superspace: Review}

In this section, one describes the $16$ currents of the large
${\cal N}=4$ linear superconformal algebra 
in ${\cal N}=4$ superspace, where $SO(4)$ symmetry 
is manifest. Then the corresponding 
large ${\cal N}=4$ linear superconformal algebra, which consists of 
$13$ nontrivial OPEs  in the component approach (in Appendix $A$), 
can be written in terms of
a single ${\cal N}=4$ (super) OPE.
Other description in other basis, where the $SU(2) \times SU(2)$ symmetry 
is manifest, is reviewed also. Finally 
the realization for  large ${\cal N}=4$ linear superconformal algebra
is also reviewed.

\subsection{The $SO(4)$-singlet ${\cal N}=4$ stress energy tensor
of super spin $0$}

The coordinates of ${\cal N}=4$ (extended) superspace 
can be described as 
$(Z, \overline{Z})$ where 
$Z=(z, \theta^i)$, $\overline{Z} =(\bar{z}, \bar{\theta}^i)$
and $i =1,2,3,4$ and the index $i$ is the $SO(4)$-vector index. 
The left covariant spinor derivative 
is given by 
$ D^i = \theta^i \frac{\pa}{\pa z } +  \frac{\pa}{\pa {\theta^i}}$
and satisfies 
the following nontrivial anticommutators 
$
\{ D^i, D^j \} = 2 \delta^{ij} \frac{\pa}{\pa z}$.
Here the Kronecker delta $\delta^{ij}$ is the rank $2$ $SO(4)$
symmetric invariant tensor.
Then the ${\cal N}=4$ stress energy tensor can be described as 
follows \cite{Schoutensnpb} (or \cite{Ahn1992}, where
the ${\cal N}=1$ and ${\cal N}=2$ superspace descriptions
are given):
\bea
{\bf J^{(4)}}(Z) & = & 
- \Delta(z) + i \theta^{j} \Gamma^{j}(z)-
i \theta^{4-jk} T^{jk}(z)-\theta^{4-j} (G^{j}-2
\alpha i \partial \Gamma^{j})(z)+\theta^{4-0}  
(2L-2\alpha\partial^2 \Delta)(z)
\nonu \\
 & = & 
-\Delta(z) + i \, \theta^{1} \:\Gamma^{1}(z)
+ i\, \theta^{2} \:\Gamma^{2}(z)+
i \, \theta^{3}\:\Gamma^{3}(z)+
i\, \theta^{4}\:\Gamma^{4}(z) \nonu \\
& - & i \, \theta^{1} \, \theta^2 \:T^{34}(z)+i\, \theta^{1} \, \theta^3
\:T^{24}(z)
- i \, \theta^{1} \, \theta^4 \:T^{23}(z)
-i \,  \theta^{2} \, \theta^3 \:T^{14}(z)+ i\, \theta^{2} \, \theta^4 \:T^{13}(z)
\nonu \\
& - &  i \, \theta^{3} \, \theta^4 \:T^{12}(z)
-  
\theta^{2} \, \theta^4 \, \theta^3 \, 
(G^{1}-2 \alpha\, i \, \partial \Gamma^{1})(z)-
\theta^{1} \, \theta^3 \, \theta^4 \, 
(G^{2}-2\alpha \, i \,\partial \Gamma^{2})(z)
\nonu \\
& - & \theta^{1} \, \theta^4 \, \theta^2 \, 
(G^{3}-2\alpha \, i \, \partial \Gamma^{3})(z)
-  
\theta^{1}\, \theta^2 \, \theta^3 \, 
(G^{4}-2\alpha \, i \, \partial \Gamma^{4})(z) 
\nonu \\
& + &
  \theta^{1}\, \theta^2 \, \theta^3 \, \theta^4 \, 
(2L-2\alpha\, \partial^2 \Delta)(z).
\label{j4}
\eea
%
In the first line of (\ref{j4}), the summation over 
repeated indices (which implies 
that the ${\cal N}=4$ stress energy tensor 
${\bf J^{(4)}}(Z)$ is $SO(4)$-singlet) 
is taken
 \footnote{
In this notation the superscript $(4)$ is nothing to do with the 
spin. It stands for the number of supersymmetry. Later one introduces
the ${\cal N}=4$ (higher spin) 
multiplet ${\bf \Phi^{(s)}}(Z)$ of super spin $s$.
Then it is more appropriate to write the ${\cal N}=4$ stress energy tensor 
${\bf J^{(4)}}(Z)$ as 
 ${\bf \Phi^{(0)}}(Z)$, but one follows the original notation in 
\cite{Schoutensnpb}. 
One uses the boldface notation for the ${\cal N}=4$ or ${\cal N}=2$ 
multiplet in order to emphasize 
the fact that the corresponding multiplet has many component currents.
For the ${\cal N}=4$ multiplet, the $16$ independent component currents arise
while for the ${\cal N}=2$ multiplet, the four independent 
components arise (if they are 
unconstrained currents). Sometimes the right hand side of the given OPE
contains too many multiplets and then we ignore 
the boldface notation for simplicity. See also section $7$.}. 
The simplified notation (in the 
multiple product) 
$\theta^{4-0}$ is used for 
$\theta^1 \,\theta^2 \,\theta^3 \,\theta^4$. 
The complement $4-i$ is defined such that $\theta^{4} = \theta^{4-i}\,
\theta^{i}$.
In the second line of (\ref{j4}), 
the complete $16$ currents for the ${\cal N}=4$ stress energy tensor 
are described in an expansion of Grassmann coordinates 
completely. The quintic- and higher-order terms in $\theta^i$
vanish due to the property of $\theta^i$. 
The $16$ currents are given by
a single spin-$0$ current $\Delta(z)$, four spin-$\frac{1}{2}$ currents
$\Gamma^i(z)$ transforming as  a vector representation under the $SO(4)$,
six spin-$1$ currents $T^{ij}(z)$
transforming as an adjoint representation under the 
$SO(4)$, four spin-$\frac{3}{2}$ currents $G^i(z)$ 
(transforming as  a vector representation under the $SO(4)$)
and 
the spin-$2$ current $L(z)$. 
In particular, the spin-$0$ and spin-$2$ 
currents are $SO(4)$-singlets.
The spin of $\theta^i$ is given by 
$-\frac{1}{2}$ (and the covariant spinor derivative
$D^i$ has spin $\frac{1}{2}$)
and therefore the ${\cal N}=4$ (super) spin of the stress energy tensor
${\bf J^{(4)}}(Z)$ 
is equal to zero.
Each term in (\ref{j4}) has spin-$0$ value. 

In the cubic- and quartic-terms in the $\theta^i$, there exist 
$\alpha$-dependent terms, 
where 
the parameter is introduced in \cite{Schoutensnpb}
as follows:
\bea
\alpha= \frac{1}{2} \frac{(k^{+}-k^{-})}{(k^{+} +k^{-})}.
\label{alpha}
\eea
The self-dual and anti self-dual combinations of the spin-$1$ current 
$T^{ij}(z)$ have the level $k^{+}$ and $k^{-}$, respectively.
That is, two commuting $SU(2)$ Kac-Moody algebras 
have  their levels $k^{\pm}$.
The parameter $\alpha$ in (\ref{alpha}) 
reflects the asymmetry between the occurrences of these two 
$SU(2)$ Kac-Moody subalgebras in the (anti)commutators  involving odd
currents.
More precisely, in Appendix $A$, 
the $\alpha$-dependence appears in the OPEs between the spin-$\frac{3}{2}$
currents and in the OPEs between 
the spin-$\frac{3}{2}$ currents and the spin-$1$ currents.
For $\alpha =\pm \frac{1}{2}$ (or $k^{\pm} \rightarrow \infty$),
one of the $SU(2)$ subalgebras with the currents $\Gamma^i(z)$ and $\Delta(z)$
decouple and the above large ${\cal N}=4$ linear superconformal algebra 
is reduced to the $SU(2)$-extended ${\cal N}=4$ superconformal algebra in 
\cite{Ademolloetal}.

Furthermore, the central charge appearing in the OPE between the 
bosonic stress energy tensor $L(z)$ is given by \cite{Schoutensnpb}
\bea
c =\frac{6 k^{+} k^{-}}{(k^{+}+k^{-})}.
\label{centralcharge}
\eea
When $k^{+} =k^{-}$ (that is, $\alpha=0$),
this central charge is the positive multiple of $3$
and can be realized by one real scalar and four Majorana fermions 
\cite{Schoutensplb}.
Note that there is no twisted anomaly, where the spin-$0$ current
$\Delta(w)$ 
transforms as a primary field under the stress energy tensor $L(z)$.
Later one sees the realization of the ${\cal N}=4$ stress energy
tensor  in the ${\cal N}=4$ coset theory and the above two parameters
are given by two parameters $(k+1)$ and $(N+1)$.
Accordingly, the central charge can be characterized by 
$k$ and $N$.

\subsection{The OPE of ${\cal N}=4$ stress energy tensor}

The ${\cal N}=4$ super OPE between the ${\cal N}=4$
stress energy tensor and itself  can be summarized by \cite{Schoutensnpb}
\bea
{\bf J^{(4)}}(Z_{1}) \, {\bf J^{(4)}}(Z_{2}) & = & 
\frac{\theta_{12}^{4}}{z_{12}^{2}}\, \frac{1}{2} (k^{+}-k^{-})
+
\frac{\theta_{12}^{4-i}}{z_{12}} \, D^{i} {\bf J^{(4)}}(Z_{2})+
\frac{\theta_{12}^{4}}{z_{12}} \, 2 \, \partial {\bf J^{(4)}}(Z_{2})
\nonu \\
& - & \frac{1}{2} (k^{+}+k^{-}) \,\log (z_{12})
+\cdots,
\label{j4j4}
\eea
where the summation over the 
repeated indices is assumed (the OPE between the 
$SO(4)$-singlet current and itself) and 
the fermionic coordinate difference for given index $i$ is defined as
 $ \theta_{12}^i = \theta_1^i-\theta_2^i$, and the bosonic
coordinate difference is given by $z_{12} = z_1 -z_2 -
\theta_1^i \theta_2^i$.
By introducing the spin-$\frac{1}{2}$ 
field in $J^i(Z) \equiv D^i {\bf J^{(4)}}(Z)$,
the OPE $J^i (Z_1) \, J^j(Z_2)$ from 
(\ref{j4j4}) does not contain the $\log (z_{12})$-term
and then the nonlocal operator associated with the central term disappears
\cite{Schoutensnpb,IKL1,IKL2} \footnote{
Recall that 
the super spin of 
${\cal N}=4$ stress energy tensor 
is zero and the OPE between the spin-$0$ current,  
$\Delta(z_1) \, \Delta(z_2)$, has a term $\log(z_1-z_2)$. 
One can see this feature
from (\ref{j4j4}) by taking $\theta^i=0$ both sides. 
Then the left hand side is given by the OPE $\Delta(z_1) \, \Delta(z_2)$
while the right hand side is given by 
$ -  \frac{1}{2} (k^{+}+k^{-}) \,\log (z_1-z_2)$.
The associated OPE is in the last equation of Appendix \ref{16SCAOPEs}, where
one introduces $U(z) \equiv - \pa \Delta(z)$.}.
Note that there is no ${\bf J^{(4)}}(Z_2)$-term in the right hand side 
of (\ref{j4j4}). On the other hand, for less supersymmetric theory 
with ${\cal N} \leq 3$, the corresponding 
stress energy tensor-term 
occurs in the right hand side of the OPE
from $\frac{\theta_{12}^{\cal N}}{z_{12}^2} \, (4-{\cal N}) 
{\bf J^{({\cal N})}}(Z_2)$.
Also the explicit component results (which will be described in
next subsection) will be given in Appendix $A$.



\subsection{The OPEs of
${\cal N}=4$ stress energy tensor in component approach}

Now the ${\cal N}=4$ stress energy tensor is given by (\ref{j4})
and its ${\cal N}=4$ OPE is given by (\ref{j4j4}). Then it is 
straightforward to reexpress (\ref{j4j4})
in component approach by substituting (\ref{j4}) into (\ref{j4j4}).
In Appendix $A$, the $SO(4)$-extended linear superconformal algebra
\cite{Schoutensnpb} is presented.
Let us read off the OPE $\Gamma^1(z_1) \, T^{12}(z_2)$ related to  
the $11$-th equation of Appendix (\ref{16SCAOPEs})
from (\ref{j4j4}). Let us multiply the differential operator 
$D^1_1 \, D_2^{3} \, D_2^4$ in the left hand side of (\ref{j4j4}).
Then one obtains  the OPE $D_1^1 \, {\bf J^{(4)}}(Z_1) \, 
D_2^3 \, D_2^4 \, {\bf J^{(4)}}(Z_2)$, where the (bosonic) operator 
$D_2^3 \, D_2^4$ can pass the (bosonic) current ${\bf J^{(4)}}(Z_1)$
because the differential operator with respect to the super coordinate
$Z_2$ commute with 
the current ${\bf J^{(4)}}(Z_1)$, which depends on the super coordinate $Z_1$
only.
Now one takes $\theta_1^i = \theta_2^i=0$ and then one has 
the component 
OPE $-\Gamma^1(z_1) \, T^{12}(z_2)$ from the left hand side \footnote{ 
On the other hand, the cubic terms
in $\theta^i$ can arise from the second term of (\ref{j4j4}).
The nontrivial terms occur from the particular term of 
$\theta_{12}^3 \, \theta_{12}^4 \, \theta_{12}^1$. It is easy to see that 
the above differential operator $D^1_1 \, D_2^{3} \, D_2^4$ acting 
on this leads to $-1$.
Finally, one obtains the OPE
$\Gamma^1(z_1) \, T^{12}(z_2) = \frac{1}{(z_1-z_2)} \,
i \, \Gamma^2(w) +\cdots $, where $z_{12}$ is reduced to $(z_1-z_2)$
after putting the condition $\theta_1^i = \theta_2^i=0$. 
Note that the $D_2^2 \, {\bf J^{(4)}}(Z_2)$ term reduces to
the expression $i \Gamma^2(z_2)$.}.
Then one identifies the above result with the $11$-th equation of Appendix
(\ref{16SCAOPEs}) by using the property in the OPE \cite{BS}.
In this way, one can check that 
the ${\cal N}=4$ OPE in (\ref{j4j4}) is equivalent to 
the component results in Appendix (\ref{16SCAOPEs}) by acting 
the differential operators $D_1^i$ and $D_2^j$ on (\ref{j4j4}) and putting 
$\theta_1^i = \theta_2^i=0$.

On the other hand, one can check the ${\cal N}=4$ OPE 
from its component results by  
using the following identity which connects the structure of the $n$-th order 
pole
in the bosonic coordinate with those (plus other poles) 
in the ${\cal N}=4$ superspace 
coordinate  
\bea
\frac{1}{(z_1-z_2)^n} & = &  \frac{1}{z_{12}^n}
-n \, \frac{\theta_1^i \theta_2^i}{z_{12}^{n+1}}
+ \frac{1}{2 !} n (n+1)\, 
\frac{\theta_1^i \theta_2^i \theta_1^j \theta_2^j }{z_{12}^{n+2}}
-\frac{1}{3!} 
n (n+1)(n+2)\, \frac{\theta_1^i \theta_2^i \theta_1^j \theta_2^j 
\theta_1^k \theta_2^k  }{z_{12}^{n+3}}
\nonu \\
& + & \frac{1}{4!} 
n (n+1)(n+2) (n+3)\, \frac{\theta_1^i \theta_2^i \theta_1^j \theta_2^j 
\theta_1^k \theta_2^k  \theta_1^l \theta_2^l }{z_{12}^{n+4}},
\qquad z_{12} \equiv z_1-z_2-\theta_1^i \theta_2^i.
\label{polerelation}
\eea 
The next higher order terms in (\ref{polerelation}) contain 
the expression $ (\theta_1^i \theta_2^i)^5
$ 
and this is identically zero due to the property of 
the fermionic coordinates. The positive integer $n$ can be $n=1,2,3,4$
for the $16$ currents of large ${\cal N}=4$ linear superconformal algebra
because the highest spin among them is given by two 
and  the highest singular term is the fourth-order pole 
in the OPE \footnote{
Let us consider the OPE $\Gamma^1(z_1) \, T^{12}(z_2)$. Then the corresponding 
term in the left hand side of (\ref{j4j4}) is given by 
$i \, \theta^1_1 \, \Gamma^1(z_1) \, (- i) \, \theta_2^{3} \, \theta_2^4 
\, T^{12}(z_2)$.
This becomes  $\theta^1_1  \, \theta_2^{3} \, \theta_2^4
\, \Gamma^1(z_1) 
\, T^{12}(z_2)$ (there is no sign change). 
Now one can use the component result 
between the spin-$\frac{1}{2}$ current and spin-$1$ current
in Appendix $A$.
Then the above expression leads to 
the singular term 
$\theta^1_1  \, \theta_2^{3} \, \theta_2^4 \, \frac{1}{(z_1-z_2)} \,
i \, \Gamma^2(w) +\cdots$. By using the relation 
(\ref{polerelation}), one has the nontrivial singular term
$\frac{\theta^1_1  \, \theta_2^{3} \, \theta_2^4 }{z_{12}} \,
i \, \Gamma^2(z_2)$, which can be written as 
$\frac{ \theta_{12}^{3} \, \theta_{12}^4 \, \theta_{12}^1}{z_{12}} \,
D^2 \, {\bf J^{(4)}}(Z_2)$ by considering other nontrivial terms also.
In this way, one can check that all the component results in Appendix 
$A$ can be rewritten in terms of a single OPE (\ref{j4j4}) 
in ${\cal N}=4$ superspace.}.

\subsection{The ${\cal N}=4$ stress energy tensor in other basis}

It is useful to write (\ref{j4}) in the basis of \cite{BCG},
where the spin-$1$ currents are represented in $SU(2) \times SU(2)$
symmetric way
rather than in $SO(4)$ symmetric way. Then one expects that 
the previous $SO(4)$ adjoint six spin-$1$ currents $T^{ij}(z)$
can be written in terms of two $SU(2)$ adjoints. 
Let us describe the final result and then explain some important aspects 
\bea
{\bf J_{bcg}^{(4)}}(Z) & = &  -\Delta(z)-\theta^{1} \, Q^{1}(z)+
\theta^{2} \, Q^{2}(z)+\theta^{3} \, Q^{3}(z)
+\theta^{4} \, Q^{4}(z) + \theta^{12}\, (-A^{+3}+A^{-3})(z) \nonu \\
& + & \theta^{13}  (A^{+2}-A^{-2})(z)-
\theta^{14} (A^{+1}+A^{-1})(z)+\theta^{23} (A^{+1}-A^{-1})(z)
+\theta^{24}  (A^{+2}+A^{-2})(z)\nonu \\
& + & \theta^{34} \, (A^{+3}+A^{-3})(z)
- 
\theta^{243} \, (G_{bcg}^{1}+2\alpha\partial Q^{1})(z)+
\theta^{134}\, (G_{bcg}^{2}
+2\alpha\partial Q^{2})(z)
\label{bcgj4}
\\
& + & 
\theta^{142}\, (G_{bcg}^{3}+2\alpha \, \partial Q^{3})(z)
+  
\theta^{123} \, (G_{bcg}^{4}+2\alpha \, \partial Q^{4})(z)
 +  \theta^{1234} \, (2L+ 2\alpha \, \partial U)(z),
\nonu
\eea
where $U(z) = -\pa \Delta(z)$.
Let us consider the spin-$\frac{1}{2}$ currents.
In Appendix (\ref{16SCAOPEs}), the OPE
$\Gamma^i(z) \, \Gamma^j(w)$ contains the positive sign in the right hand side.
In Appendix (\ref{N4linearalg}), where 
the $SU(2) \times SU(2)$ subalgebra is manifest, 
the corresponding OPE appearing in the second
line from the below has a negative sign.
This implies that the spin-$\frac{1}{2}$ current $Q^a(z)$ in \cite{STV} 
is given by $\pm i$  times the spin-$\frac{1}{2}$ current $\Gamma^i(z)$ 
in previous 
subsection.   
Furthermore the OPE $U(z) \, G^i(w)$ in Appendix (\ref{16SCAOPEs}) 
has a second-order pole with the term $- i \, \Gamma^i(w)$ and 
in the corresponding OPE of  Appendix $B$ there is  no $-i$ factor.
If the spin-$\frac{3}{2}$ current remains the same (the spin-$1$ current 
$U(z)$ is common), then 
the corresponding spin-$\frac{1}{2}$ currents have an extra 
$-i$, but this is not the case.
The reason is as follows.
In the OPE of $G^i(z) \, \Gamma^j(w)$ with $j=i$ of Appendix $A$, there exists 
a term $i \, U(w)$ in the right hand side. On the other hand, 
the corresponding OPE of Appendix $B$ does not have the complex number $i$.

Therefore, the product of $\Gamma^i(z)$ and $G^i(z)$ 
should    contain the extra $-i$. This implies that 
some of the spin-$\frac{3}{2}$ current change with minus sign. 
Once the spin-$\frac{1}{2}$ currents are fixed, then the 
spin-$\frac{3}{2}$ currents are also determined completely
from this property.
There are still other constraints in Appendix $A$.
So one can proceed by assuming that 
the spin-$\frac{1}{2}$ currents $Q^a(z)$ can be written as 
$\pm i$ times the the spin-$\frac{1}{2}$ currents $\Gamma^i(z)$ 
and then use other nontrivial relations.
It turns out that 
as in (\ref{bcgj4}), the first component of the spin-$\frac{1}{2}$
current has an extra $-i$ ($Q^1(z) = - i \Gamma^1(z)$) 
and the remaining ones have the extra 
$i$ ($Q^j(z) = i \Gamma^j(z)$ where $j=2,3,4$). 
Furthermore, the first component of the spin-$\frac{3}{2}$ current 
remains unchanged ($G^1_{bcg}(z)=G^1(z)$) 
while the others change with minus sign ($G^i_{bcg}(z) = - G^i(z)$
where $i=2, 3, 4$) as in (\ref{bcgj4}).

What about the spin-$1$ currents?
For example, let us focus on the OPE 
$G^1(z) \, \Gamma^4(w)$ in Appendix $A$,
which has a first-order pole, $- T^{23}(w)$.
The corresponding OPE from Appendix $B$ is given by 
the expression $G^1(z) \, (-i) \, Q^4(w)$, which contains the first-order pole 
$(-i) \times 2(-\frac{1}{2} A^{+1} -\frac{1}{2} A^{-1})(w)$ from Appendix 
$B$ by substituting the values of $\alpha$ tensor explicitly.
Then one concludes that the spin-$1$ current
$T^{23}(z)$ is equal to the expression
$-i (A^{+1} +A^{-1})(z)$.
In this way one can determine the precise relations, as in (\ref{bcgj4}),  
between 
the spin-$1$ currents with $SO(4)$-indices and those with 
$SU(2) \times SU(2)$-indices as follows:
\bea
T^{12}(z) & = & i(A^{+3}+A^{-3})(z),\qquad 
T^{13}(z)=-i(A^{+2}+A^{-2})(z), \nonu \\
T^{14}(z) & = & i(A^{+1}-A^{-1})(z),
\qquad T^{23}(z)=-i(A^{+1}+A^{-1})(z),
\nonu \\
T^{24}(z) & = & -i(A^{+2}-A^{-2})(z),\qquad 
T^{34}(z)=-i(A^{+3}-A^{-3})(z).
\label{spinonerelation}
\eea
Of course, with (\ref{spinonerelation}),
one can also write down the spin-$1$ currents 
$A^{\pm i}(z)$ in terms of the spin-$1$ currents $T^{ij}(z)$.
One also identifies  the spin-$2$ current $L(z)$ both sides.
Therefore, the two Appendices $A$ and $B$ are equivalent to 
each other via the explicit current identifications described above.  
 
\subsection{The realization of ${\cal N}=4$ stress energy tensor
in the ${\cal N}=4$ coset theory}

It is known that the above $16$ currents of large ${\cal N}=4$ linear
superconformal algebra can be realized by the spin-$\frac{1}{2}$ and 
the spin-$1$ currents of
the ${\cal N}=4$ coset $\frac{SU(N+2)}{SU(N)}$ theory \cite{Saulina,AK1506}. 
The two 
parameters are related to the two quantities in the 
${\cal N}=4$ coset theory as follows \cite{ST}:
\bea
 k^{+} = (k+1), \qquad k^{-} = (N+1).
\label{kn}
\eea 
Here the level $k$ appears in the central term $k \, g^{ab}$ of the OPE
between the spin-$1$ currents
$V^a(z_1) \, V^b(z_2)$, where the indices $a,b \cdots,$ are the adjoint 
indices of the group $G=SU(N+2)$. 
The metric, structure constants and almost complex structures 
in the $16$ currents 
occur in their coefficient functions in the multiple product of 
spin-$\frac{1}{2}$ and spin-$1$ currents.
Furthermore, it is natural to ask whether there exist
the ${\cal N}=4$ affine Kac-Moody current $Q^a(Z)$ 
of superspin $\frac{1}{2}$
and corresponding Sugawara construction for the ${\cal N} =4$
stress energy tensor ${\bf J^{(4)}}(Z)$. 
It would be interesting to study this direction in detail. 
The point is how one can generalize the known ${\cal N}=2$ superspace 
description with constraints in the ${\cal N}=4$ superspace. 

\section{The OPEs between the $16$ currents   and the 
$16$ higher spin currents  in 
${\cal N}=4$ superspace}

In this section, one describes the ${\cal N}=4$ (super) primary current,
in $SO(4)$ symmetric way,
under the ${\cal N}=4$ stress energy tensor explained in previous section.
Although this $16$ higher spin currents, in general,  
transform as a nontrivial 
representation under the group $SO(4)$, 
the only $SO(4)$-singlet $16$ higher spin currents are described.
Furthermore, the super spin is, in general, 
given by the positive integer $s$, but its lowest value $s=1$
will be considered later 
when the OPEs between them are calculated for generic 
$N$ and $k$. 
The OPEs between the $16$ currents   and the 
$16$ higher spin currents in component approach are also given.
Other basis, where the $SU(2) \times SU(2)$ symmetry 
is manifest, 
is compared (the higher spin-$3$ current is not a primary current
under the stress energy tensor)
and one also gives the precise relations with the ones in \cite{Ahn1504}.

\subsection{The $SO(4)$-singlet 
${\cal N}=4$ multiplet of superspin $s$}

For general superspin $s$, one can write down the ${\cal N}=4$ (higher spin) 
multiplet
as follows \footnote{In the notation of \cite{GPZ}, this corresponds to
$R^{(s)}({\bf 1},{\bf 1})(Z)$, where the representation 
$({\bf 1},{\bf 1})$ stands for 
the singlet under the $SU(2) \times SU(2)$. In our case, there is no 
$SO(4)$ index in the ${\cal N}=4$ (higher spin) multiplet
${\bf \Phi^{(s)}}(Z)$.}:
\bea
{\bf \Phi^{(s)}}(Z) & = & 
\Phi_{0}^{(s)}(z)+\theta^{i}\:\Phi_{\frac{1}{2}}^{(s),i}(z)+
\theta^{4-ij}\:\Phi_{1}^{(s),ij}(z)+\theta^{4-i}\:\Phi_{\frac{3}{2}}^{(s),i}(z)
+\theta^{4-0}\:\Phi_{2}^{(s)}(z)
\nonu \\
 & = & \Phi_{0}^{(s)}(z)+\theta^{1} \, \Phi_{\frac{1}{2}}^{(s),1}(z)+\theta^{2}
\, \Phi_{\frac{1}{2}}^{(s),2}(z)+\theta^{3}\, \Phi_{\frac{1}{2}}^{(s),3}(z)+
\theta^{4}\, \Phi_{\frac{1}{2}}^{(s),4}(z) 
+  \theta^{12} \Phi_{1}^{(s),34}(z) \nonu \\
& + & \theta^{13}  \Phi_{1}^{(s),42}(z)+
\theta^{14}  \Phi_1^{(s),23}(z)+\theta^{23}\Phi_1^{(s),14}(z)+
\theta^{24}  \Phi_1^{(s),31}(z)+\theta^{34}\Phi_1^{(s),12}(z)
\nonu \\
& + & 
\theta^{324} \, \Phi_{\frac{3}{2}}^{(s),1}(z)+\theta^{134}\, 
\Phi_{\frac{3}{2}}^{(s),2}(z)+
\theta^{142} \, \Phi_{\frac{3}{2}}^{(s),3}(z)+
\theta^{123} \, \Phi_{\frac{3}{2}}^{(s),4}(z)+
\theta^{1234} \, \Phi_{2}^{(s)}(z),
\label{phis}
\eea
where one introduces $\theta^{ij} \equiv \theta^i \, \theta^j$, 
$\theta^{ijk} \equiv \theta^i \, \theta^j \, \theta^k $ and 
$\theta^{1234} \equiv \theta^1 \, \theta^2 \, \theta^3 \, \theta^4$.
In component, 
there are
a single higher spin-$s$ current $\Phi_0^{(s)}(z)$, four higher 
spin-$(s+\frac{1}{2})$ 
currents
$\Phi_{\frac{1}{2}}^{(s),i}(z)$ 
transforming as  a vector representation under the $SO(4)$,
six higher spin-$(s+1)$ currents $\Phi_{1}^{(s),ij}(z)$
transforming as an adjoint representation under the 
$SO(4)$, four higher spin-$(s+\frac{3}{2})$ 
currents $\Phi_{\frac{3}{2}}^{(s),i}(z)$ 
and 
the higher spin-$(s+2)$ current $\Phi_2^{(s)}(z)$. 
The higher spin-$s$ and the higher spin-$(s+2)$ currents 
are $SO(4)$ singlets.
One can easily see that the subscript $(0, \frac{1}{2}, 1, \frac{3}{2}, 2)$ 
in the component currents
stands for the number of Grassmann coordinates in the ${\cal N}=4$
superspace description.
Depending on the super spin $s$, the above ${\cal N}=4$ (higher spin) multiplet 
is a bosonic higher spin current for integer spin $s$ or 
a fermionic higher spin current for half integer spin $s$. 
In general, the ${\cal N}=4$ (higher spin) multiplet has 
nontrivial $SO(4)$ representation \cite{Schoutensnpb}. 

\subsection{The ${\cal N}=4$ primary condition}

Because the superspin of ${\bf J^{(4)}}(Z_1)$ 
is zero, the right hand side 
of the OPE 
${\bf J^{(4)}}(Z_{1}) \, {\bf \Phi^{(s)}}(Z_{2})$
has a superspin $s$. The pole structure of the 
linear term in $ {\bf \Phi^{(s)}}(Z_{2})$ in the
right hand side should have the spin-$0$ without any $SO(4)$ indices.
This implies that the structure should be $\frac{\theta_{12}^{4}}{z_{12}^{2}}$,
where the spin of $\frac{1}{z_{12}}$ 
is equal to $1$ and the spin of $\theta_{12}$
is equal to $-\frac{1}{2}$. The ordinary derivative term can occur
at the singular term $\frac{\theta_{12}^{4}}{z_{12}}$. Furthermore 
the spinor derivative terms (descendant terms) 
with triple product of $\theta_{12}$ arise.
Finally,    one obtains the following ${\cal N}=4$ primary condition
for the $16$ higher spin currents in ${\cal N}=4$ superspace \cite{Schoutensnpb}
\bea
{\bf J^{(4)}}(Z_{1}) \, {\bf \Phi^{(s)}}(Z_{2}) & = & 
\frac{\theta_{12}^{4}}{z_{12}^{2}} \, 2s\, {\bf \Phi^{(s)}}(Z_{2})+
\frac{\theta_{12}^{4-i}}{z_{12}} \, D^{i}  {\bf \Phi^{(s)}}(Z_{2})+
\frac{\theta_{12}^{4}}{z_{12}} \, 2 \, \partial {\bf \Phi^{(s)}}(Z_{2})
+\cdots,
\label{jphi}
\eea
where $\theta_{12}^{4} = \theta_{12}^1 \, \theta_{12}^2 \,
\theta_{12}^3 \, \theta_{12}^4$.
One can understand the numerical factor $2$ in the first and the 
last terms of (\ref{jphi}) by taking 
$D_1^1 \, D_1^2 \, D_1^3 \, D_1^4$ on both sides and putting 
$\theta_1^i =0=\theta^i_2$. Then the left hand side is 
given by the OPE $2 L(z_1) \, \Phi_0^{(s)}(z_2)$ and the right hand side 
is given by the expression
$\frac{1}{(z_1-z_2)^2} \, 2 s \, \Phi_0^{(s)}(z_2) + \frac{1}{(z_1-z_2)}\,
2 \, \pa \Phi_0^{(s)}(z_2) + \cdots $.
Therefore, by cancelling out the factor $2$ in this OPE,
one obtains the usual primary condition for the 
higher spin current $\Phi_0^{(s)}(z_2)$ \footnote{
One used the fact that there are no singular terms in the OPE 
of $U(z_1) \,  \Phi_0^{(s)}(z_2)$ by taking 
$\theta_1^i =0=\theta^i_2$ on both sides of (\ref{jphi}).
For the $s=1$, this is true 
because the lowest higher spin-$1$ current 
commutes with the spin-$1$ current $U(z_1)$ 
(and four spin-$\frac{1}{2}$ currents):
the so-called Goddard-Schwimmer mechanism \cite{GS}.
For the arbitrary $s$, the description of \cite{BCG}
implies that this is also true.
In this calculation, one can easily see that there is no 
contribution from the second term of (\ref{jphi}) because
the action of $D_1^1 \, D_1^2 \, D_1^3 \, D_1^4$ on this second term, 
with the condition $\theta_1^i =0=\theta^i_2$,
vanishes.}.

What happens for the coefficient of the second term in (\ref{jphi})?
In this case, one can apply 
the differential operator $ D_1^1 \, D_1^2 \, D_1^3 \, D_1^4 \, D_2^1$
with the condition 
$\theta_1^i =0=\theta^i_2$
to both sides of (\ref{jphi}).
After this action, the left hand side is given by 
the OPE $2 L(z_1) \, \Phi_{\frac{1}{2}}^{(s), 1}(z_2)$, where 
one also uses the fact that 
the OPE $U(z_1) \,  \Phi_{\frac{1}{2}}^{(s), 1}(z_2)$ 
does not have any singular terms. 
Also one assumes that the super spin $s$ is an integer \footnote{
On the other hand, the 
right hand side from the first term in (\ref{jphi}) is 
given by $\frac{1}{(z_1-z_2)^2} \, 2 s \, \Phi_{\frac{1}{2}}^{(s),1}(z_2)$.
The second term gives the singular term  
 $\frac{1}{(z_1-z_2)^2}  \, \Phi_{\frac{1}{2}}^{(s),1}(z_2)$,
where the identity $ D_2^1 \, 
\frac{1}{z_{12}} = - \frac{\theta_{12}^1}{z_{12}^2}$ 
is used. The third term gives 
$\frac{1}{(z_1-z_2)} \, 2\, \pa \Phi_{\frac{1}{2}}^{(s),1}(z_2)$.
Combining all the contributions, one obtains
the following expression $L(z_1) \, \Phi_{\frac{1}{2}}^{(s), 1}(z_2) =
\frac{1}{(z_1-z_2)^2} \, (s+\frac{1}{2}) \, \Phi_{\frac{1}{2}}^{(s),1}(z_2)+
\frac{1}{(z_1-z_2)} \,  \pa \Phi_{\frac{1}{2}}^{(s),1}(z_2) +\cdots$,
which will appear in Appendix $C$.
 The coefficient of 
the second term of (\ref{jphi}) is fixed by the $\frac{1}{2}$
of the second-order pole for the primary condition.}. 

The other cases for the remaining higher spin-$(s+\frac{1}{2})$
currents can be analyzed similarly.
Let us emphasize that for the nonsinglet $SO(4)$ 
representation for the ${\cal N}=4$ higher spin multiplet, 
there exist the nontrivial extra terms 
in the above OPE \cite{Schoutensnpb}. In particular, 
the $\frac{\theta_{12}^{4-ij}}{z_{12}}$-term contracted with 
the $SO(4)$ representation $T^{ij}$ (which is not a spin-$1$ current) occurs. 
Furthermore, one of extra indices in $T^{ij}$ is contracted with the index
of ${\cal N}=4$ (higher spin) multiplet. See also \cite{IKL1,IKL2}. 

\subsection{The ${\cal N}=4$ primary condition  in the component approach }

From (\ref{jphi}), one can read off its component expressions.
For example, let us consider the OPE 
$G^1(z_1) \, \Phi_{1}^{(s),12}(z_2)$.
Then one should multiply the differential operator $D_1^{2} \, D_1^{3} \, D_1^{4}
\, D_2^3 \, D_2^4$ on both sides of (\ref{jphi}).
At the final stage, one puts the condition 
$\theta^i_1=\theta^i_2=0$. Then the left hand side of the OPE
is given by $(G^1- 2\, \alpha \, i \, \pa \Gamma^1)(z_1) \, 
\Phi_1^{(s),12}(z_2)$ while the right hand side is given by
the contribution from the second term of (\ref{jphi}). That is, 
the action of $D_1^{2} \, D_1^{3} \, D_1^{4}$ into 
$\frac{\theta_{12}^{234}}{z_{12}}$ gives the singular term $\frac{1}{z_{12}}$
and the factor $ D_2^3 \, D_2^4 \, D_2^1 \, {\bf \Phi^{(s)}}(Z_2)$
provides the expression 
$-\Phi_{\frac{3}{2}}^{(s),2}(z_2)$ after the projection of 
$\theta_1^i=0=\theta_2^i$.
Combining all the factors leads to the final expression 
$-\frac{1}{(z_1-z_2)} \, \Phi_{\frac{3}{2}}^{(s),2}(z_2) +\cdots$.
In order to see the contribution from the left hand side in the above,
one should consider the OPE $\Gamma^1(z_1) \, \Phi_1^{(s),12}(z_2)$ 
\footnote{
Again, let us multiply
$D_1^{1} 
\, D_2^3 \, D_2^4$ on both sides of (\ref{jphi}).   
Then one obtains  that 
the left hand side is given by the OPE
$-i \, \Gamma^1(z_1) \, \Phi_1^{(s),12}(z_2)$
after projection. The nontrivial contribution from the 
right hand side can be written as $-\frac{1}{(z_1-z_2)} \, 
\Phi_{\frac{1}{2}}^{(s),2}(z_2) +\cdots$.
One concludes that the corresponding OPE 
can be written as $
 \Gamma^1(z_1) \, \Phi_1^{(s),12}(z_2)= 
-\frac{1}{(z_1-z_2)} \, i \,  
\Phi_{\frac{1}{2}}^{(s),2}(z_2) +\cdots $, which can be seen from 
Appendix $C$. Furthermore, one has 
$ \pa  \Gamma^1(z_1) \, \Phi_1^{(s),12}(z_2)= 
\frac{1}{(z_1-z_2)^2} \, i \,  
\Phi_{\frac{1}{2}}^{(s),2}(z_2) +\cdots$.}.
Therefore, one obtains the following 
OPE $
G^1(z_1) \, 
\Phi_1^{(s),12}(z_2)$ as 
$- \frac{1}{(z_1-z_2)^2} \, 2 \, \alpha \,  
\Phi_{\frac{1}{2}}^{(s),2}(z_2) - \frac{1}{(z_1-z_2)}\,
\Phi_{\frac{3}{2}}^{(s),2}(z_2)  +\cdots$, which can be seen from 
Appendix $C$.
In this way, one can obtain all the component results in Appendix $C$.
 
\subsection{The ${\cal N}=4$ multiplet where all the component currents 
are primary under the bosonic stress energy tensor}


According to the OPEs in Appendix $C$, 
the higher spin currents 
$\Phi_{0}^{(s)}(w)$, $
\Phi_{\frac{1}{2}}^{(s),i}(w)$ and $\Phi_{1}^{(s),ij}(w)$
of spins $s, (s+\frac{1}{2})$ and $(s+1)$
respectively
are primary fields under the stress energy tensor $L(z)$.
However, 
the higher spin 
currents $\Phi_{\frac{3}{2}}^{(s),i}(w)$ and 
$\Phi_{2}^{(s)}(w)$ are not primary fields.
One can make them be primary ones by introducing 
other composite or derivative term as follows:
\begin{eqnarray}
{\bf \Phi^{(s)}}(Z)&=&\Phi_{0}^{(s)}(z)+\theta^{i}\:
\Phi_{\frac{1}{2}}^{(s),i}(z)+\theta^{4-ij}\:\Phi_{1}^{(s),ij}(z)
+\theta^{4-i}\,\Bigg[\widetilde\Phi_{\frac{3}{2}}^{(s),i}(z)+
\frac{2}{(2s+1)}\alpha\,\partial\Phi_{\frac{1}{2}}^{(s),i}(z)\Bigg]
\nonumber\\&+&\theta^{4-0}\,\Bigg[\widetilde\Phi_{2}^{(s)}(z)+
p_{1}\:\partial^{2}\Phi_{0}^{(s)}(z)+p_{2}\:L\,\Phi_{0}^{(s)}(z)\Bigg],
\label{phidiff}
\end{eqnarray}
where the coefficients
depend on the spin $s$, $N$ and $k$ 
\begin{eqnarray}
p_{1} & = & 
\frac{2(k - N) (3 + 3 k + 3 N + 3 k\,N + 26 s + 13 k\,s + 13 N\,s)}{(2 +
     k + N) (3 + 3 k + 3 N + 3 k\,N - 4 s + k\,s + N\,s + 6 k\,N\,s + 
    16 s^2 + 8 k\,s^2 + 8 N\,s^2)}, \nonu \\
p_{2} & = & 
-\frac{12 \,(k - N)\, s \,(1 + s)}{(3 + 3 k + 3 N + 3 k\, N - 4 s + k
\, s + N \,s + 
 6 k\, N\,s + 16 s^2 + 8 k \,s^2 + 8 N\, s^2)}\;,
\label{p1p2} 
\end{eqnarray}
For $N=k$, these coefficients are vanishing.
It is straightforward to  obtain these two coefficients
explicitly by requiring that 
$\widetilde\Phi_{2}^{(s)}(w) \equiv \Phi_{2}^{(s)}(w)
-p_{1}\:\partial^{2}\Phi_{0}^{(s)}(w)-p_{2}\:L\,\Phi_{0}^{(s)}(w)$
with (\ref{p1p2})
transform as a primary current under the stress energy tensor
$L(z)$ via Appendix $C$. For $s=1$, this observation 
was found in \cite{Ahn1504}.

%

\subsection{The ${\cal N}=4$ multiplet in other basis }

As in previous section, one can 
also compare the expression (\ref{phis}) with the
corresponding quantity in \cite{BCG} by following the procedure 
for the $16$ currents previously.
Let us write down the answer as follows:
\bea
{\bf V^{(s)}}(Z) & = & 
i\:V_{0}^{(s)}(z)-i\, \theta^{1}\:V_{\frac{1}{2}}^{(s),1}(z)+i \, \theta^{2}
\:V_{\frac{1}{2}}^{(s),2}(z)+i \, \theta^{3}\:V_{\frac{1}{2}}^{(s),3}(z)
+i \, \theta^{4}
\:V_{\frac{1}{2}}^{(s),4}(z)
\nonu \\
& + & 
\theta^{12}\, \frac{i}{2}(V_{1}^{(s),+3}+V_{1}^{(s),-3})(z)
-\theta^{13} \, \frac{i}{2} (V_{1}^{(s),+2}+V_{1}^{(s),-2})(z)+
\theta^{14}\, \frac{i}{2} (V_{1}^{(s),+1}-V_{1}^{(s),-1})(z)
\nonu \\
& + & 
\theta^{23} \frac{i}{2} (-V_{1}^{(s),+1}-V_{1}^{(s),-1})(z)-
\theta^{24}  \frac{i}{2} (V_{1}^{(s),+2}-V_{1}^{(s),-2})(z)+
\theta^{34} \frac{i}{2}(-V_{1}^{(s),+3}+V_{1}^{(s)-3})(z)
\nonu \\
& + & 
\theta^{243}\, \frac{i}{2} \left[V_{\frac{3}{2}}^{(s),1}-
\frac{4\alpha}{(2s+1)}\partial V_{\frac{1}{2}}^{(s),1} \right](z)-
\theta^{134}\, \frac{i}{2} \left[V_{\frac{3}{2}}^{(s),2}-\frac{4\alpha}{(2s+1)}
\partial V_{\frac{1}{2}}^{(s),2}\right](z)\nonu \\
& - & 
\theta^{142} \, \frac{i}{2}\left[V_{\frac{3}{2}}^{(s),3}-
\frac{4\alpha}{(2s+1)}\partial V_{\frac{1}{2}}^{(s),3} \right]-
\theta^{123} \, \frac{i}{2} \left[ V_{\frac{3}{2}}^{(s),4}-
\frac{4\alpha}{(2s+1)}\partial V_{\frac{1}{2}}^{(s),4} \right](z)
\nonu \\
& - & 
\theta^{1234} \, \frac{i}{2} \left[ V_{2}^{(s)}-\frac{4\alpha}{(2s+1)}
\partial^{2} V_{0}^{(s)}\right](z).
\label{bcghigher}
\eea
Let us consider the lowest higher spin-$1$ current.
In \cite{BCG}, 
the normalization for the OPE $V_0^{(s)} (z) \, 
V_0^{(s)}(w)$ for $s=1$ has an extra minus sign while in 
our case there is a positive sign.
Therefore, one introduces the complex number 
$i$ as in (\ref{bcghigher}).
Now move on the next higher spin-$(s+\frac{1}{2})$ current.
Let us consider the OPE $G^i(z) \, \Phi_0^{(s)}(w)$ in Appendix $C$.
For the index $i=1$, then the spin-$\frac{3}{2}$ current 
$G^1(z)$ is the same as the one in \cite{BCG}.
In other words, one obtained $G^1_{bcg}(z) =G^1(z)$ before.
By substituting $\Phi_0^{(s)}(w)$ as $i \, V_{0}^{(s)}(w)$,
the left hand side is given by $i$ times the OPE $G_{bcg}^1(z) \, V_0^{(s)}(w)$,
which is equal to the expression 
$-\Phi_{\frac{1}{2}}^{(s),1}(w)$ from Appendix $C$.
This implies that the expression $i \, \Phi_{\frac{1}{2}}^{(s),1}(w)$
is equal to the expression $V_{\frac{1}{2}}^{(s),1}(w)$ from Appendix $D$.
Therefore, one concludes that the higher spin current 
$\Phi_{\frac{1}{2}}^{(s),1}(w)= -i \,
V_{\frac{1}{2}}^{(s),1}(w)$ as in (\ref{bcghigher}).
For the indices $i=2,3,4$, the corresponding 
spin-$\frac{3}{2}$ currents have an extra minus sign (that is, 
$G^i_{bcg}(z) =-G^i(z)$)
and this reflects the signs for the higher spin-$(s+\frac{1}{2})$ 
currents with these indices
as in (\ref{bcghigher}).

Let us consider the next higher spin-$(s+1)$ currents.
Let us look at the OPE $G^i(z) \, \Phi_{\frac{1}{2}}^{(s),j}(w)$
with indices $(i,j)=(1,2)$, which contains the 
first-order pole $\Phi_1^{(s),34}(w)$. Then the left hand side
corresponds to the OPE $G_{bcg}^1(z) \, (i) V_{\frac{1}{2}}^{(s),2}(w)$.
According to Appendix $D$, the corresponding OPE is 
given by the third OPE.  The first-order pole 
is given by the expression $i (\alpha_{12}^{+,i} \, V_1^{(s),+i} + 
\alpha_{12}^{-,i} \, V_1^{(s),-i})(w) = \frac{i}{2} ( V_1^{(s),+3} + 
V_1^{(s),-3})(w)$. This is exactly the one given in (\ref{bcghigher}). 
Then one arrives at the following results for the higher spin-$(s+1)$
currents as follows: 
\bea
\Phi_{1}^{(s),14}(z) & = & 
-\frac{i}{2 } \left( V_{1}^{(s),+1}+ V_{1}^{(s),-1} \right)(z), \qquad
\Phi_{1}^{(s),23}(z)= \frac{i}{2 } \left( V_{1}^{(s),+1}-V_{1}^{(s),-1} \right)(z),
\nonu \\
\Phi_{1}^{(s),42}(z) & = & -\frac{i}{2} \left( V_{1}^{(s),+2}+V_{1}^{(s),-2} 
\right)(z),
\qquad \Phi_{1}^{(s),31}(z) =
-\frac{i}{2} \left(V_{1}^{(s),+2}-V_{1}^{(s),-2}\right)(z),
\nonu \\
\Phi_{1}^{(s),34}(z) & = & \frac{i}{2} \left(V_{1}^{(s),+3}+
V_{1}^{(s),-3}\right)(z),
\qquad \Phi_{1}^{(s),12}(z)= - \frac{i}{2} \left(V_{1}^{(s),+3}-
V_{1}^{(s),-3}\right)(z).
\label{splusone}
\eea

Let us continue to describe the next higher spin-$(s+\frac{3}{2})$
currents by  
starting from the OPE $G^1(z) \, \Phi_1^{(s),12}(w)$ in Appendix $C$.
The first-order pole gives 
$-\Phi_{\frac{3}{2}}^{(s),2}(w)$.
From the relation (\ref{splusone}), one can calculate 
the OPE $(-\frac{i}{2}) \, G_{bcg}^1(z) \, ( V_{1}^{(s),+3}- V_{1}^{(s),-3})(w)$ 
using the results of Appendix $D$.
It turns out that the first-order pole is given by
the expression $\frac{i}{2} \, (V_{\frac{3}{2}}^{(s),2}-\frac{4\alpha}{(2s+1)}
\partial V_{\frac{1}{2}}^{(s),2})$, where the $\alpha_{ab}^{\pm,i}$ are substituted.
By considering the appropriate coefficients, one arrives at the 
final expression located at the $\theta^{134}$-term in (\ref{bcghigher}). 
The similar analysis for other type of higher spin-$(s+\frac{3}{2})$ currents
can be done
\footnote{Let us describe the final higher spin-$(s+2)$ current.
As done before, one considers the OPE
$G^1(z) \, \Phi_{\frac{3}{2}}^{(s),1(w)}$ with the help of 
Appendix $C$. The first-order pole is given by $- 
\Phi_2^{(s)}(w)$.
On the other hand, the OPE 
$G_{bcg}^1(z) \, V_{\frac{3}{2}}^{(s),1(w)}$ leads to the first-order pole 
$V_2^{(s)}(w)$. The OPE 
$G_{bcg}^1(z) \, \pa V_{\frac{1}{2}}^{(s),1(w)}(w)$ has  a first-order pole 
$\pa^2 V_0^{(s)}(w)$. Combining all the coefficients, one obtains 
the final expression in (\ref{bcghigher}).}. 

\subsection{The explicit relations between the higher spin currents 
in different basis }

Starting from the following higher spin-$1$ current
\bea
V_0^{(1)}(z) & = &  -i \, T^{(1)}(z) = -i \, \Phi_0^{(1)}(z), 
\label{lowestspin1}
\eea
where the relations (\ref{phis}) and (\ref{bcghigher}) are used, 
one can write down the higher spin-$\frac{3}{2}$ currents  as follows:
\bea
V_{\frac{1}{2}}^{(1),1}(z) & = & 
i\, \left(-i \, G^2 
+ \sqrt{2} \, T_{+}^{(\frac{3}{2})} + \sqrt{2}\, 
 T_{-}^{(\frac{3}{2})} \right)(z) = i \, \Phi_{\frac{1}{2}}^{(1),1}(z),
\nonu \\
V_{\frac{1}{2}}^{(1),2}(z) & = & 
 - \left(  G^1 
+  \sqrt{2}\,  T_{+}^{(\frac{3}{2})} - \sqrt{2} \, 
 T_{-}^{(\frac{3}{2})} \right)(z) =-i \, \Phi_{\frac{1}{2}}^{(1),2}(z), 
\nonu \\
V_{\frac{1}{2}}^{(1),3}(z) & = &  
i\, \left(   i  \, G^4 
+ \sqrt{2} \, U^{(\frac{3}{2})} +  \sqrt{2}\,
 V^{(\frac{3}{2})} \right)(z) =
 -i \, \Phi_{\frac{1}{2}}^{(1),3}(z),\nonu \\
V_{\frac{1}{2}}^{(1),4}(z) & = & 
 \left(  G^3 
+ \sqrt{2}\,  U^{(\frac{3}{2})} - \sqrt{2} \, 
 V^{(\frac{3}{2})} \right)(z) =
 -i \, \Phi_{\frac{1}{2}}^{(1),4}(z). 
\label{v3half}
\eea
The first relations of (\ref{v3half}) were obtained in \cite{Ahn1504} and the 
second relations can be obtained from (\ref{phis}) and (\ref{bcghigher}).
Obviously the higher spin-$\frac{3}{2}$ currents 
$T_{\pm}^{(\frac{3}{2})}(z)$, $U^{(\frac{3}{2})}(z)$ and $V^{(\frac{3}{2})}(z)$
can be written in terms of linear combinations of 
the higher spin current $\Phi_{\frac{1}{2}}^{(1),i}(z)$.

Similarly, the higher spin-$2$ currents can be written in terms of 
linear combination of the higher spin current 
$\Phi_{1}^{(1),ij}(z)$ as follows:
\bea
V_1^{(1), \pm 1}(z) & = & 
\pm 2 \left(  U_{\mp}^{(2)} \mp 
 V_{\pm}^{(2)} \right)(z) = i\, \left( \Phi_1^{(1), 14} 
\mp \Phi_1^{(1), 23} \right)(z), 
\nonu \\
V_1^{(1), \pm 2}(z) & = & 
-2i \left(  U_{\mp}^{(2)} + 
V_{\pm}^{(2)} \right)(z) = i\, \left( \Phi_1^{(1), 42} 
\pm \Phi_1^{(1), 31} \right)(z), 
\nonu \\
V_1^{(1), \pm 3}(z) & = & 
2 \left(  T^{(2)} \mp
 W^{(2)} \right)(z)= i\, \left( \Phi_1^{(1), 34} 
\pm \Phi_1^{(1), 12} \right)(z). 
\label{vspin2}
\eea
In (\ref{vspin2}), the last relations are obtained from (\ref{bcghigher}) 
and (\ref{phis}).

Furthermore, one has the following relations for the higher 
spin-$\frac{5}{2}$ currents  
\bea
V_{\frac{3}{2}}^{(1), 1}(z)  & = &  
-2i \sqrt{2} \left(  W_{+}^{(\frac{5}{2})} - 
 W_{-}^{(\frac{5}{2})} \right)(z) = -2 i \left( 
\Phi_{\frac{3}{2}}^{(1), 1} +\frac{2\alpha}{3} \pa \Phi_{\frac{1}{2}}^{(1), 1}
\right)(z),  
\nonu \\
V_{\frac{3}{2}}^{(1), 2}(z) & = & 
2  \sqrt{2} \left(  W_{+}^{(\frac{5}{2})} + 
 W_{-}^{(\frac{5}{2})} \right)(z)
= 2 i \left( 
\Phi_{\frac{3}{2}}^{(1), 2} -\frac{2\alpha}{3} \pa \Phi_{\frac{1}{2}}^{(1), 2}
\right)(z),   
\nonu \\
V_{\frac{3}{2}}^{(1), 3}(z)  & = &  
-2i  \sqrt{2} \left( U^{(\frac{5}{2})} - 
V^{(\frac{5}{2})} \right)(z)=2 i \left( 
\Phi_{\frac{3}{2}}^{(1), 3} -\frac{2\alpha}{3} \pa \Phi_{\frac{1}{2}}^{(1), 3}
\right)(z), 
\nonu \\
V_{\frac{3}{2}}^{(1), 4}(z) & = & 
-2  \sqrt{2} \left(  U^{(\frac{5}{2})} + 
 V^{(\frac{5}{2})} \right)(z)=
2 i \left( 
\Phi_{\frac{3}{2}}^{(1), 4} -\frac{2\alpha}{3} \pa \Phi_{\frac{1}{2}}^{(1), 4}
\right)(z).  
\label{v5half}
\eea
The first relations of (\ref{v5half}) were obtained from \cite{Ahn1504}
and the last relations are obtained from the previous results 
in (\ref{phis}) and (\ref{bcghigher}).

Finally, the higher spin-$3$ current has the following relations 
\cite{AK1506}
\bea
V_{2}^{(1)}(z) & = & 
4i \left[  W^{(3)} + 
 \frac{4  (k-N)}{((4N+5)+(3N+4)k)}  \Big( \, T^{(1)} \, L -\frac{1}{2}\,\partial^{2}T^{(1)}\, \Big)  \right](z)
\nonu \\
& = & 
i \left( \Phi_{2}^{(1)} - \frac{2\alpha}{3}  \, \pa^2 \Phi_0^{(1)}
\right)(z).  
\label{bcgspin3}
\eea
In (\ref{bcgspin3}), the primary higher spin-$3$ current $W^{(3)}(z)$
under the stress energy tensor $L(z)$ can be written in terms of 
$ \Phi_{2}^{(1)}(z)$, $\pa^2 \Phi_0^{(1)}$ and $\Phi_0^{(1)} \, L(z)$ with
(\ref{lowestspin1}).

For the next higher spin ${\cal N}=4$ multiplet of super spin $2$, 
the exact relations 
between the ones in \cite{Ahn1504} and the ones in (\ref{phis}) with $s=2$,
for general $N$ and $k$, are rather complicated. 

\subsection{The realization of the lowest ${\cal N}=4$ multiplet
in the ${\cal N}=4$ coset theory}

As done in the realization of the large ${\cal N}=4$ linear superconformal 
algebra in ${\cal N}=4$ coset theory, 
one can construct the $16$ higher spin currents in ${\cal N}=4$
coset theory, where there are two fundamental currents (the spin-$1$ current
and the spin-$\frac{1}{2}$ current) \cite{AK1506}. 
See the previous subsection $2.5$.  
It would be interesting to see whether one can construct 
the $16$ higher spin currents using the ${\cal N}=4$ affine Kac-Moody current
$Q^a(Z)$ of super spin $\frac{1}{2}$ 
via the generalized Sugawara construction.

\section{The OPEs between the $16$ currents  in 
${\cal N}=2$ superspace:Review}

In this section, the description for the ${\cal N}=2$ superspace
is given.
The previous ${\cal N}=4$ stress energy tensor is written in terms of 
four ${\cal N}=2$ currents in 
${\cal N}=2$ superspace.
The ${\cal N}=4$ primary current condition will be 
 described in the ${\cal N}=2$
superspace later.
Furthermore, the different expansions in the Grassmann coordinates
are related to the ${\cal N}=3$ superconformal algebra and 
the ${\cal N}=1$
superspace description is also reviewed.

\subsection{The basic structure of ${\cal N}=2$ superspace}

The coordinates of ${\cal N}=2$ (extended) superspace 
can be described as 
$(Z, \overline{Z})$ where 
$Z=(z, \theta^i)$, $\overline{Z} =(\bar{z}, \bar{\theta}^i)$
and $i =1,2$ stands for $SO(2)$-index. 
The ``left'' covariant spinor derivatives 
are given by 
$ D = -\frac{1}{2} \overline{\theta} 
\frac{\pa}{\pa z } +  \frac{\pa}{\pa {\theta}}$
and $\overline{D} = -\frac{1}{2} \theta 
\frac{\pa}{\pa z } +  \frac{\pa}{\pa {\overline{\theta}}}$,
where one has $\theta \equiv \theta^1 + i \, \theta^2$
and $\overline{\theta} \equiv \theta^1 - i \, \theta^2$.
The  following nontrivial anticommutator 
$
\{ D, \overline{D} \} = -\frac{\pa}{\pa z}$ satisfies.
Furthermore 
one uses
the simplified notations
$\theta_{12} \equiv \theta_1 -\theta_2$, ${\overline{\theta}}_{12} \equiv 
{\overline{\theta}}_1 -{\overline{\theta}}_2$ and 
$z_{12} \equiv z_1 -z_2 + \frac{1}{2} (\theta_1 {\overline{\theta}}_2 + 
{\overline{\theta}}_1 \theta_2)$.
One can easily see that the $\theta$- and $\overline{\theta}$-dependent 
parts in $z_{12}$ are the same as the expression 
$(\theta_1^1 \theta_2^1 + \theta_1^2
\theta_2^2)$. Because one uses the ${\cal N}=2$ package by 
Krivonos and Thielemans \cite{KT}, the same notations are also
used in this paper.

\subsection{The OPE of ${\cal N}=4$ stress energy tensor in ${\cal N}=2$
superspace}

The $16$ currents of the large ${\cal N}=4$ linear superconformal algebra
can be written in terms of ${\cal N}=2$ one spin-$1$ current ${\bf T}(Z)$,
four spin-$\frac{1}{2}$ currents ${\bf G}(Z)$, ${\bf \overline{G}}(Z)$, $
{\bf H}(Z)$ and ${\bf \overline{H}}(Z)$. 
The precise relation between these currents and its components are given 
in 
Appendix $E$.
Due to the chiral current ${\bf H}(Z)$ (that is, $D {\bf H}(Z) =0$) and 
the antichiral current ${\bf \overline{H}}(Z)$
(that is, $\overline{D} {\bf \overline{H}}(Z)=0$),
in component approach, there exist four independent component currents 
(not eight).

From the component approach in Appendix $A$ and Appendix (\ref{16compn2}), 
one can rewrite them in terms of
the following ${\cal N}=2$ super OPEs \cite{RASS,Ahn1992,BO}
\bea
{\bf T}(Z_{1}) \, {\bf T}(Z_{2}) & = & 
\frac{1}{z_{12}^2} \, \frac{1}{3}\, c +
\frac{\theta_{12}\bar{\theta}_{12}}{z_{12}^{2}} \,  {\bf T}(Z_{2})-
\frac{\theta_{12}}{z_{12}} \, D {\bf T}(Z_{2})+
\frac{\bar{\theta}_{12}}{z_{12}} \, \overline{D} {\bf T}(Z_{2})
\nonu \\
& + & 
\frac{\theta_{12}\bar{\theta}_{12}}{z_{12}} \, \partial { \bf T}(Z_{2})
+\cdots,
\nonu \\
{\bf T}(Z_{1}) \, 
\left(
\begin{array}{c}
{\bf G} \nonu \\
{\bf \overline{G}} 
\end{array}
\right)(Z_{2}) & = & 
\frac{\theta_{12}\bar{\theta}_{12}}{z_{12}^{2}} \, \frac{1}{2}\, 
\left(
\begin{array}{c}
{\bf G} \nonu \\
{\bf \overline{G}} 
\end{array}
\right)(Z_{2})
 \pm \frac{1}{z_{12}} \, 2 \, \alpha \, 
\left(
\begin{array}{c}
{\bf G} \nonu \\
{\bf \overline{G}} 
\end{array}
\right)
(Z_2)
-
\frac{\theta_{12}}{z_{12}} \, D 
\left(
\begin{array}{c}
{\bf G} \nonu \\
{\bf \overline{G}} 
\end{array}
\right)(Z_{2}) \nonu \\
& + & 
\frac{\bar{\theta}_{12}}{z_{12}} \, \overline{D} 
\left(
\begin{array}{c}
{\bf G} \nonu \\
{\bf \overline{G}} 
\end{array}
\right)(Z_{2})
+  
\frac{\theta_{12}\bar{\theta}_{12}}{z_{12}} \, \partial 
\left(
\begin{array}{c}
{\bf G} \nonu \\
{\bf \overline{G}} 
\end{array}
\right)(Z_{2})
+\cdots,
\nonu \\
{\bf T}(Z_{1}) \, 
\left(
\begin{array}{c}
{\bf H} \nonu \\
{\bf \overline{H}} 
\end{array}
\right)(Z_{2}) & = & 
\frac{\theta_{12}\bar{\theta}_{12}}{z_{12}^{2}} \, \frac{1}{2}\, 
\left(
\begin{array}{c}
{\bf H} \nonu \\
{\bf \overline{H}} 
\end{array}
\right)(Z_{2})
 \pm \frac{1}{z_{12}} \, 
\left(
\begin{array}{c}
{\bf H} \nonu \\
{\bf \overline{H}} 
\end{array}
\right)
(Z_2)
\pm
\frac{1}{z_{12}} \,  
\left(
\begin{array}{c}
\bar{\theta}_{12} \overline{D} {\bf H} \nonu \\
\theta_{12} D {\bf \overline{H}} 
\end{array}
\right)(Z_{2}) \nonu \\
& + & 
\frac{\theta_{12}\bar{\theta}_{12}}{z_{12}} \, \partial 
\left(
\begin{array}{c}
{\bf H} \nonu \\
{\bf \overline{H}} 
\end{array}
\right)(Z_{2})
+\cdots,
\nonu \\
{\bf G}(Z_{1}) \, {\bf \overline{G}}(Z_{2}) & = & 
\frac{\theta_{12}\bar{\theta}_{12}}{z_{12}^{2}} \, \frac{1}{2}(k^{+}-k^{-}) -
\frac{1}{z_{12}} \, (k^{+} +k^{-})-
\frac{\theta_{12}}{z_{12}} \, {\bf H}(Z_{2})+
\frac{\bar{\theta}_{12}}{z_{12}} \, {\bf \overline{H}}(Z_{2})
\nonu \\
& - & 
\frac{\theta_{12}\bar{\theta}_{12}}{z_{12}} \, \left[
{ \bf T} +\frac{(1+2\alpha)}{2} \, \overline{D} {\bf H}
+  \frac{(1-2\alpha)}{2} \, D {\bf \overline{H}}\right](Z_{2})
+\cdots,
\nonu \\
{\bf H}(Z_{1}) \, 
\left(
\begin{array}{c}
{\bf G} \nonu \\
{\bf \overline{G}} 
\end{array}
\right)(Z_{2}) & = & 
\mp
\frac{\bar{\theta}_{12}}{z_{12}} \, 
\left(
\begin{array}{c}
{\bf G} \nonu \\
{\bf \overline{G}} 
\end{array}
\right)(Z_{2}) +\cdots,
\nonu \\
{\bf \overline{H}}(Z_{1}) \, 
\left(
\begin{array}{c}
{\bf G} \nonu \\
{\bf \overline{G}} 
\end{array}
\right)(Z_{2}) & = & 
\pm
\frac{{\theta}_{12}}{z_{12}} \, 
\left(
\begin{array}{c}
{\bf G} \nonu \\
{\bf \overline{G}} 
\end{array}
\right)(Z_{2}) +\cdots,
\nonu \\
{\bf H}(Z_{1}) \, {\bf \overline{H}}(Z_{2}) & = & 
-\frac{\theta_{12}\bar{\theta}_{12}}{z_{12}^{2}} \, \frac{1}{2}(k^{+}+k^{-}) +
\frac{1}{z_{12}} \, (k^{+} +k^{-}) +\cdots.
\label{n2N4SCA}
\eea
The parameters $c$ and $\alpha$ are given by (\ref{centralcharge}) 
and (\ref{alpha}).
Recall that the spin for any current is encoded in the singular term 
$\frac{
\theta_{12}\bar{\theta}_{12}}{z_{12}^2}$ in the OPE (\ref{n2N4SCA}) 
between ${\bf T}(Z_1)$
and the corresponding ${\cal N}=2$ current.
In component approach this implies that 
the last component of ${\bf T}(Z_1)$ is related to 
the bosonic stress energy tensor $L(z_1)$
from Appendix $E$.
Furthermore, the differential operator 
$-\frac{1}{2} [D_1, \overline{D}_1]$
acting on the singular term
$\frac{\theta_{12} \, \overline{\theta}_{12}}{z_{12}^2}$ 
provides the factor $\frac{1}{(z_1-z_2)^2}$.
The descendant terms can be obtained similarly.
The corresponding $\frac{1}{z_{12}}$-term in this OPE provides 
the $U(1)$ charge of ${\cal N}=2$ superconformal algebra (one spin-$1$
current, two spin-$\frac{3}{2}$ currents and one spin-$2$ current). 
Then the $U(1)$ charges for the ${\cal N}=2$ currents 
${\bf G}(Z)$, ${\bf \overline{G}}(Z)$,
${\bf H}(Z)$ and ${\bf \overline{H}}(Z)$
are given by $2\alpha, -2\alpha, 1, -1$ respectively.
One can interpret this observation in the component approach.
By taking $\theta = \overline{\theta}=0$ in these
OPEs, one is left with  
the singular term $\frac{1}{(z_1-z_2)}$.
Note that the first component of ${\bf T}(Z_1)$ is related to
the $U(1)$ charge of the ${\cal N}=2$ superconformal algebra mentioned before. 
In the OPEs between the ${\cal N}=2$ currents 
${\bf T}(Z_1)$ and ${\bf H}(Z_2)$ (or 
${\bf \overline{H}}(Z_2)$), the chiral and antichiral properties 
can be seen because there is no $\frac{\theta_{12}}{z_{12}}$-term (or
$\frac{\bar{\theta}_{12}}{z_{12}}$-term) 
\footnote{
For example, one can 
check the component results in Appendix $A$ 
starting from the above OPEs in (\ref{n2N4SCA}).
Let us consider the OPE $L(z_1) \, \Gamma^3(z_2)$ appearing in 
Appendix $A$.
By multiplying 
$-\frac{1}{2} [D_1, \overline{D}_1]$ into the second equation of 
(\ref{n2N4SCA}) with the condition $\theta=\bar{\theta}=0$, then 
one obtains the left hand side is given by the OPEs $L(z_1) \, \left(
\begin{array}{c}
i \Gamma^3 -\Gamma^4 \nonu \\
i \Gamma^3 +\Gamma^4
\end{array}
\right)(z_2)$. 
Using the nonzero identities, 
$-\frac{1}{2} [D_1, \overline{D}_1] 
\frac{\theta_{12}\bar{\theta}_{12}}{z_{12}^2}|_{\theta=\bar{\theta}=0}= 
\frac{1}{(z_1-z_2)^2}$ and $-\frac{1}{2} [D_1, \overline{D}_1] 
\frac{\theta_{12}\bar{\theta}_{12}}{z_{12}}|_{\theta=\bar{\theta}=0}= 
\frac{1}{(z_1-z_2)}$, where the other contributions from the 
differential operator acting on 
$\frac{1}{z_{12}}$, $\frac{\theta_{12}}{z_{12}}$ or 
$\frac{\bar{\theta}_{12}}{z_{12}}$ are zero,
the right hand side 
is given by the OPEs
$\frac{1}{(z_1-z_2)^2} \, \left(
\begin{array}{c}
i \Gamma^3 -\Gamma^4 \nonu \\
i \Gamma^3 +\Gamma^4
\end{array}
\right)(z_2) +\frac{1}{(z_1-z_2)} \, \pa \, \left(
\begin{array}{c}
i \Gamma^3 -\Gamma^4 \nonu \\
i \Gamma^3 +\Gamma^4
\end{array}
\right)(z_2) + \cdots$. Therefore, 
one sees the usual primary currents under the stress energy tensor
$L(z_1)$.}.
 
Following this procedure, one can check all the other component results 
in Appendix $A$ from its ${\cal N}=2$ version in (\ref{n2N4SCA})
or vice versa. 
In practical, one can use the package in \cite{KT}
to check them explicitly with the help of the command 
$\tt N2OPEToComponents[ope_{-},
A_{-},B_{-}]$, which calculates the $16$ OPEs of the components of
any two ${\cal N}=2$  currents ${\bf A}(Z)$ and ${\bf B}(Z)$.


One can express the ${\cal N}=4$ stress energy tensor in different form
as follows:
\bea
{\bf J^{(4)}}(Z)  & = & \Bigg[-\Delta(z) + \theta^1 \, i\, \Gamma^1(z) + 
\theta^2 \, i \, \Gamma^2(z) - \theta^1 \, \theta^2   i\, T^{34}(z) \Bigg]
\nonu \\
& + & \theta^3 \Bigg[ i \, \Gamma^3(z) - \theta^1 \, i\, T^{24}(z) 
+\theta^2 \, i\, T^{14}(z)- \theta^1 \, \theta^2
 (G^4 - 2 \, \alpha\, i \, \pa \,
\Gamma^4)(z) \Bigg] 
\nonu \\
&+& \theta^4 \Bigg[ i \, \Gamma^4(z)  +\theta^1 \, i\, T^{23}(z) 
- \theta^2 \, i\, T^{13}(z) +\theta^1 \, \theta^2  (G^3 - 
2 \, \alpha\, i \, \pa \,
\Gamma^3)(z) \Bigg] 
\nonu \\
&+& \theta^3 \, \theta^4 \Bigg[ -i \, T^{12}(z) 
-\theta^1  (G^2 - 2 \, \alpha\, i \, \pa \,
\Gamma^2)(z)+\theta^2 (G^1 - 2 \, \alpha\, i \, \pa \,
\Gamma^1)(z)  \nonu \\
& + &  \theta^1 \, 
\theta^2 (2 \, L - 2\, \alpha \, \pa^2 \Delta)(z) \Bigg]. 
\label{j4expression}
\eea
Intentionally one expands the ${\cal N}=4$ stress energy tensor 
in terms of the fermionic coordinates $\theta^3$ and $\theta^4$.
The $\theta^3$- and $\theta^4$-independent terms contain 
only the fermionic coordinates $\theta^1$, $\theta^2$ and 
the bosonic coordinate $z$.
Furthermore, by introducing $\theta$ and $\overline{\theta}$
as 
$
\theta \equiv \theta^{1}-i\,\theta^{2},
\overline{\theta} \equiv -\theta^{1}-i\,\theta^{2}$
(in order to compare the previous notations, one can 
redefine as $\overline{\theta} \rightarrow -\theta$ and $\theta \rightarrow 
\overline{\theta}$),
one can rewrite the above ${\cal N}=4$ stress energy tensor
with the help of ${\cal N}=2$ currents as follows: 
\bea
{\bf J^{(4)}}(Z) & = & 
\frac{1}{2} \, {\bf J}(z,\theta,\overline{\theta})+
\theta^{3} \, \frac{1}{2}\, \left( {\bf G}+{\bf \overline{G}} \right)(z,\theta,
\overline{\theta})+
\theta^{4}\, \frac{i}{2}\, \left(-{\bf G}+{\bf \overline{G}} \right)(z, 
\theta,\overline{\theta}) \nonu \\
& + &
\theta^{3} \, \theta^{4} \, i\, \left( {\bf T}- \alpha\,
[D,\overline{D}] \, {\bf J} \right)(z,\theta,\overline{\theta}),
\label{n4n2}
\eea
where the ${\cal N}=2$ current 
${\bf J}(z,\theta,\overline{\theta})$ is the 
$\theta^3$, $\theta^4$-independent term of (\ref{j4expression})
and the following relations hold
\bea
D {\bf J} \equiv   {\bf H}, \qquad 
\overline{D} {\bf J} \equiv {\bf \overline{H}}.
\label{iden}
\eea
Of course, using (\ref{iden}), it is easy to see that 
the 
last $\alpha$-dependent term in (\ref{n4n2}) can be written in terms of
$D {\bf \overline{H}}$ and $\overline{D} {\bf H}$
\footnote{One can also see the (\ref{n2N4SCA}) by acting the 
corresponding differential operators on (\ref{j4j4}), where the relation 
(\ref{iden}) is used. For example, 
let us multiply the differential operator 
$(D_2^3 + i D_2^4)$ both sides and put the condition 
$\theta^3 =0 =\theta^4$. Then the left hand side 
of this OPE leads to the OPE $\frac{1}{2} \, {\bf J}(Z_1) \ {\bf G}(Z_2)$,
while the right hand side is given by the expression 
$-\frac{\theta_{12} \, 
\overline{\theta}_{12}}{z_{12}} \, \frac{1}{2} \, {\bf G}(Z_2)$.
Furthermore by acting $D_1$ on this relation, 
the left hand side goes to the OPE 
$\frac{1}{2} \, {\bf H}(Z_1) \ {\bf G}(Z_2)$,
while the right hand side is given by the singular term $-\frac{ 
\overline{\theta}_{12}}{z_{12}} \, \frac{1}{2} \, {\bf G}(Z_2)$.
Therefore, one sees the expected result in (\ref{n2N4SCA}) by canceling out
the factor $\frac{1}{2}$. In this way, one can check all the other relations
in (\ref{n2N4SCA}) from (\ref{j4j4}) and (\ref{iden}) 
with the help of the differential operators $D^3$ and $D^4$
appropriately.}.

\subsection{The ${\cal N}=3$ and ${\cal N}=1$ superspace descriptions }

Moreover, one can expand the above ${\cal N}=4$ stress energy tensor
in terms of the fermionic coordinate $\theta^4$-term 
and $\theta^4$-independent 
term
\bea
{\bf J^{(4)}}(Z)  & = & \Bigg[ -\Delta(z) + \theta^1 \, i\, \Gamma^1(z) + 
\theta^2 \, i \, \Gamma^2(z) +
\theta^3 \, i \, \Gamma^3(z)
- \theta^1 \, \theta^2   i\, T^{34}(z) 
\nonu \\
& + & 
\theta^1 \, \theta^3 \, i\, T^{24}(z) 
- \theta^2 \, \theta^3 \, i\, T^{14}(z)- \theta^1 \, \theta^2\, \theta^3\,
 (G^4 - 2 \, \alpha\, i \, \pa \,
\Gamma^4)(z)\Bigg] 
\nonu \\
&+& \theta^4 \Bigg[ i \, \Gamma^4(z)  +\theta^1 \, i\, T^{23}(z) 
- \theta^2 \, i\, T^{13}(z) 
+ \theta^3 \, i\, T^{12}(z)
+\theta^1 \, \theta^2  (G^3 - 
2 \, \alpha\, i \, \pa \,
\Gamma^3)(z) 
\nonu \\
&-& \theta^1 \, 
\theta^3  (G^2 - 2 \, \alpha\, i \, \pa \,
\Gamma^2)(z)+ \theta^2 \, \theta^3 (G^1 - 2 \, \alpha\, i \, \pa \,
\Gamma^1)(z)  \nonu \\
& - & 
\theta^1 \, 
\theta^2 \, \theta^3 (2 \, L - 2\, \alpha \, \pa^2 \Delta)(z) \Bigg]. 
\label{j4relationj3}
\eea
Then one sees that the spin-$\frac{1}{2}$ current $\Gamma^4(z)$, 
three spin-$1$ currents $-\frac{1}{2} \epsilon^{ijk} \, T^{jk}(z)$, 
three spin-$\frac{3}{2}$ currents 
$( G^i - 2\alpha \, i\, \pa \Gamma^i)(z)$ and 
one spin-$2$ current $(L-\alpha\, \pa^2 \Delta)(z)$ appearing in the 
fermionic coordinate $\theta^4$-dependent terms in (\ref{j4relationj3}) consist 
of ${\cal N}=3$ superconformal algebra 
\cite{CK} with $SO(3)$ symmetry. The index $i=1,2,3$ stands for 
the $SO(3)$-vector index. The central charge is given by 
$c =\frac{3}{2}(k^{+} +k^{-})$ \footnote{One can see the 
standard ${\cal N}=3$ superconformal algebra by using (\ref{j4j4})
and (\ref{j4relationj3}). Let us multiply $D_1^4 \, D_2^4$ into 
the equation (\ref{j4j4}) and put the condition $\theta^4=0$.
Then the left hand side is given by the OPE $ (-1) {\bf J^{(3)}}(Z_1)  
\, (-1) {\bf J^{(3)}}(Z_2)$, where one introduces $ (-1) {\bf J^{(3)}}(Z)$
corresponding to the third-fifth lines of (\ref{j4relationj3}). Then 
the nontrivial contributions from the right hand side are coming from 
the second and third terms of (\ref{j4j4}) as one acts on the above
operator $D_1^4 \, D_2^4$. The former can be written as 
the expression $\frac{\theta_{12}^{3} 
}{z_{12}^2} \, {\bf J^{(3)}}(Z_2)+ \frac{\theta_{12}^{3-i}
}{z_{12}} \, D^i \, {\bf J^{(3)}}(Z_2)
$. The latter is given by 
the singular term 
$\frac{\theta_{12}^{3}}{z_{12}} \, 2\, \pa \, {\bf J^{(3)}}(Z_2)$. 
Then one arrives at the 
${\cal N}=3$ superconformal algebra in ${\cal N}=3$ superspace 
\cite{Schoutensnpb}.
\label{n3decompos} }.

One can decompose the ${\cal N}=4$ stress energy tensor as follows
\cite{RASS,Ahn1992}: 
\bea
{\bf J^{(4)}}(Z) & = & -\Delta(z) + i \, \theta^2 \, \Gamma^2(z)
+  \theta^1 \Bigg[  i \, \Gamma^1(z) - i\, \theta^2 \, T^{34}(z) \Bigg] 
+  \theta^3 \Bigg[ i \, \Gamma^3(z) + i \, \theta^2\, T^{14}(z) \Bigg]
\nonu \\
& + &  \theta^4 \Bigg[ i \, \Gamma^4(z) - i \, \theta^2 \, T^{13}(z) \Bigg]
+  \theta^1 \, \theta^3  \Bigg[ i \, T^{24}(z) + \theta^2 
\, \left( G^4 - 2 \alpha\, i\, \pa \Gamma^4\right)(z) \Bigg]
\nonu \\
& + & \theta^1 \, \theta^4 \Bigg[- i \, T^{23}(z) -  \theta^2 \,
\left( G^3 - 2 \alpha\, i\, \pa \Gamma^3\right)(z) \Bigg]
\nonu \\
& + & \theta^3\, \theta^4 \Bigg[ -i \, T^{12}(z) + \theta^2 \left( G^1 - 2 \alpha\, i\, \pa \Gamma^1 \right)(z) \Bigg]
\nonu \\
& + & \theta^1 \, \theta^3 \, \theta^4 \Bigg[ -\left(G^2 - 2\alpha \, i\, \pa 
\Gamma^2\right)(z)
+ \theta^2 \, \left( 2 L - 2\alpha \, \pa^2 \Delta \right)(z) 
\Bigg]
\nonu \\
& = &  \Bigg[ -\Delta(z) + i \, \theta \, \Gamma^2(z) \Bigg]
+  \theta^1 \, \frac{i}{(1-4\alpha^2)}\,  {\bf N^{2}}(z,\theta) 
 +  \theta^3 \, \frac{i}{(1-4\alpha^2)}\,  {\bf N^{1}}(z,\theta)
\nonu \\
& - &  \theta^4 \, \frac{i}{(1-4\alpha^2)}\, {\bf N^{3}}(z,\theta)
+  \theta^1 \, \theta^3  \left[ {\bf C^{3}}(z,\theta)+
\frac{2 \, i \, \alpha}{(1-4\alpha^2)} \, D\, {\bf N^{3}}(z,\theta) \right]
\nonu \\
& + & \theta^1 \, \theta^4 \left[{\bf C^{1}}(z,\theta)+
\frac{2 \, i \, \alpha}{(1-4\alpha^2)}
\, D \, {\bf N^{1}}(z,\theta) \right]
+  \theta^3\, \theta^4 \left[{\bf C^{2}}(z,\theta)+
\frac{2 \, i \, \alpha}{(1-4\alpha^2)}
\, D \, {\bf N^{2}}(z,\theta)  \right]
\nonu \\
& + & \theta^1 \, \theta^3 \, \theta^4 \left[ 2\,{\bf t}(z,\theta)
-
\frac{2 \, i \, \alpha}{(1-4\alpha^2)}
\,\partial {\bf h}(z,\theta)
\right],
\label{n1superspace}
\eea
where the fermionic coordinate 
$\theta^2$ is replaced by $\theta$ and one introduces
the ${\cal N}=1$ spin-$\frac{1}{2}$ current ${\bf h}(Z)$ as follows:
\bea
\partial \, ( \Delta (z)  - i \, \theta \, \Gamma^2(z) )  =
\frac{i}{(1-4\alpha^2)}\, D \, {\bf h}(Z), \qquad
D \equiv \theta \frac{\pa}{\pa z} + \frac{\pa }{\pa \theta}.
\label{notation}
\eea
Or one has the expression ${\bf h}(Z)= -(1-4\alpha^2) 
\left( \Gamma^2(z) + \theta \, i \pa \Delta(z) \right)$ 
from (\ref{notation}) explicitly. 
Recall that the defining OPEs between the 
${\cal N}=1$ super currents 
are given by  $(2.3)$ and $(3.1)$ of \cite{ST}. 
There are spin-$\frac{3}{2}$ current ${\bf t}(Z)$, 
four spin-$\frac{1}{2}$ currents ${\bf h}(Z)$ and ${\bf N^i}(Z)$ ($i=1, 2, 3$)
and three spin-$1$ currents ${\bf C^i}(Z)$ ($i=1, 2, 3$).
The central term of ${\bf N^i}(Z)$ (and ${\bf h}(Z)$) is given by 
$\frac{1}{3} \, c \, (1-4\alpha^2)$, while 
the central term of ${\bf C^i}(Z)$ is given by $-\frac{c}{3}$. 
See also \cite{OS}.
The $16$ currents of large ${\cal N}=4$ linear superconformal algebra
can be reorganized by the following ${\cal N}=1$ quantities
introduced in \cite{ST}
\bea
{\bf N^{1}}(z,\theta) & = & (1-4 \alpha^{2})(\Gamma^{3}+\theta \,T^{14}),
\qquad
{\bf N^{2}}(z,\theta)  =  (1-4 \alpha^{2})(\Gamma^{1}-\theta \,T^{34}),
\nonu\\
{\bf N^{3}}(z,\theta) & = & (1-4 \alpha^{2})(-\Gamma^{4}+\theta \,T^{13}),
\qquad
{\bf h}(z,\theta)  =  (1-4 \alpha^{2})(-\Gamma^{2}- i\,\theta \, \pa 
\Delta),\nonu\\
{\bf C^{1}}(z,\theta) & = & (-i\, T^{23}-2\,i\, \alpha \, T^{14}-\theta \,G^{3}),
\qquad
{\bf C^{2}}(z,\theta)  =  
(i\, T^{12}+2\,i\, \alpha \, T^{34}-\theta \,G^{2}),\nonu\\
{\bf C^{3}}(z,\theta) & = & (i\, T^{24}-2\,i\, \alpha \, T^{13}+\theta \,G^{4}),
\qquad
{\bf t}(z,\theta)  =  (-\frac{1}{2}\, G^{2}+\theta \,L).
\label{n1def}
\eea

One can also write down the ${\cal N}=4$ (higher spin) multiplet
in terms of its ${\cal N}=1$ currents. 
Of course, 
the ${\cal N}=1$ superconformal algebra between (\ref{n1def}) 
in ${\cal N}=1$ superspace
can be obtained from (\ref{j4j4}) and (\ref{n1superspace})
as done in the footnote \ref{n3decompos}.

\section{The OPEs between the $16$ currents   and the 
$16$  higher spin currents  in 
${\cal N}=2$ superspace}

From the
component results in Appendix $C$, one would like to 
construct them in the ${\cal N}=2$ superspace (in $SO(2)$ symmetric way) 
explicitly.
The exact relations between the component higher spin currents 
and its ${\cal N}=2$ higher spin currents are given.
Furthermore, one also writes down the ${\cal N}=4$ higher spin current 
multiplet in terms of four ${\cal N}=2$ higher spin currents 
by expansion of Grassmann coordinates.

\subsection{The ${\cal N}=4$ primary current condition in ${\cal N}=2
$ superspace }

The $16$ higher spin currents can be 
represented by ${\cal N}=2$ higher spin-$s$ current ${\bf T^{(s)}}(Z)$,
two higher spin-$(s+\frac{1}{2})$ currents $\left(
\begin{array}{c}
{\bf U^{(s+\frac{1}{2})}} \nonu \\
{\bf V^{(s+\frac{1}{2})}}
\end{array} \right)(Z)$ and 
higher spin-$(s+1)$ current 
${\bf W^{(s+1)}}(Z)$.
Each ${\cal N}=2$ higher spin currents has four component currents
and therefore the number of component currents is given by 
$16$. The $U(1)$ charges of ${\cal N}=2$ superconformal algebra 
are given by zero for both ${\bf T^{(s)}}(Z)$ and ${\bf W^{(s+1)}}(Z)$
and $2\alpha$ for ${\bf U^{(s+\frac{1}{2})}}(Z)$ and 
$-2\alpha$ for ${\bf V^{(s+\frac{1}{2})}}(Z)$.

It turns out that there exists the following 
${\cal N}=2$ version 
corresponding to (\ref{jphi})
\bea
&& {\bf T}(Z_{1}) \, {\bf T^{(s)}}(Z_{2})  = 
\frac{\theta_{12}\bar{\theta}_{12}}{z_{12}^{2}}  s {\bf T^{(s)}}(Z_{2})-
\frac{\theta_{12}}{z_{12}}  D {\bf T^{(s)}}(Z_{2})+
\frac{\bar{\theta}_{12}}{z_{12}}  \overline{D} {\bf T^{(s)}}(Z_{2})
 +  
\frac{\theta_{12}\bar{\theta}_{12}}{z_{12}}  \partial { \bf T^{(s)}}(Z_{2})
+\cdots,
\nonu \\
&& {\bf T}(Z_{1}) \, 
\left(
\begin{array}{c}
{\bf U^{(s+\frac{1}{2})}} \nonu \\
{\bf V^{(s+\frac{1}{2})}}
\end{array} \right)(Z_{2})  =  
\frac{\theta_{12}\bar{\theta}_{12}}{z_{12}^{2}} \, (\frac{1}{2}+s)\, \left(
\begin{array}{c}
{\bf U^{(s+\frac{1}{2})}} \nonu \\
{\bf V^{(s+\frac{1}{2})}}
\end{array} \right)(Z_{2})
\pm
\frac{1}{z_{12}} \, 2\, \alpha \, \left(
\begin{array}{c}
{\bf U^{(s+\frac{1}{2})}} \nonu \\
{\bf V^{(s+\frac{1}{2})}}
\end{array} \right)(Z_{2})
\nonu \\
&& \,\,\,\, \,\,\,\,\,\, -  
\frac{\theta_{12}}{z_{12}} \, D
\left(
\begin{array}{c}
{\bf U^{(s+\frac{1}{2})}} \nonu \\
{\bf V^{(s+\frac{1}{2})}}
\end{array} \right)(Z_{2})
+ 
\frac{\bar{\theta}_{12}}{z_{12}} \, \overline{D}
\left(
\begin{array}{c}
{\bf U^{(s+\frac{1}{2})}} \nonu \\
{\bf V^{(s+\frac{1}{2})}}
\end{array} \right)(Z_{2})
 +   
\frac{\theta_{12}\bar{\theta}_{12}}{z_{12}} \, \partial 
\left(
\begin{array}{c}
{\bf U^{(s+\frac{1}{2})}} \nonu \\
{\bf V^{(s+\frac{1}{2})}}
\end{array} \right)(Z_{2})
+\cdots,
\nonu \\
&& {\bf T}(Z_{1}) \, {\bf W^{(s+1)}}(Z_{2})  =  
\frac{1}{z_{12}^{2}} \, \frac{64\alpha}{3}\, { \bf T^{(s)}}(Z_{2})+
\frac{\theta_{12}\bar{\theta}_{12}}{z_{12}^{2}} \, (1+s) \, {\bf W^{(s+1)}}(Z_{2})-
\frac{\theta_{12}}{z_{12}} \, D{\bf W^{(s+1)}}(Z_{2})
\nonu \\
&&\,\,\,\, \,\,\,\,\,\, +  
\frac{\bar{\theta}_{12}}{z_{12}} \, \overline{D}{\bf W^{(s+1)}}(Z_{2})
+  
\frac{\theta_{12}\bar{\theta}_{12}}{z_{12}} \, \partial {\bf W^{(s+1)}}(Z_{2})
+\cdots,
\nonu \\
&& {\bf H}(Z_{1}) \, 
\left(
\begin{array}{c}
{\bf U^{(s+\frac{1}{2})}} \nonu \\
{\bf V^{(s+\frac{1}{2})}}
\end{array} \right)(Z_{2})
 =  \mp \frac{\bar{\theta}_{12}}{z_{12}} \, 
\left(
\begin{array}{c}
{\bf U^{(s+\frac{1}{2})}} \nonu \\
{\bf V^{(s+\frac{1}{2})}}
\end{array} \right)(Z_{2})
+\cdots,
\nonu \\
&& {\bf\overline{H}}(Z_{1}) \,
\left(
\begin{array}{c}
{\bf U^{(s+\frac{1}{2})}} \nonu \\
{\bf V^{(s+\frac{1}{2})}}
\end{array} \right)(Z_{2})
  =   \pm
\frac{\theta_{12}}{z_{12}} \, 
\left(
\begin{array}{c}
{\bf U^{(s+\frac{1}{2})}} \nonu \\
{\bf V^{(s+\frac{1}{2})}}
\end{array} \right)(Z_{2})
+\cdots,
\nonu \\
&& \left(
\begin{array}{c}
{\bf H} \nonu \\
{\bf \overline{H}}
\end{array}
\right)(Z_1) \, 
{\bf W^{(s+1)}}(Z_{2})  =  
- \frac{\theta_{12}\bar{\theta}_{12}}{z_{12}^{2}} \, 
\frac{4(1+2s)}{(1+2s^2)}\, 
\left(
\begin{array}{c}
D {\bf T^{(s)}} \nonu \\
\overline{D} { \bf T^{(s)}} 
\end{array}
\right)(Z_{2})
\nonu \\
& & \,\,\,\,\,\,\,\,\,\, \mp  \frac{1}{z_{12}^{2}} \,
\left( 
\begin{array}{c}
\bar{\theta}_{12} \nonu \\
\theta_{12}
\end{array}
\right)
 \frac{8s(1+2s)}{(1+2s^2)}\, {\bf T^{(s)}}(Z_{2})
 \pm 
\frac{1}{z_{12}} \, \frac{8(1+2s)}{(1+2s^2)} \, 
\left(
\begin{array}{c}
D {\bf T^{(s)}} \nonu \\
\overline{D} {\bf T^{(s)}} 
\end{array}
\right)(Z_{2})
+\cdots,
\nonu \\
&& \left(
\begin{array}{c}
{\bf G} \nonu \\
{\bf \overline{ G}} 
\end{array}
\right)(Z_{1}) \, {\bf T^{(s)}}(Z_{2})  = 
\pm
\frac{\theta_{12}\bar{\theta}_{12}}{z_{12}} \, \frac{1}{2} \, 
\left(
\begin{array}{c}
{\bf U^{(s+\frac{1}{2})}} \nonu \\
{\bf V^{(s+\frac{1}{2})}}
\end{array} \right)(Z_{2})
+\cdots,
\nonu \\
&& \left(
\begin{array}{c}
{\bf G} \nonu \\
{\bf \overline{ G}} 
\end{array}
\right)(Z_{1})
 \, 
\left(
\begin{array}{c}
{\bf V^{(s+\frac{1}{2})}} \nonu \\
{\bf U^{(s+\frac{1}{2})}}
\end{array} \right)(Z_{2})
  =  
\pm \frac{\theta_{12}\bar{\theta}_{12}}{z_{12}^{2}} \, 2s\, {\bf T^{(s)}}(Z_{2})
 \mp  \frac{\theta_{12}}{z_{12}} \, 2\, D {\bf T^{(s)}}(Z_{2}) \pm
\frac{\bar{\theta}_{12}}{z_{12}} \, 2\, \overline{D} {\bf T^{(s)}}(Z_{2})
\nonu \\
&& \,\,\,\,\,\,\,\,\,\, +  \frac{\theta_{12}\bar{\theta}_{12}}{z_{12}}
\left[ \frac{2}{(1+2s)} \, \alpha \,
[D,\overline{D}] \, {\bf T^{(s)}} +
\frac{(1+2s^2)}{4(1+2s)} \, {\bf W^{(s+1)}}
\pm \partial {\bf T^{(s)}}\right](Z_{2})
  +  \cdots,
\nonu \\
&& \left(
\begin{array}{c}
{\bf G} \nonu \\
{\bf \overline{ G}} 
\end{array}
\right)(Z_{1})
\, {\bf W^{(s+1)}}(Z_{2})  =  
-\frac{\theta_{12}\bar{\theta}_{12}}{z_{12}^{2}} \, \frac{2(1+2s)^2}{(1+2s^2)}\, 
\left(
\begin{array}{c}
{\bf U^{(s+\frac{1}{2})}} \nonu \\
{\bf V^{(s+\frac{1}{2})}}
\end{array} \right)(Z_{2})
\nonu \\
& & \,\,\,\,\,\,\,\,\,\,\, \pm   \frac{1}{z_{12}} \, \frac{8}{(1+2s^2)}\alpha \, 
\left(
\begin{array}{c}
{\bf U^{(s+\frac{1}{2})}} \nonu \\
{\bf V^{(s+\frac{1}{2})}}
\end{array} \right)(Z_{2})
 +  
\frac{\theta_{12}}{z_{12}} \, 
\frac{8((1+s)k^{\pm}+ s k^{\mp})}{(k^{+}+k^{-})(1+2s^2)}
\, D
\left(
\begin{array}{c}
{\bf U^{(s+\frac{1}{2})}} \nonu \\
{\bf V^{(s+\frac{1}{2})}}
\end{array} \right)(Z_{2})
\nonu \\
& & \,\,\,\,\,\,\,\,\,\, -  
\frac{\bar{\theta}_{12}}{z_{12}} \, 
\frac{8((1+s)k^{\mp}+ s k^{\pm})}{(k^{+}+k^{-})(1+2s^2)}
\,\overline{D}
\left(
\begin{array}{c}
{\bf U^{(s+\frac{1}{2})}} \nonu \\
{\bf V^{(s+\frac{1}{2})}}
\end{array} \right)(Z_{2})
\nonu \\
 && \,\,\,\,\,\,\,\,\,\, +  \frac{\theta_{12}\bar{\theta}_{12}}{z_{12}} \left[ \mp
\frac{4}{(1+2s^2)} \, \alpha\, 
[D,\overline{D}]
\left(
\begin{array}{c}
{\bf U^{(s+\frac{1}{2})}}  \\
{\bf V^{(s+\frac{1}{2})}}
\end{array} \right)
 -    \frac{2(1+2s)}{(1+2s^2)} \, \partial 
\left(
\begin{array}{c}
{\bf U^{(s+\frac{1}{2})}}  \\
{\bf V^{(s+\frac{1}{2})}}
\end{array} \right)
 \right](Z_{2})
+  \cdots.
\label{primaryn2}
\eea
As described before, the ${\cal N}=2$ spin 
of four ${\cal N}=2$ higher spin currents can be read off 
from the $\frac{\theta_{12}\bar{\theta}_{12}}{z_{12}^2}$-term in the OPEs 
with ${\cal N}=2$ stress energy tensor ${\bf T}(Z_1)$.
The ${\cal N}=2$ higher spin current ${\bf W^{(s+1)}}(Z_2)$
is not a ${\cal N}=2$ primary under the 
${\cal N}=2$ stress energy tensor 
${\bf T}(Z_1)$ because there exists a $\frac{1}{z_{12}^2}$-term in 
(\ref{primaryn2}).
As in the standard ${\cal N}=2$ superspace formalism, 
by acting the ${\cal N}=2$ currents ${\bf G}(Z_1)$ and 
${\bf \overline{G}}(Z_1)$ on the ${\cal N}=2$ higher spin current 
${\bf T^{(s)}}(Z_2)$, the other  ${\cal N}=2$ higher spin currents
${\bf U^{(s+\frac{1}{2})}}(Z_2)$ and 
 ${\bf V^{(s+\frac{1}{2})}}(Z_2)$ can be generated.
Furthermore, 
by acting the ${\cal N}=2$ currents ${\bf G}(Z_1)$ (or
${\bf \overline{G}}(Z_1)$) on the ${\cal N}=2$ higher spin currents 
${\bf V^{(s+\frac{1}{2})}}(Z_2)$ (or 
 ${\bf U^{(s+\frac{1}{2})}}(Z_2)$),
the ${\cal N}=2$ higher spin current
${\bf W^{(s+1)}}(Z_2)$ is obtained.

How one can obtain the component results from the 
above OPEs (\ref{primaryn2})? 
Let us consider the OPE $G^1(z) \, \Phi_{\frac{1}{2}}^{(1),2}(w)$.
From Appendix $C$, one expects that the first-order pole of this OPE is 
given by $\Phi_1^{(1),34}(w)$. One would like to see this from the OPEs
(\ref{primaryn2}). From Appendix $E$, one should take 
$(D_1 +\overline{D}_1) \, {\bf T}(Z_1)$. Furthermore, for the other 
superspace coordinate one should take 
$(-i)(D_2 +\overline{D}_2) \, {\bf T^{(1)}}(Z_2)$ because of the 
decompositions, which will appear in next subsection. 
In other words, let us multiply $(-i)(D_1+\overline{D}_1)(D_2+\overline{D}_2)$
both sides of the first equation of (\ref{primaryn2}) with the conditions
$\theta =\overline{\theta}=0$.  
Then the left hand side of the OPE
becomes  the OPE $G^1(z_1) \, \Phi_{\frac{1}{2}}^{(1),2}(z_2)$ \footnote{
The nonzero contributions from the right hand side arise from 
the second- and third-terms of the first equation of (\ref{primaryn2}).
The former gives $(-\frac{1}{2} [D, \overline{D}] -\frac{1}{2} \pa) 
{\bf T^{(1)}}(Z_2)|_{\theta=\overline{\theta}=0}$ with the singular term 
$\frac{(-i)}{z_{12}}$
(one uses the fact that $D_1 \frac{\theta_{12}}{z_{12}}= \frac{1}{(z_1-z_2)}$ 
after a projection) 
while the latter 
gives $(-\frac{1}{2} [D, \overline{D}] +\frac{1}{2} \pa) 
{\bf T^{(1)}}(Z_2)|_{\theta=\overline{\theta}=0}$ with the singular term 
$\frac{(-i)}{z_{12}}$,
where the property, that 
$\overline{D}_1 \frac{\overline{\theta}_{12}}{z_{12}}$ goes to 
$\frac{1}{(z_1-z_2)}$, is used. 
This leads to the expression
$ - (-i) [D, \overline{D}] 
{\bf T^{(1)}}(Z_2)|_{\theta=\overline{\theta}=0}$ in the first-order pole, 
which is equal to 
the above  $\Phi_1^{(1),34}(z_2)$, where the decompositions (which will 
appear in next subsection)
are needed.}. 
One can also easily see that other contributions from the above 
differential operator acting on the other parts in the right hand side 
are vanishing.

In this way, one can check all the component 
results in Appendix $C$ from its ${\cal N}=2$ version in (\ref{primaryn2})
and vice versa.

\subsection{The component currents in the 
four higher spin-$s$, $(s+\frac{1}{2})$, $(s+\frac{1}{2})$, 
$(s+1)$ currents in ${\cal N}=2$ superspace }

One also has the following relations between 
the ${\cal N}=2$ higher spin currents and its components as follows:
\begin{eqnarray}
\mathbf{T}^{(s)}\mid_{\theta=\overline{\theta}=0}(z)&=&\Phi_{0}^{(s)}(z),
\nonumber\\
D\mathbf{T}^{(s)}\mid_{\theta=\overline{\theta}=0}(z)&=&
\frac{1}{2}(\Phi_{\frac{1}{2}}^{(s),1}+i\,\Phi_{\frac{1}{2}}^{(s),2})(z),
\nonumber\\
\overline{D}\mathbf{T}^{(s)}
\mid_{\theta=\overline{\theta}=0}(z)&=&\frac{1}{2}(-\Phi_{\frac{1}{2}}^{(s),1}+
i\,\Phi_{\frac{1}{2}}^{(s),2})(z),
\nonumber\\
-\frac{1}{2}[D,\overline{D}]\mathbf{T}^{(s)}
\mid_{\theta=\overline{\theta}=0}(z)&=&\frac{1}{2}\, i\,\Phi_{1}^{(s),34}(z),
\nonumber\\
\mathbf{U}^{(s+\frac{1}{2})}\mid_{\theta=\overline{\theta}=0}(z)&=&
(\Phi_{\frac{1}{2}}^{(s),3}+i\,\Phi_{\frac{1}{2}}^{(s),4})(z),
\nonumber\\
D\mathbf{U}^{(s+\frac{1}{2})}\mid_{\theta=\overline{\theta}=0}(z)&=&
\frac{1}{2}(-\Phi_{1}^{(s),13}-i\,\Phi_{1}^{(s),14}-i\,\Phi_{1}^{(s),23}+
\Phi_{1}^{(s),24})(z),
\nonumber\\
\overline{D}\mathbf{U}^{(s+\frac{1}{2})}
\mid_{\theta=\overline{\theta}=0}(z)&=&\frac{1}{2}(-\Phi_{1}^{(s),13}-i\,
\Phi_{1}^{(s),14}+i\,\Phi_{1}^{(s),23}-\Phi_{1}^{(s),24})(z),
\nonumber\\
-\frac{1}{2}[D,\overline{D}]\mathbf{U}^{(s+\frac{1}{2})}
\mid_{\theta=\overline{\theta}=0}(z)&=&\frac{1}{2}
(\Phi_{\frac{3}{2}}^{(s),3}+i\,{\Phi}_{\frac{3}{2}}^{(s),4})(z),
\nonumber\\
\mathbf{V}^{(s+\frac{1}{2})}\mid_{\theta=\overline{\theta}=0}(z)&=&
(\Phi_{\frac{1}{2}}^{(s),3}-i\,\Phi_{\frac{1}{2}}^{(s),4})(z),
\nonumber\\
D\mathbf{V}^{(s+\frac{1}{2})}\mid_{\theta=\overline{\theta}=0}(z)&=&
\frac{1}{2}(\Phi_{1}^{(s),13}-i\,\Phi_{1}^{(s),14}+i\,\Phi_{1}^{(s),23}+
\Phi_{1}^{(s),24})(z),
\nonumber\\
\overline{D}\mathbf{V}^{(s+\frac{1}{2})}
\mid_{\theta=\overline{\theta}=0}(z)&=&\frac{1}{2}(\Phi_{1}^{(s),13}-i\,
\Phi_{1}^{(s),14}-i\,\Phi_{1}^{(s),23}-\Phi_{1}^{(s),24})(z),
\nonumber\\
-\frac{1}{2}[D,\overline{D}]\mathbf{V}^{(s+\frac{1}{2})}
\mid_{\theta=\overline{\theta}=0}(z)&=&\frac{1}{2}(-\Phi_{\frac{3}{2}}^{(s),3}+i\,
\Phi_{\frac{3}{2}}^{(s),4})(z),
\nonumber\\
\mathbf{W}^{(s+1)}\mid_{\theta=\overline{\theta}=0}(z)&=&
\frac{4i\,(1+2s)}{(1+2s^{2})}\Phi_{1}^{(s),12}(z)+\frac{8i\,\alpha}{(1+2s^{2})}
\Phi_{1}^{(s),34}(z),
\nonumber\\
D\mathbf{W}^{(s+1)}\mid_{\theta=\overline{\theta}=0}(z)&=&
\frac{2\,(1+2s)}{(1+2s^{2})}(\Phi_{\frac{3}{2}}^{(s),1}+i\,
\Phi_{\frac{3}{2}}^{(s),2})(z)-\frac{4\alpha}{(1+2s^{2})}\,
\partial(\Phi_{\frac{1}{2}}^{(s),1}+i\,\Phi_{\frac{1}{2}}^{(s),2})(z),
\nonumber\\
\overline{D}\mathbf{W}^{(s+1)}\mid_{\theta=\overline{\theta}=0}(z)&=&
\frac{2\,(1+2s)}{(1+2s^{2})}({\Phi}_{\frac{3}{2}}^{(s),1}-i\,
{\Phi}_{\frac{3}{2}}^{(s),2})(z)-\frac{4\alpha}{(1+2s^{2})}\,\partial
(\Phi_{\frac{1}{2}}^{(s),1}-i\,\Phi_{\frac{1}{2}}^{(s),2})(z),
\nonumber\\
-\frac{1}{2}[D,\overline{D}]\mathbf{W}^{(s+1)}
\mid_{\theta=\overline{\theta}=0}(z)&=&-\frac{2\,(1+2s)}{(1+2s^{2})}
{\Phi}_{2}^{(s)}(z)+\frac{4\alpha}{(1+2s^{2})}\partial^{2}\Phi_{0}^{(s)}(z).
\label{16highern2}
\end{eqnarray}
For $s=1$, one can also express the right hand sides of 
(\ref{16highern2}) using (\ref{v3half}), (\ref{vspin2}), (\ref{v5half}) 
and (\ref{bcgspin3})
 \footnote{
In \cite{KT}, the command $\tt N2OPEToComponents[T_{-}]$ provides
the components of ${\cal N}=2$ current ${\bf T}(Z)$.
One can write down ${\bf T}(Z) = {\bf T}|_{\theta=\overline{\theta}=0} + \theta \,
D \, {\bf T}|_{\theta=\overline{\theta}=0} + \overline{D} \,  {\bf T}|_{\theta=\overline{\theta}=0} - \theta \, \overline{\theta} \, \frac{1}{2} \, [D, \overline{D}]
 {\bf T}|_{\theta=\overline{\theta}=0}$. These components 
are presented in the first four equations of Appendix $E$.
One can also analyze the other ${\cal N}=2$ (higher spin) currents 
similarly. 
}.
The second (third) component of ${\bf T^{(1)}}(Z)$
contains the higher spin 
currents $T_{\pm}^{(\frac{3}{2})}(z)$, in addition to the other 
$\frac{3}{2}$ currents from (\ref{v3half}),
and the last component is proportional to the higher spin current 
$T^{(2)}(z)$ from (\ref{vspin2}).
The first component of ${\bf U^{(\frac{3}{2})}}(Z)$
contains the higher spin current $U^{(\frac{3}{2})}(z)$.
The second and third components are given by the linear 
combinations between the higher spin currents 
$U_{\pm}^{(2)}(z)$ and $V_{\pm}^{(2)}(z)$
from (\ref{vspin2}). The fourth component has 
the higher spin currents $U^{(\frac{5}{2})}(z)$ and  $\pa U^{(\frac{3}{2})}(z)$.
Similarly, 
the first component of ${\bf V^{(\frac{3}{2})}}(Z)$
contains the higher spin current $V^{(\frac{3}{2})}(z)$,
the second and third components are given by the linear 
combinations between the higher spin currents 
$U_{\pm}^{(2)}(z)$ and $V_{\pm}^{(2)}(z)$
and  the fourth component has 
the higher spin currents $V^{(\frac{5}{2})}(z)$ and  $\pa V^{(\frac{3}{2})}(z)$.
The first component of ${\bf W^{(2)}}(Z)$
contains the higher spin current $W^{(2)}(z)$ as well as 
the higher spin current $T^{(2)}(z)$.
The second (third) component contains 
the higher spin currents $W_{\pm}^{(\frac{5}{2})}(z)$, 
$\pa T_{\pm}^{(\frac{3}{2})}(z)$
and other spin-$\frac{3}{2}$ currents.
The fourth component contains 
the higher spin current 
$W^{(3)}(z)$ (and other nonlinear terms) and is a quasiprimary field.

One also rewrites the ${\cal N}=4$ higher spin-$s$ current 
(\ref{phidiff})
in the expansion of the Grassmann coordinates 
$\theta^3$ and $\theta^4$ as follows:
\bea
{\bf \Phi^{(s)}}(Z)  & = & 
\Phi_0^{(s)}(z) + \theta^1  \Phi_{\frac{1}{2}}^{(s),1}(z) + 
\theta^2  \Phi_{\frac{1}{2}}^{(s),2}(z) + \theta^1  \theta^2   
\Phi_1^{(s),34}(z) 
\nonu \\
& + & \theta^3 \Bigg[ \Phi_{\frac{1}{2}}^{(s),3}(z) - \theta^1 \, 
\Phi_{1}^{(s),24}(z) 
-\theta^2 \, \Phi_1^{(s),14}(z)+ \theta^1 \, \theta^2
 \, \Phi_{\frac{3}{2}}^{(s),4}(z) \Bigg] 
\nonu \\
&+& \theta^4 \Bigg[ \Phi_{\frac{1}{2}}^{(s),4}(z)  -\theta^1 \,  
\Phi_{1}^{(s),23}(z) 
- \theta^2 \,  \Phi_{1}^{(s),31}(z) -\theta^1 \, \theta^2 \, 
\Phi_{\frac{3}{2}}^{(s),3}(z) \Bigg] 
\nonu \\
&+& \theta^3 \, \theta^4 \Bigg[ \Phi_1^{(s),12}(z) 
+ \theta^1  \, \Phi_{\frac{3}{2}}^{(s),2}(z)+\theta^2 \, 
\Phi_{\frac{3}{2}}^{(s),1}(z) +   \theta^1 \, 
\theta^2 \, \Phi_2^{(s)}(z) \Bigg]. 
\label{phis1}
\eea
As done in ${\bf J^{(4)}}(Z)$ in terms of ${\cal N}=2$ currents, 
this can be rewritten without any difficulty 
by looking at (\ref{16highern2}) in next subsection.
One can obtain the higher spin currents in terms of 
${\cal N}=3$ multiplets or ${\cal N}=1$ multiplets, as done for 
the ${\cal N}=4$ stress energy tensor in (\ref{j4relationj3})
or (\ref{n1superspace}).
Furthermore, as done in the footnote \ref{n3decompos}, 
the corresponding OPEs between the ${\cal N}=3$ currents 
can be obtained from (\ref{jphi}) together with 
(\ref{j4relationj3}) and (\ref{phis1}) in different basis. 
Similar analysis for ${\cal N}=1$ superspace description 
in the OPE between the stress energy tensor and the higher spin currents
can be done as in the footnote \ref{n3decompos}. 

\subsection{The ${\cal N}=2
$ superspace description of the ${\cal N}=4$ higher spin-$s$ current 
multiplet }

One would like to rewrite (\ref{phis1}) in terms of its
${\cal N}=2$ version.  
From the explicit results in (\ref{16highern2}), 
one obtains 
\bea
{\bf \Phi^{(s)}}(Z)&=&{\bf T^{(s)}}(z,\theta,\overline{\theta})+
\theta^{3}\, \frac{1}{2} \, \left( {\bf U^{(s+\frac{1}{2})}}+
{\bf V^{(s+\frac{1}{2})}} \right)(z, \theta, \overline{\theta})+
\theta^{4}\, \frac{i}{2}\, \left(-{\bf U^{(s+\frac{1}{2})}}+
{\bf V^{(s+\frac{1}{2})}} \right)(z, \theta, \overline{\theta})
\nonu \\
&- &\theta^{3}\, \theta^{4}\, \frac{i}{4(1+2s)} \, 
\Bigg[ (1+2s^{2}){\bf W^{(s+1)}}+8\, \alpha\,[D,\overline{D}]\, {\bf T^{(s)}}
\Bigg](z, \theta, \overline{\theta}).
\label{phisn2version}
\eea

Let us try to understand the ${\cal N}=4$ primary condition (\ref{jphi})
in ${\cal N}=2$ superspace described in (\ref{primaryn2})
with the help of  (\ref{n4n2}) and (\ref{phisn2version}).
Let us consider the action of $D_1^3$ into the 
equation (\ref{jphi}) and put the condition $\theta^3 =0=\theta^4$.
Then the left hand side of this OPE is 
given by the expression 
$\frac{1}{2} ({\bf G} +{\bf \overline{G}})(Z_1) \, 
{\bf T^{(s)}}(Z_2)$ from (\ref{n4n2}) and (\ref{phisn2version}) \footnote{
On the other hand, the right hand side of this
OPE (coming from the second term) is given by 
the expression $\frac{\theta_{12}^1 \, \theta_{12}^2}{z_{12}} \, \frac{i}{2} \,
(-{\bf U^{(s+\frac{1}{2})}} + {\bf V^{(s+\frac{1}{2})}})(Z_2)$.
Similarly,  the action of $D_1^4$ into the 
equation (\ref{jphi}) with the condition $\theta^3 =0=\theta^4$
  leads to the left hand side given by 
 $\frac{i}{2} (-{\bf G} +{\bf \overline{G}})(Z_1) \, 
{\bf T^{(s)}}(Z_2)$.
Moreover, the right hand side of this 
OPE (coming from the second term) becomes 
the following result 
$-\frac{\theta_{12}^1 \, \theta_{12}^2}{z_{12}} \, \frac{1}{2} \,
({\bf U^{(s+\frac{1}{2})}} + {\bf V^{(s+\frac{1}{2})}})(Z_2)$.
By combining these equations, one obtains the OPE
$
\left(
\begin{array}{c}
{\bf G} \nonu \\
{\bf \overline{ G}} 
\end{array}
\right)(Z_{1}) \, {\bf T^{(s)}}(Z_{2})  =  
\pm
\frac{\theta_{12}\bar{\theta}_{12}}{z_{12}} \, \frac{1}{2} \, 
\left(
\begin{array}{c}
{\bf U^{(s+\frac{1}{2})}} \nonu \\
{\bf V^{(s+\frac{1}{2})}}
\end{array} \right)(Z_{2})
+\cdots$, which is what one expects exactly.
One also uses the fact that $i \theta_{12}^1 \, \theta_{12}^2 = -\frac{1}{2}
\, \theta_{12} \overline{\theta}_{12}$.}. 
Note that 
when one operates the differential operator into $\theta_{12}^{4-i}$, 
one should be careful about the sign. The convention 
in \cite{Schoutensnpb} is such that $\theta^{4} = 
\theta^{4-i} \, \theta^{i}$.

\section{The OPEs between the $16$  higher spin  currents  in 
${\cal N}=2$ superspace}

In order to obtain the complete OPEs between 
 the $16$  higher spin  currents  in 
${\cal N}=2$ superspace, one should know the complete composite fields 
appearing in the OPEs.
It is known that some of the OPEs between the 
$16$ higher spin currents  in the component approach
are found explicitly for $N=3$ \cite{Ahn1504}. One way to obtain 
the possible candidates for the composite fields 
in the OPEs (in the ${\cal N}=2$ superspace) is to find the corresponding 
OPEs in the component approach. One can proceed the method in \cite{Ahn1504}
and to obtain the unknown OPEs, 
but in this paper one uses the power of 
${\cal N}=4$ supersymmetry,
as emphasized in the introduction. 
One can determine the undetermined OPEs in \cite{Ahn1504} by moving to the 
${\cal N}=4$ superspace and arrange the known OPEs in appropriate places
in a single  OPE in the ${\cal N}=4$ superspace.
Then one can move to the ${\cal N}=2$ superspace by collecting those OPEs 
in the component approach and rearranging them in ${\cal N}=2$ supersymmetric 
way. So far, all the coefficients in the OPEs are given with fixed $N=3$
and arbitrary $k$. Now one puts these coefficients as the functions of 
$N$ and $k$ and applies to use Jacobi identities between the ${\cal N}=2$
currents or higher spin currents. Eventually, one obtains 
the complete structure constants with arbitrary $N$ and $k$ (\ref{kn}) 
appearing 
in the complete OPEs in ${\cal N}=2$ superspace.

\subsection{The ansatz from the $136$ OPEs in the component approach}

Because the spin of ${\bf \Phi^{(1)}}(Z)$ is given by
$s=1$, the OPE between 
this ${\cal N}=4$ multiplet and itself has a spin $s=2$.
Then the right hand side of this OPE should 
preserve the total spin with $s=2$.
Then one can consider the singular terms 
$\frac{\theta_{12}^4}{z_{12}^n}$, $\frac{\theta_{12}^{4-i}}{z_{12}^n}$,
$\frac{\theta_{12}^{4-ij}}{z_{12}^n}$, 
$\frac{\theta_{12}^{4-ijk}}{z_{12}^n}$, and $\frac{1}{z_{12}^n}$ in ${\cal N}=4$
superspace.
Recall that the spin of $\theta_{12}^i$ is given by $-\frac{1}{2}$ and the 
spin of $\frac{1}{z_{12}}$ is given by $1$.
The first singular term has a spin $(n-2)$ and then
the possible $n$-value is given by $4,3,2,1$. Then the right hand sides
should contain the composite fields with 
spin $s =0, 1, 2, 3$ at each singular term respectively.
For the second singular term, the right hand sides 
should contain  the composite fields with spin 
$s=\frac{1}{2}, \frac{3}{2}, \frac{5}{2}$
at the singular term respectively (in this case the spin is given by 
$(n-\frac{3}{2})$ and the possible $n$-value is given by $n=3, 2, 1$).
Note that the derivative $D^i$ has a spin $\frac{1}{2}$.
For the third singular term, 
   the right hand sides 
should contain  the composite fields with spin 
$s=0, 1, 2$
at the singular term respectively (in this case the spin is given by 
$(n-1)$ and the possible $n$-value is given by $n=3, 2, 1$).
The $n=3$ is not allowed because
the spin $s=0$ composite field does not contain any $SO(4)$-indices,
which should be contracted with two $SO(4)$-indices $ij$. 
For the fourth singular term, 
the possible $n$-value is given by $2, 1$. Then the right hand sides
should contain the composite fields with 
spin $s =\frac{1}{2}, \frac{3}{2}$.  The only $s=\frac{3}{2}$
case can have the three $SO(4)$-indices which are contracted with 
the $SO(4)$-indices $ijk$.
For the last singular term, $n$-value can be $2, 1$. 
Then the former has a central 
term and the latter has composite fields with spin $s=1$.

Then one can rearrange the $136$ OPEs in the component approach with the help 
of (\ref{phis}) and the subsection $3.6$
to rewrite the corresponding single ${\cal N}=4$ 
OPE in ${\cal N}=4$ superspace. See also Appendix $F$.  
In other words, it turns out that all the structures (the possible composite 
terms)
appearing in the right hand side of OPEs in the component approach
are determined completely. 
In next subsection, we go to the ${\cal N}=2$ superspace
by taking the ansatz from the component approach with the arbitrary 
coefficients.  

\subsection{Jacobi identities}

The general graded Jacobi identity (that is, $(3.24)$ of \cite{BS})
reads 
\bea
(-1)^{{\bf A} {\bf C}} \, [ \, 
[{\bf A}, {\bf B} \}, {\bf C}\}(Z) +\mbox{cycl.} =0,
\label{jac}
\eea
for general currents ${\bf A}(Z), {\bf B}(Z)$ and ${\bf C}(Z)$.
The command $\tt OPEJacobi[op1_{-}, op2_{-}, op3_{-}]$ in \cite{KT}
calculates the Jacobi identities (\ref{jac}) 
for the singular part of the OPEs 
of the three arguments.
In general, all different orderings have been done.
The $\tt OPEJacobi$ returns
a list of which all should be zero up to null fields 
to be associative.
The ${\cal N}=1$ example of (\ref{jac}) appears in  
the ${\cal N}=1$ extension of $W_3$ algebra studied in \cite{ASS} (this was 
done by hand).
Of course, one can check the Jacobi identities, using the 
definition of normal ordered product between any two currents 
in ${\cal N}=4$ superspace 
(for example, \cite{IKL1,IKL2}) by hand 
\footnote{ Inside of \cite{KT},
by using the command $\tt N2OPEToComponents[ope_{-}, J1_{-}, J2_{-}]$
one calculates the $16$ OPEs of the components of ${\bf J_1}(Z_1)$ 
and ${\bf J_2}(Z_2)$ which are any two ${\cal N}=2$ (higher spin) currents. 
It shows 
a double list, where the $(m,n)$-th element is the 
component OPE between the $m$-th component of ${\bf J_1}(Z_1)$ and 
the $n$-th component 
of ${\bf J_2}(Z_2)$.}.

\subsection{The determination of structure constant}

In this subsection,
the three OPEs in the ${\cal N}=2$ superspace 
are given explicitly with Appendices $G.1$-$G.3$ where the 
structure constants are presented explicitly
and the remaining OPEs will be given 
in Appendices $G.4$-$G.7$ explicitly.

Because the OPEs ${\bf T^{(1)}}(Z_1) \, {\bf T^{(1)}}(Z_2)$,
${\bf U^{(\frac{3}{2})}}(Z_1) \,{\bf U^{(\frac{3}{2})}}(Z_2)$, 
and ${\bf V^{(\frac{3}{2})}}(Z_1) \,{\bf V^{(\frac{3}{2})}}(Z_2)$ 
have no higher spin currents, 
the structure constants of these OPEs are completely 
determined by the Jacobi identities without knowing the OPEs between 
the higher spin currents and 
the next higher spin currents. 
However the remaining  seven OPEs contain the next higher spin currents
as we will see later  and 
one does not have any 
information about the  OPEs between the higher spin currents and the 
next higher 
spin current at the moment.  Consequently, 
the insufficient Jacobi identities lead to the one unknown structure 
constant.

One can fix the unknown  structure constant via  the coefficient
of energy momentum tensor $T(w)$  in the second-order  pole 
of the OPE $T^{(1)}(z) \, {\bf W^{(3)}}(w)$  \cite{Ahn1504}.
However, since the result in \cite{Ahn1504} is for $N=3$,
one need to obtain the coefficient for  arbitrary $N$.
Let us introduce  the coefficient $ c_{1}$ in front of 
$ T(w)$ in the second-order pole  of the OPE 
$T^{(1)}(z) \, {\bf W^{(3)}}(w)$.
More explicitly,
one has $ \{ T^{(1)} \, {\bf W^{(3)}}\}_{-2} = 
 -{\bf P^{(2)}}(w)  +c_1 \, T(w)  + \cdots$.
 Let us introduce the coefficient  $ c_{2}$ of $T(z)$,
where  
in the relation between ${\bf P^{(2)}}$ and $V_{0}^{(2)}$, one has 
${\bf P^{(2)}}(z) = c_{2} \, T(z) + V_0^{(2)}(z) + \cdots$.
Then one has 
$ \{ T^{(1)} \, {\bf W^{(3)}} \}_{-2} = 
 (c_1-c_2) \, T(w) + \cdots$.
One can obtain their $N$-generalization by considering 
for low $N$ values $(N =3, 5, 7, \cdots)$ as follows
\footnote{We thank H. Kim for this calculation and other related 
computations.}:
\bea
c_{1}(N, k) & = & \frac{8(k-N)(4N+4k+5)}{3(N+k+2)(3N\,k+4N+4k+5)},\nonu \\
c_{2}(N, k) & = & \frac{2(2k+N+3)(5k+8N+10)}{3(N+k+2)^2}.
\label{c1c2}
\eea
Of course, one obtains the $16$ OPEs between 
the higher spin-$1$ current and the $16$ higher spin currents 
for several $N$-values and their complete $16$ OPEs 
can be written for generic $N$ and $k$. 
In particular, there is no new primary current in the first-order pole 
of the OPE $T^{(1)}(z) \, {\bf W^{(3)}}(w)$. This implies 
that the $16$ higher spin currents from the WZW currents in the 
${\cal N}=4$ coset theory in the final ${\cal N}=4$ OPE
${\bf \Phi^{(1)}}(Z_1)\, {\bf \Phi^{(1)}}(Z_2)$
do not generate the third ${\cal N}=4$ multiplet ${\bf \Phi^{(3)}}(Z_2)$
in the right hand side.  

One  also obtains the same quantity from the Jacobi identity only 
and solve the following equation with (\ref{c1c2})
\bea
c_{1}-c_{2}&=&\frac{1}{3(k-N)(1+N)(5+4k+4N+3k\, N)}\,4(40+62k+24k^{2}+102N+114k\, N
\nonu\\&+&36k^{2}N+86N^{2}+64k\, N^{2}+12k^{2}N^{2}+24N^{3}+12k\, N^{3}+160X+448kX+496k^{2}X
\nonu\\&+&272k^{3}X+74k^{4}X+8k^{5}X+448N\, X+1128k\, N\, X+1100k^{2}N\, X+518k^{3}N\, X
\nonu\\&-&117k^{4}N\, X+10k^{5}N\, X+496N^{2}X+1100k\, N^{2}X+912k^{2}N^{2}X+347k^{3}N^{2}X
\nonu\\&-&58k^{4}N^{2}X+3k^{5}N^{2}X+272N^{3}X+518k\, N^{3}X+347k^{2}N^{3}X+96k^{3}N^{3}X+9k^{4}N^{3}X
\nonu\\&-&74N^{4}X+117k\, N^{4}X+58k^{2}N^{4}X+9k^{3}N^{4}X+8N^{5}X+10k\, N^{5}X+3k^{2}N^{5}X),
\label{c1c2X}
\eea
the unknown quantity (appeared as the $c_{51}$-coefficient in the 
next subsection $6.3.2$) from (\ref{c1c2X}) is determined by
\bea
X=-\frac{(1+N)(32+55k+18k^{2}+41N+35kN+11N^{2})}{2(2+N)(2+k+N)^{5}}.
\label{X}
\eea
Therefore, all the structure constants are determined completely.
In the following, one uses 
the following currents without a boldface notation for simplicity
\bea
{\bf H}(Z) \equiv H(Z), \quad 
{\bf \overline{H}}(Z) \equiv \overline{H}(Z),
\quad
{\bf G}(Z) \equiv G(Z), \qquad 
{\bf \overline{G}}(Z) \equiv \overline{G}(Z),
\quad
{\bf T}(Z) \equiv T(Z).
\label{redef}
\eea
These notations are used in Appendix $G$ also.

\subsubsection{The OPE between the ${\cal N}=2$ higher spin-$1$ current}


The OPE between the ${\cal N}=2$ higher spin-$1$ currents,
where the four components of ${\cal N}=2$ higher spin-$1$ current 
are given the first four equations of (\ref{16highern2}) with $s=1$,
is summarized by
\begin{eqnarray}
&&{\bf T^{(1)}}(Z_{1})\:{\bf T^{(1)}}(Z_{2})\;=\;\frac{1}{z_{12}^{2}}\, c_{1}+
\frac{\theta_{12}\overline{\theta}_{12}}{z_{12}^{2}}\Bigg[c_{2}\,\overline{D}H+
c_{3}\, D\overline{H}+c_{4}\, T+c_{5}\, H \overline{H}+
c_{6}\, G \overline{G}\Bigg](Z_{2})
\nonumber\\&&+
\frac{\theta_{12}}{z_{12}}\Bigg[c_{7}\, DT+c_{8}\,\partial H+c_{9}\, 
DG \overline{G}+c_{10}\, G  D \overline{G}+c_{11}\, D\overline{H} H+
c_{12}\, H G \overline{G}\Bigg](Z_{2})
\nonumber\\&&+
\frac{\overline{\theta}_{12}}{z_{12}}\Bigg[c_{13}\,\overline{D}T+
c_{14}\,\partial\overline{H}+c_{15}\,\overline{D}G \overline{G}+
c_{16}\, G \overline{D}\overline{G}+c_{17}\,\overline{D}H
\overline{H}+c_{18}\,\overline{H} G \overline{G}\Bigg](Z_{2})
\nonumber\\&&+
\frac{\theta_{12}\overline{\theta}_{12}}{z_{12}}\Bigg[c_{19}\,\partial T+
c_{20}\,\partial\overline{D}H+c_{21}\,\partial D\overline{H}+
c_{22}\,[D,\overline{D}]G \overline{G}+c_{23}\, G [D,\overline{D}]\overline{G}
\nonumber\\&&+
c_{24}\,\partial H \overline{H}+c_{25}\, H \partial\overline{H}+
c_{26}\,\partial G \overline{G}+c_{27}\, G \partial\overline{G}+
c_{28}\,\overline{H} DG \overline{G}+c_{29}\, H \overline{D}G \overline{G}
\nonumber\\&&+
c_{30}\,\overline{H}  G D\overline{G}+c_{31}\, H G
\overline{D}\overline{G}+c_{32}\,\overline{D}H  G 
\overline{G}+c_{33}\, D\overline{H} G \overline{G}\Bigg](Z_{2})+\cdots,
\label{t1t1}
\end{eqnarray}
where the 
coefficients are given in Appendix $G.1$ and one uses the notation in 
(\ref{redef}) \footnote{
Let us multiply $(D_2-\overline{D}_2)$ on both sides of (\ref{t1t1})
with the condition $\theta =0 =\overline{\theta}$. 
Then the left hand side is given by 
$\Phi_0^{(1)}(z_1) \, \Phi_{\frac{1}{2}}^{(1),1}(z_2)$.
The nontrivial contributions come from 
the singular terms 
$\frac{\theta_{12}}{z_{12}}$ and $\frac{\overline{\theta}_{12}}{z_{12}}$
because the $D_2$ action on the former gives the singular term 
$-\frac{1}{(z_1-z_2)}$
and the  $\overline{D}_2$ action on the latter gives 
the singular term $-\frac{1}{(z_1-z_2)}$.
Then the first-order pole of $\Phi_0^{(1)}(z_1) \, 
\Phi_{\frac{1}{2}}^{(1),1}(z_2)$ contains 
the expression $(-c_7 D T +c_{13} \overline{D} T)$ with the condition 
$\theta =0 =\overline{\theta}$. 
By substituting the coefficients, this becomes
the result $(D_2+ \overline{D}_2) T(Z_2)$ which reduces to $G^1(z_2)$.
This term is what one expects from the component result in Appendix
$H.1$. The other terms can be identified similarly.}.

The fusion rule between 
 the ${\cal N}=2$ higher spin-$1$ currents
is given by
\bea
[{\bf T^{(1)}}] \, \cdot \,  [{\bf T^{(1)}}]= [{\bf I}], 
\label{fusionone}
\eea
where $[{\bf I}]$ denotes the large ${\cal N}=4$ linear superconformal family
of the identity operator.
Note that there are also $H(Z_2)$-, $\overline{H}(Z_2)$-, $G(Z_2)$- 
and $\overline{G}(Z_2)$-dependent terms 
as well as $T(Z_2)$-dependent terms in (\ref{t1t1}).

\subsubsection{The OPE between the ${\cal N}=2$ higher spin-$1$ current
and the ${\cal N}=2$ higher spin-$\frac{3}{2}$ currents}

The OPE between the ${\cal N}=2$ higher spin-$1$ current
and the ${\cal N}=2$ higher spin-$\frac{3}{2}$ currents,
where four components of ${\cal N}=2$ higher spin-$1, \frac{3}{2}$ 
currents 
are given the first four equations of (\ref{16highern2}) and 
the next eight equations with $s=1$,
is described by
\begin{eqnarray}
&&{\bf T^{(1)}}(Z_{1})\:
\left(
\begin{array}{c}
{\bf U^{(\frac{3}{2})}}\nonu \\
{\bf V^{(\frac{3}{2})}}
\end{array}
\right)(Z_{2})
\;=\;
\frac{\theta_{12}\overline{\theta}_{12}}{z_{12}^{3}}\, c_{\pm1}\,
G_{\pm}(Z_{2})
+\frac{\theta_{12}}{z_{12}^{2}}\Bigg[
c_{\pm2}\, D
G_{\pm}
+c_{\pm3}\, H
G_{\pm}\Bigg](Z_{2})
\nonu \\
&& +\frac{\overline{\theta}_{12}}{z_{12}^{2}}\Bigg[
c_{\pm4}\,\overline{D}
G_{\pm}
+c_{\pm5}\,\overline{H} 
G_{\pm}\Bigg](Z_{2})
\nonumber\\&&
+\frac{\theta_{12}\overline{\theta}_{12}}{z_{12}^{2}}\Bigg[
c_{\pm6}\,[D,\overline{D}]G_{\pm}
+c_{\pm7}\, H \overline{D}G_{\pm}
+c_{\pm8}\,\overline{H}  DG_{\pm}
+c_{\pm9}\, T G_{\pm}
+c_{\pm10}\,\overline{D}H  G_{\pm}
\nonumber\\&&
+c_{\pm11}\, D\overline{H}  G_{\pm}+c_{\pm12}\,\partial G_{\pm}\Bigg](Z_{2})
\nonumber\\&&
+\frac{1}{z_{12}}\Bigg[c_{\pm13}\,[D,\overline{D}]G_{\pm}+
c_{\pm14}\, H \overline{D}G_{\pm}
+c_{15}\, H \overline{H}\, G_{\pm}+c_{\pm16}\,\overline{H} DG_{\pm}
+c_{\pm17}\, T G_{\pm}
\nonumber\\&&
+c_{18}\,\overline{D}H G_{\pm}+c_{\pm19}\, D\overline{H} 
G_{\pm}+c_{\pm20}\,\partial G_{\pm}\Bigg](Z_{2})
\nonumber\\&&+
\frac{\theta_{12}}{z_{12}}\Bigg[c_{\pm21}\,\partial DG_{\pm}+
c_{\pm22}\, G_{+} DG_{\pm} {G}_{-}
+c_{\pm23}\, H \overline{H} DG_{\pm}
+c_{\pm24}\, H \overline{D}H G_{\pm}
+c_{\pm25}\, H D\overline{H} G_{\pm}
\nonumber\\&&
+c_{\pm26} \, H \partial G_{\pm}+c_{\pm27} \, T DG_{\pm}
+c_{\pm28} \, T H G_{\pm}
+c_{\pm29} \, \overline{D}H DG_{\pm}
+c_{\pm30} \, D\overline{H} DG_{\pm}
+c_{\pm31} \, \partial H G_{\pm}\Bigg]
\nonumber\\
&&
+\frac{\overline{\theta}_{12}}{z_{12}}\Bigg[c_{\pm32}\,\partial
\overline{D}G_{\pm}+c_{\pm33}\, G_{+} \overline{D}G_{\pm} G_{-}+
c_{\pm34}\,H \overline{H} \overline{D}G_{\pm}
+c_{\pm35}\,\overline{H} D\overline{H} G_{\pm}
+c_{\pm36}\,\overline{H} \partial G_{\pm}
\nonumber\\
&&+c_{\pm37} \, T \overline{D}G_{\pm}+c_{\pm38} \, T \overline{H} G_{\pm}+
c_{\pm39} \, \overline{D}H \overline{D}G_{\pm}
+c_{\pm40}  \, \overline{D}H\overline{H} G_{\pm}
+c_{\pm41} \, D\overline{H} \overline{D}G_{\pm}
+c_{\pm42} \, \partial \overline{H} G_{\pm}\Bigg]
\nonumber\\&&
+\frac{\theta_{12}\overline{\theta}_{12}}{z_{12}}\Bigg[c_{\pm43}
\left(
\begin{array}{c}
{\bf U^{(\frac{5}{2})}} \nonu \\\\
{\bf V^{(\frac{5}{2})}}
\end{array}
\right)+c_{\pm44}\, {\bf T^{(1)}}\left(
\begin{array}{c}
{\bf U^{(\frac{3}{2})}} \nonu \\\\
{\bf V^{(\frac{3}{2})}}
\end{array}
\right)+c_{\pm45}\, G_{+} \overline{D}G_{\pm}  D {G}_{-}
+c_{\pm46}\, G_{+} [D,\overline{D}]G_{\pm} {G}_{-}
\nonumber\\&&+
c_{\pm47}\, D G_{+}  G_{\pm} \overline{D} {G}_{-}+
c_{\pm48}\, H \partial\overline{D}G_{\pm}
+c_{\pm49}\, H G_{+} \overline{D}G_{\pm} {G}_{-}+
c_{\pm50}\, H \overline{H} [D,\overline{D}]G_{\pm}
\nonumber\\&&+c_{\pm51}\, H \overline{H} D\overline{H} G_{\pm}
+c_{\pm52}\, H \overline{H} \partial G_{\pm}+
c_{\pm53}\, H \overline{D}H \overline{D}G_{\pm}+
c_{\pm54}\, H \overline{D}H \overline{H} G_{\pm}
\nonumber\\&&+
c_{\pm55}\, H  D\overline{H} \overline{D}G_{\pm}+
c_{\pm56}\, H \partial\overline{H}  G_{\pm}+c_{\pm57}\,
\overline{H} \partial DG_{\pm}
+c_{\pm58}\,\overline{H}  G_{+}  DG_{\pm} {G}_{-}+
c_{\pm59}\,\overline{H} D\overline{H} DG_{\pm}
\nonumber\\&&
+c_{\pm60}\, T [D,\overline{D}]G_{\pm}
+c_{\pm61}\, T  H \overline{D}G_{\pm}
+c_{\pm62}\, T  H \overline{H} G_{\pm}+
c_{\pm63}\, T \overline{H}  DG_{\pm}+c_{\pm64}\, T  T G_{\pm}
\nonumber\\&&
+c_{\pm65}\, T \overline{D}H G_{\pm}
+c_{\pm66}\, T  D\overline{H} G_{\pm}
+c_{\pm67}\, T \partial G_{\pm}+c_{\pm68}\,\partial[D,\overline{D}]G_{\pm}
+c_{\pm69}\,\overline{D} G_{+}  DG_{\pm} {G}_{-}
\nonumber\\&&
+c_{\pm70}\,\overline{D}H [D,\overline{D}]G_{\pm}
+c_{\pm71}\,\overline{D}H \overline{H}  DG_{\pm}
+c_{\pm72}\,\overline{D}H \overline{D}H G_{\pm}+
c_{\pm73}\,\overline{D}H D\overline{H} G_{\pm}
\nonumber\\&&+
c_{\pm74}\,\overline{D}H \partial G_{\pm}
+c_{\pm75}\,\overline{D}T DG_{\pm}+c_{\pm76}\,\overline{D}T H G_{\pm}
+c_{\pm77}\,\partial\overline{D}H G_{\pm}+c_{\pm78}\,[D,\overline{D}]T G_{\pm}
\nonumber\\&&
+c_{\pm79}\, D\overline{H} [D,\overline{D}] G_{\pm}+
c_{\pm80}\, D\overline{H}  D\overline{H}  G_{\pm}+
c_{\pm81}\, D\overline{H} \partial G_{\pm}
+c_{\pm82}\, DT \overline{D} G_{\pm}
+c_{\pm83}\, DT \overline{H}  G_{\pm}\,
\nonumber\\&&+c_{\pm84}\,\partial D\overline{H} G_{\pm}+
c_{\pm85}\,\partial G_{\pm} G_{+} {G}_{-}
+c_{\pm86}\,\partial H \overline{D}G_{\pm}
+c_{\pm87}\,\partial H \overline{H} G_{\pm}+
c_{\pm88}\,\partial\overline{H} DG_{\pm}
\nonumber\\&&+
c_{\pm89}\,\partial T G_{\pm}
+c_{\pm90}\,\partial^{2} G_{\pm}
\Bigg](Z_{2})
+\cdots,
\label{tuv}
\end{eqnarray}
where the 
coefficients are given in Appendix $G.2$ after substituting (\ref{X})
explicitly.
One uses the notation in 
(\ref{redef}) and 
introduces the following notations 
\bea
G \equiv G_{+}, \qquad \overline{G} \equiv G_{-}, \qquad H \equiv H_{+},
\qquad \overline{H} \equiv H_{-},
\label{simple}
\eea
in order to write down two similar equations together \footnote{
Let us put the $\theta=0=\overline{\theta}$ in (\ref{tuv}). 
Then the left hand side 
is given by $\Phi_0^{(1)}(z_1) \, 2 \, \Phi_{\frac{1}{2}}^{(1),3}(z_2)$.
The right hand side contains 
$(c_{+13} [D, \overline{D}] G_{+}+c_{-13} [D, \overline{D}] G_{-})$, 
which is equal to 
$([D, \overline{D}] G_{+}- [D, \overline{D}] G_{-})$.
This will further reduce to $2 \, G^3(z_2)$ with other terms.
Therefore, the first-order term of 
the OPE  $\Phi_0^{(1)}(z_1) \, \Phi_{\frac{1}{2}}^{(1),3}(z_2)$
contains the $G^3(z_2)$ term, as one expects in the component approach.}.

The fusion rule between 
 the ${\cal N}=2$ higher spin-$1$ current
and 
 the ${\cal N}=2$ higher spin-$\frac{3}{2}$ currents
is given by
\bea
[{\bf T^{(1)}}] \, \cdot
\left[
\begin{array}{c}
{\bf U^{(\frac{3}{2})}} \\
{\bf V^{(\frac{3}{2})}}
\end{array}
\right] = [{\bf I}] + 
\left[  
\begin{array}{c}
{\bf T^{(1)}}  \, {\bf U^{(\frac{3}{2})}} \\
{\bf T^{(1)}} \, {\bf V^{(\frac{3}{2})}}
\end{array} \right]
+ \left[
\begin{array}{c}
{\bf U^{(\frac{5}{2})}}\\
{\bf V^{(\frac{5}{2})}}
\end{array}
\right],
\label{fusion5half}
\eea
where the last term, which lives in the next $16$ higher spin currents, 
has its component expressions in 
the $5$-th to $12$-th equations of (\ref{16highern2}) with $s=2$
\footnote{One uses the nonstandard notation in (\ref{fusion5half}).
If one uses the $\pm$ notations in the higher spin currents
${\bf U^{(\frac{3}{2})}}, {\bf V^{(\frac{3}{2})}}, {\bf U^{(\frac{5}{2})}}$
and ${\bf V^{(\frac{5}{2})}}$ (that is, ${\bf U_{\pm}^{(\frac{3}{2})}}$ for the first
two and 
${\bf U_{\pm}^{(\frac{5}{2})}}$ for the last two), 
then one can write down them 
in one line rather than $2 \times 1$ matrices as in (\ref{fusion5half}).
In our notation, the currents 
${\bf U^{(\frac{3}{2})}}(Z)$ and ${\bf U^{(\frac{5}{2})}}(Z)$
are living in the different ${\cal N}=2$ multiplets. The former has $s=1$ 
(the component of the lowest ${\cal N}=4$ multiplet) and the latter has
$s=2$ (the component of the next ${\cal N}=4$ multiplet).}.

\subsubsection{The OPE between the ${\cal N}=2$ higher spin-$1$ current
and the ${\cal N}=2$ higher spin-$2$ current}

The OPE between the ${\cal N}=2$ higher spin-$1$ current
and the ${\cal N}=2$ higher spin-$2$ current,
where four components of ${\cal N}=2$ higher spin-$1, 2$ 
currents 
are given the first four equations of (\ref{16highern2}) and 
the last four equations with $s=1$,
is summarized by
\begin{eqnarray}
&&{\bf T^{(1)}}(Z_{1})\:{\bf W^{(2)}}(Z_{2})\;=\;
\frac{\theta_{12}}{z_{12}^{3}}\, 
c_{1}\, H(Z_{2})+\frac{\overline{\theta}_{12}}{z_{12}^{3}}\, 
c_{2}\,\overline{H}(Z_{2})
\nonumber\\&&+\frac{1}{z_{12}^{2}}\Bigg[
c_{3}\, T
+c_{4}\,\overline{D}H
+c_{5}\, D\overline{H}
+c_{6}\, G \overline{G}
+c_{7}\, H \overline{H}\Bigg](Z_{2})
\nonumber\\
&&+\frac{\theta_{12}}{z_{12}^{2}}\Bigg[
c_{8} \, DT
+c_{9} \, G D \overline{G}
+c_{10} \, H \overline{D}H+c_{11}  \, H 
D\overline{H}
+c_{12} \, H G \overline{G}
+c_{13} \, T  H
+c_{14} \, DG \overline{G}
+c_{15} \, \partial H\Bigg]
\nonumber\\
&&+\frac{\overline{\theta}_{12}}{z_{12}^{2}}\Bigg[
c_{16} \, \overline{D}T
+c_{17}  \, G \overline{D}\overline{G}
+c_{18} \, \overline{H}  D\overline{H}
+c_{19} \, \overline{H}  G \overline{G}+c_{20}  \, T
\overline{H}+c_{21} \, \overline{D}G \overline{G}
+c_{22} \, \overline{D}H \overline{H}
+c_{23} \, \partial\overline{H}\Bigg]
\nonumber\\&&
+\frac{\theta_{12}\overline{\theta}_{12}}{z_{12}^{2}}
\Bigg[
c_{24}\,{\bf T^{(2)}}
+c_{25}\,{\bf T^{(1)}
 T^{(1)}}
+c_{26}\,[D,
\overline{D}]T+
c_{27}\,\partial D\overline{H}+c_{28}\, G [D,\overline{D}]
\overline{G}
+c_{29}\, G \partial\overline{G}
\nonumber\\&&
+c_{30}\, H G \overline{D}
\overline{G}+c_{31}\, H \overline{H} D\overline{H}+
c_{32}\, H
\overline{H} G \overline{G}+c_{33}\, H \overline{D}G \overline{G}
+c_{34}\, H \overline{D}H \overline{H}+c_{35}\, H \partial
\overline{H}
\nonumber\\&&
+c_{36}\,\overline{H}  G  D\overline{G}+c_{37}\,\overline{H} 
DG\,\overline{G}+c_{38}\, T T
+c_{39}\, T \overline{D}H
+c_{40}\, T  D\overline{H}
+c_{41}\, T  G \overline{G}
+c_{42}\, T  H \overline{H}
\nonumber\\&&
+c_{43}\,\partial\overline{D}H
+c_{44}\,\overline{D}G  D\overline{G}
+c_{45}\,\overline{D}H
\overline{D}H
+c_{46}\,\overline{D}H D\overline{H}+c_{47}\,\overline{D}H  G
\overline{G}+c_{48}\,\overline{D}T  H
\nonumber\\&&
+c_{49}\,[D,\overline{D]}G \overline{G}
+c_{50}\, DG
\overline{D}\overline{G}+c_{51}\, D\overline{H} D\overline{H}+c_{52}\, 
D\overline{H}  G \overline{G}+c_{53}\, DT \overline{H}
+c_{54}\,\partial G \overline{G}\nonumber\\&&
+c_{55}\,\partial H
\overline{H}+c_{56}\,\partial T\Bigg](Z_{2})
\nonumber\\&&
+\frac{1}{z_{12}}\Bigg[c_{57}\,\partial\overline{D}H+c_{58}\,
\partial D\overline{H}+c_{59}\, G [D,\overline{D}]\overline{G}+c_{60}\, 
G \partial\overline{G}+c_{61}\, H  G \overline{D}\overline{G}
+c_{62}\, H \overline{H}  D\overline{H}
\nonumber\\&&+c_{63}\, H
\overline{D}G \overline{G}+c_{64}H\,\overline{D}H \overline{H}+c_{65}\, 
H \partial\overline{H}+c_{66} \, \overline{H}  G D\overline{G}
+c_{67}\,\overline{H} DG \overline{G}+c_{68}\,
\overline{D}H  G \overline{G}
\nonumber\\&&+c_{69}\, \overline{D}T  H+c_{70}\,[D,
\overline{D}]G \overline{G}+c_{71}\, D\overline{H}  G \overline{G}
+c_{72}\, DT \overline{H}+c_{73}\, \partial G \overline{G}+
c_{74}\,\partial H \overline{H}
+c_{75}\,\partial T\Bigg](Z_{2})
\nonumber\\&&
+\frac{\theta_{12}}{z_{12}}\Bigg[
c_{76}\,  D{\bf T^{(2)}}
+c_{77}\, {\bf T^{(1)}}  D{\bf T^{(1)}}
+c_{78}\, G \partial D\overline{G}+c_{79}\, H
\partial D\overline{H}
+c_{80}\, H G [D,\overline{D}]\overline{G}
\nonumber\\&&
+c_{81}\, H  G \partial\overline{G}+c_{82}\, H
\overline{H} G  D\overline{G}+c_{83}\, H \overline{H}  DG
\overline{G}
+
c_{84}\, H \overline{D}G  D\overline{G}
+c_{85}\, H \overline{D}H  D\overline{H}
\nonumber\\&&+c_{86}\, H
[D,\overline{D}]G \overline{G}+c_{87}\, H  DG \overline{D}
\overline{G}
+c_{88}\, H  D\overline{H}  D\overline{H}
+c_{89}\, H D\overline{H} G \overline{G}
+c_{90}\, 
H \partial G \overline{G}
\nonumber\\&&+c_{91}\,\overline{H}  DG  D\overline{G}
+c_{92}\, T DT+c_{93}\, T  G D\overline{G}
+c_{94}\, T H D\overline{H}
+c_{95}\, T DG
\overline{G}
+c_{96}\, T\partial H
\nonumber\\&&+c_{97}\,
\partial DT
+c_{98}\,
\overline{D}H G D\overline{G}
+c_{99}\,\overline{D}H DG\overline{G}+c_{100} \, \partial
\overline{D}H  H
+c_{101} \, [D,\overline{D}]G\, D\overline{G}
\nonumber\\&&
+c_{102}\,[D,
\overline{D}]T  H
+c_{103}\, DG [D,\overline{D}]\overline{G}+c_{104}\, DG
\partial\overline{G}
+c_{105}\, D\overline{H}  G  D\overline{G}+c_{106}\, 
D\overline{H} DG \overline{G}
\nonumber\\&&+c_{107}\, DT \overline{D}H+c_{108}\, DT D\overline{H}
+c_{109}\, DT G \overline{G}+c_{110}\, DT  H \overline{H}+c_{111}\,
\partial DG \overline{G}
+c_{112}\,\partial G  D\overline{G}
\nonumber\\&&+c_{113}\,\partial H\,
\overline{D}H
+c_{114}\,\partial H  D\overline{H}+c_{115}\,\partial H 
G \overline{G}+c_{116}\,\partial H  H \overline{H}
+c_{117}\,\partial T  H+c_{118}\,\partial^{2} H\Bigg](Z_{2})
\nonumber\\&&
+\frac{\overline{\theta}_{12}}{z_{12}^{2}}\Bigg[
c_{119}\,
\overline{D}{\bf T^{(2)}}
+c_{120}\,{\bf T^{(1)}}
\overline{D}{\bf T^{(1)}}
+c_{121}\, G \partial
\overline{D}\overline{G}+c_{122}\, H \overline{H}  G \overline{D}
\overline{G}
+c_{123}\, H \overline{H} \overline{D}G \overline{G}
\nonumber\\&&+
c_{124}\, H \overline{D}G \overline{D}\overline{G}+c_{125}\, H
\partial\overline{H} \overline{H}+c_{126}\,\overline{H}  G [D,
\overline{D}]\overline{G}
+c_{127}\,\overline{H}  G \partial\overline{G}+c_{128}\,
\overline{H} \overline{D}G  D\overline{G}
\nonumber\\&&+
c_{129}\,\overline{H}
[D,\overline{D}]G\,\overline{G}+c_{130}\,\overline{H}  DG
\overline{D}\overline{G}
+c_{131}\,\overline{H} \partial G \overline{G}+c_{132}\, 
T \overline{D}T+c_{133}\, T  G \overline{D}\overline{G}
\nonumber\\&&+c_{134}\, T
\overline{D}G \overline{G}+c_{135}\, T \overline{D}H \overline{H}
+c_{136}\, T \partial\overline{H}
+c_{137}\,\partial\overline{D}T
+c_{138}\,\overline{D}G [D,
\overline{D}]\overline{G}
+c_{139}\,\overline{D}G \partial\overline{G}
\nonumber\\&&
+c_{140}\,\overline{D}T  G \overline{G}
+c_{141}\,\overline{D}H \overline{H}  D\overline{H}
+c_{142}\,
\overline{D}H \overline{H}  G \overline{G}
+c_{143}\,\overline{D}H \overline{D}G \overline{G}
+c_{144}\,\overline{D}H \overline{D}H \overline{H}
\nonumber\\&&
+c_{145}\,\overline{D}H \partial\overline{H}+c_{146}\,\overline{D}T
\overline{D}H+c_{147}\,\overline{D}T  D\overline{H}
+c_{148}\,\overline{D}H  G \overline{D}\overline{G}
+c_{149}\,\overline{D}T  H \overline{H}
\nonumber\\&&+c_{150}\,
\partial\overline{D}G \overline{G}+c_{151} \partial\overline{D}H
\overline{H}
+c_{152}\,[D,\overline{D}]G \overline{D}\overline{G}
+c_{153}\,[D,\overline{D}]T \overline{H}+c_{154}\, 
D\overline{H} G \overline{D}\overline{G}
\nonumber\\&&
+c_{155}\, D\overline{H}
\overline{D}G \overline{G}
+c_{156}\,\partial D\overline{H} \overline{H}
+c_{157}\,\partial G \overline{D}\overline{G}+c_{158}\,
\partial\overline{H} D\overline{H}+c_{159}\,\partial\overline{H} G
\overline{G}+c_{160}\,\partial T \overline{H}
\nonumber\\&&
+c_{161}\,\partial^{2}\overline{H}\Bigg](Z_{2})
\nonumber\\&&+\frac{\theta_{12}\overline{\theta}_{12}}{z_{12}}
\Bigg[
c_{162}\,\partial {\bf T^{(2)}}
+c_{163}\,\partial 
{\bf T^{(1)}} {\bf T^{(1)}}
+c_{164}\,\partial^{2}D\overline{H}
+c_{165}\, G
\partial[D,\overline{D}]\overline{G}
+c_{166}\, G \partial^{2}\overline{G}
\nonumber\\&&+c_{167}\, H G
\partial\overline{D}\overline{G}+c_{168}\, H \overline{H} G \partial
\overline{G}+c_{169}\, H \overline{H} \partial G \overline{G}
+c_{170}\, H \overline{D}G [D,\overline{D}]\overline{G}+c_{171}
 H \overline{D}G \partial\overline{G}
\nonumber\\&&
+c_{172}\, H \overline{D}H
\partial\overline{H}+c_{173}\, H \partial\overline{D}G \overline{G}
+c_{174}\, H [D,\overline{D}]G \overline{D}\overline{G}+
c_{175}\, H \partial D\overline{H} \overline{H}+c_{176}\, H \partial 
G \overline{D}\overline{G}
\nonumber\\&&
+c_{177}\, H \partial\overline{H} D\overline{H}
+c_{178}\, H \partial\overline{H} G \overline{G}+c_{179}\, 
H \partial^{2}\overline{H}+c_{180}\,\overline{H} G \partial D
\overline{G}+c_{181}\,\overline{H} [D,\overline{D}]G  D\overline{G}
\nonumber\\&&
+c_{182}\,\overline{H} DG [D,\overline{D}]\overline{G}+
c_{183}\,\overline{H}  DG \partial\overline{G}+c_{184}\,\overline{H}
\partial DG \overline{G}+c_{185}\,\overline{H} \partial G  D\bar{G}
+c_{186}\, T \partial\overline{D}H
\nonumber\\&&
+c_{187}\, T \partial 
D\overline{H}+c_{188}\, T  G \partial\overline{G}+c_{189}\, T  H
\partial\overline{H}+c_{190}\, T \partial G \overline{G}
+c_{191}\, T \partial H \overline{H}+c_{192}\,
\overline{D}G \partial D\overline{G}
\nonumber\\&&
+c_{193}\,\overline{D}H
\partial D\overline{H}+c_{194}\,\overline{D}H  G \partial\overline{G}
+c_{195}\,\overline{D}H \partial G \overline{G}+c_{196}
\,\overline{D}T G D\overline{G}+c_{197}\,\overline{D}T DG
\overline{G}
\nonumber\\&&
+c_{198}\,\overline{D}T \partial H
+c_{199}\,\partial\overline{D}G D\overline{G}+c_{200}\,
\partial\overline{D}H \overline{D}H+c_{201}\,\partial\overline{D}H 
D\overline{H}+c_{202}\,\partial\overline{D}H G \overline{G}
\nonumber\\&&
+c_{203}\,\partial\overline{D}H  H \overline{H}+c_{204}\,
\partial\overline{D}T  H+c_{205}\,[D,\overline{D}]G \partial
\overline{G}+c_{206}\,\partial[D,\overline{D}]G \overline{G}
+c_{207}\, DG \partial\overline{D}\overline{G}
\nonumber\\&&
+c_{208}\, 
D\overline{H} G \partial\overline{G}+c_{209}\, D\overline{H}
\partial G \overline{G}+c_{210}\, DT G \overline{D}\overline{G}
+c_{211}\, DT \overline{D}G \overline{G}+c_{212}\, 
DT \partial\overline{H}
\nonumber\\&&
+c_{213}\,\partial DG \overline{D}\overline{G}+
c_{214}\,\partial D\overline{H}  D\overline{H}
+c_{215}\,\partial D\overline{H}  G \overline{G}+
c_{216}\,\partial DT \overline{H}+c_{217}\,\partial G [D,\overline{D}]
\overline{G}
\nonumber\\&&
+c_{218}\,\partial H  G \overline{D}\overline{G}
+c_{219}\,\partial H \overline{H}  D\overline{H}+
c_{220}\,\partial H \overline{H}  G \overline{G}+c_{221}\,\partial H
\overline{D}G \overline{G}+c_{222}\,\partial H \overline{D}H
\overline{H}
\nonumber\\&&
+c_{223}\,\partial H \partial\overline{H}+c_{224}\,\partial
\overline{H} G  D\overline{G}+c_{225}\,\partial\overline{H} DG
\overline{G}+c_{226}\,\partial T  T+c_{227}\,\partial T \overline{D}H
+c_{228}\,\partial T D\overline{H}
\nonumber\\&&
+c_{229}\,\partial T 
G\,\overline{G}+c_{230}\,\partial T  H \overline{H}
+c_{231}\,\partial[D,\overline{D}]T
+c_{232}\,\partial^{2}G \overline{G}
+c_{233}\,\partial^{2}H \overline{H}
+c_{234}\,\partial^{2} \overline{D}H
\nonumber\\&&
+c_{235}\,\partial^{2} T
\Bigg](Z_{2})+\cdots,
\label{tw}
\end{eqnarray}
where the 
coefficients are given in Appendix $G.3$
and one uses the notation in 
(\ref{redef}) \footnote{
Let us take $\theta=0=\overline{\theta}$ in (\ref{tw}).
Then the left hand side is given by
$\Phi_0^{(1)}(z_1) \, \left( 4 \, i \, 
\Phi_{1}^{(1),12}  +\frac{8 \,i \, \alpha}{3}
\Phi_{1}^{(1),34}\right)(z_2)$. The right hand
side contains $c_6 (G \, \overline{G})(Z_2)$
which reduces to $2  \, i\, c_6 \,  (\Gamma^3 \, \Gamma^4)(z_2)$ 
in the second-order pole.  This is what one expects from the 
component approach from Appendix $H.1$.}.

The fusion rule between 
 the ${\cal N}=2$ higher spin-$1$ current
and 
 the ${\cal N}=2$ higher spin-$2$ current
is given by
\bea
[{\bf T^{(1)}}] \, \cdot \, [{\bf W^{(2)}}] = [{\bf I}] + 
[{\bf T^{(1)}} \, {\bf T^{(1)}} ] 
+ [{\bf T^{(2)}}],
\label{fusion2}
\eea
where the last term, which lives in the next $16$ higher spin currents, 
has its component expressions in 
the first four equations of (\ref{16highern2}) with $s=2$.
Recall that the next higher ${\cal N}=4$ multiplet 
has four ${\cal N}=2$ higher spin currents 
denoted by ${\bf T^{(2)}}(Z)$, ${\bf U^{(\frac{5}{2})}}(Z)$,
 ${\bf V^{(\frac{5}{2})}}(Z)$ and 
${\bf W^{(3)}}(Z)$.
Some of the terms in the right hand side of 
(\ref{tw}) occur in (\ref{t1t1}) but most of the terms 
in (\ref{tw}) arise newly. For example, $c_{10}$- and $c_{13}$-terms 
do not appear in (\ref{t1t1}).

\subsubsection{ Other remaining OPEs in ${\cal N}=2$ superspace}

From Appendix $G.4$, 
the fusion rule between the ${\cal N}=2$ higher spin-$\frac{3}{2}$ currents,
where four components of ${\cal N}=2$ higher spin-$\frac{3}{2}$ 
currents 
are given the $5$-th to $12$-th equations of (\ref{16highern2}) with $s=1$,
is summarized by
\bea
\left[
\begin{array}{c}
{\bf U^{(\frac{3}{2})}} \\
{\bf V^{(\frac{3}{2})}}
\end{array}
\right] 
\, \cdot\, 
\left[
\begin{array}{c}
{\bf U^{(\frac{3}{2})}} \\
{\bf V^{(\frac{3}{2})}}
\end{array}
\right] = [{\bf I}].
\label{fusiontwo}
\eea

From Appendix $G.5$, 
the other 
fusion rule between the ${\cal N}=2$ higher spin-$\frac{3}{2}$ currents,
where four components of ${\cal N}=2$ higher spin-$\frac{3}{2}$ 
currents 
are given the $5$-th to $12$-th equations of (\ref{16highern2}) with $s=1$,
is summarized by
\bea
[{\bf U^{(\frac{3}{2})}}] \, 
\cdot \, 
[{\bf V^{(\frac{3}{2})}}
] =[{\bf I}] + 
[{\bf T^{(1)}} \, {\bf T^{(1)}} ] 
 +
[{\bf T^{(1)}} \, {\bf W^{(2)}} ] 
+ [{\bf U^{(\frac{3}{2})}} \,  
{\bf V^{(\frac{3}{2})}}
] + [{\bf T^{(2)}}] + [{\bf W^{(3)}}],
\label{fusion23}
\eea
where
 the last two terms, which live in the next $16$ higher spin currents, 
has its component expressions in 
the first four and the last four 
equations of (\ref{16highern2}) with $s=2$.
Therefore, one sees that the right hand sides of 
(\ref{fusion5half}), (\ref{fusion2}) and (\ref{fusion23})
contain the next $16$ higher spin currents in terms of 
its ${\cal N}=2$ version.

From Appendix $G.6$, 
the  
fusion rule between the ${\cal N}=2$ higher spin-$\frac{3}{2}$ current
and  the ${\cal N}=2$ higher spin-$2$ current,
where four components of ${\cal N}=2$ higher spin-$\frac{3}{2}$ 
currents 
are given the $5$-th-$12$-th equations of (\ref{16highern2}) 
and 
four components of ${\cal N}=2$ higher spin-$2$ 
currents 
are given the last four equations of (\ref{16highern2}) 
with $s=1$,
is summarized by
\bea
\left[
\begin{array}{c}
{\bf U^{(\frac{3}{2})}} \\
{\bf V^{(\frac{3}{2})}}
\end{array}
\right] \, \cdot \, [{\bf W^{(2)}} ]
=[{\bf I}] + 
\left[
\begin{array}{c}
 {\bf T^{(1)}} \, {\bf U^{(\frac{3}{2})}} \\
 {\bf T^{(1)}} \, {\bf V^{(\frac{3}{2})}}
\end{array}
 \right]
+ \left[
\begin{array}{c}
{\bf U^{(\frac{5}{2})}} \\
{\bf V^{(\frac{5}{2})}}
\end{array}
\right]
\label{fusionthree}
\eea
where the last term, which lives in the next $16$ higher spin currents,
has its component expressions in 
the $5$-th to $12$-th equations of (\ref{16highern2}) with $s=2$.

Finally 
from Appendix $G.7$, 
the  
fusion rule between the ${\cal N}=2$ higher spin-$2$ currents,
where 
four components of ${\cal N}=2$ higher spin-$2$ 
current 
are given the last four equations of (\ref{16highern2}) 
with $s=1$,
is described by
\bea
[{\bf W^{(2)}} ] \, \cdot \, [{\bf W^{(2)}} ]
= [{\bf I}] + 
[{\bf T^{(1)}} \, {\bf T^{(1)}} ] 
 +
[{\bf T^{(1)}} \, {\bf W^{(2)}} ] 
+ [{\bf U^{(\frac{3}{2})}} \,  
{\bf V^{(\frac{3}{2})}}
] 
+ [{\bf T^{(2)}}] +
  [{\bf W^{(3)}}]
\label{fusionfour}
\eea
where
 the last two terms, which live in the next $16$ higher spin currents, 
have its component expressions in 
the first four and the last four 
equations of (\ref{16highern2}) with $s=2$.
The presence of $[{\bf W^{(3)}}]$ in (\ref{fusionfour})
occurs also in the ${\cal N}=2$ ${\cal W}_{N+1}$ algebra studied in 
\cite{Ahn1208,Ahn1206}. 

By adding (\ref{fusionone}), (\ref{fusion5half}), (\ref{fusion2}),
(\ref{fusiontwo}), (\ref{fusion23}), (\ref{fusionthree}) and 
(\ref{fusionfour}) together, 
one obtains 
the fusion rule
for the lowest $16$ higher spin currents 
\bea
&& [{\bf I}] + 
\left( [{\bf T^{(1)}} \, {\bf T^{(1)}} ] 
 +
[
{\bf T^{(1)}} \, {\bf U^{(\frac{3}{2})}}]
+ [{\bf T^{(1)}} \, {\bf V^{(\frac{3}{2})}}]
+
[{\bf T^{(1)}} \, {\bf W^{(2)}} ] 
+ [{\bf U^{(\frac{3}{2})}} \,  
{\bf V^{(\frac{3}{2})}}
] \right) 
\nonu \\
&& + \left( [{\bf T^{(2)}}] +
[{\bf U^{(\frac{5}{2})}}]
+  [{\bf V^{(\frac{5}{2})}}]
+
  [{\bf W^{(3)}}] \right)
\label{finalfusionn2}
\eea
in the ${\cal N}=2$ superspace.
In next section, this fusion rule will be written in the ${\cal N}=4$ 
superspace. 
At the moment it is not clear why the terms 
$ [{\bf U^{(\frac{3}{2})}} \,  
{\bf U^{(\frac{3}{2})}}
] $,
$[{\bf U^{(\frac{3}{2})}} \, {\bf W^{(2)}} ]$, 
$ [{\bf V^{(\frac{3}{2})}} \,  
{\bf V^{(\frac{3}{2})}}
] $,
$[{\bf V^{(\frac{3}{2})}} \, {\bf W^{(2)}} ]$
and 
$[{\bf W^{(2)}} \, {\bf W^{(2)}} ]$
do not appear in the above fusion rule, but 
if one goes to ${\cal N}=4$ superspace, one can easily 
understand the reason as follows.
In the fusion rule 
of $[{\bf \Phi^{(1)}}] \, \cdot \, [{\bf \Phi^{(1)}}]$, 
the relation (\ref{phisn2version}) has an expansion of 
$\theta^3$ and $\theta^4$ and the normal ordered product 
$ {\bf \Phi^{(1)}}  \, {\bf \Phi^{(1)}} $ at the same point of ${\cal N}=4$
superspace coordinate  
contains 
the above normal ordered product in (\ref{finalfusionn2}) because 
those terms have linear in $\theta^3$- or $\theta^4$- or 
$\theta^{3,4}$-independent term. 
However, 
the above five normal ordered products, which are not allowed in 
(\ref{finalfusionn2}), have the quadratic terms in $\theta^{3,4}$
and therefore they vanish in the viewpoint of ${\cal N}=4$ superspace
\footnote{Let us emphasize, as noticed in \cite{Ahn1504}, 
that if one takes the basis in \cite{BCG}
where the higher spin-$3$ current is not a primary current, 
then the above nonlinear terms in ${\cal N}=2$ superspace 
are disappeared with the redefinitions of the next $16$ higher spin 
currents.}.

\section{The OPEs betweenthe $16$ higher spin  currents  in 
${\cal N}=4$ superspace}

In this final section, one would like to 
summarize what has been obtained in previous section in ${\cal N}=4$
superspace.

\subsection{The $136$ OPEs from ${\cal N}=2$ superspace results }

In previous section, the complete ${\cal N}=2$ OPEs with complete
structure constants are determined.
Again, using the package \cite{KT}, one can go to the component approach
where the $136$ OPEs are determined completely.
They are presented in Appendix $H$ with simplified notation.
One might ask whether there exists a possibility for 
having the new primary currents in these $136$ OPEs or not.
Because one does not check them from the $16$ higher spin currents 
in the ${\cal N}=4$ coset theory for generic $N$ by hand, one should 
be careful about the occurrence of the new primary currents in the OPEs.
For example, in the bosonic coset theory described in \cite{AK1308},
the higher spin-$4$ current appears as long as $N \geq 4$. Or 
for $N=3$, there is no higher spin-$4$ current because the structure
constant appearing in the higher spin-$4$ current contains the factor $(N-3)$.
Furthermore,
the higher spin-$5$ current appears as long as $N \geq 5$. Or 
for $N=4$, there is no higher spin-$5$ current because the structure
constant appearing in the higher spin-$5$ current contains the factor $(N-4)$.
Therefore as the number $N$ increases, the extra new primary currents
occur. 

However, in the present ${\cal N}=4$ coset theory, this feature
does not appear. As explained around (\ref{c1c2}) before, 
we have confirmed that there are no extra primary currents in the 
basic $16$ OPEs between the higher spin-$1$ current and $16$ higher spin 
currents using the WZW currents for several $N$-values.
We believe that these basic $16$ OPEs satisfy even if one tries to 
calculate them by hand. This behavior is rather different from the 
one from the purely bosonic coset theory.    

\subsection{The ${\cal N}=4$ superspace description}

Then the final one single ${\cal N}=4$ OPE 
between the ${\cal N}=4$ (higher spin current) multiplet 
of super spin $1$ can be described as
\bea
&& {\bf \Phi^{(1)}}(Z_{1})\,{\bf \Phi^{(1)}}(Z_{2})=
\frac{\theta_{12}^{4-0}}{z_{12}^{4}}\, c_{1}+
\frac{\theta_{12}^{4-i}}{z_{12}^{3}}\, c_{2}\, J^{i}(Z_{2})+
\frac{\theta_{12}^{4-0}}{z_{12}^{3}}\, c_{3}\,\partial J(Z_{2})+
\frac{1}{z_{12}^{2}}\, c_{4}
\nonumber\\
& & +\frac{\theta_{12}^{4-ij}}{z_{12}^{2}}\Bigg[c_{5}\, J^{i} J^{j}+
c_{6}\, J^{ij}+c_{7}\,\varepsilon^{ijkl} \, J^{k} J^{l}
+c_{8}\, J^{4-ij}\Bigg](Z_{2})
\nonumber\\
& & +\frac{\theta_{12}^{4-i}}{z_{12}^{2}}\Bigg[c_{9}\, J^{4-i}+c_{10}\, 
\partial J^{i}+
c_{11}\,\varepsilon^{ijkl} \, J^{jk} J^{l}+c_{12}\,J^{ij}
J^{j}
+c_{13}\,\partial J J^{i}
\Bigg](Z_{2})
\nonumber\\
& & +\frac{\theta_{12}^{4-0}}{z_{12}^{2}}\Bigg[
c_{14}\,{\bf \Phi^{(2)}}+
c_{15}\,{\bf \Phi^{(1)} \Phi^{(1)}}
+c_{16}\, J^{4-0}
+c_{17}\, J^{4-i} J^{i}
+\varepsilon^{ijkl}\Big\{c_{18}\,J^{i} J^{j} J^{k} J^{l}
+  c_{19}\, J^{ij} J^{kl} \Big\}
\nonu \\
&& +c_{20}\, J^{4-ij}  J^{4-ij}
+c_{21}\, J^{ij} J^{i} J^{j}
+c_{22}\, J^{4-ij} J^{i} J^{j}
 +c_{23}\,\partial^{2}J
+c_{24}\, \partial J^{i} J^{i}
+c_{25}\,\partial J \partial J
\Bigg](Z_{2})
\nonumber\\
& & +\frac{\theta_{12}^{4-jkl}}{z_{12}}\Bigg[
 c_{26}\,J^{jkl}
 +\varepsilon^{ijkl} (
c_{27}\,\partial J^{i}
 +c_{28}\,J^{ij} J^{j}
) +c_{29}\,J^{j} J^{k} J^{l}
\Bigg](Z_{2})
\nonumber\\
& & +\frac{\theta_{12}^{4-ij}}{z_{12}}\Bigg[
c_{30}\,(J^{4-i} 
J^{j}-J^{4-j} J^{i})
+c_{31}\, \partial J^{4-ij}
+c_{32}\,\partial J^{ij}
+c_{33}\,\partial (J^{i}  J^{j})
\nonumber\\
& & +c_{34}\,\varepsilon^{ijkl}\,\partial (J^{k}\,J^{l})
+c_{35}\, (J^{ik} J^{k} J^{j}-J^{jk} J^{k} J^{i})\Bigg](Z_{2})
\nonumber\\
& & +\frac{\theta_{12}^{4-i}}{z_{12}}\Bigg[
c_{36}\, D^{i} {\bf \Phi^{(2)}}
+c_{37}\, {\bf \Phi^{(1)}} D^{i} {\bf \Phi^{(1)}}+
c_{38} (J^{ij} J^{ij} J^{i}- J^{4-ij} J^{ik} J^{4-ijk})
+c_{39}\,\partial^{2} J  J^{i}
\nonu \\
&&+c_{40}\,\partial^{2} J^{i}
+c_{41}\, J^{ij} \partial J^{j}
+c_{42}\, J^{i} \partial J^{j} J^{j}
+c_{43}\, J^{ij} J^{4-j}
+c_{44}\, \partial J^{i} \, \partial J^{i}
+c_{45}\,\partial J^{4-i}
\nonu \\
&&+c_{46}\, \partial J^{ij} J^{j}
+c_{47}\, J^{4-ij} J^{4-j}
+c_{48}\, J^{ij} \partial J  J^{j}
+c_{49}\,\partial J J^{4-i}
\nonu \\
&&+\varepsilon^{ijkl}\Big\{
c_{50}\, J^{j} J^{k} J^{ijk}
+c_{51}\,\, ( J^{ij}  J^{kl}J^{i}-\partial J  J^{jk} J^{l}-J^{i} J^{j} J^{ikl})
+c_{52}\, J^{ij} J^{i} J^{k} J^{l}
+c_{53}\, J^{jk} \partial J^{l}
\nonu \\
&&c_{54}\,\partial J J^{j} J^{k} J^{l}
+c_{55}\,\partial (J^{j} J^{k} J^{l})
+c_{56}\,\partial J^{jk} J^{l}\Big\}\Bigg](Z_{2})
\nonumber\\
& & +\frac{\theta_{12}^{4-0}}{z_{12}}\Bigg[
c_{57}\,\partial {\bf \Phi^{(2)}}
+c_{58}\,\partial {\bf \Phi^{(1)} \Phi^{(1)}}
+c_{59}\, J^{4-i} \partial J^{i}
+c_{60}\, \partial J^{4-0}
+c_{61}\,\partial (J^{4-ij}  J^{4-ij})
\nonu \\
&&+c_{62}\,\partial^{2} J \partial J
+c_{63}\,\partial^{3} J
+c_{64}\,\partial^{2} J^{i} J^{i}
+c_{65}\,\partial (J^{ij} J^{i} J^{j})
+c_{66}\,J^{i} J^{j} \partial J^{ij}
+c_{67}\,\partial J  \partial J^{i} J^{i}
\nonu \\
&&+\varepsilon^{ijkl}\Big\{c_{68}\,\partial (J^{i} J^{j} J^{k} J^{l})
+c_{69}\,\partial (J^{ij} J^{kl})+c_{70}\,\partial (J^{ij} J^{k} J^{l})\Big\}\Bigg](Z_{2})+\cdots,
\label{finalPhiPhi}
\eea
where the simplified notations are introduced as follows:
\bea
{\bf J^{(4)}}(Z) & \equiv & J(Z), \qquad
D^i {\bf J^{(4)}}(Z) \equiv J^i(Z), \qquad
 D^i \, D^j {\bf J^{(4)}}(Z) \equiv J^{ij}(Z), \nonu \\
 D^i \, D^j \, D^k {\bf J^{(4)}}(Z) & \equiv & J^{ijk}(Z), \qquad
 D^1 \, D^2 \, D^3 \, D^4 {\bf J^{(4)}}(Z) \equiv J^{1234}(Z) \equiv J^{4-0}(Z).
\label{finalOPE}
\eea
Note that 
$ \theta^{4-ijk} = \varepsilon^{lijk} \theta^{l}$ and 
one uses unusual notation for the summation.
For example, see the $c_{39}$-term, where the dummy index $i$ arises four 
times.
One can easily see that there are consistent $SO(4)$ index contractions 
with $SO(4)$-invariant tensors $\varepsilon^{ijkl}$ and $\delta^{ij}$ in the 
right hand side of the OPE (\ref{finalPhiPhi})
\footnote{
Let us take some simple example, where one sees how the 
${\cal N}=4$ OPE can be reduced to the corresponding 
OPE in the component approach.
The nonlinear terms appear from the $c_5$-term to $c_8$-term in 
(\ref{finalPhiPhi}).
Let us multiply the operator $D_2^1 \, D_2^2$ both sides of (\ref{finalPhiPhi})
with the condition of $\theta_1^i=0=\theta^i_2$.
Then the left hand side is given by $\Phi_0^{(1)}(z_1) \, (-1) \, 
\Phi_1^{(1),34}(z_2)$. On the other hand, the right hand side contains
$D_2^1 \, D_2^2 \, \frac{\theta_{12}^1 \, \theta_{12}^2}{z_{12}^2}$ 
with current-dependent
terms. This leads to the singular term, $-\frac{1}{(z_1-z_2)^2}$.
Therefore, the second-order terms of the OPE 
$\Phi_0^{(1)}(z_1)  \, 
\Phi_1^{(1),34}(z_2)$
are given by $2(c_5 J^3\, J^4 + c_{6}\, J^{34}+
2 c_{7}\,J^{1}\, J^{2}
+c_{8}\, J^{12})(z_2)$ at vanishing $\theta^i$.
Then one obtains
 $2(-c_5 \Gamma^3\, \Gamma^4 +i  c_{6}\, T^{12}-
2 c_{7}\,\Gamma^{1}\, \Gamma^{2}
+ i c_{8}\, T^{34})(z_2)$ where the coefficients are given in Appendix $I$.
Then one can check that this is equal to the particular singular terms in 
the corresponding OPE in Appendix 
$H.1$.}. 

Therefore, the fusion rule for the 
lowest $16$ higher spin currents in ${\cal N}=4$ superspace
can be written as
\bea
[{\bf \Phi^{(1)}}] \, \cdot \, [ {\bf \Phi^{(1)}}] = [{\bf I}] + 
[{\bf \Phi^{(1)}} \, {\bf \Phi^{(1)}}] +[{\bf \Phi^{(2)}}],
\label{fusionfinalfinal}
\eea
where $[{\bf I}]$ denotes the large ${\cal N}=4$ linear superconformal family
of the identity operator.
This is equivalent to the previous result (\ref{finalfusionn2}) 
in ${\cal N} =2$ superspace.
Note that 
the ${\cal N}=4$ multiplet 
${\bf \Phi^{(1)}}(Z)$
stands for the 
four ${\cal N}=2$ higher spin currents 
${\bf T^{(1)}}(Z)$, ${\bf U^{(\frac{3}{2})}}(Z)$,
 ${\bf V^{(\frac{3}{2})}}(Z)$ and 
${\bf W^{(2)}}(Z)$
while 
the next ${\cal N}=4$ multiplet 
${\bf \Phi^{(2)}}(Z)$
stands for the 
four ${\cal N}=2$ next higher spin currents 
${\bf T^{(2)}}(Z)$, ${\bf U^{(\frac{5}{2})}}(Z)$,
 ${\bf V^{(\frac{5}{2})}}(Z)$ and 
${\bf W^{(3)}}(Z)$.
The explicit component expansions for the two 
${\cal N}=4$ higher spin currents with $s=1, 2$ are presented in 
(\ref{phis}) previously
\footnote{It is useful to see how one can obtain the ${\cal N}=2$
superspace description starting from its ${\cal N}=4$ version in 
(\ref{finalPhiPhi}). Let us focus on the simplest OPE given by 
(\ref{t1t1}). One can put $\theta^3 =0 = \theta^4$ in (\ref{finalPhiPhi}).
Then the left hand side gives rise to ${\bf T^{(1)}}(Z_1) \, 
{\bf T^{(1)}}(Z_2)$. Then how one can obtain the nontrivial contributions
from the right hand side with the above conditions? If one has 
$\theta_{12}^{4}$, then this does not produce the nontrivial 
contributions. For the terms $\theta^{4-i}$, one does not have any 
nonzero contributions because it contains either $\theta^3$ or $\theta^4$.
For the terms $\theta^{4-ij}$, one can have nonzero contributions for the 
indices $i, j$ being $3,4$. For the terms $\theta^{4-ijk}$, one can have 
nonzero contributions. Finally, for the term $\frac{1}{z_{12}^2}$, the nonzero
contribution occurs. Then one can easily see that the above 
coefficient $c_4$ in (\ref{finalPhiPhi}) is equal to the coefficient 
$c_1$ in (\ref{t1t1}). 
Furthermore, the expression $(c_{5}\, J^{3}\, J^{4}+
c_{6}\, J^{34}+c_{7}\,\varepsilon^{34kl}J^{k}\, J^{l}
+c_{8}\, J^{12})$ should  be reproduced to the corresponding quantity in 
(\ref{t1t1}). For example, the $c_5$-term here can be obtained 
from (\ref{n4n2}) and gives the expression 
$\frac{i}{2} G \, \overline{G}(Z_2)$.
Also one has the relation $\frac{\theta_{12}^1 \, \theta_{12}^2}{z_{12}^2}=
\frac{i}{2} \,
\frac{\theta_{12} \, \overline{\theta}_{12}}{z_{12}^2}$ and one obtains that
the coefficient of $\frac{\theta_{12} \, \overline{\theta}_{12}}{z_{12}^2}$
is given by $-\frac{1}{4} \, c_5\, G \, \overline{G}(Z_2)$, 
where the coefficient
$c_5$ is given by Appendix $I$.
This becomes $c_6 \, G \, \overline{G}(Z_2)$ 
in (\ref{t1t1}) with Appendix $G.1$, as one expects.
It is straightforward to check the remaining other terms explicitly.  }. 

\section{Conclusions and outlook }

As in the title of this paper, the 
single OPE between the ${\cal N}=4$ higher spin-$1$ multiplet
is given by (\ref{finalPhiPhi}) with the structure constants presented 
in Appendix (\ref{finalOPEcoeff}). 
From the fusion rule in (\ref{fusionfinalfinal}), 
the ${\cal N}=4$ higher spin-$2$ multiplet occurs in the right hand side
of this OPE. 

Let us make a list for the future directions as follows:

$\bullet$ Asymptotic symmetry algebra and three point functions 

As described in the introduction, the main motivation 
of this paper is to obtain the complete OPE 
between the $16$ higher spin currents in ${\cal N}=4, {\cal N}=2$
superspaces and in the component approach. One expects that 
this symmetry algebra also appears in the $AdS_3$ bulk theory
as an asymptotic symmetry algebra. So the immediate question is 
to construct the eigenvalue equations, to obtain the corresponding 
three-point functions and see the matches with the previous two 
dimensional CFT computations in \cite{AK1506}.      

$\bullet$ Other representations for the higher spin currents 
under the group $SO(4)$

In this paper, the ${\cal N}=4$ higher spin-$1$ multiplet 
is a singlet under the $SO(4)$.
In \cite{GPZ}, the ${\cal N}=4$ multiplets $R^{(s)}({\bf 2},{\bf 1}),
R^{(s)}({\bf 1},{\bf 2}), R^{(s+\frac{1}{2})}({\bf 2},{\bf 1}), 
R^{(s+\frac{3}{2})}({\bf 1},{\bf 2}), \cdots$, where 
$({\bf 2},{\bf 1})$ and $({\bf 1},{\bf 2})$ are the representations
under the $SU(2) \times SU(2)$  occur 
in the stringy chiral algebra. The above ${\cal N}=4$ higher spin multiplets 
${\bf \Phi^{(1)}}$ and ${\bf \Phi^{(2)}}$
are given by $R^{(1)}({\bf 1},{\bf 1})$ and $R^{(2)}({\bf 1},{\bf 1})$. 
At least, one can construct the ${\cal N}=4$ primary condition (\ref{jphi}) 
for these other representations by adding the extra terms \cite{Schoutensnpb}. 
The $SO(4)$ vector representation ${\bf 4}$ breaks into 
${\bf 4} \rightarrow  ({\bf 2},{\bf 1}) \oplus ({\bf 1},{\bf 2})$ under 
the $SU(2) \times SU(2)$.
It would be interesting to see the nontrivial OPEs between 
the nonsinglet ${\cal N}=4$ higher spin currents with lower spins.

$\bullet$ Orthogonal coset theory

One can also apply the present analysis of this paper to 
the ${\cal N}=4$ orthogonal coset theory. 
So far the higher spin extension  of 
the large ${\cal N}=4$ nonlinear superconformal algebra
in the realization of large ${\cal N}=4$ orthogonal coset  theory
was obtained in \cite{AP}. It is straightforward to 
construct the extension of 
the large ${\cal N}=4$ linear superconformal algebra, where
the group $G=SO(N+4)$ appears in the 
large ${\cal N}=4$ orthogonal coset  theory for $N=4$.
It would be interesting to obtain the final 
${\cal N}=4$ single OPE ${\bf \Phi^{(2)}}(Z_1)\, {\bf 
\Phi^{(2)}}(Z_2)$, where the ${\bf \Phi^{(2)}}(Z)$
is the lowest ${\cal N}=4$ higher spin multiplet of super spin $2$, 
corresponding to the equation (\ref{finalPhiPhi}).
Note that the lowest $16$ higher spin currents have the lowest higher 
spin-$2$ current, whose OPE is nontrivial, compared to the corresponding 
higher spin-$1$ current for the unitary case, because the second-order pole 
contains the nontrivial composite spin $2$ fields. 
In this direction, the work of
\cite{FG,AP1310,AP1301,Ahn1202,Ahn1106} will be useful.

$\bullet$ An extension of large ${\cal N}=4$ nonlinear superconformal 
algebra in the ${\cal N}=4$ superspace 

In this paper, the extension of the large ${\cal N}=4$ linear superconformal 
algebra was described. What about the extension of 
 the large ${\cal N}=4$ nonlinear superconformal 
algebra by decoupling the spin-$1$ current $U(z)$ and four spin-$\frac{1}{2}$
currents $\Gamma^i(z)$ \cite{AK1411,Ahn1408,Ahn1311}? Is it possible to express 
the $11$ currents in ${\cal N}=4$ superspace (or ${\cal N}=2$ superspace)?

$\bullet$ Simplification of ${\cal N}=2$ OPEs using the summation indices

In ${\cal N}=2$ superspace, the original $SO(4)$ symmetry is broken to 
$SO(2)$ symmetry. In the right hand side of the OPEs in ${\cal N}=2$ superspace,
there are too many terms in Appendix $G$. 
It would be interesting to see whether one can 
simplify them using the summation indices.  

$\bullet$ An extension of small ${\cal N}=4$ superconformal algebra

From (\ref{finalOPE}), one can take large $N$ 't Hooft limit 
with the rescales for the currents and this leads to 
an extension of small ${\cal N}=4$ superconformal algebra.
It would be interesting to study this direction because 
this algebra appears in the recent studies \cite{GG1406,GG1501}.

$\bullet$ The ${\cal N}=3$ holography

In the work of \cite{CH1506,HR1503,CHR1406}, 
the ${\cal N}=3$ holography 
was found in the context of Kazama-Suzuki model for the particular level.
See also \cite{Ahn1305,Ahn1211}.
It would be interesting to construct the higher spin currents.
One can easily see the ${\cal N}=3$ primary current condition for any 
${\cal N}=3$ multiplet, which consists of 
one higher spin-$s$ current, three higher spin-$(s+\frac{1}{2})$ currents,
three higher spin-$(s+1)$ currents and one higher spin-$(s+\frac{3}{2})$
current in the component approach. 
It is nontrivial to obtain the ${\cal N}=3$ OPE 
between the lowest ${\cal N}=3$ higher spin multiplet and itself.
Several $(N,M)$ cases will be useful to understand the structure of this 
OPE.
Therefore, the ${\cal N}=3$ description of this paper can give some hints 
for the index structure of $SO(3)$ group.

$\bullet$ The other OPEs for the next higher spin currents

One can also consider the other OPEs between the ${\cal N}=4$ higher spin
multiplets.
It is nontrivial to obtain the OPEs ${\bf \Phi^{(1)}}(Z_1) \, 
{\bf \Phi^{(2)}}(Z_2)$ and  ${\bf \Phi^{(2)}}(Z_1) \, 
{\bf \Phi^{(2)}}(Z_2)$. From the experience of this paper, 
it is enough to calculate the OPEs between the lowest higher spin-$1$
current from ${\bf \Phi^{(1)}}(Z_1)$ and the $16$ currents from 
${\bf \Phi^{(2)}}(Z_2)$ for any $N$ and $k$.
Similarly, one should calculate 
the OPEs between the higher spin-$2$ current from 
  ${\bf \Phi^{(2)}}(Z_1)$ and the $16$ higher spin currents 
from ${\bf \Phi^{(2)}}(Z_2)$ for any $N$ and $k$.
This study will help also the determination of the previous unknown 
parameter $X$ in (\ref{X}).  

$\bullet$ The ${\cal N}=3$ and ${\cal N}=1$ superspace descriptions

In this paper, the complete OPE is written in ${\cal N}=4, 2$ superspace
and in component approach. Because the two ${\cal N}=4$ multiplets
are given in terms of ${\cal N}=3$ versions (and ${\cal N}=1$ versions),
it is straightforward to write the complete OPE in the ${\cal N}=3$
and ${\cal N}=1$ superspaces.  

$\bullet$ Jacobi identity in ${\cal N}=4$ superspace

It would be interesting to see the coefficients appearing in 
(\ref{finalPhiPhi})
from the Jacobi identity in ${\cal N}=4$ superspace.
That is, one can analyze the Jacobi identity for the
three quantities ${\bf J^{(4)}}(Z)$, ${\bf \Phi^{(1)}}(Z)$ and 
${\bf \Phi^{(1)}}(Z)$ as done in the extension of ${\cal N}=1$ superconformal 
algebra.

\vspace{.7cm}

\centerline{\bf Acknowledgments}

We would like to thank H. Kim for discussions. 
This work was supported by the Mid-career Researcher Program through
the National Research Foundation of Korea (NRF) grant 
funded by the Korean government (MEST) 
(No. 2012-045385/2013-056327/2014-051185).
CA acknowledges warm hospitality from 
the School of  Liberal Arts (and Institute of Convergence Fundamental
Studies), Seoul National University of Science and Technology.

\newpage

\appendix

\renewcommand{\thesection}{\large \bf \mbox{Appendix~}\Alph{section}}
\renewcommand{\theequation}{\Alph{section}\mbox{.}\arabic{equation}}

\section{The OPEs between $16$ currents  in 
component approach}

As described in section $2$, 
the ${\cal N}=4$ OPE (\ref{j4j4}) 
can be written in terms of the following 
component results \cite{Schoutensnpb,Schoutensplb}
\bea
L(z) \, L(w) & = & 
\frac{1}{(z-w)^{4}}\, \frac{1}{2} \, c+\frac{1}{(z-w)^{2}}\, 2 L(w)+
\frac{1}{(z-w)} \, \pa L(w) + \cdots,
\nonu \\
L(z) \, G^{i}(w) & = & 
\frac{1}{(z-w)^{2}}\, \frac{3}{2}G^{i}(w)+\frac{1}{(z-w)} \, 
\partial G^{i}(w) + \cdots, \nonu \\
L(z) \, T^{ij}(w) & = & \frac{1}{(z-w)^{2}} \, T^{ij}(w)+
\frac{1}{(z-w)} \, \partial T^{ij}(w) + \cdots,
\nonu \\
L(z) \, \Gamma^{i}(w) & = & 
\frac{1}{(z-w)^{2}}\, \frac{1}{2} \, \Gamma^{i}(w)+
\frac{1}{(z-w)} \, \partial\Gamma^{i}(w) +\cdots,\nonu \\
L(z) \, U(w) & = & \frac{1}{(z-w)^{2}} \, U(w)+
\frac{1}{(z-w)} \, \pa U(w) +\cdots,
\nonu \\
G^{i}(z) \, G^{j}(w) & = & \frac{1}{(z-w)^{3}} \, \frac{2}{3} \, c \,
\delta^{ij}-
\frac{1}{(z-w)^{2}} \,  2 \,i \, \Bigg[ T^{ij} +2 \, \alpha
\, \varepsilon^{ijkl} \, T^{kl}\Bigg](w) 
\nonu \\
& + &
\frac{1}{(z-w)} \Bigg[
2 \, \delta^{ij}\, L-i \, \partial T^{ij}-2i \, \alpha\, \varepsilon^{ijkl}\,
\partial T^{kl}
\Bigg](w) + \cdots,
\nonu \\
G^{i}(z) \, T^{jk}(w) & = & 
-\frac{1}{(z-w)^{2}} \Bigg[ \varepsilon^{ijkl} \,
\Gamma^{l} + 2 \alpha (\delta^{ik}\Gamma^{j}-
\delta^{ij}\Gamma^{k})
\Bigg](w) \nonu \\
& - & \frac{1}{(z-w)} \, \Bigg[\varepsilon^{ijkl} \, \partial\Gamma^{l}
+   2 \alpha (\delta^{ik} \pa \Gamma^{j}-
\delta^{ij} \pa \Gamma^{k}) 
+ i \, \delta^{ik} \, G^{j}-i \, \delta^{ij} \, G^{k} \Bigg](w) +\cdots,
\nonu \\
G^{i}(z)\, \Gamma^{j}(w) & = & 
\frac{1}{(z-w)} \Bigg[
-\varepsilon^{ijkl} \, T^{kl}+  i\: \delta^{ij} \, U \Bigg](w) + \cdots,\nonu \\
G^{i}(z) \, U(w) & = & 
-\frac{1}{(z-w)^{2}} \, i\:\Gamma^{i}(w)-
\frac{1}{(z-w)} \, i\:\partial\Gamma^{i}(w) +\cdots,
\nonu \\
T^{ij}(z)\, T^{kl}(w) & = & 
\frac{1}{(z-w)^{2}} \, \frac{1}{2} \Bigg[(k^{+} +k^{-})(\delta^{ik}\delta^{jl}-
\delta^{il}\delta^{jk}) -\varepsilon^{ijkl}\, (k^{+}-k^{-})
\Bigg] \nonu \\
& - & 
\frac{1}{(z-w)} \, i \,
\Bigg[\delta^{ik} \, T^{jl}-\delta^{il}\, T^{jk}-\delta^{jk}\, T^{il}+
\delta^{jl}\, T^{ik} \Bigg](w) +\cdots,
\nonu \\
T^{ij}(z)\, \Gamma^{k}(w) & = & -
\frac{1}{(z-w)}\,
i\: \Bigg[\delta^{ik}\:\Gamma^{j}-\delta^{jk}\:\Gamma^{i} \Bigg](w)
+\cdots,
\nonu \\
\Gamma^{i}(z)\, \Gamma^{j}(w) & = & 
\frac{1}{(z-w)} \, \frac{1}{2}\, (k^{+} + k^{-})
\:\delta^{ij} +\cdots,\nonu \\
U(z) \, U(w) & = & -\frac{1}{(z-w)^{2}} \, \frac{1}{2}\, (k^{+} + k^{-}) 
+ \cdots,
\label{16SCAOPEs}
\eea
where one introduces
$U(z) \equiv - \pa \Delta(z)$ in order not to have $\log (z-w)$ in the 
OPE. The central charge appearing in the OPE 
$L(z) \, L(w)$ is given by (\ref{centralcharge}).
The spin-$\frac{3}{2}$ currents $G^i(w)$,  the spin-$1$ currents
 $T^{ij}(w)=-T^{ji}(w)$, the spin-$\frac{1}{2}$ currents 
$\Gamma^i(w)$ and the spin-$1$ current $U(w)$,
where $i, j=1,2,3,4$,
are primary currents of spins $\frac{3}{2}, 1, \frac{1}{2}, 1$, 
respectively 
under the stress energy tensor $L(z)$ from the second-fifth equations of 
(\ref{16SCAOPEs}). In the OPE $G^i(z) \, G^j(w)$, the parameter $\alpha$
is given by (\ref{alpha}). The totally antisymmetric 
epsilon tensor with $4$ indices 
$\varepsilon^{ijkl}$ appearing in the second order pole 
is $SO(4)$-group invariant.
Of course, the rank $2$ Kronecker delta $\delta^{ij}$ is the symmetric 
$SO(4)$-invariant tensor.
A typo in \cite{Schoutensnpb} from the OPE $T^{ij}(z) \, T^{kl}(w)$ is corrected.
Recall that in (\ref{j4}) the cubic and quartic terms in the fermionic 
coordinates $\theta^i$
contain the extra derivative terms. One can also obtain the various OPEs
by including those derivative terms in the currents $L(z)$ or $G^i(z)$.  
The above OPEs in (\ref{16SCAOPEs}) will be compared to those described 
in different basis in next Appendix $B$.
There exist two trivial OPEs 
$T^{ij}(z) \, U(w) = +\cdots$ and $\Gamma^i(z) \, U(w) = +\cdots$. 
There are no singular terms in these OPEs.



\section{The OPEs between $16$ currents  in 
component approach in different basis}

The large ${\cal N}=4$ linear superconformal algebra
can be described as \cite{STVS,STV}
\bea
 T(z) \, T(w) & = & \frac{1}{(z-w)^4} \, \frac{c}{2} +\frac{1}{(z-w)^2}
\, 2 T(w) + \frac{1}{(z-w)} \, \pa T(w) + \cdots,
\nonu \\
T (z) \, \phi (w) & = & \frac{1}{(z-w)^2}
\, h_{\phi} \, \phi(w) + \frac{1}{(z-w)} \, \pa \phi(w) + \cdots,
\nonu \\
G^{\mu}(z) \, G^{\nu}(w) &=& \frac{1}{(z-w)^3} \,
\frac{2}{3} \delta^{\mu \nu} c
- \frac{1}{(z-w)^2} \, \frac{8}{(k^{+}+k^{-})} ( k^{-} \, \alpha^{+i}_{\mu \nu} \, 
A^{+}_i      
+k^{+} \, \alpha^{-i}_{\mu \nu} \, A^{-}_i  )(w)
\nonu \\
& + & \frac{1}{(z-w)} \, \Bigg[ 2 \delta^{\mu \nu} T
-  \frac{4}{(k^{+}+k^{-})} 
\pa ( k^{-} \, \alpha^{+i}_{\mu \nu} \, A^{+}_i      
+k^{+} \, \alpha^{-i}_{\mu \nu} \, A^{-}_i  ) \Bigg](w) +\cdots,
\nonu \\
A^{\pm i}(z) \, G^{\mu }(w) & = & \mp \frac{1}{(z-w)^2} \,  
\frac{2k^{\pm}}{(k^{+}+k^{-})}  
\,  \alpha_{\mu \nu}^{\pm i } \, 
 \Gamma^{\nu }(w) + \frac{1}{(z-w)} \,  \alpha_{\mu \nu }^{\pm i} \, 
G^{\nu}(w) + \cdots,
\nonu \\
A^{\pm i}(z) \, A^{\pm j}(w) & = & - \frac{1}{(z-w)^2} \, \frac{1}{2} k^{\pm} 
\delta^{ij}  +\frac{1}{(z-w)} \, \ep^{ijk} A^{\pm k }(w) +\cdots,
\nonu \\
\Gamma^{\mu }(z) \, G^{\nu}(w) & = & \frac{1}{(z-w)} \, 
 \Bigg[ 2 ( \alpha_{\mu \nu}^{+i} A_i^{+}
- \alpha_{\mu \nu}^{-i} A_i^{-}) + \delta^{\mu \nu} \, U  \Bigg](w) 
+\cdots,
\nonu \\
A^{\pm i}(z) \, \Gamma^{\mu}(w) & = & \frac{1}{(z-w)} \,  
\alpha_{\mu \nu}^{\pm i} \, 
\Gamma^{\nu}(w)+ \cdots,
\nonu \\
U(z) \, G^{\mu}(w) & = & \frac{1}{(z-w)^2} \, \Gamma^{\mu}(w)
 + \cdots,
\nonu \\
\Gamma^{\mu}(z) \, \Gamma^{\nu}(w) & = & -\frac{1}{(z-w)} \,
 \frac{(k^{+}+k^{-})}{2} \, \delta^{\mu \nu} + \cdots, 
\nonu \\
U(z) \, U(w)  & = & -\frac{1}{(z-w)^2} \, \frac{(k^{+}+k^{-})}{2}
+ \cdots,
\label{N4linearalg}
\eea
where the conformal dimensions $h$ for 
the currents $\Gamma^{\mu}(w)$, $A^{\pm,i}(w)$, $U(w)$
and $G^{\mu}(w)$ ($i =1,2,3 $ and $\mu= 1,2,3,4$) 
are given by $\frac{1}{2}, 1, 1, \frac{3}{2}$ respectively.
The central charge is given by (\ref{centralcharge}).
The $4 \times 4$  matrices $\alpha^{\pm, i}$ is defined by
\bea
\alpha^{\pm i}_{\mu \nu} &=&
\frac{1}{2} 
\left( \pm \delta_{i \mu} \delta_{ \nu 4} \mp \delta_{i \nu} \delta_{ \mu 4}
+\ep_{i \mu \nu} \right),
\nonu
\eea
with $\ep_{123}=1$ and $\ep_{\mu \nu 4}=0$. The spin-$\frac{1}{2}$
currents $\Gamma^{\mu}(z)$ in this Appendix is different from 
those in Appendix $A$ because their OPEs are different from 
each other. The former has a minus sign and the latter has a plus sign.
The currents $T(z)$ and $U(z)$ correspond to 
the currents $L(z)$ and $U(z)$ appearing in Appendix $A$. The nontrivial 
parts in identifying the corresponding currents are given by 
the spin-$1$ currents and the spin-$\frac{3}{2}$ currents. 
They are described in the subsection $2.4$.
It is clear that the levels $k^{\pm}$ are those of two $SU(2)$ algebras
in the OPEs $A^{\pm, i}(z) \, A^{\pm, j}(w)$.

\section{The OPEs between the $16$ currents and the 
$16$ higher spin currents  in 
component approach}

From the ${\cal N}=4$ primary current condition (\ref{jphi}),
one can rewrite it in the component approach as follows:
\bea
L(z) \, \Phi_{2}^{(s)}(w) & = & 
-\frac{1}{(z-w)^{4}} \, 12\:s\:\alpha \, \Phi_{0}^{(s)}(w)+\frac{1}{(z-w)^{3}}
\, 4\:\alpha \, \partial\Phi_{0}^{(s)}(w) \nonu \\
& + & 
\frac{1}{(z-w)^{2}} \, (s+2) \, \Phi_{2}^{(s)}(w)+
\frac{1}{(z-w)} \, \partial\Phi_{2}^{(s)}(w)
+\cdots,
\nonu \\
L(z) \, \Phi_{\frac{3}{2}}^{(s),i}(w) & = & 
\frac{1}{(z-w)^{3}}\, 2\:\alpha \, \Phi_{\frac{1}{2}}^{(s),i}(w)+
\frac{1}{(z-w)^{2}}\, (s+\frac{3}{2})\, \Phi_{\frac{3}{2}}^{(s),i}(w)+
\frac{1}{(z-w)}\, \partial\Phi_{\frac{3}{2}}^{(s),i}(w) +\cdots,
\nonu \\
L(z) \, \Phi_{1}^{(s),ij}(w) & = & 
\frac{1}{(z-w)^{2}}\, (s+1) \, \Phi_{1}^{(s),ij}(w)+
\frac{1}{(z-w)} \, \partial\Phi_{1}^{(s),ij}(w)
+\cdots,
\nonu \\
L(z) \, \Phi_{\frac{1}{2}}^{(s),i}(w) & = &
\frac{1}{(z-w)^{2}}\, (s+\frac{1}{2})\, \Phi_{\frac{1}{2}}^{(s),i}(w)
+\frac{1}{(z-w)} \, \partial\Phi_{\frac{1}{2}}^{(s),i}(w)
+\cdots,
\nonu \\
L(z) \, \Phi_{0}^{(s)}(w) & = & 
\frac{1}{(z-w)^{2}} \, s\, \Phi_{0}^{(s)}(w)+
\frac{1}{(z-w)^{1}} \, \partial\Phi_{0}^{(s)}(w)
+\cdots, \nonu \\
G^{i}(z) \, \Phi_{2}^{(s)}(w) & = & 
-\frac{1}{(z-w)^{3}} \, 4(1+2s)\alpha \, \Phi_{\frac{1}{2}}^{(s),i}(w)-
\frac{1}{(z-w)^{2}} \, \Bigg[(3+2s) \Phi_{\frac{3}{2}}^{(s),i} 
-2 \, \alpha\,\partial\Phi_{\frac{1}{2}}^{(s),i}\Bigg](w)
\nonu \\
& - & \frac{1}{(z-w)}\, \partial\Phi_{\frac{3}{2}}^{(s),i}(w)
+\cdots,
\nonu \\
G^{i}(z) \, \Phi_{\frac{3}{2}}^{(s),j}(w) & = & 
\frac{1}{(z-w)^{3}} \, 8\:s\:\alpha\:\delta_{ij}\, 
\Phi_{0}^{(s)}(w)\nonu \\
& - & \frac{1}{(z-w)^{2}} \, \Bigg[2(1+s) \, \Phi_{1}^{(s),ij}+
2\, \alpha \, \Phi_{1}^{(s),4-[ij]}+
2\, \alpha \, \delta_{ij}\,  \partial\Phi_{0}^{(s)} \Bigg](w)
\nonu \\
& - & 
\frac{1}{(z-w) } \, \Bigg[ \partial\Phi_{1}^{(s),ij}+ 
 \delta_{ij} \, \Phi_{2}^{(s)}\Bigg](w)
+\cdots,
\nonu \\
G^{i}(z) \, \Phi_{1}^{(s),jk}(w) & = & 
-\frac{1}{(z-w)^{2}} \, \Bigg[ 2\:\alpha\: \Big( \delta_{ij} \, 
\Phi_{\frac{1}{2}}^{(s),k}- \delta_{ik} \, 
\Phi_{\frac{1}{2}}^{(s),j}\Big)+(1+2s)\,\Phi_{\frac{1}{2}}^{(s),4-[ijk]}
\Bigg](w)\nonu \\
& - & \frac{1}{(z-w)} \, \Bigg[ 
\Big( \delta_{ij}\, \Phi_{\frac{3}{2}}^{(s),k}- \delta_{ik} \, 
\Phi_{\frac{3}{2}}^{(s),j}\Big)+
\partial\Phi_{\frac{1}{2}}^{(s),4-[ijk]} \Bigg](w)
+\cdots,
\nonu \\
G^{i}(z) \, \Phi_{\frac{1}{2}}^{(s),j}(w) & = & 
-\frac{1}{(z-w)^{2}} \, 2\:s\:\delta_{ij}\, \Phi_{0}^{(s)}(w)-
\frac{1}{(z-w)} \, \Bigg[ \delta_{ij}\, \partial\Phi_{0}^{(s)}-
\Phi_{1}^{(s),4-[ij]} \Bigg](w)
+\cdots,
\nonu \\
G^{i}(z) \, \Phi_{0}^{(s)}(w) & = & 
-\frac{1}{(z-w)} \, \Phi_{\frac{1}{2}}^{(s),i}(w)
+\cdots, \nonu \\
T^{ij}(z) \, \Phi_{2}^{(s)}(w) & = & \frac{1}{(z-w)^{2}} \, 
2\,i\,(s+1)
\, \Phi_{1}^{(s),ij}(w)
+\cdots,
\nonu \\
T^{ij}(z) \, \Phi_{\frac{3}{2}}^{(s),k}(w) & = & 
\frac{1}{(z-w)^{2}} \, i\,(2s+1)\, 
\Phi_{\frac{1}{2}}^{(s),4-[ijk]}(w)
-\frac{1}{(z-w)} \,\Bigg[ i\delta_{ik}\, \Phi_{\frac{3}{2}}^{(s),j}
 -i\delta_{jk}\, \Phi_{\frac{3}{2}}^{(s),i}\Bigg](w)
+\cdots,
\nonu \\
T^{ij}(z) \, \Phi_{1}^{(s),kl}(w) & = & 
\frac{1}{(z-w)^{2}} \,2\:i\:s  \, 
\varepsilon_{ijkl} \, \Phi_{0}^{(s)}(w)
\nonu \\
& - &
\frac{1}{(z-w)} \, \Bigg[  
i\delta_{ik}\, \Phi_{1}^{(s),jl} -  i\delta_{il}\, \Phi_{1}^{(s),jk} - i\delta_{jk}\, \Phi_{1}^{(s),il} +  i\delta_{jl}\, \Phi_{1}^{(s),ik}\Bigg](w)
+\cdots,
\nonu \\
T^{ij}(z)\;\Phi_{\frac{1}{2}}^{(s),k}(w)
& = & 
\frac{1}{(z-w)}
\Bigg[
-i\delta^{ik}\,\Phi_{\frac{1}{2}}^{(s),j}+i\delta^{jk}\,\Phi_{\frac{1}{2}}^{(s),i}
\Bigg](w)
+\cdots,\nonu \\
U(z) \, \Phi_{2}^{(s)}(w) & = & 
-\frac{1}{(z-w)^{3}} \, 4\:s\, \Phi_{0}^{(s)}(w)+
\frac{1}{(z-w)^{2}} \, 2\, \partial\Phi_{0}^{(s)}(w)
+\cdots,
\nonu \\
U(z) \, \Phi_{\frac{3}{2}}^{(s),i}(w) & = & 
\frac{1}{(z-w)^{2}} \, \Phi_{\frac{1}{2}}^{(s),i}(w)
+\cdots,
\nonu \\
\Gamma^{i}(z) \, \Phi_{2}^{(s)}(w) & = & 
-\frac{1}{(z-w)^{2}} \, i\:(2s+1)\, \Phi_{\frac{1}{2}}^{(s),i}(w)
+\frac{i}{(z-w)} \, \partial\Phi_{\frac{1}{2}}^{(s),i}(w) +\cdots,
\nonu \\
\Gamma^{i}(z) \, \Phi_{\frac{3}{2}}^{(s),j}(w) & = & 
\frac{1}{(z-w)^{2}} \, 2\:i\:s\:\delta_{ij}\, \Phi_{0}^{(s)}(w)
-\frac{1}{(z-w)} \, \Bigg[ i \, \delta_{ij}
\, \partial\Phi_{0}^{(s)}+
i \,\Phi_{1}^{(s),4-[ij]} \Bigg](w)
+\cdots,
\nonu \\
\Gamma^{i}(z) \, \Phi_{1}^{(s),jk}(w) & = & 
-\frac{1}{(z-w)} \, \Bigg[ i \, \delta_{ij} \, \Phi_{\frac{1}{2}}^{(s),k}
- i \, \delta_{ik} \, \Phi_{\frac{1}{2}}^{(s),j}\Bigg](w) +\cdots.
\label{jphicomp}
\eea
In the first four equations of (\ref{jphicomp}), 
the higher spin currents 
are primary fields 
$\Phi_1^{(s),ij}(w)$, $\Phi_{\frac{1}{2}}^{(s),i}(w)$ and $\Phi_0^{(s)}(w)$
of spins $(s+1), (s+\frac{1}{2})$ and $s$, respectively
under the stress energy tensor $L(z)$. 
On the other hand, the currents 
$\Phi_2^{(s)}(w)$ and $\Phi_{\frac{3}{2}}^{(s),i}(w)$
are not primary fields because the right hand side of the OPEs 
contain the singular terms of order greater than $2$ with nonzero $\alpha$. 
One can easily check that by subtracting the 
expression $-\frac{2\alpha}{(2s+1)} \pa 
\Phi_{\frac{1}{2}}^{(s),i}(w)$ from  the expression 
$\Phi_{\frac{3}{2}}^{(s),i}(w)$, 
one can construct the primary current 
 $(\Phi_{\frac{3}{2}}^{(s),i}-\frac{2\alpha}{(2s+1)} \pa 
\Phi_{\frac{1}{2}}^{(s),i})(w)$, which is denoted as 
 $\widetilde{\Phi}_{\frac{3}{2}}^{(s),i}(w)$ in (\ref{phidiff}) because
the previous third-order pole disappears.
Similarly, one can easily see that 
the higher spin current $\widetilde{\Phi}_{2}^{(s)}(w)$ in (\ref{phidiff})
is a primary current under the $L(z)$ with the help of 
Appendices $A$ and $C$.

How one can obtain the higher spin-$(s+\frac{1}{2})$, $(s+1)$,
$(s+\frac{3}{2})$ and $(s+2)$ currents starting from the lowest higher
spin-$s$ current $\Phi_0^{(s)}(w)$?
By analyzing the OPEs between the spin-$\frac{3}{2}$ currents 
$G^i(z)$ and these higher spin currents as done in the subsection 
$3.5$, one can generate them systematically and completely.
The OPE $G^i(z) \, \Phi_0^{(s)}(w)$ determines 
the higher spin-$(s+\frac{1}{2})$ current $\Phi_{\frac{1}{2}}^{(s),i}(w)$. 
Now the OPE $G^i(z) \, \Phi_{\frac{1}{2}}^{(s),j}(w)$ where $j \neq i$
provides the higher spin-$(s+1)$ currents $\Phi_1^{4-ij}(w)$
from the first-order pole. Then the OPE 
 $G^i(z) \, \Phi_{1}^{(s),jk}(w)$, where $i=j$, can give us 
the higher spin-$(s+\frac{3}{2})$ current $\Phi_{\frac{3}{2}}^{(s),k}(w)$.
Finally,  the higher spin-$(s+2)$ current 
$\Phi_{2}^{(s)}(w)$ can be obtained 
from the OPE  $G^i(z) \, \Phi_{\frac{3}{2}}^{(s),j}(w)$
where $i=j$.
All the OPEs in (\ref{jphicomp}) will be compared to those in next 
Appendix $D$.

\section{The OPEs between the $16$ currents and the 
$16$ higher spin currents  in 
component approach with different basis}

Let us present the description of \cite{BCG} as follows: 
\bea
G^a(z) \, V_0^{(s)}(w)  & = & \frac{1}{(z-w)} \, V_{\frac{1}{2}}^{(s), a}(w) 
+\cdots,  
\nonu \\
A^{\pm, i}(z) \, V_{\frac{1}{2}}^{(s), a}(w)  & = & \frac{1}{(z-w)} \,
\alpha_{ab}^{\pm, i} \, V_{\frac{1}{2}}^{(s), b}(w)+\cdots,
\nonu \\
G^a(z) \, V_{\frac{1}{2}}^{(s), b}(w)  & = & \frac{1}{(z-w)^2} \, 2 s 
\, \delta^{ab} \, V_0^{(s)}(w) \nonu \\
& + &
\frac{1}{(z-w)} \, \Bigg[ \alpha_{ab}^{+, i} \, V_{1}^{(s),+, i} +
\alpha_{ab}^{-, i} \, V_{1}^{(s),-, i} +\delta^{ab} \,
\pa V_{0}^{(s)} \Bigg](w) 
+\cdots,  
\nonu \\
Q^a(z) \,  V_{1}^{(s),\pm, i}(w)  & = & \pm \frac{1}{(z-w)} \,
 2 \, \alpha_{ab}^{\pm, i} \, V_{\frac{1}{2}}^{(s),b}(w) +\cdots,
\nonu \\
A^{\pm, i}(z) \, V_{1}^{(s), \pm, j}(w)  & = & \frac{1}{(z-w)^2} \,
2s \, \delta^{ij} \, V_0^{(s)}(w) + \frac{1}{(z-w)} \,
\ep^{ijk} \, V_{1}^{(s), \pm, k}(w)+\cdots,
\nonu \\
G^a(z) \, V_1^{(s), \pm, i}(w)  & = & \frac{1}{(z-w)^2} \, 
4 \, (s + \gamma_{\mp})  \, \alpha_{ab}^{\pm, i} \, V_{\frac{1}{2}}^{(s), b}(w)  
\nonu \\
& + &
\frac{1}{(z-w)} \, \Bigg[ \frac{1}{(2s+1)} \pa (\mbox{pole-2}) \mp
\alpha_{ab}^{\pm, i} \, V_{\frac{3}{2}}^{(s), b} \Bigg](w) 
+\cdots,  
\nonu \\
U (z) \, V_{\frac{3}{2}}^{(s), a}(w)  & = & -\frac{1}{(z-w)^2} \, 
2 \, V_{\frac{1}{2}}^{(s), a}(w) +\cdots,
\nonu \\
Q^a(z) \,  V_{\frac{3}{2}}^{(s),b}(w)  & = & 
\frac{1}{(z-w)^2} \,
4 \, s \, \delta^{ab} \, V_0^{(s)}(w)
\nonu \\
& + &   \frac{1}{(z-w)} \,
2 \Bigg[ \alpha_{ab}^{+, i} \, V_{1}^{(s),+, i} 
+ \alpha_{ab}^{-, i} \, V_{1}^{(s),-, i} -\delta^{ab} \, 
\pa V_{0}^{(s)} \Bigg](w)  +\cdots,
\nonu \\
A^{\pm, i}(z) \, V_{\frac{3}{2}}^{(s), a}(w)  & = & \pm \frac{1}{(z-w)^2} \,
\left[ \frac{8s(s+1) + 4 \gamma_{\mp}}{(2s+1)} \right]  \,
\alpha_{ab}^{\pm, i} \,
V_{\frac{1}{2}}^{(s), b}(w) \nonu \\
& + &  \frac{1}{(z-w)} \,
\alpha_{ab}^{\pm, i} \, 
V_{\frac{3}{2}}^{(s), b}(w)+\cdots,
\nonu \\
G^a(z) \, V_{\frac{3}{2}}^{(s), b}(w)  & = & 
- \frac{1}{(z-w)^3} \, \Bigg[ \frac{16s(s+1)(2\gamma-1)}
{(2s+1)}\Bigg] \delta^{ab} \, V_0^{(s)}(w)  
\nonu \\
& - &  \frac{1}{(z-w)^2} \, \frac{8(s+1)}{(2s+1)} \,
\Bigg[ (s + \gamma_{+}) \,  \alpha_{ab}^{+, i} \, V_{1}^{(s), +, i}  
- (s + \gamma_{-})  \, \alpha_{ab}^{-, i} \, V_{1}^{(s), -, i}  
\Bigg](w)
\nonu \\
& + &
\frac{1}{(z-w)} \, \Bigg[ \frac{1}{2(s+1)}\, \pa (\mbox{pole-2}) +
\delta^{ab} \, V_{2}^{(s)} \Bigg](w) 
+\cdots,  
\nonu \\
U (z) \, V_{2}^{(s)}(w)  & = & \frac{1}{(z-w)^3} \,
8 \, s \, V_{0}^{(s)}(w) -\frac{1}{(z-w)^2} \, 
4 \, \pa V_{0}^{(s)}(w) +\cdots,
\nonu \\
Q^a(z) \,  V_{2}^{(s)}(w)  & = & 
- \frac{1}{(z-w)^2} \,
2  \, (2s+1)  \, V_{\frac{1}{2}}^{(s), a}(w)
+    \frac{1}{(z-w)} \,
2 \,
\pa V_{\frac{1}{2}}^{(s), a}(w)  +\cdots,
\nonu \\
A^{\pm, i}(z) \, V_{2}^{(s)}(w)  & = & \pm \frac{1}{(z-w)^2} \,
2(s+1) \, V_{1}^{(s), \pm, i}(w) +\cdots,
\nonu \\
G^a(z) \, V_{2}^{(s)}(w)  & = & 
\frac{1}{(z-w)^3} \, \left[ \frac{16s(s+1)(2\gamma-1)}
{(2s+1)}\right]  \, V_{\frac{1}{2}}^{(s), a}(w)  
\nonu \\
& + &  \frac{1}{(z-w)^2} \, (2s+3) \,
V_{\frac{3}{2}}^{(s), a}(w)
+ 
\frac{1}{(z-w)} \, \pa  V_{\frac{3}{2}}^{(s), a}(w) 
+\cdots,  
\nonu \\
T(z) \, V_{2}^{(s)}(w)  & = & 
-\frac{1}{(z-w)^4} \, \left[ \frac{24s(s+1)(2\gamma-1)}
{(2s+1)}\right]  \, V_{0}^{(s)}(w)  
\nonu \\
& + &  \frac{1}{(z-w)^2} \, (s+2) \,
V_{2}^{(s)}(w)
+ 
\frac{1}{(z-w)} \, \pa  V_{2}^{(s)}(w) 
+\cdots,  
\nonu \\
T(z) \, V_{0}^{(s)}(w)  & = & 
  \frac{1}{(z-w)^2} \, s \,
V_{0}^{(s)}(w)
+ 
\frac{1}{(z-w)} \, \pa  V_{0}^{(s)}(w) 
+\cdots,  
\nonu \\
T(z) \, V_{\frac{1}{2}}^{(s), a}(w)  & = & 
  \frac{1}{(z-w)^2} \, (s+\frac{1}{2}) \,
V_{\frac{1}{2}}^{(s), a}(w)
+ 
\frac{1}{(z-w)} \, \pa  V_{\frac{1}{2}}^{(s), a}(w) 
+\cdots,  
\nonu \\
T(z) \, V_{1}^{(s), \pm, i}(w)  & = & 
  \frac{1}{(z-w)^2} \, (s+1) \,
V_{1}^{(s), \pm, i}(w)
+ 
\frac{1}{(z-w)} \, \pa  V_{1}^{(s), \pm, i}(w) 
+\cdots,  
\nonu \\
T(z) \, V_{\frac{3}{2}}^{(s), a}(w)  & = & 
  \frac{1}{(z-w)^2} \, (s+\frac{3}{2}) \,
V_{\frac{3}{2}}^{(s), a}(w)
+ 
\frac{1}{(z-w)} \, \pa  V_{\frac{3}{2}}^{(s), a}(w) 
+\cdots.  
\label{bcgprimary}
\eea
Here the two parameters are introduced as follows:
$\gamma_{+} =\gamma= \frac{k^{-}}{(k^{+}+k^{-})}$
and $\gamma_{-} = 1-\gamma = \frac{k^{+}}{(k^{+}+k^{-})}$.
From the OPEs in (\ref{bcgprimary}), the higher spin-$s, (s+\frac{1}{2}),
(s+1), (s+\frac{3}{2})$ currents are primary fields under the stress 
energy tensor $T(z)$. 
Note that the higher spin-$(s+2)$ current $V_2^{(s)}(w)$ is not a primary
current because there is a fourth-order pole term.
One can consider the extra composite field 
$\Phi_0^{(s)} T(w)$ and make the above 
higher spin-$(s+2)$ current transforming as a primary field
as in (\ref{bcgspin3}).
As described in previous Appendix $C$, one can construct these
higher spin currents starting from the first equation of (\ref{bcgprimary})
completely. 
As explained in the subsection $3.5$ before, the linear terms in 
the fermionic coordinate $\theta^i$
and the $\theta^i$-independent term of (\ref{bcghigher}) 
can be identified without any difficulty.
The quadratic, cubic and quartic terms in the fermionic 
coordinate $\theta^i$ of (\ref{bcghigher})
can be obtained by looking at Appendices $C$ and $D$. 

\section{The four currents  in 
${\cal N}=2$ superspace and its components}

It is known that, in \cite{RASS,Ahn1992,BO},  $16$ components of the four 
${\cal N}=2$ multiplets  have  the following  
explicit forms
\bea
{\bf T}|_{\theta=\bar{\theta}=0} &= & -(T^{12} +2 \alpha \,T^{34})(z),
\nonu \\
D {\bf T}|_{\theta=\bar{\theta}=0} &= & \frac{1}{2} (G^1+ i G^2)(z),
\nonu \\
\overline{D} {\bf T}|_{\theta=\bar{\theta}=0} &= & \frac{1}{2} (G^1- i G^2)(z),
\nonu \\
-\frac{1}{2} [ D, \overline{D}] {\bf T}|_{\theta=\bar{\theta}=0} &= & L(z),
\nonu \\
{\bf G}|_{\theta=\bar{\theta}=0} &= & (i \Gamma^3 -\Gamma^4)(z), \nonu \\
 {\bf \overline{G}}|_{\theta=\bar{\theta}=0} &= & (i \Gamma^3 +\Gamma^4)(z), 
\nonu \\
D {\bf G}|_{\theta=\bar{\theta}=0} &= & \frac{1}{2} (i T^{13} -T^{14}-T^{23}-i T^{24})(z), 
\nonu \\
\overline{D} {\bf G}|_{\theta=\bar{\theta}=0} &= & \frac{1}{2} (i T^{13} -T^{14}+
T^{23}+i T^{24})(z), \nonu \\
D {\bf \overline{G}}|_{\theta=\bar{\theta}=0} &= & \frac{1}{2} (-i T^{13} -T^{14}+T^{23}-
i T^{24})(z), 
\nonu \\
\overline{D} {\bf \overline{G}}|_{\theta=\bar{\theta}=0} &= & \frac{1}{2} (-i T^{13} -T^{14}
-T^{23}+i T^{24})(z), 
\nonu \\
-\frac{1}{2} [ D, \overline{D}] {\bf G}|_{\theta=\bar{\theta}=0} &= & -\frac{1}{2}
(G^3+ i G^4- 2 i 
\, \alpha \, \pa \Gamma^3+  2 
\, \alpha \, \pa \Gamma^4)(z),
\nonu \\
-\frac{1}{2} [ D, \overline{D}] {\bf \overline{G}}|_{\theta=\bar{\theta}=0} &= & 
-\frac{1}{2}(-G^3+ i G^4+ 2 i
\, \alpha \, \pa \Gamma^3  + 2 
\, \alpha \, \pa \Gamma^4)(z),
\nonu \\
 {\bf H}|_{\theta=\bar{\theta}=0} & = & (i \Gamma^1 -\Gamma^2)(z),
\nonu \\
 {\bf \overline{H}}|_{\theta=\bar{\theta}=0} & = & (-i \Gamma^1 -\Gamma^2)(z),
\nonu \\
\overline{D} {\bf H}|_{\theta=\bar{\theta}=0} & = & (- U + T^{34})(z),
\nonu \\
D {\bf \overline{H}}|_{\theta=\bar{\theta}=0} & = & (- U - T^{34})(z),
\label{16compn2} 
\eea
where the following components are not independent
and can be written in terms of the derivative terms
of the $13$-th and the $14$-th  of  (\ref{16compn2}) 
\bea
-\frac{1}{2} [ D, \overline{D}] {\bf H}|_{\theta=\bar{\theta}=0} &= & - 
(i \pa \Gamma^1 -\pa \Gamma^2)(z),
\nonu \\
-\frac{1}{2} [ D, \overline{D}] {\bf \overline{H}}|_{\theta=\bar{\theta}=0} &= & 
(i \pa \Gamma^1 + \pa \Gamma^2)(z).
\nonu 
\eea
Recall that one has the relations 
$D \, {\bf H} =0 = \overline{D} {\bf \overline{H}}$. 
One can check that by substituting (\ref{16compn2}) into (\ref{n4n2}),
the right hand side of (\ref{n4n2}) gives rise to 
(\ref{j4}). 

\section{The OPEs between$16$ higher spin  currents  in 
component approach and its ${\cal N}=4$ version}

\subsection{The OPEs between$16$ higher spin  currents}

Let us write down the OPEs between $16$ higher spin  currents 
in the component approach by specifying 
the bosonic and fermionic coordinates at two different points 
$Z_1$ and $Z_2$
and expanding the $25(=5\times 5)$ terms  
as follows:
\bea
& & \Bigg[
\Phi_0^{(1)}(z_1) + \cdots +
\theta_1^{1}\, \theta_1^2\, \theta_1^3 \, \theta_1^4 \, \Phi_2^{(1)}(z_1)
\Bigg] 
\Bigg[
\Phi_0^{(1)}(z_2) + 
\cdots +
\theta_2^{1}\, \theta_2^2\, \theta_2^3 \, \theta_2^4 \, \Phi_2^{(1)}(z_2)
\Bigg] \nonu \\ 
 & &  =  
 \sum_{n=2}^1 \, 
\frac{1}{(z_1-z_2)^n}  \, \{  \Phi_0^{(1)} \, \Phi_0^{(1)} \}_{-n}
+ \theta^i_2 \, \sum_{n=2}^1 \, 
\frac{1}{(z_1-z_2)^n}  \, \{  \Phi_0^{(1)} \, \Phi_{\frac{1}{2}}^{(1),i} \}_{-n}
\nonu \\
&& +   \theta^{4-ij}_2 \, \sum_{n=3}^1 \, 
\frac{1}{(z_1-z_2)^n}  \, \{  \Phi_0^{(1)} \, \Phi_{1}^{(1),ij} \}_{-n}
+  \theta^{4-i}_2 \, \sum_{n=3}^1 \, 
\frac{1}{(z_1-z_2)^n}  \, \{  \Phi_0^{(1)} \, \Phi_{\frac{3}{2}}^{(1),i} \}_{-n}
\nonu \\
&& +  \theta^{4-0}_2 \, \sum_{n=4}^1 \, 
\frac{1}{(z_1-z_2)^n}  \, \{  \Phi_0^{(1)} \, \Phi_{2}^{(1)} \}_{-n}
+ \theta_1^i \, \sum_{n=2}^1 \, 
\frac{1}{(z_1-z_2)^n}  \, \{  \Phi_{\frac{1}{2}}^{(1),i} \, \Phi_0^{(1)} \}_{-n}
\nonu \\
&& - \theta_1^i \, \theta^j_2 \, \sum_{n=3}^1 \, 
\frac{1}{(z_1-z_2)^n}  \, \{  \Phi_{\frac{1}{2}}^{(1),i} \, 
\Phi_{\frac{1}{2}}^{(1),j} \}_{-n}
 +  \theta_1^i \, \theta^{4-jk}_2 \, \sum_{n=3}^1 \, 
\frac{1}{(z_1-z_2)^n}  \, \{  \Phi_{\frac{1}{2}}^{(1),i} \, \Phi_{1}^{(1),jk} \}_{-n}
\nonu \\
&& -  \theta^i_1 \, \theta^{4-j}_2 \, \sum_{n=4}^1 \, 
\frac{1}{(z_1-z_2)^n}  \, \{  \Phi_{\frac{1}{2}}^{(1),i} \, 
\Phi_{\frac{3}{2}}^{(1),j} \}_{-n}
+  \theta^i_1 \, \theta^{4-0}_2 \, \sum_{n=4}^1 \, 
\frac{1}{(z_1-z_2)^n}  \, \{  \Phi_{\frac{1}{2}}^{(1),i} \, \Phi_{2}^{(1)} \}_{-n} 
\nonu \\
&& + \theta_1^{4-ij} \, \sum_{n=3}^1 \, 
\frac{1}{(z_1-z_2)^n}  \, \{  \Phi_{1}^{(1),ij} \, \Phi_0^{(1)} \}_{-n}
 + \theta_1^{4-ij} \, \theta^k_2 \, \sum_{n=3}^1 \, 
\frac{1}{(z_1-z_2)^n}  \, \{  \Phi_{1}^{(1),ij} \, 
\Phi_{\frac{1}{2}}^{(1),k} \}_{-n}
\nonu \\
&& +  \theta_1^{4-ij} \, \theta^{4-kl}_2 \, \sum_{n=4}^1 \, 
\frac{1}{(z_1-z_2)^n}  \, \{  \Phi_{1}^{(1),ij} \, \Phi_{1}^{(1),kl} \}_{-n}
 +  \theta^{4-ij}_1 \, \theta^{4-k}_2 \, \sum_{n=4}^1 \, 
\frac{1}{(z_1-z_2)^n}  \, \{  \Phi_{1}^{(1),ij} \, 
\Phi_{\frac{3}{2}}^{(1),k} \}_{-n}
\nonu \\
&& +  \theta^{4-ij}_1 \, \theta^{4-0}_2 \, \sum_{n=5}^1 \, 
\frac{1}{(z_1-z_2)^n}  \, \{  \Phi_{1}^{(1),ij} \, \Phi_{2}^{(1)} \}_{-n} 
+ \theta_1^{4-i} \, \sum_{n=3}^1 \, 
\frac{1}{(z_1-z_2)^n}  \, \{  \Phi_{\frac{3}{2}}^{(1),i} \, \Phi_0^{(1)} \}_{-n}
\nonu \\
&& - \theta_1^{4-i} \, \theta^j_2 \, \sum_{n=3}^1 \, 
\frac{1}{(z_1-z_2)^n}  \, \{  \Phi_{\frac{3}{2}}^{(1),i} \, 
\Phi_{\frac{1}{2}}^{(1),j} \}_{-n}
+  \theta_1^{4-i} \, \theta^{4-jk}_2 \, \sum_{n=4}^1 \, 
\frac{1}{(z_1-z_2)^n}  \, \{  \Phi_{\frac{3}{2}}^{(1),i} \, \Phi_{1}^{(1),jk} \}_{-n}
\nonu \\
&& +  \theta^{4-i}_1 \, \theta^{4-j}_2 \, \sum_{n=5}^1 \, 
\frac{1}{(z_1-z_2)^n}  \, \{  \Phi_{\frac{3}{2}}^{(1),i} \, 
\Phi_{\frac{3}{2}}^{(1),j} \}_{-n}
+  \theta^{4-i}_1 \, \theta^{4-0}_2 \, \sum_{n=5}^1 \, 
\frac{1}{(z_1-z_2)^n}  \, \{  \Phi_{\frac{3}{2}}^{(1),i} \, \Phi_{2}^{(1)} \}_{-n} 
\nonu \\
&& + \theta_1^{4-0} \, \sum_{n=3}^1 \, 
\frac{1}{(z_1-z_2)^n}  \, \{  \Phi_{2}^{(1)} \, \Phi_0^{(1)} \}_{-n}
+ \theta_1^{4-0} \, \theta^i_2 \, \sum_{n=3}^1 \, 
\frac{1}{(z_1-z_2)^n}  \, \{  \Phi_{2}^{(1)} \, 
\Phi_{\frac{1}{2}}^{(1),i} \}_{-n}
\nonu \\
&& +  \theta_1^{4-0} \, \theta^{4-ij}_2 \, \sum_{n=4}^1 \, 
\frac{1}{(z_1-z_2)^n}  \, \{  \Phi_{2}^{(1)} \, \Phi_{1}^{(1),ij} \}_{-n}
+  \theta^{4-0}_1 \, \theta^{4-i}_2 \, \sum_{n=5}^1 \, 
\frac{1}{(z_1-z_2)^n}  \, \{  \Phi_{2}^{(1)} \, 
\Phi_{\frac{3}{2}}^{(1),i} \}_{-n}
\nonu \\
&& +  \theta^{4-0}_1 \, \theta^{4-0}_2 \, \sum_{n=5}^1 \, 
\frac{1}{(z_1-z_2)^n}  \, \{  \Phi_{2}^{(1)} \, \Phi_{2}^{(1)} \}_{-n}. 
\label{PhiPhiexpression} 
\eea
Now let us reexpress the singular terms $\frac{1}{(z_1-z_2)^n}$
using the identity (\ref{polerelation})
in order to express them with the coordinate $\frac{1}{z_{12}^n}$
with the definition in (\ref{polerelation}).
Then the above (\ref{PhiPhiexpression})
can be rewritten as 
\bea
 & &    
 \sum_{n=2}^1 \,
\left[ 
 \frac{1}{z_{12}^n}
-n \, \frac{\theta_1^p \theta_2^p}{z_{12}^{n+1}}
+ \frac{1}{2 !} n (n+1)\, 
\frac{\theta_1^p \theta_2^p \theta_1^q \theta_2^q }{z_{12}^{n+2}}
-\frac{1}{3!} 
n (n+1)(n+2)\, \frac{\theta_1^p \theta_2^p \theta_1^q \theta_2^q 
\theta_1^r \theta_2^r  }{z_{12}^{n+3}}
\right.
\nonu \\
& & +  \left. \frac{1}{4!} 
n (n+1)(n+2) (n+3)\, \frac{\theta_1^p \theta_2^p \theta_1^q \theta_2^q 
\theta_1^r \theta_2^r  \theta_1^s \theta_2^s }{z_{12}^{n+4}} 
\right]
\, \{  \Phi_0^{(1)} \, \Phi_0^{(1)} \}_{-n}
+ \theta^i_2 \, \sum_{n=2}^1 \, 
\left[ 
 \frac{1}{z_{12}^n}
\right. \nonu \\
&& \left. -n \, \frac{\theta_1^p \theta_2^p}{z_{12}^{n+1}}
+ \frac{1}{2 !} n (n+1)\, 
\frac{\theta_1^p \theta_2^p \theta_1^q \theta_2^q }{z_{12}^{n+2}}
-\frac{1}{3!} 
n (n+1)(n+2)\, \frac{\theta_1^p \theta_2^p \theta_1^q \theta_2^q 
\theta_1^r \theta_2^r  }{z_{12}^{n+3}}
\right]
\, \{  \Phi_0^{(1)} \, \Phi_{\frac{1}{2}}^{(1),i} \}_{-n}
\nonu \\
&& +   \theta^{4-ij}_2 \, \sum_{n=3}^1 \, 
\left[ 
 \frac{1}{z_{12}^n}
 -n \, \frac{\theta_1^p \theta_2^p}{z_{12}^{n+1}}
+ \frac{1}{2 !} n (n+1)\, 
\frac{\theta_1^p \theta_2^p \theta_1^q \theta_2^q }{z_{12}^{n+2}}
\right]
\, \{  \Phi_0^{(1)} \, \Phi_{1}^{(1),ij} \}_{-n}
\nonu \\
&& +  \theta^{4-i}_2 \, \sum_{n=3}^1 \, 
\left[ 
 \frac{1}{z_{12}^n}
 -n \, \frac{\theta_1^p \theta_2^p}{z_{12}^{n+1}} \right]
\, \{  \Phi_0^{(1)} \, \Phi_{\frac{3}{2}}^{(1),i} \}_{-n}
+  \theta^{4-0}_2 \, \sum_{n=4}^1 \, 
\frac{1}{z_{12}^n}  \, \{  \Phi_0^{(1)} \, \Phi_{2}^{(1)} \}_{-n}
\nonu \\
&& + \theta_1^i \, \sum_{n=2}^1 \, 
\left[ 
 \frac{1}{z_{12}^n}
 -n \, \frac{\theta_1^p \theta_2^p}{z_{12}^{n+1}}
+ \frac{1}{2 !} n (n+1)\, 
\frac{\theta_1^p \theta_2^p \theta_1^q \theta_2^q }{z_{12}^{n+2}}
-\frac{1}{3!} 
n (n+1)(n+2)\, \frac{\theta_1^p \theta_2^p \theta_1^q \theta_2^q 
\theta_1^r \theta_2^r  }{z_{12}^{n+3}}
\right]
\, \{  \Phi_{\frac{1}{2}}^{(1),i} \, \Phi_0^{(1)} \}_{-n}
\nonu \\
&& - \theta_1^i \, \theta^j_2 \, \sum_{n=3}^1 \, 
\left[ 
 \frac{1}{z_{12}^n}
 -n \, \frac{\theta_1^p \theta_2^p}{z_{12}^{n+1}}
+ \frac{1}{2 !} n (n+1)\, 
\frac{\theta_1^p \theta_2^p \theta_1^q \theta_2^q }{z_{12}^{n+2}}
-\frac{1}{3!} 
n (n+1)(n+2)\, \frac{\theta_1^p \theta_2^p \theta_1^q \theta_2^q 
\theta_1^r \theta_2^r  }{z_{12}^{n+3}}
\right]
\nonu \\
&& \times \, \{  \Phi_{\frac{1}{2}}^{(1),i} \, 
\Phi_{\frac{1}{2}}^{(1),j} \}_{-n}
 +  \theta_1^i \, \theta^{4-jk}_2 \, \sum_{n=3}^1 \, 
\left[ 
 \frac{1}{z_{12}^n}
 -n \, \frac{\theta_1^p \theta_2^p}{z_{12}^{n+1}}
+ \frac{1}{2 !} n (n+1)\, 
\frac{\theta_1^p \theta_2^p \theta_1^q \theta_2^q }{z_{12}^{n+2}}
\right]
\, \{  \Phi_{\frac{1}{2}}^{(1),i} \, \Phi_{1}^{(1),jk} \}_{-n}
\nonu \\
&& -  \theta^i_1 \, \theta^{4-j}_2 \, \sum_{n=4}^1 \, 
\left[ 
 \frac{1}{z_{12}^n}
 -n \, \frac{\theta_1^p \theta_2^p}{z_{12}^{n+1}}
\right]
\, \{  \Phi_{\frac{1}{2}}^{(1),i} \, 
\Phi_{\frac{3}{2}}^{(1),j} \}_{-n}
+  \theta^i_1 \, \theta^{4-0}_2 \, \sum_{n=4}^1 \, 
\frac{1}{z_{12}^n}  \, \{  \Phi_{\frac{1}{2}}^{(1),i} \, \Phi_{2}^{(1)} \}_{-n} 
\nonu \\
&& + \theta_1^{4-ij} \, \sum_{n=3}^1 \, 
\left[ 
 \frac{1}{z_{12}^n}
 -n \, \frac{\theta_1^p \theta_2^p}{z_{12}^{n+1}}
+ \frac{1}{2 !} n (n+1)\, 
\frac{\theta_1^p \theta_2^p \theta_1^q \theta_2^q }{z_{12}^{n+2}}
\right]
\, \{  \Phi_{1}^{(1),ij} \, \Phi_0^{(1)} \}_{-n}
\nonu \\ 
&& + \theta_1^{4-ij} \, \theta^k_2 \, \sum_{n=3}^1 \, 
\left[ 
 \frac{1}{z_{12}^n}
 -n \, \frac{\theta_1^p \theta_2^p}{z_{12}^{n+1}}
+ \frac{1}{2 !} n (n+1)\, 
\frac{\theta_1^p \theta_2^p \theta_1^q \theta_2^q }{z_{12}^{n+2}}
\right]
\, \{  \Phi_{1}^{(1),ij} \, 
\Phi_{\frac{1}{2}}^{(1),k} \}_{-n}
\nonu \\
&& +  \theta_1^{4-ij} \, \theta^{4-kl}_2 \, \sum_{n=4}^1 \, 
\left[ 
 \frac{1}{z_{12}^n}
 -n \, \frac{\theta_1^p \theta_2^p}{z_{12}^{n+1}}
+ \frac{1}{2 !} n (n+1)\, 
\frac{\theta_1^p \theta_2^p \theta_1^q \theta_2^q }{z_{12}^{n+2}}
\right]
\, \{  \Phi_{1}^{(1),ij} \, \Phi_{1}^{(1),kl} \}_{-n}
 \nonu \\
&& +  \theta^{4-ij}_1 \, \theta^{4-k}_2 \, \sum_{n=4}^1 \, 
\left[ 
 \frac{1}{z_{12}^n}
 -n \, \frac{\theta_1^p \theta_2^p}{z_{12}^{n+1}} \right]
\, \{  \Phi_{1}^{(1),ij} \, 
\Phi_{\frac{3}{2}}^{(1),k} \}_{-n}
 +  \theta^{4-ij}_1 \, \theta^{4-0}_2 \, \sum_{n=5}^1 \, 
\frac{1}{z_{12}^n}  \, \{  \Phi_{1}^{(1),ij} \, \Phi_{2}^{(1)} \}_{-n} 
\nonu \\
&& + \theta_1^{4-i} \, \sum_{n=3}^1 \, 
\left[ 
 \frac{1}{z_{12}^n}
 -n \, \frac{\theta_1^p \theta_2^p}{z_{12}^{n+1}} \right]
\, \{  \Phi_{\frac{3}{2}}^{(1),i} \, \Phi_0^{(1)} \}_{-n}
- \theta_1^{4-i} \, \theta^j_2 \, \sum_{n=4}^1 \, 
\left[ 
 \frac{1}{z_{12}^n}
 -n \, \frac{\theta_1^p \theta_2^p}{z_{12}^{n+1}} \right]
\, \{  \Phi_{\frac{3}{2}}^{(1),i} \, 
\Phi_{\frac{1}{2}}^{(1),j} \}_{-n}
\nonu \\
&& +  \theta_1^{4-i} \, \theta^{4-jk}_2 \, \sum_{n=4}^1 \, 
\left[ 
 \frac{1}{z_{12}^n}
 -n \, \frac{\theta_1^p \theta_2^p}{z_{12}^{n+1}} \right]
\, \{  \Phi_{\frac{3}{2}}^{(1),i} \, \Phi_{1}^{(1),jk} \}_{-n}
\nonu \\
&& +  \theta^{4-i}_1 \, \theta^{4-j}_2 \, \sum_{n=5}^1 \, 
\left[ 
 \frac{1}{z_{12}^n}
 -n \, \frac{\theta_1^p \theta_2^p}{z_{12}^{n+1}} \right]
\, \{  \Phi_{\frac{3}{2}}^{(1),i} \, 
\Phi_{\frac{3}{2}}^{(1),j} \}_{-n}
+  \theta^{4-i}_1 \, \theta^{4-0}_2 \, \sum_{n=5}^1 \, 
\frac{1}{z_{12}^n}  \, \{  \Phi_{\frac{3}{2}}^{(1),i} \, \Phi_{2}^{(1)} \}_{-n} 
\nonu \\
&& + \theta_1^{4-0} \, \sum_{n=4}^1 \, 
\frac{1}{z_{12}^n}  \, \{  \Phi_{2}^{(1)} \, \Phi_0^{(1)} \}_{-n}
+ \theta_1^{4-0} \, \theta^i_2 \, \sum_{n=4}^1 \, 
\frac{1}{z_{12}^n}  \, \{  \Phi_{2}^{(1)} \, 
\Phi_{\frac{1}{2}}^{(1),i} \}_{-n}
\nonu \\
&& +  \theta_1^{4-0} \, \theta^{4-ij}_2 \, \sum_{n=5}^1 \, 
\frac{1}{z_{12}^n}  \, \{  \Phi_{2}^{(1)} \, \Phi_{1}^{(1),ij} \}_{-n}
+  \theta^{4-0}_1 \, \theta^{4-i}_2 \, \sum_{n=5}^1 \, 
\frac{1}{z_{12}^n}  \, \{  \Phi_{2}^{(1)} \, 
\Phi_{\frac{3}{2}}^{(1),i} \}_{-n}
\nonu \\
&& +  \theta^{4-0}_1 \, \theta^{4-0}_2 \, \sum_{n=6}^1 \, 
\frac{1}{z_{12}^n}  \, \{  \Phi_{2}^{(1)} \, \Phi_{2}^{(1)} \}_{-n},
\label{PhiPhiother}
\eea
where 
one ignores the terms, which do not contribute to
due to the property of $\theta^i$. 
The remaining things one should do are
that for given each fractional expression, the denominator is already 
written in terms of the singular term 
$\frac{1}{z_{12}^n}$ and the numerator is 
the combination of the product of the fermionic coordinate 
$\theta^i$ and the specific singular term
with given order.

It is useful to see the above OPE when $\theta_2^i=0$.
It is easy to see from (\ref{PhiPhiexpression}) and it is given by
\bea
&& 
\frac{1}{(z_1-z_2)^2}  \, \{  \Phi_0^{(1)} \, \Phi_0^{(1)} \}_{-2}(z_2)
+ \theta_1^i \, 
\frac{1}{(z_1-z_2)}  \, \{  \Phi_{\frac{1}{2}}^{(1),i} \, \Phi_0^{(1)} \}_{-1}(z_2)
\nonu \\
&& + \theta_1^{4-ij} \, \sum_{n=2}^1 \, 
\frac{1}{(z_1-z_2)^n}  \, \{  \Phi_{1}^{(1),ij} \, \Phi_0^{(1)} \}_{-n}(z_2)
+ \theta_1^{4-i} \, \sum_{n=3}^1 \, 
\frac{1}{(z_1-z_2)^n}  \, \{  \Phi_{\frac{3}{2}}^{(1),i} \, \Phi_0^{(1)} \}_{-n}(z_2)
\nonu \\
&& + \theta_1^{4-0} \, \sum_{n=4}^1 \, 
\frac{1}{(z_1-z_2)^n}  \, \{  \Phi_{2}^{(1)} \, \Phi_0^{(1)} \}_{-n}(z_2).
\label{restricted}
\eea
The crucial point is  that the ${\cal N}=4$ OPE 
can be obtained by replacing all the component currents
appearing in (\ref{restricted}) 
with the corresponding ${\cal N}=4$ versions using 
(\ref{j4}) and (\ref{phis}). 
That is,
\bea
\Delta & \rightarrow & -{\bf J^{(4)}}, \qquad
\Gamma^i \rightarrow - i \, D^i \, {\bf J^{(4)}}, \qquad
T^{ij} \rightarrow - \frac{i}{2} \, 
\varepsilon^{ijkl} D^k \, D^l \, {\bf J^{(4)}},  
\nonu \\
G^i  & \rightarrow &
\left( -\frac{1}{6} \, \varepsilon^{ijkl} \, D^j \, D^k \, D^l 
+ 2 \, \alpha \, D^i \, \right) {\bf J^{(4)}}, 
\qquad
 L \rightarrow \left( \frac{1}{2} \,
D^1 \, D^2\, D^3 \, D^4  - \alpha \, \pa^2  \right) {\bf J^{(4)}}, 
\nonu \\
\Phi_0^{(s)} & \rightarrow & {\bf \Phi^{(s)}}, \qquad
\Phi_{\frac{1}{2}}^{(s),i} \rightarrow  D^i \,  {\bf \Phi^{(s)}},
\qquad
s=1,2.
\label{compsuper}
\eea
The final OPE in (\ref{finalPhiPhi})
is obtained from (\ref{restricted}) with the replacement 
(\ref{compsuper}), $(z_1-z_2) \rightarrow z_{12}$ and $\theta_1^i 
\rightarrow \theta_{12}^i$.
Let us emphasize that it is nontrivial to obtain 
the higher spin currents $\Phi_0^{(s=2)}(w)$ and 
$\Phi_{\frac{1}{2}}^{(s=2),i}(w)$ for general $N$ and $k$.
In \cite{Ahn1504}, the corresponding quantities 
are given by the higher spin currents 
${\bf P^{(2)}}(w)$, ${\bf P_{\pm}^{(\frac{5}{2})}}(w)$,
${\bf Q^{(\frac{5}{2})}}(w)$, and ${\bf R^{(\frac{5}{2})}}(w)$.
This is one of the reasons why one should resort to the Jacobi identities
to obtain $\Phi_0^{(2)}(w)$ and 
$\Phi_{\frac{1}{2}}^{(2),i}(w)$ in terms of 
 ${\bf P^{(2)}}(w)$, ${\bf P_{\pm}^{(\frac{5}{2})}}(w)$,
${\bf Q^{(\frac{5}{2})}}(w)$ and ${\bf R^{(\frac{5}{2})}}(w)$.
See also Appendix $J$.

The remaining things are to check the above statement explicitly.
One should arrange the numerators in (\ref{PhiPhiother})
in order to see its ${\cal N}=4$ superspace notation appropriately
and it will turn out that
\bea
& &  
\frac{\theta_{12}^{4-0}}{z_{12}^{4}}\, \Bigg[ 
\{  \Phi_{2}^{(1)} \, \Phi_0^{(1)} \}_{-4}
+ 
 \cdots
\Bigg] +  
\frac{\theta_{12}^{4-i}}{z_{12}^{3}}\, \Bigg[  
\{  \Phi_{\frac{3}{2}}^{(1),i} \, \Phi_0^{(1)} \}_{-3}
+ \cdots
\Bigg]
 + 
\frac{\theta_{12}^{4-0}}{z_{12}^{3}}\,\Bigg[ 
\{  \Phi_{2}^{(1)} \, \Phi_0^{(1)} \}_{-3}
+ \cdots
\Bigg]
\nonu \\
& & + 
\frac{1}{z_{12}^{2}}\, \Bigg[ 
 \{  \Phi_0^{(1)} \, \Phi_0^{(1)} \}_{-2}
+ \cdots
\Bigg]
 +  \frac{\theta_{12}^{4-ij}}{z_{12}^{2}}\, \Bigg[ 
\{  \Phi_{1}^{(1),ij} \, \Phi_0^{(1)} \}_{-2}
+ \cdots
\Bigg]
 +  \frac{\theta_{12}^{4-i}}{z_{12}^{2}} \, 
\Bigg[ 
\{  \Phi_{\frac{3}{2}}^{(1),i} \, \Phi_0^{(1)} \}_{-2}
+ \cdots
\Bigg]
\nonu \\
& & +  
\frac{\theta_{12}^{4-0}}{z_{12}^{2}}\, \Bigg[
\{  \Phi_{2}^{(1)} \, \Phi_0^{(1)} \}_{-2}
+ \cdots
\Bigg]
 +  \frac{\theta_{12}^{4-jkl}}{z_{12}}\, 
\Bigg[
\{  \Phi_{\frac{1}{2}}^{(1),i} \, \Phi_0^{(1)} \}_{-1}
+ \cdots
\Bigg]
 +\frac{\theta_{12}^{4-ij}}{z_{12}}\Bigg[
\{  \Phi_{1}^{(1),ij} \, \Phi_0^{(1)} \}_{-1}
+ \cdots
\Bigg] \nonu \\
&& +\frac{\theta_{12}^{4-i}}{z_{12}}\, \Bigg[
\{  \Phi_{\frac{3}{2}}^{(1),i} \, \Phi_0^{(1)} \}_{-1}
+ \cdots
\Bigg]  +\frac{\theta_{12}^{4-0}}{z_{12}}\, \Bigg[
\{  \Phi_{2}^{(1)} \, \Phi_0^{(1)} \}_{-1}
+ \cdots
\Bigg]+\cdots,
\label{interinter}
\eea
where the abbreviated parts in the singular terms 
depend on the coordinate
$\theta_2^i$.
One can see that when $\theta_2^i$ is equal to zero, 
this expression (\ref{interinter}) becomes the one in (\ref{finalPhiPhi})
evaluated at $\theta_2^i=0$
with the help of Appendix $H.1$ or Appendix $I$.
One should replace all the singular terms in terms of composite fields
explicitly.
Furthermore, it is straightforward to 
obtain the abbreviated $\theta_2^i$-dependent part of (\ref{interinter})  
above 
by substituting the replacement (\ref{compsuper}) into 
the reduced $\theta_2^i$-independent part of  (\ref{interinter})
with Appendix $H$ or Appendix $I$.
Note that the 
fermionic coordinate in the right hand side of 
(\ref{interinter}) is characterized by $\theta_2^i$.
In other words, the final OPE in (\ref{finalPhiPhi})
can be obtained 
1) by taking  the $\theta_2^i$-independent part of (\ref{interinter}) 
or the OPEs in Appendix $I$
and 2) by substituting  
the replacement (\ref{compsuper}) with $(z_1-z_2) \rightarrow z_{12}$ and 
$\theta_1^i \rightarrow \theta_{12}^i$. The seventy structure constants
in (\ref{finalPhiPhi}) come from those in the OPEs in Appendix $I$.  

\subsection{Various covariant spinor derivatives acting on the 
fractional super coordinates}

Sometimes one should 
calculate the spinor derivative
acting on the fractional super coordinates in order to check 
any ${\cal N}=4$ space version and its component results.
In doing this, one should calculate the following computations
\bea
D_1^1 \, \frac{\theta^{4-0}_{12}}{z_{12}^n} & = & 
-\frac{\theta_{12}^2 \, \theta_{12}^4
\, \theta_{12}^3}{z_{12}^n},
\qquad
D_1^2 \, \frac{\theta^{4-0}_{12}}{z_{12}^n}  =  -
\frac{\theta_{12}^1 \, \theta_{12}^3
\, \theta_{12}^4}{z_{12}^n},
\qquad
D_1^3 \, \frac{\theta^{4-0}_{12}}{z_{12}^n}  = 
-\frac{\theta_{12}^1 \, \theta_{12}^4
\, \theta_{12}^2}{z_{12}^n},
\nonu \\
D_1^4 \, \frac{\theta^{4-0}_{12}}{z_{12}^n}  & = & 
-\frac{\theta_{12}^1 \, \theta_{12}^2
\, \theta_{12}^3}{z_{12}^n},
\qquad
D^1_1 \, D_1^2 \, \frac{\theta^{4-0}_{12}}{z_{12}^n}  = 
-\frac{\theta_{12}^3
\, \theta_{12}^4}{z_{12}^n},
\qquad
D^1_1 \, D_1^3 \, \frac{\theta^{4-0}_{12}}{z_{12}^n}  = 
\frac{\theta_{12}^2 \, 
\, \theta_{12}^4}{z_{12}^n},
\nonu \\
D^1_1 \, D_1^4 \, \frac{\theta^{4-0}_{12}}{z_{12}^n} & = &
-\frac{\theta_{12}^2 \, \theta_{12}^3}{z_{12}^n},
\qquad
D^2_1 \, D_1^3 \, \frac{\theta^{4-0}_{12}}{z_{12}^n}  = 
-\frac{\theta_{12}^1  
\, \theta_{12}^4}{z_{12}^n},
\qquad
D^2_1 \, D_1^4 \, \frac{\theta^{4-0}_{12}}{z_{12}^n}  = 
\frac{\theta_{12}^1 \, \theta_{12}^3}{z_{12}^n},
\nonu \\
D^3_1 \, D_1^4 \, \frac{\theta^{4-0}_{12}}{z_{12}^n} & = & 
-\frac{\theta_{12}^1 \, \theta_{12}^2
}{z_{12}^n},
\qquad
D^1_1 \, D_1^2 \,  D^3_1\, \frac{\theta^{4-0}_{12}}{z_{12}^n}  = 
-\frac{\theta_{12}^4 
}{z_{12}^n},
\qquad
D^1_1 \, D_1^4 \,  D^2_1\, \frac{\theta^{4-0}_{12}}{z_{12}^n}   =   -
\frac{\theta_{12}^3
}{z_{12}^n},
\nonu \\
D^1_1 \, D_1^3 \,  D^4_1\, \frac{\theta^{4-0}_{12}}{z_{12}^n} & = &
-\frac{ \theta_{12}^2
}{z_{12}^n},
\qquad
D^2_1 \, D_1^4 \,  D^3_1\, \frac{\theta^{4-0}_{12}}{z_{12}^n}  = 
-\frac{\theta_{12}^1
}{z_{12}^n},
\qquad
D^1_1 \, D_1^2 \,  D^3_1\, D^4_1\, \frac{\theta^{4-0}_{12}}{z_{12}^n}  =  
\frac{1
}{z_{12}^n}.
\nonu 
\eea
Similarly, one can obtain the following relations
\bea
D_2^1 \, \frac{\theta^{4-0}_{12}}{z_{12}^n} & = & 
\frac{\theta_{12}^2 \, \theta_{12}^4
\, \theta_{12}^3}{z_{12}^n},
\quad
D_2^2 \, \frac{\theta^{4-0}_{12}}{z_{12}^n}  = 
\frac{\theta_{12}^1 \, \theta_{12}^3
\, \theta_{12}^4}{z_{12}^n},
\qquad
D_2^3 \, \frac{\theta^{4-0}_{12}}{z_{12}^n}  = 
\frac{\theta_{12}^1 \, \theta_{12}^4
\, \theta_{12}^2}{z_{12}^n},
\nonu \\
D_2^4 \, \frac{\theta^{4-0}_{12}}{z_{12}^n}  & = &  
\frac{\theta_{12}^1 \, \theta_{12}^2
\, \theta_{12}^3}{z_{12}^n},
\qquad
D^1_2 \, D_2^2 \, \frac{\theta^{4-0}_{12}}{z_{12}^n}  = 
-\frac{\theta_{12}^3
\, \theta_{12}^4}{z_{12}^n},
\qquad
D^1_2 \, D_2^3 \, \frac{\theta^{4-0}_{12}}{z_{12}^n}  = 
\frac{\theta_{12}^2 \, 
\, \theta_{12}^4}{z_{12}^n},
\nonu \\
D^1_2 \, D_2^4 \, \frac{\theta^{4-0}_{12}}{z_{12}^n} & = &
-\frac{\theta_{12}^2 \, \theta_{12}^3}{z_{12}^n},
\qquad
D^2_2 \, D_2^3 \, \frac{\theta^{4-0}_{12}}{z_{12}^n}  = 
-\frac{\theta_{12}^1  
\, \theta_{12}^4}{z_{12}^n},
\qquad
D^2_2 \, D_2^4 \, \frac{\theta^{4-0}_{12}}{z_{12}^n}  = 
\frac{\theta_{12}^1 \, \theta_{12}^3}{z_{12}^n},
\nonu \\
D^3_2 \, D_2^4 \, \frac{\theta^{4-0}_{12}}{z_{12}^n}  & = &  
-\frac{\theta_{12}^1 \, \theta_{12}^2
}{z_{12}^n},
\qquad
D^1_2 \, D_2^2 \,  D^3_2\, \frac{\theta^{4-0}_{12}}{z_{12}^n}  = 
\frac{\theta_{12}^4 
}{z_{12}^n},
\qquad
D^1_2 \, D_2^4 \,  D^2_2\, \frac{\theta^{4-0}_{12}}{z_{12}^n}  = 
\frac{\theta_{12}^3
}{z_{12}^n},
\nonu \\
D^1_2 \, D_2^3 \,  D^4_2\, \frac{\theta^{4-0}_{12}}{z_{12}^n} & = &
\frac{ \theta_{12}^2
}{z_{12}^n},
\qquad
D^2_2 \, D_2^4 \,  D^3_2\, \frac{\theta^{4-0}_{12}}{z_{12}^n}  = 
\frac{\theta_{12}^1
}{z_{12}^n},
\qquad
D^1_2 \, D_2^2 \,  D^3_2\, D^4_2\, \frac{\theta^{4-0}_{12}}{z_{12}^n}  =  
\frac{1
}{z_{12}^n}.
\nonu 
\eea
For the three Grassmann coordinates, one has the following 
nonzero identities
\bea
D_1^1 \, \frac{\theta^{2}_{12} \, \theta_{12}^{4} \, \theta_{12}^{3}}{z_{12}^n} & = & 
\frac{\theta_{12}^{4-0}}{z_{12}^{n+1}} \, n, 
\qquad
D_1^1 \, \frac{\theta^{1}_{12} \, \theta_{12}^{3} \, \theta_{12}^{4}}{z_{12}^n} 
=  
\frac{\theta_{12}^{3} \, \theta_{12}^4}{z_{12}^{n}},
\qquad
D_1^1 \, \frac{\theta^{1}_{12} \, \theta_{12}^{4} \, \theta_{12}^{2}}{z_{12}^n} 
  =   
-\frac{\theta_{12}^{2} \, \theta_{12}^4}{z_{12}^{n}},
\nonu \\
D_1^1 \, \frac{\theta^{1}_{12} \, \theta_{12}^{2} \, \theta_{12}^{3}}{z_{12}^n} 
& = &  
\frac{\theta_{12}^{2} \, \theta_{12}^3}{z_{12}^{n}},
\qquad
D_1^2 \, \frac{\theta^{2}_{12} \, \theta_{12}^{4} \, \theta_{12}^{3}}{z_{12}^n}  =  
-\frac{\theta_{12}^3 \, \theta_{12}^{4}}{z_{12}^{n}}, 
\qquad
D_1^2 \, \frac{\theta^{1}_{12} \, \theta_{12}^{3} \, \theta_{12}^{4}}{z_{12}^n} 
=  
\frac{\theta_{12}^{4-0}}{z_{12}^{n+1}} \, n,
\nonu \\
D_1^2 \, \frac{\theta^{1}_{12} \, \theta_{12}^{4} \, \theta_{12}^{2}}{z_{12}^n} 
 & = &  
\frac{\theta_{12}^{1} \, \theta_{12}^4}{z_{12}^{n}},
\qquad
D_1^2 \, \frac{\theta^{1}_{12} \, \theta_{12}^{2} \, \theta_{12}^{3}}{z_{12}^n} 
=  -
\frac{\theta_{12}^{1} \, \theta_{12}^3}{z_{12}^{n}},
\qquad
D_1^3 \, \frac{\theta^{2}_{12} \, \theta_{12}^{4} \, \theta_{12}^{3}}{z_{12}^n}  =  
\frac{\theta_{12}^2 \, \theta_{12}^{4}}{z_{12}^{n}}, 
\nonu \\
D_1^3 \, \frac{\theta^{1}_{12} \, \theta_{12}^{3} \, \theta_{12}^{4}}{z_{12}^n} 
& = & -  
\frac{\theta_{12}^1 \, \theta_{12}^{4}}{z_{12}^{n}},
\qquad
D_1^3 \, \frac{\theta^{1}_{12} \, \theta_{12}^{4} \, \theta_{12}^{2}}{z_{12}^n} 
  =  
\frac{\theta_{12}^{4-0}}{z_{12}^{n+1}} \, n,
\qquad
D_1^3 \, \frac{\theta^{1}_{12} \, \theta_{12}^{2} \, \theta_{12}^{3}}{z_{12}^n} 
=  
\frac{\theta_{12}^{1} \, \theta_{12}^2}{z_{12}^{n}},
\nonu \\
D_1^4 \, \frac{\theta^{2}_{12} \, \theta_{12}^{4} \, \theta_{12}^{3}}{z_{12}^n} & = &
-\frac{\theta_{12}^2 \, \theta_{12}^{3}}{z_{12}^{n}}, 
\qquad
D_1^4 \, \frac{\theta^{1}_{12} \, \theta_{12}^{3} \, \theta_{12}^{4}}{z_{12}^n} 
=  
\frac{\theta_{12}^1 \, \theta_{12}^{3}}{z_{12}^{n}},
\qquad
D_1^4 \, \frac{\theta^{1}_{12} \, \theta_{12}^{4} \, \theta_{12}^{2}}{z_{12}^n} 
  =  
-\frac{\theta_{12}^{1} \, \theta_{12}^{2}}{z_{12}^{n}},
\nonu \\
D_1^4 \, \frac{\theta^{1}_{12} \, \theta_{12}^{2} \, \theta_{12}^{3}}{z_{12}^n} 
& = &  
\frac{\theta_{12}^{4-0}}{z_{12}^{n+1}} \, n,
\nonu \\
D_1^1 \, 
D_1^2 \, \frac{\theta^{2}_{12} \, \theta_{12}^{3} \, \theta_{12}^{4}}{z_{12}^n} & = & -
\frac{\theta_{12}^1 \,
\theta_{12}^3 \, \theta_{12}^{4}}{z_{12}^{n+1}} \, n, 
\qquad
D_1^1 \, 
D_1^2 \, \frac{\theta^{1}_{12} \, \theta_{12}^{3} \, \theta_{12}^{4}}{z_{12}^n} 
=   
-\frac{\theta_{12}^2 \, \theta_{12}^4 \, \theta_{12}^{3}}{z_{12}^{n+1}} \, n,
\nonu \\
D_1^1 \,
D_1^2 \, \frac{\theta^{1}_{12} \, \theta_{12}^{4} \, \theta_{12}^{2}}{z_{12}^n} 
 & = &   
\frac{ \theta_{12}^4}{z_{12}^{n}},
\qquad
D_1^1 \, 
D_1^2 \, \frac{\theta^{1}_{12} \, \theta_{12}^{2} \, \theta_{12}^{3}}{z_{12}^n} 
=  -
\frac{ \theta_{12}^3}{z_{12}^{n}},
\nonu \\
D_1^1\, D_1^3 \, \frac{\theta^{2}_{12} \, \theta_{12}^{4} \, \theta_{12}^{3}}{z_{12}^n} & = & 
\frac{\theta_{12}^1 \, \theta_{12}^4 \, \theta_{12}^{2}}{z_{12}^{n+1}} \, n, 
\qquad
D_1^1\, 
D_1^3 \, \frac{\theta^{1}_{12} \, \theta_{12}^{3} \, \theta_{12}^{4}}{z_{12}^n} 
= -  
\frac{ \theta_{12}^{4}}{z_{12}^{n}},
\nonu \\
D_1^1 \, D_1^3 \, \frac{\theta^{1}_{12} \, \theta_{12}^{4} \, \theta_{12}^{2}}{z_{12}^n} 
 & = & 
\frac{\theta_{12}^2 \, \theta_{12}^4 \, \theta_{12}^{3}}{z_{12}^{n+1}} \, n,
\qquad
D_1^1 \, 
D_1^3 \, \frac{\theta^{1}_{12} \, \theta_{12}^{2} \, \theta_{12}^{3}}{z_{12}^n} 
=  
\frac{ \theta_{12}^2}{z_{12}^{n}},
\nonu \\
D_1^1 \, 
D_1^4 \, \frac{\theta^{2}_{12} \, \theta_{12}^{4} \, \theta_{12}^{3}}{z_{12}^n} & = &
\frac{\theta_{12}^1 \, \theta_{12}^2 \, \theta_{12}^{3}}{z_{12}^{n+1}} \, n, 
\qquad
D_1^1 \, 
D_1^4 \, \frac{\theta^{1}_{12} \, \theta_{12}^{3} \, \theta_{12}^{4}}{z_{12}^n} 
=  
\frac{ \theta_{12}^{3}}{z_{12}^{n}},
\nonu \\
D_1^1 \, 
D_1^4 \, \frac{ \theta_{12}^{2} \, \theta_{12}^{4}}{z_{12}^n} 
 & = & 
\frac{\theta_{12}^{1} \, \theta_{12}^{2}}{z_{12}^{n}},
\qquad
D_1^1 \,
D_1^4 \, \frac{\theta^{1}_{12} \, \theta_{12}^{2} \, \theta_{12}^{3}}{z_{12}^n} 
= 
-\frac{\theta_{12}^2 \, \theta_{12}^4 \, \theta_{12}^{3}}{z_{12}^{n+1}} \, n,
\nonu \\
D_1^2\,
D_1^3 \, \frac{\theta^{2}_{12} \, \theta_{12}^{4} \, \theta_{12}^{3}}{z_{12}^n} & = & 
\frac{ \theta_{12}^{4}}{z_{12}^{n}}, 
\qquad
D_1^2\,
D_1^3 \, \frac{\theta^{1}_{12} \, \theta_{12}^{3} \, \theta_{12}^{4}}{z_{12}^n} 
=   
\frac{\theta_{12}^1 \, \theta_{12}^4 \, \theta_{12}^{2}}{z_{12}^{n+1}} \, n,
\nonu \\
D_1^2\,
D_1^3 \, \frac{\theta^{1}_{12} \, \theta_{12}^{4} \, \theta_{12}^{2}}{z_{12}^n} 
 & = & 
-\frac{\theta_{12}^1 \, \theta_{12}^3 \, \theta_{12}^{4}}{z_{12}^{n+1}} \, n,
\qquad
D_1^2\,
D_1^3 \, \frac{\theta^{1}_{12} \, \theta_{12}^{2} \, \theta_{12}^{3}}{z_{12}^n} 
=  -
\frac{\theta_{12}^{1} }{z_{12}^{n}},
\nonu \\
D_1^2 \,
D_1^4 \, \frac{\theta^{2}_{12} \, \theta_{12}^{4} \, \theta_{12}^{3}}{z_{12}^n} & = & -
\frac{ \theta_{12}^{3}}{z_{12}^{n}}, 
\qquad
D_1^2 \,
D_1^4 \, \frac{\theta^{1}_{12} \, \theta_{12}^{3} \, \theta_{12}^{4}}{z_{12}^n} 
=  
\frac{\theta_{12}^1 \, \theta_{12}^{2}\, \theta_{12}^{3}}{z_{12}^{n+1}} \, n,
\nonu \\
D_1^2 \,
D_1^4 \, \frac{\theta^{1}_{12} \, \theta_{12}^{4} \, \theta_{12}^{2}}{z_{12}^n} 
 & = &
\frac{\theta_{12}^{1}}{z_{12}^{n}},
\qquad
D_1^2 \,
D_1^4 \, \frac{\theta^{1}_{12} \, \theta_{12}^{2} \, \theta_{12}^{3}}{z_{12}^n} 
=  
-\frac{\theta_{12}^1 \, \theta_{12}^3 \, \theta_{12}^{4}}{z_{12}^{n+1}} \, n,
\nonu \\
D_1^3 \,
D_1^4 \, \frac{\theta^{2}_{12} \, \theta_{12}^{4} \, \theta_{12}^{3}}{z_{12}^n} & = &
\frac{\theta_{12}^2 }{z_{12}^{n}}, 
\qquad
D_1^3 \,
D_1^4 \, \frac{\theta^{1}_{12} \, \theta_{12}^{3} \, \theta_{12}^{4}}{z_{12}^n} 
=  -
\frac{\theta_{12}^1}{z_{12}^{n}},
\nonu \\
D_1^3 \,
D_1^4 \, \frac{\theta^{1}_{12} \, \theta_{12}^{2} \, \theta_{12}^{4}}{z_{12}^n} 
 & = & -
\frac{\theta_{12}^{1} \, \theta_{12}^{2}\, \theta_{12}^3}{z_{12}^{n+1}} \, n,
\qquad
D_1^3 \,
D_1^4 \, \frac{\theta^{1}_{12} \, \theta_{12}^{2} \, \theta_{12}^{3}}{z_{12}^n} 
=  -
\frac{\theta_{12}^1 \, \theta_{12}^4 \, \theta_{12}^{2}}{z_{12}^{n+1}} \, n,
\nonu \\
D_1^1 \, D_1^2\,
D_1^3 \, \frac{\theta^{2}_{12} \, \theta_{12}^{4} \, \theta_{12}^{3}}{z_{12}^n} & = & 
-\frac{ \theta_{12}^1 \, \theta_{12}^{4}}{z_{12}^{n+1}}\, n, 
\qquad
D_1^1 \, D_1^2\,
D_1^3 \, \frac{\theta^{1}_{12} \, \theta_{12}^{3} \, \theta_{12}^{4}}{z_{12}^n} 
= -  
\frac{ \theta_{12}^2 \, \theta_{12}^{4}}{z_{12}^{n+1}} \, n,
\nonu \\
D_1^1 \, D_1^2\,
D_1^3 \, \frac{\theta^{1}_{12} \, \theta_{12}^{4} \, \theta_{12}^{2}}{z_{12}^n} 
 & = & 
-\frac{ \theta_{12}^3 \, \theta_{12}^{4}}{z_{12}^{n+1}} \, n,
\qquad
D_1^1 \, D_1^2\,
D_1^3 \, \frac{\theta^{1}_{12} \, \theta_{12}^{2} \, \theta_{12}^{3}}{z_{12}^n} 
=  -
\frac{1 }{z_{12}^{n}},
\nonu \\
D_1^1 \, D_1^4 \,
D_1^2 \, \frac{\theta^{2}_{12} \, \theta_{12}^{4} \, \theta_{12}^{3}}{z_{12}^n} & = &
-\frac{ \theta_{12}^1 \, \theta_{12}^{4}}{z_{12}^{n+1}} \, n, 
\qquad
D_1^1 \, D_1^4 \,
D_1^2 \, \frac{\theta^{1}_{12} \, \theta_{12}^{3} \, \theta_{12}^{4}}{z_{12}^n} 
=  
-\frac{ \theta_{12}^{2}\, \theta_{12}^{3}}{z_{12}^{n+1}} \, n,
\nonu \\
D_1^1 \, D_1^4 \,
D_1^2 \, \frac{\theta^{1}_{12} \, \theta_{12}^{4} \, \theta_{12}^{2}}{z_{12}^n} 
 & = &- 
\frac{1}{z_{12}^{n}},
\qquad
D_1^1 \, D_1^4 \,
D_1^2 \, \frac{\theta_{12}^1\, \theta_{12}^{2}\, \theta_{12}^3}{z_{12}^n} 
=  
\frac{ \theta_{12}^3 \, \theta_{12}^{4}}{z_{12}^{n+1}} \, n,
\nonu \\
D_1^1 \, D_1^3 \,
D_1^4 \, \frac{\theta^{2}_{12} \, \theta_{12}^{4} \, \theta_{12}^{3}}{z_{12}^n} & = & -
\frac{\theta_{12}^1 \, \theta_{12}^2 }{z_{12}^{n+1}} \, n, 
\qquad
D_1^1 \,D_1^3 \,
D_1^4 \, \frac{\theta^{1}_{12} \, \theta_{12}^{3} \, \theta_{12}^{4}}{z_{12}^n} 
=  -
\frac{1}{z_{12}^{n}},
\nonu \\
D_1^1 \,D_1^3 \,
D_1^4 \, \frac{\theta^{1}_{12} \, \theta_{12}^{4} \, \theta_{12}^{2}}{z_{12}^n} 
 & = & 
\frac{ \theta_{12}^{2}\, \theta_{12}^3}{z_{12}^{n+1}} \, n,
\qquad
D_1^1 \,D_1^3 \,
D_1^4 \, \frac{\theta^{1}_{12} \, \theta_{12}^{2} \, \theta_{12}^{3}}{z_{12}^n} 
=  
\frac{ \theta_{12}^2 \, \theta_{12}^{4}}{z_{12}^{n+1}} \, n,
\nonu \\
D_1^2\, D_1^4 \,
D_1^3 \, \frac{\theta^{2}_{12} \, \theta_{12}^{4} \, \theta_{12}^{3}}{z_{12}^n} & = &
-\frac{1 }{z_{12}^{n}}, 
\qquad
D_1^2\, D_1^4 \,
D_1^3 \, \frac{\theta^{1}_{12} \, \theta_{12}^{3} \, \theta_{12}^{4}}{z_{12}^n} 
=  
\frac{\theta_{12}^1 \, \theta_{12}^2}{z_{12}^{n+1}} \, n,
\nonu \\
 D_1^2\, D_1^4 \,
D_1^3 \, \frac{\theta^{1}_{12} \, \theta_{12}^{4} \, \theta_{12}^{2}}{z_{12}^n} 
 & = & 
\frac{\theta_{12}^{1} \,  \theta_{12}^3}{z_{12}^{n+1}} \, n,
\qquad
 D_1^2\, D_1^4 \,
D_1^3 \, \frac{\theta^{1}_{12} \, \theta_{12}^{2} \, \theta_{12}^{3}}{z_{12}^n} 
=  
\frac{\theta_{12}^1 \, \theta_{12}^{4}}{z_{12}^{n+1}} \, n,
\nonu \\
D_1^1\, D_1^2\, D_1^3 \,
D_1^4 \, \frac{\theta^{2}_{12} \, \theta_{12}^{4} \, \theta_{12}^{3}}{z_{12}^n} & = &
-\frac{\theta_{12}^1 }{z_{12}^{n+1}} \, n, 
\qquad
D_1^1\,D_1^2\, D_1^3 \,
D_1^4 \, \frac{\theta^{1}_{12} \, \theta_{12}^{3} \, \theta_{12}^{4}}{z_{12}^n} 
=  -
\frac{ \theta_{12}^2}{z_{12}^{n+1}} \, n,
\nonu \\
D_1^1\, D_1^2\, D_1^3 \,
D_1^4 \, \frac{ \theta_{12}^1 \, \theta_{12}^{4} \, \theta_{12}^{2}}{z_{12}^n} 
 & = & 
-\frac{  \theta_{12}^3}{z_{12}^{n+1}} \, n,
\qquad
D_1^1\, D_1^2\, D_1^3 \,
D_1^4 \, \frac{\theta_{12}^1\, \theta_{12}^{2} \, \theta_{12}^{3}}{z_{12}^n} 
=  
-\frac{ \theta_{12}^{4}}{z_{12}^{n+1}} \, n.
\nonu
\eea
Similarly one has the following relations
\bea
D_2^1 \, \frac{\theta^{2}_{12} \, \theta_{12}^{4} \, \theta_{12}^{3}}{z_{12}^n} & = & 
\frac{\theta_{12}^{4-0}}{z_{12}^{n+1}} \, n, 
\qquad
D_2^1 \, \frac{\theta^{1}_{12} \, \theta_{12}^{3} \, \theta_{12}^{4}}{z_{12}^n} 
=  
-\frac{\theta_{12}^{3} \, \theta_{12}^4}{z_{12}^{n}},
\qquad
D_2^1 \, \frac{\theta^{1}_{12} \, \theta_{12}^{4} \, \theta_{12}^{2}}{z_{12}^n} 
  =   
\frac{\theta_{12}^{2} \, \theta_{12}^4}{z_{12}^{n}},
\nonu \\
D_2^1 \, \frac{\theta^{1}_{12} \, \theta_{12}^{2} \, \theta_{12}^{3}}{z_{12}^n} 
& = &  
-\frac{\theta_{12}^{2} \, \theta_{12}^3}{z_{12}^{n}},
\qquad
D_2^2 \, \frac{\theta^{2}_{12} \, \theta_{12}^{4} \, \theta_{12}^{3}}{z_{12}^n}  =  
\frac{\theta_{12}^3 \, \theta_{12}^{4}}{z_{12}^{n}}, 
\qquad
D_2^2 \, \frac{\theta^{1}_{12} \, \theta_{12}^{3} \, \theta_{12}^{4}}{z_{12}^n} 
=  
\frac{\theta_{12}^{4-0}}{z_{12}^{n+1}} \, n,
\nonu \\
D_2^2 \, \frac{\theta^{1}_{12} \, \theta_{12}^{4} \, \theta_{12}^{2}}{z_{12}^n} 
 & = & -  
\frac{\theta_{12}^{1} \, \theta_{12}^4}{z_{12}^{n}},
\qquad
D_2^2 \, \frac{\theta^{1}_{12} \, \theta_{12}^{2} \, \theta_{12}^{3}}{z_{12}^n} 
=  
\frac{\theta_{12}^{1} \, \theta_{12}^3}{z_{12}^{n}},
\qquad
D_2^3 \, \frac{\theta^{2}_{12} \, \theta_{12}^{3} \, \theta_{12}^{4}}{z_{12}^n}  
=  
\frac{\theta_{12}^2 \, \theta_{12}^{4}}{z_{12}^{n}}, 
\nonu \\
D_2^3 \, \frac{\theta^{1}_{12} \, \theta_{12}^{3} \, \theta_{12}^{4}}{z_{12}^n} 
 & = &    
\frac{\theta_{12}^1 \, \theta_{12}^{4}}{z_{12}^{n}},
\qquad
D_2^3 \, \frac{\theta^{1}_{12} \, \theta_{12}^{4} \, \theta_{12}^{2}}{z_{12}^n} 
  =  
\frac{\theta_{12}^{4-0}}{z_{12}^{n+1}} \, n,
\qquad
D_2^3 \, \frac{\theta^{1}_{12} \, \theta_{12}^{2} \, \theta_{12}^{3}}{z_{12}^n} 
  =   
-\frac{\theta_{12}^{1} \, \theta_{12}^2}{z_{12}^{n}},
\nonu \\
D_2^4 \, \frac{\theta^{2}_{12} \, \theta_{12}^{3} \, \theta_{12}^{4}}{z_{12}^n} & = &
-\frac{\theta_{12}^2 \, \theta_{12}^{3}}{z_{12}^{n}}, 
\qquad
D_2^4 \, \frac{\theta^{1}_{12} \, \theta_{12}^{3} \, \theta_{12}^{4}}{z_{12}^n} 
=  -
\frac{\theta_{12}^1 \, \theta_{12}^{3}}{z_{12}^{n}},
\qquad
D_2^4 \, \frac{\theta^{1}_{12} \, \theta_{12}^{4} \, \theta_{12}^{2}}{z_{12}^n} 
  =  
\frac{\theta_{12}^{1} \, \theta_{12}^{2}}{z_{12}^{n}},
\nonu \\
D_2^4 \, \frac{\theta^{1}_{12} \, \theta_{12}^{2} \, \theta_{12}^{3}}{z_{12}^n} 
& = &   
\frac{\theta_{12}^{4-0}}{z_{12}^{n+1}} \, n,
\nonu \\
D_2^1 \, 
D_2^2 \, \frac{\theta^{2}_{12} \, \theta_{12}^{4} \, \theta_{12}^{3}}{z_{12}^n} & = & -
\frac{\theta_{12}^1 \,
\theta_{12}^3 \, \theta_{12}^{4}}{z_{12}^{n+1}} \, n, 
\qquad
D_2^1 \, 
D_2^2 \, \frac{\theta^{1}_{12} \, \theta_{12}^{3} \, \theta_{12}^{4}}{z_{12}^n} 
=   
\frac{\theta_{12}^2 \, \theta_{12}^4 \, \theta_{12}^{3}}{z_{12}^{n+1}} \, n,
\nonu \\
D_2^1 \,
D_2^2 \, \frac{\theta^{1}_{12} \, \theta_{12}^{2} \, \theta_{12}^{4}}{z_{12}^n} 
 & = & -  
\frac{ \theta_{12}^4}{z_{12}^{n}},
\qquad
D_2^1 \, 
D_2^2 \, \frac{\theta^{1}_{12} \, \theta_{12}^{2} \, \theta_{12}^{3}}{z_{12}^n} 
=  -
\frac{ \theta_{12}^3}{z_{12}^{n}},
\nonu \\
D_2^1\, D_2^3 \, \frac{\theta^{2}_{12} \, \theta_{12}^{4} \, \theta_{12}^{3}}{z_{12}^n} & = & 
-\frac{\theta_{12}^1 \, \theta_{12}^4 \, \theta_{12}^{2}}{z_{12}^{n+1}} \, n, 
\qquad
D_2^1\, 
D_2^3 \, \frac{\theta^{1}_{12} \, \theta_{12}^{3} \, \theta_{12}^{4}}{z_{12}^n} 
= -  
\frac{ \theta_{12}^{4}}{z_{12}^{n}},
\nonu \\
D_2^1 \, D_2^3 \, \frac{\theta^{1}_{12} \, \theta_{12}^{4} \, \theta_{12}^{2}}{z_{12}^n} 
 & = & 
\frac{\theta_{12}^2 \, \theta_{12}^4 \, \theta_{12}^{3}}{z_{12}^{n+1}} \, n,
\qquad
D_2^1 \, 
D_2^3 \, \frac{\theta^{1}_{12} \, \theta_{12}^{2} \, \theta_{12}^{3}}{z_{12}^n} 
=  
\frac{ \theta_{12}^2}{z_{12}^{n}},
\nonu \\
D_2^1 \, 
D_2^4 \, \frac{\theta^{2}_{12} \, \theta_{12}^{4} \, \theta_{12}^{3}}{z_{12}^n} & = &
-\frac{\theta_{12}^1 \, \theta_{12}^2 \, \theta_{12}^{3}}{z_{12}^{n+1}} \, n, 
\qquad
D_2^1 \, 
D_2^4 \, \frac{\theta^{1}_{12} \, \theta_{12}^{3} \, \theta_{12}^{4}}{z_{12}^n} 
=  
\frac{ \theta_{12}^{3}}{z_{12}^{n}},
\nonu \\
D_2^1 \, 
D_2^4 \, \frac{ \theta_{12}^{1} \, \theta_{12}^{4}\, \theta_{12}^{2}}{z_{12}^n} 
 & = & 
-\frac{\theta_{12}^{2}}{z_{12}^{n}},
\qquad
D_2^1 \,
D_2^4 \, \frac{\theta^{1}_{12} \, \theta_{12}^{2} \, \theta_{12}^{3}}{z_{12}^n} 
=  
\frac{\theta_{12}^2 \, \theta_{12}^4 \, \theta_{12}^{3}}{z_{12}^{n+1}} \, n,
\nonu \\
D_2^2\,
D_2^3 \, \frac{\theta^{2}_{12} \, \theta_{12}^{4} \, \theta_{12}^{3}}{z_{12}^n} & = & 
\frac{ \theta_{12}^{4}}{z_{12}^{n}}, 
\qquad
D_2^2\,
D_2^3 \, \frac{\theta^{1}_{12} \, \theta_{12}^{3} \, \theta_{12}^{4}}{z_{12}^n} 
=   
-\frac{\theta_{12}^1 \, \theta_{12}^4 \, \theta_{12}^{2}}{z_{12}^{n+1}} \, n,
\nonu \\
D_2^2\,
D_2^3 \, \frac{\theta^{1}_{12} \, \theta_{12}^{4} \, \theta_{12}^{2}}{z_{12}^n} 
 & = & 
\frac{\theta_{12}^1 \, \theta_{12}^3 \, \theta_{12}^{4}}{z_{12}^{n+1}} \, n,
\qquad
D_2^2\,
D_2^3 \, \frac{\theta^{1}_{12} \, \theta_{12}^{2} \, \theta_{12}^{3}}{z_{12}^n} 
=  -
\frac{\theta_{12}^{1} }{z_{12}^{n}},
\nonu \\
D_2^2 \,
D_2^4 \, \frac{\theta^{2}_{12} \, \theta_{12}^{4} \, \theta_{12}^{3}}{z_{12}^n} & = & -
\frac{ \theta_{12}^{3}}{z_{12}^{n}}, 
\qquad
D_2^2 \,
D_2^4 \, \frac{\theta^{1}_{12} \, \theta_{12}^{3} \, \theta_{12}^{4}}{z_{12}^n} 
=  
-\frac{\theta_{12}^1 \, \theta_{12}^{2}\, \theta_{12}^{3}}{z_{12}^{n+1}} \, n,
\nonu \\
D_2^2 \,
D_2^4 \, \frac{\theta^{1}_{12} \, \theta_{12}^{4} \, \theta_{12}^{2}}{z_{12}^n} 
 & = & 
\frac{\theta_{12}^{1}}{z_{12}^{n}},
\qquad
D_2^2 \,
D_2^4 \, \frac{\theta^{1}_{12} \, \theta_{12}^{2} \, \theta_{12}^{3}}{z_{12}^n} 
=  
\frac{\theta_{12}^1 \, \theta_{12}^3 \, \theta_{12}^{4}}{z_{12}^{n+1}} \, n,
\nonu \\
D_2^3 \,
D_2^4 \, \frac{\theta^{2}_{12} \, \theta_{12}^{4} \, \theta_{12}^{3}}{z_{12}^n} & = &
\frac{\theta_{12}^2 }{z_{12}^{n}}, 
\qquad
D_2^3 \,
D_2^4 \, \frac{\theta^{1}_{12} \, \theta_{12}^{3} \, \theta_{12}^{4}}{z_{12}^n} 
=  -
\frac{\theta_{12}^1}{z_{12}^{n}},
\nonu \\
D_2^3 \,
D_2^4 \, \frac{\theta^{1}_{12} \, \theta_{12}^{4} \, \theta_{12}^{2}}{z_{12}^n} 
 & = & 
-\frac{\theta_{12}^{1} \, \theta_{12}^{2}\, \theta_{12}^3}{z_{12}^{n+1}} \, n,
\qquad
D_2^3 \,
D_2^4 \, \frac{\theta^{1}_{12} \, \theta_{12}^{2} \, \theta_{12}^{3}}{z_{12}^n} 
= 
\frac{\theta_{12}^1 \, \theta_{12}^4 \, \theta_{12}^{2}}{z_{12}^{n+1}} \, n,
\nonu \\
D_2^1 \, D_2^2\,
D_2^3 \, \frac{\theta^{2}_{12} \, \theta_{12}^{4} \, \theta_{12}^{3}}{z_{12}^n} & = & 
-\frac{ \theta_{12}^1 \, \theta_{12}^{4}}{z_{12}^{n+1}}\, n, 
\qquad
D_2^1 \, D_2^2\,
D_2^3 \, \frac{\theta^{1}_{12} \, \theta_{12}^{3} \, \theta_{12}^{4}}{z_{12}^n} 
=  
-\frac{ \theta_{12}^2 \, \theta_{12}^{4}}{z_{12}^{n+1}} \, n,
\nonu \\
D_2^1 \, D_2^2\,
D_2^3 \, \frac{\theta^{1}_{12} \, \theta_{12}^{4} \, \theta_{12}^{2}}{z_{12}^n} 
 & = & 
-\frac{ \theta_{12}^3 \, \theta_{12}^{4}}{z_{12}^{n+1}} \, n,
\qquad
D_2^1 \, D_2^2\,
D_2^3 \, \frac{\theta^{1}_{12} \, \theta_{12}^{2} \, \theta_{12}^{3}}{z_{12}^n} 
= \frac{1 }{z_{12}^{n}},
\nonu \\
D_2^1 \, D_2^4 \,
D_2^2 \, \frac{\theta^{2}_{12} \, \theta_{12}^{4} \, \theta_{12}^{3}}{z_{12}^n} & = &
-\frac{ \theta_{12}^1 \, \theta_{12}^{4}}{z_{12}^{n+1}} \, n, 
\qquad
D_2^1 \, D_2^4 \,
D_2^2 \, \frac{\theta^{1}_{12} \, \theta_{12}^{3} \, \theta_{12}^{4}}{z_{12}^n} 
=  
-\frac{ \theta_{12}^{2}\, \theta_{12}^{3}}{z_{12}^{n+1}} \, n,
\nonu \\
D_2^1 \, D_2^4 \,
D_2^2 \, \frac{\theta^{1}_{12} \, \theta_{12}^{4} \, \theta_{12}^{2}}{z_{12}^n} 
 & = & 
\frac{1}{z_{12}^{n}},
\qquad
D_2^1 \, D_2^4 \,
D_2^2 \, \frac{\theta_{12}^1\, \theta_{12}^{2}\, \theta_{12}^3}{z_{12}^n} 
=  
\frac{ \theta_{12}^3 \, \theta_{12}^{4}}{z_{12}^{n+1}} \, n,
\nonu \\
D_2^1 \, D_2^3 \,
D_2^4 \, \frac{\theta^{2}_{12} \, \theta_{12}^{4} \, \theta_{12}^{3}}{z_{12}^n} & = & -
\frac{\theta_{12}^1 \, \theta_{12}^2 }{z_{12}^{n+1}} \, n, 
\qquad
D_2^1 \,D_2^3 \,
D_2^4 \, \frac{\theta^{1}_{12} \, \theta_{12}^{3} \, \theta_{12}^{4}}{z_{12}^n} 
=  
\frac{1}{z_{12}^{n}},
\nonu \\
D_2^1 \,D_2^3 \,
D_2^4 \, \frac{\theta^{1}_{12} \, \theta_{12}^{4} \, \theta_{12}^{2}}{z_{12}^n} 
 & = & -
\frac{ \theta_{12}^{2}\, \theta_{12}^3}{z_{12}^{n+1}} \, n,
\qquad
D_2^1 \,D_2^3 \,
D_2^4 \, \frac{\theta^{1}_{12} \, \theta_{12}^{2} \, \theta_{12}^{3}}{z_{12}^n} 
=  
\frac{ \theta_{12}^2 \, \theta_{12}^{4}}{z_{12}^{n+1}} \, n,
\nonu \\
D_2^2\, D_2^4 \,
D_2^3 \, \frac{\theta^{2}_{12} \, \theta_{12}^{4} \, \theta_{12}^{3}}{z_{12}^n} & = &
\frac{1 }{z_{12}^{n}}, 
\qquad
D_2^2\, D_2^4 \,
D_2^3 \, \frac{\theta^{1}_{12} \, \theta_{12}^{3} \, \theta_{12}^{4}}{z_{12}^n} 
=  
\frac{\theta_{12}^1 \, \theta_{12}^2}{z_{12}^{n+1}} \, n,
\nonu \\
 D_2^2\, D_2^4 \,
D_2^3 \, \frac{\theta^{1}_{12} \, \theta_{12}^{4} \, \theta_{12}^{2}}{z_{12}^n} 
 & = & 
\frac{\theta_{12}^{1} \,  \theta_{12}^3}{z_{12}^{n+1}} \, n,
\qquad
 D_2^2\, D_2^4 \,
D_2^3 \, \frac{\theta^{1}_{12} \, \theta_{12}^{2} \, \theta_{12}^{3}}{z_{12}^n} 
=  
\frac{\theta_{12}^1 \, \theta_{12}^{4}}{z_{12}^{n+1}} \, n,
\nonu \\
D_2^1\, D_2^2\, D_2^3 \,
D_2^4 \, \frac{\theta^{2}_{12} \, \theta_{12}^{4} \, \theta_{12}^{3}}{z_{12}^n} & = &
\frac{\theta_{12}^1 }{z_{12}^{n+1}} \, n, 
\qquad
D_2^1\,D_2^2\, D_2^3 \,
D_2^4 \, \frac{\theta^{1}_{12} \, \theta_{12}^{3} \, \theta_{12}^{4}}{z_{12}^n} 
=  
\frac{ \theta_{12}^2}{z_{12}^{n+1}} \, n,
\nonu \\
D_2^1\, D_2^2\, D_2^3 \,
D_2^4 \, \frac{ \theta_{12}^1 \, \theta_{12}^{4} \, \theta_{12}^{2}}{z_{12}^n} 
 & = & 
\frac{  \theta_{12}^3}{z_{12}^{n+1}} \, n,
\qquad
D_2^1\, D_2^2\, D_2^3 \,
D_2^4 \, \frac{\theta_{12}^1\, \theta_{12}^{2} \, \theta_{12}^{3}}{z_{12}^n} 
=  
\frac{ \theta_{12}^{4}}{z_{12}^{n+1}} \, n.
\nonu
\eea

For the two Grassmann coordinates, one calculates 
the following nonzero relations
\bea
D_1^1 \, \frac{\theta_{12}^1 \, \theta_{12}^{2}}{z_{12}^n}
& = & \frac{\theta_{12}^{2}}{z_{12}^{n}},
\qquad
D_1^1 \, \frac{\theta_{12}^1 \, \theta_{12}^{3}}{z_{12}^n}
= \frac{\theta_{12}^{3}}{z_{12}^{n}},
\qquad
D_1^1 \, \frac{\theta_{12}^1 \, \theta_{12}^{4}}{z_{12}^n}
 =  \frac{\theta_{12}^{4}}{z_{12}^{n}},
\nonu \\
D_1^1 \, \frac{\theta_{12}^2 \, \theta_{12}^{3}}{z_{12}^n}
& = & -\frac{\theta_{12}^{1} \, \theta_{12}^{2} \, \theta_{12}^{3}}{z_{12}^{n+1}}\,n ,
\qquad
D_1^1 \, \frac{\theta_{12}^2 \, \theta_{12}^{4}}{z_{12}^n}
 =  \frac{\theta_{12}^{1} \, \theta_{12}^{4} \, \theta_{12}^{2}}{z_{12}^{n+1}}\,n,
\nonu \\
D_1^1 \, \frac{\theta_{12}^3 \, \theta_{12}^{4}}{z_{12}^n}
& = & -\frac{\theta_{12}^{1} \, \theta_{12}^{3} \, \theta_{12}^{4}}{z_{12}^{n+1}}\,n,
\qquad
D_1^2 \, \frac{\theta_{12}^1 \, \theta_{12}^{2}}{z_{12}^n}
 =  -\frac{\theta_{12}^{1}}{z_{12}^{n}},
\qquad
D_1^2 \, \frac{\theta_{12}^1 \, \theta_{12}^{3}}{z_{12}^n}
= \frac{\theta_{12}^{1} \, \theta_{12}^{2} \, \theta_{12}^{3}}{z_{12}^{n+1}}\,n,
\nonu \\
D_1^2 \, \frac{\theta_{12}^1 \, \theta_{12}^{4}}{z_{12}^n}
& = & -\frac{\theta_{12}^{1} \, \theta_{12}^{4} \, \theta_{12}^{2}}{z_{12}^{n+1}}\,n,
\qquad
D_1^2 \, \frac{\theta_{12}^2 \, \theta_{12}^{3}}{z_{12}^n}
= \frac{\theta_{12}^{3}}{z_{12}^{n}},
\qquad
D_1^2 \, \frac{\theta_{12}^2 \, \theta_{12}^{4}}{z_{12}^n}
 =  \frac{\theta_{12}^{4}}{z_{12}^{n}},
\nonu \\
D_1^2 \, \frac{\theta_{12}^3 \, \theta_{12}^{4}}{z_{12}^n}
& = & \frac{\theta_{12}^{2} \, \theta_{12}^{4} \, \theta_{12}^{3}}{z_{12}^{n+1}}\,n,
\qquad
D_1^3 \, \frac{\theta_{12}^1 \, \theta_{12}^{2}}{z_{12}^n}
 =  -\frac{\theta_{12}^{1} \, \theta_{12}^{2} \, \theta_{12}^{3}}{z_{12}^{n+1}}\,n,
\qquad
D_1^3 \, \frac{\theta_{12}^1 \, \theta_{12}^{3}}{z_{12}^n}
= -\frac{\theta_{12}^{1}}{z_{12}^{n}},
\nonu \\
D_1^3 \, \frac{\theta_{12}^1 \, \theta_{12}^{4}}{z_{12}^n}
& = & \frac{\theta_{12}^{1} \, \theta_{12}^{3} \, \theta_{12}^{4}}{z_{12}^{n+1}}\,n ,
\qquad
D_1^3 \, \frac{\theta_{12}^2 \, \theta_{12}^{3}}{z_{12}^n}
= -\frac{\theta_{12}^{2}}{z_{12}^{n}},
\qquad
D_1^3 \, \frac{\theta_{12}^2 \, \theta_{12}^{4}}{z_{12}^n}
 =  -\frac{\theta_{12}^{2} \, \theta_{12}^{4} \, \theta_{12}^{3}}{z_{12}^{n+1}}\,n,
\nonu \\
D_1^3 \, \frac{\theta_{12}^3 \, \theta_{12}^{4}}{z_{12}^n}
& = & \frac{\theta_{12}^{4}}{z_{12}^{n}},
\qquad
D_1^4 \, \frac{\theta_{12}^1 \, \theta_{12}^{2}}{z_{12}^n}
 =  \frac{\theta_{12}^{1} \, \theta_{12}^{4} \, \theta_{12}^{2}}{z_{12}^{n+1}}\,n,
\qquad
D_1^4 \, \frac{\theta_{12}^1 \, \theta_{12}^{3}}{z_{12}^n}
= -\frac{\theta_{12}^{1} \, \theta_{12}^{3} \, \theta_{12}^{4}}{z_{12}^{n+1}}\,n,
\nonu \\
D_1^4 \, \frac{\theta_{12}^1 \, \theta_{12}^{4}}{z_{12}^n}
& = & -\frac{\theta_{12}^{1}}{z_{12}^{n}},
\qquad
D_1^4 \, \frac{\theta_{12}^2 \, \theta_{12}^{3}}{z_{12}^n}
= \frac{\theta_{12}^{2} \, \theta_{12}^{4} \, \theta_{12}^{3}}{z_{12}^{n+1}}\,n,
\qquad
D_1^4 \, \frac{\theta_{12}^2 \, \theta_{12}^{4}}{z_{12}^n}
 =  -\frac{\theta_{12}^{2}}{z_{12}^{n}},
\nonu \\
D_1^4 \, \frac{\theta_{12}^3 \, \theta_{12}^{4}}{z_{12}^n}
& = &  -\frac{\theta_{12}^{3}}{z_{12}^{n}},
\qquad
D_1^1 \, D_1^2 \,\frac{\theta_{12}^1 \, \theta_{12}^{2}}{z_{12}^n}
 =  -\frac{1}{z_{12}^{n}},
\qquad
D_1^1 \,  D_1^2 \,\frac{\theta_{12}^1 \, \theta_{12}^{3}}{z_{12}^n}
= \frac{\theta_{12}^{2} \, \theta_{12}^{3}}{z_{12}^{n+1}}\,n,
\nonu \\
D_1^1 \,  D_1^2 \,\frac{\theta_{12}^1 \, \theta_{12}^{4}}{z_{12}^n}
& = & \frac{\theta_{12}^{2} \, \theta_{12}^{4}}{z_{12}^{n+1}}\,n,
\qquad
D_1^1 \,  D_1^2 \,\frac{\theta_{12}^2 \, \theta_{12}^{3}}{z_{12}^n}
= -\frac{\theta_{12}^{1} \, \theta_{12}^{3}}{z_{12}^{n+1}}\,n,
\nonu \\
D_1^1 \,  D_1^2 \,\frac{\theta_{12}^2 \, \theta_{12}^{4}}{z_{12}^n}
& = & -\frac{\theta_{12}^{1} \, \theta_{12}^{4}}{z_{12}^{n+1}}\,n,
\qquad
D_1^1 \,  D_1^2 \,\frac{\theta_{12}^3 \, \theta_{12}^{4}}{z_{12}^n}
= \frac{\theta_{12}^{4-0}}{z_{12}^{n+2}}\,(n+1)n,
\nonu \\
D_1^1 \, D_1^3 \,\frac{\theta_{12}^1 \, \theta_{12}^{2}}{z_{12}^n}
& = & -\frac{\theta_{12}^{2} \, \theta_{12}^{3}}{z_{12}^{n+1}}\,n,
\qquad
D_1^1 \,  D_1^3 \,\frac{\theta_{12}^1 \, \theta_{12}^{3}}{z_{12}^n}
= -\frac{1}{z_{12}^{n+1}}\,n,
\qquad
D_1^1 \,  D_1^3 \,\frac{\theta_{12}^1 \, \theta_{12}^{4}}{z_{12}^n}
 =  \frac{\theta_{12}^{3} \, \theta_{12}^{4}}{z_{12}^{n+1}}\,n,
\nonu \\
D_1^1 \,  D_1^3 \,\frac{\theta_{12}^2 \, \theta_{12}^{3}}{z_{12}^n}
& = & \frac{\theta_{12}^{1} \, \theta_{12}^{2}}{z_{12}^{n+1}}\,n,
\nonu \\
D_1^1 \,  D_1^3 \,\frac{\theta_{12}^2 \, \theta_{12}^{4}}{z_{12}^n}
& = & -\frac{\theta_{12}^{4-0}}{z_{12}^{n+2}}\,(n+1)n,
\qquad
D_1^1 \,  D_1^3 \,\frac{\theta_{12}^3 \, \theta_{12}^{4}}{z_{12}^n}
= -\frac{\theta_{12}^{1} \, \theta_{12}^{4}}{z_{12}^{n+1}}\,n,
\nonu \\
D_1^1 \, D_1^4 \,\frac{\theta_{12}^1 \, \theta_{12}^{2}}{z_{12}^n}
& = & -\frac{\theta_{12}^{2} \, \theta_{12}^{4}}{z_{12}^{n+1}}\,n,
\qquad
D_1^1 \,  D_1^4 \,\frac{\theta_{12}^1 \, \theta_{12}^{3}}{z_{12}^n}
= -\frac{\theta_{12}^{3} \, \theta_{12}^{4}}{z_{12}^{n+1}}\,n,
\nonu \\
D_1^1 \,  D_1^4 \,\frac{\theta_{12}^1 \, \theta_{12}^{4}}{z_{12}^n}
& = & -\frac{1}{z_{12}^{n+1}}\,n,
\qquad
D_1^1 \,  D_1^4 \,\frac{\theta_{12}^2 \, \theta_{12}^{3}}{z_{12}^n}
= \frac{\theta_{12}^{4-0} }{z_{12}^{n+2}}\,(n+1)n,
\nonu \\
D_1^1 \,  D_1^4 \,\frac{\theta_{12}^2 \, \theta_{12}^{4}}{z_{12}^n}
& = & \frac{\theta_{12}^{1} \, \theta_{12}^{2}}{z_{12}^{n+1}}\,n,
\qquad
D_1^1 \,  D_1^4 \,\frac{\theta_{12}^3 \, \theta_{12}^{4}}{z_{12}^n}
= \frac{\theta_{12}^{1} \, \theta_{12}^{3}}{z_{12}^{n+1}}\,n,
\nonu \\
D_1^2 \, D_1^3 \,\frac{\theta_{12}^1 \, \theta_{12}^{2}}{z_{12}^n}
& = & \frac{\theta_{12}^{1}\,\theta_{12}^{3}}{z_{12}^{n+1}}\,n,
\qquad
D_1^2 \,  D_1^3 \,\frac{\theta_{12}^1 \, \theta_{12}^{3}}{z_{12}^n}
= -\frac{\theta_{12}^{1} \, \theta_{12}^{2}}{z_{12}^{n+1}}\,n,
\nonu \\
D_1^2 \,  D_1^3 \,\frac{\theta_{12}^1 \, \theta_{12}^{4}}{z_{12}^n}
& = & \frac{\theta_{12}^{4-0}}{z_{12}^{n+2}}\,(n+1)n,
\qquad
D_1^2 \,  D_1^3 \,\frac{\theta_{12}^2 \, \theta_{12}^{3}}{z_{12}^n}
= -\frac{1}{z_{12}^{n+1}}\,n,
\nonu \\
D_1^2 \,  D_1^3 \,\frac{\theta_{12}^2 \, \theta_{12}^{4}}{z_{12}^n}
& = & \frac{\theta_{12}^{3} \, \theta_{12}^{4}}{z_{12}^{n+1}}\,n,
\qquad
D_1^2 \,  D_1^3 \,\frac{\theta_{12}^3 \, \theta_{12}^{4}}{z_{12}^n}
= -\frac{\theta_{12}^{2} \, \theta_{12}^{4}}{z_{12}^{n+1}}\,n,
\nonu \\
D_1^2 \, D_1^4 \,\frac{\theta_{12}^1 \, \theta_{12}^{2}}{z_{12}^n}
& = & \frac{\theta_{12}^{1} \, \theta_{12}^{4}}{z_{12}^{n+1}}\,n,
\qquad
D_1^2 \,  D_1^4 \,\frac{\theta_{12}^1 \, \theta_{12}^{3}}{z_{12}^n}
= -\frac{\theta_{12}^{4-0}}{z_{12}^{n+2}}\,(n+1)n,
\nonu \\
D_1^2 \,  D_1^4 \,\frac{\theta_{12}^1 \, \theta_{12}^{4}}{z_{12}^n}
& = & -\frac{\theta_{12}^{1} \, \theta_{12}^{2}}{z_{12}^{n+1}}\,n,
\qquad
D_1^2 \,  D_1^4 \,\frac{\theta_{12}^2 \, \theta_{12}^{3}}{z_{12}^n}
= -\frac{\theta_{12}^{3} \, \theta_{12}^{4}}{z_{12}^{n+1}}\,n,
\nonu \\
D_1^2 \,  D_1^4 \,\frac{\theta_{12}^2 \, \theta_{12}^{4}}{z_{12}^n}
& = & -\frac{1}{z_{12}^{n}},
\qquad
D_1^2 \,  D_1^4 \,\frac{\theta_{12}^3 \, \theta_{12}^{4}}{z_{12}^n}
= \frac{\theta_{12}^{2} \, \theta_{12}^{3}}{z_{12}^{n+1}}\,n,
\nonu \\
D_1^3 \, D_1^4 \,\frac{\theta_{12}^1 \, \theta_{12}^{2}}{z_{12}^n}
& = & \frac{\theta_{12}^{4-0}}{z_{12}^{n+2}}\,(n+1)n,
\qquad
D_1^3 \,  D_1^4 \,\frac{\theta_{12}^1 \, \theta_{12}^{3}}{z_{12}^n}
= \frac{\theta_{12}^{1} \, \theta_{12}^{4}}{z_{12}^{n+1}}\,n,
\nonu \\
D_1^3 \,  D_1^4 \,\frac{\theta_{12}^1 \, \theta_{12}^{4}}{z_{12}^n}
& = & \frac{\theta_{12}^{1} \, \theta_{12}^{3}}{z_{12}^{n+1}}\,n,
\qquad
D_1^3 \,  D_1^4 \,\frac{\theta_{12}^2 \, \theta_{12}^{3}}{z_{12}^n}
= \frac{\theta_{12}^{2} \, \theta_{12}^{4}}{z_{12}^{n+1}}\,n,
\nonu \\
D_1^3 \,  D_1^4 \,\frac{\theta_{12}^2 \, \theta_{12}^{4}}{z_{12}^n}
& = & \frac{\theta_{12}^{2} \, \theta_{12}^{3}}{z_{12}^{n+1}}\,n,
\qquad
D_1^3 \,  D_1^4 \,\frac{\theta_{12}^3 \, \theta_{12}^{4}}{z_{12}^n}
= -\frac{1}{z_{12}^{n+1}}\,n,
\nonu \\
D_1^1 \, D_1^2 \, D_1^3 \, \frac{\theta_{12}^1 \, \theta_{12}^{2}}{z_{12}^n}
& = & \frac{\theta_{12}^{3}}{z_{12}^{n+1}}\,n,
\qquad
D_1^1 \,  D_1^2 \, D_1^3 \,\frac{\theta_{12}^1 \, \theta_{12}^{3}}{z_{12}^n}
= -\frac{\theta_{12}^{2}}{z_{12}^{n+1}}\,n,
\nonu \\
D_1^1 \,  D_1^2 \, D_1^3 \,\frac{\theta_{12}^1 \, \theta_{12}^{4}}{z_{12}^n}
& = & -\frac{\theta_{12}^{2} \, \theta_{12}^{4} \, \theta_{12}^{3}}{z_{12}^{n+2}}\,(n+1)n,
\qquad
D_1^1 \,  D_1^2 \, D_1^3 \,\frac{\theta_{12}^2 \, \theta_{12}^{3}}{z_{12}^n}
= \frac{\theta_{12}^{1}}{z_{12}^{n+1}}\,n,
\nonu \\
D_1^1 \,  D_1^2 \, D_1^3 \,\frac{\theta_{12}^2 \, \theta_{12}^{4}}{z_{12}^n}
& = & -\frac{\theta_{12}^{1} \, \theta_{12}^{3} \, \theta_{12}^{4}}{z_{12}^{n+2}}\,(n+1)n,
\qquad
D_1^1 \,  D_1^2 \, D_1^3 \,\frac{\theta_{12}^3 \, \theta_{12}^{4}}{z_{12}^n}
= -\frac{\theta_{12}^{1} \, \theta_{12}^{4} \, \theta_{12}^{2}}{z_{12}^{n+2}}\,(n+1)n,
\nonu \\
D_1^1 \, D_1^4 \, D_1^2 \, \frac{\theta_{12}^1 \, \theta_{12}^{2}}{z_{12}^n}
& = & -\frac{\theta_{12}^{4}}{z_{12}^{n+1}}\,n,
\qquad
D_1^1 \,  D_1^4 \, D_1^2 \,\frac{\theta_{12}^1 \, \theta_{12}^{3}}{z_{12}^n}
= -\frac{\theta_{12}^{2} \, \theta_{12}^{4} \, \theta_{12}^{3}}{z_{12}^{n+2}}\,(n+1)n,
\nonu \\
D_1^1 \,  D_1^4 \, D_1^2 \,\frac{\theta_{12}^1 \, \theta_{12}^{4}}{z_{12}^n}
& = & \frac{\theta_{12}^{2}}{z_{12}^{n+1}}\,n,
\qquad
D_1^1 \,  D_1^4 \, D_1^2 \,\frac{\theta_{12}^2 \, \theta_{12}^{3}}{z_{12}^n}
= -\frac{\theta_{12}^{1} \, \theta_{12}^{3} \, \theta_{12}^{4}}{z_{12}^{n+2}}\,(n+1)n,
\nonu \\
D_1^1 \,  D_1^4 \, D_1^2 \,\frac{\theta_{12}^2 \, \theta_{12}^{4}}{z_{12}^n}
& = & -\frac{\theta_{12}^{1}}{z_{12}^{n+1}}\,n,
\qquad
D_1^1 \,  D_1^4 \, D_1^2 \,\frac{\theta_{12}^3 \, \theta_{12}^{4}}{z_{12}^n}
= \frac{\theta_{12}^{1} \, \theta_{12}^{2} \, \theta_{12}^{3}}{z_{12}^{n+2}}\,(n+1)n,
\nonu \\
D_1^1 \, D_1^3 \, D_1^4 \, \frac{\theta_{12}^1 \, \theta_{12}^{2}}{z_{12}^n}
& = & -\frac{\theta_{12}^{2} \, \theta_{12}^{4} \, \theta_{12}^{3}}{z_{12}^{n+2}}\,(n+1)n,
\qquad
D_1^1 \,  D_1^3 \, D_1^4 \,\frac{\theta_{12}^1 \, \theta_{12}^{3}}{z_{12}^n}
= \frac{\theta_{12}^{4}}{z_{12}^{n+1}}\,n,
\nonu \\
D_1^1 \,  D_1^3 \, D_1^4 \,\frac{\theta_{12}^1 \, \theta_{12}^{4}}{z_{12}^n}
& = & -\frac{\theta_{12}^{3}}{z_{12}^{n+1}}\,n,
\qquad
D_1^1 \,  D_1^3 \, D_1^4 \,\frac{\theta_{12}^2 \, \theta_{12}^{3}}{z_{12}^n}
= \frac{\theta_{12}^{1} \, \theta_{12}^{4} \, \theta_{12}^{2}}{z_{12}^{n+2}}\,(n+1)n,
\nonu \\
D_1^1 \,  D_1^3 \, D_1^4 \,\frac{\theta_{12}^2 \, \theta_{12}^{4}}{z_{12}^n}
& = & \frac{\theta_{12}^{1} \, \theta_{12}^{2} \, \theta_{12}^{3}}{z_{12}^{n+2}}\,(n+1)n,
\qquad
D_1^1 \,  D_1^3 \, D_1^4 \,\frac{\theta_{12}^3 \, \theta_{12}^{4}}{z_{12}^n}
= \frac{\theta_{12}^{1}}{z_{12}^{n+1}}\,n,
\nonu \\
D_1^2 \, D_1^4 \, D_1^3 \, \frac{\theta_{12}^1 \, \theta_{12}^{2}}{z_{12}^n}
& = & \frac{\theta_{12}^{1} \, \theta_{12}^{3} \, \theta_{12}^{4}}{z_{12}^{n+2}}\,(n+1)n,
\qquad
D_1^2 \,  D_1^4 \, D_1^3 \,\frac{\theta_{12}^1 \, \theta_{12}^{3}}{z_{12}^n}
= \frac{\theta_{12}^{1} \, \theta_{12}^{4} \, \theta_{12}^{2}}{z_{12}^{n+2}}\,(n+1)n,
\nonu \\
D_1^2 \,  D_1^4 \, D_1^3 \,\frac{\theta_{12}^1 \, \theta_{12}^{4}}{z_{12}^n}
& = &\frac{\theta_{12}^{1} \, \theta_{12}^{2} \, \theta_{12}^{3}}{z_{12}^{n+2}}\,(n+1)n,
\qquad
D_1^2 \,  D_1^4 \, D_1^3 \,\frac{\theta_{12}^2 \, \theta_{12}^{3}}{z_{12}^n}
=- \frac{\theta_{12}^{4}}{z_{12}^{n+1}}\,n,
\nonu \\
D_1^2 \,  D_1^4 \, D_1^3 \,\frac{\theta_{12}^2 \, \theta_{12}^{4}}{z_{12}^n}
& = & \frac{\theta_{12}^{3}}{z_{12}^{n+1}}\,n,
\qquad
D_1^2 \,  D_1^4 \, D_1^3 \,\frac{\theta_{12}^3 \, \theta_{12}^{4}}{z_{12}^n}
= -\frac{ \theta_{12}^{2}}{z_{12}^{n+1}}\,n,
\nonu \\
D_1^1 \, D_1^2 \, D_1^3 \, D_1^4 \, \frac{\theta_{12}^1 \, \theta_{12}^{2}}{z_{12}^n}
& = & -\frac{\theta_{12}^{3}\,\theta_{12}^{4}}{z_{12}^{n+2}}\,(n+1)n,
\qquad
D_1^1 \,  D_1^2 \, D_1^3 \, D_1^4 \, \frac{\theta_{12}^1 \, \theta_{12}^{3}}{z_{12}^n}
= -\frac{\theta_{12}^{4}\,\theta_{12}^{2}}{z_{12}^{n+2}}\,(n+1)n,
\nonu \\
D_1^1 \,  D_1^2 \, D_1^3 \, D_1^4 \, \frac{\theta_{12}^1 \, \theta_{12}^{4}}{z_{12}^n}
& = & -\frac{\theta_{12}^{2}\,\theta_{12}^{3}}{z_{12}^{n+2}}\,(n+1)n,
\qquad
D_1^1 \,  D_1^2 \, D_1^3 \, D_1^4 \, \frac{\theta_{12}^2 \, \theta_{12}^{3}}{z_{12}^n}
= -\frac{\theta_{12}^{1}\,\theta_{12}^{4}}{z_{12}^{n+2}}\,(n+1)n,
\nonu \\
D_1^1 \,  D_1^2 \, D_1^3 \, D_1^4 \, \frac{\theta_{12}^2 \, \theta_{12}^{4}}{z_{12}^n}
& = & -\frac{\theta_{12}^{3}\,\theta_{12}^{1}}{z_{12}^{n+2}}\,(n+1)n ,
\qquad
D_1^1 \,  D_1^2 \, D_1^3 \, D_1^4 \,\frac{\theta_{12}^3 \, \theta_{12}^{4}}{z_{12}^n}
= -\frac{\theta_{12}^{1}\,\theta_{12}^{2}}{z_{12}^{n+2}}\,(n+1)n .
\nonu 
\eea
Similarly, one has
\bea
D_2^1 \, \frac{\theta_{12}^1 \, \theta_{12}^{2}}{z_{12}^n}
& = & -\frac{\theta_{12}^{2}}{z_{12}^{n}},
\qquad
D_2^1 \, \frac{\theta_{12}^1 \, \theta_{12}^{3}}{z_{12}^n}
= -\frac{\theta_{12}^{3}}{z_{12}^{n}},
\qquad
D_2^1 \, \frac{\theta_{12}^1 \, \theta_{12}^{4}}{z_{12}^n}
 =  -\frac{\theta_{12}^{4}}{z_{12}^{n}},
\nonu \\
D_2^1 \, \frac{\theta_{12}^2 \, \theta_{12}^{3}}{z_{12}^n}
 & = & -\frac{\theta_{12}^{1} \, \theta_{12}^{2} \, \theta_{12}^{3}}{z_{12}^{n}}\,n,
\qquad
D_2^1 \, \frac{\theta_{12}^2 \, \theta_{12}^{4}}{z_{12}^n}
 =  \frac{\theta_{12}^{1} \, \theta_{12}^{4} \, \theta_{12}^{2}}{z_{12}^{n}}\,n,
\nonu \\
D_2^1 \, \frac{\theta_{12}^3 \, \theta_{12}^{4}}{z_{12}^n}
& = & 
-\frac{\theta_{12}^{1} \, \theta_{12}^{3} \, \theta_{12}^{4}}{z_{12}^{n}}\,n,
\nonu \\
D_2^2 \, \frac{\theta_{12}^1 \, \theta_{12}^{2}}{z_{12}^n}
& = & \frac{\theta_{12}^{1}}{z_{12}^{n}},
\qquad
D_2^2 \, \frac{\theta_{12}^1 \, \theta_{12}^{3}}{z_{12}^n}
= \frac{\theta_{12}^{1} \, \theta_{12}^{2} \, \theta_{12}^{3}}{z_{12}^{n}}\,n,
\qquad
D_2^2 \, \frac{\theta_{12}^1 \, \theta_{12}^{4}}{z_{12}^n}
 = -\frac{\theta_{12}^{1} \, \theta_{12}^{4} \, \theta_{12}^{2}}{z_{12}^{n}}\,n,
\nonu \\
D_2^2 \, \frac{\theta_{12}^2 \, \theta_{12}^{3}}{z_{12}^n}
& = &  -\frac{\theta_{12}^{3}}{z_{12}^{n}},
\qquad
D_2^2 \, \frac{\theta_{12}^2 \, \theta_{12}^{4}}{z_{12}^n}
 =  -\frac{\theta_{12}^{4}}{z_{12}^{n}},
\qquad
D_2^2 \, \frac{\theta_{12}^3 \, \theta_{12}^{4}}{z_{12}^n}
= \frac{\theta_{12}^{2} \, \theta_{12}^{4} \, \theta_{12}^{3}}{z_{12}^{n}}\,n,
\nonu \\
D_2^3 \, \frac{\theta_{12}^1 \, \theta_{12}^{2}}{z_{12}^n}
& = & -\frac{\theta_{12}^{1} \, \theta_{12}^{2} \, \theta_{12}^{3}}{z_{12}^{n}}\,n,
\qquad
D_2^3 \, \frac{\theta_{12}^1 \, \theta_{12}^{3}}{z_{12}^n}
= \frac{\theta_{12}^{1}}{z_{12}^{n}},
\qquad
D_2^3 \, \frac{\theta_{12}^1 \, \theta_{12}^{4}}{z_{12}^n}
 =  -\frac{\theta_{12}^{1} \, \theta_{12}^{3} \, \theta_{12}^{4}}{z_{12}^{n}}\,n,
\nonu \\
D_2^3 \, \frac{\theta_{12}^2 \, \theta_{12}^{3}}{z_{12}^n}
& = & \frac{\theta_{12}^{2}}{z_{12}^{n}},
\qquad
D_2^3 \, \frac{\theta_{12}^2 \, \theta_{12}^{4}}{z_{12}^n}
 = -\frac{\theta_{12}^{2} \, \theta_{12}^{4} \, \theta_{12}^{3}}{z_{12}^{n}}\,n,
\qquad
D_2^3 \, \frac{\theta_{12}^3 \, \theta_{12}^{4}}{z_{12}^n}
= -\frac{\theta_{12}^{4}}{z_{12}^{n}},
\nonu \\
D_2^4 \, \frac{\theta_{12}^1 \, \theta_{12}^{2}}{z_{12}^n}
& = & -\frac{\theta_{12}^{1} \, \theta_{12}^{4} \, \theta_{12}^{2}}{z_{12}^{n}}\,n,
\qquad
D_2^4 \, \frac{\theta_{12}^1 \, \theta_{12}^{3}}{z_{12}^n}
= -\frac{\theta_{12}^{1} \, \theta_{12}^{3} \, \theta_{12}^{4}}{z_{12}^{n}}\,n,
\qquad
D_2^4 \, \frac{\theta_{12}^1 \, \theta_{12}^{4}}{z_{12}^n}
 =  \frac{\theta_{12}^{1}}{z_{12}^{n}},
\nonu \\
D_2^4 \, \frac{\theta_{12}^2 \, \theta_{12}^{3}}{z_{12}^n}
& = & \frac{\theta_{12}^{2} \, \theta_{12}^{4} \, \theta_{12}^{3}}{z_{12}^{n}}\,n,
\qquad
D_2^4 \, \frac{\theta_{12}^2 \, \theta_{12}^{4}}{z_{12}^n}
 =  \frac{\theta_{12}^{2}}{z_{12}^{n}},
\qquad
D_2^4 \, \frac{\theta_{12}^3 \, \theta_{12}^{4}}{z_{12}^n}
= \frac{\theta_{12}^{3}}{z_{12}^{n}},
\nonu \\
D_2^1 \, D_2^2 \,\frac{\theta_{12}^1 \, \theta_{12}^{2}}{z_{12}^n}
& = & -\frac{1}{z_{12}^{n}},
\qquad
D_2^1 \,  D_2^2 \,\frac{\theta_{12}^1 \, \theta_{12}^{3}}{z_{12}^n}
= -\frac{\theta_{12}^{2} \, \theta_{12}^{3}}{z_{12}^{n+1}}\,n,
\qquad
D_2^1 \,  D_2^2 \,\frac{\theta_{12}^1 \, \theta_{12}^{4}}{z_{12}^n}
 = -\frac{\theta_{12}^{2} \, \theta_{12}^{4}}{z_{12}^{n+1}}\,n,
\nonu \\
D_2^1 \,  D_2^2 \,\frac{\theta_{12}^2 \, \theta_{12}^{3}}{z_{12}^n}
& = &  \frac{\theta_{12}^{1} \, \theta_{12}^{3}}{z_{12}^{n+1}}\,n,
\qquad
D_2^1 \,  D_2^2 \,\frac{\theta_{12}^2 \, \theta_{12}^{4}}{z_{12}^n}
 =  \frac{\theta_{12}^{1} \, \theta_{12}^{4}}{z_{12}^{n+1}}\,n,
\nonu \\
D_2^1 \,  D_2^2 \,\frac{\theta_{12}^3 \, \theta_{12}^{4}}{z_{12}^n}
& = & \frac{\theta_{12}^{4-0}}{z_{12}^{n+2}}\,(n+1)n,
\nonu \\
D_2^1 \, D_2^3 \,\frac{\theta_{12}^1 \, \theta_{12}^{2}}{z_{12}^n}
& = & -\frac{\theta_{12}^{2} \, \theta_{12}^{3}}{z_{12}^{n+1}}\,n,
\qquad
D_2^1 \,  D_2^3 \,\frac{\theta_{12}^1 \, \theta_{12}^{3}}{z_{12}^n}
= \frac{1}{z_{12}^{n}},
\nonu \\
D_2^1 \,  D_2^3 \,\frac{\theta_{12}^1 \, \theta_{12}^{4}}{z_{12}^n}
& = & -\frac{\theta_{12}^{3} \, \theta_{12}^{4}}{z_{12}^{n+1}}\,n,
\qquad
D_2^1 \,  D_2^3 \,\frac{\theta_{12}^2 \, \theta_{12}^{3}}{z_{12}^n}
= -\frac{\theta_{12}^{1} \, \theta_{12}^{2}}{z_{12}^{n+1}}\,n,
\nonu \\
D_2^1 \,  D_2^3 \,\frac{\theta_{12}^2 \, \theta_{12}^{4}}{z_{12}^n}
& = & -\frac{\theta_{12}^{4-0}}{z_{12}^{n+2}}\,(n+1)n,
\qquad
D_2^1 \,  D_2^3 \,\frac{\theta_{12}^3 \, \theta_{12}^{4}}{z_{12}^n}
= \frac{\theta_{12}^{1} \, \theta_{12}^{4}}{z_{12}^{n+1}}\,n,
\nonu \\
D_2^1 \, D_2^4 \,\frac{\theta_{12}^1 \, \theta_{12}^{2}}{z_{12}^n}
& = & \frac{\theta_{12}^{2} \, \theta_{12}^{4}}{z_{12}^{n+1}}\,n,
\qquad
D_2^1 \,  D_2^4 \,\frac{\theta_{12}^1 \, \theta_{12}^{3}}{z_{12}^n}
= \frac{\theta_{12}^{3} \, \theta_{12}^{4}}{z_{12}^{n+1}}\,n,
\nonu \\
D_2^1 \,  D_2^4 \,\frac{\theta_{12}^1 \, \theta_{12}^{4}}{z_{12}^n}
& = & -\frac{1}{z_{12}^{n}},
\qquad
D_2^1 \,  D_2^4 \,\frac{\theta_{12}^2 \, \theta_{12}^{3}}{z_{12}^n}
= \frac{\theta_{12}^{4-0}}{z_{12}^{n+2}}\,(n+1)n,
\nonu \\
D_2^1 \,  D_2^4 \,\frac{\theta_{12}^2 \, \theta_{12}^{4}}{z_{12}^n}
& = & -\frac{\theta_{12}^{1} \, \theta_{12}^{2}}{z_{12}^{n+1}}\,n,
\qquad
D_2^1 \,  D_2^4 \,\frac{\theta_{12}^3 \, \theta_{12}^{4}}{z_{12}^n}
= -\frac{\theta_{12}^{1} \, \theta_{12}^{3}}{z_{12}^{n+1}}\,n,
\nonu \\
D_2^2 \, D_2^3 \,\frac{\theta_{12}^1 \, \theta_{12}^{2}}{z_{12}^n}
& = & -\frac{\theta_{12}^{1} \, \theta_{12}^{3}}{z_{12}^{n+1}}\,n,
\qquad
D_2^2 \,  D_2^3 \,\frac{\theta_{12}^1 \, \theta_{12}^{3}}{z_{12}^n}
= \frac{\theta_{12}^{1} \, \theta_{12}^{2}}{z_{12}^{n+1}}\,n,
\nonu \\
D_2^2 \,  D_2^3 \,\frac{\theta_{12}^1 \, \theta_{12}^{4}}{z_{12}^n}
& = & \frac{\theta_{12}^{4-0}}{z_{12}^{n+2}}\,(n+1)n,
\qquad
D_2^2 \,  D_2^3 \,\frac{\theta_{12}^2 \, \theta_{12}^{3}}{z_{12}^n}
= -\frac{1}{z_{12}^{n}} ,
\nonu \\
D_2^2 \,  D_2^3 \,\frac{\theta_{12}^2 \, \theta_{12}^{4}}{z_{12}^n}
& = & -\frac{\theta_{12}^{3} \, \theta_{12}^{4}}{z_{12}^{n+1}}\,n,
\qquad
D_2^2 \,  D_2^3 \,\frac{\theta_{12}^3 \, \theta_{12}^{4}}{z_{12}^n}
= \frac{\theta_{12}^{2} \, \theta_{12}^{4}}{z_{12}^{n+1}}\,n,
\nonu \\
D_2^2 \, D_2^4 \,\frac{\theta_{12}^1 \, \theta_{12}^{2}}{z_{12}^n}
& = & -\frac{\theta_{12}^{1} \, \theta_{12}^{4}}{z_{12}^{n+1}}\,n,
\qquad
D_2^2 \,  D_2^4 \,\frac{\theta_{12}^1 \, \theta_{12}^{3}}{z_{12}^n}
= -\frac{\theta_{12}^{4-0}}{z_{12}^{n+2}}\,(n+1)n ,
\nonu \\
D_2^2 \,  D_2^4 \,\frac{\theta_{12}^1 \, \theta_{12}^{4}}{z_{12}^n}
& = & \frac{\theta_{12}^{1} \, \theta_{12}^{2}}{z_{12}^{n+1}}\,n,
\qquad
D_2^2 \,  D_2^4 \,\frac{\theta_{12}^2 \, \theta_{12}^{3}}{z_{12}^n}
= \frac{\theta_{12}^{3} \, \theta_{12}^{4}}{z_{12}^{n+1}}\,n,
\nonu \\
D_2^2 \,  D_2^4 \,\frac{\theta_{12}^2 \, \theta_{12}^{4}}{z_{12}^n}
& = & -\frac{1}{z_{12}^{n}},
\qquad
D_2^2 \,  D_2^4 \,\frac{\theta_{12}^3 \, \theta_{12}^{4}}{z_{12}^n}
= -\frac{\theta_{12}^{2} \, \theta_{12}^{3}}{z_{12}^{n+1}}\,n,
\nonu \\
D_2^3 \, D_2^4 \,\frac{\theta_{12}^1 \, \theta_{12}^{2}}{z_{12}^n}
& = & \frac{\theta_{12}^{4-0}}{z_{12}^{n+2}}\,(n+1)n  ,
\qquad
D_2^3 \,  D_2^4 \,\frac{\theta_{12}^1 \, \theta_{12}^{3}}{z_{12}^n}
= -\frac{\theta_{12}^{1} \, \theta_{12}^{4}}{z_{12}^{n+1}}\,n,
\nonu \\
D_2^3 \,  D_2^4 \,\frac{\theta_{12}^1 \, \theta_{12}^{4}}{z_{12}^n}
& = & \frac{\theta_{12}^{1} \, \theta_{12}^{3}}{z_{12}^{n+1}}\,n,
\qquad
D_2^3 \,  D_2^4 \,\frac{\theta_{12}^2 \, \theta_{12}^{3}}{z_{12}^n}
= -\frac{\theta_{12}^{2} \, \theta_{12}^{4}}{z_{12}^{n+1}}\,n,
\nonu \\
D_2^3 \,  D_2^4 \,\frac{\theta_{12}^2 \, \theta_{12}^{4}}{z_{12}^n}
& = & \frac{\theta_{12}^{2} \, \theta_{12}^{3}}{z_{12}^{n+1}}\,n,
\qquad
D_2^3 \,  D_2^4 \,\frac{\theta_{12}^3 \, \theta_{12}^{4}}{z_{12}^n}
= -\frac{1}{z_{12}^{n}},
\nonu \\
D_2^1 \, D_2^2 \, D_2^3 \, \frac{\theta_{12}^1 \, \theta_{12}^{2}}{z_{12}^n}
& = & \frac{\theta_{12}^{3} }{z_{12}^{n+1}}\,n,
\qquad
D_2^1 \,  D_2^2 \, D_2^3 \,\frac{\theta_{12}^1 \, \theta_{12}^{3}}{z_{12}^n}
= -\frac{\theta_{12}^{2}}{z_{12}^{n+1}}\,n,
\nonu \\
D_2^1 \,  D_2^2 \, D_2^3 \,\frac{\theta_{12}^1 \, \theta_{12}^{4}}{z_{12}^n}
& = & \frac{\theta_{12}^{2} \, \theta_{12}^{4} \, \theta_{12}^{3}}{z_{12}^{n+1}}\,n,
\qquad
D_2^1 \,  D_2^2 \, D_2^3 \,\frac{\theta_{12}^2 \, \theta_{12}^{3}}{z_{12}^n}
= \frac{\theta_{12}^{1}}{z_{12}^{n+2}}\,(n+1)n,
\nonu \\
D_2^1 \,  D_2^2 \, D_2^3 \,\frac{\theta_{12}^2 \, \theta_{12}^{4}}{z_{12}^n}
& = & \frac{\theta_{12}^{1} \,\theta_{12}^{3} \,\theta_{12}^{4}}{z_{12}^{n+2}}\,(n+1)n,
\qquad
D_2^1 \,  D_2^2 \, D_2^3 \,\frac{\theta_{12}^3 \, \theta_{12}^{4}}{z_{12}^n}
= \frac{\theta_{12}^{1} \,\theta_{12}^{4} \,\theta_{12}^{2}}{z_{12}^{n+2}}\,(n+1)n,
\nonu \\
D_2^1 \, D_2^4 \, D_2^2 \, \frac{\theta_{12}^1 \, \theta_{12}^{2}}{z_{12}^n}
& = &  -\frac{\theta_{12}^{4}}{z_{12}^{n+1}}\,n,
\qquad
D_2^1 \,  D_2^4 \, D_2^2 \,\frac{\theta_{12}^1 \, \theta_{12}^{3}}{z_{12}^n}
= \frac{\theta_{12}^{2} \,\theta_{12}^{4} \,\theta_{12}^{3}}{z_{12}^{n+2}}\,(n+1)n,
\nonu \\
D_2^1 \,  D_2^4 \, D_2^2 \,\frac{\theta_{12}^1 \, \theta_{12}^{4}}{z_{12}^n}
& = & \frac{\theta_{12}^{2} }{z_{12}^{n+1}}\,n ,
\qquad
D_2^1 \,  D_2^4 \, D_2^2 \,\frac{\theta_{12}^2 \, \theta_{12}^{3}}{z_{12}^n}
= \frac{\theta_{12}^{1} \, \theta_{12}^{3} \, \theta_{12}^{4} }{z_{12}^{n+2}}\,(n+1)n ,
\nonu \\
D_2^1 \,  D_2^4 \, D_2^2 \,\frac{\theta_{12}^2 \, \theta_{12}^{4}}{z_{12}^n}
& = & -\frac{\theta_{12}^{1}}{z_{12}^{n+1}}\,n,
\qquad
D_2^1 \,  D_2^4 \, D_2^2 \,\frac{\theta_{12}^3 \, \theta_{12}^{4}}{z_{12}^n}
= -\frac{\theta_{12}^{1} \, \theta_{12}^{2} \, \theta_{12}^{3} }{z_{12}^{n+2}}\,(n+1)n ,
\nonu \\
D_2^1 \, D_2^3 \, D_2^4 \, \frac{\theta_{12}^1 \, \theta_{12}^{2}}{z_{12}^n}
& = & \frac{\theta_{12}^{2} \, \theta_{12}^{4} \, \theta_{12}^{3} }{z_{12}^{n+2}}\,(n+1)n ,
\qquad
D_2^1 \,  D_2^3 \, D_2^4 \,\frac{\theta_{12}^1 \, \theta_{12}^{3}}{z_{12}^n}
= \frac{\theta_{12}^{4}}{z_{12}^{n+1}}\,n,
\nonu \\
D_2^1 \,  D_2^3 \, D_2^4 \,\frac{\theta_{12}^1 \, \theta_{12}^{4}}{z_{12}^n}
& = & -\frac{\theta_{12}^{3}}{z_{12}^{n+2}}\,(n+1)n ,
\qquad
D_2^1 \,  D_2^3 \, D_2^4 \,\frac{\theta_{12}^2 \, \theta_{12}^{3}}{z_{12}^n}
= \frac{\theta_{12}^{1} \, \theta_{12}^{4} \, \theta_{12}^{2} }{z_{12}^{n+2}}\,(n+1)n ,
\nonu \\
D_2^1 \,  D_2^3 \, D_2^4 \,\frac{\theta_{12}^2 \, \theta_{12}^{4}}{z_{12}^n}
& = & -\frac{\theta_{12}^{1} \, \theta_{12}^{2} \, \theta_{12}^{3} }{z_{12}^{n+2}}\,(n+1)n ,
\qquad
D_2^1 \,  D_2^3 \, D_2^4 \,\frac{\theta_{12}^3 \, \theta_{12}^{4}}{z_{12}^n}
= \frac{\theta_{12}^{1}}{z_{12}^{n+1}}\,n,
\nonu \\
D_2^2 \, D_2^4 \, D_2^3 \, \frac{\theta_{12}^1 \, \theta_{12}^{2}}{z_{12}^n}
& = & -\frac{\theta_{12}^{1} \, \theta_{12}^{3} \, \theta_{12}^{4} }{z_{12}^{n+2}}\,(n+1)n ,
\qquad
D_2^2 \,  D_2^4 \, D_2^3 \,\frac{\theta_{12}^1 \, \theta_{12}^{3}}{z_{12}^n}
= -\frac{\theta_{12}^{1} \, \theta_{12}^{4} \, \theta_{12}^{2} }{z_{12}^{n+2}}\,(n+1)n ,
\nonu \\
D_2^2 \,  D_2^4 \, D_2^3 \,\frac{\theta_{12}^1 \, \theta_{12}^{4}}{z_{12}^n}
& = &- \frac{\theta_{12}^{1} \, \theta_{12}^{2} \, \theta_{12}^{3}}{z_{12}^{n+2}}\,(n+1)n,
\qquad
D_2^2 \,  D_2^4 \, D_2^3 \,\frac{\theta_{12}^2 \, \theta_{12}^{3}}{z_{12}^n}
= -\frac{\theta_{12}^{4}}{z_{12}^{n+1}}\,n,
\nonu \\
D_2^2 \,  D_2^4 \, D_2^3 \,\frac{\theta_{12}^2 \, \theta_{12}^{4}}{z_{12}^n}
& = & \frac{\theta_{12}^{3}}{z_{12}^{n+2}}\,(n+1)n ,
\qquad
D_2^2 \,  D_2^4 \, D_2^3 \,\frac{\theta_{12}^3 \, \theta_{12}^{4}}{z_{12}^n}
= -\frac{\theta_{12}^{2}}{z_{12}^{n+2}}\,(n+1)n ,
\nonu \\
D_2^1 \, D_2^2 \, D_2^3 \, D_2^4 \, \frac{\theta_{12}^1 \, \theta_{12}^{2}}{z_{12}^n}
& = & -\frac{\theta_{12}^{3}\,\theta_{12}^{4}}{z_{12}^{n+2}}\,(n+1)n,
\qquad
D_2^1 \,  D_2^2 \, D_2^3 \, D_2^4 \, \frac{\theta_{12}^1 \, \theta_{12}^{3}}{z_{12}^n}
= -\frac{\theta_{12}^{4}\,\theta_{12}^{2}}{z_{12}^{n+2}}\,(n+1)n,
\nonu \\
D_2^1 \,  D_2^2 \, D_2^3 \, D_2^4 \, \frac{\theta_{12}^1 \, \theta_{12}^{4}}{z_{12}^n}
& = & -\frac{\theta_{12}^{2}\,\theta_{12}^{3}}{z_{12}^{n+2}}\,(n+1)n,
\qquad
D_2^1 \,  D_2^2 \, D_2^3 \, D_2^4 \, \frac{\theta_{12}^2 \, \theta_{12}^{3}}{z_{12}^n}
= -\frac{\theta_{12}^{1}\,\theta_{12}^{4}}{z_{12}^{n+2}}\,(n+1)n,
\nonu \\
D_2^1 \,  D_2^2 \, D_2^3 \, D_2^4 \, \frac{\theta_{12}^2 \, \theta_{12}^{4}}{z_{12}^n}
& = & -\frac{\theta_{12}^{3}\,\theta_{12}^{1}}{z_{12}^{n+2}}\,(n+1)n,
\qquad
D_2^1 \,  D_2^2 \, D_2^3 \, D_2^4 \,\frac{\theta_{12}^3 \, \theta_{12}^{4}}{z_{12}^n}
= -\frac{\theta_{12}^{1}\,\theta_{12}^{2}}{z_{12}^{n+2}}\,(n+1)n.
\nonu 
\eea

For the single Grassmann coordinate, one has 
the following nonzero relations
\bea
D_1^1 \, \frac{\theta_{12}^1}{z_{12}^n} & = & \frac{1}{z_{12}^{n}},
\qquad
D_1^1 \, \frac{\theta_{12}^2}{z_{12}^n}  =  -\frac{\theta_{12}^{1} \, \theta_{12}^{2}}{z_{12}^{n}}\,n,
\qquad
D_1^1 \, \frac{\theta_{12}^3}{z_{12}^n}  =   -\frac{\theta_{12}^{1} \, \theta_{12}^{3}}{z_{12}^{n}}\,n,
\nonu \\
D_1^1 \, \frac{\theta_{12}^4}{z_{12}^n}  & = &  -\frac{\theta_{12}^{1} \, \theta_{12}^{4}}{z_{12}^{n}}\,n ,
\qquad
D_1^2 \, \frac{\theta_{12}^1}{z_{12}^n}  =   
\frac{\theta_{12}^{1} \, \theta_{12}^{2}}{z_{12}^{n}}\,n,
\qquad
D_1^2 \, \frac{\theta_{12}^2}{z_{12}^n}  =   \frac{1}{z_{12}^{n}}\,n,
\nonu \\
D_1^2 \, \frac{\theta_{12}^3}{z_{12}^n} & = &  -\frac{\theta_{12}^{2} \, \theta_{12}^{3}}{z_{12}^{n}}\,n,
\qquad
D_1^2 \, \frac{\theta_{12}^4}{z_{12}^n}  =   -\frac{\theta_{12}^{2} \, \theta_{12}^{4}}{z_{12}^{n}}\,n,
\qquad
D_1^3 \, \frac{\theta_{12}^1}{z_{12}^n}  =  \frac{\theta_{12}^{1} \, \theta_{12}^{3}}{z_{12}^{n}}\,n,
\nonu \\
D_1^3 \, \frac{\theta_{12}^2}{z_{12}^n}  & = & 
\frac{\theta_{12}^{2} \, \theta_{12}^{3}}{z_{12}^{n}}\,n ,
\qquad
D_1^3 \, \frac{\theta_{12}^3}{z_{12}^n}  =  \frac{1}{z_{12}^{n}}\,n,
\qquad
D_1^3 \, \frac{\theta_{12}^4}{z_{12}^n}  = -\frac{\theta_{12}^{3} \, \theta_{12}^{4}}{z_{12}^{n}}\,n ,
\nonu \\
D_1^4 \, \frac{\theta_{12}^1}{z_{12}^n} & = & \frac{\theta_{12}^{1} \, \theta_{12}^{4}}{z_{12}^{n}}\,n,
\qquad
D_1^4 \, \frac{\theta_{12}^2}{z_{12}^n}  =  \frac{\theta_{12}^{2} \, \theta_{12}^{4}}{z_{12}^{n}}\,n,
\qquad
D_1^4 \, \frac{\theta_{12}^3}{z_{12}^n}  =  
\frac{\theta_{12}^{3} \, \theta_{12}^{4}}{z_{12}^{n}}\,n,
\nonu \\
D_1^4 \, \frac{\theta_{12}^4}{z_{12}^n}  & = & \frac{1}{z_{12}^{n}}\,n,
\qquad
D_1^1 \, D_1^2 \, \frac{\theta_{12}^1}{z_{12}^n}  =  
\frac{\theta_{12}^{2}}{z_{12}^{n+1}}\,n,
\qquad
D_1^1 \,  D_1^2 \, \frac{\theta_{12}^2}{z_{12}^n}  = -\frac{\theta_{12}^{1}}{z_{12}^{n+1}}\,n ,
\nonu \\
D_1^1 \,  D_1^2 \, \frac{\theta_{12}^3}{z_{12}^n} & = & \frac{\theta_{12}^{1} \, \theta_{12}^{2} \, \theta_{12}^{3}}{z_{12}^{n+2}}\,(n+1)n ,
\qquad
D_1^1 \,  D_1^2 \, \frac{\theta_{12}^4}{z_{12}^n}  = -\frac{\theta_{12}^{1} \, \theta_{12}^{4} \, \theta_{12}^{2}}{z_{12}^{n+2}}\,(n+1)n  ,
\nonu \\
D_1^1 \, D_1^3 \, \frac{\theta_{12}^1}{z_{12}^n} & = & \frac{\theta_{12}^{3}}{z_{12}^{n+1}}\,n ,
\qquad
D_1^1 \,  D_1^3 \, \frac{\theta_{12}^2}{z_{12}^n}  = -\frac{\theta_{12}^{1} \, \theta_{12}^{2} \, \theta_{12}^{3}}{z_{12}^{n+2}}\,(n+1)n  ,
\nonu \\
D_1^1 \,  D_1^3 \, \frac{\theta_{12}^3}{z_{12}^n} & = & -\frac{\theta_{12}^{1}}{z_{12}^{n+1}}\,n,
\qquad
D_1^1 \,  D_1^3 \, \frac{\theta_{12}^4}{z_{12}^n}  = \frac{\theta_{12}^{1} \, \theta_{12}^{3} \, \theta_{12}^{4}}{z_{12}^{n+2}}\,(n+1)n  ,
\nonu \\
D_1^1 \, D_1^4 \, \frac{\theta_{12}^1}{z_{12}^n} & = & \frac{\theta_{12}^{4}}{z_{12}^{n+1}}\,n,
\qquad
D_1^1 \,  D_1^4 \, \frac{\theta_{12}^2}{z_{12}^n}  = \frac{\theta_{12}^{1} \, \theta_{12}^{4} \, \theta_{12}^{2}}{z_{12}^{n+2}}\,(n+1)n  ,
\nonu \\
D_1^1 \,  D_1^4 \, \frac{\theta_{12}^3}{z_{12}^n} & = & -\frac{\theta_{12}^{1} \, \theta_{12}^{3} \, \theta_{12}^{4}}{z_{12}^{n+2}}\,(n+1)n ,
\qquad
D_1^1 \,  D_1^4 \, \frac{\theta_{12}^4}{z_{12}^n}  = -\frac{\theta_{12}^{1}}{z_{12}^{n+1}}\,n ,
\nonu \\
D_1^2 \, D_1^3 \, \frac{\theta_{12}^1}{z_{12}^n} & = & \frac{\theta_{12}^{1} \, \theta_{12}^{2} \, \theta_{12}^{3}}{z_{12}^{n+2}}\,(n+1)n ,
\qquad
D_1^2 \,  D_1^3 \, \frac{\theta_{12}^2}{z_{12}^n}  = \frac{\theta_{12}^{3}}{z_{12}^{n+1}}\,n ,
\nonu \\
D_1^2 \,  D_1^3 \, \frac{\theta_{12}^3}{z_{12}^n} & = & -\frac{\theta_{12}^{2}}{z_{12}^{n+1}}\,n,
\qquad
D_1^2 \,  D_1^3 \, \frac{\theta_{12}^4}{z_{12}^n}  =  -\frac{\theta_{12}^{2} \, \theta_{12}^{4} \, \theta_{12}^{3}}{z_{12}^{n+2}}\,(n+1)n ,
\nonu \\
D_1^2 \, D_1^4 \, \frac{\theta_{12}^1}{z_{12}^n} & = & -\frac{\theta_{12}^{1} \, \theta_{12}^{4} \, \theta_{12}^{2}}{z_{12}^{n+2}}\,(n+1)n ,
\qquad
D_1^2 \,  D_1^4 \, \frac{\theta_{12}^2}{z_{12}^n}  =  \frac{\theta_{12}^{4}}{z_{12}^{n+1}}\,n,
\nonu \\
D_1^2 \,  D_1^4 \, \frac{\theta_{12}^3}{z_{12}^n} & = & \frac{\theta_{12}^{2} \, \theta_{12}^{4} \, \theta_{12}^{3}}{z_{12}^{n+2}}\,(n+1)n ,
\qquad
D_1^2 \,  D_1^4 \, \frac{\theta_{12}^4}{z_{12}^n}  = -\frac{\theta_{12}^{2}}{z_{12}^{n+1}}\,n ,
\nonu \\
D_1^3 \, D_1^4 \, \frac{\theta_{12}^1}{z_{12}^n} & = & \frac{\theta_{12}^{1} \, \theta_{12}^{3} \, \theta_{12}^{4}}{z_{12}^{n+2}}\,(n+1)n ,
\qquad
D_1^3 \,  D_1^4 \, \frac{\theta_{12}^2}{z_{12}^n}  = -\frac{\theta_{12}^{2} \, \theta_{12}^{4} \, \theta_{12}^{3}}{z_{12}^{n+2}}\,(n+1)n  ,
\nonu \\
D_1^3 \,  D_1^4 \, \frac{\theta_{12}^3}{z_{12}^n} & = & \frac{\theta_{12}^{4}}{z_{12}^{n+1}}\,n,
\qquad
D_1^3 \,  D_1^4 \, \frac{\theta_{12}^4}{z_{12}^n}  = -\frac{\theta_{12}^{3}}{z_{12}^{n+1}}\,n ,
\nonu \\
D_1^1 \, D_1^2 \, D_1^3 \, \frac{\theta_{12}^1}{z_{12}^n} & = & \frac{\theta_{12}^{2} \, \theta_{12}^{3}}{z_{12}^{n+2}}\,(n+1)n ,
\qquad
D_1^1 \,  D_1^2 \, D_1^3 \, \frac{\theta_{12}^2}{z_{12}^n}  = -\frac{\theta_{12}^{1} \, \theta_{12}^{3}}{z_{12}^{n+2}}\,(n+1)n  ,
\nonu \\
D_1^1 \,  D_1^2 \,  D_1^3 \,\frac{\theta_{12}^3}{z_{12}^n} & = & \frac{\theta_{12}^{1} \, \theta_{12}^{2}}{z_{12}^{n+2}}\,(n+1)n ,
\qquad
D_1^1 \,  D_1^2 \,  D_1^3 \, \frac{\theta_{12}^4}{z_{12}^n}  = \frac{\theta_{12}^{4-0}}{z_{12}^{n+3}}\,(n+2)(n+1)n  ,
\nonu \\
D_1^1 \, D_1^4 \, D_1^2 \, \frac{\theta_{12}^1}{z_{12}^n} & = & -\frac{\theta_{12}^{2} \, \theta_{12}^{4}}{z_{12}^{n+2}}\,(n+1)n ,
\qquad
D_1^1 \,  D_1^4 \, D_1^2 \, \frac{\theta_{12}^2}{z_{12}^n}  = \frac{\theta_{12}^{1} \, \theta_{12}^{4}}{z_{12}^{n+2}}\,(n+1)n  ,
\nonu \\
D_1^1 \,  D_1^4 \,  D_1^2 \,\frac{\theta_{12}^3}{z_{12}^n} & = & \frac{\theta_{12}^{4-0}}{z_{12}^{n+3}}\,(n+2)(n+1)n ,
\qquad
D_1^1 \,  D_1^4 \,  D_1^2 \, \frac{\theta_{12}^4}{z_{12}^n}  = -\frac{\theta_{12}^{1} \, \theta_{12}^{2}}{z_{12}^{n+2}}\,(n+1)n  ,
\nonu \\
D_1^1 \, D_1^3 \, D_1^4 \, \frac{\theta_{12}^1}{z_{12}^n} & = & \frac{\theta_{12}^{3} \, \theta_{12}^{4}}{z_{12}^{n+2}}\,(n+1)n ,
\qquad
D_1^1 \,  D_1^3 \, D_1^4 \, \frac{\theta_{12}^2}{z_{12}^n}  = \frac{\theta_{12}^{4-0}}{z_{12}^{n+3}}\,(n+2)(n+1)n  ,
\nonu \\
D_1^1 \,  D_1^3 \,  D_1^4 \,\frac{\theta_{12}^3}{z_{12}^n} & = & -\frac{\theta_{12}^{1} \, \theta_{12}^{4}}{z_{12}^{n+2}}\,(n+1)n ,
\qquad
D_1^1 \,  D_1^3 \,  D_1^4 \, \frac{\theta_{12}^4}{z_{12}^n}  = \frac{\theta_{12}^{1} \, \theta_{12}^{3}}{z_{12}^{n+2}}\,(n+1)n  ,
\nonu \\
D_1^2 \, D_1^4 \, D_1^3 \, \frac{\theta_{12}^1}{z_{12}^n} & = & \frac{\theta_{12}^{4-0}}{z_{12}^{n+3}}\,(n+2)(n+1)n ,
\qquad
D_1^2 \,  D_1^4 \, D_1^3 \, \frac{\theta_{12}^2}{z_{12}^n}  = -\frac{\theta_{12}^{3} \, \theta_{12}^{4}}{z_{12}^{n+2}}\,(n+1)n  ,
\nonu \\
D_1^2 \,  D_1^4 \,  D_1^3 \,\frac{\theta_{12}^3}{z_{12}^n} & = & \frac{\theta_{12}^{2} \, \theta_{12}^{4}}{z_{12}^{n+2}}\,(n+1)n  ,
\qquad
D_1^2 \,  D_1^4 \,  D_1^3 \, \frac{\theta_{12}^4}{z_{12}^n}  = -\frac{\theta_{12}^{2} \, \theta_{12}^{3}}{z_{12}^{n+2}}\,(n+1)n  ,
\nonu \\
D_1^1 \, D_1^2 \, D_1^3 \, D_1^4 \, \frac{\theta_{12}^1}{z_{12}^n} & = & \frac{\theta_{12}^2 \,\theta_{12}^4 \,\theta_{12}^3 }{z_{12}^{n+3}}\,(n+2)(n+1)n ,
\nonu \\
D_1^1 \, D_1^2 \,  D_1^3 \, D_1^4 \, \frac{\theta_{12}^2}{z_{12}^n} & = & \frac{\theta_{12}^1 \,\theta_{12}^3 \,\theta_{12}^4 }{z_{12}^{n+3}}\,(n+2)(n+1)n  ,
\nonu \\
D_1^1 \, D_1^2 \,  D_1^3 \,  D_1^4 \,\frac{\theta_{12}^3}{z_{12}^n} & = & \frac{\theta_{12}^1 \,\theta_{12}^4 \,\theta_{12}^2 }{z_{12}^{n+3}}\,(n+2)(n+1)n ,
\nonu \\
D_1^1 \, D_1^2 \,  D_1^3 \,  D_1^4 \, \frac{\theta_{12}^4}{z_{12}^n} & = & \frac{\theta_{12}^1 \,\theta_{12}^2 \,\theta_{12}^3 }{z_{12}^{n+3}}\,(n+2)(n+1)n  .
\nonu
\eea
Similarly one has
\bea
D_2^1 \, \frac{\theta_{12}^1}{z_{12}^n} & = & -\frac{1}{z_{12}^n},
\qquad
D_2^1 \, \frac{\theta_{12}^2}{z_{12}^n}  = -\frac{\theta_{12}^1 \, \theta_{12}^2}{z_{12}^{n+1}}\,n,
\qquad
D_2^1 \, \frac{\theta_{12}^3}{z_{12}^n}  = -\frac{\theta_{12}^1 \, \theta_{12}^3}{z_{12}^{n+1}}\,n,
\nonu \\
D_2^1 \, \frac{\theta_{12}^4}{z_{12}^n}  & = & 
-\frac{\theta_{12}^1 \, \theta_{12}^4}{z_{12}^{n+1}}\,n ,
\qquad
D_2^2 \, \frac{\theta_{12}^1}{z_{12}^n}  =  
\frac{\theta_{12}^1 \, \theta_{12}^2}{z_{12}^{n+1}}\,n,
\qquad
D_2^2 \, \frac{\theta_{12}^2}{z_{12}^n}  =  -\frac{1}{z_{12}^n} ,
\nonu \\
D_2^2 \, \frac{\theta_{12}^3}{z_{12}^n} & = & -\frac{\theta_{12}^2 \, \theta_{12}^3}{z_{12}^{n+1}}\,n,
\qquad
D_2^2 \, \frac{\theta_{12}^4}{z_{12}^n}  = -\frac{\theta_{12}^2 \, \theta_{12}^4}{z_{12}^{n+1}}\,n ,
\qquad
D_2^3 \, \frac{\theta_{12}^1}{z_{12}^n}  =  
\frac{\theta_{12}^1 \, \theta_{12}^3}{z_{12}^{n+1}}\,n,
\nonu \\
D_2^3 \, \frac{\theta_{12}^2}{z_{12}^n}  & = & 
\frac{\theta_{12}^2 \, \theta_{12}^3}{z_{12}^{n+1}}\,n,
\qquad
D_2^3 \, \frac{\theta_{12}^3}{z_{12}^n}  =   -\frac{1}{z_{12}^n},
\qquad
D_2^3 \, \frac{\theta_{12}^4}{z_{12}^n}  = -\frac{\theta_{12}^3 \, \theta_{12}^4}{z_{12}^{n+1}}\,n ,
\nonu \\
D_2^4 \, \frac{\theta_{12}^1}{z_{12}^n} & = & \frac{\theta_{12}^1 \, \theta_{12}^4}{z_{12}^{n+1}}\,n,
\qquad
D_2^4 \, \frac{\theta_{12}^2}{z_{12}^n}  = \frac{\theta_{12}^2 \, \theta_{12}^4}{z_{12}^{n+1}}\,n ,
\qquad
D_2^4 \, \frac{\theta_{12}^3}{z_{12}^n}  =  
\frac{\theta_{12}^3 \, \theta_{12}^4}{z_{12}^{n+1}}\,n,
\nonu \\
D_2^4 \, \frac{\theta_{12}^4}{z_{12}^n}  & = &   -\frac{1}{z_{12}^n},
\qquad
D_2^1 \, D_2^2 \, \frac{\theta_{12}^1}{z_{12}^n}  =  
-\frac{\theta_{12}^2 }{z_{12}^{n+1}}\,n,
\qquad
D_2^1 \,  D_2^2 \, \frac{\theta_{12}^2}{z_{12}^n}   =  
\frac{\theta_{12}^1}{z_{12}^{n+1}}\,n ,
\nonu \\
D_2^1 \,  D_2^2 \, \frac{\theta_{12}^3}{z_{12}^n} & = & \frac{\theta_{12}^{1} \, \theta_{12}^{2} \, \theta_{12}^{3}}{z_{12}^{n+2}}\,(n+1)n ,
\qquad
D_2^1 \,  D_2^2 \, \frac{\theta_{12}^4}{z_{12}^n}  =  -\frac{\theta_{12}^{1} \, \theta_{12}^{4} \, \theta_{12}^{2}}{z_{12}^{n+2}}\,(n+1)n ,
\nonu \\
D_2^1 \, D_2^3 \, \frac{\theta_{12}^1}{z_{12}^n} & = & -\frac{\theta_{12}^{3}}{z_{12}^{n+1}}\,n,
\qquad
D_2^1 \,  D_2^3 \, \frac{\theta_{12}^2}{z_{12}^n}  = -\frac{\theta_{12}^{1} \, \theta_{12}^{2} \, \theta_{12}^{3}}{z_{12}^{n+2}}\,(n+1)n  ,
\nonu \\
D_2^1 \,  D_2^3 \, \frac{\theta_{12}^3}{z_{12}^n} & = & \frac{\theta_{12}^{1}}{z_{12}^{n+1}}\,n,
\qquad
D_2^1 \,  D_2^3 \, \frac{\theta_{12}^4}{z_{12}^n}  =  \frac{\theta_{12}^{1} \, \theta_{12}^{3} \, \theta_{12}^{4}}{z_{12}^{n+2}}\,(n+1)n ,
\nonu \\
D_2^1 \, D_2^4 \, \frac{\theta_{12}^1}{z_{12}^n} & = & -\frac{\theta_{12}^4 }{z_{12}^{n+1}}\,n,
\qquad
D_2^1 \,  D_2^4 \, \frac{\theta_{12}^2}{z_{12}^n}  = \frac{\theta_{12}^{1} \, \theta_{12}^{4} \, \theta_{12}^{2}}{z_{12}^{n+2}}\,(n+1)n  ,
\nonu \\
D_2^1 \,  D_2^4 \, \frac{\theta_{12}^3}{z_{12}^n} & = & -\frac{\theta_{12}^{1} \, \theta_{12}^{3} \, \theta_{12}^{4}}{z_{12}^{n+2}}\,(n+1)n ,
\qquad
D_2^1 \,  D_2^4 \, \frac{\theta_{12}^4}{z_{12}^n}  = \frac{\theta_{12}^1 \,}{z_{12}^{n+1}}\,n,
\nonu \\
D_2^2 \, D_2^3 \, \frac{\theta_{12}^1}{z_{12}^n} & = & \frac{\theta_{12}^{1} \, \theta_{12}^{2} \, \theta_{12}^{3}}{z_{12}^{n+2}}\,(n+1)n ,
\qquad
D_2^2 \,  D_2^3 \, \frac{\theta_{12}^2}{z_{12}^n}  = -\frac{\theta_{12}^{3}}{z_{12}^{n+1}}\,n ,
\nonu \\
D_2^2 \,  D_2^3 \, \frac{\theta_{12}^3}{z_{12}^n} & = & \frac{\theta_{12}^{2}}{z_{12}^{n+1}}\,n ,
\qquad
D_2^2 \,  D_2^3 \, \frac{\theta_{12}^4}{z_{12}^n}  = -\frac{\theta_{12}^{2} \, \theta_{12}^{4} \, \theta_{12}^{3}}{z_{12}^{n+2}}\,(n+1)n  ,
\nonu \\
D_2^2 \, D_2^4 \, \frac{\theta_{12}^1}{z_{12}^n} & = & -\frac{\theta_{12}^{1} \, \theta_{12}^{4} \, \theta_{12}^{2}}{z_{12}^{n+2}}\,(n+1)n ,
\qquad
D_2^2 \,  D_2^4 \, \frac{\theta_{12}^2}{z_{12}^n}  = -\frac{\theta_{12}^{4}}{z_{12}^{n+1}}\,n,
\nonu \\
D_2^2 \,  D_2^4 \, \frac{\theta_{12}^3}{z_{12}^n} & = & \frac{\theta_{12}^{2} \, \theta_{12}^{4} \, \theta_{12}^{3}}{z_{12}^{n+2}}\,(n+1)n ,
\qquad
D_2^2 \,  D_2^4 \, \frac{\theta_{12}^4}{z_{12}^n}  = \frac{\theta_{12}^{2}}{z_{12}^{n+1}}\,n,
\nonu \\
D_2^3 \, D_2^4 \, \frac{\theta_{12}^1}{z_{12}^n} & = & \frac{\theta_{12}^{1} \, \theta_{12}^{3} \, \theta_{12}^{4}}{z_{12}^{n+2}}\,(n+1)n ,
\qquad
D_2^3 \,  D_2^4 \, \frac{\theta_{12}^2}{z_{12}^n}  =  -\frac{\theta_{12}^{2} \, \theta_{12}^{4} \, \theta_{12}^{3}}{z_{12}^{n+2}}\,(n+1)n ,
\nonu \\
D_2^3 \,  D_2^4 \, \frac{\theta_{12}^3}{z_{12}^n} & = & -\frac{\theta_{12}^{4}}{z_{12}^{n+1}}\,n,
\qquad
D_2^3 \,  D_2^4 \, \frac{\theta_{12}^4}{z_{12}^n}  = \frac{\theta_{12}^{3}}{z_{12}^{n+1}}\,n,
\nonu \\
D_2^1 \, D_2^2 \, D_2^3 \, \frac{\theta_{12}^1}{z_{12}^n} & = & -\frac{\theta_{12}^{2} \, \theta_{12}^{3}}{z_{12}^{n+2}}\,(n+1)n  ,
\qquad
D_2^1 \,  D_2^2 \, D_2^3 \, \frac{\theta_{12}^2}{z_{12}^n}  = \frac{\theta_{12}^{1} \, \theta_{12}^{3}}{z_{12}^{n+2}}\,(n+1)n   ,
\nonu \\
D_2^1 \,  D_2^2 \,  D_2^3 \,\frac{\theta_{12}^3}{z_{12}^n} & = & -\frac{\theta_{12}^{1} \, \theta_{12}^{2} }{z_{12}^{n+2}}\,(n+1)n  ,
\qquad
D_2^1 \,  D_2^2 \,  D_2^3 \, \frac{\theta_{12}^4}{z_{12}^n}  = -\frac{\theta_{12}^{4-0}}{z_{12}^{n+3}}\,(n+2)(n+1)n   ,
\nonu \\
D_2^1 \, D_2^4 \, D_2^2 \, \frac{\theta_{12}^1}{z_{12}^n} & = & \frac{\theta_{12}^{2} \, \theta_{12}^{4} }{z_{12}^{n+2}}\,(n+1)n  ,
\qquad
D_2^1 \,  D_2^4 \, D_2^2 \, \frac{\theta_{12}^2}{z_{12}^n}  =- \frac{\theta_{12}^{1} \, \theta_{12}^{4}}{z_{12}^{n+2}}\,(n+1)n  ,
\nonu \\
D_2^1 \,  D_2^4 \,  D_2^2 \,\frac{\theta_{12}^3}{z_{12}^n} & = & \frac{\theta_{12}^{4-0}}{z_{12}^{n+3}}\,(n+2)(n+1)n  ,
\qquad
D_2^1 \,  D_2^4 \,  D_2^2 \, \frac{\theta_{12}^4}{z_{12}^n}  = \frac{\theta_{12}^{1} \, \theta_{12}^{2}}{z_{12}^{n+2}}\,(n+1)n ,
\nonu \\
D_2^1 \, D_2^3 \, D_2^4 \, \frac{\theta_{12}^1}{z_{12}^n} & = & -\frac{\theta_{12}^{3} \, \theta_{12}^{4}}{z_{12}^{n+2}}\,(n+1)n  ,
\qquad
D_2^1 \,  D_2^3 \, D_2^4 \, \frac{\theta_{12}^2}{z_{12}^n}  = \frac{\theta_{12}^{4-0}}{z_{12}^{n+3}}\,(n+2)(n+1)n  ,
\nonu \\
D_2^1 \,  D_2^3 \,  D_2^4 \,\frac{\theta_{12}^3}{z_{12}^n} & = & \frac{\theta_{12}^{1} \, \theta_{12}^{4}}{z_{12}^{n+2}}\,(n+1)n  ,
\qquad
D_2^1 \,  D_2^3 \,  D_2^4 \, \frac{\theta_{12}^4}{z_{12}^n}  = -\frac{\theta_{12}^{1} \, \theta_{12}^{3}}{z_{12}^{n+2}}\,(n+1)n  ,
\nonu \\
D_2^2 \, D_2^4 \, D_2^3 \, \frac{\theta_{12}^1}{z_{12}^n} & = & \frac{\theta_{12}^{4-0}}{z_{12}^{n+3}}\,(n+2)(n+1)n  ,
\qquad
D_2^2 \,  D_2^4 \, D_2^3 \, \frac{\theta_{12}^2}{z_{12}^n}  = \frac{\theta_{12}^{3} \, \theta_{12}^{4}}{z_{12}^{n+2}}\,(n+1)n  ,
\nonu \\
D_2^2 \,  D_2^4 \,  D_2^3 \,\frac{\theta_{12}^3}{z_{12}^n} & = & -\frac{\theta_{12}^{2} \, \theta_{12}^{4}}{z_{12}^{n+2}}\,(n+1)n ,
\qquad
D_2^2 \,  D_2^4 \,  D_2^3 \, \frac{\theta_{12}^4}{z_{12}^n}  = \frac{\theta_{12}^{2} \, \theta_{12}^{3}}{z_{12}^{n+2}}\,(n+1)n ,
\nonu \\
D_2^1 \, D_2^2 \, D_2^3 \, D_2^4 \, \frac{\theta_{12}^1}{z_{12}^n} & = & -\frac{\theta_{12}^{2} \, \theta_{12}^{4} \, \theta_{12}^{3}}{z_{12}^{n+3}}\,(n+2)(n+1)n  ,
\nonu \\
D_2^1 \, D_2^2 \,  D_2^3 \, D_2^4 \, \frac{\theta_{12}^2}{z_{12}^n} & = & -\frac{\theta_{12}^{1} \, \theta_{12}^{3} \, \theta_{12}^{4}}{z_{12}^{n+3}}\,(n+2)(n+1)n  ,
\nonu \\
D_2^1 \, D_2^2 \,  D_2^3 \,  D_2^4 \,\frac{\theta_{12}^3}{z_{12}^n} & = & -\frac{\theta_{12}^{1} \, \theta_{12}^{4} \, \theta_{12}^{2}}{z_{12}^{n+3}}\,(n+2)(n+1)n ,
\nonu \\
D_2^1 \, D_2^2 \,  D_2^3 \,  D_2^4 \, \frac{\theta_{12}^4}{z_{12}^n} & = & -\frac{\theta_{12}^{1} \, \theta_{12}^{2} \, \theta_{12}^{3}}{z_{12}^{n+3}}\,(n+2)(n+1)n .
\nonu
\eea

Now one can calculate the following nonzero relations
\bea
D_1^1 \, \frac{1}{z_{12}^n} & = & -\frac{\theta_{12}^{1}}{z_{12}^{n+1}}\,n
,
\qquad
D_1^2 \, \frac{1}{z_{12}^n}  = -\frac{\theta_{12}^{2}}{z_{12}^{n+1}}\,n ,
\qquad
D_1^3 \, \frac{1}{z_{12}^n}  =  -\frac{\theta_{12}^{3}}{z_{12}^{n+1}}\,n, 
\nonu \\
D_1^4 \, \frac{1}{z_{12}^n}  & = &  -\frac{\theta_{12}^{4}}{z_{12}^{n+1}}\,n,
\qquad
D_1^1 \, D_1^2 \, \frac{1}{z_{12}^n}  =  
\frac{\theta_{12}^{1}\, \theta_{12}^{2}}{z_{12}^{n+2}}\,(n+1)n, 
\qquad
D_1^1 \, D_1^3 \, \frac{1}{z_{12}^n}  = \frac{\theta_{12}^{1}\, \theta_{12}^{3}}{z_{12}^{n+2}}\,(n+1)n ,
\nonu \\
D_1^1 \, D_1^4 \, \frac{1}{z_{12}^n} & = & \frac{\theta_{12}^{1}\, \theta_{12}^{4}}{z_{12}^{n+2}}\,(n+1)n, 
\qquad
D_1^2 \, D_1^3 \, \frac{1}{z_{12}^n}  =  \frac{\theta_{12}^{2}\, \theta_{12}^{3}}{z_{12}^{n+2}}\,(n+1)n,
\nonu \\
D_1^2 \, D_1^4 \, \frac{1}{z_{12}^n} & = & \frac{\theta_{12}^{2}\, \theta_{12}^{4}}{z_{12}^{n+2}}\,(n+1)n, 
\qquad
D_1^3 \, D_1^4 \, \frac{1}{z_{12}^n}  = \frac{\theta_{12}^{3}\, \theta_{12}^{4}}{z_{12}^{n+2}}\,(n+1)n ,
\nonu \\
D_1^1 \, D_1^2 \, D_1^3 \, \frac{1}{z_{12}^n} & = & -\frac{\theta_{12}^{1}\, \theta_{12}^{2}\, \theta_{12}^{3}}{z_{12}^{n+3}}\,(n+2)(n+1)n,
\qquad
D_1^1 \, D_1^4 \, D_1^2 \, \frac{1}{z_{12}^n}  =  -\frac{\theta_{12}^{1}\, \theta_{12}^{4}\, \theta_{12}^{2}}{z_{12}^{n+3}}\,(n+2)(n+1)n, 
\nonu \\
D_1^1 \, D_1^3 \, D_1^4 \, \frac{1}{z_{12}^n} & = & -\frac{\theta_{12}^{1}\, \theta_{12}^{3}\, \theta_{12}^{4}}{z_{12}^{n+3}}\,(n+2)(n+1)n, 
\qquad
D_1^2 \, D_1^4 \, D_1^3 \, \frac{1}{z_{12}^n}  =  -\frac{\theta_{12}^{2}\, \theta_{12}^{4}\, \theta_{12}^{3}}{z_{12}^{n+3}}\,(n+2)(n+1)n,
\nonu \\
D_1^1 \, D_1^2 \, D_1^3 \, D_1^4 \, \frac{1}{z_{12}^n}  & = &  \frac{\theta_{12}^{1}\,\theta_{12}^{2}\, \theta_{12}^{3}\, \theta_{12}^{4}}{z_{12}^{n+4}}\,(n+3)(n+2)(n+1)n.
\nonu
\eea

Similarly, one has 
\bea
D_2^1 \, \frac{1}{z_{12}^n} & = & -\frac{\theta_{12}^{1}}{z_{12}^{n+1}}\,n
,
\qquad
D_2^2 \, \frac{1}{z_{12}^n}  = -\frac{\theta_{12}^{2}}{z_{12}^{n+1}}\,n ,
\qquad
D_2^3 \, \frac{1}{z_{12}^n}  = -\frac{\theta_{12}^{3}}{z_{12}^{n+1}}\,n, 
\nonu \\
D_2^4 \, \frac{1}{z_{12}^n} & =  &-\frac{\theta_{12}^{4}}{z_{12}^{n+1}}\,n ,
\qquad
D_2^1 \, D_2^2 \, \frac{1}{z_{12}^n}  =  \frac{\theta_{12}^{1}\, \theta_{12}^{2}}{z_{12}^{n+2}}\,(n+1)n, 
\qquad
D_2^1 \, D_2^3 \, \frac{1}{z_{12}^n}  =  \frac{\theta_{12}^{1}\, \theta_{12}^{3}}{z_{12}^{n+2}}\,(n+1)n,
\nonu \\
D_2^1 \, D_2^4 \, \frac{1}{z_{12}^n} & = & \frac{\theta_{12}^{1}\, \theta_{12}^{4}}{z_{12}^{n+2}}\,(n+1)n,
\qquad
D_2^2 \, D_2^3 \, \frac{1}{z_{12}^n}  = \frac{\theta_{12}^{2}\, \theta_{12}^{3}}{z_{12}^{n+2}}\,(n+1)n ,
\nonu \\
D_2^2 \, D_2^4 \, \frac{1}{z_{12}^n} & = & \frac{\theta_{12}^{2}\, \theta_{12}^{4}}{z_{12}^{n+2}}\,(n+1)n, 
\qquad
D_2^3 \, D_2^4 \, \frac{1}{z_{12}^n}  = \frac{\theta_{12}^{3}\, \theta_{12}^{4}}{z_{12}^{n+2}}\,(n+1)n ,
\nonu \\
D_2^1 \, D_2^2 \, D_2^3 \, \frac{1}{z_{12}^n} & = & -\frac{\theta_{12}^{1}\, \theta_{12}^{2}\, \theta_{12}^{3}}{z_{12}^{n+3}}\,(n+2)(n+1)n, 
\qquad
D_2^1 \, D_2^4 \, D_2^2 \, \frac{1}{z_{12}^n}   =  
-\frac{\theta_{12}^{1}\, \theta_{12}^{4}\, \theta_{12}^{2}}{z_{12}^{n+3}}\,(n+2)(n+1)n,
\nonu \\
D_2^1 \, D_2^3 \, D_2^4 \, \frac{1}{z_{12}^n} & = & -\frac{\theta_{12}^{1}\, \theta_{12}^{3}\, \theta_{12}^{4}}{z_{12}^{n+3}}\,(n+2)(n+1)n, 
\qquad
D_2^2 \, D_2^4 \, D_2^3 \, \frac{1}{z_{12}^n}  =  
-\frac{\theta_{12}^{2}\, \theta_{12}^{4}\, \theta_{12}^{3}}{z_{12}^{n+3}}\,(n+2)(n+1)n,
\nonu \\
D_2^1 \, D_2^2 \, D_2^3 \, D_2^4 \, \frac{1}{z_{12}^n} & = &  \frac{\theta_{12}^{1}\,\theta_{12}^{2}\, \theta_{12}^{3}\, \theta_{12}^{4}}{z_{12}^{n+4}}\,(n+3)(n+2)(n+1)n.
\nonu
\eea
Furthermore, 
the product of mixed covariant derivatives
$D_1^i$ and $D_2^j$ 
acting on the fractional super coordinates
can be obtained from the above explicit results.

\section{The OPEs between$16$ higher spin  currents  in 
${\cal N}=2$ superspace}

In this Appendix, one presents the coefficients in section $6$
and some OPEs with the structure constants in ${\cal N}=2$ superspace.

\subsection{The structure constants in the subsection $6.3.1$ }

The coefficients appearing in (\ref{t1t1}) of 
the subsection $6.3.1$ are given by
\bea
c_{1} & = & \frac{2k\, N}{(2+k+N)},
\qquad
c_{2}  =  -\frac{(k-N)}{(2+k+N)^{2}},
\qquad 
c_{3}  =  \frac{(k-N)}{(2+k+N)^{2}},
\nonu \\
c_{4}  & = &   \frac{(k+N)}{(2+k+N)},
\qquad
c_{5}  =  -\frac{(k+N)}{(2+k+N)^{2}}, 
\qquad
c_{6}  =  \frac{(k-N)}{(2+k+N)^{2}},\nonu \\ 
c_{7} & = & -1,\qquad
 c_{8}   =  
\frac{1}{(2+k+N)},\qquad
 c_{9}  =  \frac{1}{(2+k+N)},\nonu \\
 c_{10}  & = &   \frac{1}{(2+k+N)},
\qquad
c_{11}  =  
\frac{-1}{(2+k+N)},\qquad
c_{12}  =  -\frac{2}{(2+k+N)^{2}},\nonu \\
c_{13} & = & 1,
\qquad 
c_{14}  =  -\frac{1}{(2+k+N)},
\qquad
c_{15}  =  \frac{1}{(2+k+N)},
\nonu \\
c_{16}  & = &   \frac{1}{(2+k+N)},
\qquad
c_{17}  =  -\frac{1}{(2+k+N)},
\qquad
 c_{18}  =  \frac{2}{(2+k+N)^{2}},
\nonu \\ 
c_{19} & = & \frac{(1+k+N)}{(2+k+N)},
\qquad
c_{20}  =  \frac{(1+N)}{(2+k+N)^{2}},
\qquad
c_{21}  =  \frac{(1+k)}{(2+k+N)^{2}},
\nonu \\
c_{22}  & = &   -\frac{1}{2(2+k+N)},
\qquad
c_{23}  =  \frac{1}{2(2+k+N)},\qquad
 c_{24}  =  -\frac{(1+k+N)}{(2+k+N)^{2}},
\nonumber\\
c_{25} & = & -\frac{(1+k+N)}{(2+k+N)^{2}},
\qquad 
c_{26}  =  \frac{(k-N)}{2(2+k+N)^{2}},
\qquad
c_{27}  =  \frac{(k-N)}{2(2+k+N)^{2}},
\nonu \\
c_{28}  & =  &  \frac{1}{(2+k+N)^{2}},
\qquad
c_{29}  =  \frac{1}{(2+k+N)^{2}},
\qquad 
c_{30}   =  -\frac{1}{(2+k+N)^{2}},
\nonu \\ 
c_{31} & = & -\frac{1}{(2+k+N)^{2}},
\qquad 
c_{32}  =  -\frac{1}{(2+k+N)^{2}},
\qquad
c_{33}  =  -\frac{1}{(2+k+N)^{2}}.
\nonumber
\eea

\subsection{The structure constants in the subsection $6.3.2$}

The coefficients appearing in (\ref{tuv}) of
the subsection $6.3.2$ are given by
\bea
c_{+1} & = & -\frac{4kN}{(2+k+N)^{2}},
\qquad
c_{+2}  =  \frac{4k}{(2+k+N)},
\qquad
c_{+3}  =  -\frac{4k}{(2+k+N)^{2}},
\nonu \\
c_{+4}  & = &   -\frac{4N}{(2+k+N)},
\qquad  
c_{+5}  =  \frac{4N}{(2+k+N)^{2}},
\qquad 
c_{+6}  =  -\frac{(k-N)}{2(2+k+N)},
\nonumber \\
 c_{+7} & = & \frac{(k-N)}{(2+k+N)^{2}},
 \qquad
 c_{+8}  =  \frac{(k-N)}{(2+k+N)^{2}},
\qquad
 c_{+9}  =  \frac{(k-N)}{(2+k+N)^{2}},
\nonu \\
c_{+10}  & = &  -\frac{(3k+k^{2}+N+3kN)}{(2+k+N)^{3}},
\qquad
 c_{+11}  =  \frac{(k+3N+3kN+N^{2})}{(2+k+N)^{3}},
\nonu \\
c_{+12}  & = &  -\frac{(4k+k^{2}+4N+6kN+N^{2})}{2(2+k+N)^{2}},
\qquad
c_{+13}  =  1,
\qquad
 c_{+14}  =  -\frac{2}{(2+k+N)},
\nonumber\\
 c_{+15} & = & -\frac{4}{(2+k+N)^{2}},
\qquad
 c_{+16}  =  -\frac{2}{(2+k+N)},
\qquad
 c_{+17}  =  \frac{2}{(2+k+N)},
\nonu \\
c_{+18}  & = &  \frac{2(1+k)}{(2+k+N)^{2}},
\qquad
c_{+19}  =  \frac{2(1+N)}{(2+k+N)^{2}},
\qquad
 c_{+20}  =  \frac{(k-N)}{(2+k+N)},
 \nonumber\\
 c_{+21} & = & \frac{2k}{(2+k+N)},
 \qquad
 c_{+22}  =  -\frac{2}{(2+k+N)^{2}},
 \qquad
 c_{+23}  =  -\frac{2}{(2+k+N)^{2}},
\nonu \\
 c_{+24}  & = &  -\frac{2(1+k)}{(2+k+N)^{3}},
 \qquad
c_{+25}  =  \frac{2(1+k)}{(2+k+N)^{3}},
  \qquad
c_{+26}  = - \frac{2k}{(2+k+N)^{2}},
  \nonumber\\
c_{+27} & = & \frac{2}{(2+k+N)},
\qquad
 c_{+28}  =  -\frac{2}{(2+k+N)^{2}},
 \qquad
 c_{+29}  =  \frac{2(1+k)}{(2+k+N)^{2}},
\nonu \\
  c_{+30}  & = &   -\frac{2(1+k)}{(2+k+N)^{2}},
\qquad
  c_{+31}  =  -\frac{2(-1+k)}{(2+k+N)^{2}},
\qquad
 c_{+32}  =  -\frac{2N}{(2+k+N)},
\nonumber\\
 c_{+33} & = & \frac{2}{(2+k+N)^{2}},
\qquad
 c_{+34}  =  -\frac{2}{(2+k+N)^{2}},
\qquad
 c_{+35}  =  \frac{2(1+N)}{(2+k+N)^{3}},
\nonu \\
c_{+36}  & = &   -\frac{2N}{(2+k+N)^{2}},
\qquad
c_{+37}  =  \frac{2}{(2+k+N)},
\qquad
c_{+38}  =  \frac{2}{(2+k+N)^{2}},
\nonumber\\
c_{+39} & = & -\frac{2(1+N)}{(2+k+N)^{2}},
\qquad
 c_{+40}  =  -\frac{2(1+N)}{(2+k+N)^{3}},
 \nonumber\\
 c_{+41} & = & \frac{2(1+N)}{(2+k+N)^{2}},
 \qquad
 c_{+42}  =  -\frac{2(-1+N)}{(2+k+N)^{2}},
 \nonumber\\
 c_{+43} & = & -\frac{1}{4},
\qquad
 c_{+44}  =  
\frac{(60+77k+22k^{2}+121N+115kN+20k^{2}N+79N^{2}+42kN^{2}+16N^{3})}{4(2+N)(2+k+N)^{2}},
 \nonumber\\
  c_{+45} & = & \frac{(3+2k+N)}{(2+k+N)^{3}},
\qquad
 c_{+46}  =  -\frac{(20+21k+6k^{2}+25N+13kN+7N^{2})}{4(2+N)(2+k+N)^{3}},
 \nonumber\\
  c_{+47} & = & \frac{(26+25k+6k^{2}+30N+15kN+8N^{2})}{(2+N)(2+k+N)^{3}},
\nonumber\\
c_{+48} & = & \frac{(5k+2k^{2}+5N+14kN+4k^{2}N+7N^{2}+7kN^{2}+2N^{3})}{(2+N)(2+k+N)^{3}},
\nonumber\\
 c_{+49} & = & -\frac{(32+55k+18k^{2}+41N+35kN+11N^{2})}{2(2+N)(2+k+N)^{4}},
 \nonumber\\
  c_{+50} & = & -\frac{(13k+6k^{2}+3N+9kN+N^{2})}{4(2+N)(2+k+N)^{3}},
\nonumber\\
 c_{+51} & = & -\frac{(1+N)(32+55k+18k^{2}+41N+35kN+11N^{2})}{2(2+N)(2+k+N)^{5}},
 \nonumber\\
 c_{+52} & = & \frac{(80+92k+11k^{2}-6k^{3}+172N+158kN+21k^{2}N+119N^{2}+64kN^{2}+25N^{3})}{4(2+N)(2+k+N)^{4}},
\nonumber\\
c_{+53} & = & \frac{(20+31k+10k^{2}+35N+41kN+8k^{2}N+21N^{2}+14kN^{2}+4N^{3})}{2(2+N)(2+k+N)^{4}},
\nonumber\\
c_{+54} & = & -\frac{(1+k)(32+55k+18k^{2}+41N+35kN+11N^{2})}{2(2+N)(2+k+N)^{5}}, 
\nonumber\\
c_{+55} & = & -\frac{(12+27k+10k^{2}+19N+35kN+8k^{2}N+11N^{2}+12kN^{2}+2N^{3})}{2(2+N)(2+k+N)^{4}},
\nonumber\\
c_{+56} & = & \frac{2(6+3k+12N+11kN+3k^{2}N+9N^{2}+6kN^{2}+2N^{3})}{(2+N)(2+k+N)^{4}},
\nonumber\\
 c_{+57} & = & \frac{(5k+2k^{2}+5N+14kN+4k^{2}N+7N^{2}+7kN^{2}+2N^{3})}{(2+N)(2+k+N)^{3}},
\nonumber\\
 c_{+58} & = & \frac{(32+55k+18k^{2}+41N+35kN+11N^{2})}{2(2+N)(2+k+N)^{4}},
\nonumber\\
 c_{+59} & = & -\frac{(20+31k+10k^{2}+35N+41kN+8k^{2}N+21N^{2}+14kN^{2}+4N^{3})}{2(2+N)(2+k+N)^{4}},
\nonumber\\
 c_{+60} & = & \frac{(13k+6k^{2}+3N+9kN+N^{2})}{4(2+N)(2+k+N)^{2}},
\qquad
c_{+61}  =  -\frac{(13k+6k^{2}+3N+9kN+N^{2})}{2(2+N)(2+k+N)^{3}},
\nonumber\\ 
c_{+62} & = & -\frac{(32+55k+18k^{2}+41N+35kN+11N^{2})}{2(2+N)(2+k+N)^{4}},
\nonumber\\
 c_{+63} & = & -\frac{(13k+6k^{2}+3N+9kN+N^{2})}{2(2+N)(2+k+N)^{3}},
\nonumber\\
 c_{+64} & = & \frac{(16+21k+6k^{2}+19N+13kN+5N^{2})}{2(2+N)(2+k+N)^{3}},
\nonumber\\
 c_{+65} & = & -\frac{(20+2k-17k^{2}-6k^{3}+48N+21kN-5k^{2}N+36N^{2}+13kN^{2}+8N^{3})}{2(2+N)(2+k+N)^{4}},
\nonumber\\
 c_{+66} & = & \frac{(20+28k+8k^{2}+54N+55kN+10k^{2}N+41N^{2}+23kN^{2}+9N^{3})}{2(2+N)(2+k+N)^{4}},
 \nonumber\\
  c_{+67} & = & -\frac{(80+72k+3k^{2}-6k^{3}+152N+102kN+5k^{2}N+91N^{2}+36kN^{2}+17N^{3})}{4(2+N)(2+k+N)^{3}},
\nonumber\\
 c_{+68} & = & -\frac{(5k+2k^{2}+5N+14kN+4k^{2}N+7N^{2}+7kN^{2}+2N^{3})}{2(2+N)(2+k+N)^{2}},
\nonumber\\
 c_{+69} & = & \frac{(13k+6k^{2}+3N+9kN+N^{2})}{(2+N)(2+k+N)^{3}},
 \nonumber\\
c_{+70} & = & -\frac{(20+31k+10k^{2}+35N+41kN+8k^{2}N+21N^{2}+14kN^{2}+4N^{3})}{4(2+N)(2+k+N)^{3}},
\nonumber\\
c_{+71} & = & -\frac{(52+89k+56k^{2}+12k^{3}+89N+97kN+30k^{2}N+47N^{2}+24kN^{2}+8N^{3})}{2(2+N)(2+k+N)^{4}},
\nonumber\\
 c_{+72} & = & -\frac{(1+k)(36+39k+10k^{2}+67N+53kN+8k^{2}N+41N^{2}+18kN^{2}+8N^{3})}{2(2+N)(2+k+N)^{5}},
\nonumber\\
 c_{+73} & = & -\frac{1}{2(2+N)(2+k+N)^{5}}(32+28k+k^{2}-2k^{3}+84N+52kN-7k^{2}N-4k^{3}N
 \nonumber\\ &+&  91N^{2}+44kN^{2}-2k^{2}N^{2}+45N^{3}+14kN^{3}+8N^{4}),
\nonumber\\
 c_{+74} & = & \frac{1}{4(2+N)(2+k+N)^{4}}(64+4k-51k^{2}-18k^{3}+116N-68kN-119k^{2}N-24k^{3}N
  \nonumber\\ &+&  83N^{2}-68kN^{2}-50k^{2}N^{2}+29N^{3}-14kN^{3}+4N^{4}),
\nonumber\\
c_{+75} & = & \frac{(18+21k+6k^{2}+22N+13kN+6N^{2})}{(2+N)(2+k+N)^{2}},
\nonumber\\
 c_{+76} & = & -\frac{(18+21k+6k^{2}+22N+13kN+6N^{2})}{(2+N)(2+k+N)^{3}},
\nonumber\\
 c_{+77} & = & -\frac{1}{2(2+N)(2+k+N)^{4}}(-68-77k-25k^{2}-2k^{3}-105N-50kN+19k^{2}N
 \nonumber\\&+&8k^{3}N-51N^{2}+13kN^{2}+18k^{2}N^{2}
 -8N^{3}+8kN^{3}),
\nonumber\\
 c_{+78} & = & \frac{1}{(2+k+N)},
\nonumber\\
 c_{+79} & = & \frac{(20+31k+10k^{2}+35N+41kN+8k^{2}N+21N^{2}+14kN^{2}+4N^{3})}{4(2+N)(2+k+N)^{3}},
\nonumber\\
 c_{+80} & = & \frac{(1+N)(4+7k+2k^{2}+19N+21kN+4k^{2}N+17N^{2}+10kN^{2}+4N^{3})}
{2(2+N)(2+k+N)^{5}},
\nonumber\\
 c_{+81} & = & -\frac{1}{4(2+N)(2+k+N)^{4}}(64+44k-15k^{2}-10k^{3}+156N+84kN-23k^{2}N
 \nonumber\\&-&8k^{3}N+159N^{2}+72kN^{2}-6k^{2}N^{2}+73N^{3}+22kN^{3}+12N^{4}),
  \nonumber\\
 c_{+82} & = & \frac{(1+N)}{(2+k+N)^{2}},
\qquad
 c_{+83}  =  \frac{(1+N)}{(2+k+N)^{3}},
 \nonumber\\
 c_{+84} & = & -\frac{1}{2(2+N)(2+k+N)^{4}}(28+35k+10k^{2}+55N+54kN+10k^{2}N+50N^{2}
 \nonumber\\&+&37kN^{2}+4k^{2}N^{2}+23N^{3}+10kN^{3}+4N^{4}),
 \nonumber\\
c_{+85} & = & \frac{(-64-84k-11k^{2}+6k^{3}-76N-36kN+15k^{2}N-17N^{2}+10kN^{2}+N^{3})}{4(2+N)(2+k+N)^{4}},
\nonumber\\
 c_{+86} & = & \frac{(1+9k+4k^{2}+5N+7kN+2N^{2})}{(2+k+N)^{3}},
\nonumber\\
 c_{+87} & = & \frac{2(7+12k+4k^{2}+10N+8kN+3N^{2})}{(2+k+N)^{4}},
\nonumber\\
 c_{+88} & = & \frac{(18+13k+2k^{2}+24N+18kN+4k^{2}N+12N^{2}+7kN^{2}+2N^{3})}{(2+N)(2+k+N)^{3}},
\nonumber\\
 c_{+89} & = & -\frac{(1+N)(20+23k+6k^{2}+23N+14kN+6N^{2})}{2(2+N)(2+k+N)^{3}},
\nonumber\\
 c_{+90} & = & \frac{1}{2(2+N)(2+k+N)^{3}}(64+69k+13k^{2}-2k^{3}+91N+45kN-14k^{2}N-4k^{3}N
 \nonumber\\&+&46N^{2}+3kN^{2}-7k^{2}N^{2}+13N^{3}+kN^{3}+2N^{4}), 
 \nonumber\\
c_{-1} & = & \frac{4kN}{(2+k+N)^{2}},
\qquad
c_{-2}  = - \frac{4N}{(2+k+N)},
\qquad
c_{-3}  =  -\frac{4N}{(2+k+N)^{2}},
\nonu \\
c_{-4}  & = &   \frac{4k}{(2+k+N)},
\qquad
 c_{-5}  =  -\frac{4k}{(2+k+N)^{2}},
 \qquad
 c_{-6}  =  -\frac{(k-N)}{2(2+k+N)},
 \nonumber\\
 c_{-7} & = & -\frac{(k-N)}{(2+k+N)^{2}},
 \qquad
 c_{-8}  =  -\frac{(k-N)}{(2+k+N)^{2}},
\qquad
c_{-9}  =  \frac{(k-N)}{(2+k+N)^{2}},
\nonu \\
c_{-10}  & = &  -\frac{(k+3N+3kN+N^{2})}{(2+k+N)^{3}},
\qquad
c_{-11}  =  \frac{(3k+k^{2}+N+3kN)}{(2+k+N)^{3}},
 \nonu \\
c_{-12}  & = &   \frac{(4k+k^{2}+4N+6kN+N^{2})}{2(2+k+N)^{2}},
\qquad
c_{-13}  =  -1,
\qquad
 c_{-14}  =  -\frac{2}{(2+k+N)},
 \nonu \\
 c_{-15}  & = &   \frac{4}{(2+k+N)^{2}},
\qquad
c_{-16}  =  -\frac{2}{(2+k+N)},
\qquad
c_{-17}  =  -\frac{2}{(2+k+N)},
\nonumber\\
c_{-18} & = & \frac{2(1+N)}{(2+k+N)^{2}},
\qquad
c_{-19}  =  \frac{2(1+k)}{(2+k+N)^{2}},
\qquad
c_{-20}  =  \frac{(k-N)}{(2+k+N)},
\nonu \\
c_{-21}  & = &  -\frac{2N}{(2+k+N)},
\qquad
c_{-22}  =  -\frac{2}{(2+k+N)^{2}},
\qquad
c_{-23}  =  \frac{2}{(2+k+N)^{2}},
\nonumber\\
c_{-24} & = & \frac{2(1+N)}{(2+k+N)^{3}},
\qquad
c_{-25}  =  -\frac{2(1+N)}{(2+k+N)^{3}},
\qquad
c_{-26}  =  -\frac{2N}{(2+k+N)^{2}},
\nonu \\
c_{-27}  & = &   -\frac{2}{(2+k+N)},
\qquad
c_{-28}  =  -\frac{2}{(2+k+N)^{2}},
\qquad
c_{-29}  =  \frac{2(1+N)}{(2+k+N)^{2}},
\nonumber\\
c_{-30} & = & -\frac{2(1+N)}{(2+k+N)^{2}},
\qquad
c_{-31}  =  -\frac{2N}{(2+k+N)^{2}},
\qquad
c_{-32}  =  \frac{2k}{(2+k+N)},
\nonu \\
c_{-33}  & = &   \frac{2}{(2+k+N)^{2}},
\qquad
c_{-34}   =   \frac{2}{(2+k+N)^{2}},
\qquad
c_{-35}  =  -\frac{2(1+k)}{(2+k+N)^{3}},
\nonumber\\
c_{-36} & = & -\frac{2k}{(2+k+N)^{2}},
\qquad
 c_{-37}  =  -\frac{2}{(2+k+N)},
 \qquad
 c_{-38}  =  \frac{2}{(2+k+N)^{2}},
 \nonu \\
 c_{-39}  & = &   -\frac{2(1+k)}{(2+k+N)^{2}},
\qquad
c_{-40}  =  \frac{2(1+k)}{(2+k+N)^{3}},
\qquad
c_{-41}  =  \frac{2(1+k)}{(2+k+N)^{2}},
\nonumber\\
c_{-42} & = & -\frac{2k}{(2+k+N)^{2}},
\qquad
c_{-43}  =  \frac{1}{4},
\nonumber\\
c_{-44} & = & -\frac{(60+77k+22k^{2}+121N+115kN+20k^{2}N+79N^{2}+42kN^{2}+16N^{3})}{4(2+N)(2+k+N)^{2}},
\nonumber\\
c_{-45} & = & -\frac{(13k+6k^{2}+3N+9kN+N^{2})}{(2+N)(2+k+N)^{3}},
\nonumber\\
 c_{-46} & = & \frac{(20+21k+6k^{2}+25N+13kN+7N^{2})}{4(2+N)(2+k+N)^{3}},
\nonumber\\
c_{-47} & = & -\frac{(26+25k+6k^{2}+30N+15kN+8N^{2})}{(2+N)(2+k+N)^{3}},
\nonumber\\
c_{-48} & = & -\frac{(5k+2k^{2}+5N+14kN+4k^{2}N+7N^{2}+7kN^{2}+2N^{3})}{(2+N)(2+k+N)^{3}},
\nonumber\\
 c_{-49} & = & \frac{(32+55k+18k^{2}+41N+35kN+11N^{2})}{2(2+N)(2+k+N)^{4}},
 \nonumber\\
  c_{-50} & = & \frac{(13k+6k^{2}+3N+9kN+N^{2})}{4(2+N)(2+k+N)^{3}},
\nonumber\\
c_{-51} & = & -\frac{(1+k)(32+55k+18k^{2}+41N+35kN+11N^{2})}{2(2+N)(2+k+N)^{5}},
\nonumber\\
c_{-52} & = & \frac{(80+92k+11k^{2}-6k^{3}+172N+158kN+21k^{2}N+119N^{2}+64kN^{2}+25N^{3})}{4(2+N)(2+k+N)^{4}},
\nonumber\\
c_{-53} & = & \frac{(20+31k+10k^{2}+35N+41kN+8k^{2}N+21N^{2}+14kN^{2}+4N^{3})}{2(2+N)(2+k+N)^{4}},
\nonumber\\
c_{-54} & = & -\frac{(1+N)(32+55k+18k^{2}+41N+35kN+11N^{2})}{2(2+N)(2+k+N)^{5}},
\nonumber\\
c_{-55} & = & \frac{(52+89k+56k^{2}+12k^{3}+89N+97kN+30k^{2}N+47N^{2}+24kN^{2}+8N^{3})}{2(2+N)(2+k+N)^{4}},
\nonumber\\
c_{-56} & = & \frac{(24+41k+14k^{2}+67N+77kN+16k^{2}N+53N^{2}+32kN^{2}+12N^{3})}{2(2+N)(2+k+N)^{4}},
\nonumber\\
c_{-57} & = & -\frac{(5k+2k^{2}+5N+14kN+4k^{2}N+7N^{2}+7kN^{2}+2N^{3})}{(2+N)(2+k+N)^{3}},
\nonumber\\
 c_{-58} & = & -\frac{(32+55k+18k^{2}+41N+35kN+11N^{2})}{2(2+N)(2+k+N)^{4}},
\nonumber\\
 c_{-59} & = & -\frac{(20+31k+10k^{2}+35N+41kN+8k^{2}N+21N^{2}+14kN^{2}+4N^{3})}{2(2+N)(2+k+N)^{4}},
\nonumber\\
c_{-60} & = & -\frac{(13k+6k^{2}+3N+9kN+N^{2})}{4(2+N)(2+k+N)^{2}},
\qquad
c_{-61}  =  -\frac{(13k+6k^{2}+3N+9kN+N^{2})}{2(2+N)(2+k+N)^{3}},
\nonumber\\
 c_{-62} & = & \frac{(32+55k+18k^{2}+41N+35kN+11N^{2})}{2(2+N)(2+k+N)^{4}},
\nonumber\\
c_{-63} & = & -\frac{(13k+6k^{2}+3N+9kN+N^{2})}{2(2+N)(2+k+N)^{3}},
\nonumber\\
 c_{-64} & = & -\frac{(16+21k+6k^{2}+19N+13kN+5N^{2})}{2(2+N)(2+k+N)^{3}},
\nonumber\\
 c_{-65} & = & \frac{(20+28k+8k^{2}+54N+55kN+10k^{2}N+41N^{2}+23kN^{2}+9N^{3})}{2(2+N)(2+k+N)^{4}},
\nonumber\\
 c_{-66} & = & -\frac{(20+2k-17k^{2}-6k^{3}+48N+21kN-5k^{2}N+36N^{2}+13kN^{2}+8N^{3})}{2(2+N)(2+k+N)^{4}},
\nonumber\\
c_{-67} & = & -\frac{(80+72k+3k^{2}-6k^{3}+152N+102kN+5k^{2}N+91N^{2}+36kN^{2}+17N^{3})}{4(2+N)(2+k+N)^{3}},
\nonumber\\
 c_{-68} & = & -\frac{(5k+2k^{2}+5N+14kN+4k^{2}N+7N^{2}+7kN^{2}+2N^{3})}{2(2+N)(2+k+N)^{2}},
\qquad
 c_{-69}  =  -\frac{(3+2k+N)}{(2+k+N)^{3}},
\nonumber\\
c_{-70} & = & \frac{(20+31k+10k^{2}+35N+41kN+8k^{2}N+21N^{2}+14kN^{2}+4N^{3})}{4(2+N)(2+k+N)^{3}},
\nonumber\\
 c_{-71} & = & \frac{(12+27k+10k^{2}+19N+35kN+8k^{2}N+11N^{2}+12kN^{2}+2N^{3})}{2(2+N)(2+k+N)^{4}},
\nonumber\\
c_{-72} & = & -\frac{(1+N)(4+7k+2k^{2}+19N+21kN+4k^{2}N+17N^{2}+10kN^{2}+4N^{3})}{2(2+N)(2+k+N)^{5}},
\nonumber\\
 c_{-73} & = & \frac{1}{2(2+N)(2+k+N)^{5}}(32+28k+k^{2}-2k^{3}+84N+52kN-7k^{2}N-4k^{3}N
 \nonumber\\&+&91N^{2}+44kN^{2}-2k^{2}N^{2}+45N^{3}+14kN^{3}+8N^{4}),
\nonumber\\
 c_{-74} & = & \frac{1}{4(2+N)(2+k+N)^{4}}(64+44k-15k^{2}-10k^{3}+156N+84kN-23k^{2}N
 \nonumber\\&-&8k^{3}N+159N^{2}+72kN^{2}-6k^{2}N^{2}+73N^{3}+22kN^{3}+12N^{4}),
\nonumber\\
 c_{-75} & = & \frac{(1+N)}{(2+k+N)^{2}},
\qquad
c_{-76}  =  \frac{(1+N)}{(2+k+N)^{3}},
\nonumber\\
c_{-77} & = & \frac{(1+N)(-8-7k-2k^{2}-3N+4kN+2k^{2}N+5N^{2}+5kN^{2}+2N^{3})}{(2+N)(2+k+N)^{4}},
\nonumber\\
c_{-78} & = & -\frac{1}{(2+k+N)},
\nonumber\\
 c_{-79} & = & -\frac{(20+31k+10k^{2}+35N+41kN+8k^{2}N+21N^{2}+14kN^{2}+4N^{3})}{4(2+N)(2+k+N)^{3}},
\nonumber\\
c_{-80} & = & \frac{(1+k)(36+39k+10k^{2}+67N+53kN+8k^{2}N+41N^{2}+18kN^{2}+8N^{3})}{2(2+N)(2+k+N)^{5}},
\nonumber\\
 c_{-81} & = & -\frac{1}{4(2+N)(2+k+N)^{4}}(64+4k-51k^{2}-18k^{3}+116N-68kN-119k^{2}N
 \nonumber\\&-&24k^{3}N+83N^{2}-68kN^{2}-50k^{2}N^{2}+29N^{3}-14kN^{3}+4N^{4}),
\nonumber\\
 c_{-82} & = & \frac{(18+21k+6k^{2}+22N+13kN+6N^{2})}{(2+N)(2+k+N)^{2}},
\nonumber\\
 c_{-83} & = & -\frac{(18+21k+6k^{2}+22N+13kN+6N^{2})}{(2+N)(2+k+N)^{3}},
\nonumber\\
 c_{-84} & = & \frac{(1+k)(8+9k+2k^{2}+21N+20kN+4k^{2}N+17N^{2}+9kN^{2}+4N^{3})}{(2+N)(2+k+N)^{4}},
\nonumber\\
 c_{-85} & = & \frac{(-64-84k-11k^{2}+6k^{3}-76N-36kN+15k^{2}N-17N^{2}+10kN^{2}+N^{3})}{4(2+N)(2+k+N)^{4}},
\nonumber\\
 c_{-86} & = & -\frac{(44+51k+14k^{2}+57N+42kN+4k^{2}N+21N^{2}+7kN^{2}+2N^{3})}{(2+N)(2+k+N)^{3}},
\nonumber\\
 c_{-87} & = & \frac{(56+67k+18k^{2}+89N+79kN+12k^{2}N+47N^{2}+24kN^{2}+8N^{3})}{2(2+N)(2+k+N)^{4}},
\nonumber\\
 c_{-88} & = & -\frac{(-4+k+2k^{2}+3N+12kN+4k^{2}N+7N^{2}+7kN^{2}+2N^{3})}{(2+N)(2+k+N)^{3}},
\nonumber\\
 c_{-89} & = & -\frac{(k+9N+12kN+3k^{2}N+11N^{2}+7kN^{2}+3N^{3})}{(2+N)(2+k+N)^{3}},
\nonumber\\
 c_{-90} & = & \frac{1}{2(2+N)(2+k+N)^{3}}(-32-27k-k^{2}+2k^{3}-53N-19kN+14k^{2}N+4k^{3}N
 \nonumber\\&-&36N^{2}-3kN^{2}+7k^{2}N^{2}-13N^{3}-kN^{3}-2N^{4}).\nonumber
\eea

\subsection{The structure constants in the subsection $6.3.3$}

The coefficients appearing in (\ref{tw}) of
the subsection $6.3.3$ are given by
\bea
c_{1} &=& \frac{16k\, N}{(2+k+N)^{2}},\qquad
c_{2} = -\frac{16k\, N}{(2+k+N)^{2}},\qquad
c_{3} = \frac{16(k-N)(3+2k+2N)}{3(2+k+N)^{2}},\nonu\\
\nonu\\
c_{4} &=& -\frac{16(3k+2k^{2}+3N+8k\, N+3k^{2}N+2N^{2}+3kN^{2})}{3(2+k+N)^{3}},
\nonu\\
c_{5} &=& \frac{16(3k+2k^{2}+3N+8k\, N+3k^{2}N+2N^{2}+3k\, N^{2})}{3(2+k+N)^{3}},\nonu\\
c_{6} &=& \frac{16(3k+2k^{2}+3N+2k\, N+2N^{2})}{3(2+k+N)^{3}},
\qquad c_{7} = -\frac{16(k-N)(3+2k+2N)}{3(2+k+N)^{3}},
\nonu\\ c_{8} &=& -\frac{16(k-N)}{3(2+k+N)^{2}},
\qquad c_{9} = \frac{16(k-N)(1+N)}{3(2+k+N)^{3}},
\qquad c_{10} = -\frac{8(k-N)}{(2+k+N)^{3}},
\nonu \\
 c_{11}  & = &  \frac{8(k-N)}{3(2+k+N)^{3}},
\qquad c_{12} = \frac{8(k-N)}{3(2+k+N)^{3}},
\qquad c_{13} = \frac{8(k+N)}{(2+k+N)^{2}},
\nonu\\c_{14} &=& \frac{16(1+k)(k-N)}{3(2+k+N)^{3}},
\qquad c_{15} = -\frac{8(4k+k^{2}+8N+8k\, N+3N^{2})}{3(2+k+N)^{3}},
\nonu\\c_{16} &=& -\frac{16(k-N)}{3(2+k+N)^{2}},
\qquad c_{17} = -\frac{16(1+k)(k-N)}{3(2+k+N)^{3}},
\qquad 
c_{18} = \frac{8(k-N)}{(2+k+N)^{3}},
\nonu \\
 c_{19} & = &  \frac{8(k-N)}{3(2+k+N)^{3}},
\qquad c_{20} = \frac{8(k+N)}{(2+k+N)^{2}},
\qquad c_{21} = -\frac{16(k-N)(1+N)}{3(2+k+N)^{3}},
\nonu\\ c_{22} &=& -\frac{8(k-N)}{3(2+k+N)^{3}},
\qquad c_{23} = \frac{8(8k+3k^{2}+4N+8k\, N+N^{2})}{3(2+k+N)^{3}},
\qquad c_{24} = -4,
\nonu\\ c_{25} &=& \frac{2(60\ +77k+22k^{2}+121N+115k\, N+20k^{2}N+79N^{2}+42k\, N^{2}+16N^{3})}{(2+N)(2+k+N)^{2}},
\nonu\\ c_{26} &=& \frac{4(30+37k+10k^{2}+32N+21k\, N+8N^{2})}{3(2+k+N)^{2}},
\nonu\\ c_{27} &=& -\frac{4(96+167k+56k^{2}+193N+218k\, N+37k^{2}N+122N^{2}+71k\, N^{2}+24N^{3})}{3(2+N)(2+k+N)^{3}},
\nonu\\c_{28} &=& \frac{2(12+24k+7k^{2}+24N+22k\, N+7N^{2})}{3(2+k+N)^{3}},
\nonu\\ c_{29} &=& \frac{2(60+49k+8k^{2}+149N+101k\, N+13k^{2}N+107N^{2}+42k\, N^{2}+23N^{3})}{3(2+N)(2+k+N)^{3}},
\nonu\\c_{30} &=& -\frac{8(42+55k+18k^{2}+56N+35k\, N+16N^{2})}{3(2+N)(2+k+N)^{3}},
\nonu\\ c_{31} &=& -\frac{4(8+11k+4k^{2}+25N+23k\, N+5k^{2}N+21N^{2}+10k\, N^{2}+5N^{3})}{(2+N)(2+k+N)^{4}},
\nonu\\ c_{32} &=& -\frac{4(32+55k+18k^{2}+41N+35k\, N+11N^{2})}{(2+N)(2+k+N)^{4}},
\qquad c_{33} = \frac{8(9+2k+7N)}{3(2+k+N)^{3}},
\nonu\\ c_{34} &=& \frac{4(8+11k+4k^{2}+25N+23k\, N+5k^{2}N+21N^{2}+10k\, N^{2}+5N^{3})}{(2+N)(2+k+N)^{4}},
\nonu\\ c_{35} &=& -\frac{2(132+85k+22k^{2}+221N+119k\, N+20k^{2}N+111N^{2}+42k\, N^{2}+16N^{3})}{3(2+N)(2+k+N)^{3}},
\nonu\\ c_{36} &=& \frac{8(9+2k+7N)}{3(2+k+N)^{3}},
\qquad 
c_{37} = -\frac{8(42+55k+18k^{2}+56N+35k\, N+16N^{2})}{3(2+N)(2+k+N)^{3}},
\nonu\\ c_{38} &=& \frac{2(20+21k+6k^{2}+25N+13k\, N+7N^{2})}{(2+N)(2+k+N)^{2}},
\nonu\\ c_{39} &=& -\frac{4(-5k-2k^{2}+5N+3k\, N+2k^{2}N+7N^{2}+4k\, N^{2}+2N^{3})}{(2+N)(2+k+N)^{3}},
\nonu\\ c_{40} &=& \frac{4(-5k-2k^{2}+5N+3k\, N+2k^{2}N+7N^{2}+4k\, N^{2}+2N^{3})}{(2+N)(2+k+N)^{3}},\nonu\\c_{41} &=& \frac{4(8+17k+6k^{2}+11N+11k\, N+3N^{2})}{(2+N)(2+k+N)^{3}},
\nonu\\ c_{42} &=& -\frac{4(20+21k+6k^{2}+25N+13k\, N+7N^{2})}{(2+N)(2+k+N)^{3}},
\nonu\\ c_{43} &=& -\frac{4(-24-23k-6k^{2}-13N-30k\, N-12k^{2}N-13k\, N^{2}+N^{3})}{3(2+N)(2+k+N)^{3}},
\nonu\\ c_{44} &=& \frac{8(30+37k+10k^{2}+35N+23k\, N+9N^{2})}{3(2+k+N)^{3}},
\nonu\\ c_{45} &=& -\frac{2(20+31k+10k^{2}+35N+41k\, N+8k^{2}N+21N^{2}+14k\, N^{2}+4N^{3})}{(2+N)(2+k+N)^{4}},
\nonu\\c_{46} &=& -\frac{1}{3(2+N)(2+k+N)^{4}}4(180+323k+198k^{2}+40k^{3}+391N+529k\, N
\nonu\\&+&214k^{2}N+20k^{3}N+317N^{2}+296k\, N^{2}+62k^{2}N^{2}+116N^{3}+58k\, N^{3}+16N^{4}),
\nonu\\ c_{47} &=& -\frac{4(60+73k+20k^{2}+125N+113k\, N+19k^{2}N+83N^{2}+42k\, N^{2}+17N^{3})}{3(2+N)(2+k+N)^{4}},
\nonu\\ c_{48} &=& \frac{4(2+3k+3N)}{(2+k+N)^{2}},
\qquad 
c_{49} = \frac{2(12+24k+7k^{2}+24N+22k\, N+7N^{2})}{3(2+k+N)^{3}},
\nonu\\ c_{50} &=& \frac{8(60+128k+87k^{2}+18k^{3}+94N+141k\, N+51k^{2}N+48N^{2}+37k\, N^{2}+8N^{3})}{3(2+N)(2+k+N)^{3}},
\nonu\\ c_{51} &=& -\frac{2(20+31k+10k^{2}+35N+41k\, N+8k^{2}N+21N^{2}+14k\, N^{2}+4N^{3})}{(2+N)(2+k+N)^{4}},
\nonu\\  c_{52} &=& \frac{4(60+73k+20k^{2}+125N+113k\, N+19k^{2}N+83N^{2}+42k\, N^{2}+17N^{3})}{3(2+N)(2+k+N)^{4}},
\nonu\\  c_{53} &=& -\frac{4(2+3k+3N)}{(2+k+N)^{2}},
\nonu\\  c_{54} &=& -\frac{2(60+49k+8k^{2}+149N+101k\, N+13k^{2}N+107N^{2}+42k\, N^{2}+23N^{3})}{3(2+N)(2+k+N)^{3})},
\nonu\\  c_{55} &=& \frac{2(228+289k+94k^{2}+353N+251k\, N+20k^{2}N+147N^{2}+42k\, N^{2}+16N^{3})}{3(2+N)(2+k+N)^{3}},
\nonu\\  c_{56} &=& -\frac{4(4+15k+6k^{2}+7N+10k\, N+2N^{2})}{(2+N)(2+k+N)^{2}},
\nonu\\  c_{57} &=& \frac{8(3+2k+N)(4+k+3N)}{3(2+k+N)^{3}},
\qquad c_{58} = -\frac{8(4+3k+N)(3+k+2N)}{3(2+k+N)^{3}},
\nonu\\  c_{59} &=& -\frac{4(k-N)}{3(2+k+N)^{2}},
\qquad c_{60} = \frac{4(k-N)^{2}}{3(2+k+N)^{3}},
\qquad  c_{61} = \frac{16(3+2k+N)}{3(2+k+N)^{3}},
\nonu \\  
c_{62} & = & -\frac{8}{(2+k+N)^{2}},
\qquad c_{63} = \frac{16(3+k+2N)}{3(2+k+N)^{3}},
\qquad  c_{64} = -\frac{8}{(2+k+N)^{2}},
\nonu\\  c_{65} &=& -\frac{16(3+k+2N)}{3(2+k+N)^{3}},
\qquad  c_{66} = -\frac{16(3+k+2N)}{3(2+k+N)^{3}},
\qquad  c_{67} = -\frac{16(3+2k+N)}{3(2+k+N)^{3}},
\nonu \\  c_{68} & = & \frac{8(k-N)}{3(2+k+N)^{3}},
\qquad  c_{69} = -\frac{8}{(2+k+N)},
\qquad c_{70} = \frac{4(k-N)}{3(2+k+N)^{2}},
\nonu\\  c_{71} &=& \frac{8(k-N)}{3(2+k+N)^{3}},
\qquad  c_{72} = -\frac{8}{(2+k+N)},
\qquad c_{73} = \frac{4(k-N)^{2}}{3(2+k+N)^{3}},
\nonu \\  c_{74} & = & \frac{16(3+2k+N)}{3(2+k+N)^{3}},
\qquad  c_{75} = -\frac{8(k-N)}{3(2+k+N)^{2}},
\qquad c_{76} = 2,
\nonu\\  c_{77} &=& \ -\frac{2(60+77k+22k^{2}+121N+115k\, N+20k^{2}N+79N^{2}+42k\, N^{2}+16N^{3})}{(2+N)(2+k+N)^{2}},
\nonu\\  c_{78} &=& -\frac{2(108+93k+22k^{2}+177N+115k\, N+20k^{2}N+103N^{2}+38k\, N^{2}+20N^{3})}{3(2+N)(2+k+N)^{3}},
\nonu\\  c_{79} &=& \frac{2(72+93k+22k^{2}+123N+123k\, N+20k^{2}N+77N^{2}+42k\, N^{2}+16N^{3})}{3(2+N)(2+k+N)^{3}},
\nonu\\ c_{80} &=& \frac{2(54+67k+18k^{2}+62N+41k\, N+16N^{2})}{3(2+N)(2+k+N)^{3}},
\nonu\\  c_{81} &=& \frac{1}{3(2+N)(2+k+N)^{4}}2(168+155k-5k^{2}-18k^{3}+313N+238k\, N+17k^{2}N
\nonu\\ &+&199N^{2}+93k\, N^{2}+40N^{3}),
\nonu\\ c_{82} &=& -\frac{2(32+55k+18k^{2}+41N+35k\, N+11N^{2})}{(2+N)(2+k+N)^{4}},
\nonu\\  c_{83} &=& \frac{2(32+55k+18k^{2}+41N+35k\, N+11N^{2})}{(2+N)(2+k+N)^{4}},
\qquad  c_{84} = -\frac{4(9+4k+5N)}{3(2+k+N)^{3}},
\nonu\\  c_{85} &=& \frac{2(32+35k+10k^{2}+61N+47k\, N+8k^{2}N+39N^{2}+16k\, N^{2}+8N^{3})}{(2+N)(2+k+N)^{4}},
\nonu\\  c_{86} &=& \frac{2(3+2k+N)}{3(2+k+N)^{3}},
\qquad  c_{87} = -\frac{4(78+79k+18k^{2}+86N+47k\, N+22N^{2})}{3(2+N)(2+k+N)^{3})},
\nonu\\  c_{88} &=& -\frac{2(3k+2k^{2}+13N+15k\, N+4k^{2}N+15N^{2}+8k\, N^{2}+4N^{3})}{(2+N)(2+k+N)^{4}},
\nonu\\  c_{89} &=& -\frac{2(32+55k+18k^{2}+41N+35k\, N+11N^{2})}{(2+N)(2+k+N)^{4}},
\nonu\\  c_{90} &=& \frac{2(72+137k+46k^{2}+187N+244k\, N+50k^{2}N+142N^{2}+99k\, N^{2}+31N^{3})}{3(2+N)(2+k+N)^{4}},
\nonu\\  c_{91} &=& -\frac{4(13k+6k^{2}+3N+9k\, N+N^{2})}{(2+N)(2+k+N)^{3}},
\nonu\\  c_{92} &=& -\frac{2(20+21k+6k^{2}+25N+13k\, N+7N^{2})}{(2+N)(2+k+N)^{2}},
\nonu\\  c_{93} &=& \frac{2(16+21k+6k^{2}+19N+13k\, N+5N^{2})}{(2+N)(2+k+N)^{3}},
\nonu\\  c_{94} &=& -\frac{2(20+21k+6k^{2}+25N+13k\, N+7N^{2})}{(2+N)(2+k+N)^{3}},
\nonu\\  c_{95} &=& -\frac{2(16+21k+6k^{2}+19N+13k\, N+5N^{2})}{(2+N)(2+k+N)^{3}},
\nonu\\  c_{96} &=& \frac{2(16+21k+6k^{2}+19N+13k\, N+5N^{2})}{(2+N)(2+k+N)^{3}},
\qquad  c_{97} = \frac{4(3+2k+N)(9+5k+8N)}{3(2+k+N)^{2}},
\nonu\\  c_{98} &=& -\frac{2(12+37k+14k^{2}+41N+71k\, N+16k^{2}N+35N^{2}+30k\, N^{2}+8N^{3})}{3(2+N)(2+k+N)^{4}},
\nonu\\  c_{99} &=& \frac{2(108+133k+38k^{2}+185N+167k\, N+28k^{2}N+107N^{2}+54k\, N^{2}+20N^{3})}{3(2+N)(2+k+N)^{4}},
\nonu\\  c_{100} &=& -\frac{8(1+k)}{(2+k+N)^{3}},
\qquad  c_{101} = -\frac{2(3+2k+N)}{3(2+k+N)^{2}},
\qquad  c_{102} = -\frac{4}{(2+k+N)},
\nonu\\  c_{103} &=& -\frac{2(54+67k+18k^{2}+62N+41k\, N+16N^{2})}{3(2+N)(2+k+N)^{2}},
\nonu\\  c_{104} &=& -\frac{1}{3(2+N)(2+k+N)^{3}}2(108+86k-27k^{2}-18k^{3}+232N+159k\, N
\nonu\\ &-&3k^{2}N+156N^{2}+67k\, N^{2}+32N^{3}),
\nonu\\  c_{105} &=& \frac{2(-4+3k+2k^{2}+3N+15k\, N+4k^{2}N+7N^{2}+8k\, N^{2}+2N^{3})}{(2+N)(2+k+N)^{4}},
\nonu\\  c_{106} &=& \frac{2(3+2k+N)(12+19k+6k^{2}+15N+12k\, N+4N^{2})}{(2+N)(2+k+N)^{4}},
\nonu\\  c_{107} &=& \frac{2(3k+2k^{2}+13N+15k\, N+4k^{2}N+15N^{2}+8k\, N^{2}+4N^{3})}{(2+N)(2+k+N)^{3}},
\nonu\\  c_{108} &=& -\frac{2(3k+2k^{2}+13N+15k\, N+4k^{2}N+15N^{2}+8k\, N^{2}+4N^{3})}{(2+N)(2+k+N)^{3}},
\nonu\\  c_{109} &=& -\frac{2(13k+6k^{2}+3N+9k\, N+N^{2})}{(2+N)(2+k+N)^{3}},
\nonu\\  c_{110} &=& \frac{2(20+21k+6k^{2}+25N+13k\, N+7N^{2})}{(2+N)(2+k+N)^{3}},
\nonu\\  c_{111} &=& -\frac{2(12+45k+14k^{2}+81N+107k\, N+16k^{2}N+71N^{2}+46k\, N^{2}+16N^{3})}{3(2+N)(2+k+N)^{3}},
\nonu\\  c_{112} &=& -\frac{2(12+20k+8k^{2}+58N+77k\, N+22k^{2}N+59N^{2}+41k\, N^{2}+15N^{3})}{3(2+N)(2+k+N)^{3}},
\nonu\\  c_{113} &=& -\frac{2(4+7k+2k^{2}+19N+21k\, N+4k^{2}N+17N^{2}+10k\, N^{2}+4N^{3})}{(2+N)(2+k+N)^{4}},
\nonu\\  c_{114} &=& \frac{1}{3(2+N)(2+k+N)^{4}}2(228+393k+218k^{2}+40k^{3}+537N+681k\, N
\nonu\\&+&242k^{2}N+20k^{3}N+433N^{2}+362k\, N^{2}+62k^{2}N^{2}+142N^{3}+58k\, N^{3}+16N^{4}),
\nonu\\  c_{115} &=& \frac{4(1+N)(20+23k+6k^{2}+23N+14k\, N+6N^{2})}{(2+N)(2+k+N)^{4}},
\nonu\\  c_{116} &=& \frac{2(3k+2k^{2}+13N+15k\, N+4k^{2}N+15N^{2}+8k\, N^{2}+4N^{3})}{(2+N)(2+k+N)^{4}},
\qquad  c_{117} = \frac{4(k+N)}{(2+k+N)^{2}},
\nonu\\  c_{118} &=& -\frac{1}{3(2+N)(2+k+N)^{3}}(348+429k+118k^{2}+633N+523k\, N+68k^{2}N
\nonu\\ &+&367N^{2}+158k\, N^{2}+68N^{3}),
\nonu\\  c_{119}&=&-2,
\nonu\\ c_{120} &=& \ \frac{2(60+77k+22k^{2}+121N+115k\, N+20k^{2}N+79N^{2}+42k\, N^{2}+16N^{3})}{(2+N)(2+k+N)^{2}},
\nonu\\ c_{121} &=& \frac{2(12+45k+14k^{2}+81N+107k\, N+16k^{2}N+71N^{2}+46k\, N^{2}+16N^{3})}{3(2+N)(2+k+N)^{3}},
\nonu\\ c_{122} &=& \frac{2(32+55k+18k^{2}+41N+35k\, N+11N^{2})}{(2+N)(2+k+N)^{4}},
\nonu\\  c_{123} &=& -\frac{2(32+55k+18k^{2}+41N+35k\, N+11N^{2})}{(2+N)(2+k+N)^{4}},
\nonu\\  c_{124} &=& \frac{4(13k+6k^{2}+3N+9k\, N+N^{2})}{(2+N)(2+k+N)^{3}},
\nonu\\  c_{125} &=& -\frac{2(3k+2k^{2}+13N+15k\, N+4k^{2}N+15N^{2}+8k\, N^{2}+4N^{3})}{(2+N)(2+k+N)^{4}},
\nonu \\
  c_{126} & = &  -\frac{2(3+2k+N)}{3(2+k+N)^{3}},
\nonu\\ c_{127} &=& \frac{2(72+137k+46k^{2}+187N+244k\, N+50k^{2}N+142N^{2}+99k\, N^{2}+31N^{3})}{3(2+N)(2+k+N)^{4}},
\nonu\\ c_{128} &=& \frac{4(9+4k+5N)}{3(2+k+N)^{3}},
\qquad  
c_{129} = -\frac{2(54+67k+18k^{2}+62N+41k\, N+16N^{2})}{3(2+N)(2+k+N)^{3}},
\nonu\\  c_{130} &=& \frac{4(78+79k+18k^{2}+86N+47k\, N+22N^{2})}{(3(2+N)(2+k+N)^{3}},
\nonu\\  c_{131} &=& \frac{2(168+155k-5k^{2}-18k^{3}+313N+238k\, N+17k^{2}N+199N^{2}+93k\, N^{2}+40N^{3})}{3(2+N)(2+k+N)^{4}},
\nonu\\ c_{132} &=& \frac{2(20+21k+6k^{2}+25N+13k\, N+7N^{2})}{(2+N)(2+k+N)^{2}},
\nonu\\  c_{133} &=& -\frac{2(16+21k+6k^{2}+19N+13k\, N+5N^{2})}{(2+N)(2+k+N)^{3}},
\nonu\\  c_{134} &=& \frac{2(16+21k+6k^{2}+19N+13k\, N+5N^{2})}{(2+N)(2+k+N)^{3}},
\nonu\\  c_{135} &=& -\frac{2(20+21k+6k^{2}+25N+13k\, N+7N^{2})}{(2+N)(2+k+N)^{3}},
\nonu\\  c_{136} &=& \frac{2(16+21k+6k^{2}+19N+13k\, N+5N^{2})}{(2+N)(2+k+N)^{3}},
\nonu\\  c_{137} &=& \frac{4(6+3k+2k^{2}+36N+36k\, N+10k^{2}N+34N^{2}+21k\, N^{2}+8N^{3})}{3(2+N)(2+k+N)^{2}},
\nonu\\  c_{138} &=& -\frac{2(3+2k+N)}{3(2+k+N)^{2}},
\nonu\\  c_{139} &=& \frac{2(12+20k+8k^{2}+58N+77k\, N+22k^{2}N+59N^{2}+41k\, N^{2}+15N^{3})}{(3(2+N)(2+k+N)^{3}},
\nonu\\  c_{140} &=& -\frac{2(3+2k+N)(12+19k+6k^{2}+15N+12k\, N+4N^{2})}{(2+N)(2+k+N)^{4}},
\nonu\\  c_{141} &=& -\frac{2(32+35k+10k^{2}+61N+47k\, N+8k^{2}N+39N^{2}+16k\, N^{2}+8N^{3})}{(2+N)(2+k+N)^{4}},
\nonu\\  c_{142} &=& -\frac{2(32+55k+18k^{2}+41N+35k\, N+11N^{2})}{(2+N)(2+k+N)^{4}},
\nonu\\  c_{143} &=& -\frac{2(-4+3k+2k^{2}+3N+15k\, N+4k^{2}N+7N^{2}+8k\, N^{2}+2N^{3})}{(2+N)(2+k+N)^{4}},
\nonu\\  c_{144} &=& \frac{2(3k+2k^{2}+13N+15k\, N+4k^{2}N+15N^{2}+8k\, N^{2}+4N^{3})}{(2+N)(2+k+N)^{4}},
\nonu\\  c_{145} &=& -\frac{1}{3(2+N)(2+k+N)^{4}}2(132+141k+44k^{2}+4k^{3}+357N+381k\, N+140k^{2}N
\nonu\\ &+&20k^{3}N+331N^{2}+278k\, N^{2}+62k^{2}N^{2}+124N^{3}+58k\, N^{3}+16N^{4}),
\nonu\\  c_{146} &=& -\frac{2(3k+2k^{2}+13N+15k\, N+4k^{2}N+15N^{2}+8k\, N^{2}+4N^{3})}{(2+N)(2+k+N)^{3}},
\nonu\\  c_{147} &=& \frac{2(3k+2k^{2}+13N+15k\, N+4k^{2}N+15N^{2}+8k\, N^{2}+4N^{3})}{(2+N)(2+k+N)^{3}},
\nonu\\  c_{148} &=& \frac{2(13k+6k^{2}+3N+9k\, N+N^{2})}{(2+N)(2+k+N)^{3}},
\nonu\\  c_{149} &=& -\frac{2(20+21k+6k^{2}+25N+13k\, N+7N^{2})}{(2+N)(2+k+N)^{3}},
\nonu\\  c_{150} &=& \frac{2(108+93k+22k^{2}+177N+115k\, N+20k^{2}N+103N^{2}+38k\, N^{2}+20N^{3})}{3(2+N)(2+k+N)^{3}},
\nonu\\  c_{151} &=& -\frac{2(-72-57k-14k^{2}-39N+33k\, N+20k^{2}N+35N^{2}+42k\, N^{2}+16N^{3})}{3(2+N)(2+k+N)^{3}},
\nonu\\  c_{152} &=& -\frac{2(54+67k+18k^{2}+62N+41k\, N+16N^{2})}{3(2+N)(2+k+N)^{2}},
\qquad  c_{153} = \frac{4}{(2+k+N)},
\nonu\\  c_{154} &=& -\frac{2(108+133k+38k^{2}+185N+167k\, N+28k^{2}N+107N^{2}+54k\, N^{2}+20N^{3})}{3(2+N)(2+k+N)^{4}},
\nonu\\  c_{155} &=& \frac{2(12+37k+14k^{2}+41N+71k\, N+16k^{2}N+35N^{2}+30k\, N^{2}+8N^{3})}{3(2+N)(2+k+N)^{4}},
\nonu\\  c_{156} &=& -\frac{8(1+N)}{(2+k+N)^{3}},
\nonu\\  c_{157} &=& \frac{2(108+86k-27k^{2}-18k^{3}+232N+159k\, N-3k^{2}N+156N^{2}+67k\, N^{2}+32N^{3})}{3(2+N)(2+k+N)^{3}},
\nonu\\  c_{158} &=& \frac{2(36+39k+10k^{2}+67N+53k\, N+8k^{2}N+41N^{2}+18k\, N^{2}+8N^{3})}{(2+N)(2+k+N)^{4}},
\nonu\\ c_{159} &=& \frac{4(1+N)(20+23k+6k^{2}+23N+14k\, N+6N^{2})}{(2+N)(2+k+N)^{4}},
\qquad  c_{160} = \frac{4(k+N)}{(2+k+N)^{2}},
\nonu\\  c_{161} &=& \frac{(228+291k+82k^{2}+375N+301k\, N+32k^{2}N+193N^{2}+74k\, N^{2}+32N^{3})}{3(2+N)(2+k+N)^{3}},
\nonu\\ c_{162} &=& -2,
\nonu\\ c_{163} &=& \ \frac{2(60+77k+22k^{2}+121N+115k\, N+20k^{2}N+79N^{2}+42k\, N^{2}+16N^{3})}{(2+N)(2+k+N)^{2}},
\nonu\\ c_{164} &=& -\frac{2(84+149k+50k^{2}+181N+203k\, N+34k^{2}N+119N^{2}+68k\, N^{2}+24N^{3})}{3(2+N)(2+k+N)^{3}},
\nonu\\  c_{165} &=& \frac{4(-6+k^{2}+4k\, N+N^{2})}{3(2+k+N)^{3}},
\nonu\\  c_{166} &=& \frac{60+61k+14k^{2}+137N+107k\, N+16k^{2}N+95N^{2}+42k\, N^{2}+20N^{3}}{3(2+N)(2+k+N)^{3}},
\nonu\\  c_{167} &=& -\frac{4(42+55k+18k^{2}+56N+35k\, N+16N^{2})}{3(2+N)(2+k+N)^{3}},
\nonu\\  c_{168} &=& -\frac{2(32+55k+18k^{2}+41N+35k\, N+11N^{2})}{(2+N)(2+k+N)^{4}},
\nonu\\  c_{169} &=& -\frac{2(32+55k+18k^{2}+41N+35k\, N+11N^{2})}{(2+N)(2+k+N)^{4}},
\nonu\\  c_{170} &=& -\frac{4}{(2+k+N)^{2}},
\qquad c_{171} = -\frac{4(-3+k-4N)}{3(2+k+N)^{3}},
\nonu\\  c_{172} &=& \frac{2(32+35k+10k^{2}+61N+47k\, N+8k^{2}N+39N^{2}+16k\, N^{2}+8N^{3})}{(2+N)(2+k+N)^{4}},
\nonu\\  c_{173} &=& \frac{4(9+2k+7N)}{3(2+k+N)^{3}},
\qquad  c_{174} = -\frac{4}{(2+k+N)^{2}},
\nonu\\  c_{175} &=& -\frac{2(3k+2k^{2}+13N+15k\, N+4k^{2}N+15N^{2}+8k\, N^{2}+4N^{3})}{(2+N)(2+k+N)^{4}},
\nonu\\  c_{176} &=& -\frac{4(54+61k+18k^{2}+68N+38k\, N+19N^{2})}
{3(2+N)(2+k+N)^{3}},
\nonu\\  c_{177} &=& -\frac{2(3k+2k^{2}+13N+15k\, N+4k^{2}N+15N^{2}+8k\, N^{2}+4N^{3})}{(2+N)(2+k+N)^{4}},
\nonu\\  c_{178} &=& -\frac{2(32+55k+18k^{2}+41N+35k\, N+11N^{2})}{(2+N)(2+k+N)^{4}},
\nonu\\  c_{179} &=& -\frac{84+61k+22k^{2}+173N+107k\, N+20k^{2}N+99N^{2}+42k\, N^{2}+16N^{3}}{3(2+N)(2+k+N)^{3}},
\nonu\\  c_{180} &=& \frac{4(9+2k+7N)}{3(2+k+N)^{3}},
\qquad c_{181} = \frac{4}{(2+k+N)^{2}},
\qquad  c_{182} = \frac{4}{(2+k+N)^{2}},
\nonu\\  c_{183} &=& -\frac{4(54+61k+18k^{2}+68N+38k\, N+19N^{2})}{3(2+N)(2+k+N)^{3}},
\nonu\\ c_{184} &=& -\frac{4(42+55k+18k^{2}+56N+35k\, N+16N^{2})}{3(2+N)(2+k+N)^{3}},
\qquad  c_{185} = -\frac{4(-3+k-4N)}{3(2+k+N)^{3}},
\nonu\\ c_{186} &=& \frac{2(16+21k+6k^{2}+19N+13k\, N+5N^{2})}{(2+N)(2+k+N)^{3}},
\nonu\\  c_{187} &=& -\frac{2(16+21k+6k^{2}+19N+13k\, N+5N^{2})}{(2+N)(2+k+N)^{3}},
\nonu\\  c_{188} &=& \frac{2(13k+6k^{2}+3N+9k\, N+N^{2})}{(2+N)(2+k+N)^{3}},
\nonu\\  c_{189} &=& -\frac{2(20+21k+6k^{2}+25N+13k\, N+7N^{2})}{(2+N)(2+k+N)^{3}},
\nonu\\  c_{190} &=& \frac{2(13k+6k^{2}+3N+9k\, N+N^{2})}{(2+N)(2+k+N)^{3}},
\nonu\\  c_{191} &=& -\frac{2(20+21k+6k^{2}+25N+13k\, N+7N^{2})}{(2+N)(2+k+N)^{3}},
\nonu\\ c_{192} &=& \frac{4(18+31k+10k^{2}+29N+23k\, N+9N^{2})}{3(2+k+N)^{3}},
\nonu\\  c_{193} &=& -\frac{1}{3(2+N)(2+k+N)^{4}}2(180+323k+198k^{2}+40k^{3}+391N+529k\, N+214k^{2}N
\nonu\\ &+&20k^{3}N+317N^{2}+296k\, N^{2}+62k^{2}N^{2}+116N^{3}+58k\, N^{3}+16N^{4}),
\nonu\\  c_{194} &=& -\frac{2(12+37k+14k^{2}+41N+71k\, N+16k^{2}N+35N^{2}+30k\, N^{2}+8N^{3})}{3(2+N)(2+k+N)^{4}},
\nonu\\ c_{195} &=& -\frac{2(108+133k+38k^{2}+185N+167k\, N+28k^{2}N+107N^{2}+54k\, N^{2}+20N^{3})}{3(2+N)(2+k+N)^{4}},
\nonu\\ c_{196} &=& \frac{8}{(2+k+N)^{2}},
\qquad c_{197} = -\frac{8}{(2+k+N)^{2}},
\qquad c_{198} = -\frac{8}{(2+k+N)^{2}},
\nonu\\  c_{199} &=& \frac{4(18+31k+10k^{2}+29N+23k\, N+9N^{2})}{3(2+k+N)^{3}},
\nonu\\  c_{200} &=& -\frac{2(20+31k+10k^{2}+35N+41k\, N+8k^{2}N+21N^{2}+14k\, N^{2}+4N^{3})}{(2+N)(2+k+N)^{4}},
\nonu\\  c_{201} &=& -\frac{1}{3(2+N)(2+k+N)^{4}}
2(180+323k+198k^{2}+40k^{3}+391N+529k\, N+214k^{2}N
\nonu\\  &+&20k^{3}N+317N^{2}+296k\, N^{2}+62k^{2}N^{2}+116N^{3}+58k\, N^{3}+16N^{4}),
\nonu\\  c_{202} &=& -\frac{2(60+61k+14k^{2}+137N+107k\, N+16k^{2}N+95N^{2}+42k\, N^{2}+20N^{3})}{3(2+N)(2+k+N)^{4}},
\nonu\\ c_{203} &=& \frac{2(3k+2k^{2}+13N+15k\, N+4k^{2}N+15N^{2}+8k\, N^{2}+4N^{3})}{(2+N)(2+k+N)^{4}},
\nonu\\  c_{204} &=& \frac{8(1+k+N)}{(2+k+N)^{2}},
\qquad c_{205} = \frac{8(3+6k+2k^{2}+6N+5k\, N+2N^{2})}{3(2+k+N)^{3}},
\nonu\\ c_{206} &=& \frac{4(-6+k^{2}+4k\, N+N^{2})}{3(2+k+N)^{3}},
\nonu\\  c_{207} &=& \frac{4(84+140k+87k^{2}+18k^{3}+118N+147k\, N+51k^{2}N+54N^{2}+37k\, N^{2}+8N^{3})}{3(2+N)(2+k+N)^{3}},
\nonu\\  c_{208} &=& \frac{2(108+133k+38k^{2}+185N+167k\, N+28k^{2}N+107N^{2}+54k\, N^{2}+20N^{3})}{3(2+N)(2+k+N)^{4}},
\nonu\\  c_{209} &=& \frac{2(12+37k+14k^{2}+41N+71k\, N+16k^{2}N+35N^{2}+30k\, N^{2}+8N^{3})}{3(2+N)(2+k+N)^{4}},
\nonu\\  c_{210} &=& \frac{8}{(2+k+N)^{2}},
\qquad c_{211} = -\frac{8}{(2+k+N)^{2}},
\qquad  c_{212} = \frac{8}{(2+k+N)^{2}},
\nonu\\ c_{213} &=& \frac{4(84+140k+87k^{2}+18k^{3}+118N+147k\, N+51k^{2}N+54N^{2}+37k\, N^{2}+8N^{3})}{3(2+N)(2+k+N)^{3}},
\nonu\\  c_{214} &=& -\frac{2(20+31k+10k^{2}+35N+41k\, N+8k^{2}N+21N^{2}+14k\, N^{2}+4N^{3})}{(2+N)(2+k+N)^{4}},
\nonu\\  c_{215} &=& \frac{2(60+61k+14k^{2}+137N+107k\, N+16k^{2}N+95N^{2}+42k\, N^{2}+20N^{3})}{3(2+N)(2+k+N)^{4}},
\nonu\\ c_{216} &=& -\frac{8(1+k+N)}{(2+k+N)^{2}},
\qquad  c_{217} = \frac{8(3+6k+2k^{2}+6N+5k\, N+2N^{2})}{3(2+k+N)^{3}},
\nonu\\  c_{218} &=& -\frac{4(30+55k+18k^{2}+38N+35k\, N+10N^{2})}{3(2+N)(2+k+N)^{3}},
\nonu\\  c_{219} &=& -\frac{2(32+35k+10k^{2}+61N+47k\, N+8k^{2}N+39N^{2}+16k\, N^{2}+8N^{3})}{(2+N)(2+k+N)^{4}},
\nonu\\  c_{220} &=& -\frac{2(32+55k+18k^{2}+41N+35k\, N+11N^{2})}{(2+N)(2+k+N)^{4}},
\qquad  c_{221} = \frac{4(15+8k+7N)}{3(2+k+N)^{3}},
\nonu\\ c_{222} &=& \frac{2(3k+2k^{2}+13N+15k\, N+4k^{2}N+15N^{2}+8k\, N^{2}+4N^{3})}{(2+N)(2+k+N)^{4}},
\nonu\\  c_{223} &=& \frac{4(13k+6k^{2}+3N+9k\, N+N^{2})}{(2+N)(2+k+N)^{3}},
\qquad  c_{224} = \frac{4(15+8k+7N)}{3(2+k+N)^{3}},
\nonu\\ c_{225} &=& -\frac{4(30+55k+18k^{2}+38N+35k\, N+10N^{2})}{3(2+N)(2+k+N)^{3}},
\nonu\\  c_{226} &=& \frac{2(20+21k+6k^{2}+25N+13k\, N+7N^{2})}{(2+N)(2+k+N)^{2}},
\nonu\\  c_{227} &=& -\frac{2(16+11k+2k^{2}+29N+19k\, N+4k^{2}N+19N^{2}+8k\, N^{2}+4N^{3})}{(2+N)(2+k+N)^{3}},
\nonu\\  c_{228} &=& \frac{2(16+11k+2k^{2}+29N+19k\, N+4k^{2}N+19N^{2}+8k\, N^{2}+4N^{3})}{(2+N)(2+k+N)^{3}},
\nonu\\  c_{229} &=& \frac{2(16+21k+6k^{2}+19N+13k\, N+5N^{2})}{(2+N)(2+k+N)^{3}},
\nonu\\  c_{230} &=& -\frac{2(20+21k+6k^{2}+25N+13k\, N+7N^{2})}{(2+N)(2+k+N)^{3}},
\nonu\\  c_{231} &=& \frac{2(30+37k+10k^{2}+32N+21k\, N+8N^{2})}{3(2+k+N)^{2}},
\nonu\\  c_{232} &=& -\frac{60+61k+14k^{2}+137N+107k\, N+16k^{2}N+95N^{2}+42k\, N^{2}+20N^{3}}{3(2+N)(2+k+N)^{3}},
\nonu\\  c_{233} &=& \frac{(276+313k+94k^{2}+401N+263k\, N+20k^{2}N+159N^{2}+42k\, N^{2}+16N^{3})}{3(2+N)(2+k+N)^{3}},
\nonu\\ c_{234} &=& \frac{2(36+29k+6k^{2}+37N+39k\, N+12k^{2}N+15N^{2}+16k\, N^{2}+2N^{3})}{3(2+N)(2+k+N)^{3}},
\nonu\\  c_{235} &=& -\frac{2(8+17k+6k^{2}+11N+11k\, N+3N^{2})}{(2+N)(2+k+N)^{2}}.\nonu
\eea

\subsection{The OPEs between the ${\cal N}=2$ higher spin-$\frac{3}{2}$ 
currents}

The remaining OPE in the section $6$ between 
 the ${\cal N}=2$ higher spin-$\frac{3}{2}$ 
currents 
with (\ref{simple})
can be summarized by \footnote{In the remaining Appendix $G$, the 
${\cal N}=2$ currents are denoted without a boldface notation for 
simplicity. One uses a boldface notation for the ${\cal N}=2$ higher spin 
currents.  }
\begin{eqnarray}
&&\left(
\begin{array}{c}
{\bf U^{(\frac{3}{2})}} \nonu \\\\
{\bf V^{(\frac{3}{2})}}
\end{array}\right)(Z_{1})
\left(
\begin{array}{c}
{\bf U^{(\frac{3}{2})}} \nonu \\\\
{\bf V^{(\frac{3}{2})}}
\end{array}\right)(Z_{2})\;=\;\frac{\theta_{12}}{z_{12}^{2}}\Bigg[ c_{{\pm}1}\, G_{\pm}\, DG_{\pm}\Bigg](Z_{2})+\frac{\overline{\theta}_{12}}{z_{12}^{2}}\Bigg[ c_{{\pm}2}\, G_{\pm}\,\overline{D}G_{\pm}\Bigg](Z_{2})\nonumber\\&&+\frac{\theta_{12}\overline{\theta}_{12}}{z_{12}^{2}}\Bigg[c_{{\pm}3}\, G_{\pm}\,[D,\overline{D}]G_{\pm}+c_{{\pm}4}\, G_{\pm}\,\partial G_{\pm}+c_{{\pm}5}\, DG_{\pm}\,\overline{D}G_{\pm}\Bigg](Z_{2})
\nonumber\\&&+\frac{1}{z_{12}}\Bigg[c_{{\pm}6}\, G_{\pm}\,\partial G_{\pm}+c_{{\pm}7}\, DG_{\pm}\,\overline{D}G_{\pm}+c_{{\pm}8}\,\overline{H}\, G_{\pm}\, DG_{\pm}
+c_{{\pm}9}\, H\, G_{\pm}\,\overline{D}G_{\pm}\Bigg](Z_{2})
\nonumber\\&&+\frac{\theta_{12}}{z_{12}}\Bigg[c_{{\pm}10}\, G_{\pm}\,\partial DG_{\pm}+c_{{\pm}11}\,\partial G_{\pm}\, DG_{\pm}+c_{{\pm}12}\, DG_{\pm}\,[D,\overline{D}]G_{\pm}
+c_{{\pm}13}\, H\, G_{\pm}\,\partial G_{\pm}
\nonumber\\&&+c_{{\pm}14}\, H\, G_{\pm}\,[D,\overline{D}]G_{\pm}+c_{{\pm}15}\, H\, DG_{\pm}\,\overline{D}G_{\pm}+c_{{\pm}16}\,\overline{H}\, DG_{\pm}\, DG_{\pm}+c_{{\pm}17}\, D\overline{H}\, G_{\pm}\, DG_{\pm}\Bigg](Z_{2})
\nonumber\\&&+\frac{\overline{\theta}_{12}}{z_{12}^{2}}\Bigg[c_{{\pm}18}\, G_{\pm}\,\partial\overline{D}G_{\pm}+c_{{\pm}19}\,\partial G_{\pm}\,\overline{D}G_{\pm}+c_{{\pm}20}\,\overline{D}G_{\pm}\,[D,\overline{D}]G_{\pm}
+c_{{\pm}21}\, H\,\overline{D}G_{\pm}\,\overline{D}G_{\pm}
\nonumber\\&&+c_{{\pm}22}\,\overline{D}H\, G_{\pm}\,\overline{D}G_{\pm}+c_{{\pm}23}\,\overline{H}\, G_{\pm}\,\partial G_{\pm}
+c_{{\pm}24}\,\overline{H}\, G_{\pm}\,[D,\overline{D}]G_{\pm}+c_{{\pm}25}\,\overline{H}\, DG_{\pm}\,\overline{D}G_{\pm}\Bigg](Z_{2})
\nonumber\\&&+\frac{\theta_{12}\overline{\theta}_{12}}{z_{12}}\Bigg[c_{{\pm}26}\, G_{\pm}\,\partial^{2}G_{\pm}+c_{{\pm}27}\, G_{\pm}\,\partial[D,\overline{D}]G_{\pm}+c_{{\pm}28}\,\partial G_{\pm}\,[D,\overline{D}]G_{\pm}
+c_{{\pm}29}\, DG_{\pm}\,\partial\overline{D}G_{\pm}
\nonumber\\&&+c_{{\pm}30}\,\partial DG_{\pm}\,\overline{D}G_{\pm}+c_{{\pm}31}\, T\, G_{\pm}\,\partial G_{\pm}
+c_{{\pm}32}\, DT\, G_{\pm}\,\overline{D}G_{\pm}+c_{{\pm}33}\,\overline{D}T\, G_{\pm}\, DG_{\pm}
\nonumber\\&&+c_{{\pm}34}\, H\, G_{\pm}\,\partial\overline{D}G_{\pm}
+c_{{\pm}35}\, H\,\partial G_{\pm}\,\overline{D}G_{\pm}+c_{{\pm}36}\, H\,\overline{D}G_{\pm}\,[D,\overline{D}]G_{\pm}
+c_{{\pm}37}\,\partial H\,G_{\pm}\,\overline{D}G_{\pm}
\nonumber\\&&+c_{{\pm}38}\,\overline{D}H\,G_{\pm}\,\partial G_{\pm}\,\overline{D}H\, DG_{\pm}\,\overline{D}G_{\pm}+c_{{\pm}40}\,\overline{H}\,G_{\pm}\,\partial DG_{\pm}
+c_{{\pm}41}\,\overline{H}\,\partial G_{\pm}\, DG_{\pm}
\nonumber\\&&+c_{{\pm}42}\,\overline{H}\, DG_{\pm}\,[D,\overline{D}]\overline{G}+c_{{\pm}43}\,\partial\overline{H}\, G_{\pm}\, DG_{\pm}
+c_{{\pm}44}\, D\overline{H}\,G_{\pm}\,\partial G_{\pm}+c_{{\pm}45}\, D\overline{H}\, DG_{\pm}\,\overline{D}G_{\pm}\Bigg](Z_{2})+\cdots,\nonumber
\end{eqnarray}
where 
the coefficients are
\bea
c_{+1} & = & -\frac {8 k} {(2 + k + N)^{2}},\qquad
c_{+2}  =  -\frac {8 N} {(2 + k + N)^{2}},\qquad
c_{+3}  =  -\frac {2 (k + N)} {(2 + k + N)^{2}},\nonu \\
c_{+4}  & = &  -\frac {2 (k - N)} {(2 + k + N)^{2}},\qquad
c_{+5}  =  -\frac {4 (k - N)} {(2 + k + N)^{2}}\qquad
c_{+6}  =  \frac {8} {(2 + k + N)^{2}}, \nonu\\
c_{+7} & = & \frac {8} {(2 + k + N)},\qquad
c_{+8}  =  \frac {8} {(2 + k + N)^{2}},\qquad
c_{+9}  =  -\frac {8} {(2 + k + N)^{2}},\nonu \\
c_{+10}  & = &   -\frac {4 (1 + k)} {(2 + k + N)^{2}},\qquad
c_{+11}  =  -\frac {2 (3 k + N)} {(2 + k + N)^{2}},\qquad
c_{+12}  =  \frac {2} {(2 + k + N)},\nonu\\
c_{+13} & = & \frac {2} {(2 + k + N)^{2}},\qquad
c_{+14}  =  -\frac {2} {(2 + k + N)^{2}},\qquad
c_{+15}  =  \frac {4} {(2 + k + N)^{2}},\nonu \\
c_{+16}  & = &   -\frac {4} {(2 + k + N)^{2}},\qquad
c_{+17}  =  \frac {4} {(2 + k + N)^{2}},\qquad
c_{+18}  =  -\frac {4 (1 + N)} {(2 + k + N)^{2}},\nonu\\
c_{+19} & = & -\frac {2 (k + 3 N)} {(2 + k + N)^{2}},\qquad
c_{+20}  =  -\frac {2} {(2 + k + N)},\qquad
c_{+21}  =  \frac {4} {(2 + k + N)^{2}},\nonu \\
c_{+22} & = &  -\frac {4} {(2 + k + N)^{2}},\qquad
c_{+23}  =  -\frac {2} {(2 + k + N)^{2}},\qquad
c_{+24}  =  -\frac {2} {(2 + k + N)^{2}},\nonu\\
c_{+25} & = & -\frac {4} {(2 + k + N)^{2}},\qquad
c_{+26}  =  -\frac {2 (k - N)} {(2 + k + N)^{2}},\qquad
c_{+27}  =  -\frac {2} {(2 + k + N)},\nonu \\
c_{+28}  & = &   \frac {4} {(2 + k + N)^{2}},\qquad
c_{+29}  =  -\frac {4 (3 + 2 k + N)} {(2 + k + N)^{2}},\qquad
c_{+30}  =  \frac {4 (3 + k + 2 N)} {(2 + k + N)^{2}},\nonu\\
c_{+31} & = & -\frac {4} {(2 + k + N)^{2}},\qquad
c_{+32}  =  \frac {4} {(2 + k + N)^{2}},\qquad
c_{+33}  =  \frac {4} {(2 + k + N)^{2}},\nonu \\
c_{+34}  & = &   \frac {4} {(2 + k + N)^{2}},\qquad
c_{+35}  =  -\frac {2} {(2 + k + N)^{2}},\qquad
c_{+36}  =  \frac {2} {(2 + k + N)^{2}},\nonu\\
c_{+37} & = & -\frac {4 (3 + 2 k + N)} {(2 + k + N)^{3}},\qquad
c_{+38}  =  -\frac {4 (1 + k)} {(2 + k + N)^{3}},\qquad
c_{+39}  =  \frac {4} {(2 + k + N)^{2}},\nonu \\
c_{+40}  & = &   \frac {4} {(2 + k + N)^{2}},\qquad
c_{+41}  =  -\frac {2} {(2 + k + N)^{2}},\qquad
c_{+42}  =  -\frac {2} {(2 + k + N)^{2}},\nonu\\
c_{+43} & = & -\frac {4 (3 + k + 2 N)} {(2 + k + N)^{3}},\qquad
c_{+44}  =  -\frac {4 (1 + N)} {(2 + k + N)^{3}},\qquad
c_{+45}  =  \frac {4} {(2 + k + N)^{2}},\nonu \\
c_{-1}  & = &   -\frac {8 N} {(2 + k + N)^{2}},\qquad
c_{-2}  =  -\frac {8 k} {(2 + k + N)^{2}},\qquad
c_{-3}  =  -\frac {2 (k + N)} {(2 + k + N)^{2}},\nonu\\
c_{-4} & = & \frac {2 (k - N)} {(2 + k + N)^{2}},\qquad
c_{-5}  =  \frac {4 (k - N)} {(2 + k + N)^{2}},\qquad
c_{-6}  =  \frac {8} {(2 + k + N)^{2}},\nonu \\
c_{-7}  & = &  \frac {8} {(2 + k + N)},\qquad
c_{-8}  =  -\frac {8} {(2 + k + N)^{2}},\qquad
c_{-9}  =  \frac {8} {(2 + k + N)^{2}},\nonu\\
c_{-10} & = & -\frac {4 (1 + N)} {(2 + k + N)^{2}},\qquad
c_{-11}  =  -\frac {2 (k + 3 N)} {(2 + k + N)^{2}},\qquad
c_{-12}  =  \frac {2} {(2 + k + N)},\nonu \\
c_{-13}  & = &   -\frac {2} {(2 + k + N)^{2}},\qquad
c_{-14}  =  \frac {2} {(2 + k + N)^{2}},\qquad
c_{-15}  =  -\frac {4} {(2 + k + N)^{2}},\nonu\\
c_{-16} & = & \frac {4} {(2 + k + N)^{2}},\qquad
c_{-17}  =  -\frac {4} {(2 + k + N)^{2}},\qquad
c_{-18}  =  -\frac {4 (1 + k)} {(2 + k + N)^{2}},\nonu \\
c_{-19}  & = &  -\frac {2 (3 k + N)} {(2 + k + N)^{2}},\qquad
c_{-20}  =  -\frac {2} {(2 + k + N)},\qquad
c_{-21}  =  -\frac {4} {(2 + k + N)^{2}},\nonu\\
c_{-22}  & = &   \frac {4} {(2 + k + N)^{2}},\qquad
c_{-23}  =  \frac {2} {(2 + k + N)^{2}},\qquad
c_{-24}  =  \frac {2} {(2 + k + N)^{2}},\nonu \\
c_{-25}  & = &   \frac {4} {(2 + k + N)^{2}},\qquad
c_{-26}  =   \frac {2 (k - N)} {(2 + k + N)^{2}},\qquad
c_{-27}  =  -\frac {2} {(2 + k + N)},\nonu\\
c_{-28}  & = &  \frac {4} {(2 + k + N)^{2}},\qquad
c_{-29}  =  -\frac {4 (3 + k + 2 N)} {(2 + k + N)^{2}},\qquad
c_{-30}  =   \frac {4 (3 + 2 k + N)} {(2 + k + N)^{2}},\nonu \\
c_{-31}  & = &  -\frac {4} {(2 + k + N)^{2}},\qquad
c_{-32}   =   \frac {4} {(2 + k + N)^{2}},\qquad
c_{-33}  =  \frac {4} {(2 + k + N)^{2}},\nonu\\
c_{-34} & = & -\frac {4} {(2 + k + N)^{2}},\qquad
c_{-35}  =  \frac {2} {(2 + k + N)^{2}},\qquad
c_{-36}  =   -\frac {2} {(2 + k + N)^{2}},\nonu \\
c_{-37}  & = &  \frac {4 (3 + k + 2 N)} {(2 + k + N)^{3}},\qquad
c_{-38}   =   \frac {4 (1 + N)} {(2 + k + N)^{3}},\qquad
c_{-39}  =  -\frac {4} {(2 + k + N)^{2}},\nonu\\
c_{-40} & = & -\frac {4} {(2 + k + N)^{2}},\qquad
c_{-41}  =  \frac {2} {(2 + k + N)^{2}},\qquad
c_{-42}  =  \frac {2} {(2 + k + N)^{2}},\nonu \\
c_{-43}  & = &  \frac {4 (3 + 2 k + N)} {(2 + k + N)^{3}},\qquad
c_{-44}  =  \frac {4 (1 + k)} {(2 + k + N)^{3}},\qquad
c_{-45}  =  -\frac {4} {(2 + k + N)^{2}}.
\nonu
\eea

\subsection{The OPEs between the ${\cal N}=2$ higher spin-$\frac{3}{2}$ 
currents}

The remaining OPE in the section $6$ between 
 the ${\cal N}=2$ higher spin-$\frac{3}{2}$ 
currents can be summarized by
\begin{eqnarray}
&&{\bf U^{(\frac{3}{2})}}(Z_{1})\:{\bf V^{(\frac{3}{2})}}(Z_{2})\;=\;\frac{\theta_{12}\overline{\theta}_{12}}{z_{12}^{4}}\, c_{1}+\frac{1}{z_{12}^{3}}\, c_{2}+\frac{\theta_{12}}{z_{12}^{3}}\,c_{3}\, H(Z_{2})+\frac{\overline{\theta}_{12}}{z_{12}^{3}}\,c_{4}\,\overline{H}(Z_{2})\nonumber\\&&+\frac{\theta_{12}\overline{\theta}_{12}}{z_{12}^{3}}\Bigg[c_{5}\, T+c_{6}\,\overline{D}H+c_{7}\, D\overline{H}+c_{8}\, G\,\overline{G}+c_{9}\, H\,\overline{H}\Bigg](Z_{2})
\nonu\\&&+\frac{1}{z_{12}^{2}}\Bigg[c_{10}\, T+c_{11}\,\overline{D}H+c_{12}\, D\overline{H}+c_{13}\, G\,\overline{G}+c_{14}\, H\,\overline{H}\Bigg](Z_{2})
\nonu\\&&+\frac{\theta_{12}}{z_{12}^{2}}\Bigg[c_{15}\, DT+c_{16}\, G\, D\overline{G}+c_{17}\, H\,\overline{D}H+c_{18}\, H\, D\overline{H}+c_{19}\, H\, G\,\overline{G}+c_{20}\, T\, H
\nonu\\&&+c_{21}\, DG\,\overline{G}+c_{22}\,\partial H\Bigg](Z_{2})
\nonu\\&&+\frac{\overline{\theta}_{12}}{z_{12}^{2}}\Bigg[c_{23}\,\overline{D}T+c_{24}\, G\,\overline{D}\overline{G}+c_{25}\,\overline{H}\, D\overline{H}+c_{26}\,\overline{H}\, G\,\overline{G}+c_{27}\, T\,\overline{H}+c_{28}\,\overline{D}G\,\overline{G}+c_{29}\,\overline{D}H\,\overline{H}
\nonu\\&&+c_{30}\,\partial\overline{H}\Bigg](Z_{2})
\nonu\\&&+\frac{\theta_{12}\overline{\theta}_{12}}{z_{12}^{2}}\Bigg[
c_{31}\, {\bf T^{(2)}}
+c_{32}\, {\bf T^{(1)}\, T^{(1)}}
+c_{33}\,[D,\overline{D}]T
+c_{34}\,\partial D\overline{H}
+c_{35}\, G\,[D,\overline{D}]\overline{G}
+c_{36}\, G\,\partial\overline{G}
\nonu\\&&+c_{37}\, H\, G\,\overline{D}\overline{G}+c_{38}\, H\,\overline{H}\, D\overline{H}+c_{39}\, H\,\overline{H}\, G\,\overline{G}+c_{40}\, H\,\overline{D}G\,\overline{G}+c_{41}\, H\,\overline{D}H\,\overline{H}+c_{42}\, H\,\partial\overline{H}
\nonu\\&&+c_{43}\,\overline{H}\, G\, D\overline{G}+c_{44}\,\overline{H}\, DG\,\overline{G}
+c_{45}\, T\, T+c_{46}\, T\,\overline{D}H
+c_{47}\, T\, D\overline{H}+c_{48}\, T\, G\,\overline{G}
+c_{49}\, T\, H\,\overline{H}
\nonu\\&&+c_{50}\,\partial\overline{D}H
+c_{51}\,\overline{D}G\, D\overline{G}
+c_{52}\,\overline{D}H\,\overline{D}H
+c_{53}\,\overline{D}H\, D\overline{H}
+c_{54}\,\overline{D}H\, G\,\overline{G}
+c_{55}\,\overline{D}T\, H
\nonu\\&&+c_{56}\, DG\,\overline{D}\overline{G}
+c_{57}\, D\overline{H}\, D\overline{H}+c_{58}\, D\overline{H}\, G\,\overline{G}+c_{59}\, DT\,\overline{H}+c_{60}\,\partial G\,\overline{G}+c_{61}\,\partial H\,\overline{H}+c_{62}\,\partial T
\Bigg](Z_{2})\nonu\\&&+\frac{1}{z_{12}}\Bigg[c_{63}\,\partial\overline{D}H+c_{64}\,[D,\overline{D}]T+c_{65}\,\partial D\overline{H}+c_{66}\, H\, G\,\overline{D}\overline{G}+c_{67}\, H\,\overline{H}\, D\overline{H}+c_{68}\, H\,\overline{D}G\,\overline{G}
\nonu\\&&+c_{69}\, H\,\overline{D}H\,\overline{H}+c_{70}\, H\,\partial\overline{H}+c_{71}\,\overline{H}\, G\, D\overline{G}+c_{72}\,\overline{H}\, DG\,\overline{G}+c_{73}\, T\, T+c_{74}\, T\,\overline{D}H+c_{75}\, T\, D\overline{H}
\nonu\\&&+c_{76}\, T\, H\,\overline{H}+c_{77}\,\overline{D}G\, D\overline{G}+c_{78}\,\overline{D}H\,\overline{D}H+c_{79}\,\overline{D}H\, D\overline{H}+c_{80}\, DG\,\overline{D}\overline{G}+c_{81}\, D\overline{H}\, D\overline{H}
\nonu\\&&+c_{82}\,\partial G\,\overline{G}+c_{83}\,\partial H\,\overline{H}+c_{84}\,\partial T
\Bigg](Z_{2})
\nonu\\&&+
\frac{\theta_{12}}{z_{12}}\Bigg[
c_{85}\,  D{\bf T^{(2)}}
+c_{109}\,{\bf T^{(1)}}\, D{\bf T^{(1)}}
+c_{87}\, G\,\partial D\overline{G}+c_{88}\, H\,\partial D\overline{H}+c_{89}\, H\, G\,[D,\overline{D}]\overline{G}
\nonu\\&&+c_{90}\, H\, G\,\partial\overline{G}+c_{91}\, H\,\overline{H}\, G\, D\overline{G}+c_{92}\, H\,\overline{H}\, DG\,\overline{G}+c_{93}\, H\,\overline{D}G\, D\overline{G}+c_{94}\, H\,\overline{D}H\,\overline{D}H
\nonu\\&&+c_{95}\, H\,\overline{D}H\, D\overline{H}+c_{96}\, H\,[D,\overline{D}]G\,\overline{G}+c_{97}\, H\, DG\,\overline{D}\overline{G}+c_{98}\, H\, D\overline{H}\, D\overline{H}+c_{99}\, H\, D\overline{H}\, G\,\overline{G}
\nonu\\&&+c_{100}\, H\,\partial G\,\overline{G}
+c_{101}\,\overline{H}\, DG\, D\overline{G}
+c_{102}\, T\, DT
+c_{103}\, T\, G\, D\overline{G}
+c_{104}\, T\, H\,\overline{D}H
+c_{105}\, T\, H\, D\overline{H}
\nonu\\&&
+c_{106}\, T\, T\, H+c_{107}\, T\, DG\,\overline{G}
+c_{108}\, T\,\partial H
+c_{109}\,\partial DT
+c_{110}\,\overline{D}H\, G\, D\overline{G}
+c_{111}\,\overline{D}H\, DG\,\overline{G}
\nonu\\&&
+c_{112}\,\partial\overline{D}H\, H
+c_{113}\,[D,\overline{D}]G\, D\overline{G}
+c_{114}\,[D,\overline{D}]T\, H+c_{115}\, DG\,[D,\overline{D}]\overline{G}+c_{116}\, DG\,\partial\overline{G}
\nonu\\&&+c_{117}\, D\overline{H}\, G\, D\overline{G}+c_{118}\, D\overline{H}\, DG\,\overline{G}+c_{119}\, DT\,\overline{D}H+c_{120}\, DT\, D\overline{H}+c_{121}\, DT\, G\,\overline{G}
\nonu\\&&+c_{122}\, DT\, H\,\overline{H}+c_{123}\,\partial DG\,\overline{G}+c_{124}\,\partial G\, D\overline{G}+c_{125}\,\partial H\,\overline{D}H+c_{126}\,\partial H\, D\overline{H}+c_{127}\,\partial H\, G\,\overline{G}
\nonu\\&&+c_{128}\,\partial H\, H\,\overline{H}+c_{129}\,\partial T\, H+c_{130}\,\partial^{2}H
\Bigg](Z_{2})
\nonu\\&&+\frac{\overline{\theta}_{12}}{z_{12}}\Bigg[
c_{131}\,\overline{D}{\bf T^{(2)}}
+c_{132}\, {\bf T^{(1)}}\,\overline{D}{\bf T^{(1)}}
+c_{133}\, G\,\partial\overline{D}\overline{G}
+c_{134}\, H\,\overline{H}\, G\,\overline{D}\overline{G}+c_{135}\, H\,\overline{H}\,\overline{D}G\,\overline{G}
\nonu\\&&+c_{136}\, H\,\overline{D}G\,\overline{D}\overline{G}+c_{137}\, H\,\partial\overline{H}\,\overline{H}+c_{138}\,\overline{H}\, G\,[D,\overline{D}]\overline{G}+c_{139}\,\overline{H}\, G\,\partial\overline{G}+c_{140}\,\overline{H}\,\overline{D}G\, D\overline{G}
\nonu\\&&+c_{141}\,\overline{H}\,[D,\overline{D}]G\,\overline{G}+c_{142}\,\overline{H}\, DG\,\overline{D}\overline{G}+c_{143}\,\overline{H}\, D\overline{H}\, D\overline{H}+c_{144}\,\overline{H}\,\partial G\,\overline{G}+c_{145}\, T\,\overline{D}T
\nonu\\&&+c_{146}\, T\, G\,\overline{D}\overline{G}+c_{147}\, T\,\overline{H}\, D\overline{H}+c_{148}\, T\, T\,\overline{H}+c_{149}\, T\,\overline{D}G\,\overline{G}+c_{150}\, T\,\overline{D}H\,\overline{H}
+c_{151}\, T\,\partial\overline{H}
\nonu\\&&+c_{152}\,\partial\overline{D}T
+c_{153}\,\overline{D}G\,[D,\overline{D}]\overline{G}
+c_{154}\,\overline{D}G\,\partial\overline{G}
+c_{155}\,\overline{D}H\, G\,\overline{D}\overline{G}
+c_{156}\,\overline{D}H\,\overline{H}\, D\overline{H}
\nonu\\&&+c_{157}\,\overline{D}H\,\overline{H}\, G\,\overline{G}+c_{158}\,\overline{D}H\,\overline{D}G\,\overline{G}+c_{159}\,\overline{D}H\,\overline{D}H\,\overline{H}+c_{160}\,\overline{D}H\,\partial\overline{H}+c_{161}\,\overline{D}T\,\overline{D}H
\nonu\\&&+c_{162}\,\overline{D}T\, D\overline{H}+c_{163}\,\overline{D}T\, G\,\overline{G}+c_{164}\,\overline{D}T\, H\,\overline{H}+c_{165}\,\partial\overline{D}G\,\overline{G}+c_{166}\,\partial\overline{D}H\,\overline{H}
\nonu\\&&+c_{167}\,[D,\overline{D}]G\,\overline{D}\overline{G}+c_{168}\,[D,\overline{D}]T\,\overline{H}+c_{169}\, D\overline{H}\, G\,\overline{D}\overline{G}+c_{170}\, D\overline{H}\,\overline{D}G\,\overline{G}+c_{171}\,\partial D\overline{H}\,\overline{H}
\nonu\\&&+c_{172}\,\partial G\,\overline{D}\overline{G}+c_{173}\,\partial\overline{H}\, D\overline{H}+c_{174}\,\partial\overline{H}\, G\,\overline{G}+c_{175}\,\partial T\,\overline{H}+c_{176}\,\partial^{2}\overline{H}\Bigg](Z_{2})
\nonu\\&&+\frac{\theta_{12}\overline{\theta}_{12}}{z_{12}}\Bigg[
c_{177}\, {\bf W^{(3)}}
+c_{178}\,\partial {\bf T^{(2)}}
+c_{179}\,[D,\overline{D}]{\bf T^{(2)}}
+c_{180}\, {\bf T^{(1)}}\,{\bf  W^{(2)}}
+c_{181}\,{\bf T^{(1)}}\,[D,\overline{D}]{\bf T^{(1)}}
\nonu\\&&+c_{182}\, {\bf U^{(\frac{3}{2})}}\, {\bf V^{(\frac{3}{2})}}
+c_{183}\,\partial {\bf T^{(1)}\, T^{(1)}}
+c_{184}\, G\, DG\,\overline{G}\,\overline{D}\overline{G}
+c_{185}\, G\,\partial^{2}\overline{G}+c_{186}\, H\, G\,\partial\overline{D}\overline{G}
\nonu\\&&+c_{187}\, H\,\overline{H}\, G\,\partial\overline{G}
+c_{188}\, H\,\overline{H}\, D\overline{H}\, D\overline{H}
+c_{189}\, H\,\overline{H}\,\partial G\,\overline{G}+c_{190}\, H\,\overline{D}G\,[D,\overline{D}]\overline{G}
\nonu\\&&+c_{191}\, H\,\overline{D}G\,\partial\overline{G}+c_{192}\, H\,\overline{D}H\, G\,\overline{D}\overline{G}
+c_{193}\, H\,\overline{D}H\,\overline{H}\, D\overline{H}
+c_{194}\, H\,\overline{D}H\,\overline{D}G\,\overline{G}
\nonu\\&&+c_{195}\, H\,\overline{D}H\,\overline{D}H\,\overline{H}
+c_{196}\, H\,\overline{D}H\,\overline{H}
+c_{197}\, H\,\partial\overline{D}G\,\overline{G}+c_{198}\, H\,[D,\overline{D}]G\,\overline{D}\overline{G}
\nonu\\&&+c_{199}\, H\, D\overline{H}\, G\,\overline{D}\overline{G}+c_{200}\, H\, D\overline{H}\,\overline{D}G\,\overline{G}+c_{201}\, H\,\partial D\overline{H}\,\overline{H}+c_{202}\, H\,\partial G\,\overline{D}\overline{G}
\nonu\\&&+c_{203}\, H\,\partial\overline{H}\,\overline{D}\overline{H}+c_{204}\, H\,\partial\overline{H}\, G\,\overline{G}+c_{205}\, H\,\partial^{2}\overline{H}+c_{206}\,\overline{H}\, G\,\partial D\overline{G}+c_{207}\,\overline{H}\,[D,\overline{D}]G\, D\overline{G}
\nonu\\&&+c_{208}\,\overline{H}\, DG\,[D,\overline{D}]\overline{G}+c_{209}\,\overline{H}\, DG\,\partial\overline{G}+c_{210}\,\overline{H}\, D\overline{H}\, G\, D\overline{G}+c_{211}\,\overline{H}\, D\overline{H}\, DG\,\overline{G}
\nonu\\&&+c_{212}\,\overline{H}\,\partial DG\,\overline{G}+c_{213}\,\overline{H}\,\partial G\, D\overline{G}+c_{214}\, T\,\partial\overline{D}H+c_{215}\, T\,[D,\overline{D}]T+c_{216}\, T\,\partial D\overline{H}
\nonu\\&&+c_{217}\, T\, G\,[D,\overline{D}]\overline{G}+c_{218}\, T\, G\,\partial\overline{G}+c_{219}\, T\, H\, G\,\overline{D}\overline{G}+c_{220}\, T\, H\,\overline{H}\, D\overline{H}+c_{221}\, T\, H\,\overline{D}G\,\overline{G}
\nonu\\&&+c_{222}\, T\, H\,\overline{D}H\,\overline{H}+c_{223}\, T\, H\,\partial\overline{H}+c_{224}\, T\,\overline{H}\, G\, D\overline{G}+c_{225}\, T\,\overline{H}\, DG\,\overline{G}+c_{226}\, T\, T\, T
\nonu\\&&+c_{227}\, T\, T\,\overline{D}H+c_{228}\, T\, T\, D\overline{H}+c_{229}\, T\, T\, H\,\overline{H}+c_{230}\, T\,\overline{D}G\, D\overline{G}+c_{231}\, T\,\overline{D}H\,\overline{D}H
\nonu\\&&+c_{232}\, T\,\overline{D}H\, D\overline{H}+c_{233}\, T\,\overline{D}T\, H+c_{234}\, T\,[D,\overline{D}]G\,\overline{G}
+c_{235}\, T\, DG\,\overline{D}\overline{G}+c_{236}\, T\, D\overline{H}\, D\overline{H}
\nonu\\&&+c_{237}\, T\, DT\,\overline{H}+c_{238}\, T\,\partial G\,\overline{G}
+c_{239}\, T\,\partial H\,\overline{H}
+c_{240}\,\partial[D,\overline{D}]T
+c_{241}\,\partial^{2}D\overline{H}
\nonu\\&&+c_{242}\, G\,\partial[D,\overline{D}]\overline{G}
+c_{243}\,\overline{D}\overline{G}\,\partial D\overline{G}+c_{244}\,\overline{D}H\,\partial D\overline{H}+c_{245}\,\overline{D}H\, G\,[D,\overline{D}]\overline{G}+c_{246}\,\overline{D}H\, G\,\partial\overline{G}
\nonu\\&&+c_{247}\,\overline{D}H\,\overline{H}\, G\, D\overline{G}+c_{248}\,\overline{D}H\,\overline{H}\, DG\,\overline{G}+c_{249}\,\overline{D}H\,\overline{D}G\, D\overline{G}
+c_{250}\,\overline{D}H\,\overline{D}H\,\overline{D}H
\nonu\\&&+c_{251}\,\overline{D}H\,\overline{D}H\, D\overline{H}+c_{252}\,\overline{D}H\,[D,\overline{D}]G\,\overline{G}
+c_{253}\,\overline{D}H\, DG\,\overline{D}\overline{G}
+c_{254}\,\overline{D}H\, D\overline{H}\, D\overline{H}
\nonu\\&&+c_{255}\,\overline{D}H\,\partial G\,\overline{G}+c_{256}\,\overline{D}T\, DT+c_{257}\,\overline{D}T\, G\, D\overline{G}+c_{258}\,\overline{D}T\, H\,\overline{D}H+c_{259}\,\overline{D}T\, H\, D\overline{H}
\nonu\\&&+c_{260}\,\overline{D}T\, H\, G\,\overline{G}+c_{261}\,\overline{D}T\, DG\,\overline{G}+c_{262}\,\overline{D}T\,\partial H+c_{263}\,\partial\overline{D}G\, D\overline{G}+c_{264}\,\partial\overline{D}H\,\overline{D}H
\nonu\\&&+c_{265}\,\partial\overline{D}H\, D\overline{H}+c_{266}\,\partial\overline{D}H\, G\,\overline{G}+c_{267}\,\partial\overline{D}H\, H\,\overline{H}+c_{268}\,\partial\overline{D}T\, H+c_{269}\,[D,\overline{D}]G\,[D,\overline{D}]\overline{G}
\nonu\\&&+c_{270}\,[D,\overline{D}]G\,\overline{G}+c_{271}\,[D,\overline{D}]T\,\overline{D}H+c_{272}\,[D,\overline{D}]T\, D\overline{H}+c_{273}\,[D,\overline{D}]T\, G\,\overline{G}
\nonu\\&&+c_{274}\,[D,\overline{D}]T\, H\,\overline{H}+c_{275}\,\partial[D,\overline{D}]G\,\overline{G}+c_{276}\, DG\,\partial\overline{D}\overline{G}+c_{277}\, D\overline{H}\, G\,[D,\overline{D}]\overline{G}
\nonu\\&&+c_{278}\, D\overline{H}\, G\,\partial\overline{G}+c_{279}\, D\overline{H}\,\overline{D}G\, D\overline{G}+c_{280}\, D\overline{H}\,[D,\overline{D}]G\,\overline{G}+c_{281}\, D\overline{H}\, DG\,\overline{D}\overline{G}
\nonu\\&&+c_{282}\, D\overline{H}\, D\overline{H}\, D\overline{H}+c_{283}\, D\overline{H}\,\partial G\,\overline{G}+c_{284}\, DT\, G\,\overline{D}\overline{G}+c_{285}\, DT\,\overline{H}\, D\overline{H}+c_{286}\, D\overline{T}\,\overline{H}\, G\,\overline{G}
\nonu\\&&+c_{287}\, DT\,\overline{D}G\,\overline{G}+c_{288}\, DT\,\overline{D}H\,\overline{H}+c_{289}\, DT\,\partial\overline{H}+c_{290}\,\partial DG\,\overline{D}\overline{G}+c_{291}\,\partial D\overline{H}\, D\overline{H}
\nonu\\&&+c_{292}\,\partial D\overline{H}\, G\,\overline{G}+c_{293}\,\partial DT\,\overline{H}+c_{294}\,\partial G\,[D,\overline{D}]\overline{G}
+c_{295}\,\partial G\,\partial\overline{G}+c_{296}\,\partial H\, G\,\overline{D}\overline{G}
\nonu\\&&+c_{297}\,\partial H\,\overline{H}\, D\overline{H}
+c_{298}\,\partial H\,\overline{H}\, G\,\overline{G}+c_{299}\,\partial H\,\overline{D}G\,\overline{G}+c_{300}\,\partial H\,\overline{D}H\,\overline{H}+c_{301}\,\partial H\,\partial\overline{H}
\nonu\\&&+c_{302}\,\partial\overline{H}\, G\, D\overline{G}+c_{303}\,\overline{H}\, DG\,\overline{G}+c_{304}\,\partial T\, T+c_{305}\,\partial T\,\overline{D}H+c_{306}\,\partial T\, D\overline{H}+c_{307}\,\partial T\, G\,\overline{G}
\nonu\\&&+c_{308}\,\partial T\, H\,\overline{H}+c_{309}\, G\,\overline{D}G\,\overline{G}\, D\overline{G}+c_{310}\,\partial^{2}G\,\overline{G}
+c_{311}\,\partial^{2}H\,\overline{H}
c_{312}\,\partial^{2}\overline{D}H
+c_{313}\,\partial^{2}T\Bigg](Z_{2})+\cdots,\nonumber
\end{eqnarray}
where the coefficients are
\begin{eqnarray}
c_ {1} & =& \frac {4 k (k - N) N} {(2 + k + N)^{2}},\qquad
c_ {2}   =  -\frac {8 k\, N} {(2 + k + N)},\nonumber\\
c_ {3}  & =&  -\frac {8 k\, N} {(2 + k + N)^{2}},\qquad
c_ {4}   =  \frac {8 k\, N} {(2 + k + N)^{2}},\nonumber\\
c_ {5}  & =&  -\frac {4 (2 k + k^{2} + 2 N + 4 k\, 
     N + N^{2})} {(2 + k + N)^{2}},\qquad
c_ {6}   =  -\frac {4 (-2 k - k^{2} + 2 N + 2 k\, 
     N + 2 k^{2} N + N^{2})} {(2 + k + N)^{3}},\nonumber\\
c_ {7}  & =&  -\frac {4 (2 k + k^{2} - 2 N + 2 k\, N - N^{2} + 2 k\, 
     N^{2})} {(2 + k + N)^{3}},\qquad
c_ {8}   =  -\frac {4 (k - N)} {(2 + k + N)^{2}},\nonumber\\
c_ {9}  & =&  \frac {4 (k + N)} {(2 + k + N)^{2}},\qquad
c_ {10}   =  -\frac {4 (k - N)} {(2 + k + N)},\qquad
c_ {11}   =  \frac {4 (k + N + 2 k\, N)} {(2 + k + N)^{2}},\nonu \\
c_ {12}   & = &  
-\frac {4 (k + N + 2 k\, N)} {(2 + k + N)^{2}},\qquad
c_ {13}   =  -\frac {4 (k + N)} {(2 + k + N)^{2}},\qquad
c_ {14}   =  \frac {4 (k - N)} {(2 + k + N)^{2}},\nonumber\\
c_ {15}  & =&  \frac {4 (1 + k + 2 N)} {(2 + k + N)},\qquad
c_ {16}   =  -\frac {4 (1 + k)} {(2 + k + N)^{2}},\nonumber\\
c_ {17}  & =&  \frac {4 (k + N + 2 k\, N)} {(2 + k + N)^{3}},\qquad
c_ {18}   =  \frac {4 (2 + 2 k + k^{2} + 4 N + k\, 
     N + 2 N^{2})} {(2 + k + N)^{3}},\nonumber\\
c_ {19}  & =&  \frac {4} {(2 + k + N)^{2}},\qquad
c_ {20}   =  -\frac {4 (k - N)} {(2 + k + N)^{2}},\qquad
c_ {21}   =  -\frac {4 (1 + k + 2 N)} {(2 + k + N)^{2}},\nonu \\
c_ {22}   & = &  -\frac {4 (1 + 3 N + 2 k\, N)} {(2 + k + N)^{2}},\qquad
c_ {23}   =  -\frac {4 (1 + 2 k + N)} {(2 + k + N)},\qquad
c_ {24}   =  -\frac {4 (1 + N)} {(2 + k + N)^{2}},\nonumber\\
c_ {25}  & =&  \frac {4 (k + N + 2 k\, N)} {(2 + k + N)^{3}},\qquad
c_ {26}   =  -\frac {4} {(2 + k + N)^{2}},\nonumber\\
c_ {27}  & =&  \frac {4 (k - N)} {(2 + k + N)^{2}},\qquad
c_ {28}   =  -\frac {4 (1 + 2 k + N)} {(2 + k + N)^{2}},\nonumber\\
c_ {29}  & =&  \frac {4 (2 + 4 k + 2 k^{2} + 2 N + k\,
     N + N^{2})} {(2 + k + N)^{3}},\qquad
c_ {30}   =  \frac {4 (1 + 3 k + 2 k\, N)} {(2 + k + N)^{2}},\qquad
c_ {31}   =  2,\nonumber\\
c_ {32}  & =&  -\frac {(60 + 77 k + 22 k^{2} + 121 N + 115 k\, 
    N + 20 k^{2} N + 79 N^{2} + 42 k\, 
    N^{2} + 16 N^{3})} {(2 + N) (2 + k + N)^{2}},\nonumber\\
c_ {33}  & =&  -\frac {(20 + 24 k + 7 k^{2} + 22 N + 14 k\,
    N + 5 N^{2})} {(2 + k + N)^{2}}, \nonumber\\
c_ {34}  & =&  -\frac{1}{(2 + N) (2 + k + N)^{3}}2 (-28 - 37 k - 12 k^{2} - 65 N - 48 k\, 
     N - 7 k^{2} N - 44 N^{2}
     \nonu\\ &-& 10 k\, 
     N^{2} + k^{2} N^{2} - 9 N^{3} + 3 k\, 
     N^{3}) ,\nonumber\\
c_ {35}  & =&  -\frac {4 (1 + k + N)} {(2 + k + N)^{2}},\nonumber\\
c_ {36}  & =&  -\frac {(20 + 31 k + 10 k^{2} + 35 N + 41 k\, 
    N + 8 k^{2} N + 21 N^{2} + 14 k\, 
    N^{2} + 4 N^{3})} {(2 + N) (2 + k + N)^{3}},\nonumber\\
c_ {37}  & =&  \frac {2 (32 + 40 k + 12 k^{2} + 40 N + 25 k\,
     N + 11 N^{2})} {(2 + N) (2 + k + N)^{3}},\nonumber\\
c_ {38}  & =&  \frac {2 (8 + 11 k + 4 k^{2} + 25 N + 23 k\, 
     N + 5 k^{2} N + 21 N^{2} + 10 k\, 
     N^{2} + 5 N^{3})} {(2 + N) (2 + k + N)^{4}},\nonumber\\
c_ {39}  & =&  \frac {2 (32 + 55 k + 18 k^{2} + 41 N + 35 k\, 
     N + 11 N^{2})} {(2 + N) (2 + k + N)^{4}},\qquad
c_ {40}   =  -\frac {2 (8 + 3 k + 5 N)} {(2 + k + N)^{3}},\nonumber\\
c_ {41}  & =&  -\frac {2 (8 + 11 k + 4 k^{2} + 25 N + 23 k\, 
     N + 5 k^{2} N + 21 N^{2} + 10 k\, 
     N^{2} + 5 N^{3})} {(2 + N) (2 + k + N)^{4}}, \nonumber\\
c_ {42}  & =&  \frac {(52 + 47 k + 18 k^{2} + 99 N + 61 k\, 
    N + 12 k^{2} N + 57 N^{2} + 20 k\, 
    N^{2} + 10 N^{3})} {(2 + N) (2 + k + N)^{3}},\nonumber\\
c_ {43}  & =&  -\frac {2 (4 + k + 3 N)} {(2 + k + N)^{3}},\qquad
c_ {44}   =  \frac {2 (24 + 36 k + 12 k^{2} + 32 N + 23 k\, 
     N + 9 N^{2})} {(2 + N) (2 + k + N)^{3}},\nonumber\\
c_ {45}  & =&  -\frac {(20 + 25 k + 6 k^{2} + 21 N + 15 k\,
    N + 5 N^{2})} {(2 + N) (2 + k + N)^{2}},\nonumber\\
c_ {46}  & =&  \frac {2 (-5 k - 4 k^{2} + 13 N + 11 k\, 
     N + k^{2} N + 13 N^{2} + 8 k\, 
     N^{2} + 3 N^{3})} {(2 + N) (2 + k + N)^{3}},\nonumber\\
c_ {47}  & =&  -\frac {2 (3 k + 5 N + 15 k\, 
     N + 3 k^{2} N + 5 N^{2} + 8 k\, 
     N^{2} + N^{3})} {(2 + N) (2 + k + N)^{3}},\nonumber\\
c_ {48}  & =&  -\frac {2 (8 + 17 k + 6 k^{2} + 11 N + 11 k\, 
     N + 3 N^{2})} {(2 + N) (2 + k + N)^{3}},\nonumber\\
c_ {49}  & =&  \frac {2 (20 + 21 k + 6 k^{2} + 25 N + 13 k\, 
     N + 7 N^{2})} {(2 + N) (2 + k + N)^{3}}, \nonumber\\
c_ {50}  & =&  -\frac {2 (12 + 5 k + 25 N + 24 k\, 
     N + 9 k^{2} N + 16 N^{2} + 14 k\, 
     N^{2} + 3 k^{2} N^{2} + 3 N^{3} + k\, 
     N^{3})} {(2 + N) (2 + k + N)^{3}},\nonumber\\
c_ {51}  & =&  -\frac {2 (10 + 7 k + 7 N)} {(2 + k + N)^{2}},\nonumber\\
c_ {52}  & =&  \frac {(20 + 35 k + 14 k^{2} + 47 N + 67 k\, 
    N + 22 k^{2} N + 31 N^{2} + 30 k\, 
    N^{2} + 6 k^{2} N^{2} + 6 N^{3} + 2 k\, 
    N^{3})} {(2 + N) (2 + k + N)^{4}}, \nonumber\\
c_ {53}  & =&  \frac{1}{(2 + N) (2 + k + N)^{4}}2 (60 + 109 k + 66 k^{2} + 14 k^{3} + 129 N + 
      177 k\, N + 68 k^{2} N + 7 k^{3} N 
      \nonu\\ &+& 105 N^{2} + 102 k\, 
     N^{2} + 19 k^{2} N^{2} + 38 N^{3} + 21 k\, 
     N^{3} + 5 N^{4}) , \nonumber\\
c_ {54}  & =&  \frac {2 (28 + 35 k + 10 k^{2} + 51 N + 47 k\, 
     N + 8 k^{2} N + 31 N^{2} + 16 k\, 
     N^{2} + 6 N^{3})} {(2 + N) (2 + k + N)^{4}},\nonumber\\
c_ {55}  & =&  -\frac {4 (1 + 2 k + N)} {(2 + k + N)^{2}},\qquad
c_ {56}   =  -\frac {2 (20 + 32 k + 12 k^{2} + 22 N + 21 k\, 
     N + 5 N^{2})} {(2 + N) (2 + k + N)^{2}}, \nonumber\\
c_ {57}  & =&  \frac {(20 + 19 k + 6 k^{2} + 31 N + 11 k\, 
    N + 2 k^{2} N + 15 N^{2} - 10 k\, 
    N^{2} - 2 k^{2} N^{2} + 2 N^{3} - 6 k\, 
    N^{3})} {(2 + N) (2 + k + N)^{4}}, \nonumber\\
c_ {58}  & =&  -\frac {2 (1 + N) (12 + 19 k + 6 k^{2} + 15 N + 12 k\, 
     N + 4 N^{2})} {(2 + N) (2 + k + N)^{4}},\qquad
c_ {59}   =  \frac {4 (1 + k + 2 N)} {(2 + k + N)^{2}},\nonumber\\
c_ {60}  & =&  \frac {(20 + 7 k - 2 k^{2} + 59 N + 29 k\, 
    N + 2 k^{2} N + 45 N^{2} + 14 k\, 
    N^{2} + 10 N^{3})} {(2 + N) (2 + k + N)^{3}}, \nonumber\\
c_ {61}  & =&  -\frac {(68 + 75 k + 22 k^{2} + 95 N + 61 k\, 
    N + 2 k^{2} N + 29 N^{2} + 8 k\, 
    N^{2})} {(2 + N) (2 + k + N)^{3}}, \nonumber\\
c_ {62}  & =&  -\frac {(-2 k - 2 k^{2} + 18 N + 22 k\, 
    N + 5 k^{2} N + 20 N^{2} + 14 k\, 
    N^{2} + 5 N^{3})} {(2 + N) (2 + k + N)^{2}},\nonumber\\
c_ {63}  & =&  \frac {4 (1 + k) (1 + N)} {(2 + k + N)^{2}},\qquad
c_ {64}   =  2,\qquad
c_ {65}   =  -\frac {4 (1 + k) (1 + N)} {(2 + k + N)^{2}},\nonumber\\
c_ {66}  & =&  -\frac {4} {(2 + k + N)^{2}},\qquad
c_ {67}   =  \frac {4 (k - N)} {(2 + k + N)^{3}},\qquad
c_ {68}   =  \frac {4} {(2 + k + N)^{2}},\nonu \\
c_ {69}   & = &  -\frac {4 (k - N)} {(2 + k + N)^{3}},\qquad
c_ {70}   =  -\frac {4 (2 + N)} {(2 + k + N)^{2}},\qquad
c_ {71}   =  \frac {4} {(2 + k + N)^{2}},\nonumber\\
c_ {72}  & =&  -\frac {4} {(2 + k + N)^{2}},\qquad
c_ {73}   =  \frac {4} {(2 + k + N)},\qquad
c_ {74}   =  \frac {4 (k - N)} {(2 + k + N)^{2}},\nonu \\
c_ {75}   & = &   -\frac {4 (k - N)} {(2 + k + N)^{2}},\qquad
c_ {76}   =  -\frac {8} {(2 + k + N)^{2}},\qquad
c_ {77}   =  \frac {4} {(2 + k + N)},\nonumber\\
c_ {78}  & =&  -\frac {4 (1 + k) (1 + N)} {(2 + k + N)^{3}},\qquad
c_ {79}   =  -\frac {4 (2 + 2 k + k^{2} + 2 N + N^{2})} {(2 + k + N)^{3}},
\qquad
c_ {80}   =  \frac {4} {(2 + k + N)},\nonu \\ 
c_ {81}   & = &  -\frac {4 (1 + k) (1 + N)} {(2 + k + N)^{3}},\qquad
c_ {82}   =  -\frac {4 (k + N)} {(2 + k + N)^{2}},\qquad
c_ {83}   =  \frac {4 (2 + k)} {(2 + k + N)^{2}},\nonumber\\
c_ {84}  & =&  -\frac {2 (k - N)} {(2 + k + N)},\qquad
c_ {85}   =  1,\nonumber\\
c_ {86}  & =&  \frac {(60 + 77 k + 22 k^{2} + 121 N + 115 k\, 
    N + 20 k^{2} N + 79 N^{2} + 42 k\, 
    N^{2} + 16 N^{3})} {(2 + N) (2 + k + N)^{2}}, \nonumber\\
c_ {87}  & =&  \frac {(4 + 7 k + 2 k^{2} + 19 N + 21 k\, 
    N + 4 k^{2} N + 17 N^{2} + 10 k\, 
    N^{2} + 4 N^{3})} {(2 + N) (2 + k + N)^{3}},\nonumber\\
c_ {88}  & =&  -\frac {(-8 + 7 k + 2 k^{2} - 7 N + 21 k\, 
    N + 4 k^{2} N - N^{2} + 10 k\, N^{2})} {(2 + N) (2 + k + N)^{3}},\nonumber\\
c_ {89}  & =&  -\frac {(18 + 21 k + 6 k^{2} + 22 N + 13 k\, 
    N + 6 N^{2})} {(2 + N) (2 + k + N)^{3}},\nonumber\\
c_ {90}  & =&  -\frac {(24 + 17 k - 11 k^{2} - 6 k^{3} + 59 N + 46 k\, 
    N + k^{2} N + 45 N^{2} + 23 k\, 
    N^{2} + 10 N^{3})} {(2 + N) (2 + k + N)^{4}},\nonumber\\
c_ {91}  & =&  \frac {(32 + 55 k + 18 k^{2} + 41 N + 35 k\, 
    N + 11 N^{2})} {(2 + N) (2 + k + N)^{4}},\nonumber\\
c_ {92}  & =&  -\frac {(32 + 55 k + 18 k^{2} + 41 N + 35 k\, 
    N + 11 N^{2})} {(2 + N) (2 + k + N)^{4}},\nonumber\\
c_ {93}  & =&  -\frac {2 (1 + N)} {(2 + k + N)^{3}},\qquad
c_ {94}   =  -\frac {4 (1 + k) (1 + N)} {(2 + k + N)^{4}},\nonumber\\
c_ {95}  & =&  -\frac {(32 + 27 k + 10 k^{2} + 69 N + 35 k\, 
    N + 8 k^{2} N + 51 N^{2} + 12 k\, 
    N^{2} + 12 N^{3})} {(2 + N) (2 + k + N)^{4}},\nonumber\\
c_ {96}  & =&  -\frac {(5 + 2 k + 3 N)} {(2 + k + N)^{3}},\qquad
c_ {97}   =  \frac {2 (26 + 25 k + 6 k^{2} + 30 N + 15 k\,
     N + 8 N^{2})} {(2 + N) (2 + k + N)^{3}},\nonumber\\
c_ {98}  & =&  \frac {(8 + 3 k + 2 k^{2} + 33 N + 15 k\, 
    N + 4 k^{2} N + 31 N^{2} + 8 k\, 
    N^{2} + 8 N^{3})} {(2 + N) (2 + k + N)^{4}},\nonumber\\
c_ {99}  & =&  \frac {(32 + 55 k + 18 k^{2} + 41 N + 35 k\, 
    N + 11 N^{2})} {(2 + N) (2 + k + N)^{4}},\nonumber\\
c_ {100}  & =&  -\frac {(8 + 27 k + 10 k^{2} + 41 N + 64 k\, 
    N + 14 k^{2} N + 38 N^{2} + 29 k\, 
    N^{2} + 9 N^{3})} {(2 + N) (2 + k + N)^{4}},\nonumber\\
c_ {101}  & =&  \frac {2 (8 + 17 k + 6 k^{2} + 11 N + 11 k\, 
     N + 3 N^{2})} {(2 + N) (2 + k + N)^{3}},\nonu \\
c_ {102}  & = &  \frac {(36 + 29 k + 6 k^{2} + 41 N + 17 k\, 
    N + 11 N^{2})} {(2 + N) (2 + k + N)^{2}},\nonumber\\
c_ {103}  & =&  -\frac {(13 k + 6 k^{2} + 3 N + 9 k\, 
    N + N^{2})} {(2 + N) (2 + k + N)^{3}},\qquad
c_ {104}   =  \frac {4 (k - N)} {(2 + k + N)^{3}},\nonumber\\
c_ {105}  & =&  \frac {(36 + 21 k + 6 k^{2} + 49 N + 13 k\, 
    N + 15 N^{2})} {(2 + N) (2 + k + N)^{3}},\qquad
c_ {106}   =  \frac {4} {(2 + k + N)^{2}},\nonumber\\
c_ {107}  & =&  \frac {(13 k + 6 k^{2} + 3 N + 9 k\, 
    N + N^{2})} {(2 + N) (2 + k + N)^{3}},\nonumber\\
c_ {108}  & =&  \frac {(-48 - 53 k - 14 k^{2} - 35 N - 29 k\, 
    N - 4 k^{2} N + 3 N^{2} + 4 N^{3})} {(2 + N) (2 + k + N)^{3}},\nonumber\\
c_ {109}  & =&  -\frac {2 (1 + k) (5 + 2 k + 3 N)} {(2 + k + N)^{2}},\nonumber\\
c_ {110}  & =&  \frac {(20 + 31 k + 10 k^{2} + 35 N + 41 k\, 
    N + 8 k^{2} N + 21 N^{2} + 14 k\, 
    N^{2} + 4 N^{3})} {(2 + N) (2 + k + N)^{4}}, \nonumber\\
c_ {111}  & =&  -\frac {(20 + 31 k + 10 k^{2} + 35 N + 41 k\, 
    N + 8 k^{2} N + 21 N^{2} + 14 k\, 
    N^{2} + 4 N^{3})} {(2 + N) (2 + k + N)^{4}},\nonumber\\
c_ {112}  & =&  \frac {4 (1 + k) (1 + N)} {(2 + k + N)^{3}},\qquad
c_ {113}   =  \frac {(5 + 2 k + 3 N)} {(2 + k + N)^{2}},\nonumber\\
c_ {114}  & =&  \frac {2} {(2 + k + N)},\qquad
c_ {115}   =  \frac {(18 + 21 k + 6 k^{2} + 22 N + 13 k\, 
    N + 6 N^{2})} {(2 + N) (2 + k + N)^{2}},\nonumber\\
c_ {116}  & =&  \frac {(20 + 10 k - 13 k^{2} - 6 k^{3} + 40 N + 25 k\, 
    N - 3 k^{2} N + 28 N^{2} + 13 k\, 
    N^{2} + 6 N^{3})} {(2 + N) (2 + k + N)^{3}}, \nonumber\\
c_ {117}  & =&  \frac {(20 + 5 k - 2 k^{2} + 29 N - 3 k\, 
    N - 4 k^{2} N + 13 N^{2} - 4 k\, 
    N^{2} + 2 N^{3})} {(2 + N) (2 + k + N)^{4}}, \nonumber\\
c_ {118}  & =&  -\frac {(52 + 89 k + 56 k^{2} + 12 k^{3} + 89 N + 97 k\, 
    N + 30 k^{2} N + 47 N^{2} + 24 k\, 
    N^{2} + 8 N^{3})} {(2 + N) (2 + k + N)^{4}}, \nonumber\\
c_ {119}  & =&  -\frac {(16 + 11 k + 2 k^{2} + 45 N + 27 k\, 
    N + 4 k^{2} N + 35 N^{2} + 12 k\, 
    N^{2} + 8 N^{3})} {(2 + N) (2 + k + N)^{3}}, \nonumber\\
c_ {120}  & =&  \frac {(16 + 11 k + 2 k^{2} + 45 N + 27 k\, 
    N + 4 k^{2} N + 35 N^{2} + 12 k\, 
    N^{2} + 8 N^{3})} {(2 + N) (2 + k + N)^{3}},\nonumber\\
c_ {121}  & =&  \frac {(16 + 21 k + 6 k^{2} + 19 N + 13 k\, 
    N + 5 N^{2})} {(2 + N) (2 + k + N)^{3}},\nonumber\\
c_ {122}  & =&  -\frac {(36 + 29 k + 6 k^{2} + 41 N + 17 k\, 
    N + 11 N^{2})} {(2 + N) (2 + k + N)^{3}},\nonumber\\
c_ {123}  & =&  \frac {(20 + 15 k + 2 k^{2} + 19 N + 17 k\, 
    N + 4 k^{2} N + 5 N^{2} + 6 k\, N^{2})} {(2 + N) (2 + k + N)^{3}},\nonumber\\
c_ {124}  & =&  \frac {(-12 - 4 k + 6 N + 23 k\, 
    N + 6 k^{2} N + 17 N^{2} + 15 k\, 
    N^{2} + 5 N^{3})} {(2 + N) (2 + k + N)^{3}}, \nonumber\\
c_ {125}  & =&  \frac{1} {(2 + N) (2 + k + N)^{4}}(36 + 39 k + 10 k^{2} + 99 N + 101 k\, 
    N + 24 k^{2} N + 73 N^{2} 
    \nonu\\ &+& 58 k\, 
    N^{2} + 8 k^{2} N^{2} + 16 N^{3} + 8 k\, 
    N^{3}), \nonumber\\
c_ {126}  & =&  \frac {1}{(2 + N) (2 + k + N)^{4}}(-76 - 115 k - 54 k^{2} - 8 k^{3} - 163 N - 203 k\, 
    N - 66 k^{2} N 
    \nonu\\ &-& 4 k^{3} N - 107 N^{2} - 102 k\, 
    N^{2} - 18 k^{2} N^{2} - 22 N^{3} - 14 k\, 
    N^{3}), \nonumber\\
c_ {127}  & =&  -\frac {2 (1 + N) (12 + 19 k + 6 k^{2} + 15 N + 12 k\, 
     N + 4 N^{2})} {(2 + N) (2 + k + N)^{4}},\nonumber\\
c_ {128}  & =&  -\frac {(-16 - 21 k - 6 k^{2} - 3 N - 5 k\, 
    N + 11 N^{2} + 4 k\, N^{2} + 4 N^{3})} {(2 + N) (2 + k + N)^{4}},
\qquad
c_ {129}   =  -\frac {2 (k - N)} {(2 + k + N)^{2}},\nonumber\\
c_ {130}  & =&  \frac{1} {2 (2 + N) (2 + k + N)^{3}} (100 + 119 k + 34 k^{2} + 131 N + 85 k\, 
    N + 4 k^{2} N + 57 N^{2} 
    \nonu\\ &-& 2 k\, 
    N^{2} - 8 k^{2} N^{2} + 8 N^{3} - 8 k\, 
    N^{3}),
    \nonumber\\
c_ {131}  & =& 1,
\nonumber\\
c_ {132}  & =&  -\frac {(60 + 77 k + 22 k^{2} + 121 N + 115 k\, 
    N + 20 k^{2} N + 79 N^{2} + 42 k\, 
    N^{2} + 16 N^{3})} {(2 + N) (2 + k + N)^{2}}, \nonumber\\
c_ {133}  & =&  -\frac {(36 + 39 k + 10 k^{2} + 67 N + 53 k\, 
    N + 8 k^{2} N + 41 N^{2} + 18 k\, 
    N^{2} + 8 N^{3})} {(2 + N) (2 + k + N)^{3}},\nonumber\\
c_ {134}  & =&  -\frac {(32 + 55 k + 18 k^{2} + 41 N + 35 k\, 
    N + 11 N^{2})} {(2 + N) (2 + k + N)^{4}},\nonumber\\
c_ {135}  & =&  \frac {(32 + 55 k + 18 k^{2} + 41 N + 35 k\, 
    N + 11 N^{2})} {(2 + N) (2 + k + N)^{4}},\nonumber\\
c_ {136}  & =&  -\frac {2 (8 + 17 k + 6 k^{2} + 11 N + 11 k\, 
     N + 3 N^{2})} {(2 + N) (2 + k + N)^{3}},\nonumber\\
c_ {137}  & =&  \frac {(-16 - 5 k + 2 k^{2} - 19 N + 3 k\, 
    N + 4 k^{2} N - 5 N^{2} + 4 k\, N^{2})} {(2 + N) (2 + k + N)^{4}},
\qquad
c_ {138}   =  \frac {(1 + N)} {(2 + k + N)^{3}},\nonumber\\
c_ {139}  & =&  -\frac {(56 + 75 k + 22 k^{2} + 113 N + 112 k\, 
    N + 20 k^{2} N + 74 N^{2} + 41 k\, 
    N^{2} + 15 N^{3})} {(2 + N) (2 + k + N)^{4}},\nonumber\\
c_ {140}  & =&  -\frac {2 (3 + 2 k + N)} {(2 + k + N)^{3}},\qquad
c_ {141}   =  \frac {(10 + 17 k + 6 k^{2} + 14 N + 11 k\, 
    N + 4 N^{2})} {(2 + N) (2 + k + N)^{3}},\nonumber\\
c_ {142}  & =&  -\frac {2 (18 + 21 k + 6 k^{2} + 22 N + 13 k\, 
     N + 6 N^{2})} {(2 + N) (2 + k + N)^{3}}, \qquad
c_ {143}   =  \frac {4 (1 + k) (1 + N)} {(2 + k + N)^{4}},\nonumber\\
c_ {144}  & =&  -\frac {(72 + 65 k + k^{2} - 6 k^{3} + 131 N + 94 k\, 
    N + 7 k^{2} N + 81 N^{2} + 35 k\, 
    N^{2} + 16 N^{3})} {(2 + N) (2 + k + N)^{4}},\nonumber\\
c_ {145}  & =&  -\frac {(4 + 13 k + 6 k^{2} + 9 N + 9 k\, 
    N + 3 N^{2})} {(2 + N) (2 + k + N)^{2}},\nonumber\\
c_ {146}  & =&  \frac {(13 k + 6 k^{2} + 3 N + 9 k\, 
    N + N^{2})} {(2 + N) (2 + k + N)^{3}},\qquad
c_ {147}   =  \frac {4 (k - N)} {(2 + k + N)^{3}},\nonumber\\
c_ {148}  & =&  -\frac {4} {(2 + k + N)^{2}},\qquad
c_ {149}   =  -\frac {(13 k + 6 k^{2} + 3 N + 9 k\, 
    N + N^{2})} {(2 + N) (2 + k + N)^{3}},\nonumber\\
c_ {150}  & =&  \frac {(4 + 5 k + 6 k^{2} + 17 N + 5 k\, 
    N + 7 N^{2})} {(2 + N) (2 + k + N)^{3}},\nonumber\\
c_ {151}  & =&  -\frac {(48 + 21 k - 2 k^{2} + 67 N + 13 k\, 
    N - 4 k^{2} N + 29 N^{2} + 4 N^{3})} {(2 + N) (2 + k + N)^{3}},\nonumber\\
c_ {152}  & =&  -\frac {2 (1 + N) (10 + 17 k + 6 k^{2} + 14 N + 11 k\, 
     N + 4 N^{2})} {(2 + N) (2 + k + N)^{2}},\qquad
c_ {153}   =  \frac {(1 + N)} {(2 + k + N)^{2}},\nonumber\\
c_ {154}  & =&  -\frac {(1 + N) (20 + 36 k + 12 k^{2} + 26 N + 23 k\, 
     N + 7 N^{2})} {(2 + N) (2 + k + N)^{3}}, \nonumber\\
c_ {155}  & =&  \frac {(20 + 57 k + 48 k^{2} + 12 k^{3} + 41 N + 65 k\, 
    N + 26 k^{2} N + 23 N^{2} + 16 k\, 
    N^{2} + 4 N^{3})} {(2 + N) (2 + k + N)^{4}}, \nonumber\\
c_ {156}  & =&  \frac {(32 + 43 k + 18 k^{2} + 53 N + 43 k\, 
    N + 12 k^{2} N + 35 N^{2} + 12 k\, 
    N^{2} + 8 N^{3})} {(2 + N) (2 + k + N)^{4}},\nonumber\\
c_ {157}  & =&  \frac {(32 + 55 k + 18 k^{2} + 41 N + 35 k\, 
    N + 11 N^{2})} {(2 + N) (2 + k + N)^{4}},\nonumber\\
c_ {158}  & =&  \frac {(12 + 27 k + 10 k^{2} + 19 N + 35 k\, 
    N + 8 k^{2} N + 11 N^{2} + 12 k\, 
    N^{2} + 2 N^{3})} {(2 + N) (2 + k + N)^{4}}, \nonumber\\
c_ {159}  & =&  -\frac {(8 + 19 k + 10 k^{2} + 17 N + 23 k\, 
    N + 8 k^{2} N + 15 N^{2} + 8 k\, 
    N^{2} + 4 N^{3})} {(2 + N) (2 + k + N)^{4}}, \nonumber\\
c_ {160}  & =&  \frac{1} {(2 + N) (2 + k + N)^{4}}(44 + 63 k + 44 k^{2} + 12 k^{3} + 135 N + 151 k\, 
    N + 72 k^{2} N + 12 k^{3} N 
    \nonu\\ &+& 137 N^{2} + 106 k\, 
    N^{2} + 26 k^{2} N^{2} + 56 N^{3} + 22 k\, 
    N^{3} + 8 N^{4}) , \nonumber\\
c_ {161}  & =&  \frac {(16 + 27 k + 10 k^{2} + 29 N + 35 k\, 
    N + 8 k^{2} N + 19 N^{2} + 12 k\, 
    N^{2} + 4 N^{3})} {(2 + N) (2 + k + N)^{3}}, \nonumber\\
c_ {162}  & =&  -\frac {(16 + 27 k + 10 k^{2} + 29 N + 35 k\, 
    N + 8 k^{2} N + 19 N^{2} + 12 k\, 
    N^{2} + 4 N^{3})} {(2 + N) (2 + k + N)^{3}},\nonumber\\
c_ {163}  & =&  -\frac {(16 + 21 k + 6 k^{2} + 19 N + 13 k\, 
    N + 5 N^{2})} {(2 + N) (2 + k + N)^{3}},\nonumber\\
c_ {164}  & =&  \frac {(4 + 13 k + 6 k^{2} + 9 N + 9 k\, 
    N + 3 N^{2})} {(2 + N) (2 + k + N)^{3}},\nonumber\\
c_ {165}  & =&  -\frac {(20 + 47 k + 18 k^{2} + 51 N + 65 k\, 
    N + 12 k^{2} N + 37 N^{2} + 22 k\, 
    N^{2} + 8 N^{3})} {(2 + N) (2 + k + N)^{3}}, \nonumber\\
c_ {166}  & =&  \frac {(8 + 13 k + 6 k^{2} + 27 N + 35 k\, 
    N + 12 k^{2} N + 29 N^{2} + 18 k\, 
    N^{2} + 8 N^{3})} {(2 + N) (2 + k + N)^{3}},\nonumber\\
c_ {167}  & =&  \frac {(10 + 17 k + 6 k^{2} + 14 N + 11 k\, 
    N + 4 N^{2})} {(2 + N) (2 + k + N)^{2}},\qquad
c_ {168}   =  -\frac {2} {(2 + k + N)},\nonumber\\
c_ {169}  & =&  \frac {(20 + 31 k + 10 k^{2} + 35 N + 41 k\, 
    N + 8 k^{2} N + 21 N^{2} + 14 k\, 
    N^{2} + 4 N^{3})} {(2 + N) (2 + k + N)^{4}}, \nonumber\\
c_ {170}  & =&  -\frac {(20 + 31 k + 10 k^{2} + 35 N + 41 k\, 
    N + 8 k^{2} N + 21 N^{2} + 14 k\, 
    N^{2} + 4 N^{3})} {(2 + N) (2 + k + N)^{4}},\nonumber\\
c_ {171}  & =&  \frac {4 (1 + k) (1 + N)} {(2 + k + N)^{3}}, \nonumber\\
c_ {172}  & =&  -\frac {(52 + 34 k - 9 k^{2} - 6 k^{3} + 96 N + 53 k\, 
    N - k^{2} N + 60 N^{2} + 21 k\, 
    N^{2} + 12 N^{3})} {(2 + N) (2 + k + N)^{3}}, \nonumber\\
c_ {173}  & =&  \frac {1} {(2 + N) (2 + k + N)^{4}}(-4 + 25 k + 14 k^{2} - 19 N + 43 k\, 
    N + 20 k^{2} N - 17 N^{2} + 30 k\, 
    N^{2} 
    \nonu\\ &+& 8 k^{2} N^{2} - 4 N^{3} + 8 k\, 
    N^{3}), \nonumber\\
c_ {174}  & =&  -\frac {2 (28 + 35 k + 10 k^{2} + 51 N + 47 k\, 
     N + 8 k^{2} N + 31 N^{2} + 16 k\, 
     N^{2} + 6 N^{3})} {(2 + N) (2 + k + N)^{4}}, \nonumber\\
c_ {175}  & =&  \frac {2 (k - N)} {(2 + k + N)^{2}}, \nonumber\\
c_ {176}  & =&  \frac{1}{2 (2 + N) (2 + k + N)^{3}}(-60 - 25 k + 2 k^{2} - 93 N + 13 k\, 
    N + 20 k^{2} N - 47 N^{2} + 30 k\, 
    N^{2} 
    \nonu\\ &+& 8 k^{2} N^{2} - 8 N^{3} + 8 k\, 
    N^{3}), \nonumber\\
c_ {177}  & =&  \frac {9} {40}, \qquad
c_ {178}   =  \frac {3} {2},\qquad
 c_ {179}   =  \frac {(k - N)} {10 (2 + k + N)},\nonumber\\
 c_ {180}  & =&  -\frac {(60 + 77 k + 22 k^{2} + 121 N + 115 k\, 
    N + 20 k^{2} N + 79 N^{2} + 42 k\, 
    N^{2} + 16 N^{3})} {8 (2 + N) (2 + k + N)^{2}}, \nonumber\\
c_ {181}  & =&  \frac {(-k + N) (60 + 77 k + 22 k^{2} + 121 N + 115 k\, 
     N + 20 k^{2} N + 79 N^{2} + 42 k\, 
     N^{2} + 16 N^{3})} {6 (2 + N) (2 + k + N)^{3}}, \nonumber\\
c_ {182}  & =&  -\frac {(60 + 77 k + 22 k^{2} + 121 N + 115 k\, 
    N + 20 k^{2} N + 79 N^{2} + 42 k\, 
    N^{2} + 16 N^{3})} {4 (2 + N) (2 + k + N)^{2}}, \nonumber\\
c_ {183}  & =&  -\frac {3 (60 + 77 k + 22 k^{2} + 121 N + 115 k\, 
     N + 20 k^{2} N + 79 N^{2} + 42 k\, 
     N^{2} + 16 N^{3})} {2 (2 + N) (2 + k + N)^{2}},\nonumber\\
c_ {184}  & =&  -\frac {(32 + 55 k + 18 k^{2} + 41 N + 35 k\, 
    N + 11 N^{2})} {(2 + N) (2 + k + N)^{4}},\nonumber\\
c_ {185}  & =&  -\frac {(5 k + 2 k^{2} + 5 N + 14 k\, 
    N + 4 k^{2} N + 7 N^{2} + 7 k\, 
    N^{2} + 2 N^{3})} {(2 + N) (2 + k + N)^{3}}, \nonumber\\
c_ {186}  & =&  \frac {(40 + 65 k + 48 k^{2} + 12 k^{3} + 83 N + 103 k\, 
    N + 38 k^{2} N + 57 N^{2} + 38 k\, 
    N^{2} + 12 N^{3})} {(2 + N) (2 + k + N)^{4}},\nonumber\\
c_ {187}  & =&  \frac {(32 + 55 k + 18 k^{2} + 41 N + 35 k\, 
    N + 11 N^{2})} {(2 + N) (2 + k + N)^{4}},\nonumber\\
c_ {188}  & =&  \frac {(1 + k) (1 + N) (32 + 55 k + 18 k^{2} + 41 N + 
      35 k\, N + 11 N^{2})} {(2 + N) (2 + k + N)^{6}},\nonumber\\
c_ {189}  & =&  \frac {2 (32 + 55 k + 18 k^{2} + 41 N + 35 k\,
     N + 11 N^{2})} {(2 + N) (2 + k + N)^{4}},\nonumber\\
c_ {190}  & =&  \frac {(16 + 21 k + 6 k^{2} + 19 N + 13 k\, 
    N + 5 N^{2})} {2 (2 + N) (2 + k + N)^{3}},\nonumber\\
c_ {191}  & =&  \frac {(32 + 52 k + 3 k^{2} - 6 k^{3} + 92 N + 114 k\, 
    N + 17 k^{2} N + 75 N^{2} + 52 k\, 
    N^{2} + 17 N^{3})} {2 (2 + N) (2 + k + N)^{4}}, \nonumber\\
c_ {192}  & =&  -\frac {(1 + k) (32 + 55 k + 18 k^{2} + 41 N + 35 k\, 
     N + 11 N^{2})} {(2 + N) (2 + k + N)^{5}},\nonumber\\
c_ {193}  & =&  \frac {(2 + 2 k + k^{2} + 2 N + N^{2}) (32 + 55 k + 
      18 k^{2} + 41 N + 35 k\, 
     N + 11 N^{2})} {(2 + N) (2 + k + N)^{6}},\nonumber\\
c_ {194}  & =&  -\frac {(1 + N) (32 + 55 k + 18 k^{2} + 41 N + 35 k\, 
     N + 11 N^{2})} {(2 + N) (2 + k + N)^{5}},\nonumber\\
c_ {195}  & =&  \frac {(1 + k) (1 + N) (32 + 55 k + 18 k^{2} + 41 N + 
      35 k\, N + 11 N^{2})} {(2 + N) (2 + k + N)^{6}}, \nonumber\\
c_ {196}  & =&  -\frac {(1 + N) (36 + 81 k + 82 k^{2} + 24 k^{3} + 57 N + 
      91 k\, N + 48 k^{2} N + 27 N^{2} + 22 k\, 
     N^{2} + 4 N^{3})} {(2 + N) (2 + k + N)^{5}},\nonumber\\
c_ {197}  & =&  \frac {2 (3 + 15 k + 6 k^{2} + 7 N + 11 k\, 
     N + 2 N^{2})} {(2 + k + N)^{4}},\nonumber\\
c_ {198}  & =&  -\frac {(13 k + 6 k^{2} + 3 N + 9 k\, 
    N + N^{2})} {2 (2 + N) (2 + k + N)^{3}},\nonumber\\
c_ {199}  & =&  -\frac {(1 + N) (32 + 55 k + 18 k^{2} + 41 N + 35 k\, 
     N + 11 N^{2})} {(2 + N) (2 + k + N)^{5}},\nonumber\\
c_ {200}  & =&  -\frac {(1 + k) (32 + 55 k + 18 k^{2} + 41 N + 35 k\, 
     N + 11 N^{2})} {(2 + N) (2 + k + N)^{5}}, \nonumber\\
c_ {201}  & =&  \frac{1}{(2 + N) (2 + k + N)^{5}} (-40 - 27 k + 14 k^{2} + 8 k^{3} - 97 N - 49 k\, 
    N + 29 k^{2} N + 10 k^{3} N
    \nonu\\ & - & 97 N^{2} - 43 k\, 
    N^{2} + 9 k^{2} N^{2} - 46 N^{3} - 15 k\, 
    N^{3} - 8 N^{4}), \nonumber\\
c_ {202}  & =&  \frac {(124 + 147 k + 42 k^{2} + 151 N + 91 k\, 
    N + 41 N^{2})} {2 (2 + N) (2 + k + N)^{3}},\nonumber\\
c_ {203}  & =&  \frac{1}{1} {(2 + N) (2 + k + N)^{5}}(60 + 141 k + 122 k^{2} + 32 k^{3} + 117 N + 
     190 k\, N + 116 k^{2} N 
     \nonu\\ &+& 16 k^{3} N + 70 N^{2} + 69 k\, 
    N^{2} + 26 k^{2} N^{2} + 13 N^{3} + 4 k\, 
    N^{3}) , \nonumber\\
c_ {204}  & =&  \frac {(3 + k + 2 N) (32 + 55 k + 18 k^{2} + 41 N + 35 k\, 
     N + 11 N^{2})} {(2 + N) (2 + k + N)^{5}}, \nonumber\\
c_ {205}  & =&  \frac{1} {2 (2 + N) (2 + k + N)^{4}}(-28 - 97 k - 38 k^{2} - 53 N - 116 k\, 
    N + 4 k^{2} N + 12 k^{3} N - 30 N^{2}
    \nonu\\ & - &29 k\, 
    N^{2} + 18 k^{2} N^{2} - 5 N^{3} + 2 k\, 
    N^{3}), \nonumber\\
c_ {206}  & =&  -\frac {(40 + 59 k + 18 k^{2} + 57 N + 41 k\, 
    N + 21 N^{2} + 2 k\, N^{2} + 2 N^{3})} {(2 + N) (2 + k + N)^{4}},\nonumber\\
c_ {207}  & =&  \frac {(13 k + 6 k^{2} + 3 N + 9 k\, 
    N + N^{2})} {2 (2 + N) (2 + k + N)^{3}},\qquad
c_ {208}   =  -\frac {(16 + 21 k + 6 k^{2} + 19 N + 13 k\, 
    N + 5 N^{2})} {(2 (2 + N) (2 + k + N)^{3})}, \nonumber\\
c_ {209}  & =&  -\frac {(-32 - 116 k - 113 k^{2} - 30 k^{3} - 28 N - 
     86 k\, N - 51 k^{2} N + 7 N^{2} - 4 k\, 
    N^{2} + 5 N^{3})} {2 (2 + N) (2 + k + N)^{4}}, \nonumber\\
c_ {210}  & =&  \frac {(1 + N) (32 + 55 k + 18 k^{2} + 41 N + 35 k\, 
     N + 11 N^{2})} {(2 + N) (2 + k + N)^{5}},\nonumber\\
c_ {211}  & =&  \frac {(1 + k) (32 + 55 k + 18 k^{2} + 41 N + 35 k\,
     N + 11 N^{2})} {(2 + N) (2 + k + N)^{5}},\nonumber\\
c_ {212}  & =&  \frac {2 (26 + 77 k + 56 k^{2} + 12 k^{3} + 36 N + 66 k\, 
     N + 24 k^{2} N + 10 N^{2} + 9 k\, 
     N^{2})} {(2 + N) (2 + k + N)^{4}},\nonumber\\
c_ {213}  & =&  -\frac {(-4 - 21 k - 6 k^{2} - N - 13 k\, 
    N + N^{2})} {2 (2 + N) (2 + k + N)^{3}}, \nonumber\\
c_ {214}  & =&  \frac{1}{(2 + N) (2 + k + N)^{4}} (24 + 71 k + 59 k^{2} + 14 k^{3} + 53 N + 121 k\, 
    N + 74 k^{2} N + 10 k^{3} N
    \nonu\\ & + &20 N^{2} + 45 k\, 
    N^{2} + 19 k^{2} N^{2} - 9 N^{3} - k\, 
    N^{3} - 4 N^{4}) , \nonumber\\
c_ {215}  & =&  \frac {(16 + 21 k + 6 k^{2} + 19 N + 13 k\, 
    N + 5 N^{2})} {2 (2 + N) (2 + k + N)^{2}}, \nonumber\\
c_ {216}  & =&  \frac{1}{(2 + N) (2 + k + N)^{4}} (56 + 21 k - 25 k^{2} - 10 k^{3} + 119 N + 7 k\, 
    N - 60 k^{2} N - 14 k^{3} N
    \nonu\\ & + & 106 N^{2} + 3 k\, 
    N^{2} - 23 k^{2} N^{2} + 47 N^{3} + 5 k\, 
    N^{3} + 8 N^{4}) , \nonumber\\
c_ {217}  & =&  -\frac {(20 + 21 k + 6 k^{2} + 25 N + 13 k\, 
    N + 7 N^{2})} {2 (2 + N) (2 + k + N)^{3}}, \nonumber\\
c_ {218}  & =&  -\frac {(64 + 136 k + 61 k^{2} + 6 k^{3} + 88 N + 104 k\, 
    N + 15 k^{2} N + 27 N^{2} + 10 k\, 
    N^{2} + N^{3})} {2 (2 + N) (2 + k + N)^{4}},\nonumber\\
c_ {219}  & =&  -\frac {(32 + 55 k + 18 k^{2} + 41 N + 35 k\, 
    N + 11 N^{2})} {(2 + N) (2 + k + N)^{4}},\nonumber\\
c_ {220}  & =&  -\frac {(-k + N) (32 + 55 k + 18 k^{2} + 41 N + 35 k\, 
     N + 11 N^{2})} {(2 + N) (2 + k + N)^{5}},\nonumber\\
c_ {221}  & =&  \frac {(32 + 55 k + 18 k^{2} + 41 N + 35 k\, 
    N + 11 N^{2})} {(2 + N) (2 + k + N)^{4}},\nonumber\\
c_ {222}  & =&  \frac {(-k + N) (32 + 55 k + 18 k^{2} + 41 N + 35 k\, 
     N + 11 N^{2})} {(2 + N) (2 + k + N)^{5}}, \nonumber\\
c_ {223}  & =&  -\frac {(16 + 47 k + 5 k^{2} - 6 k^{3} + 41 N + 62 k\, 
    N + 3 k^{2} N + 29 N^{2} + 21 k\, 
    N^{2} + 6 N^{3})} {(2 + N) (2 + k + N)^{4}},\nonumber\\
c_ {224}  & =&  \frac {(32 + 55 k + 18 k^{2} + 41 N + 35 k\, 
    N + 11 N^{2})} {(2 + N) (2 + k + N)^{4}},\nonumber\\
c_ {225}  & =&  -\frac {(32 + 55 k + 18 k^{2} + 41 N + 35 k\, 
    N + 11 N^{2})} {(2 + N) (2 + k + N)^{4}},\nonumber\\
c_ {226}  & =&  \frac {(16 + 21 k + 6 k^{2} + 19 N + 13 k\, 
    N + 5 N^{2})} {(2 + N) (2 + k + N)^{3}},\nonumber\\
c_ {227}  & =&  -\frac {(4 - 9 k - 19 k^{2} - 6 k^{3} + 35 N + 23 k\, 
    N - 3 k^{2} N + 36 N^{2} + 18 k\, 
    N^{2} + 9 N^{3})} {(2 + N) (2 + k + N)^{4}}, \nonumber\\
c_ {228}  & =&  \frac {(36 + 23 k - 11 k^{2} - 6 k^{3} + 83 N + 55 k\, 
    N + k^{2} N + 60 N^{2} + 26 k\, 
    N^{2} + 13 N^{3})} {(2 + N) (2 + k + N)^{4}},\nonumber\\
c_ {229}  & =&  -\frac {(32 + 55 k + 18 k^{2} + 41 N + 35 k\, 
    N + 11 N^{2})} {(2 + N) (2 + k + N)^{4}},\nonumber\\
c_ {230}  & =&  \frac {(12 + 21 k + 6 k^{2} + 13 N + 13 k\, 
    N + 3 N^{2})} {(2 + N) (2 + k + N)^{3}}, \nonumber\\
c_ {231}  & =&  \frac{1}{(2 + N) (2 + k + N)^{5}} (-16 - 41 k - 34 k^{2} - 8 k^{3} - 31 N - 81 k\, 
    N - 59 k^{2} N - 10 k^{3} N 
    \nonu\\ &-& 5 N^{2} - 33 k\, 
    N^{2} - 19 k^{2} N^{2} + 12 N^{3} + k\, 
    N^{3} + 4 N^{4}) , \nonumber\\
c_ {232}  & =&  \frac{1}{(2 + N) (2 + k + N)^{5}}(-32 - 34 k - 24 k^{2} - 21 k^{3} - 6 k^{4} - 
     110 N - 66 k\, N + 5 k^{2} N 
     \nonu\\ &-& k^{3} N - 150 N^{2} - 73 k\, 
    N^{2} + 5 k^{2} N^{2} - 87 N^{3} - 29 k\, 
    N^{3} - 17 N^{4}) , \nonumber\\
c_ {233}  & =&  -\frac {(4 + 13 k + 6 k^{2} + 9 N + 9 k\, 
    N + 3 N^{2})} {(2 + N) (2 + k + N)^{3}},\nonumber\\
c_ {234}  & =&  -\frac {(20 + 21 k + 6 k^{2} + 25 N + 13 k\, 
    N + 7 N^{2})} {(2 (2 + N) (2 + k + N)^{3})}, \nonumber\\
c_ {235}  & =&  \frac {(52 + 63 k + 18 k^{2} + 63 N + 39 k\, 
    N + 17 N^{2})} {(2 + N) (2 + k + N)^{3}}, \nonumber\\
c_ {236}  & =&  \frac{1}{(2 + N) (2 + k + N)^{5}}(-16 - 73 k - 66 k^{2} - 16 k^{3} + N - 97 k\, 
    N - 83 k^{2} N - 14 k^{3} N 
    \nonu\\ &+& 43 N^{2} - 25 k\, 
    N^{2} - 23 k^{2} N^{2} + 36 N^{3} + 5 k\, 
    N^{3} + 8 N^{4}) , \nonumber\\
c_ {237}  & =&  \frac {(36 + 29 k + 6 k^{2} + 41 N + 17 k\, 
    N + 11 N^{2})} {(2 + N) (2 + k + N)^{3}}, \nonumber\\
c_ {238}  & =&  -\frac {(84 k + 105 k^{2} + 30 k^{3} + 44 N + 164 k\, 
    N + 83 k^{2} N + 51 N^{2} + 66 k\, 
    N^{2} + 13 N^{3})} {2 (2 + N) (2 + k + N)^{4}}, \nonumber\\
c_ {239}  & =&  -\frac {(-16 - 63 k - 52 k^{2} - 12 k^{3} - 25 N - 40 k\, 
    N - 16 k^{2} N - 4 N^{2} + 3 k\, 
    N^{2} + N^{3})} {(2 + N) (2 + k + N)^{4}},\nonumber\\
c_ {240}  & =&  -\frac {(1 + N)} {(2 + k + N)^{2}}, \nonumber\\
c_ {241}  & =&  \frac {1}{2 (2 + N) (2 + k + N)^{4}}(200 + 296 k + 141 k^{2} + 22 k^{3} + 460 N + 
     420 k\, N + 65 k^{2} N 
     \nonu\\ &-& 12 k^{3} N + 375 N^{2} + 120 k\, 
    N^{2} - 74 k^{2} N^{2} - 16 k^{3} N^{2} + 129 N^{3} - 42 k\, 
    N^{3} - 36 k^{2} N^{3} 
    \nonu\\ &+& 16 N^{4} - 16 k\, 
    N^{4}),\nonumber\\
c_ {242}  & =&  -\frac {(-5 k - 2 k^{2} + 5 N + 3 k\, 
    N + 2 k^{2} N + 7 N^{2} + 4 k\, 
    N^{2} + 2 N^{3})} {(2 + N) (2 + k + N)^{3}},\nonumber\\
c_ {243}  & =&  -\frac {2 (18 + 21 k + 6 k^{2} + 24 N + 13 k\, 
     N + 9 N^{2} + N^{3})} {(2 + N) (2 + k + N)^{3}}, \nonumber\\
c_ {244}  & =&  \frac {1}{(2 + N) (2 + k + N)^{5}}(116 + 321 k + 285 k^{2} + 100 k^{3} + 12 k^{4} + 
     273 N + 639 k\, 
    N
    \nonu\\ & +& 418 k^{2} N + 80 k^{3} N + 248 N^{2} + 491 k\, 
    N^{2} + 227 k^{2} N^{2} + 20 k^{3} N^{2} + 103 N^{3} + 169 k\, 
    N^{3} 
    \nonu\\ &+& 46 k^{2} N^{3} + 16 N^{4} + 20 k\, 
    N^{4}) ,\nonumber\\
c_ {245}  & =&  -\frac {(1 + k) (20 + 21 k + 6 k^{2} + 25 N + 13 k\, 
     N + 7 N^{2})} {2 (2 + N) (2 + k + N)^{4}}, \nonumber\\
c_ {246}  & =&  \frac {1}{2 (2 + N) (2 + k + N)^{5}}(16 + 16 k + 7 k^{2} + 15 k^{3} + 6 k^{4} + 104 N + 
     184 k\, N + 113 k^{2} N
     \nonu\\ & +& 31 k^{3} N + 137 N^{2} + 173 k\, 
    N^{2} + 54 k^{2} N^{2} + 59 N^{3} + 37 k\, 
    N^{3} + 8 N^{4}) , \nonumber\\
c_ {247}  & =&  \frac {(1 + k) (32 + 55 k + 18 k^{2} + 41 N + 35 k\, 
     N + 11 N^{2})} {(2 + N) (2 + k + N)^{5}},\nonumber\\
c_ {248}  & =&  \frac {(1 + N) (32 + 55 k + 18 k^{2} + 41 N + 35 k\, 
     N + 11 N^{2})} {(2 + N) (2 + k + N)^{5}},\nonumber\\
c_ {249}  & =&  -\frac {(7 k + 2 k^{2} + 9 N + 21 k\, 
    N + 4 k^{2} N + 9 N^{2} + 10 k\, 
    N^{2} + 2 N^{3})} {(2 + N) (2 + k + N)^{4}}, \nonumber\\
c_ {250}  & =&  \frac {(1 + k) (1 + N) (4 + 7 k + 2 k^{2} + 19 N + 21 k\, 
     N + 4 k^{2} N + 17 N^{2} + 10 k\, 
     N^{2} + 4 N^{3})} {(2 + N) (2 + k + N)^{6}}, \nonumber\\
c_ {251}  & =&  \frac{1}{(2 + N) (2 + k + N)^{6}}(-28 - 53 k - 27 k^{2} + k^{3} + 2 k^{4} - 57 N - 
     101 k\, N - 37 k^{2} N 
     \nonu\\ &+& 11 k^{3} N + 4 k^{4} N - 32 N^{2} - 
     76 k\, N^{2} - 32 k^{2} N^{2} + 2 k^{3} N^{2} + 12 N^{3} - 
     18 k\, N^{3} - 10 k^{2} N^{3} 
     \nonu\\ &+& 17 N^{4} + 2 k\, 
    N^{4} + 4 N^{5}) , \nonumber\\
c_ {252}  & =&  \frac {(1 + N) (20 + 21 k + 6 k^{2} + 25 N + 13 k\, 
     N + 7 N^{2})} {2 (2 + N) (2 + k + N)^{4}},\nonumber\\
c_ {253}  & =&  -\frac {1}{(2 + N) (2 + k + N)^{4}}(-32 - 71 k - 52 k^{2} - 12 k^{3} - 25 N - 49 k\, 
    N - 22 k^{2} N 
    \nonu\\ &+& 5 N^{2} - 2 k\, 
    N^{2} + 4 N^{3}) , \nonumber\\
c_ {254}  & =&  \frac {1}{(2 + N) (2 + k + N)^{6}}(-68 - 139 k - 125 k^{2} - 57 k^{3} - 10 k^{4} - 
     183 N - 267 k\, 
    N 
    \nonu\\ &-& 175 k^{2} N - 63 k^{3} N - 8 k^{4} N - 216 N^{2} - 196 k\, 
    N^{2} - 68 k^{2} N^{2} - 14 k^{3} N^{2} - 144 N^{3} 
    \nonu\\ &-& 74 k\, 
    N^{3} - 6 k^{2} N^{3} - 53 N^{4} - 14 k\, 
    N^{4} - 8 N^{5}) , \nonumber\\
c_ {255}  & =&  \frac{1} {2 (2 + N) (2 + k + N)^{5}}(160 + 308 k + 209 k^{2} + 46 k^{3} + 380 N + 
     600 k\, N + 296 k^{2} N 
     \nonu\\ &+& 38 k^{3} N + 359 N^{2} + 414 k\, 
    N^{2} + 111 k^{2} N^{2} + 156 N^{3} + 98 k\, 
    N^{3} + 25 N^{4}) ,\nonumber\\
c_ {256}  & =&  \frac {2 (16 + 21 k + 6 k^{2} + 19 N + 13 k\, 
     N + 5 N^{2})} {(2 + N) (2 + k + N)^{2}},\nonumber\\
c_ {257}  & =&  -\frac {2 (8 + 17 k + 6 k^{2} + 11 N + 11 k\, 
     N + 3 N^{2})} {(2 + N) (2 + k + N)^{3}},\nonumber\\
c_ {258}  & =&  \frac {(16 + 27 k + 10 k^{2} + 29 N + 35 k\, 
    N + 8 k^{2} N + 19 N^{2} + 12 k\, 
    N^{2} + 4 N^{3})} {(2 + N) (2 + k + N)^{4}},\nonumber\\
c_ {259}  & =&  \frac {(3 + 2 k + N) (16 + 19 k + 6 k^{2} + 21 N + 12 k\, 
     N + 6 N^{2})} {(2 + N) (2 + k + N)^{4}},\nonumber\\
c_ {260}  & =&  \frac {(32 + 55 k + 18 k^{2} + 41 N + 35 k\, 
    N + 11 N^{2})} {(2 + N) (2 + k + N)^{4}},\nonumber\\
c_ {261}  & =&  -\frac {2 (16 + 21 k + 6 k^{2} + 19 N + 13 k\, 
     N + 5 N^{2})} {(2 + N) (2 + k + N)^{3}},\nonumber\\
c_ {262}  & =&  -\frac {2 (32 + 45 k + 14 k^{2} + 43 N + 37 k\, 
     N + 4 k^{2} N + 17 N^{2} + 6 k\, 
     N^{2} + 2 N^{3})} {(2 + N) (2 + k + N)^{3}}, \nonumber\\
c_ {263}  & =&  -\frac {(40 + 29 k + 6 k^{2} + 55 N + 17 k\, 
    N + 25 N^{2} + 4 N^{3})} {(2 + N) (2 + k + N)^{3}}, \nonumber\\
c_ {264}  & =&  -\frac {1}{(2 + N) (2 + k + N)^{5}}(1 + k) (36 + 49 k + 14 k^{2} + 101 N + 110 k\, 
     N + 22 k^{2} N 
     \nonu\\ &+& 122 N^{2} + 99 k\, 
     N^{2} + 12 k^{2} N^{2} + 65 N^{3} + 30 k\, 
     N^{3} + 12 N^{4}), \nonumber\\
c_ {265}  & =&  \frac {1}{(2 + N) (2 + k + N)^{5}}(84 + 165 k + 107 k^{2} + 22 k^{3} + 221 N + 
     387 k\, N + 214 k^{2} N 
     \nonu\\ &+& 34 k^{3} N + 230 N^{2} + 353 k\, 
    N^{2} + 157 k^{2} N^{2} + 16 k^{3} N^{2} + 107 N^{3} + 135 k\, 
    N^{3} + 38 k^{2} N^{3} 
    \nonu\\ &+& 18 N^{4} + 16 k\, 
    N^{4}), \nonumber\\
c_ {266}  & =&  \frac {1}{(2 + N) (2 + k + N)^{5}}(60 + 97 k + 63 k^{2} + 14 k^{3} + 129 N + 173 k\, 
    N + 82 k^{2} N + 10 k^{3} N 
    \nonu\\ &+& 114 N^{2} + 117 k\, 
    N^{2} + 31 k^{2} N^{2} + 49 N^{3} + 29 k\, 
    N^{3} + 8 N^{4}), \nonumber\\
c_ {267}  & =& \frac {1}{(2 + N) (2 + k + N)^{5}}(-40 - 105 k - 84 k^{2} - 20 k^{3} - 115 N - 
     231 k\, N - 139 k^{2} N 
     \nonu\\ &-& 22 k^{3} N - 105 N^{2} - 145 k\, 
    N^{2} - 49 k^{2} N^{2} - 36 N^{3} - 25 k\, 
    N^{3} - 4 N^{4}) , \nonumber\\
c_ {268}  & =&  -\frac {2 (1 + N) (10 + 17 k + 6 k^{2} + 14 N + 11 k\, 
     N + 4 N^{2})} {(2 + N) (2 + k + N)^{3}},\nonumber\\
c_ {269}  & =&  -\frac {(20 + 21 k + 6 k^{2} + 25 N + 13 k\, 
    N + 7 N^{2})} {4 (2 + N) (2 + k + N)^{2}},\nonumber\\
c_ {270}  & =&  -\frac {(64 + 64 k + 3 k^{2} - 6 k^{3} + 96 N + 48 k\, 
    N - 7 k^{2} N + 45 N^{2} + 6 k\, 
    N^{2} + 7 N^{3})} {4 (2 + N) (2 + k + N)^{3}}, \nonumber\\
c_ {271}  & =&  -\frac {(4 + 7 k + 2 k^{2} + 19 N + 21 k\, 
    N + 4 k^{2} N + 17 N^{2} + 10 k\, 
    N^{2} + 4 N^{3})} {2 (2 + N) (2 + k + N)^{3}}, \nonumber\\
c_ {272}  & =&  \frac {(36 + 39 k + 10 k^{2} + 67 N + 53 k\, 
    N + 8 k^{2} N + 41 N^{2} + 18 k\, 
    N^{2} + 8 N^{3})} {2 (2 + N) (2 + k + N)^{3}},\nonumber\\
c_ {273}  & =&  \frac {(20 + 21 k + 6 k^{2} + 25 N + 13 k\,
    N + 7 N^{2})} {2 (2 + N) (2 + k + N)^{3}},\nonumber\\
c_ {274}  & =&  -\frac {(13 k + 6 k^{2} + 3 N + 9 k\, 
    N + N^{2})} {2 (2 + N) (2 + k + N)^{3}},\nonumber\\
c_ {275}  & =&  -\frac {(16 + 21 k + 6 k^{2} + 19 N + 13 k\, 
    N + 5 N^{2})} {(2 + N) (2 + k + N)^{3}},\nonumber\\
c_ {276}  & =&  -\frac {2 (2 + 18 k + 21 k^{2} + 6 k^{3} + 3 N + 22 k\, 
     N + 13 k^{2} N + N^{2} + 6 k\, N^{2})} {(2 + N) (2 + k + N)^{3}},\nonumber\\
c_ {277}  & =&  -\frac {(1 + N) (20 + 21 k + 6 k^{2} + 25 N + 13 k\, 
     N + 7 N^{2})} {2 (2 + N) (2 + k + N)^{4}}, \nonumber\\
c_ {278}  & =&  \frac {1}{2 (2 + N) (2 + k + N)^{5}}(-144 - 248 k - 113 k^{2} - 14 k^{3} - 320 N - 
     416 k\, N - 130 k^{2} N 
     \nonu\\ &-& 10 k^{3} N - 247 N^{2} - 206 k\, 
    N^{2} - 29 k^{2} N^{2} - 74 N^{3} - 26 k\, 
    N^{3} - 7 N^{4}) ,\nonumber\\
c_ {279}  & =&  \frac {(32 + 39 k + 10 k^{2} + 57 N + 53 k\,
    N + 8 k^{2} N + 33 N^{2} + 18 k\, 
    N^{2} + 6 N^{3})} {(2 + N) (2 + k + N)^{4}},\nonumber\\
c_ {280}  & =&  \frac {(1 + k) (20 + 21 k + 6 k^{2} + 25 N + 13 k\, 
     N + 7 N^{2})} {2 (2 + N) (2 + k + N)^{4}}, \nonumber\\
c_ {281}  & =&  \frac {(-39 k - 44 k^{2} - 12 k^{3} + 23 N - 17 k\, 
    N - 18 k^{2} N + 29 N^{2} + 6 k\, 
    N^{2} + 8 N^{3})} {(2 + N) (2 + k + N)^{4}}, \nonumber\\
c_ {282}  & =&  -\frac {(1 + k) (1 + N) (36 + 39 k + 10 k^{2} + 67 N + 
      53 k\, N + 8 k^{2} N + 41 N^{2} + 18 k\, 
     N^{2} + 8 N^{3})} {(2 + N) (2 + k + N)^{6}}, \nonumber\\
c_ {283}  & =&  \frac{1}{2 (2 + N) (2 + k + N)^{5}} (-160 - 348 k - 219 k^{2} - 29 k^{3} + 6 k^{4} - 
     340 N - 544 k\, 
    N 
    \nonu\\ &-& 217 k^{2} N - 9 k^{3} N - 277 N^{2} - 303 k\, 
    N^{2} - 62 k^{2} N^{2} - 107 N^{3} - 63 k\, 
    N^{3} - 16 N^{4}) , \nonumber\\
c_ {284}  & =&  -\frac {2 (8 + 17 k + 6 k^{2} + 11 N + 11 k\, 
     N + 3 N^{2})} {(2 + N) (2 + k + N)^{3}},\nonumber\\
c_ {285}  & =&  \frac {(16 + 11 k + 2 k^{2} + 45 N + 27 k\, 
    N + 4 k^{2} N + 35 N^{2} + 12 k\, 
    N^{2} + 8 N^{3})} {(2 + N) (2 + k + N)^{4}},\nonumber\\
c_ {286}  & =&  -\frac {(32 + 55 k + 18 k^{2} + 41 N + 35 k\, 
    N + 11 N^{2})} {(2 + N) (2 + k + N)^{4}}, \nonumber\\
c_ {287}  & =&  -\frac {2 (16 + 21 k + 6 k^{2} + 19 N + 13 k\, 
     N + 5 N^{2})} {(2 + N) (2 + k + N)^{3}}, \nonumber\\
c_ {288}  & =&  \frac {(48 + 105 k + 64 k^{2} + 12 k^{3} + 63 N + 105 k\, 
    N + 34 k^{2} N + 23 N^{2} + 24 k\, 
    N^{2} + 2 N^{3})} {(2 + N) (2 + k + N)^{4}}, \nonumber\\
c_ {289}  & =&  \frac {2 (32 + 37 k + 10 k^{2} + 51 N + 33 k\, 
     N + 2 k^{2} N + 25 N^{2} + 6 k\, 
     N^{2} + 4 N^{3})} {(2 + N) (2 + k + N)^{3}}, \nonumber\\
c_ {290}  & =&  -\frac {(40 + 127 k + 102 k^{2} + 24 k^{3} + 53 N + 
     123 k\, N + 52 k^{2} N + 15 N^{2} + 24 k\, 
    N^{2})} {(2 + N) (2 + k + N)^{3}}, \nonumber\\
c_ {291}  & =&  -\frac{1}{(2 + N) (2 + k + N)^{5}}(1 + N) (4 + 73 k + 73 k^{2} + 18 k^{3} + 29 N + 
      190 k\, N + 141 k^{2} N 
      \nonu\\ &+& 24 k^{3} N + 31 N^{2} + 131 k\, 
     N^{2} + 54 k^{2} N^{2} + 8 N^{3} + 24 k\, 
     N^{3}) , \nonumber\\
c_ {292}  & =&  \frac {1}{(2 + N) (2 + k + N)^{5}}(-60 - 149 k - 113 k^{2} - 26 k^{3} - 141 N - 
     241 k\, N - 112 k^{2} N 
     \nonu\\ &-& 10 k^{3} N - 124 N^{2} - 137 k\, 
    N^{2} - 31 k^{2} N^{2} - 51 N^{3} - 29 k\, 
    N^{3} - 8 N^{4}), \nonumber\\
c_ {293}  & =&  -\frac {2 (1 + k) (5 + 2 k + 3 N)} {(2 + k + N)^{3}},\nonumber\\
c_ {294}  & =&  \frac {(-64 - 96 k - 45 k^{2} - 6 k^{3} - 128 N - 128 k\, 
    N - 31 k^{2} N - 83 N^{2} - 42 k\, 
    N^{2} - 17 N^{3})} {4 (2 + N) (2 + k + N)^{3}}, \nonumber\\
c_ {295}  & =&  \frac{1}{4 (2 + N) (2 + k + N)^{4}} (-144 k - 156 k^{2} - 27 k^{3} + 6 k^{4} - 16 N - 
     296 k\, N - 217 k^{2} N 
     \nonu\\ &-& 23 k^{3} N - 76 N^{2} - 261 k\, 
    N^{2} - 93 k^{2} N^{2} - 71 N^{3} - 81 k\, 
    N^{3} - 17 N^{4}) , \nonumber\\
c_ {296}  & =&  \frac {(88 + 193 k + 122 k^{2} + 24 k^{3} + 139 N + 
     185 k\, N + 56 k^{2} N + 61 N^{2} + 36 k\, 
    N^{2} + 8 N^{3})} {(2 + N) (2 + k + N)^{4}}, \nonumber\\
c_ {297}  & =&  \frac {(1 + k) (-4 + 3 k + 2 k^{2} + 3 N + 41 k\, 
     N + 16 k^{2} N + 13 N^{2} + 26 k\, 
     N^{2} + 4 N^{3})} {(2 + N) (2 + k + N)^{5}}, \nonumber\\
c_ {298}  & =&  \frac {(3 + 2 k + N) (32 + 55 k + 18 k^{2} + 41 N + 35 k\, 
     N + 11 N^{2})} {(2 + N) (2 + k + N)^{5}}, \nonumber\\
c_ {299}  & =&  \frac {(12 + 27 k + 10 k^{2} + 19 N + 35 k\, 
    N + 8 k^{2} N + 11 N^{2} + 12 k\, 
    N^{2} + 2 N^{3})} {(2 + N) (2 + k + N)^{4}}, \nonumber\\
c_ {300}  & =&  \frac {1}{(2 + N) (2 + k + N)^{5}}(-100 - 223 k - 147 k^{2} - 30 k^{3} - 207 N - 
     358 k\, N - 177 k^{2} N 
     \nonu\\ &-& 24 k^{3} N - 153 N^{2} - 175 k\, 
    N^{2} - 48 k^{2} N^{2} - 44 N^{3} - 22 k\, 
    N^{3} - 4 N^{4}), \nonumber\\
c_ {301}  & =&  \frac {1}{(2 + N) (2 + k + N)^{4}}2 (-48 - 61 k - 18 k^{2} - 75 N - 57 k\, 
     N + 7 k^{2} N + 6 k^{3} N - 37 N^{2} 
     \nonu\\ &-& 9 k\, 
     N^{2} + 9 k^{2} N^{2} - 6 N^{3} + k\, 
     N^{3}) , \nonumber\\
c_ {302}  & =&  \frac {(8 + 53 k + 22 k^{2} + 15 N + 53 k\, 
    N + 8 k^{2} N + 5 N^{2} + 12 k\, 
    N^{2})} {(2 + N) (2 + k + N)^{4}}, \nonumber\\
c_ {303}  & =&  \frac {(52 + 89 k + 56 k^{2} + 12 k^{3} + 89 N + 97 k\, 
    N + 30 k^{2} N + 47 N^{2} + 24 k\, 
    N^{2} + 8 N^{3})} {(2 + N) (2 + k + N)^{4}},\nonumber\\
c_ {304}  & =&  \frac {3 (-k + N) (16 + 21 k + 6 k^{2} + 19 N + 13 k\, 
     N + 5 N^{2})} {2 (2 + N) (2 + k + N)^{3}}, \nonumber\\
c_ {305}  & =&  \frac{1}{2 (2 + N) (2 + k + N)^{4}} (36 k + 49 k^{2} + 14 k^{3} + 28 N + 156 k\, 
    N + 141 k^{2} N + 28 k^{3} N 
    \nonu\\ &+& 19 N^{2} + 108 k\, 
    N^{2} + 58 k^{2} N^{2} - 7 N^{3} + 14 k\, 
    N^{3} - 4 N^{4}) , \nonumber\\
c_ {306}  & =&  \frac{1}{2 (2 + N) (2 + k + N)^{4}}(-68 k - 81 k^{2} - 22 k^{3} + 4 N - 172 k\, 
    N - 165 k^{2} N - 32 k^{3} N 
    \nonu\\ &+& 29 N^{2} - 100 k\, 
    N^{2} - 62 k^{2} N^{2} + 31 N^{3} - 10 k\, 
    N^{3} + 8 N^{4}) , \nonumber\\
c_ {307}  & =&  -\frac {(k + N) (32 + 55 k + 18 k^{2} + 41 N + 35 k\, 
     N + 11 N^{2})} {(2 (2 + N) (2 + k + N)^{4})},\nonumber\\
c_ {308}  & =&  -\frac {(-k + N) (32 + 55 k + 18 k^{2} + 41 N + 35 k\, 
     N + 11 N^{2})} {(2 (2 + N) (2 + k + N)^{4})},\nonumber\\
c_ {309}  & =&  \frac {(32 + 55 k + 18 k^{2} + 41 N + 35 k\, 
    N + 11 N^{2})} {(2 + N) (2 + k + N)^{4}},\nonumber\\
c_ {310}  & =&  -\frac {(1 + N) (-60 - 51 k + 3 k^{2} + 6 k^{3} - 67 N - 
      18 k\, N + 13 k^{2} N - 17 N^{2} + 7 k\, 
     N^{2})} {(2 + N) (2 + k + N)^{4}}, \nonumber\\
c_ {311}  & =&  \frac{1} {2 (2 + N) (2 + k + N)^{4}}(-68 - 175 k - 131 k^{2} - 30 k^{3} - 107 N - 
     156 k\, N - 31 k^{2} N 
     \nonu\\ &+& 12 k^{3} N - 41 N^{2} - 17 k\, 
    N^{2} + 18 k^{2} N^{2} - 4 N^{3} + 2 k\, 
    N^{3}), \nonumber\\
c_ {312}  & =&  \frac{1}{2 (2 + N) (2 + k + N)^{4}}(120 + 268 k + 181 k^{2} + 38 k^{3} + 224 N + 
     384 k\, N + 181 k^{2} N 
     \nonu\\ &+& 20 k^{3} N + 163 N^{2} + 220 k\, 
    N^{2} + 82 k^{2} N^{2} + 8 k^{3} N^{2} + 57 N^{3} + 66 k\, 
    N^{3} + 20 k^{2} N^{3} 
    \nonu\\ &+& 8 N^{4} + 8 k\, 
    N^{4}) ,\nonumber\\
c_ {313}  & =&  -\frac {(-1 + k\, N) (16 + 21 k + 6 k^{2} + 19 N + 13 k\, 
     N + 5 N^{2})} {(2 + N) (2 + k + N)^{3}}.\nonumber
\end{eqnarray}

\subsection{The OPEs between the ${\cal N} = 2$ higher spin-$\frac{3}{2}$
currents and the ${\cal N}=2$ higher spin-$2$ current }

The remaining OPE in the section $6$ 
 between the ${\cal N} = 2$ higher spin-$\frac{3}{2}$
currents and the ${\cal N}=2$ higher spin-$2$ current with (\ref{simple})
can be summarized by 
\begin{eqnarray}
&&\left(
\begin{array}{c}
{\bf U^{(\frac{3}{2})}} \nonu \\\\
{\bf V^{(\frac{3}{2})}}
\end{array}\right)(Z_{1})\; {\bf W^{(2)}}(Z_{2})\;=
\;\frac{\theta_{12}\overline{\theta}_{12}}{z_{12}^{4}}\, 
c_{{\pm}1}\, G_{\pm}(Z_{2})+\frac{1}{z_{12}^{3}}\, 
c_{{\pm}2}\, G_{\pm}(Z_{2})
 \nonu \\&&+\frac{\theta_{12}}{z_{12}^{3}}\Bigg[
c_{{\pm}3}\, DG_{\pm}
+c_{{\pm}4}\, H\, G_{\pm}\Bigg](Z_{2})+\frac{\overline{\theta}_{12}}{z_{12}^{3}}\Bigg[
c_{{\pm}5}\,\overline{D}G_{\pm}
+c_{{\pm}6}\,\overline{H}\, G_{\pm}\Bigg](Z_{2})
 \nonu \\&&+\frac{\theta_{12}\overline{\theta}_{12}}{z_{12}^{3}}\Bigg[
c_{{\pm}7}\,[D,\overline{D}]G_{\pm}
+c_{{\pm}8}\, H\,\overline{D}G_{\pm}
+c_{{\pm}9}\, H\,\overline{H}\, G_{\pm}
+c_{{\pm}10}\,\overline{H}\, DG_{\pm}
+c_{{\pm}11}\, T\, G_{\pm}
\nonu \\&&+c_{{\pm}12}\,\overline{D}H\, G_{\pm}
+c_{{\pm}13}\, D\overline{H}\, G_{\pm}
+c_{{\pm}14}\,\partial G_{\pm}\Bigg](Z_{2})
\nonu \\&&+\frac{1}{z_{12}^{2}}\Bigg[
c_{{\pm}15}\,[D,\overline{D}]G_{\pm}
+c_{{\pm}16}\, H\,\overline{D}G_{\pm}
+c_{{\pm}17}\, H\,\overline{H}\, G_{\pm}
+c_{{\pm}18}\,\overline{H}\, DG_{\pm}
+c_{{\pm}19}\, T\, G_{\pm}
\nonu \\&&+c_{{\pm}20}\,\overline{D}H\, G_{\pm}
+c_{{\pm}21}\, D\overline{H}\, G_{\pm}
+c_{{\pm}22}\,\partial G_{\pm}\Bigg](Z_{2})
\nonu \\&&+\frac{\theta_{12}}{z_{12}^{2}}\Bigg[
c_{{\pm}23}\,\partial DG_{\pm}
+c_{{\pm}24}\, G_{+}\, DG_{\pm}\,G_{-}
+c_{{\pm}25}\, H\,[D,\overline{D}]G_{\pm}
+c_{{\pm}26}\, H\,\overline{H}\, G_{\pm}
+c_{{\pm}27}\, H\,\overline{D}H\, G_{\pm}
\nonu \\&&+c_{{\pm}28}\, H\, D\overline{H}\, G_{\pm}
+c_{{\pm}29}\, H\,\partial G_{\pm}
+c_{{\pm}30}\, T\, DG_{\pm}
+c_{{\pm}31}\, T\, H\, G_{\pm}
+c_{{\pm}32}\overline{D}H\, DG_{\pm}
\nonu \\&&+  c_{{\pm}33}D\overline{H}\, DG_{\pm}
+c_{{\pm}34}\, DT\, G_{\pm}
+c_{{\pm}35}\,\partial H\, G_{\pm}\Bigg](Z_{2})
 \nonu \\&&+\frac{\overline{\theta}_{12}}{z_{12}^{2}}\Bigg[
c_{{\pm}36}\,\partial\overline{D}G_{\pm}
+c_{{\pm}37}\, G_{+}\,\overline{D}G_{\pm}\,G_{-}
+c_{{\pm}38}\, H\,\overline{H}\,\overline{D}G_{\pm}
+c_{{\pm}39}\,\overline{H}\,[D,\overline{D}]G_{\pm}
+c_{{\pm}40}\,\overline{H}\, D\overline{H}\, G_{\pm}
\nonu \\&&+c_{{\pm}41}\,\overline{H}\,\partial G_{\pm}
+c_{{\pm}42}\, T\,\overline{D}G_{\pm}
+c_{{\pm}43}\, T\,\overline{H}\, G_{\pm}
+c_{{\pm}44}\,\overline{D}H\,\overline{D}G_{\pm}
+c_{{\pm}45}\,\overline{D}H\,\overline{H}\, G_{\pm}
\nonu \\&&+c_{{\pm}46}\,\overline{D}T\, G_{\pm}
+c_{{\pm}47}\, D\overline{H}\,\overline{D}G_{\pm}
+c_{{\pm}48}\,\partial\overline{H}\, G_{\pm}\Bigg](Z_{2})
\nonu \\&&+\frac{\theta_{12}\overline{\theta}_{12}}{z_{12}^{2}}\Bigg[
c_{{\pm}49}\left(
\begin{array}{c}
{\bf U^{(\frac{5}{2})}} \nonu \\\\
{\bf V^{(\frac{5}{2})}}
\end{array}\right)
+ c_{{\pm}50}\; {\bf T^{(1)}}\left(
\begin{array}{c}
{\bf U^{(\frac{3}{2})}} \nonu \\\\
{\bf V^{(\frac{3}{2})}}
\end{array}\right)
+c_{{\pm}51}\, G_{+}\,\overline{D}G_{\pm}\, DG_{-}
+c_{{\pm}52}\, G_{+}\,[D,\overline{D}]G_{\pm}\,G_{-}
\nonu \\&&+c_{{\pm}53}\, DG_{+}\, G_{\pm}\,\overline{D}G_{-}
+c_{{\pm}54}\, H\,\partial\overline{D}G_{\pm}
+c_{{\pm}55}\, H\, G_{+}\,\overline{D}G_{\pm}\,G_{-}
+c_{{\pm}56}\, H\,\overline{H}\,[D,\overline{D}]G_{\pm}
\nonu \\&&+c_{{\pm}57}\, H\,\overline{H}\, D\overline{H}\, G_{\pm}
+c_{{\pm}58}\, H\,\overline{H}\,\partial G_{\pm}
+c_{{\pm}59}\, H\,\overline{D}H\,\overline{D}G_{\pm}
+c_{{\pm}60}\, H\,\overline{D}H\,\overline{H}\, G_{\pm}
\nonu \\&&+c_{{\pm}61}\, H\, D\overline{H}\,\overline{D}G_{\pm}
+c_{{\pm}62}\, H\,\partial\overline{H}\, G_{\pm}
+c_{{\pm}63}\,\overline{H}\,\partial DG_{\pm}
+c_{{\pm}64}\,\overline{H}\, G_{+}\, DG_{\pm}\,G_{-}
\nonu \\&&+c_{{\pm}65}\,\overline{H}\, D\overline{H}\, DG_{\pm}
+c_{{\pm}66}\, T\,[D,\overline{D}]G_{\pm}
+c_{{\pm}67}\, T\, H\,\overline{D}G_{\pm}
+c_{{\pm}68}\, T\, H\,\overline{H}\, G_{\pm}
+c_{{\pm}69}\, T\,\overline{H}\, DG_{\pm}
\nonu \\&&+c_{{\pm}70}T\, T\, G_{\pm}
+c_{{\pm}71}\, T\,\overline{D}H\, G_{\pm}
+c_{{\pm}72}\, T\, D\overline{H}\, G_{\pm}
+c_{{\pm}73}\, T\,\partial G_{\pm}
+c_{{\pm}74}\,\partial[D,\overline{D}]G_{\pm}
\nonu \\&&+c_{{\pm}75}\,\overline{D}G_{+}\, DG_{\pm}\,G_{-}
+c_{{\pm}76}\,\overline{D}H\,[D,\overline{D}]G_{\pm}
+c_{{\pm}77}\,\overline{D}H\,\overline{H}\, DG_{\pm}
+c_{{\pm}78}\,\overline{D}H\,\overline{D}H\, G_{\pm}
\nonu \\&&+c_{{\pm}79}\,\overline{D}H\, D\overline{H}\, G_{\pm}
+c_{{\pm}80}\,\overline{D}H\,\partial G_{\pm}
+c_{{\pm}81}\,\overline{D}T\, DG_{\pm}
+c_{{\pm}82}\,\overline{D}T\, H\, G_{\pm}
+c_{{\pm}83}\,\partial\overline{D}H\, G_{\pm}
\nonu \\&&+c_{{\pm}84}\,[D,\overline{D}]T\, G_{\pm}
+c_{{\pm}85}\, D\overline{H}\,[D,\overline{D}]G_{\pm}
+c_{{\pm}86}\, D\overline{H}\, D\overline{H}\, G_{\pm}
+c_{{\pm}87}\, D\overline{H}\,\partial G_{\pm}
\nonu \\&&+c_{{\pm}88}\, DT\,\overline{D}G_{\pm}
+c_{{\pm}89}\, DT\,\overline{H}\, G_{\pm}
+c_{{\pm}90}\,\partial D\overline{H}\, G_{\pm}
+c_{{\pm}91}\, G_{+}\,\partial G_{\pm}\,G_{-}
+c_{{\pm}92}\,\partial H\,\overline{D}G_{\pm}
\nonu \\&&+c_{{\pm}93}\,\partial H\,\overline{H}\, G_{\pm}
+c_{{\pm}94}\,\partial\overline{H}\, DG_{\pm}
+c_{{\pm}95}\,\partial T\, G_{\pm}
+c_{{\pm}96}\,\partial^{2}G_{\pm}
\Bigg](Z_{2})
\nonu \\&&+\frac{1}{z_{12}}\Bigg[
c_{{\pm}97}
\left(
\begin{array}{c}
{\bf U^{(\frac{5}{2})}} \nonu \\\\
{\bf V^{(\frac{5}{2})}}
\end{array}\right)
+c_{{\pm}98}\; {\bf T^{(1)}}\left(
\begin{array}{c}
{\bf U^{(\frac{3}{2})}} \nonu \\\\
{\bf V^{(\frac{3}{2})}}
\end{array}\right)
+c_{{\pm}99}\, G_{+}\,\overline{D}G_{\pm}\, DG_{-}
+c_{{\pm}100}\, G_{+}\,[D,\overline{D}]G_{\pm}\,G_{-}
\nonu \\&&+c_{{\pm}101}\, DG_{+}\, G_{\pm}\,\overline{D}G_{-}
+c_{{\pm}102}\, H\,\partial\overline{D}G_{\pm}
+c_{{\pm}103}\, H\, G_{+}\,\overline{D}G_{\pm}\,G_{-}
+c_{{\pm}104}\, H\,\overline{H}\,[D,\overline{D}]G_{\pm}
\nonu \\&&+c_{{\pm}105}\, H\,\overline{H}\, D\overline{H}\, G_{\pm}
+c_{{\pm}106}\, H\,\overline{H}\,\partial G_{\pm}
+c_{{\pm}107}\, H\,\overline{D}H\,\overline{D}G_{\pm}
+c_{{\pm}108}\, H\,\overline{D}H\,\overline{H}\, G_{\pm}
\nonu \\&&+c_{{\pm}109}\, H\, D\overline{H}\,\overline{D}G_{\pm}
+c_{{\pm}110}\, H\,\partial\overline{H}\, G_{\pm}
+c_{{\pm}111}\,\overline{H}\,\partial DG_{\pm}
+c_{{\pm}112}\,\overline{H}\, G_{+}\, DG_{\pm}\,G_{-}
\nonu \\&&+c_{{\pm}113}\,\overline{H}\, D\overline{H}\, DG_{\pm}
+c_{{\pm}114}\, T\,[D,\overline{D}]G_{\pm}
+c_{{\pm}115}\, T\, H\,\overline{D}G_{\pm}
+c_{{\pm}116}\, T\, H\,\overline{H}\, G_{\pm}
\nonu \\&&+c_{117}\, T\,\overline{H}\, DG_{\pm}
+c_{118}\, T\, T\, G_{\pm}
+c_{119}\, T\,\overline{D}H\, G_{\pm}
+c_{120}\, T\, D\overline{H}\, G_{\pm}
+c_{121}\, T\,\partial G_{\pm}
\nonu \\&&+c_{122}\,\partial[D,\overline{D}]G_{\pm}
+c_{123}\,\overline{D}G_{+}\, DG_{\pm}\,G_{-}
+c_{124}\,\overline{D}H\,[D,\overline{D}]G_{\pm}
+c_{125}\,\overline{D}H\,\overline{H}\, DG_{\pm}
\nonu \\&&+c_{126}\,\overline{D}H\,\overline{D}H\, G_{\pm}
+c_{127}\,\overline{D}H\, D\overline{H}\, G_{\pm}
+c_{128}\,\overline{D}H\,\partial G_{\pm}
+c_{129}\,\overline{D}T\, DG_{\pm}
+c_{130}\,\overline{D}T\, H\, G_{\pm}
\nonu \\&&+c_{131}\,\partial\overline{D}H\, G_{\pm}
+c_{132}\,[D,\overline{D}]T\, G_{\pm}
+c_{133}\, D\overline{H}\,[D,\overline{D}]G_{\pm}
+c_{134}\, D\overline{H}\, D\overline{H}\, G_{\pm}
\nonu \\&&+c_{135}\, D\overline{H}\,\partial G_{\pm}
+c_{136}\, DT\,\overline{D}G_{\pm}
+c_{137}\, DT\,\overline{H}\, G_{\pm}
+c_{138}\,\partial D\overline{H}\, G_{\pm}
+c_{139}\,\partial G_{\pm}\, G_{+}\,G_{-}
\nonu \\&&+c_{140}\,\partial H\,\overline{D}G_{\pm}
+c_{141}\,\partial H\,\overline{H}\, G_{\pm}
+c_{142}\,\partial\overline{H}\, DG_{\pm}
+c_{143}\,\partial T\, G_{\pm}
+c_{144}\,\partial^{2}G_{\pm}
\Bigg](Z_{2})
\nonu \\&&+\frac{\theta_{12}}{z_{12}}\Bigg[
c_{{\pm}145}\left(
\begin{array}{c}
D{\bf  U^{(\frac{5}{2})}} \nonu \\\\
D{\bf  V^{(\frac{5}{2})}}
\end{array}\right) 
+c_{{\pm}146}\,D{\bf T^{(1)}}\left(
\begin{array}{c}
{\bf U^{(\frac{3}{2})}} \nonu \\\\
{\bf V^{(\frac{3}{2})}}
\end{array}\right) 
+c_{{\pm}147}\,{\bf T^{(1)}}\left(
\begin{array}{c}
D {\bf U^{(\frac{3}{2})}} \nonu \\\\
D {\bf V^{(\frac{3}{2})}}
\end{array}\right) 
\nonu \\&&+c_{{\pm}148}\, DG_{+}\, G_{\pm}\,[D,\overline{D}]G_{-}
+c_{{\pm}149}\, DG_{+}\,G_{\pm}\, \partial G_{-}
+c_{{\pm}150}\, H\,\partial[D,\overline{D}]G_{\pm}
+c_{{\pm}151}\, H\, G_{+}\,\overline{D}G_{\pm}\, DG_{-}
\nonu \\&&+c_{{\pm}152}\, H\, G_{+}\,[D,\overline{D}]G_{\pm}\,G_{-}
+c_{{\pm}153}\, H\,\overline{H}\,\partial G_{\pm}
+c_{{\pm}154}\, H\,\overline{H}\, D\overline{H}\, DG_{\pm}
+c_{{\pm}155}\, H\, DG_{+}\,\overline{D}G_{\pm}\,G_{-}
\nonu \\&&+c_{{\pm}156}\, H\,\overline{D}H\,[D,\overline{D}]G_{\pm}
+c_{{\pm}157}\, H\,\overline{D}H\,\overline{H}\, DG_{\pm}
+c_{{\pm}158}\, H\,\overline{D}H\, D\overline{H}\, G_{\pm}
+c_{{\pm}159}\, H\,\overline{D}H\,\partial G_{\pm}
\nonu \\&&+c_{{\pm}160}\, H\, D\overline{H}\,[D,\overline{D}]G_{\pm}
+c_{{\pm}161}\, H\, D\overline{H}\, D\overline{H}\, G_{\pm}
+c_{{\pm}162}\, H\, D\overline{H}\,\partial G_{\pm}
+c_{{\pm}163}\, H\,\partial D\overline{H}\, G_{\pm}
\nonu \\&&+c_{{\pm}164}\, H\, G_{+}\,\partial G_{\pm}\,G_{-}
+c_{{\pm}165}\, H\,\partial\overline{H}\, DG_{\pm}
+c_{{\pm}166}\, H\,\partial^{2}G_{\pm}
+c_{{\pm}167}\,\overline{H}\, G_{+}\, DG_{\pm}\, DG_{-}
\nonu \\&&+c_{{\pm}168}\,\overline{H}\, DG_{+}\, DG_{\pm}\,G_{-}
+c_{{\pm}169}\, T\,\partial DG_{\pm}
+c_{{\pm}170}\, T\, H\,[D,\overline{D}]G_{\pm}
+c_{{\pm}171}T\, H\,\overline{H}\, DG_{\pm}
\nonu \\&&+c_{{\pm}172}\, T\, H\, D\overline{H}\, G_{\pm}
+c_{{\pm}173}\, T\, H\,\partial G_{\pm}
+c_{{\pm}174}\, T\, T\, DG_{\pm}
+c_{{\pm}175}\, T\,\overline{D}H\, DG_{\pm}
\nonu \\&&+c_{{\pm}176}\, T\, D\overline{H}\, DG_{\pm}
+c_{{\pm}177}\, T\, DT\, G_{\pm}
+c_{{\pm}178}\, T\,\partial H\, G_{\pm}
+c_{{\pm}179}\, G_{+}\,[D,\overline{D}]G_{\pm}\, DG_{-}
\nonu \\&&+c_{{\pm}180}\,\overline{D}G_{+}\, DG_{\pm}\, DG_{-}
+c_{{\pm}181}\,\overline{D}H\,\partial DG_{\pm}
+c_{{\pm}182}\,\overline{D}H\,\overline{D}H\, DG_{\pm}
+c_{{\pm}183}\,\overline{D}H\, D\overline{H}\, DG_{\pm}
\nonu \\&&+c_{{\pm}184}\,\overline{D}T\, H\, DG_{\pm}
+c_{{\pm}185}\,\partial\overline{D}H\, DG_{\pm}
+c_{{\pm}186}\,\partial\overline{D}H\, H\, G_{\pm}
+c_{{\pm}187}\,[D,\overline{D}]G_{+}\, DG_{\pm}\,G_{-}
\nonu \\&&+c_{{\pm}188}\,[D,\overline{D}]T\, DG_{\pm}
+c_{{\pm}189}\,[D,\overline{D}]T\, H\, G_{\pm}
+c_{{\pm}190}\, DG_{+}\, DG_{\pm}\,\overline{D}G_{-}
+c_{{\pm}191}\, D\overline{H}\,\partial DG_{\pm}
\nonu \\&&+c_{{\pm}192}\, D\overline{H}\, G_{+}\, DG_{\pm}\,G_{-}
+c_{{\pm}193}\, D\overline{H}\, D\overline{H}\, DG_{\pm}
+c_{{\pm}194}\, DT\,[D,\overline{D}]G_{\pm}
+c_{{\pm}195}\, DT\, H\,\overline{D}G_{\pm}
\nonu \\&&+c_{{\pm}196}\, DT\, H\,\overline{H}\, G_{\pm}
+c_{{\pm}197}\, DT\,\overline{H}\, DG_{\pm}
+c_{{\pm}198}\, DT\,\overline{D}H\, G_{\pm}
+c_{{\pm}199}\, DT\, D\overline{H}\, G_{\pm}
\nonu \\&&+c_{{\pm}200}\, DT\,\partial G_{\pm}
++c_{{\pm}201}\,\partial^{2}G_{\pm}
+c_{{\pm}202}\, G_{+}\, \partial DG_{\pm}\,G_{-}
+c_{{\pm}203}\,\partial D\overline{H}\, DG_{\pm}
+c_{{\pm}204}\partial DT\, G_{\pm}
\nonu \\&&+
c_{{\pm}205}\, G_{+}\,\partial G_{\pm}\, DG_{-}
+c_{{\pm}206}\,\partial G_{+}\, DG_{\pm}\,G_{-}
+c_{{\pm}207}\,\partial H\,[D,\overline{D}]G_{\pm}
+c_{{\pm}208}\,\partial H\, H\,\overline{D}G_{\pm}\nonu \\&&+
c_{{\pm}209}\,\partial H\, H\,\overline{H}\, G_{\pm}
+c_{{\pm}210}\,\partial H\,\overline{H}\, DG_{\pm}
+c_{{\pm}211}\,\partial H\,\overline{D}H\, G_{\pm}
+c_{{\pm}212}\,\partial H\, D\overline{H}\, G_{\pm}
\nonu \\&&+
c_{{\pm}213}\,\partial H\,\partial G_{\pm}
+c_{{\pm}214}\,\partial T\, DG_{\pm}
+c_{215}\,\partial T\, H\, G_{\pm}
+c_{{\pm}216}\,\partial^{2}H\, G_{\pm}\Bigg](Z_{2})
\nonu \\&&+\frac{\overline{\theta}_{12}}{z_{12}}\Bigg[
c_{{\pm}217}\left(
\begin{array}{c}
\overline{D}{\bf  U^{(\frac{5}{2})}} \nonu \\\\
\overline{D}{\bf  V^{(\frac{5}{2})}}
\end{array}\right) 
+c_{{\pm}218}\,\overline{D}{\bf T^{(1)}}\left(
\begin{array}{c}
{\bf U^{(\frac{3}{2})}} \nonu \\\\
{\bf V^{(\frac{3}{2})}}
\end{array}\right) 
+c_{{\pm}219}\,{\bf T^{(1)}}\left(
\begin{array}{c}
\overline{D} {\bf U^{(\frac{3}{2})}} \nonu \\\\
\overline{D} {\bf V^{(\frac{3}{2})}}
\end{array}\right) 
\nonu \\&&+c_{{\pm}220}\, G_{+}\,\overline{D}G_{\pm}\,\partial G_{-}
+c_{{\pm}221}\, G_{+}\,[D,\overline{D}]G_{\pm}\,\overline{D}G_{-}
+c_{{\pm}222}\, H\, G_{+}\,\overline{D}G_{\pm}\,\overline{D}G_{-}
+c_{{\pm}223}\, H\,\overline{H}\,\partial\overline{D}G_{\pm}
\nonu \\&&+c_{{\pm}224}\, H\,\overline{H}\, D\overline{H}\,\overline{D}G_{\pm}
+c_{{\pm}225}\, H\,\overline{D}G_{+}\,\overline{D}G_{\pm}\,G_{-}
+c_{{\pm}226}\, H\,\overline{D}H\,\overline{H}\,\overline{D}G_{\pm}
+c_{{\pm}227}\, H\,\partial\overline{H}\,\overline{D}G_{\pm}
\nonu \\&&+c_{{\pm}228}\, H\,\partial\overline{H}\,\overline{H}\, G_{\pm}
+c_{{\pm}229}\,\overline{H}\,\partial[D,\overline{D}]G_{\pm}
+c_{{\pm}230}\,\overline{H}\, G_{+}\,[D,\overline{D}]G_{\pm}\,G_{-}
\nonu \\&&+
c_{{\pm}231}\,\overline{H}\, G_{+}\, DG_{\pm}\,\overline{D}G_{-}
+c_{{\pm}232}\,\overline{H}\,\overline{D}G_{+}\, DG_{\pm}\,G_{-}
+c_{{\pm}233}\,\overline{H}\, D\overline{H}\,[D,\overline{D}]G_{\pm}
\nonu \\&&+
c_{{\pm}234}\,\overline{H}\, D\overline{H}\,\partial G_{\pm}
+c_{{\pm}235}\,\overline{H}\, G_{+}\,\partial G_{\pm}\,G_{-}
+c_{{\pm}236}\,\overline{H}\,\partial^{2}G_{\pm}
+c_{{\pm}237}\, T\,\partial\overline{D}G_{\pm}
\nonu \\&&+
c_{{\pm}238}\, T\, H\,\overline{H}\,\overline{D}G_{\pm}
+c_{{\pm}239}\, T\,\overline{H}\,[D,\overline{D}]G_{\pm}
+c_{{\pm}240}\, T\,\overline{H}\,\partial G_{\pm}
+c_{{\pm}241}\, T\, T\,\overline{D}G_{\pm}
\nonu \\&&+
c_{{\pm}242}\, T\,\overline{D}H\,\overline{D}G_{\pm}
+c_{{\pm}243}\, T\,\overline{D}H\,\overline{H}\, G_{\pm}
+c_{{\pm}244}\, T\,\overline{D}T\, G_{\pm}
+c_{{\pm}245}\, T\, D\overline{H}\,\overline{D}G_{\pm}
\nonu \\&&+
c_{{\pm}246}\, T\,\partial\overline{H}\, G_{\pm}
+c_{{\pm}247}\, G_{+}\,\overline{D}G_{\pm}\,[D,\overline{D}]G_{-}
+c_{{\pm}248}\,\overline{D}G_{+}\,\overline{D}G_{\pm}\, DG_{-}
+c_{{\pm}249}\,\overline{D}G_{+}\,[D,\overline{D}]G_{\pm}\,G_{-}
\nonu \\&&+
c_{{\pm}250}\,\overline{D}G_{+}\, DG_{\pm}\,\overline{D}G_{-}
+c_{{\pm}251}\,\overline{D}H\,\partial\overline{D}G_{\pm}
+c_{{\pm}252}\,\overline{D}H\, G_{+}\,\overline{D}G_{\pm}\,G_{-}
+c_{{\pm}253}\,\overline{D}H\,\overline{H}\,[D,\overline{D}]G_{\pm}
\nonu \\&&+
c_{{\pm}254}\,\overline{D}H\,\overline{H}\, D\overline{H}\, G_{\pm}
+c_{{\pm}255}\,\overline{D}H\,\overline{H}\,\partial G_{\pm}
+c_{{\pm}256}\,\overline{D}H\,\overline{D}H\,\overline{D}G_{\pm}
+c_{{\pm}257}\,\overline{D}H\,\overline{D}H\,\overline{H}\, G_{\pm}
\nonu \\&&+
c_{{\pm}258}\,\overline{D}H\, D\overline{H}\,\overline{D}G_{\pm}
+c_{{\pm}259}\,\overline{D}H\,\partial\overline{H}\, G_{\pm}
+c_{{\pm}260}\,\overline{D}T\,[D,\overline{D}]G_{\pm}
+c_{{\pm}261}\,\overline{D}T\, H\,\overline{D}G_{\pm}
\nonu \\&&+
c_{{\pm}262}\,\overline{D}T\, H\,\overline{H}\, G_{+}
+c_{{\pm}263}\,\overline{D}T\,\overline{H}\, DG_{\pm}
+c_{{\pm}264}\,\overline{D}T\,\overline{D}H\, G_{\pm}
+c_{{\pm}265}\,\overline{D}T\, D\overline{H}\, G_{\pm}
\nonu \\&&+
c_{{\pm}266}\,\overline{D}T\,\partial G_{\pm}
+c_{{\pm}267}\,\partial^{2}\overline{D}G_{\pm}
+c_{{\pm}268}\, G_{+}\,\partial\overline{D}G_{\pm}\,G_{-}
+c_{{\pm}269}\,\partial\overline{D}H\,\overline{D}G_{\pm}
+c_{{\pm}270}\,\partial\overline{D}H\,\overline{H}\, G_{\pm}
\nonu \\&&+
c_{{\pm}271}\,\partial\overline{D}T\, G_{\pm}
+c_{{\pm}272}\,[D,\overline{D}]T\,\overline{D}G_{\pm}
+c_{{\pm}273}\,[D,\overline{D}]T\,\overline{H}\, G_{\pm}
+c_{{\pm}274}\, D\overline{H}\,\partial\overline{D}G_{\pm}
\nonu \\&&+
c_{{\pm}275}\, D\overline{H}\, D\overline{H}\,\overline{D}G_{\pm}
+c_{{\pm}276}\, DT\,\overline{H}\,\overline{D}G_{\pm}
+c_{{\pm}277}\,\partial D\overline{H}\,\overline{D}G_{\pm}
+c_{{\pm}278}\,\partial D\overline{H}\,\overline{H}\, G_{\pm}
\nonu \\&&+
c_{{\pm}279}\,\partial G_{+}\, G_{\pm}\,\overline{D}G_{-}
+c_{{\pm}280}\,\partial G_{+}\,\overline{D}G_{\pm}\,G_{-}
+c_{{\pm}281}\,\partial H\,\overline{H}\,\overline{D}G_{\pm}
+c_{{\pm}282}\,\partial\overline{H}\,[D,\overline{D}]G_{\pm}
\nonu \\&&+
c_{{\pm}283}\,\partial\overline{H}\,\overline{H}\, DG_{\pm}
+c_{{\pm}284}\,\partial\overline{H}\, D\overline{H}\, G_{\pm}
+c_{{\pm}285}\,\partial\overline{H}\,\partial G_{\pm}
+c_{{\pm}286}\,\partial T\,\overline{D}G_{\pm}
\nonu \\&&+
c_{{\pm}287}\,\partial T\,\overline{H}\, G_{\pm}
+c_{{\pm}288}\,\partial^{2}\overline{H}\, G_{\pm}
\Bigg](Z_{2})
\nonu \\&&+\frac{\theta_{12}\overline{\theta}_{12}}{z_{12}}\Bigg[
c_{{\pm}289}\,[D,\overline{D}]\left(
\begin{array}{c}
{\bf  U^{(\frac{5}{2})}} \nonu \\\\
{\bf  V^{(\frac{5}{2})}}
\end{array}\right)
+c_{{\pm}290}\,\left(
\begin{array}{c}
{\bf \partial U^{(\frac{5}{2})}} \nonu \\\\
{\bf \partial  V^{(\frac{5}{2})}}
\end{array}\right)
+c_{{\pm}291}\,\partial {\bf T^{(1)}}\left(
\begin{array}{c}
{\bf U^{(\frac{3}{2})}} \nonu \\\\
{\bf V^{(\frac{3}{2})}}
\end{array}\right)
\nonu \\&&+
c_{{\pm}292}\,{\bf T^{(1)}}\left(
\begin{array}{c}
{\bf \partial U^{(\frac{3}{2})}} \nonu \\\\
{\bf \partial V^{(\frac{3}{2})}}
\end{array}\right)
+c_{{\pm}293}\,[D,\overline{D}]{\bf T^{(1)}}\left(
\begin{array}{c}
{\bf U^{(\frac{3}{2})}} \nonu \\\\
{\bf V^{(\frac{3}{2})}}
\end{array}\right)
+c_{{\pm}294}\,{\bf T^{(1)}}\left(
\begin{array}{c}
\big[D,\overline{D}\big]  {\bf U^{(\frac{3}{2})}} \nonu \\\\
\big[D,\overline{D}\big]  {\bf V^{(\frac{3}{2})}}
\end{array}\right)
\nonu \\&&+
c_{{\pm}295}\,D{\bf T^{(1)}}\left(
\begin{array}{c}
\overline{D}{\bf {\bf U^{(\frac{3}{2})}}} \nonu \\\\
\overline{D}{\bf {\bf V^{(\frac{3}{2})}}}
\end{array}\right)
+c_{{\pm}296}\,\overline{D}{\bf T^{(1)}}\left(
\begin{array}{c}
D{\bf {\bf U^{(\frac{3}{2})}}} \nonu \\\\
D{\bf {\bf V^{(\frac{3}{2})}}}
\end{array}\right)
+c_{{\pm}297}\, H\, G_{+}\,\overline{D}G_{\pm}\,\partial G_{-}
\nonu \\&&+
c_{{\pm}298}\, H\, G_{+}\,[D,\overline{D}]G_{\pm}\,\overline{D}G_{-}
+c_{{\pm}299}\, H\,\overline{H}\,\partial[D,\overline{D}]G_{\pm}
+c_{{\pm}300}\, H\,\overline{H}\, D\overline{H}\,[D,\overline{D}]G_{\pm}
\nonu \\&&+
c_{{\pm}301}\, H\,\overline{H}\, D\overline{H}\,\partial G_{\pm}
+c_{{\pm}302}\, H\,\overline{H}\,\partial^{2}G_{\pm}
+c_{{\pm}303}\, H\,\overline{D}G_{+}\,\overline{D}G_{\pm}\, DG_{-}
\nonu \\&&+
c_{{\pm}304}\, H\,[D,\overline{D}]G_{+}\,\overline{D}G_{\pm}\,G_{-}
+c_{{\pm}305}\, H\, DG_{+}\,\overline{D}G_{\pm}\,\overline{D}G_{-}
+c_{{\pm}306}\, H\,\overline{D}H\,\partial\overline{D}G_{\pm}
\nonu \\&&+
c_{{\pm}307}\, H\,\overline{D}H\,\overline{H}\,[D,\overline{D}]G_{\pm}
+c_{{\pm}308}\, H\,\overline{D}H\,\overline{H}\,\partial G_{\pm}
+c_{{\pm}309}\, H\,\overline{D}H\, D\overline{H}\,\overline{D}G_{\pm}
\nonu \\&&+
c_{{\pm}310}\, H\,\overline{D}H\,\partial\overline{H}\, G_{\pm}
+c_{{\pm}311}\, H\, G_{+}\,\partial\overline{D}G_{\pm}\, G_{-}
+c_{{\pm}312}\, H\, D\overline{H}\,\partial\overline{D}G_{\pm}
\nonu \\&&+
c_{{\pm}313}\, H\, D\overline{H}\, D\overline{H}\,\overline{D}G_{\pm}
+c_{{\pm}314}\, H\,\partial D\overline{H}\,\overline{D}G_{\pm}
+c_{{\pm}315}\, H\,\partial D\overline{H}\,\overline{H}\, G_{\pm}
\nonu \\&&+
c_{{\pm}316}\, H\,\partial G_{+}\, G_{\pm}\,\overline{D}G_{-}
+c_{{\pm}317}\, H\,\overline{D}G_{+}\,\partial G_{\pm}\,G_{-}
+c_{{\pm}318}\, H\,\partial\overline{H}\,[D,\overline{D}]G_{\pm}
\nonu \\&&+
c_{{\pm}319}\, H\,\partial\overline{H}\,\overline{H}\, DG_{\pm}
+c_{{\pm}320}\, H\,\partial\overline{H}\, D\overline{H}\, G_{\pm}
+c_{{\pm}321}\, H\,\partial\overline{H}\,\partial G_{\pm}
+c_{{\pm}322}\, H\,\partial^{2}\overline{H}\, G_{\pm}
\nonu \\&&+
c_{{\pm}323}\,\overline{H}\,\partial^{2}DG_{\pm}
+c_{{\pm}324}\,\overline{H}\,G_{+}\,[D,\overline{D}]G_{\pm}\,  DG_{-}
+c_{{\pm}325}\,\overline{H}\,  DG_{+}\,G_{\pm}\,[D,\overline{D}]G_{-}
\nonu \\&&+
c_{{\pm}326}\,\overline{H}\, G_{+}\, DG_{\pm}\,\partial G_{-}
+c_{{\pm}327}\,\overline{H}\,\overline{D}G_{+}\, DG_{\pm}\, DG_{-}
+c_{{\pm}328}\,\overline{H}\,[D,\overline{D}]G_{+}\, DG_{\pm}\,G_{-}
\nonu \\&&+
c_{{\pm}329}\,\overline{H}\, DG_{+}\, DG_{\pm}\,\overline{D}G_{-}
+c_{{\pm}330}\,\overline{H}\, D\overline{H}\,\partial DG_{\pm}
+c_{{\pm}331}\,\overline{H}\, G_{+}\,\partial DG_{\pm}\,G_{-}
\nonu \\&&+
c_{{\pm}332}\,\overline{H}\,\partial G_{+}\, G_{\pm}\, DG_{-}
+c_{{\pm}333}\,\overline{H}\, DG_{+}\, \partial G_{\pm}\, G_{-}
+c_{{\pm}334}\, T\,\partial[D,\overline{D}]G_{\pm}
+c_{{\pm}335}\, T\, H\,\partial\overline{D}G_{\pm}
\nonu \\&&+
c_{{\pm}336}\, T\, H\,\overline{H}\,[D,\overline{D}]G_{\pm}
+c_{{\pm}337}\, T\, H\,\overline{H}\,\partial G_{\pm}
+c_{{\pm}338}\, T\, H\, D\overline{H}\,\overline{D}G_{\pm}
+c_{{\pm}339}\, T\, H\,\partial\overline{H}\, G_{\pm}
\nonu \\&&+
c_{{\pm}340}\, T\,\overline{H}\,\partial DG_{\pm}
+c_{{\pm}341}\, T\, T\,[D,\overline{D}]G_{\pm}
+c_{{\pm}342}\, T\, T\,\partial G_{\pm}
+c_{{\pm}343}\, T\,\overline{D}H\,[D,\overline{D}]G_{\pm}
\nonu \\&&+
c_{{\pm}344}\, T\,\overline{D}H\,\overline{H}\, DG_{\pm}
+c_{{\pm}345}\, T\,\overline{D}H\, D\overline{H}\, G_{\pm}
+c_{{\pm}346}\, T\,\overline{D}H\,\partial G_{\pm}
+c_{{\pm}347}\, T\,\overline{D}T\, DG_{\pm}
\nonu \\&&+
c_{{\pm}348}\, T\,\partial\overline{D}H\, G_{\pm}
+c_{{\pm}349}\, T\,[D,\overline{D}]T\, G_{\pm}
+c_{{\pm}350}\, T\, D\overline{H}\,[D,\overline{D}]G_{\pm}
+c_{{\pm}351}\, T\, D\overline{H}\,\partial G_{\pm}
\nonu \\&&+
c_{{\pm}352}\, T\, DT\,\overline{D}G_{\pm}
+c_{{\pm}353}\, T\,\partial D\overline{H}\, G_{\pm}
+c_{{\pm}354}\, T\,\partial H\,\overline{D}G_{\pm}
+c_{{\pm}355}\, T\,\partial H\,\overline{H}\, G_{\pm}
\nonu \\&&+
c_{{\pm}356}\, T\,\partial\overline{H}\, DG_{\pm}
+c_{{\pm}357}\, T\,\partial^{2}G_{\pm}
+c_{{\pm}358}\, DG_{+}\, G_{\pm}\,\partial\overline{D}G_{-}
+c_{{\pm}359}\,[D,\overline{D}]G_{+}\, G_{\pm}\,[D,\overline{D}]G_{-}
\nonu \\&&+
c_{{\pm}360}\,\overline{D}G_{+}\,[D,\overline{D}]G_{\pm}\, DG_{-}
+c_{{\pm}361}\, DG_{+}\,\overline{D}G_{\pm}\,[D,\overline{D}]G_{-}
+c_{{\pm}362}\,\overline{D}G_{+}\, DG_{\pm}\,\partial G_{-}
\nonu \\&&+
c_{{\pm}363}\,\overline{D}H\,\partial[D,\overline{D}]G_{\pm}
+c_{{\pm}364}\,\overline{D}H\, G_{+}\,\overline{D}G_{\pm}\, DG_{-}
+c_{{\pm}365}\,\overline{D}H\, G_{+}\,[D,\overline{D}]G_{\pm}\,G_{-}
\nonu \\&&+
c_{{\pm}366}\,\overline{D}H\,\overline{H}\,\partial DG_{\pm}
+c_{{\pm}367}\,\overline{D}H\,\overline{H}\, D\overline{H}\, DG_{\pm}
+c_{{\pm}368}\,\overline{D}H\,DG_{+}\, \overline{D}G_{\pm}\, G_{-}
\nonu \\&&+
c_{{\pm}369}\,\overline{D}H\,\overline{D}H\,[D,\overline{D}]G_{\pm}
+c_{{\pm}370}\,\overline{D}H\,\overline{D}H\,\overline{H}\, DG_{\pm}
+c_{{\pm}371}\,\overline{D}H\,\overline{D}H\, D\overline{H}\, G_{\pm}
\nonu \\&&+
c_{{\pm}372}\,\overline{D}H\,\overline{D}H\,\partial G_{\pm}
+c_{{\pm}373}\,\overline{D}H\, D\overline{H}\,[D,\overline{D}]G_{\pm}
+c_{{\pm}374}\,\overline{D}H\, D\overline{H}\, D\overline{H}\, G_{\pm}
\nonu \\&&+
c_{{\pm}375}\,\overline{D}H\, D\overline{H}\,\partial G_{\pm}
+c_{{\pm}376}\,\overline{D}H\,\partial D\overline{H}\, G_{\pm}
+c_{{\pm}377}\,\overline{D}H\, G_{+}\, \partial G_{\pm}\, G_{-}
+c_{{\pm}378}\,\overline{D}H\,\partial\overline{H}\, DG_{\pm}
\nonu \\&&+
c_{{\pm}379}\,\overline{D}H\,\partial^{2}G_{\pm}
+c_{{\pm}380}\,\overline{D}T\,\partial DG_{\pm}
+c_{{\pm}381}\,\overline{D}T\, H\,[D,\overline{D}]G_{\pm}
+c_{{\pm}382}\,\overline{D}T\, H\,\overline{H}\, DG_{\pm}
\nonu \\&&+
c_{{\pm}383}\overline{D}T\, H\, D\overline{H}\, G_{\pm}
+c_{{\pm}384}\,\overline{D}T\, H\,\partial G_{\pm}
+c_{{\pm}385}\,\overline{D}T\,\overline{D}H\, DG_{\pm}
+c_{{\pm}386}\,\overline{D}T\, D\overline{H}\, DG_{\pm}
\nonu \\&&+
c_{{\pm}387}\,\overline{D}T\, DT\, G_{\pm}
+c_{{\pm}388}\,\overline{D}T\,\partial H\, G_{\pm}
+c_{{\pm}389}\, H\, \overline{D}G_{+}\,G_{\pm}\,[D,\overline{D}]G_{-}
+c_{{\pm}390}\,G_{+}\,\partial\overline{D}G_{\pm}\,  DG_{-}
\nonu \\&&+
c_{{\pm}391}\,\partial\overline{D}G_{+}\, DG_{\pm}\,G_{-}
+c_{{\pm}392}\,\partial\overline{D}H\,[D,\overline{D}]G_{\pm}
+c_{{\pm}393}\,\partial\overline{D}H\, H\,\overline{D}G_{\pm}
\nonu \\&&+
c_{{\pm}394}\,\partial\overline{D}H\, H\,\overline{H}\, G_{\pm}
+c_{{\pm}395}\,\partial\overline{D}H\,\overline{H}\, DG_{\pm}
+c_{{\pm}396}\,\partial\overline{D}H\,\overline{D}H\, G_{\pm}
\nonu \\&&+
c_{{\pm}397}\,\partial\overline{D}H\, D\overline{H}\, G_{\pm}
+c_{{\pm}398}\,\partial\overline{D}H\,\partial G_{\pm}
+c_{{\pm}399}\,\partial\overline{D}T\, DG_{\pm}
+c_{{\pm}400}\,\partial\overline{D}T\, H\, G_{\pm}
\nonu \\&&+
c_{{\pm}401}\,\partial^{2}\overline{D}H\, G_{\pm}
+c_{{\pm}402}\,[D,\overline{D}]G_{+}\, DG_{\pm}\,\overline{D}G_{-}
+c_{{\pm}403}\,[D,\overline{D}]T\,[D,\overline{D}]G_{\pm}
\nonu \\&&+
c_{{\pm}404}\,[D,\overline{D}]T\, H\,\overline{D}G_{\pm}
+c_{{\pm}405}\,[D,\overline{D}]T\, H\,\overline{H}\, G_{\pm}
+c_{{\pm}406}\,[D,\overline{D}]T\,\overline{H}\, DG_{\pm}
\nonu \\&&+
c_{{\pm}407}\,[D,\overline{D}]T\,\overline{D}H\, G_{\pm}
+c_{{\pm}408}\,[D,\overline{D}]T\, D\overline{H}\, G_{\pm}
+c_{{\pm}409}\,[D,\overline{D}]T\,\partial G_{\pm}
\nonu \\&&+
c_{{\pm}410}\, [D,\overline{D}]G_{+}\,G_{\pm}\,\partial G_{-}
+c_{{\pm}411}\,G_{+}\,\partial[D,\overline{D}]G_{\pm}\, G_{-}
+c_{{\pm}412}\,\partial[D,\overline{D}]T\, G_{\pm}
\nonu \\&&+
c_{{\pm}413}\, D\overline{H}\,\partial[D,\overline{D}]G_{\pm}
+c_{{\pm}414}\, D\overline{H}\, G_{+}\,[D,\overline{D}]G_{\pm}\,G_{-}
+c_{{\pm}415}\, D\overline{H}\, G_{+}\, DG_{\pm}\,\overline{D}G_{-}
\nonu \\&&+
c_{{\pm}416}\, D\overline{H}\,\overline{D}G_{+}\, DG_{\pm}\,G_{-}
+c_{{\pm}417}D\overline{H}\, D\overline{H}\,[D,\overline{D}]G_{\pm}
+c_{{\pm}418}\, D\overline{H}\, D\overline{H}\,\partial G_{\pm}
\nonu \\&&+
c_{{\pm}419}\, D\overline{H}\,G_{+}\,\partial G_{\pm}\, G_{-}
+c_{{\pm}420}\, D\overline{H}\,\partial^{2}G_{\pm}
+c_{{\pm}421}\, DT\,\partial\overline{D}G_{\pm}
+_{{\pm}422}\, DT\, H\,\overline{H}\,\overline{D}G_{\pm}
\nonu \\&&+
c_{{\pm}423}\, DT\,\overline{H}\,[D,\overline{D}]G_{\pm}
+c_{{\pm}424}\, DT\,\overline{H}\,\partial G_{\pm}
+c_{{\pm}425}\, DT\,\overline{D}H\,\overline{D}G_{\pm}
\nonu \\&&+
c_{{\pm}426}\, DT\,\overline{D}H\,\overline{H}\, G_{\pm}
+c_{{\pm}427}\, DT\, D\overline{H}\,\overline{D}G_{\pm}
+c_{{\pm}428}\, DT\,\partial\overline{H}\, G_{\pm}
+c_{{\pm}429}\, H\,\partial^{2}\overline{D}G_{\pm}
\nonu \\&&+
c_{{\pm}430}\,\partial DG_{+}\, G_{\pm}\,\overline{D}G_{-}
+c_{{\pm}431}\,\overline{D}G_{+}\,\partial DG_{\pm}\,G_{-}
+c_{{\pm}432}\,\partial D\overline{H}\,[D,\overline{D}]G_{\pm}
\nonu \\&&+
c_{{\pm}433}\,\partial D\overline{H}\,\overline{H}\, DG_{\pm}
+c_{{\pm}434}\,\partial D\overline{H}\, D\overline{H}\, G_{\pm}
+c_{{\pm}435}\,\partial D\overline{H}\,\partial G_{\pm}
+c_{{\pm}436}\,\partial DT\,\overline{D}G_{\pm}
\nonu \\&&+
c_{{\pm}437}\,\partial DT\,\overline{H}\, G_{\pm}
+c_{{\pm}438}\,\partial^{2} D\overline{H}\, G_{\pm}
+c_{{\pm}439}\, G_{+}\,\partial G_{\pm}\,[D,\overline{D}]G_{-}
+c_{{\pm}440}\,\partial G_{+}\, G_{\pm}\,\partial G_{-}
\nonu \\&&+
c_{{\pm}441}\,\partial G_{+}\,\overline{D}G_{\pm}\, DG_{-}
+c_{{\pm}442}\,\partial G_{+}\,[D,\overline{D}]G_{\pm}\,G_{-}
+c_{{\pm}443}\, DG_{+}\,\partial G_{\pm}\,\overline{D}G_{-}
+c_{{\pm}444}\,\partial H\,\partial\overline{D}G_{\pm}
\nonu \\&&+
c_{{\pm}445}\,\partial H\, G_{+}\,\overline{D}G_{\pm}\,G_{-}
+c_{{\pm}446}\,\partial H\, H\,\overline{H}\,\overline{D}G_{\pm}
+c_{{\pm}447}\,\partial H\,\overline{H}\,[D,\overline{D}]G_{\pm}
+c_{{\pm}448}\,\partial H\,\overline{H}\, D\overline{H}\, G_{\pm}
\nonu \\&&+
c_{{\pm}449}\,\partial H\,\overline{H}\,\partial G_{\pm}
+c_{{\pm}450}\,\partial H\,\overline{D}H\,\overline{D}G_{\pm}
+c_{{\pm}451}\,\partial H\,\overline{D}H\,\overline{H}\, G_{\pm}
+c_{{\pm}452}\,\partial H\, D\overline{H}\,\overline{D}G_{\pm}
\nonu \\&&+
c_{{\pm}453}\,\partial H\,\partial\overline{H}\, G_{\pm}
+c_{{\pm}454}\,\partial\overline{H}\,\partial DG_{\pm}
+c_{{\pm}455}\,\partial\overline{H}\, G_{+}\, DG_{\pm}\,G_{-}
+c_{{\pm}456}\,\partial\overline{H}\, D\overline{H}\, DG_{\pm}
\nonu \\&&+
c_{{\pm}457}\,\partial T\,[D,\overline{D}]G_{\pm}
+c_{{\pm}458}\,\partial T\, H\,\overline{D}G_{\pm}
+c_{{\pm}459}\,\partial T\, H\,\overline{H}\, G_{\pm}
+c_{{\pm}460}\,\partial T\,\overline{H}\, DG_{\pm}
\nonu \\&&+
c_{{\pm}461}\,\partial T\, T\, G_{\pm}
+c_{{\pm}462}\,\partial T\,\overline{D}H\, G_{\pm}
+c_{{\pm}463}\,\partial T\, D\overline{H}\, G_{\pm}
+c_{{\pm}464}\,\partial T\,\partial G_{\pm}
\nonu \\&&+
c_{{\pm}465}\, G_{+}\,\overline{D}G_{\pm}\,\partial DG_{-}
+c_{{\pm}466}\, G_{+}\,\partial^{2}G_{\pm}\,G_{-}
+c_{{\pm}467}\,\partial^{2}H\,\overline{D}G_{\pm}
+c_{{\pm}468}\,\partial^{2}H\,\overline{H}\, G_{\pm}
\nonu \\&&+
c_{{\pm}469}\,\partial^{2}\overline{H}\, DG_{\pm}
+c_{{\pm}470}\,\partial^{2}T\, G_{\pm}
+c_{{\pm}471}\,\partial^{2}[D,\overline{D}]G_{\pm}
+c_{{\pm}472}\,\partial^{3}G_{\pm}
\Bigg](Z_{2}),
\nonu
\end{eqnarray}
where the 
coefficients are 
\bea
c_ {+1} & = & -\frac {48 k\, N} {(2 + k + N)^{2}}, \qquad
c_ {+2}  =  \frac {32 k (k - N) N} {3 (2 + k + N)^{3}}, \nonu \\
c_ {+3} & = & \frac {64 k (3 + 2 k + N) (2 + k + 
      2 N)} {3 (2 + k + N)^{3}},\qquad 
c_ {+4}  =  -\frac {64 k (3 + 2 k + N)} {3 (2 + k + N)^{3}}, \nonu \\
c_ {+5} & = & -\frac {64 N (2 + 2 k + N) (3 + k + 
      2 N)} {3 (2 + k + N)^{3}}, \qquad
c_ {+6}  =  -\frac {64 N (3 + k + 2 N)} {3 (2 + k + N)^{3}}, \nonu \\
c_ {+7} & = & -\frac {16 (k - N) (2 + 2 k + N) (2 + k + 
      2 N)} {3 (2 + k + N)^{3}}, \qquad
c_ {+8}  =  \frac {32 (k - N) (2 + 2 k + N)} {3 (2 + k + N)^{3}}, \nonu \\
c_ {+9} & = & \frac {16 (k - N)} {3 (2 + k + N)^{3}}, \qquad
c_ {+10}  =  \frac {32 (k - N) (2 + k + 2 N)} {3 (2 + k + N)^{3}}, \nonu \\
c_ {+11} & = & \frac {16 (k - N)} {3 (2 + k + N)^{2}}, \qquad
c_ {+12}  =  -\frac {16 (5 k + 4 k^{2} + N + 2 k\, 
     N)} {3 (2 + k + N)^{3}}, \nonu \\
c_ {+13} & = & \frac {16 (k + 5 N + 2 k\, 
     N + 4 N^{2})} {3 (2 + k + N)^{3}}, \qquad
c_ {+14}  =  -\frac {16 (3 k + 2 k^{2} + 3 N + 5 k\, 
     N + 2 N^{2})} {3 (2 + k + N)^{2}}, \nonu \\
c_ {+15} & = & \frac {16 (3 + 6 k + 2 k^{2} + 6 N + 5 k\, 
     N + 2 N^{2})} {3 (2 + k + N)^{2}}, \qquad
c_ {+16}  =  -\frac {32 (3 + 6 k + 2 k^{2} + 2 k\, 
     N - N^{2})} {3 (2 + k + N)^{3}}, \nonu \\
c_ {+17} & = & -\frac {16} {(2 + k + N)^{2}}, \qquad
c_ {+18}  =  \frac {32 (-3 + k^{2} - 6 N - 2 k\, 
     N - 2 N^{2})} {3 (2 + k + N)^{3}}, \nonu \\
c_ {+19} & = & \frac {16 (6 + 12 k + 5 k^{2} + 12 N + 8 k\, 
     N + 5 N^{2})} {3 (2 + k + N)^{3}}, \nonu \\
c_ {+20} & = & \frac {16 (6 + 18 k + 15 k^{2} + 4 k^{3} + 12 N + 22 k\, 
     N + 8 k^{2} N + 5 N^{2} + 6 k\, N^{2})} {3 (2 + k + N)^{4}},\nonu \\ 
c_ {+21} & = & \frac {16 (6 + 12 k + 5 k^{2} + 18 N + 22 k\, 
     N + 6 k^{2} N + 15 N^{2} + 8 k\, 
     N^{2} + 4 N^{3})} {3 (2 + k + N)^{4}}, \nonu \\
c_ {+22} & = & \frac {16 (k - N) (3 + 5 k + 2 k^{2} + 5 N + 5 k\, 
     N + 2 N^{2})} {3 (2 + k + N)^{3}}, \nonu \\
c_ {+23} & = & \frac {16 (3 k + 9 k^{2} + 4 k^{3} + 3 N + 13 k\, 
     N + 10 k^{2} N + 2 N^{2} + 4 k\, N^{2})} {3 (2 + k + N)^{3}},\nonu \\ 
c_ {+24} & = & -\frac {16 (3 + k + 2 N)} {3 (2 + k + N)^{3}}, \qquad
c_ {+25}  =  \frac {8 (k - N) (3 + k + 2 N)} {3 (2 + k + N)^{3}}, \nonu \\
c_ {+26} & = & -\frac {16 (3 + k + 2 N)} {3 (2 + k + N)^{3}}, \qquad
c_ {+27}  =  -\frac {16 (1 + k) (3 + k + 2 N)} {3 (2 + k + N)^{4}}, \nonu \\
c_ {+28} & = & \frac {16 (1 + k) (3 + k + 2 N)} {3 (2 + k + N)^{4}}, \qquad
c_ {+29}  =  -\frac {8 (9 k + 7 k^{2} - 3 N + k\, 
     N - 2 N^{2})} {3 (2 + k + N)^{3}}, \nonu \\
c_ {+30} & = & \frac {32 (3 + 7 k + 3 k^{2} + 5 N + 4 k\, 
     N + 2 N^{2})} {3 (2 + k + N)^{3}}, \qquad
c_ {+31}  =  -\frac {16 (3 + k + 2 N)} {3 (2 + k + N)^{3}}, \nonu \\
c_ {+32} & = & \frac {32 (3 + 4 k + 3 k^{2} + k^{3} + 5 N + 4 k\, 
     N + k^{2} N + 2 N^{2} + k\, N^{2})} {3 (2 + k + N)^{4}}, \nonu \\
c_ {+33} & = & -\frac {32 (3 + 4 k + 3 k^{2} + k^{3} + 5 N + 4 k\, 
     N + k^{2} N + 2 N^{2} + k\, N^{2})} {3 (2 + k + N)^{4}}, \nonu \\
c_ {+34} & = & -\frac {16 (k - N) (3 + k + 2 N)} {3 (2 + k + N)^{3}}, \nonu \\
c_ {+35} & = & -\frac {32 (-3 + 5 k + 7 k^{2} + 2 k^{3} - 2 N + 8 k\, 
     N + 4 k^{2} N + 3 k\, N^{2})} {3 (2 + k + N)^{4}}, \nonu \\
c_ {+36} & = & -\frac {16 (3 k + 2 k^{2} + 3 N + 13 k\, 
     N + 4 k^{2} N + 9 N^{2} + 10 k\, 
     N^{2} + 4 N^{3})} {3 (2 + k + N)^{3}}, \nonu \\
c_ {+37} & = & \frac {16 (3 + 2 k + N)} {3 (2 + k + N)^{3}}, \qquad
c_ {+38}  =  -\frac {16 (3 + 2 k + N)} {3 (2 + k + N)^{3}}, \nonu \\
c_ {+39} & = & \frac {8 (k - N) (3 + 2 k + N)} {3 (2 + k + N)^{3}}, \qquad
c_ {+40}  =  \frac {16 (1 + N) (3 + 2 k + N)} {3 (2 + k + N)^{4}}, \nonu \\
c_ {+41} & = & \frac {8 (3 k + 2 k^{2} - 9 N - k\, 
     N - 7 N^{2})} {3 (2 + k + N)^{3}}, \qquad
c_ {+42}  =  \frac {32 (3 + 5 k + 2 k^{2} + 7 N + 4 k\, 
     N + 3 N^{2})} {3 (2 + k + N)^{3}}, \nonu \\
c_ {+43} & = & \frac {16 (3 + 2 k + N)} {3 (2 + k + N)^{3}}, \nonu \\
c_ {+44} & = & -\frac {32 (3 + 5 k + 2 k^{2} + 4 N + 4 k\, 
     N + k^{2} N + 3 N^{2} + k\, N^{2} + N^{3})} {3 (2 + k + N)^{4}}, \nonu \\
c_ {+45} & = & -\frac {16 (1 + N) (3 + 2 k + N)} {3 (2 + k + N)^{4}}, \qquad
c_ {+46}  =  \frac {16 (k - N) (3 + 2 k + N)} {3 (2 + k + N)^{3}}, \nonu \\
c_ {+47} & = & \frac {32 (3 + 5 k + 2 k^{2} + 4 N + 4 k\, 
     N + k^{2} N + 3 N^{2} + k\, N^{2} + N^{3})} {3 (2 + k + N)^{4}}, \nonu \\
c_ {+48} & = & -\frac {32 (-3 - 2 k + 5 N + 8 k\, 
     N + 3 k^{2} N + 7 N^{2} + 4 k\, 
     N^{2} + 2 N^{3})} {3 (2 + k + N)^{4}}, \qquad
c_ {+49}  =  -5, \nonu \\
c_ {+50} & = & \frac {5 (60 + 77 k + 22 k^{2} + 121 N + 115 k\, 
     N + 20 k^{2} N + 79 N^{2} + 42 k\, 
     N^{2} + 16 N^{3})} {(2 + N) (2 + k + N)^{2}}, \nonu \\
c_ {+51} & = & -\frac {4 (27 + 4 k + 23 N)} {3 (2 + k + N)^{3}}, \nonu \\
c_ {+52} & = & -\frac {(300 + 299 k + 90 k^{2} + 391 N + 187 k\, 
    N + 113 N^{2})} {3 (2 + N) (2 + k + N)^{3}}, \nonu \\
c_ {+53} & = & \frac {4 (246 + 299 k + 90 k^{2} + 310 N + 187 k\, 
     N + 86 N^{2})} {3 (2 + N) (2 + k + N)^{3}}, \nonu \\
c_ {+54} & = & \frac {4 (-24 + 35 k + 26 k^{2} + 7 N + 146 k\, 
     N + 58 k^{2} N + 53 N^{2} + 83 k\, 
     N^{2} + 18 N^{3})} {3 (2 + N) (2 + k + N)^{3}}, \nonu \\
c_ {+55} & = & -\frac {10 (32 + 55 k + 18 k^{2} + 41 N + 35 k\, 
     N + 11 N^{2})} {(2 + N) (2 + k + N)^{4}}, \nonu \\
c_ {+56} & = & -\frac {(32 + 81 k + 30 k^{2} + 47 N + 53 k\, 
    N + 13 N^{2})} {(2 + N) (2 + k + N)^{3}}, \nonu \\
c_ {+57} & = & -\frac {10 (1 + N) (32 + 55 k + 18 k^{2} + 41 N + 35 k\, 
     N + 11 N^{2})} {(2 + N) (2 + k + N)^{5}}, \nonu \\
c_ {+58} & = & \frac {1} {3 (2 + N) (2 + k + N)^{4}}(1200 + 1412 k + 181 k^{2} - 90 k^{3} + 2548 N + 
     2386 k\, N + 323 k^{2} N 
     \nonu \\ & + & 1753 N^{2} + 960 k\, 
    N^{2} + 367 N^{3}), \nonu \\
c_ {+59} & = & \frac {2 (300 + 409 k + 118 k^{2} + 581 N + 595 k\, 
     N + 104 k^{2} N + 367 N^{2} + 214 k\, 
     N^{2} + 72 N^{3})} {3 (2 + N) (2 + k + N)^{4}}, \nonu \\
c_ {+60} & = & -\frac {10 (1 + k) (32 + 55 k + 18 k^{2} + 41 N + 35 k\, 
     N + 11 N^{2})} {(2 + N) (2 + k + N)^{5}}, \nonu \\
c_ {+61} & = & -\frac {2 (-12 + 173 k + 78 k^{2} + 37 N + 321 k\, 
     N + 84 k^{2} N + 57 N^{2} + 136 k\, 
     N^{2} + 14 N^{3})} {3 (2 + N) (2 + k + N)^{4}}, \nonu \\
c_ {+62} & = & \frac {8 (18 - 33 k - 22 k^{2} + 78 N + 92 k\, 
     N + 34 k^{2} N + 86 N^{2} + 73 k\, 
     N^{2} + 22 N^{3})} {3 (2 + N) (2 + k + N)^{4}}, \nonu \\
c_ {+63} & = & \frac {4 (24 + 131 k + 54 k^{2} + 127 N + 282 k\, 
     N + 72 k^{2} N + 129 N^{2} + 127 k\, 
     N^{2} + 32 N^{3})} {3 (2 + N) (2 + k + N)^{3}}, \nonu \\
c_ {+64} & = & \frac {10 (32 + 55 k + 18 k^{2} + 41 N + 35 k\, 
     N + 11 N^{2})} {(2 + N) (2 + k + N)^{4}}, \nonu \\
c_ {+65} & = & -\frac {2 (300 + 409 k + 126 k^{2} + 581 N + 579 k\, 
     N + 108 k^{2} N + 375 N^{2} + 206 k\, 
     N^{2} + 76 N^{3})} {3 (2 + N) (2 + k + N)^{4}}, \nonu \\
c_ {+66} & = & -\frac {1} {3 (2 + N) (2 + k + N)^{3}}(-96 - 390 k - 343 k^{2} - 90 k^{3} - 138 N - 
     478 k\, N - 209 k^{2} N 
     \nonu \\ & - & 43 N^{2} - 134 k\, 
    N^{2} + N^{3}), \nonu \\
c_ {+67} & = & -\frac {2 (96 + 251 k + 90 k^{2} + 133 N + 163 k\, 
     N + 35 N^{2})} {3 (2 + N) (2 + k + N)^{3}}, \nonu \\
c_ {+68} & = & -\frac {10 (32 + 55 k + 18 k^{2} + 41 N + 35 k\, 
     N + 11 N^{2})} {(2 + N) (2 + k + N)^{4}}, \nonu \\
c_ {+69} & = & -\frac {2 (96 + 235 k + 90 k^{2} + 149 N + 155 k\, 
     N + 43 N^{2})} {3 (2 + N) (2 + k + N)^{3}}, \nonu \\
c_ {+70} & = & \frac {2 (48 + 89 k + 30 k^{2} + 63 N + 57 k\, 
     N + 17 N^{2})} {(2 + N) (2 + k + N)^{3}}, \nonu \\
c_ {+71} & = & -\frac{1}{3 (2 + N) (2 + k + N)^{4}}2 (204 + 14 k - 239 k^{2} - 90 k^{3} + 496 N + 
      259 k\, N - 67 k^{2} N \nonu \\ & + & 388 N^{2} + 171 k\, 
     N^{2} + 88 N^{3}) , \nonu \\
c_ {+72} & = & \frac {2 (396 + 596 k + 184 k^{2} + 874 N + 961 k\, 
     N + 182 k^{2} N + 607 N^{2} + 369 k\, 
     N^{2} + 127 N^{3})} {3 (2 + N) (2 + k + N)^{4}}, \nonu \\
c_ {+73} & = & -\frac {5 (80 + 72 k + 3 k^{2} - 6 k^{3} + 152 N + 102 k\, 
     N + 5 k^{2} N + 91 N^{2} + 36 k\, 
     N^{2} + 17 N^{3})} {(2 + N) (2 + k + N)^{3}}, \nonu \\
c_ {+74} & = & -\frac {1}{3 (2 + N) (2 + k + N)^{3}}2 (46 k + 71 k^{2} + 26 k^{3} + 254 N + 518 k\, 
     N + 332 k^{2} N + 58 k^{3} N 
     \nonu \\ & + & 401 N^{2} + 521 k\, 
     N^{2} + 167 k^{2} N^{2} + 201 N^{3} + 133 k\, 
     N^{3} + 32 N^{4}) , \nonu \\
c_ {+75} & = & \frac {4 (32 + 81 k + 30 k^{2} + 47 N + 53 k\, 
     N + 13 N^{2})} {(2 + N) (2 + k + N)^{3}}, \nonu \\
c_ {+76} & = & -\frac{1}{3 (2 + N) (2 + k + N)^{4}}(504 + 910 k + 557 k^{2} + 118 k^{3} + 1142 N + 
     1708 k\, N + 773 k^{2} N 
     \nonu \\ & + & 104 k^{3} N + 963 N^{2} + 1078 k\, 
    N^{2} + 266 k^{2} N^{2} + 347 N^{3} + 222 k\, 
    N^{3} + 44 N^{4}) , \nonu \\
c_ {+77} & = & -\frac {1}{3 (2 + N) (2 + k + N)^{4}}2 (588 + 1183 k + 808 k^{2} + 180 k^{3} + 1007 N + 
      1291 k\, N 
      \nonu \\ & + & 434 k^{2} N + 517 N^{2} + 316 k\, 
     N^{2} + 84 N^{3}), \nonu \\
c_ {+78} & = & -\frac{1}{3 (2 + N) (2 + k + N)^{5}} 2 (1 + k) (348 + 425 k + 118 k^{2} + 685 N + 
      619 k\, N + 104 k^{2} N 
      \nonu \\ & + & 439 N^{2} + 222 k\, 
     N^{2} + 88 N^{3}) , \nonu \\
c_ {+79} & = & -\frac {1} {3 (2 + N) (2 + k + N)^{5}}2 (96 - 156 k - 305 k^{2} - 94 k^{3} + 492 N - 
      20 k\, N - 393 k^{2} N 
      \nonu \\ & - & 92 k^{3} N + 757 N^{2} + 276 k\, 
     N^{2} - 94 k^{2} N^{2} + 451 N^{3} + 146 k\, 
     N^{3} + 88 N^{4}), \nonu \\
c_ {+80} & = & \frac {1}{3 (2 + N) (2 + k + N)^{4}}(384 - 164 k - 637 k^{2} - 222 k^{3} + 716 N - 
     1164 k\, N - 1593 k^{2} N 
     \nonu \\ & - & 336 k^{3} N + 589 N^{2} - 988 k\, 
    N^{2} - 686 k^{2} N^{2} + 259 N^{3} - 186 k\, 
    N^{3} + 44 N^{4}), \nonu \\
c_ {+81} & = & \frac{1}{3 (2 + N) (2 + k + N)^{3}}4 (396 + 796 k + 483 k^{2} + 90 k^{3} + 674 N + 
      903 k\, N + 279 k^{2} N 
      \nonu \\ & + & 366 N^{2} + 245 k\, 
     N^{2} + 64 N^{3}), \nonu \\
c_ {+82} & = & -\frac {4 (222 + 287 k + 90 k^{2} + 286 N + 181 k\, 
     N + 80 N^{2})} {3 (2 + N) (2 + k + N)^{3}}, \nonu \\
c_ {+83} & = & -\frac {1} {3 (2 + N) (2 + k + N)^{4}}2 (-684 - 931 k - 367 k^{2} - 38 k^{3} - 1199 N - 
      814 k\, N + 161 k^{2} N 
      \nonu \\ & + & 116 k^{3} N - 637 N^{2} + 3 k\, 
     N^{2} + 206 k^{2} N^{2} - 108 N^{3} + 68 k\, 
     N^{3}), \nonu \\
c_ {+84} & = & \frac {4 (24 + 36 k + 11 k^{2} + 36 N + 26 k\, 
     N + 11 N^{2})} {3 (2 + k + N)^{3}}, \nonu \\
c_ {+85} & = & \frac {1}{3 (2 + N) (2 + k + N)^{4}}(696 + 1390 k + 877 k^{2} + 182 k^{3} + 1718 N + 
     2652 k\, N + 1157 k^{2} N 
     \nonu \\ & + & 136 k^{3} N + 1523 N^{2} + 1654 k\, 
    N^{2} + 378 k^{2} N^{2} + 571 N^{3} + 334 k\, 
    N^{3} + 76 N^{4}) , \nonu \\
c_ {+86} & = & \frac {1}{3 (2 + N) (2 + k + N)^{5}}2 (1 + N) (252 + 329 k + 94 k^{2} + 541 N + 523 k\, 
     N + 92 k^{2} N 
     \nonu \\ & + & 367 N^{2} + 198 k\, 
     N^{2} + 76 N^{3}) , \nonu \\
c_ {+87} & = & -\frac {1}{3 (2 + N) (2 + k + N)^{4}}(384 - 76 k - 513 k^{2} - 182 k^{3} + 1828 N + 
     860 k\, N - 441 k^{2} N 
     \nonu \\ & - & 136 k^{3} N + 2401 N^{2} + 1192 k\, 
    N^{2} - 66 k^{2} N^{2} + 1207 N^{3} + 394 k\, 
    N^{3} + 204 N^{4}) , \nonu \\
c_ {+88} & = & \frac {4 (3 + 2 k + N) (34 + 13 k + 
      21 N)} {3 (2 + k + N)^{3}}, \qquad
c_ {+89}  =  \frac {4 (39 + 10 k + 29 N)} {3 (2 + k + N)^{3}}, \nonu \\
c_ {+90} & = & -\frac{1}{3 (2 + N) (2 + k + N)^{4}}2 (84 + 317 k + 126 k^{2} + 433 N + 882 k\, 
     N + 242 k^{2} N + 630 N^{2} 
     \nonu \\ & + & 771 k\, 
     N^{2} + 112 k^{2} N^{2} + 349 N^{3} + 214 k\, 
     N^{3} + 64 N^{4}), \nonu \\
c_ {+91} & = & -\frac{1} {(2 + N) (2 + k + N)^{4}}(-256 - 356 k - 39 k^{2} + 30 k^{3} - 284 N - 
     116 k\, N + 83 k^{2} N 
     \nonu \\ & - & 37 N^{2} + 66 k\, 
    N^{2} + 13 N^{3}), \nonu \\
c_ {+92} & = & \frac {1} {3 (2 + k + N)^{4}} 4 (222 + 473 k + 
      295 k^{2} + 58 k^{3} + 505 N + 668 k\, 
     N + 199 k^{2} N + 327 N^{2} 
     \nonu \\ & + & 213 k\, N^{2} + 64 N^{3}), \nonu \\
c_ {+93} & = & \frac {8 (141 + 213 k + 68 k^{2} + 189 N + 137 k\, 
     N + 56 N^{2})} {3 (2 + k + N)^{4}}, \nonu \\
c_ {+94} & = & \frac {1} {3 (2 + N) (2 + k + N)^{4}}4 (156 - 20 k - 129 k^{2} - 38 k^{3} + 422 N + 
      259 k\, N + 72 k^{2} N 
      \nonu \\ & + & 26 k^{3} N + 452 N^{2} + 344 k\, 
     N^{2} + 87 k^{2} N^{2} + 204 N^{3} + 101 k\, 
     N^{3} + 32 N^{4}), \nonu \\
c_ {+95} & = & -\frac {2 (300 + 409 k + 122 k^{2} + 581 N + 587 k\, 
     N + 106 k^{2} N + 371 N^{2} + 210 k\, 
     N^{2} + 74 N^{3})} {3 (2 + N) (2 + k + N)^{3}}, \nonu \\
c_ {+96} & = & \frac {1}{3 (2 + N) (2 + k + N)^{3}}2 (480 + 771 k + 191 k^{2} - 26 k^{3} + 861 N + 
      599 k\, N - 168 k^{2} N 
      \nonu \\ & - & 58 k^{3} N + 554 N^{2} + 117 k\, 
     N^{2} - 83 k^{2} N^{2} + 197 N^{3} + 37 k\, 
     N^{3} + 32 N^{4}) , \nonu \\
c_ {+97} & = & \frac {2 (k - N)} {3 (2 + k + N)}, \nonu \\
c_ {+98} & = & \frac {2 (-k + N) (60 + 77 k + 22 k^{2} + 121 N + 115 k\, 
     N + 20 k^{2} N + 79 N^{2} + 42 k\, 
     N^{2} + 16 N^{3})} {3 (2 + N) (2 + k + N)^{3}}, \nonu \\
c_ {+99} & = & \frac {8 (24 + 21 k + 4 k^{2} + 27 N + 13 k\, 
     N + 7 N^{2})} {3 (2 + k + N)^{4}}, \nonu \\
c_ {+100} & = & -\frac {2 (-k + N) (20 + 21 k + 6 k^{2} + 25 N + 13 k\, 
     N + 7 N^{2})} {3 (2 + N) (2 + k + N)^{4}},\nonu \\
     c_ {+101} & = & \frac {8 (-48 - 74 k - 37 k^{2} - 6 k^{3} - 46 N - 53 k\, 
       N - 15 k^{2} N - 6 N^{2} - 5 k\, 
       N^{2} + 2 N^{3})} {3 (2 + N) (2 + k + N)^{4}}, \nonu \\
  c_ {+102} & = & \frac{1}{3 (2 + N) (2 + k + N)^{4}}8 (48 + 24 k - 13 k^{2} - 6 k^{3} + 96 N + 
        28 k\, N - 24 k^{2} N - 6 k^{3} N 
        \nonu \\ & + & 81 N^{2} + 19 k\, 
       N^{2} - 7 k^{2} N^{2} + 35 N^{3} + 7 k\, 
       N^{3} + 6 N^{4}), \nonu \\
  c_ {+103} & = & -\frac {4 (-k + N) (32 + 55 k + 18 k^{2} + 41 N + 
        35 k\, N + 11 N^{2})} {3 (2 + N) (2 + k + N)^{5}},\nonu \\ 
  c_ {+104} & = & -\frac {2 (-k + N) (13 k + 6 k^{2} + 3 N + 9 k\, 
       N + N^{2})} {3 (2 + N) (2 + k + N)^{4}}, \nonu \\
  c_ {+105} & = & -\frac {4 (1 + N) (-k + N) (32 + 55 k + 18 k^{2} + 
        41 N + 35 k\, N + 11 N^{2})} {3 (2 + N) (2 + k + N)^{6}},\nonu \\ 
  c_ {+106} & = & \frac{1} {3 (2 + N) (2 + k + N)^{5}}2 (-k + N) (80 + 92 k + 11 k^{2} - 6 k^{3} + 
        172 N + 158 k\, N + 21 k^{2} N 
        \nonu \\ & + & 119 N^{2} + 64 k\, 
       N^{2} + 25 N^{3}), \nonu \\
  c_ {+107} & = & -\frac {1} {3 (2 + N) (2 + k + N)^{5}}4 (96 + 116 k + 55 k^{2} + 10 k^{3} + 220 N + 
        196 k\, N + 67 k^{2} N 
        \nonu \\ & + & 8 k^{3} N + 181 N^{2} + 100 k\, 
       N^{2} + 18 k^{2} N^{2} + 63 N^{3} + 14 k\, 
       N^{3} + 8 N^{4}), \nonu \\
  c_ {+108} & = & \frac {4 (1 + k) (k - N) (32 + 55 k + 18 k^{2} + 41 N + 
        35 k\, N + 11 N^{2})} {3 (2 + N) (2 + k + N)^{6}}, \nonu \\
  c_ {+109} & = & -\frac {1} {3 (2 + N) (2 + k + N)^{5}}4 (96 + 180 k + 93 k^{2} + 14 k^{3} + 156 N + 
        248 k\, N + 83 k^{2} N \nonu \\ & + & 4 k^{3} N + 91 N^{2} + 120 k\, 
       N^{2} + 20 k^{2} N^{2} + 23 N^{3} + 22 k\, 
       N^{3} + 2 N^{4}), \nonu \\
  c_ {+110} & = & \frac{1}{3 (2 + N) (2 + k + N)^{5}}16 (-24 - 54 k - 33 k^{2} - 6 k^{3} - 30 N - 
        69 k\, N - 38 k^{2} N - 6 k^{3} N 
        \nonu \\ & - & 6 N^{2} - 22 k\, 
       N^{2} - 9 k^{2} N^{2} + 6 N^{3} + k\, 
       N^{3} + 2 N^{4}) , \nonu \\
  c_ {+111} & = & \frac {1} {3 (2 + N) (2 + k + N)^{4}}8 (48 + 72 k + 35 k^{2} + 6 k^{3} + 48 N + 
        52 k\, N + 12 k^{2} N + 9 N^{2} 
        \nonu \\ & + & 7 k\, 
       N^{2} - k^{2} N^{2} - N^{3} + k\, 
       N^{3}), \nonu \\
  c_ {+112} & = & \frac {4 (-k + N) (32 + 55 k + 18 k^{2} + 41 N + 35 k\, 
       N + 11 N^{2})} {3 (2 + N) (2 + k + N)^{5}},  \nonu \\
  c_ {+113} & = & -\frac{1} {3 (2 + N) (2 + k + N)^{5}}4 (-96 - 212 k - 151 k^{2} - 34 k^{3} - 124 N - 
        244 k\, N - 139 k^{2} N 
        \nonu \\ & - & 20 k^{3} N - 37 N^{2} - 76 k\, 
       N^{2} - 30 k^{2} N^{2} + 9 N^{3} - 2 k\, 
       N^{3} + 4 N^{4}), \nonu \\
  c_ {+114} & = & \frac {2 (-k + N) (13 k + 6 k^{2} + 3 N + 9 k\, 
       N + N^{2})} {3 (2 + N) (2 + k + N)^{3}}, \nonu \\
  c_ {+115} & = & \frac {4 (96 + 96 k + 37 k^{2} + 6 k^{3} + 144 N + 
        86 k\, N + 15 k^{2} N + 69 N^{2} + 16 k\, 
       N^{2} + 11 N^{3})} {3 (2 + N) (2 + k + N)^{4}},\nonu \\ 
  c_ {+116} & = & -\frac {4 (-k + N) (32 + 55 k + 18 k^{2} + 41 N + 
        35 k\, N + 11 N^{2})} {3 (2 + N) (2 + k + N)^{5}}, \nonu \\
  c_ {+117} & = & -\frac {4 (96 + 96 k + 11 k^{2} - 6 k^{3} + 144 N + 
        106 k\, N + 9 k^{2} N + 75 N^{2} + 32 k\, 
       N^{2} + 13 N^{3})} {3 (2 + N) (2 + k + N)^{4}}, \nonu \\
  c_ {+118} & = & \frac {4 (-k + N) (16 + 21 k + 6 k^{2} + 19 N + 13 k\, 
       N + 5 N^{2})} {3 (2 + N) (2 + k + N)^{4}}, \nonu \\
  c_ {+119} & = & -\frac {4 (-k + N) (20 + 2 k - 17 k^{2} - 6 k^{3} + 
        48 N + 21 k\, N - 5 k^{2} N + 36 N^{2} + 13 k\, 
       N^{2} + 8 N^{3})} {3 (2 + N) (2 + k + N)^{5}}, \nonu \\
  c_ {+120} & = & \frac {4 (-k + N) (20 + 28 k + 8 k^{2} + 54 N + 55 k\, 
       N + 10 k^{2} N + 41 N^{2} + 23 k\, 
       N^{2} + 9 N^{3})} {3 (2 + N) (2 + k + N)^{5}}, \nonu \\
  c_ {+121} & = & -\frac {1}{3 (2 + N) (2 + k + N)^{4}}2 (-192 - 368 k - 232 k^{2} - 35 k^{3} + 
        6 k^{4} - 304 N - 480 k\, 
       N 
       \nonu \\ & - & 243 k^{2} N - 27 k^{3} N - 152 N^{2} - 181 k\, 
       N^{2} - 63 k^{2} N^{2} - 21 N^{3} - 13 k\, 
       N^{3} + N^{4}), \nonu \\
  c_ {+122} & = & \frac {1} {3 (2 + N) (2 + k + N)^{3}}4 (-48 - 24 k + 13 k^{2} + 6 k^{3} - 48 N + 
        20 k\, N + 36 k^{2} N + 6 k^{3} N 
        \nonu \\ & - & 9 N^{2} + 29 k\, 
       N^{2} + 13 k^{2} N^{2} + N^{3} + 5 k\, 
       N^{3}), \nonu \\
  c_ {+123} & = & \frac {8 (-k + N) (13 k + 6 k^{2} + 3 N + 9 k\, 
       N + N^{2})} {3 (2 + N) (2 + k + N)^{4}}, \nonu \\
  c_ {+124} & = & -\frac {1} {3 (2 + N) (2 + k + N)^{4}} 2 (-k + N) (20 + 31 k + 10 k^{2} + 35 N + 
        41 k\, N + 8 k^{2} N 
        \nonu \\ & + & 21 N^{2} + 14 k\, 
       N^{2} + 4 N^{3}), \nonu \\
  c_ {+125} & = & \frac {1} {3 (2 + 
        N) (2 + k + N)^{5}}4 (3 + 2 k + N) (32 + 28 k + 19 k^{2} 
        \nonu \\ & + &  6 k^{3} + 52 N + 20 k\, 
       N + 6 k^{2} N + 25 N^{2} + 4 N^{3}), \nonu \\
  c_ {+126} & = & -\frac {1} {3 (2 + N) (2 + k + N)^{6}}4 (1 + k) (-k + N) (36 + 39 k + 10 k^{2} + 
        67 N + 53 k\, N + 8 k^{2} N 
        \nonu \\ & + & 41 N^{2} + 18 k\, 
       N^{2} + 8 N^{3}), \nonu \\
  c_ {+127} & = & -\frac{1} {3 (2 + N) (2 + k + N)^{6}}4 (-k + N) (32 + 28 k + k^{2} - 2 k^{3} + 
        84 N + 52 k\, N - 7 k^{2} N 
        \nonu \\ & - & 4 k^{3} N + 91 N^{2} + 44 k\, 
       N^{2} - 2 k^{2} N^{2} + 45 N^{3} + 14 k\, 
       N^{3} + 8 N^{4}), \nonu \\
  c_ {+128} & = & \frac {1}{3 (2 + N) (2 + k + N)^{5}}2 (-192 - 160 k + 92 k^{2} + 131 k^{3} + 
        34 k^{4} - 320 N - 128 k\, 
       N 
       \nonu \\ & + & 241 k^{2} N + 189 k^{3} N + 32 k^{4} N - 156 N^{2} + 9 k\, 
       N^{2} + 117 k^{2} N^{2} + 50 k^{3} N^{2} + 3 N^{3} + 23 k\, 
       N^{3} 
       \nonu \\ & + & 4 k^{2} N^{3} + 21 N^{4} + 6 k\, 
       N^{4} + 4 N^{5}), \nonu \\
  c_ {+129} & = & -\frac {8 (48 + 66 k + 33 k^{2} + 6 k^{3} + 54 N + 
        49 k\, N + 13 k^{2} N + 14 N^{2} + 5 k\, 
       N^{2})} {3 (2 + N) (2 + k + N)^{3}}, \nonu \\
  c_ {+130} & = & \frac {8 (48 + 66 k + 33 k^{2} + 6 k^{3} + 54 N + 
        49 k\, N + 13 k^{2} N + 14 N^{2} + 5 k\, 
       N^{2})} {3 (2 + N) (2 + k + N)^{4}},  \nonu \\
  c_ {+131} & = & \frac {1}{3 (2 + N) (2 + k + N)^{5}}4 (-96 - 116 k - 29 k^{2} + 15 k^{3} + 6 k^{4} - 
        124 N - 36 k\, 
       N + 87 k^{2} N 
       \nonu \\ & + & 65 k^{3} N + 12 k^{4} N - 31 N^{2} + 79 k\, 
       N^{2} + 78 k^{2} N^{2} + 22 k^{3} N^{2} + 11 N^{3} + 39 k\, 
       N^{3} + 10 k^{2} N^{3} 
       \nonu \\ & + & 4 N^{4} + 4 k\, 
       N^{4}), \nonu \\
  c_ {+132} & = & -\frac {8 (k - N)} {3 (2 + k + N)^{2}}, \nonu \\
  c_ {+133} & = & \frac {2 (-k + N) (20 + 31 k + 10 k^{2} + 35 N + 41 k\, 
       N + 8 k^{2} N + 21 N^{2} + 14 k\, 
       N^{2} + 4 N^{3})} {3 (2 + N) (2 + k + N)^{4}}, \nonu \\
  c_ {+134} & = & \frac {4 (1 + N) (-k + N) (4 + 7 k + 2 k^{2} + 19 N + 
        21 k\, N + 4 k^{2} N + 17 N^{2} + 10 k\, 
       N^{2} + 4 N^{3})} {3 (2 + N) (2 + k + N)^{6}}, \nonu \\
  c_ {+135} & = & -\frac {1} {3 (2 + N) (2 + k + N)^{5}}2 (192 + 224 k + 84 k^{2} + 31 k^{3} + 
        10 k^{4} + 256 N - 179 k^{2} N 
        \nonu \\ & - & 27 k^{3} N + 8 k^{4} N + 
        108 N^{2} - 267 k\, 
       N^{2} - 247 k^{2} N^{2} - 26 k^{3} N^{2} + 31 N^{3} - 137 k\, 
       N^{3} - 68 k^{2} N^{3} 
       \nonu \\ & + & 17 N^{4} - 14 k\, 
       N^{4} + 4 N^{5}), \nonu \\
  c_ {+136} & = & -\frac {8 (3 + 2 k + N) (8 + 3 k + 
        5 N)} {3 (2 + k + N)^{3}}, \qquad
  c_ {+137}  =  -\frac {8 (3 + 2 k + N) (8 + 3 k + 
        5 N)} {3 (2 + k + N)^{4}}, \nonu \\
  c_ {+138} & = & \frac{1} {3 (2 + N) (2 + k + N)^{5}}4 (-96 - 116 k - 29 k^{2} + 2 k^{3} - 124 N - 
        36 k\, N + 84 k^{2} N + 30 k^{3} N 
        \nonu \\ & - & 31 N^{2} + 92 k\, 
       N^{2} + 103 k^{2} N^{2} + 16 k^{3} N^{2} + 14 N^{3} + 54 k\, 
       N^{3} + 26 k^{2} N^{3} + 5 N^{4} + 6 k\, 
       N^{4}), \nonu \\
  c_ {+139} & = & -\frac{1} {3 (2 + N) (2 + k + N)^{5}}2 (-k + N) (-64 - 84 k - 11 k^{2} + 6 k^{3} - 
        76 N - 36 k\, N + 15 k^{2} N 
        \nonu \\ & - & 17 N^{2} + 10 k\, 
       N^{2} + N^{3}), \nonu \\
  c_ {+140} & = & -\frac {8 (24 + 37 k + 25 k^{2} + 6 k^{3} + 35 N + 
        36 k\, N + 13 k^{2} N + 11 N^{2} + 5 k\, 
       N^{2})} {3 (2 + k + N)^{4}}, \nonu \\
  c_ {+141} & = & -\frac {16 (12 + 19 k + 15 k^{2} + 4 k^{3} + 17 N + 
        16 k\, N + 7 k^{2} N + 5 N^{2} + k\, 
       N^{2})} {3 (2 + k + N)^{5}}, \nonu \\
  c_ {+142} & = & -\frac {1}{3 (2 + N) (2 + k + N)^{4}}8 (48 + 90 k + 45 k^{2} + 6 k^{3} + 78 N + 
        127 k\, N + 52 k^{2} N + 6 k^{3} N 
        \nonu \\ & + & 44 N^{2} + 54 k\, 
       N^{2} + 13 k^{2} N^{2} + 8 N^{3} + 5 k\, 
       N^{3}), \nonu \\
  c_ {+143} & = & \frac{1} {3 (2 + N) (2 + k + N)^{4}}4 (96 + 164 k + 103 k^{2} + 22 k^{3} + 172 N + 
        220 k\, N + 103 k^{2} N 
        \nonu \\ & + & 14 k^{3} N + 109 N^{2} + 88 k\, 
       N^{2} + 24 k^{2} N^{2} + 27 N^{3} + 8 k\, 
       N^{3} + 2 N^{4}), \nonu \\
  c_ {+144} & = & -\frac{1}{3 (2 + N) (2 + k + N)^{4}}4 (-k + N) (-64 - 61 k - k^{2} + 6 k^{3} - 
        83 N - 17 k\, N + 32 k^{2} N 
        \nonu \\ & + & 6 k^{3} N - 30 N^{2} + 21 k\, 
       N^{2} + 13 k^{2} N^{2} - 3 N^{3} + 5 k\, 
       N^{3}) , \nonu \\
  c_ {+145} & = & \frac {4 (3 + 2 k + N)} {3 (2 + k + N)},  \nonu \\
  c_ {+146} & = & -\frac {1} {3 (2 + N) (2 + k + N)^{3}}4 (3 + 2 k + N) (60 + 77 k + 22 k^{2} + 121 N + 
        115 k\, N + 20 k^{2} N 
        \nonu \\ & + & 79 N^{2} + 42 k\, 
       N^{2} + 16 N^{3}), \nonu \\
  c_ {+147} & = & -\frac{1} {3 (2 + N) (2 + k + N)^{3}}4 (3 + 2 k + N) (60 + 77 k + 22 k^{2} + 121 N + 
        115 k\, N + 20 k^{2} N 
        \nonu \\ & + & 79 N^{2} + 42 k\, 
       N^{2} + 16 N^{3}), \nonu \\
  c_ {+148} & = & \frac {8 (54 + 103 k + 62 k^{2} + 12 k^{3} + 80 N + 
        106 k\, N + 33 k^{2} N + 36 N^{2} + 25 k\, 
       N^{2} + 5 N^{3})} {3 (2 + N) (2 + k + N)^{4}}, \nonu \\
  c_ {+149} & = & -\frac {8 (30 + 75 k + 54 k^{2} + 12 k^{3} + 48 N + 
        80 k\, N + 29 k^{2} N + 22 N^{2} + 19 k\, 
       N^{2} + 3 N^{3})} {3 (2 + N) (2 + k + N)^{4}}, \nonu \\
  c_ {+150} & = & \frac{1} {3 (2 + N) (2 + k + N)^{4}}8 (-48 - 45 k - 10 k^{2} - 69 N - 17 k\, 
       N + 21 k^{2} N + 6 k^{3} N - 30 N^{2} 
       \nonu \\ & + & 19 k\, 
       N^{2} + 14 k^{2} N^{2} - 4 N^{3} + 7 k\, 
       N^{3}), \nonu \\
  c_ {+151} & = & \frac {8 (3 + 2 k + N) (32 + 55 k + 18 k^{2} + 41 N + 
        35 k\, N + 11 N^{2})} {3 (2 + N) (2 + k + N)^{5}}, \nonu \\
  c_ {+152} & = & \frac {4 (3 + 2 k + N) (32 + 55 k + 18 k^{2} + 41 N + 
        35 k\, N + 11 N^{2})} {3 (2 + N) (2 + k + N)^{5}}, \nonu \\
  c_ {+153} & = & -\frac {1}{3 (2 + N) (2 + k + N)^{5}}8 (168 + 243 k + 73 k^{2} - 28 k^{3} - 
        12 k^{4} + 393 N + 489 k\, 
       N 
       \nonu \\ & + & 156 k^{2} N + 2 k^{3} N + 338 N^{2} + 313 k\, 
       N^{2} + 65 k^{2} N^{2} + 123 N^{3} + 61 k\, 
       N^{3} + 16 N^{4}) , \nonu \\
  c_ {+154} & = & \frac {8 (1 + N) (3 + 2 k + N) (32 + 55 k + 18 k^{2} + 
        41 N + 35 k\, N + 11 N^{2})} {3 (2 + N) (2 + k + N)^{6}},\nonu \\ 
  c_ {+155} & = & -\frac {8 (3 + 2 k + N) (32 + 55 k + 18 k^{2} + 41 N + 
        35 k\, N + 11 N^{2})} {3 (2 + N) (2 + k + N)^{5}}, \nonu \\
  c_ {+156} & = & \frac {1}{3 (2 + N) (2 + k + N)^{5}}4 (108 + 181 k + 104 k^{2} + 20 k^{3} + 245 N + 
        320 k\, N + 134 k^{2} N 
        \nonu \\ & + & 16 k^{3} N + 206 N^{2} + 185 k\, 
       N^{2} + 42 k^{2} N^{2} + 75 N^{3} + 34 k\, 
       N^{3} + 10 N^{4}), \nonu \\
  c_ {+157} & = & \frac {8 (1 + k) (3 + 2 k + N) (32 + 55 k + 18 k^{2} + 
        41 N + 35 k\, N + 11 N^{2})} {3 (2 + N) (2 + k + N)^{6}},\nonu \\ 
  c_ {+158} & = & -\frac {8 (1 + k) (3 + 2 k + N) (32 + 55 k + 18 k^{2} + 
        41 N + 35 k\, N + 11 N^{2})} {3 (2 + N) (2 + k + N)^{6}},\nonu \\ 
  c_ {+159} & = & -\frac {1}{3 (2 + N) (2 + k + N)^{5}}4 (252 + 461 k + 272 k^{2} + 52 k^{3} + 469 N + 
        684 k\, N + 290 k^{2} N 
        \nonu \\ & + & 32 k^{3} N + 322 N^{2} + 337 k\, 
       N^{2} + 78 k^{2} N^{2} + 95 N^{3} + 54 k\, 
       N^{3} + 10 N^{4}), \nonu \\
  c_ {+160} & = & -\frac {1} {3 (2 + N) (2 + k + N)^{5}}4 (84 + 231 k + 223 k^{2} + 88 k^{3} + 
        12 k^{4} + 207 N + 406 k\, 
       N 
       \nonu \\ & + & 260 k^{2} N + 52 k^{3} N + 181 N^{2} + 227 k\, 
       N^{2} + 73 k^{2} N^{2} + 67 N^{3} + 40 k\, 
       N^{3} + 9 N^{4}), \nonu \\
  c_ {+161} & = & -\frac {8 (1 + N) (3 + 2 k + N) (32 + 55 k + 18 k^{2} + 
        41 N + 35 k\, N + 11 N^{2})} {3 (2 + N) (2 + k + N)^{6}},\nonu \\ 
  c_ {+162} & = & \frac {1} {3 (2 + N) (2 + k + N)^{5}}4 (84 + 293 k + 193 k^{2} + 8 k^{3} - 12 k^{4} + 
        241 N + 676 k\, 
       N + 380 k^{2} N 
       \nonu \\ & + & 44 k^{3} N + 229 N^{2} + 461 k\, 
       N^{2} + 151 k^{2} N^{2} + 81 N^{3} + 90 k\, 
       N^{3} + 9 N^{4}), \nonu \\
  c_ {+163} & = & \frac {16 (3 + 2 k + N) (8 - 2 k^{2} + 20 N + 17 k\, 
       N + 5 k^{2} N + 17 N^{2} + 11 k\, 
       N^{2} + 4 N^{3})} {3 (2 + N) (2 + k + N)^{5}}, \nonu \\
  c_ {+164} & = & -\frac {4 (3 + 2 k + N) (32 + 55 k + 18 k^{2} + 41 N + 
        35 k\, N + 11 N^{2})} {3 (2 + N) (2 + k + N)^{5}},  \nonu \\
  c_ {+165} & = & -\frac {1} {3 (2 + N) (2 + k + N)^{5}}32 (-6 - 11 k - 8 k^{2} - 2 k^{3} - 10 N + 
        8 k\, N + 16 k^{2} N + 5 k^{3} N 
        \nonu \\ & - & 3 N^{2} + 19 k\, 
       N^{2} + 11 k^{2} N^{2} + 5 k\, 
       N^{3}), \nonu \\
  c_ {+166} & = & -\frac{1}{3 (2 + N) (2 + k + N)^{4}}8 (-24 + 3 k + 24 k^{2} + 8 k^{3} - 33 N + 
        39 k\, N + 50 k^{2} N + 10 k^{3} N 
        \nonu \\ & - & 12 N^{2} + 39 k\, 
       N^{2} + 20 k^{2} N^{2} - N^{3} + 9 k\, 
       N^{3}) , \nonu \\
  c_ {+167} & = & -\frac {8 (3 + 2 k + N) (32 + 55 k + 18 k^{2} + 41 N + 
        35 k\, N + 11 N^{2})} {3 (2 + N) (2 + k + N)^{5}}, \nonu \\
  c_ {+168} & = & \frac {8 (3 + 2 k + N) (32 + 55 k + 18 k^{2} + 41 N + 
        35 k\, N + 11 N^{2})} {3 (2 + N) (2 + k + N)^{5}}, \nonu \\
  c_ {+169} & = & \frac {1}{3 (2 + N) (2 + k + N)^{4}}8 (264 + 381 k + 141 k^{2} - 16 k^{3} - 
        12 k^{4} + 531 N + 595 k\, 
       N 
       \nonu \\ & + & 162 k^{2} N - 4 k^{3} N + 398 N^{2} + 311 k\, 
       N^{2} + 49 k^{2} N^{2} + 131 N^{3} + 53 k\, 
       N^{3} + 16 N^{4}) , \nonu \\
  c_ {+170} & = & -\frac {4 (48 + 87 k + 56 k^{2} + 12 k^{3} + 81 N + 
        94 k\, N + 30 k^{2} N + 42 N^{2} + 23 k\, 
       N^{2} + 7 N^{3})} {3 (2 + N) (2 + k + N)^{4}}, \nonu \\
  c_ {+171} & = & \frac {8 (3 + 2 k + N) (32 + 55 k + 18 k^{2} + 41 N + 
        35 k\, N + 11 N^{2})} {3 (2 + N) (2 + k + N)^{5}}, \nonu \\
  c_ {+172} & = & -\frac {8 (3 + 2 k + N) (32 + 55 k + 18 k^{2} + 41 N + 
        35 k\, N + 11 N^{2})} {3 (2 + N) (2 + k + N)^{5}}, \nonu \\
  c_ {+173} & = & -\frac {4 (96 + 49 k - 24 k^{2} - 12 k^{3} + 143 N + 
        46 k\, N - 14 k^{2} N + 74 N^{2} + 13 k\, 
       N^{2} + 13 N^{3})} {3 (2 + N) (2 + k + N)^{4}}, \nonu \\
  c_ {+174} & = & -\frac {8 (3 + 2 k + N) (16 + 21 k + 6 k^{2} + 19 N + 
        13 k\, N + 5 N^{2})} {3 (2 + N) (2 + k + N)^{4}}, \nonu \\
  c_ {+175} & = & \frac {1}{3 (2 + N) (2 + k + N)^{5}}8 (3 + 2 k + N) (20 + 2 k - 17 k^{2} - 6 k^{3} + 
        48 N + 21 k\, N - 5 k^{2} N 
        \nonu \\ & + & 36 N^{2} + 13 k\, 
       N^{2} + 8 N^{3}) , \nonu \\
  c_ {+176} & = & -\frac{1}{3 (2 + N) (2 + k + N)^{5}}8 (3 + 2 k + N) (20 + 2 k - 17 k^{2} - 
        6 k^{3} + 48 N + 21 k\, N - 5 k^{2} N 
        \nonu \\ & + & 36 N^{2} + 13 k\, 
       N^{2} + 8 N^{3}) , \nonu \\
  c_ {+177} & = & -\frac {16 (24 + 71 k + 54 k^{2} + 12 k^{3} + 37 N + 
        74 k\, N + 29 k^{2} N + 16 N^{2} + 17 k\, 
       N^{2} + 2 N^{3})} {3 (2 + N) (2 + k + N)^{4}}, \nonu \\
  c_ {+178} & = & -\frac{1}{3 (2 + N) (2 + k + N)^{5}}8 (60 + 62 k - 31 k^{2} - 48 k^{3} - 12 k^{4} + 
        148 N + 169 k\, 
       N + 22 k^{2} N 
       \nonu \\ & - & 14 k^{3} N + 132 N^{2} + 128 k\, 
       N^{2} + 23 k^{2} N^{2} + 48 N^{3} + 27 k\, 
       N^{3} + 6 N^{4}), \nonu \\
  c_ {+179} & = & -\frac {4 (-55 k - 52 k^{2} - 12 k^{3} + 7 N - 54 k\, 
       N - 28 k^{2} N + 10 N^{2} - 11 k\, 
       N^{2} + 3 N^{3})} {3 (2 + N) (2 + k + N)^{4}}, \nonu \\
  c_ {+180} & = & -\frac {16 (3 + 2 k + N) (6 + 17 k + 6 k^{2} + 8 N + 
        11 k\, N + 2 N^{2})} {3 (2 + N) (2 + k + N)^{4}}, \nonu \\
  c_ {+181} & = & \frac{1} {3 (2 + N) (2 + k + N)^{5}}8 (12 + 229 k + 359 k^{2} + 196 k^{3} + 
        36 k^{4} + 29 N + 541 k\, 
       N 
       \nonu \\ & + & 694 k^{2} N + 290 k^{3} N + 36 k^{4} N + 30 N^{2} + 
        459 k\, N^{2} + 417 k^{2} N^{2} + 96 k^{3} N^{2} + 13 N^{3} 
        \nonu \\ & + &  161 k\, N^{3} + 76 k^{2} N^{3} + 2 N^{4} + 20 k\, 
       N^{4}), \nonu \\
  c_ {+182} & = & \frac{1}{3 (2 + N) (2 + k + N)^{6}}8 (1 + k) (3 + 2 k + N) (36 + 39 k + 10 k^{2} + 
        67 N + 53 k\, N + 8 k^{2} N 
        \nonu \\ & + & 41 N^{2} + 18 k\, 
       N^{2} + 8 N^{3}), \nonu \\
  c_ {+183} & = & \frac{1}{3 (2 + N) (2 + k + N)^{6}}16 (3 + 2 k + N) (68 + 129 k + 101 k^{2} + 
        39 k^{3} + 6 k^{4} + 157 N 
        \nonu \\ & + & 212 k\,N + 103 k^{2} N + 19 k^{3} N + 137 N^{2} + 118 k\, 
       N^{2} + 26 k^{2} N^{2} + 54 N^{3} + 23 k\, 
       N^{3} + 8 N^{4}), \nonu \\
  c_ {+184} & = & \frac {16 (78 + 123 k + 66 k^{2} + 12 k^{3} + 120 N + 
        128 k\, N + 35 k^{2} N + 58 N^{2} + 31 k\, 
       N^{2} + 9 N^{3})} {3 (2 + N) (2 + k + N)^{4}}, \nonu \\
  c_ {+185} & = & -\frac{1} {3 (2 + N) (2 + k + N)^{5}}8 (348 + 711 k + 533 k^{2} + 172 k^{3} + 
        20 k^{4} + 687 N + 1049 k\, 
       N 
       \nonu \\ & + & 496 k^{2} N + 54 k^{3} N - 8 k^{4} N + 494 N^{2} + 497 k\, 
       N^{2} + 95 k^{2} N^{2} - 16 k^{3} N^{2} + 155 N^{3} + 77 k\, 
       N^{3} 
       \nonu \\ & - & 6 k^{2} N^{3} + 18 N^{4}),\nonu \\
   c_ {+186} & = & \frac {16 (1 + k) (3 + 2 k + N)} {3 (2 + k + N)^{4}}, \nonu \\
  c_ {+187} & = & -\frac {4 (-12 + 23 k + 40 k^{2} + 12 k^{3} - 5 N + 
        30 k\, N + 22 k^{2} N + 2 N^{2} + 7 k\, 
       N^{2} + N^{3})} {3 (2 + N) (2 + k + N)^{4}}, \nonu \\
  c_ {+188} & = & -\frac {8 (3 + 2 k + N) (26 + 25 k + 6 k^{2} + 30 N + 
        15 k\, N + 8 N^{2})} {3 (2 + N) (2 + k + N)^{3}}, \nonu \\
  c_ {+189} & = & \frac {8 (3 + 2 k + N) (18 + 21 k + 6 k^{2} + 22 N + 
        13 k\, N + 6 N^{2})} {3 (2 + N) (2 + k + N)^{4}}, \nonu \\
  c_ {+190} & = & -\frac {16 (3 + 2 k + N) (26 + 25 k + 6 k^{2} + 30 N + 
        15 k\, N + 8 N^{2})} {3 (2 + N) (2 + k + N)^{4}}, \nonu \\
  c_ {+191} & = & -\frac{1}{3 (2 + N) (2 + k + N)^{5}}8 (12 + 229 k + 359 k^{2} + 196 k^{3} + 
        36 k^{4} + 29 N + 541 k\, 
       N 
       \nonu \\ & + & 694 k^{2} N + 290 k^{3} N + 36 k^{4} N + 30 N^{2} + 
        459 k\, N^{2} + 417 k^{2} N^{2} + 96 k^{3} N^{2} + 13 N^{3} 
        \nonu \\ & + & 161 k\, N^{3} + 76 k^{2} N^{3} + 2 N^{4} + 20 k\, 
       N^{4}), \nonu \\
  c_ {+192} & = & -\frac {8 (3 + 2 k + N) (32 + 55 k + 18 k^{2} + 41 N + 
        35 k\, N + 11 N^{2})} {3 (2 + N) (2 + k + N)^{5}}, \nonu \\
  c_ {+193} & = & \frac{1}{3 (2 + N) (2 + k + N)^{6}}8 (1 + k) (3 + 2 k + N) (36 + 39 k + 10 k^{2} + 
        67 N + 53 k\, N + 8 k^{2} N 
        \nonu \\ & + & 41 N^{2} + 18 k\, 
       N^{2} + 8 N^{3}), \nonu \\
  c_ {+194} & = & \frac {4 (3 + 2 k + N) (4 - 13 k - 6 k^{2} + 3 N - 
        9 k\, N + N^{2})} {3 (2 + N) (2 + k + N)^{3}}, \nonu \\
  c_ {+195} & = & \frac {8 (96 + 135 k + 68 k^{2} + 12 k^{3} + 153 N + 
        142 k\, N + 36 k^{2} N + 78 N^{2} + 35 k\, 
       N^{2} + 13 N^{3})} {3 (2 + N) (2 + k + N)^{4}}, \nonu \\
  c_ {+196} & = & \frac {8 (3 + 2 k + N) (32 + 55 k + 18 k^{2} + 41 N + 
        35 k\, N + 11 N^{2})} {3 (2 + N) (2 + k + N)^{5}}, \nonu \\
  c_ {+197} & = & -\frac {8 (60 + 17 k - 32 k^{2} - 12 k^{3} + 85 N + 
        14 k\, N - 18 k^{2} N + 42 N^{2} + 5 k\, 
       N^{2} + 7 N^{3})} {3 (2 + N) (2 + k + N)^{4}}, \nonu \\
  c_ {+198} & = & \frac{1}{3 (2 + N) (2 + k + N)^{5}}8 (12 - 2 k - 59 k^{2} - 52 k^{3} - 12 k^{4} + 
        44 N + 65 k\, N - 8 k^{2} N 
        \nonu \\ & - & 16 k^{3} N + 48 N^{2} + 72 k\, 
       N^{2} + 15 k^{2} N^{2} + 18 N^{3} + 17 k\, 
       N^{3} + 2 N^{4}), \nonu \\
  c_ {+199} & = & -\frac{1}{3 (2 + N) (2 + k + N)^{5}}8 (-12 + 48 k + 60 k^{2} + 16 k^{3} + 6 N + 
        151 k\, N + 118 k^{2} N 
        \nonu \\ & + & 20 k^{3} N + 23 N^{2} + 114 k\, 
       N^{2} + 46 k^{2} N^{2} + 10 N^{3} + 23 k\, 
       N^{3} + N^{4}), \nonu \\
  c_ {+200} & = & -\frac{1}{3 (2 + N) (2 + k + N)^{4}}4 (-24 - 12 k + 39 k^{2} + 44 k^{3} + 
        12 k^{4} - 48 N - 60 k\, 
       N - 6 k^{2} N 
       \nonu \\ & + & 12 k^{3} N - 27 N^{2} - 48 k\, 
       N^{2} - 17 k^{2} N^{2} - 2 N^{3} - 8 k\, 
       N^{3} + N^{4}), \nonu \\
  c_ {+201} & = & \frac{1}{3 (2 + N) (2 + k + N)^{4}}8 (-240 - 425 k - 218 k^{2} - 16 k^{3} + 
        8 k^{4} - 403 N - 422 k\, 
       N 
       \nonu \\ & + & 8 k^{2} N +94 k^{3} N + 16 k^{4} N - 230 N^{2} - 53 k\, 
       N^{2} + 134 k^{2} N^{2} + 44 k^{3} N^{2} - 53 N^{3} + 38 k\, 
       N^{3} \nonu \\ & + & 34 k^{2} N^{3} 
       -4 N^{4} + 8 k\, 
       N^{4}), \nonu \\
  c_ {+202} & = & -\frac{1}{3 (2 + N) (2 + k + N)^{5}}8 (12 + 7 k + 39 k^{2} + 44 k^{3} + 12 k^{4} + 
        83 N + 141 k\, 
       N + 128 k^{2} N
       \nonu \\ & + & 42 k^{3} N + 114 N^{2} + 141 k\, 
       N^{2} + 55 k^{2} N^{2} + 53 N^{3} + 33 k\, 
       N^{3} + 8 N^{4}), \nonu \\
  c_ {+203} & = & -\frac{1}{3 (2 + N) (2 + k + N)^{5}}8 (3 + 2 k + N) (28 - 3 k - 16 k^{2} - 
        4 k^{3} + 45 N + 2 k^{2} N + 4 k^{3} N 
        \nonu \\ & + & 26 N^{2} + 3 k\, 
       N^{2} + 6 k^{2} N^{2} + 5 N^{3}), \nonu \\
  c_ {+204} & = & \frac {(1} {3 (2 + k + N)^{4}} 16 (27 + 49 k + 
   30 k^{2} + 6 k^{3} + 44 N + 52 k\, 
  N + 16 k^{2} N + 23 N^{2} + 13 k\, N^{2} 
  \nonu \\ & + & 4 N^{3}), \nonu \\
c_ {+205} & = & \frac{1} {3 (2 + N) (2 + k + N)^{5}}4 (-72 - 224 k - 137 k^{2} + 4 k^{3} + 12 k^{4} - 
      28 N - 82 k\, N + 34 k^{2} N 
      \nonu \\ & + & 40 k^{3} N + 87 N^{2} + 104 k\, 
     N^{2} + 55 k^{2} N^{2} + 62 N^{3} + 38 k\, 
     N^{3} + 11 N^{4}),\nonu \\ 
c_ {+206} & = & \frac{1}{3 (2 + N) (2 + k + N)^{5}}4 (-192 - 208 k + 69 k^{2} + 136 k^{3} + 
      36 k^{4} - 272 N - 56 k\, 
     N 
     \nonu \\ & + & 250 k^{2} N + 110 k^{3} N - 109 N^{2} + 104 k\, 
     N^{2} + 107 k^{2} N^{2} - 10 N^{3} + 34 k\, 
     N^{3} + N^{4}), \nonu \\
c_ {+207} & = & \frac {4 (3 + 2 k + N) (32 + 45 k + 14 k^{2} + 67 N + 
      49 k\, N + 4 k^{2} N + 41 N^{2} + 12 k\, 
     N^{2} + 8 N^{3})} {3 (2 + N) (2 + k + N)^{4}}, \nonu \\
c_ {+208} & = & \frac{1}{3 (2 + N) (2 + k + N)^{5}}8 (-36 - 59 k - 28 k^{2} - 4 k^{3} - 19 N - 16 k\, 
     N + 8 k^{2} N + 4 k^{3} N 
     \nonu \\ & + & 26 N^{2} + 29 k\, 
     N^{2} + 12 k^{2} N^{2} + 21 N^{3} + 10 k\, 
     N^{3} + 4 N^{4}), \nonu \\
c_ {+209} & = & -\frac {8 (1 + k) (3 + 2 k + N) (32 + 55 k + 18 k^{2} + 
      41 N + 35 k\, N + 11 N^{2})} {3 (2 + N) (2 + k + N)^{6}},  \nonu \\
c_ {+210} & = & -\frac{1}{3 (2 + N) (2 + k + N)^{5}}8 (324 + 739 k + 602 k^{2} + 204 k^{3} + 
      24 k^{4} + 731 N + 1190 k\, 
     N 
     \nonu \\ & + & 620 k^{2} N + 100 k^{3} N + 578 N^{2} + 607 k\, 
     N^{2} + 154 k^{2} N^{2} + 195 N^{3} + 100 k\, 
     N^{3} + 24 N^{4}),\nonu \\ 
c_ {+211} & = & -\frac{1} {3 (2 + N) (2 + k + N)^{6}}16 (1 + k) (60 + 125 k + 80 k^{2} + 16 k^{3} + 
      133 N + 228 k\, 
     N 
     \nonu \\ & + & 110 k^{2} N + 14 k^{3} N + 106 N^{2} + 133 k\, 
     N^{2} + 36 k^{2} N^{2} + 35 N^{3} + 24 k\, 
     N^{3} + 4 N^{4}), \nonu \\
c_ {+212} & = & \frac{1} {3 (2 + N) (2 + k + N)^{6}}8 (528 + 1364 k + 1309 k^{2} + 548 k^{3} + 
      84 k^{4} + 1396 N + 2844 k\, 
     N 
     \nonu \\ & + & 2056 k^{2} N + 588 k^{3} N + 48 k^{4} N + 1415 N^{2} + 
      2126 k\, 
     N^{2} + 1029 k^{2} N^{2} + 152 k^{3} N^{2} + 694 N^{3} 
     \nonu \\ & + & 676 k\, 
     N^{3} + 162 k^{2} N^{3} + 167 N^{4} + 78 k\, 
     N^{4} + 16 N^{5}),\nonu \\ 
c_ {+213} & = & -\frac{1}{3 (2 + N) (2 + k + N)^{5}}4 (-168 + 24 k + 349 k^{2} + 252 k^{3} + 
      52 k^{4} - 180 N + 550 k\, 
     N 
     \nonu \\ & + & 966 k^{2} N + 440 k^{3} N + 56 k^{4} N + 49 N^{2} + 752 k\, 
     N^{2} + 681 k^{2} N^{2} + 156 k^{3} N^{2} + 130 N^{3} 
     \nonu \\ & + & 342 k\, 
     N^{3} + 136 k^{2} N^{3} + 57 N^{4} + 52 k\, 
     N^{4} + 8 N^{5}), \nonu \\
c_ {+214} & = & \frac{1}{3 (2 + N) (2 + k + N)^{4}}8 (3 + 2 k + N) (40 + 75 k + 39 k^{2} + 6 k^{3} + 
      73 N + 94 k\, N + 25 k^{2} N 
      \nonu \\ & + & 43 N^{2} + 29 k\, 
     N^{2} + 8 N^{3}), \nonu \\
c_ {+215} & = & -\frac {8 (3 + 2 k + N) (10 + 17 k + 6 k^{2} + 14 N + 
      11 k\, N + 4 N^{2})} {3 (2 + N) (2 + k + N)^{4}},\nonu \\ 
c_ {+216} & = & -\frac{1}{3 (2 + N) (2 + k + N)^{5}}8 (-132 - 239 k - 159 k^{2} - 44 k^{3} - 
      4 k^{4} - 259 N - 265 k\, 
     N 
     \nonu\\ & - & 11 k^{2} N + 62 k^{3} N + 16 k^{4} N - 176 N^{2} - 27 k\, 
     N^{2} + 110 k^{2} N^{2} + 42 k^{3} N^{2} - 50 N^{3} + 41 k\, 
     N^{3} 
     \nonu\\ & + & 34 k^{2} N^{3} - 5 N^{4} + 10 k\, 
     N^{4}), \nonu \\
c_ {+217} & = & -\frac {4 (3 + k + 2 N)} {3 (2 + k + N)},  \nonu \\
c_ {+218} & = & \frac {4 (3 + k + 2 N) (60 + 77 k + 22 k^{2} + 121 N + 
      115 k\, N + 20 k^{2} N + 79 N^{2} + 42 k\, 
     N^{2} + 16 N^{3})} {3 (2 + N) (2 + k + N)^{3}}, \nonu \\
c_ {+219} & = & \frac {4 (3 + k + 2 N) (60 + 77 k + 22 k^{2} + 121 N + 
      115 k\, N + 20 k^{2} N + 79 N^{2} + 42 k\, 
     N^{2} + 16 N^{3})} {3 (2 + N) (2 + k + N)^{3}}, \nonu \\
c_ {+220} & = & -\frac {8 (15 + 13 k + 3 k^{2} + 17 N + 7 k\, 
     N + 5 N^{2})} {3 (2 + k + N)^{4}}, \nonu \\
c_ {+221} & = & \frac {4 (23 k + 23 k^{2} + 6 k^{3} + 25 N + 48 k\, 
     N + 17 k^{2} N + 25 N^{2} + 19 k\, 
     N^{2} + 6 N^{3})} {3 (2 + N) (2 + k + N)^{4}}, \nonu \\
c_ {+222} & = & -\frac {8 (3 + k + 2 N) (32 + 55 k + 18 k^{2} + 41 N + 
      35 k\, N + 11 N^{2})} {3 (2 + N) (2 + k + N)^{5}},\nonu \\ 
c_ {+223} & = & \frac{1} {3 (2 + N) (2 + k + N)^{5}}8 (72 + 137 k + 69 k^{2} + 10 k^{3} + 259 N + 
      383 k\, N + 144 k^{2} N 
      \nonu\\ & + & 14 k^{3} N + 304 N^{2} + 313 k\, 
     N^{2} + 63 k^{2} N^{2} + 145 N^{3} + 79 k\, 
     N^{3} + 24 N^{4}), \nonu \\
c_ {+224} & = & -\frac {8 (1 + N) (3 + k + 2 N) (32 + 55 k + 18 k^{2} + 
      41 N + 35 k\, N + 11 N^{2})} {3 (2 + N) (2 + k + N)^{6}},\nonu \\ 
c_ {+225} & = & \frac {8 (3 + k + 2 N) (32 + 55 k + 18 k^{2} + 41 N + 
      35 k\, N + 11 N^{2})} {3 (2 + N) (2 + k + N)^{5}}, \nonu \\
c_ {+226} & = & -\frac {8 (1 + k) (3 + k + 2 N) (32 + 55 k + 18 k^{2} + 
      41 N + 35 k\, N + 11 N^{2})} {3 (2 + N) (2 + k + N)^{6}},\nonu \\ 
c_ {+227} & = & \frac {1}{3 (2 + N) (2 + k + N)^{5}}8 (36 - 65 k - 77 k^{2} - 18 k^{3} + 143 N + 
      10 k\, N - 23 k^{2} N + 181 N^{2} 
      \nonu\\ & + & 87 k\, 
     N^{2} + 16 k^{2} N^{2} + 92 N^{3} + 34 k\, 
     N^{3} + 16 N^{4}), \nonu \\
c_ {+228} & = & -\frac {8 (1 + N) (3 + k + 2 N) (32 + 55 k + 18 k^{2} + 
      41 N + 35 k\, N + 11 N^{2})} {3 (2 + N) (2 + k + N)^{6}},\nonu \\ 
c_ {+229} & = & \frac {1}{3 (2 + N) (2 + k + N)^{4}}8 (48 + 75 k + 37 k^{2} + 6 k^{3} + 99 N + 131 k\, 
     N + 51 k^{2} N + 6 k^{3} N 
     \nonu\\ & + & 87 N^{2} + 86 k\, 
     N^{2} + 19 k^{2} N^{2} + 37 N^{3} + 20 k\, 
     N^{3} + 6 N^{4}), \nonu \\
c_ {+230} & = & -\frac {4 (3 + k + 2 N) (32 + 55 k + 18 k^{2} + 41 N + 
      35 k\, N + 11 N^{2})} {3 (2 + N) (2 + k + N)^{5}}, \nonu \\
c_ {+231} & = & \frac {8 (3 + k + 2 N) (32 + 55 k + 18 k^{2} + 41 N + 
      35 k\, N + 11 N^{2})} {3 (2 + N) (2 + k + N)^{5}}, \nonu \\
c_ {+232} & = & -\frac {8 (3 + k + 2 N) (32 + 55 k + 18 k^{2} + 41 N + 
      35 k\, N + 11 N^{2})} {3 (2 + N) (2 + k + N)^{5}}, \nonu \\
c_ {+233} & = & -\frac {1}{3 (2 + N) (2 + k + N)^{5}}4 (12 + 17 k + k^{2} - 2 k^{3} + 73 N + 100 k\, 
     N + 31 k^{2} N + 2 k^{3} N 
     \nonu\\ & + & 97 N^{2} + 97 k\, 
     N^{2} + 18 k^{2} N^{2} + 48 N^{3} + 26 k\, 
     N^{3} + 8 N^{4}), \nonu \\
c_ {+234} & = & \frac {1}{3 (2 + N) (2 + k + N)^{5}}4 (132 + 135 k + 39 k^{2} + 2 k^{3} + 279 N + 
      200 k\, N + 29 k^{2} N - 2 k^{3} N 
      \nonu\\ & + & 211 N^{2} + 87 k\, 
     N^{2} + 2 k^{2} N^{2} + 68 N^{3} + 10 k\, 
     N^{3} + 8 N^{4}), \nonu \\
c_ {+235} & = & -\frac {4 (3 + k + 2 N) (32 + 55 k + 18 k^{2} + 41 N + 
      35 k\, N + 11 N^{2})} {3 (2 + N) (2 + k + N)^{5}}, \nonu \\
c_ {+236} & = & \frac{1}{3 (2 + N) (2 + k + N)^{4}}8 (24 + 51 k + 31 k^{2} + 6 k^{3} + 39 N + 87 k\, 
     N + 44 k^{2} N + 6 k^{3} N 
     \nonu\\ & + &29 N^{2} + 58 k\, 
     N^{2} + 17 k^{2} N^{2} + 12 N^{3} + 14 k\, 
     N^{3} + 2 N^{4}), \nonu \\
c_ {+237} & = & -\frac{1}{3 (2 + N) (2 + k + N)^{4}}8 (-24 - 13 k - 5 k^{2} - 2 k^{3} + 13 N + 73 k\, 
     N + 30 k^{2} N + 2 k^{3} N 
     \nonu\\ & + & 58 N^{2} + 93 k\, 
     N^{2} + 19 k^{2} N^{2} + 35 N^{3} + 27 k\, 
     N^{3} + 6 N^{4}), \nonu \\
c_ {+238} & = & -\frac {8 (3 + k + 2 N) (32 + 55 k + 18 k^{2} + 41 N + 
      35 k\, N + 11 N^{2})} {3 (2 + N) (2 + k + N)^{5}}, \nonu \\
c_ {+239} & = & -\frac {4 (48 + 87 k + 43 k^{2} + 6 k^{3} + 81 N + 
      104 k\, N + 27 k^{2} N + 45 N^{2} + 31 k\, 
     N^{2} + 8 N^{3})} {3 (2 + N) (2 + k + N)^{4}}, \nonu \\
c_ {+240} & = & \frac {4 (96 + 65 k - 3 k^{2} - 6 k^{3} + 127 N + 44 k\, 
     N - 7 k^{2} N + 55 N^{2} + 5 k\, 
     N^{2} + 8 N^{3})} {3 (2 + N) (2 + k + N)^{4}}, \nonu \\
c_ {+241} & = & \frac {8 (3 + k + 2 N) (16 + 21 k + 6 k^{2} + 19 N + 
      13 k\, N + 5 N^{2})} {3 (2 + N) (2 + k + N)^{4}}, \nonu \\
c_ {+242} & = & -\frac {8 (3 + k + 2 N) (20 + 28 k + 8 k^{2} + 54 N + 
      55 k\, N + 10 k^{2} N + 41 N^{2} + 23 k\, 
     N^{2} + 9 N^{3})} {3 (2 + N) (2 + k + N)^{5}}, \nonu \\
c_ {+243} & = & -\frac {8 (3 + k + 2 N) (32 + 55 k + 18 k^{2} + 41 N + 
      35 k\, N + 11 N^{2})} {3 (2 + N) (2 + k + N)^{5}}, \nonu \\
c_ {+244} & = & \frac {16 (24 + 55 k + 33 k^{2} + 6 k^{3} + 53 N + 76 k\, 
     N + 22 k^{2} N + 35 N^{2} + 25 k\, 
     N^{2} + 7 N^{3})} {3 (2 + N) (2 + k + N)^{4}}, \nonu \\
c_ {+245} & = & \frac {8 (3 + k + 2 N) (20 + 28 k + 8 k^{2} + 54 N + 
      55 k\, N + 10 k^{2} N + 41 N^{2} + 23 k\, 
     N^{2} + 9 N^{3})} {3 (2 + N) (2 + k + N)^{5}}, \nonu \\
c_ {+246} & = & -\frac{1}{3 (2 + N) (2 + k + N)^{5}}8 (1 + N) (60 + 120 k + 68 k^{2} + 12 k^{3} + 
      126 N + 163 k\, N 
      \nonu\\ & + & 45 k^{2} N + 81 N^{2} + 53 k\, 
     N^{2} + 16 N^{3}), \nonu \\
c_ {+247} & = & -\frac {8 (3 + 3 k + k^{2} + 3 N + k\, 
     N + N^{2})} {3 (2 + k + N)^{4}}, \qquad
c_ {+248}  =  \frac {16 (3 + 2 k + N) (3 + k + 
      2 N)} {3 (2 + k + N)^{4}}, \nonu \\
c_ {+249} & = & -\frac{1}{3 (2 + N) (2 + k + N)^{4}}4 (108 + 209 k + 113 k^{2} + 18 k^{3} + 205 N + 
      266 k\, N + 73 k^{2} N 
      \nonu\\ & + & 125 N^{2} + 83 k\, 
     N^{2} + 24 N^{3}), \nonu \\
c_ {+250} & = & \frac {16 (3 + k + 2 N) (26 + 38 k + 12 k^{2} + 33 N + 
      24 k\, N + 9 N^{2})} {3 (2 + N) (2 + k + N)^{4}}, \nonu \\
c_ {+251} & = & \frac{1}{3 (2 + N) (2 + k + N)^{5}}8 (108 + 185 k + 93 k^{2} + 14 k^{3} + 313 N + 
      433 k\, N + 154 k^{2} N 
      \nonu\\ & + & 12 k^{3} N + 404 N^{2} + 441 k\, 
     N^{2} + 107 k^{2} N^{2} + 4 k^{3} N^{2} + 273 N^{3} + 215 k\, 
     N^{3} + 28 k^{2} N^{3} 
     \nonu\\ & + &92 N^{4} + 40 k\, 
     N^{4} + 12 N^{5}), \nonu \\
c_ {+252} & = & -\frac {8 (3 + k + 2 N) (32 + 55 k + 18 k^{2} + 41 N + 
      35 k\, N + 11 N^{2})} {3 (2 + N) (2 + k + N)^{5}}, \nonu \\
c_ {+253} & = & -\frac{1}{3 (2 + N) (2 + k + N)^{5}}4 (204 + 493 k + 418 k^{2} + 147 k^{3} + 
      18 k^{4} + 461 N + 838 k\, 
     N 
     \nonu\\ & + & 482 k^{2} N + 87 k^{3} N + 382 N^{2} + 464 k\, 
     N^{2} + 136 k^{2} N^{2} + 137 N^{3} + 83 k\, 
     N^{3} + 18 N^{4}), \nonu \\
c_ {+254} & = & -\frac {8 (1 + N) (3 + k + 2 N) (32 + 55 k + 18 k^{2} + 
      41 N + 35 k\, N + 11 N^{2})} {3 (2 + N) (2 + k + N)^{6}},\nonu \\ 
c_ {+255} & = & \frac{1}{3 (2 + N) (2 + k + N)^{5}}4 (276 + 365 k + 56 k^{2} - 63 k^{3} - 18 k^{4} + 
      649 N + 716 k\, 
     N + 118 k^{2} N 
     \nonu\\ & - & 27 k^{3} N + 614 N^{2} + 514 k\, 
     N^{2} + 62 k^{2} N^{2} + 265 N^{3} + 127 k\, 
     N^{3} + 42 N^{4}), \nonu \\
c_ {+256} & = & \frac {8 (1 + N) (3 + k + 2 N) (4 + 7 k + 2 k^{2} + 
      19 N + 21 k\, N + 4 k^{2} N + 17 N^{2} + 10 k\, 
     N^{2} + 4 N^{3})} {3 (2 + N) (2 + k + N)^{6}}, \nonu \\
c_ {+257} & = & -\frac {8 (1 + k) (3 + k + 2 N) (32 + 55 k + 18 k^{2} + 
      41 N + 35 k\, N + 11 N^{2})} {3 (2 + N) (2 + k + N)^{6}},\nonu \\ 
c_ {+258} & = & -\frac{1}{3 (2 + N) (2 + k + N)^{6}}16 (3 + k + 2 N) (28 + 47 k + 24 k^{2} + 
      4 k^{3} + 67 N + 84 k\, 
     N 
     \nonu\\ & + &27 k^{2} N + 2 k^{3} N + 66 N^{2} + 57 k\, 
     N^{2} + 9 k^{2} N^{2} + 30 N^{3} + 14 k\, 
     N^{3} + 5 N^{4}), \nonu \\
c_ {+259} & = & \frac{1} {3 (2 + N) (2 + k + N)^{6}}8 (-48 - 300 k - 373 k^{2} - 169 k^{3} - 
      26 k^{4} + 84 N - 196 k\, 
     N 
     \nonu\\ & - & 283 k^{2} N - 87 k^{3} N - 4 k^{4} N + 377 N^{2} + 303 k\, 
     N^{2} + 68 k^{2} N^{2} + 10 k^{3} N^{2} + 389 N^{3} 
     \nonu\\ & + & 315 k\, 
     N^{3} + 62 k^{2} N^{3} + 162 N^{4} + 76 k\, 
     N^{4} + 24 N^{5}), \nonu \\
c_ {+260} & = & \frac {4 (3 + k + 2 N) (36 + 55 k + 18 k^{2} + 47 N + 
      35 k\, N + 13 N^{2})} {3 (2 + N) (2 + k + N)^{3}}, \nonu \\
c_ {+261} & = & -\frac {8 (60 + 153 k + 97 k^{2} + 18 k^{3} + 141 N + 
      214 k\, N + 65 k^{2} N + 97 N^{2} + 71 k\, 
     N^{2} + 20 N^{3})} {3 (2 + N) (2 + k + N)^{4}}, \nonu \\
c_ {+262} & = & -\frac {8 (3 + k + 2 N) (32 + 55 k + 18 k^{2} + 41 N + 
      35 k\, N + 11 N^{2})} {3 (2 + N) (2 + k + N)^{5}}, \nonu \\
c_ {+263} & = & -\frac {8 (96 + 135 k + 55 k^{2} + 6 k^{3} + 153 N + 
      152 k\, N + 33 k^{2} N + 81 N^{2} + 43 k\, 
     N^{2} + 14 N^{3})} {3 (2 + N) (2 + k + N)^{4}}, \nonu \\
c_ {+264} & = & \frac{1}{3 (2 + N) (2 + k + N)^{5}}8 (108 + 310 k + 307 k^{2} + 125 k^{3} + 
      18 k^{4} + 260 N + 543 k\, 
     N 
     \nonu\\ & + &358 k^{2} N + 73 k^{3} N + 212 N^{2} + 293 k\, 
     N^{2} + 99 k^{2} N^{2} + 70 N^{3} + 48 k\, 
     N^{3} + 8 N^{4}), \nonu \\
c_ {+265} & = & \frac{1} {3 (2 + N) (2 + k + N)^{5}}8 (108 + 200 k + 112 k^{2} + 20 k^{3} + 274 N + 
      395 k\, N + 155 k^{2} N 
      \nonu\\ & + & 16 k^{3} N + 267 N^{2} + 268 k\, 
     N^{2} + 55 k^{2} N^{2} + 115 N^{3} + 61 k\, 
     N^{3} + 18 N^{4}), \nonu \\
c_ {+266} & = & -\frac{1}{3 (2 + N) (2 + k + N)^{4}}4 (216 + 200 k - 53 k^{2} - 85 k^{3} - 18 k^{4} + 
      532 N + 392 k\, 
     N 
     \nonu\\ & - & 57 k^{2} N - 53 k^{3} N + 477 N^{2} + 257 k\, 
     N^{2} - 12 k^{2} N^{2} + 185 N^{3} + 57 k\, 
     N^{3} + 26 N^{4}), \nonu \\
c_ {+267} & = & \frac{1}{3 (2 + N) (2 + k + N)^{4}}8 (240 + 373 k + 177 k^{2} + 26 k^{3} + 575 N + 
      664 k\, N + 201 k^{2} N 
      \nonu\\ & + & 12 k^{3} N + 545 N^{2} + 449 k\, 
     N^{2} + 84 k^{2} N^{2} + 4 k^{3} N^{2} + 266 N^{3} + 152 k\, 
     N^{3} + 18 k^{2} N^{3} 
     \nonu\\ & + & 70 N^{4} + 24 k\, 
     N^{4} + 8 N^{5}), \nonu \\
c_ {+268} & = & -\frac {1}{3 (2 + N) (2 + k + N)^{5}}8 (108 + 191 k + 95 k^{2} + 14 k^{3} + 235 N + 
      269 k\, N + 62 k^{2} N 
      \nonu\\ & - &2 k^{3} N + 170 N^{2} + 99 k\, 
     N^{2} - k^{2} N^{2} + 47 N^{3} + 5 k\, 
     N^{3} + 4 N^{4}), \nonu \\
c_ {+269} & = & \frac{1} {3 (2 + N) (2 + k + N)^{5}}8 (3 + k + 2 N) (92 + 177 k + 101 k^{2} + 
      18 k^{3} + 217 N + 276 k\, N 
     \nonu \\ & + & 79 k^{2} N + 177 N^{2} + 135 k\, 
     N^{2} + 12 k^{2} N^{2} + 62 N^{3} + 22 k\, 
     N^{3} + 8 N^{4}), \nonu \\
c_ {+270} & = & \frac {16 (3 + k + 2 N) (16 + 25 k + 8 k^{2} + 23 N + 
      17 k\, N + 7 N^{2})} {3 (2 + k + N)^{5}}, \nonu \\
c_ {+271} & = & -\frac{1}{3 (2 + N) (2 + k + N)^{4}}16 (6 + k - 5 k^{2} - 2 k^{3} + 44 N + 52 k\, 
     N + 18 k^{2} N + 2 k^{3} N
     \nonu \\ & + &  64 N^{2} + 61 k\, 
     N^{2} + 13 k^{2} N^{2} + 34 N^{3} + 18 k\, 
     N^{3} + 6 N^{4}), \nonu \\
c_ {+272} & = & \frac {8 (3 + 2 k + N) (3 + k + 2 N)} {3 (2 + k + N)^{3}},
\qquad
 c_ {+273}  =  -\frac {8 (1 + N) (3 + k + 2 N)} {3 (2 + k + N)^{4}}, \nonu \\
c_ {+274} & = & -\frac{1} {3 (2 + N) (2 + k + N)^{5}}8 (108 + 185 k + 93 k^{2} + 14 k^{3} + 313 N + 
      433 k\, N + 154 k^{2} N 
      \nonu \\ & + &  12 k^{3} N + 404 N^{2} + 441 k\, 
     N^{2} + 107 k^{2} N^{2} + 4 k^{3} N^{2} + 273 N^{3} + 215 k\, 
     N^{3} + 28 k^{2} N^{3} 
     \nonu \\ & + &  92 N^{4} + 40 k\, 
     N^{4} + 12 N^{5}), \nonu \\
c_ {+275} & = & \frac{1}{3 (2 + N) (2 + k + N)^{6}}8 (1 + N) (3 + k + 2 N) (4 + 7 k + 2 k^{2} + 
      19 N + 21 k\, N + 4 k^{2} N 
      \nonu \\ & + &  17 N^{2} + 10 k\, 
     N^{2} + 4 N^{3}), \nonu \\
c_ {+276} & = & -\frac {16 (9 + 11 k + 3 k^{2} + 7 N + 5 k\, 
     N + N^{2})} {3 (2 + k + N)^{4}}, \nonu \\
c_ {+277} & = & -\frac{1}{3 (2 + N) (2 + k + N)^{5}}8 (228 + 365 k + 185 k^{2} + 30 k^{3} + 637 N + 
      843 k\, N + 328 k^{2} N 
      \nonu \\ & + &  36 k^{3} N + 736 N^{2} + 765 k\, 
     N^{2} + 207 k^{2} N^{2} + 12 k^{3} N^{2} + 437 N^{3} + 321 k\, 
     N^{3} + 46 k^{2} N^{3} 
     \nonu \\ & + &  132 N^{4} + 52 k\, 
     N^{4} + 16 N^{5}), \nonu \\
c_ {+278} & = & -\frac {16 (1 + N) (3 + k + 2 N)} {3 (2 + k + N)^{4}}, \nonu \\
c_ {+279} & = & -\frac{1}{3 (2 + N) (2 + k + N)^{5}}4 (312 + 588 k + 347 k^{2} + 79 k^{3} + 6 k^{4} + 
      696 N + 906 k\, 
     N 
     \nonu \\ & + &  313 k^{2} N + 27 k^{3} N + 535 N^{2} + 413 k\, 
     N^{2} + 54 k^{2} N^{2} + 167 N^{3} + 51 k\, 
     N^{3} + 18 N^{4}), \nonu \\
c_ {+280} & = & -\frac {1} {3 (2 + N) (2 + k + N)^{5}}4 (-192 - 176 k + 69 k^{2} + 89 k^{3} + 
      18 k^{4} - 304 N - 40 k\, 
     N 
     \nonu \\ & + &  233 k^{2} N + 79 k^{3} N - 125 N^{2} + 151 k\, 
     N^{2} + 118 k^{2} N^{2} + 7 N^{3} + 65 k\, 
     N^{3} + 8 N^{4}),  \nonu \\
c_ {+281} & = & \frac  {32 (33 + 63 k + 35 k^{2} + 
    6 k^{3} + 60 N + 76 k\, N + 21 k^{2} N + 36 N^{2} + 23 k\, 
   N^{2} + 7 N^{3})}{3 (2 + k + N)^{5}} , \nonu \\
c_ {+282} & = & -\frac {4 (3 + k + 2 N) (32 + 61 k + 22 k^{2} + 51 N + 
      57 k\, N + 8 k^{2} N + 25 N^{2} + 12 k\, 
     N^{2} + 4 N^{3})} {3 (2 + N) (2 + k + N)^{4}}, \nonu \\
c_ {+283} & = &\frac {1}{3 (2 + N) (2 + k + N)^{5}}8 (156 + 209 k + 85 k^{2} + 10 k^{3} + 385 N + 
      412 k\, N + 121 k^{2} N 
      \nonu \\ & + & 8 k^{3} N + 349 N^{2} + 265 k\, 
     N^{2} + 42 k^{2} N^{2} + 138 N^{3} + 56 k\, 
     N^{3} + 20 N^{4}), \nonu \\
c_ {+284} & = & -\frac {1}{3 (2 + N) (2 + k + N)^{6}}16 (1 + N) (60 + 105 k + 57 k^{2} + 10 k^{3} + 
      153 N + 208 k\, 
     N 
     \nonu \\ & + & 79 k^{2} N + 8 k^{3} N + 149 N^{2} + 141 k\, 
     N^{2} + 28 k^{2} N^{2} + 64 N^{3} + 32 k\, 
     N^{3} + 10 N^{4}) , \nonu \\
c_ {+285} & = & \frac {1} {3 (2 + N) (2 + k + N)^{5}}4 (408 + 452 k + 37 k^{2} - 87 k^{3} - 22 k^{4} + 
      1216 N + 1202 k\, 
     N 
     \nonu \\ & + &  187 k^{2} N - 63 k^{3} N - 8 k^{4} N + 1533 N^{2} + 1299 k\, 
     N^{2} + 222 k^{2} N^{2} - 4 k^{3} N^{2} + 989 N^{3} 
     \nonu \\ & + & 637 k\, 
     N^{3} + 72 k^{2} N^{3} + 318 N^{4} + 116 k\, 
     N^{4} + 40 N^{5}), \nonu \\
c_ {+286} & = & -\frac {8 (3 + k + 2 N) (40 + 49 k + 14 k^{2} + 67 N + 
      60 k\, N + 10 k^{2} N + 38 N^{2} + 19 k\, 
     N^{2} + 7 N^{3})} {3 (2 + N) (2 + k + N)^{4}}, \nonu \\
c_ {+287} & = & -\frac {8 (3 + k + 2 N) (5 + 2 k + 
      3 N)} {3 (2 + k + N)^{4}}, \nonu \\
c_ {+288} & = & \frac {1}{3 (2 + N) (2 + k + N)^{5}}8 (12 + 45 k + 27 k^{2} + 4 k^{3} + 57 N + 135 k\, 
     N + 53 k^{2} N + 3 k^{3} N 
     \nonu \\ & + &  126 N^{2} + 193 k\, 
     N^{2} + 53 k^{2} N^{2} + 2 k^{3} N^{2} + 122 N^{3} + 120 k\, 
     N^{3} + 18 k^{2} N^{3} + 52 N^{4}
     \nonu \\ & + & 26 k\, 
     N^{4} + 8 N^{5}) , \nonu \\
c_ {+289} & = & -\frac {(k - N)} {3 (2 + k + N)},\qquad
c_ {+290}  =  -3, \nonu \\
c_ {+291} & = &\frac {3 (60 + 77 k + 22 k^{2} + 121 N + 
               115 k\, N + 20 k^{2} N + 79 N^{2} + 42 k\, 
              N^{2} + 16 N^{3})} {(2 + N) (2 + k + N)^{2}}, \nonu \\
c_ {+292} & = &\frac {3 (60 + 77 k + 22 k^{2} + 121 N + 
               115 k\, N + 20 k^{2} N + 79 N^{2} + 42 k\, 
              N^{2} + 16 N^{3})} {(2 + N) (2 + k + N)^{2}}, \nonu \\
c_ {+293} & = & - \frac {(-k + N) (60 + 77 k + 22 k^{2} + 
                121 N + 115 k\, N + 20 k^{2} N + 79 N^{2} + 42 k\, 
               N^{2} + 16 N^{3})} {3 (2 + N) (2 + k + N)^{3}}, \nonu \\
c_ {+294} & = & - \frac {(-k + N) (60 + 77 k + 22 k^{2} + 
                121 N + 115 k\, N + 20 k^{2} N + 79 N^{2} + 42 k\, 
               N^{2} + 16 N^{3})} {3 (2 + N) (2 + k + N)^{3}}, \nonu \\
c_ {+295} & = & - \frac {2 (-k + N) (60 + 77 k + 22 k^{2} + 
                121 N + 115 k\, N + 20 k^{2} N + 79 N^{2} + 42 k\, 
               N^{2} + 16 N^{3})} {3 (2 + N) (2 + k + N)^{3}}, \nonu \\
c_ {+296} & = &\frac {2 (-k + N) (60 + 77 k + 22 k^{2} + 
               121 N + 115 k\, N + 20 k^{2} N + 79 N^{2} + 42 k\, 
              N^{2} + 16 N^{3})} {3 (2 + N) (2 + k + N)^{3}}, \nonu \\
c_ {+297} & = & - \frac {6 (32 + 55 k + 18 k^{2} + 41 N + 
                35 k\, N + 11 N^{2})} {(2 + N) (2 + k + N)^{4}}, \nonu \\
c_ {+298} & = &\frac {2 (-k + N) (32 + 55 k + 18 k^{2} + 
               41 N + 35 k\, 
              N + 11 N^{2})} {3 (2 + N) (2 + k + N)^{5}},\nonu \\ 
 c_ {+299} & = & - \frac{1}{3 (2 + N) (2 + k + N)^{5}}2 (194 k + 296 k^{2} + 161 k^{3} + 
                30 k^{4} + 94 N + 410 k\, 
               N + 327 k^{2} N \nonu \\& + & 81 k^{3} N + 158 N^{2} + 285 k\, 
               N^{2} + 91 k^{2} N^{2} + 91 N^{3} + 69 k\, 
               N^{3} + 17 N^{4}), \nonu \\
c_ {+300} & = &\frac {2 (1 + N) (-k + N) (32 + 55 k + 
               18 k^{2} + 41 N + 35 k\, 
              N + 11 N^{2})} {3 (2 + N) (2 + k + N)^{6}}, \nonu \\
c_ {+301} & = & - \frac {6 (1 + N) (32 + 55 k + 18 k^{2} + 
                41 N + 35 k\, 
               N + 11 N^{2})} {(2 + N) (2 + k + N)^{5}}, \nonu \\
c_ {+302} & = &\frac {2 (360 + 414 k + 43 k^{2} - 30 k^{3} + 
               774 N + 716 k\, N + 93 k^{2} N + 537 N^{2} + 292 k\, 
              N^{2} + 113 N^{3})} {3 (2 + N) (2 + k + N)^{4}}, \nonu \\
c_ {+303} & = &\frac {4 (-k + N) (32 + 55 k + 18 k^{2} + 
               41 N + 35 k\, 
              N + 11 N^{2})} {3 (2 + N) (2 + k + N)^{5}}, \nonu \\
c_ {+304} & = &\frac {4 (-k + N) (32 + 55 k + 18 k^{2} + 
               41 N + 35 k\, 
              N + 11 N^{2})} {3 (2 + N) (2 + k + N)^{5}}, \nonu \\
c_ {+305} & = & - \frac {4 (-k + N) (32 + 55 k + 18 k^{2} + 
                41 N + 35 k\, 
               N + 11 N^{2})} {3 (2 + N) (2 + k + N)^{5}}, \nonu \\
c_ {+306} & = &\frac{1} {3 (2 + N) (2 + k + N)^{5}}4 (324 + 563 k + 317 k^{2} + 
               58 k^{3} + 715 N + 984 k\, 
              N + 401 k^{2} N 
              \nonu \\& + & 44 k^{3} N + 589 N^{2} + 571 k\, 
              N^{2} + 126 k^{2} N^{2} + 212 N^{3} + 108 k\, 
              N^{3} + 28 N^{4}),\nonu \\ 
c_ {+307} & = & - \frac {2 (1 + k) (k - N) (32 + 55 k + 
                18 k^{2} + 41 N + 35 k\, 
               N + 11 N^{2})} {3 (2 + N) (2 + k + N)^{6}}, \nonu \\
c_ {+308} & = & - \frac {6 (1 + k) (32 + 55 k + 18 k^{2} + 
                41 N + 35 k\, 
               N + 11 N^{2})} {(2 + N) (2 + k + N)^{5}}, \nonu \\
c_ {+309} & = &\frac {4 (1 + k) (k - N) (32 + 55 k + 
               18 k^{2} + 41 N + 35 k\, 
              N + 11 N^{2})} {3 (2 + N) (2 + k + N)^{6}}, \nonu \\
c_ {+310} & = & - \frac {4 (1 + k) (9 + 5 k + 4 N) (32 + 
                55 k + 18 k^{2} + 41 N + 35 k\, 
               N + 11 N^{2})} {3 (2 + N) (2 + k + N)^{6}}, \nonu \\
c_ {+311} & = & - \frac {4 (9 + 5 k + 4 N) (32 + 55 k + 
                18 k^{2} + 41 N + 35 k\, 
               N + 11 N^{2})} {3 (2 + N) (2 + k + N)^{5}}, \nonu \\
c_ {+312} & = &\frac {1} {3 (2 + N) (2 + k + N)^{5}}4 (36 - 111 k - 156 k^{2} - 44 k^{3} + 
               93 N - 224 k\, 
              N - 245 k^{2} N 
              \nonu \\& - & 46 k^{3} N + 110 N^{2} - 117 k\, 
              N^{2} - 83 k^{2} N^{2} + 64 N^{3} - 10 k\, 
              N^{3} + 13 N^{4}),\nonu \\ 
c_ {+313} & = & - \frac {4 (1 + N) (-k + N) (32 + 55 k + 
                18 k^{2} + 41 N + 35 k\, 
               N + 11 N^{2})} {3 (2 + N) (2 + k + N)^{6}}, \nonu \\
c_ {+314} & = &\frac {1} {3 (2 + N) (2 + k + N)^{5}}4 (36 - 83 k - 82 k^{2} - 16 k^{3} + 
               65 N - 238 k\, 
              N - 186 k^{2} N 
              \nonu\\ & - & 32 k^{3} N + 50 N^{2} - 167 k\, 
              N^{2} - 72 k^{2} N^{2} + 27 N^{3} - 28 k\, 
              N^{3} + 6 N^{4}), \nonu \\
c_ {+315} & = & - \frac {4 (1 + N) (9 + 4 k + 5 N) (32 + 
                55 k + 18 k^{2} + 41 N + 35 k\, 
               N + 11 N^{2})} {3 (2 + N) (2 + k + N)^{6}}, \nonu \\
c_ {+316} & = &\frac {2 (-k + N) (32 + 55 k + 18 k^{2} + 
               41 N + 35 k\, 
              N + 11 N^{2})} {3 (2 + N) (2 + k + N)^{5}}, \nonu \\
c_ {+317} & = & - \frac {4 (9 + 4 k + 5 N) (32 + 55 k + 
                18 k^{2} + 41 N + 35 k\, 
               N + 11 N^{2})} {3 (2 + N) (2 + k + N)^{5}}, \nonu \\
c_ {+318} & = & - \frac{1} {3 (2 + N) (2 + k + N)^{5}}2 (288 + 606 k + 538 k^{2} + 
                213 k^{3} + 30 k^{4} + 690 N + 1020 k\, 
               N 
               \nonu \\& + & 566 k^{2} N + 107 k^{3} N + 602 N^{2} + 571 k\, 
               N^{2} + 150 k^{2} N^{2} + 234 N^{3} + 111 k\, 
               N^{3} + 34 N^{4}),\nonu \\ 
c_ {+319} & = & - \frac {4 (1 + N) (-k + N) (32 + 55 k + 
                18 k^{2} + 41 N + 35 k\, 
               N + 11 N^{2})} {3 (2 + N) (2 + k + N)^{6}}, \nonu \\
c_ {+320} & = & - \frac {4 (1 + N) (3 + 2 k + N) (32 + 
                55 k + 18 k^{2} + 41 N + 35 k\, 
               N + 11 N^{2})} {(2 + N) (2 + k + N)^{6}}, \nonu \\
c_ {+321} & = & \frac {1} {3 (2 + N) (2 + k + N)^{5}}2 (576 + 824 k + 270 k^{2} - 61 k^{3} - 
          30 k^{4} + 1768 N + 2702 k\, N 
          \nonu \\& + & 1240 k^{2} N + 161 k^{3} N + 1924 N^{2} + 2291 k\, 
         N^{2} + 628 k^{2} N^{2} + 850 N^{3} + 551 k\, 
         N^{3} + 130 N^{4}), \nonu \\
c_ {+322} & = & \frac {1} {3 (2 + N) (2 + k + N)^{5}}16 (18 - k - 17 k^{2} - 6 k^{3} + 64 N + 
          90 k\, N + 52 k^{2} N + 12 k^{3} N 
          \nonu \\& + & 80 N^{2} + 105 k\, 
         N^{2} + 35 k^{2} N^{2} + 38 N^{3} + 28 k\, 
         N^{3} + 6 N^{4}), \nonu \\
c_ {+323} & = &\frac{1} {3 (2 + N) (2 + k + N)^{4}}8 (48 + 121 k + 80 k^{2} + 16 k^{3} + 
          137 N + 281 k\, 
         N + 143 k^{2} N 
         \nonu \\& + & 20 k^{3} N + 152 N^{2} + 213 k\, 
         N^{2} + 58 k^{2} N^{2} + 72 N^{3} + 51 k\, 
         N^{3} + 12 N^{4}), \nonu \\
    c_ {+324} & = & - \frac {2 (-k + N) (32 + 55 k + 18 k^{2} + 
           41 N + 35 k\, N + 11 N^{2})} {3 (2 + N) (2 + k + N)^{5}},\nonu \\ 
    c_ {+325} & = & - \frac {2 (-k + N) (32 + 55 k + 18 k^{2} + 
           41 N + 35 k\, N + 11 N^{2})} {3 (2 + N) (2 + k + N)^{5}},\nonu \\ 
    c_ {+326} & = &\frac {6 (32 + 55 k + 18 k^{2} + 41 N + 35 k\, 
         N + 11 N^{2})} {(2 + N) (2 + k + N)^{4}}, \nonu \\
    c_ {+327} & = & - \frac {4 (-k + N) (32 + 55 k + 18 k^{2} + 
           41 N + 35 k\, N + 11 N^{2})} {3 (2 + N) (2 + k + N)^{5}},\nonu \\ 
    c_ {+328} & = & - \frac {4 (-k + N) (32 + 55 k + 18 k^{2} + 
           41 N + 35 k\, N + 11 N^{2})} {3 (2 + N) (2 + k + N)^{5}},\nonu \\ 
    c_ {+329} & = &\frac {4 (-k + N) (32 + 55 k + 18 k^{2} + 41 N + 
          35 k\, N + 11 N^{2})} {3 (2 + N) (2 + k + N)^{5}}, \nonu \\
    c_ {+330} & = & - \frac{1} {3 (2 + N) (2 + k + N)^{5}}4 (36 + 111 k + 78 k^{2} + 16 k^{3} + 
           159 N + 316 k\, 
          N + 154 k^{2} N 
          \nonu \\& + & 20 k^{3} N + 200 N^{2} + 255 k\, 
          N^{2} + 64 k^{2} N^{2} + 97 N^{3} + 62 k\, 
          N^{3} + 16 N^{4}), \nonu \\
    c_ {+331} & = &\frac {4 (9 + 4 k + 5 N) (32 + 55 k + 18 k^{2} + 
          41 N + 35 k\, N + 11 N^{2})} {3 (2 + N) (2 + k + N)^{5}}, \nonu \\
    c_ {+332} & = &\frac {2 (-k + N) (32 + 55 k + 18 k^{2} + 41 N + 
          35 k\, N + 11 N^{2})} {3 (2 + N) (2 + k + N)^{5}}, \nonu \\
    c_ {+333} & = &\frac {4 (9 + 5 k + 4 N) (32 + 55 k + 18 k^{2} + 
          41 N + 35 k\, N + 11 N^{2})} {3 (2 + N) (2 + k + N)^{5}}, \nonu \\
    c_ {+334} & = & - \frac{1} {3 (2 + N) (2 + k + N)^{4}}2 (288 + 238 k - 82 k^{2} - 125 k^{3} - 
           30 k^{4} + 482 N + 222 k\, 
          N 
          \nonu \\& - & 135 k^{2} N - 69 k^{3} N + 292 N^{2} + 41 k\, 
          N^{2} - 45 k^{2} N^{2} + 75 N^{3} - 7 k\, 
          N^{3} + 7 N^{4}), \nonu \\
    c_ {+335} & = & - \frac {4 (48 + 173 k + 135 k^{2} + 30 k^{3} + 
           91 N + 200 k\, N + 77 k^{2} N + 49 N^{2} + 53 k\, 
          N^{2} + 8 N^{3})} {3 (2 + N) (2 + k + N)^{4}}, \nonu \\
    c_ {+336} & = &\frac {2 (-k + N) (32 + 55 k + 18 k^{2} + 41 N + 
          35 k\, N + 11 N^{2})} {3 (2 + N) (2 + k + N)^{5}}, \nonu \\
    c_ {+337} & = & - \frac {6 (32 + 55 k + 18 k^{2} + 41 N + 35 k\, 
          N + 11 N^{2})} {(2 + N) (2 + k + N)^{4}}, \nonu \\
    c_ {+338} & = & - \frac {4 (-k + N) (32 + 55 k + 18 k^{2} + 
           41 N + 35 k\, N + 11 N^{2})} {3 (2 + N) (2 + k + N)^{5}},\nonu \\ 
    c_ {+339} & = & - \frac {4 (9 + 5 k + 4 N) (32 + 55 k + 
           18 k^{2} + 41 N + 35 k\, 
          N + 11 N^{2})} {3 (2 + N) (2 + k + N)^{5}}, \nonu \\
    c_ {+340} & = & - \frac {4 (48 + 157 k + 114 k^{2} + 24 k^{3} + 
           107 N + 202 k\, N + 70 k^{2} N + 68 N^{2} + 61 k\, 
          N^{2} + 13 N^{3})} {3 (2 + N) (2 + k + N)^{4}}, \nonu \\
    c_ {+341} & = & - \frac {2 (-k + N) (16 + 21 k + 6 k^{2} + 19 N + 
           13 k\, N + 5 N^{2})} {3 (2 + N) (2 + k + N)^{4}}, \nonu \\
    c_ {+342} & = &\frac {2 (16 + 47 k + 18 k^{2} + 25 N + 31 k\, 
         N + 7 N^{2})} {(2 + N) (2 + k + N)^{3}}, \nonu \\
    c_ {+343} & = &\frac {2 (-k + N) (20 + 28 k + 8 k^{2} + 54 N + 
          55 k\, N + 10 k^{2} N + 41 N^{2} + 23 k\, 
         N^{2} + 9 N^{3})} {3 (2 + N) (2 + k + N)^{5}}, \nonu \\
    c_ {+344} & = & - \frac {4 (-k + N) (32 + 55 k + 18 k^{2} + 
           41 N + 35 k\, N + 11 N^{2})} {3 (2 + N) (2 + k + N)^{5}},\nonu \\ 
    c_ {+345} & = &\frac {4 (-k + N) (32 + 55 k + 18 k^{2} + 41 N + 
          35 k\, N + 11 N^{2})} {3 (2 + N) (2 + k + N)^{5}}, \nonu \\
    c_ {+346} & = & - \frac {1} {3 (2 + N) (2 + k + N)^{4}}2 (276 + 194 k - 102 k^{2} - 60 k^{3} + 
           496 N + 335 k\, N - 16 k^{2} N 
           \nonu \\& + & 319 N^{2} + 149 k\, 
          N^{2} + 65 N^{3}), \nonu \\
    c_ {+347} & = & - \frac {8 (48 + 64 k + 33 k^{2} + 6 k^{3} + 
           56 N + 46 k\, N + 13 k^{2} N + 17 N^{2} + 4 k\, 
          N^{2} + N^{3})} {3 (2 + N) (2 + k + N)^{4}}, \nonu \\
    c_ {+348} & = & - \frac{1}{3 (2 + N) (2 + k + N)^{5}}4 (180 + 150 k - 111 k^{2} - 131 k^{3} - 
           30 k^{4} + 480 N + 453 k\, 
          N 
          \nonu \\& + & 16 k^{2} N - 45 k^{3} N + 468 N^{2} + 373 k\, 
          N^{2} + 49 k^{2} N^{2} + 192 N^{3} + 88 k\, 
          N^{3} + 28 N^{4}), \nonu \\
    c_ {+349} & = & - \frac {4 (-k + N) (16 + 21 k + 6 k^{2} + 19 N + 
           13 k\, N + 5 N^{2})} {3 (2 + N) (2 + k + N)^{4}}, \nonu \\
    c_ {+350} & = & - \frac {2 (-k + N) (20 + 2 k - 17 k^{2} - 
           6 k^{3} + 48 N + 21 k\, N - 5 k^{2} N + 36 N^{2} + 13 k\, 
          N^{2} + 8 N^{3})} {3 (2 + N) (2 + k + N)^{5}}, \nonu \\
    c_ {+351} & = &\frac {2 (84 + 236 k + 75 k^{2} - 6 k^{3} + 
          262 N + 449 k\, N + 95 k^{2} N + 220 N^{2} + 191 k\, 
         N^{2} + 50 N^{3})} {3 (2 + N) (2 + k + N)^{4}}, \nonu \\
    c_ {+352} & = & - \frac {8 (48 + 32 k - 9 k^{2} - 6 k^{3} + 
           88 N + 50 k\, N - k^{2} N + 55 N^{2} + 20 k\, 
          N^{2} + 11 N^{3})} {3 (2 + N) (2 + k + N)^{4}}, \nonu \\
    c_ {+353} & = &\frac{1} {3 (2 + N) (2 + k + N)^{5}}4 (180 + 364 k + 216 k^{2} + 40 k^{3} + 
          554 N + 867 k\, 
         N + 374 k^{2} N 
         \nonu \\& + & 44 k^{3} N + 591 N^{2} + 638 k\, 
         N^{2} + 146 k^{2} N^{2} + 262 N^{3} + 147 k\, 
         N^{3} + 41 N^{4}), \nonu \\
    c_ {+354} & = & - \frac{1}{3 (2 + N) (2 + k + N)^{5}}4 (182 k + 295 k^{2} + 163 k^{3} + 
           30 k^{4} + 106 N + 398 k\, 
          N + 334 k^{2} N 
          \nonu \\& + & 85 k^{3} N + 171 N^{2} + 273 k\, 
          N^{2} + 93 k^{2} N^{2} + 94 N^{3} + 63 k\, 
          N^{3} + 17 N^{4}), \nonu \\
    c_ {+355} & = & - \frac {4 (9 + 4 k + 5 N) (32 + 55 k + 
           18 k^{2} + 41 N + 35 k\, 
          N + 11 N^{2})} {3 (2 + N) (2 + k + N)^{5}}, \nonu \\
    c_ {+356} & = & - \frac {1}{3 (2 + N) (2 + k + N)^{5}}4 (246 k + 359 k^{2} + 179 k^{3} + 
           30 k^{4} + 42 N + 430 k\, 
          N + 382 k^{2} N 
          \nonu \\& + & 93 k^{3} N + 75 N^{2} + 257 k\, 
          N^{2} + 101 k^{2} N^{2} + 46 N^{3} + 55 k\, 
          N^{3} + 9 N^{4}), \nonu \\
    c_ {+357} & = & - \frac{1} {3 (2 + N) (2 + k + N)^{3}}2 (360 + 340 k + 15 k^{2} - 30 k^{3} + 
           668 N + 472 k\, N + 25 k^{2} N 
           \nonu \\& + & 395 N^{2} + 166 k\, 
          N^{2} + 73 N^{3}), \nonu \\
    c_ {+358} & = &\frac {8 (9 + 5 k + 4 N) (18 + 21 k + 6 k^{2} + 
          22 N + 13 k\, N + 6 N^{2})} {3 (2 + N) (2 + k + N)^{4}}, \nonu \\
    c_ {+359} & = &\frac {(-k + N) (20 + 21 k + 6 k^{2} + 25 N + 
          13 k\, N + 7 N^{2})} {3 (2 + N) (2 + k + N)^{4}}, \nonu \\
    c_ {+360} & = & - \frac {2 (96 + 52 k - 39 k^{2} - 18 k^{3} + 
           188 N + 108 k\, N - 9 k^{2} N + 123 N^{2} + 50 k\, 
          N^{2} + 25 N^{3})} {3 (2 + N) (2 + k + N)^{4}}, \nonu \\
    c_ {+361} & = & - \frac {8 (-k + N) (16 + 21 k + 6 k^{2} + 19 N + 
           13 k\, N + 5 N^{2})} {3 (2 + N) (2 + k + N)^{4}}, \nonu \\
    c_ {+362} & = &\frac {8 (96 + 203 k + 126 k^{2} + 24 k^{3} + 
          181 N + 247 k\, N + 76 k^{2} N + 107 N^{2} + 72 k\, 
         N^{2} + 20 N^{3})} {3 (2 + N) (2 + k + N)^{4}}, \nonu \\
    c_ {+363} & = & - \frac {1} {3 (2 + N) (2 + k + N)^{5}}2 (456 + 1094 k + 966 k^{2} + 
           383 k^{3} + 58 k^{4} + 1054 N + 2138 k\, 
          N 
          \nonu \\& + & 1505 k^{2} N + 445 k^{3} N + 44 k^{4} N + 1024 N^{2} + 
           1641 k\, 
          N^{2} + 805 k^{2} N^{2} + 130 k^{3} N^{2} + 527 N^{3} 
          \nonu \\& + & 
           589 k\, N^{3} + 146 k^{2} N^{3} + 143 N^{4} + 84 k\, 
          N^{4} + 16 N^{5}), \nonu \\
    c_ {+364} & = & - \frac {4 (-k + N) (32 + 55 k + 18 k^{2} + 
           41 N + 35 k\, N + 11 N^{2})} {3 (2 + N) (2 + k + N)^{5}},\nonu \\ 
    c_ {+365} & = & - \frac {2 (-k + N) (32 + 55 k + 18 k^{2} + 
           41 N + 35 k\, N + 11 N^{2})} {3 (2 + N) (2 + k + N)^{5}},\nonu \\ 
    c_ {+366} & = & - \frac{1}{3 (2 + N) (2 + k + N)^{5}}4 (324 + 849 k + 797 k^{2} + 309 k^{3} + 
           42 k^{4} + 717 N + 1416 k\, 
          N 
          \nonu \\& + & 911 k^{2} N + 181 k^{3} N + 541 N^{2} + 724 k\, 
          N^{2} + 246 k^{2} N^{2} + 162 N^{3} + 109 k\, 
          N^{3} + 16 N^{4}) , \nonu \\
    c_ {+367} & = & - \frac {4 (1 + N) (-k + N) (32 + 55 k + 
           18 k^{2} + 41 N + 35 k\, 
          N + 11 N^{2})} {3 (2 + N) (2 + k + N)^{6}}, \nonu \\
    c_ {+368} & = &\frac {4 (-k + N) (32 + 55 k + 18 k^{2} + 41 N + 
          35 k\, N + 11 N^{2})} {3 (2 + N) (2 + k + N)^{5}}, \nonu \\
    c_ {+369} & = & - \frac {2 (1 + N) (-k + N) (4 + 7 k + 2 k^{2} + 
           19 N + 21 k\, N + 4 k^{2} N + 17 N^{2} + 10 k\, 
          N^{2} + 4 N^{3})} {3 (2 + N) (2 + k + N)^{6}}, \nonu \\
    c_ {+370} & = &\frac {4 (1 + k) (k - N) (32 + 55 k + 18 k^{2} + 
          41 N + 35 k\, N + 11 N^{2})} {3 (2 + N) (2 + k + N)^{6}}, \nonu \\
    c_ {+371} & = & - \frac {4 (1 + k) (k - N) (32 + 55 k + 
           18 k^{2} + 41 N + 35 k\, 
          N + 11 N^{2})} {3 (2 + N) (2 + k + N)^{6}}, \nonu \\
    c_ {+372} & = &\frac{1}{3 (2 + N) (2 + k + N)^{5}}2 (-324 - 727 k - 520 k^{2} - 116 k^{3} - 
          551 N - 1068 k\, 
         N - 604 k^{2} N 
         \nonu \\& - & 88 k^{3} N - 302 N^{2} - 479 k\, 
         N^{2} - 168 k^{2} N^{2} - 43 N^{3} - 54 k\, 
         N^{3} + 4 N^{4}), \nonu \\
    c_ {+373} & = &\frac{1} {3 (2 + N) (2 + k + N)^{6}}2 (-k + N) (160 + 376 k + 325 k^{2} + 
          125 k^{3} + 18 k^{4} + 376 N 
          \nonu \\& + & 640 k\, 
         N + 356 k^{2} N + 67 k^{3} N + 331 N^{2} + 365 k\, 
         N^{2} + 97 k^{2} N^{2} + 130 N^{3} + 71 k\, 
         N^{3} + 19 N^{4}), \nonu \\
    c_ {+374} & = &\frac {4 (1 + N) (-k + N) (32 + 55 k + 18 k^{2} + 
          41 N + 35 k\, N + 11 N^{2})} {3 (2 + N) (2 + k + N)^{6}}, \nonu \\
    c_ {+375} & = & - \frac{1}{3 (2 + N) (2 + k + N)^{5}}2 (288 + 212 k - 53 k^{2} + k^{3} + 
           18 k^{4} + 796 N + 460 k\, 
          N 
          \nonu \\& - & 194 k^{2} N - 61 k^{3} N + 889 N^{2} + 457 k\, 
          N^{2} - 63 k^{2} N^{2} + 456 N^{3} + 167 k\, 
          N^{3} + 83 N^{4}), \nonu \\
    c_ {+376} & = & - \frac{1}{3 (2 + N) (2 + k + N)^{6}}4 (-k + N) (148 + 223 k + 106 k^{2} + 
           16 k^{3} + 439 N + 592 k\, 
          N 
          \nonu \\& + & 246 k^{2} N + 32 k^{3} N + 464 N^{2} + 463 k\, 
          N^{2} + 108 k^{2} N^{2} + 205 N^{3} + 110 k\, 
          N^{3} + 32 N^{4}), \nonu \\
    c_ {+377} & = &\frac {2 (-k + N) (32 + 55 k + 18 k^{2} + 41 N + 
          35 k\, N + 11 N^{2})} {3 (2 + N) (2 + k + N)^{5}}, \nonu \\
    c_ {+378} & = & - \frac{1}{3 (2 + N) (2 + k + N)^{6}}4 (936 + 2702 k + 3099 k^{2} + 
           1778 k^{3} + 514 k^{4} + 60 k^{5} 
           \nonu \\& + & 2374 N + 5530 k\, 
          N + 4808 k^{2} N + 1854 k^{3} N + 270 k^{4} N + 
           2315 N^{2} + 4086 k\, 
          N^{2} 
          \nonu \\& + & 2405 k^{2} N^{2} + 468 k^{3} N^{2} + 1064 N^{3} + 
           1268 k\, N^{3} + 384 k^{2} N^{3} + 223 N^{4} + 134 k\, 
          N^{4} + 16 N^{5}), \nonu \\
    c_ {+379} & = &\frac{1}  {3 (2 + N) (2 + k + N)^{4}}2 (96 - 120 k - 255 k^{2} - 82 k^{3} + 
          180 N - 452 k\, 
         N - 559 k^{2} N 
         \nonu \\& - & 116 k^{3} N + 161 N^{2} - 330 k\, 
         N^{2} - 226 k^{2} N^{2} + 83 N^{3} - 52 k\, 
         N^{3} + 16 N^{4}), \nonu \\
    c_ {+380} & = &\frac{1} {3 (2 + N) (2 + k + N)^{4}}4 (456 + 1104 k + 945 k^{2} + 337 k^{3} + 
          42 k^{4} + 972 N + 1836 k\, 
         N 
         \nonu \\& + & 1089 k^{2} N + 201 k^{3} N + 759 N^{2} + 993 k\, 
         N^{2} + 308 k^{2} N^{2} + 257 N^{3} + 173 k\, 
         N^{3} + 32 N^{4}), \nonu \\
    c_ {+381} & = &\frac {2 (-96 - 132 k - 79 k^{2} - 18 k^{3} - 
          108 N - 88 k\, N - 29 k^{2} N - 25 N^{2} - 2 k\, 
         N^{2} + N^{3})} {3 (2 + N) (2 + k + N)^{4}}, \nonu \\
    c_ {+382} & = & - \frac {4 (-k + N) (32 + 55 k + 18 k^{2} + 
           41 N + 35 k\, N + 11 N^{2})} {3 (2 + N) (2 + k + N)^{5}},\nonu \\ 
    c_ {+383} & = &\frac {4 (-k + N) (32 + 55 k + 18 k^{2} + 41 N + 
          35 k\, N + 11 N^{2})} {3 (2 + N) (2 + k + N)^{5}}, \nonu \\
    c_ {+384} & = & - \frac{1} {3 (2 + N) (2 + k + N)^{4}}2 (648 + 1064 k + 573 k^{2} + 
           102 k^{3} + 1132 N + 1244 k\, 
          N 
          \nonu \\& + & 335 k^{2} N + 631 N^{2} + 350 k\, 
          N^{2} + 113 N^{3}), \nonu \\
    c_ {+385} & = &\frac{1}{3 (2 + N) (2 + k + N)^{5}}4 (-96 - 244 k - 238 k^{2} - 107 k^{3} - 
          18 k^{4} - 92 N - 198 k\, 
         N 
         \nonu \\& - & 142 k^{2} N - 37 k^{3} N + 4 N^{2} - 11 k\, 
         N^{2} - 6 k^{2} N^{2} + 20 N^{3} + 9 k\, 
         N^{3} + 4 N^{4}), \nonu \\
    c_ {+386} & = & - \frac{1}{3 (2 + N) (2 + k + N)^{5}}4 (96 + 116 k + 26 k^{2} - 17 k^{3} - 
           6 k^{4} + 220 N + 238 k\, 
          N + 74 k^{2} N 
          \nonu \\& + & k^{3} N + 168 N^{2} + 135 k\, 
          N^{2} + 30 k^{2} N^{2} + 48 N^{3} + 19 k\, 
          N^{3} + 4 N^{4}), \nonu \\
    c_ {+387} & = &\frac {8 (-k + N) (16 + 21 k + 6 k^{2} + 19 N + 
          13 k\, N + 5 N^{2})} {3 (2 + N) (2 + k + N)^{4}}, \nonu \\
    c_ {+388} & = & - \frac{1}{3 (2 + N) (2 + k + N)^{5}}4 (360 + 940 k + 844 k^{2} + 315 k^{3} + 
           42 k^{4} + 896 N + 1728 k\, 
          N 
          \nonu \\& + & 1034 k^{2} N + 193 k^{3} N + 776 N^{2} + 999 k\, 
          N^{2} + 304 k^{2} N^{2} + 280 N^{3} + 181 k\, 
          N^{3} + 36 N^{4}), \nonu \\
    c_ {+389} & = &\frac {2 (-k + N) (32 + 55 k + 18 k^{2} + 41 N + 
          35 k\, N + 11 N^{2})} {3 (2 + N) (2 + k + N)^{5}}, \nonu \\
    c_ {+390} & = & - \frac{1}{3 (2 + N) (2 + k + N)^{5}}4 (264 + 280 k + 59 k^{2} - 6 k^{3} + 
           644 N + 566 k\, 
          N + 118 k^{2} N 
          \nonu \\& + &6 k^{3} N + 563 N^{2} + 342 k\, 
          N^{2} + 39 k^{2} N^{2} + 206 N^{3} + 60 k\, 
          N^{3} + 27 N^{4}), \nonu \\
    c_ {+391} & = &\frac {1}{3 (2 + N) (2 + k + N)^{5}}4 (192 + 704 k + 781 k^{2} + 360 k^{3} + 
          60 k^{4} + 544 N + 1312 k\, 
         N \nonu \\& + & 922 k^{2} N + 208 k^{3} N + 499 N^{2} + 744 k\, 
         N^{2} + 253 k^{2} N^{2} + 182 N^{3} + 128 k\, 
         N^{3} + 23 N^{4}) , \nonu \\
    c_ {+392} & = & - \frac{1}{3 (2 + N) (2 + k + N)^{5}}4 (36 + 55 k + 6 k^{2} - 14 k^{3} - 
           4 k^{4} + 251 N + 427 k\, 
          N + 244 k^{2} N 
          \nonu \\& + & 69 k^{3} N + 10 k^{4} N + 431 N^{2} + 
           579 k\, 
          N^{2} + 242 k^{2} N^{2} + 37 k^{3} N^{2} + 307 N^{3} + 
           278 k\, N^{3} 
           \nonu \\& + & 58 k^{2} N^{3} + 99 N^{4} + 45 k\, 
          N^{4} + 12 N^{5}), \nonu \\
    c_ {+393} & = &\frac {4 (9 + 5 k + 4 N) (4 + 7 k + 2 k^{2} + 
          19 N + 21 k\, N + 4 k^{2} N + 17 N^{2} + 10 k\, 
         N^{2} + 4 N^{3})} {3 (2 + N) (2 + k + N)^{5}}, \nonu \\
    c_ {+394} & = & - \frac {4 (1 + k) (9 + 5 k + 4 N) (32 + 55 k + 
           18 k^{2} + 41 N + 35 k\, 
          N + 11 N^{2})} {3 (2 + N) (2 + k + N)^{6}}, \nonu \\
    c_ {+395} & = & - \frac{1}{3 (2 + N) (2 + k + N)^{5}}4 (324 + 917 k + 949 k^{2} + 404 k^{3} + 
           60 k^{4} + 649 N + 1378 k\, 
          N 
          \nonu \\& + & 963 k^{2} N + 206 k^{3} N + 427 N^{2} + 631 k\, 
          N^{2} + 232 k^{2} N^{2} + 108 N^{3} + 88 k\, 
          N^{3} + 8 N^{4}), \nonu \\
    c_ {+396} & = & - \frac{1}{3 (2 + N) (2 + k + N)^{6}}8 (1 + k) (180 + 323 k + 185 k^{2} + 
           34 k^{3} + 451 N + 648 k\, 
          N 
          \nonu \\& + & 275 k^{2} N + 32 k^{3} N + 409 N^{2} + 415 k\, 
          N^{2} + 96 k^{2} N^{2} + 158 N^{3} + 84 k\, 
          N^{3} + 22 N^{4}), \nonu \\
    c_ {+397} & = & - \frac{1}{3 (2 + N) (2 + k + N)^{6}}4 (-k + N) (372 + 683 k + 397 k^{2} + 
           74 k^{3} + 891 N + 1252 k\, 
          N 
          \nonu \\& + & 505 k^{2} N + 52 k^{3} N + 777 N^{2} + 747 k\, 
          N^{2} + 158 k^{2} N^{2} + 292 N^{3} + 144 k\, 
          N^{3} + 40 N^{4}), \nonu \\
    c_ {+398} & = &\frac {1} {3 (2 + N) (2 + k + N)^{5}}4 (756 + 1261 k + 658 k^{2} + 92 k^{3} - 
          8 k^{4} + 1745 N + 1807 k\, 
         N 
         \nonu \\& + & 46 k^{2} N - 375 k^{3} N - 82 k^{4} N + 1549 N^{2} + 
          757 k\, N^{2} - 492 k^{2} N^{2} - 225 k^{3} N^{2} + 
          671 N^{3} 
          \nonu \\& + & 66 k\, 
         N^{3} - 162 k^{2} N^{3} + 143 N^{4} - 11 k\, 
         N^{4} + 12 N^{5}), \nonu \\
    c_ {+399} & = &\frac {1}{3 (2 + N) (2 + k + N)^{4}}8 (276 + 674 k + 590 k^{2} + 221 k^{3} + 
          30 k^{4} + 580 N + 1078 k\, 
         N 
         \nonu \\& + & 642 k^{2} N + 123 k^{3} N + 438 N^{2} + 551 k\, 
         N^{2} + 168 k^{2} N^{2} + 140 N^{3} + 89 k\, 
         N^{3} + 16 N^{4}), \nonu \\
    c_ {+400} & = & - \frac{1} {3 (2 + N) (2 + k + N)^{4}}8 (138 + 251 k + 151 k^{2} + 30 k^{3} + 
           238 N + 285 k\, N + 85 k^{2} N 
           \nonu \\& + & 128 N^{2} + 76 k\, 
          N^{2} + 22 N^{3}), \nonu \\
    c_ {+401} & = & - \frac{1} {3 (2 + N) (2 + k + N)^{5}}4 (-324 - 673 k - 502 k^{2} - 
           159 k^{3} - 18 k^{4} - 713 N - 927 k\, 
          N 
          \nonu \\& - & 225 k^{2} N + 103 k^{3} N 
        + 36 k^{4} N - 551 N^{2} - 
           298 k\, 
          N^{2} + 189 k^{2} N^{2} + 98 k^{3} N^{2} - 182 N^{3} 
          \nonu \\& + & 
           36 k\, N^{3} + 80 k^{2} N^{3} - 22 N^{4} + 20 k\, 
          N^{4}), \nonu \\
    c_ {+402} & = & - \frac {2 (-96 - 180 k - 129 k^{2} - 30 k^{3} - 
           60 N - 92 k\, N - 47 k^{2} N + 29 N^{2} + 14 k\, 
          N^{2} + 15 N^{3})} {3 (2 + N) (2 + k + N)^{4}}, \nonu \\
    c_ {+403} & = & - \frac {(-k + N) (16 + 21 k + 6 k^{2} + 19 N + 
           13 k\, N + 5 N^{2})} {(2 + N) (2 + k + N)^{3}},\nonu \\ 
    c_ {+404} & = &\frac {2 (-96 - 132 k - 79 k^{2} - 18 k^{3} - 
          108 N - 88 k\, N - 29 k^{2} N - 25 N^{2} - 2 k\, 
         N^{2} + N^{3})} {3 (2 + N) (2 + k + N)^{4}}, \nonu \\
    c_ {+405} & = &\frac {2 (-k + N) (32 + 55 k + 18 k^{2} + 41 N + 
          35 k\, N + 11 N^{2})} {3 (2 + N) (2 + k + N)^{5}}, \nonu \\
    c_ {+406} & = &\frac {2 (96 + 100 k + 11 k^{2} - 6 k^{3} + 
          140 N + 112 k\, N + 9 k^{2} N + 69 N^{2} + 34 k\, 
         N^{2} + 11 N^{3})} {3 (2 + N) (2 + k + N)^{4}}, \nonu \\
    c_ {+407} & = & - \frac{1}{3 (2 + N) (2 + k + N)^{5}}2 (-k + N) (52 + 118 k + 83 k^{2} + 
           18 k^{3} + 76 N + 117 k\, 
          N 
          \nonu \\ & + &  43 k^{2} N + 32 N^{2} + 25 k\, 
          N^{2} + 4 N^{3}) , \nonu \\
    c_ {+408} & = & - \frac {2 (-k + N) (12 + 24 k + 8 k^{2} + 38 N + 
           49 k\, N + 10 k^{2} N + 31 N^{2} + 21 k\, 
          N^{2} + 7 N^{3})} {3 (2 + N) (2 + k + N)^{5}}, \nonu \\
    c_ {+409} & = &\frac{1} {3 (2 + N) (2 + k + N)^{4}}(384 + 768 k + 516 k^{2} + 151 k^{3} + 
         18 k^{4} + 960 N + 1312 k\, 
        N 
        \nonu \\ & + & 507 k^{2} N + 59 k^{3} N + 860 N^{2} + 737 k\, 
        N^{2} + 119 k^{2} N^{2} + 333 N^{3} + 141 k\, 
        N^{3} + 47 N^{4}), \nonu \\
    c_ {+410} & = &\frac{1}{3 (2 + N) (2 + k + N)^{4}}(360 + 590 k + 339 k^{2} + 66 k^{3} + 
         598 N + 644 k\, N + 185 k^{2} N 
         \nonu \\ & + & 313 N^{2} + 164 k\, 
        N^{2} + 53 N^{3}), \nonu \\
    c_ {+411} & = & - \frac{1}{3 (2 + N) (2 + k + N)^{5}}2 (360 + 674 k + 502 k^{2} + 189 k^{3} + 
           30 k^{4} + 874 N + 1230 k\, 
          N 
          \nonu \\ & + &  607 k^{2} N + 113 k^{3} N + 752 N^{2} + 697 k\, 
          N^{2} + 169 k^{2} N^{2} + 271 N^{3} + 121 k\, 
          N^{3} + 35 N^{4}), \nonu \\
    c_ {+412} & = &\frac { 4 (24 + 39 k + 
     17 k^{2} + 2 k^{3} + 57 N + 64 k\, 
    N + 15 k^{2} N + 39 N^{2} + 23 k\, N^{2} + 8 N^{3})}{3 (2 + k + N)^{4}}, \nonu \\
  c_ {+413} & = &\frac{1} {3 (2 + N) (2 + k + N)^{5}}2 (264 + 902 k + 970 k^{2} + 433 k^{3} + 
        70 k^{4} + 766 N + 2074 k\, 
       N 
        \nonu \\ & + & 1715 k^{2} N + 553 k^{3} N + 56 k^{4} N + 892 N^{2} + 
        1779 k\, 
       N^{2} + 987 k^{2} N^{2} + 168 k^{3} N^{2} + 513 N^{3}
       \nonu \\ & + &  
        669 k\, N^{3} + 182 k^{2} N^{3} + 145 N^{4} + 94 k\, 
       N^{4} + 16 N^{5}), \nonu \\
  c_ {+414} & = &\frac {2 (-k + N) (32 + 55 k + 18 k^{2} + 41 N + 
        35 k\, N + 11 N^{2})} {3 (2 + N) (2 + k + N)^{5}},\nonu \\ 
  c_ {+415} & = & - \frac {4 (-k + N) (32 + 55 k + 18 k^{2} + 41 N + 
         35 k\, N + 11 N^{2})} {3 (2 + N) (2 + k + N)^{5}},\nonu \\ 
  c_ {+416} & = &\frac {4 (-k + N) (32 + 55 k + 18 k^{2} + 41 N + 
        35 k\, N + 11 N^{2})} {3 (2 + N) (2 + k + N)^{5}}, \nonu \\
  c_ {+417} & = &\frac{1}{3 (2 + N) (2 + k + N)^{6}}2 (1 + k) (-k + N) (36 + 39 k + 10 k^{2} + 
        67 N + 53 k\, N + 8 k^{2} N 
        \nonu \\ & + &  41 N^{2} + 18 k\, 
       N^{2} + 8 N^{3}), \nonu \\
  c_ {+418} & = &\frac{1} {3 (2 + N) (2 + k + N)^{5}}2 (36 + 75 k + 3 k^{2} - 10 k^{3} + 195 N + 
        296 k\, N + 47 k^{2} N - 8 k^{3} N 
        \nonu \\ & + &  295 N^{2} + 315 k\, 
       N^{2} + 38 k^{2} N^{2} + 170 N^{3} + 100 k\, 
       N^{3} + 32 N^{4}), \nonu \\
  c_ {+419} & = &\frac {2 (-k + N) (32 + 55 k + 18 k^{2} + 41 N + 
        35 k\, N + 11 N^{2})} {3 (2 + N) (2 + k + N)^{5}}, \nonu \\
  c_ {+420} & = & - \frac{1}{3 (2 + N) (2 + k + N)^{4}}2 (96 - 68 k - 205 k^{2} - 70 k^{3} + 
         488 N + 168 k\, 
        N - 207 k^{2} N 
        \nonu \\ & - &  56 k^{3} N + 679 N^{2} + 324 k\, 
        N^{2} - 40 k^{2} N^{2} + 361 N^{3} + 122 k\, 
        N^{3} + 64 N^{4}) , \nonu \\
  c_ {+421} & = &\frac {1}{3 (2 + N) (2 + k + N)^{4}}4 (264 + 404 k + 187 k^{2} + 26 k^{3} + 
        616 N + 704 k\, 
       N + 204 k^{2} N 
       \nonu \\ & + & 10 k^{3} N + 537 N^{2} + 416 k\, 
       N^{2} + 57 k^{2} N^{2} + 206 N^{3} + 84 k\, 
       N^{3} + 29 N^{4}), \nonu \\
  c_ {+422} & = &\frac {4 (-k + N) (32 + 55 k + 18 k^{2} + 41 N + 
        35 k\, N + 11 N^{2})} {3 (2 + N) (2 + k + N)^{5}}, \nonu \\
  c_ {+423} & = &\frac {2 (96 + 100 k + 11 k^{2} - 6 k^{3} + 140 N + 
        112 k\, N + 9 k^{2} N + 69 N^{2} + 34 k\, 
       N^{2} + 11 N^{3})} {3 (2 + N) (2 + k + N)^{4}}, \nonu \\
  c_ {+424} & = &\frac {2 (72 + 20 k - 21 k^{2} - 6 k^{3} + 160 N + 
        56 k\, N - 7 k^{2} N + 109 N^{2} + 26 k\, 
       N^{2} + 23 N^{3})} {3 (2 + N) (2 + k + N)^{4}}, \nonu \\
  c_ {+425} & = &\frac{1}{3 (2 + N) (2 + k + N)^{5}}4 (-96 - 212 k - 148 k^{2} - 32 k^{3} - 
        124 N - 266 k\, 
       N - 155 k^{2} N 
       \nonu \\ & - &  22 k^{3} N - 18 N^{2} - 82 k\, 
       N^{2} - 37 k^{2} N^{2} + 29 N^{3} + 2 k\, 
       N^{3} + 9 N^{4}), \nonu \\
  c_ {+426} & = &\frac {4 (-k + N) (32 + 55 k + 18 k^{2} + 41 N + 
        35 k\, N + 11 N^{2})} {3 (2 + N) (2 + k + N)^{5}}, \nonu \\
  c_ {+427} & = & - \frac{1}{3 (2 + N) (2 + k + N)^{5}}4 (96 + 84 k - 8 k^{3} + 252 N + 178 k\, 
        N - 5 k^{2} N - 10 k^{3} N 
        \nonu \\ & + &  254 N^{2} + 138 k\, 
        N^{2} + k^{2} N^{2} + 115 N^{3} + 38 k\, 
        N^{3} + 19 N^{4}) , \nonu \\
  c_ {+428} & = &\frac{1} {3 (2 + N) (2 + k + N)^{5}}4 (360 + 472 k + 192 k^{2} + 24 k^{3} + 
        788 N + 764 k\, 
       N + 187 k^{2} N 
       \nonu \\ & + &  6 k^{3} N + 664 N^{2} + 434 k\, 
       N^{2} + 49 k^{2} N^{2} + 255 N^{3} + 88 k\, 
       N^{3} + 37 N^{4}), \nonu \\
  c_ {+429} & = &\frac {1}{3 (2 + N) (2 + k + N)^{4}}8 (-48 - 23 k + 13 k^{2} + 6 k^{3} - 55 N + 
        65 k\, N + 83 k^{2} N + 18 k^{3} N 
        \nonu \\ & + & 3 N^{2} + 98 k\, 
       N^{2} + 43 k^{2} N^{2} + 17 N^{3} + 28 k\, 
       N^{3} + 4 N^{4}), \nonu \\
  c_ {+430} & = &\frac {1}{3 (2 + N) (2 + k + N)^{5}}4 (456 + 1036 k + 887 k^{2} + 351 k^{3} + 
        54 k^{4} + 1136 N + 1898 k\, 
       N
       \nonu \\ & + &  1063 k^{2} N + 207 k^{3} N + 995 N^{2} + 1089 k\, 
       N^{2} + 300 k^{2} N^{2} + 365 N^{3} + 195 k\, 
       N^{3} + 48 N^{4}), \nonu \\
  c_ {+431} & = &\frac{1} {3 (2 + N) (2 + k + N)^{5}}4 (192 + 768 k + 813 k^{2} + 321 k^{3} + 
        42 k^{4} + 480 N + 1344 k\, 
       N 
       \nonu \\ & + & 929 k^{2} N + 181 k^{3} N + 435 N^{2} + 783 k\, 
       N^{2} + 268 k^{2} N^{2} + 175 N^{3} + 155 k\, 
       N^{3} + 26 N^{4}), \nonu \\
  c_ {+432} & = &\frac{1} {3 (2 + N) (2 + k + N)^{5}}4 (324 + 671 k + 492 k^{2} + 155 k^{3} + 
        18 k^{4} + 931 N + 1603 k\, 
       N 
       \nonu \\ & + &  935 k^{2} N + 222 k^{3} N + 18 k^{4} N + 1073 N^{2} + 
        1426 k\, 
       N^{2} + 575 k^{2} N^{2} + 73 k^{3} N^{2} + 616 N^{3} 
       \nonu \\ & + &  557 k\, 
       N^{3} + 114 k^{2} N^{3} + 176 N^{4} + 81 k\, 
       N^{4} + 20 N^{5}), \nonu \\
  c_ {+433} & = & - \frac {1}{3 (2 + N) (2 + k + N)^{5}}4 (9 + 4 k + 5 N) (36 + 39 k + 10 k^{2} + 
         67 N + 53 k\, N + 8 k^{2} N 
         \nonu \\ & + &  41 N^{2} + 18 k\, 
        N^{2} + 8 N^{3}) , \nonu \\
  c_ {+434} & = &\frac{1} {3 (2 + N) (2 + k + N)^{6}}8 (1 + N) (180 + 303 k + 162 k^{2} + 
        28 k^{3} + 471 N + 628 k\, 
       N 
       \nonu \\ & + &  244 k^{2} N + 26 k^{3} N + 452 N^{2} + 423 k\, 
       N^{2} + 88 k^{2} N^{2} + 187 N^{3} + 92 k\, 
       N^{3} + 28 N^{4}), \nonu \\
  c_ {+435} & = & - \frac{1}{3 (2 + N) (2 + k + N)^{5}}4 (396 + 643 k + 256 k^{2} - 19 k^{3} - 
         18 k^{4} + 1535 N + 2113 k\, 
        N 
        \nonu \\ & + &  729 k^{2} N - 2 k^{3} N - 18 k^{4} N + 2257 N^{2} + 
         2478 k\, 
        N^{2} + 649 k^{2} N^{2} + 15 k^{3} N^{2} + 1582 N^{3} 
        \nonu \\ & + &  
         1235 k\, N^{3} + 182 k^{2} N^{3} + 530 N^{4} + 221 k\, 
        N^{4} + 68 N^{5}), \nonu \\
  c_ {+436} & = &\frac  {8 (42 + 56 k + 
   21 k^{2} + 2 k^{3} + 70 N + 66 k\, 
  N + 13 k^{2} N + 39 N^{2} + 20 k\, N^{2} + 7 N^{3})} {3 (2 + k + N)^{4}}, \nonu \\
c_ {+437} & = &\frac {8 (21 + 14 k + 2 k^{2} + 28 N + 10 k\, 
     N + 9 N^{2})} {3 (2 + k + N)^{4}}, \nonu \\
c_ {+438} & = & - \frac{1}{3 (2 + N) (2 + k + N)^{5}}4 (-36 + 67 k + 82 k^{2} + 20 k^{3} + 
       131 N + 525 k\, 
      N + 330 k^{2} N 
      \nonu \\ & + &  54 k^{3} N + 437 N^{2} + 815 k\, 
      N^{2} + 322 k^{2} N^{2} + 28 k^{3} N^{2} + 419 N^{3} + 465 k\, 
      N^{3} + 92 k^{2} N^{3} 
      \nonu \\ & + &  167 N^{4} + 90 k\, 
      N^{4} + 24 N^{5}), \nonu \\
c_ {+439} & = & - \frac{1}{3 (2 + N) (2 + k + N)^{5}}(-k + N) (192 + 264 k + 93 k^{2} + 6 k^{3} + 
       280 N + 232 k\, N 
       \nonu \\ & + &  31 k^{2} N + 123 N^{2} + 42 k\, 
      N^{2} + 17 N^{3}), \nonu \\
c_ {+440} & = & - \frac{1}{3 (2 + N) (2 + k + N)^{4}}(768 + 908 k + 105 k^{2} - 66 k^{3} + 
      1012 N + 508 k\, N - 125 k^{2} N 
      \nonu \\ & + & 347 N^{2} - 30 k\, 
     N^{2} + 29 N^{3}), \nonu \\
c_ {+441} & = & - \frac{1}{3 (2 + N) (2 + k + N)^{5}}2 (-48 - 88 k - 80 k^{2} - 63 k^{3} - 
       18 k^{4} - 80 N - 124 k\, 
      N - 89 k^{2} N 
      \nonu \\ & - &  39 k^{3} N - 12 N^{2} - k\, 
      N^{2} - 5 k^{2} N^{2} + 33 N^{3} + 27 k\, 
      N^{3} + 11 N^{4}), \nonu \\
c_ {+442} & = & - \frac{1}{3 (2 + N) (2 + k + N)^{5}}4 (180 + 369 k + 267 k^{2} + 75 k^{3} + 
       6 k^{4} + 405 N + 631 k\, 
      N 
      \nonu \\ & + &  307 k^{2} N + 43 k^{3} N + 344 N^{2} + 368 k\, 
      N^{2} + 92 k^{2} N^{2} + 132 N^{3} + 74 k\, 
      N^{3} + 19 N^{4}), \nonu \\
c_ {+443} & = &\frac{1} {3 (2 + N) (2 + k + N)^{5}}2 (1488 + 3232 k + 2564 k^{2} + 861 k^{3} + 
      102 k^{4} + 3128 N + 5148 k\, 
     N 
     \nonu \\ & + &  2743 k^{2} N + 461 k^{3} N + 2440 N^{2} + 2723 k\, 
     N^{2} + 735 k^{2} N^{2} + 849 N^{3} + 487 k\, 
     N^{3} + 111 N^{4}), \nonu \\
c_ {+444} & = &\frac{1} {3 (2 + N) (2 + k + N)^{5}}4 (744 + 1796 k + 1559 k^{2} + 592 k^{3} + 
      84 k^{4} + 2392 N + 4626 k\, 
     N 
     \nonu \\ & + &  3078 k^{2} N + 832 k^{3} N + 72 k^{4} N + 2995 N^{2} + 
      4352 k\, 
     N^{2} + 1955 k^{2} N^{2} + 276 k^{3} N^{2} 
     \nonu \\ & + & 1830 N^{3} + 
      1778 k\, N^{3} + 400 k^{2} N^{3} + 547 N^{4} + 268 k\, 
     N^{4} + 64 N^{5}), \nonu \\
c_ {+445} & = & - \frac {4 (9 + 4 k + 5 N) (32 + 55 k + 18 k^{2} + 
       41 N + 35 k\, N + 11 N^{2})} {3 (2 + N) (2 + k + N)^{5}},\nonu \\ 
c_ {+446} & = &\frac {4 (1 + k) (k - N) (32 + 55 k + 18 k^{2} + 
      41 N + 35 k\, N + 11 N^{2})} {3 (2 + N) (2 + k + N)^{6}}, \nonu \\
c_ {+447} & = & - \frac{1}{3 (2 + N) (2 + k + N)^{5}}2 (288 + 606 k + 408 k^{2} + 114 k^{3} + 
       12 k^{4} + 690 N + 1120 k\, 
      N 
      \nonu \\ & + &  527 k^{2} N + 80 k^{3} N + 632 N^{2} + 690 k\, 
      N^{2} + 165 k^{2} N^{2} + 253 N^{3} + 138 k\, 
      N^{3} + 37 N^{4}), \nonu \\
c_ {+448} & = & - \frac {4 (1 + N) (9 + 4 k + 5 N) (32 + 55 k + 
       18 k^{2} + 41 N + 35 k\, 
      N + 11 N^{2})} {3 (2 + N) (2 + k + N)^{6}}, \nonu \\
c_ {+449} & = &\frac {1} {3 (2 + N) (2 + k + N)^{5}}2 (2304 + 4312 k + 2452 k^{2} + 414 k^{3} - 
      12 k^{4} + 5480 N + 7710 k\, 
     N 
     \nonu \\ & + & 2945 k^{2} N + 256 k^{3} N + 4814 N^{2} + 4584 k\, 
     N^{2} + 895 k^{2} N^{2} + 1849 N^{3} + 904 k\, 
     N^{3} + 261 N^{4}), \nonu \\
c_ {+450} & = &\frac{1} {3 (2 + N) (2 + k + N)^{6}}4 (360 + 962 k + 919 k^{2} + 381 k^{3} + 
      58 k^{4} + 946 N + 2062 k\, 
     N 
     \nonu \\ & + &  1535 k^{2} N + 463 k^{3} N + 44 k^{4} N + 1051 N^{2} + 
      1715 k\, 
     N^{2} + 862 k^{2} N^{2} + 138 k^{3} N^{2} + 617 N^{3} 
     \nonu \\ & + &  659 k\, 
     N^{3} + 162 k^{2} N^{3} + 190 N^{4} + 100 k\, 
     N^{4} + 24 N^{5}), \nonu \\
c_ {+451} & = & - \frac {4 (1 + k) (3 + k + 2 N) (32 + 55 k + 
       18 k^{2} + 41 N + 35 k\, 
      N + 11 N^{2})} {(2 + N) (2 + k + N)^{6}}, \nonu \\
c_ {+452} & = & - \frac{1}{3 (2 + N) (2 + k + N)^{6}}4 (216 + 738 k + 875 k^{2} + 429 k^{3} + 
       74 k^{4} + 522 N + 1494 k\, 
      N 
      \nonu \\ & + &  1409 k^{2} N + 503 k^{3} N + 52 k^{4} N + 511 N^{2} + 
       1135 k\, 
      N^{2} + 750 k^{2} N^{2} + 146 k^{3} N^{2} + 267 N^{3}
      \nonu \\ & + &  395 k\, 
      N^{3} + 132 k^{2} N^{3} + 78 N^{4} + 56 k\, 
      N^{4} + 10 N^{5}), \nonu \\
c_ {+453} & = &\frac{1}  {3 (2 + N) (2 + k + N)^{6}}4 (1440 + 3128 k + 2568 k^{2} + 943 k^{3} + 
      130 k^{4} + 4504 N + 8072 k\, 
     N 
     \nonu \\ & + &  5193 k^{2} N + 1377 k^{3} N + 116 k^{4} N + 5488 N^{2} + 
      7577 k\, 
     N^{2} + 3365 k^{2} N^{2} + 474 k^{3} N^{2} 
     \nonu \\ & + &  3279 N^{3} + 
      3091 k\, N^{3} + 704 k^{2} N^{3} + 965 N^{4} + 466 k\, 
     N^{4} + 112 N^{5}), \nonu \\
c_ {+454} & = &\frac{1} {3 (2 + N) (2 + k + N)^{5}}4 (-24 - 112 k - 117 k^{2} - 37 k^{3} - 
      2 k^{4} + 460 N + 958 k\, 
     N 
     \nonu \\ & + &  829 k^{2} N + 351 k^{3} N + 56 k^{4} N + 1139 N^{2} + 
      1845 k\, 
     N^{2} + 1030 k^{2} N^{2} + 204 k^{3} N^{2} + 975 N^{3} 
     \nonu \\ & + &  
      1059 k\, N^{3} + 288 k^{2} N^{3} + 358 N^{4} + 196 k\, 
     N^{4} + 48 N^{5}), \nonu \\
c_ {+455} & = &\frac {4 (9 + 5 k + 4 N) (32 + 55 k + 18 k^{2} + 
      41 N + 35 k\, N + 11 N^{2})} {3 (2 + N) (2 + k + N)^{5}}, \nonu \\
c_ {+456} & = & - \frac{1}{3 (2 + N) (2 + k + N)^{6}}4 (360 + 1002 k + 969 k^{2} + 400 k^{3} + 
       60 k^{4} + 906 N + 2134 k\, 
      N 
      \nonu \\ & + &  1648 k^{2} N + 504 k^{3} N + 48 k^{4} N + 929 N^{2} + 
       1708 k\, 
      N^{2} + 917 k^{2} N^{2} + 152 k^{3} N^{2} + 492 N^{3} 
      \nonu \\ & + &  614 k\, 
      N^{3} + 166 k^{2} N^{3} + 137 N^{4} + 86 k\, 
      N^{4} + 16 N^{5}), \nonu \\
c_ {+457} & = &\frac{1}{3 (2 + N) (2 + k + N)^{4}}(576 + 1212 k + 930 k^{2} + 319 k^{3} + 
     42 k^{4} + 1380 N + 2172 k\, 
    N 
    \nonu \\ & + &  1093 k^{2} N + 183 k^{3} N + 1218 N^{2} + 1285 k\, 
    N^{2} + 319 k^{2} N^{2} + 471 N^{3} + 253 k\, 
    N^{3} + 67 N^{4}), \nonu \\
c_ {+458} & = & - \frac{1}{3 (2 + N) (2 + k + N)^{4}}2 (192 + 406 k + 239 k^{2} + 42 k^{3} + 
       362 N + 504 k\, N + 149 k^{2} N 
       \nonu \\ & + &  217 N^{2} + 152 k\, 
      N^{2} + 41 N^{3}), \nonu \\
c_ {+459} & = & - \frac {6 (32 + 55 k + 18 k^{2} + 41 N + 35 k\, 
      N + 11 N^{2})} {(2 + N) (2 + k + N)^{4}}, \nonu \\
c_ {+460} & = & - \frac{1}{3 (2 + N) (2 + k + N)^{4}}2 (192 + 406 k + 265 k^{2} + 54 k^{3} + 
       362 N + 484 k\, N + 155 k^{2} N 
       \nonu \\ & + &  211 N^{2} + 136 k\, 
      N^{2} + 39 N^{3}), \nonu \\
c_ {+461} & = &\frac {4 (32 + 55 k + 18 k^{2} + 41 N + 35 k\, 
     N + 11 N^{2})} {(2 + N) (2 + k + N)^{3}}, \nonu \\
c_ {+462} & = & - \frac {2 (-12 - 106 k - 143 k^{2} - 42 k^{3} + 
       76 N + 15 k\, N - 47 k^{2} N + 104 N^{2} + 55 k\, 
      N^{2} + 28 N^{3})} {3 (2 + N) (2 + k + N)^{4}}, \nonu \\
c_ {+463} & = &\frac {2 (372 + 472 k + 136 k^{2} + 746 N + 697 k\, 
     N + 122 k^{2} N + 487 N^{2} + 253 k\, 
     N^{2} + 99 N^{3})} {3 (2 + N) (2 + k + N)^{4}}, \nonu \\
c_ {+464} & = & - \frac{1} {3 (2 + N) (2 + k + N)^{3}}(1080 + 1190 k + 225 k^{2} - 42 k^{3} + 
      2014 N + 1628 k\, N + 191 k^{2} N 
      \nonu \\ & + &  1207 N^{2} + 560 k\, 
     N^{2} + 227 N^{3}), \nonu \\
c_ {+465} & = & - \frac {8 (1 + N) (9 + 4 k + 
       5 N)} {3 (2 + k + N)^{4}}, \nonu \\
c_ {+466} & = & - \frac {2 (-192 - 272 k - 15 k^{2} + 30 k^{3} - 
       208 N - 64 k\, N + 83 k^{2} N - 17 N^{2} + 66 k\, 
      N^{2} + 13 N^{3})} {3 (2 + N) (2 + k + N)^{4}}, \nonu \\
c_ {+467} & = &\frac{1} {3 (2 + N) (2 + k + N)^{5}}4 (504 + 1288 k + 1115 k^{2} + 407 k^{3} + 
      54 k^{4} + 1700 N + 3310 k\, 
     N 
     \nonu \\ & + &  2095 k^{2} N + 509 k^{3} N + 36 k^{4} N + 2127 N^{2} + 
      3005 k\, 
     N^{2} + 1232 k^{2} N^{2} + 146 k^{3} N^{2} + 1261 N^{3} 
     \nonu \\ & + &  
      1157 k\, N^{3} + 228 k^{2} N^{3} + 360 N^{4} + 162 k\, 
     N^{4} + 40 N^{5}), \nonu \\
c_ {+468} & = &\frac {1} {3 (2 + k + N)^{5}} 16 (81 + 161 k + 
      96 k^{2} + 18 k^{3} + 154 N + 198 k\, 
     N + 58 k^{2} N + 93 N^{2}
     \nonu \\ & + &  59 k\, N^{2} + 18 N^{3}), \nonu \\
c_ {+469} & = &\frac{1} {3 (2 + N) (2 + k + N)^{5}}4 (216 + 180 k - 77 k^{2} - 96 k^{3} - 
      20 k^{4} + 648 N + 738 k\, 
     N 
     \nonu \\ & + & 284 k^{2} N + 84 k^{3} N + 20 k^{4} N + 851 N^{2} + 972 k\, 
     N^{2} + 399 k^{2} N^{2} + 74 k^{3} N^{2} + 568 N^{3} 
     \nonu \\ & + &  502 k\, 
     N^{3} + 116 k^{2} N^{3} + 187 N^{4} + 90 k\, 
     N^{4} + 24 N^{5}),\nonu \\
c_ {+470} & = & - \frac{1}{3 (2 + N) (2 + k + N)^{4}}4 (180 + 329 k + 192 k^{2} + 36 k^{3} + 
       445 N + 641 k\, 
      N + 269 k^{2} N 
      \nonu \\ & + &  30 k^{3} N + 409 N^{2} + 413 k\, 
      N^{2} + 93 k^{2} N^{2} + 164 N^{3} + 87 k\, 
      N^{3} + 24 N^{4}),\nonu \\
c_ {+471} & = & - \frac{1}{3 (2 + N) (2 + k + N)^{4}}2 (-12 k + 15 k^{2} + 36 k^{3} + 12 k^{4} + 
       372 N + 778 k\, 
      N 
      \nonu \\ & + &  671 k^{2} N + 268 k^{3} N + 36 k^{4} N + 755 N^{2} + 
       1210 k\, 
      N^{2} + 678 k^{2} N^{2} + 130 k^{3} N^{2} + 567 N^{3} 
      \nonu \\ & + &  616 k\, 
      N^{3} + 172 k^{2} N^{3} + 190 N^{4} + 106 k\, 
      N^{4} + 24 N^{5}),\nonu \\
c_ {+472} & = &\frac{1} {3 (2 + N) (2 + k + N)^{3}}2 (352 + 517 k + 132 k^{2} - 12 k^{3} + 603 N + 
      425 k\, N - 88 k^{2} N 
       \nonu\\ & - &  36 k^{3} N + 387 N^{2} + 110 k\, 
     N^{2} - 46 k^{2} N^{2} + 142 N^{3} + 34 k\, 
     N^{3} + 24 N^{4}), \nonu \\
c_ {-1} & = & - \frac {48 k\, N)}{(2 + k + N)^{2}}, \qquad
c_ {-2}  =  - \frac {32 k (k - N) N}{3 (2 + k + N)^{3}}, \nonu\\
c_ {-3} & = &\frac {64 N (2 + 2 k + N) (3 + k + 
       2 N)}{3 (2 + k + N)^{3}}, \qquad
c_ {-4}  = \frac {64 N (3 + k + 2 N)}{3 (2 + k + N)^{3}},\nonu\\ 
c_ {-5} & = & - \frac {64 k (3 + 2 k + N) (2 + k + 
        2 N)}{3 (2 + k + N)^{3}}, \qquad
c_ {-6}  = \frac {64 k (3 + 2 k + N)}{3 (2 + k + N)^{3}},\nonu\\ 
c_ {-7} & = &\frac {16 (k - N) (2 + 2 k + N) (2 + k + 
       2 N)}{3 (2 + k + N)^{3}}, \qquad
c_ {-8}  = \frac {32 (k - N) (2 + k + 2 N)}{3 (2 + k + N)^{3}},\nonu\\ 
c_ {-9} & = & - \frac {16 (k - N)}{3 (2 + k + N)^{3}}, \qquad
c_ {-10}  = \frac {32 (k - N) (2 + 2 k + N)}{3 (2 + k + N)^{3}},\nonu\\ 
c_ {-11} & = & - \frac {16 (k - N)}{3 (2 + k + N)^{2}}, \qquad
c_ {-12}  = \frac {16 (k + 5 N + 2 k\, 
      N + 4 N^{2})}{3 (2 + k + N)^{3}}, \nonu\\
c_ {-13} & = & - \frac {16 (5 k + 4 k^{2} + N + 2 k\, 
       N)}{3 (2 + k + N)^{3}}, \qquad
c_ {-14}  =  - \frac {16 (3 k + 2 k^{2} + 3 N + 5 k\, 
       N + 2 N^{2})}{3 (2 + k + N)^{2}}, \nonu\\
c_ {-15} & = &\frac {16 (3 + 6 k + 2 k^{2} + 6 N + 5 k\, 
      N + 2 N^{2})}{3 (2 + k + N)^{2}}, \qquad
c_ {-16}  =  - \frac {32 (-3 + k^{2} - 6 N - 2 k\, 
       N - 2 N^{2})}{3 (2 + k + N)^{3}},\nonu\\ 
c_ {-17} & = & - \frac {16}{(2 + k + N)^{2}},\qquad
c_ {-18}  = \frac {32 (3 + 6 k + 2 k^{2} + 2 k\, 
      N - N^{2})}{3 (2 + k + N)^{3}}, \nonu\\
c_ {-19} & = &\frac {16 (6 + 12 k + 5 k^{2} + 12 N + 8 k\, 
      N + 5 N^{2})}{3 (2 + k + N)^{3}}, \nonu\\
c_ {-20} & = & - \frac {16 (6 + 12 k + 5 k^{2} + 18 N + 22 k\, 
       N + 6 k^{2} N + 15 N^{2} + 8 k\, 
       N^{2} + 4 N^{3})}{3 (2 + k + N)^{4}}, \nonu\\
c_ {-21} & = & - \frac {16 (6 + 18 k + 15 k^{2} + 4 k^{3} + 12 N + 
        22 k\, N + 8 k^{2} N + 5 N^{2} + 6 k\, 
       N^{2})}{3 (2 + k + N)^{4}}, \nonu\\
c_ {-22} & = & - \frac {16 (k - N) (3 + 5 k + 2 k^{2} + 5 N + 5 k\, 
       N + 2 N^{2})}{3 (2 + k + N)^{3}}, \nonu\\
c_ {-23} & = &\frac {16 (3 k + 2 k^{2} + 3 N + 13 k\, 
      N + 4 k^{2} N + 9 N^{2} + 10 k\, 
      N^{2} + 4 N^{3})}{3 (2 + k + N)^{3}}, \nonu\\
c_ {-24} & = &\frac {16 (3 + 2 k + N)}{3 (2 + k + N)^{3}}, \qquad
c_ {-25}  = \frac {8 (k - N) (3 + 2 k + N)}{3 (2 + k + N)^{3}}, \nonu\\
c_ {-26} & = & - \frac {16 (3 + 2 k + N)}{3 (2 + k + N)^{3}}, \qquad
c_ {-27}  =  - \frac {16 (1 + N) (3 + 2 k + N)}{3 (2 + k + N)^{4}},\nonu\\
 c_ {-28} & = &\frac {16 (1 + N) (3 + 2 k + N)}{3 (2 + k + N)^{4}}, \qquad
c_ {-29}  =  - \frac {8 (3 k + 2 k^{2} - 9 N - k\, 
       N - 7 N^{2})}{3 (2 + k + N)^{3}}, \nonu\\
c_ {-30} & = &\frac {32 (3 + 5 k + 2 k^{2} + 7 N + 4 k\, 
      N + 3 N^{2})}{3 (2 + k + N)^{3}}, \qquad
c_ {-31}  = \frac {16 (3 + 2 k + N)}{3 (2 + k + N)^{3}}, \nonu\\
c_ {-32} & = & - \frac {32 (3 + 5 k + 2 k^{2} + 4 N + 4 k\, 
       N + k^{2} N + 3 N^{2} + k\, 
       N^{2} + N^{3})}{3 (2 + k + N)^{4}}, \nonu\\
c_ {-33} & = &\frac {32 (3 + 5 k + 2 k^{2} + 4 N + 4 k\, 
      N + k^{2} N + 3 N^{2} + k\, N^{2} + N^{3})}{3 (2 + k + N)^{4}},\nonu\\
 c_ {-34} & = &\frac {16 (k - N) (3 + 2 k + N)}{3 (2 + k + N)^{3}}, \nonu\\
c_ {-35} & = &\frac {16 (3 k + 2 k^{2} + 15 N + 19 k\, 
      N + 6 k^{2} N + 15 N^{2} + 8 k\, 
      N^{2} + 4 N^{3})}{3 (2 + k + N)^{4}}, \nonu\\
c_ {-36} & = & - \frac {16 (3 k + 9 k^{2} + 4 k^{3} + 3 N + 13 k\, 
       N + 10 k^{2} N + 2 N^{2} + 4 k\, N^{2})}{3 (2 + k + N)^{3}}, \nonu\\
c_ {-37} & = & - \frac {16 (3 + k + 2 N)}{3 (2 + k + N)^{3}}, \qquad
c_ {-38}  =  - \frac {16 (3 + k + 2 N)}{3 (2 + k + N)^{3}}, \nonu\\
c_ {-39} & = &\frac {8 (k - N) (3 + k + 2 N)}{3 (2 + k + N)^{3}}, \qquad
c_ {-40}  = \frac {16 (1 + k) (3 + k + 2 N)}{3 (2 + k + N)^{4}}, \nonu\\
c_ {-41} & = &\frac {8 (9 k + 7 k^{2} - 3 N + k\, 
      N - 2 N^{2})}{3 (2 + k + N)^{3}}, \qquad
c_ {-42}  = \frac {32 (3 + 7 k + 3 k^{2} + 5 N + 4 k\, 
      N + 2 N^{2})}{3 (2 + k + N)^{3}}, \nonu\\
c_ {-43} & = & - \frac {16 (3 + k + 2 N)}{3 (2 + k + N)^{3}}, \nonu\\
c_ {-44} & = &\frac {32 (3 + 4 k + 3 k^{2} + k^{3} + 5 N + 4 k\, 
      N + k^{2} N + 2 N^{2} + k\, N^{2})}{3 (2 + k + N)^{4}}, \nonu\\
c_ {-45} & = & - \frac {16 (1 + k) (3 + k + 2 N)}{3 (2 + k + N)^{4}},\qquad
 c_ {-46}  =  - \frac {16 (k - N) (3 + k + 
        2 N)}{3 (2 + k + N)^{3}}, \nonu\\
c_ {-47} & = & - \frac {32 (3 + 4 k + 3 k^{2} + k^{3} + 5 N + 4 k\, 
       N + k^{2} N + 2 N^{2} + k\, N^{2})}{3 (2 + k + N)^{4}}, \nonu\\
c_ {-48} & = &\frac {16 (15 k + 15 k^{2} + 4 k^{3} + 3 N + 19 k\, 
      N + 8 k^{2} N + 2 N^{2} + 6 k\, N^{2})}{3 (2 + k + N)^{4}}, \qquad
c_ {-49}  =  - 5, \nonu\\
c_ {-50} & = &\frac {(5 (60 + 77 k + 22 k^{2} + 121 N + 115 k\, 
       N + 20 k^{2} N + 79 N^{2} + 42 k\, N^{2} + 16 N^{3}))}{(2 + 
       N) (2 + k + N)^{2}}, \nonu\\
c_ {-51} & = &\frac {4 (32 + 81 k + 30 k^{2} + 47 N + 53 k\, 
      N + 13 N^{2})}{(2 + N) (2 + k + N)^{3}}, \nonu\\
c_ {-52} & = & - \frac {(300 + 299 k + 90 k^{2} + 391 N + 187 k\, 
      N + 113 N^{2})}{3 (2 + N) (2 + k + N)^{3}}, \nonu\\
c_ {-53} & = &\frac {4 (246 + 299 k + 90 k^{2} + 310 N + 187 k\, 
      N + 86 N^{2})}{3 (2 + N) (2 + k + N)^{3}}, \nonu\\
c_ {-54} & = &\frac {4 (24 + 131 k + 54 k^{2} + 127 N + 282 k\, 
      N + 72 k^{2} N + 129 N^{2} + 127 k\, 
      N^{2} + 32 N^{3})}{3 (2 + N) (2 + k + N)^{3}}, \nonu\\
c_ {-55} & = & - \frac {10 (32 + 55 k + 18 k^{2} + 41 N + 35 k\, 
       N + 11 N^{2})}{(2 + N) (2 + k + N)^{4}}, \nonu\\
c_ {-56} & = & - \frac {(32 + 81 k + 30 k^{2} + 47 N + 53 k\, 
      N + 13 N^{2})}{(2 + N) (2 + k + N)^{3}}, \nonu\\
c_ {-57} & = &\frac {10 (1 + k) (32 + 55 k + 18 k^{2} + 41 N + 35 k\, 
      N + 11 N^{2})}{(2 + N) (2 + k + N)^{5}}, \nonu\\
c_ {-58} & = & - \frac {1}{3 (2 + N) (2 + k + N)^{4}}(1200 + 1412 k + 181 k^{2} - 90 k^{3} + 
       2548 N + 2386 k\, N 
       \nonu\\ & + & 323 k^{2} N + 1753 N^{2} + 960 k\, 
      N^{2} + 367 N^{3}), \nonu\\
c_ {-59} & = & - \frac {2 (300 + 409 k + 126 k^{2} + 581 N + 579 k\, 
       N + 108 k^{2} N + 375 N^{2} + 206 k\, 
       N^{2} + 76 N^{3})}{3 (2 + N) (2 + k + N)^{4}}, \nonu\\
c_ {-60} & = &\frac {10 (1 + N) (32 + 55 k + 18 k^{2} + 41 N + 35 k\, 
      N + 11 N^{2})}{(2 + N) (2 + k + N)^{5}}, \nonu\\
c_ {-61} & = & - \frac{1}{3 (2 + N) (2 + k + N)^{4}}2 (588 + 1183 k + 808 k^{2} + 180 k^{3} + 
        1007 N + 1291 k\, N 
        \nonu\\ & + & 434 k^{2} N + 517 N^{2} + 316 k\, 
       N^{2} + 84 N^{3}), \nonu\\
c_ {-62} & = & - \frac{1}{3 (2 + N) (2 + k + N)^{4}}2 (648 + 879 k + 274 k^{2} + 1461 N + 
        1423 k\, N + 272 k^{2} N 
        \nonu\\ & + &  1039 N^{2} + 548 k\, 
       N^{2} + 224 N^{3}), \nonu\\
c_ {-63} & = &\frac {4 (-24 + 35 k + 26 k^{2} + 7 N + 146 k\, 
      N + 58 k^{2} N + 53 N^{2} + 83 k\, 
      N^{2} + 18 N^{3})}{3 (2 + N) (2 + k + N)^{3}}, \nonu\\
c_ {-64} & = &\frac {10 (32 + 55 k + 18 k^{2} + 41 N + 35 k\, 
      N + 11 N^{2})}{(2 + N) (2 + k + N)^{4}}, \nonu\\
c_ {-65} & = &\frac {2 (300 + 409 k + 118 k^{2} + 581 N + 595 k\, 
      N + 104 k^{2} N + 367 N^{2} + 214 k\, 
      N^{2} + 72 N^{3})}{3 (2 + N) (2 + k + N)^{4}}, \nonu\\
c_ {-66} & = & - \frac {(-96 - 390 k - 343 k^{2} - 90 k^{3} - 138 N - 
       478 k\, N - 209 k^{2} N - 43 N^{2} - 134 k\, 
      N^{2} + N^{3})}{3 (2 + N) (2 + k + N)^{3}}, \nonu\\
c_ {-67} & = &\frac {2 (96 + 235 k + 90 k^{2} + 149 N + 155 k\, 
      N + 43 N^{2})}{3 (2 + N) (2 + k + N)^{3}}, \nonu\\
c_ {-68} & = & - \frac {10 (32 + 55 k + 18 k^{2} + 41 N + 35 k\, 
       N + 11 N^{2})}{(2 + N) (2 + k + N)^{4}}, \nonu\\
c_ {-69} & = &\frac {2 (96 + 251 k + 90 k^{2} + 133 N + 163 k\, 
      N + 35 N^{2})}{3 (2 + N) (2 + k + N)^{3}}, \nonu\\
c_ {-70} & = &\frac {2 (48 + 89 k + 30 k^{2} + 63 N + 57 k\, 
      N + 17 N^{2})}{(2 + N) (2 + k + N)^{3}}, \nonu\\
c_ {-71} & = & - \frac {2 (396 + 596 k + 184 k^{2} + 874 N + 961 k\, 
       N + 182 k^{2} N + 607 N^{2} + 369 k\, 
       N^{2} + 127 N^{3})}{3 (2 + N) (2 + k + N)^{4}}, \nonu\\
c_ {-72} & = &\frac{1}{3 (2 + N) (2 + k + N)^{4}}2 (204 + 14 k - 239 k^{2} - 90 k^{3} + 496 N + 
       259 k\, N - 67 k^{2} N
       \nonu\\ & + & 388 N^{2} + 171 k\, 
      N^{2} + 88 N^{3}), \nonu\\
c_ {-73} & = &\frac {5 (80 + 72 k + 3 k^{2} - 6 k^{3} + 152 N + 
       102 k\, N + 5 k^{2} N + 91 N^{2} + 36 k\, 
      N^{2} + 17 N^{3})}{(2 + N) (2 + k + N)^{3}}, \nonu\\
c_ {-74} & = &\frac{1}{3 (2 + N) (2 + k + N)^{3}}2 (46 k + 71 k^{2} + 26 k^{3} + 254 N + 518 k\, 
      N + 332 k^{2} N + 58 k^{3} N 
      \nonu\\ & + &  401 N^{2} + 521 k\, 
      N^{2} + 167 k^{2} N^{2} + 201 N^{3} + 133 k\, 
      N^{3} + 32 N^{4}), \nonu\\
c_ {-75} & = & - \frac {4 (27 + 4 k + 23 N)}{3 (2 + k + N)^{3}}, \nonu\\
c_ {-76} & = & - \frac {1}{3 (2 + N) (2 + k + N)^{4}}(696 + 1390 k + 877 k^{2} + 182 k^{3} + 
       1718 N + 2652 k\, 
      N 
       \nonu\\ & + & 1157 k^{2} N + 136 k^{3} N + 1523 N^{2} + 1654 k\, 
      N^{2} + 378 k^{2} N^{2} + 571 N^{3} + 334 k\, 
      N^{3} + 76 N^{4}), \nonu\\
c_ {-77} & = & - \frac {2 (-12 + 173 k + 78 k^{2} + 37 N + 321 k\, 
       N + 84 k^{2} N + 57 N^{2} + 136 k\, 
       N^{2} + 14 N^{3})}{3 (2 + N) (2 + k + N)^{4}}, \nonu\\
c_ {-78} & = &\frac {2 (1 + N) (252 + 329 k + 94 k^{2} + 541 N + 
       523 k\, N + 92 k^{2} N + 367 N^{2} + 198 k\, 
      N^{2} + 76 N^{3})}{3 (2 + N) (2 + k + N)^{5}}, \nonu\\
c_ {-79} & = & - \frac {1}{3 (2 + N) (2 + k + N)^{5}}2 (96 - 156 k - 305 k^{2} - 94 k^{3} + 
        492 N - 20 k\, 
       N - 393 k^{2} N 
       \nonu\\ & - &  92 k^{3} N + 757 N^{2} + 276 k\, 
       N^{2} - 94 k^{2} N^{2} + 451 N^{3} + 146 k\, 
       N^{3} + 88 N^{4}), \nonu\\
c_ {-80} & = & - \frac {1}{3 (2 + N) (2 + k + N)^{4}}(384 - 76 k - 513 k^{2} - 182 k^{3} + 
       1828 N + 860 k\, 
      N - 441 k^{2} N 
      \nonu\\ & - &  136 k^{3} N + 2401 N^{2} + 1192 k\, 
      N^{2} - 66 k^{2} N^{2} + 1207 N^{3} + 394 k\, 
      N^{3} + 204 N^{4}), \nonu\\
c_ {-81} & = & - \frac {4 (3 + 2 k + N) (34 + 13 k + 
        21 N)}{3 (2 + k + N)^{3}}, \qquad
c_ {-82}  =  - \frac {4 (39 + 10 k + 29 N)}{3 (2 + k + N)^{3}}, \nonu\\
c_ {-83} & = & - \frac{1}{3 (2 + N) (2 + k + N)^{4}}4 (71 k + 26 k^{2} + 79 N + 275 k\, 
       N + 80 k^{2} N + 194 N^{2} + 315 k\, 
       N^{2} 
       \nonu\\ & + &  56 k^{2} N^{2} + 145 N^{3} + 107 k\, 
       N^{3} + 32 N^{4}), \nonu\\
c_ {-84} & = &\frac {4 (24 + 36 k + 11 k^{2} + 36 N + 26 k\, 
      N + 11 N^{2})}{3 (2 + k + N)^{3}}, \nonu\\
c_ {-85} & = &\frac {1}{3 (2 + N) (2 + k + N)^{4}}(504 + 910 k + 557 k^{2} + 118 k^{3} + 1142 N + 
      1708 k\, N + 773 k^{2} N 
      \nonu\\ & + &  104 k^{3} N + 963 N^{2} + 1078 k\, 
     N^{2} + 266 k^{2} N^{2} + 347 N^{3} + 222 k\, 
     N^{3} + 44 N^{4}), \nonu\\
c_ {-86} & = & - \frac {1}{3 (2 + N) (2 + k + N)^{5}}2 (1 + k) (348 + 425 k + 118 k^{2} + 685 N + 
        619 k\, N + 104 k^{2} N 
        \nonu\\ & + &  439 N^{2} + 222 k\, 
       N^{2} + 88 N^{3}), \nonu\\
c_ {-87} & = &\frac{1}{3 (2 + N) (2 + k + N)^{4}}(384 - 164 k - 637 k^{2} - 222 k^{3} + 716 N - 
      1164 k\, N - 1593 k^{2} N 
      \nonu\\ & - &  336 k^{3} N + 589 N^{2} - 988 k\, 
     N^{2} - 686 k^{2} N^{2} + 259 N^{3} - 186 k\, 
     N^{3} + 44 N^{4}), \nonu\\
c_ {-88} & = & - \frac{1}{3 (2 + N) (2 + k + N)^{3}}4 (396 + 796 k + 483 k^{2} + 90 k^{3} + 
        674 N + 903 k\, N + 279 k^{2} N 
        \nonu\\ & + & 366 N^{2} + 245 k\, 
       N^{2} + 64 N^{3}), \nonu\\
c_ {-89} & = &\frac {4 (222 + 287 k + 90 k^{2} + 286 N + 181 k\, 
      N + 80 N^{2})}{3 (2 + N) (2 + k + N)^{3}}, \nonu\\
c_ {-90} & = & - \frac{1}{3 (2 + N) (2 + k + N)^{4}}4 (71 k + 93 k^{2} + 26 k^{3} + 79 N + 
        291 k\, N + 260 k^{2} N + 58 k^{3} N
        \nonu\\ & + & 111 N^{2} + 222 k\, 
       N^{2} + 103 k^{2} N^{2} + 32 N^{3} + 34 k\, 
       N^{3}), \nonu\\
c_ {-91} & = &\frac{1}{(2 + N) (2 + k + N)^{4}}(-256 - 356 k - 39 k^{2} + 30 k^{3} - 284 N - 
      116 k\, N + 83 k^{2} N 
      \nonu\\ & - & 37 N^{2} + 66 k\, 
     N^{2} + 13 N^{3}), \nonu\\
c_ {-92} & = &\frac{1}{3 (2 + N) (2 + k + N)^{4}}4 (840 + 1406 k + 773 k^{2} + 142 k^{3} + 
       1666 N + 1944 k\, 
      N + 598 k^{2} N 
      \nonu\\ & + & 26 k^{3} N + 1153 N^{2} + 815 k\, 
      N^{2} + 87 k^{2} N^{2} + 329 N^{3} + 101 k\, 
      N^{3} + 32 N^{4}), \nonu\\
c_ {-93} & = & - \frac {2 (552 + 693 k + 182 k^{2} + 927 N + 893 k\, 
       N + 136 k^{2} N + 509 N^{2} + 292 k\, 
       N^{2} + 88 N^{3})}{3 (2 + N) (2 + k + N)^{4}}, \nonu\\
c_ {-94} & = &\frac {1}{3 (2 + N) (2 + k + N)^{4}}4 (360 + 434 k + 175 k^{2} + 26 k^{3} + 1054 N + 
       1196 k\, N + 448 k^{2} N 
       \nonu\\ & + & 58 k^{3} N + 1059 N^{2} + 923 k\, 
      N^{2} + 199 k^{2} N^{2} + 439 N^{3} + 213 k\, 
      N^{3} + 64 N^{4}), \nonu\\
c_ {-95} & = &\frac {4 (47 k + 16 k^{2} + 103 N + 196 k\, 
      N + 53 k^{2} N + 133 N^{2} + 105 k\, 
      N^{2} + 37 N^{3})}{3 (2 + N) (2 + k + N)^{3}}, \nonu\\
c_ {-96} & = &\frac{1}{3 (2 + N) (2 + k + N)^{3}}2 (192 + 237 k + 11 k^{2} - 26 k^{3} + 483 N + 
       257 k\, N - 168 k^{2} N 
       \nonu\\ & - & 58 k^{3} N + 452 N^{2} + 117 k\, 
      N^{2} - 83 k^{2} N^{2} + 197 N^{3} + 37 k\, 
      N^{3} + 32 N^{4}), \nonu\\
c_ {-97} & = & - \frac {2 (k - N)}{3 (2 + k + N)}, \nonu\\
c_ {-98} & = & - \frac {2 (-k + N) (60 + 77 k + 22 k^{2} + 121 N + 
        115 k\, N + 20 k^{2} N + 79 N^{2} + 42 k\, 
       N^{2} + 16 N^{3})}{3 (2 + N) (2 + k + N)^{3}}, \nonu\\
c_ {-99} & = & - \frac {8 (-k + N) (13 k + 6 k^{2} + 3 N + 9 k\, 
       N + N^{2})}{3 (2 + N) (2 + k + N)^{4}}, \nonu\\
c_ {-100} & = &\frac {2 (-k + N) (20 + 21 k + 6 k^{2} + 25 N + 13 k\, 
      N + 7 N^{2})}{3 (2 + N) (2 + k + N)^{4}}, \nonu\\
c_ {-101} & = & - \frac {8 (-48 - 74 k - 37 k^{2} - 6 k^{3} - 46 N - 
        53 k\, N - 15 k^{2} N - 6 N^{2} - 5 k\, 
       N^{2} + 2 N^{3})}{3 (2 + N) (2 + k + N)^{4}}, \nonu\\
c_ {-102} & = & - \frac {1}{3 (2 + N) (2 + k + N)^{4}}8 (48 + 72 k + 35 k^{2} + 6 k^{3} + 48 N + 
        52 k\, N + 12 k^{2} N + 9 N^{2} 
        \nonu\\ & + & 7 k\, 
       N^{2} - k^{2} N^{2} - N^{3} + k\, 
       N^{3}), \nonu\\
c_ {-103} & = &\frac {4 (-k + N) (32 + 55 k + 18 k^{2} + 41 N + 
       35 k\, N + 11 N^{2})}{3 (2 + N) (2 + k + N)^{5}}, \nonu\\
c_ {-104} & = &\frac {2 (-k + N) (13 k + 6 k^{2} + 3 N + 9 k\, 
      N + N^{2})}{3 (2 + N) (2 + k + N)^{4}}, \nonu\\
c_ {-105} & = &\frac {4 (1 + k) (k - N) (32 + 55 k + 18 k^{2} + 
       41 N + 35 k\, N + 11 N^{2})}{3 (2 + N) (2 + k + N)^{6}}, \nonu\\
c_ {-106} & = &\frac{1}{3 (2 + N) (2 + k + N)^{5}}2 (-k + N) (80 + 92 k + 11 k^{2} - 6 k^{3} + 
       172 N + 158 k\, N + 21 k^{2} N 
       \nonu\\ & + & 119 N^{2} + 64 k\, 
      N^{2} + 25 N^{3}), \nonu\\
c_ {-107} & = &\frac{1}{3 (2 + N) (2 + k + N)^{5}}4 (-96 - 212 k - 151 k^{2} - 34 k^{3} - 124 N - 
       244 k\, N - 139 k^{2} N 
       \nonu\\ & - & 20 k^{3} N - 37 N^{2} - 76 k\, 
      N^{2} - 30 k^{2} N^{2} + 9 N^{3} - 2 k\, 
      N^{3} + 4 N^{4}), \nonu\\
c_ {-108} & = & - \frac {4 (1 + N) (-k + N) (32 + 55 k + 18 k^{2} + 
        41 N + 35 k\, N + 11 N^{2})}{3 (2 + N) (2 + k + N)^{6}}, \nonu\\
c_ {-109} & = & - \frac {4 (3 + 2 k + N) (32 + 28 k + 19 k^{2} + 
        6 k^{3} + 52 N + 20 k\, 
       N + 6 k^{2} N + 25 N^{2} + 4 N^{3})}{3 (2 + 
        N) (2 + k + N)^{5}}, \nonu\\
c_ {-110} & = & - \frac{1}{3 (2 + N) (2 + k + N)^{5}}4 (96 + 120 k + 65 k^{2} + 14 k^{3} + 
        216 N + 218 k\, 
       N + 99 k^{2} N 
       \nonu\\ & + & 16 k^{3} N + 149 N^{2} + 96 k\, 
       N^{2} + 28 k^{2} N^{2} + 31 N^{3} + 4 k\, 
       N^{3}), \nonu\\
c_ {-111} & = & - \frac {1}{3 (2 + N) (2 + k + N)^{4}}8 (48 + 24 k - 13 k^{2} - 6 k^{3} + 96 N + 
        28 k\, N - 24 k^{2} N - 6 k^{3} N 
        \nonu\\ & + & 81 N^{2} + 19 k\, 
       N^{2} - 7 k^{2} N^{2} + 35 N^{3} + 7 k\, 
       N^{3} + 6 N^{4}), \nonu\\
c_ {-112} & = & - \frac {4 (-k + N) (32 + 55 k + 18 k^{2} + 41 N + 
        35 k\, N + 11 N^{2})}{3 (2 + N) (2 + k + N)^{5}}, \nonu\\
c_ {-113} & = &\frac{1}{3 (2 + N) (2 + k + N)^{5}}4 (96 + 116 k + 55 k^{2} + 10 k^{3} + 220 N + 
       196 k\, N + 67 k^{2} N + 8 k^{3} N 
       \nonu\\ & + & 181 N^{2} + 100 k\, 
      N^{2} + 18 k^{2} N^{2} + 63 N^{3} + 14 k\, 
      N^{3} + 8 N^{4}), \nonu\\
c_ {-114} & = & - \frac {2 (-k + N) (13 k + 6 k^{2} + 3 N + 9 k\, 
       N + N^{2})}{3 (2 + N) (2 + k + N)^{3}}, \nonu\\
c_ {-115} & = & - \frac {4 (96 + 96 k + 11 k^{2} - 6 k^{3} + 144 N + 
        106 k\, N + 9 k^{2} N + 75 N^{2} + 32 k\, 
       N^{2} + 13 N^{3})}{3 (2 + N) (2 + k + N)^{4}}, \nonu\\
c_ {-116} & = &\frac {4 (-k + N) (32 + 55 k + 18 k^{2} + 41 N + 
       35 k\, N + 11 N^{2})}{3 (2 + N) (2 + k + N)^{5}}, \nonu\\
c_ {-117} & = &\frac {4 (96 + 96 k + 37 k^{2} + 6 k^{3} + 144 N + 
       86 k\, N + 15 k^{2} N + 69 N^{2} + 16 k\, 
      N^{2} + 11 N^{3})}{3 (2 + N) (2 + k + N)^{4}}, \nonu\\
c_ {-118} & = & - \frac {4 (-k + N) (16 + 21 k + 6 k^{2} + 19 N + 
        13 k\, N + 5 N^{2})}{3 (2 + N) (2 + k + N)^{4}}, \nonu\\
c_ {-119} & = &\frac {4 (-k + N) (20 + 28 k + 8 k^{2} + 54 N + 55 k\, 
      N + 10 k^{2} N + 41 N^{2} + 23 k\, 
      N^{2} + 9 N^{3})}{3 (2 + N) (2 + k + N)^{5}}, \nonu\\
c_ {-120} & = & - \frac {4 (-k + N) (20 + 2 k - 17 k^{2} - 6 k^{3} + 
        48 N + 21 k\, N - 5 k^{2} N + 36 N^{2} + 13 k\, 
       N^{2} + 8 N^{3})}{3 (2 + N) (2 + k + N)^{5}}, \nonu\\
c_ {-121} & = & - \frac{1}{3 (2 + N) (2 + k + N)^{4}}2 (-192 - 368 k - 232 k^{2} - 35 k^{3} + 
        6 k^{4} - 304 N - 480 k\,N       
       \nonu\\ & - & 243 k^{2} N - 27 k^{3} N - 152 N^{2} - 181 k\, 
       N^{2} - 63 k^{2} N^{2} - 21 N^{3} - 13 k\, 
       N^{3} + N^{4}), \nonu\\
c_ {-122} & = &\frac{1}{3 (2 + N) (2 + k + N)^{3}}4 (-48 - 24 k + 13 k^{2} + 6 k^{3} - 48 N + 
       20 k\, N + 36 k^{2} N + 6 k^{3} N 
       \nonu\\ & - & 9 N^{2} + 29 k\, 
      N^{2} + 13 k^{2} N^{2} + N^{3} + 5 k\, 
      N^{3}), \nonu\\
c_ {-123} & = & - \frac {8 (24 + 21 k + 4 k^{2} + 27 N + 13 k\, 
       N + 7 N^{2})}{3 (2 + k + N)^{4}},\nonu\\
       c_ {-124} & = &\frac {2 (-k + N) (20 + 31 k + 10 k^{2} + 35 N + 
      41 k\, N + 8 k^{2} N + 21 N^{2} + 14 k\, 
     N^{2} + 4 N^{3})} {3 (2 + N) (2 + k + N)^{4}}, \nonu\\
c_ {-125} & = &\frac{1}{3 (2 + N) (2 + k + N)^{5}}4 (96 + 180 k + 93 k^{2} + 14 k^{3} + 156 N + 
      248 k\, N + 83 k^{2} N 
      \nonu\\ & + & 4 k^{3} N + 91 N^{2} + 120 k\, 
     N^{2} + 20 k^{2} N^{2} + 23 N^{3} + 22 k\, 
     N^{3} + 2 N^{4}), \nonu\\
c_ {-126} & = & - \frac {4 (1 + N) (-k + N) (4 + 7 k + 2 k^{2} + 
       19 N + 21 k\, N + 4 k^{2} N + 17 N^{2} + 10 k\, 
      N^{2} + 4 N^{3})} {3 (2 + N) (2 + k + N)^{6}},\nonu\\ 
c_ {-127} & = &\frac{1}{3 (2 + N) (2 + k + N)^{6}}4 (-k + N) (32 + 28 k + k^{2} - 2 k^{3} + 
      84 N + 52 k\, N - 7 k^{2} N 
      \nonu\\ & - & 4 k^{3} N + 91 N^{2} + 44 k\, 
     N^{2} - 2 k^{2} N^{2} + 45 N^{3} + 14 k\, 
     N^{3} + 8 N^{4}), \nonu\\
c_ {-128} & = &\frac {1}{3 (2 + N) (2 + k + N)^{5}}2 (192 + 224 k + 84 k^{2} + 31 k^{3} + 
      10 k^{4} + 256 N - 179 k^{2} N 
      \nonu\\ & - & 27 k^{3} N + 8 k^{4} N + 
      108 N^{2} - 267 k\, 
     N^{2} - 247 k^{2} N^{2} - 26 k^{3} N^{2} + 31 N^{3} - 137 k\, 
     N^{3} 
     \nonu\\ & - & 68 k^{2} N^{3} + 17 N^{4} - 14 k\, 
     N^{4} + 4 N^{5}), \nonu\\
c_ {-129} & = & - \frac {8 (3 + 2 k + N) (8 + 3 k + 
       5 N)} {3 (2 + k + N)^{3}}, \qquad
c_ {-130}  =  - \frac {8 (3 + 2 k + N) (8 + 3 k + 
       5 N)} {3 (2 + k + N)^{4}}, \nonu\\
c_ {-131} & = & - \frac{1}{3 (2 + N) (2 + k + N)^{5}} 8 (-48 - 32 k + 9 k^{2} + 6 k^{3} - 88 N - 
       8 k\, N + 55 k^{2} N + 16 k^{3} N
       \nonu\\ & - & 49 N^{2} + 41 k\, 
      N^{2} + 53 k^{2} N^{2} + 8 k^{3} N^{2} - 6 N^{3} + 26 k\, 
      N^{3} + 13 k^{2} N^{3} + N^{4} + 3 k\, 
      N^{4}), \nonu\\
c_ {-132} & = &\frac {8 (k - N)} {3 (2 + k + N)^{2}},\nonu\\ 
c_ {-133} & = & - \frac {2 (-k + N) (20 + 31 k + 10 k^{2} + 35 N + 
       41 k\, N + 8 k^{2} N + 21 N^{2} + 14 k\, 
      N^{2} + 4 N^{3})} {3 (2 + N) (2 + k + N)^{4}},\nonu\\ 
c_ {-134} & = &\frac {4 (1 + k) (-k + N) (36 + 39 k + 10 k^{2} + 
      67 N + 53 k\, N + 8 k^{2} N + 41 N^{2} + 18 k\, 
     N^{2} + 8 N^{3})} {3 (2 + N) (2 + k + N)^{6}}, \nonu\\
c_ {-135} & = & - \frac{1}{3 (2 + N) (2 + k + N)^{5}}2 (-192 - 160 k + 92 k^{2} + 131 k^{3} + 
       34 k^{4} - 320 N - 128 k\,N 
      \nonu\\ & + & 241 k^{2} N + 189 k^{3} N + 32 k^{4} N - 156 N^{2} + 9 k\, 
      N^{2} + 117 k^{2} N^{2} + 50 k^{3} N^{2} + 3 N^{3} + 23 k\,N^{3} 
      \nonu\\ & + &4 k^{2} N^{3} + 21 N^{4} + 6 k\, 
      N^{4} + 4 N^{5}), \nonu\\
c_ {-136} & = & - \frac {8 (48 + 66 k + 33 k^{2} + 6 k^{3} + 54 N + 
       49 k\, N + 13 k^{2} N + 14 N^{2} + 5 k\, 
      N^{2})} {3 (2 + N) (2 + k + N)^{3}}, \nonu\\
c_ {-137} & = &\frac {8 (48 + 66 k + 33 k^{2} + 6 k^{3} + 54 N + 
      49 k\, N + 13 k^{2} N + 14 N^{2} + 5 k\, 
     N^{2})} {3 (2 + N) (2 + k + N)^{4}}, \nonu\\
c_ {-138} & = & - \frac{1}{3 (2 + N) (2 + k + N)^{5}}8 (-48 - 64 k - 7 k^{2} + 19 k^{3} + 
       6 k^{4} - 56 N - 24 k\, 
      N + 50 k^{2} N 
      \nonu\\ & + &38 k^{3} N + 6 k^{4} N - 17 N^{2} + 28 k\, 
      N^{2} + 38 k^{2} N^{2} + 11 k^{3} N^{2} - N^{3} + 14 k\, 
      N^{3} + 5 k^{2} N^{3} + 2 k\,N^{4}), \nonu\\
c_ {-139} & = & - \frac {1}{3 (2 + N) (2 + k + N)^{5}}2 (-k + N) (-64 - 84 k - 11 k^{2} + 
       6 k^{3} - 76 N - 36 k\, N 
       \nonu\\ & + & 15 k^{2} N - 17 N^{2} + 10 k\, 
      N^{2} + N^{3}), \nonu\\
c_ {-140} & = &\frac {1}{3 (2 + N) (2 + k + N)^{4}}8 (96 + 164 k + 95 k^{2} + 18 k^{3} + 124 N + 
      170 k\, N + 70 k^{2} N 
      \nonu\\ & + & 6 k^{3} N + 47 N^{2} + 51 k\, 
     N^{2} + 13 k^{2} N^{2} + 5 N^{3} + 5 k\, 
     N^{3}), \nonu\\
c_ {-141} & = &\frac {1}{3 (2 + N) (2 + k + N)^{5}}4 (-96 - 248 k - 187 k^{2} - 42 k^{3} - 88 N - 
      262 k\, N - 169 k^{2} N 
      \nonu\\ & - & 24 k^{3} N + 17 N^{2} - 64 k\, 
     N^{2} - 36 k^{2} N^{2} + 35 N^{3} + 4 k\, 
     N^{3} + 8 N^{4}), \nonu\\
c_ {-142} & = &\frac{1} {3 (2 + N) (2 + k + N)^{4}}8 (96 + 116 k + 45 k^{2} + 6 k^{3} + 172 N + 
      166 k\, N + 52 k^{2} N + 6 k^{3} N 
      \nonu\\ & + & 101 N^{2} + 67 k\, 
     N^{2} + 13 k^{2} N^{2} + 19 N^{3} + 5 k\, 
     N^{3}), \nonu\\
c_ {-143} & = &\frac{1}{3 (2 + N) (2 + k + N)^{4}}8 (-48 - 24 k + 17 k^{2} + 8 k^{3} - 48 N + 
      12 k\, N + 36 k^{2} N + 7 k^{3} N 
      \nonu\\ & - & 5 N^{2} + 23 k\, 
     N^{2} + 12 k^{2} N^{2} + 5 N^{3} + 4 k\, 
     N^{3} + N^{4}), \nonu\\
c_ {-144} & = &\frac{1} {3 (2 + N) (2 + k + N)^{4}}4 (-k + N) (-32 - 19 k + 11 k^{2} + 6 k^{3} - 
      45 N + 9 k\, N + 32 k^{2} N 
      \nonu\\ & + & 6 k^{3} N - 20 N^{2} + 21 k\, 
     N^{2} + 13 k^{2} N^{2} - 3 N^{3} + 5 k\, 
     N^{3}), \nonu\\
c_ {-145} & = &\frac {4 (3 + k + 2 N)} {3 (2 + k + N)}, \nonu\\
c_ {-146} & = & - \frac{1}{3 (2 + N) (2 + k + N)^{3}}4 (3 + k + 2 N) (60 + 77 k + 22 k^{2} + 
       121 N + 115 k\, N + 20 k^{2} N 
       \nonu\\ & + & 79 N^{2} + 42 k\, 
      N^{2} + 16 N^{3}),\nonu\\ 
c_ {-147} & = & - \frac {1}{3 (2 + N) (2 + k + N)^{3}}4 (3 + k + 2 N) (60 + 77 k + 22 k^{2} + 
       121 N + 115 k\, N + 20 k^{2} N 
       \nonu\\ & + & 79 N^{2} + 42 k\, 
      N^{2} + 16 N^{3}),\nonu\\ 
c_ {-148} & = &\frac {4 (23 k + 23 k^{2} + 6 k^{3} + 25 N + 48 k\, 
     N + 17 k^{2} N + 25 N^{2} + 19 k\, 
     N^{2} + 6 N^{3})} {3 (2 + N) (2 + k + N)^{4}}, \nonu\\
c_ {-149} & = & - \frac{1}{3 (2 + N) (2 + k + N)^{5}}4 (312 + 588 k + 347 k^{2} + 79 k^{3} + 
       6 k^{4} + 696 N + 906 k\,N 
      \nonu\\ & + & 313 k^{2} N + 27 k^{3} N + 535 N^{2} + 413 k\, 
      N^{2} + 54 k^{2} N^{2} + 167 N^{3} + 51 k\, 
      N^{3} + 18 N^{4}),\nonu\\ 
c_ {-150} & = &\frac {1}{3 (2 + N) (2 + k + N)^{4}}8 (48 + 75 k + 37 k^{2} + 6 k^{3} + 99 N + 
      131 k\, N + 51 k^{2} N + 6 k^{3} N 
      \nonu\\ & + & 87 N^{2} + 86 k\, 
     N^{2} + 19 k^{2} N^{2} + 37 N^{3} + 20 k\, 
     N^{3} + 6 N^{4}), \nonu\\
c_ {-151} & = &\frac {8 (3 + k + 2 N) (32 + 55 k + 18 k^{2} + 41 N + 
      35 k\, N + 11 N^{2})} {3 (2 + N) (2 + k + N)^{5}}, \nonu\\
c_ {-152} & = &\frac {4 (3 + k + 2 N) (32 + 55 k + 18 k^{2} + 41 N + 
      35 k\, N + 11 N^{2})} {3 (2 + N) (2 + k + N)^{5}},\nonu\\ 
c_ {-153} & = &\frac{1} {3 (2 + N) (2 + k + N)^{5}}8 (72 + 137 k + 69 k^{2} + 10 k^{3} + 259 N + 
      383 k\, N + 144 k^{2} N 
      \nonu\\ & + & 14 k^{3} N + 304 N^{2} + 313 k\, 
     N^{2} + 63 k^{2} N^{2} + 145 N^{3} + 79 k\, 
     N^{3} + 24 N^{4}), \nonu\\
c_ {-154} & = & - \frac {8 (1 + k) (3 + k + 2 N) (32 + 55 k + 
       18 k^{2} + 41 N + 35 k\, 
      N + 11 N^{2})} {3 (2 + N) (2 + k + N)^{6}}, \nonu\\
c_ {-155} & = & - \frac {8 (3 + k + 2 N) (32 + 55 k + 18 k^{2} + 
       41 N + 35 k\, N + 11 N^{2})} {3 (2 + N) (2 + k + N)^{5}},\nonu\\ 
c_ {-156} & = & - \frac {1}{3 (2 + N) (2 + k + N)^{5}}4 (12 + 17 k + k^{2} - 2 k^{3} + 73 N + 
       100 k\, N + 31 k^{2} N + 2 k^{3} N 
       \nonu\\ & + &97 N^{2} + 97 k\, 
      N^{2} + 18 k^{2} N^{2} + 48 N^{3} + 26 k\, 
      N^{3} + 8 N^{4}), \nonu\\
c_ {-157} & = & - \frac {8 (1 + N) (3 + k + 2 N) (32 + 55 k + 
       18 k^{2} + 41 N + 35 k\, 
      N + 11 N^{2})} {3 (2 + N) (2 + k + N)^{6}}, \nonu\\
c_ {-158} & = &\frac {8 (1 + N) (3 + k + 2 N) (32 + 55 k + 18 k^{2} + 
      41 N + 35 k\, N + 11 N^{2})} {3 (2 + N) (2 + k + N)^{6}}, \nonu\\
c_ {-159} & = & - \frac {1} {3 (2 + N) (2 + k + N)^{5}}4 (132 + 135 k + 39 k^{2} + 2 k^{3} + 
       279 N + 200 k\, 
      N + 29 k^{2} N 
      \nonu\\ & - & 2 k^{3} N + 211 N^{2} + 87 k\, 
      N^{2} + 2 k^{2} N^{2} + 68 N^{3} + 10 k\, 
      N^{3} + 8 N^{4}), \nonu\\
c_ {-160} & = & - \frac{1} {3 (2 + N) (2 + k + N)^{5}}4 (204 + 493 k + 418 k^{2} + 147 k^{3} + 
       18 k^{4} + 461 N + 838 k\, 
      N 
      \nonu\\ & + & 482 k^{2} N + 87 k^{3} N + 382 N^{2} + 464 k\, 
      N^{2} + 136 k^{2} N^{2} + 137 N^{3} + 83 k\, 
      N^{3} + 18 N^{4}), \nonu\\
c_ {-161} & = &\frac {8 (1 + k) (3 + k + 2 N) (32 + 55 k + 18 k^{2} + 
      41 N + 35 k\, N + 11 N^{2})} {3 (2 + N) (2 + k + N)^{6}}, \nonu\\
c_ {-162} & = & - \frac {1}{3 (2 + N) (2 + k + N)^{5}}4 (276 + 365 k + 56 k^{2} - 63 k^{3} - 
       18 k^{4} + 649 N + 716 k\, 
      N 
      \nonu\\ & + & 118 k^{2} N - 27 k^{3} N + 614 N^{2} + 514 k\, 
      N^{2} + 62 k^{2} N^{2} + 265 N^{3} + 127 k\, 
      N^{3} + 42 N^{4}), \nonu\\
c_ {-163} & = & - \frac {8 (3 + k + 2 N) (32 + 45 k + 14 k^{2} + 
       83 N + 83 k\, N + 16 k^{2} N + 63 N^{2} + 34 k\, 
      N^{2} + 14 N^{3})} {3 (2 + N) (2 + k + N)^{5}}, \nonu\\
c_ {-164} & = & - \frac {4 (3 + k + 2 N) (32 + 55 k + 18 k^{2} + 
       41 N + 35 k\, N + 11 N^{2})} {3 (2 + N) (2 + k + N)^{5}},\nonu\\ 
c_ {-165} & = &\frac {1} {3 (2 + N) (2 + k + N)^{5}}8 (168 + 307 k + 171 k^{2} + 30 k^{3} + 425 N + 
      604 k\, N + 237 k^{2} N
      \nonu\\ & + & 24 k^{3} N + 413 N^{2} + 407 k\, 
     N^{2} + 84 k^{2} N^{2} + 178 N^{3} + 92 k\, 
     N^{3} + 28 N^{4}), \nonu\\
c_ {-166} & = & - \frac{1}{3 (2 + N) (2 + k + N)^{4}}8 (24 + 51 k + 31 k^{2} + 6 k^{3} + 39 N + 
       87 k\, N + 44 k^{2} N + 6 k^{3} N 
       \nonu\\ & + &29 N^{2} + 58 k\, 
      N^{2} + 17 k^{2} N^{2} + 12 N^{3} + 14 k\, 
      N^{3} + 2 N^{4}), \nonu\\
c_ {-167} & = & - \frac {8 (3 + k + 2 N) (32 + 55 k + 18 k^{2} + 
       41 N + 35 k\, N + 11 N^{2})} {3 (2 + N) (2 + k + N)^{5}},\nonu\\ 
c_ {-168} & = &\frac {8 (3 + k + 2 N) (32 + 55 k + 18 k^{2} + 41 N + 
      35 k\, N + 11 N^{2})} {3 (2 + N) (2 + k + N)^{5}}, \nonu\\
c_ {-169} & = & - \frac {1}{3 (2 + N) (2 + k + N)^{4}}8 (-24 - 13 k - 5 k^{2} - 2 k^{3} + 13 N + 
       73 k\, N + 30 k^{2} N + 2 k^{3} N 
       \nonu\\ & + & 58 N^{2} + 93 k\, 
      N^{2} + 19 k^{2} N^{2} + 35 N^{3} + 27 k\, 
      N^{3} + 6 N^{4}), \nonu\\
c_ {-170} & = &\frac {4 (48 + 87 k + 43 k^{2} + 6 k^{3} + 81 N + 
      104 k\, N + 27 k^{2} N + 45 N^{2} + 31 k\, 
     N^{2} + 8 N^{3})} {3 (2 + N) (2 + k + N)^{4}}, \nonu\\
c_ {-171} & = &\frac {8 (3 + k + 2 N) (32 + 55 k + 18 k^{2} + 41 N + 
      35 k\, N + 11 N^{2})} {3 (2 + N) (2 + k + N)^{5}}, \nonu\\
c_ {-172} & = & - \frac {8 (3 + k + 2 N) (32 + 55 k + 18 k^{2} + 
       41 N + 35 k\, N + 11 N^{2})} {3 (2 + N) (2 + k + N)^{5}},\nonu\\ 
c_ {-173} & = &\frac {4 (96 + 65 k - 3 k^{2} - 6 k^{3} + 127 N + 
      44 k\, N - 7 k^{2} N + 55 N^{2} + 5 k\, 
     N^{2} + 8 N^{3})} {3 (2 + N) (2 + k + N)^{4}}, \nonu\\
c_ {-174} & = & - \frac {8 (3 + k + 2 N) (16 + 21 k + 6 k^{2} + 
       19 N + 13 k\, N + 5 N^{2})} {3 (2 + N) (2 + k + N)^{4}}, \nonu\\
c_ {-175} & = &\frac {8 (3 + k + 2 N) (20 + 28 k + 8 k^{2} + 54 N + 
      55 k\, N + 10 k^{2} N + 41 N^{2} + 23 k\, 
     N^{2} + 9 N^{3})} {3 (2 + N) (2 + k + N)^{5}}, \nonu\\
c_ {-176} & = & - \frac {8 (3 + k + 2 N) (20 + 28 k + 8 k^{2} + 
       54 N + 55 k\, N + 10 k^{2} N + 41 N^{2} + 23 k\, 
      N^{2} + 9 N^{3})} {3 (2 + N) (2 + k + N)^{5}}, \nonu\\
c_ {-177} & = & - \frac {16 (24 + 55 k + 33 k^{2} + 6 k^{3} + 53 N + 
       76 k\, N + 22 k^{2} N + 35 N^{2} + 25 k\, 
      N^{2} + 7 N^{3})} {3 (2 + N) (2 + k + N)^{4}}, \nonu\\
c_ {-178} & = & - \frac{1}{3 (2 + N) (2 + k + N)^{5}}8 (1 + N) (60 + 120 k + 68 k^{2} + 
       12 k^{3} + 126 N + 163 k\, N 
       \nonu\\ & + & 45 k^{2} N + 81 N^{2} + 53 k\, 
      N^{2} + 16 N^{3}), \nonu\\
c_ {-179} & = &\frac {4 (108 + 209 k + 113 k^{2} + 18 k^{3} + 205 N + 
      266 k\, N + 73 k^{2} N + 125 N^{2} + 83 k\, 
     N^{2} + 24 N^{3})} {3 (2 + N) (2 + k + N)^{4}}, \nonu\\
c_ {-180} & = & - \frac {16 (3 + 2 k + N) (3 + k + 
       2 N)} {3 (2 + k + N)^{4}}, \nonu\\
c_ {-181} & = &\frac {1}{3 (2 + N) (2 + k + N)^{5}}8 (108 + 185 k + 93 k^{2} + 14 k^{3} + 313 N + 
      433 k\, N + 154 k^{2} N 
      \nonu\\ & + & 12 k^{3} N + 404 N^{2} + 441 k\, 
     N^{2} + 107 k^{2} N^{2} + 4 k^{3} N^{2} + 273 N^{3} + 215 k\, 
     N^{3} + 28 k^{2} N^{3} 
     \nonu\\ & + & 92 N^{4} + 40 k\, 
     N^{4} + 12 N^{5}), \nonu\\
c_ {-182} & = & - \frac {1}{3 (2 + N) (2 + k + N)^{6}}8 (1 + N) (3 + k + 2 N) (4 + 7 k + 2 k^{2} + 
       19 N + 21 k\, N 
       \nonu\\ & + & 4 k^{2} N + 17 N^{2} + 10 k\, 
      N^{2} + 4 N^{3}), \nonu\\
c_ {-183} & = &\frac{1}{3 (2 + N) (2 + k + N)^{6}}16 (3 + k + 2 N) (28 + 47 k + 24 k^{2} + 
      4 k^{3} + 67 N + 84 k\,N 
     \nonu\\ & + & 27 k^{2} N + 2 k^{3} N + 66 N^{2} + 57 k\, 
     N^{2} + 9 k^{2} N^{2} + 30 N^{3} + 14 k\, 
     N^{3} + 5 N^{4}), \nonu\\
c_ {-184} & = & - \frac {16 (9 + 11 k + 3 k^{2} + 7 N + 5 k\, 
      N + N^{2})} {3 (2 + k + N)^{4}}, \nonu\\
c_ {-185} & = &\frac {1}{3 (2 + N) (2 + k + N)^{5}}16 (48 + 63 k + 23 k^{2} + 2 k^{3} + 183 N + 
      232 k\, N + 87 k^{2} N 
      \nonu\\ & + & 10 k^{3} N + 258 N^{2} + 277 k\, 
     N^{2} + 81 k^{2} N^{2} + 6 k^{3} N^{2} + 177 N^{3} + 140 k\, 
     N^{3} + 23 k^{2} N^{3} + 60 N^{4}
     \nonu\\ & + & 26 k\, 
     N^{4} + 8 N^{5}), \nonu\\
c_ {-186} & = &\frac {16 (1 + N) (3 + k + 2 N)} {3 (2 + k + N)^{4}}, \qquad
c_ {-187}  = \frac {8 (3 + 3 k + k^{2} + 3 N + k\, 
     N + N^{2})} {3 (2 + k + N)^{4}}, \nonu\\
c_ {-188} & = & - \frac {8 (3 + 2 k + N) (3 + k + 
       2 N)} {3 (2 + k + N)^{3}}, \qquad
c_ {-189}  = \frac {8 (1 + N) (3 + k + 2 N)} {3 (2 + k + N)^{4}}, \nonu\\
c_ {-190} & = & - \frac {16 (3 + k + 2 N) (26 + 38 k + 12 k^{2} + 
       33 N + 24 k\, N + 9 N^{2})} {3 (2 + N) (2 + k + N)^{4}}, \nonu\\
c_ {-191} & = & - \frac{1}{3 (2 + N) (2 + k + N)^{5}}8 (108 + 185 k + 93 k^{2} + 14 k^{3} + 
       313 N + 433 k\, 
      N + 154 k^{2} N 
      \nonu\\ & + & 12 k^{3} N + 404 N^{2} + 441 k\, 
      N^{2} + 107 k^{2} N^{2} + 4 k^{3} N^{2} + 273 N^{3} + 215 k\, 
      N^{3} + 28 k^{2} N^{3} 
      \nonu\\ & + & 92 N^{4} + 40 k\, 
      N^{4} + 12 N^{5}),\nonu\\ 
c_ {-192} & = & - \frac {8 (3 + k + 2 N) (32 + 55 k + 18 k^{2} + 
       41 N + 35 k\, N + 11 N^{2})} {3 (2 + N) (2 + k + N)^{5}},\nonu\\ 
c_ {-193} & = & - \frac {1}{3 (2 + N) (2 + k + N)^{6}}8 (1 + N) (3 + k + 2 N) (4 + 7 k + 2 k^{2} + 
       19 N + 21 k\, N + 4 k^{2} N 
       \nonu\\ & + & 17 N^{2} + 10 k\, 
      N^{2} + 4 N^{3}), \nonu\\
c_ {-194} & = & - \frac {4 (3 + k + 2 N) (36 + 55 k + 18 k^{2} + 
       47 N + 35 k\, N + 13 N^{2})} {3 (2 + N) (2 + k + N)^{3}},\nonu\\ 
c_ {-195} & = & - \frac {8 (96 + 135 k + 55 k^{2} + 6 k^{3} + 153 N + 
       152 k\, N + 33 k^{2} N + 81 N^{2} + 43 k\, 
      N^{2} + 14 N^{3})} {3 (2 + N) (2 + k + N)^{4}}, \nonu\\
c_ {-196} & = &\frac {8 (3 + k + 2 N) (32 + 55 k + 18 k^{2} + 41 N + 
      35 k\, N + 11 N^{2})} {3 (2 + N) (2 + k + N)^{5}}, \nonu\\
c_ {-197} & = & - \frac {8 (60 + 153 k + 97 k^{2} + 18 k^{3} + 
       141 N + 214 k\, N + 65 k^{2} N + 97 N^{2} + 71 k\, 
      N^{2} + 20 N^{3})} {3 (2 + N) (2 + k + N)^{4}}, \nonu\\
c_ {-198} & = &\frac {1}{3 (2 + N) (2 + k + N)^{5}}8 (108 + 200 k + 112 k^{2} + 20 k^{3} + 274 N + 
      395 k\, N + 155 k^{2} N 
      \nonu\\ & + & 16 k^{3} N + 267 N^{2} + 268 k\, 
     N^{2} + 55 k^{2} N^{2} + 115 N^{3} + 61 k\, 
     N^{3} + 18 N^{4}), \nonu\\
c_ {-199} & = &\frac {1}{3 (2 + N) (2 + k + N)^{5}}8 (108 + 310 k + 307 k^{2} + 125 k^{3} + 
      18 k^{4} + 260 N + 543 k\, 
     N 
     \nonu\\ & + & 358 k^{2} N + 73 k^{3} N + 212 N^{2} + 293 k\, 
     N^{2} + 99 k^{2} N^{2} + 70 N^{3} + 48 k\, 
     N^{3} + 8 N^{4}), \nonu\\
c_ {-200} & = & - \frac {1}{3 (2 + N) (2 + k + N)^{4}}4 (216 + 200 k - 53 k^{2} - 85 k^{3} - 
       18 k^{4} + 532 N + 392 k\, 
      N 
      \nonu\\ & - & 57 k^{2} N - 53 k^{3} N + 477 N^{2} + 257 k\, 
      N^{2} - 12 k^{2} N^{2} + 185 N^{3} + 57 k\, 
      N^{3} + 26 N^{4}), \nonu\\
c_ {-201} & = & - \frac{1} {3 (2 + N) (2 + k + N)^{4}}8 (144 + 215 k + 99 k^{2} + 14 k^{3} + 
       397 N + 464 k\, 
      N + 151 k^{2} N 
      \nonu\\ & + &  12 k^{3} N + 439 N^{2} + 387 k\, 
      N^{2} + 84 k^{2} N^{2} + 4 k^{3} N^{2} + 246 N^{3} + 152 k\, 
      N^{3} + 18 k^{2} N^{3} 
      \nonu\\ & + & 70 N^{4} 
      +  24 k\, 
      N^{4} + 8 N^{5}), \nonu\\
c_ {-202} & = & - \frac{1}{3 (2 + N) (2 + k + N)^{5}}8 (108 + 191 k + 95 k^{2} + 14 k^{3} + 
       235 N + 269 k\, 
      N + 62 k^{2} N 
      \nonu\\ & - &  2 k^{3} N + 170 N^{2} + 99 k\, 
      N^{2} - k^{2} N^{2} + 47 N^{3} + 5 k\, 
      N^{3} + 4 N^{4}), \nonu\\
c_ {-203} & = & - \frac {1}{3 (2 + N) (2 + k + N)^{5}}16 (-24 - 53 k - 37 k^{2} - 8 k^{3} + 47 N + 
       62 k\, N + 27 k^{2} N 
       \nonu\\ & + & 5 k^{3} N + 164 N^{2} + 195 k\, 
      N^{2} + 66 k^{2} N^{2} + 6 k^{3} N^{2} + 149 N^{3} + 127 k\, 
      N^{3} + 23 k^{2} N^{3} 
      \nonu\\ & + & 57 N^{4} 
     +26 k\, N^{4} + 8 N^{5}), \nonu\\
c_ {-204} & = & - \frac{1}{3 (2 + N) (2 + k + N)^{4}}8 (-33 k - 37 k^{2} - 10 k^{3} + 45 N + 
       46 k\, N + 17 k^{2} N + 4 k^{3} N 
       \nonu\\ & + &  93 N^{2} + 101 k\, 
      N^{2} + 26 k^{2} N^{2} + 60 N^{3} + 36 k\, 
      N^{3} + 12 N^{4}), \nonu\\
c_ {-205} & = & - \frac {1}{3 (2 + N) (2 + k + N)^{5}}4 (-192 - 176 k + 69 k^{2} + 89 k^{3} + 
       18 k^{4} - 304 N - 40 k\, 
      N 
      \nonu\\ & + &  233 k^{2} N + 79 k^{3} N - 125 N^{2} + 151 k\, 
      N^{2} + 118 k^{2} N^{2} + 7 N^{3} + 65 k\, 
      N^{3} + 8 N^{4}), \nonu\\
c_ {-206} & = & - \frac {8 (15 + 13 k + 3 k^{2} + 17 N + 7 k\, 
      N + 5 N^{2})} {3 (2 + k + N)^{4}}, \nonu\\
c_ {-207} & = & - \frac {1}{3 (2 + N) (2 + k + N)^{4}}4 (-12 - 17 k - 9 k^{2} - 2 k^{3} - 13 N + 
       30 k\, N + 35 k^{2} N + 8 k^{3} N 
       \nonu\\ & + &  27 N^{2} + 73 k\, 
      N^{2} + 28 k^{2} N^{2} + 32 N^{3} + 28 k\, 
      N^{3} + 8 N^{4}), \nonu\\
c_ {-208} & = & - \frac{1}{3 (2 + N) (2 + k + N)^{5}}8 (156 + 209 k + 85 k^{2} + 10 k^{3} + 
       385 N + 412 k\, 
      N + 121 k^{2} N 
      \nonu\\ & + &  8 k^{3} N + 349 N^{2} + 265 k\, 
      N^{2} + 42 k^{2} N^{2} + 138 N^{3} + 56 k\, 
      N^{3} + 20 N^{4}), \nonu\\
c_ {-209} & = &\frac {8 (1 + N) (3 + k + 2 N) (32 + 55 k + 18 k^{2} + 
      41 N + 35 k\, N + 11 N^{2})} {3 (2 + N) (2 + k + N)^{6}}, \nonu\\
      c_ {-210} & = &\frac {16 (1 + N) (33 + 33 k + 8 k^{2} + 33 N + 17 k\, 
     N + 8 N^{2})} {3 (2 + k + N)^{5}}, \nonu\\
   c_ {-211} & = &\frac{1} {3 (2 + N) (2 + k + N)^{6}}16 (1 + N) (60 + 105 k + 57 k^{2} + 
         10 k^{3} + 153 N + 208 k\, 
        N 
        \nonu\\ & + &  79 k^{2} N + 8 k^{3} N + 149 N^{2} + 141 k\, 
        N^{2} + 28 k^{2} N^{2} + 64 N^{3} + 32 k\, 
        N^{3} + 10 N^{4}), \nonu\\
   c_ {-212} & = & - \frac{1}{3 (2 + N) (2 + k + N)^{6}}16 (72 + 95 k + 21 k^{2} - 12 k^{3} - 
          4 k^{4} + 277 N + 350 k\, 
         N 
         \nonu\\ & + &  112 k^{2} N + k^{3} N - 2 k^{4} N + 397 N^{2} + 418 k\, 
         N^{2} + 110 k^{2} N^{2} + 5 k^{3} N^{2} + 274 N^{3} + 
          209 k\, N^{3} 
          \nonu\\ & + & 31 k^{2} N^{3} + 92 N^{4} + 38 k\, 
         N^{4} + 12 N^{5}), \nonu\\
   c_ {-213} & = & - \frac {1}{3 (2 + N) (2 + k + N)^{5}}4 (528 + 864 k + 453 k^{2} + 81 k^{3} + 
          2 k^{4} + 1608 N + 2068 k\, 
         N 
         \nonu\\ & + & 733 k^{2} N + 43 k^{3} N - 8 k^{4} N + 1943 N^{2} + 
          1863 k\, 
         N^{2} + 394 k^{2} N^{2} - 4 k^{3} N^{2} + 1163 N^{3} 
         \nonu\\ & + &
          753 k\, N^{3} + 72 k^{2} N^{3} + 344 N^{4} + 116 k\, 
         N^{4} + 40 N^{5}), \nonu\\
   c_ {-214} & = & - \frac{1} {3 (2 + N) (2 + k + N)^{4}}8 (156 + 200 k + 76 k^{2} + 8 k^{3} + 
          310 N + 335 k\, 
         N + 105 k^{2} N 
         \nonu\\ & + & 10 k^{3} N + 249 N^{2} + 206 k\, 
         N^{2} + 39 k^{2} N^{2} + 95 N^{3} + 45 k\, 
         N^{3} + 14 N^{4}), \nonu\\
   c_ {-215} & = & - \frac {8 (3 + k + 2 N) (5 + 2 k + 
          3 N)} {3 (2 + k + N)^{4}}, \nonu\\
   c_ {-216} & = & - \frac{1}{3 (2 + N) (2 + k + N)^{5}}8 (24 + 38 k + 12 k^{2} + 58 N + 88 k\, 
         N + 19 k^{2} N - 2 k^{3} N + 98 N^{2} 
         \nonu\\ & + & 140 k\, 
         N^{2} + 37 k^{2} N^{2} + 2 k^{3} N^{2} + 99 N^{3} + 104 k\, 
         N^{3} + 18 k^{2} N^{3} + 47 N^{4} + 26 k\, 
         N^{4} + 8 N^{5}), \nonu\\
   c_ {-217} & = & - \frac {4 (3 + 2 k + N)} {3 (2 + k + N)}, \nonu\\
   c_ {-218} & = &\frac {1} {3 (2 + N) (2 + k + N)^{3}}4 (3 + 2 k + N) (60 + 77 k + 22 k^{2} + 
         121 N + 115 k\, N + 20 k^{2} N 
         \nonu\\ & + & 79 N^{2} + 42 k\, 
        N^{2} + 16 N^{3}), \nonu\\
   c_ {-219} & = &\frac{1} {3 (2 + N) (2 + k + N)^{3}}4 (3 + 2 k + N) (60 + 77 k + 22 k^{2} + 
         121 N + 115 k\, N + 20 k^{2} N 
         \nonu\\ & + &79 N^{2} + 42 k\, 
        N^{2} + 16 N^{3}), \nonu\\
   c_ {-220} & = &\frac{1}{3 (2 + N) (2 + k + N)^{5}}4 (-192 - 208 k + 69 k^{2} + 136 k^{3} + 
         36 k^{4} - 272 N - 56 k\, 
        N 
        \nonu\\ & + & 250 k^{2} N + 110 k^{3} N - 109 N^{2} + 104 k\, 
        N^{2} + 107 k^{2} N^{2} - 10 N^{3} + 34 k\, 
        N^{3} + N^{4}), \nonu\\
   c_ {-221} & = & - \frac {8 (54 + 103 k + 62 k^{2} + 12 k^{3} + 
          80 N + 106 k\, N + 33 k^{2} N + 36 N^{2} + 25 k\, 
         N^{2} + 5 N^{3})} {3 (2 + N) (2 + k + N)^{4}}, \nonu\\
   c_ {-222} & = & - \frac {8 (3 + 2 k + N) (32 + 55 k + 18 k^{2} + 
          41 N + 35 k\, N + 11 N^{2})} {3 (2 + N) (2 + k + N)^{5}}, \nonu\\
   c_ {-223} & = & - \frac{1}{3 (2 + N) (2 + k + N)^{5}}8 (168 + 243 k + 73 k^{2} - 28 k^{3} - 
          12 k^{4} + 393 N + 489 k\, 
         N 
         \nonu\\ & + & 156 k^{2} N + 2 k^{3} N + 338 N^{2} + 313 k\, 
         N^{2} + 65 k^{2} N^{2} + 123 N^{3} + 61 k\, 
         N^{3} + 16 N^{4}), \nonu\\
   c_ {-224} & = &\frac {8 (1 + k) (3 + 2 k + N) (32 + 55 k + 
         18 k^{2} + 41 N + 35 k\, 
        N + 11 N^{2})} {3 (2 + N) (2 + k + N)^{6}}, \nonu\\
   c_ {-225} & = &\frac {8 (3 + 2 k + N) (32 + 55 k + 18 k^{2} + 
         41 N + 35 k\, N + 11 N^{2})} {3 (2 + N) (2 + k + N)^{5}},\nonu\\ 
   c_ {-226} & = &\frac {8 (1 + N) (3 + 2 k + N) (32 + 55 k + 
         18 k^{2} + 41 N + 35 k\, 
        N + 11 N^{2})} {3 (2 + N) (2 + k + N)^{6}}, \nonu\\
   c_ {-227} & = & - \frac{1}{3 (2 + N) (2 + k + N)^{5}}16 (114 + 255 k + 219 k^{2} + 84 k^{3} + 
          12 k^{4} + 288 N + 474 k\, 
         N 
         \nonu\\ & + &  266 k^{2} N + 50 k^{3} N + 252 N^{2} + 275 k\, 
         N^{2} + 77 k^{2} N^{2} + 92 N^{3} + 50 k\, 
         N^{3} + 12 N^{4}), \nonu\\
   c_ {-228} & = &\frac {8 (1 + k) (3 + 2 k + N) (32 + 55 k + 
         18 k^{2} + 41 N + 35 k\, 
        N + 11 N^{2})} {3 (2 + N) (2 + k + N)^{6}}, \nonu\\
   c_ {-229} & = &\frac{1} {3 (2 + N) (2 + k + N)^{4}}8 (-48 - 45 k - 10 k^{2} - 69 N - 17 k\, 
        N + 21 k^{2} N + 6 k^{3} N - 30 N^{2} 
        \nonu\\ & + &  19 k\, 
        N^{2} + 14 k^{2} N^{2} - 4 N^{3} + 7 k\, 
        N^{3}), \nonu\\
   c_ {-230} & = & - \frac {4 (3 + 2 k + N) (32 + 55 k + 18 k^{2} + 
          41 N + 35 k\, N + 11 N^{2})} {3 (2 + N) (2 + k + N)^{5}},\nonu\\ 
   c_ {-231} & = &\frac {8 (3 + 2 k + N) (32 + 55 k + 18 k^{2} + 
         41 N + 35 k\, N + 11 N^{2})} {3 (2 + N) (2 + k + N)^{5}}, \nonu\\
   c_ {-232} & = & - \frac {8 (3 + 2 k + N) (32 + 55 k + 18 k^{2} + 
          41 N + 35 k\, N + 11 N^{2})} {3 (2 + N) (2 + k + N)^{5}},\nonu\\ 
   c_ {-233} & = &\frac {1} {3 (2 + N) (2 + k + N)^{5}}4 (108 + 181 k + 104 k^{2} + 20 k^{3} + 
         245 N + 320 k\, 
        N + 134 k^{2} N 
        \nonu\\ & + & 16 k^{3} N + 206 N^{2} + 185 k\, 
        N^{2} + 42 k^{2} N^{2} + 75 N^{3} + 34 k\, 
        N^{3} + 10 N^{4}), \nonu\\
        c_ {-234} & = &\frac{1} {3 (2 + N) (2 + k + N)^{5}}4 (252 + 461 k + 272 k^{2} + 52 k^{3} + 
         469 N + 684 k\, 
        N + 290 k^{2} N 
        \nonu\\ & + & 32 k^{3} N + 322 N^{2} + 337 k\, 
        N^{2} + 78 k^{2} N^{2} + 95 N^{3} + 54 k\, 
        N^{3} + 10 N^{4}), \nonu\\
c_ {-235} & = & - \frac {4 (3 + 2 k + N) (32 + 55 k + 18 k^{2} + 
          41 N + 35 k\, N + 11 N^{2})} {3 (2 + N) (2 + k + N)^{5}},\nonu\\ 
c_ {-236} & = &\frac{1} {3 (2 + N) (2 + k + N)^{4}}8 (-24 + 3 k + 24 k^{2} + 8 k^{3} - 33 N + 
         39 k\, N + 50 k^{2} N + 10 k^{3} N 
         \nonu\\ & - &  12 N^{2} + 39 k\, 
        N^{2} + 20 k^{2} N^{2} - N^{3} + 9 k\, 
        N^{3}), \nonu\\
c_ {-237} & = &\frac {1}{3 (2 + N) (2 + k + N)^{4}}8 (264 + 381 k + 141 k^{2} - 16 k^{3} - 
         12 k^{4} + 531 N + 595 k\, 
        N 
        \nonu\\ & + &  162 k^{2} N - 4 k^{3} N + 398 N^{2} + 311 k\, 
        N^{2} + 49 k^{2} N^{2} + 131 N^{3} + 53 k\, 
        N^{3} + 16 N^{4}), \nonu\\
c_ {-238} & = & - \frac {8 (3 + 2 k + N) (32 + 55 k + 18 k^{2} + 
          41 N + 35 k\, N + 11 N^{2})} {3 (2 + N) (2 + k + N)^{5}},\nonu\\ 
c_ {-239} & = &\frac {4 (48 + 87 k + 56 k^{2} + 12 k^{3} + 81 N + 
         94 k\, N + 30 k^{2} N + 42 N^{2} + 23 k\, 
        N^{2} + 7 N^{3})} {3 (2 + N) (2 + k + N)^{4}},\nonu\\
c_ {-240} & = & - \frac {4 (96 + 49 k - 24 k^{2} - 12 k^{3} + 143 N + 
       46 k\, N - 14 k^{2} N + 74 N^{2} + 13 k\, 
      N^{2} + 13 N^{3})} {3 (2 + N) (2 + k + N)^{4}}, \nonu\\
c_ {-241} & = &\frac {8 (3 + 2 k + N) (16 + 21 k + 6 k^{2} + 19 N + 
      13 k\, N + 5 N^{2})} {3 (2 + N) (2 + k + N)^{4}}, \nonu\\
c_ {-242} & = & - \frac{1}{3 (2 + N) (2 + k + N)^{5}}8 (3 + 2 k + N) (20 + 2 k - 17 k^{2} - 
       6 k^{3} + 48 N + 21 k\, N - 5 k^{2} N 
       \nonu\\ & + &  36 N^{2} + 13 k\, 
      N^{2} + 8 N^{3}), \nonu\\
c_ {-243} & = & - \frac {8 (3 + 2 k + N) (32 + 55 k + 18 k^{2} + 
       41 N + 35 k\, N + 11 N^{2})} {3 (2 + N) (2 + k + N)^{5}},\nonu\\ 
c_ {-244} & = &\frac {16 (24 + 71 k + 54 k^{2} + 12 k^{3} + 37 N + 
      74 k\, N + 29 k^{2} N + 16 N^{2} + 17 k\, 
     N^{2} + 2 N^{3})} {3 (2 + N) (2 + k + N)^{4}}, \nonu\\
c_ {-245} & = &\frac{1} {3 (2 + N) (2 + k + N)^{5}}8 (3 + 2 k + N) (20 + 2 k - 17 k^{2} - 
      6 k^{3} + 48 N + 21 k\, N - 5 k^{2} N 
      \nonu\\ & + &  36 N^{2} + 13 k\, 
     N^{2} + 8 N^{3}), \nonu\\
c_ {-246} & = & - \frac{1}{3 (2 + N) (2 + k + N)^{5}}8 (60 + 62 k - 31 k^{2} - 48 k^{3} - 
       12 k^{4} + 148 N + 169 k\, 
      N + 22 k^{2} N 
      \nonu\\ & - &  14 k^{3} N + 132 N^{2} + 128 k\, 
      N^{2} + 23 k^{2} N^{2} + 48 N^{3} + 27 k\, 
      N^{3} + 6 N^{4}), \nonu\\
c_ {-247} & = &\frac {4 (-12 + 23 k + 40 k^{2} + 12 k^{3} - 5 N + 
      30 k\, N + 22 k^{2} N + 2 N^{2} + 7 k\, 
     N^{2} + N^{3})} {3 (2 + N) (2 + k + N)^{4}}, \nonu\\
c_ {-248} & = &\frac {16 (3 + 2 k + N) (6 + 17 k + 6 k^{2} + 8 N + 
      11 k\, N + 2 N^{2})} {3 (2 + N) (2 + k + N)^{4}}, \nonu\\
c_ {-249} & = & - \frac {4 (-55 k - 52 k^{2} - 12 k^{3} + 7 N - 
       54 k\, N - 28 k^{2} N + 10 N^{2} - 11 k\, 
      N^{2} + 3 N^{3})} {3 (2 + N) (2 + k + N)^{4}}, \nonu\\
c_ {-250} & = &\frac {16 (3 + 2 k + N) (26 + 25 k + 6 k^{2} + 30 N + 
      15 k\, N + 8 N^{2})} {3 (2 + N) (2 + k + N)^{4}}, \nonu\\
c_ {-251} & = &\frac{1} {3 (2 + N) (2 + k + N)^{5}}8 (12 + 229 k + 359 k^{2} + 196 k^{3} + 
      36 k^{4} + 29 N + 541 k\, 
     N 
     \nonu\\ & + &  694 k^{2} N + 290 k^{3} N + 36 k^{4} N + 30 N^{2} + 459 k\, 
     N^{2} + 417 k^{2} N^{2} + 96 k^{3} N^{2} + 13 N^{3}
     \nonu\\ & + &  161 k\, 
     N^{3} + 76 k^{2} N^{3} + 2 N^{4} + 20 k\, 
     N^{4}), \nonu\\
c_ {-252} & = & - \frac {8 (3 + 2 k + N) (32 + 55 k + 18 k^{2} + 
       41 N + 35 k\, N + 11 N^{2})} {3 (2 + N) (2 + k + N)^{5}},\nonu\\ 
c_ {-253} & = & - \frac{1}{3 (2 + N) (2 + k + N)^{5}}4 (84 + 231 k + 223 k^{2} + 88 k^{3} + 
       12 k^{4} + 207 N + 406 k\, 
      N 
      \nonu\\ & + &  260 k^{2} N + 52 k^{3} N + 181 N^{2} + 227 k\, 
      N^{2} + 73 k^{2} N^{2} + 67 N^{3} + 40 k\, 
      N^{3} + 9 N^{4}), \nonu\\
c_ {-254} & = &\frac {8 (1 + k) (3 + 2 k + N) (32 + 55 k + 18 k^{2} + 
      41 N + 35 k\, N + 11 N^{2})} {3 (2 + N) (2 + k + N)^{6}}, \nonu\\
c_ {-255} & = & - \frac {1}{3 (2 + N) (2 + k + N)^{5}}4 (84 + 293 k + 193 k^{2} + 8 k^{3} - 
       12 k^{4} + 241 N + 676 k\, 
      N 
      \nonu\\ & + &  380 k^{2} N + 44 k^{3} N + 229 N^{2} + 461 k\, 
      N^{2} + 151 k^{2} N^{2} + 81 N^{3} + 90 k\, 
      N^{3} + 9 N^{4}), \nonu\\
c_ {-256} & = & - \frac{1}{3 (2 + N) (2 + k + N)^{6}}8 (1 + k) (3 + 2 k + N) (36 + 39 k + 
       10 k^{2} + 67 N + 53 k\, N 
       \nonu\\ & + &  8 k^{2} N + 41 N^{2} + 18 k\, 
      N^{2} + 8 N^{3}), \nonu\\
c_ {-257} & = &\frac {8 (1 + N) (3 + 2 k + N) (32 + 55 k + 18 k^{2} + 
      41 N + 35 k\, N + 11 N^{2})} {3 (2 + N) (2 + k + N)^{6}}, \nonu\\
c_ {-258} & = & - \frac{1}{3 (2 + N) (2 + k + N)^{6}}16 (3 + 2 k + N) (68 + 129 k + 101 k^{2} + 
       39 k^{3} + 6 k^{4} + 157 N 
       \nonu\\ & + &  212 k\, 
      N + 103 k^{2} N + 19 k^{3} N + 137 N^{2} + 118 k\, 
      N^{2} + 26 k^{2} N^{2} + 54 N^{3} + 23 k\, 
      N^{3} + 8 N^{4}), \nonu\\
c_ {-259} & = & - \frac {1}{3 (2 + N) (2 + k + N)^{6}}16 (1 + N) (168 + 405 k + 376 k^{2} + 
       156 k^{3} + 24 k^{4} + 327 N 
       \nonu\\ & + &  583 k\, 
      N + 361 k^{2} N + 76 k^{3} N + 229 N^{2} + 265 k\, 
      N^{2} + 81 k^{2} N^{2} + 70 N^{3} + 39 k\, 
      N^{3} + 8 N^{4}), \nonu\\
c_ {-260} & = & - \frac {4 (3 + 2 k + N) (4 - 13 k - 6 k^{2} + 3 N - 
       9 k\, N + N^{2})} {3 (2 + N) (2 + k + N)^{3}}, \nonu\\
c_ {-261} & = & - \frac {8 (60 + 17 k - 32 k^{2} - 12 k^{3} + 85 N + 
       14 k\, N - 18 k^{2} N + 42 N^{2} + 5 k\, 
      N^{2} + 7 N^{3})} {3 (2 + N) (2 + k + N)^{4}}, \nonu\\
c_ {-262} & = & - \frac {8 (3 + 2 k + N) (32 + 55 k + 18 k^{2} + 
       41 N + 35 k\, N + 11 N^{2})} {3 (2 + N) (2 + k + N)^{5}},\nonu\\ 
c_ {-263} & = &\frac {8 (96 + 135 k + 68 k^{2} + 12 k^{3} + 153 N + 
      142 k\, N + 36 k^{2} N + 78 N^{2} + 35 k\, 
     N^{2} + 13 N^{3})} {3 (2 + N) (2 + k + N)^{4}}, \nonu\\
c_ {-264} & = & - \frac {1}{3 (2 + N) (2 + k + N)^{5}}8 (-12 + 48 k + 60 k^{2} + 16 k^{3} + 6 N + 
       151 k\, N + 118 k^{2} N 
       \nonu\\ & + &  20 k^{3} N + 23 N^{2} + 114 k\, 
      N^{2} + 46 k^{2} N^{2} + 10 N^{3} + 23 k\, 
      N^{3} + N^{4}), \nonu\\
c_ {-265} & = &\frac{1}{3 (2 + N) (2 + k + N)^{5}}8 (12 - 2 k - 59 k^{2} - 52 k^{3} - 12 k^{4} + 
      44 N + 65 k\, N - 8 k^{2} N 
      \nonu\\ & - &  16 k^{3} N + 48 N^{2} + 72 k\, 
     N^{2} + 15 k^{2} N^{2} + 18 N^{3} + 17 k\, 
     N^{3} + 2 N^{4}), \nonu\\
c_ {-266} & = & - \frac{1}{3 (2 + N) (2 + k + N)^{4}}4 (-24 - 12 k + 39 k^{2} + 44 k^{3} 
+ 12 k^{4} - 48 N - 60 k\, N 
\nonu\\ & - &  6 k^{2} N + 12 k^{3} N - 27 N^{2} - 48 k\, 
      N^{2} - 17 k^{2} N^{2} - 2 N^{3} - 8 k\, 
      N^{3} + N^{4}), \nonu\\
c_ {-267} & = & - \frac {1}{3 (2 + N) (2 + k + N)^{4}}8 (-144 - 235 k - 98 k^{2} + 8 k^{3} + 
       8 k^{4} - 257 N - 226 k\, 
      N \nonu\\ & + &  72 k^{2} N 
      + 94 k^{3} N + 16 k^{4} N - 162 N^{2} - 7 k\, 
      N^{2} + 134 k^{2} N^{2} + 44 k^{3} N^{2} - 43 N^{3} + 38 k\, 
      N^{3} 
      \nonu\\ & + & 34 k^{2} N^{3}- 4 N^{4} + 8 k\, 
      N^{4}), \nonu\\
c_ {-268} & = & - \frac{1}{3 (2 + N) (2 + k + N)^{5}}8 (12 + 7 k + 39 k^{2} + 44 k^{3} + 
       12 k^{4} + 83 N + 141 k\, 
      N + 128 k^{2} N 
      \nonu\\ & + &  42 k^{3} N + 114 N^{2} + 141 k\, 
      N^{2} + 55 k^{2} N^{2} + 53 N^{3} + 33 k\, 
      N^{3} + 8 N^{4}), \nonu\\
c_ {-269} & = &\frac{1} {3 (2 + N) (2 + k + N)^{5}}16 (264 + 527 k + 409 k^{2} + 146 k^{3} + 
      20 k^{4} + 571 N + 866 k\, 
     N 
     \nonu\\ & + &  467 k^{2} N + 97 k^{3} N + 4 k^{4} N + 450 N^{2} + 461 k\, 
     N^{2} + 136 k^{2} N^{2} + 8 k^{3} N^{2} + 153 N^{3}
     \nonu\\ & + &  79 k\, 
     N^{3} + 3 k^{2} N^{3} + 19 N^{4}),\nonu\\ 
c_ {-270} & = & - \frac {8 (3 + 2 k + N) (48 + 55 k + 14 k^{2} + 
       81 N + 69 k\, N + 10 k^{2} N + 45 N^{2} + 22 k\, 
      N^{2} + 8 N^{3})} {3 (2 + N) (2 + k + N)^{5}}, \nonu\\
c_ {-271} & = &\frac {1}{3 (2 + N) (2 + k + N)^{4}}8 (45 k + 48 k^{2} + 12 k^{3} + 63 N + 148 k\, 
     N + 86 k^{2} N + 12 k^{3} N 
     \nonu\\ & + &  98 N^{2} + 117 k\, 
     N^{2} + 32 k^{2} N^{2} + 49 N^{3} + 26 k\, 
     N^{3} + 8 N^{4}), \nonu\\
c_ {-272} & = &\frac {8 (3 + 2 k + N) (26 + 25 k + 6 k^{2} + 30 N + 
      15 k\, N + 8 N^{2})} {3 (2 + N) (2 + k + N)^{3}}, \nonu\\
c_ {-273} & = & - \frac {8 (3 + 2 k + N) (18 + 21 k + 6 k^{2} + 
       22 N + 13 k\, N + 6 N^{2})} {3 (2 + N) (2 + k + N)^{4}}, \nonu\\
c_ {-274} & = & - \frac{1}{3 (2 + N) (2 + k + N)^{5}}8 (12 + 229 k + 359 k^{2} + 196 k^{3} + 
       36 k^{4} + 29 N + 541 k\, 
      N 
      \nonu\\ & + &  694 k^{2} N + 290 k^{3} N + 36 k^{4} N + 30 N^{2} + 459 k\, 
      N^{2} + 417 k^{2} N^{2} + 96 k^{3} N^{2} + 13 N^{3} 
      \nonu\\ & + &  161 k\, 
      N^{3} + 76 k^{2} N^{3} + 2 N^{4} + 20 k\, 
      N^{4}), \nonu\\
c_ {-275} & = & - \frac{1}{3 (2 + N) (2 + k + N)^{6}}8 (1 + k) (3 + 2 k + N) (36 + 39 k + 
       10 k^{2} + 67 N + 53 k\, N 
       \nonu\\ & + &  8 k^{2} N + 41 N^{2} + 18 k\, 
      N^{2} + 8 N^{3}), \nonu\\
c_ {-276} & = &\frac {16 (78 + 123 k + 66 k^{2} + 12 k^{3} + 120 N + 
      128 k\, N + 35 k^{2} N + 58 N^{2} + 31 k\, 
     N^{2} + 9 N^{3})} {3 (2 + N) (2 + k + N)^{4}}, \nonu\\
c_ {-277} & = & - \frac{1} {3 (2 + N) (2 + k + N)^{5}}16 (-48 - 129 k - 109 k^{2} - 36 k^{3} - 
       4 k^{4} - 57 N - 136 k\, 
      N 
      \nonu\\ & - &  67 k^{2} N + 2 k^{3} N + 4 k^{4} N - 10 N^{2} - 33 k\, 
      N^{2} + 3 k^{2} N^{2} + 8 k^{3} N^{2} + 7 N^{3} + 
       3 k^{2} N^{3} + 2 N^{4}), \nonu\\
c_ {-278} & = & - \frac {16 (1 + k) (3 + 2 k + 
       N)} {3 (2 + k + N)^{4}}, \nonu\\        
c_ {-279} & = & - \frac {8 (30 + 75 k + 54 k^{2} + 12 k^{3} + 48 N + 
       80 k\, N + 29 k^{2} N + 22 N^{2} + 19 k\, 
      N^{2} + 3 N^{3})} {3 (2 + N) (2 + k + N)^{4}}, \nonu\\
c_ {-280} & = & - \frac{1}{3 (2 + N) (2 + k + N)^{5}}4 (-72 - 224 k - 137 k^{2} + 4 k^{3} + 
       12 k^{4} - 28 N - 82 k\, 
      N 
      \nonu \\& + & 34 k^{2} N + 40 k^{3} N + 87 N^{2} + 104 k\, 
      N^{2} + 55 k^{2} N^{2} + 62 N^{3} + 38 k\, 
      N^{3} + 11 N^{4}), \nonu\\
c_ {-281} & = & - \frac{1} {3 (2 + N) (2 + k + N)^{5}}8 (72 + 185 k + 132 k^{2} + 28 k^{3} + 
       115 N + 274 k\, 
      N + 152 k^{2} N 
      \nonu \\& + & 20 k^{3} N + 62 N^{2} + 133 k\, 
      N^{2} + 44 k^{2} N^{2} + 11 N^{3} + 20 k\, 
      N^{3}), \nonu\\
c_ {-282} & = &\frac{1}{3 (2 + N) (2 + k + N)^{4}}4 (108 + 121 k + 40 k^{2} + 4 k^{3} + 245 N + 
      242 k\, N + 74 k^{2} N 
      \nonu \\& + & 8 k^{3} N + 198 N^{2} + 149 k\, 
     N^{2} + 28 k^{2} N^{2} + 67 N^{3} + 28 k\, 
     N^{3} + 8 N^{4}), \nonu\\
c_ {-283} & = & - \frac{1}{3 (2 + N) (2 + k + N)^{5}}8 (-36 - 59 k - 28 k^{2} - 4 k^{3} - 19 N - 
       16 k\, N + 8 k^{2} N + 4 k^{3} N 
       \nonu \\& + & 26 N^{2} + 29 k\, 
      N^{2} + 12 k^{2} N^{2} + 21 N^{3} + 10 k\, 
      N^{3} + 4 N^{4}), \nonu\\
c_ {-284} & = &\frac {1}{3 (2 + N) (2 + k + N)^{6}}16 (1 + k) (60 + 125 k + 80 k^{2} + 16 k^{3} + 
      133 N + 228 k\, 
     N 
     \nonu\\ & + & 110 k^{2} N 
    + 14 k^{3} N + 106 N^{2} + 133 k\, 
     N^{2} + 36 k^{2} N^{2} + 35 N^{3} + 24 k\, 
     N^{3} + 4 N^{4}), \nonu\\
c_ {-285} & = &\frac{1} {3 (2 + N) (2 + k + N)^{5}}4 (-48 + 8 k + 143 k^{2} + 120 k^{3} + 
      28 k^{4} + 64 N + 524 k\, 
     N 
     \nonu\\ & + &750 k^{2} N + 370 k^{3} N + 56 k^{4} N + 245 N^{2} + 752 k\, 
     N^{2} + 629 k^{2} N^{2} + 156 k^{3} N^{2} + 202 N^{3} 
     \nonu\\ & + & 346 k\, 
     N^{3} + 136 k^{2} N^{3} + 67 N^{4} + 52 k\, 
     N^{4} + 8 N^{5}), \nonu\\
c_ {-286} & = &\frac {1}{3 (2 + N) (2 + k + N)^{4}}8 (-36 + 106 k + 183 k^{2} + 84 k^{3} + 
      12 k^{4} + 20 N + 297 k\, 
     N 
     \nonu \\& + & 258 k^{2} N + 56 k^{3} N + 84 N^{2} + 216 k\, 
     N^{2} + 83 k^{2} N^{2} + 48 N^{3} + 45 k\, 
     N^{3} + 8 N^{4}), \nonu\\
c_ {-287} & = & - \frac {8 (3 + 2 k + N) (10 + 17 k + 6 k^{2} + 
       14 N + 11 k\, N + 4 N^{2})} {3 (2 + N) (2 + k + N)^{4}}, \nonu\\
c_ {-288} & = &\frac{1}{3 (2 + N) (2 + k + N)^{5}}8 (-24 + 14 k + 64 k^{2} + 42 k^{3} + 8 k^{4} + 
      10 N + 196 k\, 
     N + 255 k^{2} N 
     \nonu \\& + & 113 k^{3} N + 16 k^{4} N + 58 N^{2} + 238 k\, 
     N^{2} + 186 k^{2} N^{2} + 42 k^{3} N^{2} + 35 N^{3} + 89 k\,N^{3} 
     \nonu \\& + & 34 k^{2} N^{3} + 6 N^{4} + 10 k\,N^{4}), \nonu\\
c_ {-289} & = &\frac {(k - N)} {3 (2 + k + N)},\qquad
c_ {-290}  =  - 3,\nonu\\ 
c_ {-291} & = &\frac {3 (60 + 77 k + 22 k^{2} + 121 N + 115 k\, 
     N + 20 k^{2} N + 79 N^{2} + 42 k\, 
     N^{2} + 16 N^{3})} {(2 + N) (2 + k + N)^{2}},\nonu\\
 c_ {-292} & = & \frac {3 (60 + 77 k + 22 k^{2} + 121 N + 115 k\, 
     N + 20 k^{2} N + 79 N^{2} + 42 k\, 
     N^{2} + 16 N^{3})} {(2 + N) (2 + k + N)^{2}}, \nonu\\
c_ {-293} & = & \frac {(-k + N) (60 + 77 k + 22 k^{2} + 121 N + 
      115 k\, N + 20 k^{2} N + 79 N^{2} + 42 k\, 
     N^{2} + 16 N^{3})} {3 (2 + N) (2 + k + N)^{3}},\nonu\\ 
c_ {-294} & = & \frac {(-k + N) (60 + 77 k + 22 k^{2} + 121 N + 
      115 k\, N + 20 k^{2} N + 79 N^{2} + 42 k\, 
     N^{2} + 16 N^{3})} {3 (2 + N) (2 + k + N)^{3}},\nonu\\ 
c_ {-295} & = & \frac {2 (-k + N) (60 + 77 k + 22 k^{2} + 121 N + 
      115 k\, N + 20 k^{2} N + 79 N^{2} + 42 k\, 
     N^{2} + 16 N^{3})} {3 (2 + N) (2 + k + N)^{3}},\nonu\\ 
c_ {-296} & = & - \frac {2 (-k + N) (60 + 77 k + 22 k^{2} + 121 N + 
       115 k\, N + 20 k^{2} N + 79 N^{2} + 42 k\, 
      N^{2} + 16 N^{3})} {3 (2 + N) (2 + k + N)^{3}},\nonu\\ 
c_ {-297} & = & - \frac {4 (9 + 5 k + 4 N) (32 + 55 k + 18 k^{2} + 
       41 N + 35 k\, N + 11 N^{2})} {3 (2 + N) (2 + k + N)^{5}},\nonu\\ 
c_ {-298} & = & - \frac {4 (-k + N) (32 + 55 k + 18 k^{2} + 41 N + 
       35 k\, N + 11 N^{2})} {3 (2 + N) (2 + k + N)^{5}}, \nonu\\
c_ {-299} & = & - \frac{1}{3 (2 + N) (2 + k + N)^{5}}2 (194 k + 296 k^{2} + 161 k^{3} + 
       30 k^{4} + 94 N + 410 k\, 
      N + 327 k^{2} N 
      \nonu \\& + & 81 k^{3} N + 158 N^{2} + 285 k\, 
      N^{2} + 91 k^{2} N^{2} + 91 N^{3} + 69 k\, 
      N^{3} + 17 N^{4}), \nonu\\
c_ {-300} & = & - \frac {2 (1 + k) (k - N) (32 + 55 k + 18 k^{2} + 
       41 N + 35 k\, N + 11 N^{2})} {3 (2 + N) (2 + k + N)^{6}},\nonu\\ 
c_ {-301} & = & \frac {6 (1 + k) (32 + 55 k + 18 k^{2} + 41 N + 
      35 k\, N + 11 N^{2})} {(2 + N) (2 + k + N)^{5}}, \nonu\\
c_ {-302} & = & - \frac{1} {3 (2 + N) (2 + k + N)^{4}}2 (360 + 414 k + 43 k^{2} - 30 k^{3} + 
       774 N + 716 k\, N + 93 k^{2} N 
       \nonu \\& + & 537 N^{2} + 292 k\, 
      N^{2} + 113 N^{3}), \nonu\\
c_ {-303} & = & - \frac {4 (-k + N) (32 + 55 k + 18 k^{2} + 41 N + 
       35 k\, N + 11 N^{2})} {3 (2 + N) (2 + k + N)^{5}}, \nonu\\
c_ {-304} & = & - \frac {2 (-k + N) (32 + 55 k + 18 k^{2} + 41 N + 
       35 k\, N + 11 N^{2})} {3 (2 + N) (2 + k + N)^{5}}, \nonu\\
c_ {-305} & = & \frac {4 (-k + N) (32 + 55 k + 18 k^{2} + 41 N + 
      35 k\, N + 11 N^{2})} {3 (2 + N) (2 + k + N)^{5}}, \nonu\\
c_ {-306} & = & - \frac{1}{3 (2 + N) (2 + k + N)^{5}}4 (36 + 111 k + 78 k^{2} + 16 k^{3} + 
       159 N + 316 k\, 
      N + 154 k^{2} N 
      \nonu \\& + & 20 k^{3} N + 200 N^{2} + 255 k\, 
      N^{2} + 64 k^{2} N^{2} + 97 N^{3} + 62 k\, 
      N^{3} + 16 N^{4}), \nonu\\
c_ {-307} & = & \frac {2 (1 + N) (-k + N) (32 + 55 k + 18 k^{2} + 
      41 N + 35 k\, N + 11 N^{2})} {3 (2 + N) (2 + k + N)^{6}}, \nonu\\
c_ {-308} & = & \frac {6 (1 + N) (32 + 55 k + 18 k^{2} + 41 N + 
      35 k\, N + 11 N^{2})} {(2 + N) (2 + k + N)^{5}}, \nonu\\
c_ {-309} & = & - \frac {4 (1 + N) (-k + N) (32 + 55 k + 18 k^{2} + 
       41 N + 35 k\, N + 11 N^{2})} {3 (2 + N) (2 + k + N)^{6}},\nonu\\ 
c_ {-310} & = & \frac {4 (1 + N) (9 + 4 k + 5 N) (32 + 55 k + 
      18 k^{2} + 41 N + 35 k\, 
     N + 11 N^{2})} {3 (2 + N) (2 + k + N)^{6}}, \nonu\\
c_ {-311} & = & - \frac {4 (9 + 4 k + 5 N) (32 + 55 k + 18 k^{2} + 
       41 N + 35 k\, N + 11 N^{2})} {3 (2 + N) (2 + k + N)^{5}},\nonu\\ 
c_ {-312} & = & - \frac{1}{3 (2 + N) (2 + k + N)^{5}}4 (324 + 849 k + 797 k^{2} + 309 k^{3} + 
       42 k^{4} + 717 N + 1416 k\, 
      N 
      \nonu \\& + & 911 k^{2} N + 181 k^{3} N + 541 N^{2} + 724 k\, 
      N^{2} + 246 k^{2} N^{2} + 162 N^{3} + 109 k\, 
      N^{3} + 16 N^{4}), \nonu\\
c_ {-313} & = & \frac {4 (1 + k) (k - N) (32 + 55 k + 18 k^{2} + 
      41 N + 35 k\, N + 11 N^{2})} {3 (2 + N) (2 + k + N)^{6}}, \nonu\\
c_ {-314} & = & - \frac {1}{3 (2 + N) (2 + k + N)^{5}}4 (324 + 885 k + 894 k^{2} + 386 k^{3} + 
       60 k^{4} + 681 N + 1392 k\, 
      N 
      \nonu \\& + & 946 k^{2} N + 206 k^{3} N + 468 N^{2} + 655 k\, 
      N^{2} + 232 k^{2} N^{2} + 119 N^{3} + 88 k\, 
      N^{3} + 8 N^{4}), \nonu\\
c_ {-315} & = & \frac {4 (1 + k) (9 + 5 k + 4 N) (32 + 55 k + 
      18 k^{2} + 41 N + 35 k\, 
     N + 11 N^{2})} {3 (2 + N) (2 + k + N)^{6}}, \nonu\\
c_ {-316} & = & - \frac {6 (32 + 55 k + 18 k^{2} + 41 N + 35 k\, 
      N + 11 N^{2})} {(2 + N) (2 + k + N)^{4}}, \nonu\\
c_ {-317} & = & \frac {2 (-k + N) (32 + 55 k + 18 k^{2} + 41 N + 
      35 k\, N + 11 N^{2})} {3 (2 + N) (2 + k + N)^{5}}, \nonu\\
c_ {-318} & = & - \frac {1}{3 (2 + N) (2 + k + N)^{5}}2 (288 + 638 k + 463 k^{2} + 132 k^{3} + 
       12 k^{4} + 658 N + 1106 k\, 
      N 
      \nonu \\& + & 544 k^{2} N + 80 k^{3} N + 591 N^{2} + 666 k\, 
      N^{2} + 165 k^{2} N^{2} + 242 N^{3} + 138 k\, 
      N^{3} + 37 N^{4}), \nonu\\
c_ {-319} & = & \frac {4 (1 + k) (k - N) (32 + 55 k + 18 k^{2} + 
      41 N + 35 k\, N + 11 N^{2})} {3 (2 + N) (2 + k + N)^{6}}, \nonu\\
c_ {-320} & = & \frac {4 (1 + k) (3 + k + 2 N) (32 + 55 k + 
      18 k^{2} + 41 N + 35 k\, 
     N + 11 N^{2})} {(2 + N) (2 + k + N)^{6}}, \nonu\\
c_ {-321} & = & - \frac{1} {3 (2 + N) (2 + k + N)^{5}}2 (1728 + 3034 k + 1633 k^{2} + 252 k^{3} - 
       12 k^{4} + 4454 N + 6216 k\, 
      N 
      \nonu \\& + & 2468 k^{2} N + 256 k^{3} N + 4247 N^{2} + 4170 k\, 
      N^{2} + 895 k^{2} N^{2} + 1750 N^{3} + 904 k\, 
      N^{3} + 261 N^{4}), \nonu\\
c_ {-322} & = & - \frac{1}{3 (2 + N) (2 + k + N)^{5}}4 (1 + N) (360 + 665 k + 386 k^{2} + 
       72 k^{3} + 667 N + 809 k\, 
      N 
      \nonu \\& + & 232 k^{2} N + 389 N^{2} + 236 k\, 
      N^{2} + 72 N^{3}), \nonu\\
c_ {-323} & = & \frac{1}{3 (2 + N) (2 + k + N)^{4}}8 (-48 - 23 k + 13 k^{2} + 6 k^{3} - 55 N + 
      65 k\, N + 83 k^{2} N + 18 k^{3} N 
      \nonu \\& + & 3 N^{2} + 98 k\, 
     N^{2} + 43 k^{2} N^{2} + 17 N^{3} + 28 k\, 
     N^{3} + 4 N^{4}), \nonu\\
c_ {-324} & = & \frac {4 (-k + N) (32 + 55 k + 18 k^{2} + 41 N + 
      35 k\, N + 11 N^{2})} {3 (2 + N) (2 + k + N)^{5}}, \nonu\\
c_ {-325} & = & - \frac {2 (-k + N) (32 + 55 k + 18 k^{2} + 41 N + 
       35 k\, N + 11 N^{2})} {3 (2 + N) (2 + k + N)^{5}}, \nonu\\
c_ {-326} & = & \frac {4 (9 + 4 k + 5 N) (32 + 55 k + 18 k^{2} + 
      41 N + 35 k\, N + 11 N^{2})} {3 (2 + N) (2 + k + N)^{5}}, \nonu\\
c_ {-327} & = & \frac {4 (-k + N) (32 + 55 k + 18 k^{2} + 41 N + 
      35 k\, N + 11 N^{2})} {3 (2 + N) (2 + k + N)^{5}}, \nonu\\
c_ {-328} & = & \frac {2 (-k + N) (32 + 55 k + 18 k^{2} + 41 N + 
      35 k\, N + 11 N^{2})} {3 (2 + N) (2 + k + N)^{5}}, \nonu\\
c_ {-329} & = & - \frac {4 (-k + N) (32 + 55 k + 18 k^{2} + 41 N + 
       35 k\, N + 11 N^{2})} {3 (2 + N) (2 + k + N)^{5}}, \nonu\\
c_ {-330} & = & \frac{1}{3 (2 + N) (2 + k + N)^{5}}4 (324 + 563 k + 317 k^{2} + 58 k^{3} + 
      715 N + 984 k\, 
     N + 401 k^{2} N 
     \nonu \\& + & 44 k^{3} N + 589 N^{2} + 571 k\, 
     N^{2} + 126 k^{2} N^{2} + 212 N^{3} + 108 k\, 
     N^{3} + 28 N^{4}), \nonu\\
c_ {-331} & = & \frac {4 (9 + 5 k + 4 N) (32 + 55 k + 18 k^{2} + 
      41 N + 35 k\, N + 11 N^{2})} {3 (2 + N) (2 + k + N)^{5}}, \nonu\\
c_ {-332} & = & \frac {6 (32 + 55 k + 18 k^{2} + 41 N + 35 k\, 
     N + 11 N^{2})} {(2 + N) (2 + k + N)^{4}}, \nonu\\
c_ {-333} & = & \frac {2 (-k + N) (32 + 55 k + 18 k^{2} + 41 N + 
      35 k\, N + 11 N^{2})} {3 (2 + N) (2 + k + N)^{5}}, \nonu\\
c_ {-334} & = & - \frac{1} {3 (2 + N) (2 + k + N)^{4}}2 (288 + 238 k - 82 k^{2} - 125 k^{3} - 
       30 k^{4} + 482 N + 222 k\,N 
      \nonu \\& + & 135 k^{2} N - 69 k^{3} N + 292 N^{2} + 41 k\, 
      N^{2} - 45 k^{2} N^{2} + 75 N^{3} - 7 k\, 
      N^{3} + 7 N^{4}), \nonu\\
c_ {-335} & = & \frac {4 (48 + 157 k + 114 k^{2} + 24 k^{3} + 107 N + 
      202 k\, N + 70 k^{2} N + 68 N^{2} + 61 k\, 
     N^{2} + 13 N^{3})} {3 (2 + N) (2 + k + N)^{4}}, \nonu\\
c_ {-336} & = & - \frac {2 (-k + N) (32 + 55 k + 18 k^{2} + 41 N + 
       35 k\, N + 11 N^{2})} {3 (2 + N) (2 + k + N)^{5}}, \nonu\\
c_ {-337} & = & - \frac {6 (32 + 55 k + 18 k^{2} + 41 N + 35 k\, 
      N + 11 N^{2})} {(2 + N) (2 + k + N)^{4}}, \nonu\\
c_ {-338} & = & \frac {4 (-k + N) (32 + 55 k + 18 k^{2} + 41 N + 
      35 k\, N + 11 N^{2})} {3 (2 + N) (2 + k + N)^{5}}, \nonu\\
c_ {-339} & = & - \frac {4 (9 + 4 k + 5 N) (32 + 55 k + 18 k^{2} + 
       41 N + 35 k\, N + 11 N^{2})} {3 (2 + N) (2 + k + N)^{5}},\nonu\\ 
c_ {-340} & = & \frac {4 (48 + 173 k + 135 k^{2} + 30 k^{3} + 91 N + 
      200 k\, N + 77 k^{2} N + 49 N^{2} + 53 k\, 
     N^{2} + 8 N^{3})} {3 (2 + N) (2 + k + N)^{4}}, \nonu\\
c_ {-341} & = & \frac {2 (-k + N) (16 + 21 k + 6 k^{2} + 19 N + 
      13 k\, N + 5 N^{2})} {3 (2 + N) (2 + k + N)^{4}}, \nonu\\
c_ {-342} & = & \frac {2 (16 + 47 k + 18 k^{2} + 25 N + 31 k\, 
     N + 7 N^{2})} {(2 + N) (2 + k + N)^{3}}, \nonu\\
c_ {-343} & = & - \frac{1}{3 (2 + N) (2 + k + N)^{5}}2 (-k + N) (20 + 2 k - 17 k^{2} - 6 k^{3} + 
       48 N + 21 k\, N - 5 k^{2} N 
       \nonu \\& + & 36 N^{2} + 13 k\, 
      N^{2} + 8 N^{3}), \nonu\\
c_ {-344} & = & \frac {4 (-k + N) (32 + 55 k + 18 k^{2} + 41 N + 
      35 k\, N + 11 N^{2})} {3 (2 + N) (2 + k + N)^{5}}, \nonu\\
c_ {-345} & = & - \frac {4 (-k + N) (32 + 55 k + 18 k^{2} + 41 N + 
       35 k\, N + 11 N^{2})} {3 (2 + N) (2 + k + N)^{5}}, \nonu\\
c_ {-346} & = & - \frac {2 (84 + 236 k + 75 k^{2} - 6 k^{3} + 262 N + 
       449 k\, N + 95 k^{2} N + 220 N^{2} + 191 k\, 
      N^{2} + 50 N^{3})} {3 (2 + N) (2 + k + N)^{4}}, \nonu\\
c_ {-347} & = & - \frac {8 (48 + 32 k - 9 k^{2} - 6 k^{3} + 88 N + 
       50 k\, N - k^{2} N + 55 N^{2} + 20 k\, 
      N^{2} + 11 N^{3})} {3 (2 + N) (2 + k + N)^{4}}, \nonu\\
c_ {-348} & = & - \frac{1}{3 (2 + N) (2 + k + N)^{5}}4 (180 + 364 k + 216 k^{2} + 40 k^{3} + 
       554 N + 867 k\, 
      N + 374 k^{2} N 
      \nonu \\& + & 44 k^{3} N + 591 N^{2} + 638 k\, 
      N^{2} + 146 k^{2} N^{2} + 262 N^{3} + 147 k\, 
      N^{3} + 41 N^{4}), \nonu\\
c_ {-349} & = & \frac {4 (-k + N) (16 + 21 k + 6 k^{2} + 19 N + 
      13 k\, N + 5 N^{2})} {3 (2 + N) (2 + k + N)^{4}}, \nonu\\
c_ {-350} & = & \frac {2 (-k + N) (20 + 28 k + 8 k^{2} + 54 N + 
      55 k\, N + 10 k^{2} N + 41 N^{2} + 23 k\, 
     N^{2} + 9 N^{3})} {3 (2 + N) (2 + k + N)^{5}}, \nonu\\
c_ {-351} & = & \frac {1}{3 (2 + N) (2 + k + N)^{4}}2 (276 + 194 k - 102 k^{2} - 60 k^{3} + 
      496 N + 335 k\, N - 16 k^{2} N 
      \nonu \\& + & 319 N^{2} + 149 k\, 
     N^{2} + 65 N^{3}), \nonu\\
c_ {-352} & = & - \frac {8 (48 + 64 k + 33 k^{2} + 6 k^{3} + 56 N + 
       46 k\, N + 13 k^{2} N + 17 N^{2} + 4 k\, 
      N^{2} + N^{3})} {3 (2 + N) (2 + k + N)^{4}}, \nonu\\
c_ {-353} & = & \frac{1} {3 (2 + N) (2 + k + N)^{5}}4 (180 + 150 k - 111 k^{2} - 131 k^{3} - 
      30 k^{4} + 480 N + 453 k\, 
     N 
     \nonu \\& + & 16 k^{2} N - 45 k^{3} N + 468 N^{2} + 373 k\, 
     N^{2} + 49 k^{2} N^{2} + 192 N^{3} + 88 k\, 
     N^{3} + 28 N^{4}), \nonu\\
c_ {-354} & = & \frac{1} {3 (2 + N) (2 + k + N)^{5}}4 (246 k + 359 k^{2} + 179 k^{3} + 30 k^{4} + 
      42 N + 430 k\, 
     N + 382 k^{2} N 
     \nonu \\& + & 93 k^{3} N + 75 N^{2} + 257 k\, 
     N^{2} + 101 k^{2} N^{2} + 46 N^{3} + 55 k\, 
     N^{3} + 9 N^{4}), \nonu\\
c_ {-355} & = & - \frac {4 (9 + 5 k + 4 N) (32 + 55 k + 18 k^{2} + 
       41 N + 35 k\, N + 11 N^{2})} {3 (2 + N) (2 + k + N)^{5}},\nonu\\ 
c_ {-356} & = & \frac {1}{3 (2 + N) (2 + k + N)^{5}}4 (182 k + 295 k^{2} + 163 k^{3} + 30 k^{4} + 
      106 N + 398 k\, 
     N + 334 k^{2} N 
     \nonu \\& + & 85 k^{3} N + 171 N^{2} + 273 k\, 
     N^{2} + 93 k^{2} N^{2} + 94 N^{3} + 63 k\, 
     N^{3} + 17 N^{4}), \nonu\\
c_ {-357} & = & \frac {2 (360 + 340 k + 15 k^{2} - 30 k^{3} + 668 N + 
      472 k\, N + 25 k^{2} N + 395 N^{2} + 166 k\, 
     N^{2} + 73 N^{3})} {3 (2 + N) (2 + k + N)^{3}}, \nonu\\
c_ {-358} & = & \frac{1}{3 (2 + N) (2 + k + N)^{5}}4 (456 + 1036 k + 887 k^{2} + 351 k^{3} + 
      54 k^{4} + 1136 N + 1898 k\, 
     N 
     \nonu \\& + & 1063 k^{2} N + 207 k^{3} N + 995 N^{2} + 1089 k\, 
     N^{2} + 300 k^{2} N^{2} + 365 N^{3} + 195 k\, 
     N^{3} + 48 N^{4}), \nonu\\
c_ {-359} & = & - \frac {(-k + N) (20 + 21 k + 6 k^{2} + 25 N + 
       13 k\, N + 7 N^{2})} {3 (2 + N) (2 + k + N)^{4}}, \nonu\\
c_ {-360} & = & \frac {2 (96 + 52 k - 39 k^{2} - 18 k^{3} + 188 N + 
      108 k\, N - 9 k^{2} N + 123 N^{2} + 50 k\, 
     N^{2} + 25 N^{3})} {3 (2 + N) (2 + k + N)^{4}}, \nonu\\
c_ {-361} & = & \frac {2 (-96 - 180 k - 129 k^{2} - 30 k^{3} - 60 N - 
      92 k\, N - 47 k^{2} N + 29 N^{2} + 14 k\, 
     N^{2} + 15 N^{3})} {3 (2 + N) (2 + k + N)^{4}}, \nonu\\
c_ {-362} & = & - \frac{1} {3 (2 + N) (2 + k + N)^{5}}2 (-48 - 88 k - 80 k^{2} - 63 k^{3} - 
       18 k^{4} - 80 N - 124 k\, 
      N
      \nonu\\&-& 89 k^{2} N - 39 k^{3} N - 12 N^{2} - k\, 
      N^{2} - 5 k^{2} N^{2} + 33 N^{3} + 27 k\, 
      N^{3} + 11 N^{4}), \nonu\\
c_ {-363} & = & - \frac{1}{3 (2 + N) (2 + k + N)^{5}}2 (264 + 902 k + 970 k^{2} + 433 k^{3} + 
       70 k^{4} + 766 N + 2074 k\,N 
      \nonu\\ & + & 1715 k^{2} N + 553 k^{3} N + 56 k^{4} N + 892 N^{2} + 
       1779 k\, 
      N^{2} + 987 k^{2} N^{2} + 168 k^{3} N^{2} + 513 N^{3} 
      \nonu\\ & + &  669 k\, 
      N^{3} + 182 k^{2} N^{3} + 145 N^{4} + 94 k\, 
      N^{4} + 16 N^{5}), \nonu\\
c_ {-364} & = & \frac {4 (-k + N) (32 + 55 k + 18 k^{2} + 41 N + 
      35 k\, N + 11 N^{2})} {3 (2 + N) (2 + k + N)^{5}}, \nonu\\
c_ {-365} & = & \frac {2 (-k + N) (32 + 55 k + 18 k^{2} + 41 N + 
      35 k\, N + 11 N^{2})} {3 (2 + N) (2 + k + N)^{5}}, \nonu\\
c_ {-366} & = & \frac{1} {3 (2 + N) (2 + k + N)^{5}}4 (36 - 111 k - 156 k^{2} - 44 k^{3} + 93 N - 
      224 k\, N - 245 k^{2} N 
      \nonu\\ &-& 46 k^{3} N + 110 N^{2} - 117 k\, 
     N^{2} - 83 k^{2} N^{2} + 64 N^{3} - 10 k\, 
     N^{3} + 13 N^{4}), \nonu\\
c_ {-367} & = & \frac {4 (1 + k) (k - N) (32 + 55 k + 18 k^{2} + 
      41 N + 35 k\, N + 11 N^{2})} {3 (2 + N) (2 + k + N)^{6}}, \nonu\\
c_ {-368} & = & - \frac {4 (-k + N) (32 + 55 k + 18 k^{2} + 41 N + 
       35 k\, N + 11 N^{2})} {3 (2 + N) (2 + k + N)^{5}}, \nonu\\
c_ {-369} & = & - \frac{1}{3 (2 + N) (2 + k + N)^{6}}2 (1 + k) (-k + N) (36 + 39 k + 10 k^{2} + 
       67 N + 53 k\, N + 8 k^{2} N 
       \nonu\\ & + &  41 N^{2} + 18 k\, 
      N^{2} + 8 N^{3}), \nonu\\
c_ {-370} & = & - \frac {4 (1 + N) (-k + N) (32 + 55 k + 18 k^{2} + 
       41 N + 35 k\, N + 11 N^{2})} {3 (2 + N) (2 + k + N)^{6}},\nonu\\ 
c_ {-371} & = & \frac {4 (1 + N) (-k + N) (32 + 55 k + 18 k^{2} + 
      41 N + 35 k\, N + 11 N^{2})} {3 (2 + N) (2 + k + N)^{6}}, \nonu\\
c_ {-372} & = & \frac{1} {3 (2 + N) (2 + k + N)^{5}}2 (36 + 75 k + 3 k^{2} - 10 k^{3} + 195 N + 
      296 k\, N + 47 k^{2} N - 8 k^{3} N 
      \nonu\\ & + &  295 N^{2} + 315 k\, 
     N^{2} + 38 k^{2} N^{2} + 170 N^{3} + 100 k\, 
     N^{3} + 32 N^{4}), \nonu\\
c_ {-373} & = & - \frac{1}{3 (2 + N) (2 + k + N)^{6}}2 (-k + N) (160 + 376 k + 325 k^{2} + 
       125 k^{3} + 18 k^{4} + 376 N 
       \nonu\\ & + &  640 k\, 
      N + 356 k^{2} N + 67 k^{3} N + 331 N^{2} + 365 k\, 
      N^{2} + 97 k^{2} N^{2} + 130 N^{3} + 71 k\, 
      N^{3} + 19 N^{4}), \nonu\\
c_ {-374} & = & - \frac {4 (1 + k) (k - N) (32 + 55 k + 18 k^{2} + 
       41 N + 35 k\, N + 11 N^{2})} {3 (2 + N) (2 + k + N)^{6}},\nonu\\ 
c_ {-375} & = & - \frac{1}{3 (2 + N) (2 + k + N)^{5}}2 (288 + 212 k - 53 k^{2} + k^{3} + 
       18 k^{4} + 796 N + 460 k\, 
      N - 194 k^{2} N 
      \nonu\\ & - &  61 k^{3} N + 889 N^{2} + 457 k\, 
      N^{2} - 63 k^{2} N^{2} + 456 N^{3} + 167 k\, 
      N^{3} + 83 N^{4}), \nonu\\
c_ {-376} & = & - \frac {1}{3 (2 + N) (2 + k + N)^{6}}4 (-k + N) (308 + 541 k + 306 k^{2} + 
       56 k^{3} + 777 N + 1086 k\, 
      N 
      \nonu\\ & + & 452 k^{2} N + 52 k^{3} N + 714 N^{2} + 701 k\, 
      N^{2} + 158 k^{2} N^{2} + 281 N^{3} + 144 k\, 
      N^{3} + 40 N^{4}), \nonu\\
c_ {-377} & = & - \frac {2 (-k + N) (32 + 55 k + 18 k^{2} + 41 N + 
       35 k\, N + 11 N^{2})} {3 (2 + N) (2 + k + N)^{5}}, \nonu\\
c_ {-378} & = & - \frac{1}{3 (2 + N) (2 + k + N)^{6}}4 (216 + 674 k + 733 k^{2} + 338 k^{3} + 
       56 k^{4} + 586 N + 1522 k\, 
      N 
      \nonu\\ & + &  1334 k^{2} N + 468 k^{3} N + 52 k^{4} N + 625 N^{2} + 
       1238 k\, 
      N^{2} + 757 k^{2} N^{2} + 146 k^{3} N^{2} + 330 N^{3} 
      \nonu\\ & + &  430 k\, 
      N^{3} + 132 k^{2} N^{3} + 89 N^{4} + 56 k\, 
      N^{4} + 10 N^{5}), \nonu\\
c_ {-379} & = & - \frac{1}{3 (2 + N) (2 + k + N)^{4}}2 (96 - 68 k - 205 k^{2} - 70 k^{3} + 
       488 N + 168 k\, 
      N - 207 k^{2} N 
      \nonu\\ & - &  56 k^{3} N + 679 N^{2} + 324 k\, 
      N^{2} - 40 k^{2} N^{2} + 361 N^{3} + 122 k\, 
      N^{3} + 64 N^{4}), \nonu\\
c_ {-380} & = & - \frac {1}{3 (2 + N) (2 + k + N)^{4}}4 (264 + 404 k + 187 k^{2} + 26 k^{3} + 
       616 N + 704 k\, 
      N + 204 k^{2} N 
      \nonu\\ & + & 10 k^{3} N + 537 N^{2} + 416 k\, 
      N^{2} + 57 k^{2} N^{2} + 206 N^{3} + 84 k\, 
      N^{3} + 29 N^{4}), \nonu\\
c_ {-381} & = & \frac {2 (96 + 100 k + 11 k^{2} - 6 k^{3} + 140 N + 
      112 k\, N + 9 k^{2} N + 69 N^{2} + 34 k\, 
     N^{2} + 11 N^{3})} {3 (2 + N) (2 + k + N)^{4}}, \nonu\\
c_ {-382} & = & \frac {4 (-k + N) (32 + 55 k + 18 k^{2} + 41 N + 
      35 k\, N + 11 N^{2})} {3 (2 + N) (2 + k + N)^{5}}, \nonu\\
c_ {-383} & = & - \frac {4 (-k + N) (32 + 55 k + 18 k^{2} + 41 N + 
       35 k\, N + 11 N^{2})} {3 (2 + N) (2 + k + N)^{5}}, \nonu\\
c_ {-384} & = & - \frac {2 (72 + 20 k - 21 k^{2} - 6 k^{3} + 160 N + 
       56 k\, N - 7 k^{2} N + 109 N^{2} + 26 k\, 
      N^{2} + 23 N^{3})} {3 (2 + N) (2 + k + N)^{4}}, \nonu\\
c_ {-385} & = & \frac{1} {3 (2 + N) (2 + k + N)^{5}}4 (96 + 84 k - 8 k^{3} + 252 N + 178 k\, 
     N - 5 k^{2} N - 10 k^{3} N + 254 N^{2} 
     \nonu\\& + & 138 k\, 
     N^{2} + k^{2} N^{2} + 115 N^{3} + 38 k\, 
     N^{3} + 19 N^{4}), \nonu\\
c_ {-386} & = & - \frac{1}{3 (2 + N) (2 + k + N)^{5}}4 (-96 - 212 k - 148 k^{2} - 32 k^{3} - 
       124 N - 266 k\, 
      N - 155 k^{2} N 
      \nonu\\& - & 22 k^{3} N - 18 N^{2} - 82 k\, 
      N^{2} - 37 k^{2} N^{2} + 29 N^{3} + 2 k\, 
      N^{3} + 9 N^{4}), \nonu\\
c_ {-387} & = & - \frac {8 (-k + N) (16 + 21 k + 6 k^{2} + 19 N + 
       13 k\, N + 5 N^{2})} {3 (2 + N) (2 + k + N)^{4}}, \nonu\\
c_ {-388} & = & - \frac{1}{3 (2 + N) (2 + k + N)^{5}}4 (360 + 472 k + 192 k^{2} + 24 k^{3} + 
       788 N + 764 k\, 
      N + 187 k^{2} N 
      \nonu\\& + & 6 k^{3} N + 664 N^{2} + 434 k\, 
      N^{2} + 49 k^{2} N^{2} + 255 N^{3} + 88 k\, 
      N^{3} + 37 N^{4}), \nonu\\
c_ {-389} & = & \frac {2 (-k + N) (32 + 55 k + 18 k^{2} + 41 N + 
      35 k\, N + 11 N^{2})} {3 (2 + N) (2 + k + N)^{5}}, \nonu\\
c_ {-390} & = & \frac{1} {3 (2 + N) (2 + k + N)^{5}}4 (192 + 768 k + 813 k^{2} + 321 k^{3} + 
      42 k^{4} + 480 N + 1344 k\, 
     N 
     \nonu\\& + & 929 k^{2} N + 181 k^{3} N + 435 N^{2} + 783 k\, 
     N^{2} + 268 k^{2} N^{2} + 175 N^{3} + 155 k\, 
     N^{3} + 26 N^{4}), \nonu\\
c_ {-391} & = & - \frac {8 (1 + N) (9 + 4 k + 
       5 N)} {3 (2 + k + N)^{4}}, \nonu\\
c_ {-392} & = & - \frac {1}{3 (2 + N) (2 + k + N)^{5}}2 (648 + 1394 k + 1141 k^{2} + 424 k^{3} + 
       60 k^{4} + 1810 N + 3166 k\, 
      N 
      \nonu\\& + & 1958 k^{2} N + 490 k^{3} N + 36 k^{4} N + 2029 N^{2} + 
       2726 k\, 
      N^{2} + 1141 k^{2} N^{2} + 146 k^{3} N^{2} \nonu\\ & + &  1156 N^{3} 
      +1068 k\, N^{3} + 228 k^{2} N^{3} + 337 N^{4} + 162 k\, 
      N^{4} + 40 N^{5}), \nonu\\
c_ {-393} & = & - \frac {4 (9 + 4 k + 5 N) (36 + 39 k + 10 k^{2} + 
       67 N + 53 k\, N + 8 k^{2} N + 41 N^{2} + 18 k\, 
      N^{2} + 8 N^{3})} {3 (2 + N) (2 + k + N)^{5}}, \nonu\\
c_ {-394} & = & \frac {4 (1 + N) (9 + 4 k + 5 N) (32 + 55 k + 
      18 k^{2} + 41 N + 35 k\, 
     N + 11 N^{2})} {3 (2 + N) (2 + k + N)^{6}}, \nonu\\
c_ {-395} & = & \frac {1}{3 (2 + N) (2 + k + N)^{5}}4 (36 - 115 k - 137 k^{2} - 34 k^{3} + 97 N - 
      224 k\, N - 203 k^{2} N
      \nonu\\& - & 32 k^{3} N + 91 N^{2} - 143 k\, 
     N^{2} - 72 k^{2} N^{2} + 38 N^{3} - 28 k\, 
     N^{3} + 6 N^{4}), \nonu\\
c_ {-396} & = & \frac {1}{3 (2 + N) (2 + k + N)^{6}}8 (1 + N) (180 + 303 k + 162 k^{2} + 
      28 k^{3} + 471 N + 628 k\, 
     N 
     \nonu\\& + & 244 k^{2} N + 26 k^{3} N + 452 N^{2} + 423 k\, 
     N^{2} + 88 k^{2} N^{2} + 187 N^{3} + 92 k\, 
     N^{3} + 28 N^{4}), \nonu\\
c_ {-397} & = & - \frac{1}{3 (2 + N) (2 + k + N)^{6}}4 (-k + N) (212 + 365 k + 197 k^{2} + 
       34 k^{3} + 553 N + 758 k\, 
      N 
      \nonu\\& + & 299 k^{2} N + 32 k^{3} N + 527 N^{2} + 509 k\, 
      N^{2} + 108 k^{2} N^{2} + 216 N^{3} + 110 k\, 
      N^{3} + 32 N^{4}), \nonu\\
c_ {-398} & = & - \frac{1} {3 (2 + N) (2 + k + N)^{5}}2 (384 + 448 k - 129 k^{2} - 248 k^{3} - 
       60 k^{4} + 2192 N + 2880 k\, 
      N 
      \nonu\\& + & 762 k^{2} N - 122 k^{3} N - 36 k^{4} N + 3801 N^{2} + 
       4236 k\, 
      N^{2} + 1111 k^{2} N^{2} + 30 k^{3} N^{2} + 2906 N^{3} 
      \nonu\\& + & 
       2342 k\, N^{3} + 364 k^{2} N^{3} + 1025 N^{4} + 442 k\, 
      N^{4} + 136 N^{5}), \nonu\\
c_ {-399} & = & - \frac {1} {3 (2 + N) (2 + k + N)^{4}}4 (264 + 340 k + 129 k^{2} + 14 k^{3} + 
       488 N + 476 k\, 
      N + 113 k^{2} N 
      \nonu\\& + & 4 k^{3} N + 343 N^{2} + 230 k\, 
      N^{2} + 26 k^{2} N^{2} + 111 N^{3} + 40 k\, 
      N^{3} + 14 N^{4}), \nonu\\
c_ {-400} & = & - \frac {8 (21 + 14 k + 2 k^{2} + 28 N + 10 k\, 
      N + 9 N^{2})} {3 (2 + k + N)^{4}}, \nonu\\
      c_ {-401} & = & - \frac{1}{3 (2 + N) (2 + k + N)^{5}}8 (1 + N) (-24 + 5 k + 16 k^{2} + 4 k^{3} + 
       25 N + 139 k\, 
      N + 87 k^{2} N 
      \nonu\\& + & 14 k^{3} N + 94 N^{2} + 159 k\, 
      N^{2} + 46 k^{2} N^{2} + 62 N^{3} + 45 k\, 
      N^{3} + 12 N^{4}) , \nonu\\
c_ {-402} & = & \frac {8 (-k + N) (16 + 21 k + 6 k^{2} + 19 N + 
      13 k\, N + 5 N^{2})} {3 (2 + N) (2 + k + N)^{4}}, \nonu\\
c_ {-403} & = & \frac {(-k + N) (16 + 21 k + 6 k^{2} + 19 N + 13 k\, 
     N + 5 N^{2})} {(2 + N) (2 + k + N)^{3}}, \nonu\\
c_ {-404} & = & \frac {2 (96 + 100 k + 11 k^{2} - 6 k^{3} + 140 N + 
      112 k\, N + 9 k^{2} N + 69 N^{2} + 34 k\, 
     N^{2} + 11 N^{3})} {3 (2 + N) (2 + k + N)^{4}}, \nonu\\
c_ {-405} & = & - \frac {2 (-k + N) (32 + 55 k + 18 k^{2} + 41 N + 
       35 k\, N + 11 N^{2})} {3 (2 + N) (2 + k + N)^{5}}, \nonu\\
c_ {-406} & = & \frac {2 (-96 - 132 k - 79 k^{2} - 18 k^{3} - 108 N - 
      88 k\, N - 29 k^{2} N - 25 N^{2} - 2 k\, 
     N^{2} + N^{3})} {3 (2 + N) (2 + k + N)^{4}}, \nonu\\
c_ {-407} & = & - \frac {2 (-k + N) (12 + 24 k + 8 k^{2} + 38 N + 
       49 k\, N + 10 k^{2} N + 31 N^{2} + 21 k\, 
      N^{2} + 7 N^{3})} {3 (2 + N) (2 + k + N)^{5}}, \nonu\\
c_ {-408} & = & - \frac{1}{3 (2 + N) (2 + k + N)^{5}}2 (-k + N) (52 + 118 k + 83 k^{2} + 
       18 k^{3} + 76 N + 117 k\, N 
       \nonu\\& + & 43 k^{2} N + 32 N^{2} + 25 k\, 
      N^{2} + 4 N^{3}), \nonu\\
c_ {-409} & = & \frac{1}{3 (2 + N) (2 + k + N)^{4}}(384 + 768 k + 516 k^{2} + 151 k^{3} + 
     18 k^{4} + 960 N + 1312 k\, 
    N 
    \nonu\\& + & 507 k^{2} N + 59 k^{3} N + 860 N^{2} + 737 k\, 
    N^{2} + 119 k^{2} N^{2} + 333 N^{3} + 141 k\, 
    N^{3} + 47 N^{4}), \nonu\\
c_ {-410} & = & \frac{1}{3 (2 + N) (2 + k + N)^{5}}(-k + N) (192 + 264 k + 93 k^{2} + 6 k^{3} + 
      280 N + 232 k\, N + 31 k^{2} N 
      \nonu\\& + & 123 N^{2} + 42 k\, 
     N^{2} + 17 N^{3}), \nonu\\
c_ {-411} & = & - \frac {1}{3 (2 + N) (2 + k + N)^{5}}2 (360 + 674 k + 502 k^{2} + 189 k^{3} + 
       30 k^{4} + 874 N + 1230 k\, 
      N 
      \nonu\\& + & 607 k^{2} N + 113 k^{3} N + 752 N^{2} + 697 k\, 
      N^{2} + 169 k^{2} N^{2} + 271 N^{3} + 121 k\, 
      N^{3} + 35 N^{4}), \nonu\\
c_ {-412} & = & \frac {1}{3 (2 + N) (2 + k + N)^{4}}2 (96 + 208 k + 131 k^{2} + 26 k^{3} + 224 N + 
      334 k\, N + 115 k^{2} N 
      \nonu\\& + & 4 k^{3} N + 207 N^{2} + 198 k\, 
     N^{2} + 30 k^{2} N^{2} + 93 N^{3} + 46 k\, 
     N^{3} + 16 N^{4}), \nonu\\
c_ {-413} & = & \frac {1}{3 (2 + N) (2 + k + N)^{5}}2 (456 + 1094 k + 966 k^{2} + 383 k^{3} + 
      58 k^{4} + 1054 N + 2138 k\, 
     N 
     \nonu\\& + & 1505 k^{2} N + 445 k^{3} N + 44 k^{4} N + 1024 N^{2} + 
      1641 k\, 
     N^{2} + 805 k^{2} N^{2} + 130 k^{3} N^{2} + 527 N^{3}
     \nonu\\& + & 589 k\, 
     N^{3} + 146 k^{2} N^{3} + 143 N^{4} + 84 k\, 
     N^{4} + 16 N^{5}), \nonu\\
c_ {-414} & = & - \frac {2 (-k + N) (32 + 55 k + 18 k^{2} + 41 N + 
       35 k\, N + 11 N^{2})} {3 (2 + N) (2 + k + N)^{5}}, \nonu\\
c_ {-415} & = & \frac {4 (-k + N) (32 + 55 k + 18 k^{2} + 41 N + 
      35 k\, N + 11 N^{2})} {3 (2 + N) (2 + k + N)^{5}}, \nonu\\
c_ {-416} & = & - \frac {4 (-k + N) (32 + 55 k + 18 k^{2} + 41 N + 
       35 k\, N + 11 N^{2})} {3 (2 + N) (2 + k + N)^{5}}, \nonu\\
c_ {-417} & = & \frac {2 (1 + N) (-k + N) (4 + 7 k + 2 k^{2} + 19 N + 
      21 k\, N + 4 k^{2} N + 17 N^{2} + 10 k\, 
     N^{2} + 4 N^{3})} {3 (2 + N) (2 + k + N)^{6}}, \nonu\\
c_ {-418} & = & \frac {1}{3 (2 + N) (2 + k + N)^{5}}2 (-324 - 727 k - 520 k^{2} - 116 k^{3} - 
      551 N - 1068 k\, 
     N - 604 k^{2} N 
     \nonu\\& - & 88 k^{3} N - 302 N^{2} - 479 k\, 
     N^{2} - 168 k^{2} N^{2} - 43 N^{3} - 54 k\, 
     N^{3} + 4 N^{4}), \nonu\\
c_ {-419} & = & - \frac {2 (-k + N) (32 + 55 k + 18 k^{2} + 41 N + 
       35 k\, N + 11 N^{2})} {3 (2 + N) (2 + k + N)^{5}}, \nonu\\
c_ {-420} & = & \frac{1} {3 (2 + N) (2 + k + N)^{4}}2 (96 - 120 k - 255 k^{2} - 82 k^{3} + 180 N - 
      452 k\, N - 559 k^{2} N 
      \nonu\\& - & 116 k^{3} N + 161 N^{2} - 330 k\, 
     N^{2} - 226 k^{2} N^{2} + 83 N^{3} - 52 k\, 
     N^{3} + 16 N^{4}), \nonu\\
c_ {-421} & = & - \frac {1}{3 (2 + N) (2 + k + N)^{4}}4 (456 + 1104 k + 945 k^{2} + 337 k^{3} + 
       42 k^{4} + 972 N + 1836 k\, 
      N 
      \nonu\\& + & 1089 k^{2} N + 201 k^{3} N + 759 N^{2} + 993 k\, 
      N^{2} + 308 k^{2} N^{2} + 257 N^{3} + 173 k\, 
      N^{3} + 32 N^{4}), \nonu\\
c_ {-422} & = & - \frac {4 (-k + N) (32 + 55 k + 18 k^{2} + 41 N + 
       35 k\, N + 11 N^{2})} {3 (2 + N) (2 + k + N)^{5}}, \nonu\\
c_ {-423} & = & \frac {2 (-96 - 132 k - 79 k^{2} - 18 k^{3} - 108 N - 
      88 k\, N - 29 k^{2} N - 25 N^{2} - 2 k\, 
     N^{2} + N^{3})} {3 (2 + N) (2 + k + N)^{4}}, \nonu\\
c_ {-424} & = & \frac{1} {3 (2 + N) (2 + k + N)^{4}}2 (648 + 1064 k + 573 k^{2} + 102 k^{3} + 
      1132 N + 1244 k\, N + 335 k^{2} N 
      \nonu\\& + & 631 N^{2} + 350 k\, 
     N^{2} + 113 N^{3}), \nonu\\
c_ {-425} & = & \frac{1}{3 (2 + N) (2 + k + N)^{5}}4 (96 + 116 k + 26 k^{2} - 17 k^{3} - 
      6 k^{4} + 220 N + 238 k\, 
     N + 74 k^{2} N 
     \nonu\\& + & k^{3} N + 168 N^{2} + 135 k\, 
     N^{2} + 30 k^{2} N^{2} + 48 N^{3} + 19 k\, 
     N^{3} + 4 N^{4}), \nonu\\
c_ {-426} & = & - \frac {4 (-k + N) (32 + 55 k + 18 k^{2} + 41 N + 
       35 k\, N + 11 N^{2})} {3 (2 + N) (2 + k + N)^{5}}, \nonu\\
c_ {-427} & = & - \frac{1} {3 (2 + N) (2 + k + N)^{5}}4 (-96 - 244 k - 238 k^{2} - 107 k^{3} - 
       18 k^{4} - 92 N - 198 k\, 
      N 
      \nonu\\& - &142 k^{2} N - 37 k^{3} N + 4 N^{2} - 11 k\, 
      N^{2} - 6 k^{2} N^{2} + 20 N^{3} + 9 k\, 
      N^{3} + 4 N^{4}), \nonu\\
c_ {-428} & = & \frac{1}{3 (2 + N) (2 + k + N)^{5}}4 (360 + 940 k + 844 k^{2} + 315 k^{3} + 
      42 k^{4} + 896 N + 1728 k\, 
     N 
     \nonu\\& + &1034 k^{2} N + 193 k^{3} N + 776 N^{2} + 999 k\, 
     N^{2} + 304 k^{2} N^{2} + 280 N^{3} + 181 k\, 
     N^{3} + 36 N^{4}), \nonu\\
c_ {-429} & = & \frac{1}{3 (2 + N) (2 + k + N)^{4}}8 (48 + 121 k + 80 k^{2} + 16 k^{3} + 137 N + 
      281 k\, N + 143 k^{2} N 
      \nonu\\& + & 20 k^{3} N + 152 N^{2} + 213 k\, 
     N^{2} + 58 k^{2} N^{2} + 72 N^{3} + 51 k\, 
     N^{3} + 12 N^{4}), \nonu\\
c_ {-430} & = & \frac {8 (9 + 5 k + 4 N) (18 + 21 k + 6 k^{2} + 
      22 N + 13 k\, N + 6 N^{2})} {3 (2 + N) (2 + k + N)^{4}}, \nonu\\
c_ {-431} & = & - \frac{1} {3 (2 + N) (2 + k + N)^{5}}4 (264 + 280 k + 59 k^{2} - 6 k^{3} + 
       644 N + 566 k\, 
      N + 118 k^{2} N 
      \nonu\\& + & 6 k^{3} N + 563 N^{2} + 342 k\, 
      N^{2} + 39 k^{2} N^{2} + 206 N^{3} + 60 k\, 
      N^{3} + 27 N^{4}), \nonu\\
c_ {-432} & = & \frac {1}{3 (2 + N) (2 + k + N)^{5}}2 (72 + 122 k + 107 k^{2} + 53 k^{3} + 
      10 k^{4} + 490 N + 806 k\, 
     N + 537 k^{2} N 
     \nonu\\& + & 171 k^{3} N + 20 k^{4} N + 815 N^{2} + 
      1065 k\, 
     N^{2} + 474 k^{2} N^{2} + 74 k^{3} N^{2} + 577 N^{3} + 523 k\, 
     N^{3} 
     \nonu\\& + & 116 k^{2} N^{3} + 190 N^{4} + 90 k\, 
     N^{4} + 24 N^{5}), \nonu\\
     c_ {-433} & = & \frac {4 (9 + 5 k + 4 N) (4 + 7 k + 2 k^{2} + 19 N + 
      21 k\, N + 4 k^{2} N + 17 N^{2} + 10 k\, 
     N^{2} + 4 N^{3})} {3 (2 + N) (2 + k + N)^{5}}, \nonu\\
c_ {-434} & = & - \frac{1}{3 (2 + N) (2 + k + N)^{6}}8 (1 + k) (180 + 323 k + 185 k^{2} + 
       34 k^{3} + 451 N + 648 k\, 
      N 
      \nonu\\& + & 275 k^{2} N + 32 k^{3} N + 409 N^{2} + 415 k\, 
      N^{2} + 96 k^{2} N^{2} + 158 N^{3} + 84 k\, 
      N^{3} + 22 N^{4}), \nonu\\
c_ {-435} & = & \frac{1}{3 (2 + N) (2 + k + N)^{5}}2 (384 + 400 k - 129 k^{2} - 233 k^{3} - 
      58 k^{4} + 800 N - 240 k\, 
     N 
     \nonu\\& - & 1693 k^{2} N - 1015 k^{3} N - 164 k^{4} N + 729 N^{2} - 
      787 k\, N^{2} - 1528 k^{2} N^{2} - 450 k^{3} N^{2} + 
      433 N^{3} 
      \nonu\\& - & 317 k\, 
     N^{3} - 324 k^{2} N^{3} + 158 N^{4} - 22 k\, 
     N^{4} + 24 N^{5}), \nonu\\
c_ {-436} & = & - \frac {1}{3 (2 + N) (2 + k + N)^{4}}4 (456 + 1232 k + 1135 k^{2} + 436 k^{3} + 
       60 k^{4} + 1036 N + 2056 k\, 
      N 
      \nonu\\& + & 1265 k^{2} N + 246 k^{3} N + 829 N^{2} + 1084 k\, 
      N^{2} + 336 k^{2} N^{2} + 275 N^{3} + 178 k\, 
      N^{3} + 32 N^{4}), \nonu\\
c_ {-437} & = & \frac {8 (138 + 251 k + 151 k^{2} + 30 k^{3} + 
      238 N + 285 k\, N + 85 k^{2} N + 128 N^{2} + 76 k\, 
     N^{2} + 22 N^{3})} {3 (2 + N) (2 + k + N)^{4}}, \nonu\\
c_ {-438} & = & - \frac {1}{3 (2 + N) (2 + k + N)^{5}}8 (1 + k) (24 + 53 k + 33 k^{2} + 6 k^{3} + 
       97 N + 187 k\, 
      N + 105 k^{2} N 
      \nonu\\& + & 18 k^{3} N + 125 N^{2} + 164 k\, 
      N^{2} + 49 k^{2} N^{2} + 61 N^{3} + 40 k\, 
      N^{3} + 10 N^{4}), \nonu\\
c_ {-439} & = & \frac {1}{3 (2 + N) (2 + k + N)^{5}}4 (180 + 369 k + 267 k^{2} + 75 k^{3} + 
      6 k^{4} + 405 N + 631 k\, 
     N 
     \nonu\\& + & 307 k^{2} N + 43 k^{3} N + 344 N^{2} + 368 k\, 
     N^{2} + 92 k^{2} N^{2} + 132 N^{3} + 74 k\, 
     N^{3} + 19 N^{4}), \nonu\\
c_ {-440} & = & \frac {1}{3 (2 + N) (2 + k + N)^{4}}(768 + 908 k + 105 k^{2} - 66 k^{3} + 1012 N + 
     508 k\, N - 125 k^{2} N 
     \nonu\\& + & 347 N^{2} - 30 k\, 
    N^{2} + 29 N^{3}), \nonu\\
c_ {-441} & = & \frac{1}{3 (2 + N) (2 + k + N)^{4}}8 (96 + 203 k + 126 k^{2} + 24 k^{3} + 181 N + 
      247 k\, N + 76 k^{2} N 
      \nonu\\& + & 107 N^{2} + 72 k\, 
     N^{2} + 20 N^{3}), \nonu\\
c_ {-442} & = & - \frac{1}{3 (2 + N) (2 + k + N)^{4}}(360 + 590 k + 339 k^{2} + 66 k^{3} + 
      598 N + 644 k\, N + 185 k^{2} N 
      \nonu\\& + & 313 N^{2} + 164 k\, 
     N^{2} + 53 N^{3}), \nonu\\
c_ {-443} & = & \frac{1} {3 (2 + N) (2 + k + N)^{5}}2 (1488 + 3232 k + 2564 k^{2} + 861 k^{3} + 
      102 k^{4} + 3128 N + 5148 k\,N 
     \nonu\\& + & 2743 k^{2} N + 461 k^{3} N + 2440 N^{2} + 2723 k\, 
     N^{2} + 735 k^{2} N^{2} + 849 N^{3} + 487 k\, 
     N^{3} + 111 N^{4}), \nonu\\
c_ {-444} & = & \frac{1}{3 (2 + N) (2 + k + N)^{5}}4 (624 + 1692 k + 1583 k^{2} + 635 k^{3} + 
      94 k^{4} + 2076 N + 4200 k\, 
     N 
     \nonu\\& + & 2821 k^{2} N + 739 k^{3} N + 56 k^{4} N + 2569 N^{2} + 
      3717 k\, 
     N^{2} + 1598 k^{2} N^{2} + 204 k^{3} N^{2} 
     \nonu\\& + & 1515 N^{3} + 
      1409 k\, N^{3} + 288 k^{2} N^{3} + 432 N^{4} + 196 k\, 
     N^{4} + 48 N^{5}), \nonu\\
c_ {-445} & = & - \frac {4 (9 + 5 k + 4 N) (32 + 55 k + 18 k^{2} + 
       41 N + 35 k\, N + 11 N^{2})} {3 (2 + N) (2 + k + N)^{5}},\nonu\\ 
c_ {-446} & = & - \frac {4 (1 + N) (-k + N) (32 + 55 k + 18 k^{2} + 
       41 N + 35 k\, N + 11 N^{2})} {3 (2 + N) (2 + k + N)^{6}},\nonu\\ 
c_ {-447} & = & - \frac{1}{3 (2 + N) (2 + k + N)^{5}}2 (288 + 574 k + 483 k^{2} + 195 k^{3} + 
       30 k^{4} + 722 N + 1034 k\,N 
      \nonu\\& + & 549 k^{2} N + 107 k^{3} N + 643 N^{2} + 595 k\, 
      N^{2} + 150 k^{2} N^{2} + 245 N^{3} + 111 k\, 
      N^{3} + 34 N^{4}), \nonu\\
c_ {-448} & = & \frac {4 (1 + k) (9 + 5 k + 4 N) (32 + 55 k + 
      18 k^{2} + 41 N + 35 k\, 
     N + 11 N^{2})} {3 (2 + N) (2 + k + N)^{6}}, \nonu\\
c_ {-449} & = & - \frac {1}{3 (2 + N) (2 + k + N)^{5}}2 (1152 + 2102 k + 1089 k^{2} + 101 k^{3} - 
       30 k^{4} + 2794 N + 4196 k\, 
      N 
      \nonu\\& + & 1717 k^{2} N + 161 k^{3} N + 2491 N^{2} + 2705 k\, 
      N^{2} + 628 k^{2} N^{2} + 949 N^{3} + 551 k\, 
      N^{3} + 130 N^{4}), \nonu\\
c_ {-450} & = & - \frac{1}{3 (2 + N) (2 + k + N)^{6}}4 (360 + 1002 k + 969 k^{2} + 400 k^{3} + 
       60 k^{4} + 906 N + 2134 k\, 
      N 
      \nonu\\& + & 1648 k^{2} N + 504 k^{3} N + 48 k^{4} N + 929 N^{2} + 
       1708 k\, 
      N^{2} + 917 k^{2} N^{2} + 152 k^{3} N^{2} + 492 N^{3} 
      \nonu\\& + & 614 k\, 
      N^{3} + 166 k^{2} N^{3} + 137 N^{4} + 86 k\, 
      N^{4} + 16 N^{5}), \nonu\\
c_ {-451} & = & \frac {4 (1 + N) (3 + 2 k + N) (32 + 55 k + 
      18 k^{2} + 41 N + 35 k\, 
     N + 11 N^{2})}{(2 + N) (2 + k + N)^{6}}, \nonu\\
c_ {-452} & = & - \frac {1}{3 (2 + N) (2 + k + N)^{6}}4 (936 + 2766 k + 3241 k^{2} + 1869 k^{3} + 
       532 k^{4} + 60 k^{5} + 2310 N 
       \nonu\\& + & 5502 k\, 
      N + 4883 k^{2} N + 1889 k^{3} N + 270 k^{4} N + 2201 N^{2} + 
       3983 k\, 
      N^{2} + 2398 k^{2} N^{2} 
      \nonu\\& + & 468 k^{3} N^{2} + 1001 N^{3} + 
       1233 k\, N^{3} + 384 k^{2} N^{3} + 212 N^{4} + 134 k\, 
      N^{4} + 16 N^{5}), \nonu\\
c_ {-453} & = & - \frac {1}{3 (2 + N) (2 + k + N)^{6}}8 (720 + 1532 k + 1213 k^{2} + 426 k^{3} + 
       56 k^{4} + 2284 N + 4050 k\, 
      N 
      \nonu\\& + &2559 k^{2} N + 671 k^{3} N + 58 k^{4} N + 2801 N^{2} + 
       3840 k\, 
      N^{2} + 1686 k^{2} N^{2} + 237 k^{3} N^{2} 
      \nonu\\& + & 1671 N^{3} + 
       1563 k\, N^{3} + 352 k^{2} N^{3} + 488 N^{4} + 233 k\, 
      N^{4} + 56 N^{5}), \nonu\\
c_ {-454} & = & \frac {1}{3 (2 + N) (2 + k + N)^{5}}4 (816 + 1372 k + 837 k^{2} + 226 k^{3} + 
      24 k^{4} + 2492 N + 3880 k\, 
     N 
     \nonu\\& + & 2274 k^{2} N + 630 k^{3} N + 72 k^{4} N + 3059 N^{2} + 
      3950 k\, 
     N^{2} + 1741 k^{2} N^{2} + 276 k^{3} N^{2} 
     \nonu\\& + &1854 N^{3} + 
      1710 k\, N^{3} + 400 k^{2} N^{3} + 551 N^{4} + 268 k\, 
     N^{4} + 64 N^{5}), \nonu\\
c_ {-455} & = & \frac {4 (9 + 4 k + 5 N) (32 + 55 k + 18 k^{2} + 
      41 N + 35 k\, N + 11 N^{2})} {3 (2 + N) (2 + k + N)^{5}}, \nonu\\
c_ {-456} & = & \frac{1}{3 (2 + N) (2 + k + N)^{6}}4 (360 + 962 k + 919 k^{2} + 381 k^{3} + 
      58 k^{4} + 946 N + 2062 k\, 
     N 
     \nonu\\& + & 1535 k^{2} N + 463 k^{3} N + 44 k^{4} N + 1051 N^{2} + 
      1715 k\, 
     N^{2} + 862 k^{2} N^{2} + 138 k^{3} N^{2} + 617 N^{3}
     \nonu\\& + & 659 k\, 
     N^{3} + 162 k^{2} N^{3} + 190 N^{4} + 100 k\, 
     N^{4} + 24 N^{5}), \nonu\\
c_ {-457} & = & \frac{1}{3 (2 + N) (2 + k + N)^{4}}(960 + 1636 k + 1068 k^{2} + 331 k^{3} + 
     42 k^{4} + 1916 N + 2564 k\, 
    N 
    \nonu\\& + & 1155 k^{2} N + 183 k^{3} N + 1456 N^{2} + 1369 k\, 
    N^{2} + 319 k^{2} N^{2} + 505 N^{3} + 253 k\, 
    N^{3} + 67 N^{4}), \nonu\\
c_ {-458} & = & \frac{1}{3 (2 + N) (2 + k + N)^{4}}2 (192 + 406 k + 265 k^{2} + 54 k^{3} + 
      362 N + 484 k\, N + 155 k^{2} N 
      \nonu\\ & + & 211 N^{2} + 136 k\, 
     N^{2} + 39 N^{3}), \nonu\\
c_ {-459} & = & - \frac {6 (32 + 55 k + 18 k^{2} + 41 N + 35 k\, 
      N + 11 N^{2})} {(2 + N) (2 + k + N)^{4}}, \nonu\\
c_ {-460} & = & \frac {1}{3 (2 + N) (2 + k + N)^{4}}2 (192 + 406 k + 239 k^{2} + 42 k^{3} + 
      362 N + 504 k\, N + 149 k^{2} N 
      \nonu\\ & + &  217 N^{2} + 152 k\, 
     N^{2} + 41 N^{3}), \nonu\\
c_ {-461} & = & \frac {4 (32 + 55 k + 18 k^{2} + 41 N + 35 k\, 
     N + 11 N^{2})} {(2 + N) (2 + k + N)^{3}}, \nonu\\
c_ {-462} & = & - \frac {2 (372 + 472 k + 136 k^{2} + 746 N + 697 k\, 
      N + 122 k^{2} N + 487 N^{2} + 253 k\, 
      N^{2} + 99 N^{3})} {3 (2 + N) (2 + k + N)^{4}}, \nonu\\
c_ {-463} & = & \frac {2 (-12 - 106 k - 143 k^{2} - 42 k^{3} + 76 N + 
      15 k\, N - 47 k^{2} N + 104 N^{2} + 55 k\, 
     N^{2} + 28 N^{3})} {3 (2 + N) (2 + k + N)^{4}}, \nonu\\
c_ {-464} & = & \frac {1}{3 (2 + N) (2 + k + N)^{3}}(720 + 812 k + 117 k^{2} - 42 k^{3} + 1564 N + 
     1394 k\, N + 191 k^{2} N 
     \nonu\\ & + &  1081 N^{2} + 560 k\, 
    N^{2} + 227 N^{3}), \nonu\\
c_ {-465} & = &  \frac {1}{3 (2 + N) (2 + k + N)^{5}}4 (192 + 704 k + 781 k^{2} + 360 k^{3} + 
      60 k^{4} + 544 N + 1312 k\, 
     N
     \nonu\\ & + &  922 k^{2} N + 208 k^{3} N + 499 N^{2} + 744 k\, 
     N^{2} + 253 k^{2} N^{2} + 182 N^{3} + 128 k\, 
     N^{3} + 23 N^{4}), \nonu\\
c_ {-466} & = & \frac {2 (-192 - 272 k - 15 k^{2} + 30 k^{3} - 
      208 N - 64 k\, N + 83 k^{2} N - 17 N^{2} + 66 k\, 
     N^{2} + 13 N^{3})} {3 (2 + N) (2 + k + N)^{4}}, \nonu\\
c_ {-467} & = & \frac {1}{3 (2 + N) (2 + k + N)^{5}}8 (624 + 1330 k + 1058 k^{2} + 375 k^{3} + 
      50 k^{4} + 1466 N + 2414 k\, 
     N 
     \nonu\\ & + &1344 k^{2} N + 274 k^{3} N + 10 k^{4} N +1334 N^{2} + 
      1569 k\, N^{2} + 518 k^{2} N^{2} + 37 k^{3} N^{2} + 594 N^{3} 
      \nonu\\ & + & 436 k\, 
     N^{3} + 58 k^{2} N^{3} + 132 N^{4} + 45 k\, 
     N^{4} + 12 N^{5}), \nonu\\
c_ {-468} & = & - \frac{1}{3 (2 + N) (2 + k + N)^{5}}4 (360 + 651 k + 369 k^{2} + 66 k^{3} + 
       753 N + 1100 k\, 
      N + 455 k^{2} N 
      \nonu\\ & + & 48 k^{3} N + 583 N^{2} + 615 k\, 
      N^{2} + 140 k^{2} N^{2} + 196 N^{3} + 112 k\, 
      N^{3} + 24 N^{4}), \nonu\\
c_ {-469} & = & \frac{1} {3 (2 + N) (2 + k + N)^{5}}8 (96 + 174 k + 124 k^{2} + 43 k^{3} + 
      6 k^{4} + 486 N + 838 k\, 
     N + 544 k^{2} N 
     \nonu\\ & + &  161 k^{3} N + 18 k^{4} N + 748 N^{2} + 
      1033 k\, 
     N^{2} + 471 k^{2} N^{2} + 73 k^{3} N^{2} + 510 N^{3} + 489 k\, 
     N^{3} 
     \nonu\\ & + &  114 k^{2} N^{3} + 163 N^{4} + 81 k\, 
     N^{4} + 20 N^{5}), \nonu\\ 
c_ {-470} & = & \frac {1}{3 (2 + N) (2 + k + N)^{4}}2 (100 k + 87 k^{2} + 18 k^{3} + 260 N + 
      634 k\, N + 367 k^{2} N + 60 k^{3} N 
      \nonu\\ & + &  467 N^{2} + 646 k\, 
     N^{2} + 186 k^{2} N^{2} + 265 N^{3} + 174 k\, 
     N^{3} + 48 N^{4}), \nonu\\
c_ {-471} & = & \frac {1}{3 (2 + N) (2 + k + N)^{4}}2 (20 k + 57 k^{2} + 48 k^{3} + 12 k^{4} + 
      340 N + 774 k\, 
     N + 685 k^{2} N 
     \nonu\\ & + &  268 k^{3} N + 36 k^{4} N + 717 N^{2} + 
      1194 k\, 
     N^{2} + 678 k^{2} N^{2} + 130 k^{3} N^{2} + 557 N^{3} + 616 k\, 
     N^{3} 
     \nonu\\ & + &  172 k^{2} N^{3} + 190 N^{4} + 106 k\, 
     N^{4} + 24 N^{5}), \nonu\\
c_ {-472} & = & \frac{1}{3 (2 + N) (2 + k + N)^{3}}2 (128 + 171 k + 24 k^{2} - 12 k^{3} + 325 N + 
      207 k\, N - 88 k^{2} N 
      \nonu\\ & - &  36 k^{3} N + 313 N^{2} + 110 k\, 
     N^{2} - 46 k^{2} N^{2} + 142 N^{3} + 34 k\, 
     N^{3} + 24 N^{4}).\nonu
\eea

\subsection{The OPE between the ${\cal N}=2$ higher spin-$2$ current}

The remaining OPE  in section $6$ 
between the ${\cal N}=2$ higher spin-$2$ current
can be summarized by
\bea
&& {\bf W^{(2)}}(Z_{1})\, {\bf W^{(2)}}(Z_{2})\;=\;\frac{1}{z_{12}^{4}}\, c_{1}+\frac{\theta_{12}\overline{\theta}_{12}}{z_{12}^{4}}\Bigg[c_{2}\, T+c_{3}\,\overline{D}H+c_{4}\, D\overline{H}+c_{5}\, G\,\overline{G}+c_{6}\, H\,\overline{H}\Bigg](Z_{2})
\nonu \\&&+\frac{\theta_{12}}{z_{12}^{3}}\Bigg[c_{7}\, DT+c_{8}\, G\, D\overline{G}+c_{9}\, H\,\overline{D}H+c_{10}\, H\, D\overline{H}+c_{11}\, H\, G\,\overline{G}+c_{12}\, T\, H
+c_{13}\, DG\,\overline{G}
\nonu \\&&+c_{14}\,\partial H\Bigg](Z_{2})
\nonu \\&&+\frac{\overline{\theta}_{12}}{z_{12}^{3}}\Bigg[c_{15}\,\overline{D}T+c_{16}\, G\,\overline{D}\overline{G}+c_{17}\,\overline{H}\, D\overline{H}+c_{18}\,\overline{H}\, G\,\overline{G}+c_{19}\, T\,\overline{H}+c_{20}\,\overline{D}G\,\overline{G}
\nonu \\&&+c_{21}\,\overline{D}H\,\overline{H}+c_{22}\,\partial\overline{H}\Bigg](Z_{2})
\nonu \\&&+\frac{\theta_{12}\overline{\theta}_{12}}{z_{12}^{3}}\Bigg[c_{23}\,\partial\overline{D}H+c_{24}\,\partial D\overline{H}+c_{25}\, G\,[D,\overline{D}]\overline{G}+c_{26}\, G\,\partial\overline{G}+c_{27}\, H\, G\,\overline{D}\overline{G}
\nonu \\&&+c_{28}\, H\,\overline{D}G\,\overline{G}+c_{29}\, H\,\partial\overline{H}+c_{30}\,\overline{H}\, G\, D\overline{G}+c_{31}\,\overline{H}\, DG\,\overline{G}+c_{32}\, T\,\overline{D}H
+c_{33}\, T\, D\overline{H}
\nonu \\&&+c_{34}\,\overline{D}H\,\overline{D}H+c_{35}\,\overline{D}H\, G\,\overline{G}+c_{36}\,\overline{D}T\, H+c_{37}\,[D,\overline{D}]G\,\overline{G}
+c_{38}\, D\overline{H}\, D\overline{H}+c_{39}\, D\overline{H}\, G\,\overline{G}
\nonu \\&&+c_{40}\, DT\,\overline{H}+c_{41}\,\partial G\,\overline{G}+c_{42}\,\partial H\,\overline{H}+c_{43}\,\partial T
\Bigg](Z_{2})
\nonu \\&&+\frac{1}{z_{12}^{2}}\Bigg[c_{44}\,{\bf T^{(2)}}+c_{45}\, {\bf T^{(1)}}\, {\bf T^{(1)}}+c_{46}\,[D,\overline{D}]T+c_{47}\,\partial D\overline{H}+c_{48}\, G\,[D,\overline{D}]\overline{G}
+c_{49}\, G\,\partial\overline{G}
\nonu \\&&+c_{50}\, H\, G\,\overline{D}\overline{G}+c_{51}\, H\,\overline{H}\,\overline{H}+c_{52}\, H\,\overline{H}\, G\,\overline{G}+c_{53}\, H\,\overline{D}G\,\overline{G}
+c_{54}\, H\,\overline{D}H\,\overline{H}+c_{55}\, H\,\partial\overline{H}
\nonu \\&&+c_{56}\,\overline{H}\, G\, D\overline{G}+c_{57}\,\overline{H}\, DG\,\overline{G}+c_{58}\, T\, T
+c_{59}\, T\,\overline{D}H+c_{60}\, T\, D\overline{H}+c_{61}\, T\, G\,\overline{G}+c_{62}\, T\, H\,\overline{H}
\nonu \\&&+c_{63}\,\partial\overline{D}H
+c_{64}\,\overline{D}G\, D\overline{G}+c_{65}\,\overline{D}H\,\overline{D}H+c_{66}\,\overline{D}H\, D\overline{H}+c_{67}\,\overline{D}H\, G\,\overline{G}+c_{68}\,\overline{D}T\, H
\nonu \\&&+c_{69}\,[D,\overline{D}]G\,\overline{G}+c_{70}\, DG\,\overline{D}\overline{G}+c_{71}\, D\overline{H}\, D\overline{H}+c_{72}\, D\overline{H}\, G\,\overline{G}+c_{73}\, DT\,\overline{H}
+c_{74}\,\partial G\,\overline{G}
\nonu \\&&+c_{75}\,\partial H\,\overline{H}+c_{76}\,\partial T
\Bigg](Z_{2})
\nonu \\&&+\frac{\theta_{12}}{z_{12}^{2}}\Bigg[c_{77}\, D{\bf T^{(2)}}+c_{78}\, {\bf T^{(1)}}\, D{\bf T^{(1)}}+c_{79}\, G\,\partial D\overline{G}+c_{80}\, H\,\partial D\overline{H}+c_{81}\, H\, G\,[D,\overline{D}]\overline{G}
\nonu \\&&+c_{82}\, H\, G\,\partial\overline{G}+c_{83}\, H\,\overline{H}\, G\, D\overline{G}+c_{84}\, H\,\overline{H}\, DG\,\overline{G}+c_{85}\, H\,\overline{D}G\, D\overline{G}
+c_{86}\, H\,\overline{D}H\, D\overline{H}
\nonu \\&&+c_{87}\, H\,[D,\overline{D}]G\,\overline{G}+c_{88}\, H\, DG\,\overline{D}\overline{G}+c_{89}\, H\, D\overline{H}\, D\overline{H}
+c_{90}\, H\, D\overline{H}\, G\,\overline{G}+c_{91}\, H\,\partial G\,\overline{G}
\nonu \\&&+c_{92}\,\overline{H}\, DG\, D\overline{G}+c_{93}\, T\, DT+c_{94}\, T\, G\, D\overline{G}
+c_{95}\, T\, H\, D\overline{H}+c_{96}\, T\, DG\,\overline{G}+c_{97}\, T\,\partial H
\nonu \\&&+c_{98}\,\partial DT+c_{99}\,\overline{D}H\, G\, D\overline{G}
+c_{100}\,\overline{D}H\, DG\,\overline{G}+c_{101}\,\partial\overline{D}H\, H
+c_{102}\,[D,\overline{D}]G\, D\overline{G}
\nonu \\&&+c_{103}\,[D,\overline{D}]T\, H
+c_{104}\, DG\,[D,\overline{D}]\overline{G}+c_{105}\, DG\,\partial\overline{G}+c_{106}\, D\overline{H}\, G\, D\overline{G}+c_{107}\, D\overline{H}\, DG\,\overline{G}
\nonu \\&&+c_{108}\, DT\,\overline{D}H+c_{109}\, DT\, D\overline{H}+c_{110}\, DT\, G\,\overline{G}+c_{111}\, DT\, H\,\overline{H}+c_{112}\,\partial DG\,\overline{G}
+c_{113}\,\partial G\, D\overline{G}
\nonu \\&&+c_{114}\,\partial H\,\overline{D}H+c_{115}\,\partial H\, D\overline{H}+c_{116}\,\partial H\, G\,\overline{G}+c_{117}\,\partial H\, H\,\overline{H}
+c_{118}\,\partial T\, H+c_{119}\,\partial^{2}H
\Bigg](Z_{2})
\nonu \\&&+\frac{\overline{\theta}_{12}}{z_{12}^{2}}\Bigg[c_{120}\,\overline{D}{\bf T^{(2)}}+c_{121}\,{\bf  T^{(1)}}\,\overline{D}{\bf T^{(1)}}+c_{122}\, G\,\partial\overline{D}\overline{G}+c_{123}\, H\,\overline{H}\, G\,\overline{D}\overline{G}
+c_{124}\, H\,\overline{H}\,\overline{D}G\,\overline{G}
\nonu \\&&+c_{125}\, H\,\overline{D}G\,\overline{D}\overline{G}+c_{126}\, H\,\partial\overline{H}\,\overline{H}+c_{127}\,\overline{H}\, G\,[D,\overline{D}]\overline{G}
+c_{128}\,\overline{H}\, G\,\partial\overline{G}+c_{129}\,\overline{H}\,\overline{D}G\, D\overline{G}
\nonu \\&&+c_{130}\,\overline{H}\,[D,\overline{D}]G\,\overline{G}+c_{131}\,\overline{H}\, DG\,\overline{D}\overline{G}
+c_{132}\,\overline{H}\,\partial G\,\overline{G}+c_{133}\, T\,\overline{D}T+c_{134}\, T\, G\,\overline{D}\overline{G}
\nonu \\&&+c_{135}\, T\,\overline{D}G\,\overline{G}+c_{136}\, T\,\overline{D}H\,\overline{H}
+c_{137}\, T\,\partial\overline{H}+c_{138}\,\partial\overline{D}T+c_{139}\,\overline{D}G\,[D,\overline{D}]\overline{G}+c_{140}\,\overline{D}G\,\partial\overline{G}
\nonu \\&&+c_{141}\,\overline{D}H\, G\,\overline{D}\overline{G}+c_{142}\,\overline{D}H\,\overline{H}\, D\overline{H}+c_{144}\,\overline{D}H\,\overline{H}\, G\,\overline{G}+c_{144}\,\overline{D}H\,\overline{D}G\,\overline{G}
+c_{145}\,\overline{D}H\,\overline{D}H\,\overline{H}
\nonu \\&&+c_{146}\,\overline{D}H\,\partial\overline{H}+c_{147}\,\overline{D}T\,\overline{D}H+c_{148}\,\overline{D}T\, D\overline{H}
+c_{149}\,\overline{D}T\, G\,\overline{G}+c_{150}\,\overline{D}T\, H\,\overline{H}
+c_{151}\,\partial\overline{D}G\,\overline{G}
\nonu \\&&+c_{152}\,\partial\overline{D}H\,\overline{H}
+c_{153}\,[D,\overline{D}]G\,\overline{D}\overline{G}+c_{154}\,[D,\overline{D}]T\,\overline{H}+c_{155}\, D\overline{H}\, G\,\overline{D}\overline{G}
+c_{156}\, D\overline{H}\,\overline{D}G\,\overline{G}
\nonu \\&&+c_{157}\,\partial D\overline{H}\,\overline{H}+c_{158}\,\partial G\,\overline{D}\overline{G}+c_{159}\,\partial\overline{H}\, D\overline{H}+c_{160}\,\partial\overline{H}\, G\,\overline{G}+c_{161}\,\partial T\,\overline{H}
+c_{162}\partial^{2}\overline{H}
\Bigg](Z_{2})
\nonu \\&&+\frac{\theta_{12}\overline{\theta}_{12}}{z_{12}^{2}}\Bigg[c_{163}\, {\bf W^{(3)}}+c_{164}\,[D,\overline{D}]{\bf T^{(2)}}+c_{165}\,{\bf  T^{(1)}}\,{\bf  W^{(2)}}+c_{166}\,\overline{D}{\bf T^{(1)}}\, D{\bf T^{(1)}}
+c_{167}\,{\bf U^{(\frac{3}{2})}\, V^{(\frac{3}{2})}}
\nonu \\&&+c_{168}\, G\,\partial[D,\overline{D}]\overline{G}+c_{169}\, G\,\overline{D}G\,\overline{G}\, D\overline{G}+c_{170}\, G\, DG\,\overline{G}\,\overline{D}\overline{G}
+c_{171}\, G\,\partial^{2}\overline{G}+c_{172}\, H\, G\,\partial\overline{D}\overline{G}
\nonu \\&&+c_{173}\, H\,\overline{H}\, G\,[D,\overline{D}]\overline{G}+c_{174}\, H\,\overline{H}\, G\,\partial\overline{G}
+c_{175}\, H\,\overline{H}\,\overline{D}G\, D\overline{G}+c_{176}\, H\,\overline{H}\,[D,\overline{D}]G\,\overline{G}
\nonu \\&&+c_{177}\, H\,\overline{H}\, DG\,\overline{D}\overline{G}
+c_{178}\, H\,\overline{H}\, D\overline{H}\, D\overline{H}+c_{179}\, H\,\overline{H}\,\partial G\,\overline{G}+c_{180}\, H\,\overline{D}G\,[D,\overline{D}]\overline{G}
\nonu \\&&+c_{181}\, H\,\overline{D}G\,\partial\overline{G}
+c_{182}\, H\,\overline{D}H\, G\,\overline{D}\overline{G}+c_{183}\, H\,\overline{D}H\,\overline{H}\, D\overline{H}+c_{184}\, H\,\overline{D}H\,\overline{D}G\,\overline{G}
\nonu \\&&+c_{185}\, H\,\overline{D}H\,\overline{D}H\,\overline{H}+c_{186}\, H\,\overline{D}H\,\partial\overline{H}+c_{187}\, H\,\partial\overline{D}G\,\overline{G}+c_{188}\, H\,[D,\overline{D}]G\,\overline{D}\overline{G}
\nonu \\&&+c_{189}\, H\, D\overline{H}\, G\,\overline{D}\overline{G}+c_{190}\, H\, D\overline{H}\,\overline{D}G\,\overline{G}+c_{191}\, H\,\partial D\overline{H}\,\overline{H}+c_{192}\, H\,\partial G\,\overline{D}\overline{G}
\nonumber\\&&+c_{193}\, H\,\partial\overline{H}\, D\overline{H}+c_{194}\, H\,\partial\overline{H}\, G\,\overline{G}+c_{195}\, H\,\partial^{2}\overline{H}+c_{196}\,\overline{H}\, G\,\partial D\overline{G}
+c_{197}\,\overline{H}\,[D,\overline{D}]G\, D\overline{G}
\nonu \\&&+c_{198}\,\overline{H}\, DG\,[D,\overline{D}]\overline{G}+c_{199}\,\overline{H}\, DG\,\partial\overline{G}
+c_{200}\,\overline{H}\, D\overline{H}\, G\, D\overline{G}+c_{201}\,\overline{H}\, D\overline{H}\, DG\,\overline{G}
\nonu \\&&+c_{202}\,\overline{H}\,\partial DG\,\overline{G}+c_{203}\,\overline{H}\,\partial G\, D\overline{G}
+c_{204}\, T\,\partial\overline{D}H+c_{205}\, T\,[D,\overline{D}]T+c_{206}\, T\,\partial D\overline{H}
\nonu \\&&+c_{207}\, T\, G\,[D,\overline{D}]\overline{G}
+c_{208}\, T\, G\,\partial\overline{G}+c_{209}\, T\, H\, G\,\overline{D}\overline{G}+c_{210}\, T\, H\,\overline{H}\, D\overline{H}+c_{211}\, T\, H\,\overline{D}G\,\overline{G}
\nonu \\&&+c_{212}\, T\, H\,\overline{D}H\,\overline{H}+c_{213}\, T\, H\,\partial\overline{H}+c_{214}\, T\,\overline{H}\, G\, D\overline{G}+c_{215}\, T\,\overline{H}\, DG\,\overline{G}
+c_{216}\, T\, T\, T
\nonu \\&&+c_{217}\, T\, T\,\overline{D}H+c_{218}\, T\, T\, D\overline{H}+c_{219}\, T\, T\, H\,\overline{H}+c_{220}\, T\,\overline{D}G\, D\overline{G}
+c_{221}\, T\,\overline{D}H\,\overline{D}H
\nonu \\&&+c_{222}\, T\,\overline{D}H\, D\overline{H}+c_{223}\, T\,\overline{D}T\, H+c_{224}\, T\,[D,\overline{D}]G\,\overline{G}
+c_{225}\, T\, DG\,\overline{D}\overline{G}+c_{226}\, T\, D\overline{H}\, D\overline{H}
\nonu \\&&+c_{227}\, T\, DT\,\overline{H}+c_{228}\, T\,\partial G\,\overline{G}
+c_{229}\, T\,\partial H\,\overline{H}+c_{230}\,\partial^{2}\overline{D}H+c_{231}\,\partial^{2}D\overline{H}+c_{232}\,\overline{D}G\,\partial D\overline{G}
\nonu \\&&+c_{233}\,\overline{D}H\,\partial D\overline{H}+c_{234}\,\overline{D}H\, G\,[D,\overline{D}]\overline{G}+c_{235}\,\overline{D}H\, G\,\partial\overline{G}+c_{236}\,\overline{D}H\,\overline{H}\, G\, D\overline{G}
\nonu \\&&+c_{237}\,\overline{D}H\,\overline{H}\, DG\,\overline{G}+c_{238}\,\overline{D}H\,\overline{D}G\, D\overline{G}+c_{239}\,\overline{D}H\,\overline{D}H\,\overline{D}H
+c_{240}\,\overline{D}H\,\overline{D}H\, D\overline{H}
\nonu \\&&+c_{241}\,\overline{D}H\,[D,\overline{D}]G\,\overline{G}+c_{242}\,\overline{D}H\, DG\,\overline{D}\overline{G}
+c_{243}\,\overline{D}H\, D\overline{H}\, D\overline{H}+c_{244}\,\overline{D}H\, D\overline{H}\, G\,\overline{G}
\nonu \\&&+c_{245}\,\overline{D}H\,\partial G\,\overline{G}+c_{246}\,\overline{D}T\, DT
+c_{247}\,\overline{D}T\, G\, D\overline{G}+c_{248}\,\overline{D}T\, H\,\overline{D}H+c_{249}\,\overline{D}T\, H\, D\overline{H}
\nonu \\&&+c_{250}\,\overline{D}T\, H\, G\,\overline{G}
+c_{251}\,\overline{D}T\, DG\,\overline{G}+c_{252}\,\overline{D}T\,\partial H+c_{253}\,\partial[D,\overline{D}]T+c_{254}\,\partial\overline{D}G\, D\overline{G}
\nonu \\&&+c_{255}\,\partial\overline{D}H\,\overline{D}H+c_{256}\,\partial\overline{D}H\, D\overline{H}+c_{257}\,\partial\overline{D}H\, G\,\overline{G}+c_{258}\,\partial\overline{D}H\, H\,\overline{H}+c_{259}\,\partial\overline{D}T\, H
\nonu \\&&+c_{260}\,[D,\overline{D}]G\,[D,\overline{D}]\overline{G}+c_{261}\,[D,\overline{D}]G\,\partial\overline{G}+c_{262}\,[D,\overline{D}]T\,\overline{D}H
+c_{263}\,[D,\overline{D}]T\, D\overline{H}
\nonu \\&&+c_{264}\,[D,\overline{D}]T\, G\,\overline{G}+c_{265}\,[D,\overline{D}]T\, H\,\overline{H}
+c_{266}\,\partial[D,\overline{D}]G\,\overline{G}+c_{267}\, DG\,\partial\overline{D}\overline{G}
\nonu \\&&+c_{268}\, D\overline{H}\, G\,[D,\overline{D}]\overline{G}+c_{269}\, D\overline{H}\, G\,\partial\overline{G}
+c_{270}\, D\overline{H}\,\overline{D}G\, D\overline{G}+c_{271}\, D\overline{H}\,[D,\overline{D}]G\,\overline{G}
\nonu \\&&+c_{272}\, D\overline{H}\, DG\,\overline{D}\overline{G}
+c_{273}\, D\overline{H}\, D\overline{H}\, D\overline{H}+c_{274}\, D\overline{H}\,\partial G\,\overline{G}+c_{275}\, DT\, G\,\overline{D}\overline{G}
+c_{276}\, DT\,\overline{H}\, D\overline{H}
\nonu \\&&+c_{277}\, DT\,\overline{H}\, G\,\overline{G}+c_{278}\, DT\,\overline{D}G\,\overline{G}+c_{279}\, DT\,\overline{D}H\,\overline{H}+c_{282}\, DT\,\partial\overline{H}
+c_{281}\,\partial DG\,\overline{D}\overline{G}
\nonu \\&&+c_{282}\,\partial D\overline{H}\, D\overline{H}+c_{283}\,\partial D\overline{H}\, G\,\overline{G}+c_{284}\,\partial DT\,\overline{H}
+c_{285}\,\partial G\,[D,\overline{D}]\overline{G}
+c_{286}\,\partial G\,\partial\overline{G}
\nonu \\&&+c_{287}\,\partial H\, G\,\overline{D}\overline{G}+c_{288}\,\partial H\,\overline{H}\, D\overline{H}
+c_{289}\,\partial H\,\overline{H}\, G\,\overline{G}+c_{290}\,\partial H\,\overline{D}G\,\overline{G}
+c_{291}\,\partial H\,\overline{D}H\,\overline{H}
\nonu \\&&+c_{292}\,\partial H\,\partial\overline{H}
+c_{293}\,\partial\overline{H}\, G\, D\overline{G}+c_{294}\,\partial\overline{H}\, DG\,\overline{G}+c_{295}\,\partial T\, T
+c_{296}\,\partial T\,\overline{D}H
+c_{297}\,\partial T\, D\overline{H}
\nonu \\&&+c_{298}\,\partial T\, G\,\overline{G}+c_{299}\,\partial T\, H\,\overline{H}+c_{300}\,\partial^{2}G\,\overline{G}+c_{301}\,\partial^{2}H\,\overline{H}
+c_{302}\,\partial^{2}T
\Bigg](Z_{2})
\nonu \\&&+\frac{1}{z_{12}}\Bigg[c_{303}\,\partial {\bf T^{(2)}}+c_{304}\,\partial {\bf  T^{(1)}}\, {\bf T^{(1)}}+c_{305}\,\partial^{2}D\overline{H}+c_{306}\, G\,\partial[D,\overline{D}]\overline{G}
+c_{307}\, G\,\partial^{2}\overline{G}
\nonu \\&&+c_{308}\, H\, G\,\partial\overline{D}\overline{G}+c_{309}\, H\,\overline{H}\, G\,\partial\overline{G}+c_{310}\, H\,\overline{H}\,\partial G\,\overline{G}
+c_{311}\, H\,\overline{D}G\,\partial\overline{G}+c_{312}\, H\,\overline{D}H\,\partial\overline{H}
\nonu \\&&+c_{313}\, H\,\partial\overline{D}G\,\overline{G}+c_{314}\, H\,\partial D\overline{H}\,\overline{H}
+c_{315}\, H\,\partial G\,\overline{D}\overline{G}+c_{316}\, H\,\partial\overline{H}\, D\overline{H}+c_{317}\, H\,\partial\overline{H}\, G\,\overline{G}
\nonu \\&&+c_{318}\, H\,\partial^{2}\overline{H}
+c_{319}\,\overline{H}\, G\,\partial D\overline{G}+c_{320}\,\overline{H}\, DG\,\partial\overline{G}+c_{321}\,\overline{H}\,\partial DG\,\overline{G}+c_{322}\,\overline{H}\,\partial G\, D\overline{G}
\nonu \\&&+c_{333}\,\overline{D}T\,\partial H+c_{334}\,\partial\overline{D}G\, D\overline{G}+c_{335}\,\partial\overline{D}H\,\overline{D}H+c_{336}\,\partial\overline{D}H\, D\overline{H}
+c_{337}\,\partial\overline{D}H\, G\,\overline{G}
\nonu \\&&+c_{338}\,\partial\overline{D}H\, H\,\overline{H}+c_{339}\,\partial\overline{D}T\, H+c_{340}\,[D,\overline{D}]G\,\partial\overline{G}
+c_{341}\,\partial[D,\overline{D}]G\,\overline{G}+c_{342}\, DG\,\partial\overline{D}\overline{G}
\nonu \\&&+c_{343}\, D\overline{H}\, G\,\partial\overline{G}+c_{344}\, D\overline{H}\,\partial G\,\overline{G}
+c_{345}\, DT\,\partial\overline{H}+c_{346}\,\partial DG\,\overline{D}\overline{G}+c_{347}\,\partial D\overline{H}\, D\overline{H}
\nonu \\&&+c_{348}\,\partial D\overline{H}\, G\,\overline{G}
+c_{349}\,\partial DT\,\overline{H}+c_{350}\,\partial G\,[D,\overline{D}]\overline{G}+c_{351}\,\partial H\, G\,\overline{D}\overline{G}+c_{352}\,\partial H\,\overline{H}\, D\overline{H}
\nonu \\&&+c_{353}\,\partial H\,\overline{H}\, G\,\overline{G}+c_{354}\,\partial H\,\overline{D}G\,\overline{G}+c_{355}\,\partial H\,\overline{D}H\,\overline{H}+c_{356}\,\partial H\,\partial\overline{H}
+c_{357}\,\partial\overline{H}\, G\, D\overline{G}
\nonu \\&&+c_{358}\,\partial\overline{H}\, DG\,\overline{G}+c_{359}\,\partial T\, T+c_{360}\,\partial T\,\overline{D}H+c_{361}\,\partial T\, D\overline{H}
+c_{362}\,\partial T\, G\,\overline{G}+c_{363}\,\partial T\, H\,\overline{H}
\nonu \\&&+c_{364}\,\partial^{2}\overline{D}H+c_{365}\,\partial^{2}G\,\overline{G}+c_{366}\,\partial^{2}H\,\overline{H}
+c_{367}\,\partial[D,\overline{D}]T+c_{368}\,\partial^{2}T
\Bigg](Z_{2})
\nonu \\&&+\frac{\theta_{12}}{z_{12}}\Bigg[
c_{369}\, D{\bf W^{(3)}}
+c_{370}\,\partial D{\bf T^{(2)}}
+c_{371}\, D{\bf T^{(1)}}\, {\bf W^{(2)}}
+c_{372}\, {\bf T^{(1)}}\,  D{\bf W^{(2)}}
+c_{373}\,\partial D{\bf T^{(1)}}\, {\bf T^{(1)}}
\nonu \\&&+c_{374}\,\partial {\bf T^{(1)}}\, D{\bf T^{(1)}}
+c_{375}\,[D,\overline{D}]{\bf T^{(1)}}\, D{\bf T^{(1)}}
+c_{376}\, D{\bf U^{(\frac{3}{2})}}\, {\bf V^{(\frac{3}{2})}}
+c_{377}\,{\bf  U^{(\frac{3}{2})}}\, D{\bf V^{(\frac{3}{2})}}
\nonu \\&&+c_{378}\, H\,\partial^{2}D\overline{H}
+c_{379}\, H\, G\,\partial[D,\overline{D}]\overline{G}
+c_{380}\, H\, G\,\partial^{2}\overline{G}
+c_{381}\, H\,\overline{H}\, G\,\partial D\overline{G}
\nonu \\&&+c_{382}\, H\,\overline{H}\, DG\,\partial\overline{G}
+c_{383}\, H\,\overline{H}\,\partial DG\,\overline{G}
+c_{384}\, H\,\overline{H}\,\partial G\, D\overline{G}
+c_{385}\, H\,\overline{D}G\,\partial D\overline{G}
\nonu \\&&+c_{386}\, H\,\overline{D}H\,\partial D\overline{H}
+c_{387}\, H\,\overline{D}H\, G\,[D,\overline{D}]\overline{G}
+c_{388}\, H\,\overline{D}H\, G\,\partial\overline{G}
+c_{389}\, H\,\overline{D}H\,\overline{D}G\, D\overline{G}
\nonu \\&&+c_{390}\, H\,\overline{D}H\,\overline{D}H\, D\overline{H}
+c_{391}\, H\,\overline{D}H\,[D,\overline{D}]G\,\overline{G}
+c_{392}\, H\,\overline{D}H\, DG\,\overline{D}\overline{G}
+c_{393}\, H\,\overline{D}H\, D\overline{H}\, D\overline{H}
\nonu \\&&+c_{394}\, H\,\overline{D}H\,\partial G\,\overline{G}
+c_{395}\, H\,\partial\overline{D}G\, D\overline{G}
+c_{396}\, H\,[D,\overline{D}]G\,[D,\overline{D}]\overline{G}
+c_{397}\, H\,[D,\overline{D}]G\,\partial\overline{G}
\nonu \\&&
+c_{398}\, H\,\partial[D,\overline{D}]G\,\overline{G}
+c_{399}\, H\, DG\,\partial\overline{D}\overline{G}
+c_{400}\, H\, D\overline{H}\, G\,[D,\overline{D}]\overline{G}
+c_{401}\, H\, D\overline{H}\, G\,\partial\overline{G}
\nonu \\&&+c_{402}\, H\, D\overline{H}\,\overline{D}G\, D\overline{G}
+c_{403}\, H\, D\overline{H}\,[D,\overline{D}]G\,\overline{G}
+c_{404}\, H\, D\overline{H}\, DG\,\overline{D}\overline{G}
\nonu \\&&+c_{405}\, H\, D\overline{H}\, D\overline{H}\, D\overline{H}
+c_{406}\, H\, D\overline{H}\,\partial G\,\overline{G}
+c_{407}\, H\,\partial DG\,\overline{D}\overline{G}
+c_{408}\, H\,\partial D\overline{H}\, D\overline{H}
\nonu \\&&+
c_{409}\, H\,\partial D\overline{H}\, G\,\overline{G}
+c_{410}\, H\,\partial G\,[D,\overline{D}]\overline{G}
+c_{411}\, H\,\partial G\,\partial\overline{G}
+c_{412}\, H\,\partial\overline{H}\, G\, D\overline{G}
\nonu \\&&+c_{413}\, H\,\partial\overline{H}\, DG\,\overline{G}
+c_{414}\, H\,\partial^{2}G\,\overline{G}
+c_{415}\,\overline{H}\, DG\,\partial D\overline{G}
+c_{416}\,\overline{H}\, D\overline{H}\, DG\, D\overline{G}
\nonu \\&&+c_{417}\,\overline{H}\,\partial DG\, D\overline{G}
+c_{418}\, T\, G\,\partial D\overline{G}
+c_{419}\, T\, H\,\partial D\overline{H}
+c_{420}\, T\, H\, G\,[D,\overline{D}]\overline{G}
\nonu \\&&+c_{421}\, T\, H\, G\,\partial\overline{G}
+c_{422}\, T\, H\,\overline{D}G\, D\overline{G}
+c_{423}\, T\, H\,\overline{D}H\, D\overline{H}
+c_{424}\, T\, H\,[D,\overline{D}]G\,\overline{G}
\nonu \\&&+c_{425}\, T\, H\, DG\,\overline{D}\overline{G}
+c_{426}\, T\, H\, D\overline{H}\, D\overline{H}
+c_{427}\, T\, H\,\partial G\,\overline{G}
+c_{428}\, T\, T\, DT
+c_{429}\, T\, T\, H\, D\overline{H}
\nonu \\&&+
c_{430}\, T\, T\,\partial H
+c_{431}\, T\,\partial\overline{D}H\, H
+c_{432}\, T\,[D,\overline{D}]G\, D\overline{G}
+c_{433}\, T\,[D,\overline{D}]T\, H
\nonu \\&&+
c_{434}\, T\, DG\,[D,\overline{D}]\overline{G}
+c_{435}\, T\, DG\,\partial\overline{G}
+c_{436}\, T\, D\overline{H}\, G\, D\overline{G}
+c_{437}\, T\, D\overline{H}\, DG\,\overline{G}
\nonu \\&&+
c_{438}\, T\, DT\,\overline{D}H
+c_{439}\, T\, DT\, D\overline{H}
+c_{440}\, T\, DT\, H\,\overline{H}
+c_{441}T\,\partial DG\,\overline{G}
+c_{442}\, T\,\partial G\, D\overline{G}
\nonu \\&&+c_{443}\, T\,\partial H\,\overline{D}H
+c_{444}\, T\,\partial H\, D\overline{H}
+c_{445}\, T\,\partial H\, H\,\overline{H}
+c_{446}\, T\,\partial^{2}H
+c_{447}\, G\,\partial^{2}D\overline{G}
\nonu \\&&
+c_{448}\, G\, DG\,\partial\overline{G}\,\overline{G}
+c_{449}\,\overline{D}G\, DG\,\overline{G}\, D\overline{G}
+c_{450}\,\overline{D}H\, G\,\partial D\overline{G}
+c_{451}\,\overline{D}H\,\overline{H}\, DG\, D\overline{G}
\nonu \\&&+c_{452}\,\overline{D}H\,[D,\overline{D}]G\, D\overline{G}
+c_{453}\,\overline{D}H\, DG\,[D,\overline{D}]\overline{G}
+c_{454}\,\overline{D}H\, DG\,\partial\overline{G}
+c_{455}\,\overline{D}H\, D\overline{H}\, G\, D\overline{G}
\nonu \\&&+
c_{456}\,\overline{D}H\, D\overline{H}\, DG\,\overline{G}
+c_{457}\,\overline{D}H\,\partial DG\,\overline{G}
+c_{458}\,\overline{D}H\,\partial G\, D\overline{G}
+c_{459}\,\overline{D}T\, H\, G\, D\overline{G}
\nonu \\&&+
c_{460}\,\overline{D}T\, H\, DG\,\overline{G}
+c_{461}\,\overline{D}T\, DG\, D\overline{G}
+c_{462}\,\overline{D}T\, DT\, H
+c_{463}\,\overline{D}T\,\partial H\, H
\nonu \\&&+
c_{464}\,\partial\overline{D}H\, G\, D\overline{G}
+c_{465}\,\partial\overline{D}H\, H\, D\overline{H}
+c_{466}\,\partial\overline{D}H\, DG\,\overline{G}
+c_{467}\,\partial\overline{D}H\,\partial H
\nonu \\&&+
c_{468}\,\partial^{2}\overline{D}H\, H
+c_{469}\,[D,\overline{D}]G\,\partial D\overline{G}
+c_{470}\,[D,\overline{D}]T\, DT
+c_{471}\,[D,\overline{D}]T\, G\, D\overline{G}
\nonu \\&&+
c_{472}\,[D,\overline{D}]T\, H\,\overline{D}H
+c_{473}\,[D,\overline{D}]T\, H\, D\overline{H}
+c_{474}\,[D,\overline{D}]T\, H\, G\,\overline{G}
+c_{475}\,[D,\overline{D}]T\, DG\,\overline{G}
\nonu \\&&+c_{476}\,[D,\overline{D}]T\,\partial H
+c_{477}\, G\, DG\,\overline{G}\,[D,\overline{D}]\overline{G}
+c_{478}\,\partial[D,\overline{D}]G\, D\overline{G}
+c_{479}\,\partial[D,\overline{D}]T\, H
\nonu \\&&+
c_{480}\, DG\,\partial[D,\overline{D}]\overline{G}
+c_{481}\, DG\, DG\,\overline{G}\,\overline{D}\overline{G}
+c_{482}\, DG\,\partial^{2}\overline{G}
+c_{483}\, D\overline{H}\, G\,\partial D\overline{G}
\nonu \\&&+
c_{484}\, D\overline{H}\,[D,\overline{D}]G\, D\overline{G}
+c_{485}\, D\overline{H}\, DG\,[D,\overline{D}]\overline{G}
+c_{486}\, D\overline{H}\, DG\,\partial\overline{G}
\nonu \\&&+
c_{487}\, D\overline{H}\, D\overline{H}\, G\, D\overline{G}
+c_{488}\, D\overline{H}\, D\overline{H}\, DG\,\overline{G}
+c_{489}\, D\overline{H}\,\partial DG\,\overline{G}
+c_{490}\, D\overline{H}\,\partial G\, D\overline{G}
\nonu \\&&+c_{491}\, DT\,\partial\overline{D}H
+c_{492}\, DT\,\partial D\overline{H}
+c_{493}\, DT\, G\,[D,\overline{D}]\overline{G}
+c_{494}\, DT\, G\,\partial\overline{G}
\nonu \\&&+
c_{495}\, DT\, H\, G\,\overline{D}\overline{G}
+c_{496}\, DT\, H\,\overline{H}\, D\overline{H}
+c_{497}\, DT\, H\,\overline{D}G\,\overline{G}
+c_{498}\, DT\, H\,\overline{D}H\,\overline{H}
\nonu \\&&+
c_{499}\, DT\, H\,\partial\overline{H}
+c_{500}\, DT\,\overline{H}\, G\, D\overline{G}
+c_{501}\, DT\,\overline{H}\, DG\,\overline{G}
+c_{502}\, DT\,\overline{D}G\, D\overline{G}
\nonu \\&&+
c_{503}\, DT\,\overline{D}H\,\overline{D}H
+c_{504}\, DT\,\overline{D}H\, D\overline{H}
+c_{505}\, DT\,[D,\overline{D}]G\,\overline{G}
+c_{506}\, DT\, DG\,\overline{D}\overline{G}
\nonu \\&&+
c_{507}\, DT\, D\overline{H}\, D\overline{H}
+c_{508}\, DT\, D\overline{H}\, G\,\overline{G}
+c_{509}\, DT\,\partial G\,\overline{G}
+c_{510}\, DT\,\partial H\,\overline{H}
+c_{511}\,\partial^{2}DT
\nonu \\&&+
c_{512}\, G\, DG\,\overline{D}\overline{G}\, D\overline{G}
+c_{513}\,\partial DG\,[D,\overline{D}]\overline{G}
+c_{514}\,\partial DG\,\partial\overline{G}
+c_{515}\,\partial D\overline{H}\, G\, D\overline{G}
\nonu \\&&+
c_{516}\,\partial D\overline{H}\, DG\,\overline{G}
+c_{517}\,\partial DT\, T
+c_{518}\,\partial DT\,\overline{D}H
+c_{519}\,\partial DT\, D\overline{H}+c_{520}\,\partial DT\, G\,\overline{G}
\nonu \\&&+
c_{521}\,\partial DT\, H\,\overline{H}
+c_{522}\, G\,\overline{D}G\, D\overline{G}\, D\overline{G}
+c_{523}\,\partial^{2}DG\,\overline{G}
+c_{524}\,\partial G\,\partial D\overline{G}
+c_{525}\,\partial G\, G\,\overline{G}\, D\overline{G}
\nonu \\&&+
c_{526}\,\partial H\,\partial D\overline{H}
+c_{527}\,\partial H\, G\,[D,\overline{D}]\overline{G}
+c_{528}\,\partial H\, G\,\overline{G}
+c_{529}\,\partial H\, H\, G\,\overline{D}\overline{G}
\nonu \\&&+
c_{530}\,\partial H\, H\,\overline{H}\, D\overline{H}
+c_{531}\,\partial H\, H\,\overline{D}G\,\overline{G}
+c_{532}\,\partial H\, H\,\overline{D}H\,\overline{H}
+c_{533}\,\partial H\, H\,\partial\overline{H}
\nonu \\&&+
c_{534}\,\partial H\,\overline{H}\, G\, D\overline{G}
+c_{535}\,\partial H\,\overline{H}\, DG\,\overline{G}
+c_{536}\,\partial H,\overline{D}G\, D\overline{G}
+c_{537}\,\partial H\,\overline{D}H,\overline{D}H
\nonu \\&&+
c_{538}\,\partial H\,\overline{D}H\, D\overline{H}
+c_{539}\,\partial H\,[D,\overline{D}]G\,\overline{G}
+c_{540}\,\partial H\, DG\,\overline{D}\overline{G}
+c_{541}\,\partial H\, D\overline{H}\, D\overline{H}
\nonu \\&&+
c_{542}\,\partial H\,\partial G\,\overline{G}
+c_{543}\,\partial\overline{H}\, DG\, D\overline{G}
+c_{544}\,\partial T\, DT
+c_{545}\,\partial T\, G\, D\overline{G}
+c_{546}\,\partial T\, H\,\overline{D}H
\nonu \\&&+c_{547}\,\partial T\, H\, D\overline{H}
+c_{548}\,\partial T\, H\, G\,\overline{G}
+c_{549}\,\partial T\, T\, H
+c_{550}\,\partial T\, DG\,\overline{G}
+c_{551}\,\partial T\,\partial H
\nonu \\&&+
c_{552}\, G\,[D,\overline{D}]G\,\overline{G}\, D\overline{G}
+c_{553}\,\partial^{2}G\, D\overline{G}
+c_{554}\,\partial^{2}H\,\overline{D}H
+c_{555}\,\partial^{2}H\, D\overline{H}
+c_{556}\,\partial^{2}H\, G\,\overline{G}
\nonu \\&&
+c_{557}\,\partial^{2}H\, H\,\overline{H}
+c_{558}\,\partial^{2}T\, H
+c_{559}\,\partial^{3}H
\Bigg](Z_{2})
\nonu \\&&+\frac{\overline{\theta}_{12}}{z_{12}}\Bigg[
c_{560}\, \overline{D}{\bf W^{(3)}}
+c_{561}\,\partial \overline{D}{\bf T^{(2)}}
+c_{562}\,\overline{D}{\bf T^{(1)}\, W^{(2)}}
+c_{563}\, {\bf T^{(1)}}\,\overline{D}{\bf W^{(2)}}
+c_{564}\,\partial\overline{D}{\bf T^{(1)}\, T^{(1)}}
\nonu \\&&+c_{565}\,\partial {\bf T^{(1)}}\,\overline{D}{\bf T^{(1)}}
+c_{566}\,\overline{D}{\bf T^{(1)}}\,[D,\overline{D}]{\bf T^{(1)}}
+c_{567}\,\overline{D}{\bf U^{(\frac{3}{2})}\, V^{(\frac{3}{2})}}
+c_{568}\, {\bf U^{(\frac{3}{2})}}\,\overline{D}{\bf V^{(\frac{3}{2})}}
\nonu \\&&+c_{569}\, H\, \overline{H}\, G \,\partial \overline{D}\overline{G}
+c_{570}\, H\,\overline{H}\,\overline{D}G\,\partial \overline{G}
+c_{571}\, H\,\overline{H}\,\overline{D}\partial G\,\overline{G}
+c_{572}\, H\,\overline{H}\,\partial G\,\overline{D}\overline{G}
\nonu \\&&+c_{573}\, H\,\overline{D}G\,\partial \overline{D}\overline{G}
+c_{574}\, H\,\overline{D}H\,\overline{D}G\,\overline{D}\overline{G}+c_{575}\, H\,\overline{D}H\,\partial\overline{H}\,\overline{H}
+c_{576}\, H\,\partial\overline{D} G\,\overline{D}\overline{G}]
\nonu \\&&+c_{577}\, H\, D\overline{H}\,\overline{D}G\,\overline{D}\overline{G}+c_{578}\, H\,\partial\overline{H}\, G\,\overline{D}\overline{G}
+c_{579}\, H\,\partial\overline{H}\,\overline{H}\, D\overline{H}
+c_{580}\, H\,\partial\overline{H}\,\overline{D}G\,\overline{G}
\nonu \\&&+c_{581}\, H\,\partial^{2}\overline{H}\,\overline{H}+c_{582}\,\overline{H}\, G\,\partial[D,\overline{D}]\overline{G}+c_{583}\,\overline{H}\, G\,\partial^{2}\overline{G}
+c_{584}\,\overline{H}\,\overline{D}G\,\partial D\overline{G}
\nonu \\&&+c_{585}\,\overline{H}\,\partial\overline{D}G\, D\overline{G}+c_{586}\,\overline{H}\,[D,\overline{D}]G\, [D ,\overline{D}] \overline{G}
+c_{587}\,\overline{H}\,[D,\overline{D}]G\,\partial\overline{G}
+c_{588}\,\overline{H}\,\partial[D,\overline{D}]G\,\overline{G}
\nonu \\&&+c_{589}\,\overline{H}\, DG\,\partial\overline{D}\overline{G}
+c_{590}\,\overline{H}\, D\overline{H}\, G\,[D,\overline{D}]\overline{G}+c_{591}\,\overline{H}\, D\overline{H}\, G\,\partial\overline{G}
+c_{592}\,\overline{H}\, D\overline{H}\,\overline{D}G\, D\overline{G}
\nonu \\&&+c_{593}\,\overline{H}\, D\overline{H}\,[D,\overline{D}]G\,\overline{G}+c_{594}\,\overline{H}\, D\overline{H}\, DG\,\overline{D}\overline{G}+c_{595}\,\overline{H}\, D\overline{H}\,\partial G\,\overline{G}
+c_{596}\,\overline{H}\,\partial DG\,\overline{D}\overline{G}
\nonu \\&&+c_{597}\,\overline{H}\,\partial G\,[D,\overline{D}]\overline{G}
+c_{598}\,\overline{H}\,\partial G\,\partial\overline{G}+c_{599}\,\overline{H}\,\partial^{2}G\,\overline{G}
+c_{600}\, T\, G\,\partial\overline{D}\overline{G}+c_{601}\, T\, H\,\partial\overline{H}\,\overline{H}
\nonu \\&&+c_{602}\, T\,\overline{H}\, G\,[D,\overline{D}]\overline{G}+c_{603}\, T\,\overline{H}\, G\,\partial\overline{G}
+c_{604}\, T\,\overline{H}\,\overline{D}G\, D\overline{G}+c_{605}\, T\,\overline{H}\,[D,\overline{D}]G\,\overline{G}
\nonu \\&&+c_{606}\, T\,\overline{H}\, DG\,\overline{D}\overline{G}
+c_{607}\, T\,\overline{H}\,\partial G\,\overline{G}
+c_{608}\, T\, T\,\overline{D}T
+c_{609}\, T\, T\,\overline{D}H\,\overline{H}+c_{610}\, T\, T\,\partial\overline{H}
\nonu \\&&+c_{611}\, T\,\overline{D}G\,[D,\overline{D}]\overline{G}+c_{612}\, T\,\overline{D}G\,\partial\overline{G}
+c_{613}\, T\,\overline{D}H\, G\,\overline{D}\overline{G}
+c_{614}\, T\,\overline{D}H\,\overline{H}\, D\overline{H}
\nonu \\&&+c_{615}\, T\,\overline{D}H\,\overline{D}G\,\overline{G}+c_{616}\, T\,\overline{D}H\,\overline{D}H\,\overline{H}
+c_{617}\, T\,\overline{D}H\,\partial\overline{H}+c_{618}\, T\,\overline{D}T\,\overline{D}H
+c_{618}\, T\,\overline{D}T\,\overline{D}H
\nonu \\&&+c_{619}\, T\,\overline{D}T\, D\overline{H}
+c_{620}\, T\,\overline{D}T\, H\,\overline{H}
+c_{621}\, T\,\partial\overline{D}G\,\overline{G}+c_{622}\, T\,\partial\overline{D}H\,\overline{H}+c_{623}\, T\,[D,\overline{D}]G\,\overline{D}\overline{G}
\nonu \\&&+c_{624}\, T\, D\,\overline{D}T\,\overline{H}+c_{625}\, T\,\partial D\overline{H}\,\overline{H}
+c_{626}\, T\,\partial G\,\overline{D}\overline{G}+c_{627}\, T\,\partial\overline{H}\, D\overline{H}
+c_{628}\, T\,\partial^{2}\overline{H}
\nonu \\&&+c_{629}\, G\,\partial^{2}\overline{D}\overline{G}+c_{630}\, G\, DG\,\overline{D}\overline{G}\,\overline{D}\overline{G}
+c_{631}\,\overline{D}G\,\partial[D,\overline{D}]\overline{G}
+c_{632}\,\overline{D}G\,\overline{D}G\,\overline{G}\, D\overline{G}
\nonu \\&&+c_{633}\,\overline{D}G\, DG\,\overline{G}\,\overline{D}\overline{G}+c_{634}\,\overline{D}G\,\partial^{2}\overline{G}
+c_{635}\,\overline{D}H\, G\,\partial\overline{D}\overline{G}\
+c_{636}\,\overline{D}H\,\overline{H}\, G\,[D,\overline{D}]\,\overline{G}
\nonu \\&&+c_{637}\,\overline{D}H\,\overline{H}\, G\,\partial\overline{G}+c_{638}\,\overline{D}H\,\overline{H}\,\overline{D}G\, D\overline{G}
+c_{639}\,\overline{D}H\,\overline{H}\,[D,\overline{D}]G\,\overline{G}+c_{640}\,\overline{D}H\,\overline{H}\, DG\,\overline{D}\overline{G}
\nonu \\&&+c_{641}\,\overline{D}H\,\overline{H}\, D\overline{H}\, D\overline{H}
+c_{642}\,\overline{D}H\,\overline{H}\,\partial G\,\overline{G}
+c_{643}\,\overline{D}H\,\overline{D}G\,[D,\overline{D}]\overline{G}+c_{644}\,\overline{D}H\,\overline{D}G\,\partial\overline{G}
\nonu \\&&+c_{645}\,\overline{D}H\,\overline{D}H\, G\,\overline{D}\overline{G}+c_{646}\,\overline{D}H\,\overline{D}H\,\overline{H}\, D\overline{H}
+c_{647}\,\overline{D}H\,\overline{D}H\,\overline{D}G\,\overline{G}
+c_{648}\,\overline{D}H\,\overline{D}H\,\overline{D}H\,\overline{H}
\nonu \\&&+c_{649}\,\overline{D}H\,\overline{D}H\,\partial\overline{H}
+c_{650}\,\overline{D}H\,\partial\overline{D}G\,\overline{G}
+c_{651}\,\overline{D}H\,[D,\overline{D}]G\,\overline{D}\overline{G}+c_{652}\,\overline{D}H\, D\overline{H}\, G\,\overline{D}\overline{G}
\nonu \\&&+c_{653}\,\overline{D}H\, D\overline{H}\,\overline{D}G\,\overline{G}
+c_{654}\,\overline{D}H\,\partial D\overline{H}\,\overline{H}
+c_{655}\,\overline{D}H\,\partial G\,\overline{D}\overline{G}+c_{656}\,\overline{D}H\,\partial\overline{H}\, D\overline{H}
\nonu \\&&+c_{657}\,\overline{D}H\,\partial^{2}\overline{H}
+c_{658}\,\overline{D}T\,\partial\overline{D}H+c_{659}\,\overline{D}T\, D\,\overline{D}T
+c_{660}\,\overline{D}T\,\partial D\overline{H}+c_{661}\,\overline{D}T\, G\,[D,\overline{D}]\overline{G}
\nonu \\&&+c_{662}\,\overline{D}T\, G\,\partial\overline{G}+c_{663}\,\overline{D}T\, H\, G\,\overline{D}\overline{G}
+c_{664}\,\overline{D}T\, H\,\overline{H}\, D\overline{H}+c_{665}\,\overline{D}T\, H\,\overline{D}G\,\overline{G}
\nonu \\&&+c_{666}\,\overline{D}T\, H\,\overline{D}H\,\overline{H}+c_{667}\,\overline{D}T\, H\,\partial\overline{H}
+c_{668}\,\overline{D}T\,\overline{H}\, G\, D\overline{G}+c_{669}\,\overline{D}T\,\overline{H}\, DG\,\overline{G}
\nonu \\&&+c_{670}\,\overline{D}T\,\overline{D}G\, D\overline{G}+c_{671}\,\overline{D}T\,\overline{D}H\,\overline{D}H
+c_{672}\,\overline{D}T\,\overline{D}H\, D\overline{H}
+c_{673}\,\overline{D}T\,\overline{D}H\, G\,\overline{G}
\nonu \\&&+c_{674}\,\overline{D}T\,[D,\overline{D}]G\,\overline{G}+c_{675}\,\overline{D}T\, DG\,\overline{D}\overline{G}
+c_{676}\,\overline{D}T\, D\overline{H}\, D\overline{H}+c_{677}\,\overline{D}T\, DT\,\overline{H}
\nonu \\&&+c_{678}\,\overline{D}T\,\partial G\,\overline{G}+c_{679}\,\overline{D}T\,\partial H\,\overline{H}
+c_{680}\,\partial^{2}\overline{D}T
+c_{681}\, G\,\overline{D}G\, \partial \overline{G}\,\overline{G}+c_{682}\, G\, [D,\overline{D}]G\,\overline{G}\, \overline{D}\overline{G}
\nonu \\&&+c_{683}\,\partial\overline{D}G\,[D,\overline{D}]\overline{G}\ +c_{684}\,\partial\overline{D}G\,\partial\overline{G}
+c_{685}\,\partial\overline{D}H\, G\,\overline{D}\overline{G}+c_{686}\,\partial\overline{D}H\,\overline{H}\, D\overline{H}
\nonu \\&&+c_{687}\,\partial\overline{D}H\,\overline{H}\, G\,\overline{G}+c_{688}\,\partial\overline{D}H\,\overline{D}G\,\overline{G}
+c_{689}\,\partial\overline{D}H\,\overline{D}H\,\overline{H}\ +c_{690}\,\partial\overline{D}H\,\partial\overline{H}
+c_{691}\,\partial\overline{D}T\, T
\nonu \\&&+c_{692}\,\partial\overline{D}T\,\overline{D}H+c_{693}\,\partial\overline{D}T\, D\overline{H}
+c_{694}\,\partial\overline{D}T\, G\,\overline{G}
+c_{695}\,\partial\overline{D}T\, H\,\overline{H}+c_{696}\,  G\,\overline{D}\overline{G}\, \overline{G}\, [D,\overline{D}]\overline{G}
\nonu \\&&+c_{697}\,\partial^{2}\overline{D}G\,\overline{G}+c_{698}\,\partial^{2}\overline{D}H\,\overline{H}
+c_{699}\,[D,\overline{D}]G\,\partial\overline{D}\overline{G}+c_{700}\,[D,\overline{D}]T\, G\,\overline{D}\overline{G}
\nonu \\&&+c_{701}\,[D,\overline{D}]T\,\overline{H}\, D\overline{H}
+c_{702}\,[D,\overline{D}]T\,\overline{H}\, G\,\overline{G}
+c_{703}\,[D,\overline{D}]T\,\overline{D}G\,\overline{G}+c_{704}\,[D,\overline{D}]T\,\overline{D}H\,\overline{H}
\nonu \\&&+c_{705}\,[D,\overline{D}]T\,\partial\overline{H}+c_{706}\,[D,\overline{D}]\partial G\,\overline{D}\overline{G}\ 
+c_{707}\, D\,\partial\overline{D}T\,\overline{H}
+c_{708}\, D\overline{H}\, G\,\partial\overline{D}\overline{G}
\nonu \\&&+c_{709}\, D\overline{H}\,\overline{D}G\,[D,\overline{D}]\overline{G}+c_{710}\, D\overline{H}\,\overline{D}G\,\partial\overline{G}
+c_{711}\, D\overline{H}\,\partial\overline{D}G\,\overline{G}+c_{712}\, D\overline{H}\,[D,\overline{D}]G\,\overline{D}\overline{G}
\nonu \\&&+c_{713}\, D\overline{H}\,\partial G\,\overline{D}\overline{G}
+c_{714}\, DT\,\overline{H}\, G\,\overline{D}\overline{G}
+c_{715}\, DT\,\overline{H}\,\overline{D}G\,\overline{G}+c_{716}\, DT\,\overline{D}G\,\overline{D}\overline{G}
\nonu \\&&+c_{717}\, DT\,\partial\overline{H}\,\overline{H}
+c_{718}\,\partial D\overline{H}\, G\,\overline{D}\overline{G}
+c_{719}\,\partial D\overline{H}\,\overline{D}G\,\overline{G}+c_{720}\,\partial D\overline{H}\,\partial\overline{H}
+c_{721}\,\partial^{2}D\overline{H}\,\overline{H}
\nonu \\&&+c_{722}\,\partial G\,\partial\overline{D}\overline{G}+c_{723}\,\partial G\, G\,\overline{G}\,\overline{D}\overline{G}
+c_{724}\,\partial H\,\overline{H}\, G\,\overline{D}\overline{G}+c_{725}\,\partial H\,\overline{H}\,\overline{D}G\,\overline{G}
\nonu \\&&+c_{726}\,\partial H\,\overline{D}G\,\overline{D}\overline{G}+c_{727}\,\partial H\,\partial\overline{H}\,\overline{H}
+c_{728}\,\partial\overline{H}\, G\,[D,\overline{D}]\overline{G}+c_{729}\,\partial\overline{H}\, G\,\partial\overline{G}
\nonu \\&&+c_{730}\,\partial\overline{H}\,\overline{H}\, G\, D\overline{G}
+c_{731}\,\partial\overline{H}\,\overline{H}\, DG\,\overline{G}
+c_{732}\,\partial\overline{H}\,\overline{D}G\, D\overline{G}
+c_{733}\,\partial\overline{H}\,[D,\overline{D}]G\,\overline{G}
\nonu \\&&+c_{734}\,\partial\overline{H}\, DG\,\overline{D}\overline{G}+c_{735}\,\partial\overline{H}\, D\overline{H}\, D\overline{H}
+c_{736}\,\partial\overline{H}\,\partial G\,\overline{G}
+c_{737}\,\partial T\,\overline{D}T+c_{738}\,\partial T\, G\,\overline{D}\overline{G}
\nonu \\&&+c_{739}\,\partial T\,\overline{H}\, D\overline{H}+c_{740}\,\partial T\,\overline{H}\, G\,\overline{G}
+c_{741}\,\partial T\, T\,\overline{H}+c_{742}\,\partial T\,\overline{D}G\,\overline{G}+c_{743}\,\partial T\,\overline{D}H\,\overline{H}
\nonu \\&&+c_{744}\,\partial T\,\partial\overline{H}
+c_{745}\, G\,\overline{D}G\, \overline{D}\overline{G}\, D\overline{G}
+c_{746}\,\partial^{2}G\,\overline{D}\overline{G}+c_{747}\,\partial^{2}\overline{H}\, D\overline{H}+c_{748}\,\partial^{2}\overline{H}\, G\,\overline{G}
\nonu \\&&+c_{749}\,\partial^{2}T\,\overline{H}+c_{750}\partial^{3}\overline{H}
\Bigg](Z_{2})
\nonu \\&&+\frac{\theta_{12}\overline{\theta}_{12}}{z_{12}}\Bigg[
c_{751}\,\partial {\bf W^{(3)}}
+c_{752}\,\partial[D,\overline{D}]{\bf T^{(2)}}
+c_{753}\,\partial {\bf T^{(1)}}\,{\bf  W^{(2)}}
+c_{754}\, {\bf T^{(1)}}\,\partial{\bf  W^{(2)}}
\nonu \\&&+c_{755}\,\partial D{\bf T^{(1)}}\,\overline{D}{\bf T^{(1)}}
+c_{756}\,\partial\overline{D}{\bf T^{(1)}}\, D{\bf T^{(1)}}
+c_{757}\,{\bf  U^{(\frac{3}{2})}}\,\partial {\bf V^{(\frac{3}{2})}}
+c_{758}\,\partial {\bf  U^{(\frac{3}{2})}\, V^{(\frac{3}{2})}}
\nonu \\&&+c_{759}\, G\, DG\,\partial\overline{G}\,\overline{D}\overline{G}
+c_{760}\, G\,\partial^{3}\overline{G}
+c_{761}\, H\, G\,\partial^{2}\overline{D}\overline{G}
+c_{762}\, H\,\overline{H}\, G\,\partial[D,\overline{D}]\overline{G}
\nonu \\&&+c_{763}\, H\,\overline{H}\, G\,\partial^{2}\overline{G}+c_{764}\, H\,\overline{H}\,\overline{D}G\,\partial D\overline{G}
+c_{765}\, H\,\overline{H}\,\partial\overline{D}G\, D\overline{G}+c_{766}\, H\,\overline{H}\,[D,\overline{D}]G\,\partial\overline{G}
\nonu \\&&+c_{767}\, H\,\overline{H}\,\partial[D,\overline{D}]G\,\overline{G}
+c_{768}\, H\,\overline{H}\, DG\,\partial\overline{D}\overline{G}+c_{769}\, H\,\overline{H}\,\partial DG\,\overline{D}\overline{G}+c_{770}\, H\,\overline{H}\,\partial G\,[D,\overline{D}]\overline{G}
\nonu \\&&+c_{771}\, H\,\overline{H}\,\partial^{2}G\,\overline{G}+c_{772}\, H\,\overline{D}G\,\partial[D,\overline{D}]\overline{G}+c_{773}\, H\,\overline{D}G\,\partial^{2}\overline{G}
+c_{774}\, H\,\overline{D}H\, G\,\partial\overline{D}\overline{G}
\nonu \\&&+c_{775}\, H\,\overline{D}H\,\overline{D}G\,\partial\overline{G}+c_{776}\, H\,\overline{D}H\,\overline{D}H\,\partial\overline{H}
+c_{777}\, H\,\overline{D}H\,\partial\overline{D}G\,\overline{G}+c_{778}\, H\,\overline{D}H\,\partial D\overline{H}\,\overline{H}
\nonu \\&&+c_{779}\, H\,\overline{D}H\,\partial G\,\overline{D}\overline{G}
+c_{780}\, H\,\overline{D}H\,\partial\overline{H}\, D\overline{H}+c_{781}\, H\,\overline{D}H\,\partial^{2}\overline{H}+c_{782}\, H\,\partial\overline{D}G\,[D,\overline{D}]\overline{G}
\nonu \\&&+c_{783}\, H\,\partial\overline{D}G\,\partial\overline{G}+c_{784}\, H\,\partial^{2}\overline{D}G\,\overline{G}+c_{785}\, H\,[D,\overline{D}]G\,\partial\overline{D}\overline{G}
+c_{786}\, H\,\partial[D,\overline{D}]G\,\overline{D}\overline{G}
\nonu \\&&+c_{787}\, H\, D\overline{H}\, G\,\partial\overline{D}\overline{G}+c_{788}\, H\, D\overline{H}\,\overline{D}G\,\partial\overline{G}
+c_{789}\, H\, D\overline{H}\,\partial\overline{D}G\,\overline{G}+c_{790}\, H\, D\overline{H}\,\partial G\,\overline{D}\overline{G}
\nonu \\&&+c_{791}\, H\,\partial D\overline{H}\, G\,\overline{D}\overline{G}
+c_{792}\, H\,\partial D\overline{H}\,\overline{H}\, D\overline{H}+c_{793}\, H\,\partial D\overline{H}\,\overline{D}G\,\overline{G}+c_{794}\, H\,\partial D\overline{H}\,\partial\overline{H}
\nonu \\&&+c_{795}\, H\,\partial^{2}D\overline{H}\,\overline{H}+c_{796}\, H\,\partial G\,\partial\overline{D}\overline{G}+c_{797}\, H\,\partial\overline{H}\, G\,[D,\overline{D}]\overline{G}
+c_{798}\, H\,\partial\overline{H}\, G\,\partial\overline{G}
\nonu \\&&+c_{799}\, H\,\partial\overline{H}\,\overline{D}G\, D\overline{G}+c_{800}\, H\,\partial\overline{H}\,[D,\overline{D}]G\,\overline{G}
+c_{801}\, H\,\partial\overline{H}\, DG\,\overline{D}\overline{G}+c_{802}\, H\,\partial\overline{H}\, D\overline{H}\, D\overline{H}
\nonu \\&&+c_{803}\, H\,\partial\overline{H}\,\partial G\,\overline{G}
+c_{804}\, H\,\partial^{2}G\,\overline{D}\overline{G}+c_{805}\, H\,\partial^{2}\overline{H}\, D\overline{H}+c_{806}\, H\,\partial^{2}\overline{H}\, G\,\overline{G}+c_{807}\, H\,\partial^{3}\overline{H}
\nonu \\&&+c_{808}\,\overline{H}\, G\,\partial^{2}D\overline{G}+c_{809}\,\overline{H}\,[D,\overline{D}]G\,\partial D\overline{G}+c_{810}\,\overline{H}\,\partial[D,\overline{D}]G\, D\overline{G}
+c_{811}\,\overline{H}\, DG\,\partial[D,\overline{D}]\overline{G}
\nonu \\&&+c_{812}\,\overline{H}\, DG\,\partial^{2}\overline{G}+c_{813}\,\overline{H}\, D\overline{H}\, G\,\partial D\overline{G}
+c_{814}\,\overline{H}\, D\overline{H}\, DG\,\partial\overline{G}+c_{815}\,\overline{H}\, D\overline{H}\,\partial DG\,\overline{G}
\nonu \\&&+c_{816}\,\overline{H}\, D\overline{H}\,\partial G\, D\overline{G}
+c_{817}\,\overline{H}\,\partial DG\,[D,\overline{D}]\overline{G}+c_{818}\,\overline{H}\,\partial DG\,\partial\overline{G}+c_{819}\,\overline{H}\,\partial^{2}DG\,\overline{G}
\nonu \\&&+c_{820}\,\overline{H}\,\partial G\,\partial D\overline{G}+c_{821}\,\overline{H}\,\partial^{2}G\, D\overline{G}+c_{822}\, T\,\partial^{2}\overline{D}H+c_{823}\, T\,\partial^{2}D\overline{H}
+c_{824}\, T\, G\,\partial[D,\overline{D}]\overline{G}
\nonu \\&&+c_{825}\, T\, G\,\partial^{2}\overline{G}+c_{826}\, T\, H\, G\,\partial\overline{D}\overline{G}+c_{827}\, T\, H\,\overline{D}G\,\partial\overline{G}
+c_{828}\, T\, H\,\overline{D}H\,\partial\overline{H}
\nonu \\&&+c_{829}\, T\, H\,\partial\overline{D}G\,\overline{G}+c_{830}\, T\, H\,\partial D\overline{H}\,\overline{H}+c_{831}\, T\, H\,\partial G\,\overline{D}\overline{G}
+c_{832}\, T\, H\,\partial\overline{H}\, D\overline{H}
\nonu \\&&+c_{833}\, T\, H\,\partial^{2}\overline{H}+c_{834}\, T\,\overline{H}\, G\,\partial D\overline{G}+c_{835}\, T\,\overline{H}\, DG\,\partial\overline{G}
+c_{836}\, T\,\overline{H}\,\partial DG\,\overline{G}
\nonu \\&&+c_{837}\, T\,\overline{H}\,\partial G\, D\overline{G}+c_{838}\, T\, T\,\partial\overline{D}H+c_{839}\, T\, T\,\partial D\overline{H}
+c_{840}\, T\, T\, H\,\partial\overline{H}+c_{841}\, T\, T\,\partial H\,\overline{H}
\nonu \\&&+c_{842}\, T\,\overline{D}G\,\partial D\overline{G}+c_{843}\, T\,\overline{D}H\,\partial D\overline{H}
+c_{844}\, T\,\overline{D}T\,\partial H+c_{845}\, T\,\partial\overline{D}G\, D\overline{G}
\nonu \\&&+c_{846}\, T\,\partial\overline{D}H\,\overline{D}H+c_{847}\, T\,\partial\overline{D}H\, D\overline{H}
+c_{848}\, T\,\partial\overline{D}H\, H\,\overline{H}+c_{849}\, T\,[D,\overline{D}]G\,\partial\overline{G}
\nonu \\&&+c_{850}\, T\,\partial[D,\overline{D}]G\,\overline{G}
+c_{851}\, T\, DG\,\partial\overline{D}\overline{G}+c_{852}\, T\, DT\,\partial\overline{H}+c_{853}\, T\,\partial DG\,\overline{D}\overline{G}
\nonu \\&&+c_{854}\, T\,\partial D\overline{H}\, D\overline{H}
+c_{855}\, T\,\partial G\,[D,\overline{D}]\overline{G}+c_{856}\, T\,\partial H\, G\,\overline{D}\overline{G}+c_{857}\, T\,\partial H\,\overline{H}\, D\overline{H}
\nonu \\&&+c_{858}\, T\,\partial H\,\overline{D}G\,\overline{G}+c_{859}\, T\,\partial H\,\overline{D}H\,\overline{H}+c_{860}\, T\,\partial H\,\partial\overline{H}+c_{861}\, T\,\partial\overline{H}\, G\, D\overline{G}
\nonu \\&&+c_{862}\, T\,\partial\overline{H}\, DG\,\overline{G}+c_{863}\, T\,\partial^{2}G\,\overline{G}+c_{864}\, T\,\partial^{2}H\,\overline{H}
+c_{865}\,\partial^{3}D\overline{H}
+c_{866}\, G\,\overline{D}G\,\partial\overline{G}\, D\overline{G}
\nonu \\&&+c_{867}\,\overline{D}G\,\partial^{2}D\overline{G}+c_{868}\,\overline{D}H\,\partial^{2}D\overline{H}+c_{869}\,\overline{D}H\, G\,\partial[D,\overline{D}]\overline{G}
+c_{870}\,\overline{D}H\, G\,\partial^{2}\overline{G}
\nonu \\&&+c_{871}\,\overline{D}H\,\overline{H}\, G\,\partial D\overline{G}+c_{872}\,\overline{D}H\,\overline{H}\, DG\,\partial\overline{G}
+c_{873}\,\overline{D}H\,\overline{H}\,\partial DG\,\overline{G}+c_{874}\,\overline{D}H\,\overline{H}\,\partial G\, D\overline{G}
\nonu \\&&+c_{875}\,\overline{D}H\,\overline{D}G\,\partial D\overline{G}
+c_{876}\,\overline{D}H\,\overline{D}H\,\partial D\overline{H}+c_{877}\,\overline{D}H\,\partial\overline{D}G\, D\overline{G}+c_{878}\,\overline{D}H\,[D,\overline{D}]G\,\partial\overline{G}
\nonu \\&&+c_{879}\,\overline{D}H\,\partial[D,\overline{D}]G\,\overline{G}+c_{880}\,\overline{D}H\, DG\,\partial\overline{D}\overline{G}+c_{881}\,\overline{D}H\, D\overline{H}\, G\,\partial\overline{G}
+c_{882}\,\overline{D}H\, D\overline{H}\,\partial G\,\overline{G}
\nonu \\&&+c_{883}\,\overline{D}H\,\partial DG\,\overline{D}\overline{G}+c_{884}\,\overline{D}H\,\partial D\overline{H}\, D\overline{H}
+c_{885}\,\overline{D}H\,\partial D\overline{H}\, G\,\overline{G}+c_{886}\,\overline{D}H\,\partial G\,[D,\overline{D}]\overline{G}
\nonu \\&&+c_{887}\,\overline{D}H\,\partial G\,\partial\overline{G}
+c_{888}\,\overline{D}H\,\partial\overline{H}\, G\, D\overline{G}+c_{889}\,\overline{D}H\,\partial\overline{H}\, DG\,\overline{G}
+c_{890}\,\overline{D}H\,\partial^{2}G\,\overline{G}
\nonu \\&&+c_{891}\,\overline{D}T\, G\,\partial D\overline{G}+c_{892}\,\overline{D}T\, H\,\partial D\overline{H}+c_{893}\,\overline{D}T\, H\, G\,\partial\overline{G}+c_{894}\,\overline{D}T\, H\,\partial G\,\overline{G}
\nonu \\&&+c_{895}\,\overline{D}T\,\partial\overline{D}H\, H+c_{896}\,\overline{D}T\, DG\,\partial\overline{G}+c_{897}\,\overline{D}T\,\partial DG\,\overline{G}+c_{898}\,\overline{D}T\,\partial G\, D\overline{G}
\nonu \\&&+c_{899}\,\overline{D}T\,\partial H\,\overline{D}H+c_{900}\,\overline{D}T\,\partial H\, D\overline{H}+c_{901}\,\overline{D}T\,\partial H\, G\,\overline{G}+c_{902}\,\overline{D}T\,\partial^{2}H
\nonu \\&&+c_{903}\,\partial\overline{D}G\,\partial D\overline{G}+c_{904}\,\partial\overline{D}G\, G\,\overline{G}\, D\overline{G}+c_{905}\,\partial\overline{D}H\,\partial\overline{D}H+c_{906}\,\overline{D}H\,\partial D\overline{H}
\nonu \\&&+c_{907}\,\partial\overline{D}H\, G\,[D,\overline{D}]\overline{G}+c_{908}\,\partial\overline{D}H\, G\,\partial\overline{G}+c_{909}\,\partial\overline{D}H\, H\, G\,\overline{D}\overline{G}
+c_{910}\,\partial\overline{D}H\, H\,\overline{H}\, D\overline{H}
\nonu \\&&+c_{911}\,\partial\overline{D}H\, H\,\overline{D}G\,\overline{G}+c_{912}\,\partial\overline{D}H\, H\,\overline{D}H\,\overline{H}
+c_{913}\,\partial\overline{D}H\, H\,\partial\overline{H}+c_{914}\,\partial\overline{D}H\,\overline{H}\, G\, D\overline{G}
\nonu \\&&+c_{915}\,\partial\overline{D}H\,\overline{H}\, DG\,\overline{G}
+c_{916}\,\partial\overline{D}H\,\overline{D}G\, D\overline{G}+c_{917}\,\partial\overline{D}H\,\overline{D}H\,\overline{D}H+c_{918}\,\partial\overline{D}H\,\overline{D}H\, D\overline{H}
\nonu \\&&+c_{919}\,\partial\overline{D}H\,[D,\overline{D}]G\,\overline{G}+c_{920}\,\partial\overline{D}H\, DG\,\overline{D}\overline{G}+c_{921}\,\partial\overline{D}H\, D\overline{H}\, D\overline{H}
+c_{922}\,\partial\overline{D}H\, D\overline{H}\, G\,\overline{G}
\nonu \\&&+c_{923}\,\partial\overline{D}H\,\partial G\,\overline{G}+c_{924}\,\partial\overline{D}H\,\partial H\,\overline{H}+c_{925}\,\partial\overline{D}T\, DT
+c_{926}\,\partial\overline{D}T\, G\,\overline{G}+c_{927}\,\partial\overline{D}T\, H\,\overline{D}H
\nonu \\&&+c_{928}\,\partial\overline{D}T\, H\, D\overline{H}+c_{929}\,\partial\overline{D}T\, H\, G\,\overline{G}
+c_{930}\,\partial\overline{D}T\, T\, H+c_{931}\,\partial\overline{D}T\, DG\,\overline{G}+c_{932}\,\partial\overline{D}T\,\partial H
\nonu \\&&+c_{933}\, G\,\overline{D}G\,\partial D\overline{G}\,\overline{G}
+c_{934}\,\partial^{2}\overline{D}G\, D\overline{G}+c_{935}\,\partial^{2}\overline{D}H\,\overline{D}H+c_{936}\,\partial^{2}\overline{D}H\, D\overline{H}
\nonu \\&&+c_{937}\,\partial^{2}\overline{D}H\, G\,\overline{G}
+c_{938}\,\partial^{2}\overline{D}H\, H\,\overline{H}+c_{939}\,\partial^{2}\overline{D}T\, H+c_{940}\,[D,\overline{D}]G\,\partial[D,\overline{D}]\overline{G}
\nonu \\&&+c_{941}\,[D,\overline{D}]G\,\partial^{2}\overline{G}+c_{942}\,[D,\overline{D}]T\,\partial\overline{D}H+c_{943}\,[D,\overline{D}]T\,\partial D\overline{H}
+c_{944}\,[D,\overline{D}]T\, G\,\partial\overline{G}
\nonu \\&&+c_{945}\,[D,\overline{D}]T\, H\,\partial\overline{H}+c_{946}\,[D,\overline{D}]T\,\partial G\,\overline{G}
+c_{947}\,[D,\overline{D}]T\,\partial H\,\overline{H}+c_{948}\,\partial[D,\overline{D}]G\,[D,\overline{D}]\overline{G}
\nonu \\&&+c_{949}\,\partial[D,\overline{D}]G\,\partial\overline{G}
+c_{950}\,\partial[D,\overline{D}]T\, T+c_{951}\,[D,\overline{D}]T\,\overline{D}H+c_{952}\,\partial[D,\overline{D}]T\, D\overline{H}
\nonu \\&&+c_{953}\,\partial[D,\overline{D}]T\, G\,\overline{G}+c_{954}\,\partial[D,\overline{D}]T\, H\,\overline{H}+c_{955}\,\partial^{2}[D,\overline{D}]G\,\overline{G}
+c_{956}\, DG\,\partial^{2}\overline{D}\overline{G}
\nonu \\&&+c_{957}\, D\overline{H}\, G\,\partial[D,\overline{D}]\overline{G}+c_{958}\, D\overline{H}\, G\,\partial^{2}\overline{G}
+c_{959}\, D\overline{H}\,\overline{D}G\,\partial D\overline{G}+c_{960}\, D\overline{H}\,\partial\overline{D}G\, D\overline{G}
\nonu \\&&+c_{961}\, D\overline{H}\,[D,\overline{D}]G\,\partial\overline{G}
+c_{962}\, D\overline{H}\,[D,\overline{D}]G\,\overline{G}+c_{963}\, D\overline{H}\, DG\,\partial\overline{D}\overline{G}+c_{964}\, D\overline{H}\,\partial DG\,\overline{D}\overline{G}
\nonu \\&&+c_{965}\, D\overline{H}\,\partial G\,[D,\overline{D}]\overline{G}+c_{966}\, D\overline{H}\,\partial G\,\partial\overline{G}+c_{967}\, D\overline{H}\,\partial^{2}G\,\overline{G}
+c_{968}\, DT\, G\,\partial\overline{D}\overline{G}
\nonu \\&&+c_{969}\, DT\,\overline{H}\, G\,\partial\overline{G}+c_{970}\, DT\,\overline{H}\,\partial G\,\overline{G}+c_{971}\, DT\,\overline{D}G\,\partial\overline{G}
+c_{972}\, DT\,\overline{D}H\,\partial\overline{H}
\nonu \\&&+c_{973}\, DT\,\partial\overline{D}G\,\overline{G}+c_{974}\, DT\,\partial\overline{D}H\,\overline{H}+c_{975}\, DT\,\partial D\overline{H}\,\overline{H}
+c_{976}\, DT\,\partial G\,\overline{D}\overline{G}
\nonu \\&&+c_{977}\, DT\,\partial\overline{H}\, D\overline{H}+c_{978}\, DT\,\partial\overline{H}\, G\,\overline{G}+c_{979}\, DT\,\partial^{2}\overline{H}
+c_{980}\,\partial DG\,\partial\overline{D}\overline{G}
\nonu \\&&+c_{981}\,\partial DG\, G\,\overline{G}\,\overline{D}\overline{G}+c_{982}\,\partial D\overline{H}\,\partial D\overline{H}
+c_{983}\,\partial D\overline{H}\, G\,[D,\overline{D}]\overline{G}+c_{984}\,\partial D\overline{H}\, G\,\partial\overline{G}
\nonu \\&&+c_{985}\,\partial D\overline{H}\,\overline{H}\, G\, D\overline{G}
+c_{986}\,\partial\overline{H}\,\overline{H}\, DG\,\overline{G}+c_{987}\,\partial D\overline{H}\,\overline{D}G\, D\overline{G}+c_{988}\,\partial D\overline{H}\,[D,\overline{D}]G\,\overline{G}
\nonu \\&&+c_{989}\,\partial D\overline{H}\, DG\,\overline{D}\overline{G}+c_{990}\,\partial D\overline{H}\, D\overline{H}\, D\overline{H}+c_{991}\,\partial D\overline{H}\,\partial G\,\overline{G}+c_{992}\,\partial DT\,\overline{D}T
\nonu \\&&+c_{993}\,\partial DT\, G\,\overline{D}\overline{G}+c_{994}\,\partial DT\,\overline{H}\, D\overline{H}+c_{995}\,\partial DT\,\overline{H}\, G\,\overline{G}+c_{996}\,\partial DT\, T\,\overline{H}
\nonu \\&&+c_{997}\,\partial DT\,\overline{D}G\,\overline{G}+c_{998}\,\partial DT\,\overline{D}H\,\overline{H}+c_{999}\,\partial DT\,\partial\overline{H}+
c_{1000}\, G\,\partial^{2}[D,\overline{D}]\overline{G}
\nonu \\&&+c_{1001}\,\partial^{2}DG\,\overline{D}\overline{G}+c_{1002}\,\partial^{2}D\overline{H}\, D\overline{H}+c_{1003}\,\partial^{2}D\overline{H}\, G\,\overline{G}+c_{1004}\,\partial^{2}DT\,\overline{H}
+c_{1005}\,\partial G\,[D,\overline{D}]\overline{G}
\nonu \\&&+c_{1006}\,\partial G\,\overline{D}G\,\overline{G}\, D\overline{G}+c_{1007}\,\partial G\, DG\,\overline{G}\,\overline{D}\overline{G}
+c_{1008}\,\partial G\,\partial^{2}\overline{G}+c_{1009}\,\partial H\, G\,\partial\overline{D}\overline{G}
\nonu \\&&+c_{1010}\,\partial H\,\overline{H}\, G\,[D,\overline{D}]\overline{G}
+c_{1011}\,\partial H\,\overline{H}\, G\,\partial\overline{G}+c_{1012}\,\partial H\,\overline{H}\,\overline{D}G\, D\overline{G}+c_{1013}\,\partial H\,\overline{H}\,[D,\overline{D}]G\,\overline{G}
\nonu \\&&+c_{1014}\,\partial H\,\overline{H}\, DG\,\overline{D}\overline{G}+c_{1015}\,\partial H\,\overline{H}\, D\overline{H}\, D\overline{H}+c_{1016}\,\partial H\,\overline{H}\,\partial G\,\overline{G}
+c_{1017}\,\partial H\,\overline{D}G\,[D,\overline{D}]\overline{G}
\nonu \\&&+c_{1018}\,\partial H\,\overline{D}G\,\partial\overline{G}+c_{1019}\,\partial H\,\overline{D}H\, G\,\overline{D}\overline{G}
+c_{1020}\,\partial H\,\overline{D}H\,\overline{H}\, D\overline{H}+c_{1021}\,\partial H\,\overline{D}H\,\overline{D}G\,\overline{G}
\nonu \\&&+c_{1022}\,\partial H\,\overline{D}H\,\overline{D}H\,\overline{H}
+c_{1023}\,\partial H\,\overline{D}H\,\partial\overline{H}+c_{1024}\,\partial H\,\partial\overline{D}G\,\overline{G}+c_{1025}\,\partial H\,[D,\overline{D}]G\,\overline{D}\overline{G}
\nonu \\&&+c_{1026}\,\partial H\, D\overline{H}\, G\,\overline{D}\overline{G}+c_{1027}\,\partial H\, D\overline{H}\,\overline{D}G\,\overline{G}+c_{1028}\,\partial H\,\partial D\overline{H}\,\overline{H}
+c_{1029}\,\partial H\,\partial G\,\overline{D}\overline{G}
\nonu \\&&+c_{1030}\,\partial H\,\partial\overline{H}\, D\overline{H}+c_{1031}\,\partial H\,\partial^{2}\overline{H}+c_{1032}\,\partial\overline{H}\, G\,\partial D\overline{G}
+c_{1033}\,\partial\overline{H}\,[D,\overline{D}]G\, D\overline{G}
\nonu \\&&+c_{1034}\,\partial\overline{H}\, DG\,[D,\overline{D}]\overline{G}+c_{1035}\,\partial\overline{H}\, DG\,\partial\overline{G}
+c_{1036}\,\partial\overline{H}\, D\overline{H}\, G\, D\overline{G}+c_{1037}\,\partial\overline{H}\, D\overline{H}\, DG\,\overline{G}
\nonu \\&&+c_{1038}\,\partial\overline{H}\,\partial DG\,\overline{G}
+c_{1039}\,\partial\overline{H}\,\partial G\, D\overline{G}+c_{1040}\,\partial T\,\partial\overline{D}H+c_{1041}\,\partial T\,[D,\overline{D}]T+c_{1042}\,\partial T\,\partial D\overline{H}
\nonu \\&&+c_{1043}\,\partial T\, G\,[D,\overline{D}]\overline{G}+c_{1044}\,\partial T\, G\,\partial\overline{G}+c_{1045}\,\partial T\, H\, G\,\overline{D}\overline{G}
+c_{1046}\,\partial T\, H\,\overline{H}\, D\overline{H}
\nonu \\&&+c_{1047}\,\partial T\, H\,\overline{D}G\,\overline{G}+c_{1048}\,\partial T\, H\,\overline{D}H\,\overline{H}
+c_{1049}\,\partial T\, H\,\partial\overline{H}+c_{1050}\,\partial T\,\overline{H}\, G\, D\overline{G}
\nonu \\&&+c_{1051}\,\partial T\,\overline{H}\, DG\,\overline{G}+c_{1052}\,\partial T\, T\, T
+c_{1053}\,\partial T\, T\,\overline{D}H+c_{1054}\,\partial T\, T\, D\overline{H}+c_{1055}\,\partial T\, T\, H\,\overline{H}
\nonu \\&&+c_{1056}\,\partial T\,\overline{D}G\, D\overline{G}
+c_{1057}\,\partial T\,\overline{D}H\,\overline{D}H+c_{1058}\,\partial T\,\overline{D}H\, D\overline{H}+c_{1059}\,\partial T\,\overline{D}T\, H
\nonu \\&&+c_{1060}\,\partial T\,[D,\overline{D}]G\,\overline{G}+c_{1061}\,\partial T\, DG\,\overline{D}\overline{G}+c_{1062}\,\partial T\, D\overline{H}\, D\overline{H}
+c_{1063}\,\partial T\, DT\,\overline{H}
\nonu \\&&+c_{1064}\,\partial T\,\partial G\,\overline{G}+c_{1065}\,\partial T\,\partial H\,\overline{H}+c_{1066}\,\partial T\,\partial T
+c_{1067}\,\partial^{2}[D,\overline{D}]T
+c_{1068}\, G\, DG\,\partial\overline{D}\overline{G}\,\overline{G}
\nonu \\&&+c_{1069}\,\partial^{2}G\,[D,\overline{D}]\overline{G}+c_{1070}\,\partial^{2}G\,\partial\overline{G}
+c_{1071}\,\partial^{2}H\, G\,\overline{D}\overline{G}+c_{1072}\,\partial^{2}H\,\overline{H}\, D\overline{H}
\nonu \\&&+c_{1073}\,\partial^{2}H\,\overline{H}\, G\,\overline{G}+c_{1074}\,\partial^{2}H\,\overline{D}G\,\overline{G}
+c_{1075}\,\partial^{2}H\,\overline{D}H\,\overline{H}+c_{1076}\,\partial^{2}H\,\partial\overline{H}
\nonu \\&&+c_{1077}\,\partial^{2}\overline{H}\, G\, D\overline{G}+c_{1078}\,\partial^{2}\overline{H}\, DG\,\overline{G}
+c_{1079}\,\partial^{2}T\, T+c_{1080}\,\partial^{2}T\,\overline{D}H+c_{1081}\,\partial^{2}T\, D\overline{H}
\nonu \\&&+c_{1082}\,\partial^{2}T\, G\,\overline{G}
+c_{1083}\,\partial^{2}T\, H\,\overline{H}+c_{1084}\,\partial^{3}G\,\overline{G}+c_{1085}\,\partial^{3}H\,\overline{H}+
c_{1086}\,\partial^{3}\overline{D}H
+c_{1087}\,\partial^{3}T
\Bigg](Z_{2})\nonumber,
\eea
where the coefficients are 
\bea
c_ {1} & = &\frac {256 k\, 
   N (3 + 2 k + N) (3 + k + 2 N)} {3 (2 + k + N)^{3}}, \nonu\\
c_ {2} & = &\frac {128 (9 k + 12 k^{2} + 4 k^{3} + 9 N + 30 k\, 
     N + 14 k^{2} N + 12 N^{2} + 14 k\, 
     N^{2} + 4 N^{3})} {3 (2 + k + N)^{3}}, \nonu\\
c_ {3} & = & - \frac {128 (k - N) (9 + 12 k + 4 k^{2} + 12 N + 7 k\, 
      N + 4 N^{2})} {3 (2 + k + N)^{4}}, \nonu\\
c_ {4} & = &\frac {128 (k - N) (9 + 12 k + 4 k^{2} + 12 N + 7 k\, 
     N + 4 N^{2})} {3 (2 + k + N)^{4}}, \nonu\\
c_ {5} & = &\frac {128 (k - N) (9 + 12 k + 4 k^{2} + 12 N + 7 k\, 
     N + 4 N^{2})} {3 (2 + k + N)^{4}}, \nonu\\
c_ {6} & = & - \frac {128 (9 k + 12 k^{2} + 4 k^{3} + 9 N + 12 k\, 
      N + 5 k^{2} N + 12 N^{2} + 5 k\, 
      N^{2} + 4 N^{3})} {3 (2 + k + N)^{4}}, \nonu\\
c_ {7} & = & - \frac {128 (18 + 63 k + 52 k^{2} + 12 k^{3} + 63 N + 
       112 k\, N + 42 k^{2} N + 52 N^{2} + 42 k\, 
      N^{2} + 12 N^{3})} {9 (2 + k + N)^{3}}, \nonu\\
c_ {8} & = &\frac {128 (18 + 63 k + 52 k^{2} + 12 k^{3} + 27 N + 
      58 k\, N + 25 k^{2} N + 16 N^{2} + 13 k\, 
     N^{2} + 4 N^{3})} {9 (2 + k + N)^{4}}, \nonu\\
c_ {9} & = & - \frac {256 (3 k + 2 k^{2} + 3 N + 8 k\, 
      N + 3 k^{2} N + 2 N^{2} + 3 k\, N^{2})} {3 (2 + k + N)^{4}}, \nonu\\
c_ {10} & = & - \frac {128 (18 + 45 k + 40 k^{2} + 12 k^{3} + 45 N + 
       46 k\, N + 15 k^{2} N + 40 N^{2} + 15 k\, 
      N^{2} + 12 N^{3})} {9 (2 + k + N)^{4}}, \nonu\\
c_ {11} & = & - \frac {256 (3 + 2 k + N) (3 + k + 
       2 N)} {9 (2 + k + N)^{4}}, \qquad
c_ {12}  = \frac {256 (k - N) (3 + 2 k + 2 N)} {3 (2 + k + N)^{3}},\nonu\\ 
c_ {13} & = &\frac {128 (18 + 27 k + 16 k^{2} + 4 k^{3} + 63 N + 
      58 k\, N + 13 k^{2} N + 52 N^{2} + 25 k\, 
     N^{2} + 12 N^{3})} {9 (2 + k + N)^{4}}, \nonu\\
c_ {14} & = & - \frac {128 (-18 + 9 k + 26 k^{2} + 8 k^{3} - 99 N - 
       58 k\, N - k^{2} N - 94 N^{2} - 37 k\, 
      N^{2} - 24 N^{3})} {9 (2 + k + N)^{4}}, \nonu\\
c_ {15} & = &\frac {128 (18 + 63 k + 52 k^{2} + 12 k^{3} + 63 N + 
      112 k\, N + 42 k^{2} N + 52 N^{2} + 42 k\, 
     N^{2} + 12 N^{3})} {9 (2 + k + N)^{3}}, \nonu\\
c_ {16} & = &\frac {128 (18 + 27 k + 16 k^{2} + 4 k^{3} + 63 N + 
      58 k\, N + 13 k^{2} N + 52 N^{2} + 25 k\, 
     N^{2} + 12 N^{3})} {9 (2 + k + N)^{4}}, \nonu\\
c_ {17} & = & - \frac {256 (3 k + 2 k^{2} + 3 N + 8 k\, 
      N + 3 k^{2} N + 2 N^{2} + 3 k\, N^{2})} {3 (2 + k + N)^{4}}, \nonu\\
c_ {18} & = &\frac {256 (3 + 2 k + N) (3 + k + 
      2 N)} {9 (2 + k + N)^{4}}, \qquad
c_ {19}  =  - \frac {256 (k - N) (3 + 2 k + 
       2 N)} {3 (2 + k + N)^{3}}, \nonu\\
c_ {20} & = &\frac {128 (18 + 63 k + 52 k^{2} + 12 k^{3} + 27 N + 
      58 k\, N + 25 k^{2} N + 16 N^{2} + 13 k\, 
     N^{2} + 4 N^{3})} {9 (2 + k + N)^{4}}, \nonu\\
c_ {21} & = & - \frac {128 (18 + 45 k + 40 k^{2} + 12 k^{3} + 45 N + 
       46 k\, N + 15 k^{2} N + 40 N^{2} + 15 k\, 
      N^{2} + 12 N^{3})} {9 (2 + k + N)^{4}}, \nonu\\
c_ {22} & = & - \frac {128 (18 + 99 k + 94 k^{2} + 24 k^{3} - 9 N + 
       58 k\, N + 37 k^{2} N - 26 N^{2} + k\, 
      N^{2} - 8 N^{3})} {9 (2 + k + N)^{4}}, \nonu\\
c_ {23} & = & - \frac {128 (-9 + 18 k + 31 k^{2} + 10 k^{3} - 63 N - 
       29 k\, N + 4 k^{2} N - 65 N^{2} - 23 k\, 
      N^{2} - 18 N^{3})} {9 (2 + k + N)^{4}}, \nonu\\
c_ {24} & = &\frac {128 (9 + 63 k + 65 k^{2} + 18 k^{3} - 18 N + 
      29 k\, N + 23 k^{2} N - 31 N^{2} - 4 k\, 
     N^{2} - 10 N^{3})} {9 (2 + k + N)^{4}}, \nonu\\
c_ {25} & = &\frac {64 (9 + 18 k + 8 k^{2} + 18 N + 11 k\, 
     N + 8 N^{2})} {9 (2 + k + N)^{3}}, \nonu\\
c_ {26} & = &\frac {64 (k - N) (45 + 54 k + 16 k^{2} + 54 N + 31 k\, 
     N + 16 N^{2})} {9 (2 + k + N)^{4}}, \nonu\\
c_ {27} & = & - \frac {128 (3 + 2 k + N) (3 + k + 
       2 N)} {9 (2 + k + N)^{4}}, \qquad
c_ {28}  = \frac {128 (3 + 2 k + N) (3 + k + 
      2 N)} {9 (2 + k + N)^{4}}, \nonu\\
c_ {29} & = & - \frac {128 (9 + 36 k + 38 k^{2} + 12 k^{3} + 36 N + 
       41 k\, N + 15 k^{2} N + 38 N^{2} + 15 k\, 
      N^{2} + 12 N^{3})} {9 (2 + k + N)^{4}}, \nonu\\
c_ {30} & = & - \frac {128 (3 + 2 k + N) (3 + k + 
       2 N)} {9 (2 + k + N)^{4}}, \qquad
c_ {31}  = \frac {128 (3 + 2 k + N) (3 + k + 
      2 N)} {9 (2 + k + N)^{4}}, \nonu\\
c_ {32} & = &\frac {128 (k - N) (3 + 2 k + 2 N)} {3 (2 + k + N)^{3}},\qquad
c_ {33}  = \frac {128 (k - N) (3 + 2 k + 2 N)} {3 (2 + k + N)^{3}},\nonu\\ 
c_ {34} & = & - \frac {128 (3 k + 2 k^{2} + 3 N + 8 k\, 
      N + 3 k^{2} N + 2 N^{2} + 3 k\, N^{2})} {3 (2 + k + N)^{4}}, \nonu\\
c_ {35} & = & - \frac {128 (3 + 2 k + N) (3 + k + 
       2 N)} {9 (2 + k + N)^{4}}, \qquad
c_ {36}  = \frac {128 (k - N) (3 + 2 k + 2 N)} {3 (2 + k + N)^{3}},\nonu\\ 
c_ {37} & = & - \frac {64 (9 + 18 k + 8 k^{2} + 18 N + 11 k\, 
      N + 8 N^{2})} {9 (2 + k + N)^{3}}, \nonu\\
c_ {38} & = &\frac {128 (3 k + 2 k^{2} + 3 N + 8 k\, 
     N + 3 k^{2} N + 2 N^{2} + 3 k\, N^{2})} {3 (2 + k + N)^{4}}, \nonu\\
c_ {39} & = & - \frac {128 (3 + 2 k + N) (3 + k + 
       2 N)} {9 (2 + k + N)^{4}}, \qquad
c_ {40}  = \frac {128 (k - N) (3 + 2 k + 2 N)} {3 (2 + k + N)^{3}},\nonu\\ 
c_ {41} & = &\frac {64 (k - N) (45 + 54 k + 16 k^{2} + 54 N + 31 k\, 
     N + 16 N^{2})} {9 (2 + k + N)^{4}}, \nonu\\
c_ {42} & = & - \frac {128 (9 + 36 k + 38 k^{2} + 12 k^{3} + 36 N + 
       41 k\, N + 15 k^{2} N + 38 N^{2} + 15 k\, 
      N^{2} + 12 N^{3})} {9 (2 + k + N)^{4}}, \nonu\\
c_ {43} & = &\frac {128 (9 + 45 k + 44 k^{2} + 12 k^{3} + 45 N + 
      101 k\, N + 42 k^{2} N + 44 N^{2} + 42 k\, 
     N^{2} + 12 N^{3})} {9 (2 + k + N)^{3}}, \nonu\\
c_ {44} & = & - \frac {64 (k - N)} {3 (2 + k + N)}, \nonu\\
c_ {45} & = & - \frac {32 (-k + N) (60 + 77 k + 22 k^{2} + 121 N + 
       115 k\, N + 20 k^{2} N + 79 N^{2} + 42 k\, 
      N^{2} + 16 N^{3})} {3 (2 + N) (2 + k + N)^{3}}, \nonu\\
c_ {46} & = &\frac {64 (-36 - 42 k - 7 k^{2} + 2 k^{3} - 102 N - 
      97 k\, N - 17 k^{2} N - 76 N^{2} - 41 k\, 
     N^{2} - 16 N^{3})} {9 (2 + k + N)^{3}}, \nonu\\
c_ {47} & = &\frac {1} {9 (2 + N) (2 + k + N)^{4}} 64 (72 + 84 k - 
    25 k^{2} - 22 k^{3} + 168 N + 140 k\, 
   N - 51 k^{2} N 
   \nonu\\ & - & 20 k^{3} N + 209 N^{2} + 132 k\, 
   N^{2} - 14 k^{2} N^{2} + 121 N^{3} + 46 k\, N^{3} + 24 N^{4}), \nonu\\
c_ {48} & = &\frac {128 (k - N) (3 + 6 k + 2 k^{2} + 6 N + 5 k\, 
     N + 2 N^{2})} {9 (2 + k + N)^{4}}, \nonu\\
c_ {49} & = & - \frac {1} {9 (2 + N) (2 + k + N)^{4}} 32 (84 k + 
     91 k^{2} + 26 k^{3} + 204 N + 268 k\, 
    N + 87 k^{2} N + 4 k^{3} N 
     \nonu\\ & + & 361 N^{2} + 252 k\, 
    N^{2} + 26 k^{2} N^{2} + 211 N^{3} + 74 k\, N^{3} + 40 N^{4}), \nonu\\
c_ {50} & = &\frac {128 (36 - 6 k - 47 k^{2} - 18 k^{3} + 96 N + 
      37 k\, N - 13 k^{2} N + 82 N^{2} + 29 k\, 
     N^{2} + 20 N^{3})} {9 (2 + N) (2 + k + N)^{4}}, \nonu\\
c_ {51} & = &\frac {64 (-k + N) (32 + 35 k + 10 k^{2} + 61 N + 47 k\, 
     N + 8 k^{2} N + 39 N^{2} + 16 k\, 
     N^{2} + 8 N^{3})} {3 (2 + N) (2 + k + N)^{5}}, \nonu\\
c_ {52} & = &\frac {64 (-k + N) (32 + 55 k + 18 k^{2} + 41 N + 35 k\, 
     N + 11 N^{2})} {3 (2 + N) (2 + k + N)^{5}}, \nonu\\
c_ {53} & = & - \frac {128 (18 + 9 k + 2 k^{2} + 27 N + 5 k\, 
      N + 11 N^{2})} {9 (2 + k + N)^{4}}, \nonu\\
c_ {54} & = & - \frac {64 (-k + N) (32 + 35 k + 10 k^{2} + 61 N + 
       47 k\, N + 8 k^{2} N + 39 N^{2} + 16 k\, 
      N^{2} + 8 N^{3})} {3 (2 + N) (2 + k + N)^{5}}, \nonu\\
c_ {55} & = &\frac {1} {9 (2 + N) (2 + k + N)^{4}} 32 (288 + 228 k + 
    91 k^{2} + 10 k^{3} + 636 N + 268 k\, 
   N + 31 k^{2} N 
    \nonu\\ & - & 4 k^{3} N + 577 N^{2} + 160 k\, 
   N^{2} - 2 k^{2} N^{2} + 231 N^{3} + 46 k\, N^{3} + 32 N^{4}), \nonu\\
c_ {56} & = & - \frac {128 (18 + 9 k + 2 k^{2} + 27 N + 5 k\, 
      N + 11 N^{2})} {9 (2 + k + N)^{4}}, \nonu\\
c_ {57} & = &\frac {128 (36 - 6 k - 47 k^{2} - 18 k^{3} + 96 N + 
      37 k\, N - 13 k^{2} N + 82 N^{2} + 29 k\, 
     N^{2} + 20 N^{3})} {9 (2 + N) (2 + k + N)^{4}}, \nonu\\
c_ {58} & = & - \frac {32 (48 + 52 k + 3 k^{2} - 6 k^{3} + 116 N + 
       80 k\, N + 5 k^{2} N + 85 N^{2} + 30 k\, 
      N^{2} + 19 N^{3})} {3 (2 + N) (2 + k + N)^{3}}, \nonu\\
c_ {59} & = &\frac {64 (-k + N) (24 + 31 k + 10 k^{2} + 53 N + 45 k\, 
     N + 8 k^{2} N + 37 N^{2} + 16 k\, 
     N^{2} + 8 N^{3})} {3 (2 + N) (2 + k + N)^{4}}, \nonu\\
c_ {60} & = & - \frac {64 (-k + N) (24 + 31 k + 10 k^{2} + 53 N + 
       45 k\, N + 8 k^{2} N + 37 N^{2} + 16 k\, 
      N^{2} + 8 N^{3})} {3 (2 + N) (2 + k + N)^{4}}, \nonu\\
c_ {61} & = & - \frac {64 (-k + N) (8 + 17 k + 6 k^{2} + 11 N + 
       11 k\, N + 3 N^{2})} {3 (2 + N) (2 + k + N)^{4}}, \nonu\\
c_ {62} & = &\frac {64 (48 + 28 k - 9 k^{2} - 6 k^{3} + 92 N + 44 k\, 
     N - k^{2} N + 61 N^{2} + 18 k\, 
     N^{2} + 13 N^{3})} {3 (2 + N) (2 + k + N)^{4}}, \nonu\\
c_ {63} & = & - \frac {1} {9 (2 + N) (2 + k + N)^{4}} 64 (72 + 12 k - 
     25 k^{2} - 6 k^{3} + 240 N + 104 k\, 
    N - 27 k^{2} N 
     \nonu\\ & - & 12 k^{3} N + 245 N^{2} + 108 k\, 
    N^{2} - 2 k^{2} N^{2} + 105 N^{3} + 34 k\, N^{3} + 16 N^{4}), \nonu\\
c_ {64} & = &\frac {128 (-36 - 42 k - 7 k^{2} + 2 k^{3} - 66 N - 
      58 k\, N - 7 k^{2} N - 43 N^{2} - 22 k\, 
     N^{2} - 9 N^{3})} {9 (2 + k + N)^{4}}, \nonu\\
c_ {65} & = &\frac {1} {3 (2 + N) (2 + k + N)^{5}} 32 (48 + 100 k + 
    65 k^{2} + 14 k^{3} + 164 N + 296 k\, 
   N + 161 k^{2} N 
    \nonu\\ & + & 28 k^{3} N + 191 N^{2} + 284 k\, 
   N^{2} + 114 k^{2} N^{2} + 12 k^{3} N^{2} + 93 N^{3} + 106 k\, 
   N^{3} + 24 k^{2} N^{3} 
    \nonu\\ & + & 16 N^{4} + 12 k\, N^{4}), \nonu\\
c_ {66} & = &\frac {1} {9 (2 + N) (2 + k + N)^{5}} 64 (144 + 180 k + 
    29 k^{2} - 30 k^{3} - 8 k^{4} + 612 N + 704 k\, 
   N
    \nonu\\ & + & 181 k^{2} N - 18 k^{3} N - 4 k^{4} N + 923 N^{2} + 844 k\, 
   N^{2} + 166 k^{2} N^{2} - 6 k^{3} N^{2} + 661 N^{3} 
   \nonu\\ & + & 420 k\, 
   N^{3} + 44 k^{2} N^{3} + 232 N^{4} + 78 k\, N^{4} + 32 N^{5}), \nonu\\
c_ {67} & = &\frac {64 (-k + N) (60 + 77 k + 22 k^{2} + 121 N + 
      115 k\, N + 20 k^{2} N + 79 N^{2} + 42 k\, 
     N^{2} + 16 N^{3})} {9 (2 + N) (2 + k + N)^{5}}, \nonu\\
c_ {68} & = & - \frac {256 (k - N)} {3 (2 + k + N)^{3}}, \qquad
c_ {69}  = \frac {128 (k - N) (3 + 6 k + 2 k^{2} + 6 N + 5 k\, 
     N + 2 N^{2})} {9 (2 + k + N)^{4}}, \nonu\\
c_ {70} & = & - \frac {1} {9 (2 + N) (2 + k + N)^{4}} 128 (72 + 
     12 k - 112 k^{2} - 87 k^{3} - 18 k^{4} + 240 N + 182 k\, 
    N 
     \nonu\\ & - & 30 k^{2} N - 33 k^{3} N + 254 N^{2} + 189 k\, 
    N^{2} + 22 k^{2} N^{2} + 108 N^{3} + 49 k\, N^{3} + 16 N^{4}), \nonu\\
c_ {71} & = &\frac {1} {3 (2 + N) (2 + k + N)^{5}} 32 (48 + 100 k + 
    65 k^{2} + 14 k^{3} + 164 N + 296 k\, 
   N + 161 k^{2} N 
    \nonu\\ & + & 28 k^{3} N + 191 N^{2} + 284 k\, 
   N^{2} + 114 k^{2} N^{2} 
   \nonu\\ & + & 12 k^{3} N^{2} + 93 N^{3} + 106 k\, 
   N^{3} + 24 k^{2} N^{3} + 16 N^{4} + 12 k\, N^{4}), \nonu\\
c_ {72} & = & - \frac {64 (-k + N) (60 + 77 k + 22 k^{2} + 121 N + 
       115 k\, N + 20 k^{2} N + 79 N^{2} + 42 k\, 
      N^{2} + 16 N^{3})} {9 (2 + N) (2 + k + N)^{5}}, \nonu\\
c_ {73} & = &\frac {256 (k - N)} {3 (2 + k + N)^{3}}, \nonu\\
c_ {74} & = &\frac {1} {9 (2 + N) (2 + k + N)^{4}} 32 (84 k + 
    91 k^{2} + 26 k^{3} + 204 N + 268 k\, 
   N + 87 k^{2} N + 4 k^{3} N 
    \nonu\\ & + & 361 N^{2} + 252 k\, 
   N^{2} + 26 k^{2} N^{2} + 211 N^{3} + 74 k\, N^{3} + 40 N^{4}), \nonu\\
c_ {75} & = & - \frac {1} {9 (2 + N) (2 + k + N)^{4}} 32 (288 + 
     132 k - 113 k^{2} - 62 k^{3} + 732 N + 340 k\, 
    N - 29 k^{2} N 
    \nonu\\ & + &4 k^{3} N + 709 N^{2} + 256 k\, 
    N^{2} - 2 k^{2} N^{2} + 267 N^{3} + 46 k\, N^{3} + 32 N^{4}), \nonu\\
c_ {76} & = &\frac {64 (-k + N) (16 + 21 k + 6 k^{2} + 19 N + 13 k\, 
     N + 5 N^{2})} {3 (2 + N) (2 + k + N)^{3}}, \qquad
c_ {77}  =  - \frac {32 (k - N)} {3 (2 + k + N)}, \nonu\\
c_ {78} & = & - \frac {32 (-k + N) (60 + 77 k + 22 k^{2} + 121 N + 
       115 k\, N + 20 k^{2} N + 79 N^{2} + 42 k\, 
      N^{2} + 16 N^{3})} {3 (2 + N) (2 + k + N)^{3}}, \nonu\\
c_ {79} & = &\frac {1} {9 (2 + N) (2 + k + N)^{4}} 32 (72 + 372 k + 
    345 k^{2} + 86 k^{3} + 168 N + 552 k\, 
   N + 363 k^{2} N 
   \nonu\\ & + & 52 k^{3} N + 147 N^{2} + 252 k\, 
   N^{2} + 90 k^{2} N^{2} + 55 N^{3} + 30 k\, N^{3} + 8 N^{4}), \nonu\\
c_ {80} & = & - \frac {1} {9 (2 + N) (2 + k + N)^{4}} 32 (216 + 
     384 k + 321 k^{2} + 86 k^{3} + 444 N + 474 k\, 
    N + 295 k^{2} N 
    \nonu\\ & + &52 k^{3} N + 357 N^{2} + 178 k\, 
    N^{2} + 62 k^{2} N^{2} + 125 N^{3} + 14 k\, N^{3} + 16 N^{4}), \nonu\\
c_ {81} & = &\frac {32 (36 - 6 k - 47 k^{2} - 18 k^{3} + 96 N + 
      37 k\, N - 13 k^{2} N + 82 N^{2} + 29 k\, 
     N^{2} + 20 N^{3})} {9 (2 + N) (2 + k + N)^{4}}, \nonu\\
c_ {82} & = &\frac {1} {9 (2 + N) (2 + k + N)^{5}} 32 (-216 - 420 k - 
    235 k^{2} + k^{3} + 18 k^{4} - 336 N - 628 k\, 
   N 
   \nonu\\ & - & 351 k^{2} N - 37 k^{3} N - 109 N^{2} - 237 k\, 
   N^{2} - 110 k^{2} N^{2} + 47 N^{3} + k\, N^{3} + 20 N^{4}), \nonu\\
c_ {83} & = & - \frac {32 (-k + N) (32 + 55 k + 18 k^{2} + 41 N + 
       35 k\, N + 11 N^{2})} {3 (2 + N) (2 + k + N)^{5}}, \nonu\\
c_ {84} & = &\frac {32 (-k + N) (32 + 55 k + 18 k^{2} + 41 N + 35 k\, 
     N + 11 N^{2})} {3 (2 + N) (2 + k + N)^{5}}, \nonu\\
c_ {85} & = &\frac {64 (18 + 9 k + 2 k^{2} + 27 N + 5 k\, 
     N + 11 N^{2})} {9 (2 + k + N)^{4}}, \nonu\\
c_ {86} & = &\frac {32 (-k + N) (32 + 35 k + 10 k^{2} + 61 N + 47 k\, 
     N + 8 k^{2} N + 39 N^{2} + 16 k\, 
     N^{2} + 8 N^{3})} {3 (2 + N) (2 + k + N)^{5}}, \nonu\\
c_ {87} & = &\frac {32 (18 + 9 k + 2 k^{2} + 27 N + 5 k\, 
     N + 11 N^{2})} {9 (2 + k + N)^{4}}, \nonu\\
c_ {88} & = & - \frac {64 (36 - 6 k - 47 k^{2} - 18 k^{3} + 96 N + 
       37 k\, N - 13 k^{2} N + 82 N^{2} + 29 k\, 
      N^{2} + 20 N^{3})} {9 (2 + N) (2 + k + N)^{4}}, \nonu\\
c_ {89} & = & - \frac {32 (-k + N) (32 + 35 k + 10 k^{2} + 61 N + 
       47 k\, N + 8 k^{2} N + 39 N^{2} + 16 k\, 
      N^{2} + 8 N^{3})} {3 (2 + N) (2 + k + N)^{5}}, \nonu\\
c_ {90} & = & - \frac {32 (-k + N) (32 + 55 k + 18 k^{2} + 41 N + 
       35 k\, N + 11 N^{2})} {3 (2 + N) (2 + k + N)^{5}}, \nonu\\
c_ {91} & = &\frac {1} {9 (2 + N) (2 + k + N)^{5}} 32 (-216 - 468 k - 
    361 k^{2} - 86 k^{3} - 288 N - 592 k\, 
   N 
   -414 k^{2} N 
   \nonu\\ & - & 70 k^{3} N - 19 N^{2} - 138 k\, 
   N^{2} - 101 k^{2} N^{2} + 98 N^{3} + 34 k\, N^{3} + 29 N^{4}), \nonu\\
c_ {92} & = & - \frac {64 (-k + N) (8 + 17 k + 6 k^{2} + 11 N + 
       11 k\, N + 3 N^{2})} {3 (2 + N) (2 + k + N)^{4}}, \nonu\\
c_ {93} & = & - \frac {32 (48 + 52 k + 3 k^{2} - 6 k^{3} + 116 N + 
       80 k\, N + 5 k^{2} N + 85 N^{2} + 30 k\, 
      N^{2} + 19 N^{3})} {3 (2 + N) (2 + k + N)^{3}}, \nonu\\
c_ {94} & = &\frac {32 (-k + N) (8 + 17 k + 6 k^{2} + 11 N + 11 k\, 
     N + 3 N^{2})} {3 (2 + N) (2 + k + N)^{4}}, \nonu\\
c_ {95} & = & - \frac {32 (48 + 28 k - 9 k^{2} - 6 k^{3} + 92 N + 
       44 k\, N - k^{2} N + 61 N^{2} + 18 k\, 
      N^{2} + 13 N^{3})} {3 (2 + N) (2 + k + N)^{4}}, \nonu\\
c_ {96} & = & - \frac {32 (-k + N) (8 + 17 k + 6 k^{2} + 11 N + 
       11 k\, N + 3 N^{2})} {3 (2 + N) (2 + k + N)^{4}}, \nonu\\
c_ {97} & = & - \frac {32 (-k + N) (72 + 87 k + 26 k^{2} + 133 N + 
       105 k\, N + 16 k^{2} N + 81 N^{2} + 32 k\, 
      N^{2} + 16 N^{3})} {3 (2 + N) (2 + k + N)^{4}}, \nonu\\
c_ {98} & = & - \frac {64 (18 + 99 k + 99 k^{2} + 26 k^{3} + 45 N + 
       138 k\, N + 67 k^{2} N + 33 N^{2} + 43 k\, 
      N^{2} + 8 N^{3})} {9 (2 + k + N)^{3}}, \nonu\\
c_ {99} & = & - \frac {32 (-k + N) (60 + 77 k + 22 k^{2} + 121 N + 
       115 k\, N + 20 k^{2} N + 79 N^{2} + 42 k\, 
      N^{2} + 16 N^{3})} {9 (2 + N) (2 + k + N)^{5}}, \nonu\\
c_ {100} & = &\frac {32 (-k + N) (60 + 77 k + 22 k^{2} + 121 N + 
      115 k\, N + 20 k^{2} N + 79 N^{2} + 42 k\, 
     N^{2} + 16 N^{3})} {9 (2 + N) (2 + k + N)^{5}},\nonu\\
c_ {100} & = & \frac {32 (-k + N) (60 + 77 k + 22 k^{2} + 121 N + 
      115 k\, N + 20 k^{2} N + 79 N^{2} + 42 k\, 
     N^{2} + 16 N^{3})} {9 (2 + N) (2 + k + N)^{5}}, \nonu\\
c_ {101} & = & - \frac {128 (3 k + 2 k^{2} + 3 N + 8 k\, 
      N + 3 k^{2} N + 2 N^{2} + 3 k\, N^{2})} {3 (2 + k + N)^{4}}, \nonu\\
c_ {102} & = & - \frac {32 (3 + 2 k + N) (6 + k + 
       5 N)} {9 (2 + k + N)^{3}}, \qquad
c_ {103}  =  - \frac {64 (k - N)} {3 (2 + k + N)^{3}}, \nonu\\
c_ {104} & = & - \frac {32 (36 - 18 k - 59 k^{2} - 18 k^{3} + 108 N + 
       43 k\, N - 19 k^{2} N + 88 N^{2} + 35 k\, 
      N^{2} + 20 N^{3})} {9 (2 + N) (2 + k + N)^{3}}, \nonu\\
c_ {105} & = & - \frac {1} {9 (2 + N) (2 + k + N)^{4}} 32 (-144 - 
     168 k - 22 k^{2} + 51 k^{3} + 18 k^{4} - 336 N - 400 k\, 
    N 
    \nonu\\ & - & 150 k^{2} N - 3 k^{3} N - 226 N^{2} - 225 k\, 
    N^{2} - 68 k^{2} N^{2} - 36 N^{3} - 23 k\, N^{3} + 4 N^{4}),\nonu\\ 
c_ {106} & = & - \frac {1} {3 (2 + N) (2 + k + N)^{5}} 32 (48 + 
     68 k + 43 k^{2} + 10 k^{3} + 100 N + 100 k\, 
    N + 49 k^{2} N 
    \nonu\\ & + & 8 k^{3} N + 73 N^{2} + 40 k\, 
    N^{2} + 12 k^{2} N^{2} + 21 N^{3} + 2 k\, N^{3} + 2 N^{4}), \nonu\\
c_ {107} & = & \frac {1} {3 (2 + N) (2 + k + N)^{5}} 32 (48 + 36 k - 
    41 k^{2} - 48 k^{3} - 12 k^{4} + 132 N + 124 k\, 
   N + 7 k^{2} N 
   \nonu\\ & - &  14 k^{3} N + 133 N^{2} + 106 k\, 
   N^{2} + 18 k^{2} N^{2} + 55 N^{3} + 24 k\, N^{3} + 8 N^{4}), \nonu\\
c_ {108} & = & \frac {32 (-k + N) (24 + 31 k + 10 k^{2} + 53 N + 
      45 k\, N + 8 k^{2} N + 37 N^{2} + 16 k\, 
     N^{2} + 8 N^{3})} {3 (2 + N) (2 + k + N)^{4}}, \nonu\\
c_ {109} & = & - \frac {32 (-k + N) (8 + 23 k + 10 k^{2} + 37 N + 
       41 k\, N + 8 k^{2} N + 33 N^{2} + 16 k\, 
      N^{2} + 8 N^{3})} {3 (2 + N) (2 + k + N)^{4}}, \nonu\\
c_ {110} & = & - \frac {32 (-k + N) (8 + 17 k + 6 k^{2} + 11 N + 
       11 k\, N + 3 N^{2})} {3 (2 + N) (2 + k + N)^{4}}, \nonu\\
c_ {111} & = & \frac {32 (48 + 28 k - 9 k^{2} - 6 k^{3} + 92 N + 
      44 k\, N - k^{2} N + 61 N^{2} + 18 k\, 
     N^{2} + 13 N^{3})} {3 (2 + N) (2 + k + N)^{4}}, \nonu\\
c_ {112} & = & \frac {1} {9 (2 + N) (2 + k + N)^{4}} 32 (72 + 252 k + 
    249 k^{2} + 70 k^{3} + 288 N + 492 k\, 
   N + 291 k^{2} N 
   \nonu\\ & + &  44 k^{3} N + 303 N^{2} + 276 k\, 
   N^{2} + 78 k^{2} N^{2} + 119 N^{3} + 42 k\, N^{3} + 16 N^{4}),\nonu\\ 
c_ {113} & = & - \frac {1} {9 (2 + N) (2 + k + N)^{4}} 32 (-144 - 
     384 k - 292 k^{2} - 64 k^{3} - 120 N - 448 k\, 
    N - 321 k^{2} N 
    \nonu\\ & - &  50 k^{3} N + 92 N^{2} - 90 k\, 
    N^{2} - 77 k^{2} N^{2} + 115 N^{3} + 28 k\, N^{3} + 27 N^{4}),\nonu\\ 
c_ {114} & = & - \frac {1} {3 (2 + N) (2 + k + N)^{5}} 32 (48 + 
     148 k + 121 k^{2} + 30 k^{3} + 212 N + 496 k\, 
    N + 317 k^{2} N 
    \nonu\\ & + &  60 k^{3} N + 271 N^{2} + 500 k\, 
    N^{2} + 226 k^{2} N^{2} + 24 k^{3} N^{2} + 137 N^{3} + 194 k\, 
    N^{3} + 48 k^{2} N^{3} 
    \nonu\\ & + &  24 N^{4} + 24 k\, N^{4}), \nonu\\
c_ {115} & = & - \frac {1} {9 (2 + N) (2 + k + N)^{5}} 32 (516 k + 
     897 k^{2} + 526 k^{3} + 104 k^{4} + 60 N + 1008 k\, 
    N 
    \nonu\\ & + &  1271 k^{2} N + 500 k^{3} N + 52 k^{4} N + 111 N^{2} + 668 k\, 
    N^{2} + 558 k^{2} N^{2} + 114 k^{3} N^{2} + 127 N^{3} 
    \nonu\\ & + &  202 k\, 
    N^{3} + 76 k^{2} N^{3} + 76 N^{4} + 30 k\, N^{4} + 16 N^{5}), \nonu\\
c_ {116} & = & \frac {1} {3 (2 + N) (2 + k + N)^{5}} 64 (-24 - 56 k - 
    43 k^{2} - 10 k^{3} - 28 N - 70 k\, 
   N - 49 k^{2} N - 8 k^{3} N 
   \nonu\\ & + &  5 N^{2} - 16 k\, 
   N^{2} - 12 k^{2} N^{2} + 15 N^{3} + 4 k\, N^{3} + 4 N^{4}), \nonu\\
c_ {117} & = & \frac {32 (-k + N) (32 + 35 k + 10 k^{2} + 61 N + 
      47 k\, N + 8 k^{2} N + 39 N^{2} + 16 k\, 
     N^{2} + 8 N^{3})} {3 (2 + N) (2 + k + N)^{5}}, \nonu\\
c_ {118} & = & \frac {64 (k - N) (7 + 4 k + 4 N)} {3 (2 + k + N)^{3}},\nonu\\
 c_ {119} & = & \frac {1} {9 (2 + N) (2 + k + N)^{4}} 16 (288 + 
    132 k - 27 k^{2} - 10 k^{3} + 1164 N + 768 k\, 
   N + 129 k^{2} N 
   \nonu\\ & + &  4 k^{3} N + 1275 N^{2} + 636 k\, 
   N^{2} + 66 k^{2} N^{2} + 541 N^{3} + 138 k\, N^{3} + 80 N^{4}), \nonu\\
c_ {120} & = & - \frac {32 (k - N)} {3 (2 + k + N)},  \nonu\\
c_ {121} & = & - \frac {32 (-k + N) (60 + 77 k + 22 k^{2} + 121 N + 
       115 k\, N + 20 k^{2} N + 79 N^{2} + 42 k\, 
      N^{2} + 16 N^{3})} {3 (2 + N) (2 + k + N)^{3}}, \nonu\\
c_ {122} & = & \frac {1} {9 (2 + N) (2 + k + N)^{4}} 32 (72 + 252 k + 
    249 k^{2} + 70 k^{3} + 288 N + 492 k\, 
   N + 291 k^{2} N 
   \nonu\\ & + &  44 k^{3} N + 303 N^{2} + 276 k\, 
   N^{2} + 78 k^{2} N^{2} + 119 N^{3} + 42 k\, N^{3} + 16 N^{4}), \nonu\\
c_ {123} & = & - \frac {32 (-k + N) (32 + 55 k + 18 k^{2} + 41 N + 
       35 k\, N + 11 N^{2})} {3 (2 + N) (2 + k + N)^{5}}, \nonu\\
c_ {124} & = & \frac {32 (-k + N) (32 + 55 k + 18 k^{2} + 41 N + 
      35 k\, N + 11 N^{2})} {3 (2 + N) (2 + k + N)^{5}}, \nonu\\
c_ {125} & = & - \frac {64 (-k + N) (8 + 17 k + 6 k^{2} + 11 N + 
       11 k\, N + 3 N^{2})} {3 (2 + N) (2 + k + N)^{4}}, \nonu\\
c_ {126} & = & \frac {32 (-k + N) (32 + 35 k + 10 k^{2} + 61 N + 
      47 k\, N + 8 k^{2} N + 39 N^{2} + 16 k\, 
     N^{2} + 8 N^{3})} {3 (2 + N) (2 + k + N)^{5}}, \nonu\\
c_ {127} & = & \frac {32 (18 + 9 k + 2 k^{2} + 27 N + 5 k\, 
     N + 11 N^{2})} {9 (2 + k + N)^{4}}, \nonu\\
c_ {128} & = & - \frac {1} {9 (2 + N) (2 + k + N)^{5}} 32 (-216 - 
     468 k - 361 k^{2} - 86 k^{3} - 288 N - 592 k\, 
    N - 414 k^{2} N
    \nonu\\ & - &  70 k^{3} N - 19 N^{2} - 138 k\, 
    N^{2} - 101 k^{2} N^{2} + 98 N^{3} + 34 k\, N^{3} + 29 N^{4}), \nonu\\
c_ {129} & = & \frac {64 (18 + 9 k + 2 k^{2} + 27 N + 5 k\, 
     N + 11 N^{2})} {9 (2 + k + N)^{4}}, \nonu\\
c_ {130} & = & \frac {32 (36 - 6 k - 47 k^{2} - 18 k^{3} + 96 N + 
      37 k\, N - 13 k^{2} N + 82 N^{2} + 29 k\, 
     N^{2} + 20 N^{3})} {9 (2 + N) (2 + k + N)^{4}}, \nonu\\
c_ {131} & = & - \frac {64 (36 - 6 k - 47 k^{2} - 18 k^{3} + 96 N + 
       37 k\, N - 13 k^{2} N + 82 N^{2} + 29 k\, 
      N^{2} + 20 N^{3})} {9 (2 + N) (2 + k + N)^{4}}, \nonu\\
c_ {132} & = & - \frac {1} {9 (2 + N) (2 + k + N)^{5}} 32 (-216 - 
     420 k - 235 k^{2} + k^{3} + 18 k^{4} - 336 N - 628 k\, 
    N 
    \nonu\\ & - &  351 k^{2} N - 37 k^{3} N - 109 N^{2} - 237 k\, 
    N^{2} - 110 k^{2} N^{2} + 47 N^{3} + k\, N^{3} + 20 N^{4}), \nonu\\
c_ {133} & = & - \frac {32 (48 + 52 k + 3 k^{2} - 6 k^{3} + 116 N + 
       80 k\, N + 5 k^{2} N + 85 N^{2} + 30 k\, 
      N^{2} + 19 N^{3})} {3 (2 + N) (2 + k + N)^{3}}, \nonu\\
c_ {134} & = & \frac {32 (-k + N) (8 + 17 k + 6 k^{2} + 11 N + 11 k\, 
     N + 3 N^{2})} {3 (2 + N) (2 + k + N)^{4}},  \nonu\\
c_ {135} & = & - \frac {32 (-k + N) (8 + 17 k + 6 k^{2} + 11 N + 
       11 k\, N + 3 N^{2})} {3 (2 + N) (2 + k + N)^{4}}, \nonu\\
c_ {136} & = & \frac {32 (48 + 28 k - 9 k^{2} - 6 k^{3} + 92 N + 
      44 k\, N - k^{2} N + 61 N^{2} + 18 k\, 
     N^{2} + 13 N^{3})} {3 (2 + N) (2 + k + N)^{4}}, \nonu\\
c_ {137} & = & \frac {32 (-k + N) (72 + 87 k + 26 k^{2} + 133 N + 
      105 k\, N + 16 k^{2} N + 81 N^{2} + 32 k\, 
     N^{2} + 16 N^{3})} {3 (2 + N) (2 + k + N)^{4}}, \nonu\\
c_ {138} & = & \frac {1} {9 (2 + N) (2 + k + N)^{3}} 64 (36 + 150 k + 
    135 k^{2} + 34 k^{3} + 156 N + 381 k\, 
   N + 212 k^{2} N
   \nonu\\ & + &  26 k^{3} N + 168 N^{2} + 248 k\, 
   N^{2} + 67 k^{2} N^{2} + 64 N^{3} + 43 k\, N^{3} + 8 N^{4}), \nonu\\
c_ {139} & = & \frac {32 (3 + 2 k + N) (6 + k + 
      5 N)} {9 (2 + k + N)^{3}}, \nonu\\
c_ {140} & = & - \frac {1} {9 (2 + N) (2 + k + N)^{4}} 32 (-144 - 
     384 k - 292 k^{2} - 64 k^{3} - 120 N - 448 k\, 
    N - 321 k^{2} N 
    \nonu\\ & - &  50 k^{3} N + 92 N^{2} - 90 k\, 
    N^{2} - 77 k^{2} N^{2} + 115 N^{3} + 28 k\, N^{3} + 27 N^{4}), \nonu\\
c_ {141} & = & \frac {1} {3 (2 + N) (2 + k + N)^{5}} 32 (48 + 36 k - 
    41 k^{2} - 48 k^{3} - 12 k^{4} + 132 N + 124 k\, 
   N + 7 k^{2} N 
   \nonu\\ & - &  14 k^{3} N + 133 N^{2} + 106 k\, 
   N^{2} + 18 k^{2} N^{2} + 55 N^{3} + 24 k\, N^{3} + 8 N^{4}), \nonu\\
c_ {142} & = & \frac {32 (-k + N) (32 + 35 k + 10 k^{2} + 61 N + 
      47 k\, N + 8 k^{2} N + 39 N^{2} + 16 k\, 
     N^{2} + 8 N^{3})} {3 (2 + N) (2 + k + N)^{5}}, \nonu\\
c_ {143} & = & \frac {32 (-k + N) (32 + 55 k + 18 k^{2} + 41 N + 
      35 k\, N + 11 N^{2})} {3 (2 + N) (2 + k + N)^{5}}, \nonu\\
c_ {144} & = & - \frac {1} {3 (2 + N) (2 + k + N)^{5}} 32 (48 + 
     68 k + 43 k^{2} + 10 k^{3} + 100 N + 100 k\, 
    N + 49 k^{2} N 
     \nonu\\ & + & 8 k^{3} N + 73 N^{2} + 40 k\, 
    N^{2} + 12 k^{2} N^{2} + 21 N^{3} + 2 k\, N^{3} + 2 N^{4}), \nonu\\
c_ {145} & = & - \frac {32 (-k + N) (32 + 35 k + 10 k^{2} + 61 N + 
       47 k\, N + 8 k^{2} N + 39 N^{2} + 16 k\, 
      N^{2} + 8 N^{3})} {3 (2 + N) (2 + k + N)^{5}}, \nonu\\
c_ {146} & = & - \frac {1} {9 (2 + N) (2 + k + N)^{5}} 32 (420 k + 
     645 k^{2} + 352 k^{3} + 68 k^{4} + 156 N + 1080 k\, 
    N  
    \nonu\\ & + & 1145 k^{2} N + 434 k^{3} N + 52 k^{4} N + 291 N^{2} + 866 k\, 
    N^{2} + 576 k^{2} N^{2} + 114 k^{3} N^{2} + 229 N^{3}
    \nonu\\ & + &  268 k\, N^{3} +76 k^{2} N^{3} + 94 N^{4} + 30 k\, N^{4} + 16 N^{5}), \nonu\\
c_ {147} & = & \frac {32 (-k + N) (8 + 23 k + 10 k^{2} + 37 N + 
      41 k\, N + 8 k^{2} N + 33 N^{2} + 16 k\, 
     N^{2} + 8 N^{3})} {3 (2 + N) (2 + k + N)^{4}}, \nonu\\
c_ {148} & = & - \frac {32 (-k + N) (24 + 31 k + 10 k^{2} + 53 N + 
       45 k\, N + 8 k^{2} N + 37 N^{2} + 16 k\, 
      N^{2} + 8 N^{3})} {3 (2 + N) (2 + k + N)^{4}}, \nonu\\
c_ {149} & = & - \frac {32 (-k + N) (8 + 17 k + 6 k^{2} + 11 N + 
       11 k\, N + 3 N^{2})} {3 (2 + N) (2 + k + N)^{4}}, \nonu\\
c_ {150} & = & \frac {32 (48 + 28 k - 9 k^{2} - 6 k^{3} + 92 N + 
      44 k\, N - k^{2} N + 61 N^{2} + 18 k\, 
     N^{2} + 13 N^{3})} {3 (2 + N) (2 + k + N)^{4}}, \nonu\\
c_ {151} & = & \frac {1} {9 (2 + N) (2 + k + N)^{4}} 32 (72 + 372 k + 
    345 k^{2} + 86 k^{3} + 168 N + 552 k\, 
   N + 363 k^{2} N 
   \nonu\\ & + &  52 k^{3} N + 147 N^{2} + 252 k\, 
   N^{2} + 90 k^{2} N^{2} + 55 N^{3} + 30 k\, N^{3} + 8 N^{4}), \nonu\\
c_ {152} & = & - \frac {1} {9 (2 + N) (2 + k + N)^{4}} 32 (216 + 
     336 k + 219 k^{2} + 50 k^{3} + 492 N + 510 k\, 
    N + 265 k^{2} N 
    \nonu\\ & + &  52 k^{3} N + 423 N^{2} + 226 k\, 
    N^{2} + 62 k^{2} N^{2} + 143 N^{3} + 14 k\, N^{3} + 16 N^{4}), \nonu\\
c_ {153} & = & \frac {32 (36 - 18 k - 59 k^{2} - 18 k^{3} + 108 N + 
      43 k\, N - 19 k^{2} N + 88 N^{2} + 35 k\, 
     N^{2} + 20 N^{3})} {9 (2 + N) (2 + k + N)^{3}}, \nonu\\
c_ {154} & = & - \frac {64 (k - N)} {3 (2 + k + N)^{3}}, \nonu\\
c_ {155} & = & \frac {32 (-k + N) (60 + 77 k + 22 k^{2} + 121 N + 
      115 k\, N + 20 k^{2} N + 79 N^{2} + 42 k\, 
     N^{2} + 16 N^{3})} {9 (2 + N) (2 + k + N)^{5}}, \nonu\\
c_ {156} & = & - \frac {32 (-k + N) (60 + 77 k + 22 k^{2} + 121 N + 
       115 k\, N + 20 k^{2} N + 79 N^{2} + 42 k\, 
      N^{2} + 16 N^{3})} {9 (2 + N) (2 + k + N)^{5}}, \nonu\\
c_ {157} & = & - \frac {128 (3 k + 2 k^{2} + 3 N + 8 k\, 
      N + 3 k^{2} N + 2 N^{2} + 3 k\, N^{2})} {3 (2 + k + N)^{4}}, \nonu\\
c_ {158} & = & - \frac {1} {9 (2 + N) (2 + k + N)^{4}} 32 (-144 - 
     168 k - 22 k^{2} + 51 k^{3} + 18 k^{4} - 336 N - 400 k\, 
    N 
    \nonu\\ & - &  150 k^{2} N - 3 k^{3} N - 226 N^{2} - 225 k\, 
    N^{2} - 68 k^{2} N^{2} - 36 N^{3} - 23 k\, N^{3} + 4 N^{4}), \nonu\\
c_ {159} & = & - \frac {1} {3 (2 + N) (2 + k + N)^{5}} 32 (48 + 
     148 k + 121 k^{2} + 30 k^{3} + 212 N + 496 k\, 
    N + 317 k^{2} N 
    \nonu\\ & + &  60 k^{3} N + 271 N^{2} + 500 k\, 
    N^{2} + 226 k^{2} N^{2} + 24 k^{3} N^{2} + 137 N^{3} + 194 k\, 
    N^{3} + 48 k^{2} N^{3} 
    \nonu\\ & + & 24 N^{4} + 24 k\, N^{4}), \nonu\\
c_ {160} & = & - \frac {1} {3 (2 + N) (2 + k + N)^{5}} 64 (-24 - 
     56 k - 43 k^{2} - 10 k^{3} - 28 N - 70 k\, 
    N - 49 k^{2} N 
    \nonu\\ & - &  8 k^{3} N + 5 N^{2} - 16 k\, 
    N^{2} - 12 k^{2} N^{2} + 15 N^{3} + 4 k\, N^{3} + 4 N^{4}), \nonu\\
c_ {161} & = & - \frac {64 (k - N) (7 + 4 k + 
       4 N)} {3 (2 + k + N)^{3}}, \nonu\\
c_ {162} & = & \frac {1} {9 (2 + N) (2 + k + N)^{4}} 16 (-288 - 
    1140 k - 933 k^{2} - 214 k^{3} - 156 N - 1272 k\, 
   N 
   \nonu\\ & - &  849 k^{2} N - 116 k^{3} N + 189 N^{2} - 396 k\, 
   N^{2} - 186 k^{2} N^{2} + 163 N^{3} - 18 k\, N^{3} + 32 N^{4}), \nonu\\
c_ {163} & = & - \frac {54} {5}, \qquad
c_ {164}  =  \frac {16 (k - N)} {5 (2 + k + N)}, \nonu\\
c_ {165} & = & \frac {6 (60 + 77 k + 22 k^{2} + 121 N + 115 k\, 
     N + 20 k^{2} N + 79 N^{2} + 42 k\, 
     N^{2} + 16 N^{3})} {(2 + N) (2 + k + N)^{2}}, \nonu\\
c_ {166} & = & - \frac {16 (-k + N) (60 + 77 k + 22 k^{2} + 121 N + 
       115 k\, N + 20 k^{2} N + 79 N^{2} + 42 k\, 
      N^{2} + 16 N^{3})} {(2 + N) (2 + k + N)^{3}}, \nonu\\
c_ {167} & = & \frac {12 (60 + 77 k + 22 k^{2} + 121 N + 115 k\, 
     N + 20 k^{2} N + 79 N^{2} + 42 k\, 
     N^{2} + 16 N^{3})} {(2 + N) (2 + k + N)^{2}}, \nonu\\
c_ {168} & = & - \frac {1} {3 (2 + N) (2 + k + N)^{4}} 8 (336 + 
     720 k + 413 k^{2} + 70 k^{3} + 648 N + 884 k\, 
    N + 189 k^{2} N 
    \nonu\\ & - &  28 k^{3} N + 407 N^{2} + 288 k\, 
    N^{2} - 26 k^{2} N^{2} + 101 N^{3} + 22 k\, N^{3} + 8 N^{4}), \nonu\\
c_ {169} & = & - \frac {48 (32 + 55 k + 18 k^{2} + 41 N + 35 k\, 
      N + 11 N^{2})} {(2 + N) (2 + k + N)^{4}}, \nonu\\
c_ {170} & = & \frac {48 (32 + 55 k + 18 k^{2} + 41 N + 35 k\, 
     N + 11 N^{2})} {(2 + N) (2 + k + N)^{4}}, \nonu\\
c_ {171} & = & - \frac {1} {3 (2 + N) (2 + k + N)^{4}} 8 (540 + 
     613 k + 179 k^{2} + 6 k^{3} + 1349 N + 1136 k\, 
    N + 157 k^{2} N 
    \nonu\\ & - &  24 k^{3} N + 1223 N^{2} + 699 k\, 
    N^{2} + 36 k^{2} N^{2} + 488 N^{3} + 150 k\, N^{3} + 72 N^{4}), \nonu\\
c_ {172} & = & \frac {1} {(2 + N) (2 + k + N)^{5}} 32 (156 + 355 k + 
    273 k^{2} + 92 k^{3} + 12 k^{4} + 335 N + 506 k\, 
   N 
   \nonu\\ & + &  219 k^{2} N + 30 k^{3} N + 259 N^{2} + 229 k\, 
   N^{2} + 36 k^{2} N^{2} + 90 N^{3} + 36 k\, N^{3} + 12 N^{4}), \nonu\\
c_ {173} & = & - \frac {8 (-k + N) (32 + 55 k + 18 k^{2} + 41 N + 
       35 k\, N + 11 N^{2})} {(2 + N) (2 + k + N)^{5}}, \nonu\\
c_ {174} & = & \frac {24 (32 + 55 k + 18 k^{2} + 41 N + 35 k\, 
     N + 11 N^{2})} {(2 + N) (2 + k + N)^{4}}, \nonu\\
c_ {175} & = & - \frac {16 (-k + N) (32 + 55 k + 18 k^{2} + 41 N + 
       35 k\, N + 11 N^{2})} {(2 + N) (2 + k + N)^{5}}, \nonu\\
c_ {176} & = & - \frac {8 (-k + N) (32 + 55 k + 18 k^{2} + 41 N + 
       35 k\, N + 11 N^{2})} {(2 + N) (2 + k + N)^{5}}, \nonu\\
c_ {177} & = & \frac {16 (-k + N) (32 + 55 k + 18 k^{2} + 41 N + 
      35 k\, N + 11 N^{2})} {(2 + N) (2 + k + N)^{5}}, \nonu\\
c_ {178} & = & - \frac {48 (1 + k) (1 + N) (32 + 55 k + 18 k^{2} + 
       41 N + 35 k\, N + 11 N^{2})} {(2 + N) (2 + k + N)^{6}}, \nonu\\
c_ {179} & = & - \frac {24 (32 + 55 k + 18 k^{2} + 41 N + 35 k\, 
      N + 11 N^{2})} {(2 + N) (2 + k + N)^{4}}, \nonu\\
c_ {180} & = & - \frac {1}{3 (2 + N) (2 + k + N)^{4}}8 (144 + 322 k + 155 k^{2} + 18 k^{3} + 
       326 N + 482 k\, N + 121 k^{2} N 
        \nonu\\ & + & 227 N^{2} + 174 k\, 
      N^{2} + 47 N^{3}), \nonu\\
c_ {181} & = & - \frac {1} {3 (2 + N) (2 + k + N)^{5}} 8 (1680 + 
     2440 k + 952 k^{2} + 49 k^{3} - 18 k^{4} + 4592 N + 5432 k\, 
    N 
    \nonu\\ & + &  1623 k^{2} N + 89 k^{3} N + 4632 N^{2} + 3959 k\, 
    N^{2} + 659 k^{2} N^{2} + 2025 N^{3} + 943 k\, N^{3} + 319 N^{4}),\nonu\\
 c_ {182} & = & \frac {48 (1 + k) (32 + 55 k + 18 k^{2} + 41 N + 
      35 k\, N + 11 N^{2})} {(2 + N) (2 + k + N)^{5}}, \nonu\\
c_ {183} & = & - \frac {48 (2 + 2 k + k^{2} + 2 N + N^{2}) (32 + 
       55 k + 18 k^{2} + 41 N + 35 k\, 
      N + 11 N^{2})} {(2 + N) (2 + k + N)^{6}}, \nonu\\
c_ {184} & = & \frac {48 (1 + N) (32 + 55 k + 18 k^{2} + 41 N + 
      35 k\, N + 11 N^{2})} {(2 + N) (2 + k + N)^{5}}, \nonu\\
c_ {185} & = & - \frac {48 (1 + k) (1 + N) (32 + 55 k + 18 k^{2} + 
       41 N + 35 k\, N + 11 N^{2})} {(2 + N) (2 + k + N)^{6}}, \nonu\\
c_ {186} & = & - \frac {1} {3 (2 + N) (2 + k + N)^{5}} 16 (108 - 
     64 k - 343 k^{2} - 130 k^{3} + 442 N - 44 k\, 
    N - 537 k^{2} N 
    \nonu\\ & - &  128 k^{3} N + 585 N^{2} + 52 k\, 
    N^{2} - 200 k^{2} N^{2} + 309 N^{3} + 38 k\, N^{3} + 56 N^{4}), \nonu\\
c_ {187} & = & - \frac {1} {(2 + N) (2 + k + N)^{5}} 16 (432 + 
     860 k + 499 k^{2} + 90 k^{3} + 1036 N + 1488 k\, 
    N + 531 k^{2} N 
    \nonu\\ & + &  36 k^{3} N + 917 N^{2} + 868 k\, 
    N^{2} + 146 k^{2} N^{2} + 359 N^{3} + 174 k\, N^{3} + 52 N^{4}), \nonu\\
c_ {188} & = & \frac {8 (144 + 434 k + 367 k^{2} + 90 k^{3} + 214 N + 
      418 k\, N + 185 k^{2} N + 79 N^{2} + 78 k\, 
     N^{2} + 7 N^{3})} {3 (2 + N) (2 + k + N)^{4}}, \nonu\\
c_ {189} & = & \frac {16 (3 + k + 2 N) (32 + 55 k + 18 k^{2} + 41 N + 
      35 k\, N + 11 N^{2})} {(2 + N) (2 + k + N)^{5}}, \nonu\\
c_ {190} & = & \frac {16 (3 + 2 k + N) (32 + 55 k + 18 k^{2} + 41 N + 
      35 k\, N + 11 N^{2})} {(2 + N) (2 + k + N)^{5}}, \nonu\\
c_ {191} & = & \frac {1} {3 (2 + N) (2 + k + N)^{5}} 16 (360 + 
    356 k - 4 k^{2} - 40 k^{3} + 1192 N + 1063 k\, 
   N + 81 k^{2} N 
   \nonu\\ & - &  38 k^{3} N + 1425 N^{2} + 1000 k\, 
   N^{2} + 79 k^{2} N^{2} + 723 N^{3} + 299 k\, N^{3} + 128 N^{4}), \nonu\\
c_ {192} & = & \frac {1} {3 (2 + N) (2 + k + N)^{5}} 8 (-480 - 
    1444 k - 1414 k^{2} - 505 k^{3} - 54 k^{4} - 812 N - 2168 k\, 
   N 
   \nonu\\ & - &  1641 k^{2} N - 329 k^{3} N - 306 N^{2} - 863 k\, 
   N^{2} - 431 k^{2} N^{2} + 81 N^{3} - 43 k\, N^{3} + 41 N^{4}), \nonu\\
c_ {193} & = & \frac {1} {3 (2 + N) (2 + k + N)^{5}} 32 (-54 - 
    334 k - 379 k^{2} - 106 k^{3} + 145 N - 104 k\, 
   N - 210 k^{2} N 
   \nonu\\ & - &  26 k^{3} N + 384 N^{2} + 232 k\, 
   N^{2} + k^{2} N^{2} + 237 N^{3} + 98 k\, N^{3} + 44 N^{4}), \nonu\\
c_ {194} & = & - \frac {32 (-k + N) (32 + 55 k + 18 k^{2} + 41 N + 
       35 k\, N + 11 N^{2})} {(2 + N) (2 + k + N)^{5}}, \nonu\\
c_ {195} & = & \frac {1} {3 (2 + N) (2 + k + N)^{4}} 16 (816 + 
    1263 k + 483 k^{2} + 38 k^{3} + 1839 N + 2112 k\, 
   N + 424 k^{2} N 
   \nonu\\ & - &  26 k^{3} N + 1476 N^{2} + 1129 k\, 
   N^{2} + 77 k^{2} N^{2} + 509 N^{3} + 200 k\, N^{3} + 64 N^{4}), \nonu\\
c_ {196} & = & - \frac {16 (-36 - 121 k - 46 k^{2} - 17 N - 55 k\, 
      N + 4 k^{2} N + 29 N^{2} + 14 k\, 
      N^{2} + 12 N^{3})} {(2 + N) (2 + k + N)^{4}}, \nonu\\
c_ {197} & = & - \frac{1} {3 (2 + N) (2 + k + N)^{4}}8 (144 + 322 k + 155 k^{2} + 18 k^{3} + 
       326 N + 482 k\, N + 121 k^{2} N 
        \nonu\\ & + &
         227 N^{2} + 174 k\, 
      N^{2} + 47 N^{3}), \nonu\\
c_ {198} & = & \frac {8 (144 + 434 k + 367 k^{2} + 90 k^{3} + 214 N + 
      418 k\, N + 185 k^{2} N + 79 N^{2} + 78 k\, 
     N^{2} + 7 N^{3})} {3 (2 + N) (2 + k + N)^{4}}, \nonu\\
c_ {199} & = & \frac{1} {3 (2 + N) (2 + k + N)^{4}}8 (1320 + 1808 k + 661 k^{2} + 54 k^{3} + 
      2644 N + 2578 k\, N + 515 k^{2} N
       \nonu\\ & + & 1705 N^{2} + 900 k\, 
     N^{2} + 343 N^{3}), \nonu\\
c_ {200} & = & - \frac {48 (1 + N) (32 + 55 k + 18 k^{2} + 41 N + 
       35 k\, N + 11 N^{2})} {(2 + N) (2 + k + N)^{5}}, \nonu\\
c_ {201} & = & - \frac {48 (1 + k) (32 + 55 k + 18 k^{2} + 41 N + 
       35 k\, N + 11 N^{2})} {(2 + N) (2 + k + N)^{5}}, \nonu\\
c_ {202} & = & \frac {64 (6 - 19 k - 23 k^{2} - 6 k^{3} + 34 N + 
      21 k\, N + k^{2} N + 38 N^{2} + 22 k\, 
     N^{2} + 10 N^{3})} {(2 + N) (2 + k + N)^{4}}, \nonu\\
c_ {203} & = & - \frac{1} {3 (2 + N) (2 + k + N)^{4}}8 (240 + 406 k + 179 k^{2} + 18 k^{3} + 
       482 N + 560 k\, N + 133 k^{2} N 
        \nonu\\ & + & 317 N^{2} + 192 k\, 
      N^{2} + 65 N^{3}), \nonu\\
c_ {204} & = & \frac {1} {3 (2 + N) (2 + k + N)^{4}} 16 (-552 - 
    1118 k - 772 k^{2} - 168 k^{3} - 838 N - 1261 k\, 
   N  \nonu\\ & - &  623 k^{2} N 
   -66 k^{3} N - 223 N^{2} - 192 k\, 
   N^{2} - 75 k^{2} N^{2} + 131 N^{3} + 93 k\, N^{3} + 48 N^{4}), \nonu\\
c_ {205} & = & - \frac {16 (32 + 81 k + 56 k^{2} + 12 k^{3} + 47 N + 
       85 k\, N + 30 k^{2} N + 19 N^{2} + 20 k\, 
      N^{2} + 2 N^{3})} {(2 + N) (2 + k + N)^{3}}, \nonu\\
c_ {206} & = & - \frac {1} {3 (2 + N) (2 + k + N)^{4}} 16 (168 - 
     274 k - 467 k^{2} - 138 k^{3} + 694 N - 131 k\, 
    N - 532 k^{2} N 
     \nonu\\ & - & 114 k^{3} N + 934 N^{2} + 240 k\, 
    N^{2} - 123 k^{2} N^{2} + 514 N^{3} + 141 k\, N^{3} + 96 N^{4}), \nonu\\
c_ {207} & = & \frac {16 (180 + 263 k + 121 k^{2} + 18 k^{3} + 
      331 N + 331 k\, N + 77 k^{2} N + 196 N^{2} + 102 k\, 
     N^{2} + 37 N^{3})} {3 (2 + N) (2 + k + N)^{4}}, \nonu\\
c_ {208} & = & - \frac {48 (-16 - 13 k + 11 k^{2} + 6 k^{3} - 11 N + 
       15 k\, N + 17 k^{2} N + 6 N^{2} + 14 k\, 
      N^{2} + 3 N^{3})} {(2 + N) (2 + k + N)^{4}}, \nonu\\
c_ {209} & = & \frac {48 (32 + 55 k + 18 k^{2} + 41 N + 35 k\, 
     N + 11 N^{2})} {(2 + N) (2 + k + N)^{4}}, \nonu\\
c_ {210} & = & \frac {48 (-k + N) (32 + 55 k + 18 k^{2} + 41 N + 
      35 k\, N + 11 N^{2})} {(2 + N) (2 + k + N)^{5}}, \nonu\\
c_ {211} & = & - \frac {48 (32 + 55 k + 18 k^{2} + 41 N + 35 k\, 
      N + 11 N^{2})} {(2 + N) (2 + k + N)^{4}}, \nonu\\
c_ {212} & = & - \frac {48 (-k + N) (32 + 55 k + 18 k^{2} + 41 N + 
       35 k\, N + 11 N^{2})} {(2 + N) (2 + k + N)^{5}}, \nonu\\
c_ {213} & = & \frac {32 (146 + 235 k + 89 k^{2} + 6 k^{3} + 274 N + 
      282 k\, N + 50 k^{2} N + 161 N^{2} + 83 k\, 
     N^{2} + 30 N^{3})} {(2 + N) (2 + k + N)^{4}}, \nonu\\
c_ {214} & = & - \frac {48 (32 + 55 k + 18 k^{2} + 41 N + 35 k\, 
      N + 11 N^{2})} {(2 + N) (2 + k + N)^{4}}, \nonu\\
c_ {215} & = & \frac {48 (32 + 55 k + 18 k^{2} + 41 N + 35 k\, 
     N + 11 N^{2})} {(2 + N) (2 + k + N)^{4}}, \nonu\\
c_ {216} & = & - \frac {16 (32 + 55 k + 18 k^{2} + 41 N + 35 k\, 
      N + 11 N^{2})} {(2 + N) (2 + k + N)^{3}}, \nonu\\
c_ {217} & = & \frac {48 (20 + 15 k - 11 k^{2} - 6 k^{3} + 51 N + 
      43 k\, N + k^{2} N + 40 N^{2} + 22 k\, 
     N^{2} + 9 N^{3})} {(2 + N) (2 + k + N)^{4}}, \nonu\\
c_ {218} & = & - \frac {48 (20 + 15 k - 11 k^{2} - 6 k^{3} + 51 N + 
       43 k\, N + k^{2} N + 40 N^{2} + 22 k\, 
      N^{2} + 9 N^{3})} {(2 + N) (2 + k + N)^{4}}, \nonu\\
c_ {219} & = & \frac {48 (32 + 55 k + 18 k^{2} + 41 N + 35 k\, 
     N + 11 N^{2})} {(2 + N) (2 + k + N)^{4}}, \nonu\\
c_ {220} & = & \frac {32 (-12 - 163 k - 152 k^{2} - 36 k^{3} - 11 N - 
      167 k\, N - 82 k^{2} N + 7 N^{2} - 36 k\, 
     N^{2} + 4 N^{3})} {3 (2 + N) (2 + k + N)^{4}}, \nonu\\
c_ {221} & = & - \frac {1} {(2 + N) (2 + k + N)^{5}} 16 (-32 - 
     147 k - 150 k^{2} - 40 k^{3} - 13 N - 207 k\, 
    N - 201 k^{2} N 
    \nonu\\ & - &  38 k^{3} N + 69 N^{2} - 51 k\, 
    N^{2} - 57 k^{2} N^{2} + 68 N^{3} + 15 k\, N^{3} + 16 N^{4}), \nonu\\
c_ {222} & = & \frac {1} {(2 + N) (2 + k + N)^{5}} 32 (32 + 47 k + 
    43 k^{2} + 40 k^{3} + 12 k^{4} + 113 N + 91 k\, 
   N + 4 k^{3} N 
   \nonu\\ & + &  154 N^{2} + 93 k\, 
   N^{2} - 7 k^{2} N^{2} + 91 N^{3} + 37 k\, N^{3} + 18 N^{4}), \nonu\\
c_ {223} & = & \frac {48 (20 + 21 k + 6 k^{2} + 25 N + 13 k\, 
     N + 7 N^{2})} {(2 + N) (2 + k + N)^{3}}, \nonu\\
c_ {224} & = & \frac {1} {3 (2 + N) (2 + k + N)^{4}}16 (180 + 263 k + 121 k^{2} + 18 k^{3} + 
      331 N + 331 k\, N + 77 k^{2} N 
       \nonu\\ & + & 196 N^{2} + 102 k\, 
     N^{2} + 37 N^{3}), \nonu\\
c_ {225} & = & - \frac{1}{3 (2 + N) (2 + k + N)^{4}}32 (372 + 689 k + 394 k^{2} + 72 k^{3} + 
       673 N + 829 k\, N + 236 k^{2} N 
       \nonu\\ & + &  385 N^{2} + 240 k\, 
      N^{2} + 70 N^{3}), \nonu\\
c_ {226} & = & - \frac {1} {(2 + N) (2 + k + N)^{5}} 16 (-32 - 
     147 k - 150 k^{2} - 40 k^{3} - 13 N - 207 k\, 
    N - 201 k^{2} N 
    \nonu\\ & - &  38 k^{3} N + 69 N^{2} - 51 k\, 
    N^{2} - 57 k^{2} N^{2} + 68 N^{3} + 15 k\, N^{3} + 16 N^{4}), \nonu\\
c_ {227} & = & - \frac {48 (20 + 21 k + 6 k^{2} + 25 N + 13 k\, 
      N + 7 N^{2})} {(2 + N) (2 + k + N)^{3}}, \nonu\\
c_ {228} & = & \frac {48 (-16 - 13 k + 11 k^{2} + 6 k^{3} - 11 N + 
      15 k\, N + 17 k^{2} N + 6 N^{2} + 14 k\, 
     N^{2} + 3 N^{3})} {(2 + N) (2 + k + N)^{4}}, \nonu\\
c_ {229} & = & \frac {32 (34 - 4 k - 23 k^{2} - 6 k^{3} + 89 N + 
      63 k\, N + 10 k^{2} N + 76 N^{2} + 43 k\, 
     N^{2} + 18 N^{3})} {(2 + N) (2 + k + N)^{4}}, \nonu\\
c_ {230} & = & - \frac {1} {3 (2 + N) (2 + k + N)^{4}} 16 (384 + 
     793 k + 510 k^{2} + 104 k^{3} + 947 N + 1692 k\, 
    N + 892 k^{2} N
    \nonu\\ & + &  136 k^{3} N + 912 N^{2} + 1383 k\, 
    N^{2} + 578 k^{2} N^{2} + 60 k^{3} N^{2} + 399 N^{3} + 476 k\, 
    N^{3} + 126 k^{2} N^{3} 
    \nonu\\ & + &  64 N^{4} + 48 k\, N^{4}), \nonu\\
c_ {231} & = & \frac {1} {3 (2 + N) (2 + k + N)^{4}} 8 (672 + 
    1466 k + 1007 k^{2} + 218 k^{3} + 1966 N + 3680 k\, 
   N 
   \nonu\\ & + &  2075 k^{2} N + 340 k^{3} N + 2117 N^{2} + 3198 k\, 
   N^{2} + 1322 k^{2} N^{2} + 120 k^{3} N^{2} + 977 N^{3} 
   \nonu\\ & + &  1070 k\, 
   N^{3} + 252 k^{2} N^{3} + 160 N^{4} + 96 k\, N^{4}), \nonu\\
c_ {232} & = & - \frac {1} {(2 + N) (2 + k + N)^{4}} 16 (176 + 
     232 k + 91 k^{2} + 10 k^{3} + 448 N + 524 k\, 
    N + 183 k^{2} N 
    \nonu\\ & + &  20 k^{3} N + 433 N^{2} + 384 k\, 
    N^{2} + 78 k^{2} N^{2} + 183 N^{3} + 90 k\, N^{3} + 28 N^{4}), \nonu\\
c_ {233} & = & \frac {1} {3 (2 + N) (2 + k + N)^{5}} 16 (432 + 
    506 k + 227 k^{2} + 107 k^{3} + 30 k^{4} + 1546 N + 1597 k\, 
   N 
   \nonu\\ & + &  645 k^{2} N + 277 k^{3} N + 60 k^{4} N + 2298 N^{2} + 1891 k\, 
   N^{2} + 454 k^{2} N^{2} + 90 k^{3} N^{2} + 1731 N^{3} 
   \nonu\\ & + &  1016 k\, 
   N^{3} + 96 k^{2} N^{3} + 653 N^{4} + 216 k\, N^{4} + 96 N^{5}), \nonu\\
c_ {234} & = & \frac {1} {3 (2 + N) (2 + k + N)^{5}} 8 (360 + 794 k + 
    604 k^{2} + 183 k^{3} + 18 k^{4} + 754 N + 1326 k\, 
   N 
   \nonu\\ & + &  720 k^{2} N + 117 k^{3} N + 554 N^{2} + 697 k\, 
   N^{2} + 206 k^{2} N^{2} + 164 N^{3} + 111 k\, N^{3} + 16 N^{4}), \nonu\\
c_ {235} & = & \frac {1} {3 (2 + N) (2 + k + N)^{5}} 8 (744 + 
    1414 k + 844 k^{2} + 121 k^{3} - 18 k^{4} + 1478 N + 1544 k\, 
   N 
   \nonu\\ & + &  138 k^{2} N - 145 k^{3} N + 1248 N^{2} + 719 k\, 
   N^{2} - 88 k^{2} N^{2} + 594 N^{3} + 223 k\, N^{3} + 112 N^{4}), \nonu\\
c_ {236} & = & - \frac {16 (3 + 2 k + N) (32 + 55 k + 18 k^{2} + 
       41 N + 35 k\, N + 11 N^{2})} {(2 + N) (2 + k + N)^{5}}, \nonu\\
c_ {237} & = & - \frac {16 (3 + k + 2 N) (32 + 55 k + 18 k^{2} + 
       41 N + 35 k\, N + 11 N^{2})} {(2 + N) (2 + k + N)^{5}}, \nonu\\
c_ {238} & = & \frac {1} {3 (2 + N) (2 + k + N)^{5}} 32 (96 + 257 k + 
    186 k^{2} + 40 k^{3} + 223 N + 497 k\, 
   N + 273 k^{2} N 
   \nonu\\ & + &  38 k^{3} N + 181 N^{2} + 301 k\, 
   N^{2} + 93 k^{2} N^{2} + 58 N^{3} + 55 k\, N^{3} + 6 N^{4}), \nonu\\
c_ {239} & = & - \frac{1}{(2 + N) (2 + k + N)^{6}}16 (1 + k) (1 + N) (60 + 77 k + 22 k^{2} + 
       121 N + 115 k\, N 
       \nonu\\ & + & 20 k^{2} N+79 N^{2} + 42 k\, 
      N^{2} + 16 N^{3}), \nonu\\
c_ {240} & = & - \frac {1} {3 (2 + N) (2 + k + N)^{6}} 16 (180 + 
     475 k + 591 k^{2} + 366 k^{3} + 80 k^{4} + 479 N + 851 k\, 
    N 
    \nonu\\ & + &  816 k^{2} N + 448 k^{3} N + 76 k^{4} N + 574 N^{2} + 519 k\, 
    N^{2} + 231 k^{2} N^{2} + 98 k^{3} N^{2} + 423 N^{3}
     \nonu\\ & + &  173 k\, 
    N^{3} - 18 k^{2} N^{3} + 184 N^{4} + 46 k\, N^{4} + 32 N^{5}), \nonu\\
c_ {241} & = & - \frac {1} {3 (2 + N) (2 + k + N)^{5}} 8 (360 + 
     554 k + 264 k^{2} + 40 k^{3} + 994 N + 1214 k\, 
    N + 417 k^{2} N
    \nonu\\ & + &  38 k^{3} N + 1006 N^{2} + 868 k\, 
    N^{2} + 159 k^{2} N^{2} + 439 N^{3} + 202 k\, N^{3} + 69 N^{4}), \nonu\\
c_ {242} & = & \frac {1} {3 (2 + N) (2 + k + N)^{5}} 32 (-96 - 
    385 k - 470 k^{2} - 222 k^{3} - 36 k^{4} - 95 N - 441 k\, 
   N 
   \nonu\\ & - &  423 k^{2} N - 108 k^{3} N + 47 N^{2} - 95 k\, 
   N^{2} - 79 k^{2} N^{2} + 68 N^{3} + 15 k\, N^{3} + 16 N^{4}), \nonu\\
c_ {243} & = & \frac {1} {3 (2 + N) (2 + k + N)^{6}} 16 (180 + 
    475 k + 591 k^{2} + 366 k^{3} + 80 k^{4} + 479 N + 851 k\, 
   N 
   \nonu\\ & + &  816 k^{2} N + 448 k^{3} N + 76 k^{4} N + 574 N^{2} + 519 k\, 
   N^{2} + 231 k^{2} N^{2} + 98 k^{3} N^{2} + 423 N^{3} 
    \nonu\\ & + &  173 k\, 
   N^{3} - 18 k^{2} N^{3} + 184 N^{4} + 46 k\, N^{4} + 32 N^{5}), \nonu\\
c_ {244} & = & - \frac {16 (-k + N) (32 + 55 k + 18 k^{2} + 41 N + 
       35 k\, N + 11 N^{2})} {(2 + N) (2 + k + N)^{5}}, \nonu\\
c_ {245} & = & \frac {1} {3 (2 + N) (2 + k + N)^{5}} 8 (24 - 382 k - 
    730 k^{2} - 244 k^{3} + 178 N - 800 k\, 
   N - 1155 k^{2} N 
    \nonu\\ & - & 230 k^{3} N + 198 N^{2} - 620 k\, 
   N^{2} - 467 k^{2} N^{2} + 63 N^{3} - 160 k\, N^{3} + 5 N^{4}), \nonu\\
c_ {246} & = & - \frac {16 (96 + 232 k + 153 k^{2} + 30 k^{3} + 
       200 N + 296 k\, N + 95 k^{2} N + 127 N^{2} + 90 k\, 
      N^{2} + 25 N^{3})} {(2 + N) (2 + k + N)^{3}}, \nonu\\
c_ {247} & = & \frac {16 (112 + 260 k + 161 k^{2} + 30 k^{3} + 
      212 N + 322 k\, N + 99 k^{2} N + 125 N^{2} + 96 k\, 
     N^{2} + 23 N^{3})} {(2 + N) (2 + k + N)^{4}}, \nonu\\
c_ {248} & = & - \frac {16 (32 + 41 k + 14 k^{2} + 87 N + 77 k\, 
      N + 16 k^{2} N + 69 N^{2} + 32 k\, 
      N^{2} + 16 N^{3})} {(2 + N) (2 + k + N)^{4}}, \nonu\\
c_ {249} & = & - \frac {16 (96 + 223 k + 147 k^{2} + 30 k^{3} + 
       161 N + 251 k\, N + 83 k^{2} N + 82 N^{2} + 66 k\, 
      N^{2} + 13 N^{3})} {(2 + N) (2 + k + N)^{4}}, \nonu\\
c_ {250} & = & - \frac {48 (32 + 55 k + 18 k^{2} + 41 N + 35 k\, 
      N + 11 N^{2})} {(2 + N) (2 + k + N)^{4}}, \nonu\\
c_ {251} & = & \frac {1}{(2 + N) (2 + k + N)^{4}}16 (112 + 276 k + 195 k^{2} + 42 k^{3} + 
      196 N + 310 k\, N + 109 k^{2} N 
       \nonu\\ & + &  103 N^{2} + 80 k\, 
     N^{2} + 17 N^{3}), \nonu\\
c_ {252} & = & \frac {1}{3 (2 + N) (2 + k + N)^{4}}16 (336 + 700 k + 443 k^{2} + 90 k^{3} + 
      716 N + 926 k\, N + 277 k^{2} N 
       \nonu\\ & + &  455 N^{2} + 288 k\, 
     N^{2} + 89 N^{3}), \nonu\\
c_ {253} & = & - \frac {16 (3 + k + 2 N) (24 + 33 k + 10 k^{2} + 
       30 N + 21 k\, N + 8 N^{2})} {(2 + k + N)^{3}}, \nonu\\
c_ {254} & = & - \frac {1} {(2 + N) (2 + k + N)^{4}} 16 (184 + 
     350 k + 195 k^{2} + 34 k^{3} + 518 N + 716 k\, 
    N + 255 k^{2} N 
     \nonu\\ & + &  20 k^{3} N + 525 N^{2} + 486 k\, 
    N^{2} + 86 k^{2} N^{2} + 229 N^{3} + 110 k\, N^{3} + 36 N^{4}), \nonu\\
c_ {255} & = & \frac {1} {3 (2 + N) (2 + k + N)^{5}} 16 (912 + 
    1782 k + 1150 k^{2} + 236 k^{3} + 2526 N + 4267 k\, 
   N 
    \nonu\\ & + &  2319 k^{2} N + 370 k^{3} N + 2593 N^{2} + 3594 k\, 
   N^{2} + 1505 k^{2} N^{2} + 144 k^{3} N^{2} + 1153 N^{3} 
    \nonu\\ & + &  1187 k\, 
   N^{3} + 306 k^{2} N^{3} + 184 N^{4} + 108 k\, N^{4}), \nonu\\
c_ {256} & = & \frac {1} {3 (2 + N) (2 + k + N)^{5}} 16 (1008 + 
    2306 k + 1958 k^{2} + 776 k^{3} + 120 k^{4} + 2914 N 
    \nonu\\ & + &  5029 k\,  N+2955 k^{2} N + 748 k^{3} N + 60 k^{4} N + 3471 N^{2} + 
    4228 k\, 
   N^{2} + 1447 k^{2} N^{2} 
   \nonu\\ & + &  162 k^{3} N^{2} + 2175 N^{3}
+ 1703 k\,N^{3} + 240 k^{2} N^{3} + 716 N^{4} + 288 k\, N^{4} + 96 N^{5}), \nonu\\
c_ {257} & = & \frac {1} {3 (2 + N) (2 + k + N)^{5}} 16 (240 + 
    300 k - 31 k^{2} - 50 k^{3} + 828 N + 791 k\, 
   N + 6 k^{2} N 
    \nonu\\ & - &  34 k^{3} N + 896 N^{2} + 552 k\, 
   N^{2} - 11 k^{2} N^{2} + 380 N^{3} + 109 k\, N^{3} + 56 N^{4}), \nonu\\
c_ {258} & = & - \frac {1} {3 (2 + N) (2 + k + N)^{5}} 16 (-360 - 
     896 k - 689 k^{2} - 158 k^{3} - 652 N - 1459 k\, 
    N 
     \nonu\\ & - &  918 k^{2} N - 142 k^{3} N - 336 N^{2} - 676 k\, 
    N^{2} - 277 k^{2} N^{2} - 12 N^{3} - 65 k\, N^{3} + 16 N^{4}), \nonu\\
c_ {259} & = & \frac {32 (54 + 69 k + 22 k^{2} + 102 N + 138 k\, 
     N + 38 k^{2} N + 74 N^{2} + 63 k\, 
     N^{2} + 16 N^{3})} {3 (2 + N) (2 + k + N)^{3}}, \nonu\\
c_ {260} & = & \frac {4 (120 + 170 k + 65 k^{2} + 6 k^{3} + 226 N + 
      228 k\, N + 47 k^{2} N + 139 N^{2} + 76 k\, 
     N^{2} + 27 N^{3})} {(2 + N) (2 + k + N)^{3}}, \nonu\\
c_ {261} & = & \frac {1} {3 (2 + N) (2 + k + N)^{4}} 4 (96 - 48 k - 
    266 k^{2} - 139 k^{3} - 18 k^{4} - 728 k\, 
   N - 837 k^{2} N 
   \nonu\\ & - & 203 k^{3} N - 110 N^{2} - 669 k\, 
   N^{2} - 367 k^{2} N^{2} - 35 N^{3} - 133 k\, N^{3} + N^{4}), \nonu\\
c_ {262} & = & \frac {1} {3 (2 + N) (2 + k + N)^{4}} 16 (180 + 
    337 k + 202 k^{2} + 40 k^{3} + 437 N + 665 k\, 
   N + 297 k^{2} N
   \nonu\\ & + & 38 k^{3} N + 375 N^{2} + 413 k\, 
   N^{2} + 101 k^{2} N^{2} + 132 N^{3} + 79 k\, N^{3} + 16 N^{4}), \nonu\\
c_ {263} & = & - \frac {1} {3 (2 + N) (2 + k + N)^{4}} 16 (180 + 
     337 k + 202 k^{2} + 40 k^{3} + 437 N + 665 k\, 
    N + 297 k^{2} N 
    \nonu\\ & + & 38 k^{3} N + 375 N^{2} + 413 k\, 
    N^{2} + 101 k^{2} N^{2} + 132 N^{3} + 79 k\, N^{3} + 16 N^{4}), \nonu\\
c_ {264} & = & - \frac {16 (180 + 287 k + 172 k^{2} + 36 k^{3} + 
       307 N + 313 k\, N + 92 k^{2} N + 163 N^{2} + 78 k\, 
      N^{2} + 28 N^{3})} {3 (2 + N) (2 + k + N)^{4}}, \nonu\\
c_ {265} & = & \frac {16 (16 + 65 k + 52 k^{2} + 12 k^{3} + 23 N + 
      69 k\, N + 28 k^{2} N + 7 N^{2} + 16 k\, 
     N^{2})} {(2 + N) (2 + k + N)^{4}}, \nonu\\
c_ {266} & = & \frac {1} {3 (2 + N) (2 + k + N)^{4}} 8 (336 + 720 k + 
    413 k^{2} + 70 k^{3} + 648 N + 884 k\, 
   N + 189 k^{2} N 
   \nonu\\ & - & 28 k^{3} N + 407 N^{2} + 288 k\, 
   N^{2} - 26 k^{2} N^{2} + 101 N^{3} + 22 k\, N^{3} + 8 N^{4}), \nonu\\
c_ {267} & = & - \frac {1} {(2 + N) (2 + k + N)^{4}} 64 (76 + 173 k + 
     137 k^{2} + 47 k^{3} + 6 k^{4} + 173 N + 282 k\, 
    N 
    \nonu\\ & + & 141 k^{2} N + 23 k^{3} N + 147 N^{2} + 154 k\, 
    N^{2} + 36 k^{2} N^{2} + 56 N^{3} + 29 k\, N^{3} + 8 N^{4}), \nonu\\
c_ {268} & = & \frac {1} {3 (2 + N) (2 + k + N)^{5}} 8 (360 + 554 k + 
    264 k^{2} + 40 k^{3} + 994 N + 1214 k\, 
   N + 417 k^{2} N 
   \nonu\\ & + & 38 k^{3} N + 1006 N^{2} + 868 k\, 
   N^{2} + 159 k^{2} N^{2} + 439 N^{3} + 202 k\, N^{3} + 69 N^{4}), \nonu\\
c_ {269} & = & - \frac {1} {3 (2 + N) (2 + k + N)^{5}} 8 (-24 + 
     22 k + 268 k^{2} + 112 k^{3} + 182 N + 536 k\, 
    N + 597 k^{2} N 
    \nonu\\ & + & 110 k^{3} N + 528 N^{2} + 836 k\, 
    N^{2} + 335 k^{2} N^{2} + 411 N^{3} + 316 k\, N^{3} + 91 N^{4}), \nonu\\
c_ {270} & = & - \frac {1} {3 (2 + N) (2 + k + N)^{5}} 32 (96 + 
     257 k + 186 k^{2} + 40 k^{3} + 223 N + 497 k\, 
    N + 273 k^{2} N 
    \nonu\\ & + & 38 k^{3} N + 181 N^{2} + 301 k\, 
    N^{2} + 93 k^{2} N^{2} + 58 N^{3} + 55 k\, N^{3} + 6 N^{4}), \nonu\\
c_ {271} & = & - \frac {1} {3 (2 + N) (2 + k + N)^{5}} 8 (360 + 
     794 k + 604 k^{2} + 183 k^{3} + 18 k^{4} + 754 N + 1326 k\, 
    N 
    \nonu\\ & + & 720 k^{2} N + 117 k^{3} N + 554 N^{2} + 697 k\, 
    N^{2} + 206 k^{2} N^{2} + 164 N^{3} + 111 k\, N^{3} + 16 N^{4}), \nonu\\
c_ {272} & = & - \frac {1} {3 (2 + N) (2 + k + N)^{5}} 32 (-96 - 
     385 k - 470 k^{2} - 222 k^{3} - 36 k^{4} - 95 N - 441 k\, 
    N 
    \nonu\\ & - & 423 k^{2} N - 108 k^{3} N + 47 N^{2} - 95 k\, 
    N^{2} - 79 k^{2} N^{2} + 68 N^{3} + 15 k\, N^{3} + 16 N^{4}), \nonu\\
c_ {273} & = & \frac {1}{(2 + N) (2 + k + N)^{6}}16 (1 + k) (1 + N) (60 + 77 k + 22 k^{2} + 
      121 N + 115 k\, N + 20 k^{2} N 
       \nonu\\ & + & 79 N^{2} + 42 k\, 
     N^{2} + 16 N^{3}), \nonu\\
c_ {274} & = & \frac {1} {3 (2 + N) (2 + k + N)^{5}} 8 (744 + 
    1774 k + 1306 k^{2} + 253 k^{3} - 18 k^{4} + 1118 N + 1808 k\, 
   N 
   \nonu\\ & + & 696 k^{2} N - 25 k^{3} N + 522 N^{2} + 503 k\, 
   N^{2} + 44 k^{2} N^{2} + 120 N^{3} + 67 k\, N^{3} + 16 N^{4}), \nonu\\
c_ {275} & = & \frac{1} {(2 + N) (2 + k + N)^{4}}16 (112 + 276 k + 195 k^{2} + 42 k^{3} + 
      196 N + 310 k\, N + 109 k^{2} N 
       \nonu\\ & + &  103 N^{2} + 80 k\, 
     N^{2} + 17 N^{3}), \nonu\\
c_ {276} & = & - \frac {16 (32 + 41 k + 14 k^{2} + 87 N + 77 k\, 
      N + 16 k^{2} N + 69 N^{2} + 32 k\, 
      N^{2} + 16 N^{3})} {(2 + N) (2 + k + N)^{4}}, \nonu\\
c_ {277} & = & \frac {48 (32 + 55 k + 18 k^{2} + 41 N + 35 k\, 
     N + 11 N^{2})} {(2 + N) (2 + k + N)^{4}}, \nonu\\
c_ {278} & = & \frac {16 (112 + 260 k + 161 k^{2} + 30 k^{3} + 
      212 N + 322 k\, N + 99 k^{2} N + 125 N^{2} + 96 k\, 
     N^{2} + 23 N^{3})} {(2 + N) (2 + k + N)^{4}}, \nonu\\
c_ {279} & = & - \frac {16 (96 + 223 k + 147 k^{2} + 30 k^{3} + 
       161 N + 251 k\, N + 83 k^{2} N + 82 N^{2} + 66 k\, 
      N^{2} + 13 N^{3})} {(2 + N) (2 + k + N)^{4}}, \nonu\\
c_ {280} & = & - \frac {1}{3 (2 + N) (2 + k + N)^{4}}16 (336 + 908 k + 625 k^{2} + 126 k^{3} + 
       508 N + 970 k\, N + 347 k^{2} N 
       \nonu\\ & + &  229 N^{2} + 240 k\, 
      N^{2} + 31 N^{3}), \nonu\\
c_ {281} & = & - \frac {1} {(2 + N) (2 + k + N)^{4}} 32 (28 - 55 k - 
     131 k^{2} - 72 k^{3} - 12 k^{4} + 137 N + 56 k\, 
    N - 63 k^{2} N 
    \nonu\\ & - & 26 k^{3} N + 185 N^{2} + 127 k\, 
    N^{2} + 10 k^{2} N^{2} + 94 N^{3} + 42 k\, N^{3} + 16 N^{4}), \nonu\\
c_ {282} & = & \frac {1} {3 (2 + N) (2 + k + N)^{5}} 16 (528 + 
    1326 k + 995 k^{2} + 226 k^{3} + 1638 N + 3731 k\, 
   N 
    \nonu\\ & + &  2424 k^{2} N+446 k^{3} N + 1844 N^{2} + 3666 k\, 
   N^{2} + 1867 k^{2} N^{2} + 216 k^{3} N^{2} + 890 N^{3} 
   \nonu\\ & + &1453 k\, N^{3} +450 k^{2} N^{3} +152 N^{4} + 180 k\, N^{4}), \nonu\\
c_ {283} & = & \frac {1} {3 (2 + N) (2 + k + N)^{5}} 16 (240 + 
    768 k + 695 k^{2} + 178 k^{3} + 360 N + 797 k\, 
   N + 438 k^{2} N 
   \nonu\\ & + & 26 k^{3} N + 164 N^{2} + 228 k\, 
   N^{2} + 55 k^{2} N^{2} + 44 N^{3} + 31 k\, N^{3} + 8 N^{4}), \nonu\\
c_ {284} & = & \frac {32 (63 + 111 k + 38 k^{2} + 69 N + 63 k\, 
     N + 16 N^{2})} {3 (2 + k + N)^{3}}, \nonu\\
c_ {285} & = & - \frac {1} {3 (2 + N) (2 + k + N)^{4}} 4 (96 - 48 k - 
     266 k^{2} - 139 k^{3} - 18 k^{4} - 728 k\, 
    N - 837 k^{2} N 
    \nonu\\ & - & 203 k^{3} N - 110 N^{2} - 669 k\, 
    N^{2} - 367 k^{2} N^{2} - 35 N^{3} - 133 k\, N^{3} + N^{4}), \nonu\\
c_ {286} & = & \frac {1} {3 (2 + N) (2 + k + N)^{4}} 4 (752 k + 
    1048 k^{2} + 273 k^{3} - 18 k^{4} + 688 N + 2272 k\, 
   N + 1631 k^{2} N 
   \nonu\\ & + & 201 k^{3} N + 1432 N^{2} + 2295 k\, 
   N^{2} + 747 k^{2} N^{2} + 985 N^{3} + 735 k\, N^{3} + 207 N^{4}), \nonu\\
c_ {287} & = & \frac {1} {3 (2 + N) (2 + k + N)^{5}} 16 (-264 - 
    1234 k - 1345 k^{2} - 538 k^{3} - 72 k^{4} + 22 N - 812 k\, 
   N 
   \nonu\\ & - & 765 k^{2} N - 164 k^{3} N + 681 N^{2} + 418 k\, 
   N^{2} + 34 k^{2} N^{2} + 513 N^{3} + 254 k\, N^{3} + 104 N^{4}), \nonu\\
c_ {288} & = & \frac {1} {3 (2 + N) (2 + k + N)^{5}} 16 (468 + 
    802 k + 469 k^{2} + 94 k^{3} + 1124 N + 1142 k\, 
   N + 177 k^{2} N 
   \nonu\\ & - & 52 k^{3} N + 1071 N^{2} + 686 k\, 
   N^{2} + 2 k^{2} N^{2} + 501 N^{3} + 196 k\, N^{3} + 88 N^{4}), \nonu\\
c_ {289} & = & \frac {32 (-k + N) (32 + 55 k + 18 k^{2} + 41 N + 
      35 k\, N + 11 N^{2})} {(2 + N) (2 + k + N)^{5}}, \nonu\\
c_ {290} & = & - \frac {1} {3 (2 + N) (2 + k + N)^{5}} 16 (672 + 
     910 k + 373 k^{2} + 46 k^{3} + 1730 N + 1838 k\, 
    N + 519 k^{2} N 
    \nonu\\ & + & 32 k^{3} N + 1677 N^{2} + 1262 k\, 
    N^{2} + 188 k^{2} N^{2} + 717 N^{3} + 292 k\, N^{3} + 112 N^{4}), \nonu\\
c_ {291} & = & - \frac {1} {3 (2 + N) (2 + k + N)^{5}} 32 (-234 - 
     575 k - 377 k^{2} - 74 k^{3} - 388 N - 841 k\, 
    N - 447 k^{2} N 
    \nonu\\ & - & 64 k^{3} N - 123 N^{2} - 277 k\, 
    N^{2} - 100 k^{2} N^{2} + 69 N^{3} + 19 k\, N^{3} + 28 N^{4}), \nonu\\
c_ {292} & = & \frac {1} {3 (2 + N) (2 + k + N)^{5}} 16 (96 - 972 k - 
    1653 k^{2} - 874 k^{3} - 152 k^{4} + 156 N - 2340 k\, 
   N 
   \nonu\\ & - & 2993 k^{2} N - 1112 k^{3} N - 112 k^{4} N + 393 N^{2} - 
    1544 k\, 
   N^{2} - 1530 k^{2} N^{2} - 312 k^{3} N^{2} 
   \nonu\\ & + & 467 N^{3} -250 k\,N^{3} - 208 k^{2} N^{3} + 212 N^{4} + 24 k\, N^{4} + 32 N^{5}), \nonu\\
c_ {293} & = & - \frac {1} {3 (2 + N) (2 + k + N)^{5}} 32 (492 + 
     1057 k + 652 k^{2} + 124 k^{3} + 1097 N + 1688 k\, 
    N 
    \nonu\\ & + & 636 k^{2} N + 44 k^{3} N + 882 N^{2} + 881 k\, 
    N^{2} + 152 k^{2} N^{2} + 309 N^{3} + 154 k\, N^{3} + 40 N^{4}), \nonu\\
c_ {294} & = & \frac {1} {3 (2 + N) (2 + k + N)^{5}} 32 (384 + 
    851 k + 692 k^{2} + 254 k^{3} + 36 k^{4} + 925 N + 1519 k\, 
   N 
   \nonu\\ & + &
   801 k^{2} N + 142 k^{3} N + 813 N^{2} + 889 k\, 
   N^{2} + 229 k^{2} N^{2} + 312 N^{3} + 173 k\, N^{3} + 44 N^{4}), \nonu\\
c_ {295} & = & - \frac {48 (60 + 77 k + 22 k^{2} + 121 N + 115 k\, 
      N + 20 k^{2} N + 79 N^{2} + 42 k\, 
      N^{2} + 16 N^{3})} {(2 + N) (2 + k + N)^{3}}, \nonu\\
c_ {296} & = & \frac {1} {3 (2 + N) (2 + k + N)^{4}} 16 (96 - 65 k - 
    214 k^{2} - 72 k^{3} + 449 N + 47 k\, 
   N - 341 k^{2} N 
   \nonu\\ & - & 90 k^{3} N + 647 N^{2} + 231 k\, 
   N^{2} - 99 k^{2} N^{2} + 374 N^{3} + 117 k\, N^{3} + 72 N^{4}), \nonu\\
c_ {297} & = & - \frac {1} {3 (2 + N) (2 + k + N)^{4}} 16 (96 - 
     49 k - 206 k^{2} - 72 k^{3} + 433 N + 55 k\, 
    N - 337 k^{2} N 
    \nonu\\ & - & 90 k^{3} N + 631 N^{2} + 231 k\, 
    N^{2} - 99 k^{2} N^{2} + 370 N^{3} + 117 k\, N^{3} + 72 N^{4}), \nonu\\
c_ {298} & = & - \frac {48 (32 + 55 k + 18 k^{2} + 41 N + 35 k\, 
      N + 11 N^{2})} {(2 + N) (2 + k + N)^{4}}, \nonu\\
c_ {299} & = & \frac {48 (60 + 77 k + 22 k^{2} + 121 N + 115 k\, 
     N + 20 k^{2} N + 79 N^{2} + 42 k\, 
     N^{2} + 16 N^{3})} {(2 + N) (2 + k + N)^{4}}, \nonu\\
c_ {300} & = & \frac {1} {3 (2 + N) (2 + k + N)^{4}} 8 (-540 - 
    253 k + 283 k^{2} + 126 k^{3} - 629 N + 514 k\, 
   N + 797 k^{2} N 
   \nonu\\ & + & 144 k^{3} N + 229 N^{2} + 1155 k\, 
   N^{2} + 456 k^{2} N^{2} + 460 N^{3} + 450 k\, N^{3} + 120 N^{4}), \nonu\\
c_ {301} & = & - \frac {1} {3 (2 + N) (2 + k + N)^{4}} 8 (1248 + 
     1530 k + 591 k^{2} + 74 k^{3} + 2850 N + 2916 k\, 
    N 
    \nonu\\ & + &  1075 k^{2} N+172 k^{3} N + 2283 N^{2} + 1588 k\, 
    N^{2} + 338 k^{2} N^{2} + 695 N^{3} + 200 k\, N^{3} + 64 N^{4}), \nonu\\
c_ {302} & = & - \frac {1} {3 (2 + N) (2 + k + N)^{3}} 32 (-108 k - 
     133 k^{2} - 38 k^{3} + 12 N - 196 k\, 
    N - 177 k^{2} N
     \nonu\\ & - &  28 k^{3} N+ 41 N^{2} - 84 k\, 
    N^{2} - 50 k^{2} N^{2} + 35 N^{3} - 2 k\, N^{3} + 8 N^{4}), \nonu\\
c_ {303} & = & - \frac {32 (k - N)} {3 (2 + k + N)},  \nonu\\
c_ {304} & = & - \frac {32 (-k + N) (60 + 77 k + 22 k^{2} + 121 N + 
       115 k\, N + 20 k^{2} N + 79 N^{2} + 42 k\, 
      N^{2} + 16 N^{3})} {3 (2 + N) (2 + k + N)^{3}}, \nonu\\
c_ {305} & = & \frac {1} {9 (2 + N) (2 + k + N)^{4}} 32 (72 + 84 k - 
    25 k^{2} - 22 k^{3} + 168 N + 140 k\, 
   N - 51 k^{2} N 
   \nonu\\ & - & 20 k^{3} N +209 N^{2} + 132 k\, 
   N^{2} - 14 k^{2} N^{2} + 121 N^{3} + 46 k\, N^{3} + 24 N^{4}), \nonu\\
c_ {306} & = & \frac {64 (k - N) (3 + 6 k + 2 k^{2} + 6 N + 5 k\, 
     N + 2 N^{2})} {9 (2 + k + N)^{4}}, \nonu\\
c_ {307} & = & - \frac {1} {9 (2 + N) (2 + k + N)^{4}} 16 (84 k + 
     91 k^{2} + 26 k^{3} + 204 N + 268 k\, 
    N + 87 k^{2} N + 4 k^{3} N 
    \nonu\\ & + & 361 N^{2} + 252 k\, 
    N^{2} + 26 k^{2} N^{2} + 211 N^{3} + 74 k\, N^{3} + 40 N^{4}), \nonu\\
c_ {308} & = & \frac {64 (36 - 6 k - 47 k^{2} - 18 k^{3} + 96 N + 
      37 k\, N - 13 k^{2} N + 82 N^{2} + 29 k\, 
     N^{2} + 20 N^{3})} {9 (2 + N) (2 + k + N)^{4}}, \nonu\\
c_ {309} & = & \frac {32 (-k + N) (32 + 55 k + 18 k^{2} + 41 N + 
      35 k\, N + 11 N^{2})} {3 (2 + N) (2 + k + N)^{5}}, \nonu\\
c_ {310} & = & \frac {32 (-k + N) (32 + 55 k + 18 k^{2} + 41 N + 
      35 k\, N + 11 N^{2})} {3 (2 + N) (2 + k + N)^{5}}, \nonu\\
c_ {311} & = & - \frac {64 (18 + 9 k + 2 k^{2} + 27 N + 5 k\, 
      N + 11 N^{2})} {9 (2 + k + N)^{4}}, \nonu\\
c_ {312} & = & - \frac {32 (-k + N) (32 + 35 k + 10 k^{2} + 61 N + 
       47 k\, N + 8 k^{2} N + 39 N^{2} + 16 k\, 
      N^{2} + 8 N^{3})} {3 (2 + N) (2 + k + N)^{5}}, \nonu\\
c_ {313} & = & - \frac {64 (18 + 9 k + 2 k^{2} + 27 N + 5 k\, 
      N + 11 N^{2})} {9 (2 + k + N)^{4}}, \nonu\\
c_ {314} & = & \frac {32 (-k + N) (32 + 35 k + 10 k^{2} + 61 N + 
      47 k\, N + 8 k^{2} N + 39 N^{2} + 16 k\, 
     N^{2} + 8 N^{3})} {3 (2 + N) (2 + k + N)^{5}}, \nonu\\
c_ {315} & = & \frac {64 (36 - 6 k - 47 k^{2} - 18 k^{3} + 96 N + 
      37 k\, N - 13 k^{2} N + 82 N^{2} + 29 k\, 
     N^{2} + 20 N^{3})} {9 (2 + N) (2 + k + N)^{4}}, \nonu\\
c_ {316} & = & \frac {32 (-k + N) (32 + 35 k + 10 k^{2} + 61 N + 
      47 k\, N + 8 k^{2} N + 39 N^{2} + 16 k\, 
     N^{2} + 8 N^{3})} {3 (2 + N) (2 + k + N)^{5}}, \nonu\\
c_ {317} & = & \frac {32 (-k + N) (32 + 55 k + 18 k^{2} + 41 N + 
      35 k\, N + 11 N^{2})} {3 (2 + N) (2 + k + N)^{5}}, \nonu\\
c_ {318} & = & \frac {1} {9 (2 + N) (2 + k + N)^{4}} 16 (288 + 
    228 k + 91 k^{2} + 10 k^{3} + 636 N + 268 k\, 
   N + 31 k^{2} N 
   \nonu\\ & - & 4 k^{3} N + 577 N^{2} + 160 k\, 
   N^{2} - 2 k^{2} N^{2} + 231 N^{3} + 46 k\, N^{3} + 32 N^{4}), \nonu\\
c_ {319} & = & - \frac {64 (18 + 9 k + 2 k^{2} + 27 N + 5 k\, 
      N + 11 N^{2})} {9 (2 + k + N)^{4}}, \nonu\\
c_ {320} & = & \frac {64 (36 - 6 k - 47 k^{2} - 18 k^{3} + 96 N + 
      37 k\, N - 13 k^{2} N + 82 N^{2} + 29 k\, 
     N^{2} + 20 N^{3})} {9 (2 + N) (2 + k + N)^{4}}, \nonu\\
c_ {321} & = & \frac {64 (36 - 6 k - 47 k^{2} - 18 k^{3} + 96 N + 
      37 k\, N - 13 k^{2} N + 82 N^{2} + 29 k\, 
     N^{2} + 20 N^{3})} {9 (2 + N) (2 + k + N)^{4}}, \nonu\\
c_ {322} & = & - \frac {64 (18 + 9 k + 2 k^{2} + 27 N + 5 k\, 
      N + 11 N^{2})} {9 (2 + k + N)^{4}}, \nonu\\
c_ {323} & = & \frac {32 (-k + N) (24 + 31 k + 10 k^{2} + 53 N + 
      45 k\, N + 8 k^{2} N + 37 N^{2} + 16 k\, 
     N^{2} + 8 N^{3})} {3 (2 + N) (2 + k + N)^{4}}, \nonu\\
c_ {324} & = & - \frac {32 (-k + N) (24 + 31 k + 10 k^{2} + 53 N + 
       45 k\, N + 8 k^{2} N + 37 N^{2} + 16 k\, 
      N^{2} + 8 N^{3})} {3 (2 + N) (2 + k + N)^{4}}, \nonu\\
c_ {325} & = & - \frac {32 (-k + N) (8 + 17 k + 6 k^{2} + 11 N + 
       11 k\, N + 3 N^{2})} {3 (2 + N) (2 + k + N)^{4}}, \nonu\\
c_ {326} & = & \frac {32 (48 + 28 k - 9 k^{2} - 6 k^{3} + 92 N + 
      44 k\, N - k^{2} N + 61 N^{2} + 18 k\, 
     N^{2} + 13 N^{3})} {3 (2 + N) (2 + k + N)^{4}}, \nonu\\
c_ {327} & = & - \frac {32 (-k + N) (8 + 17 k + 6 k^{2} + 11 N + 
       11 k\, N + 3 N^{2})} {3 (2 + N) (2 + k + N)^{4}}, \nonu\\
c_ {328} & = & \frac {32 (48 + 28 k - 9 k^{2} - 6 k^{3} + 92 N + 
      44 k\, N - k^{2} N + 61 N^{2} + 18 k\, 
     N^{2} + 13 N^{3})} {3 (2 + N) (2 + k + N)^{4}}, \nonu\\
c_ {329} & = & \frac {64 (-36 - 42 k - 7 k^{2} + 2 k^{3} - 66 N - 
      58 k\, N - 7 k^{2} N - 43 N^{2} - 22 k\, 
     N^{2} - 9 N^{3})} {9 (2 + k + N)^{4}}, \nonu\\
c_ {330} & = & \frac {1} {9 (2 + N) (2 + k + N)^{5}} 32 (144 + 
    180 k + 29 k^{2} - 30 k^{3} - 8 k^{4} + 612 N + 704 k\, 
   N \nonu\\ & + & 181 k^{2} N 
  -18 k^{3} N - 4 k^{4} N + 923 N^{2} + 844 k\, 
   N^{2} + 166 k^{2} N^{2} - 6 k^{3} N^{2} + 661 N^{3} 
   \nonu\\ & + &  420 k\, N^{3}+44 k^{2} N^{3}+232 N^{4} + 78 k\, N^{4} + 32 N^{5}), \nonu\\ 
c_ {331} & = & \frac {32 (-k + N) (60 + 77 k + 22 k^{2} + 121 N + 
      115 k\, N + 20 k^{2} N + 79 N^{2} + 42 k\, 
     N^{2} + 16 N^{3})} {9 (2 + N) (2 + k + N)^{5}}, \nonu\\
c_ {332} & = & \frac {32 (-k + N) (60 + 77 k + 22 k^{2} + 121 N + 
      115 k\, N + 20 k^{2} N + 79 N^{2} + 42 k\, 
     N^{2} + 16 N^{3})} {9 (2 + N) (2 + k + N)^{5}}, \nonu\\
c_ {333} & = & - \frac {128 (k - N)} {3 (2 + k + N)^{3}}, \nonu\\
c_ {334} & = & \frac {64 (-36 - 42 k - 7 k^{2} + 2 k^{3} - 66 N - 
      58 k\, N - 7 k^{2} N - 43 N^{2} - 22 k\, 
     N^{2} - 9 N^{3})} {9 (2 + k + N)^{4}}, \nonu\\
c_ {335} & = & \frac {1} {3 (2 + N) (2 + k + N)^{5}} 32 (48 + 100 k + 
    65 k^{2} + 14 k^{3} + 164 N + 296 k\, 
   N + 161 k^{2} N 
   \nonu\\ & + & 28 k^{3} N + 191 N^{2} + 284 k\, 
   N^{2} + 114 k^{2} N^{2} 
   \nonu\\ & + & 12 k^{3} N^{2} + 93 N^{3} + 106 k\, 
   N^{3} + 24 k^{2} N^{3} + 16 N^{4} + 12 k\, N^{4}), \nonu\\
c_ {336} & = & \frac {1} {9 (2 + N) (2 + k + N)^{5}} 32 (144 + 
    180 k + 29 k^{2} - 30 k^{3} - 8 k^{4} + 612 N + 704 k\, 
   N  \nonu\\ & + & 181 k^{2} N 
   -18 k^{3} N - 4 k^{4} N + 923 N^{2} + 844 k\, 
   N^{2} + 166 k^{2} N^{2} - 6 k^{3} N^{2} + 661 N^{3}
    \nonu\\ & + &  420 k\, N^{3} +44 k^{2} N^{3} 
   + 232 N^{4} + 78 k\, N^{4} + 32 N^{5}), \nonu\\
c_ {337} & = & \frac {32 (-k + N) (60 + 77 k + 22 k^{2} + 121 N + 
      115 k\, N + 20 k^{2} N + 79 N^{2} + 42 k\, 
     N^{2} + 16 N^{3})} {9 (2 + N) (2 + k + N)^{5}}, \nonu\\
c_ {338} & = & - \frac {32 (-k + N) (32 + 35 k + 10 k^{2} + 61 N + 
       47 k\, N + 8 k^{2} N + 39 N^{2} + 16 k\, 
      N^{2} + 8 N^{3})} {3 (2 + N) (2 + k + N)^{5}}, \nonu\\
c_ {339} & = & - \frac {128 (k - N)} {3 (2 + k + N)^{3}}, \qquad
c_ {340}  =  \frac {64 (k - N) (3 + 6 k + 2 k^{2} + 6 N + 5 k\, 
     N + 2 N^{2})} {9 (2 + k + N)^{4}}, \nonu\\
c_ {341} & = & \frac {64 (k - N) (3 + 6 k + 2 k^{2} + 6 N + 5 k\, 
     N + 2 N^{2})} {9 (2 + k + N)^{4}}, \nonu\\
c_ {342} & = & - \frac {1} {9 (2 + N) (2 + k + N)^{4}} 64 (72 + 
     12 k - 112 k^{2} - 87 k^{3} - 18 k^{4} + 240 N + 182 k\, 
    N 
    \nonu\\ & - & 30 k^{2} N - 33 k^{3} N + 254 N^{2} + 189 k\, 
    N^{2} + 22 k^{2} N^{2} + 108 N^{3} + 49 k\, N^{3} + 16 N^{4}),\nonu\\ 
c_ {343} & = & - \frac {32 (-k + N) (60 + 77 k + 22 k^{2} + 121 N + 
       115 k\, N + 20 k^{2} N + 79 N^{2} + 42 k\, 
      N^{2} + 16 N^{3})} {9 (2 + N) (2 + k + N)^{5}}, \nonu\\
c_ {344} & = & - \frac {32 (-k + N) (60 + 77 k + 22 k^{2} + 121 N + 
       115 k\, N + 20 k^{2} N + 79 N^{2} + 42 k\, 
      N^{2} + 16 N^{3})} {9 (2 + N) (2 + k + N)^{5}}, \nonu\\
c_ {345} & = & \frac {128 (k - N)} {3 (2 + k + N)^{3}}, \nonu\\
c_ {346} & = & - \frac {1} {9 (2 + N) (2 + k + N)^{4}} 64 (72 + 
     12 k - 112 k^{2} - 87 k^{3} - 18 k^{4} + 240 N + 182 k\, 
    N 
    \nonu\\ & - & 30 k^{2} N - 33 k^{3} N + 254 N^{2} + 189 k\, 
    N^{2} + 22 k^{2} N^{2} + 108 N^{3} + 49 k\, N^{3} + 16 N^{4}),\nonu\\ 
c_ {347} & = & \frac {1} {3 (2 + N) (2 + k + N)^{5}} 32 (48 + 100 k + 
    65 k^{2} + 14 k^{3} + 164 N + 296 k\, 
   N + 161 k^{2} N 
   \nonu\\ & + & 28 k^{3} N + 191 N^{2} + 284 k\, 
   N^{2} + 114 k^{2} N^{2} + 12 k^{3} N^{2} + 93 N^{3} + 106 k\, 
   N^{3} + 24 k^{2} N^{3} 
   \nonu\\ & + & 16 N^{4} 
   + 12 k\, N^{4}), \nonu\\
c_ {348} & = & - \frac {32 (-k + N) (60 + 77 k + 22 k^{2} + 121 N + 
       115 k\, N + 20 k^{2} N + 79 N^{2} + 42 k\, 
      N^{2} + 16 N^{3})} {9 (2 + N) (2 + k + N)^{5}}, \nonu\\
c_ {349} & = & \frac {128 (k - N)} {3 (2 + k + N)^{3}}, \qquad
c_ {350}  =  \frac {64 (k - N) (3 + 6 k + 2 k^{2} + 6 N + 5 k\, 
     N + 2 N^{2})} {9 (2 + k + N)^{4}}, \nonu\\
c_ {351} & = & \frac {64 (36 - 6 k - 47 k^{2} - 18 k^{3} + 96 N + 
      37 k\, N - 13 k^{2} N + 82 N^{2} + 29 k\, 
     N^{2} + 20 N^{3})} {9 (2 + N) (2 + k + N)^{4}}, \nonu\\
c_ {352} & = & \frac {32 (-k + N) (32 + 35 k + 10 k^{2} + 61 N + 
      47 k\, N + 8 k^{2} N + 39 N^{2} + 16 k\, 
     N^{2} + 8 N^{3})} {3 (2 + N) (2 + k + N)^{5}}, \nonu\\
c_ {353} & = & \frac {32 (-k + N) (32 + 55 k + 18 k^{2} + 41 N + 
      35 k\, N + 11 N^{2})} {3 (2 + N) (2 + k + N)^{5}}, \nonu\\
c_ {354} & = & - \frac {64 (18 + 9 k + 2 k^{2} + 27 N + 5 k\, 
      N + 11 N^{2})} {9 (2 + k + N)^{4}}, \nonu\\
c_ {355} & = & - \frac {32 (-k + N) (32 + 35 k + 10 k^{2} + 61 N + 
       47 k\, N + 8 k^{2} N + 39 N^{2} + 16 k\, 
      N^{2} + 8 N^{3})} {3 (2 + N) (2 + k + N)^{5}}, \nonu\\
c_ {356} & = & - \frac {64 (-k + N) (8 + 17 k + 6 k^{2} + 11 N + 
       11 k\, N + 3 N^{2})} {3 (2 + N) (2 + k + N)^{4}}, \nonu\\
c_ {357} & = & - \frac {64 (18 + 9 k + 2 k^{2} + 27 N + 5 k\, 
      N + 11 N^{2})} {9 (2 + k + N)^{4}}, \nonu\\
c_ {358} & = & \frac {64 (36 - 6 k - 47 k^{2} - 18 k^{3} + 96 N + 
      37 k\, N - 13 k^{2} N + 82 N^{2} + 29 k\, 
     N^{2} + 20 N^{3})} {9 (2 + N) (2 + k + N)^{4}}, \nonu\\
c_ {359} & = & - \frac {32 (48 + 52 k + 3 k^{2} - 6 k^{3} + 116 N + 
       80 k\, N + 5 k^{2} N + 85 N^{2} + 30 k\, 
      N^{2} + 19 N^{3})} {3 (2 + N) (2 + k + N)^{3}}, \nonu\\
c_ {360} & = & \frac {32 (-k + N) (24 + 31 k + 10 k^{2} + 53 N + 
      45 k\, N + 8 k^{2} N + 37 N^{2} + 16 k\, 
     N^{2} + 8 N^{3})} {3 (2 + N) (2 + k + N)^{4}}, \nonu\\
c_ {361} & = & - \frac {32 (-k + N) (24 + 31 k + 10 k^{2} + 53 N + 
       45 k\, N + 8 k^{2} N + 37 N^{2} + 16 k\, 
      N^{2} + 8 N^{3})} {3 (2 + N) (2 + k + N)^{4}}, \nonu\\
c_ {362} & = & - \frac {32 (-k + N) (8 + 17 k + 6 k^{2} + 11 N + 
       11 k\, N + 3 N^{2})} {3 (2 + N) (2 + k + N)^{4}}, \nonu\\
c_ {363} & = & \frac {32 (48 + 28 k - 9 k^{2} - 6 k^{3} + 92 N + 
      44 k\, N - k^{2} N + 61 N^{2} + 18 k\, 
     N^{2} + 13 N^{3})} {3 (2 + N) (2 + k + N)^{4}}, \nonu\\
c_ {364} & = & - \frac{1}{9 (2 + N) (2 + k + N)^{4}}32 (72 + 12 k - 25 k^{2} - 6 k^{3} + 240 N + 
       104 k\, N - 27 k^{2} N 
       \nonu\\ & - & 12 k^{3} N + 245 N^{2} + 108 k\, 
      N^{2} - 2 k^{2} N^{2} + 105 N^{3} + 34 k\, 
      N^{3} + 16 N^{4}), \nonu\\
c_ {365} & = & \frac {1}{9 (2 + N) (2 + k + N)^{4}}16 (84 k + 91 k^{2} + 26 k^{3} + 204 N + 
      268 k\, N + 87 k^{2} N + 4 k^{3} N 
      \nonu\\ & + & 361 N^{2} + 252 k\, 
     N^{2} + 26 k^{2} N^{2} + 211 N^{3} + 74 k\, 
     N^{3} + 40 N^{4}), \nonu\\
c_ {366} & = & - \frac {1}{9 (2 + N) (2 + k + N)^{4}}16 (288 + 132 k - 113 k^{2} - 62 k^{3} + 
       732 N + 340 k\, 
      N - 29 k^{2} N 
      \nonu\\ & - & 4 k^{3} N + 709 N^{2} + 256 k\, 
      N^{2} - 2 k^{2} N^{2} + 267 N^{3} + 46 k\, 
      N^{3} + 32 N^{4}), \nonu\\
c_ {367} & = & \frac {32 (-36 - 42 k - 7 k^{2} + 2 k^{3} - 102 N - 
      97 k\, N - 17 k^{2} N - 76 N^{2} - 41 k\, 
     N^{2} - 16 N^{3})} {9 (2 + k + N)^{3}}, \nonu\\
c_ {368} & = & \frac {32 (-k + N) (16 + 21 k + 6 k^{2} + 19 N + 
      13 k\, N + 5 N^{2})} {3 (2 + N) (2 + k + N)^{3}}, \nonu\\
c_ {369} & = & \frac {18} {5}, \qquad
c_ {370}  =  - \frac {32 (k - N)} {5 (2 + k + N)}, \nonu\\
c_ {371} & = & - \frac {2 (60 + 77 k + 22 k^{2} + 121 N + 115 k\, 
      N + 20 k^{2} N + 79 N^{2} + 42 k\, 
      N^{2} + 16 N^{3})} {(2 + N) (2 + k + N)^{2}}, \nonu\\
c_ {372} & = & - \frac {2 (60 + 77 k + 22 k^{2} + 121 N + 115 k\, 
      N + 20 k^{2} N + 79 N^{2} + 42 k\, 
      N^{2} + 16 N^{3})} {(2 + N) (2 + k + N)^{2}}, \nonu\\
c_ {373} & = & - \frac {1} {3 (2 + N) (2 + k + N)^{3}}16 (-k + N) (60 + 77 k + 22 k^{2} + 121 N + 
       115 k\, N + 20 k^{2} N 
       \nonu\\ & + & 79 N^{2} + 42 k\, 
      N^{2} + 16 N^{3}), \nonu\\
c_ {374} & = & - \frac {8 (-k + N) (60 + 77 k + 22 k^{2} + 121 N + 
       115 k\, N + 20 k^{2} N + 79 N^{2} + 42 k\, 
      N^{2} + 16 N^{3})} {(2 + N) (2 + k + N)^{3}}, \nonu\\
c_ {375} & = & \frac {8 (-k + N) (60 + 77 k + 22 k^{2} + 121 N + 
      115 k\, N + 20 k^{2} N + 79 N^{2} + 42 k\, 
     N^{2} + 16 N^{3})} {3 (2 + N) (2 + k + N)^{3}}, \nonu\\
c_ {376} & = & - \frac {4 (60 + 77 k + 22 k^{2} + 121 N + 115 k\, 
      N + 20 k^{2} N + 79 N^{2} + 42 k\, 
      N^{2} + 16 N^{3})} {(2 + N) (2 + k + N)^{2}}, \nonu\\
c_ {377} & = & \frac {4 (60 + 77 k + 22 k^{2} + 121 N + 115 k\, 
     N + 20 k^{2} N + 79 N^{2} + 42 k\, 
     N^{2} + 16 N^{3})} {(2 + N) (2 + k + N)^{2}}, \nonu\\
c_ {378} & = & \frac{1} {3 (2 + N) (2 + k + N)^{4}}8 (528 + 846 k + 313 k^{2} + 22 k^{3} + 
      1134 N + 1396 k\, 
     N + 283 k^{2} N
     \nonu\\ & - &  16 k^{3} N + 853 N^{2} + 718 k\, 
     N^{2} + 52 k^{2} N^{2} + 273 N^{3} + 118 k\, 
     N^{3} + 32 N^{4}), \nonu\\
c_ {379} & = & - \frac {1} {3 (2 + N) (2 + k + N)^{5}}16 (156 + 391 k + 333 k^{2} + 125 k^{3} + 
       18 k^{4} + 299 N + 508 k\, 
      N 
      \nonu\\ & + &  255 k^{2} N + 43 k^{3} N + 197 N^{2} + 194 k\, 
      N^{2} + 36 k^{2} N^{2} + 56 N^{3} + 23 k\, 
      N^{3} + 6 N^{4}), \nonu\\
c_ {380} & = & \frac{1}{3 (2 + N) (2 + k + N)^{5}}16 (60 + 231 k + 245 k^{2} + 109 k^{3} + 
      18 k^{4} + 123 N + 284 k\, 
     N 
     \nonu\\ & + &  171 k^{2} N + 35 k^{3} N + 77 N^{2} + 90 k\, 
     N^{2} + 16 k^{2} N^{2} + 20 N^{3} + 7 k\, 
     N^{3} + 2 N^{4}), \nonu\\
c_ {381} & = & \frac {16 (1 + k) (32 + 55 k + 18 k^{2} + 41 N + 
      35 k\, N + 11 N^{2})} {(2 + N) (2 + k + N)^{5}}, \nonu\\
c_ {382} & = & - \frac {16 (1 + k) (32 + 55 k + 18 k^{2} + 41 N + 
       35 k\, N + 11 N^{2})} {(2 + N) (2 + k + N)^{5}}, \nonu\\
c_ {383} & = & \frac {16 (1 + N) (32 + 55 k + 18 k^{2} + 41 N + 
      35 k\, N + 11 N^{2})} {(2 + N) (2 + k + N)^{5}}, \nonu\\
c_ {384} & = & - \frac {16 (1 + N) (32 + 55 k + 18 k^{2} + 41 N + 
       35 k\, N + 11 N^{2})} {(2 + N) (2 + k + N)^{5}}, \nonu\\
c_ {385} & = & - \frac {1}{3 (2 + N) (2 + k + N)^{5}}16 (72 + 194 k + 108 k^{2} + 16 k^{3} + 
       346 N + 626 k\, 
      N + 255 k^{2} N 
      \nonu\\ & + & 26 k^{3} N + 454 N^{2} + 544 k\, 
      N^{2} + 117 k^{2} N^{2} + 229 N^{3} + 142 k\, 
      N^{3} + 39 N^{4}), \nonu\\
c_ {386} & = & \frac {1} {(2 + N) (2 + k + N)^{5}}32 (-6 - 16 k - k^{2} + 2 k^{3} - 5 N - 9 k\, 
     N + 13 k^{2} N + 4 k^{3} N - N^{2} 
     \nonu\\ & + &  2 k\, 
     N^{2} + 8 k^{2} N^{2} + k\, N^{3}),\nonu\\ 
c_ {387} & = & - \frac {8 (1 + k) (32 + 55 k + 18 k^{2} + 41 N + 
       35 k\, N + 11 N^{2})} {(2 + N) (2 + k + N)^{5}}, \nonu\\
c_ {388} & = & \frac {8 (1 + k) (32 + 55 k + 18 k^{2} + 41 N + 35 k\, 
     N + 11 N^{2})} {(2 + N) (2 + k + N)^{5}}, \nonu\\
c_ {389} & = & \frac {16 (1 + N) (32 + 55 k + 18 k^{2} + 41 N + 
      35 k\, N + 11 N^{2})} {(2 + N) (2 + k + N)^{5}}, \nonu\\
c_ {390} & = & - \frac {16 (1 + k) (1 + N) (32 + 55 k + 18 k^{2} + 
       41 N + 35 k\, N + 11 N^{2})} {(2 + N) (2 + k + N)^{6}}, \nonu\\
c_ {391} & = & \frac {8 (1 + N) (32 + 55 k + 18 k^{2} + 41 N + 35 k\, 
     N + 11 N^{2})} {(2 + N) (2 + k + N)^{5}}, \nonu\\
c_ {392} & = & \frac {16 (1 + k) (32 + 55 k + 18 k^{2} + 41 N + 
      35 k\, N + 11 N^{2})} {(2 + N) (2 + k + N)^{5}}, \nonu\\
c_ {393} & = & - \frac {16 (2 + 2 k + k^{2} + 2 N + N^{2}) (32 + 
       55 k + 18 k^{2} + 41 N + 35 k\, 
      N + 11 N^{2})} {(2 + N) (2 + k + N)^{6}}, \nonu\\
c_ {394} & = & - \frac {8 (1 + N) (32 + 55 k + 18 k^{2} + 41 N + 
       35 k\, N + 11 N^{2})} {(2 + N) (2 + k + N)^{5}}, \nonu\\
c_ {395} & = & - \frac{1}{3 (2 + N) (2 + k + N)^{5}}16 (528 + 996 k + 559 k^{2} + 98 k^{3} + 
       1236 N + 1708 k\, 
      N + 597 k^{2} N 
      \nonu\\ & + &  40 k^{3} N + 1069 N^{2} + 984 k\, 
      N^{2} + 164 k^{2} N^{2} + 409 N^{3} + 194 k\, 
      N^{3} + 58 N^{4}), \nonu\\
c_ {396} & = & - \frac {8 (16 + 21 k + 6 k^{2} + 19 N + 13 k\, 
      N + 5 N^{2})} {(2 + N) (2 + k + N)^{3}}, \nonu\\
c_ {397} & = & - \frac {1}{3 (2 + N) (2 + k + N)^{5}}8 (168 + 198 k - 32 k^{2} - 79 k^{3} - 
       18 k^{4} + 486 N + 550 k\, 
      N 
      \nonu\\ & + &  72 k^{2} N - 29 k^{3} N + 526 N^{2} + 471 k\, 
      N^{2} + 62 k^{2} N^{2} + 244 N^{3} + 125 k\, 
      N^{3} + 40 N^{4}), \nonu\\
c_ {398} & = & - \frac{1}{3 (2 + N) (2 + k + N)^{5}}8 (432 + 868 k + 503 k^{2} + 90 k^{3} + 
       1028 N + 1500 k\, 
      N + 537 k^{2} N 
      \nonu\\ & + &  36 k^{3} N + 901 N^{2} + 872 k\, 
      N^{2} + 148 k^{2} N^{2} + 349 N^{3} + 174 k\, 
      N^{3} + 50 N^{4}), \nonu\\
c_ {399} & = & \frac{1} {3 (2 + N) (2 + k + N)^{5}}32 (204 + 471 k + 377 k^{2} + 133 k^{3} + 
      18 k^{4} + 387 N + 620 k\, 
     N 
     \nonu\\ & + & 297 k^{2} N + 47 k^{3} N + 257 N^{2} + 246 k\, 
     N^{2} + 46 k^{2} N^{2} + 74 N^{3} + 31 k\, 
     N^{3} + 8 N^{4}), \nonu\\
c_ {400} & = & - \frac {8 (1 + N) (32 + 55 k + 18 k^{2} + 41 N + 
       35 k\, N + 11 N^{2})} {(2 + N) (2 + k + N)^{5}},
\nonu\\
c_ {401} & = &\frac {8 (3 + 2 k + N) (32 + 55 k + 18 k^{2} + 41 N + 
      35 k\, N + 11 N^{2})} {(2 + N) (2 + k + N)^{5}}, \nonu\\
c_ {402} & = &\frac {16 (1 + k) (32 + 55 k + 18 k^{2} + 41 N + 35 k\, 
     N + 11 N^{2})} {(2 + N) (2 + k + N)^{5}}, \nonu\\
c_ {403} & = &\frac {8 (1 + k) (32 + 55 k + 18 k^{2} + 41 N + 35 k\, 
     N + 11 N^{2})} {(2 + N) (2 + k + N)^{5}}, \nonu\\
c_ {404} & = &\frac {16 (1 + N) (32 + 55 k + 18 k^{2} + 41 N + 35 k\, 
     N + 11 N^{2})} {(2 + N) (2 + k + N)^{5}}, \nonu\\
c_ {405} & = & - \frac {16 (1 + k) (1 + N) (32 + 55 k + 18 k^{2} + 
       41 N + 35 k\, N + 11 N^{2})} {(2 + N) (2 + k + N)^{6}}, \nonu\\
c_ {406} & = & - \frac {8 (3 + k + 2 N) (32 + 55 k + 18 k^{2} + 
       41 N + 35 k\, N + 11 N^{2})} {(2 + N) (2 + k + N)^{5}}, \nonu\\
c_ {407} & = & - \frac {1}{3 (2 + N) (2 + k + N)^{5}}16 (288 + 416 k + 145 k^{2} - 33 k^{3} - 
       18 k^{4} + 592 N + 702 k\, 
      N 
      \nonu\\ & + & 225 k^{2} N + 3 k^{3} N + 449 N^{2} + 385 k\, 
      N^{2} + 80 k^{2} N^{2} + 143 N^{3} + 63 k\, 
      N^{3} + 16 N^{4}), \nonu\\
c_ {408} & = &\frac{1} {3 (2 + N) (2 + k + N)^{5}}16 (84 - 40 k - 184 k^{2} - 64 k^{3} + 430 N + 
      337 k\, N - 39 k^{2} N 
      \nonu\\ & + & 14 k^{3} N + 609 N^{2} + 472 k\, 
     N^{2} + 43 k^{2} N^{2} + 321 N^{3} + 149 k\, 
     N^{3} + 56 N^{4}), \nonu\\
c_ {409} & = & - \frac {16 (-k + N) (32 + 55 k + 18 k^{2} + 41 N + 
       35 k\, N + 11 N^{2})} {(2 + N) (2 + k + N)^{5}}, \nonu\\
c_ {410} & = &\frac{1} {3 (2 + N) (2 + k + N)^{5}} 8 (192 + 444 k + 317 k^{2} + 70 k^{3} + 420 N + 
      794 k\, N + 432 k^{2} N 
      \nonu\\ & + & 62 k^{3} N + 329 N^{2} + 450 k\, 
     N^{2} + 139 k^{2} N^{2} + 104 N^{3} + 76 k\, 
     N^{3} + 11 N^{4}), \nonu\\
c_ {411} & = &\frac {1} {3 (2 + N) (2 + k + N)^{5}}8 (552 + 774 k + 311 k^{2} + 34 k^{3} + 1254 N + 
      1352 k\, N + 339 k^{2} N
      \nonu\\ & + & 8 k^{3} N + 1109 N^{2} + 846 k\, 
     N^{2} + 106 k^{2} N^{2} + 449 N^{3} + 190 k\, 
     N^{3} + 68 N^{4}), \nonu\\
c_ {412} & = & - \frac {16 (-k + N) (32 + 55 k + 18 k^{2} + 41 N + 
       35 k\, N + 11 N^{2})} {(2 + N) (2 + k + N)^{5}}, \nonu\\
c_ {413} & = &\frac {16 (-k + N) (32 + 55 k + 18 k^{2} + 41 N + 
      35 k\, N + 11 N^{2})} {(2 + N) (2 + k + N)^{5}}, \nonu\\
c_ {414} & = &\frac {1} {3 (2 + N) (2 + k + N)^{5}}8 (240 + 612 k + 391 k^{2} + 74 k^{3} + 612 N + 
      1084 k\, N + 417 k^{2} N 
      \nonu\\ & + & 28 k^{3} N + 565 N^{2} + 648 k\, 
     N^{2} + 116 k^{2} N^{2} + 229 N^{3} + 134 k\, 
     N^{3} + 34 N^{4}), \nonu\\
c_ {415} & = &\frac {1} {3 (2 + N) (2 + k + N)^{4}}16 (216 + 436 k + 230 k^{2} + 36 k^{3} + 392 N + 
      473 k\, N + 112 k^{2} N 
      \nonu\\ & + & 209 N^{2} + 117 k\, 
     N^{2} + 35 N^{3}), \nonu\\
c_ {416} & = & - \frac {16 (32 + 55 k + 18 k^{2} + 41 N + 35 k\, 
      N + 11 N^{2})} {(2 + N) (2 + k + N)^{4}}, \nonu\\
c_ {417} & = &\frac {16 (24 - 128 k - 103 k^{2} - 18 k^{3} + 92 N - 
      43 k\, N - 23 k^{2} N + 98 N^{2} + 27 k\, 
     N^{2} + 26 N^{3})} {3 (2 + N) (2 + k + N)^{4}}, \nonu\\
c_ {418} & = &\frac {16 (36 + 36 k - 5 k^{2} - 6 k^{3} + 54 N + 
      22 k\, N - 11 k^{2} N + 23 N^{2} + 3 N^{3})} {(2 + 
      N) (2 + k + N)^{4}}, \nonu\\
c_ {419} & = &\frac {16 (60 + 126 k + 57 k^{2} + 6 k^{3} + 120 N + 
      152 k\, N + 31 k^{2} N + 71 N^{2} + 44 k\, 
     N^{2} + 13 N^{3})} {(2 + N) (2 + k + N)^{4}},    \nonu\\   
c_ {420} & = & - \frac {8 (32 + 55 k + 18 k^{2} + 41 N + 35 k\, 
      N + 11 N^{2})} {(2 + N) (2 + k + N)^{4}},  \nonu\\
c_ {421} & = &\frac {8 (32 + 55 k + 18 k^{2} + 41 N + 35 k\, 
     N + 11 N^{2})} {(2 + N) (2 + k + N)^{4}},  \nonu\\
c_ {422} & = & - \frac {16 (32 + 55 k + 18 k^{2} + 41 N + 35 k\, 
      N + 11 N^{2})} {(2 + N) (2 + k + N)^{4}},  \nonu\\
c_ {423} & = & - \frac {16 (-k + N) (32 + 55 k + 18 k^{2} + 41 N + 
       35 k\, N + 11 N^{2})} {(2 + N) (2 + k + N)^{5}}, \nonu\\ 
c_ {424} & = & - \frac {8 (32 + 55 k + 18 k^{2} + 41 N + 35 k\, 
      N + 11 N^{2})} {(2 + N) (2 + k + N)^{4}},  \nonu\\
c_ {425} & = &\frac {16 (32 + 55 k + 18 k^{2} + 41 N + 35 k\, 
     N + 11 N^{2})} {(2 + N) (2 + k + N)^{4}},  \nonu\\
c_ {426} & = &\frac {16 (-k + N) (32 + 55 k + 18 k^{2} + 41 N + 
      35 k\, N + 11 N^{2})} {(2 + N) (2 + k + N)^{5}},  \nonu\\
c_ {427} & = &\frac {8 (32 + 55 k + 18 k^{2} + 41 N + 35 k\, 
     N + 11 N^{2})} {(2 + N) (2 + k + N)^{4}},  \nonu\\
c_ {428} & = &\frac {16 (32 + 55 k + 18 k^{2} + 41 N + 35 k\, 
     N + 11 N^{2})} {(2 + N) (2 + k + N)^{3}},  \nonu\\
c_ {429} & = &\frac {16 (32 + 55 k + 18 k^{2} + 41 N + 35 k\, 
     N + 11 N^{2})} {(2 + N) (2 + k + N)^{4}},  \nonu\\
c_ {430} & = &\frac {16 (20 + 15 k - 11 k^{2} - 6 k^{3} + 51 N + 
      43 k\, N + k^{2} N + 40 N^{2} + 22 k\, 
     N^{2} + 9 N^{3})} {(2 + N) (2 + k + N)^{4}},  \nonu\\
c_ {431} & = & - \frac {64} {(2 + k + N)^{2}},  \qquad
c_ {432} = \frac {16 (8 + 17 k + 6 k^{2} + 11 N + 11 k\, 
     N + 3 N^{2})} {(2 + N) (2 + k + N)^{3}},  \nonu\\
c_ {433} & = & - \frac {8 (20 + 21 k + 6 k^{2} + 25 N + 13 k\, 
      N + 7 N^{2})} {(2 + N) (2 + k + N)^{3}},  \nonu\\
c_ {434} & = &\frac {16 (8 + 17 k + 6 k^{2} + 11 N + 11 k\, 
     N + 3 N^{2})} {(2 + N) (2 + k + N)^{3}},  \nonu\\
c_ {435} & = & - \frac {32 (10 + 23 k + 8 k^{2} + 18 N + 20 k\, 
      N + k^{2} N + 8 N^{2} + 3 k\, 
      N^{2} + N^{3})} {(2 + N) (2 + k + N)^{4}},  \nonu\\
c_ {436} & = &\frac {16 (32 + 55 k + 18 k^{2} + 41 N + 35 k\, 
     N + 11 N^{2})} {(2 + N) (2 + k + N)^{4}},  \nonu\\
c_ {437} & = & - \frac {16 (32 + 55 k + 18 k^{2} + 41 N + 35 k\, 
      N + 11 N^{2})} {(2 + N) (2 + k + N)^{4}},  \nonu\\
c_ {438} & = & - \frac {32 (20 + 15 k - 11 k^{2} - 6 k^{3} + 51 N + 
       43 k\, N + k^{2} N + 40 N^{2} + 22 k\, 
      N^{2} + 9 N^{3})} {(2 + N) (2 + k + N)^{4}},  \nonu\\
c_ {439} & = &\frac {16 (-k + N) (32 + 55 k + 18 k^{2} + 41 N + 
      35 k\, N + 11 N^{2})} {(2 + N) (2 + k + N)^{4}},  \nonu\\
c_ {440} & = & - \frac {32 (32 + 55 k + 18 k^{2} + 41 N + 35 k\, 
      N + 11 N^{2})} {(2 + N) (2 + k + N)^{4}},  \nonu\\
c_ {441} & = & - \frac {16 (4 + 10 k + 17 k^{2} + 6 k^{3} + 32 N + 
       52 k\, N + 23 k^{2} N + 35 N^{2} + 28 k\, 
      N^{2} + 9 N^{3})} {(2 + N) (2 + k + N)^{4}},  \nonu\\
c_ {442} & = &\frac {32 (10 - 2 k^{2} + 25 N + 17 k\, 
     N + 5 k^{2} N + 21 N^{2} + 11 k\, 
     N^{2} + 5 N^{3})} {(2 + N) (2 + k + N)^{4}},  \nonu\\
c_ {443} & = & - \frac {32 (1 + N) (-33 k - 42 k^{2} - 12 k^{3} + 
       17 N - 12 k\, N - 17 k^{2} N + 22 N^{2} + 7 k\, 
      N^{2} + 6 N^{3})} {(2 + N) (2 + k + N)^{5}},  \nonu\\
c_ {444} & = &\frac {1} {(2 + N) (2 + k + N)^{5}} 16 (-8 - 50 k - 
    40 k^{2} + 5 k^{3} + 6 k^{4} + 54 N + 36 k\, 
   N - 3 k^{2} N 
   \nonu\\ & + & 5 k^{3} N + 128 N^{2} + 117 k\, 
   N^{2} + 17 k^{2} N^{2} + 85 N^{3} + 47 k\, N^{3} + 17 N^{4}), \nonu\\ 
c_ {445} & = & - \frac {16 (-k + N) (32 + 55 k + 18 k^{2} + 41 N + 
       35 k\, N + 11 N^{2})} {(2 + N) (2 + k + N)^{5}},  \nonu\\
c_ {446} & = &\frac {1} {3 (2 + N) (2 + k + N)^{4}} 16 (-120 - 
    226 k - 149 k^{2} - 30 k^{3} - 266 N - 305 k\, 
   N - 118 k^{2} N 
   \nonu\\ & - & 6 k^{3} N - 170 N^{2} - 72 k\, 
   N^{2} - 9 k^{2} N^{2} - 32 N^{3} + 15 k\, N^{3}),  \nonu\\
c_ {447} & = & - \frac {1} {3 (2 + N) (2 + k + N)^{4}} 8 (180 + 
     307 k + 151 k^{2} + 22 k^{3} + 347 N + 396 k\, 
    N + 65 k^{2} N 
    \nonu\\ & - & 16 k^{3} N + 203 N^{2} + 133 k\, 
    N^{2} - 12 k^{2} N^{2} + 38 N^{3} + 10 k\, N^{3}),  \nonu\\
c_ {448} & = & - \frac {8 (32 + 55 k + 18 k^{2} + 41 N + 35 k\, 
      N + 11 N^{2})} {(2 + N) (2 + k + N)^{4}},  \nonu\\
c_ {449} & = &\frac {16 (32 + 55 k + 18 k^{2} + 41 N + 35 k\, 
     N + 11 N^{2})} {(2 + N) (2 + k + N)^{4}},  \nonu\\
c_ {450} & = &\frac {1} {3 (2 + N) (2 + k + N)^{5}} 16 (168 + 380 k + 
    267 k^{2} + 58 k^{3} + 304 N + 473 k\, 
   N + 174 k^{2} N 
   \nonu\\ & + & 2 k^{3} N + 208 N^{2} + 208 k\, 
   N^{2} + 27 k^{2} N^{2} + 76 N^{3} + 43 k\, N^{3} + 12 N^{4}),  \nonu\\
c_ {451} & = & - \frac {16 (32 + 55 k + 18 k^{2} + 41 N + 35 k\, 
      N + 11 N^{2})} {(2 + N) (2 + k + N)^{4}},  \nonu\\
c_ {452} & = & - \frac {8 (36 + 52 k + 16 k^{2} + 70 N + 75 k\, 
      N + 14 k^{2} N + 45 N^{2} + 27 k\, 
      N^{2} + 9 N^{3})} {(2 + N) (2 + k + N)^{4}},  \nonu\\
c_ {453} & = & - \frac {8 (4 - 6 k - 17 k^{2} - 6 k^{3} + 16 N + 
       9 k\, N - 5 k^{2} N + 16 N^{2} + 9 k\, 
      N^{2} + 4 N^{3})} {(2 + N) (2 + k + N)^{4}},  \nonu\\
c_ {454} & = & - \frac {1} {3 (2 + N) (2 + k + N)^{5}} 8 (696 + 
     1296 k + 878 k^{2} + 235 k^{3} + 18 k^{4} + 1332 N + 1694 k\, 
    N 
    \nonu\\ & + & 642 k^{2} N + 53 k^{3} N + 944 N^{2} + 735 k\, 
    N^{2} + 106 k^{2} N^{2} + 320 N^{3} + 127 k\, N^{3} + 44 N^{4}), \nonu\\ 
c_ {455} & = &\frac {16 (1 + k) (32 + 55 k + 18 k^{2} + 41 N + 35 k\, 
     N + 11 N^{2})} {(2 + N) (2 + k + N)^{5}},  \nonu\\
c_ {456} & = &\frac {16 (1 + N) (32 + 55 k + 18 k^{2} + 41 N + 35 k\, 
     N + 11 N^{2})} {(2 + N) (2 + k + N)^{5}},  \nonu\\
c_ {457} & = &\frac {1} {3 (2 + N) (2 + k + N)^{5}} 16 (72 + 176 k + 
    172 k^{2} + 48 k^{3} + 172 N + 357 k\, 
   N + 255 k^{2} N 
   \nonu\\ & + & 42 k^{3} N + 179 N^{2} + 268 k\, 
   N^{2} + 101 k^{2} N^{2} + 89 N^{3} + 69 k\, N^{3} + 16 N^{4}),  \nonu\\
c_ {458} & = &\frac {1} {3 (2 + N) (2 + k + N)^{5}} 8 (456 + 708 k + 
    260 k^{2} + 16 k^{3} + 888 N + 998 k\, 
   N + 183 k^{2} N 
   \nonu\\ & - & 10 k^{3} N + 602 N^{2} + 402 k\, 
   N^{2} + k^{2} N^{2} + 155 N^{3} + 34 k\, N^{3} + 11 N^{4}),  \nonu\\
c_ {459} & = & - \frac {16 (32 + 55 k + 18 k^{2} + 41 N + 35 k\, 
      N + 11 N^{2})} {(2 + N) (2 + k + N)^{4}},  \nonu\\
c_ {460} & = &\frac {16 (32 + 55 k + 18 k^{2} + 41 N + 35 k\, 
     N + 11 N^{2})} {(2 + N) (2 + k + N)^{4}},  \nonu\\
c_ {461} & = &\frac {64 (16 + 21 k + 6 k^{2} + 19 N + 13 k\, 
     N + 5 N^{2})} {(2 + N) (2 + k + N)^{3}},  \nonu\\
c_ {462} & = &\frac {16 (20 + 21 k + 6 k^{2} + 25 N + 13 k\, 
     N + 7 N^{2})} {(2 + N) (2 + k + N)^{3}},  \nonu\\
c_ {463} & = &\frac {16 (64 + 67 k + 18 k^{2} + 109 N + 79 k\, 
     N + 12 k^{2} N + 63 N^{2} + 24 k\, 
     N^{2} + 12 N^{3})} {(2 + N) (2 + k + N)^{4}},  \nonu\\
c_ {464} & = &\frac {1} {3 (2 + N) (2 + k + N)^{5}} 16 (48 + 40 k - 
    38 k^{2} - 20 k^{3} + 224 N + 185 k\, 
   N - 15 k^{2} N 
   \nonu\\ & - & 10 k^{3} N + 261 N^{2} + 140 k\, 
   N^{2} - 7 k^{2} N^{2} + 111 N^{3} + 25 k\, N^{3} + 16 N^{4}),  \nonu\\
c_ {465} & = & - \frac {1} {3 (2 + N) (2 + k + N)^{5}} 16 (-120 - 
     352 k - 280 k^{2} - 64 k^{3} - 164 N - 503 k\, 
    N 
    \nonu\\ & - & 333 k^{2} N - 50 k^{3} N - 45 N^{2} - 212 k\, 
    N^{2} - 95 k^{2} N^{2} + 21 N^{3} - 19 k\, N^{3} + 8 N^{4}),  \nonu\\
c_ {466} & = & - \frac {1} {3 (2 + N) (2 + k + N)^{5}} 16 (48 + 
     104 k + 26 k^{2} - 4 k^{3} + 160 N + 217 k\, 
    N + 33 k^{2} N 
    \nonu\\ & - & 2 k^{3} N + 165 N^{2} + 124 k\, 
    N^{2} + k^{2} N^{2} + 63 N^{3} + 17 k\, N^{3} + 8 N^{4}),  \nonu\\
c_ {467} & = &\frac {1} {3 (2 + N) (2 + k + N)^{5}} 16 (768 + 
    1230 k + 665 k^{2} + 118 k^{3} + 1926 N + 2657 k\, 
   N 
   \nonu\\ & + &1212 k^{2} N + 170 k^{3} N + 1772 N^{2} + 1986 k\, 
   N^{2} + 697 k^{2} N^{2} + 60 k^{3} N^{2} + 710 N^{3} \nonu \\
& + & 583 k\, 
   N^{3} + 126 k^{2} N^{3} + 104 N^{4} + 48 k\, N^{4}),  \nonu\\
c_ {468} & = & - \frac {64 (1 + N) (9 + 10 k + 3 k^{2} + 5 N + 3 k\, 
      N)} {3 (2 + k + N)^{4}},  \nonu\\
c_ {469} & = &\frac {1} {(2 + N) (2 + k + N)^{4}} 8 (88 + 114 k + 
    42 k^{2} + 4 k^{3} + 194 N + 198 k\, 
   N + 49 k^{2} N + 2 k^{3} N 
   \nonu\\ & + & 172 N^{2} + 128 k\, 
   N^{2} + 17 k^{2} N^{2} + 71 N^{3} + 30 k\, N^{3} + 11 N^{4}),  \nonu\\
c_ {470} & = &\frac {24 (16 + 21 k + 6 k^{2} + 19 N + 13 k\, 
     N + 5 N^{2})} {(2 + N) (2 + k + N)^{2}},  \nonu\\
c_ {471} & = & - \frac {8 (108 + 173 k + 54 k^{2} + 133 N + 109 k\, 
      N + 35 N^{2})} {3 (2 + N) (2 + k + N)^{3}},  \nonu\\
c_ {472} & = &\frac {8 (3 k + 2 k^{2} + 13 N + 15 k\, 
     N + 4 k^{2} N + 15 N^{2} + 8 k\, 
     N^{2} + 4 N^{3})} {(2 + N) (2 + k + N)^{4}},  \nonu\\
c_ {473} & = &\frac {8 (64 + 139 k + 89 k^{2} + 18 k^{3} + 101 N + 
      151 k\, N + 49 k^{2} N + 48 N^{2} + 38 k\, 
     N^{2} + 7 N^{3})} {(2 + N) (2 + k + N)^{4}},  \nonu\\
c_ {474} & = &\frac {8 (32 + 55 k + 18 k^{2} + 41 N + 35 k\, 
     N + 11 N^{2})} {(2 + N) (2 + k + N)^{4}},  \nonu\\
c_ {475} & = & - \frac {8 (-12 + 31 k + 18 k^{2} - N + 23 k\, 
      N + N^{2})} {3 (2 + N) (2 + k + N)^{3}},  \nonu\\
c_ {476} & = &\frac {8 (12 - 33 k - 14 k^{2} + 63 N + 45 k\, 
     N + 20 k^{2} N + 65 N^{2} + 42 k\, 
     N^{2} + 16 N^{3})} {3 (2 + N) (2 + k + N)^{3}},  \nonu\\
c_ {477} & = & - \frac {8 (32 + 55 k + 18 k^{2} + 41 N + 35 k\, 
      N + 11 N^{2})} {(2 + N) (2 + k + N)^{4}},  \nonu\\
c_ {478} & = &\frac {1} {3 (2 + N) (2 + k + N)^{4}} 16 (252 + 431 k + 
    227 k^{2} + 38 k^{3} + 547 N + 676 k\, 
   N + 210 k^{2} N 
   \nonu\\ & + & 10 k^{3} N + 447 N^{2} + 361 k\, 
   N^{2} + 49 k^{2} N^{2} + 165 N^{3} + 68 k\, N^{3} + 23 N^{4}),  \nonu\\
c_ {479} & = & - \frac {8 (-48 - 41 k - 14 k^{2} - 19 N + 41 k\, 
      N + 20 k^{2} N + 39 N^{2} + 42 k\, 
      N^{2} + 16 N^{3})} {3 (2 + N) (2 + k + N)^{3}},  \nonu\\
c_ {480} & = &\frac {1} {3 (2 + N) (2 + k + N)^{4}} 8 (624 + 1268 k + 
    923 k^{2} + 296 k^{3} + 36 k^{4} + 1204 N + 1756 k\, 
   N 
   \nonu\\ & + & 795 k^{2} N + 118 k^{3} N + 849 N^{2} + 778 k\, 
   N^{2} + 154 k^{2} N^{2} + 267 N^{3} + 116 k\, N^{3} + 32 N^{4}),  \nonu\\
c_ {481} & = & - \frac {16 (32 + 55 k + 18 k^{2} + 41 N + 35 k\, 
      N + 11 N^{2})} {(2 + N) (2 + k + N)^{4}},  \nonu\\
c_ {482} & = & - \frac {1} {3 (2 + N) (2 + k + N)^{4}} 8 (-36 + 
     241 k + 404 k^{2} + 210 k^{3} + 36 k^{4} - 103 N + 296 k\, 
    N 
\nonu\\ & + & 370 k^{2} N + 102 k^{3} N - 82 N^{2} + 117 k\, 
    N^{2} + 78 k^{2} N^{2} - 19 N^{3} + 18 k\, N^{3}),  \nonu\\
c_ {483} & = &\frac {1} {3 (2 + N) (2 + k + N)^{5}} 16 (-120 - 
    332 k - 278 k^{2} - 68 k^{3} - 184 N - 379 k\, 
   N - 204 k^{2} N 
   \nonu\\ & - & 16 k^{3} N - 75 N^{2} - 118 k\, 
   N^{2} - 34 k^{2} N^{2} - 6 N^{3} - 11 k\, N^{3} + N^{4}),  \nonu\\
c_ {484} & = &\frac {8 (84 + 128 k + 40 k^{2} + 178 N + 199 k\, 
     N + 38 k^{2} N + 121 N^{2} + 75 k\, 
     N^{2} + 25 N^{3})} {3 (2 + N) (2 + k + N)^{4}},  \nonu\\
c_ {485} & = & - \frac {8 (204 + 394 k + 257 k^{2} + 54 k^{3} + 
       308 N + 395 k\, N + 133 k^{2} N + 140 N^{2} + 87 k\, 
      N^{2} + 20 N^{3})} {3 (2 + N) (2 + k + N)^{4}},  \nonu\\
c_ {486} & = & - \frac {1} {(2 + N) (2 + k + N)^{5}} 8 (328 + 600 k + 
     344 k^{2} + 77 k^{3} + 6 k^{4} + 740 N + 1030 k\, 
    N 
    \nonu\\ & + & 400 k^{2} N + 47 k^{3} N + 614 N^{2} + 573 k\, 
    N^{2} + 110 k^{2} N^{2} + 218 N^{3} + 101 k\, N^{3} + 28 N^{4}), \nonu\\ 
c_ {487} & = &\frac {16 (1 + N) (32 + 55 k + 18 k^{2} + 41 N + 35 k\, 
     N + 11 N^{2})} {(2 + N) (2 + k + N)^{5}},  \nonu\\
c_ {488} & = &\frac {16 (1 + k) (32 + 55 k + 18 k^{2} + 41 N + 35 k\, 
     N + 11 N^{2})} {(2 + N) (2 + k + N)^{5}},  \nonu\\
c_ {489} & = & - \frac {1} {3 (2 + N) (2 + k + N)^{5}} 16 (-118 k - 
     166 k^{2} - 91 k^{3} - 18 k^{4} + 118 N + 19 k\, 
    N - 75 k^{2} N 
    \nonu\\ & - & 35 k^{3} N + 243 N^{2} + 199 k\, 
    N^{2} + 31 k^{2} N^{2} + 159 N^{3} + 86 k\, N^{3} + 32 N^{4}),  \nonu\\
c_ {490} & = &\frac {1} {(2 + N) (2 + k + N)^{5}} 8 (168 + 304 k + 
    192 k^{2} + 40 k^{3} + 348 N + 414 k\, 
   N + 133 k^{2} N 
   \nonu\\ & + & 2 k^{3} N + 278 N^{2} + 202 k\, 
   N^{2}    + 23 k^{2} N^{2} + 109 N^{3} + 42 k\, N^{3} + 17 N^{4}),  \nonu\\
c_ {491} & = & - \frac {1} {3 (2 + N) (2 + k + N)^{4}} 16 (-264 - 
     554 k - 373 k^{2} - 78 k^{3} - 298 N - 541 k\, 
    N 
    \nonu\\ & - & 278 k^{2} N - 30 k^{3} N + 2 N^{2} - 60 k\, 
    N^{2} - 33 k^{2} N^{2} + 92 N^{3} + 39 k\, N^{3} + 24 N^{4}),  \nonu\\
c_ {492} & = &\frac {1} {(2 + N) (2 + k + N)^{4}} 16 (-40 - 150 k - 
    119 k^{2} - 26 k^{3} - 14 N - 139 k\, 
   N - 90 k^{2} N 
   \nonu\\ & - & 10 k^{3} N + 50 N^{2} - 8 k\, 
   N^{2} - 11 k^{2} N^{2} + 40 N^{3} + 13 k\, N^{3} + 8 N^{4}),  \nonu\\
c_ {493} & = & - \frac {8 (108 + 173 k + 54 k^{2} + 133 N + 109 k\, 
      N + 35 N^{2})} {3 (2 + N) (2 + k + N)^{3}},  \nonu\\
c_ {494} & = &\frac {8 (-32 + 42 k + 93 k^{2} + 30 k^{3} - 26 N + 
      96 k\, N + 71 k^{2} N + 3 N^{2} + 40 k\, 
     N^{2} + 3 N^{3})} {(2 + N) (2 + k + N)^{4}},  \nonu\\
c_ {495} & = & - \frac {16 (32 + 55 k + 18 k^{2} + 41 N + 35 k\, 
      N + 11 N^{2})} {(2 + N) (2 + k + N)^{4}},  \nonu\\
c_ {496} & = & - \frac {16 (-k + N) (32 + 55 k + 18 k^{2} + 41 N + 
       35 k\, N + 11 N^{2})} {(2 + N) (2 + k + N)^{5}},  \nonu\\
c_ {497} & = &\frac {16 (32 + 55 k + 18 k^{2} + 41 N + 35 k\, 
     N + 11 N^{2})} {(2 + N) (2 + k + N)^{4}},  \nonu\\
c_ {498} & = &\frac {16 (-k + N) (32 + 55 k + 18 k^{2} + 41 N + 
      35 k\, N + 11 N^{2})} {(2 + N) (2 + k + N)^{5}},  \nonu\\
c_ {499} & = & - \frac {16 (124 + 190 k + 73 k^{2} + 6 k^{3} + 
       216 N + 216 k\, N + 39 k^{2} N + 119 N^{2} + 60 k\, 
      N^{2} + 21 N^{3})} {(2 + N) (2 + k + N)^{4}},  \nonu\\
c_ {500} & = &\frac {32 (32 + 55 k + 18 k^{2} + 41 N + 35 k\, 
     N + 11 N^{2})} {(2 + N) (2 + k + N)^{4}},  \nonu\\
c_ {501} & = & - \frac {32 (32 + 55 k + 18 k^{2} + 41 N + 35 k\, 
      N + 11 N^{2})} {(2 + N) (2 + k + N)^{4}},  \nonu\\
c_ {502} & = &\frac {16 (132 + 181 k + 54 k^{2} + 161 N + 113 k\, 
     N + 43 N^{2})} {3 (2 + N) (2 + k + N)^{3}},  \nonu\\
c_ {503} & = &\frac {1} {(2 + N) (2 + k + N)^{5}} 16 (-32 - 81 k - 
    66 k^{2} - 16 k^{3} - 47 N - 117 k\, 
   N - 83 k^{2} N
   \nonu\\ & - & 14 k^{3} N - 9 N^{2} - 41 k\, 
   N^{2} - 23 k^{2} N^{2} + 12 N^{3} + k\, N^{3} + 4 N^{4}),  \nonu\\
c_ {504} & = & - \frac {1} {(2 + N) (2 + k + N)^{5}} 16 (128 + 
     276 k + 233 k^{2} + 101 k^{3} + 18 k^{4} + 300 N + 452 k\, 
    N 
    \nonu\\ & + & 220 k^{2} N + 43 k^{3} N + 275 N^{2} + 265 k\, 
    N^{2} + 53 k^{2} N^{2} + 118 N^{3} + 59 k\, N^{3} + 19 N^{4}),  \nonu\\
c_ {505} & = &\frac {8 (-12 + 31 k + 18 k^{2} - N + 23 k\, 
     N + N^{2})} {3 (2 + N) (2 + k + N)^{3}},  \nonu\\
c_ {506} & = &\frac {16 (252 + 323 k + 90 k^{2} + 295 N + 199 k\, 
     N + 77 N^{2})} {3 (2 + N) (2 + k + N)^{3}},  \nonu\\
c_ {507} & = & - \frac {16 (1 + k) (1 + N) (32 + 55 k + 18 k^{2} + 
       41 N + 35 k\, N + 11 N^{2})} {(2 + N) (2 + k + N)^{5}},  \nonu\\
c_ {508} & = &\frac {16 (32 + 55 k + 18 k^{2} + 41 N + 35 k\, 
     N + 11 N^{2})} {(2 + N) (2 + k + N)^{4}},  \nonu\\
c_ {509} & = & - \frac {8 (-32 + 10 k + 51 k^{2} + 18 k^{3} + 6 N + 
       100 k\, N + 57 k^{2} N + 41 N^{2} + 56 k\, 
      N^{2} + 13 N^{3})} {(2 + N) (2 + k + N)^{4}},  \nonu\\
c_ {510} & = & - \frac {16 (-4 - 65 k - 56 k^{2} - 12 k^{3} + 7 N - 
       35 k\, N - 18 k^{2} N + 19 N^{2} + 6 k\, 
      N^{2} + 6 N^{3})} {(2 + N) (2 + k + N)^{4}},  \nonu\\
c_ {511} & = &\frac {1} {3 (2 + N) (2 + k + N)^{3}} 16 (96 + 124 k + 
    39 k^{2} + 2 k^{3} + 254 N + 249 k\, 
   N + 31 k^{2} N 
   \nonu\\ & - & 8 k^{3} N + 249 N^{2} + 162 k\, 
   N^{2} + 2 k^{2} N^{2} + 105 N^{3} + 35 k\, N^{3} + 16 N^{4}),  \nonu\\
c_ {512} & = &\frac {16 (32 + 55 k + 18 k^{2} + 41 N + 35 k\, 
     N + 11 N^{2})} {(2 + N) (2 + k + N)^{4}},  \nonu\\
c_ {513} & = & - \frac {1} {(2 + N) (2 + k + N)^{4}} 8 (32 + 132 k + 
     137 k^{2} + 51 k^{3} + 6 k^{4} + 76 N + 226 k\, 
    N 
    \nonu\\ & + & 161 k^{2} N + 31 k^{3} N + 53 N^{2} + 113 k\, 
    N^{2} + 44 k^{2} N^{2} + 11 N^{3} + 15 k\, N^{3}),  \nonu\\
c_ {514} & = & - \frac {1} {3 (2 + N) (2 + k + N)^{4}} 8 (312 + 
     434 k + 143 k^{2} - 37 k^{3} - 18 k^{4} + 706 N + 812 k\, 
    N 
    \nonu\\ & + & 209 k^{2} N - 17 k^{3} N + 713 N^{2} + 639 k\, 
    N^{2} + 110 k^{2} N^{2} + 353 N^{3} + 185 k\, N^{3} + 64 N^{4}), \nonu\\ 
c_ {515} & = &\frac {1} {3 (2 + N) (2 + k + N)^{5}} 16 (360 + 914 k + 
    634 k^{2} + 132 k^{3} + 730 N + 1323 k\, 
   N + 537 k^{2} N 
   \nonu\\ & + & 30 k^{3} N + 527 N^{2} + 634 k\, 
   N^{2} + 113 k^{2} N^{2} + 173 N^{3} + 111 k\, N^{3} + 22 N^{4}),  \nonu\\
c_ {516} & = & - \frac {1} {3 (2 + N) (2 + k + N)^{5}} 16 (288 + 
     766 k + 702 k^{2} + 266 k^{3} + 36 k^{4} + 626 N + 1165 k\, 
    N 
    \nonu\\ & + & 657 k^{2} N + 112 k^{3} N + 485 N^{2} + 584 k\, 
    N^{2} + 153 k^{2} N^{2} + 173 N^{3} + 107 k\, N^{3} + 24 N^{4}), \nonu\\ 
c_ {517} & = &\frac {16 (-20 - 11 k + 11 k^{2} + 6 k^{3} - 7 N + 
      39 k\, N + 23 k^{2} N + 14 N^{2} + 28 k\, 
     N^{2} + 5 N^{3})} {(2 + N) (2 + k + N)^{3}},  \nonu\\
c_ {518} & = & - \frac {1} {3 (2 + N) (2 + k + N)^{4}} 16 (48 + 
     131 k + 108 k^{2} + 28 k^{3} + 97 N + 169 k\, 
    N + 52 k^{2} N 
    \nonu\\ & - & 4 k^{3} N + 113 N^{2} + 134 k\, 
    N^{2} + 14 k^{2} N^{2} + 74 N^{3} + 52 k\, N^{3} + 16 N^{4}),  \nonu\\
c_ {519} & = &\frac {1} {3 (2 + N) (2 + k + N)^{4}} 16 (264 + 503 k + 
    320 k^{2} + 68 k^{3} + 577 N + 787 k\, 
   N + 282 k^{2} N 
   \nonu\\ & + & 16 k^{3} N + 495 N^{2} + 466 k\, 
   N^{2} + 76 k^{2} N^{2} + 204 N^{3} + 110 k\, N^{3} + 32 N^{4}),  \nonu\\
c_ {520} & = &\frac {16 (52 + 86 k + 41 k^{2} + 6 k^{3} + 76 N + 
      66 k\, N + 11 k^{2} N + 29 N^{2} + 6 k\, 
     N^{2} + 3 N^{3})} {(2 + N) (2 + k + N)^{4}},  \nonu\\
c_ {521} & = & - \frac {16 (-4 + 13 k + 19 k^{2} + 6 k^{3} + 25 N + 
       67 k\, N + 27 k^{2} N + 34 N^{2} + 36 k\, 
      N^{2} + 9 N^{3})} {(2 + N) (2 + k + N)^{4}},  \nonu\\
c_ {522} & = & - \frac {16 (32 + 55 k + 18 k^{2} + 41 N + 35 k\, 
      N + 11 N^{2})} {(2 + N) (2 + k + N)^{4}},  \nonu\\
c_ {523} & = & - \frac {1} {3 (2 + N) (2 + k + N)^{4}} 8 (-180 - 
     67 k + 79 k^{2} + 30 k^{3} - 227 N + 146 k\, 
    N + 193 k^{2} N 
    \nonu\\ & + & 24 k^{3} N + 21 N^{2} + 317 k\, 
    N^{2} + 112 k^{2} N^{2} + 114 N^{3} + 126 k\, N^{3} + 32 N^{4}), \nonu\\ 
c_ {524} & = & - \frac {1} {3 (2 + N) (2 + k + N)^{4}} 8 (432 + 
     376 k - 12 k^{2} - 40 k^{3} + 800 N + 432 k\, 
    N - 113 k^{2} N 
    \nonu\\ & - & 38 k^{3} N + 492 N^{2} + 78 k\, 
    N^{2} - 79 k^{2} N^{2} + 99 N^{3} - 28 k\, N^{3} + N^{4}),  \nonu\\
c_ {525} & = & - \frac {8 (32 + 55 k + 18 k^{2} + 41 N + 35 k\, 
      N + 11 N^{2})} {(2 + N) (2 + k + N)^{4}},  \nonu\\
c_ {526} & = &\frac {1} {3 (2 + N) (2 + k + N)^{5}} 16 (288 + 78 k - 
    361 k^{2} - 261 k^{3} - 50 k^{4} + 630 N - 61 k\, 
   N 
   \nonu\\ & - & 701 k^{2} N - 273 k^{3} N - 16 k^{4} N + 764 N^{2} + 55 k\, 
   N^{2} - 368 k^{2} N^{2} - 72 k^{3} N^{2} + 565 N^{3} 
   \nonu\\ & + & 156 k\, 
   N^{3} - 52 k^{2} N^{3} + 217 N^{4} + 54 k\, N^{4} + 32 N^{5}),  \nonu\\
c_ {527} & = & - \frac {1} {3 (2 + N) (2 + k + N)^{5}} 8 (-288 - 
     708 k - 583 k^{2} - 185 k^{3} - 18 k^{4} - 492 N - 862 k\, 
    N 
    \nonu\\ & - & 471 k^{2} N - 73 k^{3} N - 235 N^{2} - 243 k\, 
    N^{2} - 68 k^{2} N^{2} - 13 N^{3} + 7 k\, N^{3} + 8 N^{4}),  \nonu\\
c_ {528} & = &\frac {1} {(2 + N) (2 + k + N)^{5}} 8 (-16 - 20 k - 
    27 k^{2} - 23 k^{3} - 6 k^{4} - 36 N - 44 k\, 
   N - 57 k^{2} N 
   \nonu\\ & - & 23 k^{3} N - N^{2} + 15 k\, 
   N^{2} - 10 k^{2} N^{2} + 25 N^{3} + 23 k\, N^{3} + 8 N^{4}),  \nonu\\
c_ {529} & = &\frac {16 (1 + k) (32 + 55 k + 18 k^{2} + 41 N + 35 k\, 
     N + 11 N^{2})} {(2 + N) (2 + k + N)^{5}},  \nonu\\
c_ {530} & = & - \frac {16 (2 + 2 k + k^{2} + 2 N + N^{2}) (32 + 
       55 k + 18 k^{2} + 41 N + 35 k\, 
      N + 11 N^{2})} {(2 + N) (2 + k + N)^{6}},  \nonu\\
c_ {531} & = &\frac {16 (1 + N) (32 + 55 k + 18 k^{2} + 41 N + 35 k\, 
     N + 11 N^{2})} {(2 + N) (2 + k + N)^{5}},  \nonu\\
c_ {532} & = & - \frac {32 (1 + k) (1 + N) (32 + 55 k + 18 k^{2} + 
       41 N + 35 k\, N + 11 N^{2})} {(2 + N) (2 + k + N)^{6}},  \nonu\\
c_ {533} & = & - \frac {1} {3 (2 + N) (2 + k + N)^{5}} 32 (18 - 
     80 k - 125 k^{2} - 38 k^{3} + 143 N - 37 k\, 
    N - 135 k^{2} N 
     \nonu\\ & - & 28 k^{3} N + 195 N^{2} + 26 k\, 
    N^{2} - 40 k^{2} N^{2} + 96 N^{3} + 13 k\, N^{3} + 16 N^{4}),  \nonu\\
c_ {534} & = & - \frac {16 (1 + k) (32 + 55 k + 18 k^{2} + 41 N + 
       35 k\, N + 11 N^{2})} {(2 + N) (2 + k + N)^{5}},  \nonu\\
c_ {535} & = & - \frac {16 (1 + N) (32 + 55 k + 18 k^{2} + 41 N + 
       35 k\, N + 11 N^{2})} {(2 + N) (2 + k + N)^{5}},  \nonu\\
c_ {536} & = & - \frac {1} {3 (2 + N) (2 + k + N)^{5}} 16 (336 + 
     452 k + 175 k^{2} + 18 k^{3} + 724 N + 732 k\, 
    N + 165 k^{2} N 
     \nonu\\ & + & 605 N^{2} + 424 k\, 
    N^{2} + 44 k^{2} N^{2} + 233 N^{3} + 90 k\, N^{3} + 34 N^{4}),  \nonu\\
c_ {537} & = & - \frac{1}{(2 + N) (2 + k + N)^{6}}16 (1 + k) (1 + N) (60 + 77 k + 22 k^{2} + 
       121 N + 115 k\, N 
       \nonu\\ & + & 20 k^{2} N + 79 N^{2} + 42 k\, 
      N^{2} + 16 N^{3}),  \nonu\\
c_ {538} & = & - \frac {1} {3 (2 + N) (2 + k + N)^{6}} 16 (-192 - 
     542 k - 390 k^{2} - 51 k^{3} + 14 k^{4} - 418 N - 1144 k\, 
    N 
     \nonu\\ & - & 729 k^{2} N - 71 k^{3} N + 16 k^{4} N - 194 N^{2} - 717 k\, 
    N^{2} - 417 k^{2} N^{2} - 28 k^{3} N^{2} + 153 N^{3} 
     \nonu\\ & - &61 k\, 
    N^{3} - 66 k^{2} N^{3} + 151 N^{4} + 46 k\, N^{4} + 32 N^{5}),  \nonu\\
c_ {539} & = & - \frac {1} {3 (2 + N) (2 + k + N)^{5}} 8 (264 + 
     360 k + 131 k^{2} + 10 k^{3} + 756 N + 860 k\, 
    N + 246 k^{2} N 
     \nonu\\ & + & 14 k^{3} N + 773 N^{2} + 642 k\, 
    N^{2} + 103 k^{2} N^{2} + 338 N^{3} + 154 k\, N^{3} + 53 N^{4}), \nonu\\ 
c_ {540} & = &\frac {1} {3 (2 + N) (2 + k + N)^{5}} 16 (24 - 380 k - 
    567 k^{2} - 256 k^{3} - 36 k^{4} + 272 N - 170 k\, 
   N 
    \nonu\\ & - & 369 k^{2} N - 98 k^{3} N + 461 N^{2} + 230 k\, 
   N^{2} - 12 k^{2} N^{2} + 263 N^{3} + 110 k\, N^{3} + 48 N^{4}),  \nonu\\
c_ {541} & = &\frac {1} {3 (2 + N) (2 + k + N)^{6}} 16 (372 + 913 k + 
    879 k^{2} + 396 k^{3} + 68 k^{4} + 1001 N + 1739 k\, 
   N 
    \nonu\\ & + & 1062 k^{2} N + 268 k^{3} N + 16 k^{4} N + 1126 N^{2} + 
    1317 k\, 
   N^{2} + 411 k^{2} N^{2} + 32 k^{3} N^{2} + 693 N^{3} 
    \nonu\\ & + & 521 k\, 
   N^{3} + 60 k^{2} N^{3} + 232 N^{4} + 94 k\, N^{4} + 32 N^{5}),  \nonu\\
c_ {542} & = &\frac {1} {(2 + N) (2 + k + N)^{5}} 8 (8 - 56 k - 
    103 k^{2} - 34 k^{3} + 84 N - 34 k\, 
   N - 120 k^{2} N - 26 k^{3} N 
    \nonu\\ & + & 109 N^{2} - 2 k\, 
   N^{2} - 41 k^{2} N^{2} + 48 N^{3} + 7 N^{4}),  \nonu\\
c_ {543} & = & - \frac {32 (18 - 4 k - 47 k^{2} - 18 k^{3} + 49 N + 
       k\, N - 31 k^{2} N + 34 N^{2} + 3 k\, 
      N^{2} + 7 N^{3})} {3 (2 + N) (2 + k + N)^{4}},  \nonu\\
c_ {544} & = &\frac {8 (184 + 170 k + 23 k^{2} - 6 k^{3} + 290 N + 
      206 k\, N + 21 k^{2} N + 155 N^{2} + 66 k\, 
     N^{2} + 27 N^{3})} {(2 + N) (2 + k + N)^{3}},  \nonu\\
c_ {545} & = &\frac {8 (-64 - 58 k + k^{2} + 6 k^{3} - 70 N + 8 k\, 
     N + 27 k^{2} N - 9 N^{2} + 28 k\, 
     N^{2} + 3 N^{3})} {(2 + N) (2 + k + N)^{4}},  \nonu\\
c_ {546} & = &\frac {8 (64 + 61 k + 14 k^{2} + 83 N + 49 k\, 
     N + 4 k^{2} N + 33 N^{2} + 8 k\, 
     N^{2} + 4 N^{3})} {(2 + N) (2 + k + N)^{4}},  \nonu\\
c_ {547} & = &\frac {8 (120 + 125 k + 17 k^{2} - 6 k^{3} + 223 N + 
      181 k\, N + 21 k^{2} N + 138 N^{2} + 66 k\, 
     N^{2} + 27 N^{3})} {(2 + N) (2 + k + N)^{4}},  \nonu\\
c_ {548} & = & - \frac {8 (32 + 55 k + 18 k^{2} + 41 N + 35 k\, 
      N + 11 N^{2})} {(2 + N) (2 + k + N)^{4}},  \nonu\\
c_ {549} & = &\frac {8 (20 + 21 k + 6 k^{2} + 25 N + 13 k\, 
     N + 7 N^{2})} {(2 + N) (2 + k + N)^{3}},  \nonu\\
c_ {550} & = &\frac {8 (64 + 162 k + 99 k^{2} + 18 k^{3} + 94 N + 
      128 k\, N + 33 k^{2} N + 29 N^{2} + 12 k\, 
     N^{2} + N^{3})} {(2 + N) (2 + k + N)^{4}},  \nonu\\
c_ {551} & = &\frac {1} {3 (2 + N) (2 + k + N)^{4}} 8 (-192 - 266 k - 
    157 k^{2} - 30 k^{3} + 74 N - 52 k\, 
   N - 209 k^{2} N 
   \nonu\\ & - & 60 k^{3} N + 401 N^{2} + 174 k\, 
   N^{2} - 66 k^{2} N^{2} + 257 N^{3} + 78 k\, N^{3} + 48 N^{4}),  \nonu\\
c_ {552} & = & - \frac {8 (32 + 55 k + 18 k^{2} + 41 N + 35 k\, 
      N + 11 N^{2})} {(2 + N) (2 + k + N)^{4}},  \nonu\\
c_ {553} & = &\frac {1} {3 (2 + N) (2 + k + N)^{4}} 8 (-204 - 209 k - 
    7 k^{2} + 18 k^{3} - 361 N - 142 k\, 
   N + 133 k^{2} N 
    \nonu\\ & + & 36 k^{3} N - 145 N^{2} + 123 k\, 
   N^{2} + 102 k^{2} N^{2} + 32 N^{3} + 78 k\, N^{3} + 18 N^{4}),  \nonu\\
c_ {554} & = &\frac {1} {3 (2 + N) (2 + k + N)^{5}} 16 (144 + 310 k + 
    217 k^{2} + 46 k^{3} + 374 N + 677 k\, 
   N + 396 k^{2} N 
    \nonu\\ & + & 62 k^{3} N + 384 N^{2} + 554 k\, 
   N^{2} + 257 k^{2} N^{2} + 24 k^{3} N^{2} + 174 N^{3} + 175 k\, 
   N^{3} + 54 k^{2} N^{3} 
    \nonu\\ & + & 28 N^{4} + 12 k\, N^{4}),  \nonu\\
c_ {555} & = &\frac {1} {3 (2 + N) (2 + k + N)^{5}} 16 (48 + 160 k + 
    194 k^{2} + 106 k^{3} + 20 k^{4} + 104 N + 153 k\, 
   N 
    \nonu\\ & + & 38 k^{2} N - 10 k^{3} N - 8 k^{4} N + 157 N^{2} + 91 k\, 
   N^{2} - 83 k^{2} N^{2} - 30 k^{3} N^{2} + 173 N^{3} + 103 k\, 
   N^{3}
    \nonu\\ & - & 17 k^{2} N^{3} + 90 N^{4} + 39 k\, N^{4} + 16 N^{5}),  \nonu\\
c_ {556} & = &\frac {1} {3 (2 + N) (2 + k + N)^{5}} 16 (96 + 144 k + 
    34 k^{2} - 4 k^{3} + 288 N + 301 k\, 
   N + 45 k^{2} N - 2 k^{3} N 
    \nonu\\ & + & 289 N^{2} + 180 k\, 
   N^{2} + 5 k^{2} N^{2} + 115 N^{3} + 29 k\, N^{3} + 16 N^{4}),  \nonu\\
c_ {557} & = &\frac {1} {3 (2 + N) (2 + k + N)^{5}} 16 (120 + 224 k + 
    152 k^{2} + 32 k^{3} + 292 N + 439 k\, 
   N + 237 k^{2} N 
    \nonu\\ & + & 34 k^{3} N + 237 N^{2} + 244 k\, 
   N^{2} + 79 k^{2} N^{2} + 75 N^{3} + 35 k\, N^{3} + 8 N^{4}),  \nonu\\
c_ {558} & = &\frac {16 (10 + 17 k + 6 k^{2} + 8 N + 20 k\, 
     N + 6 k^{2} N + 2 N^{2} + 7 k\, 
     N^{2})} {(2 + N) (2 + k + N)^{3}},  \nonu\\
c_ {559} & = & - \frac {1} {3 (2 + N) (2 + k + N)^{4}} 16 (80 + 
     263 k + 197 k^{2} + 42 k^{3} + 133 N + 480 k\, 
    N + 313 k^{2} N 
     \nonu\\ & + & 52 k^{3} N + 105 N^{2} + 373 k\, 
    N^{2} + 192 k^{2} N^{2} + 20 k^{3} N^{2} + 46 N^{3} + 134 k\, 
    N^{3} + 42 k^{2} N^{3} 
     \nonu\\ & + & 8 N^{4} + 16 k\, N^{4}),  \nonu\\
c_ {560} & = & - \frac {18} {5},  \qquad
c_ {561}  =  - \frac {32 (k - N)} {5 (2 + k + N)},  \nonu\\
c_ {562} & = &\frac {2 (60 + 77 k + 22 k^{2} + 121 N + 115 k\, 
     N + 20 k^{2} N + 79 N^{2} + 42 k\, 
     N^{2} + 16 N^{3})} {(2 + N) (2 + k + N)^{2}},  \nonu\\
c_ {563} & = &\frac {2 (60 + 77 k + 22 k^{2} + 121 N + 115 k\, 
     N + 20 k^{2} N + 79 N^{2} + 42 k\, 
     N^{2} + 16 N^{3})} {(2 + N) (2 + k + N)^{2}},  \nonu\\
c_ {564} & = & - \frac {16 (-k + N) (60 + 77 k + 22 k^{2} + 121 N + 
       115 k\, N + 20 k^{2} N + 79 N^{2} + 42 k\, 
      N^{2} + 16 N^{3})} {3 (2 + N) (2 + k + N)^{3}},  \nonu\\
c_ {565} & = & - \frac {8 (-k + N) (60 + 77 k + 22 k^{2} + 121 N + 
       115 k\, N + 20 k^{2} N + 79 N^{2} + 42 k\, 
      N^{2} + 16 N^{3})} {(2 + N) (2 + k + N)^{3}},  \nonu\\
c_ {566} & = & - \frac {8 (-k + N) (60 + 77 k + 22 k^{2} + 121 N + 
       115 k\, N + 20 k^{2} N + 79 N^{2} + 42 k\, 
      N^{2} + 16 N^{3})} {3 (2 + N) (2 + k + N)^{3}},  \nonu\\
c_ {567} & = &\frac {4 (60 + 77 k + 22 k^{2} + 121 N + 115 k\, 
     N + 20 k^{2} N + 79 N^{2} + 42 k\, 
     N^{2} + 16 N^{3})} {(2 + N) (2 + k + N)^{2}},  \nonu\\
c_ {568} & = & - \frac {4 (60 + 77 k + 22 k^{2} + 121 N + 115 k\, 
      N + 20 k^{2} N + 79 N^{2} + 42 k\, 
      N^{2} + 16 N^{3})} {(2 + N) (2 + k + N)^{2}},  \nonu\\
c_ {569} & = & - \frac {16 (1 + N) (32 + 55 k + 18 k^{2} + 41 N + 
       35 k\, N + 11 N^{2})} {(2 + N) (2 + k + N)^{5}},  \nonu\\
c_ {570} & = &\frac {16 (1 + N) (32 + 55 k + 18 k^{2} + 41 N + 35 k\, 
     N + 11 N^{2})} {(2 + N) (2 + k + N)^{5}},  \nonu\\
c_ {571} & = & - \frac {16 (1 + k) (32 + 55 k + 18 k^{2} + 41 N + 
       35 k\, N + 11 N^{2})} {(2 + N) (2 + k + N)^{5}},  \nonu\\
c_ {572} & = &\frac {16 (1 + k) (32 + 55 k + 18 k^{2} + 41 N + 35 k\, 
     N + 11 N^{2})} {(2 + N) (2 + k + N)^{5}},  \nonu\\
c_ {573} & = &\frac {16 (24 - 128 k - 103 k^{2} - 18 k^{3} + 92 N - 
      43 k\, N - 23 k^{2} N + 98 N^{2} + 27 k\, 
     N^{2} + 26 N^{3})} {3 (2 + N) (2 + k + N)^{4}},  \nonu\\
c_ {574} & = & - \frac {16 (32 + 55 k + 18 k^{2} + 41 N + 35 k\, 
      N + 11 N^{2})} {(2 + N) (2 + k + N)^{4}},  \nonu\\
c_ {575} & = & - \frac {16 (2 + 2 k + k^{2} + 2 N + N^{2}) (32 + 
       55 k + 18 k^{2} + 41 N + 35 k\, 
      N + 11 N^{2})} {(2 + N) (2 + k + N)^{6}},  \nonu\\
c_ {576} & = &\frac {1} {3 (2 + N) (2 + k + N)^{4}}16 (216 + 436 k + 230 k^{2} + 36 k^{3} + 392 N + 
      473 k\, N + 112 k^{2} N 
      \nonu\\ & + & 209 N^{2} + 117 k\, 
     N^{2} + 35 N^{3}),  \nonu\\
c_ {577} & = & - \frac {16 (32 + 55 k + 18 k^{2} + 41 N + 35 k\, 
      N + 11 N^{2})} {(2 + N) (2 + k + N)^{4}},  \nonu\\
c_ {578} & = &\frac {16 (1 + N) (32 + 55 k + 18 k^{2} + 41 N + 35 k\, 
     N + 11 N^{2})} {(2 + N) (2 + k + N)^{5}},  \nonu\\
c_ {579} & = & - \frac {32 (1 + k) (1 + N) (32 + 55 k + 18 k^{2} + 
       41 N + 35 k\, N + 11 N^{2})} {(2 + N) (2 + k + N)^{6}},  \nonu\\
c_ {580} & = &\frac {16 (1 + k) (32 + 55 k + 18 k^{2} + 41 N + 35 k\, 
     N + 11 N^{2})} {(2 + N) (2 + k + N)^{5}},  \nonu\\
c_ {581} & = &\frac {1} {3 (2 + N) (2 + k + N)^{5}} 16 (120 + 44 k - 
    79 k^{2} - 34 k^{3} + 472 N + 307 k\, 
   N - 42 k^{2} N 
   \nonu\\ & - & 26 k^{3} N + 600 N^{2} + 352 k\, 
   N^{2} + 13 k^{2} N^{2} + 312 N^{3} + 113 k\, N^{3} + 56 N^{4}),  \nonu\\
c_ {582} & = &\frac {8 (-36 - 125 k - 46 k^{2} - 13 N - 61 k\, 
     N + 4 k^{2} N + 35 N^{2} + 12 k\, 
     N^{2} + 14 N^{3})} {3 (2 + N) (2 + k + N)^{4}},  \nonu\\
c_ {583} & = &\frac {1} {3 (2 + N) (2 + k + N)^{5}} 16 (60 + 15 k - 
    14 k^{2} - 4 k^{3} + 147 N + 100 k\, 
   N + 57 k^{2} N 
   \nonu\\ & + &  16 k^{3} N + 136 N^{2} + 93 k\, 
   N^{2} + 35 k^{2} N^{2} + 52 N^{3} + 20 k\, N^{3} + 7 N^{4}),  \nonu\\
c_ {584} & = &\frac {16 (-28 - 55 k - 18 k^{2} - 31 N - 35 k\, 
     N - 3 N^{2} + 2 N^{3})} {(2 + N) (2 + k + N)^{4}},  \nonu\\
c_ {585} & = &\frac {16 (72 + 61 k + 14 k^{2} + 83 N + 33 k\, 
     N + 25 N^{2})} {3 (2 + k + N)^{4}},  \nonu\\
c_ {586} & = &\frac {8 (16 + 21 k + 6 k^{2} + 19 N + 13 k\, 
     N + 5 N^{2})} {(2 + N) (2 + k + N)^{3}},  \nonu\\
c_ {587} & = &\frac {8 (92 + 125 k + 38 k^{2} + 185 N + 184 k\, 
     N + 34 k^{2} N + 122 N^{2} + 67 k\, 
     N^{2} + 25 N^{3})} {(2 + N) (2 + k + N)^{4}},  \nonu\\
c_ {588} & = &\frac {16 (12 - 56 k - 67 k^{2} - 18 k^{3} + 86 N + 
      41 k\, N - 5 k^{2} N + 98 N^{2} + 51 k\, 
     N^{2} + 26 N^{3})} {3 (2 + N) (2 + k + N)^{4}},  \nonu\\
c_ {589} & = & - \frac {16 (324 + 337 k + 47 k^{2} - 18 k^{3} + 
       617 N + 494 k\, N + 61 k^{2} N + 395 N^{2} + 183 k\, 
      N^{2} + 80 N^{3})} {3 (2 + N) (2 + k + N)^{4}},  \nonu\\
c_ {590} & = &\frac {8 (1 + N) (32 + 55 k + 18 k^{2} + 41 N + 35 k\, 
     N + 11 N^{2})} {(2 + N) (2 + k + N)^{5}},  \nonu\\
c_ {591} & = &\frac {8 (1 + N) (32 + 55 k + 18 k^{2} + 41 N + 35 k\, 
     N + 11 N^{2})} {(2 + N) (2 + k + N)^{5}},  \nonu\\
c_ {592} & = &\frac {16 (1 + N) (32 + 55 k + 18 k^{2} + 41 N + 35 k\, 
     N + 11 N^{2})} {(2 + N) (2 + k + N)^{5}},  \nonu\\
c_ {593} & = & - \frac {8 (1 + k) (32 + 55 k + 18 k^{2} + 41 N + 
       35 k\, N + 11 N^{2})} {(2 + N) (2 + k + N)^{5}},  \nonu\\
c_ {594} & = &\frac {16 (1 + k) (32 + 55 k + 18 k^{2} + 41 N + 35 k\, 
     N + 11 N^{2})} {(2 + N) (2 + k + N)^{5}},  \nonu\\
c_ {595} & = & - \frac {8 (1 + k) (32 + 55 k + 18 k^{2} + 41 N + 
       35 k\, N + 11 N^{2})} {(2 + N) (2 + k + N)^{5}},  \nonu\\
c_ {596} & = & - \frac {32 (-4 - 28 k - 25 k^{2} - 6 k^{3} + 18 N + 
       5 k\, N - 3 k^{2} N + 28 N^{2} + 15 k\, 
      N^{2} + 8 N^{3})} {(2 + N) (2 + k + N)^{4}},  \nonu\\
c_ {597} & = &\frac {8 (32 + 48 k + 29 k^{2} + 6 k^{3} + 64 N + 
      59 k\, N + 17 k^{2} N + 40 N^{2} + 17 k\, 
     N^{2} + 8 N^{3})} {(2 + N) (2 + k + N)^{4}},  \nonu\\
c_ {598} & = &\frac {1} {3 (2 + N) (2 + k + N)^{5}} 8 (168 + 510 k + 
    415 k^{2} + 98 k^{3} + 558 N + 1216 k\, 
   N + 723 k^{2} N 
   \nonu\\ & + &  112 k^{3} N + 565 N^{2} + 822 k\, 
   N^{2} + 266 k^{2} N^{2} + 217 N^{3} + 158 k\, N^{3} + 28 N^{4}),  \nonu\\
c_ {599} & = &\frac {1} {3 (2 + N) (2 + k + N)^{5}} 8 (240 + 180 k - 
    127 k^{2} - 152 k^{3} - 36 k^{4} + 660 N + 716 k\, 
   N 
   \nonu\\ & + &  189 k^{2} N - 10 k^{3} N + 683 N^{2} + 654 k\, 
   N^{2} + 154 k^{2} N^{2} + 293 N^{3} + 160 k\, N^{3} + 44 N^{4}),  \nonu\\
c_ {600} & = &\frac {16 (4 + 10 k + 17 k^{2} + 6 k^{3} + 32 N + 
      52 k\, N + 23 k^{2} N + 35 N^{2} + 28 k\, 
     N^{2} + 9 N^{3})} {(2 + N) (2 + k + N)^{4}},  \nonu\\
c_ {601} & = &\frac {16 (-k + N) (32 + 55 k + 18 k^{2} + 41 N + 
      35 k\, N + 11 N^{2})} {(2 + N) (2 + k + N)^{5}},  \nonu\\
c_ {602} & = &\frac {8 (32 + 55 k + 18 k^{2} + 41 N + 35 k\, 
     N + 11 N^{2})} {(2 + N) (2 + k + N)^{4}},  \nonu\\
c_ {603} & = &\frac {8 (32 + 55 k + 18 k^{2} + 41 N + 35 k\, 
     N + 11 N^{2})} {(2 + N) (2 + k + N)^{4}},  \nonu\\
c_ {604} & = &\frac {16 (32 + 55 k + 18 k^{2} + 41 N + 35 k\, 
     N + 11 N^{2})} {(2 + N) (2 + k + N)^{4}},  \nonu\\
c_ {605} & = &\frac {8 (32 + 55 k + 18 k^{2} + 41 N + 35 k\, 
     N + 11 N^{2})} {(2 + N) (2 + k + N)^{4}},  \nonu\\
c_ {606} & = & - \frac {16 (32 + 55 k + 18 k^{2} + 41 N + 35 k\, 
      N + 11 N^{2})} {(2 + N) (2 + k + N)^{4}},  \nonu\\
c_ {607} & = &\frac {8 (32 + 55 k + 18 k^{2} + 41 N + 35 k\, 
     N + 11 N^{2})} {(2 + N) (2 + k + N)^{4}},  \nonu\\
c_ {608} & = & - \frac {16 (32 + 55 k + 18 k^{2} + 41 N + 35 k\, 
      N + 11 N^{2})} {(2 + N) (2 + k + N)^{3}},  \nonu\\
c_ {609} & = &\frac {16 (32 + 55 k + 18 k^{2} + 41 N + 35 k\, 
     N + 11 N^{2})} {(2 + N) (2 + k + N)^{4}},  \nonu\\
c_ {610} & = &\frac {16 (20 + 15 k - 11 k^{2} - 6 k^{3} + 51 N + 
      43 k\, N + k^{2} N + 40 N^{2} + 22 k\, 
     N^{2} + 9 N^{3})} {(2 + N) (2 + k + N)^{4}},  \nonu\\
c_ {611} & = &\frac {16 (8 + 17 k + 6 k^{2} + 11 N + 11 k\, 
     N + 3 N^{2})} {(2 + N) (2 + k + N)^{3}},  \nonu\\
c_ {612} & = & - \frac {32 (10 - 2 k^{2} + 25 N + 17 k\, 
      N + 5 k^{2} N + 21 N^{2} + 11 k\, 
      N^{2} + 5 N^{3})} {(2 + N) (2 + k + N)^{4}},  \nonu\\
c_ {613} & = &\frac {16 (32 + 55 k + 18 k^{2} + 41 N + 35 k\, 
     N + 11 N^{2})} {(2 + N) (2 + k + N)^{4}},  \nonu\\
c_ {614} & = &\frac {16 (-k + N) (32 + 55 k + 18 k^{2} + 41 N + 
      35 k\, N + 11 N^{2})} {(2 + N) (2 + k + N)^{5}},  \nonu\\
c_ {615} & = & - \frac {16 (32 + 55 k + 18 k^{2} + 41 N + 35 k\, 
      N + 11 N^{2})} {(2 + N) (2 + k + N)^{4}},  \nonu\\
c_ {616} & = & - \frac {16 (-k + N) (32 + 55 k + 18 k^{2} + 41 N + 
       35 k\, N + 11 N^{2})} {(2 + N) (2 + k + N)^{5}},  \nonu\\
c_ {617} & = &\frac {1} {(2 + N) (2 + k + N)^{5}} 16 (248 + 478 k + 
    282 k^{2} + 39 k^{3} - 6 k^{4} + 550 N + 820 k\, 
   N 
   \nonu\\ & + & 357 k^{2} N + 35 k^{3} N + 430 N^{2} + 439 k\, 
   N^{2} + 107 k^{2} N^{2} + 137 N^{3} + 69 k\, N^{3} + 15 N^{4}), \nonu\\ 
c_ {618} & = &\frac {16 (-k + N) (32 + 55 k + 18 k^{2} + 41 N + 
      35 k\, N + 11 N^{2})} {(2 + N) (2 + k + N)^{4}},  \nonu\\
c_ {619} & = & - \frac {32 (20 + 15 k - 11 k^{2} - 6 k^{3} + 51 N + 
       43 k\, N + k^{2} N + 40 N^{2} + 22 k\, 
      N^{2} + 9 N^{3})} {(2 + N) (2 + k + N)^{4}},  \nonu\\
c_ {620} & = &\frac {32 (32 + 55 k + 18 k^{2} + 41 N + 35 k\, 
     N + 11 N^{2})} {(2 + N) (2 + k + N)^{4}},  \nonu\\
c_ {621} & = & - \frac {16 (36 + 36 k - 5 k^{2} - 6 k^{3} + 54 N + 
       22 k\, N - 11 k^{2} N + 23 N^{2} + 3 N^{3})} {(2 + 
       N) (2 + k + N)^{4}},  \nonu\\
c_ {622} & = &\frac {16 (60 + 28 k - 13 k^{2} - 6 k^{3} + 122 N + 
      78 k\, N + 9 k^{2} N + 87 N^{2} + 40 k\, 
     N^{2} + 19 N^{3})} {(2 + N) (2 + k + N)^{4}},  \nonu\\
c_ {623} & = &\frac {16 (8 + 17 k + 6 k^{2} + 11 N + 11 k\, 
     N + 3 N^{2})} {(2 + N) (2 + k + N)^{3}},  \nonu\\
c_ {624} & = &\frac {8 (20 + 21 k + 6 k^{2} + 25 N + 13 k\, 
     N + 7 N^{2})} {(2 + N) (2 + k + N)^{3}},  \qquad
c_ {625}  = \frac {64} {(2 + k + N)^{2}},  \nonu\\
c_ {626} & = &\frac {32 (10 + 23 k + 8 k^{2} + 18 N + 20 k\, 
     N + k^{2} N + 8 N^{2} + 3 k\, 
     N^{2} + N^{3})} {(2 + N) (2 + k + N)^{4}},  \nonu\\
c_ {627} & = &\frac {32 (1 + N) (-33 k - 42 k^{2} - 12 k^{3} + 17 N - 
      12 k\, N - 17 k^{2} N + 22 N^{2} + 7 k\, 
     N^{2} + 6 N^{3})} {(2 + N) (2 + k + N)^{5}},  \nonu\\
c_ {628} & = &\frac {1} {3 (2 + N) (2 + k + N)^{4}} 16 (120 - 110 k - 
    232 k^{2} - 72 k^{3} + 410 N - 31 k\, 
   N - 251 k^{2} N 
   \nonu\\ & - &  54 k^{3} N + 503 N^{2} + 120 k\, 
   N^{2} - 57 k^{2} N^{2} + 263 N^{3} + 63 k\, N^{3} + 48 N^{4}),  \nonu\\
c_ {629} & = &\frac {4 (180 + 127 k + 50 k^{2} + 347 N + 189 k\, 
     N + 52 k^{2} N + 229 N^{2} + 74 k\, 
     N^{2} + 48 N^{3})} {3 (2 + N) (2 + k + N)^{3}},  \nonu\\
c_ {630} & = & - \frac {16 (32 + 55 k + 18 k^{2} + 41 N + 35 k\, 
      N + 11 N^{2})} {(2 + N) (2 + k + N)^{4}},  \nonu\\
c_ {631} & = &\frac {16 (-36 - 52 k - 16 k^{2} - 14 N + 5 k\, 
     N + 10 k^{2} N + 23 N^{2} + 23 k\, 
     N^{2} + 9 N^{3})} {3 (2 + N) (2 + k + N)^{3}},  \nonu\\
c_ {632} & = & - \frac {16 (32 + 55 k + 18 k^{2} + 41 N + 35 k\, 
      N + 11 N^{2})} {(2 + N) (2 + k + N)^{4}},  \nonu\\
c_ {633} & = &\frac {16 (32 + 55 k + 18 k^{2} + 41 N + 35 k\, 
     N + 11 N^{2})} {(2 + N) (2 + k + N)^{4}},  \nonu\\
c_ {634} & = &\frac {1} {3 (2 + N) (2 + k + N)^{4}} 4 (312 + 566 k + 
    327 k^{2} + 58 k^{3} + 670 N + 1140 k\, 
   N + 599 k^{2} N 
   \nonu\\ & + &  92 k^{3} N + 537 N^{2} + 714 k\, 
   N^{2} + 226 k^{2} N^{2} + 177 N^{3} + 130 k\, N^{3} + 20 N^{4}), \nonu\\ 
c_ {635} & = &\frac {1} {3 (2 + N) (2 + k + N)^{5}} 16 (360 + 760 k + 
    529 k^{2} + 157 k^{3} + 18 k^{4} + 788 N + 1265 k\, 
   N 
   \nonu\\ & + &  606 k^{2} N + 95 k^{3} N + 594 N^{2} + 635 k\, 
   N^{2} + 155 k^{2} N^{2} + 174 N^{3} + 88 k\, N^{3} + 16 N^{4}),  \nonu\\
c_ {636} & = &\frac {8 (1 + k) (32 + 55 k + 18 k^{2} + 41 N + 35 k\, 
     N + 11 N^{2})} {(2 + N) (2 + k + N)^{5}},  \nonu\\
c_ {637} & = &\frac {8 (3 + k + 2 N) (32 + 55 k + 18 k^{2} + 41 N + 
      35 k\, N + 11 N^{2})} {(2 + N) (2 + k + N)^{5}},  \nonu\\
c_ {638} & = &\frac {16 (1 + k) (32 + 55 k + 18 k^{2} + 41 N + 35 k\, 
     N + 11 N^{2})} {(2 + N) (2 + k + N)^{5}},  \nonu\\
c_ {639} & = & - \frac {8 (1 + N) (32 + 55 k + 18 k^{2} + 41 N + 
       35 k\, N + 11 N^{2})} {(2 + N) (2 + k + N)^{5}},  \nonu\\
c_ {640} & = &\frac {16 (1 + N) (32 + 55 k + 18 k^{2} + 41 N + 35 k\, 
     N + 11 N^{2})} {(2 + N) (2 + k + N)^{5}},  \nonu\\
c_ {641} & = & - \frac {16 (1 + k) (1 + N) (32 + 55 k + 18 k^{2} + 
       41 N + 35 k\, N + 11 N^{2})} {(2 + N) (2 + k + N)^{6}},  \nonu\\
c_ {642} & = & - \frac {8 (3 + 2 k + N) (32 + 55 k + 18 k^{2} + 
       41 N + 35 k\, N + 11 N^{2})} {(2 + N) (2 + k + N)^{5}},  \nonu\\
c_ {643} & = & - \frac {8 (84 + 128 k + 40 k^{2} + 178 N + 199 k\, 
      N + 38 k^{2} N + 121 N^{2} + 75 k\, 
      N^{2} + 25 N^{3})} {3 (2 + N) (2 + k + N)^{4}},  \nonu\\
c_ {644} & = & - \frac {1} {(2 + N) (2 + k + N)^{5}} 8 (72 + 124 k + 
     50 k^{2} + 4 k^{3} + 256 N + 442 k\, 
    N + 221 k^{2} N
    \nonu\\ & + &  38 k^{3} N + 280 N^{2} + 354 k\, 
    N^{2} + 101 k^{2} N^{2} + 113 N^{3} + 74 k\, N^{3} + 15 N^{4}), \nonu\\ 
c_ {645} & = &\frac {16 (1 + k) (32 + 55 k + 18 k^{2} + 41 N + 35 k\, 
     N + 11 N^{2})} {(2 + N) (2 + k + N)^{5}},  \nonu\\
c_ {646} & = & - \frac {16 (2 + 2 k + k^{2} + 2 N + N^{2}) (32 + 
       55 k + 18 k^{2} + 41 N + 35 k\, 
      N + 11 N^{2})} {(2 + N) (2 + k + N)^{6}},  \nonu\\
c_ {647} & = &\frac {16 (1 + N) (32 + 55 k + 18 k^{2} + 41 N + 35 k\, 
     N + 11 N^{2})} {(2 + N) (2 + k + N)^{5}},  \nonu\\
c_ {648} & = & - \frac {16 (1 + k) (1 + N) (32 + 55 k + 18 k^{2} + 
       41 N + 35 k\, N + 11 N^{2})} {(2 + N) (2 + k + N)^{6}},  \nonu\\
c_ {649} & = & - \frac {1} {3 (2 + N) (2 + k + N)^{6}} 16 (12 - 
     229 k - 549 k^{2} - 378 k^{3} - 80 k^{4} + 235 N - 239 k\, 
    N 
    \nonu\\ & - &  894 k^{2} N - 520 k^{3} N - 76 k^{4} N + 500 N^{2} + 93 k\, 
    N^{2} - 417 k^{2} N^{2} - 158 k^{3} N^{2} + 399 N^{3} 
    \nonu\\ & + &  151 k\,N^{3} - 48 k^{2} N^{3} + 134 N^{4} + 32 k\, N^{4} + 16 N^{5}),  \nonu\\
c_ {650} & = & - \frac {1} {3 (2 + N) (2 + k + N)^{5}} 16 (480 + 
     974 k + 641 k^{2} + 134 k^{3} + 1090 N + 1663 k\, 
    N 
    \nonu\\ & + &  735 k^{2} N + 76 k^{3} N + 912 N^{2} + 952 k\, 
    N^{2} + 220 k^{2} N^{2} + 339 N^{3} + 185 k\, N^{3} + 47 N^{4}), \nonu\\ 
c_ {651} & = &\frac {8 (204 + 394 k + 257 k^{2} + 54 k^{3} + 308 N + 
      395 k\, N + 133 k^{2} N + 140 N^{2} + 87 k\, 
     N^{2} + 20 N^{3})} {3 (2 + N) (2 + k + N)^{4}},  \nonu\\
c_ {652} & = &\frac {16 (1 + N) (32 + 55 k + 18 k^{2} + 41 N + 35 k\, 
     N + 11 N^{2})} {(2 + N) (2 + k + N)^{5}},  \nonu\\
c_ {653} & = &\frac {16 (1 + k) (32 + 55 k + 18 k^{2} + 41 N + 35 k\, 
     N + 11 N^{2})} {(2 + N) (2 + k + N)^{5}},  \nonu\\
c_ {654} & = &\frac {1} {3 (2 + N) (2 + k + N)^{5}} 16 (120 + 172 k + 
    49 k^{2} - 2 k^{3} + 344 N + 371 k\, 
   N + 54 k^{2} N 
   \nonu\\ & - &  10 k^{3} N + 408 N^{2} + 320 k\, 
   N^{2} + 29 k^{2} N^{2} + 216 N^{3} + 97 k\, N^{3} + 40 N^{4}),  \nonu\\
c_ {655} & = &\frac {1} {(2 + N) (2 + k + N)^{5}} 8 (-88 - 172 k - 
    102 k^{2} - 33 k^{3} - 6 k^{4} - 136 N - 174 k\, 
   N - 46 k^{2} N 
   \nonu\\ & - & 7 k^{3} N - 56 N^{2} - 17 k\, 
   N^{2} + 14 k^{2} N^{2} + 4 N^{3} + 15 k\, N^{3} + 4 N^{4}),  \nonu\\
c_ {656} & = &\frac {1} {3 (2 + N) (2 + k + N)^{6}} 16 (-192 - 
    862 k - 1224 k^{2} - 693 k^{3} - 134 k^{4} - 98 N - 884 k\, 
   N 
   \nonu\\ & - &  1281 k^{2} N - 607 k^{3} N - 76 k^{4} N + 380 N^{2} + 201 k\, 
   N^{2} - 183 k^{2} N^{2} - 98 k^{3} N^{2} + 429 N^{3} 
   \nonu\\ & + &  391 k\, 
   N^{3} + 78 k^{2} N^{3} + 149 N^{4} + 80 k\, N^{4} + 16 N^{5}),  \nonu\\
c_ {657} & = & - \frac {1} {3 (2 + N) (2 + k + N)^{5}} 4 (-2112 - 
     4568 k - 3106 k^{2} - 681 k^{3} - 6 k^{4} - 4744 N 
     \nonu\\ & - &  8340 k\, 
    N - 3831 k^{2} N - 111 k^{3} N + 132 k^{4} N - 3818 N^{2} - 
     5383 k\,N^{2} - 1553 k^{2} N^{2} 
    \nonu\\ & + & 90 k^{3} N^{2} - 1145 N^{3} - 1329 k\, 
    N^{3} - 216 k^{2} N^{3} + 23 N^{4} - 54 k\, N^{4} + 48 N^{5}), \nonu\\ 
c_ {658} & = &\frac {1} {(2 + N) (2 + k + N)^{4}} 16 (-120 - 226 k - 
    148 k^{2} - 32 k^{3} - 202 N - 269 k\, 
   N - 113 k^{2} N 
   \nonu\\ & - &  10 k^{3} N - 79 N^{2} - 56 k\, 
   N^{2} - 11 k^{2} N^{2} + 13 N^{3} + 13 k\, N^{3} + 8 N^{4}),  \nonu\\
c_ {659} & = & - \frac {24 (16 + 21 k + 6 k^{2} + 19 N + 13 k\, 
      N + 5 N^{2})} {(2 + N) (2 + k + N)^{2}},  \nonu\\
c_ {660} & = & - \frac {1} {3 (2 + N) (2 + k + N)^{4}} 16 (-24 - 
     70 k - 80 k^{2} - 24 k^{3} + 10 N - 83 k\, 
    N - 127 k^{2} N \nonu \\
& - & 30 k^{3} N + 115 N^{2} + 36 k\, 
    N^{2} - 33 k^{2} N^{2} + 103 N^{3} + 39 k\, N^{3} + 24 N^{4}), \nonu\\ 
c_ {661} & = & - \frac {8 (-12 + 31 k + 18 k^{2} - N + 23 k\, 
      N + N^{2})} {3 (2 + N) (2 + k + N)^{3}},  \nonu\\
c_ {662} & = & - \frac {8 (-32 + 10 k + 51 k^{2} + 18 k^{3} + 6 N + 
       100 k\, N + 57 k^{2} N + 41 N^{2} + 56 k\, 
      N^{2} + 13 N^{3})} {(2 + N) (2 + k + N)^{4}},  \nonu\\
c_ {663} & = &\frac {32 (32 + 55 k + 18 k^{2} + 41 N + 35 k\, 
     N + 11 N^{2})} {(2 + N) (2 + k + N)^{4}},  \nonu\\
c_ {664} & = &\frac {16 (-k + N) (32 + 55 k + 18 k^{2} + 41 N + 
      35 k\, N + 11 N^{2})} {(2 + N) (2 + k + N)^{5}},  \nonu\\
c_ {665} & = & - \frac {32 (32 + 55 k + 18 k^{2} + 41 N + 35 k\, 
      N + 11 N^{2})} {(2 + N) (2 + k + N)^{4}},  \nonu\\
c_ {666} & = & - \frac {16 (-k + N) (32 + 55 k + 18 k^{2} + 41 N + 
       35 k\, N + 11 N^{2})} {(2 + N) (2 + k + N)^{5}},  \nonu\\
c_ {667} & = &\frac {16 (124 + 219 k + 100 k^{2} + 12 k^{3} + 235 N + 
      265 k\, N + 58 k^{2} N + 139 N^{2} + 78 k\, 
     N^{2} + 26 N^{3})} {(2 + N) (2 + k + N)^{4}},  \nonu\\
c_ {668} & = & - \frac {16 (32 + 55 k + 18 k^{2} + 41 N + 35 k\, 
      N + 11 N^{2})} {(2 + N) (2 + k + N)^{4}},  \nonu\\
c_ {669} & = &\frac {16 (32 + 55 k + 18 k^{2} + 41 N + 35 k\, 
     N + 11 N^{2})} {(2 + N) (2 + k + N)^{4}},  \nonu\\
c_ {670} & = & - \frac {16 (132 + 181 k + 54 k^{2} + 161 N + 113 k\, 
      N + 43 N^{2})} {3 (2 + N) (2 + k + N)^{3}},  \nonu\\
c_ {671} & = &\frac {16 (1 + k) (1 + N) (32 + 55 k + 18 k^{2} + 
      41 N + 35 k\, N + 11 N^{2})} {(2 + N) (2 + k + N)^{5}},  \nonu\\
c_ {672} & = &\frac {1} {(2 + N) (2 + k + N)^{5}} 16 (128 + 276 k + 
    233 k^{2} + 101 k^{3} + 18 k^{4} + 300 N + 452 k\,N 
   \nonu\\ & + & 220 k^{2} N + 43 k^{3} N + 275 N^{2} + 265 k\, 
   N^{2} + 53 k^{2} N^{2} + 118 N^{3} + 59 k\, N^{3} + 19 N^{4}),  \nonu\\
c_ {673} & = &\frac {16 (32 + 55 k + 18 k^{2} + 41 N + 35 k\, 
     N + 11 N^{2})} {(2 + N) (2 + k + N)^{4}},  \nonu\\
c_ {674} & = &\frac {8 (108 + 173 k + 54 k^{2} + 133 N + 109 k\, 
     N + 35 N^{2})} {3 (2 + N) (2 + k + N)^{3}},  \nonu\\
c_ {675} & = & - \frac {16 (252 + 323 k + 90 k^{2} + 295 N + 199 k\, 
      N + 77 N^{2})} {3 (2 + N) (2 + k + N)^{3}},  \nonu\\
c_ {676} & = & - \frac {1} {(2 + N) (2 + k + N)^{5}} 16 (-32 - 81 k - 
     66 k^{2} - 16 k^{3} - 47 N - 117 k\, 
    N - 83 k^{2} N 
    \nonu\\ & - &  14 k^{3} N - 9 N^{2} - 41 k\, 
    N^{2} - 23 k^{2} N^{2} + 12 N^{3} + k\, N^{3} + 4 N^{4}),  \nonu\\
c_ {677} & = & - \frac {16 (20 + 21 k + 6 k^{2} + 25 N + 13 k\, 
      N + 7 N^{2})} {(2 + N) (2 + k + N)^{3}},  \nonu\\
c_ {678} & = &\frac {8 (-32 + 42 k + 93 k^{2} + 30 k^{3} - 26 N + 
      96 k\, N + 71 k^{2} N + 3 N^{2} + 40 k\, 
     N^{2} + 3 N^{3})} {(2 + N) (2 + k + N)^{4}},  \nonu\\
c_ {679} & = &\frac {16 (-4 - 36 k - 29 k^{2} - 6 k^{3} + 26 N + 
      14 k\, N + k^{2} N + 39 N^{2} + 24 k\, 
     N^{2} + 11 N^{3})} {(2 + N) (2 + k + N)^{4}},  \nonu\\
c_ {680} & = &\frac {1} {3 (2 + N) (2 + k + N)^{3}} 8 (432 + 731 k + 
    376 k^{2} + 60 k^{3} + 961 N + 1297 k\, 
   N + 527 k^{2} N 
   \nonu\\ & + &  66 k^{3} N + 745 N^{2} + 717 k\, 
   N^{2} + 171 k^{2} N^{2} + 232 N^{3} + 117 k\, N^{3} + 24 N^{4}), \nonu\\ 
c_ {681} & = & - \frac {8 (32 + 55 k + 18 k^{2} + 41 N + 35 k\, 
      N + 11 N^{2})} {(2 + N) (2 + k + N)^{4}},  \nonu\\
c_ {682} & = &\frac {8 (32 + 55 k + 18 k^{2} + 41 N + 35 k\, 
     N + 11 N^{2})} {(2 + N) (2 + k + N)^{4}},  \nonu\\
c_ {683} & = &\frac {8 (48 + 96 k + 32 k^{2} + 168 N + 191 k\, 
     N + 34 k^{2} N + 137 N^{2} + 79 k\, 
     N^{2} + 31 N^{3})} {3 (2 + N) (2 + k + N)^{3}},  \nonu\\
c_ {684} & = &\frac {1} {3 (2 + N) (2 + k + N)^{4}} 8 (-72 + 266 k + 
    375 k^{2} + 106 k^{3} + 106 N + 852 k\, 
   N + 644 k^{2} N 
   \nonu\\ & + & 98 k^{3} N + 345 N^{2} + 756 k\, 
   N^{2} + 265 k^{2} N^{2} + 234 N^{3} + 202 k\, N^{3} + 47 N^{4}),  \nonu\\
c_ {685} & = &\frac {1} {3 (2 + N) (2 + k + N)^{5}} 16 (-120 - 
    454 k - 447 k^{2} - 200 k^{3} - 36 k^{4} - 158 N - 421 k\, 
   N 
   \nonu\\ & - & 234 k^{2} N - 52 k^{3} N + 40 N^{2} + 40 k\, 
   N^{2} + 33 k^{2} N^{2} + 94 N^{3} + 67 k\, N^{3} + 24 N^{4}),  \nonu\\
c_ {686} & = &\frac{1} {(2 + N) (2 + k + N)^{5}}32 (26 + 25 k + 6 k^{2} + 82 N + 52 k\, 
     N - 7 k^{2} N - 6 k^{3} N + 91 N^{2} 
     \nonu\\ & + &  43 k\,N^{2} - 3 k^{2} N^{2} + 45 N^{3} + 14 k\, 
     N^{3} + 8 N^{4}),  \nonu\\
c_ {687} & = &\frac {16 (-k + N) (32 + 55 k + 18 k^{2} + 41 N + 
      35 k\, N + 11 N^{2})} {(2 + N) (2 + k + N)^{5}},  \nonu\\
c_ {688} & = & - \frac {1} {3 (2 + N) (2 + k + N)^{5}} 16 (192 + 
     250 k + 167 k^{2} + 42 k^{3} + 422 N + 417 k\, 
    N + 204 k^{2} N 
    \nonu\\ & + & 30 k^{3} N + 376 N^{2} + 266 k\, 
    N^{2} + 73 k^{2} N^{2} + 160 N^{3} + 63 k\, N^{3} + 26 N^{4}),  \nonu\\
c_ {689} & = & - \frac {1} {3 (2 + N) (2 + k + N)^{5}} 16 (-276 - 
     746 k - 551 k^{2} - 122 k^{3} - 412 N - 1099 k\, 
    N 
    \nonu\\ & - & 678 k^{2} N - 106 k^{3} N - 72 N^{2} - 394 k\, 
    N^{2} - 175 k^{2} N^{2} + 120 N^{3} + 7 k\, N^{3} + 40 N^{4}), \nonu\\ 
c_ {690} & = & - \frac {1} {3 (2 + N) (2 + k + N)^{5}} 8 (384 + 
     2052 k + 2398 k^{2} + 1085 k^{3} + 174 k^{4} + 1380 N 
     \nonu\\ & + & 4894 k\, 
    N + 4197 k^{2} N + 1339 k^{3} N + 132 k^{4} N + 1744 N^{2} + 
     3999 k\, 
    N^{2} + 2171 k^{2} N^{2} 
    \nonu\\ & + &354 k^{3} N^{2} + 1075 N^{3} + 
     1379 k\, N^{3} + 324 k^{2} N^{3} + 349 N^{4} + 186 k\, 
    N^{4} + 48 N^{5}),  \nonu\\
c_ {691} & = & - \frac {16 (140 + 165 k + 33 k^{2} - 6 k^{3} + 
       249 N + 191 k\, N + 17 k^{2} N + 144 N^{2} + 56 k\, 
      N^{2} + 27 N^{3})} {(2 + N) (2 + k + N)^{3}},  \nonu\\
c_ {692} & = &\frac {1} {3 (2 + N) (2 + k + N)^{4}} 16 (24 - 161 k - 
    254 k^{2} - 80 k^{3} + 89 N - 325 k\, 
   N - 390 k^{2} N 
   \nonu\\ & - & 76 k^{3} N + 129 N^{2} - 196 k\, 
   N^{2} - 142 k^{2} N^{2} + 78 N^{3} - 32 k\, N^{3} + 16 N^{4}),  \nonu\\
c_ {693} & = & - \frac {1} {3 (2 + N) (2 + k + N)^{4}} 16 (48 - 
     137 k - 240 k^{2} - 76 k^{3} + 245 N - 103 k\, 
    N - 274 k^{2} N 
    \nonu\\ & - &56 k^{3} N + 337 N^{2} + 28 k\, 
    N^{2} - 80 k^{2} N^{2} + 178 N^{3} + 26 k\, N^{3} + 32 N^{4}), \nonu\\ 
c_ {694} & = &\frac {16 (-12 - 24 k + 5 k^{2} + 6 k^{3} - 6 N - 4 k\, 
     N + 11 k^{2} N + 7 N^{2} + 6 k\, 
     N^{2} + 3 N^{3})} {(2 + N) (2 + k + N)^{4}},  \nonu\\
c_ {695} & = &\frac {16 (124 + 141 k + 25 k^{2} - 6 k^{3} + 217 N + 
      163 k\, N + 13 k^{2} N + 124 N^{2} + 48 k\, 
     N^{2} + 23 N^{3})} {(2 + N) (2 + k + N)^{4}},  \nonu\\
c_ {696} & = &\frac {8 (32 + 55 k + 18 k^{2} + 41 N + 35 k\, 
     N + 11 N^{2})} {(2 + N) (2 + k + N)^{4}},  \nonu\\
c_ {697} & = &\frac {1} {3 (2 + N) (2 + k + N)^{4}} 4 (-360 - 314 k + 
    83 k^{2} + 66 k^{3} - 274 N + 352 k\, 
   N + 599 k^{2} N
   \nonu\\ & + & 132 k^{3} N + 441 N^{2} + 934 k\, 
   N^{2} + 374 k^{2} N^{2} + 477 N^{3} + 354 k\, N^{3} + 112 N^{4}), \nonu\\ 
c_ {698} & = & - \frac {1} {3 (2 + N) (2 + k + N)^{4}} 4 (864 + 
     1008 k + 353 k^{2} + 30 k^{3} + 2088 N + 2060 k\, 
    N + 769 k^{2} N 
    \nonu\\ & + & 132 k^{3} N + 1751 N^{2} + 1156 k\, 
    N^{2} + 246 k^{2} N^{2} + 541 N^{3} + 138 k\, N^{3} + 48 N^{4}), \nonu\\ 
c_ {699} & = &\frac {8 (228 + 375 k + 161 k^{2} + 18 k^{3} + 483 N + 
      530 k\, N + 115 k^{2} N + 317 N^{2} + 181 k\, 
     N^{2} + 64 N^{3})} {3 (2 + N) (2 + k + N)^{3}},  \nonu\\
c_ {700} & = & - \frac {8 (-12 + 31 k + 18 k^{2} - N + 23 k\, 
      N + N^{2})} {3 (2 + N) (2 + k + N)^{3}},  \nonu\\
c_ {701} & = &\frac {8 (3 k + 2 k^{2} + 13 N + 15 k\, 
     N + 4 k^{2} N + 15 N^{2} + 8 k\, 
     N^{2} + 4 N^{3})} {(2 + N) (2 + k + N)^{4}},  \nonu\\
c_ {702} & = & - \frac {8 (32 + 55 k + 18 k^{2} + 41 N + 35 k\, 
      N + 11 N^{2})} {(2 + N) (2 + k + N)^{4}},  \nonu\\
c_ {703} & = & - \frac {8 (108 + 173 k + 54 k^{2} + 133 N + 109 k\, 
      N + 35 N^{2})} {3 (2 + N) (2 + k + N)^{3}},  \nonu\\
c_ {704} & = &\frac {8 (64 + 139 k + 89 k^{2} + 18 k^{3} + 101 N + 
      151 k\, N + 49 k^{2} N + 48 N^{2} + 38 k\, 
     N^{2} + 7 N^{3})} {(2 + N) (2 + k + N)^{4}},  \nonu\\
c_ {705} & = &\frac {8 (108 + 171 k + 58 k^{2} + 195 N + 177 k\, 
     N + 20 k^{2} N + 101 N^{2} + 42 k\, 
     N^{2} + 16 N^{3})} {3 (2 + N) (2 + k + N)^{3}},  \nonu\\
c_ {706} & = &\frac {8 (-132 - 277 k - 180 k^{2} - 36 k^{3} - 53 N - 
      93 k\, N - 42 k^{2} N + 81 N^{2} + 52 k\, 
     N^{2} + 32 N^{3})} {3 (2 + N) (2 + k + N)^{3}},  \nonu\\
c_ {707} & = & - \frac {16 (27 + 37 k + 10 k^{2} + 29 N + 21 k\, 
      N + 8 N^{2})} {3 (2 + k + N)^{3}},  \nonu\\
c_ {708} & = &\frac {1} {3 (2 + N) (2 + k + N)^{5}} 16 (72 + 116 k + 
    95 k^{2} + 26 k^{3} + 232 N + 313 k\, 
   N + 162 k^{2} N \nonu\\ & + & 22 k^{3} N + 300 N^{2} + 304 k\, 
   N^{2} + 79 k^{2} N^{2} + 168 N^{3} + 95 k\, N^{3} + 32 N^{4}),  \nonu\\
c_ {709} & = &\frac {8 (36 + 52 k + 16 k^{2} + 70 N + 75 k\, 
     N + 14 k^{2} N + 45 N^{2} + 27 k\, 
     N^{2} + 9 N^{3})} {(2 + N) (2 + k + N)^{4}},  \nonu\\
c_ {710} & = & - \frac {1} {(2 + N) (2 + k + N)^{5}} 8 (-152 - 
     276 k - 138 k^{2} - 20 k^{3} - 256 N - 362 k\, 
    N - 123 k^{2} N 
    \nonu\\ & - & 10 k^{3} N - 120 N^{2} - 110 k\, 
    N^{2} - 15 k^{2} N^{2} + N^{3} + 6 k\, N^{3} + 7 N^{4}),  \nonu\\
c_ {711} & = & - \frac {1} {3 (2 + N) (2 + k + N)^{5}} 16 (-168 - 
     440 k - 344 k^{2} - 80 k^{3} - 244 N - 517 k\, 
    N - 267 k^{2} N 
    \nonu\\ & - &22 k^{3} N - 87 N^{2} - 172 k\, 
    N^{2} - 49 k^{2} N^{2} + 3 N^{3} - 17 k\, N^{3} + 4 N^{4}),  \nonu\\
c_ {712} & = &\frac {8 (4 - 6 k - 17 k^{2} - 6 k^{3} + 16 N + 9 k\, 
     N - 5 k^{2} N + 16 N^{2} + 9 k\, 
     N^{2} + 4 N^{3})} {(2 + N) (2 + k + N)^{4}},  \nonu\\
c_ {713} & = & - \frac {1} {(2 + N) (2 + k + N)^{5}} 8 (232 + 472 k + 
     344 k^{2} + 93 k^{3} + 6 k^{4} + 404 N + 594 k\, 
    N 
    \nonu\\ & + & 276 k^{2} N + 31 k^{3} N + 234 N^{2} + 221 k\, 
    N^{2} + 50 k^{2} N^{2} + 54 N^{3} + 25 k\, N^{3} + 4 N^{4}),  \nonu\\
c_ {714} & = & - \frac {16 (32 + 55 k + 18 k^{2} + 41 N + 35 k\, 
      N + 11 N^{2})} {(2 + N) (2 + k + N)^{4}},  \nonu\\
c_ {715} & = &\frac {16 (32 + 55 k + 18 k^{2} + 41 N + 35 k\, 
     N + 11 N^{2})} {(2 + N) (2 + k + N)^{4}},  \nonu\\
c_ {716} & = & - \frac {64 (16 + 21 k + 6 k^{2} + 19 N + 13 k\, 
      N + 5 N^{2})} {(2 + N) (2 + k + N)^{3}},  \nonu\\
c_ {717} & = & - \frac {16 (64 + 67 k + 18 k^{2} + 109 N + 79 k\, 
      N + 12 k^{2} N + 63 N^{2} + 24 k\, 
      N^{2} + 12 N^{3})} {(2 + N) (2 + k + N)^{4}},  \nonu\\
c_ {718} & = &\frac {1} {3 (2 + N) (2 + k + N)^{5}} 16 (-48 - 260 k - 
    268 k^{2} - 72 k^{3} - 4 N - 219 k\, 
   N - 177 k^{2} N 
   \nonu\\ & - & 18 k^{3} N + 79 N^{2} - 16 k\, 
   N^{2} - 23 k^{2} N^{2} + 49 N^{3} + 9 k\, N^{3} + 8 N^{4}),  \nonu\\
c_ {719} & = &\frac {16 (1 + k) (48 + 148 k + 56 k^{2} + 68 N + 
      119 k\, N + 10 k^{2} N + 17 N^{2} + 15 k\, 
     N^{2} - N^{3})} {3 (2 + N) (2 + k + N)^{5}},  \nonu\\
c_ {720} & = & - \frac {1} {3 (2 + N) (2 + k + N)^{5}} 16 (-288 - 
     450 k - 206 k^{2} - 28 k^{3} - 282 N - 119 k\, 
    N 
    \nonu\\ & + & 177 k^{2} N + 70 k^{3} N + 91 N^{2} + 498 k\, 
    N^{2} + 395 k^{2} N^{2} + 60 k^{3} N^{2} + 163 N^{3} + 329 k\, 
    N^{3} 
    \nonu\\ & + & 126 k^{2} N^{3} + 40 N^{4} + 48 k\, N^{4}),  \nonu\\
c_ {721} & = & - \frac {64 (1 + k) (9 + 5 k + 10 N + 3 k\, 
      N + 3 N^{2})} {3 (2 + k + N)^{4}},  \nonu\\
c_ {722} & = & - \frac {1} {3 (2 + N) (2 + k + N)^{4}} 8 (-48 - 
     208 k - 220 k^{2} - 103 k^{3} - 18 k^{4} - 200 N - 472 k\, 
    N 
    \nonu\\ & - & 322 k^{2} N - 77 k^{3} N - 124 N^{2} - 195 k\, 
    N^{2} - 76 k^{2} N^{2} + 20 N^{3} + 11 k\, N^{3} + 16 N^{4}),  \nonu\\
c_ {723} & = & - \frac {8 (32 + 55 k + 18 k^{2} + 41 N + 35 k\, 
      N + 11 N^{2})} {(2 + N) (2 + k + N)^{4}},  \nonu\\
c_ {724} & = & - \frac {16 (-k + N) (32 + 55 k + 18 k^{2} + 41 N + 
       35 k\, N + 11 N^{2})} {(2 + N) (2 + k + N)^{5}},  \nonu\\
c_ {725} & = &\frac {16 (-k + N) (32 + 55 k + 18 k^{2} + 41 N + 
      35 k\, N + 11 N^{2})} {(2 + N) (2 + k + N)^{5}},  \nonu\\
c_ {726} & = & - \frac {32 (-78 - 169 k - 101 k^{2} - 18 k^{3} - 
       74 N - 104 k\, N - 31 k^{2} N + N^{2} + 3 k\, 
      N^{2} + 7 N^{3})} {3 (2 + N) (2 + k + N)^{4}},  \nonu\\
c_ {727} & = &\frac {1} {3 (2 + N) (2 + k + N)^{5}} 32 (78 + 203 k + 
    146 k^{2} + 32 k^{3} + 118 N + 220 k\, 
   N + 75 k^{2} N 
   \nonu\\ & - & 2 k^{3} N + 81 N^{2} + 97 k\, 
   N^{2} + 7 k^{2} N^{2} + 39 N^{3} + 26 k\, N^{3} + 8 N^{4}),  \nonu\\
c_ {728} & = &\frac {8 (144 + 252 k + 88 k^{2} + 204 N + 211 k\, 
     N + 26 k^{2} N + 85 N^{2} + 35 k\, 
     N^{2} + 11 N^{3})} {3 (2 + N) (2 + k + N)^{4}},  \nonu\\
c_ {729} & = &\frac {1} {3 (2 + N) (2 + k + N)^{5}} 8 (528 + 1176 k + 
    868 k^{2} + 200 k^{3} + 1056 N + 1840 k\, 
   N + 957 k^{2} N 
   \nonu\\ & + & 118 k^{3} N + 820 N^{2} + 1014 k\, 
   N^{2} + 287 k^{2} N^{2} + 301 N^{3} + 200 k\, N^{3} + 43 N^{4}),  \nonu\\
c_ {730} & = &\frac {16 (1 + N) (32 + 55 k + 18 k^{2} + 41 N + 35 k\, 
     N + 11 N^{2})} {(2 + N) (2 + k + N)^{5}},  \nonu\\
c_ {731} & = &\frac {16 (1 + k) (32 + 55 k + 18 k^{2} + 41 N + 35 k\, 
     N + 11 N^{2})} {(2 + N) (2 + k + N)^{5}},  \nonu\\
c_ {732} & = &\frac {32 (27 + 56 k + 20 k^{2} + 46 N + 40 k\, 
     N + 15 N^{2})} {3 (2 + k + N)^{4}},  \nonu\\
c_ {733} & = &\frac {8 (228 + 318 k + 139 k^{2} + 18 k^{3} + 444 N + 
      421 k\, N + 95 k^{2} N + 280 N^{2} + 137 k\, 
     N^{2} + 56 N^{3})} {3 (2 + N) (2 + k + N)^{4}},  \nonu\\
c_ {734} & = & - \frac {32 (36 + 116 k + 87 k^{2} + 18 k^{3} + 70 N + 
       135 k\, N + 51 k^{2} N + 42 N^{2} + 37 k\, 
      N^{2} + 8 N^{3})} {3 (2 + N) (2 + k + N)^{4}},  \nonu\\
c_ {735} & = & - \frac{1}{(2 + N) (2 + k + N)^{6}}16 (1 + k) (1 + N) (60 + 77 k + 22 k^{2} + 
       121 N + 115 k\, N + 20 k^{2} N
       \nonu\\ & + & 79 N^{2} + 42 k\, 
      N^{2} + 16 N^{3}),  \nonu\\
c_ {736} & = &\frac {1} {3 (2 + N) (2 + k + N)^{5}} 8 (216 + 216 k + 
    94 k^{2} + 59 k^{3} + 18 k^{4} + 732 N + 874 k\, 
   N 
   \nonu\\ & + & 450 k^{2} N + 109 k^{3} N + 772 N^{2} + 687 k\, 
   N^{2} + 194 k^{2} N^{2} + 304 N^{3} + 131 k\, N^{3} + 40 N^{4}),  \nonu\\
c_ {737} & = & - \frac {8 (56 + 138 k + 65 k^{2} + 6 k^{3} + 194 N + 
       254 k\, N + 59 k^{2} N + 161 N^{2} + 102 k\, 
      N^{2} + 37 N^{3})} {(2 + N) (2 + k + N)^{3}},  \nonu\\
c_ {738} & = & - \frac {8 (-64 - 58 k + 27 k^{2} + 18 k^{3} - 70 N - 
       12 k\, N + 33 k^{2} N - 15 N^{2} + 12 k\, 
      N^{2} + N^{3})} {(2 + N) (2 + k + N)^{4}},  \nonu\\
c_ {739} & = & - \frac {8 (64 + 61 k + 14 k^{2} + 83 N + 49 k\, 
      N + 4 k^{2} N + 33 N^{2} + 8 k\, 
      N^{2} + 4 N^{3})} {(2 + N) (2 + k + N)^{4}},  \nonu\\
c_ {740} & = & - \frac {8 (32 + 55 k + 18 k^{2} + 41 N + 35 k\, 
      N + 11 N^{2})} {(2 + N) (2 + k + N)^{4}},  \nonu\\
c_ {741} & = &\frac {8 (20 + 21 k + 6 k^{2} + 25 N + 13 k\, 
     N + 7 N^{2})} {(2 + N) (2 + k + N)^{3}},  \nonu\\
c_ {742} & = & - \frac {8 (64 + 162 k + 73 k^{2} + 6 k^{3} + 94 N + 
       148 k\, N + 27 k^{2} N + 35 N^{2} + 28 k\, 
      N^{2} + 3 N^{3})} {(2 + N) (2 + k + N)^{4}},  \nonu\\
c_ {743} & = &\frac {8 (120 + 183 k + 71 k^{2} + 6 k^{3} + 261 N + 
      279 k\, N + 59 k^{2} N + 178 N^{2} + 102 k\, 
     N^{2} + 37 N^{3})} {(2 + N) (2 + k + N)^{4}},  \nonu\\
c_ {744} & = &\frac {1} {3 (2 + N) (2 + k + N)^{4}} 8 (-192 - 10 k + 
    49 k^{2} + 6 k^{3} - 182 N + 16 k\, 
   N - 127 k^{2} N 
   \nonu\\ & - & 60 k^{3} N + 127 N^{2} + 126 k\, 
   N^{2} - 66 k^{2} N^{2} + 187 N^{3} + 78 k\, N^{3} + 48 N^{4}),  \nonu\\
c_ {745} & = &\frac {16 (32 + 55 k + 18 k^{2} + 41 N + 35 k\, 
     N + 11 N^{2})} {(2 + N) (2 + k + N)^{4}},  \nonu\\
c_ {746} & = & - \frac {1} {3 (2 + N) (2 + k + N)^{4}} 4 (-792 - 
     502 k + 467 k^{2} + 398 k^{3} + 72 k^{4} - 1598 N - 832 k\, 
    N 
    \nonu\\ & + & 407 k^{2} N + 184 k^{3} N - 991 N^{2} - 234 k\, 
    N^{2} + 134 k^{2} N^{2} - 151 N^{3} + 62 k\, N^{3} + 16 N^{4}),  \nonu\\
c_ {747} & = & - \frac {1} {3 (2 + N) (2 + k + N)^{5}} 16 (336 + 
     790 k + 562 k^{2} + 124 k^{3} + 950 N + 2021 k\,N 
    \nonu\\ & + & 1233 k^{2} N + 218 k^{3} N + 999 N^{2} + 1850 k\, 
    N^{2} + 875 k^{2} N^{2} + 96 k^{3} N^{2} + 459 N^{3} + 697 k\, 
    N^{3} 
    \nonu\\ & + & 198 k^{2} N^{3} + 76 N^{4} + 84 k\, N^{4}),  \nonu\\
c_ {748} & = & - \frac {1} {3 (2 + N) (2 + k + N)^{5}} 8 (192 + 
     540 k + 475 k^{2} + 122 k^{3} + 324 N + 562 k\, 
    N + 285 k^{2} N 
    \nonu\\ & + & 16 k^{3} N + 211 N^{2} + 180 k\, 
    N^{2} + 32 k^{2} N^{2} + 85 N^{3} + 32 k\, N^{3} + 16 N^{4}),  \nonu\\
c_ {749} & = & - \frac {16 (5 + 15 k + 6 k^{2} + 3 N + 7 k\, 
      N)} {(2 + k + N)^{3}},  \nonu\\
c_ {750} & = & - \frac {1} {3 (2 + N) (2 + k + N)^{4}} 4 (640 + 
     1512 k + 1003 k^{2} + 202 k^{3} + 1512 N + 3068 k\, 
    N 
    \nonu\\ & + & 1659 k^{2} N + 252 k^{3} N + 1489 N^{2} + 2384 k\, 
    N^{2} + 938 k^{2} N^{2} + 80 k^{3} N^{2} + 675 N^{3} + 754 k\, 
    N^{3}
    \nonu\\ & + & 168 k^{2} N^{3} + 112 N^{4} + 64 k\, N^{4}),  \nonu\\
c_ {751} & = & - \frac {36} {5},  \qquad
c_ {752}  = \frac {32 (k - N)} {15 (2 + k + N)}, \nonu\\ 
c_ {753} & = &\frac {4 (60 + 77 k + 22 k^{2} + 121 N + 115 k\, 
     N + 20 k^{2} N + 79 N^{2} + 42 k\, 
     N^{2} + 16 N^{3})} {(2 + N) (2 + k + N)^{2}},  \nonu\\
c_ {754} & = &\frac {4 (60 + 77 k + 22 k^{2} + 121 N + 115 k\, 
     N + 20 k^{2} N + 79 N^{2} + 42 k\, 
     N^{2} + 16 N^{3})} {(2 + N) (2 + k + N)^{2}},  \nonu\\
c_ {755} & = &\frac {32 (-k + N) (60 + 77 k + 22 k^{2} + 121 N + 
      115 k\, N + 20 k^{2} N + 79 N^{2} + 42 k\, 
     N^{2} + 16 N^{3})} {3 (2 + N) (2 + k + N)^{3}},  \nonu\\
c_ {756} & = & - \frac {32 (-k + N) (60 + 77 k + 22 k^{2} + 121 N + 
       115 k\, N + 20 k^{2} N + 79 N^{2} + 42 k\, 
      N^{2} + 16 N^{3})} {3 (2 + N) (2 + k + N)^{3}},  \nonu\\
c_ {757} & = &\frac {8 (60 + 77 k + 22 k^{2} + 121 N + 115 k\, 
     N + 20 k^{2} N + 79 N^{2} + 42 k\, 
     N^{2} + 16 N^{3})} {(2 + N) (2 + k + N)^{2}},  \nonu\\
c_ {758} & = &\frac {8 (60 + 77 k + 22 k^{2} + 121 N + 115 k\, 
     N + 20 k^{2} N + 79 N^{2} + 42 k\, 
     N^{2} + 16 N^{3})} {(2 + N) (2 + k + N)^{2}},  \nonu\\
c_ {759} & = &\frac {32 (32 + 55 k + 18 k^{2} + 41 N + 35 k\, 
     N + 11 N^{2})} {(2 + N) (2 + k + N)^{4}},  \nonu\\
c_ {760} & = & - \frac {1} {9 (2 + N) (2 + k + N)^{5}} 16 (1080 + 
     2106 k + 1578 k^{2} + 547 k^{3} + 74 k^{4} + 2898 N 
     \nonu\\ & + & 4386 k\,N + 2337 k^{2} N + 499 k^{3} N + 28 k^{4} N + 3036 N^{2} + 
     3315 k\,N^{2} + 1083 k^{2} N^{2} 
    \nonu\\ & + & 100 k^{3} N^{2} + 1577 N^{3} + 
     1099 k\, N^{3} + 156 k^{2} N^{3} + 413 N^{4} + 140 k\, 
    N^{4} + 44 N^{5}),  \nonu\\
c_ {761} & = &\frac {1} {3 (2 + N) (2 + k + N)^{5}} 16 (624 + 
    1480 k + 1169 k^{2} + 390 k^{3} + 48 k^{4} + 1280 N + 2068 k\, 
   N 
   \nonu\\ & + & 969 k^{2} N + 140 k^{3} N + 915 N^{2} + 880 k\, 
   N^{2} + 166 k^{2} N^{2} + 281 N^{3} + 118 k\, N^{3} + 32 N^{4}),  \nonu\\
c_ {762} & = & - \frac {16 (-k + N) (32 + 55 k + 18 k^{2} + 41 N + 
       35 k\, N + 11 N^{2})} {3 (2 + N) (2 + k + N)^{5}},  \nonu\\
c_ {763} & = &\frac {16 (32 + 55 k + 18 k^{2} + 41 N + 35 k\, 
     N + 11 N^{2})} {(2 + N) (2 + k + N)^{4}},  \nonu\\
c_ {764} & = & - \frac {32 (-k + N) (32 + 55 k + 18 k^{2} + 41 N + 
       35 k\, N + 11 N^{2})} {3 (2 + N) (2 + k + N)^{5}},  \nonu\\
c_ {765} & = & - \frac {32 (-k + N) (32 + 55 k + 18 k^{2} + 41 N + 
       35 k\, N + 11 N^{2})} {3 (2 + N) (2 + k + N)^{5}},  \nonu\\
c_ {766} & = & - \frac {16 (-k + N) (32 + 55 k + 18 k^{2} + 41 N + 
       35 k\, N + 11 N^{2})} {3 (2 + N) (2 + k + N)^{5}},  \nonu\\
c_ {767} & = & - \frac {16 (-k + N) (32 + 55 k + 18 k^{2} + 41 N + 
       35 k\, N + 11 N^{2})} {3 (2 + N) (2 + k + N)^{5}},  \nonu\\
c_ {768} & = &\frac {32 (-k + N) (32 + 55 k + 18 k^{2} + 41 N + 
      35 k\, N + 11 N^{2})} {3 (2 + N) (2 + k + N)^{5}},  \nonu\\
c_ {769} & = &\frac {32 (-k + N) (32 + 55 k + 18 k^{2} + 41 N + 
      35 k\, N + 11 N^{2})} {3 (2 + N) (2 + k + N)^{5}},  \nonu\\
c_ {770} & = & - \frac {16 (-k + N) (32 + 55 k + 18 k^{2} + 41 N + 
       35 k\, N + 11 N^{2})} {3 (2 + N) (2 + k + N)^{5}},  \nonu\\
c_ {771} & = & - \frac {16 (32 + 55 k + 18 k^{2} + 41 N + 35 k\, 
      N + 11 N^{2})} {(2 + N) (2 + k + N)^{4}},  \nonu\\
c_ {772} & = & - \frac {16 (62 k + 41 k^{2} + 6 k^{3} + 34 N + 
       114 k\, N + 35 k^{2} N + 37 N^{2} + 46 k\, 
      N^{2} + 9 N^{3})} {3 (2 + N) (2 + k + N)^{4}},  \nonu\\
c_ {773} & = & - \frac {1} {3 (2 + N) (2 + k + N)^{5}} 16 (528 + 
     820 k + 365 k^{2} + 33 k^{3} - 6 k^{4} + 1412 N + 1780 k\,N 
    \nonu\\ & + & 606 k^{2} N + 47 k^{3} N + 1383 N^{2} + 1249 k\, 
    N^{2} + 235 k^{2} N^{2} + 584 N^{3} + 283 k\, N^{3} + 89 N^{4}), \nonu\\ 
c_ {774} & = &\frac {32 (1 + k) (32 + 55 k + 18 k^{2} + 41 N + 35 k\, 
     N + 11 N^{2})} {(2 + N) (2 + k + N)^{5}},  \nonu\\
c_ {775} & = &\frac {32 (1 + N) (32 + 55 k + 18 k^{2} + 41 N + 35 k\, 
     N + 11 N^{2})} {(2 + N) (2 + k + N)^{5}},  \nonu\\
c_ {776} & = & - \frac {32 (1 + k) (1 + N) (32 + 55 k + 18 k^{2} + 
       41 N + 35 k\, N + 11 N^{2})} {(2 + N) (2 + k + N)^{6}},  \nonu\\
c_ {777} & = &\frac {32 (1 + N) (32 + 55 k + 18 k^{2} + 41 N + 35 k\, 
     N + 11 N^{2})} {(2 + N) (2 + k + N)^{5}},  \nonu\\
c_ {778} & = & - \frac {32 (2 + 2 k + k^{2} + 2 N + N^{2}) (32 + 
       55 k + 18 k^{2} + 41 N + 35 k\, 
      N + 11 N^{2})} {(2 + N) (2 + k + N)^{6}},  \nonu\\
c_ {779} & = &\frac {32 (1 + k) (32 + 55 k + 18 k^{2} + 41 N + 35 k\, 
     N + 11 N^{2})} {(2 + N) (2 + k + N)^{5}},  \nonu\\
c_ {780} & = & - \frac {32 (2 + 2 k + k^{2} + 2 N + N^{2}) (32 + 
       55 k + 18 k^{2} + 41 N + 35 k\, 
      N + 11 N^{2})} {(2 + N) (2 + k + N)^{6}},  \nonu\\
c_ {781} & = & - \frac {1} {3 (2 + N) (2 + k + N)^{5}} 32 (36 - 
     93 k^{2} - 38 k^{3} + 126 N - 4 k\, 
    N - 163 k^{2} N - 40 k^{3} N 
    \nonu\\ & + & 163 N^{2} + 12 k\,N^{2} - 64 k^{2} N^{2} + 87 N^{3} + 10 k\, N^{3} + 16 N^{4}),  \nonu\\
c_ {782} & = & - \frac {16 (96 + 158 k + 65 k^{2} + 6 k^{3} + 178 N + 
       210 k\, N + 47 k^{2} N + 109 N^{2} + 70 k\, 
      N^{2} + 21 N^{3})} {3 (2 + N) (2 + k + N)^{4}},  \nonu\\
c_ {783} & = & - \frac {1} {3 (2 + N) (2 + k + N)^{5}} 16 (1584 + 
     2888 k + 1584 k^{2} + 259 k^{3} - 6 k^{4} + 3808 N 
     \nonu\\ & + & 5232 k\,N + 1905 k^{2} N + 151 k^{3} N + 3384 N^{2} + 3165 k\, 
    N^{2} + 585 k^{2} N^{2} + 1319 N^{3} 
    \nonu\\ & + & 641 k\, N^{3} + 189 N^{4}), \nonu\\
 c_ {784} & = & - \frac {1} {3 (2 + N) (2 + k + N)^{5}} 16 (864 + 
     1780 k + 1075 k^{2} + 202 k^{3} + 2012 N + 3020 k\, 
    N 
    \nonu\\ & + &1155 k^{2} N + 92 k^{3} N + 1713 N^{2} + 1700 k\, 
    N^{2} + 314 k^{2} N^{2} + 639 N^{3} + 322 k\, N^{3} + 88 N^{4}),  \nonu\\
c_ {785} & = &\frac {16 (96 + 190 k + 133 k^{2} + 30 k^{3} + 146 N + 
      186 k\, N + 67 k^{2} N + 65 N^{2} + 38 k\, 
     N^{2} + 9 N^{3})} {3 (2 + N) (2 + k + N)^{4}},  \nonu\\
c_ {786} & = & - \frac {16 (-94 k - 109 k^{2} - 30 k^{3} - 2 N - 
       90 k\, N - 55 k^{2} N + 7 N^{2} - 14 k\, 
      N^{2} + 3 N^{3})} {3 (2 + N) (2 + k + N)^{4}},  \nonu\\
c_ {787} & = &\frac {32 (3 + k + 2 N) (32 + 55 k + 18 k^{2} + 41 N + 
      35 k\, N + 11 N^{2})} {3 (2 + N) (2 + k + N)^{5}},  \nonu\\
c_ {788} & = &\frac {32 (3 + 2 k + N) (32 + 55 k + 18 k^{2} + 41 N + 
      35 k\, N + 11 N^{2})} {3 (2 + N) (2 + k + N)^{5}},  \nonu\\
c_ {789} & = &\frac {32 (3 + 2 k + N) (32 + 55 k + 18 k^{2} + 41 N + 
      35 k\, N + 11 N^{2})} {3 (2 + N) (2 + k + N)^{5}},  \nonu\\
c_ {790} & = &\frac {32 (3 + k + 2 N) (32 + 55 k + 18 k^{2} + 41 N + 
      35 k\, N + 11 N^{2})} {3 (2 + N) (2 + k + N)^{5}},  \nonu\\
c_ {791} & = &\frac {32 (3 + k + 2 N) (32 + 55 k + 18 k^{2} + 41 N + 
      35 k\, N + 11 N^{2})} {3 (2 + N) (2 + k + N)^{5}},  \nonu\\
c_ {792} & = & - \frac {64 (1 + k) (1 + N) (32 + 55 k + 18 k^{2} + 
       41 N + 35 k\, N + 11 N^{2})} {(2 + N) (2 + k + N)^{6}},  \nonu\\
c_ {793} & = &\frac {32 (3 + 2 k + N) (32 + 55 k + 18 k^{2} + 41 N + 
      35 k\, N + 11 N^{2})} {3 (2 + N) (2 + k + N)^{5}},  \nonu\\
c_ {794} & = &\frac {1} {3 (2 + N) (2 + k + N)^{5}} 32 (84 - 104 k - 
    254 k^{2} - 84 k^{3} + 494 N + 285 k\, 
   N - 113 k^{2} N 
   \nonu\\ & - & 30 k^{3} N + 731 N^{2} + 488 k\, 
   N^{2} + 27 k^{2} N^{2} + 399 N^{3} + 165 k\, N^{3} + 72 N^{4}),  \nonu\\
c_ {795} & = &\frac {1} {3 (2 + N) (2 + k + N)^{5}} 32 (120 + 108 k - 
    12 k^{2} - 16 k^{3} + 408 N + 349 k\, 
   N + 19 k^{2} N 
   \nonu\\ & - & 14 k^{3} N + 491 N^{2} + 336 k\, 
   N^{2} + 25 k^{2} N^{2} + 249 N^{3} + 101 k\, N^{3} + 44 N^{4}),  \nonu\\
c_ {796} & = &\frac {1} {3 (2 + N) (2 + k + N)^{5}} 16 (624 + 
    1304 k + 900 k^{2} + 265 k^{3} + 30 k^{4} + 1264 N + 1752 k\, 
   N 
   \nonu\\ & + & 639 k^{2} N + 61 k^{3} N + 924 N^{2} + 735 k\, 
   N^{2} + 75 k^{2} N^{2} + 305 N^{3} + 107 k\, N^{3} + 39 N^{4}),  \nonu\\
c_ {797} & = & - \frac {16 (-k + N) (32 + 55 k + 18 k^{2} + 41 N + 
       35 k\, N + 11 N^{2})} {3 (2 + N) (2 + k + N)^{5}},  \nonu\\
c_ {798} & = & - \frac {16 (-6 - 7 k + N) (32 + 55 k + 18 k^{2} + 
       41 N + 35 k\, N + 11 N^{2})} {3 (2 + N) (2 + k + N)^{5}},  \nonu\\
c_ {799} & = & - \frac {32 (-k + N) (32 + 55 k + 18 k^{2} + 41 N + 
       35 k\, N + 11 N^{2})} {3 (2 + N) (2 + k + N)^{5}},     \nonu\\
c_ {800} & = & - \frac {16 (-k + N) (32 + 55 k + 18 k^{2} + 41 N + 
       35 k\, N + 11 N^{2})} {3 (2 + N) (2 + k + N)^{5}}, \nonu\\
c_ {801} & = & \frac {32 (-k + N) (32 + 55 k + 18 k^{2} + 41 N + 
      35 k\, N + 11 N^{2})} {3 (2 + N) (2 + k + N)^{5}}, \nonu\\
c_ {802} & = & - \frac {32 (1 + k) (1 + N) (32 + 55 k + 18 k^{2} + 
       41 N + 35 k\, N + 11 N^{2})} {(2 + N) (2 + k + N)^{6}},\nonu\\ 
c_ {803} & = & - \frac {16 (6 - k + 7 N) (32 + 55 k + 18 k^{2} + 
       41 N + 35 k\, N + 11 N^{2})} {3 (2 + N) (2 + k + N)^{5}},\nonu\\ 
c_ {804} & = & - \frac {1} {3 (2 + N) (2 + k + N)^{5}} 16 (192 + 
     464 k + 413 k^{2} + 149 k^{3} + 18 k^{4} + 400 N + 748 k\, 
    N   
     \nonu\\ & + & 474 k^{2} N + 91 k^{3} N + 279 N^{2} + 361 k\, 
    N^{2} + 127 k^{2} N^{2} + 72 N^{3} + 47 k\, N^{3} + 5 N^{4}),\nonu\\ 
c_ {805} & = & \frac {1} {3 (2 + N) (2 + k + N)^{5}} 64 (-18 - 
    106 k - 121 k^{2} - 34 k^{3} + 43 N - 32 k\, 
   N - 66 k^{2} N 
    \nonu\\ & - & 8 k^{3} N + 120 N^{2} + 76 k\, 
   N^{2} + k^{2} N^{2} + 75 N^{3} + 32 k\, N^{3} + 14 N^{4}), \nonu\\
c_ {806} & = & - \frac {64 (-k + N) (32 + 55 k + 18 k^{2} + 41 N + 
       35 k\, N + 11 N^{2})} {3 (2 + N) (2 + k + N)^{5}}, \nonu\\
c_ {807} & = & \frac {1} {9 (2 + N) (2 + k + N)^{5}} 16 (3456 + 
    7296 k + 5217 k^{2} + 1522 k^{3} + 152 k^{4} + 9336 N 
     \nonu\\ & + & 15624 k\, 
   N + 7955 k^{2} N + 1274 k^{3} N + 4 k^{4} N + 9723 N^{2} + 
    12086 k\, 
   N^{2} + 3846 k^{2} N^{2} 
    \nonu\\ & + & 222 k^{3} N^{2} + 4897 N^{3} + 4012 k\, 
   N^{3} + 586 k^{2} N^{3} + 1192 N^{4} + 480 k\, N^{4} + 112 N^{5}), \nonu\\
c_ {808} & = & - \frac {1} {3 (2 + N) (2 + k + N)^{5}} 16 (-144 - 
     496 k - 349 k^{2} - 70 k^{3} - 200 N - 452 k\, 
    N - 93 k^{2} N 
     \nonu\\ & + & 28 k^{3} N - 39 N^{2} - 32 k\, 
    N^{2} + 58 k^{2} N^{2} + 27 N^{3} + 26 k\, N^{3} + 8 N^{4}), \nonu\\
c_ {809} & = & - \frac {16 (96 + 158 k + 65 k^{2} + 6 k^{3} + 178 N + 
       210 k\, N + 47 k^{2} N + 109 N^{2} + 70 k\, 
      N^{2} + 21 N^{3})} {3 (2 + N) (2 + k + N)^{4}}, \nonu\\
c_ {810} & = & - \frac {16 (62 k + 41 k^{2} + 6 k^{3} + 34 N + 
       114 k\, N + 35 k^{2} N + 37 N^{2} + 46 k\, 
      N^{2} + 9 N^{3})} {3 (2 + N) (2 + k + N)^{4}}, \nonu\\
c_ {811} & = & - \frac {16 (-94 k - 109 k^{2} - 30 k^{3} - 2 N - 
       90 k\, N - 55 k^{2} N + 7 N^{2} - 14 k\, 
      N^{2} + 3 N^{3})} {3 (2 + N) (2 + k + N)^{4}}, \nonu\\
c_ {812} & = & \frac {1} {3 (2 + N) (2 + k + N)^{5}} 16 (912 + 
    1748 k + 1139 k^{2} + 281 k^{3} + 18 k^{4} + 2212 N + 3316 k\, 
   N 
   \nonu\\ & + &1536 k^{2} N + 211 k^{3} N + 1953 N^{2} + 2029 k\, 
   N^{2} + 499 k^{2} N^{2} + 738 N^{3} + 395 k\, N^{3} + 101 N^{4}),\nonu\\ 
c_ {813} & = & - \frac {32 (1 + N) (32 + 55 k + 18 k^{2} + 41 N + 
       35 k\, N + 11 N^{2})} {(2 + N) (2 + k + N)^{5}}, \nonu\\
c_ {814} & = & - \frac {32 (1 + k) (32 + 55 k + 18 k^{2} + 41 N + 
       35 k\, N + 11 N^{2})} {(2 + N) (2 + k + N)^{5}}, \nonu\\
c_ {815} & = & - \frac {32 (1 + k) (32 + 55 k + 18 k^{2} + 41 N + 
       35 k\, N + 11 N^{2})} {(2 + N) (2 + k + N)^{5}}, \nonu\\
c_ {816} & = & - \frac {32 (1 + N) (32 + 55 k + 18 k^{2} + 41 N + 
       35 k\, N + 11 N^{2})} {(2 + N) (2 + k + N)^{5}}, \nonu\\
c_ {817} & = & \frac {16 (96 + 190 k + 133 k^{2} + 30 k^{3} + 146 N + 
      186 k\, N + 67 k^{2} N + 65 N^{2} + 38 k\, 
     N^{2} + 9 N^{3})} {3 (2 + N) (2 + k + N)^{4}}, \nonu\\
c_ {818} & = & \frac {1} {3 (2 + N) (2 + k + N)^{5}} 16 (816 + 
    1264 k + 552 k^{2} - k^{3} - 30 k^{4} + 2360 N + 3384 k\, 
   N  \nonu\\ & + & 1485 k^{2} N + 179 k^{3} N + 2424 N^{2} + 2601 k\, 
   N^{2} + 669 k^{2} N^{2} + 1027 N^{3} + 589 k\, N^{3} + 153 N^{4}), \nonu\\
c_ {819} & = & \frac {1} {3 (2 + N) (2 + k + N)^{5}} 16 (96 - 196 k - 
    443 k^{2} - 258 k^{3} - 48 k^{4} + 532 N + 500 k\, 
   N 
   \nonu\\ & + & 93 k^{2} N - 20 k^{3} N + 759 N^{2} + 788 k\, 
   N^{2} + 206 k^{2} N^{2} + 385 N^{3} + 230 k\, N^{3} + 64 N^{4}), \nonu\\
c_ {820} & = & - \frac {1} {3 (2 + N) (2 + k + N)^{5}} 16 (-144 - 
     320 k - 132 k^{2} + 5 k^{3} + 6 k^{4} - 184 N - 96 k\, 
    N 
    \nonu\\ & + & 219 k^{2} N + 89 k^{3} N - 36 N^{2} + 171 k\, 
    N^{2} + 159 k^{2} N^{2} + 13 N^{3} + 55 k\, N^{3} + 3 N^{4}), \nonu\\
c_ {821} & = & - \frac {1} {3 (2 + N) (2 + k + N)^{5}} 16 (192 + 
     464 k + 361 k^{2} + 99 k^{3} + 6 k^{4} + 400 N + 788 k\, 
    N
    \nonu\\ & + & 456 k^{2} N + 73 k^{3} N + 291 N^{2} + 419 k\, 
    N^{2} + 137 k^{2} N^{2} + 82 N^{3} + 65 k\, N^{3} + 7 N^{4}), \nonu\\
c_ {822} & = & \frac {1} {3 (2 + N) (2 + k + N)^{4}} 32 (-168 - 
    354 k - 252 k^{2} - 56 k^{3} - 258 N - 403 k\, 
   N - 205 k^{2} N 
   \nonu\\ & - & 22 k^{3} N - 65 N^{2} - 60 k\, 
   N^{2} - 25 k^{2} N^{2} + 45 N^{3} + 31 k\, N^{3} + 16 N^{4}), \nonu\\
c_ {823} & = & - \frac {1} {3 (2 + N) (2 + k + N)^{4}} 32 (72 - 
     78 k - 153 k^{2} - 46 k^{3} + 258 N - 29 k\, 
    N - 176 k^{2} N 
    \nonu\\ & - & 38 k^{3} N + 326 N^{2} + 84 k\, 
    N^{2} - 41 k^{2} N^{2} + 174 N^{3} + 47 k\, N^{3} + 32 N^{4}), \nonu\\
c_ {824} & = & \frac {32 (60 + 85 k + 39 k^{2} + 6 k^{3} + 113 N + 
      109 k\, N + 25 k^{2} N + 68 N^{2} + 34 k\, 
     N^{2} + 13 N^{3})} {3 (2 + N) (2 + k + N)^{4}}, \nonu\\
c_ {825} & = & - \frac {32 (-16 - 13 k + 11 k^{2} + 6 k^{3} - 11 N + 
       15 k\, N + 17 k^{2} N + 6 N^{2} + 14 k\, 
      N^{2} + 3 N^{3})} {(2 + N) (2 + k + N)^{4}}, \nonu\\
c_ {826} & = & \frac {32 (32 + 55 k + 18 k^{2} + 41 N + 35 k\, 
     N + 11 N^{2})} {(2 + N) (2 + k + N)^{4}}, \nonu\\
c_ {827} & = & - \frac {32 (32 + 55 k + 18 k^{2} + 41 N + 35 k\, 
      N + 11 N^{2})} {(2 + N) (2 + k + N)^{4}}, \nonu\\
c_ {828} & = & - \frac {32 (-k + N) (32 + 55 k + 18 k^{2} + 41 N + 
       35 k\, N + 11 N^{2})} {(2 + N) (2 + k + N)^{5}}, \nonu\\
c_ {829} & = & - \frac {32 (32 + 55 k + 18 k^{2} + 41 N + 35 k\, 
      N + 11 N^{2})} {(2 + N) (2 + k + N)^{4}}, \nonu\\
c_ {830} & = & \frac {32 (-k + N) (32 + 55 k + 18 k^{2} + 41 N + 
      35 k\, N + 11 N^{2})} {(2 + N) (2 + k + N)^{5}}, \nonu\\
c_ {831} & = & \frac {32 (32 + 55 k + 18 k^{2} + 41 N + 35 k\, 
     N + 11 N^{2})} {(2 + N) (2 + k + N)^{4}}, \nonu\\
c_ {832} & = & \frac {32 (-k + N) (32 + 55 k + 18 k^{2} + 41 N + 
      35 k\, N + 11 N^{2})} {(2 + N) (2 + k + N)^{5}}, \nonu\\
c_ {833} & = & \frac {64 (138 + 227 k + 87 k^{2} + 6 k^{3} + 262 N + 
      274 k\, N + 49 k^{2} N + 155 N^{2} + 81 k\, 
     N^{2} + 29 N^{3})} {3 (2 + N) (2 + k + N)^{4}}, \nonu\\
c_ {834} & = & - \frac {32 (32 + 55 k + 18 k^{2} + 41 N + 35 k\, 
      N + 11 N^{2})} {(2 + N) (2 + k + N)^{4}}, \nonu\\
c_ {835} & = & \frac {32 (32 + 55 k + 18 k^{2} + 41 N + 35 k\, 
     N + 11 N^{2})} {(2 + N) (2 + k + N)^{4}}, \nonu\\
c_ {836} & = & \frac {32 (32 + 55 k + 18 k^{2} + 41 N + 35 k\, 
     N + 11 N^{2})} {(2 + N) (2 + k + N)^{4}}, \nonu\\
c_ {837} & = & - \frac {32 (32 + 55 k + 18 k^{2} + 41 N + 35 k\, 
      N + 11 N^{2})} {(2 + N) (2 + k + N)^{4}}, \nonu\\
c_ {838} & = & \frac {32 (20 + 15 k - 11 k^{2} - 6 k^{3} + 51 N + 
      43 k\, N + k^{2} N + 40 N^{2} + 22 k\, 
     N^{2} + 9 N^{3})} {(2 + N) (2 + k + N)^{4}}, \nonu\\
c_ {839} & = & - \frac {32 (20 + 15 k - 11 k^{2} - 6 k^{3} + 51 N + 
       43 k\, N + k^{2} N + 40 N^{2} + 22 k\, 
      N^{2} + 9 N^{3})} {(2 + N) (2 + k + N)^{4}}, \nonu\\
c_ {840} & = & \frac {32 (32 + 55 k + 18 k^{2} + 41 N + 35 k\, 
     N + 11 N^{2})} {(2 + N) (2 + k + N)^{4}}, \nonu\\
c_ {841} & = & \frac {32 (32 + 55 k + 18 k^{2} + 41 N + 35 k\, 
     N + 11 N^{2})} {(2 + N) (2 + k + N)^{4}}, \nonu\\
c_ {842} & = & \frac {64 (12 - 41 k - 48 k^{2} - 12 k^{3} + 23 N - 
      41 k\, N - 26 k^{2} N + 17 N^{2} - 8 k\, 
     N^{2} + 4 N^{3})} {3 (2 + N) (2 + k + N)^{4}}, \nonu\\
c_ {843} & = & \frac {1} {3 (2 + N) (2 + k + N)^{5}} 64 (-k + 
    19 k^{2} + 36 k^{3} + 12 k^{4} + 49 N + 19 k\, 
   N - 24 k^{2} N + 2 k^{3} N 
   \nonu\\ & + & 106 N^{2} + 57 k\, 
   N^{2} - 13 k^{2} N^{2} + 75 N^{3} + 31 k\, N^{3} + 16 N^{4}), \nonu\\
c_ {844} & = & \frac {32 (20 + 21 k + 6 k^{2} + 25 N + 13 k\, 
     N + 7 N^{2})} {(2 + N) (2 + k + N)^{3}}, \nonu\\
c_ {845} & = & \frac {64 (12 - 41 k - 48 k^{2} - 12 k^{3} + 23 N - 
      41 k\, N - 26 k^{2} N + 17 N^{2} - 8 k\, 
     N^{2} + 4 N^{3})} {3 (2 + N) (2 + k + N)^{4}}, \nonu\\
c_ {846} & = & - \frac {64 (1 + N) (-33 k - 42 k^{2} - 12 k^{3} + 
       17 N - 12 k\, N - 17 k^{2} N + 22 N^{2} + 7 k\, 
      N^{2} + 6 N^{3})} {(2 + N) (2 + k + N)^{5}}, \nonu\\
c_ {847} & = & \frac {1} {3 (2 + N) (2 + k + N)^{5}} 64 (-k + 
    19 k^{2} + 36 k^{3} + 12 k^{4} + 49 N + 19 k\, 
   N - 24 k^{2} N + 2 k^{3} N 
   \nonu\\ & + & 106 N^{2} + 57 k\, 
   N^{2} - 13 k^{2} N^{2} + 75 N^{3} + 31 k\, N^{3} + 16 N^{4}), \nonu\\
c_ {848} & = & - \frac {32 (-k + N) (32 + 55 k + 18 k^{2} + 41 N + 
       35 k\, N + 11 N^{2})} {(2 + N) (2 + k + N)^{5}}, \nonu\\
c_ {849} & = & \frac {32 (60 + 85 k + 39 k^{2} + 6 k^{3} + 113 N + 
      109 k\, N + 25 k^{2} N + 68 N^{2} + 34 k\, 
     N^{2} + 13 N^{3})} {3 (2 + N) (2 + k + N)^{4}}, \nonu\\
c_ {850} & = & \frac {32 (60 + 85 k + 39 k^{2} + 6 k^{3} + 113 N + 
      109 k\, N + 25 k^{2} N + 68 N^{2} + 34 k\, 
     N^{2} + 13 N^{3})} {3 (2 + N) (2 + k + N)^{4}}, \nonu\\
c_ {851} & = & - \frac{1}{3 (2 + N) (2 + k + N)^{4}}64 (108 + 211 k + 126 k^{2} + 24 k^{3} + 
       203 N + 259 k\, N + 76 k^{2} N 
       \nonu\\ & + & 119 N^{2} + 76 k\, 
      N^{2} + 22 N^{3}), \nonu\\
c_ {852} & = & - \frac {32 (20 + 21 k + 6 k^{2} + 25 N + 13 k\, 
      N + 7 N^{2})} {(2 + N) (2 + k + N)^{3}}, \nonu\\
c_ {853} & = & - \frac{1} {3 (2 + N) (2 + k + N)^{4}}64 (108 + 211 k + 126 k^{2} + 24 k^{3} + 
       203 N + 259 k\, N + 76 k^{2} N 
       \nonu\\ & + & 119 N^{2} + 76 k\, 
      N^{2} + 22 N^{3}), \nonu\\
c_ {854} & = & - \frac {64 (1 + N) (-33 k - 42 k^{2} - 12 k^{3} + 
       17 N - 12 k\, N - 17 k^{2} N + 22 N^{2} + 7 k\, 
      N^{2} + 6 N^{3})} {(2 + N) (2 + k + N)^{5}}, \nonu\\
c_ {855} & = & \frac {32 (60 + 85 k + 39 k^{2} + 6 k^{3} + 113 N + 
      109 k\, N + 25 k^{2} N + 68 N^{2} + 34 k\, 
     N^{2} + 13 N^{3})} {3 (2 + N) (2 + k + N)^{4}}, \nonu\\
c_ {856} & = & \frac {32 (32 + 55 k + 18 k^{2} + 41 N + 35 k\, 
     N + 11 N^{2})} {(2 + N) (2 + k + N)^{4}}, \nonu\\
c_ {857} & = & \frac {32 (-k + N) (32 + 55 k + 18 k^{2} + 41 N + 
      35 k\, N + 11 N^{2})} {(2 + N) (2 + k + N)^{5}}, \nonu\\
c_ {858} & = & - \frac {32 (32 + 55 k + 18 k^{2} + 41 N + 35 k\, 
      N + 11 N^{2})} {(2 + N) (2 + k + N)^{4}}, \nonu\\
c_ {859} & = & - \frac {32 (-k + N) (32 + 55 k + 18 k^{2} + 41 N + 
       35 k\, N + 11 N^{2})} {(2 + N) (2 + k + N)^{5}}, \nonu\\
c_ {860} & = & \frac {64 (60 + 77 k + 22 k^{2} + 121 N + 115 k\, 
     N + 20 k^{2} N + 79 N^{2} + 42 k\, 
     N^{2} + 16 N^{3})} {(2 + N) (2 + k + N)^{4}}, \nonu\\
c_ {861} & = & - \frac {32 (32 + 55 k + 18 k^{2} + 41 N + 35 k\, 
      N + 11 N^{2})} {(2 + N) (2 + k + N)^{4}}, \nonu\\
c_ {862} & = & \frac {32 (32 + 55 k + 18 k^{2} + 41 N + 35 k\, 
     N + 11 N^{2})} {(2 + N) (2 + k + N)^{4}}, \nonu\\
c_ {863} & = & \frac {32 (-16 - 13 k + 11 k^{2} + 6 k^{3} - 11 N + 
      15 k\, N + 17 k^{2} N + 6 N^{2} + 14 k\, 
     N^{2} + 3 N^{3})} {(2 + N) (2 + k + N)^{4}}, \nonu\\
c_ {864} & = & \frac {64 (42 + 4 k - 21 k^{2} - 6 k^{3} + 101 N + 
      71 k\, N + 11 k^{2} N + 82 N^{2} + 45 k\, 
     N^{2} + 19 N^{3})} {3 (2 + N) (2 + k + N)^{4}}, \nonu\\
c_ {865} & = & \frac {1} {9 (2 + N) (2 + k + N)^{5}} 16 (1344 + 
    3624 k + 3483 k^{2} + 1418 k^{3} + 208 k^{4} + 4488 N 
    \nonu\\ & + & 10776 k\, 
   N + 8946 k^{2} N + 3014 k^{3} N + 344 k^{4} N + 5877 N^{2} + 
    12012 k\, 
   N^{2} + 7989 k^{2} N^{2} 
   \nonu\\ & + &1928 k^{3} N^{2} + 120 k^{4} N^{2} + 
    3736 N^{3} + 6122 k\, 
   N^{3} + 2898 k^{2} N^{3} + 372 k^{3} N^{3} + 1147 N^{4} 
   \nonu\\ & + & 1366 k\, 
   N^{4} + 348 k^{2} N^{4} + 136 N^{5} + 96 k\, N^{5}), \nonu\\
c_ {866} & = & - \frac {32 (32 + 55 k + 18 k^{2} + 41 N + 35 k\, 
      N + 11 N^{2})} {(2 + N) (2 + k + N)^{4}}, \nonu\\
c_ {867} & = & - \frac {1} {3 (2 + N) (2 + k + N)^{4}} 16 (288 + 
     460 k + 243 k^{2} + 42 k^{3} + 740 N + 1028 k\, 
    N + 451 k^{2} N 
    \nonu\\ & + & 60 k^{3} N + 697 N^{2} + 716 k\, 
    N^{2} + 178 k^{2} N^{2} + 279 N^{3} + 154 k\, N^{3} + 40 N^{4}),\nonu\\ 
c_ {868} & = & \frac {1} {3 (2 + N) (2 + k + N)^{5}} 16 (288 + 
    436 k + 344 k^{2} + 187 k^{3} + 42 k^{4} + 932 N + 1142 k\, 
   N 
   \nonu\\ & + & 721 k^{2} N + 337 k^{3} N + 60 k^{4} N + 1262 N^{2} + 1117 k\, 
   N^{2} + 401 k^{2} N^{2} + 102 k^{3} N^{2} + 891 N^{3}
   \nonu\\ & + & 513 k\, 
   N^{3} + 60 k^{2} N^{3} + 327 N^{4} + 102 k\, N^{4} + 48 N^{5}), \nonu\\
c_ {869} & = & \frac {1} {3 (2 + N) (2 + k + N)^{5}} 16 (120 + 
    270 k + 204 k^{2} + 61 k^{3} + 6 k^{4} + 246 N + 450 k\, 
   N 
   \nonu\\ & + &244 k^{2} N + 39 k^{3} N + 174 N^{2} + 235 k\, 
   N^{2} + 70 k^{2} N^{2} + 48 N^{3} + 37 k\, N^{3} + 4 N^{4}), \nonu\\
c_ {870} & = & \frac {1} {3 (2 + N) (2 + k + N)^{5}} 16 (216 + 
    478 k + 329 k^{2} + 57 k^{3} - 6 k^{4} + 374 N + 484 k\, 
   N 
   \nonu\\ & + & 111 k^{2} N - 31 k^{3} N + 255 N^{2} + 169 k\, 
   N^{2} - 14 k^{2} N^{2} + 107 N^{3} + 43 k\, N^{3} + 20 N^{4}), \nonu\\
c_ {871} & = & - \frac {32 (3 + 2 k + N) (32 + 55 k + 18 k^{2} + 
       41 N + 35 k\, N + 11 N^{2})} {3 (2 + N) (2 + k + N)^{5}}, \nonu\\
c_ {872} & = & - \frac {32 (3 + k + 2 N) (32 + 55 k + 18 k^{2} + 
       41 N + 35 k\, N + 11 N^{2})} {3 (2 + N) (2 + k + N)^{5}}, \nonu\\
c_ {873} & = & - \frac {32 (3 + k + 2 N) (32 + 55 k + 18 k^{2} + 
       41 N + 35 k\, N + 11 N^{2})} {3 (2 + N) (2 + k + N)^{5}}, \nonu\\
c_ {874} & = & - \frac {32 (3 + 2 k + N) (32 + 55 k + 18 k^{2} + 
       41 N + 35 k\, N + 11 N^{2})} {3 (2 + N) (2 + k + N)^{5}}, \nonu\\
c_ {875} & = & \frac {1} {3 (2 + N) (2 + k + N)^{5}} 64 (48 + 115 k + 
    78 k^{2} + 16 k^{3} + 101 N + 207 k\, 
   N + 107 k^{2} N 
   \nonu\\ & + & 14 k^{3} N + 75 N^{2} + 119 k\, 
   N^{2} + 35 k^{2} N^{2} + 22 N^{3} + 21 k\, N^{3} + 2 N^{4}), \nonu\\
c_ {876} & = & - \frac {1} {3 (2 + N) (2 + k + N)^{6}} 32 (60 + 
     201 k + 261 k^{2} + 154 k^{3} + 32 k^{4} + 117 N + 305 k\, 
    N 
    \nonu\\ & + & 336 k^{2} N + 176 k^{3} N + 28 k^{4} N + 106 N^{2} + 141 k\, 
    N^{2} + 93 k^{2} N^{2} + 38 k^{3} N^{2} + 77 N^{3} 
    \nonu\\ & + & 31 k\, 
    N^{3} - 6 k^{2} N^{3} + 40 N^{4} + 10 k\, N^{4} + 8 N^{5}), \nonu\\
c_ {877} & = & \frac {1} {3 (2 + N) (2 + k + N)^{5}} 64 (48 + 115 k + 
    78 k^{2} + 16 k^{3} + 101 N + 207 k\, 
   N + 107 k^{2} N 
   \nonu\\ & + & 14 k^{3} N + 75 N^{2} + 119 k\, 
   N^{2} + 35 k^{2} N^{2} + 22 N^{3} + 21 k\, N^{3} + 2 N^{4}), \nonu\\
c_ {878} & = & - \frac {1} {3 (2 + N) (2 + k + N)^{5}} 16 (120 + 
     190 k + 96 k^{2} + 16 k^{3} + 326 N + 402 k\, 
    N + 143 k^{2} N 
    \nonu\\ & + &14 k^{3} N + 330 N^{2} + 284 k\, 
    N^{2} + 53 k^{2} N^{2} + 145 N^{3} + 66 k\, N^{3} + 23 N^{4}), \nonu\\
c_ {879} & = & - \frac {1} {3 (2 + N) (2 + k + N)^{5}} 16 (120 + 
     190 k + 96 k^{2} + 16 k^{3} + 326 N + 402 k\, 
    N + 143 k^{2} N 
    \nonu\\ & + & 14 k^{3} N + 330 N^{2} + 284 k\, 
    N^{2} + 53 k^{2} N^{2} + 145 N^{3} + 66 k\, N^{3} + 23 N^{4}), \nonu\\
c_ {880} & = & \frac {1} {3 (2 + N) (2 + k + N)^{5}} 64 (-48 - 
    147 k - 162 k^{2} - 74 k^{3} - 12 k^{4} - 69 N - 183 k\, 
   N 
   \nonu\\ & - & 149 k^{2} N - 36 k^{3} N - 15 N^{2} - 53 k\, 
   N^{2} - 29 k^{2} N^{2} + 12 N^{3} + k\, N^{3} + 4 N^{4}), \nonu\\
c_ {881} & = & - \frac {32 (-k + N) (32 + 55 k + 18 k^{2} + 41 N + 
       35 k\, N + 11 N^{2})} {3 (2 + N) (2 + k + N)^{5}}, \nonu\\
c_ {882} & = & - \frac {32 (-k + N) (32 + 55 k + 18 k^{2} + 41 N + 
       35 k\, N + 11 N^{2})} {3 (2 + N) (2 + k + N)^{5}}, \nonu\\
c_ {883} & = & \frac {1} {3 (2 + N) (2 + k + N)^{5}} 64 (-48 - 
    147 k - 162 k^{2} - 74 k^{3} - 12 k^{4} - 69 N - 183 k\, 
   N 
   \nonu\\ & - &149 k^{2} N - 36 k^{3} N - 15 N^{2} - 53 k\, 
   N^{2} - 29 k^{2} N^{2} + 12 N^{3} + k\, N^{3} + 4 N^{4}), \nonu\\
c_ {884} & = & \frac {1} {3 (2 + N) (2 + k + N)^{6}} 64 (1 + N) (60 + 
    137 k + 165 k^{2} + 106 k^{3} + 24 k^{4} + 121 N 
    \nonu\\ & + & 136 k\, 
   N + 75 k^{2} N + 30 k^{3} N + 113 N^{2} + 53 k\, 
   N^{2} - 6 k^{2} N^{2} + 60 N^{3} + 18 k\, N^{3} + 12 N^{4}), \nonu\\
c_ {885} & = & - \frac {32 (-k + N) (32 + 55 k + 18 k^{2} + 41 N + 
       35 k\, N + 11 N^{2})} {3 (2 + N) (2 + k + N)^{5}}, \nonu\\
c_ {886} & = & \frac {1} {3 (2 + N) (2 + k + N)^{5}} 16 (120 + 
    270 k + 204 k^{2} + 61 k^{3} + 6 k^{4} + 246 N + 450 k\, 
   N 
   \nonu\\ & + & 244 k^{2} N + 39 k^{3} N + 174 N^{2} + 235 k\, 
   N^{2} + 70 k^{2} N^{2} + 48 N^{3} + 37 k\, N^{3} + 4 N^{4}), \nonu\\
c_ {887} & = & \frac {1} {3 (2 + N) (2 + k + N)^{5}} 16 (576 + 
    944 k + 432 k^{2} + 43 k^{3} - 6 k^{4} + 1072 N + 1056 k\, 
   N 
   \nonu\\ & + &87 k^{2} N - 65 k^{3} N + 720 N^{2} + 321 k\, 
   N^{2} - 81 k^{2} N^{2} + 221 N^{3} + 29 k\, N^{3} + 27 N^{4}), \nonu\\
c_ {888} & = & - \frac {32 (3 + 2 k + N) (32 + 55 k + 18 k^{2} + 
       41 N + 35 k\, N + 11 N^{2})} {3 (2 + N) (2 + k + N)^{5}}, \nonu\\
c_ {889} & = & - \frac {32 (3 + k + 2 N) (32 + 55 k + 18 k^{2} + 
       41 N + 35 k\, N + 11 N^{2})} {3 (2 + N) (2 + k + N)^{5}}, \nonu\\
c_ {890} & = & - \frac {1} {3 (2 + N) (2 + k + N)^{5}} 16 (24 + 
     110 k + 185 k^{2} + 62 k^{3} + 70 N + 292 k\, 
    N + 312 k^{2} N 
    \nonu\\ & + & 58 k^{3} N + 111 N^{2} + 280 k\, 
    N^{2} + 139 k^{2} N^{2} + 78 N^{3} + 86 k\, N^{3} + 17 N^{4}), \nonu\\
c_ {891} & = & \frac {32 (96 + 236 k + 153 k^{2} + 30 k^{3} + 196 N + 
      302 k\, N + 95 k^{2} N + 121 N^{2} + 92 k\, 
     N^{2} + 23 N^{3})} {3 (2 + N) (2 + k + N)^{4}}, \nonu\\
c_ {892} & = & - \frac {32 (96 + 223 k + 147 k^{2} + 30 k^{3} + 
       161 N + 251 k\, N + 83 k^{2} N + 82 N^{2} + 66 k\, 
      N^{2} + 13 N^{3})} {3 (2 + N) (2 + k + N)^{4}},\nonu\\ 
c_ {893} & = & - \frac {32 (32 + 55 k + 18 k^{2} + 41 N + 35 k\, 
      N + 11 N^{2})} {(2 + N) (2 + k + N)^{4}}, \nonu\\
c_ {894} & = & - \frac {32 (32 + 55 k + 18 k^{2} + 41 N + 35 k\, 
      N + 11 N^{2})} {(2 + N) (2 + k + N)^{4}}, \nonu\\
c_ {895} & = & - \frac {32 (32 + 35 k + 10 k^{2} + 61 N + 47 k\, 
      N + 8 k^{2} N + 39 N^{2} + 16 k\, 
      N^{2} + 8 N^{3})} {(2 + N) (2 + k + N)^{4}}, \nonu\\
c_ {896} & = & \frac {32 (12 + 7 k + 5 N) (16 + 21 k + 6 k^{2} + 
      19 N + 13 k\, N + 5 N^{2})} {3 (2 + N) (2 + k + N)^{4}}, \nonu\\
c_ {897} & = & \frac {32 (96 + 268 k + 195 k^{2} + 42 k^{3} + 164 N + 
      298 k\, N + 109 k^{2} N + 83 N^{2} + 76 k\, 
     N^{2} + 13 N^{3})} {3 (2 + N) (2 + k + N)^{4}}, \nonu\\
c_ {898} & = & \frac {32 (12 + 5 k + 7 N) (16 + 21 k + 6 k^{2} + 
      19 N + 13 k\, N + 5 N^{2})} {3 (2 + N) (2 + k + N)^{4}}, \nonu\\
c_ {899} & = & - \frac {32 (3 k + 2 k^{2} + 13 N + 15 k\, 
      N + 4 k^{2} N + 15 N^{2} + 8 k\, 
      N^{2} + 4 N^{3})} {(2 + N) (2 + k + N)^{4}}, \nonu\\
c_ {900} & = & - \frac {32 (96 + 223 k + 147 k^{2} + 30 k^{3} + 
       161 N + 251 k\, N + 83 k^{2} N + 82 N^{2} + 66 k\, 
      N^{2} + 13 N^{3})} {3 (2 + N) (2 + k + N)^{4}}, \nonu\\
c_ {901} & = & - \frac {32 (32 + 55 k + 18 k^{2} + 41 N + 35 k\, 
      N + 11 N^{2})} {(2 + N) (2 + k + N)^{4}}, \nonu\\
c_ {902} & = & \frac {32 (144 + 268 k + 157 k^{2} + 30 k^{3} + 
      284 N + 342 k\, N + 97 k^{2} N + 173 N^{2} + 104 k\, 
     N^{2} + 33 N^{3})} {3 (2 + N) (2 + k + N)^{4}}, \nonu\\
c_ {903} & = & - \frac {32 (60 + 77 k + 22 k^{2} + 121 N + 115 k\, 
      N + 20 k^{2} N + 79 N^{2} + 42 k\, 
      N^{2} + 16 N^{3})} {(2 + N) (2 + k + N)^{3}}, \nonu\\
c_ {904} & = & - \frac {32 (32 + 55 k + 18 k^{2} + 41 N + 35 k\, 
      N + 11 N^{2})} {(2 + N) (2 + k + N)^{4}}, \nonu\\
c_ {905} & = & \frac {1} {3 (2 + N) (2 + k + N)^{5}} 32 (432 + 
    802 k + 498 k^{2} + 100 k^{3} + 1178 N + 1905 k\, 
   N + 1001 k^{2} N
   \nonu\\ & + & 158 k^{3} N + 1179 N^{2} + 1574 k\, 
   N^{2} + 635 k^{2} N^{2} + 60 k^{3} N^{2} + 511 N^{3} + 513 k\, 
   N^{3} + 126 k^{2} N^{3} 
   \nonu\\ & + & 80 N^{4} + 48 k\, N^{4}), \nonu\\
c_ {906} & = & \frac {1} {3 (2 + N) (2 + k + N)^{5}} 32 (480 + 
    1100 k + 985 k^{2} + 426 k^{3} + 72 k^{4} + 1324 N + 2318 k\, 
   N 
   \nonu\\ & + & 1539 k^{2} N + 502 k^{3} N + 60 k^{4} N + 1557 N^{2} + 
    1880 k\, 
   N^{2} + 740 k^{2} N^{2} + 126 k^{3} N^{2} + 991 N^{3} 
   \nonu\\ & + & 734 k\, 
   N^{3} + 108 k^{2} N^{3} + 340 N^{4} + 126 k\, N^{4} + 48 N^{5}),\nonu\\ 
c_ {907} & = & \frac {1} {3 (2 + N) (2 + k + N)^{5}} 16 (120 + 
    254 k + 196 k^{2} + 61 k^{3} + 6 k^{4} + 262 N + 426 k\, 
   N 
   \nonu\\ & + & 232 k^{2} N + 39 k^{3} N + 206 N^{2} + 227 k\, 
   N^{2} + 66 k^{2} N^{2} + 68 N^{3} + 37 k\, N^{3} + 8 N^{4}), \nonu\\
c_ {908} & = & \frac {1} {3 (2 + N) (2 + k + N)^{5}} 16 (408 + 
    802 k + 420 k^{2} + 51 k^{3} - 6 k^{4} + 914 N + 1146 k\, 
   N 
   \nonu\\ & + & 244 k^{2} N - 31 k^{3} N + 750 N^{2} + 541 k\, 
   N^{2} + 10 k^{2} N^{2} + 280 N^{3} + 95 k\, N^{3} + 40 N^{4}), \nonu\\
c_ {909} & = & \frac {32 (1 + k) (32 + 55 k + 18 k^{2} + 41 N + 
      35 k\, N + 11 N^{2})} {(2 + N) (2 + k + N)^{5}}, \nonu\\
c_ {910} & = & - \frac {32 (2 + 2 k + k^{2} + 2 N + N^{2}) (32 + 
       55 k + 18 k^{2} + 41 N + 35 k\, 
      N + 11 N^{2})} {(2 + N) (2 + k + N)^{6}}, \nonu\\
c_ {911} & = & \frac {32 (1 + N) (32 + 55 k + 18 k^{2} + 41 N + 
      35 k\, N + 11 N^{2})} {(2 + N) (2 + k + N)^{5}}, \nonu\\
c_ {912} & = & - \frac {64 (1 + k) (1 + N) (32 + 55 k + 18 k^{2} + 
       41 N + 35 k\, N + 11 N^{2})} {(2 + N) (2 + k + N)^{6}}, \nonu\\
c_ {913} & = & - \frac {1} {3 (2 + N) (2 + k + N)^{5}} 32 (-84 - 
     384 k - 408 k^{2} - 112 k^{3} - 6 N - 533 k\, 
    N - 533 k^{2} N  
    \nonu\\ & - &98 k^{3} N + 179 N^{2} - 192 k\, 
    N^{2} - 167 k^{2} N^{2} + 147 N^{3} - k\, N^{3} + 32 N^{4}), \nonu\\
c_ {914} & = & - \frac {32 (3 + 2 k + N) (32 + 55 k + 18 k^{2} + 
       41 N + 35 k\, N + 11 N^{2})} {3 (2 + N) (2 + k + N)^{5}}, \nonu\\
c_ {915} & = & - \frac {32 (3 + k + 2 N) (32 + 55 k + 18 k^{2} + 
       41 N + 35 k\, N + 11 N^{2})} {3 (2 + N) (2 + k + N)^{5}}, \nonu\\
c_ {916} & = & \frac {64 (3 + 2 k + N) (9 k + 4 k^{2} + 7 N + 20 k\, 
     N + 5 k^{2} N + 8 N^{2} + 9 k\, 
     N^{2} + 2 N^{3})} {3 (2 + N) (2 + k + N)^{5}}, \nonu\\
c_ {917} & = & - \frac {1}{(2 + N) (2 + k + N)^{6}}32 (1 + k) (1 + N) (60 + 77 k + 22 k^{2} + 
       121 N + 115 k\, N + 20 k^{2} N 
       \nonu\\ & + & 79 N^{2} + 42 k\, 
      N^{2} + 16 N^{3}), \nonu\\
c_ {918} & = & - \frac {1} {3 (2 + N) (2 + k + N)^{6}} 64 (1 + 
     N) (60 + 137 k + 165 k^{2} + 106 k^{3} + 24 k^{4} + 121 N 
     \nonu\\ & + & 
     136 k\, N + 75 k^{2} N + 30 k^{3} N + 113 N^{2} + 53 k\, 
    N^{2} - 6 k^{2} N^{2} + 60 N^{3} + 18 k\, N^{3} + 12 N^{4}), \nonu\\
c_ {919} & = & - \frac {1} {3 (2 + N) (2 + k + N)^{5}} 16 (120 + 
     174 k + 72 k^{2} + 8 k^{3} + 342 N + 410 k\, 
    N + 131 k^{2} N
    \nonu\\ & + & 10 k^{3} N + 346 N^{2} + 300 k\, 
    N^{2} + 53 k^{2} N^{2} + 149 N^{3} + 70 k\, N^{3} + 23 N^{4}), \nonu\\
c_ {920} & = & \frac {1} {3 (2 + N) (2 + k + N)^{5}} 64 (-91 k - 
    146 k^{2} - 74 k^{3} - 12 k^{4} + 43 N - 75 k\, 
   N - 125 k^{2} N 
   \nonu\\ & - & 36 k^{3} N + 77 N^{2} + 11 k\, 
   N^{2} - 21 k^{2} N^{2} + 44 N^{3} + 13 k\, N^{3} + 8 N^{4}), \nonu\\
c_ {921} & = & \frac {1} {3 (2 + N) (2 + k + N)^{6}} 32 (60 + 201 k + 
    261 k^{2} + 154 k^{3} + 32 k^{4} + 117 N + 305 k\, 
   N 
   \nonu\\ & + & 336 k^{2} N + 176 k^{3} N + 28 k^{4} N + 106 N^{2} + 141 k\, 
   N^{2} + 93 k^{2} N^{2} + 38 k^{3} N^{2} + 77 N^{3} + 31 k\, 
   N^{3} 
   \nonu\\ & - & 6 k^{2} N^{3} + 40 N^{4} + 10 k\, N^{4} + 8 N^{5}), \nonu\\
c_ {922} & = & - \frac {32 (-k + N) (32 + 55 k + 18 k^{2} + 41 N + 
       35 k\, N + 11 N^{2})} {3 (2 + N) (2 + k + N)^{5}}, \nonu\\ 
c_ {923} & = & \frac {1} {3 (2 + N) (2 + k + N)^{5}} 16 (168 + 
    182 k - 126 k^{2} - 76 k^{3} + 502 N + 354 k\, 
   N - 203 k^{2} N 
   \nonu\\ & - & 62 k^{3} N + 432 N^{2} + 100 k\, 
   N^{2} - 119 k^{2} N^{2} + 119 N^{3} - 30 k\, N^{3} + 7 N^{4}),\nonu\\ 
c_ {924} & = & - \frac {1} {3 (2 + N) (2 + k + N)^{5}} 32 (-276 - 
     682 k - 481 k^{2} - 102 k^{3} - 476 N - 1047 k\, 
    N 
    \nonu\\ & - & 604 k^{2} N - 90 k^{3} N - 194 N^{2} - 410 k\, 
    N^{2} - 159 k^{2} N^{2} + 42 N^{3} - 9 k\, N^{3} + 24 N^{4}),\nonu\\ 
c_ {925} & = & - \frac{1} {3 (2 + N) (2 + k + N)^{3}}32 (96 + 232 k + 153 k^{2} + 30 k^{3} + 
       200 N + 296 k\, N + 95 k^{2} N 
       \nonu\\ & + &127 N^{2} + 90 k\, 
      N^{2} + 25 N^{3}), \nonu\\
c_ {926} & = & \frac {32 (96 + 252 k + 161 k^{2} + 30 k^{3} + 180 N + 
      310 k\, N + 99 k^{2} N + 105 N^{2} + 92 k\, 
     N^{2} + 19 N^{3})} {3 (2 + N) (2 + k + N)^{4}}, \nonu\\
c_ {927} & = & - \frac {32 (3 k + 2 k^{2} + 13 N + 15 k\, 
      N + 4 k^{2} N + 15 N^{2} + 8 k\, 
      N^{2} + 4 N^{3})} {(2 + N) (2 + k + N)^{4}}, \nonu\\
c_ {928} & = & - \frac {32 (96 + 223 k + 147 k^{2} + 30 k^{3} + 
       161 N + 251 k\, N + 83 k^{2} N + 82 N^{2} + 66 k\, 
      N^{2} + 13 N^{3})} {3 (2 + N) (2 + k + N)^{4}}, \nonu\\
c_ {929} & = & - \frac {32 (32 + 55 k + 18 k^{2} + 41 N + 35 k\, 
      N + 11 N^{2})} {(2 + N) (2 + k + N)^{4}}, \nonu\\
c_ {930} & = & \frac {32 (20 + 21 k + 6 k^{2} + 25 N + 13 k\, 
     N + 7 N^{2})} {(2 + N) (2 + k + N)^{3}}, \nonu\\
c_ {931} & = & \frac {32 (96 + 252 k + 187 k^{2} + 42 k^{3} + 180 N + 
      290 k\, N + 105 k^{2} N + 99 N^{2} + 76 k\, 
     N^{2} + 17 N^{3})} {3 (2 + N) (2 + k + N)^{4}}, \nonu\\
c_ {932} & = & \frac {1} {3 (2 + N) (2 + k + N)^{4}} 32 (120 + 
    260 k + 167 k^{2} + 34 k^{3} + 352 N + 524 k\, 
   N + 211 k^{2} N 
   \nonu\\ & + &20 k^{3} N + 329 N^{2} + 316 k\, 
   N^{2} + 62 k^{2} N^{2} + 123 N^{3} + 58 k\, N^{3} + 16 N^{4}),\nonu\\ 
c_ {933} & = & - \frac {32 (32 + 55 k + 18 k^{2} + 41 N + 35 k\, 
      N + 11 N^{2})} {(2 + N) (2 + k + N)^{4}}, \nonu\\
c_ {934} & = & - \frac {1} {3 (2 + N) (2 + k + N)^{4}} 16 (432 + 
     824 k + 483 k^{2} + 90 k^{3} + 1072 N + 1540 k\, 
    N + 611 k^{2} N 
    \nonu\\ & + & 60 k^{3} N + 977 N^{2} + 952 k\, 
    N^{2} + 194 k^{2} N^{2} + 387 N^{3} + 194 k\, N^{3} + 56 N^{4}),\nonu\\ 
c_ {935} & = & \frac {1} {3 (2 + N) (2 + k + N)^{5}} 32 (240 + 
    514 k + 354 k^{2} + 76 k^{3} + 698 N + 1281 k\, 
   N + 737 k^{2} N 
   \nonu\\ & + &122 k^{3} N + 747 N^{2} + 1118 k\, 
   N^{2} + 491 k^{2} N^{2} + 48 k^{3} N^{2} + 343 N^{3} + 381 k\, 
   N^{3} + 102 k^{2} N^{3} 
   \nonu\\ & + & 56 N^{4} + 36 k\, N^{4}), \nonu\\
c_ {936} & = & \frac {1} {3 (2 + N) (2 + k + N)^{5}} 16 (672 + 
    1636 k + 1498 k^{2} + 633 k^{3} + 102 k^{4} + 1844 N + 3430 k\, 
   N 
   \nonu\\ & + & 2261 k^{2} N + 651 k^{3} N + 60 k^{4} N + 2044 N^{2} + 
    2675 k\, 
   N^{2} + 1063 k^{2} N^{2} + 150 k^{3} N^{2} + 1187 N^{3} 
   \nonu\\ & + &971 k\, 
   N^{3} + 156 k^{2} N^{3} + 369 N^{4} + 150 k\, N^{4} + 48 N^{5}), \nonu\\
c_ {937} & = & \frac {1} {3 (2 + N) (2 + k + N)^{5}} 16 (192 + 
    308 k + 83 k^{2} - 6 k^{3} + 556 N + 638 k\, 
   N + 121 k^{2} N 
   \nonu\\ & + & 527 N^{2} + 364 k\, 
   N^{2} + 20 k^{2} N^{2} + 193 N^{3} + 52 k\, N^{3} + 24 N^{4}), \nonu\\
c_ {938} & = & - \frac {1} {3 (2 + N) (2 + k + N)^{5}} 32 (-120 - 
     288 k - 219 k^{2} - 50 k^{3} - 228 N - 481 k\, 
    N - 298 k^{2} N 
    \nonu\\ & - & 46 k^{3} N - 128 N^{2} - 228 k\, 
    N^{2} - 91 k^{2} N^{2} - 12 N^{3} - 23 k\, N^{3} + 4 N^{4}), \nonu\\
c_ {939} & = & \frac {16 (4 + 7 k + 2 k^{2} + 23 N + 47 k\, 
     N + 16 k^{2} N + 29 N^{2} + 28 k\, 
     N^{2} + 8 N^{3})} {(2 + N) (2 + k + N)^{3}}, \nonu\\
c_ {940} & = & \frac {8 (120 + 170 k + 65 k^{2} + 6 k^{3} + 226 N + 
      228 k\, N + 47 k^{2} N + 139 N^{2} + 76 k\, 
     N^{2} + 27 N^{3})} {3 (2 + N) (2 + k + N)^{3}}, \nonu\\
c_ {941} & = & - \frac {8 (6 k + k^{2} + 6 N + 10 k\, 
      N + N^{2}) (16 + 21 k + 6 k^{2} + 19 N + 13 k\, 
      N + 5 N^{2})} {3 (2 + N) (2 + k + N)^{4}}, \nonu\\
c_ {942} & = & \frac {32 (3 + 2 k + N) (20 + 17 k + 4 k^{2} + 49 N + 
      32 k\, N + 5 k^{2} N + 36 N^{2} + 13 k\, 
     N^{2} + 8 N^{3})} {3 (2 + N) (2 + k + N)^{4}},\nonu\\ 
c_ {943} & = & - \frac {32 (3 + 2 k + N) (20 + 17 k + 4 k^{2} + 
       49 N + 32 k\, N + 5 k^{2} N + 36 N^{2} + 13 k\, 
      N^{2} + 8 N^{3})} {3 (2 + N) (2 + k + N)^{4}},\nonu\\ 
c_ {944} & = & - \frac {32 (60 + 101 k + 60 k^{2} + 12 k^{3} + 97 N + 
       107 k\, N + 32 k^{2} N + 49 N^{2} + 26 k\, 
      N^{2} + 8 N^{3})} {3 (2 + N) (2 + k + N)^{4}},\nonu\\ 
c_ {945} & = & \frac {32 (48 + 97 k + 60 k^{2} + 12 k^{3} + 71 N + 
      101 k\, N + 32 k^{2} N + 31 N^{2} + 24 k\, 
     N^{2} + 4 N^{3})} {3 (2 + N) (2 + k + N)^{4}}, \nonu\\
c_ {946} & = & - \frac {32 (60 + 101 k + 60 k^{2} + 12 k^{3} + 97 N + 
       107 k\, N + 32 k^{2} N + 49 N^{2} + 26 k\, 
      N^{2} + 8 N^{3})} {3 (2 + N) (2 + k + N)^{4}},\nonu\\ 
c_ {947} & = & \frac {32 (48 + 97 k + 60 k^{2} + 12 k^{3} + 71 N + 
      101 k\, N + 32 k^{2} N + 31 N^{2} + 24 k\, 
     N^{2} + 4 N^{3})} {3 (2 + N) (2 + k + N)^{4}}, \nonu\\
c_ {948} & = & \frac {8 (120 + 170 k + 65 k^{2} + 6 k^{3} + 226 N + 
      228 k\, N + 47 k^{2} N + 139 N^{2} + 76 k\, 
     N^{2} + 27 N^{3})} {3 (2 + N) (2 + k + N)^{3}},\nonu\\ 
c_ {949} & = & \frac {1} {3 (2 + N) (2 + k + N)^{4}} 8 (576 + 944 k + 
    432 k^{2} + 43 k^{3} - 6 k^{4} + 1072 N + 1056 k\, 
   N \nonu\\ & + &  87 k^{2} N - 65 k^{3} N + 720 N^{2} + 321 k\, 
   N^{2} - 81 k^{2} N^{2} + 221 N^{3} + 29 k\, N^{3} + 27 N^{4}), \nonu\\
c_ {950} & = & \frac {32 (-49 k - 48 k^{2} - 12 k^{3} + N - 53 k\, 
     N - 26 k^{2} N + 5 N^{2} - 12 k\, 
     N^{2} + 2 N^{3})} {3 (2 + N) (2 + k + N)^{3}}, \nonu\\
c_ {951} & = & \frac {1} {3 (2 + N) (2 + k + N)^{4}} 32 (60 + 123 k + 
    78 k^{2} + 16 k^{3} + 135 N + 227 k\, 
   N + 107 k^{2} N 
   \nonu\\ & + &  14 k^{3} N + 109 N^{2} + 135 k\, 
   N^{2} + 35 k^{2} N^{2} + 36 N^{3} + 25 k\, N^{3} + 4 N^{4}), \nonu\\
c_ {952} & = & - \frac {1} {3 (2 + N) (2 + k + N)^{4}} 32 (60 + 
     123 k + 78 k^{2} + 16 k^{3} + 135 N + 227 k\, 
    N + 107 k^{2} N 
    \nonu\\ & + &  14 k^{3} N + 109 N^{2} + 135 k\, 
    N^{2} + 35 k^{2} N^{2} + 36 N^{3} + 25 k\, N^{3} + 4 N^{4}), \nonu\\
c_ {953} & = & - \frac {32 (60 + 93 k + 56 k^{2} + 12 k^{3} + 105 N + 
       103 k\, N + 30 k^{2} N + 57 N^{2} + 26 k\, 
      N^{2} + 10 N^{3})} {3 (2 + N) (2 + k + N)^{4}}, \nonu\\
c_ {954} & = & - \frac {32 (-49 k - 48 k^{2} - 12 k^{3} + N - 53 k\, 
      N - 26 k^{2} N + 5 N^{2} - 12 k\, 
      N^{2} + 2 N^{3})} {3 (2 + N) (2 + k + N)^{4}}, \nonu\\
c_ {955} & = & \frac {1} {3 (2 + N) (2 + k + N)^{4}} 16 (96 + 232 k + 
    143 k^{2} + 26 k^{3} + 200 N + 296 k\, 
   N + 71 k^{2} N 
   \nonu\\ & - &  8 k^{3} N + 137 N^{2} + 104 k\, 
   N^{2} - 6 k^{2} N^{2} + 39 N^{3} + 10 k\, N^{3} + 4 N^{4}), \nonu\\
c_ {956} & = & - \frac {1} {3 (2 + N) (2 + k + N)^{4}} 16 (672 + 
     1508 k + 1189 k^{2} + 398 k^{3} + 48 k^{4} + 1420 N + 2364 k\, 
    N 
    \nonu\\ & + & 1229 k^{2} N + 204 k^{3} N + 1103 N^{2} + 1212 k\, 
    N^{2} + 310 k^{2} N^{2} + 377 N^{3} + 206 k\, N^{3} + 48 N^{4}),\nonu\\ 
c_ {957} & = & \frac {1} {3 (2 + N) (2 + k + N)^{5}} 16 (120 + 
    190 k + 96 k^{2} + 16 k^{3} + 326 N + 402 k\, 
   N + 143 k^{2} N 
   \nonu\\ & + &  14 k^{3} N + 330 N^{2} + 284 k\, 
   N^{2} + 53 k^{2} N^{2} + 145 N^{3} + 66 k\, N^{3} + 23 N^{4}), \nonu\\
c_ {958} & = & - \frac {1} {3 (2 + N) (2 + k + N)^{5}} 16 (24 + 
     110 k + 185 k^{2} + 62 k^{3} + 70 N + 292 k\, 
    N + 312 k^{2} N 
    \nonu\\ & + &  58 k^{3} N + 111 N^{2} + 280 k\, 
    N^{2} + 139 k^{2} N^{2} + 78 N^{3} + 86 k\, N^{3} + 17 N^{4}), \nonu\\
c_ {959} & = & - \frac {1} {3 (2 + N) (2 + k + N)^{5}} 64 (48 + 
     115 k + 78 k^{2} + 16 k^{3} + 101 N + 207 k\, 
    N + 107 k^{2} N 
    \nonu\\ & + & 14 k^{3} N + 75 N^{2} + 119 k\, 
    N^{2} + 35 k^{2} N^{2} + 22 N^{3} + 21 k\, N^{3} + 2 N^{4}), \nonu\\
c_ {960} & = & - \frac {1} {3 (2 + N) (2 + k + N)^{5}} 64 (48 + 
     115 k + 78 k^{2} + 16 k^{3} + 101 N + 207 k\, 
    N + 107 k^{2} N 
    \nonu\\ & + &  14 k^{3} N + 75 N^{2} + 119 k\, 
    N^{2} + 35 k^{2} N^{2} + 22 N^{3} + 21 k\, N^{3} + 2 N^{4}), \nonu\\
c_ {961} & = & - \frac {1} {3 (2 + N) (2 + k + N)^{5}} 16 (120 + 
     270 k + 204 k^{2} + 61 k^{3} + 6 k^{4} + 246 N + 450 k\, 
    N 
    \nonu\\ & + &  244 k^{2} N + 39 k^{3} N + 174 N^{2} + 235 k\, 
    N^{2} + 70 k^{2} N^{2} + 48 N^{3} + 37 k\, N^{3} + 4 N^{4}), \nonu\\
c_ {962} & = & - \frac {1} {3 (2 + N) (2 + k + N)^{5}} 16 (120 + 
     270 k + 204 k^{2} + 61 k^{3} + 6 k^{4} + 246 N + 450 k\, 
    N 
    \nonu\\ & + & 244 k^{2} N + 39 k^{3} N + 174 N^{2} + 235 k\, 
    N^{2} + 70 k^{2} N^{2} + 48 N^{3} + 37 k\, N^{3} + 4 N^{4}), \nonu\\
c_ {963} & = & - \frac {1} {3 (2 + N) (2 + k + N)^{5}} 64 (-48 - 
     147 k - 162 k^{2} - 74 k^{3} - 12 k^{4} - 69 N - 183 k\, 
    N 
    \nonu\\ & - &  149 k^{2} N - 36 k^{3} N - 15 N^{2} - 53 k\, 
    N^{2} - 29 k^{2} N^{2} + 12 N^{3} + k\, N^{3} + 4 N^{4}), \nonu\\
c_ {964} & = & - \frac {1} {3 (2 + N) (2 + k + N)^{5}} 64 (-48 - 
     147 k - 162 k^{2} - 74 k^{3} - 12 k^{4} - 69 N - 183 k\, 
    N 
    \nonu\\ & - &  149 k^{2} N - 36 k^{3} N - 15 N^{2} - 53 k\, 
    N^{2} - 29 k^{2} N^{2} + 12 N^{3} + k\, N^{3} + 4 N^{4}), \nonu\\
c_ {965} & = & \frac {1} {3 (2 + N) (2 + k + N)^{5}} 16 (120 + 
    190 k + 96 k^{2} + 16 k^{3} + 326 N + 402 k\, 
   N + 143 k^{2} N 
   \nonu\\ & + &  14 k^{3} N + 330 N^{2} + 284 k\, 
   N^{2} + 53 k^{2} N^{2} + 145 N^{3} + 66 k\, N^{3} + 23 N^{4}), \nonu\\
c_ {966} & = & \frac {1} {3 (2 + N) (2 + k + N)^{5}} 16 (576 + 
    944 k + 432 k^{2} + 43 k^{3} - 6 k^{4} + 1072 N + 1056 k\, 
   N 
   \nonu\\ & + &  87 k^{2} N - 65 k^{3} N + 720 N^{2} + 321 k\, 
   N^{2} - 81 k^{2} N^{2} + 221 N^{3} + 29 k\, N^{3} + 27 N^{4}), \nonu\\
c_ {967} & = & \frac {1} {3 (2 + N) (2 + k + N)^{5}} 16 (216 + 
    478 k + 329 k^{2} + 57 k^{3} - 6 k^{4} + 374 N + 484 k\, 
   N 
   \nonu\\ & + &  111 k^{2} N - 31 k^{3} N + 255 N^{2} + 169 k\, 
   N^{2} - 14 k^{2} N^{2} + 107 N^{3} + 43 k\, N^{3} + 20 N^{4}), \nonu\\
c_ {968} & = & \frac {32 (96 + 268 k + 195 k^{2} + 42 k^{3} + 164 N + 
      298 k\, N + 109 k^{2} N + 83 N^{2} + 76 k\, 
     N^{2} + 13 N^{3})} {3 (2 + N) (2 + k + N)^{4}}, \nonu\\
c_ {969} & = & \frac {32 (32 + 55 k + 18 k^{2} + 41 N + 35 k\, 
     N + 11 N^{2})} {(2 + N) (2 + k + N)^{4}}, \nonu\\
c_ {970} & = & \frac {32 (32 + 55 k + 18 k^{2} + 41 N + 35 k\, 
     N + 11 N^{2})} {(2 + N) (2 + k + N)^{4}}, \nonu\\
c_ {971} & = & \frac {32 (12 + 5 k + 7 N) (16 + 21 k + 6 k^{2} + 
      19 N + 13 k\, N + 5 N^{2})} {3 (2 + N) (2 + k + N)^{4}}, \nonu\\
c_ {972} & = & - \frac {32 (96 + 223 k + 147 k^{2} + 30 k^{3} + 
       161 N + 251 k\, N + 83 k^{2} N + 82 N^{2} + 66 k\, 
      N^{2} + 13 N^{3})} {3 (2 + N) (2 + k + N)^{4}}, \nonu\\
c_ {973} & = & \frac {32 (96 + 236 k + 153 k^{2} + 30 k^{3} + 196 N + 
      302 k\, N + 95 k^{2} N + 121 N^{2} + 92 k\, 
     N^{2} + 23 N^{3})} {3 (2 + N) (2 + k + N)^{4}}, \nonu\\
c_ {974} & = & - \frac {32 (96 + 223 k + 147 k^{2} + 30 k^{3} + 
       161 N + 251 k\, N + 83 k^{2} N + 82 N^{2} + 66 k\, 
      N^{2} + 13 N^{3})} {3 (2 + N) (2 + k + N)^{4}}, \nonu\\
c_ {975} & = & - \frac {32 (32 + 35 k + 10 k^{2} + 61 N + 47 k\, 
      N + 8 k^{2} N + 39 N^{2} + 16 k\, 
      N^{2} + 8 N^{3})} {(2 + N) (2 + k + N)^{4}}, \nonu\\
c_ {976} & = & \frac {32 (12 + 7 k + 5 N) (16 + 21 k + 6 k^{2} + 
      19 N + 13 k\, N + 5 N^{2})} {3 (2 + N) (2 + k + N)^{4}},\nonu\\ 
c_ {977} & = & - \frac {32 (3 k + 2 k^{2} + 13 N + 15 k\, 
      N + 4 k^{2} N + 15 N^{2} + 8 k\, 
      N^{2} + 4 N^{3})} {(2 + N) (2 + k + N)^{4}}, \nonu\\
c_ {978} & = & \frac {32 (32 + 55 k + 18 k^{2} + 41 N + 35 k\, 
     N + 11 N^{2})} {(2 + N) (2 + k + N)^{4}}, \nonu\\
c_ {979} & = & - \frac {1} {3 (2 + N) (2 + k + N)^{4}}32 (144 + 332 k + 215 k^{2} + 42 k^{3} + 
       220 N + 354 k\, N + 119 k^{2} N 
       \nonu\\ & + & 103 N^{2} + 88 k\, 
      N^{2} + 15 N^{3}), \nonu\\
c_ {980} & = & - \frac {32 (60 + 77 k + 22 k^{2} + 121 N + 115 k\, 
      N + 20 k^{2} N + 79 N^{2} + 42 k\, 
      N^{2} + 16 N^{3})} {(2 + N) (2 + k + N)^{3}}, \nonu\\
c_ {981} & = & \frac {32 (32 + 55 k + 18 k^{2} + 41 N + 35 k\, 
     N + 11 N^{2})} {(2 + N) (2 + k + N)^{4}}, \nonu\\
c_ {982} & = & \frac {1} {3 (2 + N) (2 + k + N)^{5}} 32 (48 + 170 k + 
    153 k^{2} + 38 k^{3} + 274 N + 729 k\, 
   N + 532 k^{2} N 
   \nonu\\ & + & 106 k^{3} N + 396 N^{2} + 862 k\, 
   N^{2} + 481 k^{2} N^{2} + 60 k^{3} N^{2} + 218 N^{3} + 375 k\, 
   N^{3} + 126 k^{2} N^{3} 
   \nonu\\ & + & 40 N^{4} + 48 k\, N^{4}), \nonu\\
c_ {983} & = & \frac {1} {3 (2 + N) (2 + k + N)^{5}} 16 (120 + 
    174 k + 72 k^{2} + 8 k^{3} + 342 N + 410 k\, 
   N + 131 k^{2} N 
   \nonu\\ & + & 10 k^{3} N + 346 N^{2} + 300 k\, 
   N^{2} + 53 k^{2} N^{2} + 149 N^{3} + 70 k\, N^{3} + 23 N^{4}),\nonu\\ 
c_ {984} & = & \frac {1} {3 (2 + N) (2 + k + N)^{5}} 16 (168 + 
    374 k + 204 k^{2} + 32 k^{3} + 310 N + 270 k\, 
   N - 101 k^{2} N 
   \nonu\\ & - & 62 k^{3} N + 186 N^{2} - 44 k\, 
   N^{2} - 119 k^{2} N^{2} + 53 N^{3} - 30 k\, N^{3} + 7 N^{4}), \nonu\\
c_ {985} & = & - \frac {32 (1 + N) (32 + 55 k + 18 k^{2} + 41 N + 
       35 k\, N + 11 N^{2})} {(2 + N) (2 + k + N)^{5}}, \nonu\\
c_ {986} & = & - \frac {32 (1 + k) (32 + 55 k + 18 k^{2} + 41 N + 
       35 k\, N + 11 N^{2})} {(2 + N) (2 + k + N)^{5}}, \nonu\\
c_ {987} & = & - \frac {64 (3 + 2 k + N) (9 k + 4 k^{2} + 7 N + 
       20 k\, N + 5 k^{2} N + 8 N^{2} + 9 k\, 
      N^{2} + 2 N^{3})} {3 (2 + N) (2 + k + N)^{5}}, \nonu\\
c_ {988} & = & - \frac {1} {3 (2 + N) (2 + k + N)^{5}} 16 (120 + 
     254 k + 196 k^{2} + 61 k^{3} + 6 k^{4} + 262 N + 426 k\, 
    N 
    \nonu\\ & + & 232 k^{2} N + 39 k^{3} N + 206 N^{2} + 227 k\, 
    N^{2} + 66 k^{2} N^{2} + 68 N^{3} + 37 k\, N^{3} + 8 N^{4}), \nonu\\
c_ {989} & = & - \frac {1} {3 (2 + N) (2 + k + N)^{5}} 64 (-91 k - 
     146 k^{2} - 74 k^{3} - 12 k^{4} + 43 N - 75 k\, 
    N - 125 k^{2} N 
    \nonu\\ & - & 36 k^{3} N + 77 N^{2} + 11 k\, 
    N^{2} - 21 k^{2} N^{2} + 44 N^{3} + 13 k\, N^{3} + 8 N^{4}), \nonu\\
c_ {990} & = & \frac{1} {(2 + N) (2 + k + N)^{6}}32 (1 + k) (1 + N) (60 + 77 k + 22 k^{2} + 
      121 N + 115 k\, N + 20 k^{2} N 
      \nonu\\ & + & 79 N^{2} + 42 k\, 
     N^{2} + 16 N^{3}), \nonu\\
c_ {991} & = & \frac {1} {3 (2 + N) (2 + k + N)^{5}} 16 (408 + 
    994 k + 750 k^{2} + 159 k^{3} - 6 k^{4} + 722 N + 1062 k\, 
   N 
   \nonu\\ & + &346 k^{2} N - 31 k^{3} N + 504 N^{2} + 397 k\, 
   N^{2} + 10 k^{2} N^{2} + 214 N^{3} + 95 k\, N^{3} + 40 N^{4}),\nonu\\ 
c_ {992} & = & \frac {32 (96 + 232 k + 153 k^{2} + 30 k^{3} + 200 N + 
      296 k\, N + 95 k^{2} N + 127 N^{2} + 90 k\, 
     N^{2} + 25 N^{3})} {3 (2 + N) (2 + k + N)^{3}}, \nonu\\
c_ {993} & = & \frac {32 (96 + 252 k + 187 k^{2} + 42 k^{3} + 180 N + 
      290 k\, N + 105 k^{2} N + 99 N^{2} + 76 k\, 
     N^{2} + 17 N^{3})} {3 (2 + N) (2 + k + N)^{4}}, \nonu\\
c_ {994} & = & - \frac {32 (3 k + 2 k^{2} + 13 N + 15 k\, 
      N + 4 k^{2} N + 15 N^{2} + 8 k\, 
      N^{2} + 4 N^{3})} {(2 + N) (2 + k + N)^{4}}, \nonu\\
c_ {995} & = & \frac {32 (32 + 55 k + 18 k^{2} + 41 N + 35 k\, 
     N + 11 N^{2})} {(2 + N) (2 + k + N)^{4}}, \nonu\\
c_ {996} & = & - \frac {32 (20 + 21 k + 6 k^{2} + 25 N + 13 k\, 
      N + 7 N^{2})} {(2 + N) (2 + k + N)^{3}}, \nonu\\
c_ {997} & = & \frac {32 (96 + 252 k + 161 k^{2} + 30 k^{3} + 180 N + 
      310 k\, N + 99 k^{2} N + 105 N^{2} + 92 k\, 
     N^{2} + 19 N^{3})} {3 (2 + N) (2 + k + N)^{4}}, \nonu\\
c_ {998} & = & - \frac {32 (96 + 223 k + 147 k^{2} + 30 k^{3} + 
       161 N + 251 k\, N + 83 k^{2} N + 82 N^{2} + 66 k\, 
      N^{2} + 13 N^{3})} {3 (2 + N) (2 + k + N)^{4}}, \nonu\\
c_ {999} & = & \frac {1} {3 (2 + N) (2 + k + N)^{4}} 32 (120 + 
    104 k + 17 k^{2} - 2 k^{3} + 316 N + 320 k\, 
   N + 121 k^{2} N 
   \nonu\\ & + & 20 k^{3} N + 299 N^{2} + 256 k\, 
   N^{2} + 62 k^{2} N^{2} + 117 N^{3} + 58 k\, N^{3} + 16 N^{4}),\nonu\\ 
c_ {1000} & = & - \frac {1} {3 (2 + N) (2 + k + N)^{4}} 16 (96 + 
     232 k + 143 k^{2} + 26 k^{3} + 200 N + 296 k\, 
    N + 71 k^{2} N 
    \nonu\\ & - & 8 k^{3} N + 137 N^{2} + 104 k\, 
    N^{2} - 6 k^{2} N^{2} + 39 N^{3} + 10 k\, N^{3} + 4 N^{4}), \nonu\\
c_ {1001} & = & - \frac {1} {3 (2 + N) (2 + k + N)^{4}} 16 (48 - 
     224 k - 463 k^{2} - 266 k^{3} - 48 k^{4} + 392 N + 204 k\, 
    N 
    \nonu\\ & - & 167 k^{2} N - 84 k^{3} N + 571 N^{2} + 456 k\, 
    N^{2} + 62 k^{2} N^{2} + 289 N^{3} + 142 k\, N^{3} + 48 N^{4}),\nonu\\ 
c_ {1002} & = & \frac {1} {3 (2 + N) (2 + k + N)^{5}} 32 (240 + 
    554 k + 393 k^{2} + 86 k^{3} + 658 N + 1401 k\, 
   N + 868 k^{2} N 
   \nonu\\ & + &  154 k^{3} N + 684 N^{2} + 1294 k\, 
   N^{2} + 637 k^{2} N^{2} + 72 k^{3} N^{2} + 314 N^{3} + 495 k\, 
   N^{3} + 150 k^{2} N^{3} 
   \nonu\\ & + & 52 N^{4} + 60 k\, N^{4}), \nonu\\
c_ {1003} & = & \frac {1} {3 (2 + N) (2 + k + N)^{5}} 16 (192 + 
    500 k + 413 k^{2} + 102 k^{3} + 364 N + 554 k\, 
   N + 223 k^{2} N
   \nonu\\ & + & 281 N^{2} + 220 k\, 
   N^{2} + 20 k^{2} N^{2} + 127 N^{3} + 52 k\, N^{3} + 24 N^{4}), \nonu\\
c_ {1004} & = & \frac {16 (36 + 75 k + 26 k^{2} + 67 N + 91 k\, 
     N + 16 k^{2} N + 41 N^{2} + 28 k\, 
     N^{2} + 8 N^{3})} {(2 + N) (2 + k + N)^{3}}, \nonu\\
c_ {1005} & = & - \frac {1} {3 (2 + N) (2 + k + N)^{4}} 8 (576 + 
     944 k + 432 k^{2} + 43 k^{3} - 6 k^{4} + 1072 N + 1056 k\, 
    N 
    \nonu\\ & + & 87 k^{2} N - 65 k^{3} N + 720 N^{2} + 321 k\, 
    N^{2} - 81 k^{2} N^{2} + 221 N^{3} + 29 k\, N^{3} + 27 N^{4}), \nonu\\
c_ {1006} & = & - \frac {32 (32 + 55 k + 18 k^{2} + 41 N + 35 k\, 
      N + 11 N^{2})} {(2 + N) (2 + k + N)^{4}}, \nonu\\
c_ {1007} & = & \frac {32 (32 + 55 k + 18 k^{2} + 41 N + 35 k\, 
     N + 11 N^{2})} {(2 + N) (2 + k + N)^{4}}, \nonu\\
c_ {1008} & = & \frac {1} {3 (2 + N) (2 + k + N)^{5}} 8 (-720 - 
    988 k - 292 k^{2} + 30 k^{3} - k^{4} - 6 k^{5} - 1388 N - 892 k\, 
   N 
   \nonu\\ & + & 498 k^{2} N + 314 k^{3} N + 21 k^{4} N - 688 N^{2} + 750 k\, 
   N^{2} + 1128 k^{2} N^{2} + 232 k^{3} N^{2} + 162 N^{3} 
   \nonu\\ & + & 954 k\, 
   N^{3} + 422 k^{2} N^{3} + 197 N^{4} + 230 k\, N^{4} + 37 N^{5}),\nonu\\ 
c_ {1009} & = & \frac {128 (3 + k + 2 N) (20 + 27 k + 8 k^{2} + 
      39 N + 39 k\, N + 7 k^{2} N + 25 N^{2} + 14 k\, 
     N^{2} + 5 N^{3})} {3 (2 + N) (2 + k + N)^{5}}, \nonu\\
c_ {1010} & = & - \frac {16 (-k + N) (32 + 55 k + 18 k^{2} + 41 N + 
       35 k\, N + 11 N^{2})} {3 (2 + N) (2 + k + N)^{5}}, \nonu\\
c_ {1011} & = & \frac {16 (6 - k + 7 N) (32 + 55 k + 18 k^{2} + 
      41 N + 35 k\, N + 11 N^{2})} {3 (2 + N) (2 + k + N)^{5}}, \nonu\\
c_ {1012} & = & - \frac {32 (-k + N) (32 + 55 k + 18 k^{2} + 41 N + 
       35 k\, N + 11 N^{2})} {3 (2 + N) (2 + k + N)^{5}}, \nonu\\
c_ {1013} & = & - \frac {16 (-k + N) (32 + 55 k + 18 k^{2} + 41 N + 
       35 k\, N + 11 N^{2})} {3 (2 + N) (2 + k + N)^{5}}, \nonu\\
c_ {1014} & = & \frac {32 (-k + N) (32 + 55 k + 18 k^{2} + 41 N + 
      35 k\, N + 11 N^{2})} {3 (2 + N) (2 + k + N)^{5}}, \nonu\\
c_ {1015} & = & - \frac {32 (1 + k) (1 + N) (32 + 55 k + 18 k^{2} + 
       41 N + 35 k\, N + 11 N^{2})} {(2 + N) (2 + k + N)^{6}}, \nonu\\
c_ {1016} & = & \frac {16 (-6 - 7 k + N) (32 + 55 k + 18 k^{2} + 
      41 N + 35 k\, N + 11 N^{2})} {3 (2 + N) (2 + k + N)^{5}}, \nonu\\
c_ {1017} & = & - \frac {16 (6 + k + 5 N) (16 + 21 k + 6 k^{2} + 
       19 N + 13 k\, N + 5 N^{2})} {3 (2 + N) (2 + k + N)^{4}}, \nonu\\
c_ {1018} & = & - \frac {1} {3 (2 + N) (2 + k + N)^{5}} 16 (1008 + 
     1620 k + 792 k^{2} + 107 k^{3} - 6 k^{4} + 2484 N +3180 k\, 
    N 
    \nonu\\ & + & 1133 k^{2} N + 99 k^{3} N + 2292 N^{2} + 2081 k\, 
    N^{2} + 401 k^{2} N^{2} + 927 N^{3} + 449 k\, N^{3} + 137 N^{4}),\nonu\\ 
c_ {1019} & = & \frac {32 (1 + k) (32 + 55 k + 18 k^{2} + 41 N + 
      35 k\, N + 11 N^{2})} {(2 + N) (2 + k + N)^{5}}, \nonu\\
c_ {1020} & = & - \frac {32 (2 + 2 k + k^{2} + 2 N + N^{2}) (32 + 
       55 k + 18 k^{2} + 41 N + 35 k\, 
      N + 11 N^{2})} {(2 + N) (2 + k + N)^{6}}, \nonu\\
c_ {1021} & = & \frac {32 (1 + N) (32 + 55 k + 18 k^{2} + 41 N + 
      35 k\, N + 11 N^{2})} {(2 + N) (2 + k + N)^{5}}, \nonu\\
c_ {1022} & = & - \frac {32 (1 + k) (1 + N) (32 + 55 k + 18 k^{2} + 
       41 N + 35 k\, N + 11 N^{2})} {(2 + N) (2 + k + N)^{6}}, \nonu\\
c_ {1023} & = & - \frac {1} {3 (2 + N) (2 + k + N)^{5}} 32 (-120 - 
     362 k - 323 k^{2} - 82 k^{3} - 154 N - 554 k\, 
    N - 445 k^{2} N 
    \nonu\\ & - &80 k^{3} N + 49 N^{2} - 178 k\, 
    N^{2} - 128 k^{2} N^{2} + 117 N^{3} + 20 k\, N^{3} + 32 N^{4}), \nonu\\
c_ {1024} & = & - \frac {1} {3 (2 + N) (2 + k + N)^{5}} 32 (672 + 
     1226 k + 687 k^{2} + 122 k^{3} + 1606 N + 2178 k\, 
    N 
    \nonu\\ & + & 793 k^{2} N + 64 k^{3} N + 1407 N^{2} + 1282 k\, 
    N^{2} + 232 k^{2} N^{2} + 539 N^{3} + 252 k\, N^{3} + 76 N^{4}), \nonu\\
c_ {1025} & = & \frac {16 (6 + 5 k + N) (16 + 21 k + 6 k^{2} + 19 N + 
      13 k\, N + 5 N^{2})} {3 (2 + N) (2 + k + N)^{4}}, \nonu\\
c_ {1026} & = & \frac {32 (3 + k + 2 N) (32 + 55 k + 18 k^{2} + 
      41 N + 35 k\, N + 11 N^{2})} {3 (2 + N) (2 + k + N)^{5}}, \nonu\\
c_ {1027} & = & \frac {32 (3 + 2 k + N) (32 + 55 k + 18 k^{2} + 
      41 N + 35 k\, N + 11 N^{2})} {3 (2 + N) (2 + k + N)^{5}}, \nonu\\
c_ {1028} & = & \frac {1} {3 (2 + N) (2 + k + N)^{5}} 32 (276 + 
    450 k + 219 k^{2} + 34 k^{3} + 708 N + 767 k\, 
   N + 134 k^{2} N
   \nonu\\ & - & 22 k^{3} N + 736 N^{2} + 546 k\, 
   N^{2} + 35 k^{2} N^{2} + 360 N^{3} + 157 k\, N^{3} + 64 N^{4}), \nonu\\
c_ {1029} & = & \frac {1} {3 (2 + N) (2 + k + N)^{5}} 16 (-336 - 
    1264 k - 1302 k^{2} - 507 k^{3} - 66 k^{4} - 296 N - 1200 k\, 
   N 
   \nonu\\ & - &931 k^{2} N - 191 k^{3} N + 222 N^{2} - 49 k\, 
   N^{2} - 85 k^{2} N^{2} + 263 N^{3} + 115 k\, N^{3} + 59 N^{4}), \nonu\\
c_ {1030} & = & \frac {1} {3 (2 + N) (2 + k + N)^{5}} 32 (120 + 2 k - 
    139 k^{2} - 50 k^{3} + 514 N + 290 k\, 
   N - 113 k^{2} N
   \nonu\\ & - &40 k^{3} N + 677 N^{2} + 394 k\, 
   N^{2} - 4 k^{2} N^{2} + 357 N^{3} + 136 k\, N^{3} + 64 N^{4}), \nonu\\
c_ {1031} & = & \frac {1} {3 (2 + N) (2 + k + N)^{5}} 16 (1344 + 
    2136 k + 993 k^{2} + 89 k^{3} - 22 k^{4} + 3504 N + 4240 k\, 
   N 
   \nonu\\ & + & 1110 k^{2} N - 161 k^{3} N - 56 k^{4} N + 3803 N^{2} + 
    3395 k\, 
   N^{2} + 457 k^{2} N^{2} - 96 k^{3} N^{2} + 2114 N^{3} 
   \nonu\\ & + & 1287 k\, 
   N^{3} + 82 k^{2} N^{3} + 587 N^{4} + 186 k\, N^{4} + 64 N^{5}),\nonu\\ 
c_ {1032} & = & - \frac {128 (3 + 2 k + N) (20 + 27 k + 8 k^{2} + 
       39 N + 39 k\, N + 7 k^{2} N + 25 N^{2} + 14 k\, 
      N^{2} + 5 N^{3})} {3 (2 + N) (2 + k + N)^{5}}, \nonu\\
c_ {1033} & = & - \frac {16 (6 + k + 5 N) (16 + 21 k + 6 k^{2} + 
       19 N + 13 k\, N + 5 N^{2})} {3 (2 + N) (2 + k + N)^{4}}, \nonu\\
c_ {1034} & = & \frac {16 (6 + 5 k + N) (16 + 21 k + 6 k^{2} + 19 N + 
      13 k\, N + 5 N^{2})} {3 (2 + N) (2 + k + N)^{4}}, \nonu\\
c_ {1035} & = & \frac {1} {3 (2 + N) (2 + k + N)^{5}} 16 (1392 + 
    2980 k + 2208 k^{2} + 663 k^{3} + 66 k^{4} + 3236 N + 5340 k\, 
   N 
   \nonu\\ & + & 2737 k^{2} N + 431 k^{3} N + 2748 N^{2} + 3109 k\, 
   N^{2} + 829 k^{2} N^{2} + 1003 N^{3} + 581 k\, N^{3} + 133 N^{4}),\nonu\\ 
c_ {1036} & = & - \frac {32 (1 + N) (32 + 55 k + 18 k^{2} + 41 N + 
       35 k\, N + 11 N^{2})} {(2 + N) (2 + k + N)^{5}}, \nonu\\
c_ {1037} & = & - \frac {32 (1 + k) (32 + 55 k + 18 k^{2} + 41 N + 
       35 k\, N + 11 N^{2})} {(2 + N) (2 + k + N)^{5}}, \nonu\\
c_ {1038} & = & \frac {1} {3 (2 + N) (2 + k + N)^{5}} 32 (288 + 
    454 k + 249 k^{2} + 46 k^{3} + 842 N + 1230 k\, 
   N + 599 k^{2} N 
   \nonu\\ & + &92 k^{3} N + 873 N^{2} + 950 k\, 
   N^{2} + 260 k^{2} N^{2} + 373 N^{3} + 216 k\, N^{3} + 56 N^{4}), \nonu\\
c_ {1039} & = & - \frac {1} {3 (2 + N) (2 + k + N)^{5}} 16 (816 + 
     1800 k + 1206 k^{2} + 265 k^{3} + 6 k^{4} + 1824 N + 2952 k\, 
    N 
    \nonu\\ & + & 1309 k^{2} N + 141 k^{3} N + 1434 N^{2} + 1531 k\, 
    N^{2} + 343 k^{2} N^{2} + 471 N^{3} + 247 k\, N^{3} + 55 N^{4}), \nonu\\
c_ {1040} & = & \frac {1} {3 (2 + N) (2 + k + N)^{4}} 32 (-216 - 
    517 k - 406 k^{2} - 96 k^{3} - 167 N - 498 k\, 
   N - 376 k^{2} N 
   \nonu\\ & - &60 k^{3} N + 184 N^{2} + 13 k\, 
   N^{2} - 66 k^{2} N^{2} + 207 N^{3} + 78 k\, N^{3} + 48 N^{4}), \nonu\\
c_ {1041} & = & - \frac {32 (3 + 2 k + N) (32 + 27 k + 6 k^{2} + 
       37 N + 16 k\, N + 10 N^{2})} {3 (2 + N) (2 + k + N)^{3}}, \nonu\\
c_ {1042} & = & - \frac {1} {3 (2 + N) (2 + k + N)^{4}} 32 (24 - 
     113 k - 179 k^{2} - 54 k^{3} + 221 N 
     \nonu\\ & - & 60 k\, 
    N - 251 k^{2} N - 60 k^{3} N + 383 N^{2} + 125 k\, 
    N^{2} - 66 k^{2} N^{2} + 240 N^{3} + 78 k\, N^{3} + 48 N^{4}), \nonu\\
c_ {1043} & = & \frac {32 (60 + 93 k + 43 k^{2} + 6 k^{3} + 105 N + 
      113 k\, N + 27 k^{2} N + 60 N^{2} + 34 k\, 
     N^{2} + 11 N^{3})} {3 (2 + N) (2 + k + N)^{4}}, \nonu\\
c_ {1044} & = & - \frac {32 (8 + 17 k + 6 k^{2} + 11 N + 11 k\, 
      N + 3 N^{2})} {(2 + N) (2 + k + N)^{3}}, \nonu\\
c_ {1045} & = & \frac {32 (32 + 55 k + 18 k^{2} + 41 N + 35 k\, 
     N + 11 N^{2})} {(2 + N) (2 + k + N)^{4}}, \nonu\\
c_ {1046} & = & \frac {32 (-k + N) (32 + 55 k + 18 k^{2} + 41 N + 
      35 k\, N + 11 N^{2})} {(2 + N) (2 + k + N)^{5}}, \nonu\\
c_ {1047} & = & - \frac {32 (32 + 55 k + 18 k^{2} + 41 N + 35 k\, 
      N + 11 N^{2})} {(2 + N) (2 + k + N)^{4}}, \nonu\\
c_ {1048} & = & - \frac {32 (-k + N) (32 + 55 k + 18 k^{2} + 41 N + 
       35 k\, N + 11 N^{2})} {(2 + N) (2 + k + N)^{5}}, \nonu\\
c_ {1049} & = & \frac {1} {3 (2 + N) (2 + k + N)^{4}}32 (504 + 733 k + 252 k^{2} + 12 k^{3} + 
      959 N + 941 k\, N + 164 k^{2} N
       \nonu\\ & + & 583 N^{2} + 300 k\, 
     N^{2} + 112 N^{3}), \nonu\\
c_ {1050} & = & - \frac {32 (32 + 55 k + 18 k^{2} + 41 N + 35 k\, 
      N + 11 N^{2})} {(2 + N) (2 + k + N)^{4}}, \nonu\\
c_ {1051} & = & \frac {32 (32 + 55 k + 18 k^{2} + 41 N + 35 k\, 
     N + 11 N^{2})} {(2 + N) (2 + k + N)^{4}}, \nonu\\
c_ {1052} & = & - \frac {32 (32 + 55 k + 18 k^{2} + 41 N + 35 k\, 
      N + 11 N^{2})} {(2 + N) (2 + k + N)^{3}}, \nonu\\
c_ {1053} & = & \frac {64 (20 + 15 k - 11 k^{2} - 6 k^{3} + 51 N + 
      43 k\, N + k^{2} N + 40 N^{2} + 22 k\, 
     N^{2} + 9 N^{3})} {(2 + N) (2 + k + N)^{4}}, \nonu\\
c_ {1054} & = & - \frac {64 (20 + 15 k - 11 k^{2} - 6 k^{3} + 51 N + 
       43 k\, N + k^{2} N + 40 N^{2} + 22 k\, 
      N^{2} + 9 N^{3})} {(2 + N) (2 + k + N)^{4}}, \nonu\\
c_ {1055} & = & \frac {64 (32 + 55 k + 18 k^{2} + 41 N + 35 k\, 
     N + 11 N^{2})} {(2 + N) (2 + k + N)^{4}}, \nonu\\
c_ {1056} & = & - \frac {64 (3 + 2 k + N) (12 + 19 k + 6 k^{2} + 
       15 N + 12 k\, N + 4 N^{2})} {3 (2 + N) (2 + k + N)^{4}}, \nonu\\
c_ {1057} & = & - \frac {1}{(2 + N) (2 + k + N)^{5}}32 (-32 - 81 k - 66 k^{2} - 16 k^{3} - 
       47 N - 117 k\, N - 83 k^{2} N 
        \nonu\\ & - &14 k^{3} N - 9 N^{2} - 41 k\, 
      N^{2} - 23 k^{2} N^{2} + 12 N^{3} + k\, 
      N^{3} + 4 N^{4}), \nonu\\
c_ {1058} & = & \frac {1}{3 (2 + N) (2 + k + N)^{5}}64 (96 + 143 k + 91 k^{2} + 48 k^{3} + 
      12 k^{4} + 241 N + 235 k\, 
     N + 48 k^{2} N 
      \nonu\\ & + & 8 k^{3} N + 250 N^{2} + 165 k\, 
     N^{2} + 5 k^{2} N^{2} + 123 N^{3} + 49 k\, 
     N^{3} + 22 N^{4}), \nonu\\
c_ {1059} & = & \frac {32 (20 + 21 k + 6 k^{2} + 25 N + 13 k\, 
     N + 7 N^{2})} {(2 + N) (2 + k + N)^{3}}, \nonu\\
c_ {1060} & = & \frac {32 (60 + 93 k + 43 k^{2} + 6 k^{3} + 105 N + 
      113 k\, N + 27 k^{2} N + 60 N^{2} + 34 k\, 
     N^{2} + 11 N^{3})} {3 (2 + N) (2 + k + N)^{4}}, \nonu\\
c_ {1061} & = & - \frac{1}{3 (2 + N) (2 + k + N)^{4}}64 (156 + 267 k + 142 k^{2} + 24 k^{3} + 
       267 N + 311 k\, N + 84 k^{2} N 
        \nonu\\ & + & 147 N^{2} + 88 k\, 
      N^{2} + 26 N^{3}), \nonu\\
c_ {1062} & = & - \frac {1}{(2 + N) (2 + k + N)^{5}}32 (-32 - 81 k - 66 k^{2} - 16 k^{3} - 
       47 N - 117 k\, N - 83 k^{2} N
        \nonu\\ & - & 14 k^{3} N - 9 N^{2} - 41 k\, 
      N^{2} - 23 k^{2} N^{2} + 12 N^{3} + k\, 
      N^{3} + 4 N^{4}), \nonu\\
c_ {1063} & = & - \frac {32 (20 + 21 k + 6 k^{2} + 25 N + 13 k\, 
      N + 7 N^{2})} {(2 + N) (2 + k + N)^{3}}, \nonu\\
c_ {1064} & = & \frac {32 (-48 - 68 k - 7 k^{2} + 6 k^{3} - 52 N - 
      20 k\, N + 17 k^{2} N - 5 N^{2} + 14 k\, 
     N^{2} + 3 N^{3})} {(2 + N) (2 + k + N)^{4}}, \nonu\\
c_ {1065} & = & \frac{1}{3 (2 + N) (2 + k + N)^{4}}32 (216 + 191 k + 12 k^{2} - 12 k^{3} + 
      493 N + 439 k\, N + 76 k^{2} N 
      \nonu\\ & + & 365 N^{2} + 204 k\, 
     N^{2} + 80 N^{3}), \nonu\\
c_ {1066} & = & - \frac {32 (60 + 77 k + 22 k^{2} + 121 N + 115 k\, 
      N + 20 k^{2} N + 79 N^{2} + 42 k\, 
      N^{2} + 16 N^{3})} {(2 + N) (2 + k + N)^{3}}, \nonu\\
c_ {1067} & = & - \frac{1}{3 (2 + N) (2 + k + N)^{3}}16 (240 + 406 k + 214 k^{2} + 36 k^{3} + 
       566 N + 756 k\, 
      N + 289 k^{2} N 
      \nonu\\ & + &  30 k^{3} N + 488 N^{2} + 455 k\, 
      N^{2} + 93 k^{2} N^{2} + 180 N^{3} + 87 k\, 
      N^{3} + 24 N^{4}), \nonu\\
c_ {1068} & = & \frac {32 (32 + 55 k + 18 k^{2} + 41 N + 35 k\, 
     N + 11 N^{2})} {(2 + N) (2 + k + N)^{4}}, \nonu\\
c_ {1069} & = & \frac {8 (6 k + k^{2} + 6 N + 10 k\, 
     N + N^{2}) (16 + 21 k + 6 k^{2} + 19 N + 13 k\, 
     N + 5 N^{2})} {3 (2 + N) (2 + k + N)^{4}}, \nonu\\
c_ {1070} & = & \frac {1} {3 (2 + N) (2 + k + N)^{5}} 8 (-720 - 
    268 k + 992 k^{2} + 756 k^{3} + 131 k^{4} - 6 k^{5} - 668 N
    \nonu\\ & + &  2204 k\, 
   N + 3792 k^{2} N + 1508 k^{3} N + 141 k^{4} N + 1124 N^{2} + 
    4992 k\, 
   N^{2} + 3858 k^{2} N^{2} 
   \nonu\\ & + &  724 k^{3} N^{2} + 1836 N^{3} + 3288 k\, 
   N^{3} + 1142 k^{2} N^{3} + 863 N^{4} + 674 k\, N^{4} + 133 N^{5}),\nonu\\ 
c_ {1071} & = & \frac {1}{3 (2 + N) (2 + k + N)^{5}}16 (-144 - 704 k - 777 k^{2} - 326 k^{3} - 
      48 k^{4} + 8 N - 436 k\, 
     N 
     \nonu\\ & - &  385 k^{2} N - 84 k^{3} N + 373 N^{2} + 264 k\, 
     N^{2} + 50 k^{2} N^{2} + 279 N^{3} + 146 k\, 
     N^{3} + 56 N^{4}), \nonu\\
c_ {1072} & = & \frac {1}{3 (2 + N) (2 + k + N)^{5}}32 (156 + 246 k + 135 k^{2} + 26 k^{3} + 
      396 N + 370 k\, 
     N + 43 k^{2} N 
     \nonu\\ & - &  20 k^{3} N + 389 N^{2} + 234 k\, 
     N^{2} - 2 k^{2} N^{2} + 183 N^{3} + 68 k\, 
     N^{3} + 32 N^{4}), \nonu\\
c_ {1073} & = & \frac {64 (-k + N) (32 + 55 k + 18 k^{2} + 41 N + 
      35 k\, N + 11 N^{2})} {3 (2 + N) (2 + k + N)^{5}}, \nonu\\
c_ {1074} & = & - \frac {1}{3 (2 + N) (2 + k + N)^{5}}16 (480 + 704 k + 347 k^{2} + 58 k^{3} + 
       1168 N + 1320 k\, 
      N 
      \nonu\\ & + & 455 k^{2} N 44 k^{3} N + 1069 N^{2} + 832 k\, 
      N^{2} + 150 k^{2} N^{2} + 431 N^{3} + 174 k\, 
      N^{3} + 64 N^{4}), \nonu\\
c_ {1075} & = & - \frac{1}{3 (2 + N) (2 + k + N)^{5}}64 (-78 - 197 k - 131 k^{2} - 26 k^{3} - 
       124 N - 283 k\, 
      N - 153 k^{2} N 
      \nonu\\ & - &  22 k^{3} N - 33 N^{2} - 91 k\, 
      N^{2} - 34 k^{2} N^{2} + 27 N^{3} + 7 k\, 
      N^{3} + 10 N^{4}), \nonu\\
c_ {1076} & = & \frac {1} {3 (2 + N) (2 + k + N)^{5}} 16 (-576 - 
    1776 k - 1739 k^{2} - 683 k^{3} - 94 k^{4} - 1560 N - 4200 k\, 
   N 
    \nonu\\ & - & 3474 k^{2} N 
  - 1103 k^{3} N - 116 k^{4} N - 1369 N^{2} -3113 k\, 
   N^{2} - 1911 k^{2} N^{2} - 342 k^{3} N^{2} 
   \nonu\\ & - &  398 N^{3} - 783 k\, 
   N^{3} - 278 k^{2} N^{3} + 15 N^{4} - 36 k\, N^{4} + 16 N^{5}), \nonu\\
c_ {1077} & = & - \frac {1}{3 (2 + N) (2 + k + N)^{5}}16 (624 + 1432 k + 925 k^{2} + 182 k^{3} + 
       1328 N + 2244 k\, 
      N 
      \nonu\\ & + &  925 k^{2} N + 76 k^{3} N + 983 N^{2} + 1112 k\, 
      N^{2} + 222 k^{2} N^{2} + 301 N^{3} + 174 k\, 
      N^{3} + 32 N^{4}), \nonu\\
c_ {1078} & = & \frac {1}{3 (2 + N) (2 + k + N)^{5}}16 (480 + 1136 k + 957 k^{2} + 350 k^{3} + 
      48 k^{4} + 1120 N + 2008 k\, 
     N 
     \nonu\\ & + &  1129 k^{2} N + 204 k^{3} N + 923 N^{2} + 1128 k\, 
     N^{2} + 322 k^{2} N^{2} + 321 N^{3} + 202 k\, 
     N^{3} + 40 N^{4}), \nonu\\
c_ {1079} & = & - \frac {32 (60 + 77 k + 22 k^{2} + 121 N + 115 k\, 
      N + 20 k^{2} N + 79 N^{2} + 42 k\, 
      N^{2} + 16 N^{3})} {(2 + N) (2 + k + N)^{3}}, \nonu\\
c_ {1080} & = & \frac {1}{(2 + N) (2 + k + N)^{4}}32 (16 - k - 22 k^{2} - 8 k^{3} + 57 N + 
      11 k\, N - 37 k^{2} N - 10 k^{3} N 
      \nonu\\ & + &  75 N^{2} + 27 k\, 
     N^{2} - 11 k^{2} N^{2} + 42 N^{3} + 13 k\, 
     N^{3} + 8 N^{4}), \nonu\\
c_ {1081} & = & - \frac{1} {(2 + N) (2 + k + N)^{4}}32 (16 - k - 22 k^{2} - 8 k^{3} + 57 N + 
       11 k\, N - 37 k^{2} N - 10 k^{3} N 
       \nonu\\ & + &  75 N^{2} + 27 k\, 
      N^{2} - 11 k^{2} N^{2} + 42 N^{3} + 13 k\, 
      N^{3} + 8 N^{4}), \nonu\\
c_ {1082} & = & - \frac {32 (32 + 55 k + 18 k^{2} + 41 N + 35 k\, 
      N + 11 N^{2})} {(2 + N) (2 + k + N)^{4}}, \nonu\\
c_ {1083} & = & \frac {32 (60 + 77 k + 22 k^{2} + 121 N + 115 k\, 
     N + 20 k^{2} N + 79 N^{2} + 42 k\, 
     N^{2} + 16 N^{3})} {(2 + N) (2 + k + N)^{4}}, \nonu\\
c_ {1084} & = & \frac{1}{9 (2 + N) (2 + k + N)^{5}}32 (-540 - 513 k + 174 k^{2} + 271 k^{3} + 
      62 k^{4} - 909 N + 129 k\, 
     N 
     \nonu\\ & + & 1302 k^{2} N + 646 k^{3} N + 76 k^{4} N - 159 N^{2} + 
      1524 k\, 
     N^{2} + 1506 k^{2} N^{2} + 319 k^{3} N^{2} + 467 N^{3} 
     \nonu\\ & + & 
      1201 k\, N^{3} + 462 k^{2} N^{3} + 293 N^{4} + 263 k\, 
     N^{4} + 50 N^{5}), \nonu\\
c_ {1085} & = & - \frac {1} {9 (2 + N) (2 + k + N)^{5}} 16 (2304 + 
     4056 k + 2595 k^{2} + 698 k^{3} + 64 k^{4} + 6240 N
     \nonu\\ & + & 9504 k\, 
    N + 5509 k^{2} N + 1504 k^{3} N + 176 k^{4} N + 6369 N^{2} + 
     7534 k\, 
    N^{2} + 3210 k^{2} N^{2} 
    \nonu\\ & + & 516 k^{3} N^{2} + 2927 N^{3} + 
     2246 k\, N^{3} + 494 k^{2} N^{3} + 572 N^{4} + 186 k\, 
    N^{4} + 32 N^{5}), \nonu\\
c_ {1086} & = & - \frac {1} {9 (2 + N) (2 + k + N)^{5}} 16 (1536 + 
     4056 k + 3789 k^{2} + 1499 k^{3} + 214 k^{4} + 4536 N 
     \nonu\\ & + &
     10536 k\, 
    N + 8361 k^{2} N + 2669 k^{3} N + 284 k^{4} N + 5427 N^{2} + 
     10887 k\, 
    N^{2} + 7098 k^{2} N^{2}
    \nonu\\ & + & 1718 k^{3} N^{2} + 120 k^{4} N^{2} + 
     3253 N^{3} + 5423 k\, 
    N^{3} + 2646 k^{2} N^{3} + 372 k^{3} N^{3} + 964 N^{4} 
    \nonu\\ & + & 1240 k\, 
    N^{4} + 348 k^{2} N^{4} + 112 N^{5} + 96 k\, N^{5}), \nonu\\
c_ {1087} & = & - \frac {1}{9 (2 + N) (2 + k + N)^{4}}16 (-96 - 756 k - 1125 k^{2} - 584 k^{3} - 
       100 k^{4} + 36 N - 1356 k\, 
      N 
      \nonu\\ & - & 1933 k^{2} N - 781 k^{3} N - 86 k^{4} N + 513 N^{2} - 
       544 k\, N^{2} - 912 k^{2} N^{2} - 219 k^{3} N^{2} + 
       613 N^{3} 
       \nonu\\ & + & 133 k\, 
      N^{3} - 98 k^{2} N^{3} + 268 N^{4} + 75 k\, 
      N^{4} + 40 N^{5}).
\nonu
\eea
It would be interesting to see how one can simplify the 
above OPEs using the summation indices. 

\section{The complete $136$ OPEs in component approach
corresponding to the single OPE (\ref{finalPhiPhi}) 
in ${\cal N}=4$ superspace in section $7$ }

In this Appendix, one rewrites the complete OPEs, which were 
presented in ${\cal N}=2$ superspace in section $6$,
in the component approach. 

\subsection{The OPEs between the higher spin-$1$ current and the other 
$16$ higher spin currents}

The OPE between the higher spin-$1$ current is given by
\bea
\Phi_{0}^{(1)}(z)\;\Phi_{0}^{(1)}(w)&=&\frac{1}{(z-w)^{2}}\,\frac{2\, k\, N}
{(2+k+N)}+\cdots.
\nonu
\eea

The OPEs between the higher spin-$1$ current and the four 
higher spin-$\frac{3}{2}$ currents are given by
\bea
&& \Phi_{0}^{(1)}(z)\;\Phi_{\frac{1}{2}}^{(1),i}(w)
=\frac{1}{(z-w)}\Bigg[c_{1}\, G^{i}+c_{2}\, U\,\Gamma^{i}+c_{3}\,\varepsilon^{ijkl}\,\Gamma^{j}\,\Gamma^{k}\,\Gamma^{l}+c_{4}\,\varepsilon^{ijkl}\, T^{jk}\,\Gamma^{l}\Bigg](w)+\cdots
\nonu,
\eea
where the coefficients are
\bea
c_{1}&=&1,\qquad
c_{2} =-\frac{2\, i}{(2+k+N)},\qquad
c_{3}=-\frac{8\, i}{6(2+k+N)},\qquad
c_{4}=\frac{1}{(2+k+N)}.
\nonu
\eea
The fusion rule is given by
\bea
[\Phi_{0}^{(1)}] \, \cdot \, [\Phi_{\frac{1}{2}}^{(1),i}]
=[I^i], 
\nonu
\eea
where 
$[I]$ denotes the large ${\cal N}=4$ linear superconformal family of 
identity operator.

The OPEs between the higher spin-$1$ current and the six 
higher spin-$2$ currents are given by
\bea
&& \Phi_{0}^{(1)}(z)\;\Phi_{1}^{(1),ij}(w)
=\frac{1}{(z-w)^{2}}\Bigg[c_{1}\,\Gamma^{i}\,\Gamma^{j}+c_{2}\,\varepsilon^{ijkl}\,\Gamma^{k}\,\Gamma^{l}+c_{3}\, T^{ij}+c_{4}\,\varepsilon^{ijkl}\, T^{kl}\Bigg](w)
\nonu\\
& & +\frac{1}{(z-w)}\Bigg[c_{5}\, U\,\Gamma^{i}\,\Gamma^{j}+c_{6}\,(G^{i}\,\Gamma^{j}-G^{j}\,\Gamma^{i})
\nonu\\
& &+\varepsilon^{ijkl}\Big\{c_{7}\,\partial(\Gamma^{k}\,\Gamma^{l})
 +c_{8}\,(T^{ik}\,\Gamma^{i}\,\Gamma^{l}+
 T^{jk}\,\Gamma^{j}\,
\Gamma^{l})+c_{9}\,\partial(T^{kl})\Big\}\Bigg](w)+\cdots,
\nonu
\eea
where \footnote{One assumes that there is no summation over 
the repeated indices with 
more than two in this Appendix. For example, the whole $c_8$-term 
contains the index $i$ which appears three times. For fixed index $i$, this
has only one single term.}
the coefficients are
\bea
c_{1}&=&\frac{4\,(k-N)}{(2+k+N)^{2}},\qquad
c_{2}=\frac{2\,(k+N)}{(2+k+N)^{2}},\qquad
c_{3}=\frac{2\, i\,(k-N)}{(2+k+N)},\qquad
c_{4}=\frac{ i\,(k+N)}{(2+k+N)},\nonu\\
c_{5}&=&-\frac{8}{(2+k+N)^{2}},\qquad
c_{6}=-\frac{2i}{(2+k+N)},\qquad
c_{7}=-\frac{4}{(2+k+N)^{2}},\qquad
c_{8}=\frac{4\, i}{(2+k+N)^{2}},
\nonu\\
c_{9}&=&-\frac{2\, i}{(2+k+N)}.
\nonu
\eea
Again the fusion rule is given by
\bea
[\Phi_{0}^{(1)}]\, \cdot \, [\Phi_{1}^{(1),ij}]
=[I^{ij}].  
\nonu
\eea

The OPEs between the higher spin-$1$ current and the four 
higher spin-$\frac{5}{2}$ currents are given by \footnote{
In Appendix $H$, all the higher spin currents are bosonic (or fermionic) 
component currents.  We use a boldface notation for the higher spin currents
appearing in the right hand side of the OPE in order to emphasize that 
they are the higher spin currents. They are not super currents.}
\bea
&& \Phi_{0}^{(1)}(z)\;\widetilde{\Phi}_{\frac{3}{2}}^{(1),i}(w)
=\frac{1}{(z-w)^{3}}c_{1}\,\Gamma^{i}(w)+\frac{1}{(z-w)^{2}}\Bigg[c_{2}\, G^{i}+c_{3}\, U\,\Gamma^{i}
\nonu\\
& & +\varepsilon^{ijkl}(c_{4}\,\Gamma^{j}\,\Gamma^{k}\,\Gamma^{l}+c_{5}\, T^{jk}\,\Gamma^{l})+c_{6}\, T^{ij}\,\Gamma^{j}+c_{7}\,\Gamma^{i}\Bigg](w)
\nonu\\
& & +\frac{1}{(z-w)}\Bigg[ c_{8}\,{\bf \Phi_{\frac{1}{2}}^{(2),i}}+
c_{9}\,{\bf \Phi_{0}^{(1)}\,\Phi_{\frac{1}{2}}^{(1),i}}+c_{10}\, U\, U\,\Gamma^{i}+c_{11}\, U\,\partial\Gamma^{i}+c_{12}\, G^{i}\, U
\nonu\\
& & +c_{13}\,\partial G^{i}+c_{14}\,\partial^{2}\Gamma^{i}+c_{15}\,\partial U\,\Gamma^{i}+c_{16}\,\Gamma^{i}\,\partial\Gamma^{j}\,\Gamma^{j}+c_{17}\, T^{ij}\, U\,\Gamma^{j}+c_{18}\, G^{j}\,\Gamma^{i}\,\Gamma^{j}
\nonu\\
& & +c_{19}\, G^{j}\, T^{ij}+c_{20}\,\partial T^{ij}\,\Gamma^{j}+c_{21}\,T^{jk}\,\Gamma^{i}\,\Gamma^{j}\,\Gamma^{k}
 +\varepsilon^{ijkl} \left(c_{22}\, G^{j}\,\Gamma^{k}\,\Gamma^{l}+
c_{23}\, U\,\Gamma^{j}\,\Gamma^{k}\,\Gamma^{l} \right.
\nonu \\
&& + \left.
c_{24}\,\partial(\Gamma^{j}\,\Gamma^{k}\,\Gamma^{l})+c_{25}\, G^{j}\, T^{kl}
+c_{26}\,\partial T^{jk}\,\Gamma^{l}+c_{27}\, T^{jk}\, U\,\Gamma^{l}+c_{28}\, T^{jk}\,\partial\Gamma^{l} \right)\nonu \\
&& +c_{29}\,\widetilde{T}^{ij}\,\widetilde{T}^{ij}\,\Gamma^{i}
 +c_{30}\,\varepsilon^{ijkl}\, T^{ij}\, T^{kl}\,\Gamma^{i}+c_{31}\, T^{ij}\, T^{jk}\,\Gamma^{k}(1-\delta^{ik})
\Bigg](w)
+\cdots,
\nonu
\eea
where the coefficients are
\bea
c_ {1} & = & - \frac {8 i\, k\, N} {(2 + k + N)^{2}}, \qquad
c_ {2}  = \frac {8 (k - N)} {3 (2 + k + N)}, \qquad
c_ {3}  =  - \frac {16 i (k - N)} {3 (2 + k + N)^{2}},
\nonu \\ 
c_ {4}  & = &   \frac {4 i (k - N)} {9 (2 + k + N)^{3}}, \qquad
c_ {5}  =   \frac {2 (k - N)} {3 (2 + k + N)^{2}}, \qquad
c_ {6}  =  - \frac {4 (k + N)} {(2 + k + N)^{2}}, \nonu\\
c_ {7} & = &\frac {4 i (3 k + 3 N + 2 k\, N)} {(2 + k + N)^{2}},\qquad 
c_ {8}  =  - \frac {1} {2}, \nonu\\
c_ {9} & = &\frac {(60 + 77 k + 22 k^{2} + 121 N + 115 k\, 
    N + 20 k^{2} N + 79 N^{2} + 42 k\, 
    N^{2} + 16 N^{3})} {2 (2 + N) (2 + k + N)^{2}}, \nonu\\
c_ {10} & = &\frac {4 i} {(2 + k + N)^{2}}, \qquad
c_ {11}  = \frac {2 i (3 + 2 k + N) (10 + 5 k + 
      8 N)} {3 (2 + k + N)^{3}}, \qquad
c_ {12}  =  - \frac {2} {(2 + k + N)}, \nonu\\
c_ {13} & = & - \frac {(-60 - 49 k - 14 k^{2} - 29 N + 37 k\, 
     N + 20 k^{2} N + 37 N^{2} + 42 k\, 
     N^{2} + 16 N^{3})} {6 (2 + N) (2 + k + N)^{2}},\nonu\\ 
c_ {14} & = &\frac {4 i (1 + 2 k + 2 N)} {(2 + k + N)^{2}},\nonu\\ 
c_ {15} & = &\frac {i (-60 - 49 k - 14 k^{2} - 29 N + 37 k\, 
     N + 20 k^{2} N + 37 N^{2} + 42 k\, 
     N^{2} + 16 N^{3})} {3 (2 + N) (2 + k + N)^{3}}, \nonu\\
c_ {16} & = &\frac {4 i (13 k + 6 k^{2} + 3 N + 9 k\, 
     N + N^{2})} {(2 + N) (2 + k + N)^{4}}, \nonu\\
c_ {17} & = & - \frac {(20 + 21 k + 6 k^{2} + 25 N + 13 k\, 
     N + 7 N^{2})} {(2 + N) (2 + k + N)^{3}}, \nonu\\
c_ {18} & = & - \frac {(20 + 21 k + 6 k^{2} + 25 N + 13 k\, 
     N + 7 N^{2})} {(2 + N) (2 + k + N)^{3}}, \nonu\\
c_ {19} & = & - \frac {i (20 + 21 k + 6 k^{2} + 25 N + 13 k\, 
      N + 7 N^{2})} {2 (2 + N) (2 + k + N)^{2}}, \qquad
c_ {20}  = \frac {8} {(2 + k + N)^{2}}, \nonu\\
c_ {21} & = &\frac { (32 + 55 k + 18 k^{2} + 41 N + 35 k\, 
     N + 11 N^{2})} {(2 + N) (2 + k + N)^{4}}, \nonu\\
c_ {22} & = &-\frac {(13 k + 6 k^{2} + 3 N + 9 k\, 
    N + N^{2})} {2(2 + N) (2 + k + N)^{3}}, \nonu\\
c_ {23} & = &  \frac { i (32 + 55 k + 18 k^{2} + 41 N + 35 k\, 
      N + 11 N^{2})} {3(2 + N) (2 + k + N)^{4}}, \nonu\\
c_ {24} & = &  \frac {2 i (60 + 77 k + 22 k^{2} + 121 N + 115 k\, 
      N + 20 k^{2} N + 79 N^{2} + 42 k\, 
      N^{2} + 16 N^{3})} {9 (2 + N) (2 + k + N)^{4}}, \nonu\\
c_ {25} & = &-\frac {i (16 + 21 k + 6 k^{2} + 19 N + 13 k\, 
     N + 5 N^{2})} {4 (2 + N) (2 + k + N)^{2}}, \nonu\\
c_ {26} & = &-\frac { (7 k + 2 k^{2} + 23 N + 38 k\, 
     N + 10 k^{2} N + 29 N^{2} + 21 k\, 
     N^{2} + 8 N^{3})} {3 (2 + N) (2 + k + N)^{3}}, \nonu\\
c_ {27} & = &-\frac {(32 + 29 k + 6 k^{2} + 35 N + 17 k\, 
    N + 9 N^{2})} {2(2 + N) (2 + k + N)^{3}}, \nonu\\
c_ {28} & = &-\frac { (3 + 2 k + N) (10 + 5 k + 
      8 N)} {3 (2 + k + N)^{3}}, \nonu\\
c_ {29} & = &\frac {i (16 + 21 k + 6 k^{2} + 19 N + 13 k\, 
     N + 5 N^{2})} {(2 + N) (2 + k + N)^{3}}, \nonu\\
c_ {30} & = &\frac {i (20 + 21 k + 6 k^{2} + 25 N + 13 k\, 
     N + 7 N^{2})} {6(2 + N) (2 + k + N)^{3}}, \nonu\\
c_ {31} & = &\frac {i (16 + 21 k + 6 k^{2} + 19 N + 13 k\, 
     N + 5 N^{2})} {(2 + N) (2 + k + N)^{3}}.
\nonu
\eea   
We also introduce 
\bea
\widetilde{T}^{ij}(w) = \frac{1}{2!} \, \varepsilon^{ijkl} \, T^{kl}(w).
\nonu
\eea
The fusion rule is given by 
\bea
[\Phi_{0}^{(1)}] \, \cdot \, [\widetilde{\Phi}_{\frac{3}{2}}^{(1),i}]
= [I^i]  + [\Phi_{0}^{(1)}\,\Phi_{\frac{1}{2}}^{(1),i}]
+ [\Phi_{\frac{1}{2}}^{(2),i}],
\nonu 
\eea
where the last term belongs to the next $16$ higher spin currents with 
spin $s=2$.

The OPE between the higher spin-$1$ current and the 
higher spin-$3$ current is given by
\bea
&& \Phi_{0}^{(1)}(z)\;\widetilde{\Phi}_{2}^{(1)}(w)
=\frac{1}{(z-w)^{3}}\, c_{1}\, U(w)+\frac{1}{(z-w)^{2}}\Bigg[
c_{2}\,{\bf \Phi_{0}^{(2)}}+c_{3}\,{\bf \Phi_{0}^{(1)}\,\Phi_{0}^{(1)}}+
c_{4}\, L+c_{5}\, U\, U
\nonu\\
& & +c_{6}\, G^{i}\,\Gamma^{i}+\varepsilon^{ijkl}(c_{7}\,\Gamma^{i}\,\Gamma^{j}\,\Gamma^{k}\,\Gamma^{l}+c_{8}\, T^{ij}\, T^{kl}+c_{9}\, T^{ij}\,\Gamma^{k}\,\Gamma^{l})
\nonu\\
& & +\delta^{ik}\delta^{jl}(c_{10}\,T^{ij}\, T^{kl}+c_{11}\,T^{ij}\,\Gamma^{k}\,\Gamma^{l})+c_{12}\,\partial\Gamma^{i}\,\Gamma^{i}+c_{13}\,\partial U\Bigg](w)
\nonu\\
& & +\frac{1}{(z-w)}\Bigg[c_{14}\, U\,\partial\Gamma^{i}\,\Gamma^{i}+c_{15}\, G^{i}\,\partial\Gamma^{i}+c_{16}\, \partial G^{i}\,\Gamma^{i}
\nonu\\
& & +c_{17}\,\varepsilon^{ijkl}(T^{ij}\,\partial(\Gamma^{k}\,\Gamma^{l})-\partial T^{ij}\,\Gamma^{k}\,\Gamma^{l})+c_{18}\,\partial^{2}U\Bigg](w)
+\cdots,
\nonu
\eea
where the coefficients 
are
\bea
c_ {1} &=& -\frac {16 k\,N} {(2 + k + N)^{2}}, \qquad
c_ {2} = 2, \nonu\\
c_ {3} &=& -\frac{1} {(2 + 
      N) (2 + k + N)^{2} (5 + 4 k + 4 N + 3 k\,N)}3 (100 + 187 k + 118 k^{2} + 24 k^{3} + 303 N
      \nonu\\&+& 
      505 k\,N + 277 k^{2} N + 46 k^{3} N + 325 N^{2} + 
      455 k\,N^{2} + 195 k^{2} N^{2} + 20 k^{3} N^{2} + 148 N^{3} 
      \nonu\\&+& 
      159 k\,N^{3} + 42 k^{2} N^{3} + 24 N^{4} + 16 k\,N^{4}), \nonu\\
c_ {4} &=& \frac {1}{(2 + k + N)^{2} (5 + 4 k + 4 N + 3 k\,N)}4 (50 + 85 k + 46 k^{2} + 8 k^{3} + 110 N + 
       157 k\,N 
       \nonu\\&+&  71 k^{2} N + 10 k^{3} N+ 76 N^{2} + 78 k\,N^{2} + 
       21 k^{2} N^{2} + 16 N^{3} + 
       8 k\,N^{3}), \nonu\\
c_ {5} &=& \frac {4 (10 + 9 k + 2 k^{2} + 14 N + 7 k\,N + 
       4 N^{2})} {(2 + k + N)^{3}}, \qquad
c_ {6} = -\frac {2 i} {(2 + k + N)}, \nonu\\
c_ {7} &=& \frac { (32 + 55 k + 18 k^{2} + 41 N + 35 k\,N + 
       11 N^{2})} {3(2 + N) (2 + k + N)^{4}}, \nonu\\
c_ {8} &=& -\frac { (13 k + 6 k^{2} + 3 N + 9 k\,N + N^{2})} {4(2 + 
      N) (2 + k + N)^{2}}, \nonu\\
c_ {9} &=& \frac {i (8 + 17 k + 6 k^{2} + 11 N + 11 k\,N + 
       3 N^{2})} {(2 + N) (2 + k + N)^{3}}, \nonu\\
c_ {10} &=& -\frac {(20 + 21 k + 6 k^{2} + 25 N + 13 k\,N + 
     7 N^{2})} {2(2 + N) (2 + k + N)^{2}}, \nonu\\
c_ {11} &=& \frac {2 i (20 + 21 k + 6 k^{2} + 25 N + 13 k\,N + 
       7 N^{2})} {(2 + N) (2 + k + N)^{3}}, \nonu\\
c_ {12} &=& -\frac {2 (100 + 99 k + 26 k^{2} + 151 N + 85 k\,N + 
      4 k^{2} N + 65 N^{2} + 14 k\,N^{2} + 8 N^{3})} {(2 + 
      N) (2 + k + N)^{3}}, \nonu\\
c_ {13} &=& \frac {8 (-2 - k - N + k\,N)} {(2 + k + N)^{2}}, \qquad
c_ {14} = -\frac {16} {(2 + k + N)^{2}}, \qquad
c_ {15} = \frac {6 i} {(2 + k + N)}, \nonu \\
c_ {16} & = &  -\frac {2 i} {(2 + k + N)}, \qquad
c_ {17} = -\frac {2 i} {(2 + k + N)^{2}}, \qquad
c_ {18} = \frac {8} {(2 + k + N)^{2}}.
\nonu
\eea
The fusion rule is given by
\bea
[\Phi_{0}^{(1)}] \, \cdot \, [\widetilde{\Phi}_{2}^{(1)}]
= [I]  + [\Phi_{0}^{(1)}\,\Phi_{0}^{(1)}]
+ [\Phi_{0}^{(2)}],
\nonu 
\eea
where the last term belongs to the next $16$ higher spin currents with 
spin $s=2$.

The OPEs between the higher spin-$1$ current and the four 
higher spin-$\frac{5}{2}$ currents in different basis are given by
\bea
&& \Phi_{0}^{(1)}(z)\;\Phi_{\frac{3}{2}}^{(1),i}(w)
=\frac{1}{(z-w)^{3}}c_{1}\,\Gamma^{i}(w)+\frac{1}{(z-w)^{2}}\Bigg[c_{2}\, G^{i}+c_{3}\, U\,\Gamma^{i}
\nonu\\
& & +\varepsilon^{ijkl}\,c_{4}\, T^{jk}\,\Gamma^{l}+c_{5}\, T^{ij}\,\Gamma^{j}+c_{6}\,\Gamma^{i}\Bigg](w)
\nonu\\
& & +\frac{1}{(z-w)}\Bigg[  c_{7}\,{\bf \Phi_{\frac{1}{2}}^{(2),i}}+
c_{8}\,{\bf \Phi_{0}^{(1)}\,\Phi_{\frac{1}{2}}^{(1),i}}+
c_{9}\, U\, U\,\Gamma^{i}+c_{10}\, U\,\partial\Gamma^{i}+c_{11}\, G^{i}\, U
\nonu\\
& & +c_{12}\,\partial G^{i}+c_{13}\,\partial^{2}\Gamma^{i}+c_{14}\,\partial U\,\Gamma^{i}+c_{15}\,\Gamma^{i}\,\partial\Gamma^{j}\,\Gamma^{j}+c_{16}\, T^{ij}\, U\,\Gamma^{j}+c_{17}\, G^{j}\,\Gamma^{i}\,\Gamma^{j}
\nonu\\
& & +c_{18}\, G^{j}\, T^{ij}+c_{19}\,\partial T^{ij}\,\Gamma^{j}+c_{20}\,T^{jk}\,\Gamma^{i}\,\Gamma^{j}\,\Gamma^{k}
 +\varepsilon^{ijkl} \left(c_{21}\, G^{j}\,\Gamma^{k}\,\Gamma^{l}+
c_{22}\, U\,\Gamma^{j}\,\Gamma^{k}\,\Gamma^{l} \right.
\nonu \\
&& + \left.
c_{23}\,\partial(\Gamma^{j}\,\Gamma^{k}\,\Gamma^{l})+c_{24}\, G^{j}\, T^{kl}
+c_{25}\,\partial T^{jk}\,\Gamma^{l}+c_{26}\, T^{jk}\, U\,\Gamma^{l}+c_{27}\, T^{jk}\,\partial\Gamma^{l} \right)\nonu \\
&& +c_{28}\,\widetilde{T}^{ij}\,\widetilde{T}^{ij}\,\Gamma^{i}
 +c_{29}\,\varepsilon^{ijkl}\, T^{ij}\, T^{kl}\,\Gamma^{i}+c_{30}\, T^{ij}\, T^{jk}\,\Gamma^{k}(1-\delta^{ik})
\Bigg](w)
+\cdots,
\nonu
\eea
where the coefficients are
\bea
c_ {1} & = & - \frac {8 i\, k\, N} {(2 + k + N)^{2}}, \qquad
c_ {2}  = \frac {3 (k - N)} {(2 + k + N)},  \qquad
c_ {3}  =  - \frac {6 i (k - N)} {(2 + k + N)^{2}},  \nonu\\
c_ {4} & = &\frac {(k - N)} {(2 + k + N)^{2}}, \qquad
c_ {5}  =  - \frac {4 (k + N)} {(2 + k + N)^{2}},\qquad
c_ {6}  = \frac {4 i (3 k + 3 N + 2 k\, N)} {(2 + k + N)^{2}}, \nonu\\
c_ {7} & = & - \frac {1} {2}, \nonu\\
c_ {8} & = &\frac {(60 + 77 k + 22 k^{2} + 121 N + 115 k\, 
    N + 20 k^{2} N + 79 N^{2} + 42 k\, 
    N^{2} + 16 N^{3})} {2 (2 + N) (2 + k + N)^{2}}, \nonu\\
c_ {9} & = &\frac {4 i} {(2 + k + N)^{2}},\qquad
c_ {10}  = \frac {2 i (10 + 11 k + 3 k^{2} + 12 N + 7 k\, 
     N + 3 N^{2})} {(2 + k + N)^{3}}, \nonu\\
c_ {11} & = & - \frac {2} {(2 + k + N)}, \nonu\\
c_ {12} & = & - \frac {(-20 - 19 k - 6 k^{2} - 7 N + 11 k\, 
     N + 6 k^{2} N + 15 N^{2} + 14 k\, 
     N^{2} + 6 N^{3})} {(2 (2 + N) (2 + k + N)^{2})}, \nonu\\
c_ {13} & = &\frac {4 i (1 + 2 k + 2 N)} {(2 + k + N)^{2}}, \nonu\\
c_ {14} & = &\frac {i (-20 - 19 k - 6 k^{2} - 7 N + 11 k\, 
     N + 6 k^{2} N + 15 N^{2} + 14 k\, 
     N^{2} + 6 N^{3})} {(2 + N) (2 + k + N)^{3}}, \nonu\\
c_ {15} & = &\frac {4 i (13 k + 6 k^{2} + 3 N + 9 k\, 
     N + N^{2})} {(2 + N) (2 + k + N)^{4}}, \nonu\\
c_ {16} & = & - \frac {(20 + 21 k + 6 k^{2} + 25 N + 13 k\, 
     N + 7 N^{2})} {(2 + N) (2 + k + N)^{3}}, \nonu\\
c_ {17} & = & - \frac {(20 + 21 k + 6 k^{2} + 25 N + 13 k\, 
     N + 7 N^{2})} {(2 + N) (2 + k + N)^{3}}, \nonu\\
c_ {18} & = & - \frac {i (20 + 21 k + 6 k^{2} + 25 N + 13 k\, 
      N + 7 N^{2})} {2 (2 + N) (2 + k + N)^{2}}, \nonu\\
c_ {19} & = &\frac {8} {(2 + k + N)^{2}}, \qquad
c_ {20}  = \frac {(32 + 55 k + 18 k^{2} + 41 N + 35 k\, 
    N + 11 N^{2})} {(2 + N) (2 + k + N)^{4}}, \nonu\\
c_ {21} & = & - \frac {(13 k + 6 k^{2} + 3 N + 9 k\, 
     N + N^{2})} {2 (2 + N) (2 + k + N)^{3}}, \nonu\\
c_ {22} & = &\frac {i (32 + 55 k + 18 k^{2} + 41 N + 35 k\, 
     N + 11 N^{2})} {3 (2 + N) (2 + k + N)^{4}}, \nonu\\
c_ {23} & = &\frac {2 i (1 + N) (20 + 23 k + 6 k^{2} + 23 N + 14 k\, 
     N + 6 N^{2})} {3 (2 + N) (2 + k + N)^{4}}, \nonu\\
c_ {24} & = & - \frac {i (16 + 21 k + 6 k^{2} + 19 N + 13 k\, 
      N + 5 N^{2})} {4 (2 + N) (2 + k + N)^{2}}, \nonu\\
c_ {25} & = & - \frac {(k + 9 N + 12 k\, 
     N + 3 k^{2} N + 11 N^{2} + 7 k\, 
     N^{2} + 3 N^{3})} {(2 + N) (2 + k + N)^{3}}, \nonu\\
c_ {26} & = & - \frac {(32 + 29 k + 6 k^{2} + 35 N + 17 k\, 
     N + 9 N^{2})} {2 (2 + N) (2 + k + N)^{3}}, \nonu\\
c_ {27} & = & - \frac {(10 + 11 k + 3 k^{2} + 12 N + 7 k\, 
     N + 3 N^{2})} {(2 + k + N)^{3}}, \nonu\\
c_ {28} & = &\frac {i (16 + 21 k + 6 k^{2} + 19 N + 13 k\, 
     N + 5 N^{2})} {(2 + N) (2 + k + N)^{3}}, \nonu\\
c_ {29} & = &\frac {i (20 + 21 k + 6 k^{2} + 25 N + 13 k\, 
     N + 7 N^{2})} {2 (2 + N) (2 + k + N)^{3}}, \nonu\\
c_ {30} & = &\frac {i (16 + 21 k + 6 k^{2} + 19 N + 13 k\, 
     N + 5 N^{2})} {(2 + N) (2 + k + N)^{3}}.
\nonu
\eea

Furthermore, 
the OPE between the higher spin-$1$ current and the 
higher spin-$3$ current in different basis is given by
\bea
&& \Phi_{0}^{(1)}(z)\;\Phi_{2}^{(1)}(w)
=\frac{1}{(z-w)^{4}}\, c_{1}+\frac{1}{(z-w)^{3}}\, c_{2}\, U(w)
\nonu\\
& &+\frac{1}{(z-w)^{2}}\Bigg[ c_{3}\, {\bf \Phi_{0}^{(2)}}+
c_{4}\,{\bf \Phi_{0}^{(1)}\,\Phi_{0}^{(1)}}+c_{5}\, L+c_{6}\, U\, U
+c_{7}\, G^{i}\,\Gamma^{i}
\nonu\\
& &+\varepsilon^{ijkl}(c_{8}\,\Gamma^{i}\,\Gamma^{j}\,\Gamma^{k}\,\Gamma^{l}+c_{9}\, T^{ij}\, T^{kl}+c_{10}\, T^{ij}\,\Gamma^{k}\,\Gamma^{l})
+\delta^{ik}\delta^{jl}(c_{11}\,T^{ij}\, T^{kl}+c_{12}\,T^{ij}\,\Gamma^{k}\,\Gamma^{l})
\nonu\\
& &+c_{13}\,\partial\Gamma^{i}\,\Gamma^{i}+c_{14}\,\partial U\Bigg](w)
\nonu\\
& & +\frac{1}{(z-w)}\Bigg[c_{15}\, U\,\partial\Gamma^{i}\,\Gamma^{i}+c_{16}\, G^{i}\,\partial\Gamma^{i}+c_{17}\, \partial G^{i}\,\Gamma^{i}
\nonu\\
& & +c_{18}\,\varepsilon^{ijkl}(T^{ij}\,\partial(\Gamma^{k}\,\Gamma^{l})-\partial T^{ij}\,\Gamma^{k}\,\Gamma^{l})+c_{19}\,\partial^{2}U\Bigg](w)
+\cdots,
\nonu
\eea
where the coefficients are
\bea
c_ {1} & = &\frac {4 k (k - N) N} {(2 + k + N)^{2}}, \qquad
c_ {2}  =  - \frac {16 k\, N} {(2 + k + N)^{2}}, \qquad
c_ {3}  =   2, \nonu\\
c_ {4} & = & - \frac {(60 + 77 k + 22 k^{2} + 121 N + 115 k\, 
     N + 20 k^{2} N + 79 N^{2} + 42 k\, 
     N^{2} + 16 N^{3})} {(2 + N) (2 + k + N)^{2}}, \nonu\\
c_ {5} & = &\frac {4 (10 + 9 k + 2 k^{2} + 14 N + 7 k\, 
     N + 4 N^{2})} {(2 + k + N)^{2}}, \nonu\\
c_ {6} & = &\frac {4 (10 + 9 k + 2 k^{2} + 14 N + 7 k\, 
     N + 4 N^{2})} {(2 + k + N)^{3}}, \qquad
c_ {7}  =  - \frac {2 i} {(2 + k + N)}, \nonu\\
c_ {8} & = &\frac {(32 + 55 k + 18 k^{2} + 41 N + 35 k\, 
    N + 11 N^{2})} {3 (2 + N) (2 + k + N)^{4}}, \nonu\\
c_ {9} & = & - \frac {(13 k + 6 k^{2} + 3 N + 9 k\, 
     N + N^{2})} {4 (2 + N) (2 + k + N)^{2}}, \nonu\\
c_ {10} & = &\frac {4 i (8 + 17 k + 6 k^{2} + 11 N + 11 k\, 
     N + 3 N^{2})} {(2 + N) (2 + k + N)^{3}}, \nonu\\
c_ {11} & = & - \frac {(20 + 21 k + 6 k^{2} + 25 N + 13 k\, 
     N + 7 N^{2})} {2 (2 + N) (2 + k + N)^{2}}, \nonu\\
c_ {12} & = &\frac {2 i (20 + 21 k + 6 k^{2} + 25 N + 13 k\, 
     N + 7 N^{2})} {(2 + N) (2 + k + N)^{3}}, \nonu\\
c_ {13} & = & - \frac {2 (100 + 99 k + 26 k^{2} + 151 N + 85 k\, 
      N + 4 k^{2} N + 65 N^{2} + 14 k\, 
      N^{2} + 8 N^{3})} {(2 + N) (2 + k + N)^{3}}, \nonu\\
c_ {14} & = &\frac {8 (-2 - k - N + k\, N)} {(2 + k + N)^{2}}, \qquad
c_ {15}  =  - \frac {16} {(2 + k + N)^{2}}, \qquad
c_ {16}  = \frac {6 i} {(2 + k + N)}, \nonu\\
c_ {17} & = & - \frac {2 i} {(2 + k + N)},\qquad
c_ {18}  =  - \frac {2 i} {(2 + k + N)^{2}}, \qquad
c_ {19}  = \frac {8} {(2 + k + N)}.
\nonu
\eea

One can rewrite the above OPEs by changing the order of the 
operators. For example,  
the OPE between the higher spin-$\frac{3}{2}$ current and the 
higher spin-$1$ current is given by
\bea
\Phi_{\frac{1}{2}}^{(1),i}(z)\:\Phi_{0}^{(1)}(w)=
\frac{1}{(z-w)}\Bigg[c_{1}\,G^{i}+c_{2}\,U\,\Gamma^{i}+\varepsilon^{ijkl}\Big(\,c_{3}\,T^{jk}\,\Gamma^{l}+c_{4}\,\Gamma^{j}\,\Gamma^{k}\,\Gamma^{l}\Big)\Bigg](w)+\cdots,
\nonu
\eea
where the coefficients are 
\bea
c_ {1} &= & -1, \qquad 
c_ {2} = \frac{2i} {(2 + k + N)}, \qquad
c_ {3} = -\frac {1} {(2 + k + N)}, \qquad
c_ {4} = \frac {4 i} {3 (2 + k + N)^{2}}.
\nonu
\eea

The OPE between the higher spin-$2$ current and 
the higher spin-$1$ current is given by
\bea
\Phi_{1}^{(1),ij}(z)\:\Phi_{0}^{(1)}(w)&=&
\frac{1}{(z-w)^{2}}\Bigg[c_{1}\,\Gamma^{i}\,\Gamma^{j}+\varepsilon^{ijkl}\,\Big(\,c_{2}\,T^{kl}+c_{3}\,\Gamma^{k}\,\Gamma^{l}\Big)+c_{4}\,T^{ij}\Bigg](w)
\nonu\\
&+&\frac{1}{(z-w)}\Bigg[c_{5}\,(G^{i}\,\Gamma^{j}-G^{j}\,\Gamma^{i})+c_{6}\,\partial T^{ij}+\varepsilon^{ijkl}\,c_{7}\,\partial T^{kl}+c_{8}\,\partial(\Gamma^{i}\,\Gamma^{j})
\nonu\\
&+&\varepsilon^{ijkl}\,\Big(\,c_{9}\,\partial(\Gamma^{k}\,\Gamma^{l})+c_{10}\,(T^{ki}\,\Gamma^{i}\,\Gamma^{l}+T^{kj}\,\Gamma^{j}\,\Gamma^{l})\Big)+c_{11}\,U\,\Gamma^{i}\,\Gamma^{j}\Bigg](w)+\cdots,
\nonu
\eea
where the coefficients are given by
\bea
c_ {1} &=& \frac {4 (k - N)} {(2 + k + N)^{2}}, \qquad
c_ {2} = \frac {i (k + N)} {(2 + k + N)}, \qquad
c_ {3} = \frac {2 (k + N)} {(2 + k + N)^{2}}, \qquad
c_ {4} = \frac {2 i (k - N)} {(2 + k + N)}, \nonu\\
c_ {5} &=& \frac {2 i} {(2 + k + N)}, \qquad
c_ {6} = \frac {2 i (k - N)} {(2 + k + N)}, \qquad
c_ {7} = i, \qquad
c_ {8} = \frac {4 (k - N)} {(2 + k + N)^{2}}, \nonu\\
c_ {9} &=& \frac {2} {(2 + k + N)}, \qquad
c_ {10} = \frac {4 i} {(2 + k + N)^{2}}, \qquad
c_ {11} = \frac {8} {(2 + k + N)^{2}}.
\nonu
\eea

The OPE between the higher spin-$\frac{5}{2}$ current and the 
higher spin-$1$ current 
is given by
\bea
\Phi_{\frac{3}{2}}^{(1),i}(z)\:\Phi_{0}^{(1)}(w)&=&
\frac{1}{(z-w)^{3}}\,c_{1}\,\Gamma^{i}(w)
\nonu\\
&+&\frac{1}{(z-w)^{2}}\Bigg[\,c_{2}\,G^{i}+c_{3}\,\partial\Gamma^{i}+c_{4}\,T^{ij}\,\Gamma^{j}+\varepsilon^{ijkl}\,c_{5}\,T^{jk}\,\Gamma^{l}+c_{6}\,U\,\Gamma^{i}\Bigg](w)
\nonu\\
&+&\frac{1}{(z-w)}\Bigg[c_{7}\, {\bf \Phi_{\frac{1}{2}}^{(2),i}}+
c_{8}\, {\bf \Phi_{0}^{(1)}\,\Phi_{\frac{1}{2}}^{(1),i}}+
c_{9}\,T^{jk}\,T^{jk}\,\Gamma^{i}+c_{10}\,T^{ij}\,T^{jk}\,\Gamma^{k}+c_{11}\,\partial U\,\Gamma^{i}
\nonu\\
&+&c_{12}\,\partial^{2}\Gamma^{i}+\varepsilon^{ijkl}\,c_{13}\,T^{jk}\,\partial\Gamma^{l}+c_{14}\,\Gamma^{i}\,\partial\Gamma^{j}\,\Gamma^{j}+\varepsilon^{ijkl}\,c_{15}\,T^{jk}\,G^{l}+c_{16}\,U\,U\,\Gamma^{i}
\nonu\\
&+&c_{17}\,U\,\partial\Gamma^{i}+c_{18}\,\partial G^{i}+c_{19}\,\partial T^{jk}\,\Gamma^{l}+c_{20}\,T^{ij}\,G^{j}
+\varepsilon^{ijkl}\Big(\,c_{21}\,T^{jk}\,U\,\Gamma^{l}+c_{22}\,\Gamma^{j}\,\Gamma^{k}\,G^{l}\Big)
\nonu\\
&+&c_{23}\,U\,G^{i}+\varepsilon^{ijkl}\,c_{24}\,T^{ij}\,T^{kl}\,\Gamma^{i}+c_{25}\,(U\,T^{ij}\,\Gamma^{j}-\Gamma^{j}\,\Gamma^{i}\,G^{j})+c_{26}\,T^{jk}\,\Gamma^{i}\,\Gamma^{j}\,\Gamma^{k}
\nonu\\
&+&c_{27}\,T^{ij}\,\partial\Gamma^{j}+\varepsilon^{ijkl}\,c_{28}\,U\,\Gamma^{j}\,\Gamma^{k}\,\Gamma^{l}+c_{29}\,\partial(\Gamma^{j}\,\Gamma^{k}\,\Gamma^{l})+c_{30}\,\partial T^{ij}\,\Gamma^{j}\Bigg](w)+\cdots,
\nonu
\eea
where the 
coefficients
are given by
\bea
c_ {1} &=& \frac {8 i k\, N} {(2 + k + N)^{2}}, \qquad
c_ {2} = \frac {3 (k - N)} {(2 + k + N)}, \qquad
c_ {3} = \frac {4 i (3 k + 3 N + 4 k\, N)} {(2 + k + N)^{2}}, \nonu\\
c_ {4} &=& -\frac {4 (k + N)} {(2 + k + N)^{2}}, \qquad
c_ {5} = \frac {(k - N)} {(2 + k + N)^{2}}, \qquad
c_ {6} = -\frac {6 i (k - N)} {(2 + k + N)^{2}}, \qquad
c_ {7} = \frac {1} {2}, \nonu\\
c_ {8} &=& -\frac {(60 + 77 k + 22 k^{2} + 121 N + 115 k\, 
    N + 20 k^{2} N + 79 N^{2} + 42 k\, 
    N^{2} + 16 N^{3})} {2 (2 + N) (2 + k + N)^{2}}, \nonu\\
c_ {9} &=& -\frac {i (16 + 21 k + 6 k^{2} + 19 N + 13 k\, 
     N + 5 N^{2})} {2 (2 + N) (2 + k + N)^{3}}, \nonu\\
c_ {10} &=& -\frac {i (16 + 21 k + 6 k^{2} + 19 N + 13 k\, 
     N + 5 N^{2})} {(2 + N) (2 + k + N)^{3}}, \nonu\\
c_ {11} &=& -\frac {2 i (10 + 17 k + 6 k^{2} + 6 N + 7 k\, 
     N)} {(2 + k + N)^{3}}, \qquad
c_ {12} = \frac {12 i (k + N + k\, N)} {(2 + k + N)^{2}}, \nonu\\
c_ {13} &=& \frac {(10 + 13 k + 4 k^{2} + 10 N + 7 k\, 
    N + 2 N^{2})} {(2 + k + N)^{3}}, \nonu\\
c_ {14} &=& -\frac {4 i (13 k + 6 k^{2} + 3 N + 9 k\, 
     N + N^{2})} {(2 + N) (2 + k + N)^{4}}, \nonu\\
c_ {15} &=& \frac {i (16 + 21 k + 6 k^{2} + 19 N + 13 k\, 
     N + 5 N^{2})} {4 (2 + N) (2 + k + N)^{2}}, \qquad
c_ {16} = -\frac {4 i} {(2 + k + N)^{2}}, \nonu\\
c_ {17} &=& -\frac {2 i (10 + 17 k + 6 k^{2} + 6 N + 7 k\, 
     N)} {(2 + k + N)^{3}}, \qquad
c_ {18} = \frac {(10 + 17 k + 6 k^{2} + 6 N + 7 k\, 
    N)} {(2 + k + N)^{2}}, \nonu\\
c_ {19} &=& \frac {(20 + 31 k + 10 k^{2} + 35 N + 41 k\, 
    N + 8 k^{2} N + 21 N^{2} + 14 k\, 
    N^{2} + 4 N^{3})} {2 (2 + N) (2 + k + N)^{3}}, \nonu\\
c_ {20} &=& \frac {i (20 + 21 k + 6 k^{2} + 25 N + 13 k\, 
     N + 7 N^{2})} {2 (2 + N) (2 + k + N)^{2}}, \nonu\\
c_ {21} &=& \frac {(32 + 29 k + 6 k^{2} + 35 N + 17 k\, 
    N + 9 N^{2})} {2 (2 + N) (2 + k + N)^{3}}, \qquad
c_ {22} = \frac {(13 k + 6 k^{2} + 3 N + 9 k\, 
    N + N^{2})} {2 (2 + N) (2 + k + N)^{3}}, \nonu\\
c_ {23} &=& \frac {2} {(2 + k + N)}, \qquad
c_ {24} = -\frac {i (20 + 21 k + 6 k^{2} + 25 N + 13 k\, 
     N + 7 N^{2})} {2 (2 + N) (2 + k + N)^{3}}, \nonu\\
c_ {25} &=& \frac {(20 + 21 k + 6 k^{2} + 25 N + 13 k\, 
    N + 7 N^{2})} {(2 + N) (2 + k + N)^{3}}, \nonu\\
c_ {26} &=& -\frac {(32 + 55 k + 18 k^{2} + 41 N + 35 k\, 
    N + 11 N^{2})} {(2 + N) (2 + k + N)^{4}}, \nonu\\
c_ {27} &=& -\frac {4 (k + N)} {(2 + k + N)^{2}}, \qquad
c_ {28} = -\frac {i (32 + 55 k + 18 k^{2} + 41 N + 35 k\, 
     N + 11 N^{2})} {3 (2 + N) (2 + k + N)^{4}}, \nonu\\
c_ {29} &=& -\frac {2 i (1 + N) (20 + 23 k + 6 k^{2} + 23 N + 14 k\, 
     N + 6 N^{2})} {3 (2 + N) (2 + k + N)^{4}}, \nonu\\
c_ {30} &=& -\frac {2 (16 + 29 k + 10 k^{2} + 27 N + 25 k\, 
     N + 2 k^{2} N + 13 N^{2} + 4 k\, 
     N^{2} + 2 N^{3})} {(2 + N) (2 + k + N)^{3}}.
\nonu
\eea

The OPE between the higher spin-$3$ current and the higher spin-$1$
current is 
\bea
\Phi_{2}^{(1)}(z)\:\Phi_{0}^{(1)}(w)&=&
\frac{1}{(z-w)^{4}}\,c_{1}\,\Gamma^{i}(w)+\frac{1}{(z-w)^{3}}\,c_{2}\,U(w)
\nonu\\
&+&\frac{1}{(z-w)^{2}}\Bigg[\,c_{3}\, {\bf \Phi_{0}^{(2)}}+
c_{4}\, {\bf \Phi_{0}^{(1)}\,\Phi_{0}^{(1)}}+
c_{5}\,L+c_{6}\,G^{i}\,\Gamma^{i}
\nonu \\
& + &
\varepsilon^{ijkl}\Big(\,c_{7}\,\Gamma^{i}\,\Gamma^{j}\,\Gamma^{k}\,\Gamma^{l}+c_{8}\,T^{ij}\,T^{kl}\Big)
\nonu\\
&+&c_{9}\,T^{ij}\,T^{ij}+\varepsilon^{ijkl}\,c_{10}\,T^{ij}\,\Gamma^{k}\,\Gamma^{l}+c_{11}\,T^{ij}\,\Gamma^{i}\,\Gamma^{j}+c_{12}\,\partial U+c_{13}\,\partial\Gamma^{i}\,\Gamma^{i}+c_{14}\,U\,U\Bigg](w)
\nonu\\
&+&\frac{1}{(z-w)}\Bigg[c_{15}\, {\bf \partial\Phi_{0}^{(2)}}+
c_{16}\, { \bf \partial\Phi_{0}^{(1)}\,\Phi_{0}^{(1)}}+
c_{17}\,G^{i}\,\partial\Gamma^{i}+c_{18}\,\partial L+c_{19}\,\partial(T^{ij}\,T^{ij})
\nonu\\
&+&c_{20}\,\partial U\,U+c_{21}\,\partial^{2}U+c_{22}\,\partial^{2}\Gamma^{i}\,\Gamma^{i}+\varepsilon^{ijkl}\Big(\,c_{23}\,\partial(T^{ij}\,\Gamma^{k}\,\Gamma^{l})+c_{24}\,\Gamma^{i}\,\Gamma^{j}\,\partial T^{kl}\Big)
\nonu\\
&+&c_{25}\,U\,\partial\Gamma^{i}\,\Gamma^{i}
+\varepsilon^{ijkl}\Big(\,c_{26}\,\partial(\Gamma^{i}\,\Gamma^{j}\,\Gamma^{k}\,\Gamma^{l})+c_{27}\,\partial(T^{ij}\,T^{kl})\Big)+c_{28}\,\partial(T^{ij}\,\Gamma^{i}\,\Gamma^{j})\Bigg](w) \nonu \\
& + & \cdots,
\nonu
\eea
where the coefficients are
given by
\bea
c_ {1} & = &\frac {4 k (k - N) N} {(2 + k + N)^{2}}, \qquad
c_ {2}  =\frac {16 k\, N} {(2 + k + N)^{2}},\qquad
c_ {3}  =  2, \nonu\\
c_ {4} & = & - \frac {(60 + 77 k + 22 k^{2} + 121 N + 115 k\, 
     N + 20 k^{2} N + 79 N^{2} + 42 k\, 
     N^{2} + 16 N^{3})} {(2 + N) (2 + k + N)^{2}}, \nonu\\
c_ {5} & = &\frac {4 (10 + 9 k + 2 k^{2} + 14 N + 7 k\, 
     N + 4 N^{2})} {(2 + k + N)^{2}}, \qquad
c_ {6}  =  - \frac {2 i} {(2 + k + N)}, \nonu\\
c_ {7} & = &\frac {(32 + 55 k + 18 k^{2} + 41 N + 35 k\, 
    N + 11 N^{2})} {3 (2 + N) (2 + k + N)^{4}}, \nonu\\
c_ {8} & = & - \frac {(13 k + 6 k^{2} + 3 N + 9 k\, 
     N + N^{2})} {4 (2 + N) (2 + k + N)^{2}}, \nonu\\
c_ {9} & = & - \frac {(20 + 21 k + 6 k^{2} + 25 N + 13 k\, 
     N + 7 N^{2})} {2 (2 + N) (2 + k + N)^{2}}, \nonu\\
c_ {10} & = &\frac {4 i (8 + 17 k + 6 k^{2} + 11 N + 11 k\, 
     N + 3 N^{2})} {(2 + N) (2 + k + N)^{3}}, \nonu\\
c_ {11} & = &\frac {4 i (20 + 21 k + 6 k^{2} + 25 N + 13 k\, 
     N + 7 N^{2})} {(2 + N) (2 + k + N)^{3}}, \qquad
c_ {12}  = \frac {8 (-2 - k - N + 3 k\, N)} {(2 + k + N)^{2}}, \nonu\\
c_ {13} & = & - \frac {2 (100 + 99 k + 26 k^{2} + 151 N + 85 k\, 
      N + 4 k^{2} N + 65 N^{2} + 14 k\, 
      N^{2} + 8 N^{3})} {(2 + N) (2 + k + N)^{3}}, \nonu\\
c_ {14} & = &\frac {4 (10 + 9 k + 2 k^{2} + 14 N + 7 k\, 
     N + 4 N^{2})} {(2 + k + N)^{3}},\qquad
c_ {15}  =  2, \nonu\\
c_ {16} & = & - \frac {2 (60 + 77 k + 22 k^{2} + 121 N + 115 k\, 
      N + 20 k^{2} N + 79 N^{2} + 42 k\, 
      N^{2} + 16 N^{3})} {(2 + N) (2 + k + N)^{2}}, \nonu\\
c_ {17} & = & - \frac {8 i} {(2 + k + N)}, \qquad
c_ {18}  = \frac {4 (10 + 9 k + 2 k^{2} + 14 N + 7 k\, 
     N + 4 N^{2})} {(2 + k + N)^{2}}, \nonu\\
c_ {19} & = & - \frac {(20 + 21 k + 6 k^{2} + 25 N + 13 k\, 
     N + 7 N^{2})} {2 (2 + N) (2 + k + N)^{2}}, \nonu\\
c_ {20} & = &\frac {8 (10 + 9 k + 2 k^{2} + 14 N + 7 k\, 
     N + 4 N^{2})} {(2 + k + N)^{3}}, \nonu\\
c_ {21} & = &\frac {16 (-2 - k - N + k\, N)} {(2 + k + N)^{2}}, \nonu\\
c_ {22} & = & - \frac {2 (100 + 99 k + 26 k^{2} + 151 N + 85 k\, 
      N + 4 k^{2} N + 65 N^{2} + 14 k\, 
      N^{2} + 8 N^{3})} {(2 + N) (2 + k + N)^{3}}, \nonu\\
c_ {23} & = &\frac {i (16 + 21 k + 6 k^{2} + 19 N + 13 k\, 
     N + 5 N^{2})} {(2 + N) (2 + k + N)^{3}}, \qquad
c_ {24}  =  - \frac {4 i} {(2 + k + N)^{2}}, \nonu\\
c_ {25} & = &\frac {16} {(2 + k + N)^{2}}, \qquad
c_ {26}  = \frac {(32 + 55 k + 18 k^{2} + 41 N + 35 k\, 
    N + 11 N^{2})} {3 (2 + N) (2 + k + N)^{4}}, \nonu\\
c_ {27} & = & - \frac {(13 k + 6 k^{2} + 3 N + 9 k\, 
     N + N^{2})} {4 (2 + N) (2 + k + N)^{2}}, \nonu\\
c_ {28} & = &\frac {2 i (20 + 21 k + 6 k^{2} + 25 N + 13 k\, 
     N + 7 N^{2})} {(2 + N) (2 + k + N)^{3}}.\nonu
\eea
In Appendix $J$, the same OPEs with rearrangement of 
composite fields are presented.

\subsection{The OPEs between the higher spin-$\frac{3}{2}$ currents and 
the other 
$15$ higher spin currents}

The OPEs between the higher spin-$\frac{3}{2}$ currents are given by
\bea
&& \Phi_{\frac{1}{2}}^{(1),i}(z)\;\Phi_{\frac{1}{2}}^{(1),j}(w) =
\frac{1}{(z-w)^{3}}\, c_{1}\,\delta^{ij}
+\frac{1}{(z-w)^{2}}\Bigg[c_{2}\,\Gamma^{i}\,\Gamma^{j}+c_{3}\,\varepsilon^{ijkl}\,\Gamma^{k}\,\Gamma^{l}+c_{4}\, T^{ij}+c_{5}\,\widetilde{T}^{ij}\Bigg](w)
\nonu\\
& & +
\frac{1}{(z-w)}\Bigg[(1-\delta^{ij})\Big(c_{6}\,\Gamma^{i}\,\partial\Gamma^{j}+c_{7}\,\partial\Gamma^{i}\,\Gamma^{j}
+c_{8}\,(T^{ik}\,\Gamma^{k}\,\Gamma^{j}+T^{jk}\,\Gamma^{k}\,\Gamma^{i})+c_{9}\,(T^{ik}\, T^{kj})\Big)
\nonu\\
& & +c_{10}\, \partial T^{ij}+c_{11}\,\varepsilon^{ijkl}\,\partial(\Gamma^{k}\,\Gamma^{l})+c_{12}\,\partial\tilde{T}^{ij}
+\delta^{ij}\Big(c_{13}\, L+c_{14}\, U\, U
+ c_{15}\,\partial\Gamma^{k}\,\Gamma^{k}
\nonu\\
& & +c_{16}\,\delta^{ik}\,\partial\Gamma^{k}\,\Gamma^{k}
+c_{17}\,\varepsilon^{iabc}\,\widetilde{T}^{ia}\,\Gamma^{b}\,\Gamma^{c}
 + c_{18}\,\widetilde{T}^{ik}\,\widetilde{T}^{ik}\Big)\Bigg](w)+\cdots,
\nonu
\eea
where the coefficients are 
\bea
c_{1}&=&-\frac{4\, k\, N}{(2+k+N)},\qquad
c_{2}=\frac{4(k+N)}{(2+k+N)^{2}},\qquad
c_{3}=\frac{2(k-N)}{(2+k+N)^{2}},\qquad
c_{4}=\frac{2i(k+N)}{(2+k+N)},\nonu\\
c_{5}&=&\frac{2i(k-N)}{(2+k+N)}, \qquad
c_{6}=\frac{2}{(2+k+N)},\qquad
c_{7}=\frac{2(-2+k+N)}{(2+k+N)},\qquad
c_{8}=-\frac{2i}{(2+k+N)^{2}},\nonu\\
c_{9} &=&\frac{1}{(2+k+N)},\qquad
c_{10}=\frac{i(-2+k+N)}{2+k+N},\qquad
c_{11} = \frac{(k-N)}{(2+k+N)^{2}},\nonu \\
c_{12}&=&\frac{i}{2+k+N},\qquad
c_{13}=-2,\qquad
c_{14}=-\frac{2}{(2+k+N)},\qquad
c_{15}=\frac{2(6+k+N)}{(2+k+N)^{2}},\nonu\\
c_{16}&=&-\frac{8}{(2+k+N)^{2}},\qquad
c_{17}=-\frac{4i}{(2+k+N)^{2}},\qquad
c_{18}=\frac{2}{(2+k+N)}.
\nonu
\eea
Again the fusion rule is given by
\bea
[\Phi_{\frac{1}{2}}^{(1),i}] \, \cdot \, [\Phi_{\frac{1}{2}}^{(1),j}] = [I].
\nonu 
\eea

The OPEs between the higher spin-$\frac{3}{2}$ currents and 
the higher spin-$2$ currents are given by
\bea
&& \Phi_{\frac{1}{2}}^{(1),i}(z)\;\Phi_{1}^{(1),jk}(w) =
\frac{1}{(z-w)^{3}}\,c_{1}\,\delta^{ij}\,\Gamma^{k}(w)
+\frac{1}{(z-w)^{2}}\Bigg[\delta^{ij}\Big(c_{2}\,G^{k}+c_{3}\,U\,\Gamma^{k}
\nonu\\
& &+c_{4}\,T^{ik}\,\Gamma^{i}
+ c_{5}\,\widetilde{T}^{ik}\,\Gamma^{k}
+c_{6}\,\varepsilon^{ikab}\,T^{ia}\,\Gamma^{b}+c_{7}\,\partial\Gamma^{k}\Big)+\varepsilon^{ijkl}\Big(c_{8}\,G^{l}+c_{9}\,U\,\Gamma^{l}+c_{10}\,\Gamma^{i}\,\Gamma^{j}\,\Gamma^{k}
\nonu\\
& & +c_{11}(T^{ij}\Gamma^{k}-T^{ik}\Gamma^{j})+c_{12}\,T^{jk}\,\Gamma^{i}+c_{13}\,\partial\Gamma^{k}
+c_{14}(T^{jl}\,\Gamma^{j}+T^{kl}\,\Gamma^{k})\Big)\Bigg](w)
\nonu\\
& & +\frac{1}{(z-w)}\Bigg[\delta^{ij} \left\{
 c_{15}\,{\bf \Phi_{\frac{1}{2}}^{(2),k}}+
c_{16}\,{\bf \Phi_{0}^{(1)}\,\Phi_{\frac{1}{2}}^{(1),k}}+
c_{17}\,L\,\Gamma^{k}+c_{18}\,U\,U\,\Gamma^{k}+c_{19}\,U\,\partial\Gamma^{k}
\right.
\nonu\\
& & +c_{20}\,\partial U\,\Gamma^{k}+c_{21}\,\partial^{2}\Gamma^{k}+c_{22}\,\partial\Gamma^{i}\,\Gamma^{i}\,\Gamma^{k}+c_{23}\,G^{i}\,\Gamma^{i}\,\Gamma^{k}+c_{24}\,G^{i}\,T^{ik}
+c_{25}\,T^{ik}\,U\,\Gamma^{i}
\nonu\\
& & +c_{26}\partial\,T^{ik}\,\Gamma^{i}+c_{27}\,\partial G^{k}+\varepsilon^{iabk}\Bigg(c_{28}\,\partial\Gamma^{i}\,\Gamma^{a}\,\Gamma^{b}
+c_{29}\,U\,\Gamma^{i}\,\Gamma^{a}\,\Gamma^{b}
\nonu\\
& & +c_{30}\,\Gamma^{i}\,\partial(\Gamma^{a}\,\Gamma^{b})+c_{31}\,G^{i}\,T^{ab}+c_{32}\,G^{i}\,\Gamma^{a}\,\Gamma^{b}
+c_{33}\,\partial T^{ab}\,\Gamma^{i}+c_{34}\,T^{ab}\,U\,\Gamma^{i}
\nonu\\
& & +c_{35}\,T^{ia}\,U\,\Gamma^{b}
+c_{36}\,G^{a}\,\Gamma^{b}\,\Gamma^{i}+c_{37}\,\partial T^{ia}\,\Gamma^{b}
+c_{38}\,G^{a}T^{ib}+c_{39}\,T^{ab}\,\Gamma^{k}\,\Gamma^{a}\,\Gamma^{b}\Bigg)
\nonu\\
& & +c_{40}(T^{kl}\,\partial\Gamma^{l}-\delta^{il}T^{kl}\,\partial\Gamma^{l})
+c_{41}(\Gamma^{k}\,\partial\Gamma^{l}\,\Gamma^{l}-\delta^{il}\Gamma^{k}\,\partial\Gamma^{l}\,\Gamma^{l})
\nonu\\
& & +c_{42}\,(T^{kl}\,U\,\Gamma^{l}-\delta^{il}T^{kl}\,U\,\Gamma^{l})
+c_{43}(G^{l}\,\Gamma^{k}\,\Gamma^{l}-\delta^{il}G^{l}\,\Gamma^{k}\,\Gamma^{l})
\nonu\\
& & +c_{44}(T^{kl}\,T^{kl}\,\Gamma^{k}-\delta^{il}\,T^{kl}\,T^{kl}\,\Gamma^{k})
+c_{45}(G^{l}\,T^{kl}-\delta^{il}G^{l}\,T^{kl})+c_{46}\,\widetilde{T}^{ik}\,\widetilde{T}^{ik}\,\Gamma^{k}
\nonu\\
& & +c_{47}(T^{ik}\,T^{il}\,\Gamma^{l}-T^{il}\,T^{il}\,\Gamma^{k})
+c_{48}(T^{kl}\,T^{li}\,\Gamma^{i}+\varepsilon^{ikab}\,T^{ka}\,\widetilde{T}^{ik}\,\Gamma^{b})
\nonu\\ 
& & +
\left.
c_{49}\,\varepsilon^{abck}\,T^{ab}\,\partial\Gamma^{c}+c_{50}\,T^{il}\,\Gamma^{i}\,\Gamma^{k}\,\Gamma^{l}
+c_{51}\,\varepsilon^{abcd}\,T^{ab}\,T^{cd}\,\Gamma^{k} \right\}
\nonu\\
& & +\varepsilon^{ijkl}\Big(c_{52}\,\partial U\,\Gamma^{l}+c_{53}\,\partial\Gamma^{l}\,U+c_{54}\,\partial G^{l}+c_{55}\,\partial^{2}\Gamma^{l}\Big)
+c_{56}\,\Gamma^{i}\,\partial(\Gamma^{j}\,\Gamma^{k})
\nonu\\
& & +c_{57}(\widetilde{T}^{ij}\,U\,\Gamma^{k}-\widetilde{T}^{ik}\,U\,\Gamma^{j})
+c_{58}(\partial\Gamma^{j}\,\Gamma^{j}\,\Gamma^{4-ijk}+\,\partial\Gamma^{k}\,\Gamma^{k}\,\Gamma^{4-ijk})
\nonu\\
& & +c_{59}\,(G^{j}\,\Gamma^{j}\,\Gamma^{4-ijk}+\,G^{k}\,\Gamma^{k}\,\Gamma^{4-ijk})+c_{60}(-G^{j}\,\widetilde{T}^{ik}+\,G^{k}\,\widetilde{T}^{ij})
\nonu\\
& & +c_{61}(-\partial\widetilde{T}^{ik}\,\Gamma^{j}+\partial\widetilde{T}^{ij}\,\Gamma^{k})+c_{62}\,T^{jk}\,\partial\Gamma^{i}+c_{63}\,\partial T^{jk}\,\Gamma^{i}
\nonu\\
& & +c_{64}(T^{ij}\,\widetilde{T}^{ij}\,\Gamma^{4-ijk}+T^{ik}\,\widetilde{T}^{ik}\,\Gamma^{4-ijk})+c_{65}\,(T^{ij}\,\partial\Gamma^{k}-T^{ik}\,\partial\Gamma^{j})
\nonu\\
& & +c_{66}(T^{ij}\,T^{jk}\,\Gamma^{j}+T^{ik}\,T^{jk}\,\Gamma^{k})
\Bigg](w)-\delta^{ik}\,
\sum_{n=3}^1 \, \frac{1}{(z-w)^n} \left(j\;\leftrightarrow \;k\right)(w)
+\cdots,
\nonu
\eea
where the coefficients are 
\bea
c_ {1} & = &\frac {8 i\, k\, N} {(2 + k + N)^{2}}, \qquad
c_ {2}  =  - \frac {(k - N)} {(2 + k + N)}, \qquad
c_ {3}  = \frac {2 i (k - N)} {(2 + k + N)^{2}}, \qquad
c_ {4}  =  - \frac {4 (k + N)} {(2 + k + N)^{2}}, \nonu\\
c_ {5} & = & - \frac {2 (k - N)} {(2 + k + N)^{2}}, \qquad
c_ {6}  = \frac {2 (k - N)} {(2 + k + N)^{2}}, \qquad
c_ {7}  =  - \frac {4 i (k + N)} {(2 + k + N)^{2}},\qquad
c_ {8}  = \frac {(2 + 3 k + 3 N)} {(2 + k + N)}, \nonu\\
c_ {9} & = & - \frac {2 i (2 + 3 k + 3 N)} {(2 + k + N)^{2}},\qquad 
c_ {10}  = \frac {8 i} {(2 + k + N)^{2}}, \qquad
c_ {11}  =  - \frac {2} {2 + k + N}, \nonu \\
c_ {12}  & = &  - \frac {2 (2 + 3 k + 3 N)} {(2 + k + N)^{2}},\qquad
c_ {13}  = \frac {8 i (k - N)} {(2 + k + N)^{2}}, \qquad
c_ {14}  = \frac {4 (k - N)} {(2 + k + N)^{2}}, \qquad
c_ {15}  = \frac {1} {2}, \nonu\\
c_ {16} & = & - \frac {(60 + 77 k + 22 k^{2} + 121 N + 115 k\, 
     N + 20 k^{2} N + 79 N^{2} + 42 k\, 
     N^{2} + 16 N^{3})} {2 (2 + N) (2 + k + N)^{2}}, \nonu\\
c_ {17} & = &\frac {4 i} {(2 + k + N)}, \qquad
c_ {18}  = \frac {4 i} {(2 + k + N)^{2}}, \nonu\\
c_ {19} & = & - \frac {2 i (10 + 11 k + 3 k^{2} + 12 N + 7 k\, 
      N + 3 N^{2})} {(2 + k + N)^{3}}, \nonu\\
c_ {20} & = & - \frac {i (-20 - 19 k - 6 k^{2} - 7 N + 11 k\, 
       N + 6 k^{2} N + 15 N^{2} + 14 k\, 
       N^{2} + 6 N^{3})} {(2 + N) (2 + k + N)^{3}}, \nonu\\
c_ {21} & = & - \frac {4 i (5 + 3 k + 3 N)} {(2 + k + N)^{2}},\nonu\\ 
c_ {22} & = & - \frac {4 i (16 + 29 k + 10 k^{2} + 27 N + 25 k\, 
       N + 2 k^{2} N + 13 N^{2} + 4 k\, 
       N^{2} + 2 N^{3})} {(2 + N) (2 + k + N)^{4}}, \nonu\\
c_ {23} & = & - \frac {(20 + 21 k + 6 k^{2} + 25 N + 13 k\, 
     N + 7 N^{2})} {(2 + N) (2 + k + N)^{3}}, \nonu\\
c_ {24} & = & - \frac {i (20 + 21 k + 6 k^{2} + 25 N + 13 k\, 
      N + 7 N^{2})} {2 (2 + N) (2 + k + N)^{2}}, \nonu\\
c_ {25} & = & - \frac {(20 + 21 k + 6 k^{2} + 25 N + 13 k\, 
     N + 7 N^{2})} {(2 + N) (2 + k + N)^{3}}, \qquad
c_ {26}  =  - \frac {2} {(2 + k + N)}, \nonu\\
c_ {27} & = &\frac {(-20 - 19 k - 6 k^{2} - 7 N + 11 k\, 
    N + 6 k^{2} N + 15 N^{2} + 14 k\, 
    N^{2} + 6 N^{3})} {2 (2 + N) (2 + k + N)^{2}}, \nonu\\
c_ {28} & = &\frac {2 i (1 + N) (20 + 23 k + 6 k^{2} + 23 N + 14 kN + 
      6 N^{2})} {(2 + N) (2 + k + N)^{4}}, \nonu\\
c_ {29} & = &\frac { i (32 + 55 k + 18 k^{2} + 41 N + 35 k\, 
      N + 11 N^{2})} {(2 + N) (2 + k + N)^{4}}, \nonu\\
c_ {30} & = &\frac {2 i (20 + 27 k + 8 k^{2} + 39 N + 39 k\, 
      N + 7 k^{2} N + 25 N^{2} + 14 k\, 
      N^{2} + 5 N^{3})} {(2 + N) (2 + k + N)^{4}}, \nonu\\
c_ {31} & = & -\frac {i (13 k + 6 k^{2} + 3 N + 9 k\, 
      N + N^{2})} {4 (2 + N) (2 + k + N)^{2}}, \nonu\\
c_ {32} & = & -\frac {(13 k + 6 k^{2} + 3 N + 9 k\, 
     N + N^{2})} {2(2 + N) (2 + k + N)^{3}}, \nonu\\
c_ {33} & = & -\frac { (5 k + 2 k^{2} + 5 N + 14 k\, 
       N + 4 k^{2} N + 7 N^{2} + 7 k\, 
       N^{2} + 2 N^{3})} {(2 + N) (2 + k + N)^{3}}, \nonu\\
c_ {34} & = & -\frac {(13 k + 6 k^{2} + 3 N + 9 k\, 
     N + N^{2})} {2(2 + N) (2 + k + N)^{3}}, \nonu\\
c_ {35} & = &-\frac {(16 + 21 k + 6 k^{2} + 19 N + 13 k\, 
     N + 5 N^{2})} {(2 + N) (2 + k + N)^{3}}, \nonu\\
c_ {36} & = & -\frac {(16 + 21 k + 6 k^{2} + 19 N + 13 k\, 
     N + 5 N^{2})} {(2 + N) (2 + k + N)^{3}}, \nonu\\
c_ {37} & = & -\frac {2 (k + 9 N + 12 k\, 
       N + 3 k^{2} N + 11 N^{2} + 7 k\, 
       N^{2} + 3 N^{3})} {(2 + N) (2 + k + N)^{3}}, \nonu\\
c_ {38} & = & -\frac {i (16 + 21 k + 6 k^{2} + 19 N + 13 k\, 
      N + 5 N^{2})} {2 (2 + N) (2 + k + N)^{2}}, \nonu\\
c_ {39} & = & -\frac {(32 + 55 k + 18 k^{2} + 41 N + 35 k\, 
       N + 11 N^{2})} {(2 + N) (2 + k + N)^{4}}, \qquad
c_ {40}  = \frac {12} {(2 + k + N)^{2}}, \nonu\\
c_ {41} & = & - \frac {4 i (8 + 21 k + 8 k^{2} + 15 N + 17 k\, 
       N + k^{2} N + 7 N^{2} + 2 k\, 
       N^{2} + N^{3})} {(2 + N) (2 + k + N)^{4}}, \nonu\\
c_ {42} & = &\frac {(20 + 21 k + 6 k^{2} + 25 N + 13 k\, 
    N + 7 N^{2})} {(2 + N) (2 + k + N)^{3}}, \nonu\\
c_ {43} & = &\frac {(20 + 21 k + 6 k^{2} + 25 N + 13 k\, 
    N + 7 N^{2})} {(2 + N) (2 + k + N)^{3}}, \qquad
c_ {44}  =  - \frac {4 i} {(2 + k + N)^{2}}, \nonu\\
c_ {45} & = &\frac {i (20 + 21 k + 6 k^{2} + 25 N + 13 k\, 
     N + 7 N^{2})} {2 (2 + N) (2 + k + N)^{2}}, \nonu\\
c_ {46} & = & - \frac {i (16 + 21 k + 6 k^{2} + 19 N + 13 k\, 
      N + 5 N^{2})} {(2 + N) (2 + k + N)^{3}}, \nonu\\
c_ {47} & = &\frac {i (16 + 21 k + 6 k^{2} + 19 N + 13 k\, 
     N + 5 N^{2})} {(2 + N) (2 + k + N)^{3}}, \nonu\\
c_ {48} & = & - \frac {i (13 k + 6 k^{2} + 3 N + 9 k\, 
      N + N^{2})} {(2 + N) (2 + k + N)^{3}}, \nonu\\
c_ {49} & = & - \frac { (10 + 11 k + 3 k^{2} + 12 N + 7 k\, 
       N + 3 N^{2})} {(2 + k + N)^{3}}, \nonu\\
c_ {50} & = &\frac {2 (32 + 55 k + 18 k^{2} + 41 N + 35 k\, 
     N + 11 N^{2})} {(2 + N) (2 + k + N)^{4}}, \nonu\\
c_ {51} & = & - \frac {i (20 + 21 k + 6 k^{2} + 25 N + 13 k\, 
      N + 7 N^{2})} {8(2 + N) (2 + k + N)^{3}}, \nonu\\
c_ {52} & = & - \frac {2 i (-2 + k + N)} {(2 + k + N)^{2}},  \qquad
c_ {53}  =  - \frac {2 i} {(2 + k + N)},  \qquad
c_ {54}  = \frac {(-2 + k + N)} {(2 + k + N)},  \nonu \\
c_ {55}  & = &  \frac {4 i (k - N)} {(2 + k + N)^{2}},  \qquad
c_ {56}  = \frac {4 i} {(2 + k + N)^{2}},  \qquad
c_ {57}  =  - \frac {4} {(2 + k + N)^{2}},  \nonu\\
c_ {58} & = &\frac {4 i (k - N)} {(2 + k + N)^{3}},  \qquad
c_ {59}  =  - \frac {4} {(2 + k + N)^{2}},  \qquad
c_ {60}  =  - \frac {2 i} {(2 + k + N)},  \nonu \\
c_ {61}  & = &  \frac {2 (k - N)} {(2 + k + N)^{2}},  \qquad
c_ {62}  =  - \frac {2 (-2 + k + N)} {(2 + k + N)^{2}},  \qquad
c_ {63}  =  - \frac {2 (k + N)} {(2 + k + N)^{2}},  \nonu\\
c_ {64} & = & - \frac {4 i} {(2 + k + N)^{2}},  \qquad
c_ {65}  =  - \frac {2 (4 + k + N)} {(2 + k + N)^{2}},  \qquad
c_ {66}  = \frac {4 i} {(2 + k + N)^{2}}. 
\nonu
\eea
The fusion rule is 
\bea
[\Phi_{\frac{1}{2}}^{(1),i}] \, \cdot \, [\Phi_{1}^{(1),jk}]
= [I^{ijk}] +  \delta^{ij} \, [\Phi_{0}^{(1)} \, \Phi_{\frac{1}{2}}^{(1),k}]
+ \delta^{ij} \, [\Phi_{\frac{1}{2}}^{(2),k}],
\nonu 
\eea
where the last term belongs to the next $16$ higher spin currents with 
spin $s=2$.

The OPEs between the higher spin-$\frac{3}{2}$ currents and 
the higher spin-$\frac{5}{2}$ currents are given by
\bea
&& \Phi_{\frac{1}{2}}^{(1),i}(z)\;\widetilde{\Phi}_{\frac{3}{2}}^{(1),j}(w)=
\frac{1}{(z-w)^{3}}\Bigg[c_{1}\,\delta^{ij}U+c_{2}\,\Gamma^{i}\,\Gamma^{j}+c_{3}\,T^{ij}+\varepsilon^{ijkl}(c_{4}\,\Gamma^{k}\,\Gamma^{l}+c_{5}\,T^{kl})\Bigg](w)
\nonu\\
&& +\frac{1}{(z-w)^{2}}\Bigg[\delta^{ij}
\left(
 c_{6}\,{\bf \Phi_{0}^{(2)}}+c_{7}\,{\bf \Phi_{0}^{(1)}\,\Phi_{0}^{(1)}}+
c_{8}\,L+c_{9}\,U\,U+c_{10}\,\partial U
\right.
\nonu\\
&& +\varepsilon^{abcd}(c_{11}\,\Gamma^{a}\,\Gamma^{b}\,\Gamma^{c}\,\Gamma^{d}+c_{12}\,T^{ab}\,\Gamma^{c}\,\Gamma^{d}+c_{13}\,T^{ab}\,T^{cd})+c_{14}\,G^{k}\,\Gamma^{k}+c_{15}\,G^{i}\,\Gamma^{i}
\nonu\\
&& +c_{16}\,\partial\Gamma^{k}\,\Gamma^{k}+c_{17}\,\partial\Gamma^{i}\,\Gamma^{i}+c_{18}\,T^{ik}\,T^{ik}+c_{19}\,\widetilde{T}^{ik}\,\widetilde{T}^{ik}+c_{20}\,T^{ik}\,\Gamma^{i}\,\Gamma^{k}
\nonu\\
&& + \left.
c_{21}\,\varepsilon^{iabc}\,\widetilde{T}^{ia}\,\Gamma^{b}\,\Gamma^{c}
\right)
+c_{22}\,U\,\Gamma^{i}\,\Gamma^{j}+c_{23}\,G^{i}\,\Gamma^{j}+c_{24}\,G^{j}\,\Gamma^{i}
\nonu\\
& & +c_{25}(\Gamma^{i}\,\partial\Gamma^{j}-\partial\Gamma^{i}\,\Gamma^{j})+c_{26}\,T^{ij}\,U+c_{27}\,\partial T^{ij}
+\varepsilon^{ijkl}
\left(
c_{28}(G^{k}\,\Gamma^{l})+c_{29}\,\partial(\Gamma^{k}\,\Gamma^{l})
\right.
\nonu\\
& & +
c_{30}\,T^{kl}\,U+c_{31}\,\partial T^{kl}
\left.
+c_{32}(T^{ik}\,\Gamma^{i}\,\Gamma^{l}+T^{jk}\,\Gamma^{i}\,\Gamma^{l}) \right)
\nonu\\
&& +(1-\delta^{ij})\Big(c_{33}\,T^{ik}\,T^{jk}+
c_{34}(T^{ik}\,\Gamma^{j}\,\Gamma^{k}+T^{jk}\,\Gamma^{i}\,\Gamma^{k})\Big)
\Bigg](w)
\nonu\\
& & +\frac{1}{(z-w)}\Bigg[\delta^{ij}
\left\{
 c_{35}\, {\bf \partial\Phi_{0}^{(2)}}+
c_{36}\,{ \bf \partial\Phi_{0}^{(1)}\,\partial\Phi_{0}^{(1)}}+
c_{37}\,\partial L+c_{38}\,L\,U+c_{39}\,U\,U\,U
\right.
\nonu\\
& & 
+c_{40}\,\partial U\,U+c_{41}\,\partial^{2}U+c_{42}\,U\,\partial\Gamma^{i}\,\Gamma^{i}+c_{43}\,G^{i}\,\partial\Gamma^{i}+c_{44}\,\partial G^{i}\,\Gamma^{i}+c_{45}\,\partial^{2}\Gamma^{i}\,\Gamma^{i}
\nonu\\
& & +c_{46}\,\partial G^{k}\,\Gamma^{k}+c_{47}\,G^{k}\,\partial\Gamma^{k}+c_{48}\,\partial^{2}\Gamma^{k}\,\Gamma^{k}\Big)
+\varepsilon^{iabc}\Big(c_{49}\,\Gamma^{i}\,\partial(\Gamma^{a}\,\Gamma^{b}\,\Gamma^{c})
\nonu\\
& & +c_{50}\,\partial\Gamma^{i}\,\Gamma^{a}\,\Gamma^{b}\,\Gamma^{c}+c_{51}\,T^{ia}\,\partial(\Gamma^{b}\,\Gamma^{c})
+c_{52}\,\partial T^{ia}\,\Gamma^{b}\,\Gamma^{c}+c_{53}\,T^{ia}\,\partial T^{bc}
\nonu\\
& & +c_{54}\,\partial T^{ia}\,T^{bc}+c_{55}\,T^{ab}\,\Gamma^{i}\,\partial\Gamma^{c}
+c_{56}\,\partial T^{ab}\,\Gamma^{i}\,\Gamma^{c}+c_{57}\,T^{ab}\,\partial\Gamma^{i}\,\Gamma^{c}\Big)
+c_{58}\,\partial T^{ik}\,T^{ik}
\nonu\\
& & +c_{59}\,\partial T^{ik}\,\Gamma^{i}\,\Gamma^{k}
+c_{60}\,T^{ik}\,\Gamma^{i}\,\partial\Gamma^{k}+c_{61}\,T^{ik}\,\partial\Gamma^{i}\,\Gamma^{k}+c_{62}\,\widetilde{T}^{ik}\,\widetilde{T}^{ik}\,U+c_{63}\,\partial\widetilde{T}^{ik}\,\widetilde{T}^{ik}
\nonu\\
& & +
\left.
\varepsilon^{iabc}\Big(c_{64}\,\partial\widetilde{T}^{ia}\,\Gamma^{b}\,\Gamma^{c}+c_{65}\,G^{a}\,\widetilde{T}^{ib}\,\Gamma^{c}+c_{66}\,\widetilde{T}^{ia}\,\Gamma^{b}\,\partial\Gamma^{c}+c_{67}\,\widetilde{T}^{ia}\,\partial\Gamma^{b}\,\Gamma^{c}
\Big)\right\}
\nonu\\
& & +\varepsilon^{ijkl}\Big(  
c_{68}\,{\bf \Phi_{1}^{(2),kl}}+c_{69}\,{\bf \Phi_{0}^{(1)}\,\Phi_{1}^{(1),kl}}\Big)+
 c_{70}\,{\bf \Phi_{\frac{1}{2}}^{(1),i}\,\Phi_{\frac{1}{2}}^{(1),j}}+c_{71}\,L\,T^{ij}+c_{72}\,L\,\Gamma^{i}\,\Gamma^{j}
\nonu\\
& & +c_{73}\,G^{i}\,G^{j}+c_{74}\,U\,\Gamma^{i}\,\partial\Gamma^{j}+c_{75}\,U\,\partial\Gamma^{i}\,\Gamma^{j}+c_{76}\,\partial^{2}T^{ij}+c_{77}\,T^{ij}\,U\,U
\nonu\\
& & +c_{78}\,\partial^{2}\Gamma^{i}\,\Gamma^{j}
+c_{79}\,\Gamma^{i}\,\partial^{2}\Gamma^{j}+c_{80}\,\partial\Gamma^{i}\,\partial\Gamma^{j}+c_{81}\,\partial T^{ij}\,U+c_{82}\,\partial U\,\Gamma^{i}\,\Gamma^{j}
\nonu\\
& & +c_{83}\,G^{i}\,U\,\Gamma^{j}
+c_{84}\,G^{j}\,U\,\Gamma^{i}+c_{85}\,G^{i}\,\partial\Gamma^{j}+c_{86}\,\partial G^{i}\,\Gamma^{j}+c_{87}\,\partial G^{j}\,\Gamma^{i}
\nonu\\
& & +c_{88}\,T^{ij}\,\partial\Gamma^{i}\,\Gamma^{i}
+c_{89}\,T^{ij}\,\partial\Gamma^{j}\,\Gamma^{j}+\varepsilon^{ijkl}
\left( c_{90}\,L\,T^{kl}+c_{91}\,L\,\Gamma^{k}\,\Gamma^{l}+c_{92}\,G^{k}\,G^{l}
\right.
\nonu\\
& & +c_{93}\,G^{k}\,U\,\Gamma^{l}
+c_{94}\,\partial G^{k}\,\Gamma^{l}+c_{95}\,T^{kl}\,T^{kl}\,T^{kl}+c_{96}\,T^{kl}\,U\,U+c_{97}\,T^{kl}\,\partial U
\nonu\\
& & +c_{98}\,\partial T^{kl}\,U
+c_{99}\,\partial^{2}T^{kl}+c_{100}\,T^{kl}\,T^{kl}\,\Gamma^{k}\,\Gamma^{l}+c_{101}\,\partial U\,\Gamma^{k}\,\Gamma^{l}+c_{102}\,\partial^{2}\Gamma^{k}\,\Gamma^{l}
\nonu\\
& & +c_{103}\,\partial\Gamma^{k}\,\partial\Gamma^{l}+c_{104}\,\Gamma^{k}\,\partial^{2}\Gamma^{l}+c_{105}\,U\,U\,\Gamma^{k}\,\Gamma^{l}+c_{106}\,U\,\partial(\Gamma^{k}\,\Gamma^{l})
\nonu\\
& & +c_{107}(1+\varepsilon^{ijkl})(T^{ij}\,T^{kl}\,T^{kl}+T^{ik}\,T^{lj}\,T^{kl}+T^{il}\,T^{jk}\,T^{kl})+c_{108}\,\widetilde{T}^{kl}\,\partial\Gamma^{k}\,\Gamma^{k}
\nonu\\
& & +c_{109}\,G^{i}\,\widetilde{T}^{il}\,\Gamma^{k}+c_{110}\,G^{j}\,\widetilde{T}^{lj}\,\Gamma^{k}+c_{111}(\partial\Gamma^{i}\,\Gamma^{i}\,\Gamma^{k}\,\Gamma^{l}+\partial\Gamma^{j}\,\Gamma^{j}\,\Gamma^{k}\,\Gamma^{l})
\nonu\\
& & +c_{112}\,G^{k}\,\widetilde{T}^{jl}\,\Gamma^{j}+c_{113}\,G^{k}\,\widetilde{T}^{il}\,\Gamma^{i}+c_{114}\,G^{k}\,\widetilde{T}^{kl}\,\Gamma^{k}
+c_{115}(G^{i}\,T^{kl}\,\Gamma^{i}+G^{j}\,T^{kl}\,\Gamma^{j})
\nonu\\
& & +c_{116}(\partial T^{ik}\,T^{il}+T^{ik}\,\partial T^{il})+c_{117}(\partial T^{jk}\,T^{jl}+T^{jk}\,\partial T^{jl})+c_{118}\,\widetilde{T}^{lj}\,\Gamma^{j}\,\partial\Gamma^{k}
\nonu\\
& & +c_{119}\,\widetilde{T}^{lj}\,\partial\Gamma^{j}\,\Gamma^{k}
+c_{120}\,T^{ik}\,T^{kl}\,\Gamma^{i}\,\Gamma^{k}+c_{121}\,T^{jk}\,T^{kl}\,\Gamma^{j}\,\Gamma^{k}+c_{122}\,T^{kl}\,\partial\Gamma^{i}\,\Gamma^{i}
\nonu\\
& & +c_{123}\,T^{kl}\,\partial\Gamma^{j}\,\Gamma^{j}+c_{124}\,T^{kl}\,\partial\Gamma^{k}\,\Gamma^{k}+c_{125}\,\widetilde{T}^{il}\,\Gamma^{i}\,\partial\Gamma^{k}+c_{126}\,\widetilde{T}^{il}\,\partial\Gamma^{i}\,\Gamma^{k}
\nonu\\
& & +c_{127}(T^{ik}\,T^{ik}\,T^{kl}+T^{jk}\,T^{jk}\,T^{kl})
+c_{128}\,T^{ik}\,T^{jl}\,\Gamma^{i}\,\Gamma^{j}+c_{129}\,G^{k}\,T^{kl}\,\Gamma^{k}
\nonu\\
& & +c_{130}\,\partial T^{ik}\,\partial\Gamma^{i}\,\Gamma^{l}+c_{131}\,T^{ik}\,\partial\Gamma^{i}\,\Gamma^{l}+c_{132}\,T^{ik}\,\Gamma^{i}\,\partial\Gamma^{l}+c_{133}\,T^{ij}\,T^{ik}\,T^{jl}
\nonu\\
& & +
\left.
c_{134}\,\partial T^{jk}\,\Gamma^{j}\,\Gamma^{l}+c_{135}\,T^{jk}\,\partial\Gamma^{j}\,\Gamma^{l}+c_{136}\,T^{jk}\,\Gamma^{j}\,\partial\Gamma^{l}
\right)
+c_{137}\,\Gamma^{i}\,\Gamma^{j}\,\partial\Gamma^{k}\,\Gamma^{k}
\nonu\\
& & +c_{138}\,\partial T^{ik}\,T^{jk}+c_{139}\,T^{ik}\,\partial T^{jk}+c_{140}\,T^{ik}\,T^{jk}\,U+c_{141}\,\partial T^{jk}\,\Gamma^{i}\,\Gamma^{k}
\nonu\\
& & +c_{142}(T^{ik}\,U\,\Gamma^{j}\,\Gamma^{k}+T^{jk}\,U\,\Gamma^{i}\,\Gamma^{k})
+c_{143}\,G^{k}\,\Gamma^{i}\,\Gamma^{j}\,\Gamma^{k}+c_{144}\,\partial T^{ik}\,\Gamma^{j}\,\Gamma^{k}
\Bigg](w)+\cdots,
\nonu
\eea
where the coefficients are 
\bea
c_{1}&=&\frac{8\,k\,N}{(2+k+N)^{2}},\qquad
c_{2}=-\frac{16(k-N)(3+2k+2N)}{3(2+k+N)^{3}},\nonu\\
c_{3}&=&-\frac{8i(k-N)(3+2k+2N)}{3(2+k+N)^{2}},\qquad
c_{4}=-\frac{8(3k+2k^{2}+3N+2k\,N+2N^{2})}{3(2+k+N)^{2}},\nonu\\
c_{5}&=&-\frac{4i(3k+2k^{2}+3N+5k\,N+2N^{2})}{3(2+k+N)^{2}},\qquad
c_{6}=-2,\nonu\\
c_{7}&=&\frac{60+77k+22k^{2}+121N+115kN+20k^{2}N+79N^{2}+42kN^{2}+16N^{3}}{(2+N)(2+k+N)^{2}},\nonu\\
c_{8}&=&-\frac{4(3+2k+N)(10+5k+8N)}{3(2+k+N)^{2}},\qquad
c_{9}=-\frac{4(3+2k+N)(10+5k+8N)}{3(2+k+N)^{2}},\nonu\\
c_{10}&=&\frac{4(4+5k+5N)}{(2+k+N)^{2}},\qquad
c_{11}=-\frac{(32+55k+18k^{2}+41N+35kN+11N^{2})}{3(2+N)(2+k+N)^{4}},\nonu\\
c_{12}&=&-\frac{i(8+17k+6k^{2}+11N+11kN+3N^{2})}{(2+N)(2+k+N)^{3}},\qquad
c_{13}=-\frac{(-9k-6k^{2}+N-7kN+N^{2})}{4(2+N)(2+k+N)^{2}},\nonu\\
c_{14}&=&\frac{2i(2+3k+3N)}{(2+k+N)^{2}},\qquad
c_{15}=-\frac{4i(k+N)}{(2+k+N)^{2}},\nonu\\
c_{16}&=&\frac{2(300+297k+94k^{2}+453N+255kN+20k^{2}N+179N^{2}+42kN^{2}+16N^{3})}{3(2+N)(2+k+N)^{3}},\nonu\\
c_{17}&=&\frac{32(k-N)}{3(2+k+N)},\qquad
c_{18}=\frac{(20+21k+6k^{2}+25N+13kN+7N^{2})}{(2+N)(2+k+N)^{2}},\nonu\\
c_{19}&=&\frac{(60+71k+18k^{2}+67N+43kN+17N^{2})}{(2+N)(2+k+N)^{2}},\nonu\\
c_ {20} & = & - \frac {4 i (20 + 21 k + 6 k^{2} + 25 N + 13 kN + 
       7 N^{2})} {(2 + N) (2 + k + N)^{3}},\nonu\\
c_ {21} & = & - \frac {2 i (60 + 59 k + 18 k^{2} + 79 N + 37 kN + 
       23 N^{2})} {3(2 + N) (2 + k + N)^{3}},\qquad 
c_ {22}  = \frac {8} {(2 + k + N)^{2}},\nonu\\ 
c_ {23} & = &\frac {2 i} {(2 + k + N)},\qquad
c_ {24}  =  - \frac {2 i (2 + 3 k + 3 N)} {(2 + k + N)^{2}},\qquad
c_ {25}  =  - \frac {16 (k - N)} {3 (2 + k + N)^{3}},\nonu \\
c_ {26}  & = &  - \frac {4 i (k + N)} {(2 + k + N)^{2}},\qquad
c_ {27}  = \frac {8 i (k - N)} {3 (2 + k + N)^{2}},\qquad 
c_ {28}  = \frac {2 i (k - N)} {(2 + k + N)^{2}},\nonu\\ 
c_ {29} & = &\frac {4} {(2 + k + N)^{2}},\qquad
c_ {30}  =  - \frac {2 i (k - N)} {(2 + k + N)^{2}},\qquad
c_ {31}  = \frac {4 i (1 + k + N)} {(2 + k + N)^{2}},
\nonu \\
c_ {32} & = &  - \frac  {4 (k - N)} {3 (2 + k + N)^{2}},\qquad
c_ {33}  =  - \frac{4 i} {(2 + k + N)^{2}},\qquad 
c_ {34}  = \frac {4 i (k - N)} {3 (2 + k + N)^{3}},\qquad 
c_ {35}  =  - \frac {1} {2},\nonu\\ 
c_ {36} & = &\frac {(60 + 77 k + 22 k^{2} + 121 N + 115 k\, 
    N + 20 k^{2} N + 79 N^{2} + 42 k\, 
    N^{2} + 16 N^{3})} {2 (2 + N) (2 + k + N)^{2}},\nonu\\ 
c_ {37} & = & - \frac {(3 + 2 k + N) (10 + 5 k + 
       8 N)} {3 (2 + k + N)^{2}},\qquad
c_ {38}  = \frac {4} {(2 + k + N)},\qquad
c_ {39}  = \frac {4} {(2 + k + N)^{2}},\nonu \\
c_ {40}  & = &  - \frac {2 (3 + 2 k + N) (10 + 5 k + 
       8 N)} {3 (2 + k + N)^{3}},\qquad 
c_ {41}  =  - \frac {12} {(2 + k + N)^{2}},\qquad
c_ {42}  = \frac {8} {(2 + k + N)^{2}},\nonu\\ 
c_ {43} & = & - \frac {2 i (2 + 3 k + 3 N)} {(2 + k + N)^{2}},\qquad 
c_ {44}  = \frac {2 i (6 + k + N)} {(2 + k + N)^{2}},\qquad 
c_ {45}  =  - \frac {20 (-k + N)} {3 (2 + k + N)^{3}},\nonu \\
c_ {46}  & = & \frac {2 i (-2 + k + N)} {(2 + k + N)^{2}},\qquad 
c_ {47}  =  - \frac {8 i} {(2 + k + N)^{2}},\nonu \\
c_ {48} & = & \frac {(300 + 249 k + 94 k^{2} + 501 N + 231 k\, 
    N + 20 k^{2} N + 203 N^{2} + 42 k\, 
    N^{2} + 16 N^{3})} {6 (2 + N) (2 + k + N)^{3}},\nonu\\ 
c_ {49} & = & - \frac {(32 + 55 k + 18 k^{2} + 41 N + 35 k\, 
      N + 11 N^{2})} {3(2 + N) (2 + k + N)^{4}},\nonu\\ 
c_ {50} & = & - \frac { (32 + 55 k + 18 k^{2} + 41 N + 35 k\, 
      N + 11 N^{2})} {3(2 + N) (2 + k + N)^{4}},\nonu\\ 
c_ {51} & = &\frac {i (16 - 5 k - 6 k^{2} + 13 N - 5 k\, 
     N + 3 N^{2})} {3(2 + N) (2 + k + N)^{3}},\nonu\\ 
c_ {52} & = & - \frac {i (32 + 29 k + 6 k^{2} + 35 N + 17 k\, 
      N + 9 N^{2})} {2(2 + N) (2 + k + N)^{3}},\nonu\\ 
c_ {53} & = & - \frac {(16 + 3 k - 6 k^{2} + 21 N - k\, 
     N + 7 N^{2})} {4 (2 + N) (2 + k + N)^{2}},\nonu\\ 
c_ {54} & = &\frac {(48 + 21 k + 6 k^{2} + 35 N + 13 k\, 
    N + 5 N^{2})} {4 (2 + N) (2 + k + N)^{2}},\nonu\\
c_ {55} & = & - \frac {i (32 + 29 k + 6 k^{2} + 35 N + 17 k\, 
      N + 9 N^{2})} {2(2 + N) (2 + k + N)^{3}},\nonu\\ 
c_ {56} &=& -\frac {i (13 k + 6 k^{2} + 3 N + 9 k\, 
     N + N^{2})} {2(2 + N) (2 + k + N)^{3}},\nonu\\ 
c_ {57} & = & - \frac {i (13 k + 6 k^{2} + 3 N + 9 k\, 
      N + N^{2})} {2(2 + N) (2 + k + N)^{3}},\nonu\\ 
c_ {58} & = &\frac {(20 + 21 k + 6 k^{2} + 25 N + 13 k\, 
    N + 7 N^{2})} {2 (2 + N) (2 + k + N)^{2}},\nonu\\ 
c_ {59} & = & - \frac {i (20 + 21 k + 6 k^{2} + 25 N + 13 k\, 
      N + 7 N^{2})} {(2 + N) (2 + k + N)^{3}},\nonu\\ 
c_ {60} & = & - \frac {i (20 + 21 k + 6 k^{2} + 25 N + 13 k\, 
      N + 7 N^{2})} {(2 + N) (2 + k + N)^{3}},\nonu\\ 
c_ {61} & = & - \frac {i (20 + 21 k + 6 k^{2} + 25 N + 13 k\, 
      N + 7 N^{2})} {(2 + N) (2 + k + N)^{3}},\qquad
c_ {62}  =  - \frac {2} {(2 + k + N)^{2}},\nonu\\ 
c_ {63} & = &\frac {(60 + 71 k + 18 k^{2} + 67 N + 43 k\, 
    N + 17 N^{2})} {12 (2 + N) (2 + k + N)^{2}},\nonu\\ 
c_ {64} & = & - \frac {i (60 + 47 k + 18 k^{2} + 91 N + 31 k\, 
      N + 29 N^{2})} {6(2 + N) (2 + k + N)^{3}},\qquad
c_ {65}  = \frac {4} {(2 + k + N)^{2}},\nonu\\ 
c_ {66} & = & - \frac {i (60 + 71 k + 18 k^{2} + 67 N + 43 k\, 
      N + 17 N^{2})} {6 (2 + N) (2 + k + N)^{3}},\nonu\\ 
c_ {67} & = & - \frac {i (60 + 71 k + 18 k^{2} + 67 N + 43 k\, 
      N + 17 N^{2})} {6 (2 + N) (2 + k + N)^{3}},\qquad
c_ {68}  = \frac {1} {4},\nonu\\ 
c_ {69} & = & - \frac {(60 + 77 k + 22 k^{2} + 121 N + 115 k\; 
     N + 20 k^{2} N + 79 N^{2} + 42 k\; 
     N^{2} + 16 N^{3})} {4 (2 + N) (2 + k + N)^{2}},\nonu\\ 
c_ {70} & = &\frac {(60 + 77 k + 22 k^{2} + 121 N + 115 k\; 
    N + 20 k^{2} N + 79 N^{2} + 42 k\; 
    N^{2} + 16 N^{3})} {2 (2 + N) (2 + k + N)^{2}},\nonu\\ 
c_ {71} & = & - \frac {i (20 + 21 k + 6 k^{2} + 25 N + 13 k\; 
      N + 7 N^{2})} {(2 + N) (2 + k + N)^{2}},\nonu\\ 
c_ {72} & = & - \frac {2 (20 + 21 k + 6 k^{2} + 25 N + 13 k\; 
      N + 7 N^{2})} {(2 + N) (2 + k + N)^{3}},\nonu\\ 
c_ {73} & = & - \frac {(20 + 21 k + 6 k^{2} + 25 N + 13 k\; 
     N + 7 N^{2})} {2 (2 + N) (2 + k + N)^{2}},\nonu\\ 
c_ {74} & = & - \frac {4 (16 + 29 k + 10 k^{2} + 27 N + 25 k\; 
      N + 2 k^{2} N + 13 N^{2} + 4 k\; 
      N^{2} + 2 N^{3})} {(2 + N) (2 + k + N)^{4}},\nonu\\ 
c_ {75} & = & - \frac {4 (13 k + 6 k^{2} + 3 N + 9 k\; 
      N + N^{2})} {(2 + N) (2 + k + N)^{4}},\nonu\\ 
c_ {76} & = & - \frac {1}{3 (2 + N) (2 + k + N)^{3}}i (-60 - 71 k - 15 k^{2} + 2 k^{3} - 67 N - 
       13 k\; N + 38 k^{2} N 
       \nonu\\ &+&10 k^{3} N + 10 N^{2} + 67 k\; 
      N^{2} + 31 k^{2} N^{2} + 31 N^{3} + 29 k\; 
      N^{3} + 8 N^{4}),\nonu\\ 
c_ {77} & = & - \frac {i (20 + 21 k + 6 k^{2} + 25 N + 13 k\; 
      N + 7 N^{2})} {(2 + N) (2 + k + N)^{3}},\nonu\\
c_ {78} & = & - \frac {(-60 - 65 k - 14 k^{2} - 13 N + 29 k\; 
     N + 20 k^{2} N + 45 N^{2} + 42 k\; 
     N^{2} + 16 N^{3})} {3 (2 + N) (2 + k + N)^{3}},\nonu\\
c_ {79} & = & - \frac {1}{3 (2 + N) (2 + k + N)^{4}}(240 + 352 k + 79 k^{2} - 14 k^{3} + 560 N + 
      710 k\; N + 195 k^{2} N 
      \nonu\\ &+& 20 k^{3} N + 471 N^{2} + 410 k\; 
     N^{2} + 62 k^{2} N^{2} + 153 N^{3} + 58 k\; 
     N^{3} + 16 N^{4}),\nonu\\ 
c_ {80} & = & - \frac{1}{3 (2 + N) (2 + k + N)^{4}}4 (180 + 261 k + 124 k^{2} + 20 k^{3} + 
       393 N + 429 k\; 
      N + 133 k^{2} N 
      \nonu\\ &+& 10 k^{3} N + 293 N^{2} + 211 k\; 
      N^{2} + 31 k^{2} N^{2} + 86 N^{3} + 29 k\; 
      N^{3} + 8 N^{4}),\nonu\\ 
c_ {81} & = & - \frac {2 i (8 + 21 k + 8 k^{2} + 15 N + 17 k\; 
      N + k^{2} N + 7 N^{2} + 2 k\; 
      N^{2} + N^{3})} {(2 + N) (2 + k + N)^{3}},\nonu\\
c_ {82} & = &\frac {4 (16 + 29 k + 10 k^{2} + 27 N + 25 k\; 
     N + 2 k^{2} N + 13 N^{2} + 4 k\; 
     N^{2} + 2 N^{3})} {(2 + N) (2 + k + N)^{4}},\nonu\\ 
c_ {83} & = &\frac {i (20 + 21 k + 6 k^{2} + 25 N + 13 k\; 
     N + 7 N^{2})} {(2 + N) (2 + k + N)^{3}},\nonu\\ 
c_ {84} & = & - \frac {i (20 + 21 k + 6 k^{2} + 25 N + 13 k\; 
      N + 7 N^{2})} {(2 + N) (2 + k + N)^{3}},\qquad 
c_ {85}  =  - \frac {4 i (k + N)} {(2 + k + N)^{2}},\nonu\\ 
c_ {86} & = &\frac {2 i (24 + 29 k + 8 k^{2} + 31 N + 21 k\; 
     N + k^{2} N + 11 N^{2} + 2 k\; 
     N^{2} + N^{3})} {(2 + N) (2 + k + N)^{3}},\nonu\\ 
c_ {87} & = & - \frac {2 i (8 + 21 k + 8 k^{2} + 15 N + 17 k\; 
      N + k^{2} N + 7 N^{2} + 2 k\; 
      N^{2} + N^{3})} {(2 + N) (2 + k + N)^{3}},\nonu\\ 
c_ {88} & = &\frac {i (20 + 21 k + 6 k^{2} + 25 N + 13 k\; 
     N + 7 N^{2})} {(2 + N) (2 + k + N)^{3}},\nonu\\ 
c_ {89} & = &\frac {i (20 + 21 k + 6 k^{2} + 25 N + 13 k\; 
     N + 7 N^{2})} {(2 + N) (2 + k + N)^{3}},\nonu\\ 
c_ {90} & = & - \frac {i (16 + 21 k + 6 k^{2} + 19 N + 13 k\; 
      N + 5 N^{2})} {2(2 + N) (2 + k + N)^{2}},\nonu\\ 
c_ {91} & = & - \frac { (13 k + 6 k^{2} + 3 N + 9 k\; 
      N + N^{2})} {(2 + N) (2 + k + N)^{3}},\nonu\\ 
c_ {92} & = &\frac {(16 + 21 k + 6 k^{2} + 19 N + 13 k\; 
    N + 5 N^{2})} {2(2 + N) (2 + k + N)^{2}},\nonu\\ 
c_ {93} & = & - \frac {2 i (16 + 21 k + 6 k^{2} + 19 N + 13 k\; 
      N + 5 N^{2})} {(2 + N) (2 + k + N)^{3}},\nonu\\ 
c_ {94} & = & - \frac {2 i (k + 9 N + 12 k\; 
      N + 3 k^{2} N + 11 N^{2} + 7 k\; 
      N^{2} + 3 N^{3})} {(2 + N) (2 + k + N)^{3}},\nonu\\ 
c_ {95} & = &\frac {i (16 + 21 k + 6 k^{2} + 19 N + 13 k\; 
     N + 5 N^{2})} {2(2 + N) (2 + k + N)^{3}},\nonu\\ 
c_ {96} & = & - \frac {i (16 + 21 k + 6 k^{2} + 19 N + 13 k\; 
      N + 5 N^{2})} {2(2 + N) (2 + k + N)^{3}},\nonu\\ 
c_ {97} & = & - \frac {3 i (20 + 21 k + 6 k^{2} + 25 N + 13 k\; 
      N + 7 N^{2})} {2(2 + N) (2 + k + N)^{3}},\nonu\\ 
c_ {98} & = &\frac {i (20 + 13 k + 2 k^{2} + 33 N + 9 k\; 
     N - 2 k^{2} N + 15 N^{2} + 2 N^{3})} {2(2 + N) (2 + k + N)^{3}},\nonu\\ 
c_ {99} & = &\frac{1} {6 (2 + N) (2 + k + N)^{3}}i (144 + 180 k + 47 k^{2} - 2 k^{3} + 252 N + 
      182 k\; N - 9 k^{2} N 
      \nonu\\ &-& 10 k^{3} N + 167 N^{2} + 63 k\; 
     N^{2} - 11 k^{2} N^{2} + 56 N^{3} + 13 k\; 
     N^{3} + 8 N^{4}),\nonu\\ 
c_ {100} & = &\frac { (32 + 55 k + 18 k^{2} + 41 N + 35 k\; 
     N + 11 N^{2})} {(2 + N) (2 + k + N)^{4}},\nonu\\ 
c_ {101} & = & - \frac {2 (1 + N) (20 + 23 k + 6 k^{2} + 23 N + 
       14 k\; N + 6 N^{2})} {(2 + N) (2 + k + N)^{4}},\nonu\\ 
c_ {102} & = &\frac{1}{6 (2 + N) (2 + k + N)^{4}}(96 + 252 k + 121 k^{2} + 14 k^{3} + 180 N + 
     220 k\; N - 15 k^{2} N 
     \nonu\\ &-& 20 k^{3} N + 139 N^{2} + 72 k\; 
    N^{2} - 22 k^{2} N^{2} + 73 N^{3} + 26 k\; 
    N^{3} + 16 N^{4}),\nonu\\ 
c_ {103} & = &\frac{1} {3 (2 + N) (2 + k + N)^{4}}2(96 - 42 k - 82 k^{2} - 20 k^{3} + 186 N + 
      14 k\; N - 45 k^{2} N - 10 k^{3} N 
      \nonu\\ &+& 164 N^{2} + 51 k\; 
     N^{2} - 11 k^{2} N^{2} + 62 N^{3} + 13 k\; 
     N^{3} + 8 N^{4}),\nonu\\ 
c_ {104} & = &\frac{1} {6 (2 + N) (2 + k + N)^{4}}(96 + 252 k + 121 k^{2} + 14 k^{3} + 180 N + 
     220 k\; N - 15 k^{2} N 
     \nonu\\ &-& 20 k^{3} N + 139 N^{2} + 72 k\; 
    N^{2} - 22 k^{2} N^{2} + 73 N^{3} + 26 k\; 
    N^{3} + 16 N^{4}),\nonu\\ 
c_ {105} & = & - \frac { (32 + 55 k + 18 k^{2} + 41 N + 35 k\; 
      N + 11 N^{2})} {(2 + N) (2 + k + N)^{4}},\nonu\\ 
c_ {106} & = & - \frac { (20 + 31 k + 10 k^{2} + 35 N + 41 k\; 
      N + 8 k^{2} N + 21 N^{2} + 14 k\; 
      N^{2} + 4 N^{3})} {(2 + N) (2 + k + N)^{4}},\nonu\\ 
c_ {107} & = &\frac {i (20 + 21 k + 6 k^{2} + 25 N + 13 k\; 
     N + 7 N^{2})} {2(2 + N) (2 + k + N)^{3}},\nonu\\ 
c_ {108} & = &\frac {i (k - N) (16 + 21 k + 6 k^{2} + 19 N + 13 k\; 
     N + 5 N^{2})} {2(2 + N) (2 + k + N)^{4}},\nonu\\ 
c_ {109} & = &\frac {(13 k + 6 k^{2} + 3 N + 9 k\; 
    N + N^{2})} {2(2 + N) (2 + k + N)^{3}},\nonu\\ 
c_ {110} & = & - \frac {(16 + 21 k + 6 k^{2} + 19 N + 13 k\; 
     N + 5 N^{2})} {2(2 + N) (2 + k + N)^{3}},\nonu\\ 
c_ {111} & = &\frac { (32 + 55 k + 18 k^{2} + 41 N + 35 k\; 
     N + 11 N^{2})} {(2 + N) (2 + k + N)^{4}},\nonu\\ 
c_ {112} & = & - \frac {2 (16 + 21 k + 6 k^{2} + 19 N + 13 k\; 
      N + 5 N^{2})} {(2 + N) (2 + k + N)^{3}},\nonu\\ 
c_ {113} & = & - \frac {2 (8 + 17 k + 6 k^{2} + 11 N + 11 k\; 
      N + 3 N^{2})} {(2 + N) (2 + k + N)^{3}},\nonu\\ 
c_ {114} & = & - \frac {(16 + 21 k + 6 k^{2} + 19 N + 13 k\; 
     N + 5 N^{2})} {(2 + N) (2 + k + N)^{3}},\nonu\\
c_ {115} & = &\frac {(20 + 21 k + 6 k^{2} + 25 N + 13 k\; 
    N + 7 N^{2})} {2(2 + N) (2 + k + N)^{3}},\nonu\\ 
c_ {116} & = &\frac {(16 + 21 k + 6 k^{2} + 19 N + 13 k\; 
    N + 5 N^{2})} {4 (2 + N) (2 + k + N)^{2}},\nonu\\ 
c_ {117} & = & - \frac {(48 + 3 k - 6 k^{2} + 37 N - k\; 
     N + 7 N^{2})} {4 (2 + N) (2 + k + N)^{2}},\nonu\\ 
c_ {118} & = & - \frac {i (120 + 106 k - 19 k^{2} - 18 k^{3} + 
       290 N + 236 k\; N + 19 k^{2} N + 215 N^{2} + 108 k\; 
      N^{2} + 47 N^{3})} {3 (2 + N) (2 + k + N)^{4}},\nonu\\ 
c_ {119} & = &\frac {i (k - N) (32 + 55 k + 18 k^{2} + 41 N + 35 k\; 
     N + 11 N^{2})} {3 (2 + N) (2 + k + N)^{4}},\nonu\\ 
c_ {120} & = &\frac {2 (32 + 55 k + 18 k^{2} + 41 N + 35 k\; 
     N + 11 N^{2})} {(2 + N) (2 + k + N)^{4}},\nonu\\
c_ {121} & = &\frac {2 (32 + 55 k + 18 k^{2} + 41 N + 35 k\; 
     N + 11 N^{2})} {(2 + N) (2 + k + N)^{4}},\nonu\\ 
c_ {122} & = &\frac {i (64 + 142 k + 65 k^{2} + 6 k^{3} + 114 N + 
      134 k\; N + 23 k^{2} N + 57 N^{2} + 26 k\; 
     N^{2} + 9 N^{3})} {2(2 + N) (2 + k + N)^{4}},\nonu\\ 
c_ {123} & = &\frac {i (128 + 206 k + 81 k^{2} + 6 k^{3} + 210 N + 
      198 k\; N + 31 k^{2} N + 105 N^{2} + 42 k\; 
     N^{2} + 17 N^{3})} {2(2 + N) (2 + k + N)^{4}},\nonu\\ 
c_ {124} & = &\frac {i (96 + 220 k + 93 k^{2} + 6 k^{3} + 148 N + 
      172 k\; N + 19 k^{2} N + 55 N^{2} + 18 k\; 
     N^{2} + 5 N^{3})} {(2 + N) (2 + k + N)^{4}},\nonu\\
c_ {125} & = & - \frac{1}{3 (2 + N) (2 + k + N)^{4}}i (240 + 340 k + 143 k^{2} + 18 k^{3} + 
       452 N + 446 k\; N + 97 k^{2} N 
       \nonu\\ &+& 275 N^{2} + 144 k\; 
      N^{2} + 53 N^{3}),\nonu\\ 
c_ {126} & = & - \frac{1}{3 (2 + N) (2 + k + N)^{4}}i (360 + 478 k + 179 k^{2} + 18 k^{3} + 
       710 N + 668 k\; N + 133 k^{2} N 
       \nonu\\ &+& 449 N^{2} + 228 k\; 
      N^{2} + 89 N^{3}),\nonu\\ 
c_ {127} & = &\frac {i (16 + 21 k + 6 k^{2} + 19 N + 13 k\; 
     N + 5 N^{2})} {(2 + N) (2 + k + N)^{3}},\nonu\\ 
c_ {128} & = &\frac {2 (32 + 55 k + 18 k^{2} + 41 N + 35 k\; 
     N + 11 N^{2})} {(2 + N) (2 + k + N)^{4}},\nonu\\ 
c_ {129} & = &\frac {(20 + 21 k + 6 k^{2} + 25 N + 13 k\; 
    N + 7 N^{2})} {(2 + N) (2 + k + N)^{3}},\nonu\\ 
c_ {130} & = & - \frac {i (16 + 21 k + 6 k^{2} + 19 N + 13 k\; 
      N + 5 N^{2})} {(2 + N) (2 + k + N)^{3}},\nonu\\ 
c_ {131} & = &\frac {i (16 + 21 k + 6 k^{2} + 19 N + 13 k\; 
     N + 5 N^{2})} {(2 + N) (2 + k + N)^{3}},\nonu\\ 
c_ {132} & = &\frac {i (64 + 136 k + 61 k^{2} + 6 k^{3} + 88 N + 
      104 k\; N + 15 k^{2} N + 27 N^{2} + 10 k\; 
     N^{2} + N^{3})} {(2 + N) (2 + k + N)^{4}},\nonu\\ 
c_ {133} & = &\frac {i (16 + 21 k + 6 k^{2} + 19 N + 13 k\; 
     N + 5 N^{2})} {(2 + N) (2 + k + N)^{3}},\nonu\\ 
c_ {134} & = &\frac {i (16 - 5 k - 6 k^{2} + 13 N - 5 k\; 
     N + 3 N^{2})} {(2 + N) (2 + k + N)^{3}},\nonu\\ 
c_ {135} & = &\frac {i (13 k + 6 k^{2} + 3 N + 9 k\; 
     N + N^{2})} {(2 + N) (2 + k + N)^{3}},\nonu\\ 
c_ {136} & = & - \frac {i (32 - 40 k - 37 k^{2} - 6 k^{3} + 56 N - 
       8 k\; N - 3 k^{2} N + 45 N^{2} + 14 k\; 
      N^{2} + 11 N^{3})} {(2 + N) (2 + k + N)^{4}},\nonu\\ 
c_ {137} & = &\frac {2 (k - N) (32 + 55 k + 18 k^{2} + 41 N + 35 k\; 
     N + 11 N^{2})} {(2 + N) (2 + k + N)^{5}},\nonu\\ 
c_ {138} & = &\frac {(120 + 214 k + 107 k^{2} + 18 k^{3} + 302 N + 
     368 k\; N + 97 k^{2} N + 233 N^{2} + 144 k\; 
    N^{2} + 53 N^{3})} {6 (2 + N) (2 + k + N)^{3}},\nonu\\ 
c_ {139} & = &\frac {(240 + 232 k + 17 k^{2} - 18 k^{3} + 440 N + 
     314 k\; N + 19 k^{2} N + 257 N^{2} + 108 k\; 
    N^{2} + 47 N^{3})} {6 (2 + N) (2 + k + N)^{3}},\nonu\\ 
c_ {140} & = &\frac {4} {(2 + k + N)^{2}},\nonu\\ 
c_ {141} & = & - \frac {i (240 + 292 k + 41 k^{2} - 18 k^{3} + 
       500 N + 482 k\; N + 67 k^{2} N + 341 N^{2} + 192 k\; 
      N^{2} + 71 N^{3})} {3 (2 + N) (2 + k + N)^{4}},\nonu\\ 
c_ {142} & = &\frac {2 i (32 + 55 k + 18 k^{2} + 41 N + 35 k\; 
     N + 11 N^{2})} {(2 + N) (2 + k + N)^{4}},\nonu\\
c_ {143} & = &\frac {2 i (32 + 55 k + 18 k^{2} + 41 N + 35 k\; 
     N + 11 N^{2})} {(2 + N) (2 + k + N)^{4}},\nonu\\ 
c_ {144} & = & - \frac {i (60 + 71 k + 18 k^{2} + 67 N + 43 k\; 
      N + 17 N^{2})} {3 (2 + N) (2 + k + N)^{3}}.
\nonu
\eea
The fusion rule is 
\bea
[\Phi_{\frac{1}{2}}^{(1),i}] \, \cdot \, [\widetilde{\Phi}_{\frac{3}{2}}^{(1),j}]
= [I^{ij}] +  \delta^{ij} \, [\Phi_{0}^{(1)} \, \Phi_{0}^{(1)}]
+\varepsilon^{ijkl} \, [\Phi_0^{(1)} \, \Phi_1^{(1),kl}]
+ [\Phi_{\frac{1}{2}}^{(1),i} \, \Phi_{\frac{1}{2}}^{(1),j}]
+ \delta^{ij} \, [\Phi_{0}^{(2)}] +
\varepsilon^{ijkl} \, [\Phi_1^{(2),kl}],
\nonu 
\eea
where the last two terms belong to the 
next $16$ higher spin currents with 
spin $s=2$.

The OPEs between the higher spin-$\frac{3}{2}$ currents and 
the higher spin-$3$ current are given by
\bea
&&\Phi_{\frac{1}{2}}^{(1),i}(z)\:\widetilde{\Phi}_{2}^{(1)}(w)\;=\;\frac{1}{(z-w)^{4}}\, c_{1}\,\Gamma^{i}(w)
\nonu\\
&&+\frac{1}{(z-w)^{3}}\,\Bigg[c_{2}\, G^{i}+c_{3}\, U\,\Gamma^{i}+\varepsilon^{ijkl}\, c_{4}\,\Gamma^{j}\,\Gamma^{k}\,\Gamma^{l}+c_{5}\, T^{iq}\,\Gamma^{q}+\varepsilon^{ijkl}\, c_{6}\, T^{jk}\,\Gamma^{l}+c_{7}\,\partial\Gamma^{i}\Bigg](w)
\nonu\\
&&+\frac{1}{(z-w)^{2}}\,\Bigg[  c_{8}\, {\bf \Phi_{\frac{1}{2}}^{(2),i}}+
c_{9}\,{\bf \Phi_{0}^{(1)}\,\Phi_{\frac{1}{2}}^{(1),i}}+
c_{10}\, L\,\Gamma^{i}+c_{11}\, U\, U\,\Gamma^{i}+c_{12}\, U\,\partial\Gamma^{i}+c_{13}\,\partial U\,\Gamma^{i}
\nonu\\
&&+c_{14}\, G^{i}\, U+c_{15}\,\partial G^{i}+c_{16}\,\partial^{2}\Gamma^{i}+c_{17}G^{j}\, T^{ij}+c_{18}\, G^{j}\,\Gamma^{i}\,\Gamma^{j}+c_{19}\, T^{ij}\, U\,\Gamma^{j}+c_{20}\, T^{ij}\,\partial\Gamma^{j}
\nonu\\
&&+c_{21}\,\partial T^{ij}\,\Gamma^{j}+c_{22}\, T^{ij}\, T^{jk}\,\Gamma^{k}+
c_{23}\,\widetilde{T}^{ij} \, \widetilde{T}^{ij}\,\Gamma^{i}+
c_{24}\, T^{jk}\,\Gamma^{i}\,\Gamma^{j}\,\Gamma^{k}+c_{25}\,\Gamma^{i}\,\partial\Gamma^{j}\,\Gamma^{j}
\nonu\\
&&+\varepsilon^{ijkl}\Bigg\{c_{26}\, T^{ij}\, T^{kl}\,\Gamma^{i}+c_{27}\, U\,\Gamma^{j}\,\Gamma^{k}\,\Gamma^{l}+c_{28}\, G^{j}\,\Gamma^{k}\,\Gamma^{l}+c_{29}\, G^{j}\,\Gamma^{k}\,\Gamma^{l}+c_{30}\,\partial T^{jk}\,\Gamma^{l}
\nonu\\
&&+c_{31}\, T^{jk}\,\partial\Gamma^{l}+c_{32}\,\partial(\Gamma^{j}\,\Gamma^{k}\,\Gamma^{l})+c_{33}\, T^{jk}\, U\,\Gamma^{l}
\Bigg\}\Bigg](w)
\nonu\\
&&+\frac{1}{(z-w)}\,\Bigg[ c_{34}\, {\bf \partial\Phi_{\frac{1}{2}}^{(2),i}}+
c_{35}\, {\bf \partial\Phi_{0}^{(1)}\,\Phi_{\frac{1}{2}}^{(1),i}}+
c_{36}\, {\bf \Phi_{0}^{(1)}\,\partial\Phi_{\frac{1}{2}}^{(1),i}}+
c_{37}\, L\, G^{i}+c_{38}\, L\, U\,\Gamma^{i}+c_{39}\, L\,\partial\Gamma^{i}
\nonu\\
&&+c_{40}\,\partial L\,\Gamma^{i}+c_{41}\,\partial^{2}G^{i}+c_{42}\,(\partial G^{i}\, U-G^{i}\,\partial U)+c_{43}\,U\,U\,\partial\Gamma^{i}+c_{44}\, U\,\partial^{2}\Gamma^{i}+c_{45}\,\partial U\,\partial\Gamma^{i}
\nonu\\
&&+c_{46}\,\partial^{2}U\,\Gamma^{i}+c_{47}\,\partial U\, U\,\Gamma^{i}+c_{48}\,\partial^{3}\Gamma^{i}+c_{49}\,\partial(G^{j}\, T^{ij})+c_{50}\,\partial(G^{j}\,\Gamma^{i}\,\Gamma^{j})+c_{51}\,\partial\widetilde{T}^{ij}\,\widetilde{T}^{ij}\,\Gamma^{i}
\nonu\\
&&+c_{52}\,\widetilde{T}^{ij}\,\widetilde{T}^{ij}\,\partial\Gamma^{i}+c_{53}\, T^{ij}\, T^{jk}\,\partial\Gamma^{i}+c_{54}\,\partial(T^{ij}\, U\,\Gamma^{j})+c_{55}\,\partial^{2}T^{ij}\,\Gamma^{j}+c_{56}\,\partial T^{ij}\,\partial\Gamma^{j}
\nonu\\
&&+c_{57}\, T^{ij}\,\partial^{2}\Gamma^{j}+c_{58}\,(\Gamma^{i}\,\partial^{2}\Gamma^{j}\,\Gamma^{j}+\partial\Gamma^{i}\,\partial\Gamma^{j}\,\Gamma^{j})+c_{59}\, T^{jk}\,\partial(\Gamma^{i}\,\Gamma^{j}\,\Gamma^{k})+c_{60}\, T^{ij}\,\partial T^{jk}\,\Gamma^{k}
\nonu\\
&&+c_{61}\,\partial T^{ij}\, T^{jk}\,\Gamma^{k}+c_{62}\,\partial T^{jk}\,\Gamma^{i}\,\Gamma^{j}\,\Gamma^{k}+\varepsilon^{ijkl}\Bigg\{c_{63}\, L\, T^{jk}\,\Gamma^{l}+c_{64}\, L\,\Gamma^{j}\,\Gamma^{k}\,\Gamma^{l}+c_{65}\,\partial G^{j}\, T^{kl}
\nonu\\
&&+c_{66}\, G^{j}\,\partial T^{kl}+c_{67}\,\partial G^{j}\,\Gamma^{k}\,\Gamma^{l}+c_{68}\, G^{j}\,\partial(\Gamma^{k}\,\Gamma^{l})+c_{69}\,\partial(T^{ij}\, T^{kl}\,\Gamma^{i})+c_{70}\,\partial U\,\Gamma^{j}\,\Gamma^{k}\,\Gamma^{l}
\nonu\\
&&+c_{71}\,\partial^{2}T^{jk}\,\Gamma^{l}+c_{72}\,\partial T^{jk}\,\partial\Gamma^{l}+c_{73}\, T^{jk}\,\partial^{2}\Gamma^{l}+c_{74}\,\partial(T^{jk}\, U)\,\Gamma^{l}+c_{75}\, T^{jk}\, U\,\partial\Gamma^{l}
\nonu\\
&&+c_{76}\, U\,\partial(\Gamma^{j}\,\Gamma^{k}\,\Gamma^{l})+c_{77}\,(\partial^{2}\Gamma^{j}\,\Gamma^{k}\,\Gamma^{l}+\Gamma^{j}\,\partial^{2}\Gamma^{k}\,\Gamma^{l}+\Gamma^{j}\,\Gamma^{k}\,\partial^{2}\Gamma^{l})
\nonu\\
&&+c_{78}\,(\partial\Gamma^{j}\,\partial\Gamma^{k}\,\Gamma^{l}+\Gamma^{j}\,\partial\Gamma^{k}\,\partial\Gamma^{l}+\partial\Gamma^{j}\,\Gamma^{k}\,\partial\Gamma^{l})\Bigg\}
\Bigg](w)+\cdots,
\nonu 
\eea
where
the coefficients are
\bea
c_ {1} & = &\frac {24\, i\, k\, N} {(2 + k + N)^{2}}, \qquad
c_ {2}  =  - \frac {4 (k - N) (38 + 25 k + 25 N + 12 k\, 
      N)} {3 (2 + k + N) (5 + 4 k + 4 N + 3 k\, N)}, \nonu \\
c_ {3} & = &\frac {8 i (k - N) (38 + 25 k + 25 N + 12 k\, 
     N)} {3 (2 + k + N)^{2} (5 + 4 k + 4 N + 3 k\, N)}, \qquad
c_ {4}  = -\frac {8 i (k - N) (-31 - 14 k - 14 N + 3 k\, 
     N)} {9 (2 + k + N)^{3} (5 + 4 k + 4 N + 3 k\, N)}, \nonu \\
c_ {5} & = &\frac {8 (k + N)} {(2 + k + N)^{2}}, \qquad
c_ {6}  = -\frac {4 (k - N) (23 + 13 k + 13 N + 3 k\, 
     N)} {3 (2 + k + N)^{2} (5 + 4 k + 4 N + 3 k\, N)}, \nonu \\
c_ {7} & = & - \frac {24 i (k + N + k\, 
      N)} {(2 + k + N)^{2}}, \qquad
c_ {8}  = \frac {5} {2}, \nonu \\
c_ {9} & = & - \frac {1}{2 (2 + N) (2 + k + N)^{2} (5 + 4 k + 4 N + 3 k\, 
      N)}(1500 + 2933 k + 1898 k^{2} + 392 k^{3}
      \nonu \\&+& 
      4417 N + 7639 k\, 
     N + 4251 k^{2} N + 706 k^{3} N + 4683 N^{2} + 6793 k\, 
     N^{2} + 2941 k^{2} N^{2} 
     \nonu \\&+& 300 k^{3} N^{2} + 2124 N^{3} + 
      2369 k\, N^{3} + 630 k^{2} N^{3} + 344 N^{4} + 240 k\, 
     N^{4}), \nonu \\
c_ {10} & = &\frac {4 i} {(2 + k + N)}, \qquad
c_ {11}  =  - \frac {4 i} {(2 + k + N)^{2}}, \nonu \\
c_ {12} & = & - \frac{1}{3 (2 + k + N)^{3} (5 + 4 k + 4 N + 3 k\, N)}2 i (750 + 1427 k + 902 k^{2} + 188 k^{3} + 1498 N 
\nonu \\&+& 2355 k\,N + 1133 k^{2} N + 150 k^{3} N + 928 N^{2} + 1102 k\, 
      N^{2} + 315 k^{2} N^{2} + 172 N^{3} + 120 k\, 
      N^{3}), \nonu \\
c_ {13} & = & - \frac {1} {3 (2 + N) (2 + k + N)^{3} (5 + 4 k + 4 N + 3 k\, 
      N)}i (-1500 - 2617 k - 1522 k^{2} - 328 k^{3} 
      \nonu \\&-& 
       1733 N - 1631 k\, 
      N + 81 k^{2} N + 166 k^{3} N + 633 N^{2} + 2143 k\, 
      N^{2} + 1771 k^{2} N^{2} + 300 k^{3} N^{2}
      \nonu \\&+& 1284 N^{3} + 
       1739 k\, N^{3} + 630 k^{2} N^{3} + 344 N^{4} + 240 k\, 
      N^{4}), \nonu \\
c_ {14} & = &\frac {4} {(2 + k + N)}, \nonu \\
c_ {15} & = &\frac {1}{6 (2 + N) (2 + k + N)^{2} (5 + 4 k + 4 N + 3 k\, N)}(-1500 - 2617 k - 1522 k^{2} - 328 k^{3} 
\nonu \\&-& 
     1733 N - 1631 k\, 
    N + 81 k^{2} N + 166 k^{3} N + 633 N^{2} + 2143 k\, 
    N^{2} + 1771 k^{2} N^{2} + 300 k^{3} N^{2} 
    \nonu \\&+& 1284 N^{3} + 
     1739 k\, N^{3} + 630 k^{2} N^{3} + 344 N^{4} + 240 k\, 
    N^{4})
       , \nonu \\
c_ {16} & = & - \frac {4 i (5 + 7 k + 7 N)} {(2 + k + 
       N)^{2}}, \qquad
c_ {17}  = \frac {5 i (20 + 21 k + 6 k^{2} + 25 N + 13 k\, 
     N + 7 N^{2})} {2 (2 + N) (2 + k + N)^{2}}, \nonu \\
c_ {18} & = &\frac {5 (20 + 21 k + 6 k^{2} + 25 N + 13 k\, 
     N + 7 N^{2})} {(2 + N) (2 + k + N)^{3}}, \nonu \\
c_ {19} & = &\frac {5 (20 + 21 k + 6 k^{2} + 25 N + 13 k\, 
     N + 7 N^{2})} {(2 + N) (2 + k + N)^{3}}, \qquad
c_ {20}  =  - \frac {4 (2 + 3 k + 3 N)} {(2 + k + N)^{2}}, \nonu \\
c_ {21} & = &\frac {2 (-2 + 3 k + 3 N)} {(2 + k + N)^{2}}, \qquad
c_ {22}  =  - \frac {i (48 + 89 k + 30 k^{2} + 63 N + 57 k\, 
      N + 17 N^{2})} {(2 + N) (2 + k + N)^{3}}, \nonu \\
c_ {23} & = & - \frac {i (48 + 89 k + 30 k^{2} + 63 N + 57 k\, 
      N + 17 N^{2})} {(2 + N) (2 + k + N)^{3}}, \nonu \\
c_ {24} & = & - \frac {5 (32 + 55 k + 18 k^{2} + 41 N + 35 k\, 
      N + 11 N^{2})} {(2 + N) (2 + k + N)^{4}}, \nonu \\
c_ {25} & = & - \frac {20 i (16 + 29 k + 10 k^{2} + 27 N + 25 k\, 
      N + 2 k^{2} N + 13 N^{2} + 4 k\, 
      N^{2} + 2 N^{3})} {(2 + N) (2 + k + N)^{4}}, \nonu \\
c_ {26} & = & - \frac {5 i (20 + 21 k + 6 k^{2} + 25 N + 13 k\, 
      N + 7 N^{2})} {2(2 + N) (2 + k + N)^{3}}, \nonu \\
c_ {27} & = & - \frac {5 i (32 + 55 k + 18 k^{2} + 41 N + 35 k\, 
      N + 11 N^{2})} {3(2 + N) (2 + k + N)^{4}}, \nonu \\
c_ {28} & = &\frac {(32 + 81 k + 30 k^{2} + 47 N + 53 k\, 
    N + 13 N^{2})} {2(2 + N) (2 + k + N)^{3}}, \nonu \\
c_ {29} & = &\frac {i (32 + 81 k + 30 k^{2} + 47 N + 53 k\, 
     N + 13 N^{2})} {4 (2 + N) (2 + k + N)^{2}}, \nonu \\
c_ {30} & = &\frac {1} {3 (2 + N) (2 + k + N)^{3} (5 + 4 k + 4 N + 3 k\, 
     N)} (259 k + 328 k^{2} + 88 k^{3} + 491 N + 
      1592 k\, N 
      \nonu \\&+& 1380 k^{2} N + 308 k^{3} N + 1005 N^{2} + 2054 k\, 
     N^{2} + 1268 k^{2} N^{2} + 177 k^{3} N^{2} + 663 N^{3} + 883 k\, 
     N^{3} 
     \nonu \\&+& 315 k^{2} N^{3} + 136 N^{4} + 93 k\, 
     N^{4}), \nonu \\
c_ {31} & = &\frac{1} {3 (2 + k + N)^{3} (5 + 4 k + 4 N + 3 k\, N)} (750 + 1247 k + 668 k^{2} + 116 k^{3} + 
      1678 N + 2355 k\, 
     N 
     \nonu \\&+& 953 k^{2} N + 96 k^{3} N + 1162 N^{2} + 1282 k\, 
     N^{2} + 315 k^{2} N^{2} + 244 N^{3} + 174 k\, 
     N^{3}), \nonu \\
c_ {32} & = & - \frac{1} {9 (2 + N) (2 + k + N)^{4} (5 + 4 k + 4 N + 3 k\, 
      N)}2 i (1500 + 2933 k + 1898 k^{2} + 392 k^{3} 
      \nonu \\&+& 
       4417 N + 7639 k\, 
      N + 4251 k^{2} N + 706 k^{3} N + 4683 N^{2} + 6793 k\, 
      N^{2} + 2941 k^{2} N^{2} + 300 k^{3} N^{2} 
      \nonu \\&+& 2124 N^{3} + 
       2369 k\, N^{3} + 630 k^{2} N^{3} + 344 N^{4} + 240 k\, 
      N^{4}), \nonu \\
c_ {33} & = &\frac {(64 + 97 k + 30 k^{2} + 79 N + 61 k\, 
    N + 21 N^{2})} {2(2 + N) (2 + k + N)^{3}}, \qquad
c_ {34}  = \frac {1} {2}, \nonu \\
c_ {35} & = & - \frac {(60 + 77 k + 22 k^{2} + 121 N + 115 k\, 
     N + 20 k^{2} N + 79 N^{2} + 42 k\, 
     N^{2} + 16 N^{3})} {2 (2 + N) (2 + k + N)^{2}}, \nonu \\
c_ {36} & = & - \frac{1}{2 (2 + N) (2 + k + N)^{2} (5 + 4 k + 4 N + 3 k\, N)}3 (100 + 187 k + 118 k^{2} + 24 k^{3} + 303 N 
       \nonu \\&+& 505 k\, 
      N + 277 k^{2} N + 46 k^{3} N + 325 N^{2} + 455 k\, 
      N^{2} + 195 k^{2} N^{2} + 20 k^{3} N^{2} + 148 N^{3} 
      \nonu \\&+& 159 k\, 
      N^{3} + 42 k^{2} N^{3} + 24 N^{4} + 16 k\, 
      N^{4})
        , \nonu \\
c_ {37} & = & - \frac {8 (k - N)} {(5 + 4 k + 4 N + 3 k\, 
     N)}, \qquad
c_ {38}  = \frac {16 i (k - N)} {(2 + k + N) (5 + 4 k + 4 N + 3 k\, 
     N)}, \nonu \\
c_ {39} & = & - \frac {12 i} {(2 + k + N)}, \qquad
c_ {40}  = \frac {4 i} {(2 + k + N)}, \nonu \\
c_ {41} & = &\frac{1}{6 (2 + N) (2 + k + N)^{2} (5 + 4 k + 4 N + 3 k\, 
     N)}(-300 - 389 k - 170 k^{2} - 32 k^{3} - 481 N 
     \nonu \\&-&
     259 k\, N + 117 k^{2} N + 50 k^{3} N - 75 N^{2} + 395 k\, 
    N^{2} + 371 k^{2} N^{2} + 60 k^{3} N^{2} + 156 N^{3} 
    \nonu \\&+& 331 k\, 
    N^{3} + 126 k^{2} N^{3} + 52 N^{4} + 48 k\, 
    N^{4}), \nonu \\
c_ {42} & = &\frac {4} {(2 + k + N)}, \qquad
c_ {43}  =  - \frac {20 i} {(2 + k + N)^{2}}, \nonu \\
c_ {44} & = & - \frac {1}{3 (2 + k + N)^{3} (5 + 4 k + 4 N + 3 k\, N)}2 i (150 + 319 k + 214 k^{2} + 46 k^{3} + 266 N + 471 k\, N 
\nonu \\&+& 235 k^{2} N + 30 k^{3} N + 152 N^{2} + 212 k\, 
      N^{2} + 63 k^{2} N^{2} + 26 N^{3} + 24 k\, 
      N^{3}) , \nonu \\
c_ {45} & = & - \frac{1} {3 (2 + N) (2 + k + N)^{3} (5 + 4 k + 4 N + 3 k\, 
      N)}i (300 + 887 k + 686 k^{2} + 152 k^{3} + 
       883 N 
       \nonu \\&+& 2263 k\, 
      N + 1485 k^{2} N + 262 k^{3} N + 1065 N^{2} + 2185 k\, 
      N^{2} + 1093 k^{2} N^{2} + 120 k^{3} N^{2} + 564 N^{3} 
      \nonu \\&+& 
       851 k\, N^{3} + 252 k^{2} N^{3} + 104 N^{4} + 96 k\, 
      N^{4}), \nonu \\
c_ {46} & = & - \frac {1}{3 (2 + N) (2 + k + N)^{3} (5 + 4 k + 4 N + 3 k\, 
      N)}i (-300 - 389 k - 170 k^{2} - 32 k^{3} - 
       481 N 
       \nonu \\&-& 259 k\, 
      N + 117 k^{2} N + 50 k^{3} N - 75 N^{2} + 395 k\, 
      N^{2} + 371 k^{2} N^{2} + 60 k^{3} N^{2} + 156 N^{3} 
      \nonu \\&+& 331 k\, 
      N^{3} + 126 k^{2} N^{3} + 52 N^{4} + 48 k\, 
      N^{4}), \nonu \\
c_ {47} & = &\frac {8 i} {(2 + k + N)^{2}}, \qquad
c_ {48}  = \frac {12 i} {(2 + k + N)^{2}}, \nonu \\
c_ {49} & = &\frac {i (20 + 21 k + 6 k^{2} + 25 N + 13 k\, 
     N + 7 N^{2})} {2 (2 + N) (2 + k + N)^{2}}, \nonu \\
c_ {50} & = &\frac {(20 + 21 k + 6 k^{2} + 25 N + 13 k\, 
    N + 7 N^{2})} {(2 + N) (2 + k + N)^{3}}, \nonu \\
c_ {51} & = & - \frac {2 i (16 + 21 k + 6 k^{2} + 19 N + 13 k\, 
      N + 5 N^{2})} {(2 + N) (2 + k + N)^{3}}, \nonu \\
c_ {52} & = &\frac {i (16 - 5 k - 6 k^{2} + 13 N - 5 k\, 
     N + 3 N^{2})} {(2 + N) (2 + k + N)^{3}}, \nonu \\
c_ {53} & = &\frac {i (16 - 5 k - 6 k^{2} + 13 N - 5 k\, 
     N + 3 N^{2})} {(2 + N) (2 + k + N)^{3}}, \nonu \\
c_ {54} & = &\frac {(20 + 21 k + 6 k^{2} + 25 N + 13 k\, 
    N + 7 N^{2})} {(2 + N) (2 + k + N)^{3}}, \qquad
c_ {55}  = \frac {2 (-2 + k + N)} {(2 + k + N)^{2}}, \nonu \\
c_ {56} & = & - \frac {2 (14 + 3 k + 3 N)} {(2 + k + N)^{2}}, \qquad
c_ {57}  = \frac {8} {(2 + k + N)^{2}}, \nonu \\
c_ {58} & = & - \frac {4 i (16 + 29 k + 10 k^{2} + 27 N + 25 k\, 
      N + 2 k^{2} N + 13 N^{2} + 4 k\, 
      N^{2} + 2 N^{3})} {(2 + N) (2 + k + N)^{4}}, \nonu \\
c_ {59} & = & - \frac { (32 + 55 k + 18 k^{2} + 41 N + 35 k\, 
      N + 11 N^{2})} {(2 + N) (2 + k + N)^{4}}, \nonu \\
c_ {60} & = &\frac {i (16 - 5 k - 6 k^{2} + 13 N - 5 k\, 
     N + 3 N^{2})} {(2 + N) (2 + k + N)^{3}}, \nonu \\
c_ {61} & = & - \frac {i (48 + 37 k + 6 k^{2} + 51 N + 21 k\, 
      N + 13 N^{2})} {(2 + N) (2 + k + N)^{3}}, \nonu \\
c_ {62} & = & - \frac { (32 + 55 k + 18 k^{2} + 41 N + 35 k\, 
      N + 11 N^{2})} {(2 + N) (2 + k + N)^{4}}, \nonu \\
c_ {63} & = & - \frac {8(k - N)} {(2 + k + N) (5 + 4 k + 4 N + 
       3 k\, N)}, \nonu \\
c_ {64} & = &\frac {32 i (k - N)} {3(2 + k + N)^{2} (5 + 4 k + 4 N + 
      3 k\, N)}, \nonu \\
c_ {65} & = &\frac {i (32 + 29 k + 6 k^{2} + 35 N + 17 k\, 
     N + 9 N^{2})} {4 (2 + N) (2 + k + N)^{2}}, \nonu \\
c_ {66} & = & - \frac {i (32 + 3 k - 6 k^{2} + 29 N - k\, 
      N + 7 N^{2})} {4 (2 + N) (2 + k + N)^{2}}, \nonu \\
c_ {67} & = &\frac {(32 + 29 k + 6 k^{2} + 35 N + 17 k\, 
    N + 9 N^{2})} {2(2 + N) (2 + k + N)^{3}}, \nonu \\
c_ {68} & = & - \frac {(32 + 3 k - 6 k^{2} + 29 N - k\, 
     N + 7 N^{2})} {2(2 + N) (2 + k + N)^{3}}, \nonu \\
c_ {69} & = &\frac {i (20 + 21 k + 6 k^{2} + 25 N + 13 k\, 
     N + 7 N^{2})} {(2 + N) (2 + k + N)^{3}}, \nonu \\
c_ {70} & = & - \frac { i (32 + 55 k + 18 k^{2} + 41 N + 35 k\, 
      N + 11 N^{2})} {3(2 + N) (2 + k + N)^{4}}, \nonu \\
c_ {71} & = &\frac {1} {3 (2 + N) (2 + k + N)^{3} (5 + 4 k + 4 N + 3 k\, 
     N)} (143 k + 164 k^{2} + 44 k^{3} + 7 N + 364 k\, 
     N 
     \nonu \\&+& 366 k^{2} N + 82 k^{3} N + 57 N^{2} + 370 k\, 
     N^{2} + 274 k^{2} N^{2} + 39 k^{3} N^{2} + 57 N^{3} + 149 k\, 
     N^{3} 
     \nonu \\&+& 63 k^{2} N^{3} + 14 N^{4} + 15 k\, 
     N^{4}), \nonu \\
c_ {72} & = &\frac{1}  {3 (2 + N) (2 + k + N)^{3} (5 + 4 k + 4 N + 3 k\, 
     N)} (300 + 541 k + 280 k^{2} + 40 k^{3} + 929 N 
     \nonu \\&+&
      1505 k\, N + 654 k^{2} N + 68 k^{3} N + 1059 N^{2} + 1505 k\, 
     N^{2} + 515 k^{2} N^{2} + 33 k^{3} N^{2} + 513 N^{3} 
     \nonu \\&+& 601 k\, 
     N^{3} + 126 k^{2} N^{3} + 88 N^{4} + 75 k\, 
     N^{4}), \nonu \\
c_ {73} & = &\frac{1} {3 (2 + k + N)^{3} (5 + 4 k + 4 N + 3 k\, N)}  (150 + 319 k + 214 k^{2} + 46 k^{3} + 266 N + 471 k\, N 
      \nonu \\&+& 235 k^{2} N + 30 k^{3} N + 152 N^{2} + 212 k\, 
     N^{2} + 63 k^{2} N^{2} + 26 N^{3} + 24 k\, 
     N^{3}), \nonu \\
c_ {74} & = &\frac {(13 k + 6 k^{2} + 3 N + 9 k\, 
    N + N^{2})} {2(2 + N) (2 + k + N)^{3}}, \nonu \\
c_ {75} & = &\frac {(64 + 45 k + 6 k^{2} + 67 N + 25 k\, 
    N + 17 N^{2})} {2(2 + N) (2 + k + N)^{3}}, \nonu \\
c_ {76} & = & - \frac { i (32 + 55 k + 18 k^{2} + 41 N + 35 k\, 
      N + 11 N^{2})} {3(2 + N) (2 + k + N)^{4}}, \nonu \\
c_ {77} & = &\frac{1}{9 (2 + N) (2 + k + N)^{4} (5 + 4 k + 4 N + 3 k\, N)}
       2 i (300 + 841 k + 670 k^{2} + 160 k^{3} + 
      629 N 
      \nonu \\&+& 1655 k\, 
     N + 1149 k^{2} N + 218 k^{3} N + 519 N^{2} + 1205 k\, 
     N^{2} + 665 k^{2} N^{2} + 78 k^{3} N^{2} + 198 N^{3} 
     \nonu \\&+& 361 k\, 
     N^{3} + 126 k^{2} N^{3} + 28 N^{4} + 30 k\, 
     N^{4}), \nonu \\
c_ {78} & = & - \frac {1}{9 (2 + N) (2 + k + N)^{4} (5 + 4 k + 4 N + 3 k\, 
       N)}4 i (300 + 601 k + 358 k^{2} + 64 k^{3} + 
       869 N 
       \nonu \\&+& 1535 k\, 
      N + 753 k^{2} N + 98 k^{3} N + 951 N^{2} + 1445 k\, 
      N^{2} + 545 k^{2} N^{2} + 42 k^{3} N^{2} + 450 N^{3} 
      \nonu \\&+& 553 k\, 
      N^{3} + 126 k^{2} N^{3} + 76 N^{4} + 66 k\, 
      N^{4}).
\nonu
\eea
The fusion rule is 
\bea
[\Phi_{\frac{1}{2}}^{(1),i}] \, \cdot \, [\widetilde{\Phi}_{2}^{(1)}]=
[I^i]+ [\Phi_{0}^{(1)}\,\Phi_{\frac{1}{2}}^{(1),i}]
+ [\Phi_{\frac{1}{2}}^{(2),i}].
\nonu
\eea

\subsection{The OPEs between the higher spin-$2$ currents and 
the other 
$11$ higher spin currents}

The OPEs between the higher spin-$2$ currents are given by
\bea
&&\Phi_{1}^{(1),ij}(z) \; \Phi_{1}^{(1),kl}(w) \; = \; 
\frac{1}{(z-w)^{4}}\,\Bigg[(\,\delta^{ik}\delta^{jl}-\delta^{il}\delta^{jk}\,)\, c_{1}+\varepsilon^{ijkl}\, c_{2}\,\Bigg]
\nonu\\
&&+
\frac{1}{(z-w)^{3}}\,\Bigg[\,\delta^{ik}\,\Big(\,\varepsilon^{jlab}\,(c_{3}\,\Gamma^{a}\,\Gamma^{b}+c_{4}\, T^{ab})+c_{5}\,\Gamma^{j}\,\Gamma^{l}+c_{6}\, T^{jl}\,\Big)\,\Bigg](w)
\nonu\\
&&+
\frac{1}{(z-w)^{2}}\,\Bigg[(\delta^{ik}\delta^{jl}-\delta^{il}\delta^{jk})\Bigg\{ c_{7}\, L+c_{8}\, U\, U+c_{9}\,\partial U+c_{10}(G^{i}\,\Gamma^{i}+G^{j}\,\Gamma^{j})
+c_{11}\,T^{ij}\,T^{ij}
\nonu\\
&&+c_{12}\, T^{ij}\,\Gamma^{i}\,\Gamma^{j}+\varepsilon^{iajb}\,c_{14}\, T^{ia}\, T^{jb}+c_{15}\,\Big[T^{ia}\,T^{ia}+T^{ja}\,T^{ja}-\delta^{ja}T^{iq}\,T^{iq}-\delta^{ia}T^{ja}\,T^{ja}\Big]
\nonu\\
&&+c_{16}\,(T^{ia}\,\Gamma^{i}\,\Gamma^{a}+T^{ja}\,\Gamma^{j}\,\Gamma^{a}-2\,T^{ij}\,\Gamma^{i}\,\Gamma^{j})+c_{17}(\partial\Gamma^{i}\,\Gamma^{i}+\partial\Gamma^{j}\,\Gamma^{j})+c_{18}\,\partial\Gamma^{4-ija}\,\Gamma^{4-ija}\Bigg\}
\nonu\\
&&+\delta^{ik}(1-\delta^{jl})\Bigg\{c_{19}\,(G^{j}\,\Gamma^{l}+G^{l}\,\Gamma^{j})+c_{20}\, T^{jl}+c_{21}\,\Gamma^{j}\,\partial\Gamma^{l}+c_{22}\,\partial\Gamma^{j}\,\Gamma^{l}
+\varepsilon^{ijla}\,\Big(c_{23}\,\partial(\Gamma^{i}\,\Gamma^{a})
\nonu\\
&&+c_{24}\,\partial T^{ia}+c_{25}\,(T^{ij}\, T^{ja}-T^{il}\, T^{la})\Big)+c_{26}\, T^{ij}\, T^{il}+c_{27}\, T^{ja}\, T^{la}
+c_{28}\,(T^{ja}\,\Gamma^{l}\,\Gamma^{a}+T^{la}\,\Gamma^{j}\,\Gamma^{a})\Bigg\}
\nonu\\
&&+\varepsilon^{ijkl}\Bigg\{  c_{29}\, {\bf \Phi_{0}^{(2)}}+
c_{30}\, {\bf \Phi_{0}^{(1)}\,\Phi_{0}^{(1)}}+
c_{31}\, L+c_{32}\, U\, U+c_{33}\, G^{a}\,\Gamma^{a}+c_{34}\,\partial U+c_{35}\,\partial\Gamma^{a}\,\Gamma^{a}
\nonu\\
&&+c_{36}\, T^{ab}\,\Gamma^{a}\,\Gamma^{b}+c_{37}\,(T^{ij}\,T^{ij}+T^{kl}\,T^{kl})+c_{38}\,(T^{ik}\,T^{ik}+T^{il}\,T^{il}+T^{jk}\,T^{jk}+T^{jl}\,T^{jl})
\nonu\\
&&+\varepsilon^{abcd}\,\Big(c_{39}\,\Gamma^{a}\,\Gamma^{b}\,\Gamma^{c}\,\Gamma^{d}+c_{40}(T^{ab}\,\Gamma^{c}\,\Gamma^{d})+c_{41}\, T^{ab}\, T^{cd}\Big)\Bigg\}+(\varepsilon^{ijkl})^{2}\, c_{42}\, T^{ij}\, T^{kl}\Bigg](w)
\nonu\\
&&+\frac{1}{(z-w)}\Bigg[\,(\delta^{ik}\delta^{jl}-\delta^{il}\delta^{jk})\,\Bigg\{ \, c_{43}\, L+c_{44}\,\partial U\, U+c_{45}\,\partial^{2}U+c_{46}\,\partial(G^{i}\,\Gamma^{i})+c_{47}\,\partial(G^{j}\,\Gamma^{j})
\nonu\\
&&+c_{48}\,\partial(T^{ij}\,\Gamma^{i}\,\Gamma^{j})+c_{49}\,\partial T^{ij}\, T^{ij}+c_{50}\,(\partial T^{ia}\, T^{ia}+\partial T^{ja}\, T^{ja})+c_{51}\,(\partial^{2}\Gamma^{i}\,\Gamma^{i}+\partial^{2}\Gamma^{j}\,\Gamma^{j})
\nonu\\
&&+c_{52}\,(\partial^{2}\Gamma^{a}\,\Gamma^{a})+\varepsilon^{ijab}\,\Big(c_{53}\,\Big[\partial(\widetilde{T}^{ja}\,\Gamma^{i}\,\Gamma^{b})+\partial(\widetilde{T}^{ia}\,\Gamma^{j}\,\Gamma^{b})\Big]+c_{54}\,\partial(T^{ib}\, T^{ja})\Big)\Bigg\}
\nonu\\
&&+\delta^{ik}\Bigg\{  c_{55}\, {\bf \Phi_{1}^{(2),jl}}+
c_{56}\, {\bf \Phi_{0}^{(1)}\,\Phi_{1}^{(1),jl}}+\varepsilon^{ijla}
c_{57}\, {\bf \Phi_{\frac{1}{2}}^{(1),i}\,
\Phi_{\frac{1}{2}}^{(1),a}}+c_{58}\, L\,\Gamma^{j}\,\Gamma^{l}+c_{59}\, L\, T^{jl}+c_{60}\, U\, U\,\Gamma^{j}\,\Gamma^{l}
\nonu\\
&&+c_{61}\, U\,\partial(\Gamma^{j}\,\Gamma^{l})+c_{62}\, G^{j}\, G^{l}+c_{63}\,(G^{j}\, U\,\Gamma^{l}-G^{l}\, U\,\Gamma^{j})+c_{64}\,\partial G^{j}\,\Gamma^{l}+c_{65}\, G^{l}\,\partial\Gamma^{j}+c_{66}\,\partial G^{l}\,\Gamma^{j}
\nonu\\
&&+c_{67}\, T^{jl}\, U\, U+c_{68}\,\Gamma^{j}\,\partial^{2}\Gamma^{l}+c_{69}\,\partial\Gamma^{j}\,\partial\Gamma^{l}+c_{70}\,\partial^{2}\Gamma^{j}\,\Gamma^{l}+c_{71}\, T^{jl}\,\partial U+c_{72}\,\partial T^{jl}\, U+c_{73}\,\partial U\,\Gamma^{j}\,\Gamma^{l}
\nonu\\
&&+c_{74}\,\partial^{2}T^{jl}+c_{75}\,(T^{jl}\,\partial\Gamma^{j}\,\Gamma^{j}+T^{jl}\,\partial\Gamma^{l}\,\Gamma^{l})+c_{76}\, T^{jl}\,\partial\Gamma^{i}\,\Gamma^{i}+c_{77}\, T^{jl}\,\partial\Gamma^{a}\,\Gamma^{a}
+c_{78}\,\Gamma^{j}\,\Gamma^{l}\,\partial\Gamma^{a}\,\Gamma^{a}
\nonu\\
&&+c_{79}\,(T^{ij}\,T^{ij}\, T^{jl}+T^{il}\,T^{il}\, T^{jl})+c_{80}\,(T^{ij}\,\widetilde{T}^{jl}\,\widetilde{T}^{ij}+T^{il}\,\widetilde{T}^{jl}\,\widetilde{T}^{li})+c_{81}\, G^{a}\, T^{jl}\,\Gamma^{a}+c_{82}\,\widetilde{T}^{jl}\,\widetilde{T}^{jl}\, T^{jl}
\nonu\\
&&+c_{83}\, T^{jl}\,\widetilde{T}^{ia}\,\widetilde{T}^{ia}+c_{84}\, T^{ij}\,\Gamma^{i}\,\partial\Gamma^{l}+c_{85}\, T^{il}\,\Gamma^{i}\,\partial\Gamma^{j}+c_{86}\,\widetilde{T}^{li}\,\Gamma^{l}\,\partial\Gamma^{4-ijl}+c_{87}\,\widetilde{T}^{ij}\,\Gamma^{j}\,\partial\Gamma^{4-ijl}
\nonu\\
&&+c_{88}\, T^{ij}\,\partial\Gamma^{i}\,\Gamma^{l}+c_{89}\, T^{il}\,\partial\Gamma^{i}\,\Gamma^{j}+c_{90}\,\widetilde{T}^{li}\,\partial\Gamma^{l}\,\Gamma^{4-ijl}+c_{91}\,\widetilde{T}^{ij}\,\partial\Gamma^{j}\,\Gamma^{4-ijl}+c_{92}\,\partial T^{ij}\,\Gamma^{i}\,\Gamma^{l}
\nonu\\
&&+c_{93}\,\partial T^{il}\,\Gamma^{i}\,\Gamma^{j}+c_{94}\,\partial\widetilde{T}^{li}\,\Gamma^{l}\,\Gamma^{4-ijl}+c_{95}\,\partial\widetilde{T}^{ij}\,\Gamma^{j}\,\Gamma^{4-ijl}+c_{96}\,\partial T^{ij}\, T^{il}+c_{97}\,\partial\widetilde{T}^{li}\,\widetilde{T}^{ij}
\nonu\\
&&+c_{98}\, T^{ij}\,\partial T^{il}+c_{99}\,\widetilde{T}^{li}\,\partial\widetilde{T}^{ij}+c_{100}\,\Big(T^{ij}\, T^{al}\,\Gamma^{i}\,\Gamma^{a}+T^{il}\, T^{ja}\,\Gamma^{i}\,\Gamma^{a}+T^{jl}\,T^{jl}\,\Gamma^{j}\,\Gamma^{l}
\nonu\\
&&+T^{jl}\,\widetilde{T}^{il}\,\Gamma^{4-ilj}+T^{jl}\,\widetilde{T}^{ij}\,\Gamma^{4-ijl})+\varepsilon^{ijla}\Bigg(c_{101}\, L\,\Gamma^{i}\,\Gamma^{a}+c_{102}\, L\, T^{ia}+c_{103}\, U\,\Gamma^{i}\,\partial\Gamma^{a}
\nonu\\
&&+c_{104}\, U\,\partial\Gamma^{i}\,\Gamma^{a}
+c_{105}\, G^{i}\, G^{a}+c_{106}\, G^{i}\, U\,\Gamma^{a}+c_{107}\, G^{a}\, U\,\Gamma^{i}+c_{108}\, G^{i}\,\partial\Gamma^{a}+c_{109}\, G^{a}\,\partial\Gamma^{i}
\nonu\\
&&+c_{110}\, T^{ia}\,U\,U
+c_{111}\,\Gamma^{i}\,\partial^{2}\Gamma^{a}
+c_{112}\,\partial\Gamma^{i}\,\partial\Gamma^{a}+c_{113}\,\partial^{2}\Gamma^{i}\,\Gamma^{a}
+c_{114}\, T^{ia}\,\partial U+c_{115}\,\partial T^{ia}\, U
\nonu\\
&&+c_{116}\,\partial U\,\Gamma^{i}\,\Gamma^{a}+c_{117}\,\partial^{2}T^{ia}+c_{118}\,\partial G^{i}\,\Gamma^{a}+c_{119}\,\partial G^{a}\,\Gamma^{i}+c_{120}\, T^{ia}\,\partial\Gamma^{i}\,\Gamma^{i}+c_{121}\, T^{ia}\,\partial\Gamma^{a}\,\Gamma^{a}
\nonu\\
&&+c_{122}\,(T^{ia}\,\partial\Gamma^{j}\,\Gamma^{j}+T^{ia}\,\partial\Gamma^{l}\,\Gamma^{l})+c_{123}\,(\Gamma^{i}\,\Gamma^{j}\,\partial\Gamma^{j}\,\Gamma^{a}+\Gamma^{i}\,\Gamma^{l}\,\partial\Gamma^{l}\,\Gamma^{a})+c_{124}\, G^{i}\, T^{ia}\,\Gamma^{i}
\nonu\\
&&+c_{125}\,(G^{j}\,\Gamma^{i}\,\Gamma^{j}\,\Gamma^{a}+G^{l}\,\Gamma^{i}\,\Gamma^{l}\,\Gamma^{a})+c_{126}\,(G^{i}\, T^{ja}\,\Gamma^{j}+G^{i}\, T^{la}\,\Gamma^{l}+G^{j}\, T^{ia}\,\Gamma^{j}+G^{l}\, T^{ia}\,\Gamma^{l})
\nonu\\
&&+c_{127}\,(G^{j}\, T^{ij}\,\Gamma^{a}+G^{l}\, T^{il}\,\Gamma^{a})+c_{128}\,(G^{j}\, T^{ja}\,\Gamma^{i}+G^{l}\, T^{la}\,\Gamma^{i})+c_{129}\,(G^{a}\, T^{ij}\,\Gamma^{j}+G^{a}\, T^{il}\,\Gamma^{l})
\nonu\\
&&+c_{130}\,(T^{ij}\, U\,\Gamma^{j}\,\Gamma^{a}+T^{il}\, U\,\Gamma^{l}\,\Gamma^{a}-T^{ja}\, U\,\Gamma^{i}\,\Gamma^{j}-T^{la}\, U\,\Gamma^{i}\,\Gamma^{l})
\nonu\\
&&+c_{131}\,(T^{ij}\, T^{jl}\, T^{la}-T^{il}\, T^{jl}\, T^{ja}+T^{ia}\,T^{jl}\,T^{jl}+c_{132}\,(T^{ij}\,\Gamma^{j}\,\partial\Gamma^{a}+T^{il}\,\Gamma^{l}\,\partial\Gamma^{a})
\nonu\\
&&+c_{133}\,(T^{ij}\,\partial\Gamma^{j}\,\Gamma^{a}+T^{il}\,\partial\Gamma^{l}\,\Gamma^{a})+c_{134}\,(\partial T^{ij}\,\Gamma^{j}\,\Gamma^{a}+\partial T^{il}\,\Gamma^{l}\,\Gamma^{a})+c_{135}\,(T^{ja}\,\Gamma^{i}\,\partial\Gamma^{j}+T^{la}\,\Gamma^{i}\,\partial\Gamma^{l})
\nonu\\
&&+c_{136}\,(T^{ja}\,\partial\Gamma^{i}\,\Gamma^{j}+T^{la}\,\partial\Gamma^{i}\,\Gamma^{l})+c_{137}\,(\partial T^{ja}\,\Gamma^{i}\,\Gamma^{j}+\partial T^{la}\,\Gamma^{i}\,\Gamma^{l})+c_{138}\, T^{ij}\,\partial T^{ja}+c_{139}\,\partial T^{ij}\, T^{ja}
\nonu\\
&&+c_{140}\, T^{il}\,\partial T^{la}+c_{141}\,\partial T^{il}\, T^{la}\Bigg)\Bigg\}
\nonu\\
&&+\varepsilon^{ijkl}\Bigg\{  
c_{142}\, {\bf \partial\Phi_{0}^{(2)}}+
c_{143}\, {\bf \partial\Phi_{0}^{(1)}\,\Phi_{0}^{(1)}}+
c_{144}\,\partial L+c_{145}\,\partial^{2}U+c_{146}\,\partial U\, U+c_{147}\,(\partial G^{i}\,\Gamma^{i}+\partial G^{j}\,\Gamma^{j})
\nonu\\
&&+c_{148}\,(G^{i}\,\partial\Gamma^{i}+G^{j}\,\partial\Gamma^{j})+c_{149}\,(\partial G^{k}\,\Gamma^{k}+\partial G^{l}\,\Gamma^{l})+c_{150}\,(G^{k}\,\partial\Gamma^{k}+G^{l}\,\partial\Gamma^{l})
\nonu\\
&&+c_{151}\,\varepsilon^{abcd}(G^{a}\,\widetilde{T}^{bc}\,\Gamma^{d})(1-\delta^{ia}\delta^{jd}-\delta^{id}\delta^{ja})+c_{152}\,\partial^{2}\Gamma^{a}\,\Gamma^{a}+c_{153}\, U\,\partial\Gamma^{a}\,\Gamma^{a}
\nonu\\
&&+c_{154}\,(\partial T^{ij}\, T^{ij}+\partial T^{kl}\, T^{kl})+c_{155}\,\partial T^{ab}\, T^{ab}+c_{156}\,\partial(T^{ij}\,\Gamma^{i}\,\Gamma^{j}+T^{kl}\,\Gamma^{k}\,\Gamma^{l})
\nonu\\
&&+(\delta^{ia}\delta^{kb}+\delta^{ia}\delta^{lb}+\delta^{ja}\delta^{kb}+\delta^{ja}\delta^{lb})\Big(c_{157}\,\partial T^{ab}\,\Gamma^{a}\,\Gamma^{b}+c_{158}\, T^{ab}\,\partial\Gamma^{a}\,\Gamma^{b}+c_{159}\, T^{ab}\,\Gamma^{a}\,\partial\Gamma^{b}\Big)\Bigg\}
\nonu\\
&&+c_{160}\,\partial(\Gamma^{i}\,\Gamma^{j}\,\Gamma^{k}\,\Gamma^{l})+(\varepsilon^{ijkl})^{2}\Bigg\{ c_{161}\, T^{ij}\,\partial T^{kl}+c_{162}\,\partial T^{ij}\, T^{kl}+c_{163}\,\varepsilon^{abcd}(\partial T^{ab}\, T^{cd})
\nonu\\
&&+c_{164}\,\partial T^{ij}\,\Gamma^{k}\,\Gamma^{l}+c_{165}\,\partial T^{kl}\,\Gamma^{i}\,\Gamma^{j}+c_{166}\, T^{ij}\,\partial(\Gamma^{k}\,\Gamma^{l})
+c_{167}\, T^{kl}\,\partial(\Gamma^{i}\,\Gamma^{j})
\nonu\\
&&+(\delta^{ia}\delta^{kb}+\delta^{ia}\delta^{lb}+\delta^{ja}\delta^{kb}+\delta^{ja}\delta^{lb})\Big(c_{168}\,\partial\widetilde{T}^{ab}\,\Gamma^{a}\,\Gamma^{b}+c_{169}\,\widetilde{T}^{ab}\,\partial\Gamma^{a}\,\Gamma^{b}+c_{170}\,\widetilde{T}^{ab}\,\Gamma^{a}\,\partial\Gamma^{b}\Big)\Bigg\}\Bigg](w)
\nonu\\
&&-\delta^{il}\,
\sum_{n=3}^1 \, \frac{1}{(z-w)^n} \, \left(k\;\leftrightarrow \;l\right) (w)
-\delta^{jk}\, \sum_{n=3}^1 \, \frac{1}{(z-w)^n} \,
\left(i\;\leftrightarrow \;j\right)(w)
\nonu\\
&&+\delta^{jl}\, 
\sum_{n=3}^1 \, \frac{1}{(z-w)^n} \,
\left(i\;\leftrightarrow \;j,\;\;k\;\leftrightarrow \;l\right)(w)
+\cdots,
\nonu
\eea
where the coefficients are
\bea
c_ {1} & = &-\frac {12\,k\,N}{(2 + k + N)},\qquad
c_ {2}  = \frac {4\,k(k-N)}{(2 + k + N)^{2}},\qquad
c_ {3}  = \frac {4 (k - N)} {(2 + k + N)^{2}}, \nonu \\
c_ {4}  & = & \frac {2 i (k - N)} {(2 + k + N)}, \qquad
c_ {5}  = \frac {8 (k + N)} {(2 + k + N)^{2}}, \qquad
c_ {6}  = \frac {4 i (2 k + k^{2} + 2 N + 4 kN + N^{2})} {(2 + k + 
      N)^{2}},\nonu\\
c_ {7} & = & - \frac {8 (1 + k + N)} {(2 + k + N)}, \qquad
c_ {8}  =  - \frac {8 (1 + k + N)} {(2 + k + N)^{2}},\qquad 
c_ {9}  = \frac {8 (k - N)} {(2 + k + N)^{2}}, \nonu \\
c_ {10}  & = &  \frac {4 i (k - N)} {(2 + k + N)^{2}}, \qquad
c_ {11}  = \frac {8 (1 + k + N)} {(2 + k + N)^{2}}, \qquad
c_ {12}  =  - \frac {16 i} {(2 + k + N)^{2}}, \nonu\\
c_ {14} & = & - \frac {4 (k - N)} {(2 + k + N)^{2}}, \qquad
c_ {15}  = \frac {2} {(2 + k + N)}, \qquad
c_ {16}  =  - \frac {8 i} {(2 + k + N)^{2}}, \nonu \\
c_ {17}  & = & \frac {8 (6 + 4 k + k^{2} + 4 N + N^{2})} {(2 + k + 
      N)^{3}},\qquad
c_ {18}  = \frac {8} {(2 + k + N)}, \qquad
c_ {19}  = \frac {2 i (k - N)} {(2 + k + N)^{2}}, \nonu\\
c_ {20} & = &\frac {2 i (2 + 4 k + k^{2} + 4 N + 4 kN + N^{2})} {(2 + 
      k + N)^{2}}, \qquad
c_ {21}  = \frac {4 (-2 + 2 k + k^{2} + 2 N + 4 kN + N^{2})} {(2 + 
      k + N)^{3}}, \nonu\\
c_ {22} & = &\frac {4 (2 + 2 k + k^{2} + 2 N + N^{2})} {(2 + k + 
      N)^{3}},\qquad
c_ {23}  = \frac {4 (k - N)} {(2 + k + N)^{2}}, \qquad
c_ {24}  = \frac {2 i (k - N)} {(2 + k + N)}, \nonu \\
c_ {25}  & = &   - \frac {2 (k - N)} {(2 + k + N)^{2}},\qquad
c_ {26}  = \frac {4 ( k + N)} {(2 + k + N)^{2}},\qquad
c_ {27}  = -\frac {4(k+N)} {(2 + k + N)^2}, \nonu\\
c_ {28} & = & - \frac {4 i} {(2 + k + N)^{2}}, \qquad
c_ {29}  = \frac {1} {2}, \nonu\\
c_ {30} & = & - \frac {(60 + 77 k + 22 k^{2} + 121 N + 115 kN + 
      20 k^{2} N + 79 N^{2} + 42 kN^{2} + 
      16 N^{3})} {(2 + N) (2 + k + N)^{2}}, \nonu\\
c_ {31} & = &\frac {4 (10 + 13 k + 4 k^{2} + 10 N + 7 kN + 
      2 N^{2})} {(2 + k + N)^{2}}, \nonu\\
c_ {32} & = &\frac {4 (10 + 13 k + 4 k^{2} + 10 N + 7 kN + 
      2 N^{2})} {(2 + k + N)^{3}}, \nonu\\
c_ {33} & = & - \frac {2 i (2 + 3 k + 3 N)} {(2 + k + N)^{2}},\qquad
c_ {34}  =  - \frac {8 (2 + 3 k + 3 N)} {(2 + k + N)^{2}}, \nonu\\
c_ {35} & = & - \frac {2 (100 + 107 k + 34 k^{2} + 143 N + 89 kN + 
       8 k^{2} N + 53 N^{2} + 14 kN^{2} + 4 N^{3})} {(2 + 
       N) (2 + k + N)^{3}}, \nonu\\
c_ {36} & = &\frac {2 i (20 + 21 k + 6 k^{2} + 25 N + 13 kN + 
      7 N^{2})} {(2 + N) (2 + k + N)^{3}}, \nonu\\
c_ {37} & = & - \frac {(20 + 21 k + 6 k^{2} + 25 N + 13 kN + 
      7 N^{2})} {(2 + N) (2 + k + N)^{2}}, \nonu\\
c_ {38} & = & - \frac {(20 + 25 k + 6 k^{2} + 21 N + 15 kN + 
      5 N^{2})} {(2 + N) (2 + k + N)^{2}}, \nonu\\
c_ {39} & = &\frac { (32 + 55 k + 18 k^{2} + 41 N + 35 kN + 
      11 N^{2})} {3(2 + N) (2 + k + N)^{4}}, \nonu\\
c_ {40} & = &\frac { i (8 + 17 k + 6 k^{2} + 11 N + 11 kN + 
      3 N^{2})} {(2 + N) (2 + k + N)^{3}}, \nonu\\
c_ {41} & = &\frac { (-9 k - 6 k^{2} + N - 7 kN + N^{2})} {4(2 + 
      N) (2 + k + N)^{2}}, \nonu\\
c_ {42} & = &\frac {4 (k + N)} {(2 + k + N)^{2}}, \qquad
c_ {43}  =  - \frac {4 (1 + k + N)} {(2 + k + N)}, \qquad
c_ {44}  =  - \frac {8 (1 + k + N)} {(2 + k + N)^{2}}, \nonu \\
c_ {45}  & = &  \frac {4 (k - N)} {(2 + k + N)^{2}}, \qquad
c_ {46}  = \frac {2 i (k - N)} {(2 + k + N)^{2}}, \qquad
c_ {47}  = \frac {2 i (k - N)} {(2 + k + N)^{2}}, \nonu\\
c_ {48} & = & - \frac {8 i} {(2 + k + N)^{2}}, \qquad
c_ {49}  = \frac {4 (k + N)} {(2 + k + N)^{2}}, \qquad
c_ {50}  = \frac {2} {(2 + k + N)}, \nonu \\
c_ {51}  & = &  - \frac {8 (-1 + k N)} {(2 + k + N)^{3}}, \qquad
c_ {52}  = \frac {4} {(2 + k + N)}, \qquad
c_ {53}  =  - \frac {4 i} {(2 + k + N)^{2}}, \nonu\\
c_ {54} & = & - \frac {2 (k - N)} {(2 + k + N)^{2}}, \qquad
c_ {55}  =  - \frac {1} {2}, \nonu\\
c_ {56} & = &\frac {(60 + 77 k + 22 k^{2} + 121 N + 115 kN + 
     20 k^{2} N + 79 N^{2} + 42 kN^{2} + 
     16 N^{3})} {2 (2 + N) (2 + k + N)^{2}}, \nonu\\
c_ {57} & = & - \frac {(60 + 77 k + 22 k^{2} + 121 N + 115 kN + 
      20 k^{2} N + 79 N^{2} + 42 kN^{2} + 
      16 N^{3})} {2 (2 + N) (2 + k + N)^{2}}, \nonu\\
c_ {58} & = &\frac {2 (16 + 21 k + 6 k^{2} + 19 N + 13 kN + 
      5 N^{2})} {(2 + N) (2 + k + N)^{3}}, \nonu\\
c_ {59} & = &\frac {i (32 + 29 k + 6 k^{2} + 35 N + 17 kN + 
      9 N^{2})} {(2 + N) (2 + k + N)^{2}}, \nonu\\
c_ {60} & = &\frac {2 (32 + 55 k + 18 k^{2} + 41 N + 35 kN + 
      11 N^{2})} {(2 + N) (2 + k + N)^{4}}, \nonu\\
c_ {61} & = &\frac {2 (1 + N) (20 + 23 k + 6 k^{2} + 23 N + 14 kN + 
      6 N^{2})} {(2 + N) (2 + k + N)^{4}}, \nonu\\
c_ {62} & = & - \frac {(8 + 17 k + 6 k^{2} + 11 N + 11 kN + 
      3 N^{2})} {(2 + N) (2 + k + N)^{2}}, \nonu\\
c_ {63} & = &\frac {2 i (8 + 17 k + 6 k^{2} + 11 N + 11 kN + 
      3 N^{2})} {(2 + N) (2 + k + N)^{3}}, \nonu\\
c_ {64} & = &\frac {2 i (5 k + 2 k^{2} + 5 N + 14 kN + 4 k^{2} N + 
      7 N^{2} + 7 kN^{2} + 2 N^{3})} {(2 + N) (2 + k + N)^{3}}, \qquad
c_ {65}  = \frac {2 i (k - N)} {(2 + k + N)^{2}}, \nonu\\
c_ {66} & = & - \frac {2 i (k + 9 N + 12 kN + 3 k^{2} N + 11 N^{2} + 
       7 kN^{2} + 3 N^{3})} {(2 + N) (2 + k + N)^{3}}, \nonu\\
c_ {67} & = &\frac {i (32 + 29 k + 6 k^{2} + 35 N + 17 kN + 
      9 N^{2})} {(2 + N) (2 + k + N)^{3}}, \nonu\\
c_ {68} & = & - \frac {1} {(2 + N) (2 + k + N)^{4}} (64 + 100 k + 
     35 k^{2} + 2 k^{3} + 92 N + 60 kN - 21 k^{2} N - 8 k^{3} N 
     \nonu\\ & + & 
     49 N^{2} - 14 k^{2} N^{2} + 19 N^{3} + 2 kN^{3} + 4 N^{4}),\nonu\\ 
c_ {69} & = & - \frac {1} {(2 + N) (2 + k + N)^{4}} 4 (-54 k - 
     42 k^{2} - 8 k^{3} + 6 N - 50 kN - 29 k^{2} N - 4 k^{3} N + 
     20 N^{2} 
     \nonu\\ & - & 7 kN^{2} 
     -  7 k^{2} N^{2} + 12 N^{3} + kN^{3} + 
     2 N^{4}), \nonu\\
c_ {70} & = & - \frac {1} {(2 + N) (2 + k + N)^{4}} (32 + 84 k + 
     35 k^{2} + 2 k^{3} + 60 N + 84 kN - 5 k^{2} N - 8 k^{3} N + 
     41 N^{2} 
     \nonu\\ & + & 32 kN^{2} - 6 k^{2} N^{2} + 19 N^{3} + 10 kN^{3} + 
     4 N^{4}), \nonu\\
c_ {71} & = &\frac {3 i (20 + 21 k + 6 k^{2} + 25 N + 13 kN + 
      7 N^{2})} {(2 + N) (2 + k + N)^{3}}, \nonu\\
c_ {72} & = & - \frac {i (20 + 21 k + 6 k^{2} + 25 N + 13 kN + 
       7 N^{2})} {(2 + N) (2 + k + N)^{3}}, \nonu\\
c_ {73} & = &\frac {4 (20 + 27 k + 8 k^{2} + 39 N + 39 kN + 
      7 k^{2} N + 25 N^{2} + 14 kN^{2} + 5 N^{3})} {(2 + 
      N) (2 + k + N)^{4}}, \nonu\\
c_ {74} & = & - \frac{1} {2 (2 + N) (2 + k + N)^{3}} i (96 + 75 k + 4 k^{2} - 4 k^{3} + 133 N + 
       43 kN - 30 k^{2} N - 8 k^{3} N 
       \nonu\\ & + & 73 N^{2} + 2 kN^{2} - 
       14 k^{2} N^{2} + 24 N^{3} + 2 kN^{3} + 
       4 N^{4}), \nonu\\
c_ {75} & = & - \frac {2 i (32 + 55 k + 18 k^{2} + 41 N + 35 kN + 
       11 N^{2})} {(2 + N) (2 + k + N)^{4}}, \qquad
c_ {76}  =  - \frac {8 i} {(2 + k + N)^{2}}, \nonu\\
c_ {77} & = & - \frac {1} {(2 + N) (2 + k + N)^{4}} i (64 + 142 k + 
     65 k^{2} + 6 k^{3} + 114 N + 134 kN + 23 k^{2} N + 57 N^{2} 
     \nonu\\ & + &
     26 kN^{2} + 9 N^{3}), \nonu\\
c_ {78} & = & - \frac {2 (32 + 55 k + 18 k^{2} + 41 N + 35 kN + 
       11 N^{2})} {(2 + N) (2 + k + N)^{4}}, \nonu\\
c_ {79} & = & - \frac {i (16 + 21 k + 6 k^{2} + 19 N + 13 kN + 
       5 N^{2})} {(2 + N) (2 + k + N)^{3}}, \nonu\\
c_ {80} & = &\frac {i (13 k + 6 k^{2} + 3 N + 9 kN + N^{2})} {(2 + 
      N) (2 + k + N)^{3}}, \nonu\\
c_ {81} & = & - \frac {(20 + 21 k + 6 k^{2} + 25 N + 13 kN + 
      7 N^{2})} {(2 + N) (2 + k + N)^{3}}, \qquad
c_ {82}  =  - \frac {4 i} {(2 + k + N)^{2}}, \nonu\\
c_ {83} & = & - \frac {i (16 + 21 k + 6 k^{2} + 19 N + 13 kN + 
       5 N^{2})} {(2 + N) (2 + k + N)^{3}}, \nonu\\
c_ {84} & = &\frac {i (-32 - 104 k - 53 k^{2} - 6 k^{3} - 40 N - 
      72 kN - 11 k^{2} N - 3 N^{2} - 2 kN^{2} + 3 N^{3})} {(2 + 
      N) (2 + k + N)^{4}}, \nonu\\
c_ {85} & = & - \frac {i (-72 k - 45 k^{2} - 6 k^{3} + 8 N - 40 kN - 
       7 k^{2} N + 21 N^{2} + 6 kN^{2} + 7 N^{3})} {(2 + 
       N) (2 + k + N)^{4}}, \nonu\\
c_ {86} & = & - \frac {i (16 + 21 k + 6 k^{2} + 19 N + 13 kN + 
       5 N^{2})} {(2 + N) (2 + k + N)^{3}}, \nonu\\
c_ {87} & = &\frac {i (13 k + 6 k^{2} + 3 N + 9 kN + N^{2})} {(2 + 
      N) (2 + k + N)^{3}}, \nonu\\
c_ {88} & = & - \frac {i (16 + 21 k + 6 k^{2} + 19 N + 13 kN + 
       5 N^{2})} {(2 + N) (2 + k + N)^{3}}, \nonu\\
c_ {89} & = &\frac {i (13 k + 6 k^{2} + 3 N + 9 kN + N^{2})} {(2 + 
      N) (2 + k + N)^{3}}, \nonu\\
c_ {90} & = & - \frac {1} {(2 + N) (2 + k + N)^{4}} i (96 + 168 k + 
     69 k^{2} + 6 k^{3} + 136 N + 136 kN + 19 k^{2} N + 51 N^{2} 
     \nonu\\ & + &
     18 kN^{2} + 5 N^{3}), \nonu\\
c_ {91} & = &\frac {i (64 + 136 k + 61 k^{2} + 6 k^{3} + 88 N + 
      104 kN + 15 k^{2} N + 27 N^{2} + 10 kN^{2} + N^{3})} {(2 + 
      N) (2 + k + N)^{4}}, \nonu\\
c_ {92} & = & - \frac {i (16 - 5 k - 6 k^{2} + 13 N - 5 kN + 
       3 N^{2})} {(2 + N) (2 + k + N)^{3}}, \nonu\\
c_ {93} & = & - \frac {i (13 k + 6 k^{2} + 3 N + 9 kN + N^{2})} {(2 + 
       N) (2 + k + N)^{3}}, \nonu\\
c_ {94} & = &\frac {i (64 + 136 k + 61 k^{2} + 6 k^{3} + 88 N + 
      104 kN + 15 k^{2} N + 27 N^{2} + 10 kN^{2} + N^{3})} {(2 + 
      N) (2 + k + N)^{4}}, \nonu\\
c_ {95} & = & - \frac {1} {(2 + N) (2 + k + N)^{4}} i (96 + 168 k + 
     69 k^{2} + 6 k^{3} + 136 N + 136 kN + 19 k^{2} N + 51 N^{2} 
     \nonu\\ & + & 
     18 kN^{2} + 5 N^{3}), \nonu\\
c_ {96} & = &\frac {(32 + 11 k - 6 k^{2} + 37 N + 3 kN + 
     11 N^{2})} {2 (2 + N) (2 + k + N)^{2}}, \nonu\\
c_ {97} & = & - \frac {(64 + 136 k + 61 k^{2} + 6 k^{3} + 88 N + 
      104 kN + 15 k^{2} N + 27 N^{2} + 10 kN^{2} + 
      N^{3})} {2 (2 + N) (2 + k + N)^{3}}, \nonu\\
c_ {98} & = &\frac {(-16 + 13 k + 6 k^{2} - 5 N + 9 kN + 
     N^{2})} {2 (2 + N) (2 + k + N)^{2}}, \nonu\\
c_ {99} & = &\frac {(96 + 168 k + 69 k^{2} + 6 k^{3} + 136 N + 
     136 kN + 19 k^{2} N + 51 N^{2} + 18 kN^{2} + 
     5 N^{3})} {2 (2 + N) (2 + k + N)^{3}}, \nonu\\
c_ {100} & = & - \frac {2 (32 + 55 k + 18 k^{2} + 41 N + 35 kN + 
       11 N^{2})} {(2 + N) (2 + k + N)^{4}}, \nonu\\
c_ {101} & = &\frac {2 (20 + 21 k + 6 k^{2} + 25 N + 13 kN + 
      7 N^{2})} {(2 + N) (2 + k + N)^{3}}, \nonu\\
c_ {102} & = &\frac {i (20 + 21 k + 6 k^{2} + 25 N + 13 kN + 
      7 N^{2})} {(2 + N) (2 + k + N)^{2}}, \nonu\\
c_ {103} & = & - \frac {4 (8 - 5 k - 4 k^{2} + 9 N - kN + k^{2} N + 
       5 N^{2} + 2 kN^{2} + N^{3})} {(2 + N) (2 + k + N)^{4}}, \nonu\\
c_ {104} & = &\frac {4 (24 + 37 k + 12 k^{2} + 39 N + 33 kN + 
      3 k^{2} N + 19 N^{2} + 6 kN^{2} + 3 N^{3})} {(2 + 
      N) (2 + k + N)^{4}}, \nonu\\
c_ {105} & = &\frac {(20 + 21 k + 6 k^{2} + 25 N + 13 kN + 
     7 N^{2})} {2 (2 + N) (2 + k + N)^{2}}, \nonu\\
c_ {106} & = & - \frac {i (20 + 21 k + 6 k^{2} + 25 N + 13 kN + 
       7 N^{2})} {(2 + N) (2 + k + N)^{3}}, \nonu\\
c_ {107} & = &\frac {i (20 + 21 k + 6 k^{2} + 25 N + 13 kN + 
      7 N^{2})} {(2 + N) (2 + k + N)^{3}}, \nonu\\
c_ {108} & = & - \frac {12 i} {(2 + k + N)^{2}}, \qquad
c_ {109}  =  - \frac {2 i (4 + 3 k + 3 N)} {(2 + k + N)^{2}}, \nonu\\
c_ {110} & = &\frac {i (20 + 21 k + 6 k^{2} + 25 N + 13 kN + 
      7 N^{2})} {(2 + N) (2 + k + N)^{3}}, \nonu\\
c_ {111} & = &\frac {1} {(2 + N) (2 + k + N)^{4}} (80 + 112 k + 
    29 k^{2} - 2 k^{3} + 192 N + 234 kN + 69 k^{2} N + 8 k^{3} N 
    \nonu\\ & + & 
    157 N^{2} + 134 kN^{2} + 22 k^{2} N^{2} + 47 N^{3} + 18 kN^{3} + 
    4 N^{4}), \nonu\\
c_ {112} & = &\frac {1} {(2 + N) (2 + k + N)^{4}} 4 (60 + 87 k + 
    44 k^{2} + 8 k^{3} + 131 N + 143 kN + 47 k^{2} N + 4 k^{3} N 
    \nonu\\ & + & 
    95 N^{2} + 69 kN^{2} + 11 k^{2} N^{2} + 26 N^{3} + 9 kN^{3} + 
    2 N^{4}), \nonu\\
c_ {113} & = &\frac {(-20 - 11 k - 2 k^{2} - 15 N + 15 kN + 
     8 k^{2} N + 7 N^{2} + 14 kN^{2} + 
     4 N^{3})} {(2 + N) (2 + k + N)^{3}}, \nonu\\
c_ {114} & = &\frac {12 i} {(2 + k + N)^{2}}, \qquad
c_ {115}  = \frac {2 i (8 + 17 k + 6 k^{2} + 11 N + 11 kN + 
      3 N^{2})} {(2 + N) (2 + k + N)^{3}}, \nonu\\
c_ {116} & = & - \frac {4 (8 + 21 k + 8 k^{2} + 15 N + 17 kN + 
       k^{2} N + 7 N^{2} + 2 kN^{2} + N^{3})} {(2 + 
       N) (2 + k + N)^{4}}, \nonu\\
c_ {117} & = &\frac {1} {(2 + N) (2 + k + N)^{3}} i (-20 - 17 k + 
    k^{2} + 2 k^{3} - 29 N - kN + 17 k^{2} N + 4 k^{3} N - 6 N^{2} 
    \nonu\\ & + & 
    21 kN^{2} + 11 k^{2} N^{2} + 6 N^{3} + 9 kN^{3} + 2 N^{4}), \nonu\\
c_ {118} & = & - \frac {2 i (8 + 17 k + 6 k^{2} + 11 N + 11 kN + 
       3 N^{2})} {(2 + N) (2 + k + N)^{3}}, \nonu\\
c_ {119} & = &\frac {2 i (32 + 33 k + 8 k^{2} + 39 N + 23 kN + 
      k^{2} N + 13 N^{2} + 2 kN^{2} + N^{3})} {(2 + 
      N) (2 + k + N)^{3}}, \nonu\\
c_ {120} & = & - \frac {i (20 + 5 k + 6 k^{2} + 41 N + 5 kN + 
       15 N^{2})} {(2 + N) (2 + k + N)^{3}}, \nonu\\
c_ {121} & = & - \frac {i (20 + 21 k + 6 k^{2} + 25 N + 13 kN + 
       7 N^{2})} {(2 + N) (2 + k + N)^{3}}, \nonu\\
c_ {122} & = & - \frac {i (k - N) (13 k + 6 k^{2} + 3 N + 9 kN + 
       N^{2})} {(2 + N) (2 + k + N)^{4}}, \nonu\\
c_ {123} & = & - \frac {2 (k - N) (32 + 55 k + 18 k^{2} + 41 N + 
       35 kN + 11 N^{2})} {(2 + N) (2 + k + N)^{5}}, \nonu\\
c_ {124} & = & - \frac {8} {(2 + k + N)^{2}}, \qquad
c_ {125}  = \frac {2 i (32 + 55 k + 18 k^{2} + 41 N + 35 kN + 
      11 N^{2})} {(2 + N) (2 + k + N)^{4}}, \nonu\\
c_ {126} & = &\frac {(13 k + 6 k^{2} + 3 N + 9 kN + 
     N^{2})} {(2 + N) (2 + k + N)^{3}}, \nonu\\
c_ {127} & = & - \frac {2 (8 + 17 k + 6 k^{2} + 11 N + 11 kN + 
       3 N^{2})} {(2 + N) (2 + k + N)^{3}}, \nonu\\
c_ {128} & = & - \frac {2 (16 + 21 k + 6 k^{2} + 19 N + 13 kN + 
       5 N^{2})} {(2 + N) (2 + k + N)^{3}}, \nonu\\
c_ {129} & = &\frac {(16 + 21 k + 6 k^{2} + 19 N + 13 kN + 
     5 N^{2})} {(2 + N) (2 + k + N)^{3}}, \nonu\\
c_ {130} & = &\frac {2 i (32 + 55 k + 18 k^{2} + 41 N + 35 kN + 
      11 N^{2})} {(2 + N) (2 + k + N)^{4}}, \nonu\\
c_ {131} & = & - \frac {i (20 + 21 k + 6 k^{2} + 25 N + 13 kN + 
       7 N^{2})} {(2 + N) (2 + k + N)^{3}}, \nonu\\
c_ {132} & = &\frac {i (k - N) (16 + 21 k + 6 k^{2} + 19 N + 13 kN + 
      5 N^{2})} {(2 + N) (2 + k + N)^{4}}, \nonu\\
c_ {133} & = & - \frac {i (40 + 46 k - k^{2} - 6 k^{3} + 86 N + 
       84 kN + 9 k^{2} N + 61 N^{2} + 36 kN^{2} + 13 N^{3})} {(2 + 
       N) (2 + k + N)^{4}}, \nonu\\
c_ {134} & = & - \frac {i (20 + 21 k + 6 k^{2} + 25 N + 13 kN + 
       7 N^{2})} {(2 + N) (2 + k + N)^{3}}, \nonu\\
c_ {135} & = & - \frac {1} {(2 + N) (2 + k + N)^{4}} i (80 + 124 k + 
     53 k^{2} + 6 k^{3} + 140 N + 154 kN + 35 k^{2} N + 81 N^{2} 
     \nonu\\ & + & 
     48 kN^{2} + 15 N^{3}), \nonu\\
c_ {136} & = & - \frac {1} {(2 + N) (2 + k + N)^{4}} i (120 + 154 k + 
     57 k^{2} + 6 k^{3} + 242 N + 220 kN + 43 k^{2} N + 155 N^{2} 
     \nonu\\ & + & 
     76 kN^{2} + 31 N^{3}), \nonu\\
c_ {137} & = & - \frac {1} {(2 + N) (2 + k + N)^{4}} i (80 + 92 k + 
     11 k^{2} - 6 k^{3} + 172 N + 158 kN + 21 k^{2} N + 119 N^{2} 
     \nonu\\ & + & 
     64 kN^{2} + 25 N^{3}), \nonu\\
c_ {138} & = &\frac {(40 + 30 k - 9 k^{2} - 6 k^{3} + 102 N + 76 kN + 
     5 k^{2} N + 77 N^{2} + 36 kN^{2} + 
     17 N^{3})} {2 (2 + N) (2 + k + N)^{3}}, \nonu\\
c_ {139} & = &\frac {(40 + 66 k + 33 k^{2} + 6 k^{3} + 106 N + 
     120 kN + 31 k^{2} N + 83 N^{2} + 48 kN^{2} + 
     19 N^{3})} {2 (2 + N) (2 + k + N)^{3}}, \nonu\\
c_ {140} & = &\frac {(40 + 46 k - k^{2} - 6 k^{3} + 86 N + 84 kN + 
     9 k^{2} N + 61 N^{2} + 36 kN^{2} + 
     13 N^{3})} {2 (2 + N) (2 + k + N)^{3}}, \nonu\\
c_ {141} & = &\frac {(40 + 82 k + 41 k^{2} + 6 k^{3} + 90 N + 
     128 kN + 35 k^{2} N + 67 N^{2} + 48 kN^{2} + 
     15 N^{3})} {2 (2 + N) (2 + k + N)^{3}},\nonu\\
c_ {142} & = & 1, \nonu\\
c_ {143} & = & - \frac {(60 + 77 k + 22 k^{2} + 121 N + 115 kN + 
      20 k^{2} N + 79 N^{2} + 42 kN^{2} + 
      16 N^{3})} {(2 + N) (2 + k + N)^{2}}, \nonu\\
c_ {144} & = &\frac {2 (10 + 13 k + 4 k^{2} + 10 N + 7 kN + 
      2 N^{2})} {(2 + k + N)^{2}}, \qquad
c_ {145}  =  - \frac {4 (2 + 3 k + 3 N)} {(2 + k + N)^{2}}, \nonu\\
c_ {146} & = &\frac {4 (10 + 13 k + 4 k^{2} + 10 N + 7 kN + 
      2 N^{2})} {(2 + k + N)^{3}}, \qquad
c_ {147}  =  - \frac {2 i (-2 + k + N)} {(2 + k + N)^{2}}, \nonu\\
c_ {148} & = & - \frac {6 i} {(2 + k + N)}, \qquad
c_ {149}  =  - \frac {4 i} {(2 + k + N)}, \nonu\\
c_ {150} & = &\frac {8 i} {(2 + k + N)^{2}}, \qquad
c_ {151}  = \frac {2} {(2 + k + N)^{2}}, \nonu\\
c_ {152} & = & - \frac {(100 + 107 k + 34 k^{2} + 143 N + 89 kN + 
      8 k^{2} N + 53 N^{2} + 14 kN^{2} + 
      4 N^{3})} {(2 + N) (2 + k + N)^{3}}, \nonu\\
c_ {153} & = &\frac {8} {(2 + k + N)^{2}}, \qquad
c_ {154}  = \frac {2 (k - N)} {(2 + k + N)^{2}}, \nonu\\
c_ {155} & = & - \frac {(20 + 25 k + 6 k^{2} + 21 N + 15 kN + 
      5 N^{2})} {2(2 + N) (2 + k + N)^{2}}, \nonu\\
c_ {156} & = &\frac {2 i (20 + 21 k + 6 k^{2} + 25 N + 13 kN + 
      7 N^{2})} {(2 + N) (2 + k + N)^{3}}, \nonu\\
c_ {157} & = &\frac {2 i (20 + 21 k + 6 k^{2} + 25 N + 13 kN + 
      7 N^{2})} {(2 + N) (2 + k + N)^{3}}, \nonu\\
c_ {158} & = &\frac {2 i (20 + 17 k + 6 k^{2} + 29 N + 11 kN + 
      9 N^{2})} {(2 + N) (2 + k + N)^{3}}, \nonu\\
c_ {159} & = &\frac {2 i (20 + 25 k + 6 k^{2} + 21 N + 15 kN + 
      5 N^{2})} {(2 + N) (2 + k + N)^{3}}, \nonu\\
c_ {160} & = &\frac {4 (32 + 55 k + 18 k^{2} + 41 N + 35 kN + 
      11 N^{2})} {(2 + N) (2 + k + N)^{4}}, \nonu\\
c_ {161} & = & - \frac {2 (4 + k + N)} {(2 + k + N)^{2}}, \qquad
c_ {162}  = \frac {2 (4 + 3 k + 3 N)} {(2 + k + N)^{2}}, \nonu\\
c_ {163} & = &\frac {(-9 k - 6 k^{2} + N - 7 kN + 
     N^{2})} {4(2 + N) (2 + k + N)^{2}}, \qquad
c_ {164}  = \frac {2 i (13 k + 6 k^{2} + 3 N + 9 kN + N^{2})} {(2 + 
      N) (2 + k + N)^{3}}, \nonu\\
c_ {165} & = &\frac {2 i (16 + 21 k + 6 k^{2} + 19 N + 13 kN + 
      5 N^{2})} {(2 + N) (2 + k + N)^{3}}, \nonu\\
c_ {166} & = &\frac {2 i (16 + 21 k + 6 k^{2} + 19 N + 13 kN + 
      5 N^{2})} {(2 + N) (2 + k + N)^{3}}, \nonu\\
c_ {167} & = &\frac {2 i (13 k + 6 k^{2} + 3 N + 9 kN + N^{2})} {(2 + 
      N) (2 + k + N)^{3}}, \nonu\\
c_ {168} & = &\frac {2 i (8 + 17 k + 6 k^{2} + 11 N + 11 kN + 
      3 N^{2})} {(2 + N) (2 + k + N)^{3}}, \nonu\\
c_ {169} & = &\frac {2 i (16 + 21 k + 6 k^{2} + 19 N + 13 kN + 
      5 N^{2})} {(2 + N) (2 + k + N)^{3}}, \qquad
c_ {170}  = \frac {2 i (13 k + 6 k^{2} + 3 N + 9 kN + N^{2})} {(2 + 
      N) (2 + k + N)^{3}}.
\nonu
\eea
The fusion rule is 
\bea
[\Phi_{1}^{(1),ij}] \, \cdot \, [\Phi_{1}^{(1),kl}]
& = & [I^{ijkl}]+ \varepsilon^{ijkl} [\Phi_0^{(1)} \, \Phi_0^{(1)}] + \delta^{ik} [
\Phi_0^{(1)} \, \Phi_1^{(1),jl}] + \delta^{ik} \,
\varepsilon^{ijla}[\Phi_{\frac{1}{2}}^{(1),i}\,
\Phi_{\frac{1}{2}}^{(1),a}]
\nonu \\
& + &  \varepsilon^{ijkl} \, [\Phi_0^{(2)}]
+ \delta^{ik} \,  [\Phi_1^{(2),jl}] .
\nonu
\eea

The OPEs between the higher spin-$2$ currents and
the higher spin-$\frac{5}{2}$ currents  are given by
\bea
&&\Phi_{1}^{(1),ij}(z)\:\widetilde{\Phi}_{\frac{3}{2}}^{(1),k}(w)\;=\;\frac{1}{(z-w)^{4}}\Bigg[\delta^{ik}\, c_{1}\,\Gamma^{j}+\varepsilon^{ijkl}c_{2}\,\Gamma^{l}\Bigg](w)
+\frac{1}{(z-w)^{3}}\,\Bigg[\delta^{ik}\,\Bigg\{ c_{3}\, G^{j}+c_{4}\, U\,\Gamma^{j}
\nonu\\
&&+c_{5}\,\partial\Gamma^{j}+c_{6}\, T^{ij}\,\Gamma^{i}
+\varepsilon^{iabj}\,\Big(c_{7}\,\Gamma^{i}\,\Gamma^{a}\,\Gamma^{b}+c_{8}\, T^{ia}\,\Gamma^{b}+c_{9}\, T^{ab}\,\Gamma^{i}\Big)+c_{10}\,(T^{jl}\,\Gamma^{l})(1-\delta^{il})\Bigg\}
\nonu\\
&&+\varepsilon^{ijkl}\,\Bigg\{ c_{11}\, G^{l}+c_{12}\, U\,\Gamma^{l}+c_{13}\,\partial\Gamma^{l}+c_{14}\, T^{kl}\,\Gamma^{k}+c_{15}\,(T^{il}\,\Gamma^{i}+T^{jl}\,\Gamma^{l})\Bigg\}
\nonu\\
&&+(\varepsilon^{ijkl})^{2}\,\Bigg\{ c_{16}\,(T^{jk}\,\Gamma^{i}+T^{ki}\,\Gamma^{j})+c_{17}\, T^{ij}\,\Gamma^{k}\Bigg\}+c_{18}\,\Gamma^{i}\,\Gamma^{j}\,\Gamma^{k}\Bigg](w)
\nonu\\
&&+\frac{1}{(z-w)^{2}}\,\Bigg[\delta^{ik}\,\Bigg\{  
c_{19}\, {\bf \Phi_{\frac{1}{2}}^{(2),j}}+c_{20}\,
{\bf \Phi_{0}^{(1)}\,\Phi_{\frac{1}{2}}^{(1),j}}+
c_{21}\, L\,\Gamma^{j}+c_{22}\, G^{j}\, U+c_{23}\,\partial G^{j}+c_{24}\, U\, U\,\Gamma^{j}
\nonu\\
&&+c_{25}\, U\,\partial\Gamma^{j}+c_{26}\,\partial U\,\Gamma^{j}+c_{27}\,\partial^{2}\Gamma^{j}+c_{28}\, G^{i}\, T^{ij}+c_{29}\, G^{i}\,\Gamma^{i}\,\Gamma^{j}+c_{30}\, T^{ij}\, U\,\Gamma^{i}+c_{31}\, T^{ij}\,\partial\Gamma^{i}
\nonu\\
&&+c_{32}\,\partial T^{ij}\,\Gamma^{i}+c_{33}\,\partial\Gamma^{i}\,\Gamma^{i}\,\Gamma^{j}+(1-\delta^{il})\,\Big(c_{34}\,G^{l}\,\Gamma^{j}\,\Gamma^{l}+c_{35}\,G^{l}\, T^{jl}+c_{36}\,\Gamma^{j}\,\partial\Gamma^{l}\,\Gamma^{l}
\nonu\\
&&+c_{37}\,T^{jl}\,T^{jl}\,\Gamma^{j}+(1-\delta^{ia}-\delta^{ja})(c_{38}\,T^{jl}\, T^{la}\,\Gamma^{a}+c_{39}\, T^{al}\,\Gamma^{j}\,\Gamma^{a}\,\Gamma^{l}\Big)+c_{40}\,(T^{ij}\, T^{il}\,\Gamma^{l}-\widetilde{T}^{jl}\,\widetilde{T}^{jl}\,\Gamma^{j}
\nonu\\
&&-T^{ij}\, T^{ij}\,\Gamma^{j})+c_{41}\, T^{il}\,\Gamma^{i}\,\Gamma^{j}\,\Gamma^{l}+c_{42}\, T^{il}\, T^{jl}\,\Gamma^{i}+c_{43}\, T^{jl}\, U\,\Gamma^{l}+c_{44}\,T^{jl}\,T^{jl}\,\Gamma^{l}+c_{45}\,\partial T^{jl}\,\Gamma^{l}
\nonu\\
&&+\varepsilon^{ijab}\,\Big(c_{46}\, U\,\Gamma^{i}\,\Gamma^{a}\,\Gamma^{b}+c_{47}\, G^{i}\, T^{ab}+c_{48}\, G^{i}\,\Gamma^{a}\,\Gamma^{b}+c_{49}\,\partial\Gamma^{i}\,\Gamma^{a}\,\Gamma^{b}+c_{50}\,\Gamma^{i}\,\partial\Gamma^{a}\,\Gamma^{b}
\nonu\\
&&+c_{51}\, G^{a}\, T^{ib}+c_{52}\, G^{a}\,\Gamma^{i}\,\Gamma^{b}+c_{53}\, T^{ij}\, T^{ab}\,\Gamma^{j}+c_{54}\, T^{ia}\, U\,\Gamma^{b}+c_{55}\, T^{ia}\, T^{jb}\,\Gamma^{j}+c_{56}\, T^{ia}\, T^{ab}\,\Gamma^{a}
\nonu\\
&&+c_{57}\, T^{ia}\,\partial\Gamma^{b}+c_{58}\, T^{ab}\, U\,\Gamma^{i}+c_{59}\, T^{ab}\,\partial\Gamma^{i}+c_{60}\,\partial T^{ab}\,\Gamma^{i}+c_{61}\,\partial T^{ia}\,\Gamma^{b}\Big)\Bigg\}
\nonu\\
&&+\varepsilon^{ijkl}\,\Bigg\{  
c_{62}\, {\bf \Phi_{\frac{1}{2}}^{(2),l}}+
c_{63}\, {\bf \Phi_{0}^{(1)}\,\Phi_{\frac{1}{2}}^{(1),l}}+
c_{64}\, L\,\Gamma^{l}+c_{65}\,\partial G^{l}+c_{66}\, G^{l}\, U+c_{67}\,U\,U\,\Gamma^{l}+c_{68}\, U\,\partial\Gamma^{l}
\nonu\\
&&+c_{69}\,\partial U\,\Gamma^{l}+c_{70}\,\partial^{2}\Gamma^{l}+c_{71}\,(G^{i}\, T^{il}+G^{j}\, T^{jl})+c_{74}\, G^{k}\, T^{kl}+c_{75}\,(\partial T^{il}\,\Gamma^{i}+\partial T^{jl}\,\Gamma^{j})
\nonu\\
&&+c_{76}\,\partial T^{kl}\,\Gamma^{k}
+c_{77}\,(T^{il}\,\partial\Gamma^{i}+T^{jl}\,\partial\Gamma^{j})+c_{78}\, T^{kl}\,\partial\Gamma^{l}+c_{79}\,(T^{il}\, U\,\Gamma^{i}+T^{jl}\, U\,\Gamma^{j})+c_{80}\, T^{kl}\, U\,\Gamma^{k}
\nonu\\
&&+c_{81}\,\widetilde{T}^{la}\,\widetilde{T}^{la}\,\Gamma^{l}
+c_{82}\,(T^{ij}\,\Gamma^{i}\,\Gamma^{j}\,\Gamma^{l}+T^{ik}\,\Gamma^{i}\,\Gamma^{k}\,\Gamma^{l}+T^{jk}\,\Gamma^{j}\,\Gamma^{k}\,\Gamma^{l})+c_{83}\,(T^{ik}\, T^{kl}\,\Gamma^{i}+T^{jk}\, T^{kl}\,\Gamma^{j})
\nonu\\
&&+c_{84}\,(T^{ij}\, T^{jl}\,\Gamma^{i}+T^{ji}\, T^{il}\,\Gamma^{j})+c_{85}\, T^{ak}\, T^{al}\,\Gamma^{k}+c_{86}\,(T^{il}\,T^{il}\,\Gamma^{l}+T^{jl}\,T^{jl}\,\Gamma^{l})
\nonu\\
&&+c_{87}\,(\partial\Gamma^{i}\,\Gamma^{i}\,\Gamma^{l}+\partial\Gamma^{j}\,\Gamma^{j}\,\Gamma^{l})+c_{88}\,\partial\Gamma^{k}\,\Gamma^{k}\,\Gamma^{l}\Bigg\}+(\varepsilon^{ijkl})^{2}\,\Bigg\{ c_{89}\,(G^{i}\,\Gamma^{l}\,\Gamma^{k}+G^{j}\,\Gamma^{k}\,\Gamma^{i})
\nonu\\
&&+c_{90}\, G^{k}\,\Gamma^{i}\,\Gamma^{j}
+c_{91}\,(G^{i}\, T^{jk}+G^{j}\, T^{ki})+c_{92}\, G^{k}\, T^{ij}+c_{93}\,(\partial T^{ki}\,\Gamma^{j}+\partial T^{jk}\,\Gamma^{i})+c_{94}\, T^{ij}\, U\,\Gamma^{k}
\nonu\\
&&+c_{95}\,(T^{jk}\, U\,\Gamma^{i}+T^{ki}\, U\,\Gamma^{j})+c_{96}\,(T^{ij}\, T^{ik}\,\Gamma^{i}+T^{ij}\, T^{jk}\,\Gamma^{j})+c_{97}\, U\,\Gamma^{i}\,\Gamma^{j}\,\Gamma^{k}+c_{98}\,\partial T^{ij}\,\Gamma^{k}
\nonu\\
&&+c_{99}\, T^{ij}\,\partial\Gamma^{k}+c_{100}\,(T^{ki}\,\partial\Gamma^{j}+T^{jk}\,\partial\Gamma^{i})+c_{101}\,\Gamma^{i}\,\Gamma^{j}\,\partial\Gamma^{k}+c_{102}\,\partial(\Gamma^{i}\,\Gamma^{j})\,\Gamma^{k}+c_{103}\, T^{ij}\, T^{kl}\,\Gamma^{l}
\nonu\\
&&+c_{104}\,(T^{ik}\, T^{lj}\,\Gamma^{l}+T^{il}\, T^{jk}\,\Gamma^{l})\Bigg\}\Bigg](w)
\nonu\\
&&+\frac{1}{(z-w)}\,\Bigg[\delta^{ik}\,\Bigg\{  
c_{105}\, {\bf \widetilde{\Phi}_{\frac{3}{2}}^{(2),j}}+
c_{106}\, {\bf \partial\Phi_{\frac{1}{2}}^{(2),j}}+
c_{107}\, {\bf \Phi_{0}^{(1)}\,\widetilde{\Phi}_{\frac{3}{2}}^{(1),j}}+
c_{108}\, {\bf \Phi_{\frac{1}{2}}^{(1),l}\,\Phi{}_{1}^{(1),jl}}+
c_{109}\, {\bf \Phi_{0}^{(1)}\,\partial\Phi{}_{\frac{1}{2}}^{(1),j}}
\nonu\\
&& +c_{110}\,{\bf \partial\Phi_{0}^{(1)}\,\Phi{}_{\frac{1}{2}}^{(1),j}}+
c_{111}\, L\, G^{j}+c_{112}\, L\, T^{ij}\,\Gamma^{i}+c_{113}\, L\, U\,\Gamma^{j}+c_{114}\,\partial L\,\Gamma^{j}+c_{115}\, L\,\partial\Gamma^{j}
\nonu\\
&&+c_{116}\, G^{j}\, G^{l}\,\Gamma^{l}+c_{117}\,\partial G^{i}\, T^{ji}+c_{118}\, G^{i}\,\partial T^{ji}+c_{119}\, G^{j}\,\widetilde{T}^{jl}\,\widetilde{T}^{jl}+c_{120}\, G^{j}\,T^{ij}\,T^{ij}+c_{121}\, G^{l}\, T^{jl}\, U
\nonu\\
&&+c_{122}\, G^{j}\, T^{jl}\,\Gamma^{l}\,\Gamma^{j}+c_{123}\,(\frac{1}{2}\,G^{j}\, T^{ab}\,\Gamma^{a}\,\Gamma^{b}+G^{a}\, T^{jb}\,\Gamma^{a}\,\Gamma^{b})+c_{124}\, G^{j}\, U\, U+c_{125}\,\partial G^{j}\, U
\nonu\\
&&+c_{126}\, G^{j}\,\partial U+c_{127}\,\partial^{2}G^{j}+c_{128}\,\partial G^{i}\,\Gamma^{i}\,\Gamma^{j}+c_{129}\, G^{i}\,\partial\Gamma^{i}\,\Gamma^{j}+c_{130}\, G^{i}\,\Gamma^{i}\,\partial\Gamma^{j}+c_{131}\, G^{j}\,\partial\Gamma^{j}\,\Gamma^{j}
\nonu\\
&&+c_{132}\, G^{j}\,\partial\Gamma^{i}\,\Gamma^{i}+c_{133}\,\partial T^{ij}\, T^{ij}\,\Gamma^{j}+c_{134}\,\partial\widetilde{T}^{jl}\,\widetilde{T}^{jl}\,\Gamma^{j}+c_{135}\,\partial T^{il}\, T^{jl}\,\Gamma^{i}+c_{136}\, T^{il}\,\partial T^{jl}\,\Gamma^{i}
\nonu\\
&&+c_{137}\,\widetilde{T}^{jl}\,\widetilde{T}^{jl}\,\partial\Gamma^{j}+c_{138}\,T^{ij}\,T^{ij}\,\partial\Gamma^{j}+c_{139}\,T^{ab}\,T^{jb}\, U\,\Gamma^{a}+c_{140}\, T^{ij}\,\partial U\,\Gamma^{i}
+c_{141}\, T^{ij}\, U\,\partial\Gamma^{i}
\nonu\\
&&+c_{142}\,\partial T^{ij}\, U\,\Gamma^{i}+c_{143}\, T^{ij}\,\Gamma^{i}\,\partial\Gamma^{l}\,\Gamma^{l}(1+\delta^{jl})+c_{144}\,\partial^{2}T^{ij}\,\Gamma^{i}+c_{145}\,\partial T^{ij}\,\partial\Gamma^{i}
+c_{146}\, T^{ij}\,\partial^{2}\Gamma^{i}
\nonu\\
&&+c_{147}\,\partial(T^{ab}\,\Gamma^{a}\,\Gamma^{j}\,\Gamma^{b})+c_{148}\, T^{ab}\,\Gamma^{a}\,\partial\Gamma^{j}\,\Gamma^{b}+c_{149}\,U\,U\,\Gamma^{j}+c_{150}\,\partial U\, U\,\Gamma^{j}
+c_{151}\, U\, U\,\partial\Gamma^{j}
\nonu\\
&&+c_{152}\,\partial^{2}U\,\Gamma^{j}+c_{153}\,\partial U\,\partial\Gamma^{j}+c_{154}\, U\,\partial^{2}\Gamma^{j}+c_{155}\, U\,\Gamma^{j}\,\Gamma^{l}\,\partial\Gamma^{l}+c_{156}\,\Gamma^{j}\,\partial^{2}\Gamma^{i}\,\Gamma^{i}
\nonu\\
&&+c_{157}\,\partial\Gamma^{j}\,\partial\Gamma^{i}\,\Gamma^{i}+c_{158}\,\partial^{3}\Gamma^{j}+(1-\delta^{il})\Bigg(c_{159}\, L\, T^{jl}\,\Gamma^{l}+c_{160}\,\partial G^{l}\, T^{jl}+c_{161}\, G^{l}\,\partial T^{jl}
\nonu\\
&&+c_{162}\, G^{a}\, T^{al}\, T^{jl}+c_{163}\, G^{j}\,T^{jl}\,T^{jl}+c_{164}\,\partial G^{l}\,\Gamma^{j}\,\Gamma^{l}+c_{165}\,T^{jl}\,T^{jl}\,\partial\Gamma^{j}+c_{166}\,(T^{al}\, T^{lj}\,\partial\Gamma^{a})
\nonu\\
&&+c_{167}\,\partial T^{jl}\, T^{jl}\,\Gamma^{j}+c_{168}\,\partial T^{jl}\, U\,\Gamma^{l}+c_{169}\, T^{jl}\,\partial U\,\Gamma^{l}+c_{170}\, T^{jl}\, U\,\partial\Gamma^{l}+c_{171}\,\partial^{2}T^{jl}\,\Gamma^{l}
\nonu\\
&&+c_{172}\,\partial T^{jl}\,\partial\Gamma^{l}+c_{173}\, T^{jl}\,\partial^{2}\Gamma^{l}+c_{174}\,(T^{jl}\,\partial\Gamma^{a}\,\Gamma^{a}\,\Gamma^{l})(1+\delta^{ja})+c_{175}\,\Gamma^{j}\,\partial^{2}\Gamma^{l}\,\Gamma^{l}\Bigg)
\nonu\\
&&+(1-\delta^{jl})\Big(c_{176}\, G^{l}\, T^{ij}\, T^{il}+c_{177}\,\partial T^{ij}\, T^{il}\,\Gamma^{l}+c_{178}\,(T^{ij}\,\partial T^{il}\,\Gamma^{l})+c_{179}\, T^{ij}\, T^{il}\,\partial\Gamma^{l}\Big)
\nonu\\
&&+(1-\delta^{il}-\delta^{jl})\Big(c_{180}\, G^{j}\,\partial\Gamma^{l}\,\Gamma^{l}+c_{181}\, G^{l}\,\partial\Gamma^{j}\,\Gamma^{l}+c_{182}\, G^{l}\,\Gamma^{j}\,\partial\Gamma^{l}+(1-\delta^{ia})(c_{183}\,\partial T^{ja}\, T^{al}\,\Gamma^{q}
\nonu\\
&&+c_{184}\, T^{ja}\,\partial T^{al}\,\Gamma^{l})+c_{185}\,\partial\Gamma^{j}\,\partial\Gamma^{l}\,\Gamma^{l}\Big)+\varepsilon^{ijab}\,\Bigg(c_{186}\, L\, T^{ia}\,\Gamma^{b}+c_{187}\, L\, T^{ab}\,\Gamma^{b}+c_{188}\, L\,\Gamma^{i}\,\Gamma^{a}\,\Gamma^{b}
\nonu\\
&&+c_{189}\, G^{i}\, G^{a}\,\Gamma^{b}+c_{190}\, G^{a}\, G^{b}\,\Gamma^{i}+c_{191}\,\partial G^{i}\, T^{ab}+c_{192}\,\partial G^{a}\, T^{bi}+c_{193}\,\Big(G^{j}\, T^{ij}\, T^{ab}+G^{j}\, T^{ia}\, T^{bj}
\nonu\\
&&+G^{j}\, T^{ib}\, T^{ja}\Big)+c_{194}\, G^{i}\,\partial T^{ab}+c_{195}\, G^{a}\,\partial T^{bi}+c_{196}\, G^{i}\, T^{ab}\, U+c_{197}\, G^{a}\, T^{bi}\, U+c_{198}\,\partial G^{i}\,\Gamma^{a}\,\Gamma^{b}
\nonu\\
&&+c_{199}\,\partial G^{a}\,\Gamma^{b}\,\Gamma^{i}+c_{200}\, G^{i}\,\Gamma^{a}\,\partial\Gamma^{b}+c_{201}\, G^{a}\,\Gamma^{b}\,\partial\Gamma^{i}+c_{202}\, G^{a}\,\partial\Gamma^{b}\,\Gamma^{i}
+c_{203}\,(\frac{1}{2}\,G^{i}\, U\,\Gamma^{a}\,\Gamma^{b}
\nonu\\
&&+G^{a}\, U\,\Gamma^{b}\,\Gamma^{i})
+c_{204}\, T^{qj}\,\Big(T^{ij}\, T^{ab}+T^{ia}\, T^{bj}+T^{ib}\, T^{ja}\Big)\,\Gamma^{q}+c_{205}\,\partial T^{ij}\, T^{ja}\,\Gamma^{b}
+c_{206}\,\partial T^{ij}\, T^{ab}\,\Gamma^{j}
\nonu\\
&&+c_{207}\,\partial T^{ia}\, T^{jb}\,\Gamma^{j}+c_{208}\,\partial T^{ia}\, T^{ab}\,\Gamma^{a}+c_{209}\,\partial T^{ja}\, T^{jb}\,\Gamma^{i}+c_{210}\, T^{ij}\,\partial T^{ja}\,\Gamma^{b}
+c_{211}\, T^{ij}\,\partial T^{ab}\,\Gamma^{j}
\nonu\\
&&+c_{212}\, T^{ia}\,\partial T^{jb}\,\Gamma^{j}+c_{213}\, T^{ia}\,\partial T^{ib}\,\Gamma^{i}+(1+\varepsilon^{ijab})(c_{214}\, T^{ia}\,\partial T^{ab}\,\Gamma^{a}+c_{215}\, T^{ib}\,\partial T^{ab}\,\Gamma^{b})
\nonu\\
&&+c_{216}\, T^{ia}\, T^{bj}\,\partial\Gamma^{j}+c_{217}\, T^{ij}\, T^{ja}\partial\Gamma^{b}+c_{218}\, T^{ij}\, T^{ab}\,\partial\Gamma^{j}+c_{219}\, T^{ij}\,\Gamma^{j}\,\Gamma^{a}\,\partial\Gamma^{b}
\nonu\\
&&+c_{220}\,\Big(\widetilde{T}^{bj}\,(\partial\Gamma^{i}\,\Gamma^{i}+\partial\Gamma^{a}\,\Gamma^{a})\,\Gamma^{b}+T^{ab}\,\partial \Gamma^{a}\,\Gamma^{a}\,\Gamma^{b} +T^{ja}\,\partial(\Gamma^{b}\,\Gamma^{i})\,\Gamma^{j}\Big)+c_{221}\,(\frac{1}{2}\,\partial T^{ji}\,\Gamma^{j}\,\Gamma^{a}\,\Gamma^{b}
\nonu\\
&&+\partial T^{ja}\,\Gamma^{j}\,\Gamma^{i}\,\Gamma^{b})
+c_{222}\,(T^{ia}\,U\,U\,\Gamma^{b}+\frac{1}{2}\,T^{ab}\,U\,U\,\Gamma^{i})+c_{223}\,\partial^{2}T^{ia}\,\Gamma^{b}+c_{224}\,\partial^{2}T^{ab}\,\Gamma^{i}
\nonu\\
&&+c_{225}\, T^{ia}\,\partial^{2}\Gamma^{b}+c_{226}\, T^{ab}\,\partial^{2}\Gamma^{i}+c_{227}\,\partial T^{ia}\,\partial\Gamma^{b}+c_{228}\,\partial T^{ab}\,\partial\Gamma^{i}+c_{229}\, T^{ia}\, U\,\partial\Gamma^{b}+c_{230}\, T^{ab}\, U\,\partial\Gamma^{i}
\nonu\\
&&+c_{231}\, T^{ia}\,\partial U\,\Gamma^{b}+c_{232}\, T^{ab}\,\partial U\,\Gamma^{i}+c_{233}\,\partial T^{ia}\, U\,\Gamma^{b}+c_{234}\,\partial T^{ab}\, U\,\Gamma^{i}
+c_{235}\,\partial U\,\Gamma^{i}\,\Gamma^{a}\,\Gamma^{b}
\nonu\\
&&+c_{236}\, U\,\partial(\Gamma^{i}\,\Gamma^{a}\,\Gamma^{b})+c_{237}\,\partial^{2}\Gamma^{i}\,\Gamma^{a}\,\Gamma^{b}+c_{238}\,(\Gamma^{i}\,\partial^{2}\Gamma^{a}\,\Gamma^{b}+\Gamma^{i}\,\Gamma^{a}\,\partial^{2}\Gamma^{b})+c_{239}\,\Gamma^{i}\,\partial\Gamma^{a}\,\partial\Gamma^{b}
\nonu\\
&&+c_{240}\,\partial\Gamma^{i}\,\partial(\Gamma^{a}\,\Gamma^{b})\Bigg)\Bigg\}
+\varepsilon^{ijkl}\,\Bigg\{  
c_{241}\, {\bf \partial\Phi_{\frac{1}{2}}^{(2),l}}+
c_{242}\, {\bf \partial(\Phi_{0}^{(1)}\,\Phi_{\frac{1}{2}}^{(1),l})}+
c_{243}\,(L\, T^{il}\,\Gamma^{i}-L\, T^{jl}\,\Gamma^{j})   
\nonu\\
&&+c_{244}\,\partial L\,\Gamma^{l}+c_{245}\, L\,\partial\Gamma^{l}+c_{246}\,(G^{i}\, G^{l}\,\Gamma^{i}+G^{j}\, G^{l}\,\Gamma^{j})+c_{247}\,\Big(\partial(G^{i}\, T^{il})+\partial(G^{j}\, T^{jl})\Big)
\nonu\\
&&+c_{248}\,\partial(G^{k}\, T^{kl})+c_{249}\,(G^{i}\, T^{il}+G^{j}\, T^{jl})\, U+c_{250}\,\partial G^{l}\, U+c_{251}\, G^{l}\,\partial U+c_{252}\,(G^{i}\,\Gamma^{i}+G^{j}\,\Gamma^{j})\,\partial\Gamma^{l}
\nonu\\
&&+c_{253}\,(G^{i}\,\partial\Gamma^{i}+G^{j}\,\partial\Gamma^{j})\,\Gamma^{l}+c_{254}\, G^{k}\,\partial(\Gamma^{k}\,\Gamma^{l})+c_{255}\, G^{l}\,(\partial\Gamma^{i}\,\Gamma^{i}+\partial\Gamma^{j}\,\Gamma^{j})+c_{256}\,\partial U\, U\,\Gamma^{l}
\nonu\\
&&+c_{257}\, U\, U\,\partial\Gamma^{l}+c_{258}\,\partial^{2}U\,\Gamma^{l}+c_{259}\,\partial U\,\partial\Gamma^{l}+c_{260}\, U\,\partial^{2}\Gamma^{l}+c_{261}\,\partial^{2}G^{l}+c_{262}\,(\partial G^{i}\,\Gamma^{i}+\partial G^{j}\,\Gamma^{j})\,\Gamma^{l}
\nonu\\
&&+c_{263}\,\partial G^{k}\,\Gamma^{k}\,\Gamma^{l}+c_{264}\,\Big(\partial(T^{ij}\,\Gamma^{i}\,\Gamma^{j}\,\Gamma^{l})+\partial(T^{ik}\,\Gamma^{i}\,\Gamma^{k}\,\Gamma^{l})+\partial(T^{jk}\,\Gamma^{j}\,\Gamma^{k}\,\Gamma^{l})\Big)
\nonu\\
&&+c_{265}\,(\partial T^{il}\, U\,\Gamma^{i}+\partial T^{jl}\, U\,\Gamma^{j})+c_{266}\,(T^{il}\,\partial U\,\Gamma^{i}+T^{jl}\,\partial U\,\Gamma^{j})+c_{267}\,(T^{il}\,\partial U\,\Gamma^{i}+T^{jl}\,\partial U\,\Gamma^{j})
\nonu\\
&&+c_{268}\,\partial(T^{kl}\, U\,\Gamma^{k})+c_{269}\,(T^{ij}\,\partial T^{il}\,\Gamma^{j}+T^{ij}\,\partial T^{jl}\,\Gamma^{i})
+c_{270}\,\Big(T^{ik}\,\partial(T^{il}\,\Gamma^{k})+T^{jk}\,\partial(T^{jl}\,\Gamma^{k})
\nonu\\
&&-(T^{ik}\,T^{ik}+T^{jk}\,T^{jk})\,\partial\Gamma^{l}-T^{kl}\,\Big[\partial(T^{ik}\,\Gamma^{i}+T^{jk}\,\Gamma^{j})\Big]+\partial T^{ij}\,(T^{il}\,\Gamma^{j}-T^{jl}\,\Gamma^{i})\Big)
\nonu\\
&&+c_{271}\,(T^{ik}\,\partial T^{kl}\,\Gamma^{i}+T^{jk}\,\partial T^{kl}\,\Gamma^{j})+c_{272}\,\Big(T^{ij}\,\widetilde{T}^{ka}\,\Gamma^{a}+(\partial T^{ik}\, T^{il}+\partial T^{jk}\, T^{jl})\,\Gamma^{k}\Big)
\nonu\\
&&+c_{273}\,\partial T^{ij}\, T^{ij}\,\Gamma^{l}+c_{274}\,(\partial T^{ik}\, T^{ik}+\partial T^{jk}\, T^{jk})\,\Gamma^{l}+c_{275}\,(\partial T^{il}\, T^{il}+\partial T^{jl}\, T^{jl})\,\Gamma^{l}
\nonu\\
&&+c_{276}\,(\partial T^{il}\,\partial\Gamma^{i}+\partial T^{jl}\,\partial\Gamma^{j})+c_{277}\,\partial T^{kl}\,\partial\Gamma^{k}+c_{278}\,(\partial^{2}\Gamma^{i}\,\Gamma^{i}+\partial^{2}\Gamma^{j}\,\Gamma^{j})\,\Gamma^{l}+c_{279}\,\partial^{2}\Gamma^{k}\,\Gamma^{k}\,\Gamma^{l}
\nonu\\
&&+c_{280}\,(\partial\Gamma^{i}\,\Gamma^{i}+\partial\Gamma^{j}\,\Gamma^{j})\,\Gamma^{l}+c_{281}\,\partial\Gamma^{k}\,\Gamma^{k}\,\Gamma^{l}+c_{282}\,(\partial^{2}T^{il}\,\Gamma^{i}+\partial^{2}T^{jl}\,\Gamma^{j})+c_{283}\,\partial^{2}T^{kl}\,\Gamma^{k}
\nonu\\
&&+c_{284}\,(T^{il}\,\partial^{2}\Gamma^{i}+T^{jl}\,\partial^{2}\Gamma^{j})+c_{285}\, T^{kl}\,\partial^{2}\Gamma^{k}+c_{286}\,\partial^{3}\Gamma^{l}\Bigg\}+(\varepsilon^{ijkl})^{2}\,\Bigg\{ c_{287}\,(\partial G^{i}\, T^{jk}-\partial G^{j}\, T^{ik})
\nonu\\
&&+c_{288}\,\partial G^{k}\, T^{ij}+c_{289}\,(G^{i}\,\partial T^{jk}-G^{j}\,\partial T^{ik})+c_{290}\, G^{k}\,\partial T^{ij}+c_{291}\, G^{l}\,(T^{ik}\, T^{lj}+T^{il}\, T^{jk})
\nonu\\
&&+c_{292}\,(G^{i}\, T^{ij}\, T^{ik}+G^{j}\, T^{ij}\, T^{jk})+c_{293}\,(G^{i}\,\Gamma^{j}-G^{j}\,\Gamma^{i})\,\partial\Gamma^{k}+c_{294}\,(G^{i}\,\partial\Gamma^{j}-G^{j}\,\partial\Gamma^{i})\,\Gamma^{k}
\nonu\\
&&+c_{295}\, G^{k}\,\partial(\Gamma^{i}\,\Gamma^{j})+c_{296}\,\partial(U\,\Gamma^{i}\,\Gamma^{j}\,\Gamma^{k})+c_{297}\,(\partial G^{i}\,\Gamma^{j}-\partial G^{j}\,\Gamma^{i})\,\Gamma^{k}+c_{298}\,\partial G^{k}\,\Gamma^{i}\,\Gamma^{j}
\nonu\\
&&+c_{299}\,\partial T^{ij}\, U\,\Gamma^{k}+c_{300}\, T^{ij}\,\partial U\,\Gamma^{k}+c_{301}\, T^{ij}\, U\,\partial\Gamma^{k}+c_{302}\,(\partial T^{ki}\, U\,\Gamma^{j}+\partial T^{kj}\, U\,\Gamma^{i})
\nonu\\
&&+c_{303}\,\Big(T^{ki}\,\partial(U\,\Gamma^{j})+T^{kj}\,\partial(U\,\Gamma^{i})+c_{304}\,\Big(\partial(T^{ij}\, T^{ik})\,\Gamma^{i}+\partial(T^{ij}\, T^{jk})\,\Gamma^{j}\Big)+c_{305}\,\partial(T^{ij}\, T^{kl}\,\Gamma^{l})
\nonu\\
&&+c_{306}\,\Big(\partial(T^{ik}\, T^{jl})+\partial(T^{il}\, T^{jk})\Big)\,\Gamma^{l}+c_{307}\,\Big(T^{ik}\, T^{jl}+T^{il}\, T^{jk}\Big)\,\partial\Gamma^{l}+c_{308}\, T^{ij}\,\Big(T^{ik}\,\partial\Gamma^{i}+T^{jk}\,\partial\Gamma^{j}\Big)
\nonu\\
&&+c_{309}\,\partial T^{ij}\,\partial\Gamma^{k}+c_{310}\,\Big(\partial T^{ik}\,\partial\Gamma^{j}+\partial T^{jk}\,\partial\Gamma^{i}\Big)+c_{311}\,\Gamma^{i}\,\Gamma^{j}\,\partial^{2}\Gamma^{k}+c_{312}\,\Big(\partial^{2}\Gamma^{i}\,\Gamma^{j}+\Gamma^{i}\,\partial^{2}\Gamma^{j}\Big)\,\Gamma^{k}
\nonu\\
&&+c_{313}\,\partial(\Gamma^{i}\,\Gamma^{j})\,\partial\Gamma^{k}+c_{314}\,\partial\Gamma^{i}\,\partial\Gamma^{j}\,\Gamma^{k}+c_{315}\,\partial^{2}T^{ij}\,\Gamma^{k}+c_{316}\,(\partial^{2}T^{ik}\,\Gamma^{j}-\partial^{2}T^{jk}\,\Gamma^{i})
\nonu\\
&&+c_{317}\, T^{ij}\,\partial^{2}\Gamma^{k}
+c_{318}\,(T^{ik}\,\partial^{2}\Gamma^{j}-T^{jk}\,\partial^{2}\Gamma^{i})\Bigg\}\Bigg](w)-\delta^{ik}\,
\sum_{n=4}^1 \, \frac{1}{(z-w)^n} \,
\left(j\;\leftrightarrow \;k\right)(w)
+\cdots,
\nonu
\eea
where the coefficients
are
\bea
c_{1}&=&\frac{8i\, k\, N(k-N)}{(2+k+N)^{3}},\qquad
c_{2}=-\frac{24i\, k\, N}{(2+k+N)^{2}},\nonu\\
c_ {3} & = &\frac {8 (3 + 9 k + 4 k^{2} + 9 N + 10 k\, 
     N + 4 N^{2})} {3 (2 + k + N)^{2}}, \nonu\\
c_ {4} & = & - \frac {16 i (3 + 9 k + 4 k^{2} + 9 N + 7 k\, 
      N + 4 N^{2})} {3 (2 + k + N)^{3}}, \nonu\\
c_ {5} & = &\frac {8 i (k - N) (24 + 13 k + 
      13 N)} {3 (2 + k + N)^{3}}, \qquad
c_ {6}  = \frac {8 (k - N) (12 + 7 k + 7 N)} {3 (2 + k + N)^{3}}, \nonu\\
c_ {7} & = &\frac {8 i} {(2 + k + N)^{2}}, \qquad
c_ {8}  =  - \frac {8 (6 + 12 k + 5 k^{2} + 12 N + 8 k\, 
      N + 5 N^{2})} {3 (2 + k + N)^{3}}, \nonu\\
c_ {9} & = & - \frac {4 (6 + 6 k + k^{2} + 6 N + 4 k\, 
      N + N^{2})} {3 (2 + k + N)^{3}}, \qquad
c_ {10}  =  - \frac {8 (k - N)} {(2 + k + N)^{2}}, \nonu\\
c_ {11} & = &\frac {8 (k - N)} {3 (2 + k + N)^{2}}, \qquad
c_ {12}  =  - \frac {16 i (k - N)} {3 (2 + k + N)^{3}}, \nonu\\
c_ {13} & = &\frac {8 i (6 k + k^{2} + 6 N + 10 k\, 
     N + N^{2})} {3 (2 + k + N)^{3}}, \qquad
c_ {14}  = \frac {8 (k + N)} {(2 + k + N)^{2}}, \nonu \\
c_ {15}  & = &   - \frac {8 (k - N)^{2}} {3 (2 + k + N)^{3}}, \qquad
c_ {16}  =  - \frac {8 (k - N)} {3 (2 + k + N)^{2}}, \qquad
c_ {17}  = \frac {8 (k - N) (4 + 3 k + 3 N)} {3 (2 + k + N)^{3}}, \nonu\\
c_ {18} & = & - \frac {16 i (k - N)} {3 (2 + k + N)^{3}}, \qquad
c_ {19}  = \frac {(k - N)} {6 (2 + k + N)}, \nonu\\
c_ {20} & = & - \frac {(k - N) (60 + 77 k + 22 k^{2} + 121 N + 
       115 k\, N + 20 k^{2} N + 79 N^{2} + 42 k\, 
      N^{2} + 16 N^{3})} {6 (2 + N) (2 + k + N)^{3}}, \nonu\\
c_ {21} & = &\frac {16 i (k - N)} {3 (2 + k + N)^{2}}, \qquad
c_ {22}  = \frac {2 (k - N)} {(2 + k + N)^{2}}, \nonu\\
c_ {23} & = &\frac {1} {6 (2 + N) (2 + k + N)^{3}}(-192 - 116 k + 69 k^{2} + 38 k^{3} - 172 N + 
     172 k\, N + 209 k^{2} N 
\nonu\\
&+& 28 k^{3} N + 47 N^{2} + 256 k\, 
    N^{2} + 82 k^{2} N^{2} + 73 N^{3} + 66 k\, 
    N^{3} + 16 N^{4}), \nonu\\
c_ {24} & = &\frac {4 i (k - N)} {3 (2 + k + N)^{3}}, \nonu\\
c_ {25} & = & - \frac {2 i (48 + 82 k + 57 k^{2} + 14 k^{3} + 62 N + 
       53 k\, N + 17 k^{2} N + 34 N^{2} + 9 k\, 
      N^{2} + 8 N^{3})} {3 (2 + k + N)^{4}}, \nonu\\
c_ {26} & = & - \frac{1} {3 (2 + N) (2 + k + N)^{4}}i (-192 - 20 k + 117 k^{2} + 38 k^{3} - 76 N + 
       316 k\, N + 233 k^{2} N 
\nonu\\
&+& 28 k^{3} N + 143 N^{2} + 304 k\, 
      N^{2} + 82 k^{2} N^{2} + 97 N^{3} + 66 k\, 
      N^{3} + 16 N^{4}), \nonu\\
c_ {27} & = &\frac {4 i (k - N) (13 + 8 k + 8 N)} {3 (2 + k + N)^{3}},\nonu\\
 c_ {28} & = & - \frac {i (192 + 308 k + 117 k^{2} + 6 k^{3} + 
       364 N + 340 k\, N + 55 k^{2} N + 215 N^{2} + 90 k\, 
      N^{2} + 41 N^{3})} {6 (2 + N) (2 + k + N)^{3}}, \nonu\\
c_ {29} & = & - \frac {(192 + 212 k + 69 k^{2} + 6 k^{3} + 268 N + 
      196 k\, N + 31 k^{2} N + 119 N^{2} + 42 k\, 
     N^{2} + 17 N^{3})} {3 (2 + N) (2 + k + N)^{4}}, \nonu\\
c_ {30} & = & - \frac {(192 + 212 k + 69 k^{2} + 6 k^{3} + 268 N + 
      196 k\, N + 31 k^{2} N + 119 N^{2} + 42 k\, 
     N^{2} + 17 N^{3})} {3 (2 + N) (2 + k + N)^{4}}, \nonu\\
c_ {31} & = &\frac {8 (k - N) (3 + k + N)} {3 (2 + k + N)^{3}}, \nonu\\
c_ {32} & = &\frac {2 (k - N) (64 + 67 k + 18 k^{2} + 109 N + 79 k\, 
     N + 12 k^{2} N + 63 N^{2} + 24 k\, 
     N^{2} + 12 N^{3})} {3 (2 + N) (2 + k + N)^{4}}, \nonu\\
c_ {33} & = &\frac {4 i (k - N) (32 + 19 k + 2 k^{2} + 45 N + 23 k\, 
     N + 4 k^{2} N + 23 N^{2} + 8 k\, 
     N^{2} + 4 N^{3})} {3 (2 + N) (2 + k + N)^{5}}, \nonu\\
c_ {34} & = &\frac {(96 + 116 k + 45 k^{2} + 6 k^{3} + 124 N + 
     100 k\, N + 19 k^{2} N + 47 N^{2} + 18 k\, 
    N^{2} + 5 N^{3})} {3 (2 + N) (2 + k + N)^{4}}, \nonu\\
c_ {35} & = &\frac {i (96 + 116 k + 45 k^{2} + 6 k^{3} + 124 N + 
      100 k\, N + 19 k^{2} N + 47 N^{2} + 18 k\, 
     N^{2} + 5 N^{3})} {6 (2 + N) (2 + k + N)^{3}}, \nonu\\
c_ {36} & = &\frac {4 i (k - N) (16 + 3 k - 2 k^{2} + 21 N + 7 k\, 
     N + 2 k^{2} N + 11 N^{2} + 4 k\, 
     N^{2} + 2 N^{3})} {3 (2 + N) (2 + k + N)^{5}}, \nonu\\
c_ {37} & = & - \frac {4 i (k - N)} {3 (2 + k + N)^{3}}, \qquad
c_ {38}  =  - \frac {i (k - N) (13 k + 6 k^{2} + 3 N + 9 k\, 
      N + N^{2})} {3 (2 + N) (2 + k + N)^{4}}, \nonu\\
c_ {39} & = & - \frac { (k - N) (32 + 55 k + 18 k^{2} + 41 N + 
       35 k\, N + 11 N^{2})} {3 (2 + N) (2 + k + N)^{5}}, \nonu\\
c_ {40} & = &\frac {i (k - N) (16 + 21 k + 6 k^{2} + 19 N + 13 k\, 
     N + 5 N^{2})} {3 (2 + N) (2 + k + N)^{4}}, \nonu\\
c_ {41} & = &\frac {2 (k - N) (32 + 55 k + 18 k^{2} + 41 N + 35 k\, 
     N + 11 N^{2})} {3 (2 + N) (2 + k + N)^{5}}, \nonu\\
c_ {42} & = &\frac {i (k - N) (13 k + 6 k^{2} + 3 N + 9 k\, 
     N + N^{2})} {3 (2 + N) (2 + k + N)^{4}}, \nonu\\
c_ {43} & = &\frac {(96 + 116 k + 45 k^{2} + 6 k^{3} + 124 N + 
     100 k\, N + 19 k^{2} N + 47 N^{2} + 18 k\, 
    N^{2} + 5 N^{3})} {3 (2 + N) (2 + k + N)^{4}}, \nonu\\
c_ {44} & = &\frac {4 (k - N)} {(2 + k + N)^{3}}, \qquad
c_ {45}  =  - \frac {4 (k - N)} {(2 + k + N)^{2}}, \nonu\\
c_ {46} & = &\frac { i (k - N) (32 + 55 k + 18 k^{2} + 41 N + 35 k\, 
     N + 11 N^{2})} {3 (2 + N) (2 + k + N)^{5}}, \nonu\\
c_ {47} & = & - \frac {i (k - N) (48 + 37 k + 6 k^{2} + 51 N + 21 k\, 
      N + 13 N^{2})} {12 (2 + N) (2 + k + N)^{3}}, \nonu\\
c_ {48} & = & - \frac {(k - N) (13 k + 6 k^{2} + 3 N + 9 k\, 
      N + N^{2})} {6 (2 + N) (2 + k + N)^{4}}, \nonu\\
c_ {49} & = &\frac{1} {3 (2 + N) (2 + k + N)^{5}}2 i (96 + 164 k + 95 k^{2} + 18 k^{3} + 172 N + 
      236 k\, N + 103 k^{2} N 
\nonu\\
&+& 12 k^{3} N + 101 N^{2} + 100 k\, 
     N^{2} + 26 k^{2} N^{2} + 19 N^{3} + 10 k\, 
     N^{3}), \nonu\\
c_ {50} & = &\frac {1}{3 (2 + 
      N) (2 + k + N)^{5}}4 i (48 + 92 k + 63 k^{2} + 14 k^{3} + 76 N + 
      120 k\, N + 67 k^{2} N + 10 k^{3} N 
\nonu\\
&+& 33 N^{2} + 40 k\, 
     N^{2} + 16 k^{2} N^{2} - N^{3} - 2 N^{4}), \nonu\\
c_ {51} & = &\frac {i (k - N) (-32 - 3 k + 6 k^{2} - 29 N + k\, 
     N - 7 N^{2})} {6 (2 + N) (2 + k + N)^{3}}, \nonu\\
c_ {52} & = &\frac {(k - N) (16 + 21 k + 6 k^{2} + 19 N + 13 k\, 
     N + 5 N^{2})} {3 (2 + N) (2 + k + N)^{4}}, \nonu\\
c_ {53} & = & - \frac {i (192 + 212 k + 69 k^{2} + 6 k^{3} + 268 N + 
       196 k\, N + 31 k^{2} N + 119 N^{2} + 42 k\, 
      N^{2} + 17 N^{3})} {6 (2 + N) (2 + k + N)^{4}}, \nonu\\
c_ {54} & = & - \frac {(k - N) (16 + 21 k + 6 k^{2} + 19 N + 13 k\, 
      N + 5 N^{2})} {3 (2 + N) (2 + k + N)^{4}}, \nonu\\
c_ {55} & = &\frac {i (96 + 116 k + 45 k^{2} + 6 k^{3} + 124 N + 
      100 k\, N + 19 k^{2} N + 47 N^{2} + 18 k\, 
     N^{2} + 5 N^{3})} {3 (2 + N) (2 + k + N)^{4}}, \nonu\\
c_ {56} & = & - \frac {4 i} {(2 + k + N)^{2}}, \nonu\\
c_ {57} & = & - \frac {2 (24 + 82 k + 63 k^{2} + 14 k^{3} + 62 N + 
       113 k\, N + 41 k^{2} N + 40 N^{2} + 33 k\, 
      N^{2} + 8 N^{3})} {3 (2 + k + N)^{4}}, \nonu\\
c_ {58} & = & - \frac {(k - N) (13 k + 6 k^{2} + 3 N + 9 k\, 
      N + N^{2})} {6 (2 + N) (2 + k + N)^{4}}, \nonu\\
c_ {59} & = & - \frac {(96 + 130 k + 61 k^{2} + 10 k^{3} + 110 N + 
       93 k\, N + 21 k^{2} N + 38 N^{2} + 13 k\, 
      N^{2} + 4 N^{3})} {3 (2 + k + N)^{4}}, \nonu\\
c_ {60} & = & - \frac {1}{3 (2 + N) (2 + k + N)^{4}} (-48 - 48 k - 15 k^{2} - 2 k^{3} - 72 N - 
       32 k\, N + 6 k^{2} N + 2 k^{3} N 
\nonu\\
&-& 49 N^{2} - 7 k\, 
      N^{2} + 5 k^{2} N^{2} - 21 N^{3} - 3 k\, 
      N^{3} - 4 N^{4}), \nonu\\
c_ {61} & = & - \frac {1}{3 (2 + N) (2 + k + N)^{4}}2 (-48 - 24 k + 21 k^{2} + 10 k^{3} - 48 N + 
       4 k\, N + 36 k^{2} N + 8 k^{3} N 
\nonu\\
&-& N^{2} + 17 k\, 
      N^{2} + 11 k^{2} N^{2} + 9 N^{3} + 3 k\, 
      N^{3} + 2 N^{4}), \nonu\\
c_ {62} & = & - \frac {5} {2}, \nonu\\
c_ {63} & = &\frac {5 (60 + 77 k + 22 k^{2} + 121 N + 115 k\, 
     N + 20 k^{2} N + 79 N^{2} + 42 k\, 
     N^{2} + 16 N^{3})} {2 (2 + N) (2 + k + N)^{2}}, \nonu\\
c_ {64} & = & - \frac {16 i (1 + k + N)} {(2 + k + N)^{2}}, \nonu\\
c_ {65} & = & - \frac {(-300 - 309 k - 70 k^{2} - 81 N + 153 k\, 
     N + 100 k^{2} N + 217 N^{2} + 210 k\, 
     N^{2} + 80 N^{3})} {6 (2 + N) (2 + k + N)^{2}}, \nonu\\
c_ {66} & = & - \frac {2 (2 + 3 k + 3 N)} {(2 + k + N)^{2}}, \qquad
c_ {67}  =  - \frac {4 i} {(2 + k + N)^{2}}, \nonu\\
c_ {68} & = &\frac {10 i (30 + 31 k + 10 k^{2} + 38 N + 21 k\, 
     N + 8 N^{2})} {3 (2 + k + N)^{3}}, \nonu\\
c_ {69} & = &\frac {i (-300 - 213 k - 70 k^{2} - 177 N + 201 k\, 
     N + 100 k^{2} N + 169 N^{2} + 210 k\, 
     N^{2} + 80 N^{3})} {3 (2 + N) (2 + k + N)^{3}}, \nonu\\
c_ {70} & = &\frac {4 i (18 + 27 k + 8 k^{2} + 27 N + 20 k\, 
     N + 8 N^{2})} {(2 + k + N)^{3}}, \nonu\\
c_ {71} & = &\frac {5 i (60 + 71 k + 18 k^{2} + 67 N + 43 k\, 
     N + 17 N^{2})} {6 (2 + N) (2 + k + N)^{2}}, \nonu\\
c_ {72} & = &\frac {5 i (20 + 21 k + 6 k^{2} + 25 N + 13 k\, 
     N + 7 N^{2})} {2 (2 + N) (2 + k + N)^{2}}, \nonu\\
c_ {73} & = &\frac {(300 + 307 k + 90 k^{2} + 383 N + 191 k\, 
    N + 109 N^{2})} {3 (2 + N) (2 + k + N)^{3}}, \nonu\\
c_ {74} & = &\frac {5 (20 + 21 k + 6 k^{2} + 25 N + 13 k\, 
     N + 7 N^{2})} {(2 + N) (2 + k + N)^{3}}, \nonu\\
c_ {75} & = &\frac {2 (48 + 101 k + 34 k^{2} + 75 N + 79 k\, 
     N + 2 k^{2} N + 27 N^{2} + 8 k\, 
     N^{2} + 2 N^{3})} {(2 + N) (2 + k + N)^{3}}, \nonu\\
c_ {76} & = &\frac {2 (48 + 105 k + 38 k^{2} + 79 N + 81 k\, 
     N + 4 k^{2} N + 33 N^{2} + 8 k\, 
     N^{2} + 4 N^{3})} {(2 + N) (2 + k + N)^{3}}, \nonu\\
c_ {77} & = & - \frac {4 (-6 + 15 k + 8 k^{2} + 15 N + 20 k\, 
      N + 8 N^{2})} {3 (2 + k + N)^{3}}, \qquad
c_ {78} =  - \frac {8 (1 + k + N)} {(2 + k + N)^{2}}, \nonu\\
c_ {79} & = &\frac {(300 + 307 k + 90 k^{2} + 383 N + 191 k\, 
    N + 109 N^{2})} {3 (2 + N) (2 + k + N)^{3}}, \nonu\\
c_ {80} & = &\frac {5 (20 + 21 k + 6 k^{2} + 25 N + 13 k\, 
     N + 7 N^{2})} {(2 + N) (2 + k + N)^{3}}, \nonu\\
c_ {81} & = &\frac {i (48 + 89 k + 30 k^{2} + 63 N + 57 k\, 
     N + 17 N^{2})} {(2 + N) (2 + k + N)^{3}}, \nonu\\
c_ {82} & = &\frac {10 (32 + 55 k + 18 k^{2} + 41 N + 35 k\, 
     N + 11 N^{2})} {(2 + N) (2 + k + N)^{4}}, \nonu\\
c_ {83} & = &\frac {i (48 + 89 k + 30 k^{2} + 63 N + 57 k\, 
     N + 17 N^{2})} {(2 + N) (2 + k + N)^{3}}, \nonu\\
c_ {84} & = &\frac {i (32 + 81 k + 30 k^{2} + 47 N + 53 k\, 
     N + 13 N^{2})} {(2 + N) (2 + k + N)^{3}}, \nonu\\
c_ {85} & = & - \frac {i (32 + 81 k + 30 k^{2} + 47 N + 53 k\, 
      N + 13 N^{2})} {(2 + N) (2 + k + N)^{3}}, \qquad
c_ {86}  = \frac {4 i} {(2 + k + N)^{2}}, \nonu\\
c_ {87} & = &\frac {4 i (264 + 459 k + 158 k^{2} + 441 N + 395 k\, 
     N + 34 k^{2} N + 215 N^{2} + 64 k\, 
     N^{2} + 34 N^{3})} {3 (2 + N) (2 + k + N)^{4}}, \nonu\\
c_ {88} & = &\frac {4 i (96 + 161 k + 54 k^{2} + 159 N + 141 k\, 
     N + 12 k^{2} N + 77 N^{2} + 24 k\, 
     N^{2} + 12 N^{3})} {(2 + N) (2 + k + N)^{4}}, \nonu\\
c_ {89} & = &\frac {(48 + 89 k + 30 k^{2} + 63 N + 57 k\, 
    N + 17 N^{2})} {(2 + N) (2 + k + N)^{3}}, \nonu\\
c_ {90} & = &\frac {(32 + 81 k + 30 k^{2} + 47 N + 53 k\, 
    N + 13 N^{2})} {(2 + N) (2 + k + N)^{3}}, \nonu\\
c_ {91} & = &\frac {i (32 + 81 k + 30 k^{2} + 47 N + 53 k\, 
     N + 13 N^{2})} {2 (2 + N) (2 + k + N)^{2}}, \nonu\\
c_ {92} & = &\frac {i (16 + 57 k + 30 k^{2} + 15 N + 41 k\, 
     N + N^{2})} {2 (2 + N) (2 + k + N)^{2}}, \nonu\\
c_ {93} & = &\frac {2 (63 k + 22 k^{2} + 87 N + 204 k\, 
     N + 56 k^{2} N + 119 N^{2} + 105 k\, 
     N^{2} + 34 N^{3})} {3 (2 + N) (2 + k + N)^{3}}, \nonu\\
c_ {94} & = &\frac {(32 + 81 k + 30 k^{2} + 47 N + 53 k\, 
    N + 13 N^{2})} {(2 + N) (2 + k + N)^{3}}, \nonu\\
c_ {95} & = &\frac {(48 + 89 k + 30 k^{2} + 63 N + 57 k\, 
    N + 17 N^{2})} {(2 + N) (2 + k + N)^{3}}, \qquad
c_ {96}  =  - \frac {4 i (k - N)} {3 (2 + k + N)^{3}}, \nonu\\
c_ {97} & = & - \frac {10 i (32 + 55 k + 18 k^{2} + 41 N + 35 k\, 
      N + 11 N^{2})} {(2 + N) (2 + k + N)^{4}}, \nonu\\
c_ {98} & = &\frac {2 (71 k + 34 k^{2} + 79 N + 208 k\, 
     N + 62 k^{2} N + 103 N^{2} + 105 k\, 
     N^{2} + 28 N^{3})} {3 (2 + N) (2 + k + N)^{3}}, \nonu\\
c_ {99} & = &\frac {2 (150 + 159 k + 38 k^{2} + 186 N + 105 k\, 
     N + 52 N^{2})} {3 (2 + k + N)^{3}}, \nonu\\
c_ {100} & = &\frac {2 (150 + 141 k + 34 k^{2} + 204 N + 105 k\, 
     N + 56 N^{2})} {3 (2 + k + N)^{3}}, \nonu\\
c_ {101} & = & - \frac {20 i (60 + 77 k + 22 k^{2} + 121 N + 115 k\, 
      N + 20 k^{2} N + 79 N^{2} + 42 k\, 
      N^{2} + 16 N^{3})} {3 (2 + N) (2 + k + N)^{4}}, \nonu\\
c_ {102} & = & - \frac {4 i (100 + 131 k + 38 k^{2} + 199 N + 193 k\, 
      N + 34 k^{2} N + 129 N^{2} + 70 k\, 
      N^{2} + 26 N^{3})} {(2 + N) (2 + k + N)^{4}}, \nonu\\
c_ {103} & = &\frac {5 i (20 + 21 k + 6 k^{2} + 25 N + 13 k\, 
     N + 7 N^{2})} {(2 + N) (2 + k + N)^{3}}, \nonu\\
c_ {104} & = &\frac {i (300 + 307 k + 90 k^{2} + 383 N + 191 k\, 
     N + 109 N^{2})} {3 (2 + N) (2 + k + N)^{3}}, \qquad
c_ {105}  =  - \frac {1} {2}, \nonu\\
c_ {106} & = & - \frac {(900 + 1139 k + 322 k^{2} + 1831 N + 1717 k\, 
     N + 296 k^{2} N + 1201 N^{2} + 630 k\, 
     N^{2} + 244 N^{3})} {(60 (2 + N) (2 + k + N)^{2})}, \nonu\\
c_ {107} & = &\frac {(60 + 77 k + 22 k^{2} + 121 N + 115 k\, 
    N + 20 k^{2} N + 79 N^{2} + 42 k\, 
    N^{2} + 16 N^{3})} {2 (2 + N) (2 + k + N)^{2}}, \nonu\\
c_ {108} & = & - \frac {(60 + 77 k + 22 k^{2} + 121 N + 115 k\, 
     N + 20 k^{2} N + 79 N^{2} + 42 k\, 
     N^{2} + 16 N^{3})} {2 (2 + N) (2 + k + N)^{2}}, \nonu\\
c_ {109} & = &\frac {(60 + 77 k + 22 k^{2} + 121 N + 115 k\, 
     N + 20 k^{2} N + 79 N^{2} + 42 k\, 
     N^{2} + 16 N^{3})^{2}} {4 (2 + N)^{2} (2 + k + N)^{4}}, \nonu\\
c_ {110} & = &\frac{1} {12 (2 + N)^{2} (2 + k + N)^{4}}(60 + 77 k + 22 k^{2} + 121 N + 115 k\, 
     N + 20 k^{2} N + 79 N^{2} 
\nonu\\
&+& 42 k\, 
     N^{2} + 16 N^{3}) (180 + 223 k + 62 k^{2} + 371 N + 341 k\, 
     N + 58 k^{2} N + 245 N^{2} + 126 k\, 
     N^{2} 
\nonu\\
&+& 50 N^{3}), \nonu\\
c_ {111} & = &\frac {3 (16 + 21 k + 6 k^{2} + 19 N + 13 k\, 
     N + 5 N^{2})} {(2 + N) (2 + k + N)^{2}}, \nonu\\
c_ {112} & = &\frac {2 (20 + 21 k + 6 k^{2} + 25 N + 13 k\, 
     N + 7 N^{2})} {(2 + N) (2 + k + N)^{3}}, \nonu\\
c_ {113} & = & - \frac {2 i (32 + 55 k + 18 k^{2} + 41 N + 35 k\, 
      N + 11 N^{2})} {(2 + N) (2 + k + N)^{3}}, \nonu\\
c_ {114} & = & - \frac {i (420 + 521 k + 142 k^{2} + 925 N + 865 k\, 
      N + 152 k^{2} N + 643 N^{2} + 336 k\, 
      N^{2} + 136 N^{3})} {3 (2 + N) (2 + k + N)^{3}}, \nonu\\
c_ {115} & = & - \frac {2 i (60 + 91 k + 26 k^{2} + 167 N + 191 k\, 
      N + 40 k^{2} N + 137 N^{2} + 84 k\, 
      N^{2} + 32 N^{3})} {3 (2 + N) (2 + k + N)^{3}}, \nonu\\
c_ {116} & = &\frac {i (20 + 21 k + 6 k^{2} + 25 N + 13 k\, 
     N + 7 N^{2})} {(2 + N) (2 + k + N)^{3}}, \nonu\\
c_ {117} & = & - \frac{1} {12 (2 + N)^{2} (2 + k + N)^{4}}i (2064 + 2960 k + 2707 k^{2} + 1476 k^{3} + 
       300 k^{4} + 7984 N 
\nonu\\
&+& 13070 k\, 
      N + 9788 k^{2} N + 3236 k^{3} N + 312 k^{4} N + 11907 N^{2} + 
       16874 k\, 
      N^{2} + 8603 k^{2} N^{2} 
\nonu\\
&+& 1384 k^{3} N^{2} + 8354 N^{3} + 
       8494 k\, N^{3} + 2194 k^{2} N^{3} + 2755 N^{4} + 1466 k\, 
      N^{4} + 344 N^{5}), \nonu\\
c_ {118} & = &\frac {1} {12 (2 + N)^{2} (2 + k + N)^{4}}i (-240 + 3728 k + 5947 k^{2} + 3048 k^{3} + 
      516 k^{4} + 2608 N 
\nonu\\
&+& 14030 k\, 
     N + 13724 k^{2} N + 4538 k^{3} N + 420 k^{4} N + 6363 N^{2} + 
      16550 k\, 
     N^{2} + 9941 k^{2} N^{2} 
\nonu\\
&+& 1642 k^{3} N^{2} + 5282 N^{3} + 
      7864 k\, N^{3} + 2284 k^{2} N^{3} + 1873 N^{4} + 1304 k\, 
     N^{4} + 242 N^{5}), \nonu\\
c_ {119} & = & - \frac {(16 + 21 k + 6 k^{2} + 19 N + 13 k\, 
     N + 5 N^{2})} {(2 + N) (2 + k + N)^{3}}, \nonu\\
c_ {120} & = & - \frac {(32 + 55 k + 18 k^{2} + 41 N + 35 k\, 
     N + 11 N^{2})} {(2 + N) (2 + k + N)^{3}}, \nonu\\
c_ {121} & = &\frac {i (20 + 21 k + 6 k^{2} + 25 N + 13 k\, 
     N + 7 N^{2})} {(2 + N) (2 + k + N)^{3}}, \nonu\\
c_ {122} & = & - \frac {4 i (32 + 55 k + 18 k^{2} + 41 N + 35 k\, 
      N + 11 N^{2})} {(2 + N) (2 + k + N)^{4}}, \nonu\\
c_ {123} & = &\frac {2 i (32 + 55 k + 18 k^{2} + 41 N + 35 k\, 
     N + 11 N^{2})} {(2 + N) (2 + k + N)^{4}}, \nonu\\
c_ {124} & = &\frac {3 (16 + 21 k + 6 k^{2} + 19 N + 13 k\, 
     N + 5 N^{2})} {(2 + N) (2 + k + N)^{3}}, \nonu\\
c_ {125} & = &\frac {(-20 - 11 k - 2 k^{2} - 15 N + 15 k\, 
    N + 8 k^{2} N + 7 N^{2} + 14 k\, 
    N^{2} + 4 N^{3})} {(2 + N) (2 + k + N)^{3}}, \nonu\\
c_ {126} & = &\frac {3 (20 + 21 k + 6 k^{2} + 25 N + 13 k\, 
     N + 7 N^{2})} {(2 + N) (2 + k + N)^{3}}, \nonu\\
c_ {127} & = & - \frac {1}{12 (2 + N)^{2} (2 + k + N)^{4}}(624 - 512 k - 2265 k^{2} - 1620 k^{3} - 
      340 k^{4} + 1232 N - 158 k\, 
     N 
\nonu\\
&-& 1216 k^{2} N - 364 k^{3} N + 56 k^{4} N + 3827 N^{2} + 
      8364 k\, 
     N^{2} + 7741 k^{2} N^{2} + 2942 k^{3} N^{2}
\nonu\\
&+& 356 k^{4} N^{2} + 
      7072 N^{3} + 13854 k\, 
     N^{3} + 8560 k^{2} N^{3} + 1562 k^{3} N^{3} + 5749 N^{4} + 
      7614 k\, N^{4} 
\nonu\\
&+& 2336 k^{2} N^{4} + 2068 N^{5} + 1366 k\, 
     N^{5} + 272 N^{6}), \nonu\\
c_ {128} & = &\frac{1} {6 (2 + N)^{2} (2 + k + N)^{5}}(2064 + 4208 k + 3907 k^{2} + 1764 k^{3} + 
     300 k^{4} + 8272 N 
\nonu\\
&+& 15326 k\, 
    N + 11108 k^{2} N + 3380 k^{3} N + 312 k^{4} N + 12291 N^{2} + 
     18170 k\, 
    N^{2} + 8963 k^{2} N^{2} 
\nonu\\
&+& 1384 k^{3} N^{2} + 8522 N^{3} + 
     8734 k\, N^{3} + 2194 k^{2} N^{3} + 2779 N^{4} + 1466 k\, 
    N^{4} + 344 N^{5}), \nonu\\
c_ {129} & = &\frac {1} {6 (2 + N)^{2} (2 + k + N)^{5}}(3600 + 9488 k + 8827 k^{2} + 3528 k^{3} + 
     516 k^{4} + 12208 N 
\nonu\\
&+& 25550 k\, 
    N + 18044 k^{2} N + 5018 k^{3} N + 420 k^{4} N + 15963 N^{2} + 
     25190 k\, 
    N^{2} + 12101 k^{2} N^{2} 
\nonu\\
&+& 1762 k^{3} N^{2} + 10082 N^{3} + 
     10744 k\, N^{3} + 2644 k^{2} N^{3} + 3073 N^{4} + 1664 k\, 
    N^{4} + 362 N^{5}), \nonu\\
c_ {130} & = &\frac {1} {6 (2 + N)^{2} (2 + k + N)^{5}}(2064 + 8144 k + 8539 k^{2} + 3528 k^{3} + 
     516 k^{4} + 7408 N 
\nonu\\
&+& 20846 k\, 
    N + 16268 k^{2} N + 4730 k^{3} N + 420 k^{4} N + 9819 N^{2} + 
     19718 k\, 
    N^{2} + 10421 k^{2} N^{2} 
\nonu\\
&+& 1618 k^{3} N^{2} + 6194 N^{3} + 
     8152 k\, N^{3} + 2212 k^{2} N^{3} + 1873 N^{4} + 1232 k\, 
    N^{4} + 218 N^{5}), \nonu\\
c_ {131} & = & - \frac {3 (64 + 136 k + 61 k^{2} + 6 k^{3} + 88 N + 
       104 k\, N + 15 k^{2} N + 27 N^{2} + 10 k\, 
      N^{2} + N^{3})} {(2 + N) (2 + k + N)^{4}}, \nonu\\
c_ {132} & = & - \frac {(256 + 432 k + 183 k^{2} + 18 k^{3} + 400 N + 
      400 k\, N + 69 k^{2} N + 185 N^{2} + 78 k\, 
     N^{2} + 27 N^{3})} {(2 + N) (2 + k + N)^{4}}, \nonu\\
c_ {133} & = &\frac {1} {(2 + N) (2 + k + N)^{4}} i (200 + 246 k + 
    55 k^{2} - 6 k^{3} + 414 N + 388 k\, 
   N + 61 k^{2} N + 277 N^{2} 
\nonu\\
&+& 148 k\, N^{2} + 57 N^{3}), \nonu\\
c_ {134} & = &\frac {1} {3 (2 + N)^{2} (2 + k + N)^{5}}i (16 + 21 k + 6 k^{2} + 19 N + 13 k\, 
     N + 5 N^{2}) (180 + 223 k + 62 k^{2} 
\nonu\\
&+& 371 N + 341 k\, 
     N + 58 k^{2} N + 245 N^{2} + 126 k\, 
     N^{2} + 50 N^{3}), \nonu\\
c_ {135} & = & - \frac {1}{6 (2 + N)^{2} (2 + k + N)^{5}}i (960 + 4308 k + 5515 k^{2} + 2684 k^{3} + 
       444 k^{4} + 3180 N 
\nonu\\
&+& 11408 k\, 
      N + 11336 k^{2} N + 3928 k^{3} N + 384 k^{4} N + 4149 N^{2} + 
       11238 k\, 
      N^{2} + 7739 k^{2} N^{2} 
\nonu\\
&+&1428 k^{3} N^{2} + 2630 N^{3} + 
       4804 k\, N^{3} + 1726 k^{2} N^{3} + 797 N^{4} + 738 k\, 
      N^{4} + 92 N^{5}), \nonu\\
c_ {136} & = & - \frac {1}{6 (2 + N)^{2} (2 + k + N)^{5}}i (13 k + 6 k^{2} + 3 N + 9 k\, 
      N + N^{2}) (180 + 223 k + 62 k^{2} + 371 N 
\nonu\\
&+& 341 k\, 
      N + 58 k^{2} N + 245 N^{2} + 126 k\, 
      N^{2} + 50 N^{3}), \nonu\\
c_ {137} & = &\frac{1} {6 (2 + N)^{2} (2 + k + N)^{5}}i (16 + 21 k + 6 k^{2} + 19 N + 13 k\, 
     N + 5 N^{2}) (180 + 247 k + 74 k^{2} 
\nonu\\
&+& 347 N + 353 k\, 
     N + 64 k^{2} N + 221 N^{2} + 126 k\, 
     N^{2} + 44 N^{3}), \nonu\\
c_ {138} & = &\frac {2 i (20 + 21 k + 6 k^{2} + 25 N + 13 k\, 
     N + 7 N^{2})} {(2 + N) (2 + k + N)^{3}}, \nonu\\
c_ {139} & = &\frac {2 i (32 + 55 k + 18 k^{2} + 41 N + 35 k\, 
     N + 11 N^{2})} {(2 + N) (2 + k + N)^{4}}, \nonu\\
c_ {140} & = &\frac {1} {6 (2 + N)^{2} (2 + k + N)^{5}}(2064 + 2960 k + 2707 k^{2} + 1476 k^{3} + 
     300 k^{4} + 7984 N + 13070 k\, 
    N 
\nonu\\
&+& 9788 k^{2} N + 3236 k^{3} N + 312 k^{4} N + 11907 N^{2} + 
     16874 k\, 
    N^{2} + 8603 k^{2} N^{2} + 1384 k^{3} N^{2}
\nonu\\
&+& 8354 N^{3} + 
     8494 k\, N^{3} + 2194 k^{2} N^{3} + 2755 N^{4} + 1466 k\, 
    N^{4} + 344 N^{5}), \nonu\\
c_ {141} & = &\frac{1} {6 (2 + N)^{2} (2 + k + N)^{5}}(6672 + 14576 k + 10627 k^{2} + 3108 k^{3} + 
     300 k^{4} + 19408 N 
\nonu\\
&+& 33566 k\, 
    N + 18884 k^{2} N + 4148 k^{3} N + 312 k^{4} N + 22275 N^{2} + 
     29114 k\, 
    N^{2} + 11459 k^{2} N^{2}
\nonu\\
&+& 1432 k^{3} N^{2} + 12650 N^{3} + 
     11230 k\, N^{3} + 2338 k^{2} N^{3} + 3547 N^{4} + 1610 k\, 
    N^{4} + 392 N^{5}), \nonu\\
c_ {142} & = &\frac {1} {6 (2 + N)^{2} (2 + k + N)^{5}}(528 + 4880 k + 6523 k^{2} + 3144 k^{3} + 
     516 k^{4} + 4528 N + 16334 k\, 
    N 
\nonu\\
&+& 14588 k^{2} N + 4634 k^{3} N + 420 k^{4} N + 8283 N^{2} + 
     18278 k\, 
    N^{2} + 10373 k^{2} N^{2} + 1666 k^{3} N^{2} 
\nonu\\
&+& 6242 N^{3} + 
     8440 k\, N^{3} + 2356 k^{2} N^{3} + 2113 N^{4} + 1376 k\, 
    N^{4} + 266 N^{5}), \nonu\\
c_ {143} & = & - \frac {2 (k - N) (32 + 55 k + 18 k^{2} + 41 N + 
       35 k\, N + 11 N^{2})} {(2 + N) (2 + k + N)^{5}}, \nonu\\
c_ {144} & = &\frac {1}{3 (2 + N)^{2} (2 + k + N)^{5}}(1440 + 5468 k + 6371 k^{2} + 2922 k^{3} + 
     464 k^{4} + 4804 N + 15004 k\, 
    N 
\nonu\\
&+& 13798 k^{2} N + 4617 k^{3} N + 458 k^{4} N + 6345 N^{2} + 
     15412 k\, 
    N^{2} + 10078 k^{2} N^{2} + 1925 k^{3} N^{2} 
\nonu\\
&+& 32 k^{4} N^{2} + 
     4108 N^{3} + 7055 k\, 
    N^{3} + 2627 k^{2} N^{3} + 106 k^{3} N^{3} + 1306 N^{4} + 
     1297 k\, N^{4} 
\nonu\\
&+& 120 k^{2} N^{4} + 173 N^{5} + 50 k\, 
    N^{5} + 4 N^{6}), \nonu\\
c_ {145} & = & - \frac {1}{3 (2 + N)^{2} (2 + k + N)^{5}}(-6240 - 12572 k - 9147 k^{2} - 2830 k^{3} - 
      312 k^{4} - 20260 N 
\nonu\\
&-& 33916 k\, 
     N - 19446 k^{2} N - 4299 k^{3} N - 270 k^{4} N - 25793 N^{2} - 
      33504 k\, 
     N^{2} - 13308 k^{2} N^{2} 
\nonu\\
&-& 1481 k^{3} N^{2} + 24 k^{4} N^{2} - 
      16196 N^{3} - 14477 k\, 
     N^{3} - 2937 k^{2} N^{3} + 48 k^{3} N^{3} - 5116 N^{4} 
\nonu\\
&-&  2401 k\, N^{4} - 711 N^{5} - 48 k\, 
     N^{5} - 24 N^{6}), \nonu\\
c_ {146} & = &\frac{1} {3 (2 + N) (2 + k + N)^{4}}(720 + 972 k + 235 k^{2} - 22 k^{3} + 1404 N + 
     1494 k\, N + 289 k^{2} N 
\nonu\\
&+& 16 k^{3} N + 863 N^{2} + 544 k\, 
    N^{2} + 16 k^{2} N^{2} + 125 N^{3} - 16 k\, 
    N^{3} - 16 N^{4}), \nonu\\
c_ {147} & = & - \frac{1}{6 (2 + N)^{2} (2 + k + N)^{6}}(32 + 55 k + 18 k^{2} + 41 N + 35 k\, 
      N + 11 N^{2}) (180 + 223 k + 62 k^{2} 
\nonu\\
&+& 371 N + 341 k\, 
      N + 58 k^{2} N + 245 N^{2} + 126 k\, 
      N^{2} + 50 N^{3}), \nonu\\
c_ {148} & = & - \frac {2 (k - N) (32 + 55 k + 18 k^{2} + 41 N + 
       35 k\, N + 11 N^{2})} {(2 + N) (2 + k + N)^{5}}, \nonu\\
c_ {149} & = & - \frac {2 i (32 + 55 k + 18 k^{2} + 41 N + 35 k\, 
      N + 11 N^{2})} {(2 + N) (2 + k + N)^{4}}, \nonu\\
c_ {150} & = & - \frac {2 i (540 + 677 k + 190 k^{2} + 1105 N + 
       1027 k\, N + 176 k^{2} N + 727 N^{2} + 378 k\, 
      N^{2} + 148 N^{3})} {3 (2 + N) (2 + k + N)^{4}}, \nonu\\
c_ {151} & = & - \frac {2 i (180 + 247 k + 74 k^{2} + 347 N + 353 k\, 
      N + 64 k^{2} N + 221 N^{2} + 126 k\, 
      N^{2} + 44 N^{3})} {3 (2 + N) (2 + k + N)^{4}}, \nonu\\
c_ {152} & = &\frac {1} {6 (2 + N)^{2} (2 + k + N)^{5}}i (-2064 - 3992 k - 3297 k^{2} - 1386 k^{3} - 
      232 k^{4} - 5656 N 
\nonu\\
&-& 7106 k\, 
     N - 2626 k^{2} N - 85 k^{3} N + 110 k^{4} N - 3265 N^{2} + 
      3126 k\, 
     N^{2} + 7096 k^{2} N^{2} 
+3023 k^{3} N^{2} 
\nonu\\
&+&  356 k^{4} N^{2} + 
      3406 N^{3} + 12087 k\, 
     N^{3} + 8461 k^{2} N^{3} + 1562 k^{3} N^{3} + 4798 N^{4} + 
      7389 k\, N^{4} 
\nonu\\
&+&2336 k^{2} N^{4} + 1969 N^{5} + 1366 k\, 
     N^{5} + 272 N^{6}), \nonu\\
c_ {153} & = &\frac{1} {6 (2 + N)^{2} (2 + k + N)^{5}}i (5136 + 4832 k + 1431 k^{2} + 144 k^{3} + 
      4 k^{4} + 22528 N + 34358 k\, 
     N 
\nonu\\
&+& 23800 k^{2} N + 7846 k^{3} N + 940 k^{4} N + 41431 N^{2} + 
      65154 k\, 
     N^{2} + 39797 k^{2} N^{2} + 9730 k^{3} N^{2} 
\nonu\\
&+& 712 k^{4} N^{2} + 
      39242 N^{3} + 52884 k\, 
     N^{3} + 23468 k^{2} N^{3} + 3124 k^{3} N^{3} + 20033 N^{4} + 
      19644 k\, N^{4} 
\nonu\\
&+& 4672 k^{2} N^{4} + 5222 N^{5} + 2732 k\, 
     N^{5} + 544 N^{6}), \nonu\\
c_ {154} & = &\frac {1} {(6 (2 + N) (2 + k + N)^{5})}i (4752 + 12620 k + 11220 k^{2} + 4149 k^{3} + 
      550 k^{4} + 13876 N 
\nonu\\
&+& 28846 k\, 
     N + 19327 k^{2} N + 4903 k^{3} N + 356 k^{4} N + 15842 N^{2} + 
      24991 k\, 
     N^{2} + 11655 k^{2} N^{2} 
\nonu\\
&+&  1610 k^{3} N^{2} + 8953 N^{3} + 
      9739 k\, N^{3} + 2432 k^{2} N^{3} + 2493 N^{4} + 1414 k\, 
     N^{4} + 272 N^{5}), \nonu\\
c_ {155} & = & - \frac {4 i (32 + 55 k + 18 k^{2} + 41 N + 35 k\, 
      N + 11 N^{2})} {(2 + N) (2 + k + N)^{4}}, \nonu\\
c_ {156} & = &\frac {1} {3 (2 + N)^{2} (2 + k + N)^{6}}2 i (2400 + 7564 k + 8155 k^{2} + 3602 k^{3} + 
      560 k^{4} + 9236 N 
\nonu\\
&+& 23316 k\, 
     N + 19586 k^{2} N + 6317 k^{3} N + 626 k^{4} N + 14153 N^{2} + 
      27388 k\, 
     N^{2} + 16430 k^{2} N^{2} 
\nonu\\
&+&  3181 k^{3} N^{2} + 92 k^{4} N^{2} + 
      11056 N^{3} + 15195 k\, 
     N^{3} + 5539 k^{2} N^{3} + 394 k^{3} N^{3} + 4650 N^{4} 
\nonu\\
&+&  3953 k\, N^{4} + 600 k^{2} N^{4} + 1005 N^{5} + 386 k\, 
     N^{5} + 88 N^{6}), \nonu\\
c_ {157} & = &\frac {1} {3 (2 + N)^{2} (2 + k + N)^{6}}2 i (1920 + 8260 k + 10243 k^{2} + 4916 k^{3} + 
      812 k^{4} + 5756 N 
\nonu\\
&+& 21576 k\, 
     N + 21620 k^{2} N + 7724 k^{3} N + 824 k^{4} N + 7157 N^{2} + 
      22126 k\, 
     N^{2} + 16379 k^{2} N^{2} 
\nonu\\
&+&  3616 k^{3} N^{2} + 128 k^{4} N^{2} + 
      4618 N^{3} + 10848 k\, 
     N^{3} + 5026 k^{2} N^{3} + 424 k^{3} N^{3} + 1581 N^{4} 
\nonu\\
&+&  2462 k\, N^{4} + 480 k^{2} N^{4} + 264 N^{5} + 200 k\, 
     N^{5} + 16 N^{6}), \nonu\\
c_ {158} & = &\frac {i (1020 + 1379 k + 410 k^{2} + 2047 N + 2059 k\, 
     N + 376 k^{2} N + 1341 N^{2} + 756 k\, 
     N^{2} + 272 N^{3})} {3 (2 + N) (2 + k + N)^{3}}, \nonu\\
c_ {159} & = & - \frac {2 (20 + 21 k + 6 k^{2} + 25 N + 13 k\, 
      N + 7 N^{2})} {(2 + N) (2 + k + N)^{3}}, \nonu\\
c_ {160} & = & - \frac {1} {12 (2 + N)^{2} (2 + k + N)^{4}}i (-1776 - 1648 k + 979 k^{2} + 1284 k^{3} + 
       300 k^{4} - 464 N 
\nonu\\
&+& 5006 k\, 
      N + 7484 k^{2} N + 3044 k^{3} N + 312 k^{4} N + 4611 N^{2} + 
       11690 k\, 
      N^{2} + 7595 k^{2} N^{2} 
\nonu\\
&+&  1336 k^{3} N^{2} + 5282 N^{3} + 
       7054 k\, N^{3} + 2050 k^{2} N^{3} + 2131 N^{4} + 1322 k\, 
      N^{4} + 296 N^{5}), \nonu\\
c_ {161} & = & - \frac{1}{12 (2 + N)^{2} (2 + k + N)^{4}}i (5136 + 12656 k + 10795 k^2 + 3912 k^3 + 516 k^4 + 15952 N 
\nonu\\
&+&
 31262 k N + 20420 k^2 N + 5258 k^3 N + 420 k^4 N + 19419 N^2 + 
 28790 k N^2 + 12941 k^2 N^2 
 \nonu\\
&+& 1786 k^3 N^2 + 11594 N^3 + 
 11656 k N^3 + 2716 k^2 N^3 + 3385 N^4 + 1736 k N^4 + 386 N^5), \nonu\\
c_ {162} & = & - \frac {2 (16 + 21 k + 6 k^{2} + 19 N + 13 k\, 
      N + 5 N^{2})} {(2 + N) (2 + k + N)^{3}}, \nonu\\
c_ {163} & = & - \frac {3 (16 + 21 k + 6 k^{2} + 19 N + 13 k\, 
      N + 5 N^{2})} {(2 + N) (2 + k + N)^{3}}, \nonu\\
c_ {164} & = & - \frac{1}{6 (2 + N)^{2} (2 + k + N)^{5}}(1296 + 3056 k + 3331 k^{2} + 1668 k^{3} + 
      300 k^{4} + 6352 N 
\nonu\\
&+& 13022 k\, 
     N + 10244 k^{2} N + 3284 k^{3} N + 312 k^{4} N + 10371 N^{2} + 
      16442 k\, 
     N^{2} + 8531 k^{2} N^{2} 
\nonu\\
&+& 1360 k^{3} N^{2} + 7562 N^{3} + 
      8158 k\, N^{3} + 2122 k^{2} N^{3} + 2539 N^{4} + 1394 k\, 
     N^{4} + 320 N^{5}), \nonu\\
c_ {165} & = &\frac {2 i (60 + 71 k + 18 k^{2} + 67 N + 43 k\, 
     N + 17 N^{2})} {3 (2 + N) (2 + k + N)^{3}}, \nonu\\
c_ {166} & = &\frac {1} {6 (2 + N)^{2} (2 + k + N)^{5}}i (960 + 3540 k + 4123 k^{2} + 1892 k^{3} + 
      300 k^{4} + 3948 N + 11120 k\, 
     N 
\nonu\\
&+& 9848 k^{2} N + 3220 k^{3} N + 312 k^{4} N + 5829 N^{2} + 
      12174 k\, 
     N^{2} + 7367 k^{2} N^{2} + 1272 k^{3} N^{2} 
\nonu\\
&+&3974 N^{3} + 
      5560 k\, N^{3} + 1738 k^{2} N^{3} + 1265 N^{4} + 894 k\, 
     N^{4} + 152 N^{5}), \nonu\\
c_ {167} & = & - \frac{1} {3 (2 + N) (2 + k + N)^{4}}i (-600 - 706 k - 149 k^{2} + 18 k^{3} - 
       1274 N - 1148 k\, N - 175 k^{2} N 
\nonu\\
&-& 863 N^{2} - 444 k\, 
      N^{2} - 179 N^{3}),\nonu\\ 
c_ {168} & = & - \frac{1}{6 (2 + N)^{2} (2 + k + N)^{5}}(4368 + 10640 k + 9403 k^2 + 3624 k^3 + 516 k^4 + 14128 N 
\nonu\\
&+& 
 27854 k N + 18908 k^2 N + 5114 k^3 N + 420 k^4 N + 17883 N^2 + 
 26918 k N^2 + 12533 k^2 N^2 
 \nonu\\
&+& 1786 k^3 N^2 + 11042 N^3 + 
 11320 k N^3 + 2716 k^2 N^3 + 3313 N^4 + 1736 k N^4 + 386 N^5), \nonu\\
c_ {169} & = & - \frac {1}{6 (2 + N)^{2} (2 + k + N)^{5}}(528 + 656 k + 1555 k^{2} + 1284 k^{3} + 
      300 k^{4} + 4144 N + 8462 k\, 
     N 
\nonu\\
&+& 8060 k^{2} N + 3044 k^{3} N + 312 k^{4} N + 8067 N^{2} + 
      13418 k\, 
     N^{2} + 7739 k^{2} N^{2} + 1336 k^{3} N^{2} 
\nonu\\
&+& 6434 N^{3} + 
      7342 k\, N^{3} + 2050 k^{2} N^{3} + 2275 N^{4} + 1322 k\, 
     N^{4} + 296 N^{5}), \nonu\\
c_ {170} & = & - \frac {1}{6 (2 + N)^{2} (2 + k + N)^{5}}(3600 + 9968 k + 8323 k^{2} + 2724 k^{3} + 
      300 k^{4} + 11728 N 
\nonu\\
&+& 24350 k\, 
     N + 15428 k^{2} N + 3764 k^{3} N + 312 k^{4} N + 14595 N^{2} + 
      22202 k\, 
     N^{2} + 9731 k^{2} N^{2} 
\nonu\\
&+& 1336 k^{3} N^{2} + 8810 N^{3} + 
      8926 k\, N^{3} + 2050 k^{2} N^{3} + 2587 N^{4} + 1322 k\, 
     N^{4} + 296 N^{5}), \nonu\\
c_ {171} & = & - \frac{1}{(2 + N) (2 + k + N)^{4}}2 (40 + 60 k + 23 k^{2} + 2 k^{3} + 92 N + 
       114 k\, N + 37 k^{2} N + 4 k^{3} N 
\nonu\\
&+& 73 N^{2} + 63 k\, 
      N^{2} + 11 k^{2} N^{2} + 22 N^{3} + 9 k\, 
      N^{3} + 2 N^{4}), \nonu\\
c_ {172} & = & - \frac{1}{6 (2 + N)^{2} (2 + k + N)^{5}}(11520 + 22884 k + 16219 k^{2} + 4856 k^{3} + 
      516 k^{4} + 35292 N 
\nonu\\
&+& 57992 k\, 
     N + 32180 k^{2} N + 6850 k^{3} N + 420 k^{4} N + 42429 N^{2} + 
      54066 k\, 
     N^{2} + 20849 k^{2} N^{2} 
\nonu\\
&+& 2346 k^{3} N^{2} + 24962 N^{3} + 
      21922 k\, N^{3} + 4408 k^{2} N^{3} + 7175 N^{4} + 3252 k\, 
     N^{4} + 806 N^{5}), \nonu\\
c_ {173} & = &\frac {(-240 - 260 k - 25 k^{2} + 18 k^{3} - 532 N - 
     466 k\, N - 59 k^{2} N - 373 N^{2} - 192 k\, 
    N^{2} - 79 N^{3})} {(2 + N) (2 + k + N)^{4}}, \nonu\\
c_ {174} & = &\frac {2 (k - N) (32 + 55 k + 18 k^{2} + 41 N + 35 k\, 
     N + 11 N^{2})} {(2 + N) (2 + k + N)^{5}}, \nonu\\
c_ {175} & = &\frac {1} {3 (2 + N)^{2} (2 + k + N)^{6}}2 i (2400 + 7820 k + 8539 k^{2} + 3794 k^{3} + 
      592 k^{4} + 8980 N 
\nonu\\
&+& 23572 k\, 
     N + 20162 k^{2} N + 6573 k^{3} N + 658 k^{4} N + 13513 N^{2} + 
      27260 k\, 
     N^{2} + 16718 k^{2} N^{2} 
\nonu\\
&+& 3293 k^{3} N^{2} + 100 k^{4} N^{2} + 
      10416 N^{3} + 14939 k\, 
     N^{3} + 5587 k^{2} N^{3} + 410 k^{3} N^{3} + 4330 N^{4}
\nonu\\
&+&  3841 k\, N^{4} + 600 k^{2} N^{4} + 925 N^{5} + 370 k\, 
     N^{5} + 80 N^{6}), \nonu\\
c_ {176} & = & - \frac {2 (8 + 17 k + 6 k^{2} + 11 N + 11 k\, 
      N + 3 N^{2})} {(2 + N) (2 + k + N)^{3}}, \nonu\\
c_ {177} & = & - \frac{1}{6 (2 + N)^{2} (2 + k + N)^{5}}i (2212 k + 3851 k^{2} + 2112 k^{3} + 
       372 k^{4} + 668 N + 6984 k\, 
      N 
\nonu\\
&+& 8692 k^{2} N + 3326 k^{3} N + 348 k^{4} N + 1549 N^{2} + 
       7770 k\, 
      N^{2} + 6349 k^{2} N^{2} + 1270 k^{3} N^{2} 
\nonu\\
&+& 1298 N^{3} + 
       3606 k\, N^{3} + 1484 k^{2} N^{3} + 459 N^{4} + 584 k\, 
      N^{4} + 58 N^{5}), \nonu\\
c_ {178} & = & - \frac{1} {6 (2 + N)^{2} (2 + k + N)^{5}}i (960 + 4180 k + 5387 k^{2} + 2652 k^{3} + 
       444 k^{4} + 3308 N + 11280 k\, 
      N 
\nonu\\
&+& 11176 k^{2} N + 3896 k^{3} N + 384 k^{4} N + 4405 N^{2} + 
       11238 k\, 
      N^{2} + 7675 k^{2} N^{2} + 1420 k^{3} N^{2} 
\nonu\\
&+& 2822 N^{3} + 
       4836 k\, N^{3} + 1718 k^{2} N^{3} + 861 N^{4} + 746 k\, 
      N^{4} + 100 N^{5}), \nonu\\
c_ {179} & = & - \frac {1}{6 (2 + N)^{2} (2 + k + N)^{5}}i (960 + 3796 k + 4379 k^{2} + 1956 k^{3} + 
       300 k^{4} + 3692 N + 11376 k\, 
      N 
\nonu\\
&+& 10168 k^{2} N + 3284 k^{3} N + 312 k^{4} N + 5317 N^{2} + 
       12174 k\, 
      N^{2} + 7495 k^{2} N^{2} + 1288 k^{3} N^{2} 
\nonu\\
&+& 3590 N^{3} + 
       5496 k\, N^{3} + 1754 k^{2} N^{3} + 1137 N^{4} + 878 k\, 
      N^{4} + 136 N^{5}), \nonu\\
c_ {180} & = & - \frac {(160 + 336 k + 159 k^{2} + 18 k^{3} + 256 N + 
      304 k\, N + 57 k^{2} N + 113 N^{2} + 54 k\, 
     N^{2} + 15 N^{3})} {(2 + N) (2 + k + N)^{4}}, \nonu\\
c_ {181} & = & - \frac {1}{6 (2 + N)^{2} (2 + k + N)^{5}}(5136 + 12752 k + 10843 k^{2} + 3912 k^{3} + 
      516 k^{4} + 15088 N 
\nonu\\
&+& 30062 k\, 
     N + 19724 k^{2} N + 5114 k^{3} N + 420 k^{4} N + 17499 N^{2} + 
      26630 k\, 
     N^{2} + 12149 k^{2} N^{2} 
\nonu\\
&+& 1714 k^{3} N^{2} + 10034 N^{3} + 
      10456 k\, N^{3} + 2500 k^{2} N^{3} + 2833 N^{4} + 1520 k\, 
     N^{4} + 314 N^{5}), \nonu\\
c_ {182} & = & - \frac {1}{6 (2 + N)^{2} (2 + k + N)^{5}}(4368 + 10640 k + 9403 k^{2} + 3624 k^{3} + 
      516 k^{4} + 14128 N 
\nonu\\
&+& 27854 k\, 
     N + 18908 k^{2} N + 5114 k^{3} N + 420 k^{4} N + 17883 N^{2} + 
      26918 k\, 
     N^{2} + 12533 k^{2} N^{2} 
\nonu\\
&+& 1786 k^{3} N^{2} + 11042 N^{3} + 
      11320 k\, N^{3} + 2716 k^{2} N^{3} + 3313 N^{4} + 1736 k\, 
     N^{4} + 386 N^{5}), \nonu\\
c_ {183} & = &\frac {1} {6 (2 + N)^{2} (2 + k + N)^{5}}i (13 k + 6 k^{2} + 3 N + 9 k\, 
     N + N^{2}) (180 + 223 k + 62 k^{2} + 371 N 
\nonu\\
&+& 341 k\, 
     N + 58 k^{2} N + 245 N^{2} + 126 k\, 
     N^{2} + 50 N^{3}), \nonu\\
c_ {184} & = &\frac{1} {6 (2 + N)^{2} (2 + k + N)^{5}}i (960 + 4308 k + 5515 k^{2} + 2684 k^{3} + 
      444 k^{4} + 3180 N + 11408 k\, 
     N 
\nonu\\
&+& 11336 k^{2} N + 3928 k^{3} N + 384 k^{4} N + 4149 N^{2} + 
      11238 k\, 
     N^{2} + 7739 k^{2} N^{2} + 1428 k^{3} N^{2} 
\nonu\\
&+& 2630 N^{3} + 
      4804 k\, N^{3} + 1726 k^{2} N^{3} + 797 N^{4} + 738 k\, 
     N^{4} + 92 N^{5}), \nonu\\
c_ {185} & = &\frac{1} {3 (2 + N)^{2} (2 + k + N)^{6}}2 i (1920 + 6980 k + 8323 k^{2} 
+3956 k^{3} + 
      652 k^{4} + 7036 N 
\nonu\\
&+& 20296 k\, 
     N + 18740 k^{2} N + 6444 k^{3} N + 664 k^{4} N + 10357 N^{2} + 
      22766 k\, 
     N^{2} + 14939 k^{2} N^{2} 
\nonu\\
&+& 3056 k^{3} N^{2} + 88 k^{4} N^{2} + 
      7818 N^{3} + 12128 k\, 
     N^{3} + 4786 k^{2} N^{3} + 344 k^{3} N^{3} + 3181 N^{4} 
\nonu\\
&+&  3022 k\, N^{4} + 480 k^{2} N^{4} + 664 N^{5} + 280 k\, 
     N^{5} + 56 N^{6}), \nonu\\
c_ {186} & = & - \frac {4 (8 + 17 k + 6 k^{2} + 11 N + 11 k\, 
      N + 3 N^{2})} {(2 + N) (2 + k + N)^{3}}, \nonu\\
c_ {187} & = & - \frac {2 (16 + 21 k + 6 k^{2} + 19 N + 13 k\, 
      N + 5 N^{2})} {(2 + N) (2 + k + N)^{3}}, \nonu\\
c_ {188} & = &\frac {2 i (32 + 55 k + 18 k^{2} + 41 N + 35 k\, 
     N + 11 N^{2})} {(2 + N) (2 + k + N)^{4}}, \nonu\\
c_ {189} & = &\frac {2 i (8 + 17 k + 6 k^{2} + 11 N + 11 k\, 
     N + 3 N^{2})} {(2 + N) (2 + k + N)^{3}}, \nonu\\
c_ {190} & = &\frac { i (16 + 21 k + 6 k^{2} + 19 N + 13 k\, 
     N + 5 N^{2})} {(2 + N) (2 + k + N)^{3}}, \nonu\\
c_ {191} & = &\frac {1} {24 (2 + N)^{2} (2 + k + N)^{4}}i (960 + 3540 k + 4123 k^{2} + 1892 k^{3} + 
      300 k^{4} + 3948 N 
\nonu\\
&+& 11120 k\, 
     N + 9848 k^{2} N + 3220 k^{3} N + 312 k^{4} N + 5829 N^{2} + 
      12174 k\, 
     N^{2} + 7367 k^{2} N^{2} 
\nonu\\
&+& 1272 k^{3} N^{2} + 3974 N^{3} + 
      5560 k\, N^{3} + 1738 k^{2} N^{3} + 1265 N^{4} + 894 k\, 
     N^{4} + 152 N^{5}), \nonu\\
c_ {192} & = &\frac {1} {12 (2 + N)^{2} (2 + k + N)^{4}}i (960 + 3796 k + 4379 k^{2} + 1956 k^{3} + 
      300 k^{4} + 3692 N 
\nonu\\
&+& 11376 k\, 
     N + 10168 k^{2} N + 3284 k^{3} N + 312 k^{4} N + 5317 N^{2} + 
      12174 k\, 
     N^{2} + 7495 k^{2} N^{2} 
\nonu\\
&+& 1288 k^{3} N^{2} + 3590 N^{3} + 
      5496 k\, N^{3} + 1754 k^{2} N^{3} + 1137 N^{4} + 878 k\, 
     N^{4} + 136 N^{5}), \nonu\\
c_ {193} & = & - \frac {(20 + 21 k + 6 k^{2} + 25 N + 13 k\, 
     N + 7 N^{2})} {2(2 + N) (2 + k + N)^{3}}, \nonu\\
c_ {194} & = &\frac{1} {24 (2 + N)^{2} (2 + k + N)^{4}}i (2880 + 7860 k + 7891 k^{2} + 3368 k^{3} + 
      516 k^{4} + 8844 N 
\nonu\\
&+& 19856 k\, 
     N + 15560 k^{2} N + 4714 k^{3} N + 420 k^{4} N + 10701 N^{2} + 
      18642 k\, 
     N^{2} + 10205 k^{2} N^{2}
\nonu\\
&+& 1650 k^{3} N^{2} + 6350 N^{3} + 
      7642 k\, N^{3} + 2200 k^{2} N^{3} + 1835 N^{4} + 1140 k\, 
     N^{4} + 206 N^{5}), \nonu\\
c_ {195} & = &\frac{1} {12 (2 + N)^{2} (2 + k + N)^{4}}i (16 + 21 k + 6 k^{2} + 19 N + 13 k\, 
     N + 5 N^{2}) (180 + 271 k + 86 k^{2} 
\nonu\\
&+& 323 N + 365 k\, 
     N + 70 k^{2} N + 197 N^{2} + 126 k\, 
     N^{2} + 38 N^{3}), \nonu\\
c_ {196} & = & - \frac {i (13 k + 6 k^{2} + 3 N + 9 k\, 
      N + N^{2})} {2(2 + N) (2 + k + N)^{3}}, \nonu\\
c_ {197} & = & - \frac {i (16 + 21 k + 6 k^{2} + 19 N + 13 k\, 
      N + 5 N^{2})} {(2 + N) (2 + k + N)^{3}}, \nonu\\
c_ {198} & = &\frac{1}{12 (2 + N)^{2} (2 + k + N)^{5}}(1920 + 5508 k + 5347 k^{2} + 2132 k^{3} + 
     300 k^{4} + 6588 N 
\nonu\\
&+& 15656 k\, 
    N + 12068 k^{2} N + 3532 k^{3} N + 312 k^{4} N + 8757 N^{2} + 
     16110 k\, 
    N^{2} + 8699 k^{2} N^{2} 
\nonu\\
&+& 1368 k^{3} N^{2} + 5594 N^{3} + 
     7072 k\, N^{3} + 2002 k^{2} N^{3} + 1709 N^{4} + 1110 k\, 
    N^{4} + 200 N^{5}), \nonu\\
c_ {199} & = &\frac{1} {6 (2 + N)^{2} (2 + k + N)^{5}}(1920 + 5380 k + 5219 k^{2} + 2100 k^{3} + 
     300 k^{4} + 6716 N + 15528 k\, 
    N 
\nonu\\
&+& 11908 k^{2} N + 3500 k^{3} N + 312 k^{4} N + 9013 N^{2} + 
     16110 k\, 
    N^{2} + 8635 k^{2} N^{2} + 1360 k^{3} N^{2} 
\nonu\\
&+& 5786 N^{3} + 
     7104 k\, N^{3} + 1994 k^{2} N^{3} + 1773 N^{4} + 1118 k\, 
    N^{4} + 208 N^{5}), \nonu\\
c_ {200} & = &\frac {1} {6 (2 + N)^{2} (2 + k + N)^{5}}(2880 + 7860 k + 7891 k^{2} + 3368 k^{3} + 
     516 k^{4} + 8844 N + 19856 k\, 
    N 
\nonu\\
&+& 15560 k^{2} N + 4714 k^{3} N + 420 k^{4} N + 10701 N^{2} + 
     18642 k\, 
    N^{2} + 10205 k^{2} N^{2} + 1650 k^{3} N^{2}
\nonu\\
&+& 6350 N^{3} + 
     7642 k\, N^{3} + 2200 k^{2} N^{3} + 1835 N^{4} + 1140 k\, 
    N^{4} + 206 N^{5}), \nonu\\
c_ {201} & = &\frac{1} {6 (2 + N)^{2} (2 + k + N)^{5}}(2880 + 7732 k + 7763 k^{2} + 3336 k^{3} + 
     516 k^{4} + 8972 N + 19728 k\, 
    N 
\nonu\\
&+& 15400 k^{2} N + 4682 k^{3} N + 420 k^{4} N + 10957 N^{2} + 
     18642 k\, 
    N^{2} + 10141 k^{2} N^{2} + 1642 k^{3} N^{2}
\nonu\\
&+& 6542 N^{3} + 
     7674 k\, N^{3} + 2192 k^{2} N^{3} + 1899 N^{4} + 1148 k\, 
    N^{4} + 214 N^{5}), \nonu\\
c_ {202} & = &\frac{1}{6 (2 + N)^{2} (2 + k + N)^{5}}(16 + 21 k + 6 k^{2} + 19 N + 13 k\, 
     N + 5 N^{2}) (180 + 271 k + 86 k^{2} 
\nonu\\
&+& 323 N + 365 k\, 
     N + 70 k^{2} N + 197 N^{2} + 126 k\, 
     N^{2} + 38 N^{3}), \nonu\\
c_ {203} & = &\frac {2 (32 + 55 k + 18 k^{2} + 41 N + 35 k\, 
     N + 11 N^{2})} {(2 + N) (2 + k + N)^{4}}, \nonu\\
c_ {204} & = &\frac { (32 + 55 k + 18 k^{2} + 41 N + 35 k\, 
     N + 11 N^{2})} {(2 + N) (2 + k + N)^{4}}, \qquad
c_ {205}  = \frac {2 i (13 k + 6 k^{2} + 3 N + 9 k\, 
     N + N^{2})} {(2 + N) (2 + k + N)^{3}}, \nonu\\
c_ {206} & = &\frac{1} {12 (2 + N)^{2} (2 + k + N)^{5}}i (3600 + 8864 k + 7915 k^{2} + 3084 k^{3} + 
      444 k^{4} + 12064 N 
\nonu\\
&+& 24038 k\, 
     N + 16520 k^{2} N + 4544 k^{3} N + 384 k^{4} N + 15699 N^{2} + 
      23882 k\, 
     N^{2} + 11267 k^{2} N^{2} 
\nonu\\
&+&1636 k^{3} N^{2} + 9902 N^{3} + 
      10258 k\, N^{3} + 2494 k^{2} N^{3} + 3019 N^{4} + 1598 k\, 
     N^{4} + 356 N^{5}), \nonu\\
c_ {207} & = & - \frac {1}{6 (2 + N)^{2} (2 + k + N)^{5}}i (2064 + 3584 k + 2995 k^{2} + 1320 k^{3} + 
       228 k^{4} + 8128 N
\nonu\\
&+& 13814 k\, 
      N + 9584 k^{2} N + 2906 k^{3} N + 276 k^{4} N + 12027 N^{2} + 
       16862 k\, 
      N^{2} + 8129 k^{2} N^{2} 
\nonu\\
&+&1258 k^{3} N^{2} + 8342 N^{3} + 
       8248 k\, N^{3} + 2044 k^{2} N^{3} + 2725 N^{4} + 1400 k\, 
      N^{4} + 338 N^{5}),\nonu\\
c_ {208} & = &\frac {i (64 + 136 k + 61 k^{2} + 6 k^{3} + 88 N + 
      104 k\, N + 15 k^{2} N + 27 N^{2} + 10 k\, 
     N^{2} + N^{3})} {(2 + N) (2 + k + N)^{4}}, \nonu\\
c_ {209} & = & - \frac {2 i (64 + 113 k + 51 k^{2} + 6 k^{3} + 95 N + 
       101 k\, N + 19 k^{2} N + 40 N^{2} + 18 k\, 
      N^{2} + 5 N^{3})} {(2 + N) (2 + k + N)^{4}}, \nonu\\
c_ {210} & = & - \frac {2 i (48 + 97 k + 47 k^{2} + 6 k^{3} + 71 N + 
       85 k\, N + 17 k^{2} N + 28 N^{2} + 14 k\, 
      N^{2} + 3 N^{3})} {(2 + N) (2 + k + N)^{4}}, \nonu\\
c_ {211} & = &\frac {1}{12 (2 + N)^{2} (2 + k + N)^{5}}i (528 + 1280 k + 1843 k^{2} + 1128 k^{3} + 
      228 k^{4} + 4288 N + 9206 k\, 
     N 
\nonu\\
&+& 7856 k^{2} N + 2714 k^{3} N + 276 k^{4} N + 8187 N^{2} + 
      13406 k\, 
     N^{2} + 7265 k^{2} N^{2} + 1210 k^{3} N^{2} 
\nonu\\
&+& 6422 N^{3} + 
      7096 k\, N^{3} + 1900 k^{2} N^{3} + 2245 N^{4} + 1256 k\, 
     N^{4} + 290 N^{5}), \nonu\\
c_ {212} & = & - \frac {1}{6 (2 + N)^{2} (2 + k + N)^{5}}i (3600 + 8864 k + 7915 k^{2} + 3084 k^{3} + 
       444 k^{4} + 12064 N 
\nonu\\
&+& 24038 k\, 
      N + 16520 k^{2} N + 4544 k^{3} N + 384 k^{4} N + 15699 N^{2} + 
       23882 k\, 
      N^{2} + 11267 k^{2} N^{2}
\nonu\\
&+& 1636 k^{3} N^{2} + 9902 N^{3} + 
       10258 k\, N^{3} + 2494 k^{2} N^{3} + 3019 N^{4} + 1598 k\, 
      N^{4} + 356 N^{5}), \nonu\\
c_ {213} & = &\frac {i (32 + 3 k - 6 k^{2} + 29 N - k\, 
     N + 7 N^{2})} {(2 + N) (2 + k + N)^{3}}, \nonu\\
c_ {214} & = & - \frac {i (16 + 21 k + 6 k^{2} + 19 N + 13 k\, 
      N + 5 N^{2})} {2(2 + N) (2 + k + N)^{3}}, \nonu\\
c_ {215} & = & - \frac {i (96 + 168 k + 69 k^{2} + 6 k^{3} + 136 N + 
       136 k\, N + 19 k^{2} N + 51 N^{2} + 18 k\, 
      N^{2} + 5 N^{3})} {2(2 + N) (2 + k + N)^{4}}, \nonu\\
c_ {216} & = &\frac{1} {6 (2 + N)^{2} (2 + k + N)^{5}}i (5136 + 14480 k + 12787 k^{2} + 4596 k^{3} + 
      588 k^{4} + 14896 N 
\nonu\\
&+& 32654 k\, 
     N + 21884 k^{2} N + 5612 k^{3} N + 456 k^{4} N + 16803 N^{2} + 
      27578 k\, 
     N^{2} + 12755 k^{2} N^{2} 
\nonu\\
&+& 1792 k^{3} N^{2} + 9314 N^{3} + 
      10342 k\, N^{3} + 2506 k^{2} N^{3} + 2539 N^{4} + 1442 k\, 
     N^{4} + 272 N^{5}), \nonu\\
c_ {217} & = &\frac {8 i} {(2 + k + N)^{2}}, \nonu\\
c_ {218} & = &\frac{1} {12 (2 + N)^{2} (2 + k + N)^{5}}i (6672 + 16784 k + 13939 k^{2} + 4788 k^{3} + 
      588 k^{4} + 18736 N 
\nonu\\
&+& 37262 k\, 
     N + 23612 k^{2} N + 5804 k^{3} N + 456 k^{4} N + 20643 N^{2} + 
      31034 k\, 
     N^{2} + 13619 k^{2} N^{2} 
\nonu\\
&+& 1840 k^{3} N^{2} + 11234 N^{3} + 
      11494 k\, N^{3} + 2650 k^{2} N^{3} + 3019 N^{4} + 1586 k\, 
     N^{4} + 320 N^{5}), \nonu\\
c_ {219} & = & - \frac {2 (32 + 55 k + 18 k^{2} + 41 N + 35 k\, 
      N + 11 N^{2})} {(2 + N) (2 + k + N)^{4}}, \nonu\\
c_ {220} & = &\frac {2 (32 + 55 k + 18 k^{2} + 41 N + 35 k\, 
     N + 11 N^{2})} {(2 + N) (2 + k + N)^{4}}, \nonu\\
c_ {221} & = &\frac {2 (32 + 55 k + 18 k^{2} + 41 N + 35 k\, 
     N + 11 N^{2})} {(2 + N) (2 + k + N)^{4}}, \nonu\\
c_ {222} & = & - \frac {2 (32 + 55 k + 18 k^{2} + 41 N + 35 k\, 
      N + 11 N^{2})} {(2 + N) (2 + k + N)^{4}}, \nonu\\
c_ {223} & = &\frac{1}{3 (2 + N)^{2} (2 + k + N)^{5}}(384 + 1620 k + 1847 k^{2} + 786 k^{3} + 
     112 k^{4} + 2484 N + 7626 k\, 
    N 
\nonu\\
&+&7103 k^{2} N + 2519 k^{3} N + 298 k^{4} N + 5695 N^{2} + 
     13519 k\, 
    N^{2} + 9888 k^{2} N^{2} + 2622 k^{3} N^{2} 
\nonu\\
&+&202 k^{4} N^{2} + 
     6354 N^{3} + 11327 k\, 
    N^{3} + 5806 k^{2} N^{3} + 847 k^{3} N^{3} + 3712 N^{4} + 
     4477 k\, N^{4} 
\nonu\\
&+& 1204 k^{2} N^{4} + 1083 N^{5} + 665 k\, 
    N^{5} + 124 N^{6}), \nonu\\
c_ {224} & = &\frac{1}{6 (2 + N)^{2} (2 + k + N)^{5}}(1536 + 3540 k + 3031 k^{2} + 1106 k^{3} + 
     144 k^{4} + 5556 N + 11786 k\, 
    N 
\nonu\\
&+& 9119 k^{2} N + 2935 k^{3} N + 330 k^{4} N + 9087 N^{2} + 
     17071 k\, 
    N^{2} + 11144 k^{2} N^{2} + 2798 k^{3} N^{2}
\nonu\\
&+&210 k^{4} N^{2} + 
     8338 N^{3} + 12815 k\, 
    N^{3} + 6142 k^{2} N^{3} + 871 k^{3} N^{3} + 4360 N^{4} + 
     4781 k\, N^{4} 
\nonu\\
&+& 1236 k^{2} N^{4} + 1195 N^{5} + 689 k\, 
    N^{5} + 132 N^{6}), \nonu\\
c_ {225} & = &\frac {1}{6 (2 + N) (2 + k + N)^{5}}(3216 + 8636 k + 7992 k^{2} + 3123 k^{3} + 
     442 k^{4} + 10564 N 
\nonu\\
&+& 22462 k\, 
    N + 15889 k^{2} N + 4357 k^{3} N + 356 k^{4} N + 13358 N^{2} + 
     21601 k\, 
    N^{2} + 10605 k^{2} N^{2} 
\nonu\\
&+& 1562 k^{3} N^{2} + 8167 N^{3} + 
     9037 k\, N^{3} + 2336 k^{2} N^{3} + 2403 N^{4} + 1366 k\, 
    N^{4} + 272 N^{5}), \nonu\\
c_ {226} & = &\frac{1} {12 (2 + N) (2 + k + N)^{5}}(4368 + 10940 k + 9688 k^{2} + 3667 k^{3} + 
     506 k^{4} + 13444 N 
\nonu\\
&+& 27134 k\, 
    N + 18497 k^{2} N + 4917 k^{3} N + 388 k^{4} N + 16206 N^{2} + 
     25121 k\, 
    N^{2} + 11933 k^{2} N^{2} 
\nonu\\
&+& 1706 k^{3} N^{2} + 9559 N^{3} + 
     10205 k\, N^{3} + 2560 k^{2} N^{3} + 2739 N^{4} + 1510 k\, 
    N^{4} + 304 N^{5}), \nonu\\
c_ {227} & = &\frac{1} {3 (2 + N)^{2} (2 + k + N)^{5}}(4368 + 8120 k + 5993 k^{2} + 2028 k^{3} + 
     260 k^{4} + 16408 N 
\nonu\\
&+& 29858 k\, 
    N + 20874 k^{2} N + 6336 k^{3} N + 680 k^{4} N + 26393 N^{2} + 
     43156 k\, 
    N^{2} + 25133 k^{2} N^{2} 
\nonu\\
&+& 5650 k^{3} N^{2} + 356 k^{4} N^{2} + 
     22694 N^{3} + 30374 k\, 
    N^{3} + 12768 k^{2} N^{3} + 1562 k^{3} N^{3} + 10833 N^{4} 
\nonu\\
&+&  10366 k\, N^{4} + 2336 k^{2} N^{4} + 2696 N^{5} + 1366 k\, 
    N^{5} + 272 N^{6}), \nonu\\
c_ {228} & = &\frac{1}{6 (2 + N)^{2} (2 + k + N)^{5}}(528 + 1592 k + 1961 k^{2} + 972 k^{3} + 
     164 k^{4} + 6040 N + 15266 k\, 
    N 
\nonu\\
&+& 13674 k^{2} N + 4896 k^{3} N + 584 k^{4} N + 14873 N^{2} + 
     30292 k\, 
    N^{2} + 20381 k^{2} N^{2} + 5002 k^{3} N^{2} 
\nonu\\
&+& 332 k^{4} N^{2} + 
     15974 N^{3} + 24806 k\, 
    N^{3} + 11400 k^{2} N^{3} + 1466 k^{3} N^{3} + 8673 N^{4} + 
     9190 k\, N^{4} 
\nonu\\
&+& 2192 k^{2} N^{4} + 2336 N^{5} + 1270 k\, 
    N^{5} + 248 N^{6}), \nonu\\
c_ {229} & = &\frac {1}{6 (2 + N)^{2} (2 + k + N)^{5}}(1828 k + 3155 k^{2} + 1716 k^{3} + 300 k^{4} + 
     1052 N + 6840 k\, 
    N 
\nonu\\
&+& 7948 k^{2} N + 2972 k^{3} N + 312 k^{4} N + 2389 N^{2} + 
     8238 k\, 
    N^{2} + 6163 k^{2} N^{2} + 1192 k^{3} N^{2}
\nonu\\
&+& 1970 N^{3} + 
     3984 k\, N^{3} + 1490 k^{2} N^{3} + 693 N^{4} + 662 k\, 
    N^{4} + 88 N^{5}),\nonu\\ 
c_ {230} & = &\frac{1}{12 (2 + N)^{2} (2 + k + N)^{5}}(13 k + 6 k^{2} + 3 N + 9 k\, 
     N + N^{2}) (180 + 199 k + 50 k^{2} + 395 N 
\nonu\\
&+& 329 k\, 
     N + 52 k^{2} N + 269 N^{2} + 126 k\, 
     N^{2} + 56 N^{3}), \nonu\\
c_ {231} & = &\frac{1} {6 (2 + N)^{2} (2 + k + N)^{5}}(960 + 3796 k + 4379 k^{2} + 1956 k^{3} + 
     300 k^{4} + 3692 N + 11376 k\, 
    N 
\nonu\\
&+& 10168 k^{2} N + 3284 k^{3} N + 312 k^{4} N + 5317 N^{2} + 
     12174 k\, 
    N^{2} + 7495 k^{2} N^{2} + 1288 k^{3} N^{2} 
\nonu\\
&+& 3590 N^{3} + 
     5496 k\, N^{3} + 1754 k^{2} N^{3} + 1137 N^{4} + 878 k\, 
    N^{4} + 136 N^{5}), \nonu\\
c_ {232} & = &\frac{1} {12 (2 + N)^{2} (2 + k + N)^{5}}(960 + 3540 k + 4123 k^{2} + 1892 k^{3} + 
     300 k^{4} + 3948 N 
\nonu\\
&+& 11120 k\, 
    N + 9848 k^{2} N + 3220 k^{3} N + 312 k^{4} N + 5829 N^{2} + 
     12174 k\, 
    N^{2} + 7367 k^{2} N^{2} 
\nonu\\
&+& 1272 k^{3} N^{2} + 3974 N^{3} + 
     5560 k\, N^{3} + 1738 k^{2} N^{3} + 1265 N^{4} + 894 k\, 
    N^{4} + 152 N^{5}), \nonu\\
c_ {233} & = &\frac {1}{6 (2 + N)^{2} (2 + k + N)^{5}}(1920 + 6148 k + 6923 k^{2} + 3192 k^{3} + 
     516 k^{4} + 5948 N 
\nonu\\
&+& 15576 k\, 
    N + 13660 k^{2} N + 4466 k^{3} N + 420 k^{4} N + 7261 N^{2} + 
     14706 k\, 
    N^{2} + 9001 k^{2} N^{2} 
\nonu\\
&+& 1570 k^{3} N^{2} + 4346 N^{3} + 
     6066 k\, N^{3} + 1952 k^{2} N^{3} + 1263 N^{4} + 908 k\, 
    N^{4} + 142 N^{5}), \nonu\\
c_ {234} & = &\frac{1} {12 (2 + N)^{2} (2 + k + N)^{5}}(1920 + 6276 k + 7051 k^{2} + 3224 k^{3} + 
     516 k^{4} + 5820 N 
     \nonu\\
&+& 15704 k\, 
    N + 13820 k^{2} N + 4498 k^{3} N + 420 k^{4} N + 7005 N^{2} + 
     14706 k\, 
    N^{2} + 9065 k^{2} N^{2} 
\nonu\\
&+& 1578 k^{3} N^{2} + 4154 N^{3} + 
     6034 k\, N^{3} + 1960 k^{2} N^{3} + 1199 N^{4} + 900 k\, 
    N^{4} + 134 N^{5}),\nonu\\ 
c_ {235} & = & - \frac{1}{6 (2 + N)^{2} (2 + k + N)^{6}}i (32 + 55 k + 18 k^{2} + 41 N + 35 k\, 
      N + 11 N^{2}) (180 + 199 k + 50 k^{2} 
\nonu\\
&+& 395 N + 329 k\, 
      N + 52 k^{2} N + 269 N^{2} + 126 k\, 
      N^{2} + 56 N^{3}), \nonu\\
c_ {236} & = & - \frac {1}{6 (2 + N)^{2} (2 + k + N)^{6}}i (32 + 55 k + 18 k^{2} + 41 N + 35 k\, 
      N + 11 N^{2}) (180 + 247 k + 74 k^{2} 
\nonu\\
&+& 347 N + 353 k\, 
      N + 64 k^{2} N + 221 N^{2} + 126 k\, 
      N^{2} + 44 N^{3}), \nonu\\
c_ {237} & = & - \frac {1}{3 (2 + N)^{2} (2 + k + N)^{6}} i (4368 + 11480 k + 10797 k^{2} + 
       4318 k^{3} + 624 k^{4} + 16888 N 
\nonu\\
&+& 37842 k\, 
      N + 29538 k^{2} N + 9355 k^{3} N + 990 k^{4} N + 27045 N^{2} + 
       50050 k\, 
      N^{2} + 30714 k^{2} N^{2} 
\nonu\\
&+& 6947 k^{3} N^{2} + 
       420 k^{4} N^{2} + 22894 N^{3} + 33011 k\, 
      N^{3} + 14265 k^{2} N^{3} + 1742 k^{3} N^{3} + 10752 N^{4}
\nonu\\
&+&   10773 k\, N^{4} + 2472 k^{2} N^{4} + 2641 N^{5} + 1378 k\, 
      N^{5} + 264 N^{6}), \nonu\\
c_ {238} & = & - \frac {1}{3 (2 + N)^{2} (2 + k + N)^{6}} i (3984 + 10712 k + 10189 k^{2} + 
       4094 k^{3} + 592 k^{4} + 15736 N 
\nonu\\
&+& 35986 k\, 
      N + 28386 k^{2} N + 9035 k^{3} N + 958 k^{4} N + 25573 N^{2} + 
       48226 k\, 
      N^{2} + 29890 k^{2} N^{2} 
\nonu\\
&+& 6795 k^{3} N^{2} + 
       412 k^{4} N^{2} + 21870 N^{3} + 32099 k\, 
      N^{3} + 14001 k^{2} N^{3} + 1718 k^{3} N^{3} + 10344 N^{4}
\nonu\\
&+&   10541 k\, N^{4} + 2440 k^{2} N^{4} + 2553 N^{5} + 1354 k\, 
      N^{5} + 256 N^{6}),\nonu\\ 
c_ {239} & = & - \frac{1} {3 (2 + N)^{2} (2 + k + N)^{6}}2 i (3600 + 8456 k + 7393 k^{2} + 
       2804 k^{3} + 388 k^{4} + 14536 N 
\nonu\\
&+& 30682 k\, 
      N + 23358 k^{2} N + 7400 k^{3} N + 808 k^{4} N + 24361 N^{2} + 
       43804 k\, 
      N^{2} + 26809 k^{2} N^{2} 
\nonu\\
&+& 6168 k^{3} N^{2} + 
       388 k^{4} N^{2} + 21474 N^{3} + 30602 k\, 
      N^{3} + 13296 k^{2} N^{3} + 1652 k^{3} N^{3} + 10425 N^{4} 
\nonu\\
&+&    10412 k\, N^{4} + 2404 k^{2} N^{4} + 2628 N^{5} + 1372 k\, 
      N^{5} + 268 N^{6}), \nonu\\
c_ {240} & = & - \frac {1}{3 (2 + N)^{2} (2 + k + N)^{6}}2 i (2448 + 6152 k + 5633 k^{2} + 
       2196 k^{3} + 308 k^{4} + 11080 N 
\nonu\\
&+& 24986 k\, 
      N + 19902 k^{2} N + 6504 k^{3} N + 728 k^{4} N + 20009 N^{2} + 
       38140 k\, 
      N^{2} + 24257 k^{2} N^{2} 
\nonu\\
&+& 5728 k^{3} N^{2} + 
       368 k^{4} N^{2} + 18530 N^{3} + 27770 k\, 
      N^{3} + 12456 k^{2} N^{3} + 1580 k^{3} N^{3} + 9297 N^{4} 
\nonu\\
&+&   9700 k\, N^{4} + 2300 k^{2} N^{4} + 2396 N^{5} + 1300 k\, 
      N^{5} + 248 N^{6}), \nonu\\
c_ {241} & = &\frac {(60 + 77 k + 22 k^{2} + 121 N + 115 k\, 
    N + 20 k^{2} N + 79 N^{2} + 42 k\, 
    N^{2} + 16 N^{3})} {(2 + N) (2 + k + N)^{2}}, \nonu\\
c_ {242} & = &\frac {8} {(2 + k + N)^{2}}, \qquad
c_ {243}  =  - \frac {8 i (1 + k + N)} {(2 + k + N)^{2}}, \qquad
c_ {244}  = \frac {16 i} {(2 + k + N)^{2}}, \nonu \\
c_ {245}  & = &  - \frac {4 i} {(2 + k + N)^{2}}, \qquad
c_ {246}  = \frac {i (60 + 71 k + 18 k^{2} + 67 N + 43 k\, 
     N + 17 N^{2})} {3 (2 + N) (2 + k + N)^{2}}, \nonu\\
c_ {247} & = &\frac {i (20 + 21 k + 6 k^{2} + 25 N + 13 k\, 
     N + 7 N^{2})} {(2 + N) (2 + k + N)^{2}}, \nonu\\
c_ {248} & = & - \frac {4 i} {(2 + k + N)^{2}}, \qquad
c_ {249}  =  - \frac {4} {(2 + k + N)}, \qquad
c_ {250}  = \frac {8} {(2 + k + N)^{2}}, \nonu\\
c_ {251} & = &\frac {2 (60 + 71 k + 18 k^{2} + 67 N + 43 k\, 
     N + 17 N^{2})} {3 (2 + N) (2 + k + N)^{3}}, \nonu\\
c_ {252} & = &\frac {2 (60 + 59 k + 18 k^{2} + 79 N + 37 k\, 
     N + 23 N^{2})} {3 (2 + N) (2 + k + N)^{3}}, \nonu\\
c_ {253} & = &\frac {2 (20 + 21 k + 6 k^{2} + 25 N + 13 k\, 
     N + 7 N^{2})} {(2 + N) (2 + k + N)^{3}}, \nonu\\
c_ {254} & = &\frac {4 (k - N)} {(2 + k + N)^{3}}, \qquad
c_ {255}  =  - \frac {8 i} {(2 + k + N)^{2}}, \qquad
c_ {256}  = \frac {8 i} {(2 + k + N)^{2}}, \nonu\\
c_ {257} & = &\frac {2 i (-60 - 37 k - 14 k^{2} - 41 N + 43 k\, 
     N + 20 k^{2} N + 31 N^{2} + 42 k\, 
     N^{2} + 16 N^{3})} {3 (2 + N) (2 + k + N)^{3}}, \nonu\\
c_ {258} & = &\frac {2 i (60 + 83 k + 26 k^{2} + 175 N + 187 k\, 
     N + 40 k^{2} N + 141 N^{2} + 84 k\, 
     N^{2} + 32 N^{3})} {3 (2 + N) (2 + k + N)^{3}}, \nonu\\
c_ {259} & = &\frac {4 i (30 + 30 k + 10 k^{2} + 39 N + 21 k\, 
     N + 8 N^{2})} {3 (2 + k + N)^{3}}, \nonu\\
c_ {260} & = & - \frac {(-60 - 61 k - 14 k^{2} - 17 N + 31 k\, 
     N + 20 k^{2} N + 43 N^{2} + 42 k\, 
     N^{2} + 16 N^{3})} {3 (2 + N) (2 + k + N)^{2}}, \nonu\\
c_ {261} & = &\frac {2 (60 + 59 k + 18 k^{2} + 79 N + 37 k\, 
     N + 23 N^{2})} {3 (2 + N) (2 + k + N)^{3}}, \nonu\\
c_ {262} & = &\frac {2 (20 + 21 k + 6 k^{2} + 25 N + 13 k\, 
     N + 7 N^{2})} {(2 + N) (2 + k + N)^{3}}, \nonu\\
c_ {263} & = &\frac {4 (32 + 55 k + 18 k^{2} + 41 N + 35 k\, 
     N + 11 N^{2})} {(2 + N) (2 + k + N)^{4}}, \nonu\\
c_ {264} & = &\frac {2 (60 + 59 k + 18 k^{2} + 79 N + 37 k\, 
     N + 23 N^{2})} {3 (2 + N) (2 + k + N)^{3}}, \nonu\\
c_ {265} & = &\frac {2 (60 + 71 k + 18 k^{2} + 67 N + 43 k\, 
     N + 17 N^{2})} {3 (2 + N) (2 + k + N)^{3}}, \nonu\\
c_ {266} & = &\frac {2 (60 + 47 k + 18 k^{2} + 91 N + 31 k\, 
     N + 29 N^{2})} {3 (2 + N) (2 + k + N)^{3}}, \nonu\\
c_ {267} & = &\frac {2 (20 + 21 k + 6 k^{2} + 25 N + 13 k\, 
     N + 7 N^{2})} {(2 + N) (2 + k + N)^{3}}, \nonu\\
c_ {268} & = & - \frac {2 i (8 + 17 k + 6 k^{2} + 11 N + 11 k\, 
      N + 3 N^{2})} {(2 + N) (2 + k + N)^{3}}, \nonu\\
c_ {269} & = & - \frac {2 i (13 k + 6 k^{2} + 3 N + 9 k\, 
      N + N^{2})} {(2 + N) (2 + k + N)^{3}}, \nonu\\
c_ {270} & = &\frac {2 i (24 + 25 k + 6 k^{2} + 27 N + 15 k\, 
     N + 7 N^{2})} {(2 + N) (2 + k + N)^{3}}, \nonu\\
c_ {271} & = &\frac {2 i (16 + 21 k + 6 k^{2} + 19 N + 13 k\, 
     N + 5 N^{2})} {(2 + N) (2 + k + N)^{3}}, \nonu\\
c_ {272} & = &\frac {4 i (8 + 17 k + 6 k^{2} + 11 N + 11 k\, 
     N + 3 N^{2})} {(2 + N) (2 + k + N)^{3}}, \nonu\\
c_ {273} & = &\frac {4 i (12 + 19 k + 6 k^{2} + 15 N + 12 k\, 
     N + 4 N^{2})} {(2 + N) (2 + k + N)^{3}}, \qquad
c_ {274}  = \frac {4 i} {(2 + k + N)^{2}}, \nonu\\
c_ {275} & = &\frac {4 (16 + 21 k + 6 k^{2} + 19 N + 13 k\, 
     N + 5 N^{2})} {(2 + N) (2 + k + N)^{3}}, \nonu\\
c_ {276} & = & - \frac {4 (24 + 7 k - 2 k^{2} + 29 N + 9 k\, 
      N + 2 k^{2} N + 13 N^{2} + 4 k\, 
      N^{2} + 2 N^{3})} {(2 + N) (2 + k + N)^{3}}, \nonu\\
c_ {277} & = &\frac {8 i (60 + 99 k + 34 k^{2} + 99 N + 85 k\, 
     N + 8 k^{2} N + 49 N^{2} + 14 k\, 
     N^{2} + 8 N^{3})} {3 (2 + N) (2 + k + N)^{4}}, \nonu\\
c_ {278} & = &\frac {8 i (16 + 29 k + 10 k^{2} + 27 N + 25 k\, 
     N + 2 k^{2} N + 13 N^{2} + 4 k\, 
     N^{2} + 2 N^{3})} {(2 + N) (2 + k + N)^{4}}, \nonu\\
c_ {279} & = &\frac {8 i (24 + 63 k + 22 k^{2} + 45 N + 55 k\, 
     N + 2 k^{2} N + 19 N^{2} + 8 k\, 
     N^{2} + 2 N^{3})} {3 (2 + N) (2 + k + N)^{4}}, \nonu\\
c_ {280} & = &\frac {8 i (32 + 45 k + 14 k^{2} + 51 N + 41 k\, 
     N + 4 k^{2} N + 25 N^{2} + 8 k\, 
     N^{2} + 4 N^{3})} {(2 + N) (2 + k + N)^{4}}, \nonu\\
c_ {281} & = &\frac {4 (12 + 51 k + 20 k^{2} + 27 N + 41 k\, 
     N + k^{2} N + 11 N^{2} + 4 k\, 
     N^{2} + N^{3})} {3 (2 + N) (2 + k + N)^{3}}, \nonu\\
c_ {282} & = &\frac {4 (16 + 25 k + 8 k^{2} + 23 N + 19 k\, 
     N + k^{2} N + 9 N^{2} + 2 k\, 
     N^{2} + N^{3})} {(2 + N) (2 + k + N)^{3}}, \nonu\\
c_ {283} & = & - \frac {8 (3 + 6 k + 2 k^{2} + 6 N + 5 k\, 
      N + 2 N^{2})} {3 (2 + k + N)^{3}}, \qquad
c_ {284}  = \frac {8} {(2 + k + N)^{2}}, \nonu\\
c_ {285} & = &\frac {16 i (3 + 6 k + 2 k^{2} + 6 N + 5 k\, 
     N + 2 N^{2})} {3 (2 + k + N)^{3}},\qquad
 c_ {286}  =  - 1, \nonu\\
c_ {287} & = &\frac {i (-8 + 13 k + 6 k^{2} - N + 9 k\, 
     N + N^{2})} {(2 + N) (2 + k + N)^{2}}, \nonu\\
c_ {288} & = &\frac {i (32 + 21 k + 6 k^{2} + 27 N + 13 k\, 
     N + 5 N^{2})} {(2 + N) (2 + k + N)^{2}}, \nonu\\
c_ {289} & = &\frac {i (16 + 21 k + 6 k^{2} + 19 N + 13 k\, 
     N + 5 N^{2})} {(2 + N) (2 + k + N)^{2}}, \nonu\\
c_ {290} & = & - \frac {i (24 + 3 k - 6 k^{2} + 25 N - k\, 
      N + 7 N^{2})} {(2 + N) (2 + k + N)^{2}}, \qquad
c_ {291}  =  - \frac {4} {(2 + k + N)^{2}}, \nonu \\
c_ {292} & = &  - \frac {4} {(2 + k + N)^{2}}, \qquad
c_ {293}  = \frac {2 (13 k + 6 k^{2} + 3 N + 9 k\, 
     N + N^{2})} {(2 + N) (2 + k + N)^{3}}, \nonu\\
c_ {294} & = &\frac {2 (24 + 25 k + 6 k^{2} + 27 N + 15 k\, 
     N + 7 N^{2})} {(2 + N) (2 + k + N)^{3}}, \nonu\\
c_ {295} & = & - \frac {2 (8 - 9 k - 6 k^{2} + 5 N - 7 k\, 
      N + N^{2})} {(2 + N) (2 + k + N)^{3}}, \nonu\\
c_ {296} & = & - \frac {4 i (32 + 55 k + 18 k^{2} + 41 N + 35 k\, 
      N + 11 N^{2})} {(2 + N) (2 + k + N)^{4}}, \nonu\\
c_ {297} & = &\frac {2 (8 + 17 k + 6 k^{2} + 11 N + 11 k\, 
     N + 3 N^{2})} {(2 + N) (2 + k + N)^{3}}, \nonu\\
c_ {298} & = &\frac {2 (16 + 21 k + 6 k^{2} + 19 N + 13 k\, 
     N + 5 N^{2})} {(2 + N) (2 + k + N)^{3}}, \nonu\\
c_ {299} & = & - \frac {2 (16 - 5 k - 6 k^{2} + 13 N - 5 k\, 
      N + 3 N^{2})} {(2 + N) (2 + k + N)^{3}}, \nonu\\
c_ {300} & = &\frac {2 (16 + 21 k + 6 k^{2} + 19 N + 13 k\, 
     N + 5 N^{2})} {(2 + N) (2 + k + N)^{3}}, \nonu\\
c_ {301} & = &\frac {2 (32 + 29 k + 6 k^{2} + 35 N + 17 k\, 
     N + 9 N^{2})} {(2 + N) (2 + k + N)^{3}}, \nonu\\
c_ {302} & = &\frac {2 (24 + 25 k + 6 k^{2} + 27 N + 15 k\, 
     N + 7 N^{2})} {(2 + N) (2 + k + N)^{3}}, \nonu\\
c_ {303} & = &\frac {2 (13 k + 6 k^{2} + 3 N + 9 k\, 
     N + N^{2})} {(2 + N) (2 + k + N)^{3}}, \qquad
c_ {304}  =  - \frac {4 i (k - N)} {3 (2 + k + N)^{3}}, \nonu\\
c_ {305} & = &\frac {2 i (20 + 21 k + 6 k^{2} + 25 N + 13 k\, 
     N + 7 N^{2})} {(2 + N) (2 + k + N)^{3}}, \nonu\\
c_ {306} & = &\frac {2 i (60 + 59 k + 18 k^{2} + 79 N + 37 k\, 
     N + 23 N^{2})} {3 (2 + N) (2 + k + N)^{3}}, \nonu\\
c_ {307} & = &\frac {2 i (60 + 71 k + 18 k^{2} + 67 N + 43 k\, 
     N + 17 N^{2})} {3 (2 + N) (2 + k + N)^{3}}, \qquad
c_ {308}  = \frac {8 i (k - N)} {3 (2 + k + N)^{3}}, \nonu\\
c_ {309} & = &\frac {4 (60 + 57 k + 10 k^{2} + 141 N + 105 k\, 
     N + 14 k^{2} N + 101 N^{2} + 42 k\, 
     N^{2} + 22 N^{3})} {3 (2 + N) (2 + k + N)^{3}}, \nonu\\
c_ {310} & = & - \frac {4 (60 + 77 k + 22 k^{2} + 121 N + 115 k\, 
      N + 20 k^{2} N + 79 N^{2} + 42 k\, 
      N^{2} + 16 N^{3})} {3 (2 + N) (2 + k + N)^{3}}, \nonu\\
c_ {311} & = & - \frac {8 i (60 + 85 k + 26 k^{2} + 113 N + 119 k\, 
      N + 22 k^{2} N + 71 N^{2} + 42 k\, 
      N^{2} + 14 N^{3})} {3 (2 + N) (2 + k + N)^{4}}, \nonu\\
c_ {312} & = & - \frac {8 i (60 + 77 k + 22 k^{2} + 121 N + 115 k\, 
      N + 20 k^{2} N + 79 N^{2} + 42 k\, 
      N^{2} + 16 N^{3})} {3 (2 + N) (2 + k + N)^{4}}, \nonu\\
c_ {313} & = & - \frac {16 i (1 + N) (20 + 23 k + 6 k^{2} + 23 N + 
       14 k\, N + 6 N^{2})} {(2 + N) (2 + k + N)^{4}}, \nonu\\
c_ {314} & = & - \frac {16 i (60 + 89 k + 28 k^{2} + 109 N + 121 k\, 
      N + 23 k^{2} N + 67 N^{2} + 42 k\, 
      N^{2} + 13 N^{3})} {3 (2 + N) (2 + k + N)^{4}}, \nonu\\
c_ {315} & = &\frac {4 (17 k + 8 k^{2} + 13 N + 43 k\, 
     N + 13 k^{2} N + 18 N^{2} + 21 k\, 
     N^{2} + 5 N^{3})} {3 (2 + N) (2 + k + N)^{3}}, \nonu\\
c_ {316} & = & - \frac {2 (20 k + 8 k^{2} + 40 N + 79 k\, 
      N + 22 k^{2} N + 51 N^{2} + 42 k\, 
      N^{2} + 14 N^{3})} {3 (2 + N) (2 + k + N)^{3}}, \nonu\\
c_ {317} & = &\frac {4 (30 + 38 k + 10 k^{2} + 31 N + 21 k\, 
     N + 8 N^{2})} {3 (2 + k + N)^{3}}, \nonu\\
c_ {318} & = & - \frac {4 (30 + 26 k + 6 k^{2} + 43 N + 21 k\, 
      N + 12 N^{2})} {3 (2 + k + N)^{3}}.
\nonu
\eea

The fusion rule is
\bea
[\Phi_{1}^{(1),ij}] \, \cdot \, [\widetilde{\Phi}_{\frac{3}{2}}^{(1),k}]
& = & [I^{ijk}]+ \delta^{ik} [\Phi_0^{(1)} \, \Phi_{\frac{1}{2}}^{(1),j}]
+\varepsilon^{ijkl} [\Phi_0^{(1)} \, \Phi_{\frac{1}{2}}^{(1),l}] 
+ \delta^{ik} [
\Phi_{0}^{(1)} \, \widetilde{\Phi}_{\frac{3}{2}}^{(1),j}] + \delta^{ik} \,
 [\Phi_{\frac{1}{2}}^{(1),a} \, \Phi_{1}^{(1),ja}]
\nonu \\
& + &  \delta^{ik} \, [\Phi_{\frac{1}{2}}^{(2),j}]+
 \varepsilon^{ijkl} \, [\Phi_{\frac{1}{2}}^{(2),l}]
+ \delta^{ik} \,  [\widetilde{\Phi}_{\frac{3}{2}}^{(2),j}]
+ (j \leftrightarrow k).
\nonu
\eea

The OPEs between the higher spin-$2$ currents and
the higher spin-$3$ current  are given by
\bea
&&\Phi_{1}^{(1),ij}(z)\:\widetilde{\Phi}_{2}^{(1)}(w)\;=\;\frac{1}{(z-w)^{4}}\Bigg[c_{1}\, T^{ij}+c_{2}\,\Gamma^{i}\,\Gamma^{j}+\varepsilon^{ijkl}\,\Big(c_{3}\, T^{kl}+c_{4}\,\Gamma^{k}\,\Gamma^{l}\Big)\Bigg](w)
\nonu\\ & &
+\frac{1}{(z-w)^{3}}\Bigg[c_{5}\,(G^{i}\,\Gamma^{j}-G^{j}\,\Gamma^{i})+c_{6}\,\partial T^{ij}+c_{7}\, T^{ij}\, U+c_{8}\,(T^{ik}\,\Gamma^{j}\,\Gamma^{k}-T^{jk}\,\Gamma^{i}\,\Gamma^{k})
\nonu\\ & &
+c_{9}\, U\,\Gamma^{i}\,\Gamma^{j}
+c_{10}\,\partial(\Gamma^{i}\,\Gamma^{j})+\varepsilon^{ijkl}\,\Bigg\{c_{11}\, U\,\Gamma^{k}\,\Gamma^{l}+c_{12}\, G^{k}\,\Gamma^{l}+c_{13}\,\partial T^{ki}+c_{14}\, T^{kl}\, U
\nonu\\ & &
+c_{15}\,(T^{ik}\,\Gamma^{i}\,\Gamma^{l}+T^{jk}\,\Gamma^{j}\,\Gamma^{l}\Big)+c_{16}\,\partial(\Gamma^{k}\,\Gamma^{l})\Bigg\}\Bigg](w)
\nonu\\ & & 
+\frac{1}{(z-w)^{2}}\Bigg[ c_{17}\,{\bf \Phi_{1}^{(2),ij}}+
c_{18}\, {\bf \Phi_{0}^{(1)}\,\Phi_{1}^{(1),ij}}+
c_{19}\, L\, T^{ij}+c_{20}\, L\,\Gamma^{i}\,\Gamma^{j}+c_{21}\, G^{i}\, G^{j}+c_{22}\, G^{k}\, T^{ij}\,\Gamma^{k}
\nonu\\ & &
+c_{23}\,(G^{i}\, U\,\Gamma^{j}-G^{i}\, U\,\Gamma^{j})+c_{24}\,(\partial G^{i}\,\Gamma^{j}-\partial G^{i}\,\Gamma)+c_{25}\,(G^{i}\,\partial\Gamma^{j}-G^{j}\,\partial\Gamma^{i})+c_{26}\,\partial^{2}T^{ij}
\nonu\\ & &
+c_{27}\,(\partial T^{ik}\, T^{jk}-T^{ik}\,\partial T^{jk})+c_{28}\,\Big(T^{ij}\,T^{kl}\,T^{kl}-\varepsilon^{ijkl}\, T^{ij}\,T^{kl}\,T^{kl}+2\, T^{ik}\, T^{jl}\, T^{kl}\Big)
\nonu\\ & &
+c_{29}\, T^{ij}\,\partial\Gamma^{4-ijk}\,\Gamma^{4-ijk}+c_{30}\,\partial T^{ij}\, U+c_{31}\, T^{ij}\,\partial U+c_{32}\,\Big(T^{ij}\, T^{kl}\,\Gamma^{k}\,\Gamma^{l}-\varepsilon^{ijkl}\, T^{kl}\,\Gamma^{k}\,\Gamma^{l}
\nonu\\ & &+2\, T^{ik}\, T^{jl}\,\Gamma^{k}\,\Gamma^{l}\Big)+c_{33}\, T^{ij}\,U\,U+c_{34}\,(T^{ij}\,\partial\Gamma^{i}\,\Gamma^{i}+T^{ij}\,\partial\Gamma^{j}\,\Gamma^{j})+c_{35}\,(\partial T^{ik}\,\Gamma^{j}\,\Gamma^{k}-\partial T^{jk}\,\Gamma^{i}\,\Gamma^{k})
\nonu\\ & &+c_{36}\,(T^{ik}\,\partial\Gamma^{j}\,\Gamma^{k}-T^{jk}\,\partial\Gamma^{i}\,\Gamma^{k})+c_{37}\,(T^{ik}\,\Gamma^{j}\,\partial\Gamma^{k}-T^{jk}\,\Gamma^{i}\,\partial\Gamma^{k})+c_{38}\,U\,U\,\Gamma^{i}\,\Gamma^{j}
\nonu\\ & &+c_{39}\, U\,\partial(\Gamma^{i}\,\Gamma^{j})
+c_{40}\,\partial U\,\Gamma^{i}\,\Gamma^{j}+c_{41}\,(\partial^{2}\Gamma^{i}\,\Gamma^{j}+\Gamma^{i}\,\partial^{2}\Gamma^{j})+c_{42}\,\partial\Gamma^{i}\,\partial\Gamma^{j}+c_{43}\,\Gamma^{i}\,\Gamma^{j}\,\partial\Gamma^{k}\,\Gamma^{k}
\nonu\\ & &+\varepsilon^{ijkl}\,\Bigg\{
 c_{44}\, {\bf \Phi_{\frac{1}{2}}^{(1),k}\,\Phi_{\frac{1}{2}}^{(1),l}}+
c_{45}\, L\, T^{kl}+c_{46}\, L\,\Gamma^{k}\,\Gamma^{l}+c_{47}\, G^{k}\, G^{l}+c_{48}\,\partial G^{k}\,\Gamma^{l}+c_{49}\, G^{k}\,\partial\Gamma^{l}
\nonu\\ & &+c_{50}\,(G^{i}\, T^{kl}\,\Gamma^{i}-G^{i}\, T^{kl}\,\Gamma^{i})+c_{51}\,(G^{i}\, T^{ik}\,\Gamma^{l}+G^{j}\, T^{jk}\,\Gamma^{l})+c_{52}\,(G^{k}\, T^{il}\,\Gamma^{i}+G^{k}\, T^{jl}\,\Gamma^{j})
\nonu\\ & &+c_{53}\, G^{k}\, U\,\Gamma^{l}+c_{54}\,(G^{i}\,\Gamma^{i}\,\Gamma^{k}\,\Gamma^{l}+G^{j}\,\Gamma^{j}\,\Gamma^{k}\,\Gamma^{l})+c_{55}\,\partial^{2}T^{kl}+c_{56}\,(\partial T^{ik}\, T^{il}+\partial T^{jk}\, T^{jl})
\nonu\\ & &+c_{57}\,(T^{ik}\,\partial T^{il}+T^{jk}\,\partial T^{jl})+c_{58}\,\partial T^{kl}\, U+c_{59}\, T^{kl}\,\partial U+c_{60}\, T^{kl}\,U\,U
\nonu\\ & &+c_{61}\,(T^{kl}\,\partial\Gamma^{i}\,\Gamma^{i}+T^{kl}\,\partial\Gamma^{j}\,\Gamma^{j})+c_{62}\, T^{kl}\,\partial\Gamma^{k}\,\Gamma^{k}+c_{63}\,(T^{ik}\, U\,\Gamma^{i}\,\Gamma^{l}+T^{jk}\, U\,\Gamma^{j}\,\Gamma^{l})
\nonu\\ & &+c_{64}\,\partial U\,\Gamma^{k}\,\Gamma^{l}+c_{65}\, U\,\partial(\Gamma^{k}\,\Gamma^{l})+c_{66}\,(\partial\Gamma^{i}\,\Gamma^{i}\,\Gamma^{k}\,\Gamma^{l}+\partial\Gamma^{i}\,\Gamma^{i}\,\Gamma^{k}\,\Gamma^{l})
\nonu\\ & &+c_{67}\,(\partial T^{ik}\,\Gamma^{i}\,\Gamma^{l}+\partial T^{jk}\,\Gamma^{j}\,\Gamma^{l})+c_{68}\,(T^{ik}\,\partial\Gamma^{i}\,\Gamma^{l}+T^{jk}\,\partial\Gamma^{j}\,\Gamma^{l})+c_{69}\,(T^{ik}\,\Gamma^{i}\,\partial\Gamma^{l}+T^{jk}\,\Gamma^{j}\,\partial\Gamma^{l})
\nonu\\ & &+c_{70}\,(\partial T^{il}\,\Gamma^{i}\,\Gamma^{k}+\partial T^{jl}\,\Gamma^{j}\,\Gamma^{k})+c_{71}\,(T^{il}\,\partial\Gamma^{i}\,\Gamma^{k}+T^{jl}\,\partial\Gamma^{j}\,\Gamma^{k})+c_{72}\,(T^{il}\,\Gamma^{i}\,\partial\Gamma^{k}+T^{jl}\,\Gamma^{j}\,\partial\Gamma^{k})
\nonu\\ & &+c_{73}\,\partial^{2}\Gamma^{k}\,\Gamma^{l}+c_{74}\,\partial\Gamma^{k}\,\partial\Gamma^{l}+c_{75}\,\Gamma^{k}\,\partial^{2}\Gamma^{l}\Bigg\}+\varepsilon^{abcd}\,c_{76}\, T^{ij}\, T^{ab}\, T^{cd}\Bigg](w)
\nonu\\ & &+\frac{1}{(z-w)}\Bigg[ 
c_{77}\, {\bf \partial\Phi_{\frac{1}{2}}^{(2),ij}}+
c_{78}\, {\bf \partial\Phi_{0}^{(1)}\,\Phi_{1}^{(1),ij}}+
c_{79}\, {\bf \Phi_{0}^{(1)}\,\partial\Phi_{1}^{(1),ij}}+
c_{80}\,(L\, G^{i}\,\Gamma^{j}-L\, G^{j}\,\Gamma^{i})
\nonu\\ & &+c_{81}\,\partial L\, T^{ij}+c_{82}\, L\,\partial T^{ij}+c_{83}\, L\, U\,\Gamma^{i}\,\Gamma^{j}+c_{84}\,\partial L\,\Gamma^{i}\,\Gamma^{j}+c_{85}\, L\,\partial(\Gamma^{i}\,\Gamma^{j})+c_{86}\,\partial(G^{i}\, G^{j})
\nonu\\ & &+c_{87}\,\partial(G^{k}\, T^{ij}\,\Gamma^{k})+c_{88}\,\Big(\partial(G^{i}\,\Gamma^{j})\, U-\partial(G^{j}\,\Gamma^{i})\, U\Big)+c_{89}\,(G^{i}\,\partial U\,\Gamma^{j}-G^{i}\,\partial U\,\Gamma^{j})
\nonu\\ & &+c_{90}\,(G^{i}\,\partial^{2}\Gamma^{j}-G^{i}\,\partial^{2}\Gamma^{j})+c_{91}\,(\partial G^{i}\,\partial\Gamma^{j}-\partial G^{i}\,\partial\Gamma^{j})+c_{92}\,(\partial^{2}G^{i}\,\Gamma^{j}-\partial^{2}G^{i}\,\Gamma^{j})
\nonu\\ & &+c_{93}\,(\partial^{2}T^{ik}\, T^{jk}-T^{ik}\,\partial^{2}T^{jk})+c_{94}\,\partial T^{ij}\,T^{ij}\,T^{ij}
\nonu\\ & &+(1-\delta^{ik}-\delta^{jk})\Bigg(c_{95}\,\Big[\partial T^{ij}\,T^{ik}\,T^{ik}+\partial T^{ij}\,T^{jk}\,T^{jk}+\partial(T^{il}\, T^{jk})\, T^{lk}\Big]
\nonu\\ & &
+c_{96}\,(T^{ij}\,\partial T^{ik}\, T^{ik}+T^{ij}\,\partial T^{jk}\, T^{jk})+c_{97}\,\Big[\partial^{2}(T^{ij}\,\Gamma^{k})\,\Gamma^{k}-\partial^{2}T^{ij}\,\Gamma^{i}\,\Gamma^{i}\Big]
\nonu\\ & &+c_{98}\,(\partial T^{ik}\,\partial\Gamma^{j}\,\Gamma^{k}-\partial T^{ik}\,\partial\Gamma^{j}\,\Gamma^{k})+c_{99}\,(T^{ik}\,\partial\Gamma^{j}\,\partial\Gamma^{k}-T^{ik}\,\partial\Gamma^{j}\,\partial\Gamma^{k})
\nonu\\ & &+c_{100}\,\Big[\partial T^{ik}\,\Gamma^{j}\,\partial\Gamma^{k}-\partial T^{ik}\,\Gamma^{j}\,\partial\Gamma^{k}\Big]\Bigg)+c_{101}\, T^{ik}\, T^{jl}\,\partial T^{kl}
\nonu\\ & &+c_{102}\,\Big[\,\frac{1}{2}\partial(T^{ij}\, T^{kl}\,\Gamma^{k}\,\Gamma^{l}+T^{ij}\, T^{kl}\,\Gamma^{4-ijl}\,\Gamma^{4-ijk})+\partial(T^{ik}\, T^{jl}\,\Gamma^{k}\,\Gamma^{l})\,\Big]+c_{103}\,\partial^{3}T^{ij}\,
\nonu\\ & &+c_{104}\,\partial^{2}T^{ij}\, U+c_{105}\,\partial T^{ij}\,U\,U+c_{106}\, T^{ij}\,\partial U\, U+c_{107}\, T^{ij}\,\partial^{2}U+c_{108}\,\partial T^{ij}\,\partial U
\nonu\\ & &+c_{109}\,(\partial^{2}T^{ik}\,\Gamma^{j}\,\Gamma^{k}-\partial^{2}T^{jk}\,\Gamma^{i}\,\Gamma^{k})+c_{110}\,(T^{ik}\,\partial^{2}\Gamma^{j}\,\Gamma^{k}-T^{jk}\,\partial^{2}\Gamma^{i}\,\Gamma^{k})
\nonu\\ & &+c_{111}\,(T^{ik}\,\Gamma^{j}\,\partial^{2}\Gamma^{k}-T^{jk}\,\Gamma^{i}\,\partial^{2}\Gamma^{k})+c_{112}\,(T^{ij}\,\partial^{2}\Gamma^{i}\,\Gamma^{i}+T^{ij}\,\partial^{2}\Gamma^{j}\,\Gamma^{j})
\nonu\\ & &
+c_{113}\,(\partial T^{ij}\,\partial\Gamma^{i}\,\Gamma^{i}+\partial T^{ij}\,\partial\Gamma^{j}\,\Gamma^{j})+c_{114}\,\partial^{2}U\,\Gamma^{i}\,\Gamma^{j}+c_{115}\,\partial(U\,U\,\Gamma^{i}\,\Gamma^{j})+c_{116}\,\partial U\,\partial(\Gamma^{i}\,\Gamma^{j})
\nonu\\ & &+c_{117}\, U\,\partial^{2}(\Gamma^{i}\,\Gamma^{j})+c_{118}\,\partial(\Gamma^{i}\,\Gamma^{j}\,\partial\Gamma^{k})\,\Gamma^{k}+c_{119}\,(\partial^{3}\Gamma^{i}\,\Gamma^{j}+\Gamma^{i}\,\partial^{3}\Gamma^{j})+c_{120}\,(\partial^{2}\Gamma^{i}\,\partial\Gamma^{j}+\partial\Gamma^{i}\,\partial^{2}\Gamma^{j})
\nonu\\ & &
+\varepsilon^{ijkl}\,\Bigg\{ c_{121}\, {\bf \partial(\Phi_{\frac{1}{2}}^{(1),k}\,
\Phi_{\frac{1}{2}}^{(1),l})}+
c_{122}\,\partial L\, T^{kl}+c_{123}\, L\,\partial T^{kl}+c_{124}\, L\, T^{ak}\,\Gamma^{a}\,\Gamma^{l}+c_{125}\,\partial L\,\Gamma^{k}\,\Gamma^{l}
\nonu\\ & &+c_{126}\, L\,\partial(\Gamma^{k}\,\Gamma^{l})+c_{127}\,\partial(G^{k}\, G^{l})+c_{128}\,\Big(\partial(G^{i}\, T^{ik})\,\Gamma^{l}+\partial(G^{j}\, T^{jk})\,\Gamma^{l}\Big)
\nonu\\ & &
+c_{129\,}(G^{i}\, T^{ik}\,\partial\Gamma^{l}+G^{j}\, T^{jk}\,\partial\Gamma^{l})
+c_{130}\,(\partial G^{i}\, T^{kl}\,\Gamma^{i}+\partial G^{j}\, T^{kl}\,\Gamma^{j}+G^{i}\, T^{kl}\,\partial\Gamma^{i}+G^{j}\, T^{kl}\,\partial\Gamma^{j})
\nonu\\ & &
+c_{131}\,(G^{i}\,\partial T^{kl}\,\Gamma^{i}+G^{j}\,\partial T^{kl}\,\Gamma^{j})
+c_{132}\,\Big(G^{k}\,\partial(T^{il}\,\Gamma^{i})+G^{k}\,\partial(T^{jl}\,\Gamma^{j})\Big)
\nonu\\ & &
+c_{133}\,(\partial G^{k}\, T^{il}\,\Gamma^{i}+\partial G^{k}\, T^{jl}\,\Gamma^{j})+c_{134}\, G^{k}\,\partial(U\,\Gamma^{l})
+c_{135}\,\partial(G^{a}\,\Gamma^{a}\,\Gamma^{k}\,\Gamma^{l})
+c_{136}\, G^{k}\,\partial^{2}\Gamma^{l}
\nonu\\ & &
+c_{137}\,\partial G^{k}\,\partial\Gamma^{l}+c_{138}\,\partial^{2}G^{k}\,\Gamma^{l}+c_{139}\,\partial^{3}T^{kl}+c_{140}\,(\partial^{2}T^{ik}\, T^{il}+\partial^{2}T^{jk}\, T^{jl})
\nonu\\ & &+c_{141}\,(T^{ik}\,\partial^{2}T^{il}+T^{jk}\,\partial^{2}T^{jl})+c_{142}\,\Big(\partial T^{ij}\, T^{ij}\, T^{kl}-\partial(T^{ij}\, T^{ik}\, T^{jl})\Big)
\nonu\\ & &+c_{143}\,(\partial T^{ik}\,\partial T^{il}+\partial T^{jk}\,\partial T^{jl})+c_{144}\,\partial^{2}T^{kl}\, U+c_{145}\,\partial T^{kl}\,\partial U+c_{146}\, T^{kl}\,\partial^{2}U
+c_{147}\,\partial(T^{kl}\,U\,U)
\nonu\\ & &
+c_{148}\,\partial^{2}T^{ak}\,\Gamma^{a}\,\Gamma^{l}+c_{149}\,\partial^{2}T^{al}\,\Gamma^{a}\,\Gamma^{k}+c_{150}\, T^{ak}\,\partial^{2}\Gamma^{a}\,\Gamma^{l}+c_{151}\, T^{al}\,\partial^{2}\Gamma^{a}\,\Gamma^{k}
+c_{152}\, T^{ak}\,\Gamma^{a}\,\partial^{2}\Gamma^{l}
\nonu\\ & &
+c_{153}\, T^{al}\,\Gamma^{a}\,\partial^{2}\Gamma^{k}+c_{154}\,(T^{kl}\,\partial^{2}\Gamma^{i}\,\Gamma^{i}+T^{kl}\,\partial^{2}\Gamma^{j}\,\Gamma^{j})+c_{155}\,(T^{kl}\,\partial^{2}\Gamma^{k}\,\Gamma^{k})
\nonu\\ & &+c_{156}\,(\partial T^{ik}\,\partial\Gamma^{i}\,\Gamma^{l}+\partial T^{jk}\,\partial\Gamma^{j}\,\Gamma^{l})+c_{157}\,(\partial T^{il}\,\partial\Gamma^{i}\,\Gamma^{k}+\partial T^{jl}\,\partial\Gamma^{j}\,\Gamma^{k})
\nonu\\ & &+c_{158}\,(T^{ik}\,\partial\Gamma^{i}\,\partial\Gamma^{l}+T^{jk}\,\partial\Gamma^{j}\,\partial\Gamma^{l})+c_{159}\,(T^{il}\,\partial\Gamma^{i}\,\partial\Gamma^{k}+T^{jl}\,\partial\Gamma^{j}\,\partial\Gamma^{k})
\nonu\\ & &+c_{160}\,(\partial T^{ik}\,\Gamma^{i}\,\partial\Gamma^{l}+\partial T^{jk}\,\Gamma^{j}\,\partial\Gamma^{l})+c_{161}\,(\partial T^{il}\,\Gamma^{i}\,\partial\Gamma^{k}+\partial T^{jl}\,\Gamma^{j}\,\partial\Gamma^{k})
\nonu\\ & &+c_{162}\,(\partial T^{kl}\,\partial\Gamma^{i}\,\Gamma^{i}+\partial T^{kl}\,\partial\Gamma^{j}\,\Gamma^{j})+c_{163}\,\partial T^{kl}\,\partial\Gamma^{k}\,\Gamma^{k}+c_{164}\,\partial(T^{ik}\, U\,\Gamma^{i}\,\Gamma^{l}+T^{jk}\, U\,\Gamma^{j}\,\Gamma^{l})
\nonu\\ & &+c_{165}\,\partial U\,\partial(\Gamma^{k}\,\Gamma^{l})+c_{166}\, U\,\partial^{2}\Gamma^{k}\,\Gamma^{l}+c_{167}\, U\,\partial\Gamma^{k}\,\partial\Gamma^{l}+c_{168}\,\partial^{2}U\,\Gamma^{k}\,\Gamma^{l}+c_{169}\,\partial^{3}\Gamma^{k}\,\Gamma^{l}
\nonu\\ & &+c_{170}\,\partial^{2}\Gamma^{k}\,\partial\Gamma^{l}+c_{171}\,\partial\Gamma^{k}\,\partial^{2}\Gamma^{l}+c_{172}\,\Gamma^{k}\,\partial^{3}\Gamma^{l}+c_{173}\,\partial\Gamma^{i}\,\Gamma^{i}\,\partial(\Gamma^{k}\,\Gamma^{l})
\nonu\\ & &+c_{174}\,\Big(\partial\Gamma^{j}\,\Gamma^{j}\,\partial(\Gamma^{k}\,\Gamma^{l})+\partial^{2}\Gamma^{a}\,\Gamma^{a}\,\Gamma^{k}\,\Gamma^{l}\Big)\Bigg\}\Bigg](w)+\cdots,
\nonu
\eea
where the coefficients are
\bea
c_ {1} & = & - \frac{1}{(2 + k + N)^{2} (5 + 4 k + 4 N + 3 k\, N)}24 i (5 k + 6 k^{2} + 2 k^{3} + 5 N + 24 k\, 
      N + 17 k^{2} N 
\nonu\\&+& 2 k^{3} N + 6 N^{2} + 17 k\, 
      N^{2} + 8 k^{2} N^{2} + 2 N^{3} + 2 k\, 
      N^{3}), \nonu\\
c_ {2} & = & - \frac {1}{(2 + k + N)^{3} (5 + 4 k + 4 N + 3 k\, N)}48 (5 k + 6 k^{2} + 2 k^{3} + 5 N + 14 k\, 
      N + 9 k^{2} N 
\nonu\\&+& 2 k^{3} N + 6 N^{2} + 9 k\, 
      N^{2} + 2 k^{2} N^{2} + 2 N^{3} + 2 k\, 
      N^{3}), \nonu\\
c_ {3} & = & - \frac {12 i (k - N) (5 + 6 k + 2 k^{2} + 6 N + 7 k\, 
      N + 2 k^{2} N + 2 N^{2} + 2 k\, 
      N^{2})} {(2 + k + N)^{2} (5 + 4 k + 4 N + 3 k\, N)}, \nonu\\
c_ {4} & = & - \frac {24 (k - N) (5 + 6 k + 2 k^{2} + 6 N + 7 k\, 
      N + 2 k^{2} N + 2 N^{2} + 2 k\, 
      N^{2})} {(2 + k + N)^{3} (5 + 4 k + 4 N + 3 k\, N)}, \nonu\\
c_ {5} & = &\frac {4 i (k - N) (73 + 44 k + 44 N + 15 k\, 
     N)} {3 (2 + k + N)^{2} (5 + 4 k + 4 N + 3 k\, N)}, \nonu\\
c_ {6} & = &\frac {8 i (2 + 3 k + 3 N)} {(2 + k + N)^{2}}, \qquad
c_ {7}  =  - \frac {16 i (k - N)} {(2 + k + N)^{2}}, \nonu\\
c_ {8} & = & - \frac {8 i} {(2 + k + N)^{2}}, \qquad
c_ {9}  =  - \frac {16 (k - N) (-43 - 20 k - 20 N + 3 k\, 
      N)} {3 (2 + k + N)^{3} (5 + 4 k + 4 N + 3 k\, N)}, \nonu\\
c_ {10} & = &\frac {16} {(2 + k + N)^{2}}, \qquad
c_ {11}  = \frac {16} {(2 + k + N)^{2}}, \nonu\\
c_ {12} & = &\frac {2 i (2 + 3 k + 3 N)} {(2 + k + N)^{2}}, \qquad
c_ {13}  = \frac {4 i (k - N) (73 + 44 k + 44 N + 15 k\, 
     N)} {3 (2 + k + N)^{2} (5 + 4 k + 4 N + 3 k\, N)}, \nonu\\
c_ {14} & = & - \frac {8 i (k + N)} {(2 + k + N)^{2}}, \qquad
c_ {15}  = \frac {8 i (k - N) (-43 - 20 k - 20 N + 3 k\, 
     N)} {3 (2 + k + N)^{3} (5 + 4 k + 4 N + 3 k\, N)}, \nonu\\
c_ {16} & = & - \frac {8 (k - N) (-43 - 20 k - 20 N + 3 k\, 
      N)} {3 (2 + k + N)^{3} (5 + 4 k + 4 N + 3 k\, N)}, \qquad
c_ {17}  =   3, \nonu\\
c_ {18} & = & - \frac{1}{(2 + N) (2 + k + N)^{2} (5 + 4 k + 4 N + 3 k\, N)}(900 + 1747 k + 1126 k^{2} + 232 k^{3} + 
      2663 N 
\nonu\\&+& 4577 k\, 
     N + 2541 k^{2} N + 422 k^{3} N + 2829 N^{2} + 4079 k\, 
     N^{2} + 1763 k^{2} N^{2} + 180 k^{3} N^{2} + 1284 N^{3} 
\nonu\\&+& 
      1423 k\, N^{3} + 378 k^{2} N^{3} + 208 N^{4} + 144 k\, 
     N^{4}), \nonu\\
c_ {19} & = &\frac{1}{(2 + N) (2 + k + N)^{2} (5 + 4 k + 4 N + 
      3 k\, N)}2 i (-160 - 403 k - 278 k^{2} - 56 k^{3} - 
      333 N 
\nonu\\&-& 719 k\, 
     N - 377 k^{2} N - 46 k^{3} N - 187 N^{2} - 355 k\, 
     N^{2} - 113 k^{2} N^{2} - 12 N^{3} - 41 k\, 
     N^{3} + 8 N^{4}), \nonu\\
c_ {20} & = &\frac{1} {(2 + N) (2 + k + N)^{3} (5 + 4 k + 4 N + 
      3 k\, N)}4 (-80 - 299 k - 246 k^{2} - 56 k^{3} - 189 N 
\nonu\\&-&
      555 k\, N - 337 k^{2} N - 46 k^{3} N - 103 N^{2} - 275 k\, 
     N^{2} - 101 k^{2} N^{2} + 4 N^{3} - 29 k\, 
     N^{3} + 8 N^{4}), \nonu\\
c_ {21} & = &\frac {6 (8 + 17 k + 6 k^{2} + 11 N + 11 k\, 
     N + 3 N^{2})} {(2 + N) (2 + k + N)^{2}}, \nonu\\
c_ {22} & = &\frac {6 (20 + 21 k + 6 k^{2} + 25 N + 13 k\, 
     N + 7 N^{2})} {(2 + N) (2 + k + N)^{3}}, \nonu\\
c_ {23} & = & - \frac {4 i (32 + 55 k + 18 k^{2} + 41 N + 35 k\, 
      N + 11 N^{2})} {(2 + N) (2 + k + N)^{3}}, \nonu\\
c_ {24} & = & - \frac {1}{3 (2 + N) (2 + k + N)^{3} (5 + 4 k + 4 N + 3 k\, N)}2 i (118 k + 166 k^{2} + 40 k^{3} + 782 N 
\nonu\\&+&
       1814 k\, 
      N + 1407 k^{2} N + 302 k^{3} N + 1530 N^{2} + 2600 k\, 
      N^{2} + 1454 k^{2} N^{2} + 195 k^{3} N^{2} + 975 N^{3} 
\nonu\\&+& 
       1162 k\, N^{3} + 378 k^{2} N^{3} + 196 N^{4} + 129 k\, 
      N^{4}), \nonu\\
c_ {25} & = &\frac {2 i (k - N) (173 + 124 k + 124 N + 75 k\, 
     N)} {3 (2 + k + N)^{2} (5 + 4 k + 4 N + 3 k\, N)}, \nonu\\
c_ {26} & = &\frac {1}{(2 + N) (2 + k + N)^{3} (5 + 4 k + 4 N + 3 k\, N)}2 i (360 + 828 k + 545 k^{2} + 74 k^{3} - 
      16 k^{4} 
\nonu\\&+&
 1008 N + 1838 k\, 
     N + 806 k^{2} N - 59 k^{3} N - 50 k^{4} N + 1109 N^{2} + 
      1613 k\, 
     N^{2} + 415 k^{2} N^{2} 
\nonu\\&-& 138 k^{3} N^{2} - 30 k^{4} N^{2} + 
      639 N^{3} + 788 k\, 
     N^{3} + 151 k^{2} N^{3} - 33 k^{3} N^{3} + 204 N^{4} + 225 k\, 
     N^{4} 
\nonu\\&+& 39 k^{2} N^{4} + 28 N^{5} + 24 k\, 
     N^{5}), \nonu\\
c_ {27} & = &\frac {(176 + 400 k + 183 k^{2} + 18 k^{3} + 248 N + 
     308 k\, N + 45 k^{2} N + 77 N^{2} + 30 k\, 
    N^{2} + 3 N^{3})} {(2 + N) (2 + k + N)^{3}}, \nonu\\
c_ {28} & = &\frac { i (32 + 55 k + 18 k^{2} + 41 N + 35 k\, 
     N + 11 N^{2})} {(2 + N) (2 + k + N)^{3}}, \nonu\\
c_ {29} & = &\frac {6 i (96 + 174 k + 73 k^{2} + 6 k^{3} + 162 N + 
      166 k\, N + 27 k^{2} N + 81 N^{2} + 34 k\, 
     N^{2} + 13 N^{3})} {(2 + N) (2 + k + N)^{4}}, \nonu\\
c_ {30} & = &\frac {2 i (60 + 31 k + 2 k^{2} + 107 N + 23 k\, 
     N - 8 k^{2} N + 53 N^{2} + 8 N^{3})} {(2 + N) (2 + k + N)^{3}}, \nonu\\
c_ {31} & = & - \frac {2 i (180 + 173 k + 46 k^{2} + 241 N + 109 k\, 
      N - 4 k^{2} N + 79 N^{2} + 4 N^{3})} {(2 + N) (2 + k + N)^{3}}, \nonu\\
c_ {32} & = &\frac {6 (32 + 55 k + 18 k^{2} + 41 N + 35 k\, 
     N + 11 N^{2})} {(2 + N) (2 + k + N)^{4}}, \nonu\\
c_ {33} & = & - \frac {2 i (32 + 55 k + 18 k^{2} + 41 N + 35 k\, 
      N + 11 N^{2})} {(2 + N) (2 + k + N)^{3}}, \nonu\\
c_ {34} & = &\frac {1} {(2 + N) (2 + k + N)^{4}} 2 i (416 + 788 k + 
    311 k^{2} + 18 k^{3} + 636 N + 644 k\, 
   N + 73 k^{2} N + 261 N^{2}
\nonu\\&+& 86 k\, N^{2} + 31 N^{3}), \nonu\\
c_ {35} & = & - \frac {2 i (224 + 440 k + 191 k^{2} + 18 k^{3} + 
       312 N + 344 k\, N + 49 k^{2} N + 105 N^{2} + 38 k\, 
      N^{2} + 7 N^{3})} {(2 + N) (2 + k + N)^{4}}, \nonu\\
c_ {36} & = &\frac {2 i (176 + 392 k + 179 k^{2} + 18 k^{3} + 240 N + 
      296 k\, N + 43 k^{2} N + 69 N^{2} + 26 k\, 
     N^{2} + N^{3})} {(2 + N) (2 + k + N)^{4}}, \nonu\\
c_ {37} & = &\frac {6 i (8 + 17 k + 6 k^{2} + 11 N + 11 k\, 
     N + 3 N^{2})} {(2 + N) (2 + k + N)^{3}}, \nonu\\
c_ {38} & = & - \frac {12 (32 + 55 k + 18 k^{2} + 41 N + 35 k\, 
      N + 11 N^{2})} {(2 + N) (2 + k + N)^{4}}, \nonu\\
c_ {39} & = & - \frac{1}{3 (2 + N) (2 + k + N)^{4} (5 + 4 k + 4 N + 3 k\, N)}4 (900 + 1571 k + 974 k^{2} + 200 k^{3} + 
       2839 N 
\nonu\\&+& 4489 k\, 
      N + 2481 k^{2} N + 430 k^{3} N + 3069 N^{2} + 4063 k\, 
      N^{2} + 1771 k^{2} N^{2} + 192 k^{3} N^{2} + 1392 N^{3} 
\nonu\\&+&
       1391 k\, N^{3} + 378 k^{2} N^{3} + 224 N^{4} + 132 k\, 
      N^{4}), \nonu\\
c_ {40} & = & - \frac {1}{3 (2 + N) (2 + k + N)^{4} (5 + 4 k + 4 N + 3 k\, N)}8 (900 + 1663 k + 1036 k^{2} + 208 k^{3} 
\nonu\\&+& 
       2747 N + 4535 k\, 
      N + 2460 k^{2} N + 404 k^{3} N + 2961 N^{2} + 4115 k\, 
      N^{2} + 1745 k^{2} N^{2} + 177 k^{3} N^{2} 
\nonu\\&+& 1353 N^{3} + 
       1447 k\, N^{3} + 378 k^{2} N^{3} + 220 N^{4} + 147 k\, 
      N^{4}), \nonu\\
c_ {41} & = &\frac {1}{(2 + N) (2 + k + N)^{4} (5 + 4 k + 4 N + 3 k\, N)}2 (800 + 2340 k + 2041 k^{2} + 602 k^{3} + 
      40 k^{4} 
\nonu\\&+& 2140 N + 5008 k\, 
     N + 3235 k^{2} N + 463 k^{3} N - 46 k^{4} N + 2191 N^{2} + 
      4084 k\, 
     N^{2} + 1802 k^{2} N^{2} 
\nonu\\&-& 51 k^{3} N^{2} - 60 k^{4} N^{2} + 
      1183 N^{3} + 1795 k\, 
     N^{3} + 572 k^{2} N^{3} - 30 k^{3} N^{3} + 380 N^{4} + 477 k\, 
     N^{4} 
\nonu\\&+& 114 k^{2} N^{4} + 56 N^{5} + 48 k\, 
     N^{5}), \nonu\\
c_ {42} & = &\frac {1}{(2 + N) (2 + k + N)^{4} (5 + 4 k + 4 N + 3 k\, N)}8 (-320 - 1386 k - 1606 k^{2} - 668 k^{3} - 
      88 k^{4} 
\nonu\\&-& 646 N - 2494 k\, 
     N - 2593 k^{2} N - 928 k^{3} N - 104 k^{4} N - 244 N^{2} - 
      1255 k\, 
     N^{2} - 1349 k^{2} N^{2} 
\nonu\\&-& 417 k^{3} N^{2} - 30 k^{4} N^{2} + 
      212 N^{3} + 29 k\, 
     N^{3} - 227 k^{2} N^{3} - 69 k^{3} N^{3} + 160 N^{4} + 144 k\, 
     N^{4} 
\nonu\\&+& 3 k^{2} N^{4} + 28 N^{5} + 24 k\, 
     N^{5}), \nonu\\
c_ {43} & = &\frac {12 (32 + 55 k + 18 k^{2} + 41 N + 35 k\, 
     N + 11 N^{2})} {(2 + N) (2 + k + N)^{4}}, \nonu\\
c_ {44} & = &\frac {3 (60 + 77 k + 22 k^{2} + 121 N + 115 k\, 
     N + 20 k^{2} N + 79 N^{2} + 42 k\, 
     N^{2} + 16 N^{3})} {2(2 + N) (2 + k + N)^{2}}, \nonu\\
c_ {45} & = & - \frac {1}{(2 + N) (2 + k + N)^{2} (5 + 4 k + 4 N + 
       3 k\, N)}i (300 + 555 k + 310 k^{2} + 56 k^{3} + 
       615 N 
\nonu\\&+& 927 k\, 
      N + 385 k^{2} N + 46 k^{3} N + 437 N^{2} + 481 k\, 
      N^{2} + 109 k^{2} N^{2} + 116 N^{3} + 71 k\, 
      N^{3} + 8 N^{4}), \nonu\\
c_ {46} & = & - \frac {1}{(2 + N) (2 + k + N)^{3} (5 + 4 k + 4 N + 
       3 k\, N)}2 (300 + 555 k + 310 k^{2} + 56 k^{3} + 
       615 N
\nonu\\&+& 927 k\, 
      N + 385 k^{2} N + 46 k^{3} N + 437 N^{2} + 481 k\, 
      N^{2} + 109 k^{2} N^{2} + 116 N^{3} + 71 k\, 
      N^{3} + 8 N^{4}), \nonu\\
c_ {47} & = & - \frac {3 (20 + 21 k + 6 k^{2} + 25 N + 13 k\, 
      N + 7 N^{2})} {2(2 + N) (2 + k + N)^{2}}, \nonu\\
c_ {48} & = &\frac {2 i (72 + 134 k + 46 k^{2} + 110 N + 102 k\, 
     N + 5 k^{2} N + 44 N^{2} + 10 k\, 
     N^{2} + 5 N^{3})} {(2 + N) (2 + k + N)^{3}}, \nonu\\
c_ {49} & = & - \frac {6 i (2 + 3 k + 3 N)} {(2 + k + N)^{2}}, \qquad
c_ {50}  =  - \frac { (32 + 55 k + 18 k^{2} + 41 N + 35 k\, 
      N + 11 N^{2})} {(2 + N) (2 + k + N)^{3}}, \nonu\\
c_ {51} & = & - \frac {4 (28 + 53 k + 18 k^{2} + 37 N + 34 k\, 
      N + 10 N^{2})} {(2 + N) (2 + k + N)^{3}}, \nonu\\
c_ {52} & = & - \frac {6 (8 + 17 k + 6 k^{2} + 11 N + 11 k\, 
      N + 3 N^{2})} {(2 + N) (2 + k + N)^{3}}, \nonu\\
c_ {53} & = &\frac {6 i (20 + 21 k + 6 k^{2} + 25 N + 13 k\, 
     N + 7 N^{2})} {(2 + N) (2 + k + N)^{3}}, \nonu\\
c_ {54} & = &\frac {6 i (32 + 55 k + 18 k^{2} + 41 N + 35 k\, 
     N + 11 N^{2})} {(2 + N) (2 + k + N)^{4}}, \nonu\\
c_ {55} & = & - \frac {1}{3 (2 + N) (2 + k + N)^{3} (5 + 4 k + 4 N + 3 k\, N)}i (-900 - 2329 k - 1573 k^{2} - 214 k^{3} + 
       48 k^{4} 
\nonu\\&-& 1181 N - 2663 k\, 
      N - 813 k^{2} N + 463 k^{3} N + 150 k^{4} N + 114 N^{2} + 
       526 k\, N^{2} + 1528 k^{2} N^{2} 
\nonu\\&+& 834 k^{3} N^{2} + 
       90 k^{4} N^{2} + 897 N^{3} + 1781 k\, 
      N^{3} + 1323 k^{2} N^{3} + 279 k^{3} N^{3} + 500 N^{4} + 
       723 k\, N^{4} 
\nonu\\&+& 261 k^{2} N^{4} + 84 N^{5} + 72 k\, 
      N^{5}), \nonu\\
c_ {56} & = &\frac {3 (80 + 124 k + 53 k^{2} + 6 k^{3} + 140 N + 
      154 k\, N + 35 k^{2} N + 81 N^{2} + 48 k\, 
     N^{2} + 15 N^{3})} {2(2 + N) (2 + k + N)^{3}}, \nonu\\
c_ {57} & = &\frac {3 (40 + 30 k - 9 k^{2} - 6 k^{3} + 102 N + 76 k\, 
     N + 5 k^{2} N + 77 N^{2} + 36 k\, 
     N^{2} + 17 N^{3})} {2(2 + N) (2 + k + N)^{3}}, \nonu\\
c_ {58} & = & - \frac {2 i (40 + 75 k + 26 k^{2} + 65 N + 61 k\, 
      N + 4 k^{2} N + 29 N^{2} + 8 k\, 
      N^{2} + 4 N^{3})} {(2 + N) (2 + k + N)^{3}}, \nonu\\
c_ {59} & = &\frac {4 i (1 + k + N)} {(2 + k + N)^{2}}, \qquad
c_ {60}  =  - \frac {3 i (20 + 21 k + 6 k^{2} + 25 N + 13 k\, 
      N + 7 N^{2})} {(2 + N) (2 + k + N)^{3}}, \nonu\\
c_ {61} & = &\frac { i (k - N) (32 + 55 k + 18 k^{2} + 41 N + 35 k\, 
     N + 11 N^{2})} {(2 + N) (2 + k + N)^{4}}, \nonu\\
c_ {62} & = &\frac {6 i (20 + 21 k + 6 k^{2} + 25 N + 13 k\, 
     N + 7 N^{2})} {(2 + N) (2 + k + N)^{3}}, \nonu\\
c_ {63} & = &\frac {12 i (32 + 55 k + 18 k^{2} + 41 N + 35 k\, 
     N + 11 N^{2})} {(2 + N) (2 + k + N)^{4}}, \nonu\\
c_ {64} & = &\frac {4 (56 + 95 k + 32 k^{2} + 93 N + 83 k\, 
     N + 7 k^{2} N + 45 N^{2} + 14 k\, 
     N^{2} + 7 N^{3})} {(2 + N) (2 + k + N)^{4}}, \nonu\\
c_ {65} & = & - \frac {12 (16 + 29 k + 10 k^{2} + 27 N + 25 k\, 
      N + 2 k^{2} N + 13 N^{2} + 4 k\, 
      N^{2} + 2 N^{3})} {(2 + N) (2 + k + N)^{4}}, \nonu\\
c_ {66} & = &\frac {6 (k - N) (32 + 55 k + 18 k^{2} + 41 N + 35 k\, 
     N + 11 N^{2})} {(2 + N) (2 + k + N)^{5}}, \nonu\\
c_ {67} & = & - \frac {1}{3 (2 + N) (2 + k + N)^{3} (5 + 4 k + 4 N + 3 k\, N)} i (900 + 1817 k + 1090 k^{2} + 216 k^{3} 
\nonu\\&+& 
       1693 N + 2857 k\, 
      N + 1259 k^{2} N + 162 k^{3} N + 1075 N^{2} + 1419 k\, 
      N^{2} + 339 k^{2} N^{2} + 220 N^{3} 
\nonu\\&+& 201 k\, N^{3}), \nonu\\
c_ {68} & = &\frac {1}{3 (2 + N) (2 + k + N)^{4} (5 + 4 k + 4 N + 3 k\, N)} i (-1800 - 3694 k - 1735 k^{2} + 290 k^{3} 
\nonu\\&+& 
      216 k^{4} - 5126 N - 9704 k\, 
     N - 4593 k^{2} N - 29 k^{3} N + 162 k^{4} N - 5625 N^{2} - 
      9266 k\, 
     N^{2} 
\nonu\\&-& 3836 k^{2} N^{2} - 213 k^{3} N^{2} - 2703 N^{3} - 
      3523 k\, N^{3} - 972 k^{2} N^{3} - 460 N^{4} - 381 k\, 
     N^{4}), \nonu\\
c_ {69} & = &\frac{1}{3 (2 + N) (2 + k + N)^{4} (5 + 4 k + 4 N + 3 k\, N)} i (k - N) (56 + 773 k + 818 k^{2} + 216 k^{3} 
\nonu\\&+& 
      351 N + 1499 k\, 
     N + 1063 k^{2} N + 162 k^{3} N + 327 N^{2} + 785 k\, 
     N^{2} + 309 k^{2} N^{2} + 76 N^{3} + 93 k\, 
     N^{3}), \nonu\\
c_ {70} & = & - \frac{1}{3 (2 + N) (2 + k + N)^{4} (5 + 4 k + 4 N + 3 k\, N)} i (3600 + 6716 k + 3527 k^{2} + 62 k^{3} - 
       216 k^{4} 
\nonu\\&+& 10924 N + 18622 k\, 
      N + 9357 k^{2} N + 829 k^{3} N - 162 k^{4} N + 11979 N^{2} + 
       17512 k\, 
      N^{2} 
\nonu\\&+& 7318 k^{2} N^{2} + 579 k^{3} N^{2} + 5613 N^{3} + 
       6401 k\, N^{3} + 1728 k^{2} N^{3} + 932 N^{4} + 663 k\, 
      N^{4}), \nonu\\
c_ {71} & = & - \frac {1}{3 (2 + N) (2 + k + N)^{4} (5 + 4 k + 4 N + 3 k\, N)} i (3600 + 7556 k + 5789 k^{2} + 1874 k^{3} 
\nonu\\&+& 
       216 k^{4} + 10084 N + 18142 k\, 
      N + 11229 k^{2} N + 2599 k^{3} N + 162 k^{4} N + 10197 N^{2} + 
       15016 k\, 
      N^{2} 
\nonu\\&+& 6838 k^{2} N^{2} + 867 k^{3} N^{2} + 4425 N^{3} + 
       4919 k\, N^{3} + 1296 k^{2} N^{3} + 692 N^{4} + 483 k\, 
      N^{4}), \nonu\\
c_ {72} & = & - \frac{1} {3 (2 + N) (2 + k + N)^{4} (5 + 4 k + 4 N + 3 k\, N)} i (5400 + 11306 k + 8297 k^{2} + 
       2402 k^{3} 
\nonu\\&+& 216 k^{4} + 15154 N + 27424 k\, 
      N + 16503 k^{2} N + 3475 k^{3} N + 162 k^{4} N + 15471 N^{2}
\nonu\\&+& 
       23110 k\, 
      N^{2} + 10396 k^{2} N^{2} + 1227 k^{3} N^{2} + 6801 N^{3} + 
       7733 k\, N^{3} + 2052 k^{2} N^{3} + 1076 N^{4} 
\nonu\\&+& 771 k\, 
      N^{4}), \nonu\\
c_ {73} & = & - \frac {1}{3 (2 + N) (2 + k + N)^{3} (5 + 4 k + 4 N + 3 k\, N)} (-900 - 1699 k - 782 k^{2} - 120 k^{3} - 
       911 N 
\nonu\\&-& 1043 k\, 
      N + 275 k^{2} N + 138 k^{3} N + 313 N^{2} + 1125 k\, 
      N^{2} + 1143 k^{2} N^{2} + 180 k^{3} N^{2} + 628 N^{3} 
\nonu\\&+& 
       963 k\, N^{3} + 378 k^{2} N^{3} + 168 N^{4} + 144 k\, 
      N^{4}), \nonu\\
c_ {74} & = & - \frac{1}{3 (2 + N) (2 + k + N)^{4} (5 + 4 k + 4 N + 3 k\, N)}4 (2700 + 5771 k + 4808 k^{2} + 1820 k^{3}
\nonu\\&+& 
       264 k^{4} + 8359 N + 15751 k\, 
      N + 10968 k^{2} N + 3190 k^{3} N + 312 k^{4} N + 9447 N^{2} + 
       15286 k\, 
      N^{2} 
\nonu\\&+& 8494 k^{2} N^{2} + 1725 k^{3} N^{2} + 90 k^{4} N^{2} + 
       4866 N^{3} + 6524 k\, 
      N^{3} + 2619 k^{2} N^{3} + 279 k^{3} N^{3}
\nonu\\&+& 1112 N^{4} + 
       1182 k\, N^{4} + 261 k^{2} N^{4} + 84 N^{5} + 72 k\, 
      N^{5}), \nonu\\
c_ {75} & = & - \frac {1} {3 (2 + N) (2 + k + N)^{4} (5 + 4 k + 4 N + 3 k\, N)} (3600 + 6952 k + 4261 k^{2} + 562 k^{3} - 
       120 k^{4} 
\nonu\\&+& 12488 N + 23150 k\, 
      N + 14547 k^{2} N + 3059 k^{3} N + 138 k^{4} N + 15537 N^{2} + 
       25802 k\, 
      N^{2} 
\nonu\\&+& 14360 k^{2} N^{2} + 2721 k^{3} N^{2} + 
       180 k^{4} N^{2} + 8697 N^{3} + 12121 k\, 
      N^{3} + 5130 k^{2} N^{3} + 558 k^{3} N^{3}
\nonu\\&+& 2116 N^{4} + 
       2283 k\, N^{4} + 522 k^{2} N^{4} + 168 N^{5} + 144 k\, 
      N^{5}), \nonu\\
c_ {76} & = &\frac {3 i (20 + 21 k + 6 k^{2} + 25 N + 13 k\, 
     N + 7 N^{2})} {4(2 + N) (2 + k + N)^{3}},\qquad c_ {77}  =  1, \nonu\\
c_ {78} & = & - \frac {(60 + 77 k + 22 k^{2} + 121 N + 115 k\, 
     N + 20 k^{2} N + 79 N^{2} + 42 k\, 
     N^{2} + 16 N^{3})} {(2 + N) (2 + k + N)^{2}}, \nonu\\
c_ {79} & = & - \frac {1} {(2 + N) (2 + k + N)^{2} (5 + 4 k + 4 N + 3 k\, N)}3 (100 + 187 k + 118 k^{2} + 24 k^{3} + 
       303 N 
\nonu\\&+& 505 k\, 
      N + 277 k^{2} N + 46 k^{3} N + 325 N^{2} + 455 k\, 
      N^{2} + 195 k^{2} N^{2} + 20 k^{3} N^{2} + 148 N^{3} 
\nonu\\&+& 159 k\, 
      N^{3} + 42 k^{2} N^{3} + 24 N^{4} + 16 k\, 
      N^{4}), \nonu\\
c_ {80} & = &\frac {16 i (k - N)} {(2 + k + N) (5 + 4 k + 4 N + 3 k\, 
     N)},\nonu\\
c_ {81} & = & - \frac {2 i (32 + 29 k + 6 k^{2} + 35 N + 
       17 k\, N + 9 N^{2})} {(2 + N) (2 + k + N)^{2}}, \nonu\\
c_ {82} & = &\frac{1} {(2 + N) (2 + k + N)^{2} (5 + 4 k + 4 N + 
      3 k\, N)}2 i (160 + 143 k + 14 k^{2} - 8 k^{3} + 273 N 
\nonu\\&+& 
      155 k\, N - 19 k^{2} N - 10 k^{3} N + 183 N^{2} + 63 k\, 
     N^{2} - 11 k^{2} N^{2} + 60 N^{3} + 13 k\, 
     N^{3} + 8 N^{4}), \nonu\\
c_ {83} & = &\frac {64 (k - N)} {(2 + k + N)^{2} (5 + 4 k + 4 N + 
      3 k\, N)}, \nonu\\
c_ {84} & = & - \frac {4 (16 + 21 k + 6 k^{2} + 19 N + 13 k\, 
      N + 5 N^{2})} {(2 + N) (2 + k + N)^{3}}, \nonu\\
c_ {85} & = &\frac {1} {(2 + N) (2 + k + N)^{3} (5 + 4 k + 4 N + 
      3 k\, N)}4 (80 + 39 k - 18 k^{2} - 8 k^{3} + 129 N - 9 k\, 
     N 
\nonu\\&-& 59 k^{2} N - 10 k^{3} N + 99 N^{2} - 17 k\, 
     N^{2} - 23 k^{2} N^{2} + 44 N^{3} + k\, 
     N^{3} + 8 N^{4}), \nonu\\
c_ {86} & = &\frac {2 (8 + 17 k + 6 k^{2} + 11 N + 11 k\, 
     N + 3 N^{2})} {(2 + N) (2 + k + N)^{2}}, \nonu\\
c_ {87} & = &\frac {2 (20 + 21 k + 6 k^{2} + 25 N + 13 k\, 
     N + 7 N^{2})} {(2 + N) (2 + k + N)^{3}}, \nonu\\
c_ {88} & = & - \frac {4 i (16 + 21 k + 6 k^{2} + 19 N + 13 k\, 
      N + 5 N^{2})} {(2 + N) (2 + k + N)^{3}}, \nonu\\
c_ {89} & = &\frac {4 i (13 k + 6 k^{2} + 3 N + 9 k\, 
     N + N^{2})} {(2 + N) (2 + k + N)^{3}}, \qquad
c_ {90}  = \frac {16 i (k - N)} {3 (2 + k + N)^{2}}, \nonu\\
c_ {91} & = & - \frac {4 i (-5 k - 4 k^{2} + 35 N + 32 k\, 
      N + 7 k^{2} N + 41 N^{2} + 21 k\, 
      N^{2} + 11 N^{3})} {3 (2 + N) (2 + k + N)^{3}}, \nonu\\
c_ {92} & = & - \frac {4 i (11 k + 4 k^{2} + 19 N + 40 k\, 
      N + 11 k^{2} N + 25 N^{2} + 21 k\, 
      N^{2} + 7 N^{3})} {3 (2 + N) (2 + k + N)^{3}}, \nonu\\
c_ {93} & = &\frac {(48 + 128 k + 61 k^{2} + 6 k^{3} + 72 N + 100 k\, 
    N + 15 k^{2} N + 23 N^{2} + 10 k\, 
    N^{2} + N^{3})} {(2 + N) (2 + k + N)^{3}}, \nonu\\
c_ {94} & = &\frac {2 i (32 + 55 k + 18 k^{2} + 41 N + 35 k\, 
     N + 11 N^{2})} {(2 + N) (2 + k + N)^{3}}, \nonu\\
c_ {95} & = &\frac {2 i (13 k + 6 k^{2} + 3 N + 9 k\, 
     N + N^{2})} {(2 + N) (2 + k + N)^{3}}, \nonu\\
c_ {96} & = &\frac {4 i (16 + 21 k + 6 k^{2} + 19 N + 13 k\, 
     N + 5 N^{2})} {(2 + N) (2 + k + N)^{3}}, \nonu\\
c_ {97} & = &\frac {2 i (96 + 174 k + 73 k^{2} + 6 k^{3} + 162 N + 
      166 k\, N + 27 k^{2} N + 81 N^{2} + 34 k\, 
     N^{2} + 13 N^{3})} {(2 + N) (2 + k + N)^{4}}, \nonu\\
c_ {98} & = & - \frac {4 i} {(2 + k + N)^{2}}, \nonu\\
c_ {99} & = &\frac {4 i (32 + 81 k + 43 k^{2} + 6 k^{3} + 47 N + 
      69 k\, N + 15 k^{2} N + 16 N^{2} + 10 k\, 
     N^{2} + N^{3})} {(2 + N) (2 + k + N)^{4}}, \nonu\\
c_ {100} & = &\frac {4 i (-8 - 31 k - 12 k^{2} - 5 N - 11 k\, 
     N + 3 k^{2} N + 7 N^{2} + 6 k\, 
     N^{2} + 3 N^{3})} {(2 + N) (2 + k + N)^{4}}, \nonu\\
c_ {101} & = &\frac {2 i (32 + 29 k + 6 k^{2} + 35 N + 17 k\, 
     N + 9 N^{2})} {(2 + N) (2 + k + N)^{3}}, \nonu\\
c_ {102} & = &\frac {4 (32 + 55 k + 18 k^{2} + 41 N + 35 k\, 
     N + 11 N^{2})} {(2 + N) (2 + k + N)^{4}}, \nonu\\
c_ {103} & = &\frac {1}{(2 + N) (2 + k + N)^{3} (5 + 4 k + 4 N + 3 k\, N)}2 i (40 + 132 k + 87 k^{2} - 2 k^{3} - 8 k^{4} + 
      152 N 
\nonu\\&+& 346 k\, 
     N + 136 k^{2} N - 45 k^{3} N - 18 k^{4} N + 203 N^{2} + 355 k\, 
     N^{2} + 77 k^{2} N^{2} - 52 k^{3} N^{2} 
\nonu\\&-& 10 k^{4} N^{2} + 
      139 N^{3} + 208 k\, 
     N^{3} + 41 k^{2} N^{3} - 11 k^{3} N^{3} + 52 N^{4} + 69 k\, 
     N^{4} + 13 k^{2} N^{4} + 8 N^{5} 
\nonu\\&+& 8 k\, 
     N^{5}), \nonu\\
c_ {104} & = &\frac {2 i (20 + 5 k - 2 k^{2} + 41 N + 5 k\, 
     N - 4 k^{2} N + 23 N^{2} + 4 N^{3})} {(2 + N) (2 + k + N)^{3}}, \nonu\\
c_ {105} & = &\frac {2 i (32 + 3 k - 6 k^{2} + 29 N - k\, 
     N + 7 N^{2})} {(2 + N) (2 + k + N)^{3}}, \nonu\\
c_ {106} & = & - \frac {4 i (32 + 29 k + 6 k^{2} + 35 N + 17 k\, 
      N + 9 N^{2})} {(2 + N) (2 + k + N)^{3}}, \nonu\\
c_ {107} & = & - \frac {6 i (20 + 21 k + 6 k^{2} + 25 N + 13 k\, 
      N + 7 N^{2})} {(2 + N) (2 + k + N)^{3}}, \nonu\\
c_ {108} & = & - \frac {4 i (20 + 13 k + 2 k^{2} + 33 N + 9 k\, 
      N - 2 k^{2} N + 15 N^{2} + 2 N^{3})} {(2 + N) (2 + k + N)^{3}}, \nonu\\
c_ {109} & = & - \frac {2 i (80 + 152 k + 65 k^{2} + 6 k^{3} + 
       112 N + 120 k\, N + 17 k^{2} N + 39 N^{2} + 14 k\, 
      N^{2} + 3 N^{3})} {(2 + N) (2 + k + N)^{4}}, \nonu\\
c_ {110} & = & - \frac {2 i (64 + 136 k + 61 k^{2} + 6 k^{3} + 88 N + 
       104 k\, N + 15 k^{2} N + 27 N^{2} + 10 k\, 
      N^{2} + N^{3})} {(2 + N) (2 + k + N)^{4}}, \nonu\\
c_ {111} & = & - \frac {2 i (13 k + 6 k^{2} + 3 N + 9 k\, 
      N + N^{2})} {(2 + N) (2 + k + N)^{3}}, \nonu\\
c_ {112} & = &\frac {1} {(2 + N) (2 + k + N)^{4}} 2 i (160 + 284 k + 
    109 k^{2} + 6 k^{3} + 244 N + 236 k\, 
   N + 27 k^{2} N + 103 N^{2} 
\nonu\\&+& 34 k\, N^{2} + 13 N^{3}), \nonu\\
c_ {113} & = &\frac {2 i (96 + 220 k + 93 k^{2} + 6 k^{3} + 148 N + 
      172 k\, N + 19 k^{2} N + 55 N^{2} + 18 k\, 
     N^{2} + 5 N^{3})} {(2 + N) (2 + k + N)^{4}}, \nonu\\
c_ {114} & = & - \frac {8 (60 + 85 k + 26 k^{2} + 113 N + 119 k\, 
      N + 22 k^{2} N + 71 N^{2} + 42 k\, 
      N^{2} + 14 N^{3})} {3 (2 + N) (2 + k + N)^{4}}, \nonu\\
c_ {115} & = & - \frac {4 (32 + 55 k + 18 k^{2} + 41 N + 35 k\, 
      N + 11 N^{2})} {(2 + N) (2 + k + N)^{4}}, \nonu\\
c_ {116} & = & - \frac {4 (180 + 175 k + 38 k^{2} + 419 N + 317 k\, 
      N + 46 k^{2} N + 293 N^{2} + 126 k\, 
      N^{2} + 62 N^{3})} {3 (2 + N) (2 + k + N)^{4}}, \nonu\\
c_ {117} & = & - \frac {4 (60 + 101 k + 34 k^{2} + 97 N + 127 k\, 
      N + 26 k^{2} N + 55 N^{2} + 42 k\, 
      N^{2} + 10 N^{3})} {3 (2 + N) (2 + k + N)^{4}}, \nonu\\
c_ {118} & = &\frac {4 (32 + 55 k + 18 k^{2} + 41 N + 35 k\, 
     N + 11 N^{2})} {(2 + N) (2 + k + N)^{4}}, \nonu\\
c_ {119} & = &\frac{1}{3 (2 + N) (2 + k + N)^{4} (5 + 4 k + 4 N + 3 k\, N)}2 (640 + 2052 k + 1809 k^{2} + 506 k^{3} + 
      24 k^{4} 
\nonu\\&+& 1772 N + 4560 k\, 
     N + 2991 k^{2} N + 391 k^{3} N - 54 k^{4} N + 1815 N^{2} + 
      3804 k\, 
     N^{2} 
+ 1722 k^{2} N^{2} 
\nonu\\&-& 63 k^{3} N^{2} - 60 k^{4} N^{2} + 
      971 N^{3} + 1707 k\, 
     N^{3} + 564 k^{2} N^{3} - 30 k^{3} N^{3} + 316 N^{4} + 465 k\, 
     N^{4} 
\nonu\\&+& 114 k^{2} N^{4} + 48 N^{5} + 48 k\, 
     N^{5}), \nonu\\
c_ {120} & = &\frac{1}{(2 + N) (2 + k + N)^{4} (5 + 4 k + 4 N + 3 k\, N)}2 (-320 - 1356 k - 1693 k^{2} - 786 k^{3} - 
      120 k^{4} 
\nonu\\&-& 516 N - 2104 k\, 
     N - 2623 k^{2} N - 1155 k^{3} N - 162 k^{4} N + 29 N^{2} - 
      592 k\, N^{2} - 1278 k^{2} N^{2} 
\nonu\\&-& 585 k^{3} N^{2} - 
      60 k^{4} N^{2} + 465 N^{3} + 549 k\, 
     N^{3} - 120 k^{2} N^{3} - 102 k^{3} N^{3} + 276 N^{4} + 339 k\, 
     N^{4} 
\nonu\\&+& 42 k^{2} N^{4} + 48 N^{5} + 48 k\, 
     N^{5}),\nonu\\ 
c_ {121} & = &\frac {(60 + 77 k + 22 k^{2} + 121 N + 115 k\, 
    N + 20 k^{2} N + 79 N^{2} + 42 k\, 
    N^{2} + 16 N^{3})} {2(2 + N) (2 + k + N)^{2}}, \nonu\\
c_ {122} & = & - \frac { i (20 + 21 k + 6 k^{2} + 25 N + 13 k\, 
      N + 7 N^{2})} {(2 + N) (2 + k + N)^{2}}, \nonu\\
c_ {123} & = & - \frac{1}{(2 + N) (2 + k + N)^{2} (5 + 4 k + 4 N + 
       3 k\, N)} i (100 + 121 k + 50 k^{2} + 8 k^{3} + 
       269 N 
\nonu\\&+& 277 k\, 
      N + 91 k^{2} N + 10 k^{3} N + 231 N^{2} + 171 k\, 
      N^{2} + 31 k^{2} N^{2} + 76 N^{3} + 29 k\, 
      N^{3} + 8 N^{4}), \nonu\\
c_ {124} & = & - \frac {32 i (k - N)} {(2 + k + N)^{2} (5 + 4 k + 
       4 N + 3 k\, N)}, \nonu\\
c_ {125} & = & - \frac {2 (20 + 21 k + 6 k^{2} + 25 N + 13 k\, 
      N + 7 N^{2})} {(2 + N) (2 + k + N)^{3}}, \nonu\\
c_ {126} & = & - \frac{1}{(2 + N) (2 + k + N)^{3} (5 + 4 k + 4 N + 
       3 k\, N)}2 (100 + 121 k + 50 k^{2} + 8 k^{3} + 269 N 
\nonu\\&+& 
       277 k\, N + 91 k^{2} N + 10 k^{3} N + 231 N^{2} + 171 k\, 
      N^{2} + 31 k^{2} N^{2} + 76 N^{3} + 29 k\, 
      N^{3} + 8 N^{4}), \nonu\\
c_ {127} & = & - \frac {(20 + 21 k + 6 k^{2} + 25 N + 13 k\, 
     N + 7 N^{2})} {2(2 + N) (2 + k + N)^{2}}, \nonu\\
c_ {128} & = & - \frac {4 (12 + 19 k + 6 k^{2} + 15 N + 12 k\, 
      N + 4 N^{2})} {(2 + N) (2 + k + N)^{3}}, \nonu\\
c_ {129} & = & - \frac {4 (-4 + 11 k + 6 k^{2} - N + 8 k\, 
      N)} {(2 + N) (2 + k + N)^{3}}, \nonu\\
c_ {130} & = & - \frac { (13 k + 6 k^{2} + 3 N + 9 k\, 
      N + N^{2})} {(2 + N) (2 + k + N)^{3}}, \nonu\\
c_ {131} & = & - \frac { (32 + 29 k + 6 k^{2} + 35 N + 17 k\, 
      N + 9 N^{2})} {(2 + N) (2 + k + N)^{3}}, \nonu\\
c_ {132} & = &\frac {2 (8 - 9 k - 6 k^{2} + 5 N - 7 k\, 
     N + N^{2})} {(2 + N) (2 + k + N)^{3}}, \nonu\\
c_ {133} & = & - \frac {2 (24 + 25 k + 6 k^{2} + 27 N + 15 k\, 
      N + 7 N^{2})} {(2 + N) (2 + k + N)^{3}}, \nonu\\
c_ {134} & = &\frac {2 i (20 + 21 k + 6 k^{2} + 25 N + 13 k\, 
     N + 7 N^{2})} {(2 + N) (2 + k + N)^{3}}, \nonu\\
c_ {135} & = &\frac {2 i (32 + 55 k + 18 k^{2} + 41 N + 35 k\, 
     N + 11 N^{2})} {(2 + N) (2 + k + N)^{4}}, \qquad
c_ {136}  = \frac {16 i} {(2 + k + N)^{2}}, \nonu\\
c_ {137} & = & - \frac {4 i (32 + 15 k + 41 N + 17 k\, 
      N + 3 k^{2} N + 19 N^{2} + 6 k\, 
      N^{2} + 3 N^{3})} {(2 + N) (2 + k + N)^{3}}, \nonu\\
c_ {138} & = &\frac {4 i (16 + 25 k + 8 k^{2} + 23 N + 19 k\, 
     N + k^{2} N + 9 N^{2} + 2 k\, 
     N^{2} + N^{3})} {(2 + N) (2 + k + N)^{3}}, \nonu\\
c_ {139} & = & - \frac {1}{3 (2 + N) (2 + k + N)^{3} (5 + 4 k + 4 N + 3 k\, N)} i (-300 - 635 k - 347 k^{2} - 2 k^{3} + 
       24 k^{4} 
\nonu\\&-& 535 N - 817 k\, 
      N - 105 k^{2} N + 209 k^{3} N + 54 k^{4} N - 210 N^{2} + 98 k\, 
      N^{2} + 548 k^{2} N^{2} 
\nonu\\&+& 288 k^{3} N^{2} + 30 k^{4} N^{2} + 
       141 N^{3} + 535 k\, 
      N^{3} + 441 k^{2} N^{3} + 93 k^{3} N^{3} + 124 N^{4} + 231 k\, 
      N^{4} 
\nonu\\&+& 87 k^{2} N^{4} + 24 N^{5} + 24 k\, 
      N^{5}), \nonu\\
c_ {140} & = &\frac {(80 + 124 k + 53 k^{2} + 6 k^{3} + 140 N + 
     154 k\, N + 35 k^{2} N + 81 N^{2} + 48 k\, 
    N^{2} + 15 N^{3})} {2(2 + N) (2 + k + N)^{3}}, \nonu\\
c_ {141} & = &\frac {(40 + 30 k - 9 k^{2} - 6 k^{3} + 102 N + 76 k\, 
    N + 5 k^{2} N + 77 N^{2} + 36 k\, 
    N^{2} + 17 N^{3})} {2(2 + N) (2 + k + N)^{3}}, \nonu\\
c_ {142} & = &\frac {2 i (20 + 21 k + 6 k^{2} + 25 N + 13 k\, 
     N + 7 N^{2})} {(2 + N) (2 + k + N)^{3}}, \nonu\\
c_ {143} & = &\frac { (60 + 77 k + 22 k^{2} + 121 N + 115 k\, 
     N + 20 k^{2} N + 79 N^{2} + 42 k\, 
     N^{2} + 16 N^{3})} {(2 + N) (2 + k + N)^{3}}, \nonu\\
c_ {144} & = & - \frac {2 i (24 + 33 k + 10 k^{2} + 35 N + 27 k\, 
      N + 2 k^{2} N + 15 N^{2} + 4 k\, 
      N^{2} + 2 N^{3})} {(2 + N) (2 + k + N)^{3}}, \nonu\\
c_ {145} & = &\frac {2 i (32 + 11 k - 2 k^{2} + 37 N + 11 k\, 
     N + 2 k^{2} N + 15 N^{2} + 4 k\, 
     N^{2} + 2 N^{3})} {(2 + N) (2 + k + N)^{3}}, \nonu\\
c_ {146} & = & - \frac {4 i} {(2 + k + N)^{2}}, \qquad
c_ {147}  =  - \frac {2 i (20 + 21 k + 6 k^{2} + 25 N + 13 k\, 
      N + 7 N^{2})} {(2 + N) (2 + k + N)^{3}}, \nonu\\
c_ {148} & = & - \frac { i (60 + 59 k + 18 k^{2} + 79 N + 37 k\, 
      N + 23 N^{2})} {3 (2 + N) (2 + k + N)^{3}}, \nonu\\
c_ {149} & = & - \frac {1}{3 (2 + N) (2 + k + N)^{4}} i (240 + 284 k + 37 k^{2} - 18 k^{3} + 
       508 N + 478 k\, N + 65 k^{2} N
\nonu\\&+& 349 N^{2} + 192 k\, 
      N^{2} + 73 N^{3}), \nonu\\
c_ {150} & = &\frac{1}{3 (2 + N) (2 + k + N)^{4}} i (-120 - 106 k + 19 k^{2} + 18 k^{3} - 
      290 N - 236 k\, N - 19 k^{2} N 
\nonu\\&-& 215 N^{2} - 108 k\, 
     N^{2} - 47 N^{3}), \nonu\\
c_ {151} & = & - \frac {1}{3 (2 + N) (2 + k + N)^{4}} i (240 + 356 k + 151 k^{2} + 18 k^{3} + 
       436 N + 454 k\, N + 101 k^{2} N 
\nonu\\&+& 259 N^{2} + 144 k\, 
      N^{2} + 49 N^{3}), \nonu\\
c_ {152} & = &\frac { i (k - N) (80 + 79 k + 18 k^{2} + 89 N + 
      47 k\, N + 23 N^{2})} {3 (2 + N) (2 + k + N)^{4}}, \nonu\\
c_ {153} & = & - \frac{1} {3 (2 + N) (2 + k + N)^{4}} i (360 + 542 k + 211 k^{2} + 18 k^{3} + 
       646 N + 700 k\, N + 149 k^{2} N 
\nonu\\&+& 385 N^{2} + 228 k\, 
      N^{2} + 73 N^{3}), \nonu\\
c_ {154} & = &\frac { i (k - N) (13 k + 6 k^{2} + 3 N + 9 k\, 
     N + N^{2})} {(2 + N) (2 + k + N)^{4}}, \nonu\\
c_ {155} & = &\frac {2 i (20 + 21 k + 6 k^{2} + 25 N + 13 k\, 
     N + 7 N^{2})} {(2 + N) (2 + k + N)^{3}}, \nonu\\
c_ {156} & = & - \frac {2 i (120 + 142 k + 38 k^{2} + 254 N + 224 k\, 
      N + 37 k^{2} N + 170 N^{2} + 84 k\, 
      N^{2} + 35 N^{3})} {3 (2 + N) (2 + k + N)^{4}}, \nonu\\
c_ {157} & = & - \frac {2 i (240 + 320 k + 94 k^{2} + 472 N + 466 k\, 
      N + 83 k^{2} N + 304 N^{2} + 168 k\, 
      N^{2} + 61 N^{3})} {3 (2 + N) (2 + k + N)^{4}}, \nonu\\
c_ {158} & = &\frac {2 i (-60 - 109 k + k^{2} + 18 k^{3} - 89 N - 
      161 k\, N - 19 k^{2} N - 56 N^{2} - 66 k\, 
     N^{2} - 11 N^{3})} {3 (2 + N) (2 + k + N)^{4}}, \nonu\\
c_ {159} & = & - \frac {1}{3 (2 + N) (2 + k + N)^{4}}2 i (300 + 353 k + 133 k^{2} + 18 k^{3} + 
       637 N + 529 k\, N + 101 k^{2} N 
\nonu\\&+& 418 N^{2} + 186 k\, 
      N^{2} + 85 N^{3}), \nonu\\
c_ {160} & = & - \frac {2 i (60 + 97 k + 32 k^{2} + 101 N + 125 k\, 
      N + 25 k^{2} N + 59 N^{2} + 42 k\, 
      N^{2} + 11 N^{3})} {3 (2 + N) (2 + k + N)^{4}}, \nonu\\
c_ {161} & = & - \frac {10 i (60 + 73 k + 20 k^{2} + 125 N + 113 k\, 
      N + 19 k^{2} N + 83 N^{2} + 42 k\, 
      N^{2} + 17 N^{3})} {3 (2 + N) (2 + k + N)^{4}}, \nonu\\
c_ {162} & = &\frac { i (k - N) (32 + 29 k + 6 k^{2} + 35 N + 17 k\, 
     N + 9 N^{2})} {(2 + N) (2 + k + N)^{4}}, \nonu\\
c_ {163} & = &\frac {2 i (20 + 21 k + 6 k^{2} + 25 N + 13 k\, 
     N + 7 N^{2})} {(2 + N) (2 + k + N)^{3}}, \nonu\\
c_ {164} & = &\frac {4 i (32 + 55 k + 18 k^{2} + 41 N + 35 k\, 
     N + 11 N^{2})} {(2 + N) (2 + k + N)^{4}}, \qquad
c_ {165}  = \frac {4} {(2 + k + N)^{2}}, \nonu\\
c_ {166} & = & - \frac {8 (8 + 21 k + 8 k^{2} + 15 N + 17 k\, 
      N + k^{2} N + 7 N^{2} + 2 k\, 
      N^{2} + N^{3})} {(2 + N) (2 + k + N)^{4}}, \nonu\\
c_ {167} & = & - \frac {8 (40 + 53 k + 16 k^{2} + 63 N + 49 k\, 
      N + 5 k^{2} N + 31 N^{2} + 10 k\, 
      N^{2} + 5 N^{3})} {(2 + N) (2 + k + N)^{4}}, \nonu\\
c_ {168} & = &\frac {4 (16 + 29 k + 10 k^{2} + 27 N + 25 k\, 
     N + 2 k^{2} N + 13 N^{2} + 4 k\, 
     N^{2} + 2 N^{3})} {(2 + N) (2 + k + N)^{4}}, \nonu\\
c_ {169} & = & - \frac{1}{3 (2 + N) (2 + k + N)^{3} (5 + 4 k + 4 N + 3 k\, N)} (-300 - 425 k - 154 k^{2} - 24 k^{3} - 
       445 N 
\nonu\\&-& 277 k\, 
      N + 169 k^{2} N + 54 k^{3} N - 73 N^{2} + 351 k\, 
      N^{2} + 393 k^{2} N^{2} + 60 k^{3} N^{2} + 140 N^{3} 
\nonu\\&+& 309 k\, 
      N^{3} + 126 k^{2} N^{3} + 48 N^{4} + 48 k\, 
      N^{4}), \nonu\\
c_ {170} & = & - \frac {1}{(2 + N) (2 + k + N)^{4} (5 + 4 k + 4 N + 3 k\, N)} (1000 + 2370 k + 2129 k^{2} + 834 k^{3} + 
       120 k^{4} 
\nonu\\&+& 3130 N + 6620 k\, 
      N + 5065 k^{2} N + 1575 k^{3} N + 162 k^{4} N + 3671 N^{2} + 
       6808 k\, 
      N^{2} + 4262 k^{2} N^{2} 
\nonu\\&+& 969 k^{3} N^{2} + 60 k^{4} N^{2} + 
       2021 N^{3} + 3191 k\, 
      N^{3} + 1482 k^{2} N^{3} + 186 k^{3} N^{3} + 516 N^{4} + 
       663 k\, N^{4} 
\nonu\\&+& 174 k^{2} N^{4} + 48 N^{5} + 48 k\, 
      N^{5}), \nonu\\
c_ {171} & = & - \frac{1}{(2 + N) (2 + k + N)^{4} (5 + 4 k + 4 N + 3 k\, N)} (1600 + 3620 k + 2965 k^{2} + 1010 k^{3} + 
       120 k^{4} 
\nonu\\&+& 4820 N + 9714 k\, 
      N + 6823 k^{2} N + 1867 k^{3} N + 162 k^{4} N + 5429 N^{2} + 
       9506 k\, 
      N^{2} + 5448 k^{2} N^{2} 
\nonu\\&+& 1089 k^{3} N^{2} + 60 k^{4} N^{2} + 
       2813 N^{3} + 4129 k\, 
      N^{3} + 1734 k^{2} N^{3} + 186 k^{3} N^{3} + 644 N^{4} + 
       759 k\, N^{4} 
\nonu\\&+& 174 k^{2} N^{4} + 48 N^{5} + 48 k\, 
      N^{5}), \nonu\\
c_ {172} & = & - \frac {1}{3 (2 + N) (2 + k + N)^{4} (5 + 4 k + 4 N + 3 k\, N)} (1200 + 2600 k + 1775 k^{2} + 326 k^{3} - 
       24 k^{4} 
\nonu\\&+& 3880 N + 7858 k\, 
      N + 5181 k^{2} N + 1129 k^{3} N + 54 k^{4} N + 4683 N^{2} + 
       8446 k\, 
      N^{2} + 4864 k^{2} N^{2} 
\nonu\\&+& 927 k^{3} N^{2} + 60 k^{4} N^{2} + 
       2583 N^{3} + 3923 k\, 
      N^{3} + 1710 k^{2} N^{3} + 186 k^{3} N^{3} + 620 N^{4} + 
       741 k\, N^{4} 
\nonu\\&+& 174 k^{2} N^{4} + 48 N^{5} + 48 k\, 
      N^{5}),\nonu\\ 
c_ {173} & = &\frac {2 (k - N) (32 + 55 k + 18 k^{2} + 41 N + 35 k\, 
     N + 11 N^{2})} {(2 + N) (2 + k + N)^{5}}, \nonu\\
c_ {174} & = &\frac {2 (k - N) (32 + 55 k + 18 k^{2} + 41 N + 35 k\, 
     N + 11 N^{2})} {(2 + N) (2 + k + N)^{5}}.\nonu
\eea

The fusion rule is 
\bea
[\Phi_1^{(1),ij}] \, \cdot \, [\widetilde{\Phi}_2^{(1)}]
= [I^{ij}] + [\Phi_0^{(1)} \, \Phi_1^{(1),ij}] + \varepsilon^{ijkl} \,
[\Phi_{\frac{1}{2}}^{(1),k} \, \Phi_{\frac{1}{2}}^{(1),l}] + [\Phi_1^{(2),ij}].
\nonu
\eea

\subsection{The OPEs between the higher spin-$\frac{5}{2}$ currents and 
other 
$5$ higher spin currents}

The OPEs between the higher spin-$\frac{5}{2}$ currents   are given by
\bea
&&\widetilde{\Phi}_{\frac{3}{2}}^{(1),i}(z)\:\widetilde{\Phi}_{\frac{3}{2}}^{(1),j}(w)\;=\;\frac{1}{(z-w)^{5}}\,\delta^{ij}\,c_{1}+\frac{1}{(z-w)^{4}}\Bigg[\,c_{2}\,T^{ij}+c_{3}\,\Gamma^{i}\,\Gamma^{j}+\varepsilon^{ijkl}\Big(c_{4}\,T^{ij}+c_{5}\,\Gamma^{k}\,\Gamma^{l}\Big)\Bigg](w)
\nonu\\&&+\frac{1}{(z-w)^{3}}\Bigg[\,\delta^{ij}\Bigg\{\, 
c_{6}\, {\bf \Phi_{0}^{(2)}}+c_{7}\,{\bf \Phi_{0}^{(1)}\,\Phi_{0}^{(1)}}+
c_{8}\,L+c_{9}\,G^{i}\,\Gamma^{i}+c_{10}\,\partial\Gamma^{i}\,\Gamma^{i}+c_{11}\,T^{ik}\,T^{ik}+c_{12}\,\widetilde{T}^{ik}\,\widetilde{T}^{ik}
\nonu\\&&+c_{13}\,T^{ik}\,\Gamma^{i}\,\Gamma^{k}+c_{14}\,U\,U+c_{15}\,\partial U+(1-\delta^{ik})\,\Big(c_{16}\,G^{k}\,\Gamma^{k}+c_{17}\,\partial\Gamma^{k}\,\Gamma^{k}\Big)+\varepsilon^{iabc}\,\Big(c_{18}\,\widetilde{T}^{ia}\,\Gamma^{b}\,\Gamma^{c}
\nonu\\&&+c_{19}\,(T^{ia}\,\Gamma^{b}\,\Gamma^{c}+T^{ab}\,\Gamma^{i}\,\Gamma^{c})+c_{20}\,T^{ia}\,T^{bc}+c_{21}\,\Gamma^{i}\,\Gamma^{a}\,\Gamma^{b}\,\Gamma^{c}\Big)\Bigg\}+c_{22}\,(G^{i}\,\Gamma^{j}-G^{j}\,\Gamma^{i})+c_{23}\,T^{ij}
\nonu\\&&+(1-\delta^{ij})\Bigg(\,c_{24}\,T^{ik}\,T^{jk}+c_{25}\,\partial\Gamma^{i}\,\Gamma^{j}+c_{26}\,\Gamma^{i}\,\partial\Gamma^{j}+c_{27}\,(T^{ik}\,\Gamma^{j}+T^{jk}\,\Gamma^{i})\,\Gamma^{k}\Bigg)\,
\nonu\\&&+\varepsilon^{ijkl}\,\Bigg(\,c_{28}\,\partial T^{kl}+c_{29}\,\partial(\Gamma^{k}\,\Gamma^{l})\Bigg)\Bigg](w)
\nonu\\&&+\frac{1}{(z-w)^{2}}\,\Bigg[\,\delta^{ij}\,\frac{1}{2}\,
\partial\Big(\mbox{ pole-3}\Big)+
c_{30}\, {\bf \Phi_{1}^{(2),ij}}+c_{31}\, {\bf \Phi_{0}^{(1)}\,\Phi_{1}^{(1),ij}}
+c_{32}\,L\,T^{ij}+c_{33}\,L\,\Gamma^{i}\,\Gamma^{j}
\nonu\\&&+c_{34}\,(G^{i}\,T^{ij}\,\Gamma^{i}+G^{j}\,T^{ij}\,\Gamma^{j})+(1-\delta^{ik}-\delta^{jk})\Big(c_{35}\,(G^{i}\,T^{jk}-G^{j}\,T^{ik})\,\Gamma^{k}+c_{36}\,(G^{k}\,T^{ij}\,\Gamma^{k})
\nonu\\&&+c_{37}\,(G^{k}\,T^{ik}\,\Gamma^{j}-G^{k}\,T^{jk}\,\Gamma^{i})+c_{38}\,T^{ij}\,\partial\Gamma^{k}\,\Gamma^{k}\Big)+c_{39}\,(G^{i}\,U\,\Gamma^{j}-G^{j}\,U\,\Gamma^{i})+c_{40}\,G^{k}\,\Gamma^{i}\,\Gamma^{j}\,\Gamma^{k}
\nonu\\&&+c_{41}\,T^{ij}\,T^{kl}\,T^{kl}+\varepsilon^{abcd}\,c_{42}\,T^{ab}\,T^{cd}\,\widetilde{T}^{ij}+c_{43}\,\Big[-T^{ij}\,T^{ij}\,\Gamma^{i}\,\Gamma^{j}+T^{ij}\,T^{ik}\,\Gamma^{i}\,\Gamma^{k}+T^{ij}\,T^{jk}\,\Gamma^{j}\,\Gamma^{k}
\nonu\\&&+\varepsilon^{ijkl}\,T^{ik}\,T^{jl}\,\Gamma^{k}\,\Gamma^{l}\Big]+c_{44}\,T^{ij}\,U\,U+c_{45}\,\partial T^{ij}\,U+c_{46}\,T^{ij}\,\partial U+c_{47}\,\partial^{2}T^{ij}
\nonu\\&&+c_{48}\,T^{ij}\,(\partial\Gamma^{i}\,\Gamma^{i}+\partial\Gamma^{j}\,\Gamma^{j})+c_{49}\,U\,U\,\Gamma^{i}\,\Gamma^{j}+c_{50}\,\partial U\,\Gamma^{i}\,\Gamma^{j}+c_{51}\,U\,\partial(\Gamma^{i}\,\Gamma^{j})+c_{52}\,\Gamma^{i}\,\Gamma^{j}\,\partial\Gamma^{k}\,\Gamma^{k}
\nonu\\&&+c_{53}\,\partial\Gamma^{i}\,\partial\Gamma^{j}+
(1-\delta^{ij})\Bigg\{\, c_{54}\, {\bf \Phi_{\frac{1}{2}}^{(1),i}\,
\Phi_{\frac{1}{2}}^{(1),j}}+c_{55}\,G^{i}\,G^{j}+c_{56}\,\partial G^{i}\,\Gamma^{j}+c_{57}\,\partial G^{j}\,\Gamma^{i}+c_{58}\,G^{i}\,\partial\Gamma^{j}
\nonu\\&&+c_{59}\,G^{j}\,\partial\Gamma^{i}+c_{60}\,\partial T^{ik}\,T^{jk}+c_{61}\,T^{ik}\,\partial T^{jk}+c_{62}\,(T^{ik}\,U\,\Gamma^{j}\,\Gamma^{k}-T^{jk}\,U\,\Gamma^{i}\,\Gamma^{k})
\nonu\\&&+(1-\delta^{jk})\Big(c_{63}\,\partial T^{ik}\,\Gamma^{j}\,\Gamma^{k}+c_{64}\,T^{ik}\,\partial\Gamma^{j}\,\Gamma^{k}+c_{65}\,T^{ik}\,\Gamma^{j}\,\partial\Gamma^{k}\Big)+(1-\delta^{ik})\Big(c_{66}\,\partial T^{jk}\,\Gamma^{i}\,\Gamma^{k}
\nonu\\&&+c_{67}\,T^{jk}\,\partial\Gamma^{i}\,\Gamma^{k}+c_{68}\,T^{jk}\,\Gamma^{i}\,\partial\Gamma^{k}\Big)+c_{69}\,\partial^{2}\Gamma^{i}\,\Gamma^{j}+c_{70}\,\Gamma^{i}\,\partial^{2}\Gamma^{j}\Bigg\}
\nonu\\&&+\varepsilon^{ijkl}\Bigg\{\,  
c_{71}\, {\bf \Phi_{1}^{(2),kl}}+c_{72}\,{\bf \Phi_{0}^{(1)}\,\Phi_{1}^{(1),kl}}+
c_{73}\, {\bf \Phi_{\frac{1}{2}}^{(1),k}\,\Phi_{\frac{1}{2}}^{(1),l}}+c_{74}\,L\,T^{kl}+c_{75}\,L\,\Gamma^{k}\,\Gamma^{l}
+c_{76}\,G^{k}\,G^{l}\nonu\\&&+c_{77}\,(G^{i}\,T^{kl}\,\Gamma^{i}+G^{j}\,T^{kl}\,\Gamma^{j})+c_{78}\,(G^{i}\,T^{ik}\,\Gamma^{l}+G^{j}\,T^{jk}\,\Gamma^{l})+c_{79}\,(G^{k}\,T^{il}\,\Gamma^{i}+G^{k}\,T^{jl}\,\Gamma^{j})
\nonu\\&&+c_{80}\,G^{k}\,T^{kl}\,\Gamma^{l}+c_{81}\,G^{k}\,U\,\Gamma^{l}+c_{82}\,\partial G^{k}\,\Gamma^{l}+c_{83}\,G^{k}\,\partial\Gamma^{l}+c_{84}\,G^{a}\,\Gamma^{a}\,\Gamma^{k}\,\Gamma^{l}+c_{85}\,\partial^{2}T^{kl}
\nonu\\&&+c_{86}\,(\partial T^{ik}\,T^{il}+\partial T^{jk}\,T^{jl})+c_{87}\,(T^{ik}\,\partial T^{il}+T^{jk}\,\partial T^{jl})+c_{88}\,T^{ij}\,T^{ij}\,T^{kl}+c_{89}\,T^{ij}\,T^{ik}\,T^{jl}
\nonu\\&&+c_{90}\,(T^{ik}\,T^{ik}\,T^{kl}+T^{jk}\,T^{jk}\,T^{kl}+\frac{1}{2}\,T^{kl}\,T^{kl}\,T^{kl})+c_{91}\,\partial T^{kl}\,U+c_{92}\,T^{kl}\,\partial U+c_{93}\,T^{kl}\,U\,U
\nonu\\&&+c_{94}\,T^{ak}\,U\,\Gamma^{a}\,\Gamma^{l}+c_{95}\,(T^{ik}\,T^{jl}\,\Gamma^{i}\,\Gamma^{j}+T^{ik}\,T^{kl}\,\Gamma^{i}\,\Gamma^{k}+T^{jk}\,T^{kl}\,\Gamma^{j}\,\Gamma^{k}+\frac{1}{2}T^{kl}\,T^{kl}\,\Gamma^{k}\,\Gamma^{l})
\nonu\\&&+c_{96}\,(T^{ik}\,\Gamma^{i}\,\partial\Gamma^{l}+T^{jk}\,\Gamma^{j}\,\partial\Gamma^{l})+c_{97}\,(T^{ik}\,\partial\Gamma^{i}\,\Gamma^{l}+T^{jk}\,\partial\Gamma^{j}\,\Gamma^{l})
+c_{98}\,(T^{il}\,\Gamma^{i}\,\partial\Gamma^{k}+T^{jl}\,\Gamma^{j}\,\partial\Gamma^{k})
\nonu\\&&+c_{99}\,(T^{il}\,\partial\Gamma^{i}\,\Gamma^{k}+T^{jl}\,\partial\Gamma^{j}\,\Gamma^{k})
+c_{100}\,(T^{kl}\,\partial\Gamma^{i}\,\Gamma^{i}+T^{kl}\,\partial\Gamma^{j}\,\Gamma^{j})+c_{101}\,T^{kl}\,\partial\Gamma^{k}\,\Gamma^{k}
\nonu\\&&+c_{102}\,(\partial T^{ik}\,\Gamma^{i}\,\Gamma^{l}+\partial T^{jk}\,\Gamma^{j}\,\Gamma^{l})
+c_{103}\,(\partial T^{il}\,\Gamma^{i}\,\Gamma^{k}+\partial T^{jl}\,\Gamma^{j}\,\Gamma^{k})+c_{104}\,\partial U\,\Gamma^{k}\,\Gamma^{l}+c_{105}\,U\,\Gamma^{k}\,\partial\Gamma^{l}
\nonu\\&&+c_{106}\,U\,U\,\Gamma^{k}\,\Gamma^{l}
+c_{107}\,\partial\Gamma^{a}\,\Gamma^{a}\,\Gamma^{k}\,\Gamma^{l}+c_{108}\,\partial^{2}\Gamma^{k}\,\Gamma^{l}+c_{109}\,\partial\Gamma^{k}\,\partial\Gamma^{l}+c_{110}\,\Gamma^{k}\,\partial^{2}\Gamma^{l}\Bigg\}\Bigg](w)
\nonu\\&&+\frac{1}{(z-w)}\,\Bigg[\,\delta^{ij}\Bigg\{\, c_{111}\,{\bf 
\widetilde{\Phi}_{2}^{(2)}}+c_{112}\, {\bf \partial^{2}\Phi_{0}^{(2)}}+
c_{113}\, {\bf (\Phi_{\frac{1}{2}}^{(1),k}\,\widetilde{\Phi}_{\frac{3}{2}}^{(1),k}+
\Phi_{0}^{(1)}\,\widetilde{\Phi}_{2}^{(1)})}+
c_{114}\, {\bf \partial\Phi_{\frac{1}{2}}^{(1),k}\,\Phi_{\frac{1}{2}}^{(1),k}}
\nonu\\&&  +c_{115}\, {\bf \partial^{2}\Phi_{0}^{(1)}\,\Phi_{0}^{(1)}}+
c_{116}\,{\bf \partial\Phi_{0}^{(1)}\,\partial\Phi_{0}^{(1)}}+
c_{117}\,L\, {\bf \Phi_{0}^{(2)}}+c_{118}\,L\, {\bf \Phi_{0}^{(1)}\,
\Phi_{0}^{(1)}}+c_{119}\,\partial^{2}L+c_{120}\,L\,L
\nonu\\&&+c_{121}\,\partial L\,U+c_{122}\,L\,U\,U+c_{123}\,L\,T^{ik}\,T^{ik}+c_{124}\,L\,\widetilde{T}^{ik}\,\widetilde{T}^{ik}
+c_{125}\,L\,T^{kl}\,\Gamma^{k}\,\Gamma^{l}+c_{126}\,L\,\partial\Gamma^{i}\,\Gamma^{i}
\nonu\\&&+c_{127}\,L\,\partial U+c_{128}\,\partial G^{i}\,G^{i}+c_{129}\,\partial^{2}G^{i}\,\Gamma^{i}+c_{130}\,\partial G^{i}\,\partial\Gamma^{i}+c_{131}\,G^{i}\,\partial^{2}\Gamma^{i}+c_{132}\,G^{i}\,G^{k}\,T^{ik}
\nonu\\&&+c_{133}\,G^{k}\,G^{l}\,\Gamma^{k}\,\Gamma^{l}+c_{134}\,\partial(G^{i}\,T^{ik})\,\Gamma^{k}+c_{135}\,G^{i}\,T^{ik}\,\partial\Gamma^{k}+c_{136}\,\partial(G^{k}\,T^{ik})\,\Gamma^{i}+c_{137}\,G^{k}\,T^{ik}\,\partial\Gamma^{i}
\nonu\\&&+c_{138}\,G^{i}\,T^{ik}\,U\,\Gamma^{k}+c_{139}\,\partial G^{i}\,U\,\Gamma^{i}+c_{140}\,G^{i}\,\partial U\,\Gamma^{i}+c_{141}\,G^{i}\,U\,\partial\Gamma^{i}+c_{142}\,G^{k}\,\Gamma^{k}\,\partial\Gamma^{l}\,\Gamma^{l}
\nonu\\&&+c_{143}\,\partial^{2}T^{ik}\,T^{ik}+c_{144}\,\partial^{2}\widetilde{T}^{ik}\,\widetilde{T}^{ik}+c_{145}\,\partial T^{ik}\,\partial T^{ik}
+c_{146}\,\partial\widetilde{T}^{ik}\,\partial\widetilde{T}^{ik}
+c_{147}\,(T^{ia}\,T^{ib}\,\partial T^{ab}
\nonu\\&&+\partial T^{ab}\,T^{ac}\,T^{bc})
+c_{148}\,\partial T^{ik}\,T^{il}\,T^{lk}\,
+c_{149}\,\Big[T^{ik}\,T^{il}\,\widetilde{T}^{il}\,\widetilde{T}^{ik}+\partial T^{ab}\,T^{ac}\,\Gamma^{b}\,\Gamma^{c}+T^{ab}\,T^{ac}\,\partial \Gamma^{b}\,\Gamma^{c}
\nonu\\&&+2\,T^{ab}\,T^{ab}\,\partial \Gamma^{a}\,\Gamma^{b}\Big]
+c_{150}\,\partial T^{ik}\,T^{ik}\,U
+c_{151}\,\partial\widetilde{T}^{ik}\,\widetilde{T}^{ik}\,U
+c_{152}\,T^{ik}\,T^{ik}\,\partial U
+c_{153}\,\widetilde{T}^{ik}\,\widetilde{T}^{ik}\,\partial U
\nonu\\&&+c_{154}\,T^{kl}\,T^{kl}\,U\,U+c_{155}\,\Big[T^{kl}\,\partial U\,\Gamma^{k}\,\Gamma^{l}-\,T^{kl}\,U\,\partial(\Gamma^{k}\,\Gamma^{l})\Big]
+c_{156}\,\partial^{2}T^{ik}\,\Gamma^{i}\,\Gamma^{k}+c_{157}\,\partial T^{ik}\,\partial\Gamma^{i}\,\Gamma^{k}
\nonu\\&&+c_{158}\,\partial T^{ik}\,\Gamma^{i}\,\partial\Gamma^{k}+c_{159}\,T^{ik}\,\partial\Gamma^{i}\,\partial\Gamma^{k}+c_{160}\,T^{ik}\,\partial^{2}\Gamma^{i}\,\Gamma^{k}
+c_{161}\,T^{ik}\,\Gamma^{i}\,\partial^{2}\Gamma^{k}
\nonu\\&&+\varepsilon^{iabc}\,\Big(c_{162}\,\partial^{2}\widetilde{T}^{ia}\,\Gamma^{b}\,\Gamma^{c}
+c_{163}\,\partial\widetilde{T}^{ia}\,\partial\Gamma^{b}\,\Gamma^{c}
+c_{164}\,\partial\widetilde{T}^{ia}\,\Gamma^{b}\,\partial\Gamma^{c}
+c_{165}\,\widetilde{T}^{ia}\,\partial\Gamma^{b}\,\partial\Gamma^{c}
\nonu\\&&+c_{166}\,\widetilde{T}^{ia}\,\partial^{2}\Gamma^{b}\,\Gamma^{c}\Big)+c_{167}\,\partial^{3}U+c_{168}\,\partial^{2}U\,U+c_{169}\,\partial U\,U\,U+c_{170}\,U\,U\,U\,U
+c_{171}\,\partial U\,\partial U
\nonu\\&&+c_{172}\,U\,\partial^{2}\Gamma^{i}\,\Gamma^{i}+c_{173}\,\partial U\,\partial\Gamma^{i}\,\Gamma^{i}+c_{174}\,U\,U\,\partial\Gamma^{k}\,\Gamma^{k}+c_{175}\,\partial^{2}\Gamma^{i}\,\partial\Gamma^{i}+c_{176}\,\partial^{3}\Gamma^{i}\,\Gamma^{i}
\nonu\\&&+c_{177}\,\partial\Gamma^{k}\,\Gamma^{k}\,\partial\Gamma^{l}\,\Gamma^{l}+(1-\delta^{ik})\Big(\,c_{178}\,L\,\partial\Gamma^{k}\,\Gamma^{k}+c_{179}\,\partial G^{k}\,G^{k}+c_{180}\,\partial^{2}G^{k}\,\Gamma^{k}+c_{181}\,\partial G^{k}\,\partial\Gamma^{k}
\nonu\\&&+c_{182}\,G^{k}\,\partial^{2}\Gamma^{k}+c_{183}\,G^{k}\,T^{kl}\,U\,\Gamma^{l}+c_{184}\,\partial G^{k}\,U\,\Gamma^{k}+c_{185}\,G^{k}\,\partial U\,\Gamma^{k}+c_{186}\,G^{k}\,U\,\partial\Gamma^{k}
\nonu\\&&+c_{187}\,(U\,\partial^{2}\Gamma^{k}\,\Gamma^{k}+\partial U\,\partial\Gamma^{k}\,\Gamma^{k})+c_{188}\,\partial^{2}\Gamma^{k}\,\partial\Gamma^{k}+c_{189}\,\partial^{3}\Gamma^{k}\,\Gamma^{k}
\nonu\\&&+(1-\delta^{ik}-\delta^{il})\Big(c_{190}\,G^{k}\,G^{l}\,T^{kl}
+c_{191}\,\partial(G^{k}\,T^{kl})\,\Gamma^{l}+c_{192}\,G^{k}\,T^{kl}\,\partial\Gamma^{l}\Big)
\nonu\\&&+\varepsilon^{iabc}\Bigg(\, c_{193}\, {\bf \Phi_{1}^{(1),ia}\,
\Phi_{1}^{(1),bc}}+c_{194}\,\partial G^{i}\,T^{ab}\,\Gamma^{c}
+c_{195}\,G^{i}\,\partial T^{ab}\,\Gamma^{c}+c_{196}\,G^{i}\,T^{ab}\,\partial\Gamma^{c}
\nonu\\&&+c_{197}\,(\partial G^{a}\,T^{bc}\,\Gamma^{i}+G^{a}\,T^{bc}\,\partial\Gamma^{i})+c_{198}\,G^{a}\,\partial T^{bc}\,\Gamma^{i}
+c_{199}\,\partial G^{a}\,T^{ib}\,\Gamma^{c}+c_{200}\,G^{a}\,\partial T^{ib}\,\Gamma^{c}
\nonu\\&&+c_{201}\,G^{a}\,T^{ib}\,\partial\Gamma^{c}+c_{202}\,\partial^{2}T^{ia}\,T^{bc}+c_{203}\,T^{ia}\,\partial^{2}T^{bc}
+c_{204}\,\partial T^{ia}\,T^{bc}\,U+c_{205}\,T^{ia}\,\partial T^{bc}\,U
\nonu\\&&+c_{206}\,T^{ia}\,T^{bc}\,\partial U+c_{207}\,\partial^{2}T^{ia}\,\Gamma^{b}\,\Gamma^{c}+c_{208}\,\partial T^{ia}\,\partial\Gamma^{b}\,\Gamma^{c}
+c_{209}\,T^{ia}\,\partial\Gamma^{b}\,\partial\Gamma^{c}+c_{210}\,T^{ia}\,\partial^{2}\Gamma^{b}\,\Gamma^{c}
\nonu\\&&+c_{211}\,\partial^{2}T^{ab}\,\Gamma^{i}\,\Gamma^{c}+c_{212}\,\partial T^{ab}\,\partial\Gamma^{i}\,\Gamma^{c}+c_{213}\,\partial T^{ab}\,\Gamma^{i}\,\partial\Gamma^{c}
+c_{214}\,T^{ab}\,\partial\Gamma^{i}\,\partial\Gamma^{c}+c_{215}\,T^{ab}\,\partial^{2}\Gamma^{i}\,\Gamma^{c}
\nonu\\&&+c_{216}\,T^{ab}\,\Gamma^{i}\,\partial^{2}\Gamma^{c}+c_{217}\,(\partial^{2}\Gamma^{i}\,\Gamma^{a}\,\Gamma^{b}\,\Gamma^{c}+\Gamma^{i}\,\partial^{2}\Gamma^{a}\,\Gamma^{b}\,\Gamma^{c}
+\Gamma^{i}\,\Gamma^{a}\,\partial^{2}\Gamma^{b}\,\Gamma^{c}+\Gamma^{i}\,\Gamma^{a}\,\Gamma^{b}\,\partial^{2}\Gamma^{c})
\nonu\\&&+c_{218}\,\partial^{2}(\Gamma^{i}\,\Gamma^{a}\,\Gamma^{b}\,\Gamma^{c})\Bigg)
+\varepsilon^{abcd}\Bigg(c_{219}\,G^{q}\,T^{ab}\,T^{cd}\,\Gamma^{q}
+c_{220}\,\Big[\partial G^{a}\,\Gamma^{b}\,\Gamma^{c}\,\Gamma^{d}-G^{a}\,\partial(\Gamma^{b}\,\Gamma^{c}\,\Gamma^{d})\Big]
\nonu\\&&
+c_{221}\,T^{ab}\,T^{cd}\,\partial\Gamma^{q}\,\Gamma^{q}
+c_{222}\,\Big[\partial T^{ab}\,U\,\Gamma^{c}\,\Gamma^{d}+T^{ab}\,U\,\partial(\Gamma^{c}\,\Gamma^{d})\Big]
+c_{223}\,\partial T^{ab}\,\partial T^{cd}\Bigg)\Bigg\}
\nonu\\&&+ c_{224}\, {\bf \partial\Phi_{1}^{(1),ij}}+
c_{225}\, {\bf \partial(\Phi_{0}^{(1)}\,\Phi_{1}^{(1),ij})}+
c_{226}\,\partial L\,T^{ij}+c_{227}\,L\,\partial T^{ij}+c_{228}\,\partial L\,\Gamma^{i}\,\Gamma^{j}
\nonu\\&&+c_{229}\,\Big[\partial(G^{i}\,T^{ij}\,\Gamma^{i})+\partial(G^{j}\,T^{ij}\,\Gamma^{j})\Big]+c_{230}\,\partial^{3}T^{ij}+c_{231}\,\partial T^{ij}\,T^{ij}\,T^{ij}+c_{232}\,\partial(T^{ij}\,\widetilde{T}^{ij}\,\widetilde{T}^{ij})
\nonu\\&&+c_{233}\,\partial T^{ij}\,\partial\Gamma^{i}\,\Gamma^{i}+c_{234}\,\partial T^{ij}\,\partial\Gamma^{j}\,\Gamma^{j}+c_{235}\,T^{ij}\,\partial^{2}\Gamma^{i}\,\Gamma^{i}+c_{236}\,T^{ij}\,\partial^{2}\Gamma^{j}\,\Gamma^{j}+c_{237}\,\partial^{2}T^{ij}\,U
\nonu\\&&+c_{238}\,\partial T^{ij}\,\partial U+c_{239}\,T^{ij}\,\partial^{2}U+c_{240}\,\partial(T^{ij}\,U\,U)+c_{241}\,\Big(\partial(T^{ik}\,U\,\Gamma^{j}\,\Gamma^{k})-\partial(T^{jk}\,U\,\Gamma^{i}\,\Gamma^{k})\Big)
\nonu\\&&+c_{242}\,\partial^{2}U\,\Gamma^{i}\,\Gamma^{j}+c_{243}\,U\,\partial^{2}\Gamma^{i}\,\Gamma^{j}+c_{244}\,U\,\Gamma^{i}\,\partial^{2}\Gamma^{j}+c_{245}\,\partial U\,U\,\Gamma^{i}\,\Gamma^{j}
+c_{246}\,U\,U\,\partial(\Gamma^{i}\,\Gamma^{j})
\nonu\\&&+(1-\delta^{ij})(1-\delta^{ik}-\delta^{jk})\Bigg\{\,
 c_{247}\, {\bf \partial(\Phi_{\frac{1}{2}}^{(1),i}\,\Phi_{\frac{1}{2}}^{(1),j})}
+c_{248}\,L\,T^{ik}\,T^{jk}+c_{249}\,L\,\partial\Gamma^{i}\,\Gamma^{j}+c_{250}\,L\,\Gamma^{i}\,\partial\Gamma^{j}
\nonu\\&&+c_{251}\,\partial G^{i}\,G^{j}+c_{252}\,G^{i}\,\partial G^{j}+c_{253}\,\partial^{2}G^{i}\,\Gamma^{j}
+c_{254}\,\partial^{2}G^{j}\,\Gamma^{i}+c_{255}\,\partial G^{i}\,\partial\Gamma^{j}+c_{256}\,\partial G^{j}\,\partial\Gamma^{i}
\nonu\\&&+c_{257}\,G^{i}\,\partial^{2}\Gamma^{j}+c_{258}\,G^{j}\,\partial^{2}\Gamma^{i}+c_{259}\,(G^{i}\,G^{k}\,T^{jk}+G^{j}\,G^{k}\,T^{ik})+c_{260}\,\partial(G^{i}\,T^{jk})\,\Gamma^{k}
\nonu\\&&+c_{261}\,\partial(G^{j}\,T^{ik})\,\Gamma^{k}+c_{262}\,\partial(G^{k}\,T^{ij}\,\Gamma^{k})+c_{263}\,\partial(G^{k}\,T^{ik})\,\Gamma^{j}+c_{264}\,\partial(G^{k}\,T^{jk})\,\Gamma^{i}+c_{265}\,G^{i}\,T^{jk}\,\partial\Gamma^{k}
\nonu\\&&+c_{266}\,G^{j}\,T^{ik}\,\partial\Gamma^{k}+c_{267}\,G^{k}\,T^{ik}\,\partial\Gamma^{j}+c_{268}\,G^{k}\,T^{jk}\,\partial\Gamma^{i}+c_{269}\,\Big(G^{i}\,\partial(U\,\Gamma^{j})+\partial G^{j}\,U\,\Gamma^{i}\Big)
\nonu\\&&+c_{270}\,\Big(G^{j}\,\partial(U\,\Gamma^{i})+\partial G^{i}\,U\,\Gamma^{j}\Big)+c_{271}\,\partial(G^{k}\,\Gamma^{i}\,\Gamma^{j}\,\Gamma^{k})+c_{272}\,\partial^{2}T^{ik}\,T^{jk}+c_{273}\,T^{ik}\,\partial^{2}T^{jk}
\nonu\\&&+c_{274}\,\partial T^{ik}\,\partial T^{jk}+c_{275}\,\partial T^{ij}\,T^{ik}\,T^{ik}+c_{276}\,\partial T^{ij}\,T^{ik}\,T^{ik}+c_{277}\,T^{ij}\,\partial T^{ik}\,T^{ik}+c_{278}\,T^{ij}\,\partial T^{jk}\,T^{jk}
\nonu\\&&+c_{279}\,\partial T^{ik}\,T^{jl}\,T^{kl}+c_{280}\,T^{ik}\,\partial T^{jl}\,T^{kl}+c_{281}\,T^{ik}\,T^{jl}\,\partial T^{kl}
\nonu\\&&+c_{282}\,\Big[\partial(T^{ik}\,T^{jl}\,\Gamma^{k}\,\Gamma^{l})+\partial(T^{ij}\,T^{ik}\,\Gamma^{i}\,\Gamma^{k})+\partial(T^{ij}\,T^{jk}\,\Gamma^{j}\,\Gamma^{k})+\partial(T^{ij}\,T^{ji}\,\Gamma^{j}\,\Gamma^{i})\Big]
\nonu\\&&+c_{283}\,\partial^{2}T^{ik}\,\Gamma^{j}\,\Gamma^{k}+c_{284}\,\partial^{2}T^{jk}\,\Gamma^{i}\,\Gamma^{k}+c_{285}\,T^{ik}\,\partial^{2}\Gamma^{j}\,\Gamma^{k}+c_{286}\,T^{jk}\,\partial^{2}\Gamma^{i}\,\Gamma^{k}+c_{287}\,T^{ik}\,\Gamma^{j}\,\partial^{2}\Gamma^{k}
\nonu\\&&+c_{288}\,T^{jk}\,\Gamma^{i}\,\partial^{2}\Gamma^{k}+c_{289}\,\partial T^{ik}\,\partial\Gamma^{j}\,\Gamma^{k}+c_{290}\,\partial T^{jk}\,\partial\Gamma^{i}\,\Gamma^{k}+c_{291}\,\partial T^{ik}\,\Gamma^{j}\,\partial\Gamma^{k}+c_{292}\,\partial T^{jk}\,\Gamma^{i}\,\partial\Gamma^{k}
\nonu\\&&+c_{293}\,T^{ik}\,\partial\Gamma^{j}\,\partial\Gamma^{k}+c_{294}\,T^{jk}\,\partial\Gamma^{i}\,\partial\Gamma^{k}+c_{295}\,T^{ij}\,\partial^{2}\Gamma^{k}\,\Gamma^{k}
+c_{296}\,\partial(T^{ik}\,T^{jk})\,U+c_{297}\,T^{ik}\,T^{jk}\,\partial U
\nonu\\&&+c_{298}\,\partial U\,\partial\Gamma^{i}\,\Gamma^{j}+c_{299}\,\partial U\,\Gamma^{i}\,\partial\Gamma^{j}
+c_{300}\,U\,\partial\Gamma^{i}\,\partial\Gamma^{j}+c_{301}\,\partial^{3}\Gamma^{i}\,\Gamma^{j}
+c_{302}\,\partial^{2}\Gamma^{i}\,\partial\Gamma^{j}
\nonu\\&&+c_{303}\,\partial\Gamma^{i}\,\partial^{2}\Gamma^{j}
+c_{304}\,\Gamma^{i}\,\partial^{3}\Gamma^{j}+c_{305}\,(\partial\Gamma^{i}\,\Gamma^{j}\,\partial\Gamma^{k}\,\Gamma^{k}+\Gamma^{i}\,\Gamma^{j}\,\partial^{2}\Gamma^{k}\,\Gamma^{k}+\Gamma^{i}\,\partial\Gamma^{j}\,\partial\Gamma^{k}\,\Gamma^{k})\Bigg\}
\nonu\\&&+\varepsilon^{ijkl}\Bigg\{\, c_{306}\, {\bf 
\partial\Phi_{1}^{(2),kl}}+c_{307}\, {\bf \partial(\Phi_{0}^{(1)}\,\Phi_{1}^{(1),kl})}+
c_{308}\, {\bf 
\partial(\Phi_{\frac{1}{2}}^{(1),k}\,\Phi_{\frac{1}{2}}^{(1),l})}+c_{309}\,\partial(L\,T^{kl})
+c_{310}\,\partial(L\,\Gamma^{k}\,\Gamma^{l})
\nonu\\&&+c_{311}\,\partial(G^{k}\,G^{l})+c_{312}\,\partial^{2}G^{k}\,\Gamma^{l}+c_{313}\,\partial G^{k}\,\partial\Gamma^{l}+c_{314}\,G^{k}\,\partial^{2}\Gamma^{l}
+c_{315}\,\partial(G^{i}\,T^{ik})\,\Gamma^{l}
\nonu\\&&+c_{316}\,G^{i}\,T^{ik}\,\partial\Gamma^{l}
+c_{317}\,\Big[\partial G^{i}\,T^{kl}\,\Gamma^{i}+G^{j}\,\partial(T^{kl}\,\Gamma^{j})\Big]
+c_{318}\,\Big[\partial G^{j}\,T^{kl}\,\Gamma^{j}+G^{i}\,\partial(T^{kl}\,\Gamma^{i})\Big]
\nonu\\&&+c_{319}\,\Big[\partial(G^{j}\,T^{jk})\,\Gamma^{l}+\frac{1}{2}\partial G^{k}\,T^{il}\,\Gamma^{i}\Big]+c_{320}\,G^{j}\,T^{jk}\,\partial\Gamma^{l}+c_{321}\,\partial G^{k}\,T^{jl}\,\Gamma^{j}
+c_{322}\,\partial(G^{k}\,T^{kl}\,\Gamma^{l})
\nonu\\&&+c_{323}\,(G^{k}\,T^{il}\,\partial\Gamma^{i}+G^{k}\,\partial T^{jl}\,\Gamma^{j})+c_{324}\,(G^{k}\,\partial T^{il}\,\Gamma^{i}+G^{k}\,T^{jl}\,\partial\Gamma^{j})
+c_{325}\,\partial(G^{k}\,U\,\Gamma^{l})
\nonu\\&&+c_{326}\,\Big[\partial(G^{i}\,\Gamma^{i}\,\Gamma^{k}\,\Gamma^{l})+\partial(G^{j}\,\Gamma^{j}\,\Gamma^{k}\,\Gamma^{l})\Big]+c_{327}\,\partial^{3}T^{kl}+c_{328}\,\partial^{2}T^{ik}\,T^{il}
+c_{329}\,\partial^{2}T^{jk}\,T^{jl}
\nonu\\&&+c_{330}\,T^{ik}\,\partial^{2}T^{il}+c_{331}\,T^{jk}\,\partial^{2}T^{jl}+c_{332}\,\partial(T^{ij}\,T^{ij}\,T^{kl})+c_{333}\,\partial(T^{ij}\,T^{ik}\,T^{jl})
\nonu\\&&+c_{334}\,\partial(T^{ik}\,T^{ik}\,T^{kl})+c_{335}\,(\partial T^{ik}\,\partial T^{il}-\partial T^{jk}\,\partial T^{jl})+c_{336}\,\partial T^{kl}\,T^{kl}\,T^{kl}
+c_{337}\,\Big[\partial(T^{ik}\,T^{kl}\,\Gamma^{i}\,\Gamma^{k})
\nonu\\&&+\partial(T^{jk}\,T^{kl}\,\Gamma^{j}\,\Gamma^{k})+\partial(T^{ik}\,T^{jl}\,\Gamma^{i}\,\Gamma^{j})+\frac{1}{2}\partial(T^{kl}\,T^{kl}\,\Gamma^{k}\,\Gamma^{l})\Big]
+c_{338}\,\partial^{2}T^{kl}\,U+c_{339}\,\partial T^{kl}\,\partial U
\nonu\\&&+c_{340}\,T^{kl}\,\partial^{2}U+c_{341}\,\partial(T^{kl}\,U\,U)
+c_{342}\,\Big[T^{ik}\,\partial T^{il}\,U+\partial T^{jk}\,T^{jl}\,U-\partial T^{ik}\,T^{il}\,U-T^{jk}\,\partial T^{jl}\,U\Big]
\nonu\\&&+c_{343}\,\Big[\partial(T^{ik}\,U\,\Gamma^{i}\,\Gamma^{l})+\partial(T^{jk}\,U\,\Gamma^{j}\,\Gamma^{l})\Big]+c_{344}\,\partial^{2}T^{ik}\,\Gamma^{i}\,\Gamma^{l}+c_{345}\,\partial^{2}T^{il}\,\Gamma^{i}\,\Gamma^{k}+c_{346}\,\partial^{2}T^{jk}\,\Gamma^{j}\,\Gamma^{l}
\nonu\\&&+c_{347}\,\partial^{2}T^{jl}\,\Gamma^{j}\,\Gamma^{k}+c_{348}\,\partial T^{ik}\,\partial\Gamma^{i}\,\Gamma^{l}+c_{349}\,\partial T^{il}\,\partial\Gamma^{i}\,\Gamma^{k}+c_{350}\,\partial T^{jk}\,\partial\Gamma^{j}\,\Gamma^{l}+c_{351}\,\partial T^{jl}\,\partial\Gamma^{j}\,\Gamma^{k}
\nonu\\&&+c_{352}\,\partial T^{ik}\,\Gamma^{i}\,\partial\Gamma^{l}+c_{353}\,\partial T^{il}\,\Gamma^{i}\,\partial\Gamma^{k}+c_{354}\,\partial T^{jk}\,\Gamma^{j}\,\partial\Gamma^{l}+c_{355}\,\partial T^{jl}\,\Gamma^{j}\,\partial\Gamma^{k}+c_{356}\,T^{ik}\,\partial\Gamma^{i}\,\partial\Gamma^{l}
\nonu\\&&+c_{357}\,T^{il}\,\partial\Gamma^{i}\,\partial\Gamma^{k}+c_{358}\,T^{jk}\,\partial\Gamma^{j}\,\partial\Gamma^{l}+c_{359}\,T^{jl}\,\partial\Gamma^{j}\,\partial\Gamma^{k}+c_{360}\,T^{ik}\,\partial^{2}\Gamma^{i}\,\Gamma^{l}
+c_{361}\,T^{il}\,\partial^{2}\Gamma^{i}\,\Gamma^{k}
\nonu\\&&+c_{362}\,T^{jk}\,\partial^{2}\Gamma^{j}\,\Gamma^{l}+c_{363}\,T^{jl}\,\partial^{2}\Gamma^{j}\,\Gamma^{k}+c_{364}\,(T^{ik}\,\Gamma^{i}\,\partial^{2}\Gamma^{l}+T^{jk}\,\Gamma^{j}\,\partial^{2}\Gamma^{l})
\nonu\\&&+c_{365}\,(T^{il}\,\Gamma^{i}\,\partial^{2}\Gamma^{k}+T^{jl}\,\Gamma^{j}\,\partial^{2}\Gamma^{k})
+c_{366}\,T^{kl}\,\partial^{2}\Gamma^{i}\,\Gamma^{i}+c_{367}\,T^{kl}\,\partial^{2}\Gamma^{j}\,\Gamma^{j}+c_{368}\,\partial T^{kl}\,\partial\Gamma^{i}\,\Gamma^{i}
\nonu\\&&+c_{369}\,\partial T^{kl}\,\partial\Gamma^{j}\,\Gamma^{j}+c_{370}\,(\partial T^{kl}\,\partial\Gamma^{k}\,\Gamma^{k}+T^{kl}\,\partial^{2}\Gamma^{k}\,\Gamma^{k})+c_{371}\,\partial^{2}U\,\Gamma^{k}\,\Gamma^{l}+c_{372}\,U\,\partial^{2}(\Gamma^{k}\,\Gamma^{l})
\nonu\\&&+c_{373}\,\partial U\,\Gamma^{k}\,\partial\Gamma^{l}+c_{374}\,\partial(U\,U\,\Gamma^{k}\,\Gamma^{l})+c_{375}\,\partial^{3}\Gamma^{k}\,\Gamma^{l}+c_{376}\,\partial^{2}\Gamma^{k}\,\partial\Gamma^{l}+c_{377}\,\partial\Gamma^{k}\,\partial^{2}\Gamma^{l}
\nonu\\&&+c_{378}\,\Gamma^{k}\,\partial^{3}\Gamma^{l}+c_{379}\,\Big[\partial\Gamma^{i}\,\Gamma^{i}\,\partial(\Gamma^{k}\,\Gamma^{l})+\partial\Gamma^{j}\,\Gamma^{j}\,\partial(\Gamma^{k}\,\Gamma^{l})+\partial^{2}\Gamma^{i}\,\Gamma^{i}\,\Gamma^{k}\,\Gamma^{l}+\partial^{2}\Gamma^{j}\,\Gamma^{j}\,\Gamma^{k}\,\Gamma^{l}\Big]\Bigg\}\Bigg](w)+\cdots,
\nonu
\eea
where the coefficients 
are
\bea
c_{1} & =& \frac{64\,k\,N(3+2k+N)(3+k+2N)}{3(2+k+N)^{3}},\nonu\\ 
c_{2}& =&-\frac{16i(9k+12k^{2}+4k^{3}+9N+30k\,N+14k^{2}N+12N^{2}+14k\,N^{2}+4N^{3})}{3(2+k+N)^{3}},\nonu\\ 
c_{3}& =&-\frac{32(9k+12k^{2}+4k^{3}+9N+12k\,N+5k^{2}N+12N^{2}+5k\,N^{2}+4N^{3})}{3(2+k+N)^{4}},\nonu\\ 
c_{4}& =&-\frac{16i(k-N)(9+12k+4k^{2}+12N+10k\,N+4N^{2})}{3(2+k+N)^{3}},\nonu\\ 
c_{5}& =&-\frac{32(k-N)(9+12k+4k^{2}+12N+7k\,N+4N^{2})}{3(2+k+N)^{4}},\qquad
c_ {6}  =  - \frac {8 (k - N)} {3 (2 + k + N)},\nonu\\ 
c_ {7} & = &\frac {4 (k - N) (60 + 77 k + 22 k^{2} + 121 N + 115 k\, 
     N + 20 k^{2} N + 79 N^{2} + 42 k\, 
     N^{2} + 16 N^{3})} {3 (2 + N) (2 + k + N)^{3}}, \nonu\\
c_ {8} & = &\frac {1} {9 (2 + k + N)^{3}} 16 (54 + 105 k + 59 k^{2} + 
    10 k^{3} + 165 N + 209 k\, N + 59 k^{2} N + 128 N^{2} 
    \nonu\\ &+& 83 k\, 
   N^{2} + 28 N^{3}), \nonu\\
c_ {9} & = & - \frac {64 i (k - N)} {3 (2 + k + N)^{2}}, \nonu\\
c_ {10} & = & - \frac {1}{9 (2 + N) (2 + k + N)^{4}}8 (648 + 456 k + 47 k^{2} - 14 k^{3} + 
       1380 N + 614 k\, 
      N + 39 k^{2} N 
      \nonu\\ &+&20 k^{3} N + 1175 N^{2} + 300 k\, 
      N^{2} - 8 k^{2} N^{2} + 431 N^{3} + 40 k\, 
      N^{3} + 56 N^{4}), \nonu\\
c_ {11} & = &\frac {4 (-48 - 52 k - 3 k^{2} + 6 k^{3} - 116 N - 
      80 k\, N - 5 k^{2} N - 85 N^{2} - 30 k\, 
     N^{2} - 19 N^{3})} {3 (2 + N) (2 + k + N)^{3}}, \nonu\\
c_ {12} & = &\frac {4 (-72 - 12 k + 55 k^{2} + 18 k^{3} - 168 N - 
      80 k\, N + 17 k^{2} N - 119 N^{2} - 46 k\, 
     N^{2} - 25 N^{3})} {9 (2 + N) (2 + k + N)^{3}}, \nonu\\
c_ {13} & = & - \frac {16 i (-48 - 28 k + 9 k^{2} + 6 k^{3} - 92 N - 
       44 k\, N + k^{2} N - 61 N^{2} - 18 k\, 
      N^{2} - 13 N^{3})} {3 (2 + N) (2 + k + N)^{4}}, \nonu\\
c_ {14} & = &\frac {16 (3 + k + 2 N) (18 + 29 k + 10 k^{2} + 43 N + 
      30 k\, N + 14 N^{2})} {9 (2 + k + N)^{4}}, \nonu\\
c_ {15} & = & - \frac {32 (k - N) (7 + 2 k + 
       2 N)} {3 (2 + k + N)^{3}}, \qquad
c_ {16}  =  - \frac {32 i (k - N)} {3 (2 + k + N)^{3}}, \nonu\\
c_ {17} & = & - \frac {1}{9 (2 + N) (2 + k + N)^{4}}8 (504 + 456 k + 111 k^{2} - 14 k^{3} + 
       1308 N + 918 k\, 
      N + 215 k^{2} N 
      \nonu\\ &+& 20 k^{3} N + 1239 N^{2} + 596 k\, 
      N^{2} + 64 k^{2} N^{2} + 463 N^{3} + 112 k\, 
      N^{3} + 56 N^{4}), \nonu\\
c_ {18} & = & - \frac {8 i (-72 - 12 k + 43 k^{2} + 18 k^{3} - 
       168 N - 56 k\, N + 11 k^{2} N - 131 N^{2} - 34 k\, 
      N^{2} - 31 N^{3})} {9 (2 + N) (2 + k + N)^{4}}, \nonu\\
c_ {19} & = & - \frac {8 i (k - N) (8 + 17 k + 6 k^{2} + 11 N + 
       11 k\, N + 3 N^{2})} {3 (2 + N) (2 + k + N)^{4}}, \nonu\\
c_ {20} & = &\frac {4 (k - N) (24 + 21 k + 6 k^{2} + 23 N + 13 k\, 
     N + 5 N^{2})} {3 (2 + N) (2 + k + N)^{3}}, \nonu\\
c_ {21} & = & - \frac {16 (k - N) (32 + 55 k + 18 k^{2} + 41 N + 
       35 k\, N + 11 N^{2})} {9 (2 + N) (2 + k + N)^{5}}, \nonu\\
c_ {22} & = & - \frac {16 i (k - N) (3 + 2 k + 
       2 N)} {3 (2 + k + N)^{3}}, \nonu\\
c_ {23} & = & - \frac {8 i (18 + 63 k + 52 k^{2} + 12 k^{3} + 63 N + 
       112 k\, N + 42 k^{2} N + 52 N^{2} + 42 k\, 
      N^{2} + 12 N^{3})} {(9 (2 + k + N)^{3})}, \nonu\\
c_ {24} & = & - \frac {16 (9 + 18 k + 8 k^{2} + 18 N + 11 k\, 
      N + 8 N^{2})} {(9 (2 + k + N)^{3})}, \nonu\\
c_ {25} & = & - \frac {16 (18 + 27 k + 28 k^{2} + 12 k^{3} + 27 N - 
       2 k\, N - 3 k^{2} N + 28 N^{2} - 3 k\, 
      N^{2} + 12 N^{3})} {9 (2 + k + N)^{4}}, \nonu\\
c_ {26} & = & - \frac {16 (-18 + 27 k + 44 k^{2} + 12 k^{3} + 27 N + 
       74 k\, N + 33 k^{2} N + 44 N^{2} + 33 k\, 
      N^{2} + 12 N^{3})} {9 (2 + k + N)^{4}}, \nonu\\
c_ {27} & = &\frac {32 i (3 + 2 k + N) (3 + k + 
      2 N)} {9 (2 + k + N)^{4}}, \nonu\\
c_ {28} & = & - \frac {4 i (k - N) (9 + 12 k + 4 k^{2} + 12 N + 
       10 k\, N + 4 N^{2})} {3 (2 + k + N)^{3}}, \nonu\\
c_ {29} & = & - \frac {8 (k - N) (9 + 12 k + 4 k^{2} + 12 N + 7 k\, 
      N + 4 N^{2})} {3 (2 + k + N)^{4}},\nonu\\ c_ {30} & = & 3, \nonu\\
c_ {31} & = & - \frac {3 (60 + 77 k + 22 k^{2} + 121 N + 115 k\, 
      N + 20 k^{2} N + 79 N^{2} + 42 k\, 
      N^{2} + 16 N^{3})} {(2 + N) (2 + k + N)^{2}}, \nonu\\
c_ {32} & = & - \frac {4 i (144 + 295 k + 171 k^{2} + 30 k^{3} + 
       257 N + 335 k\, N + 95 k^{2} N + 142 N^{2} + 90 k\, 
      N^{2} + 25 N^{3})} {3 (2 + N) (2 + k + N)^{3}}, \nonu\\
c_ {33} & = & - \frac {8 (96 + 223 k + 147 k^{2} + 30 k^{3} + 161 N + 
       251 k\, N + 83 k^{2} N + 82 N^{2} + 66 k\, 
      N^{2} + 13 N^{3})} {3 (2 + N) (2 + k + N)^{4}}, \nonu\\
c_ {34} & = &\frac {6 (20 + 21 k + 6 k^{2} + 25 N + 13 k\, 
     N + 7 N^{2})} {(2 + N) (2 + k + N)^{3}}, \nonu\\
c_ {35} & = &\frac {2 (k - N) (8 + 17 k + 6 k^{2} + 11 N + 11 k\, 
     N + 3 N^{2})} {3 (2 + N) (2 + k + N)^{4}}, \nonu\\
c_ {36} & = &\frac {4 (180 + 271 k + 138 k^{2} + 24 k^{3} + 323 N + 
      325 k\, N + 82 k^{2} N + 185 N^{2} + 94 k\, 
     N^{2} + 34 N^{3})} {3 (2 + N) (2 + k + N)^{4}}, \nonu\\
c_ {37} & = &\frac {4 (k - N) (12 + 19 k + 6 k^{2} + 15 N + 12 k\, 
     N + 4 N^{2})} {3 (2 + N) (2 + k + N)^{4}}, \nonu\\
c_ {38} & = &\frac {1} {3 (2 + N) (2 + k + N)^{5}}4 i (864 + 1998 k + 1448 k^{2} + 393 k^{3} + 
      30 k^{4} + 1890 N + 2990 k\, 
     N 
     \nonu\\ &+& 1307 k^{2} N + 149 k^{3} N + 1466 N^{2} + 1409 k\, 
     N^{2} + 267 k^{2} N^{2} + 491 N^{3} + 213 k\, 
     N^{3} + 61 N^{4}), \nonu\\
c_ {39} & = & - \frac {1} {3 (2 + N) (2 + k + N)^{4}}2 i (288 + 736 k + 501 k^{2} + 102 k^{3} + 
       560 N + 896 k\, N + 299 k^{2} N 
       \nonu\\ &+& 331 N^{2} + 258 k\, 
      N^{2} + 61 N^{3}), \nonu\\
c_ {40} & = &\frac {4 i (k - N) (32 + 55 k + 18 k^{2} + 41 N + 35 k\, 
     N + 11 N^{2})} {3 (2 + N) (2 + k + N)^{5}}, \nonu\\
c_ {41} & = &\frac { i (32 + 55 k + 18 k^{2} + 41 N + 35 k\, 
     N + 11 N^{2})} {(2 + N) (2 + k + N)^{3}}, \nonu\\
c_ {42} & = & - \frac { i (48 + 155 k + 114 k^{2} + 24 k^{3} + 
       109 N + 199 k\, N + 70 k^{2} N + 71 N^{2} + 60 k\, 
      N^{2} + 14 N^{3})} {6 (2 + N) (2 + k + N)^{4}}, \nonu\\
c_ {43} & = &\frac {12 (32 + 55 k + 18 k^{2} + 41 N + 35 k\, 
     N + 11 N^{2})} {(2 + N) (2 + k + N)^{4}}, \nonu\\
c_ {44} & = & - \frac{1} {3 (2 + N) (2 + k + N)^{4}}4 i (144 + 271 k + 159 k^{2} + 30 k^{3} + 
       233 N + 299 k\, N + 89 k^{2} N 
       \nonu\\ &+& 118 N^{2} + 78 k\, 
      N^{2} + 19 N^{3}), \nonu\\
c_ {45} & = & - \frac{1} {3 (2 + N) (2 + k + N)^{4}}2 i (-360 - 382 k - 119 k^{2} - 10 k^{3} - 
       806 N - 580 k\, 
      N - 57 k^{2} N 
      \nonu\\ &+& 16 k^{3} N - 597 N^{2} - 228 k\, 
      N^{2} + 16 k^{2} N^{2} - 173 N^{3} - 16 k\, 
      N^{3} - 16 N^{4}), \nonu\\
c_ {46} & = &\frac {2 i (-540 - 495 k - 130 k^{2} - 747 N - 315 k\, 
     N + 16 k^{2} N - 257 N^{2} - 16 N^{3})} {3 (2 + 
      N) (2 + k + N)^{3}}, \nonu\\
c_ {47} & = & - \frac {1}{9 (2 + N) (2 + k + N)^{4}}2 i (-1296 - 2220 k - 997 k^{2} + 52 k^{4} - 
       2748 N - 2860 k\, 
      N 
      \nonu\\ &+& 127 k^{2} N + 660 k^{3} N + 116 k^{4} N - 2479 N^{2} - 
       1442 k\, 
      N^{2} + 592 k^{2} N^{2} + 282 k^{3} N^{2} - 1349 N^{3} 
      \nonu\\ &-& 
       570 k\, N^{3} + 92 k^{2} N^{3} - 446 N^{4} - 138 k\, 
      N^{4} - 64 N^{5}), \nonu\\
c_ {48} & = &\frac{1} {3 (2 + N) (2 + k + N)^{4}}4 i (672 + 1240 k + 489 k^{2} + 30 k^{3} + 
      1016 N + 1016 k\, N + 119 k^{2} N 
      \nonu\\ &+&415 N^{2} + 138 k\, 
     N^{2} + 49 N^{3}), \nonu\\
c_ {49} & = & - \frac {12 (32 + 55 k + 18 k^{2} + 41 N + 35 k\, 
      N + 11 N^{2})} {(2 + N) (2 + k + N)^{4}}, \nonu\\
c_ {50} & = & - \frac {1}{3 (2 + N) (2 + k + N)^{5}}16 (180 + 321 k + 175 k^{2} + 30 k^{3} + 
       453 N + 647 k\, 
      N + 264 k^{2} N 
      \nonu\\ &+& 30 k^{3} N + 420 N^{2} + 421 k\, 
      N^{2} + 93 k^{2} N^{2} + 167 N^{3} + 87 k\, 
      N^{3} + 24 N^{4}), \nonu\\
c_ {51} & = & - \frac {1}{3 (2 + N) (2 + k + N)^{5}}4 (360 + 642 k + 389 k^{2} + 78 k^{3} + 
       906 N + 1264 k\, 
      N + 537 k^{2} N 
      \nonu\\ &+& 60 k^{3} N + 831 N^{2} + 818 k\, 
      N^{2} + 186 k^{2} N^{2} + 331 N^{3} + 174 k\, 
      N^{3} + 48 N^{4}), \nonu\\
c_ {52} & = &\frac {8 (18 + 18 k + 5 k^{2} + 18 N + 8 k\, 
     N + 5 N^{2}) (32 + 55 k + 18 k^{2} + 41 N + 35 k\, 
     N + 11 N^{2})} {3 (2 + N) (2 + k + N)^{6}}, \nonu\\
c_ {53} & = & - \frac {1} {9 (2 + N) (2 + k + N)^{5}}8 (1728 + 5832 k + 5427 k^{2} + 1934 k^{3} + 
       232 k^{4} + 3096 N + 8664 k\, 
      N 
      \nonu\\ &+& 6127 k^{2} N + 1522 k^{3} N + 116 k^{4} N + 1173 N^{2} + 
       3688 k\, 
      N^{2} + 2106 k^{2} N^{2} + 354 k^{3} N^{2} - 517 N^{3} 
      \nonu\\ &+& 
       284 k\, N^{3} + 236 k^{2} N^{3} - 400 N^{4} - 66 k\, 
      N^{4} - 64 N^{5}), \nonu\\
c_ {54} & = &\frac {(k - N) (60 + 77 k + 22 k^{2} + 121 N + 115 k\, 
     N + 20 k^{2} N + 79 N^{2} + 42 k\, 
     N^{2} + 16 N^{3})} {3 (2 + N) (2 + k + N)^{3}}, \nonu\\
c_ {55} & = &\frac {(192 + 592 k + 453 k^{2} + 102 k^{3} + 368 N + 
     728 k\, N + 275 k^{2} N + 211 N^{2} + 210 k\, 
    N^{2} + 37 N^{3})} {3 (2 + N) (2 + k + N)^{3}}, \nonu\\
c_ {56} & = & - \frac{1} {3 (2 + N) (2 + k + N)^{4}}4 i (82 k + 72 k^{2} + 16 k^{3} + 98 N + 
       348 k\, N + 219 k^{2} N + 38 k^{3} N 
       \nonu\\ &+& 174 N^{2} + 319 k\, 
      N^{2} + 101 k^{2} N^{2} + 94 N^{3} + 79 k\, 
      N^{3} + 16 N^{4}), \nonu\\
c_ {57} & = &\frac {1} {3 (2 + N) (2 + k + N)^{4}}4 i (34 k + 16 k^{2} + 146 N + 324 k\, 
     N + 175 k^{2} N + 30 k^{3} N + 254 N^{2} 
     \nonu\\ &+& 335 k\, 
     N^{2} + 93 k^{2} N^{2} + 138 N^{3} + 87 k\, 
     N^{3} + 24 N^{4}), \nonu\\
c_ {58} & = &\frac {32 i (k - N)} {3 (2 + k + N)^{2}}, \qquad
c_ {59}  =  - \frac {16 i (k - N) (7 + 4 k + 
       4 N)} {3 (2 + k + N)^{3}}, \nonu\\
c_ {60} & = &\frac{1}{9 (2 + N) (2 + k + N)^{4}}2 (1296 + 3876 k + 3178 k^{2} + 959 k^{3} + 
      90 k^{4} + 2244 N + 4756 k\, 
     N 
     \nonu\\ &+& 2349 k^{2} N + 283 k^{3} N + 994 N^{2} + 1383 k\, 
     N^{2} + 257 k^{2} N^{2} - 11 N^{3} - k\, 
     N^{3} - 53 N^{4}), \nonu\\
c_ {61} & = & - \frac {1}{9 (2 + N) (2 + k + N)^{4}}2 (1584 + 4416 k + 3491 k^{2} + 1021 k^{3} + 
       90 k^{4} + 3288 N + 5912 k\, 
      N 
      \nonu\\ &+& 2646 k^{2} N + 287 k^{3} N + 2261 N^{2} + 2259 k\, 
      N^{2} + 343 k^{2} N^{2} + 626 N^{3} + 229 k\, 
      N^{3} + 59 N^{4}), \nonu\\
c_ {62} & = &\frac {4 i (k - N) (32 + 55 k + 18 k^{2} + 41 N + 35 k\, 
     N + 11 N^{2})} {3 (2 + N) (2 + k + N)^{5}}, \nonu\\
c_ {63} & = & - \frac {1}{9 (2 + N) (2 + k + N)^{4}}4 i (936 + 1938 k + 875 k^{2} + 90 k^{3} + 
       1266 N + 1478 k\, N + 223 k^{2} N 
       \nonu\\ &+& 383 N^{2} + 142 k\, 
      N^{2} + 13 N^{3}), \nonu\\
c_ {64} & = &\frac {1}{9 (2 + N) (2 + k + N)^{5}}4 i (2016 + 4968 k + 3707 k^{2} + 1045 k^{3} + 
      90 k^{4} + 3816 N + 6464 k\, 
     N 
     \nonu\\ &+& 2814 k^{2} N + 299 k^{3} N + 2357 N^{2} + 2361 k\, 
     N^{2} + 373 k^{2} N^{2} + 548 N^{3} + 211 k\, 
     N^{3} + 35 N^{4}), \nonu\\
c_ {65} & = &\frac{1}{9 (2 + N) (2 + k + N)^{5}}4 i (576 + 1506 k + 1397 k^{2} + 577 k^{3} + 
      90 k^{4} + 1374 N + 2570 k\, 
     N 
     \nonu\\ &+& 1506 k^{2} N + 299 k^{3} N + 1217 N^{2} + 1467 k\, 
     N^{2} + 403 k^{2} N^{2} + 482 N^{3} + 289 k\, 
     N^{3} + 71 N^{4}), \nonu\\
c_ {66} & = &\frac {1}{9 (2 + N) (2 + k + N)^{5}}4 i (2160 + 5064 k + 3665 k^{2} + 1021 k^{3} + 
      90 k^{4} + 4224 N + 6692 k\, 
     N 
     \nonu\\ &+& 2736 k^{2} N + 269 k^{3} N + 2819 N^{2} + 2589 k\, 
     N^{2} + 355 k^{2} N^{2} + 782 N^{3} + 289 k\, 
     N^{3} + 77 N^{4}), \nonu\\
c_ {67} & = & - \frac {1}{9 (2 + N) (2 + k + N)^{5}}4 i (1728 + 4716 k + 3730 k^{2} + 1079 k^{3} + 
       90 k^{4} + 3060 N + 5932 k\, 
      N 
      \nonu\\ &+& 2877 k^{2} N + 343 k^{3} N + 1570 N^{2} + 1917 k\, 
      N^{2} + 383 k^{2} N^{2} + 175 N^{3} + 77 k\, 
      N^{3} - 29 N^{4}), \nonu\\
c_ {68} & = & - \frac {1}{9 (2 + N) (2 + k + N)^{5}}4 i (288 + 1254 k + 1420 k^{2} + 611 k^{3} + 
       90 k^{4} + 618 N + 2038 k\, 
      N 
      \nonu\\ &+& 1569 k^{2} N + 343 k^{3} N + 430 N^{2} + 1023 k\, 
      N^{2} + 413 k^{2} N^{2} + 109 N^{3} + 155 k\, 
      N^{3} + 7 N^{4}), \nonu\\
c_ {69} & = & - \frac {1} {9 (2 + N) (2 + k + N)^{4}}4 (-720 - 1560 k - 653 k^{2} - 38 k^{3} - 
       1320 N - 1730 k\, 
      N - 89 k^{2} N 
      \nonu\\ &+& 116 k^{3} N - 893 N^{2} - 740 k\, 
      N^{2} + 40 k^{2} N^{2} - 357 N^{3} - 200 k\, 
      N^{3} - 64 N^{4}), \nonu\\
c_ {70} & = & - \frac {1}{9 (2 + N) (2 + k + N)^{5}}4 (-1728 - 3804 k - 2507 k^{2} - 599 k^{3} - 
       38 k^{4} - 3828 N - 5672 k\, 
      N 
      \nonu\\ &-&1562 k^{2} N + 341 k^{3} N + 116 k^{4} N - 3413 N^{2} - 
       3251 k\, 
      N^{2} - 67 k^{2} N^{2} + 228 k^{3} N^{2} - 1716 N^{3} 
      \nonu\\ &-&
       1103 k\, N^{3} - 16 k^{2} N^{3} - 501 N^{4} - 192 k\, 
      N^{4} - 64 N^{5}), \nonu\\
c_ {71} & = &\frac {(k - N)} {6 (2 + k + N)}, \nonu\\
c_ {72} & = & - \frac {(k - N) (60 + 77 k + 22 k^{2} + 121 N + 
       115 k\, N + 20 k^{2} N + 79 N^{2} + 42 k\, 
      N^{2} + 16 N^{3})} {6 (2 + N) (2 + k + N)^{3}}, \nonu\\
c_ {73} & = &\frac {3 (60 + 77 k + 22 k^{2} + 121 N + 115 k\, 
     N + 20 k^{2} N + 79 N^{2} + 42 k\, 
     N^{2} + 16 N^{3})} {2(2 + N) (2 + k + N)^{2}}, \nonu\\
c_ {74} & = & - \frac {2 i (180 + 311 k + 171 k^{2} + 30 k^{3} + 
       283 N + 335 k\, N + 95 k^{2} N + 142 N^{2} + 86 k\, 
      N^{2} + 23 N^{3})} {3 (2 + N) (2 + k + N)^{3}}, \nonu\\
c_ {75} & = & - \frac {4 (180 + 279 k + 155 k^{2} + 30 k^{3} + 
       315 N + 319 k\, N + 87 k^{2} N + 174 N^{2} + 86 k\, 
      N^{2} + 31 N^{3})} {3 (2 + N) (2 + k + N)^{4}}, \nonu\\
c_ {76} & = & - \frac {(360 + 574 k + 279 k^{2} + 42 k^{3} + 614 N + 
      676 k\, N + 169 k^{2} N + 341 N^{2} + 196 k\, 
     N^{2} + 61 N^{3})} {6 (2 + N) (2 + k + N)^{3}}, \nonu\\
c_ {77} & = & - \frac {2 (48 + 155 k + 114 k^{2} + 24 k^{3} + 109 N + 
       199 k\, N + 70 k^{2} N + 71 N^{2} + 60 k\, 
      N^{2} + 14 N^{3})} {3 (2 + N) (2 + k + N)^{4}}, \nonu\\
c_ {78} & = & - \frac {4 (28 + 53 k + 18 k^{2} + 37 N + 34 k\, 
      N + 10 N^{2})} {(2 + N) (2 + k + N)^{3}}, \nonu\\
c_ {79} & = & - \frac {6 (8 + 17 k + 6 k^{2} + 11 N + 11 k\, 
      N + 3 N^{2})} {(2 + N) (2 + k + N)^{3}}, \nonu\\
c_ {80} & = &\frac {2 (96 + 116 k + 45 k^{2} + 6 k^{3} + 124 N + 
      100 k\, N + 19 k^{2} N + 47 N^{2} + 18 k\, 
     N^{2} + 5 N^{3})} {3 (2 + N) (2 + k + N)^{4}}, \nonu\\
c_ {81} & = &\frac {2 i (360 + 526 k + 255 k^{2} + 42 k^{3} + 662 N + 
      652 k\, N + 157 k^{2} N + 389 N^{2} + 196 k\, 
     N^{2} + 73 N^{3})} {3 (2 + N) (2 + k + N)^{4}}, \nonu\\
c_ {82} & = &\frac {1}{3 (2 + N) (2 + k + N)^{4}}4 i (240 + 522 k + 328 k^{2} + 64 k^{3} + 462 N + 
      666 k\, N + 213 k^{2} N 
      \nonu\\ &+& 2 k^{3} N + 302 N^{2} + 241 k\, 
     N^{2} + 15 k^{2} N^{2} + 82 N^{3} + 23 k\, 
     N^{3} + 8 N^{4}), \nonu\\
c_ {83} & = & - \frac {8 i (6 + 21 k + 8 k^{2} + 21 N + 20 k\, 
      N + 8 N^{2})} {3 (2 + k + N)^{3}}, \nonu\\
c_ {84} & = &\frac {6 i (32 + 55 k + 18 k^{2} + 41 N + 35 k\, 
     N + 11 N^{2})} {(2 + N) (2 + k + N)^{4}}, \nonu\\
c_ {85} & = & - \frac{1}{9 (2 + N) (2 + k + N)^{4}} i (-1080 - 1938 k - 945 k^{2} - 16 k^{3} + 
       52 k^{4} - 1626 N - 1536 k\, 
      N 
      \nonu\\ &+&531 k^{2} N + 692 k^{3} N + 116 k^{4} N - 327 N^{2} + 
       1032 k\, 
      N^{2} + 1512 k^{2} N^{2} + 410 k^{3} N^{2} + 613 N^{3} 
      \nonu\\ &+& 
       1250 k\, N^{3} + 516 k^{2} N^{3} + 382 N^{4} + 298 k\, 
      N^{4} + 64 N^{5}), \nonu\\
c_ {86} & = &\frac { (360 + 538 k + 237 k^{2} + 30 k^{3} + 650 N + 
      678 k\, N + 155 k^{2} N + 381 N^{2} + 212 k\, 
     N^{2} + 71 N^{3})} {3 (2 + N) (2 + k + N)^{3}}, \nonu\\
c_ {87} & = & - \frac {1}{3 (2 + N) (2 + k + N)^{3}} (-180 - 155 k + 39 k^{2} + 30 k^{3} - 
       439 N - 357 k\, N - 25 k^{2} N 
       \nonu\\ &-& 330 N^{2} - 166 k\, 
      N^{2} - 73 N^{3}), \nonu\\
c_ {88} & = &\frac {3 i (20 + 21 k + 6 k^{2} + 25 N + 13 k\, 
     N + 7 N^{2})} {(2 + N) (2 + k + N)^{3}}, \nonu\\
c_ {89} & = & - \frac {4 i (180 + 271 k + 138 k^{2} + 24 k^{3} + 
       323 N + 325 k\, N + 82 k^{2} N + 185 N^{2} + 94 k\, 
      N^{2} + 34 N^{3})} {3 (2 + N) (2 + k + N)^{4}}, \nonu\\
c_ {90} & = &\frac {2 i (k - N) (16 + 21 k + 6 k^{2} + 19 N + 13 k\, 
     N + 5 N^{2})} {3 (2 + N) (2 + k + N)^{4}}, \nonu\\
c_ {91} & = & - \frac{1} {3 (2 + N) (2 + k + N)^{4}} i (384 + 976 k + 661 k^{2} + 134 k^{3} + 
       848 N + 1368 k\, 
      N + 491 k^{2} N 
      \nonu\\ &+& 16 k^{3} N + 611 N^{2} + 546 k\, 
      N^{2} + 56 k^{2} N^{2} + 173 N^{3} + 56 k\, 
      N^{3} + 16 N^{4}), \nonu\\
c_ {92} & = &\frac {1}{3 (2 + N) (2 + k + N)^{4}} i (-192 - 108 k + 25 k^{2} + 14 k^{3} - 84 N + 
      172 k\, N + 135 k^{2} N 
      \nonu\\ &+& 16 k^{3} N + 139 N^{2} + 234 k\, 
     N^{2} + 56 k^{2} N^{2} + 97 N^{3} + 56 k\, 
     N^{3} + 16 N^{4}), \nonu\\
c_ {93} & = & - \frac {2 i (180 + 287 k + 159 k^{2} + 30 k^{3} + 
       307 N + 323 k\, N + 89 k^{2} N + 166 N^{2} + 86 k\, 
      N^{2} + 29 N^{3})} {3 (2 + N) (2 + k + N)^{4}}, \nonu\\
c_ {94} & = &\frac {12 i (32 + 55 k + 18 k^{2} + 41 N + 35 k\, 
     N + 11 N^{2})} {(2 + N) (2 + k + N)^{4}}, \nonu\\
c_ {95} & = &\frac {4 (k - N) (32 + 55 k + 18 k^{2} + 41 N + 35 k\, 
     N + 11 N^{2})} {3 (2 + N) (2 + k + N)^{5}}, \nonu\\
c_ {96} & = &\frac{1}{3 (2 + N) (2 + k + N)^{5}}2 i (k - N) (96 + 249 k + 159 k^{2} + 30 k^{3} + 
      167 N + 269 k\, N + 83 k^{2} N 
      \nonu\\ &+& 84 N^{2} + 66 k\, 
     N^{2} + 13 N^{3}), \nonu\\
c_ {97} & = &\frac {2 i (-180 - 183 k + 23 k^{2} + 30 k^{3} - 411 N - 
      371 k\, N - 33 k^{2} N - 300 N^{2} - 166 k\, 
     N^{2} - 65 N^{3})} {3 (2 + N) (2 + k + N)^{4}}, \nonu\\
c_ {98} & = & - \frac {1}{3 (2 + N) (2 + k + N)^{5}}2 i (1080 + 2022 k + 1338 k^{2} + 357 k^{3} + 
       30 k^{4} + 2622 N + 3770 k\, 
      N 
      \nonu\\ &+& 1703 k^{2} N + 233 k^{3} N + 2344 N^{2} + 2317 k\, 
      N^{2} + 541 k^{2} N^{2} + 915 N^{3} + 469 k\, 
      N^{3} + 131 N^{4}), \nonu\\
c_ {99} & = & - \frac {1}{3 (2 + N) (2 + k + N)^{4}}2 i (360 + 510 k + 221 k^{2} + 30 k^{3} + 
       678 N + 664 k\, N + 147 k^{2} N
       \nonu\\ &+& 411 N^{2} + 212 k\, 
      N^{2} + 79 N^{3}), \nonu\\
c_ {100} & = &\frac {1}{3 (2 + N) (2 + k + N)^{5}}2 i (k - N) (96 + 252 k + 161 k^{2} + 30 k^{3} + 
      180 N + 284 k\, N + 87 k^{2} N 
      \nonu\\ &+& 99 N^{2} + 74 k\, 
     N^{2} + 17 N^{3}), \nonu\\
c_ {101} & = &\frac {1} {3 (2 + N) (2 + k + N)^{5}}4 i (360 + 738 k + 638 k^{2} + 237 k^{3} + 
      30 k^{4} + 810 N + 1182 k\, 
     N 
     \nonu\\ &+& 647 k^{2} N + 113 k^{3} N + 664 N^{2} + 633 k\, 
     N^{2} + 169 k^{2} N^{2} + 247 N^{3} + 121 k\, 
     N^{3} + 35 N^{4}), \nonu\\
c_ {102} & = & - \frac {2 i (180 + 271 k + 151 k^{2} + 30 k^{3} + 
       323 N + 315 k\, N + 85 k^{2} N + 182 N^{2} + 86 k\, 
      N^{2} + 33 N^{3})} {3 (2 + N) (2 + k + N)^{4}}, \nonu\\
c_ {103} & = &\frac{1}{3 (2 + N) (2 + k + N)^{4}}2 i (-360 - 422 k - 47 k^{2} + 30 k^{3} - 
      766 N - 720 k\, N - 95 k^{2} N 
      \nonu\\ &-& 529 N^{2} - 292 k\, 
     N^{2} - 111 N^{3}), \nonu\\
c_ {104} & = &\frac{1}{3 (2 + N) (2 + k + N)^{5}}8 (144 + 323 k + 209 k^{2} + 42 k^{3} + 325 N + 
      467 k\, N + 160 k^{2} N 
      \nonu\\ &+& 6 k^{3} N + 260 N^{2} + 211 k\, 
     N^{2} + 23 k^{2} N^{2} + 91 N^{3} + 31 k\, 
     N^{3} + 12 N^{4}), \nonu\\
c_ {105} & = & - \frac {1}{3 (2 + N) (2 + k + N)^{5}}4 (576 + 1352 k + 913 k^{2} + 190 k^{3} + 
       1240 N + 1912 k\, 
      N + 733 k^{2} N 
      \nonu\\ &+& 44 k^{3} N + 919 N^{2} + 808 k\, 
      N^{2} + 114 k^{2} N^{2} + 285 N^{3} + 98 k\, 
      N^{3} + 32 N^{4}), \nonu\\
c_ {106} & = & - \frac {2 (k - N) (32 + 55 k + 18 k^{2} + 41 N + 
       35 k\, N + 11 N^{2})} {3 (2 + N) (2 + k + N)^{5}}, \nonu\\
c_ {107} & = &\frac {20 (k - N) (32 + 55 k + 18 k^{2} + 41 N + 35 k\, 
     N + 11 N^{2})} {3 (2 + N) (2 + k + N)^{5}}, \nonu\\
c_ {108} & = & - \frac {1}{9 (2 + N) (2 + k + N)^{5}} 2 (-1080 - 1698 k - 1110 k^{2} - 337 k^{3} - 
       38 k^{4} 
       \nonu\\ &-& 1866 N - 1086 k\, 
      N + 519 k^{2} N + 557 k^{3} N + 116 k^{4} N - 612 N^{2} + 
       1491 k\, 
      N^{2} + 1587 k^{2} N^{2} 
      \nonu\\ &+& 392 k^{3} N^{2} + 487 N^{3} + 
       1421 k\, N^{3} + 516 k^{2} N^{3} + 361 N^{4} + 316 k\, 
      N^{4} + 64 N^{5}), \nonu\\
c_ {109} & = & - \frac {1}{9 (2 + N) (2 + k + N)^{5}} 4 (3240 + 6318 k + 4863 k^{2} + 1718 k^{3} + 
       232 k^{4} + 8694 N + 13428 k\, 
      N 
      \nonu\\ &+& 7587 k^{2} N + 1742 k^{3} N + 116 k^{4} N + 8709 N^{2} + 
       9954 k\, 
      N^{2} + 3654 k^{2} N^{2} + 410 k^{3} N^{2} + 4069 N^{3} 
      \nonu\\ &+& 
       3008 k\, N^{3} + 516 k^{2} N^{3} + 868 N^{4} + 298 k\, 
      N^{4} + 64 N^{5}), \nonu\\
c_ {110} & = & - \frac {1}{9 (2 + N) (2 + k + N)^{5}} 2 (2160 + 4080 k + 2157 k^{2} + 257 k^{3} - 
       38 k^{4} + 6288 N + 10470 k\, 
      N 
      \nonu\\ &+& 5298 k^{2} N + 1097 k^{3} N + 116 k^{4} N + 6921 N^{2} + 
       8997 k\, 
      N^{2} + 3261 k^{2} N^{2} + 392 k^{3} N^{2} + 3484 N^{3} 
      \nonu\\ &+& 
       2987 k\, N^{3} + 516 k^{2} N^{3} + 793 N^{4} + 316 k\, 
      N^{4} + 64 N^{5}), \nonu\\
c_ {111} & = &\frac {1} {2}, \nonu\\
c_ {112} & = &\frac {1}{4 (2 + N) (2 + k + N)^{2} (59 + 37 k + 37 N + 15 k\, N)}(3540 + 6731 k + 4067 k^{2} + 782 k^{3} 
\nonu\\ &+& 
     9391 N + 14995 k\, 
    N + 7236 k^{2} N + 1006 k^{3} N + 9234 N^{2} + 11599 k\, 
    N^{2} + 3955 k^{2} N^{2} 
    \nonu\\ &+& 276 k^{3} N^{2} + 3939 N^{3} + 
     3443 k\, N^{3} + 630 k^{2} N^{3} + 608 N^{4} + 264 k\, 
    N^{4}),\nonu\\
 c_ {113} & = & - \frac {(60 + 77 k + 22 k^{2} + 121 N + 115 k\, 
     N + 20 k^{2} N + 79 N^{2} + 42 k\, 
     N^{2} + 16 N^{3})} {2 (2 + N) (2 + k + N)^{2}}, \nonu\\
c_ {114} & = &\frac {(k - N) (60 + 77 k + 22 k^{2} + 121 N + 115 k\, 
     N + 20 k^{2} N + 79 N^{2} + 42 k\, 
     N^{2} + 16 N^{3})} {6 (2 + N) (2 + k + N)^{3}}, \nonu\\
c_ {115} & = & - \frac{1}{12 (2 + N)^{2} (2 + k + N)^{4} (5 + 4 k + 4 N + 3 k\, 
      N)}(60 + 77 k + 22 k^{2} + 121 N + 115 k\, 
      N 
      \nonu\\ &+& 20 k^{2} N + 79 N^{2} + 42 k\, 
      N^{2} + 16 N^{3}) (900 + 2179 k + 1534 k^{2} + 328 k^{3} + 
       2231 N + 4793 k\, 
      N 
      \nonu\\ &+& 2793 k^{2} N + 446 k^{3} N + 2205 N^{2} + 4031 k\, 
      N^{2} + 1787 k^{2} N^{2} + 168 k^{3} N^{2} + 984 N^{3} + 
       1423 k\, N^{3} 
       \nonu\\ &+& 378 k^{2} N^{3} + 160 N^{4} + 156 k\, 
      N^{4}),\nonu\\ 
c_ {116} & = & - \frac{1}{4 (2 + N)^{2} (2 + k + N)^{4}}3 (1 + N) (20 + 23 k + 6 k^{2} + 
       23 N + 14 k\, 
      N + 6 N^{2}) (60 + 77 k 
      \nonu\\ &+& 22 k^{2} + 121 N + 115 k\, 
      N + 20 k^{2} N + 79 N^{2} + 42 k\, 
      N^{2} + 16 N^{3}), \nonu\\
c_ {117} & = & - \frac {36 (k - N)} {(59 + 37 k + 37 N + 15 k\, N)}, \nonu\\
c_ {118} & = &\frac {4 (k - N) (60 + 77 k + 22 k^{2} + 121 N + 
      115 k\, N + 20 k^{2} N + 79 N^{2} + 42 k\, 
     N^{2} + 16 N^{3})} {(2 + N) (2 + k + N)^{2} (5 + 4 k + 4 N + 
      3 k\, N)}, \nonu\\
c_ {119} & = &\frac {1}{2 (2 + N) (2 + k + N)^{4}}(-936 - 1546 k - 475 k^{2} + 184 k^{3} + 
     76 k^{4} - 854 N + 401 k\, 
    N 
    \nonu\\ &+& 1960 k^{2} N + 964 k^{3} N + 104 k^{4} N + 656 N^{2} + 
     2752 k\, 
    N^{2} + 2125 k^{2} N^{2} + 384 k^{3} N^{2} + 930 N^{3} 
    \nonu\\ &+& 1505 k\, 
    N^{3} + 470 k^{2} N^{3} + 316 N^{4} + 216 k\, 
    N^{4} + 32 N^{5}), \nonu\\
c_ {120} & = &\frac {6 (16 + 21 k + 6 k^{2} + 19 N + 13 k\, 
     N + 5 N^{2})} {(2 + N) (2 + k + N)^{2}}, \nonu\\
c_ {121} & = &\frac {4 (120 + 161 k + 46 k^{2} + 265 N + 268 k\, 
     N + 50 k^{2} N + 187 N^{2} + 105 k\, 
     N^{2} + 40 N^{3})} {3 (2 + N) (2 + k + N)^{3}}, \nonu\\
c_ {122} & = &\frac {4 (40 + 59 k + 18 k^{2} + 49 N + 37 k\, 
     N + 13 N^{2})} {(2 + N) (2 + k + N)^{3}}, \nonu\\
c_ {123} & = & - \frac {2 (32 + 55 k + 18 k^{2} + 41 N + 35 k\, 
      N + 11 N^{2})} {(2 + N) (2 + k + N)^{3}}, \nonu\\
c_ {124} & = & - \frac {6 (16 + 21 k + 6 k^{2} + 19 N + 13 k\, 
      N + 5 N^{2})} {(2 + N) (2 + k + N)^{3}}, \nonu\\
c_ {125} & = &\frac {4 i (32 + 55 k + 18 k^{2} + 41 N + 35 k\, 
     N + 11 N^{2})} {(2 + N) (2 + k + N)^{4}}, \nonu\\
c_ {126} & = & - \frac {12 (64 + 113 k + 51 k^{2} + 6 k^{3} + 95 N + 
       101 k\, N + 19 k^{2} N + 40 N^{2} + 18 k\, 
      N^{2} + 5 N^{3})} {(2 + N) (2 + k + N)^{4}}, \nonu\\
c_ {127} & = &\frac {8 (3 + 2 k + N) (10 + 5 k + 
      8 N)} {3 (2 + k + N)^{3}}, \nonu\\
c_ {128} & = & - \frac {(-32 - 248 k - 143 k^{2} - 18 k^{3} - 24 N - 
      152 k\, N - 25 k^{2} N + 39 N^{2} + 10 k\, 
     N^{2} + 17 N^{3})} {2 (2 + N) (2 + k + N)^{3}}, \nonu\\
c_ {129} & = &\frac {i (100 + 67 k + 10 k^{2} + 183 N + 69 k\, 
     N - 4 k^{2} N + 97 N^{2} + 14 k\, 
     N^{2} + 16 N^{3})} {2 (2 + N) (2 + k + N)^{3}}, \nonu\\
c_ {130} & = & - \frac {i (340 + 361 k + 94 k^{2} + 621 N + 453 k\, 
      N + 56 k^{2} N + 355 N^{2} + 140 k\, 
      N^{2} + 64 N^{3})} {(2 + N) (2 + k + N)^{3}}, \nonu\\
c_ {131} & = & - \frac {1}{2 (2 + N) (2 + k + N)^{4}}i (1560 + 2734 k + 1411 k^{2} + 226 k^{3} + 
       3494 N + 4928 k\, 
      N \nonu \\
& + &  1911 k^{2} N + 212 k^{3} N + 2841 N^{2} + 2822 k\, 
      N^{2} + 590 k^{2} N^{2} + 957 N^{3} + 490 k\, 
      N^{3} + 112 N^{4}),\nonu\\ 
c_ {132} & = &\frac {2 i (32 + 55 k + 18 k^{2} + 41 N + 35 k\, 
     N + 11 N^{2})} {(2 + N) (2 + k + N)^{3}}, \nonu\\
c_ {133} & = &\frac {2 (32 + 55 k + 18 k^{2} + 41 N + 35 k\, 
     N + 11 N^{2})} {(2 + N) (2 + k + N)^{4}}, \nonu\\
c_ {134} & = &\frac {(200 + 246 k + 55 k^{2} - 6 k^{3} + 414 N + 
     388 k\, N + 61 k^{2} N + 277 N^{2} + 148 k\, 
    N^{2} + 57 N^{3})} {(2 + N) (2 + k + N)^{4}}, \nonu\\
c_ {135} & = &\frac {5 (20 + 21 k + 6 k^{2} + 25 N + 13 k\, 
     N + 7 N^{2})} {(2 + N) (2 + k + N)^{3}}, \nonu\\
c_ {136} & = &\frac {(20 + 21 k + 6 k^{2} + 25 N + 13 k\, 
    N + 7 N^{2})} {(2 + N) (2 + k + N)^{3}}, \nonu\\
c_ {137} & = &\frac {(40 - 2 k - 77 k^{2} - 30 k^{3} + 134 N + 
     100 k\, N - 15 k^{2} N + 121 N^{2} + 68 k\, 
    N^{2} + 29 N^{3})} {(2 + N) (2 + k + N)^{4}}, \nonu\\
c_ {138} & = &\frac {4 (32 + 55 k + 18 k^{2} + 41 N + 35 k\, 
     N + 11 N^{2})} {(2 + N) (2 + k + N)^{4}}, \nonu\\
c_ {139} & = &\frac {i (-64 - 280 k - 151 k^{2} - 18 k^{3} - 72 N - 
      184 k\, N - 29 k^{2} N + 15 N^{2} + 2 k\, 
     N^{2} + 13 N^{3})} {(2 + N) (2 + k + N)^{4}}, \nonu\\
c_ {140} & = &\frac {3 i (72 k + 45 k^{2} + 6 k^{3} - 8 N + 40 k\, 
     N + 7 k^{2} N - 21 N^{2} - 6 k\, 
     N^{2} - 7 N^{3})} {(2 + N) (2 + k + N)^{4}}, \nonu\\
c_ {141} & = &\frac {3 i (-64 - 84 k - 11 k^{2} + 6 k^{3} - 76 N - 
      36 k\, N + 15 k^{2} N - 17 N^{2} + 10 k\, 
     N^{2} + N^{3})} {(2 + N) (2 + k + N)^{4}}, \nonu\\
c_ {142} & = &\frac {4 i (k - N) (32 + 55 k + 18 k^{2} + 41 N + 
      35 k\, N + 11 N^{2})} {(2 + N) (2 + k + N)^{5}}, \nonu\\
c_ {143} & = & - \frac {1}{12 (2 + N)^{2} (2 + k + N)^{4}}(5136 + 8208 k + 5747 k^{2} + 1948 k^{3} + 
      252 k^{4} + 15024 N 
      \nonu\\ &+& 23246 k\, 
     N + 14788 k^{2} N + 3880 k^{3} N + 288 k^{4} N + 17891 N^{2} + 
      23794 k\, 
     N^{2} + 11263 k^{2} N^{2} 
     \nonu\\ &+& 1588 k^{3} N^{2} + 10554 N^{3} + 
      10330 k\, N^{3} + 2654 k^{2} N^{3} + 3039 N^{4} + 1598 k\, 
     N^{4} + 340 N^{5}), \nonu\\
c_ {144} & = & - \frac{1}{12 (2 + N)^{2} (2 + k + N)^{4}}(-3312 - 3264 k + 1171 k^2 + 1436 k^3 + 252 k^4 - 3456 N 
\nonu\\ &+& 
 4246 k N + 9236 k^2 N + 3368 k^3 N + 288 k^4 N + 2107 N^2 + 
 12170 k N^2 + 8927 k^2 N^2 
 \nonu\\ &+& 1460 k^3 N^2 + 3858 N^3 + 
 7130 k N^3 + 2302 k^2 N^3 + 1599 N^4 + 1246 k N^4 + 212 N^5), \nonu\\
c_ {146} & = & - \frac {1}{12 (2 + N)^{2} (2 + k + N)^{4}}(1296 + 9936 k + 11003 k^{2} + 4096 k^{3} + 
      468 k^{4} + 7152 N 
      \nonu\\ &+& 26414 k\, 
     N + 20980 k^{2} N + 5470 k^{3} N + 396 k^{4} N + 10811 N^{2} + 
      24910 k\, 
     N^{2} + 13225 k^{2} N^{2} 
     \nonu\\ &+& 1846 k^{3} N^{2} + 7050 N^{3} + 
      9988 k\, N^{3} + 2744 k^{2} N^{3} + 2109 N^{4} + 1436 k\, 
     N^{4} + 238 N^{5}), \nonu\\
c_ {147} & = & - \frac{1}{12 (2 + N)^{2} (2 + k + N)^{4}}(7440 + 17232 k + 13627 k^{2} + 4352 k^{3} + 
      468 k^{4} + 20592 N 
      \nonu\\ &+& 39214 k\, 
     N + 24500 k^{2} N + 5726 k^{3} N + 396 k^{4} N + 22267 N^{2} + 
      33134 k\, 
     N^{2} + 14777 k^{2} N^{2} 
     \nonu\\ &+& 1910 k^{3} N^{2} + 11754 N^{3} + 
      12260 k\, N^{3} + 2968 k^{2} N^{3} + 3021 N^{4} + 1660 k\, 
     N^{4} + 302 N^{5}), \nonu\\
c_ {148} & = & - \frac { i (32 + 55 k + 18 k^{2} + 41 N + 35 k\, 
      N + 11 N^{2})} {(2 + N) (2 + k + N)^{4}}, \nonu\\
c_ {149} & = &\frac {2i (64 + 87 k + 26 k^{2} + 89 N + 67 k\, 
     N + 4 k^{2} N + 35 N^{2} + 8 k\, 
     N^{2} + 4 N^{3})} {(2 + N) (2 + k + N)^{4}}, \nonu\\
c_ {150} & = &\frac {2 (32 + 55 k + 18 k^{2} + 41 N + 35 k\, 
     N + 11 N^{2})} {(2 + N) (2 + k + N)^{4}}, \nonu\\
c_ {151} & = & - \frac {2 (80 + 92 k + 11 k^{2} - 6 k^{3} + 172 N + 
       158 k\, N + 21 k^{2} N + 119 N^{2} + 64 k\, 
      N^{2} + 25 N^{3})} {(2 + N) (2 + k + N)^{4}}, \nonu\\
c_ {152} & = & - \frac {2 (240 + 260 k + 25 k^{2} - 18 k^{3} + 
       532 N + 466 k\, N + 59 k^{2} N + 373 N^{2} + 192 k\, 
      N^{2} + 79 N^{3})} {3 (2 + N) (2 + k + N)^{4}}, \nonu\\
c_ {153} & = & - \frac {2 (20 + 21 k + 6 k^{2} + 25 N + 13 k\, 
      N + 7 N^{2})} {(2 + N) (2 + k + N)^{3}}, \nonu\\
c_ {154} & = & - \frac {2 (60 + 71 k + 18 k^{2} + 67 N + 43 k\, 
      N + 17 N^{2})} {3 (2 + N) (2 + k + N)^{3}}, \nonu\\
c_ {155} & = & - \frac { (32 + 55 k + 18 k^{2} + 41 N + 35 k\, 
      N + 11 N^{2})} {(2 + N) (2 + k + N)^{4}}, \nonu\\
c_ {156} & = &\frac {2 i (k - N) (32 + 55 k + 18 k^{2} + 41 N + 
      35 k\, N + 11 N^{2})} {(2 + N) (2 + k + N)^{5}}, \nonu\\
c_ {157} & = &\frac {1}{6 (2 + N)^{2} (2 + k + N)^{5}}i (5136 + 10560 k + 7979 k^2 + 2476 k^3 + 252 k^4 + 14976 N 
 \nonu\\ &+& 
 26582 k N + 16744 k^2 N + 4048 k^3 N + 288 k^4 N + 17315 N^2 + 
 24730 k N^2 + 11395 k^2 N^2 
  \nonu\\ &+& 1540 k^3 N^2 + 9822 N^3 + 9970 k N^3 + 
 2510 k^2 N^3 + 2715 N^4 + 1454 k N^4 + 292 N^5), \nonu\\
c_ {158} & = &\frac{1}{3 (2 + N)^{2} (2 + k + N)^{5}}i (7440 + 16032 k + 12059 k^{2} + 3904 k^{3} + 
      468 k^{4} + 21792 N 
      \nonu\\ &+& 37286 k\, 
     N + 21364 k^{2} N + 4990 k^{3} N + 396 k^{4} N + 25379 N^{2} + 
      32806 k\, 
     N^{2} + 12913 k^{2} N^{2} 
     \nonu\\ &+& 1654 k^{3} N^{2} + 14706 N^{3} + 
      12892 k\, N^{3} + 2624 k^{2} N^{3} + 4221 N^{4} + 1892 k\, 
     N^{4} + 478 N^{5}), \nonu\\
c_ {159} & = &\frac {1}{3 (2 + N)^{2} (2 + k + N)^{5}}i (4368 + 12864 k + 11603 k^2 + 4048 k^3 + 468 k^4 + 12672 N 
 \nonu\\ &+&
 29846 k N + 21064 k^2 N + 5350 k^3 N + 396 k^4 N + 14075 N^2 + 
 25558 k N^2 + 12829 k^2 N^2 
  \nonu\\ &+& 1798 k^3 N^2 + 7518 N^3 + 9532 k N^3 + 
 2600 k^2 N^3 + 1929 N^4 + 1292 k N^4 + 190 N^5), \nonu\\
c_ {160} & = &\frac {1}{3 (2 + N)^{2} (2 + k + N)^{5}}i (8208 + 17664 k + 12683 k^{2} + 3436 k^{3} + 
      252 k^{4} + 23232 N 
      \nonu\\ &+& 40310 k\, 
     N + 22840 k^{2} N + 4720 k^{3} N + 288 k^{4} N + 25763 N^{2} + 
      34234 k\, 
     N^{2} + 13843 k^{2} N^{2} 
     \nonu\\ &+& 1636 k^{3} N^{2} + 13998 N^{3} + 
      12754 k\, N^{3} + 2798 k^{2} N^{3} + 3723 N^{4} + 1742 k\, 
     N^{4} + 388 N^{5}), \nonu\\
c_ {161} & = &\frac{1} {6 (2 + N)^{2} (2 + k + N)^{5}}i (528 - 1584 k - 1405 k^{2} + 316 k^{3} + 
      252 k^{4} + 6384 N + 5870 k\, 
     N 
     \nonu\\ &+& 3292 k^{2} N + 1816 k^{3} N + 288 k^{4} N + 13619 N^{2} + 
      14098 k\, 
     N^{2} + 5383 k^{2} N^{2} + 964 k^{3} N^{2}
     \nonu\\ &+&11490 N^{3} + 
      8986 k\, N^{3} + 1694 k^{2} N^{3} + 4263 N^{4} + 1790 k\, 
     N^{4} + 580 N^{5}), \nonu\\
c_ {162} & = &\frac {1}{6 (2 + N)^{2} (2 + k + N)^{5}}i (3600 + 5904 k + 4595 k^{2} + 1756 k^{3} + 
      252 k^{4} + 11184 N
      \nonu\\ &+& 18638 k\,  N 
     13060 k^{2} N + 3688 k^{3} N + 288 k^{4} N + 14051 N^{2} + 
      20338 k\, 
     N^{2} + 10399 k^{2} N^{2}
     \nonu\\ &+& 1540 k^{3} N^{2}+ 8634 N^{3} + 9178 k\, N^{3} + 2510 k^{2} N^{3} + 2559 N^{4} + 1454 k\, 
     N^{4} + 292 N^{5}), \nonu\\
c_ {163} & = &\frac{1} {12(2 + N)^{2} (2 + k + N)^{5}}i (3600 + 8256 k + 6827 k^2 + 2284 k^3 + 252 k^4 + 11136 N 
 \nonu\\ &+& 21974 k N + 
 15016 k^2 N + 3856 k^3 N + 288 k^4 N + 13475 N^2 + 21274 k N^2 + 
 10531 k^2 N^2 
  \nonu\\ &+& 1492 k^3 N^2 + 7902 N^3 + 8818 k N^3 + 
 2366 k^2 N^3 + 2235 N^4 + 1310 k N^4 + 244 N^5), \nonu\\
c_ {164} & = &\frac {1}{6 (2 + N)^{2} (2 + k + N)^{5}}i (7056 + 16176 k + 12727 k^{2} + 4136 k^{3} + 
      468 k^{4} + 20112 N 
      \nonu\\ &+&36958 k\, 
     N + 22526 k^{2} N + 5330 k^{3} N + 396 k^{4} N + 22639 N^{2} + 
      31694 k\, 
     N^{2} + 13511 k^{2} N^{2}
     \nonu\\ &+& 1766 k^{3} N^{2} + 12600 N^{3} + 
      12044 k\, N^{3} + 2716 k^{2} N^{3} + 3459 N^{4} + 1696 k\, 
     N^{4} + 374 N^{5}), \nonu\\
c_ {165} & = &\frac{1}{6 (2 + N)^{2} (2 + k + N)^{5}}i (7056 + 16176 k + 12727 k^2 + 4136 k^3 + 468 k^4 + 20112 N 
 \nonu\\ &+& 
 36958 k N + 22526 k^2 N + 5330 k^3 N + 396 k^4 N + 22639 N^2 + 
 31694 k N^2 + 13511 k^2 N^2 
  \nonu\\ &+&1766 k^3 N^2 + 12600 N^3 + 
 12044 k N^3 + 2716 k^2 N^3 + 3459 N^4 + 1696 k N^4 + 374 N^5), \nonu\\
c_ {166} & = &\frac {1} {2(2 + N)^{2} (2 + k + N)^{5}} i (1712 + 
    4352 k + 3417 k^{2} + 996 k^{3} + 84 k^{4} + 5184 N + 10450 k\, 
   N 
   \nonu\\ &+& 6440 k^{2} N + 1424 k^{3} N + 96 k^{4} N + 5985 N^{2} + 
    9214 k\, 
   N^{2} + 4049 k^{2} N^{2} + 508 k^{3} N^{2} + 3322 N^{3} 
   \nonu\\ &+& 3526 k\, 
   N^{3} + 842 k^{2} N^{3} + 889 N^{4} + 490 k\, N^{4} + 92 N^{5}), \nonu\\
c_ {167} & = &\frac{1}  {2 (2 + N)^{2} (2 + k + N)^{5}}i (432 + 336 k + 361 k^{2} + 324 k^{3} + 
      84 k^{4} + 2288 N + 3274 k\, 
     N 
     \nonu\\ &+& 2448 k^{2} N + 896 k^{3} N + 96 k^{4} N + 3993 N^{2} + 
      5110 k\, 
     N^{2} + 2481 k^{2} N^{2} + 412 k^{3} N^{2} + 3066 N^{3} 
     \nonu\\ &+& 
      2814 k\, N^{3} + 674 k^{2} N^{3} + 1073 N^{4} + 514 k\, 
     N^{4} + 140 N^{5}), \nonu\\
c_ {168} & = & - \frac {(1020 + 1411 k + 410 k^{2} + 2015 N + 
      2075 k\, N + 376 k^{2} N + 1325 N^{2} + 756 k\, 
     N^{2} + 272 N^{3})} {3 (2 + N) (2 + k + N)^{3}}, \nonu\\
c_ {169} & = &\frac {1} {(2 + N) (2 + k + N)^{5}} (408 + 926 k + 
    953 k^{2} + 448 k^{3} + 76 k^{4} + 1570 N + 3385 k\, 
   N 
   \nonu\\ &+& 2844 k^{2} N + 972 k^{3} N + 104 k^{4} N + 2132 N^{2} + 
    3688 k\, 
   N^{2} + 2133 k^{2} N^{2} + 376 k^{3} N^{2} + 1278 N^{3} 
   \nonu\\ &+& 1529 k\, 
   N^{3} + 454 k^{2} N^{3} + 340 N^{4} + 208 k\, N^{4} + 32 N^{5}), \nonu\\
c_ {170} & = &\frac {8 (60 + 77 k + 22 k^{2} + 121 N + 115 k\, 
     N + 20 k^{2} N + 79 N^{2} + 42 k\, 
     N^{2} + 16 N^{3})} {(2 + N) (2 + k + N)^{4}}, \nonu\\
c_ {171} & = &\frac {2 (32 + 55 k + 18 k^{2} + 41 N + 35 k\, 
     N + 11 N^{2})} {(2 + N) (2 + k + N)^{4}}, \nonu\\
c_ {172} & = &\frac {1} {(2 + N) (2 + k + N)^{5}} (600 + 2294 k + 
    2429 k^{2} + 1012 k^{3} + 148 k^{4} + 1834 N + 5105 k\, 
   N 
   \nonu\\ &+& 4060 k^{2} N + 1192 k^{3} N + 104 k^{4} N + 2072 N^{2} + 
    4228 k\, 
   N^{2} + 2353 k^{2} N^{2} + 384 k^{3} N^{2} + 1134 N^{3}
   \nonu\\ &+& 1565 k\, 
   N^{3} + 470 k^{2} N^{3} + 304 N^{4} + 216 k\, N^{4} + 32 N^{5}), \nonu\\
c_ {173} & = &\frac {2 (540 + 821 k + 262 k^{2} + 961 N + 1099 k\, 
     N + 212 k^{2} N + 583 N^{2} + 378 k\, 
     N^{2} + 112 N^{3})} {3 (2 + N) (2 + k + N)^{4}}, \nonu\\
c_ {174} & = &\frac {2 (540 + 437 k + 70 k^{2} + 1345 N + 907 k\, 
     N + 116 k^{2} N + 967 N^{2} + 378 k\, 
     N^{2} + 208 N^{3})} {3 (2 + N) (2 + k + N)^{4}}, \nonu\\
c_ {175} & = & - \frac {6 (32 + 55 k + 18 k^{2} + 41 N + 35 k\, 
      N + 11 N^{2})} {(2 + N) (2 + k + N)^{4}}, \nonu\\
c_ {176} & = & - \frac{1}{4 (2 + N)^{2} (2 + k + N)^{5}}(16752 + 40904 k + 33511 k^{2} + 
      11156 k^{3} + 1276 k^{4} + 47656 N 
      \nonu\\ &+& 96974 k\, 
     N + 63702 k^{2} N + 15916 k^{3} N + 1216 k^{4} N + 53703 N^{2} + 
      88812 k\, 
     N^{2} + 44215 k^{2} N^{2} 
     \nonu\\ &+& 7224 k^{3} N^{2} + 208 k^{4} N^{2} + 
      30538 N^{3} + 39002 k\, 
     N^{3} + 13020 k^{2} N^{3} + 960 k^{3} N^{3} + 9103 N^{4} 
     \nonu\\ &+& 
      8100 k\, N^{4} + 1324 k^{2} N^{4} + 1304 N^{5} + 624 k\, 
     N^{5} + 64 N^{6}), \nonu\\
c_ {177} & = & - \frac {1}{12 (2 + N)^{2} (2 + k + N)^{5}}(16976 + 40728 k + 37461 k^{2} + 
      14812 k^{3} + 2100 k^{4} + 56504 N 
      \nonu\\ &+& 116490 k\, 
     N + 88914 k^{2} N + 27508 k^{3} N + 2784 k^{4} N + 75413 N^{2} + 
      125316 k\, 
     N^{2} + 72069 k^{2} N^{2} 
     \nonu\\ &+& 14488 k^{3} N^{2} + 
      624 k^{4} N^{2} + 50734 N^{3} + 62046 k\, 
     N^{3} + 22980 k^{2} N^{3} + 2016 k^{3} N^{3} + 17869 N^{4} 
     \nonu\\ &+& 
      13692 k\, N^{4} + 2244 k^{2} N^{4} + 3064 N^{5} + 1008 k\, 
     N^{5} + 192 N^{6}), \nonu\\
c_ {178} & = &\frac {4 (2 + 2 k + k^{2} + 2 N + N^{2}) (32 + 55 k + 
      18 k^{2} + 41 N + 35 k\, 
     N + 11 N^{2})} {(2 + N) (2 + k + N)^{6}}, \nonu\\
c_ {179} & = & - \frac {4 (128 + 275 k + 137 k^{2} + 18 k^{3} + 
       189 N + 239 k\, N + 49 k^{2} N + 72 N^{2} + 38 k\, 
      N^{2} + 7 N^{3})} {(2 + N) (2 + k + N)^{4}}, \nonu\\
c_ {180} & = &\frac {(352 + 536 k + 207 k^{2} + 18 k^{3} + 472 N + 
     424 k\, N + 57 k^{2} N + 169 N^{2} + 54 k\, 
    N^{2} + 15 N^{3})} {2 (2 + N) (2 + k + N)^{3}}, \nonu\\
c_ {181} & = & - \frac {1}{6 (2 + N) (2 + k + N)^{4}}i (1560 + 2174 k + 737 k^{2} + 38 k^{3} + 
       3574 N + 4036 k\, 
      N + 1137 k^{2} N 
      \nonu\\ &+& 100 k^{3} N + 2823 N^{2} + 2158 k\, 
      N^{2} + 310 k^{2} N^{2} + 855 N^{3} + 290 k\, 
      N^{3} + 80 N^{4}), \nonu\\
c_ {182} & = & - \frac {i (660 + 797 k + 214 k^{2} + 961 N + 757 k\, 
      N + 80 k^{2} N + 439 N^{2} + 168 k\, 
      N^{2} + 64 N^{3})} {3 (2 + N) (2 + k + N)^{3}}, \nonu\\
c_ {183} & = & - \frac {1}{2 (2 + N) (2 + k + N)^{4}}i (360 + 578 k + 253 k^{2} + 30 k^{3} + 
       970 N + 1312 k\, 
      N + 497 k^{2} N 
      \nonu\\ &+& 60 k^{3} N + 919 N^{2} + 882 k\, 
      N^{2} + 186 k^{2} N^{2} + 355 N^{3} + 174 k\, 
      N^{3} + 48 N^{4}), \nonu\\
c_ {184} & = & - \frac {4 (32 + 55 k + 18 k^{2} + 41 N + 35 k\, 
      N + 11 N^{2})} {(2 + N) (2 + k + N)^{4}}, \nonu\\
c_ {185} & = & - \frac {i (8 + k + N) (32 + 55 k + 18 k^{2} + 41 N + 
       35 k\, N + 11 N^{2})} {(2 + N) (2 + k + N)^{4}}, \nonu\\
c_ {186} & = &\frac {3 i (64 + 136 k + 61 k^{2} + 6 k^{3} + 88 N + 
      104 k\, N + 15 k^{2} N + 27 N^{2} + 10 k\, 
     N^{2} + N^{3})} {(2 + N) (2 + k + N)^{4}}, \nonu\\
c_ {187} & = &\frac {3 i (-20 k + 5 k^{2} + 6 k^{3} + 20 N + 28 k\, 
     N + 23 k^{2} N + 31 N^{2} + 26 k\, 
     N^{2} + 9 N^{3})} {(2 + N) (2 + k + N)^{4}}, \nonu\\
c_ {188} & = & - \frac {2 (60 + 77 k + 22 k^{2} + 121 N + 115 k\, 
      N + 20 k^{2} N + 79 N^{2} + 42 k\, 
      N^{2} + 16 N^{3})} {(2 + N) (2 + k + N)^{4}}, \nonu\\
c_ {189} & = & - \frac {1}{12 (2 + N)^{2} (2 + k + N)^{5}}(24144 + 78936 k + 76469 k^{2} + 
      29116 k^{3} + 3828 k^{4} + 73080 N 
      \nonu\\ &+& 189290 k\, 
     N + 144514 k^{2} N + 41092 k^{3} N + 3648 k^{4} N + 
      86741 N^{2} + 175108 k\, 
     N^{2} + 99493 k^{2} N^{2}
     \nonu\\ &+& 18280 k^{3} N^{2} + 
      624 k^{4} N^{2} + 52254 N^{3} + 77710 k\, 
     N^{3} + 28820 k^{2} N^{3} + 2304 k^{3} N^{3} + 16941 N^{4} 
     \nonu\\ &+& 
      16364 k\, N^{4} + 2820 k^{2} N^{4} + 2824 N^{5} + 1296 k\, 
     N^{5} + 192 N^{6}), \nonu\\
c_ {190} & = & - \frac {1}{12 (2 + N)^{2} (2 + k + N)^{5}}(19536 + 46104 k + 41045 k^{2} + 
      15580 k^{3} + 2100 k^{4} + 64440 N 
      \nonu\\ &+& 131594 k\, 
     N + 97618 k^{2} N + 29044 k^{3} N + 2784 k^{4} N + 85013 N^{2} + 
      141508 k\, 
     N^{2} + 79621 k^{2} N^{2} 
     \nonu\\ &+& 15448 k^{3} N^{2} + 
      624 k^{4} N^{2} + 56430 N^{3} + 70366 k\, 
     N^{3} + 25796 k^{2} N^{3} + 2208 k^{3} N^{3} + 19533 N^{4} 
     \nonu\\ &+& 
      15740 k\, N^{4} + 2628 k^{2} N^{4} + 3256 N^{5} + 1200 k\, 
     N^{5} + 192 N^{6}), \nonu\\
c_ {191} & = &\frac {3 i (16 + 21 k + 6 k^{2} + 19 N + 13 k\, 
     N + 5 N^{2})} {(2 + N) (2 + k + N)^{3}}, \nonu\\
c_ {192} & = &\frac {(240 + 260 k + 25 k^{2} - 18 k^{3} + 532 N + 
     466 k\, N + 59 k^{2} N + 373 N^{2} + 192 k\, 
    N^{2} + 79 N^{3})} {6 (2 + N) (2 + k + N)^{4}}, \nonu\\
c_ {193} & = & - \frac {1}{6 (2 + N) (2 + k + N)^{4}}(-240 - 548 k - 403 k^{2} - 90 k^{3} - 
      244 N - 430 k\, N - 185 k^{2} N 
      \nonu\\ &-& 31 N^{2} - 48 k\, 
     N^{2} + 11 N^{3}), \nonu\\
c_ {194} & = & - \frac {(60 + 77 k + 22 k^{2} + 121 N + 115 k\, 
     N + 20 k^{2} N + 79 N^{2} + 42 k\, 
     N^{2} + 16 N^{3})} {4 (2 + N) (2 + k + N)^{2}}, \nonu\\
c_ {195} & = &\frac { (32 + 81 k + 43 k^{2} + 6 k^{3} + 47 N + 
      69 k\, N + 15 k^{2} N + 16 N^{2} + 10 k\, 
     N^{2} + N^{3})} {(2 + N) (2 + k + N)^{4}}, \nonu\\
c_ {196} & = & - \frac { (64 + 139 k + 76 k^{2} + 12 k^{3} + 101 N + 
       135 k\, N + 34 k^{2} N + 45 N^{2} + 28 k\, 
      N^{2} + 6 N^{3})} {(2 + N) (2 + k + N)^{4}}, \nonu\\
c_ {197} & = &\frac { (96 + 145 k + 59 k^{2} + 6 k^{3} + 143 N + 
      133 k\, N + 23 k^{2} N + 64 N^{2} + 26 k\, 
     N^{2} + 9 N^{3})} {(2 + N) (2 + k + N)^{4}}, \nonu\\
c_ {198} & = &\frac { (64 + 113 k + 51 k^{2} + 6 k^{3} + 95 N + 
      101 k\, N + 19 k^{2} N + 40 N^{2} + 18 k\, 
     N^{2} + 5 N^{3})} {(2 + N) (2 + k + N)^{4}}, \nonu\\
c_ {199} & = & - \frac { (96 + 171 k + 84 k^{2} + 12 k^{3} + 149 N + 
       167 k\, N + 38 k^{2} N + 69 N^{2} + 36 k\, 
      N^{2} + 10 N^{3})} {(2 + N) (2 + k + N)^{4}}, \nonu\\
c_ {200} & = &\frac {2 (48 + 97 k + 47 k^{2} + 6 k^{3} + 71 N + 
      85 k\, N + 17 k^{2} N + 28 N^{2} + 14 k\, 
     N^{2} + 3 N^{3})} {(2 + N) (2 + k + N)^{4}}, \nonu\\
c_ {201} & = & - \frac {2 (48 + 123 k + 72 k^{2} + 12 k^{3} + 77 N + 
       119 k\, N + 32 k^{2} N + 33 N^{2} + 24 k\, 
      N^{2} + 4 N^{3})} {(2 + N) (2 + k + N)^{4}}, \nonu\\
c_ {202} & = & - \frac {2 (-16 - 65 k - 39 k^{2} - 6 k^{3} - 23 N - 
       53 k\, N - 13 k^{2} N - 4 N^{2} - 6 k\, 
      N^{2} + N^{3})} {(2 + N) (2 + k + N)^{4}}, \nonu\\
c_ {203} & = & - \frac{1}{8 (2 + N)^{2} (2 + k + N)^{4}}(-2240 - 3524 k - 1367 k^{2} + 84 k^{3} + 
      84 k^{4} - 5948 N - 7264 k\, 
     N 
     \nonu\\ &-& 1664 k^{2} N + 392 k^{3} N + 96 k^{4} N - 6297 N^{2} - 
      5502 k\, 
     N^{2} - 487 k^{2} N^{2} + 220 k^{3} N^{2} - 3318 N^{3} 
     \nonu\\ &-& 
      1820 k\, N^{3} + 2 k^{2} N^{3} - 873 N^{4} - 226 k\, 
     N^{4} - 92 N^{5}), \nonu\\
c_ {204} & = & - \frac{1}{24 (2 + N)^{2} (2 + k + N)^{4}}(-960 - 44 k + 1835 k^{2} + 1308 k^{3} + 
      252 k^{4} - 724 N + 4144 k\, 
     N 
     \nonu\\ &+& 6536 k^{2} N + 2664 k^{3} N + 288 k^{4} N + 1365 N^{2} + 
      7174 k\, 
     N^{2} + 5795 k^{2} N^{2} + 1140 k^{3} N^{2} 
     \nonu\\ &+& 1878 N^{3} + 
      3996 k\, N^{3} + 1494 k^{2} N^{3} + 781 N^{4} + 714 k\, 
     N^{4} + 108 N^{5}), \nonu\\
c_ {205} & = &\frac { (72 k + 45 k^{2} + 6 k^{3} - 8 N + 40 k\, 
     N + 7 k^{2} N - 21 N^{2} - 6 k\, 
     N^{2} - 7 N^{3})} {(2 + N) (2 + k + N)^{4}}, \nonu\\
c_ {206} & = &\frac { (96 + 168 k + 69 k^{2} + 6 k^{3} + 136 N + 
      136 k\, N + 19 k^{2} N + 51 N^{2} + 18 k\, 
     N^{2} + 5 N^{3})} {(2 + N) (2 + k + N)^{4}}, \nonu\\
c_ {207} & = &\frac {2 (-48 - 120 k - 57 k^{2} - 6 k^{3} - 64 N - 
      88 k\, N - 13 k^{2} N - 15 N^{2} - 6 k\, 
     N^{2} + N^{3})} {(2 + N) (2 + k + N)^{4}}, \nonu\\
c_ {208} & = &\frac {1}{4 (2 + N)^{2} (2 + k + N)^{5}}i (-320 - 132 k + 305 k^{2} + 324 k^{3} + 
      84 k^{4} - 444 N + 976 k\, 
     N
     \nonu\\ &+& 1708 k^{2} N + 800 k^{3} N + 96 k^{4} N - 49 N^{2} + 
      1786 k\, 
     N^{2} + 1629 k^{2} N^{2} + 364 k^{3} N^{2} + 134 N^{3} 
     \nonu\\ &+& 940 k\, 
     N^{3} + 426 k^{2} N^{3} + 51 N^{4} + 150 k\, 
     N^{4} + 4 N^{5}), \nonu\\
c_ {209} & = &\frac {1} {(2 + N)^{2} (2 + k + N)^{5}} i (480 + 
    1764 k + 2169 k^{2} + 1008 k^{3} + 156 k^{4} + 1500 N + 4412 k\, 
   N 
   \nonu\\ &+& 4196 k^{2} N + 1394 k^{3} N + 132 k^{4} N + 1891 N^{2} + 
    4226 k\, 
   N^{2} + 2771 k^{2} N^{2} + 490 k^{3} N^{2} + 1202 N^{3} 
   \nonu\\ &+& 1806 k\, 
   N^{3} + 608 k^{2} N^{3} + 377 N^{4} + 284 k\, N^{4} + 46 N^{5}), \nonu\\
c_ {210} & = &\frac {1} {2(2 + N)^{2} (2 + k + N)^{5}} i (1280 + 
    2892 k + 2089 k^{2} + 660 k^{3} + 84 k^{4} + 4212 N + 8280 k\, 
   N 
   \nonu\\ &+& 5088 k^{2} N + 1256 k^{3} N + 96 k^{4} N + 5343 N^{2} + 
    8346 k\, 
   N^{2} + 3721 k^{2} N^{2} + 508 k^{3} N^{2} + 3218 N^{3} 
   \nonu\\ &+& 3524 k\, 
   N^{3} + 850 k^{2} N^{3} + 919 N^{4} + 526 k\, N^{4} + 100 N^{5}), \nonu\\
c_ {211} & = &\frac {1}{2 (2 + N)^{2} (2 + k + N)^{5}}i (1280 + 2892 k + 2273 k^{2} + 740 k^{3} + 
      84 k^{4} + 4212 N + 8040 k\, 
     N
     \nonu\\ &+& 5124 k^{2} N + 1264 k^{3} N + 96 k^{4} N + 5399 N^{2} + 
      8098 k\, 
     N^{2} + 3661 k^{2} N^{2} + 492 k^{3} N^{2} + 3350 N^{3} 
     \nonu\\ &+& 
      3492 k\, N^{3} + 834 k^{2} N^{3} + 1003 N^{4} + 542 k\, 
     N^{4} + 116 N^{5}), \nonu\\
c_ {212} & = &\frac {1}{12 (2 + N)^{2} (2 + k + N)^{5}}i (32 + 55 k + 18 k^{2} + 41 N + 35 k\, 
     N + 11 N^{2}) (60 + 37 k + 14 k^{2} 
     \nonu\\ &+&161 N + 95 k\, 
     N + 16 k^{2} N + 107 N^{2} + 42 k\, 
     N^{2} + 20 N^{3}), \nonu\\
c_ {213} & = &\frac {1} {6 (2 + N)^{2} (2 + k + N)^{5}}i (2880 + 7988 k + 8087 k^{2} + 3320 k^{3} + 
      468 k^{4} + 8716 N + 19784 k\, 
     N 
     \nonu\\ &+& 15598 k^{2} N + 4586 k^{3} N + 396 k^{4} N + 10577 N^{2} + 
      18582 k\, 
     N^{2} + 10183 k^{2} N^{2} + 1598 k^{3} N^{2} 
     \nonu\\ &+& 6420 N^{3} + 
      7750 k\, N^{3} + 2204 k^{2} N^{3} + 1925 N^{4} + 1192 k\, 
     N^{4} + 226 N^{5}), \nonu\\
c_ {214} & = &\frac {1}{6 (2 + N)^{2} (2 + k + N)^{5}}i (3364 k + 5695 k^{2} + 2920 k^{3} + 
      468 k^{4} - 484 N + 7456 k\, 
     N 
     \nonu\\ &+& 10538 k^{2} N + 3970 k^{3} N + 396 k^{4} N - 767 N^{2} + 
      6774 k\, 
     N^{2} + 6827 k^{2} N^{2} + 1390 k^{3} N^{2} - 360 N^{3} 
     \nonu\\ &+& 
      2894 k\, N^{3} + 1492 k^{2} N^{3} - 47 N^{4} + 464 k\, 
     N^{4} + 2 N^{5}), \nonu\\
c_ {215} & = &\frac {1}{6 (2 + N)^{2} (2 + k + N)^{5}}i (960 + 2260 k + 2083 k^{2} + 1132 k^{3} + 
      252 k^{4} + 5228 N + 10720 k\, 
     N 
     \nonu\\ &+& 7964 k^{2} N + 2704 k^{3} N + 288 k^{4} N + 8269 N^{2} + 
      13230 k\, 
     N^{2} + 6911 k^{2} N^{2} + 1204 k^{3} N^{2} 
     \nonu\\ &+& 5562 N^{3} + 
      6164 k\, N^{3} + 1726 k^{2} N^{3} + 1681 N^{4} + 962 k\, 
     N^{4} + 188 N^{5}), \nonu\\
c_ {216} & = &\frac {1} {12 (2 + N)^{2} (2 + k + N)^{5}}i (3840 + 9188 k + 7331 k^{2} + 2348 k^{3} + 
      252 k^{4} + 12124 N + 24632 k\, 
     N 
     \nonu\\ &+& 16012 k^{2} N + 3920 k^{3} N + 288 k^{4} N + 15173 N^{2} + 
      24294 k\, 
     N^{2} + 11239 k^{2} N^{2} + 1508 k^{3} N^{2} 
     \nonu\\ &+& 9282 N^{3} + 
      10348 k\, N^{3} + 2534 k^{2} N^{3} + 2753 N^{4} + 1594 k\, 
     N^{4} + 316 N^{5}), \nonu\\
c_ {217} & = &\frac {1}{4 (2 + N)^{2} (2 + k + N)^{5}}i (640 + 532 k - 337 k^{2} - 388 k^{3} - 
      84 k^{4} + 1580 N + 520 k\, 
     N 
     \nonu\\ &-& 1324 k^{2} N - 768 k^{3} N - 96 k^{4} N + 1481 N^{2} - 
      226 k\, 
     N^{2} - 1253 k^{2} N^{2} - 332 k^{3} N^{2} + 658 N^{3} 
     \nonu\\ &-& 340 k\, 
     N^{3} - 338 k^{2} N^{3} + 141 N^{4} - 78 k\, 
     N^{4} + 12 N^{5}), \nonu\\
c_ {218} & = & - \frac {2 (k - N) (32 + 55 k + 18 k^{2} + 41 N + 
       35 k\, N + 11 N^{2})} {3(2 + N) (2 + k + N)^{5}}, \nonu\\
c_ {219} & = &\frac {1} {6(2 + N)^{2} (2 + k + N)^{6}} (32 + 55 k + 
    18 k^{2} + 41 N + 35 k\, 
   N + 11 N^{2}) (60 + 77 k + 22 k^{2} 
   \nonu\\ &+& 121 N + 115 k\, 
   N + 20 k^{2} N + 79 N^{2} + 42 k\, N^{2} + 16 N^{3}), \nonu\\
c_ {220} & = & - \frac { i (32 + 55 k + 18 k^{2} + 41 N + 35 k\, 
      N + 11 N^{2})} {2(2 + N) (2 + k + N)^{4}}, \nonu\\
c_ {221} & = & - \frac { i (32 + 55 k + 18 k^{2} + 41 N + 35 k\, 
      N + 11 N^{2})} {3(2 + N) (2 + k + N)^{4}}, \nonu\\
c_ {222} & = & - \frac {2 (k - N) (32 + 55 k + 18 k^{2} + 41 N + 
       35 k\, N + 11 N^{2})} {3(2 + N) (2 + k + N)^{5}}, \nonu\\
c_ {223} & = & - \frac {i (32 + 55 k + 18 k^{2} + 41 N + 35 k\, 
      N + 11 N^{2})} {2(2 + N) (2 + k + N)^{4}}, \nonu\\
c_ {224} & = & - \frac {1}{48 (2 + N)^{2} (2 + k + N)^{4}}(-3840 - 2860 k + 2635 k^{2} + 2448 k^{3} + 
      468 k^{4} - 10292 N 
      \nonu\\ &-& 5440 k\, 
     N + 5740 k^{2} N + 3462 k^{3} N + 396 k^{4} N - 10443 N^{2} - 
      2878 k\, 
     N^{2} + 4417 k^{2} N^{2} 
     \nonu\\ &+& 1254 k^{3} N^{2} - 4902 N^{3} - 
      150 k\, N^{3} + 1104 k^{2} N^{3} - 1045 N^{4} + 132 k\, 
     N^{4} - 78 N^{5}), \nonu\\
c_ {225} & = &\frac {3} {2}, \nonu\\
c_ {226} & = & - \frac {3 (60 + 77 k + 22 k^{2} + 121 N + 115 k\, 
      N + 20 k^{2} N + 79 N^{2} + 42 k\, 
      N^{2} + 16 N^{3})} {2 (2 + N) (2 + k + N)^{2}}, \nonu\\
c_ {227} & = & - \frac {2 i (144 + 295 k + 171 k^{2} + 30 k^{3} + 
       257 N + 335 k\, N + 95 k^{2} N + 142 N^{2} + 90 k\, 
      N^{2} + 25 N^{3})} {3 (2 + N) (2 + k + N)^{3}}, \nonu\\
c_ {228} & = & - \frac {2 i (96 + 271 k + 171 k^{2} + 30 k^{3} + 
       209 N + 323 k\, N + 95 k^{2} N + 130 N^{2} + 90 k\, 
      N^{2} + 25 N^{3})} {3 (2 + N) (2 + k + N)^{3}}, \nonu\\
c_ {229} & = & - \frac {4 (96 + 223 k + 147 k^{2} + 30 k^{3} + 
       161 N + 251 k\, N + 83 k^{2} N + 82 N^{2} + 66 k\, 
      N^{2} + 13 N^{3})} {3 (2 + N) (2 + k + N)^{4}}, \nonu\\
c_ {230} & = &\frac {3 (20 + 21 k + 6 k^{2} + 25 N + 13 k\, 
     N + 7 N^{2})} {(2 + N) (2 + k + N)^{3}}, \nonu\\
c_ {231} & = & - \frac {1} {3 (2 + N) (2 + k + N)^{4}} i (-432 - 
     812 k - 427 k^{2} - 40 k^{3} + 12 k^{4} - 988 N - 1184 k\, 
    N \nonu\\ &-&
    133 k^{2} N + 176 k^{3} N + 36 k^{4} N - 957 N^{2} - 706 k\, 
    N^{2} + 96 k^{2} N^{2} + 82 k^{3} N^{2} - 537 N^{3} 
    \nonu\\ &-& 278 k\, 
    N^{3} + 12 k^{2} N^{3} - 174 N^{4} - 58 k\, N^{4} - 24 N^{5}), \nonu\\
c_ {232} & = &\frac {3 i (32 + 55 k + 18 k^{2} + 41 N + 35 k\, 
     N + 11 N^{2})} {(2 + N) (2 + k + N)^{3}}, \nonu\\
c_ {233} & = &\frac {i (96 + 116 k + 45 k^{2} + 6 k^{3} + 124 N + 
      100 k\, N + 19 k^{2} N + 47 N^{2} + 18 k\, 
     N^{2} + 5 N^{3})} {3 (2 + N) (2 + k + N)^{4}}, \nonu\\
c_ {234} & = &\frac {2 i (480 + 1048 k + 441 k^{2} + 30 k^{3} + 
      728 N + 824 k\, N + 95 k^{2} N + 271 N^{2} + 90 k\, 
     N^{2} + 25 N^{3})} {3 (2 + N) (2 + k + N)^{4}}, \nonu\\
c_ {235} & = &\frac{1} {3 (2 + N) (2 + k + N)^{4}}2 i (864 + 1432 k + 537 k^{2} + 30 k^{3} + 
      1304 N + 1208 k\, N + 143 k^{2} N 
      \nonu\\ &+& 559 N^{2} + 186 k\, 
     N^{2} + 73 N^{3}), \nonu\\
c_ {236} & = &\frac {1}{3 (2 + N) (2 + k + N)^{4}}2 i (768 + 1336 k + 513 k^{2} + 30 k^{3} + 
      1160 N + 1112 k\, N + 131 k^{2} N 
      \nonu\\ &+& 487 N^{2} + 162 k\, 
     N^{2} + 61 N^{3}), \nonu\\
c_ {237} & = &\frac {1}{3 (2 + N) (2 + k + N)^{4}}2 i (576 + 1144 k + 465 k^{2} + 30 k^{3} + 
      872 N + 920 k\, N + 107 k^{2} N 
      \nonu\\ &+& 343 N^{2} + 114 k\, 
     N^{2} + 37 N^{3}), \nonu\\
c_ {238} & = & - \frac {1} {3 (2 + N) (2 + k + N)^{4}} i (-360 - 
     366 k - 111 k^{2} - 10 k^{3} - 822 N - 572 k\, 
    N - 53 k^{2} N 
    \nonu\\ &+& 16 k^{3} N - 613 N^{2} - 228 k\, 
    N^{2} + 16 k^{2} N^{2} - 177 N^{3} - 16 k\, N^{3} - 16 N^{4}), \nonu\\
c_ {239} & = & - \frac {4 i (180 + 279 k + 155 k^{2} + 30 k^{3} + 
       315 N + 319 k\, N + 87 k^{2} N + 174 N^{2} + 86 k\, 
      N^{2} + 31 N^{3})} {3 (2 + N) (2 + k + N)^{4}}, \nonu\\
c_ {240} & = &\frac {i (-540 - 495 k - 130 k^{2} - 747 N - 315 k\, 
     N + 16 k^{2} N - 257 N^{2} - 16 N^{3})} {3 (2 + 
      N) (2 + k + N)^{3}}, \nonu\\
c_ {241} & = & - \frac {2 i (144 + 271 k + 159 k^{2} + 30 k^{3} + 
       233 N + 299 k\, N + 89 k^{2} N + 118 N^{2} + 78 k\, 
      N^{2} + 19 N^{3})} {3 (2 + N) (2 + k + N)^{4}}, \nonu\\
c_ {242} & = &\frac {2 i (k - N) (32 + 55 k + 18 k^{2} + 41 N + 
      35 k\, N + 11 N^{2})} {3 (2 + N) (2 + k + N)^{5}}, \nonu\\
c_ {243} & = & - \frac {1} {3 (2 + N) (2 + k + N)^{5}} 8 (180 + 
     321 k + 175 k^{2} + 30 k^{3} + 453 N + 647 k\, 
    N + 264 k^{2} N 
    \nonu\\ &+& 30 k^{3} N + 420 N^{2} + 421 k\, 
    N^{2} + 93 k^{2} N^{2} + 167 N^{3} + 87 k\, N^{3} + 24 N^{4}), \nonu\\
c_ {244} & = & - \frac {1} {3 (2 + N) (2 + k + N)^{5}} 2 (360 + 
     514 k + 261 k^{2} + 46 k^{3} + 1034 N + 1200 k\, 
    N + 441 k^{2} N 
    \nonu\\ &+&44 k^{3} N + 1023 N^{2} + 850 k\, 
    N^{2} + 170 k^{2} N^{2} + 427 N^{3} + 190 k\, N^{3} + 64 N^{4}), \nonu\\
c_ {245} & = & - \frac {1} {3 (2 + N) (2 + k + N)^{5}} 2 (360 + 
     770 k + 517 k^{2} + 110 k^{3} + 778 N + 1328 k\, 
    N + 633 k^{2} N 
    \nonu\\ &+& 76 k^{3} N + 639 N^{2} + 786 k\, 
    N^{2} + 202 k^{2} N^{2} + 235 N^{3} + 158 k\, N^{3} + 32 N^{4}),\nonu\\ 
c_ {246} & = & - \frac {12 (32 + 55 k + 18 k^{2} + 41 N + 35 k\, 
      N + 11 N^{2})} {(2 + N) (2 + k + N)^{4}}, \nonu\\
c_ {247} & = & - \frac {6 (32 + 55 k + 18 k^{2} + 41 N + 35 k\, 
      N + 11 N^{2})} {(2 + N) (2 + k + N)^{4}}, \nonu\\
c_ {248} & = &\frac {(k - N) (60 + 77 k + 22 k^{2} + 121 N + 115 k\, 
     N + 20 k^{2} N + 79 N^{2} + 42 k\, 
     N^{2} + 16 N^{3})} {6 (2 + N) (2 + k + N)^{3}}, \nonu\\
c_ {249} & = &\frac {8} {(2 + k + N)^{2}}, \nonu\\
c_ {250} & = & - \frac {4 (192 + 319 k + 171 k^{2} + 30 k^{3} + 
       305 N + 347 k\, N + 95 k^{2} N + 154 N^{2} + 90 k\, 
      N^{2} + 25 N^{3})} {3 (2 + N) (2 + k + N)^{4}}, \nonu\\
c_ {251} & = & - \frac {4 (127 k + 123 k^{2} + 30 k^{3} + 17 N + 
       155 k\, N + 71 k^{2} N + 10 N^{2} + 42 k\, 
      N^{2} + N^{3})} {3 (2 + N) (2 + k + N)^{4}}, \nonu\\
c_ {252} & = &\frac{1}{6 (2 + N) (2 + k + N)^{3}}(-288 + 160 k + 357 k^{2} + 102 k^{3} - 304 N + 
     320 k\, N + 227 k^{2} N 
     \nonu\\ &-& 101 N^{2} + 114 k\, 
    N^{2} - 11 N^{3}), \nonu\\
c_ {253} & = &\frac{1}{6 (2 + N) (2 + k + N)^{3}}(672 + 1024 k + 549 k^{2} + 102 k^{3} + 1040 N + 
     1136 k\, N + 323 k^{2} N 
     \nonu\\ &+& 523 N^{2} + 306 k\, 
    N^{2} + 85 N^{3}), \nonu\\
c_ {254} & = & - \frac {1} {3 (2 + N) (2 + k + N)^{4}} 2 i (90 k + 
     76 k^{2} + 16 k^{3} + 90 N + 352 k\, 
    N + 221 k^{2} N + 38 k^{3} N 
    \nonu\\ &+& 166 N^{2} + 319 k\, 
    N^{2} + 101 k^{2} N^{2} + 92 N^{3} + 79 k\, N^{3} + 16 N^{4}), \nonu\\
c_ {255} & = &\frac {1} {3 (2 + N) (2 + k + N)^{4}} 2 i (26 k + 
    12 k^{2} + 154 N + 320 k\, 
   N + 173 k^{2} N + 30 k^{3} N + 262 N^{2} 
   \nonu\\ &+& 335 k\, 
   N^{2} + 93 k^{2} N^{2} + 140 N^{3} + 87 k\, N^{3} + 24 N^{4}), \nonu\\
c_ {256} & = & - \frac {1} {3 (2 + N) (2 + k + N)^{4}} 2 i (-118 k - 
     124 k^{2} - 32 k^{3} + 298 N + 248 k\, 
    N + 73 k^{2} N + 14 k^{3} N 
    \nonu\\ &+& 470 N^{2} + 367 k\, 
    N^{2} + 77 k^{2} N^{2} + 240 N^{3} + 103 k\, N^{3} + 40 N^{4}), \nonu\\
c_ {257} & = &\frac {1} {3 (2 + N) (2 + k + N)^{4}} 2 i (58 k + 
    28 k^{2} + 122 N + 336 k\, 
   N + 181 k^{2} N + 30 k^{3} N + 230 N^{2}
   \nonu\\ &+& 335 k\, 
   N^{2} + 93 k^{2} N^{2} + 132 N^{3} + 87 k\, N^{3} + 24 N^{4}), \nonu\\
c_ {258} & = & - \frac {4 i (k - N)} {3 (2 + k + N)^{3}}, \qquad
c_ {259}  =  - \frac {4 i (k - N) (23 + 12 k + 
       12 N)} {3 (2 + k + N)^{3}}, \qquad
c_ {260}  =  - \frac {4 i} {(2 + k + N)^{2}}, \nonu\\
c_ {261} & = &\frac {(k - N) (16 + 21 k + 6 k^{2} + 19 N + 13 k\, 
     N + 5 N^{2})} {3 (2 + N) (2 + k + N)^{4}}, \nonu\\
c_ {262} & = & - \frac {(k - N) (13 k + 6 k^{2} + 3 N + 9 k\, 
      N + N^{2})} {3 (2 + N) (2 + k + N)^{4}}, \nonu\\
c_ {263} & = &\frac {2 (180 + 271 k + 138 k^{2} + 24 k^{3} + 323 N + 
      325 k\, N + 82 k^{2} N + 185 N^{2} + 94 k\, 
     N^{2} + 34 N^{3})} {3 (2 + N) (2 + k + N)^{4}}, \nonu\\
c_ {264} & = &\frac {2 (k - N) (8 + 17 k + 6 k^{2} + 11 N + 11 k\, 
     N + 3 N^{2})} {3 (2 + N) (2 + k + N)^{4}}, \nonu\\
c_ {265} & = & - \frac {2 (k - N) (16 + 21 k + 6 k^{2} + 19 N + 
       13 k\, N + 5 N^{2})} {3 (2 + N) (2 + k + N)^{4}}, \nonu\\
c_ {266} & = &\frac {(k - N) (-32 - 3 k + 6 k^{2} - 29 N + k\, 
     N - 7 N^{2})} {3 (2 + N) (2 + k + N)^{4}}, \nonu\\
c_ {267} & = & - \frac {(k - N) (48 + 37 k + 6 k^{2} + 51 N + 21 k\, 
      N + 13 N^{2})} {3 (2 + N) (2 + k + N)^{4}}, \nonu\\
c_ {268} & = &\frac {2 (k - N) (32 + 29 k + 6 k^{2} + 35 N + 17 k\, 
     N + 9 N^{2})} {3 (2 + N) (2 + k + N)^{4}}, \nonu\\
c_ {269} & = & - \frac {2 (k - N) (-8 + 9 k + 6 k^{2} - 5 N + 7 k\, 
      N - N^{2})} {3 (2 + N) (2 + k + N)^{4}}, \nonu\\
c_ {270} & = & - \frac{1}{3 (2 + N) (2 + k + N)^{4}}i (576 + 1024 k + 573 k^{2} + 102 k^{3} + 
       992 N + 1184 k\, N + 335 k^{2} N 
       \nonu\\ &+& 547 N^{2} + 330 k\, 
      N^{2} + 97 N^{3}), \nonu\\
c_ {271} & = &\frac {i (448 k + 429 k^{2} + 102 k^{3} + 128 N + 
      608 k\, N + 263 k^{2} N + 115 N^{2} + 186 k\, 
     N^{2} + 25 N^{3})} {3 (2 + N) (2 + k + N)^{4}}, \nonu\\
c_ {272} & = &\frac {2 i (k - N) (32 + 55 k + 18 k^{2} + 41 N + 
      35 k\, N + 11 N^{2})} {3 (2 + N) (2 + k + N)^{5}}, \nonu\\
c_ {273} & = &\frac {1} {3 (2 + N) (2 + k + N)^{4}} (96 + 908 k + 
    930 k^{2} + 309 k^{3} + 30 k^{4} + 196 N + 1124 k\, 
   N + 681 k^{2} N 
   \nonu\\ &+& 89 k^{3} N + 10 N^{2} + 289 k\, 
   N^{2} + 67 k^{2} N^{2} - 79 N^{3} - 19 k\, N^{3} - 23 N^{4}), \nonu\\
c_ {274} & = & - \frac {1} {3 (2 + N) (2 + k + N)^{4}} (864 + 
     1856 k + 1293 k^{2} + 351 k^{3} + 30 k^{4} + 1648 N + 2432 k\, 
    N 
    \nonu\\ &+& 984 k^{2} N + 101 k^{3} N + 1075 N^{2} + 925 k\, 
    N^{2} + 133 k^{2} N^{2} + 284 N^{3} + 95 k\, N^{3} + 25 N^{4}), \nonu\\
c_ {275} & = &\frac {1} {(2 + N) (2 + k + N)^{4}} (256 + 324 k + 
    135 k^{2} + 18 k^{3} + 412 N + 396 k\, 
   N + 123 k^{2} N + 12 k^{3} N 
   \nonu\\ &+& 221 N^{2} + 140 k\, 
   N^{2} + 26 k^{2} N^{2} + 39 N^{3} + 10 k\, N^{3}), \nonu\\
c_ {276} & = &\frac {3 i (16 + 21 k + 6 k^{2} + 19 N + 13 k\, 
     N + 5 N^{2})} {(2 + N) (2 + k + N)^{3}}, \nonu\\
c_ {277} & = &\frac {i (16 + 47 k + 18 k^{2} + 25 N + 31 k\, 
     N + 7 N^{2})} {(2 + N) (2 + k + N)^{3}}, \nonu\\
c_ {278} & = &\frac {6 i (8 + 17 k + 6 k^{2} + 11 N + 11 k\, 
     N + 3 N^{2})} {(2 + N) (2 + k + N)^{3}}, \nonu\\
c_ {279} & = &\frac {2 i (40 + 59 k + 18 k^{2} + 49 N + 37 k\, 
     N + 13 N^{2})} {(2 + N) (2 + k + N)^{3}}, \nonu\\
c_ {280} & = &\frac {2 i (107 k + 102 k^{2} + 24 k^{3} + 37 N + 
      151 k\, N + 64 k^{2} N + 35 N^{2} + 48 k\, 
     N^{2} + 8 N^{3})} {3 (2 + N) (2 + k + N)^{4}}, \nonu\\
c_ {281} & = &\frac {2 i (96 + 203 k + 126 k^{2} + 24 k^{3} + 181 N + 
      247 k\, N + 76 k^{2} N + 107 N^{2} + 72 k\, 
     N^{2} + 20 N^{3})} {3 (2 + N) (2 + k + N)^{4}}, \nonu\\
c_ {282} & = &\frac {2 i (48 + 155 k + 114 k^{2} + 24 k^{3} + 109 N + 
      199 k\, N + 70 k^{2} N + 71 N^{2} + 60 k\, 
     N^{2} + 14 N^{3})} {3 (2 + N) (2 + k + N)^{4}}, \nonu\\
c_ {283} & = &\frac {6 (32 + 55 k + 18 k^{2} + 41 N + 35 k\, 
     N + 11 N^{2})} {(2 + N) (2 + k + N)^{4}}, \nonu\\
c_ {284} & = & - \frac {2 i (288 + 622 k + 285 k^{2} + 30 k^{3} + 
       386 N + 470 k\, N + 71 k^{2} N + 109 N^{2} + 42 k\, 
      N^{2} + N^{3})} {3 (2 + N) (2 + k + N)^{4}}, \nonu\\
c_ {285} & = &\frac {1} {3 (2 + N) (2 + k + N)^{5}} 2 i (768 + 
    1760 k + 1259 k^{2} + 347 k^{3} + 30 k^{4} + 1504 N + 2336 k\, 
   N 
   \nonu\\ &+& 948 k^{2} N + 93 k^{3} N + 1013 N^{2} + 915 k\, 
   N^{2} + 127 k^{2} N^{2} + 286 N^{3} + 105 k\, N^{3} + 29 N^{4}), \nonu\\
c_ {286} & = &\frac {1} {3 (2 + N) (2 + k + N)^{5}} 2 i (576 + 
    1512 k + 1161 k^{2} + 335 k^{3} + 30 k^{4} + 1080 N + 1944 k\, 
   N 
   \nonu\\ &+& 866 k^{2} N + 93 k^{3} N + 639 N^{2} + 683 k\, 
   N^{2} + 107 k^{2} N^{2} + 132 N^{3} + 53 k\, N^{3} + 5 N^{4}), \nonu\\
c_ {287} & = & - \frac {1} {3 (2 + N) (2 + k + N)^{5}} 2 i (672 + 
     1716 k + 1318 k^{2} + 373 k^{3} + 30 k^{4} + 1212 N + 2188 k\, 
    N 
    \nonu\\ &+&1031 k^{2} N + 121 k^{3} N + 670 N^{2} + 743 k\, 
    N^{2} + 145 k^{2} N^{2} + 109 N^{3} + 43 k\, N^{3} - 3 N^{4}), \nonu\\
c_ {288} & = &\frac {1} {3 (2 + N) (2 + k + N)^{5}} 2 i (288 + 
    646 k + 535 k^{2} + 203 k^{3} + 30 k^{4} + 650 N + 1078 k\, 
   N + 574 k^{2} N 
   \nonu\\ &+& 105 k^{3} N + 547 N^{2} + 601 k\, 
   N^{2} + 153 k^{2} N^{2} + 206 N^{3} + 115 k\, N^{3} + 29 N^{4}), \nonu\\
c_ {289} & = & - \frac {1} {3 (2 + N) (2 + k + N)^{5}} 2 i (274 k + 
     404 k^{2} + 193 k^{3} + 30 k^{4} + 14 N + 458 k\, 
    N + 451 k^{2} N 
    \nonu\\ &+& 109 k^{3} N + 2 N^{2} + 229 k\, 
    N^{2} + 119 k^{2} N^{2} - 9 N^{3} + 33 k\, N^{3} - 3 N^{4}),\nonu\\ 
c_ {290} & = & - \frac {2 i (k - N) (20 + 31 k + 10 k^{2} + 35 N + 
       41 k\, N + 8 k^{2} N + 21 N^{2} + 14 k\, 
      N^{2} + 4 N^{3})} {3 (2 + N) (2 + k + N)^{5}}, \nonu\\
c_ {291} & = & - \frac {1} {3 (2 + N) (2 + k + N)^{5}} 2 i (-96 - 
     44 k + 59 k^{2} + 26 k^{3} - 292 N - 148 k\, 
    N + 83 k^{2} N + 28 k^{3} N 
    \nonu\\ &-& 343 N^{2} - 172 k\, 
    N^{2} + 18 k^{2} N^{2} - 177 N^{3} - 62 k\, N^{3} - 32 N^{4}), \nonu\\
c_ {292} & = & - \frac {1} {3 (2 + N) (2 + k + N)^{5}} 2 i (864 + 
     1750 k + 1089 k^{2} + 214 k^{3} + 1562 N + 2166 k\, 
    N + 755 k^{2} N 
    \nonu\\ &+& 32 k^{3} N + 921 N^{2} + 710 k\, 
    N^{2} + 68 k^{2} N^{2} + 193 N^{3} + 36 k\, N^{3} + 8 N^{4}), \nonu\\
c_ {293} & = & - \frac {1} {3 (2 + N) (2 + k + N)^{5}} 2 i (-192 - 
     622 k - 423 k^{2} - 82 k^{3} - 338 N - 582 k\, 
    N - 65 k^{2} N 
    \nonu\\ &+& 52 k^{3} N - 147 N^{2} - 38 k\, 
    N^{2} + 100 k^{2} N^{2} - 7 N^{3} + 36 k\, N^{3} + 4 N^{4}), \nonu\\
c_ {294} & = &\frac {1} {3 (2 + N) (2 + k + N)^{5}} 4 i (576 + 
    1295 k + 968 k^{2} + 293 k^{3} + 30 k^{4} + 1153 N + 1811 k\, 
   N 
   \nonu\\ &+& 828 k^{2} N + 111 k^{3} N + 821 N^{2} + 786 k\, 
   N^{2} + 154 k^{2} N^{2} + 253 N^{3} + 108 k\, N^{3} + 29 N^{4}), \nonu\\
c_ {295} & = & - \frac {1} {3 (2 + N) (2 + k + N)^{5}} 4 i (192 + 
     779 k + 741 k^{2} + 259 k^{3} + 30 k^{4} + 325 N + 1023 k\, 
    N 
    \nonu\\ &+& 633 k^{2} N + 103 k^{3} N + 108 N^{2} + 342 k\, 
    N^{2} + 108 k^{2} N^{2} - 34 N^{3} + 14 k\, N^{3} - 15 N^{4}), \nonu\\
c_ {296} & = &\frac {1} {3 (2 + N) (2 + k + N)^{5}} 2 i (864 + 
    1998 k + 1448 k^{2} + 393 k^{3} + 30 k^{4} + 1890 N + 2990 k\, 
   N
   \nonu\\ &+& 1307 k^{2} N + 149 k^{3} N + 1466 N^{2} + 1409 k\, 
   N^{2} + 267 k^{2} N^{2} + 491 N^{3} + 213 k\, N^{3} + 61 N^{4}), \nonu\\
c_ {297} & = & - \frac {4 (k - N)} {3 (2 + k + N)^{3}}, \qquad
c_ {298}  = \frac {8 (k - N)} {3 (2 + k + N)^{3}}, \nonu\\
c_ {299} & = & - \frac {2 (540 + 821 k + 262 k^{2} + 961 N + 1099 k\, 
      N + 212 k^{2} N + 583 N^{2} + 378 k\, 
      N^{2} + 112 N^{3})} {3 (2 + N) (2 + k + N)^{4}}, \nonu\\
c_ {300} & = & - \frac {2 (540 + 565 k + 134 k^{2} + 1217 N + 971 k\, 
      N + 148 k^{2} N + 839 N^{2} + 378 k\, 
      N^{2} + 176 N^{3})} {3 (2 + N) (2 + k + N)^{4}}, \nonu\\
c_ {301} & = & - \frac {1} {3 (2 + N) (2 + k + N)^{5}} 4 (360 + 
     642 k + 389 k^{2} + 78 k^{3} + 906 N + 1264 k\, 
    N + 537 k^{2} N 
    \nonu\\ &+& 60 k^{3} N + 831 N^{2} + 818 k\, 
    N^{2} + 186 k^{2} N^{2} + 331 N^{3} + 174 k\, N^{3} + 48 N^{4}), \nonu\\
c_ {302} & = & - \frac {1} {3 (2 + N) (2 + k + N)^{4}} 2 (-304 - 
     596 k - 247 k^{2} - 18 k^{3} - 548 N - 668 k\, 
    N - 51 k^{2} N 
    \nonu\\ &+& 36 k^{3} N - 365 N^{2} - 280 k\, 
    N^{2} + 10 k^{2} N^{2} - 139 N^{3} - 70 k\, N^{3} - 24 N^{4}), \nonu\\
c_ {303} & = & - \frac {1} {3 (2 + N) (2 + k + N)^{5}} 2 (1440 + 
     3880 k + 3304 k^{2} + 1113 k^{3} + 126 k^{4} + 2600 N + 6040 k\, 
    N 
    \nonu\\ &+& 4229 k^{2} N + 1143 k^{3} N + 108 k^{4} N + 1024 N^{2} + 
     2639 k\, 
    N^{2} + 1645 k^{2} N^{2} + 318 k^{3} N^{2} - 493 N^{3} 
    \nonu\\ &+& 89 k\, 
    N^{3} + 180 k^{2} N^{3} - 411 N^{4} - 102 k\, N^{4} - 72 N^{5}), \nonu\\
c_ {304} & = & - \frac {1} {3 (2 + N) (2 + k + N)^{5}} 2 (-192 + 
     1204 k + 1877 k^{2} + 863 k^{3} + 126 k^{4} - 1012 N + 1100 k\, 
    N 
    \nonu\\ &+& 2162 k^{2} N + 883 k^{3} N + 108 k^{4} N - 1969 N^{2} - 
     613 k\, N^{2} + 675 k^{2} N^{2} + 246 k^{3} N^{2} - 1596 N^{3} 
     \nonu\\ &-&
     785 k\, N^{3} + 36 k^{2} N^{3} - 563 N^{4} - 174 k\, 
    N^{4} - 72 N^{5}), \nonu\\
c_ {305} & = & - \frac {1} {3 (2 + N) (2 + k + N)^{5}} 2 (-448 - 
     1100 k - 789 k^{2} - 213 k^{3} - 18 k^{4} - 1044 N - 1660 k\, 
    N 
    \nonu\\ &-& 492 k^{2} N + 95 k^{3} N + 36 k^{4} N - 1007 N^{2} - 1005 k\, 
    N^{2} - 33 k^{2} N^{2} + 70 k^{3} N^{2} - 562 N^{3} 
    \nonu\\ &-& 377 k\, 
    N^{3} - 12 k^{2} N^{3}- 179 N^{4} - 70 k\, N^{4} - 24 N^{5}), \nonu\\
c_ {306} & = &\frac {4 (18 + 18 k + 5 k^{2} + 18 N + 8 k\, 
     N + 5 N^{2}) (32 + 55 k + 18 k^{2} + 41 N + 35 k\, 
     N + 11 N^{2})} {3 (2 + N) (2 + k + N)^{6}}, \nonu\\
c_ {307} & = &\frac {(k - N)} {12 (2 + k + N)}, \nonu\\
c_ {308} & = & - \frac {(k - N) (60 + 77 k + 22 k^{2} + 121 N + 
       115 k\, N + 20 k^{2} N + 79 N^{2} + 42 k\, 
      N^{2} + 16 N^{3})} {12 (2 + N) (2 + k + N)^{3}}, \nonu\\
c_ {309} & = &\frac {3 (60 + 77 k + 22 k^{2} + 121 N + 115 k\, 
     N + 20 k^{2} N + 79 N^{2} + 42 k\, 
     N^{2} + 16 N^{3})} {4 (2 + N) (2 + k + N)^{2}}, \nonu\\
c_ {310} & = & - \frac { i (180 + 311 k + 171 k^{2} + 30 k^{3} + 
       283 N + 335 k\, N + 95 k^{2} N + 142 N^{2} + 86 k\, 
      N^{2} + 23 N^{3})} {3 (2 + N) (2 + k + N)^{3}}, \nonu\\
c_ {311} & = & - \frac {2 (180 + 279 k + 155 k^{2} + 30 k^{3} + 
       315 N + 319 k\, N + 87 k^{2} N + 174 N^{2} + 86 k\, 
      N^{2} + 31 N^{3})} {3 (2 + N) (2 + k + N)^{4}}, \nonu\\
c_ {312} & = & - \frac {(360 + 574 k + 279 k^{2} + 42 k^{3} + 614 N + 
      676 k\, N + 169 k^{2} N + 341 N^{2} + 196 k\, 
     N^{2} + 61 N^{3})} {12 (2 + N) (2 + k + N)^{3}}, \nonu\\
c_ {313} & = &\frac {1} {3 (2 + N) (2 + k + N)^{4}} 2 i (240 + 
    522 k + 328 k^{2} + 64 k^{3} + 462 N + 666 k\, 
   N + 213 k^{2} N 
   \nonu\\ &+& 2 k^{3} N + 302 N^{2} + 241 k\, 
   N^{2} + 15 k^{2} N^{2} + 82 N^{3} + 23 k\, N^{3} + 8 N^{4}), \nonu\\
c_ {314} & = & - \frac {1} {3 (2 + N) (2 + k + N)^{4}} 2 i (-192 - 
     330 k - 180 k^{2} - 32 k^{3} - 246 N - 242 k\, 
    N - 27 k^{2} N 
    \nonu\\ &+& 14 k^{3} N - 58 N^{2} + 35 k\, 
    N^{2} + 41 k^{2} N^{2} + 24 N^{3} + 33 k\, N^{3} + 8 N^{4}), \nonu\\
c_ {315} & = & - \frac {4 i (6 + 21 k + 8 k^{2} + 21 N + 20 k\, 
      N + 8 N^{2})} {3 (2 + k + N)^{3}}, \nonu\\
c_ {316} & = & - \frac {6 (8 + 17 k + 6 k^{2} + 11 N + 11 k\, 
      N + 3 N^{2})} {(2 + N) (2 + k + N)^{3}}, \nonu\\
c_ {317} & = & - \frac {6 (16 + 21 k + 6 k^{2} + 19 N + 13 k\, 
      N + 5 N^{2})} {(2 + N) (2 + k + N)^{3}}, \nonu\\
c_ {318} & = & - \frac { (107 k + 102 k^{2} + 24 k^{3} + 37 N + 
       151 k\, N + 64 k^{2} N + 35 N^{2} + 48 k\, 
      N^{2} + 8 N^{3})} {3 (2 + N) (2 + k + N)^{4}}, \nonu\\
c_ {319} & = & - \frac { (96 + 203 k + 126 k^{2} + 24 k^{3} + 
       181 N + 247 k\, N + 76 k^{2} N + 107 N^{2} + 72 k\, 
      N^{2} + 20 N^{3})} {3 (2 + N) (2 + k + N)^{4}}, \nonu\\
c_ {320} & = & - \frac {2 (32 + 55 k + 18 k^{2} + 41 N + 35 k\, 
      N + 11 N^{2})} {(2 + N) (2 + k + N)^{3}}, \nonu\\
c_ {321} & = & - \frac {2 (8 + 43 k + 18 k^{2} + 17 N + 29 k\, 
      N + 5 N^{2})} {(2 + N) (2 + k + N)^{3}}, \nonu\\
c_ {322} & = & - \frac {(16 + 47 k + 18 k^{2} + 25 N + 31 k\, 
     N + 7 N^{2})} {(2 + N) (2 + k + N)^{3}}, \nonu\\
c_ {323} & = &\frac {(96 + 116 k + 45 k^{2} + 6 k^{3} + 124 N + 
     100 k\, N + 19 k^{2} N + 47 N^{2} + 18 k\, 
    N^{2} + 5 N^{3})} {3 (2 + N) (2 + k + N)^{4}}, \nonu\\
c_ {324} & = & - \frac {3 (16 + 21 k + 6 k^{2} + 19 N + 13 k\, 
      N + 5 N^{2})} {(2 + N) (2 + k + N)^{3}}, \nonu\\
c_ {325} & = & - \frac {3 (13 k + 6 k^{2} + 3 N + 9 k\, 
      N + N^{2})} {(2 + N) (2 + k + N)^{3}}, \nonu\\
c_ {326} & = &\frac {i (360 + 526 k + 255 k^{2} + 42 k^{3} + 662 N + 
      652 k\, N + 157 k^{2} N + 389 N^{2} + 196 k\, 
     N^{2} + 73 N^{3})} {3 (2 + N) (2 + k + N)^{4}}, \nonu\\
c_ {327} & = &\frac {3 i (32 + 55 k + 18 k^{2} + 41 N + 35 k\, 
     N + 11 N^{2})} {(2 + N) (2 + k + N)^{4}}, \nonu\\
c_ {328} & = & - \frac {1} {6 (2 + N) (2 + k + N)^{4}} i (-360 - 
     710 k - 395 k^{2} - 40 k^{3} + 12 k^{4} - 478 N - 544 k\, 
    N 
    \nonu\\ &+& 97 k^{2} N + 200 k^{3} N + 36 k^{4} N + 3 N^{2} + 384 k\, 
    N^{2} + 484 k^{2} N^{2} + 130 k^{3} N^{2} + 279 N^{3} + 450 k\, 
    N^{3} 
    \nonu\\ &+& 172 k^{2} N^{3} + 150 N^{4} + 106 k\, N^{4} + 24 N^{5}), \nonu\\
c_ {329} & = &\frac {(360 + 546 k + 237 k^{2} + 30 k^{3} + 642 N + 
     682 k\, N + 155 k^{2} N + 377 N^{2} + 212 k\, 
    N^{2} + 71 N^{3})} {6 (2 + N) (2 + k + N)^{3}}, \nonu\\
c_ {330} & = &\frac {(360 + 530 k + 237 k^{2} + 30 k^{3} + 658 N + 
     674 k\, N + 155 k^{2} N + 385 N^{2} + 212 k\, 
    N^{2} + 71 N^{3})} {6 (2 + N) (2 + k + N)^{3}}, \nonu\\
c_ {331} & = & - \frac {(-180 - 147 k + 39 k^{2} + 30 k^{3} - 447 N - 
      353 k\, N - 25 k^{2} N - 334 N^{2} - 166 k\, 
     N^{2} - 73 N^{3})} {6 (2 + N) (2 + k + N)^{3}}, \nonu\\
c_ {332} & = & - \frac {(-180 - 163 k + 39 k^{2} + 30 k^{3} - 431 N - 
      361 k\, N - 25 k^{2} N - 326 N^{2} - 166 k\, 
     N^{2} - 73 N^{3})} {6 (2 + N) (2 + k + N)^{3}}, \nonu\\
c_ {333} & = &\frac {3 i (20 + 21 k + 6 k^{2} + 25 N + 13 k\, 
     N + 7 N^{2})} {2(2 + N) (2 + k + N)^{3}}, \nonu\\
c_ {334} & = & - \frac {2 i (180 + 271 k + 138 k^{2} + 24 k^{3} + 
       323 N + 325 k\, N + 82 k^{2} N + 185 N^{2} + 94 k\, 
      N^{2} + 34 N^{3})} {3 (2 + N) (2 + k + N)^{4}}, \nonu\\
c_ {335} & = &\frac {i (k - N) (16 + 21 k + 6 k^{2} + 19 N + 13 k\, 
     N + 5 N^{2})} {3 (2 + N) (2 + k + N)^{4}}, \nonu\\
c_ {336} & = &\frac {3 (60 + 77 k + 22 k^{2} + 121 N + 115 k\, 
     N + 20 k^{2} N + 79 N^{2} + 42 k\, 
     N^{2} + 16 N^{3})} {2(2 + N) (2 + k + N)^{3}}, \nonu\\
c_ {337} & = &\frac {i (k - N) (16 + 21 k + 6 k^{2} + 19 N + 13 k\, 
     N + 5 N^{2})} {2(2 + N) (2 + k + N)^{4}}, \nonu\\
c_ {338} & = &\frac {2 (k - N) (32 + 55 k + 18 k^{2} + 41 N + 35 k\, 
     N + 11 N^{2})} {3 (2 + N) (2 + k + N)^{5}}, \nonu\\
c_ {339} & = & - \frac {1} {6 (2 + N) (2 + k + N)^{4}} i (384 + 
     976 k + 661 k^{2} + 134 k^{3} + 848 N + 1368 k\, 
    N + 491 k^{2} N 
    \nonu\\ &+& 16 k^{3} N + 611 N^{2} + 546 k\, 
    N^{2} + 56 k^{2} N^{2} + 173 N^{3} + 56 k\, N^{3} + 16 N^{4}), \nonu\\
c_ {340} & = & - \frac {2 i (144 + 271 k + 159 k^{2} + 30 k^{3} + 
       233 N + 299 k\, N + 89 k^{2} N + 118 N^{2} + 78 k\, 
      N^{2} + 19 N^{3})} {3 (2 + N) (2 + k + N)^{4}}, \nonu\\
c_ {341} & = &\frac {1} {6 (2 + N) (2 + k + N)^{4}} i (-192 - 108 k + 
    25 k^{2} + 14 k^{3} - 84 N + 172 k\, 
   N + 135 k^{2} N + 16 k^{3} N 
   \nonu\\ &+& 139 N^{2} + 234 k\, 
   N^{2} + 56 k^{2} N^{2} + 97 N^{3} + 56 k\, N^{3} + 16 N^{4}),\nonu\\ 
c_ {342} & = & - \frac { i (180 + 287 k + 159 k^{2} + 30 k^{3} + 
       307 N + 323 k\, N + 89 k^{2} N + 166 N^{2} + 86 k\, 
      N^{2} + 29 N^{3})} {3 (2 + N) (2 + k + N)^{4}}, \nonu\\
c_ {343} & = &\frac {6} {(2 + k + N)^{2}}, \qquad
c_ {344}  = \frac {6 i (32 + 55 k + 18 k^{2} + 41 N + 35 k\, 
     N + 11 N^{2})} {(2 + N) (2 + k + N)^{4}}, \nonu\\
c_ {345} & = & - \frac { i (180 + 263 k + 147 k^{2} + 30 k^{3} + 
       331 N + 311 k\, N + 83 k^{2} N + 190 N^{2} + 86 k\, 
      N^{2} + 35 N^{3})} {3 (2 + N) (2 + k + N)^{4}}, \nonu\\
c_ {346} & = & - \frac { i (360 + 430 k + 51 k^{2} - 30 k^{3} + 
       758 N + 724 k\, N + 97 k^{2} N + 521 N^{2} + 292 k\, 
      N^{2} + 109 N^{3})} {3 (2 + N) (2 + k + N)^{4}}, \nonu\\
c_ {347} & = & - \frac { i (180 + 279 k + 155 k^{2} + 30 k^{3} + 
       315 N + 319 k\, N + 87 k^{2} N + 174 N^{2} + 86 k\, 
      N^{2} + 31 N^{3})} {3 (2 + N) (2 + k + N)^{4}}, \nonu\\
c_ {348} & = & - \frac { i (360 + 414 k + 43 k^{2} - 30 k^{3} + 
       774 N + 716 k\, N + 93 k^{2} N + 537 N^{2} + 292 k\, 
      N^{2} + 113 N^{3})} {3 (2 + N) (2 + k + N)^{4}}, \nonu\\
c_ {349} & = & - \frac {2 i (60 + 77 k + 22 k^{2} + 121 N + 115 k\, 
      N + 20 k^{2} N + 79 N^{2} + 42 k\, 
      N^{2} + 16 N^{3})} {(2 + N) (2 + k + N)^{4}}, \nonu\\
c_ {350} & = & - \frac {4 i (60 + 77 k + 22 k^{2} + 121 N + 115 k\, 
      N + 20 k^{2} N + 79 N^{2} + 42 k\, 
      N^{2} + 16 N^{3})} {(2 + N) (2 + k + N)^{4}}, \nonu\\
c_ {351} & = & - \frac {2 i (180 + 223 k + 62 k^{2} + 371 N + 341 k\, 
      N + 58 k^{2} N + 245 N^{2} + 126 k\, 
      N^{2} + 50 N^{3})} {3 (2 + N) (2 + k + N)^{4}}, \nonu\\
c_ {352} & = & - \frac {4 i (180 + 235 k + 68 k^{2} + 359 N + 347 k\, 
      N + 61 k^{2} N + 233 N^{2} + 126 k\, 
      N^{2} + 47 N^{3})} {3 (2 + N) (2 + k + N)^{4}}, \nonu\\
c_ {353} & = & - \frac {1} {3 (2 + N) (2 + k + N)^{5}} 2 i (180 + 
     353 k + 202 k^{2} 
     \nonu\\ &+& 36 k^{3} + 421 N + 673 k\, 
    N + 293 k^{2} N + 36 k^{3} N + 367 N^{2} + 417 k\, 
    N^{2} + 99 k^{2} N^{2} + 136 N^{3} 
    \nonu\\ &+& 81 k\, N^{3} + 18 N^{4}),\nonu\\ 
c_ {354} & = & - \frac {1} {3 (2 + N) (2 + k + N)^{5}} 2 i (900 + 
     1573 k + 887 k^{2} + 162 k^{3} + 2297 N + 3179 k\, 
    N \nonu\\ &+& 1300 k^{2} N 
    + 144 k^{3} N +2144 N^{2} + 2085 k\, 
    N^{2} + 459 k^{2} N^{2} + 863 N^{3} + 441 k\, N^{3} + 126 N^{4}), \nonu\\
c_ {355} & = & - \frac {1} {3 (2 + N) (2 + k + N)^{5}} 2i (180 + 
     273 k + 122 k^{2} + 16 k^{3} + 501 N + 633 k\, 
    N + 233 k^{2} N 
    \nonu\\ &+& 26 k^{3} N + 487 N^{2} + 437 k\, 
    N^{2} + 89 k^{2} N^{2} + 196 N^{3} + 91 k\, N^{3} + 28 N^{4}), \nonu\\
c_ {356} & = & - \frac {1} {3 (2 + N) (2 + k + N)^{5}} 2 i (900 + 
     1653 k + 967 k^{2} + 182 k^{3} + 2217 N + 3219 k\, 
    N \nonu\\ &+& 1360 k^{2} N 
   +154 k^{3} N +2024 N^{2} + 2065 k\, 
    N^{2} + 469 k^{2} N^{2} + 803 N^{3} + 431 k\, N^{3} + 116 N^{4}), \nonu\\
c_ {357} & = &\frac {1} {3 (2 + N) (2 + k + N)^{5}} 2 i (-180 - 
    145 k + 136 k^{2} + 141 k^{3} + 30 k^{4} - 629 N - 669 k\, 
   N 
   \nonu\\ &-& 92 k^{2} N 
  +35 k^{3} N - 709 N^{2} - 614 k\, 
   N^{2} - 98 k^{2} N^{2} - 317 N^{3} - 152 k\, N^{3} - 49 N^{4}), \nonu\\
c_ {358} & = & - \frac {1} {3 (2 + N) (2 + k + N)^{5}} 2 i (900 + 
     1781 k + 1225 k^{2} + 339 k^{3} + 30 k^{4} + 2089 N + 3183 k\, 
    N 
    \nonu\\ &+& 1501 k^{2} N + 215 k^{3} N + 1802 N^{2} + 1888 k\, 
    N^{2} + 460 k^{2} N^{2} + 682 N^{3} + 370 k\, N^{3} + 95 N^{4}), \nonu\\
c_ {359} & = &\frac {1} {3 (2 + N) (2 + k + N)^{5}} 2 i (-180 - 
    305 k - 24 k^{2} + 101 k^{3} + 30 k^{4} - 469 N - 749 k\, 
   N 
   \nonu\\ &-& 212 k^{2} N + 15 k^{3} N - 469 N^{2} - 574 k\, 
   N^{2} - 118 k^{2} N^{2} - 197 N^{3} - 132 k\, N^{3} - 29 N^{4}), \nonu\\
c_ {360} & = & - \frac {1} {3 (2 + N) (2 + k + N)^{5}} 2 i (900 + 
     1621 k + 1065 k^{2} + 299 k^{3} + 30 k^{4} + 2249 N + 3103 k\, 
    N 
    \nonu\\ &+& 1381 k^{2} N + 195 k^{3} N + 2042 N^{2} + 1928 k\, 
    N^{2} + 440 k^{2} N^{2} + 802 N^{3} + 390 k\, N^{3} + 115 N^{4}), \nonu\\
c_ {361} & = &\frac { i (-180 - 199 k + 15 k^{2} + 30 k^{3} - 
      395 N - 379 k\, N - 37 k^{2} N - 284 N^{2} - 166 k\, 
     N^{2} - 61 N^{3})} {3 (2 + N) (2 + k + N)^{4}}, \nonu\\
c_ {362} & = & - \frac {1}{3 (2 + N) (2 + k + N)^{4}} i (360 + 494 k + 213 k^{2} + 30 k^{3} + 
       694 N + 656 k\, N + 143 k^{2} N 
       \nonu\\ &+& 427 N^{2} + 212 k\, 
      N^{2} + 83 N^{3}), \nonu\\
c_ {363} & = &\frac { i (-180 - 167 k + 31 k^{2} + 30 k^{3} - 
      427 N - 363 k\, N - 29 k^{2} N - 316 N^{2} - 166 k\, 
     N^{2} - 69 N^{3})} {3 (2 + N) (2 + k + N)^{4}}, \nonu\\
c_ {364} & = & - \frac{1}{3 (2 + N) (2 + k + N)^{4}} i (360 + 526 k + 229 k^{2} + 30 k^{3} + 
       662 N + 672 k\, N + 151 k^{2} N 
       \nonu\\ &+& 395 N^{2} + 212 k\, 
      N^{2} + 75 N^{3}), \nonu\\
c_ {365} & = &\frac{1}{3 (2 + N) (2 + k + N)^{5}} i (k - N) (96 + 249 k + 159 k^{2} + 30 k^{3} + 
      167 N + 269 k\, N + 83 k^{2} N 
      \nonu\\ &+& 84 N^{2} + 66 k\, 
     N^{2} + 13 N^{3}), \nonu\\
c_ {366} & = & - \frac {1} {3 (2 + N) (2 + k + N)^{5}}  i (1080 + 
     2022 k + 1338 k^{2} + 357 k^{3} + 30 k^{4} + 2622 N + 3770 k\, 
    N 
    \nonu\\ &+& 1703 k^{2} N + 233 k^{3} N + 2344 N^{2} + 2317 k\, 
    N^{2} + 541 k^{2} N^{2} + 915 N^{3} + 469 k\, N^{3} + 131 N^{4}),\nonu\\ 
c_ {367} & = &\frac {1}{3 (2 + N) (2 + k + N)^{5}} i (k - N) (64 + 220 k + 153 k^{2} + 30 k^{3} + 
      132 N + 252 k\, N + 83 k^{2} N 
      \nonu\\ &+& 75 N^{2} + 66 k\, 
     N^{2} + 13 N^{3}), \nonu\\
c_ {368} & = &\frac {1}{3 (2 + N) (2 + k + N)^{5}} i (k - N) (128 + 284 k + 169 k^{2} + 
      30 k^{3} + 228 N + 316 k\, N + 91 k^{2} N 
      \nonu\\ &+& 123 N^{2} + 82 k\, 
     N^{2} + 21 N^{3}), \nonu\\
c_ {369} & = &\frac {1}{3 (2 + N) (2 + k + N)^{5}} i (k - N) (160 + 316 k + 177 k^{2} + 
      30 k^{3} + 276 N + 348 k\, N + 95 k^{2} N 
      \nonu\\ &+& 147 N^{2} + 90 k\, 
     N^{2} + 25 N^{3}), \nonu\\
c_ {370} & = &\frac { i (k - N) (32 + 188 k + 145 k^{2} + 30 k^{3} + 
      84 N + 220 k\, N + 79 k^{2} N + 51 N^{2} + 58 k\, 
     N^{2} + 9 N^{3})} {3 (2 + N) (2 + k + N)^{5}}, \nonu\\
c_ {371} & = &\frac {1} {3 (2 + N) (2 + k + N)^{5}} 2 i (360 + 
    738 k + 638 k^{2} + 237 k^{3} + 30 k^{4} + 810 N + 1182 k\, 
   N 
   \nonu\\ &+& 647 k^{2} N + 113 k^{3} N + 664 N^{2} + 633 k\, 
   N^{2} + 169 k^{2} N^{2} + 247 N^{3} + 121 k\, N^{3} + 35 N^{4}), \nonu\\
c_ {372} & = &\frac {1} {3 (2 + N) (2 + k + N)^{5}} 4 (144 + 323 k + 
    209 k^{2} + 42 k^{3} + 325 N + 467 k\, 
   N + 160 k^{2} N + 6 k^{3} N 
   \nonu\\ &+& 260 N^{2} + 211 k\, 
   N^{2} + 23 k^{2} N^{2} + 91 N^{3} + 31 k\, N^{3} + 12 N^{4}), \nonu\\
c_ {373} & = & - \frac {1} {3 (2 + N) (2 + k + N)^{5}}  (576 + 
     1352 k + 913 k^{2} + 190 k^{3} + 1240 N + 1912 k\, 
    N + 733 k^{2} N 
    \nonu\\ &+& 44 k^{3} N + 919 N^{2} + 808 k\, 
    N^{2} + 114 k^{2} N^{2} + 285 N^{3} + 98 k\, N^{3} + 32 N^{4}), \nonu\\
c_ {374} & = & - \frac {(k - N) (60 + 77 k + 22 k^{2} + 121 N + 
       115 k\, N + 20 k^{2} N + 79 N^{2} + 42 k\, 
      N^{2} + 16 N^{3})} {3 (2 + N) (2 + k + N)^{5}}, \nonu\\
c_ {375} & = & - \frac { (k - N) (32 + 55 k + 18 k^{2} + 41 N + 
       35 k\, N + 11 N^{2})} {3 (2 + N) (2 + k + N)^{5}}, \nonu\\
c_ {376} & = & - \frac {1} {3 (2 + N) (2 + k + N)^{5}}  (-360 - 
     590 k - 414 k^{2} - 139 k^{3} - 18 k^{4} - 598 N - 374 k\, 
    N 
    \nonu\\ &+& 127 k^{2} N + 163 k^{3} N + 36 k^{4} N - 148 N^{2} + 521 k\, 
    N^{2} + 517 k^{2} N^{2} + 126 k^{3} N^{2} + 211 N^{3}
    \nonu\\ &+& 495 k\, 
    N^{3} + 172 k^{2} N^{3} + 139 N^{4} + 110 k\, N^{4} + 24 N^{5}), \nonu\\
c_ {377} & = & - \frac {1} {3 (2 + N) (2 + k + N)^{5}}  (1800 + 
     3574 k + 2740 k^{2} + 953 k^{3} + 126 k^{4} + 5246 N + 8554 k\, 
    N 
    \nonu\\ &+& 5093 k^{2} N + 1279 k^{3} N + 108 k^{4} N + 5770 N^{2} + 
     7205 k\, 
    N^{2} + 2929 k^{2} N^{2} + 390 k^{3} N^{2} + 3021 N^{3}
    \nonu\\ &+&
     2543 k\, N^{3} + 516 k^{2} N^{3} + 755 N^{4} + 318 k\, 
    N^{4} + 72 N^{5}), \nonu\\
c_ {378} & = & - \frac {1} {3 (2 + N) (2 + k + N)^{5}}  (2880 + 
     5500 k + 3829 k^{2} + 1151 k^{3} + 126 k^{4} + 7964 N + 
     12406 k\, 
    N 
    \nonu\\ &+& 6686 k^{2} N + 1459 k^{3} N + 108 k^{4} N + 8281 N^{2} + 
     9707 k\, 
    N^{2} + 3487 k^{2} N^{2} + 390 k^{3} N^{2} + 4020 N^{3}
    \nonu\\ &+& 
     3065 k\, N^{3} + 516 k^{2} N^{3} + 899 N^{4} + 318 k\, 
    N^{4} + 72 N^{5}), \nonu\\
c_ {379} & = & - \frac {1} {3 (2 + N) (2 + k + N)^{5}}  (720 + 
     1336 k + 675 k^{2} + 59 k^{3} - 18 k^{4} + 2120 N + 3478 k\, 
    N 
    \nonu\\ &+& 1720 k^{2} N + 343 k^{3} N + 36 k^{4} N + 2363 N^{2} + 
     3023 k\, 
    N^{2} + 1075 k^{2} N^{2} + 126 k^{3} N^{2} + 1210 N^{3} 
    \nonu\\ &+& 
     1017 k\, N^{3} + 172 k^{2} N^{3} + 283 N^{4} + 110 k\, 
    N^{4} + 24 N^{5}), \nonu\\
c_ {380} & = &\frac {10 (k - N) (32 + 55 k + 18 k^{2} + 41 N + 35 k\, 
     N + 11 N^{2})} {3 (2 + N) (2 + k + N)^{5}}.
\nonu
\eea

The fusion rule is
\bea
[\widetilde{\Phi}_{\frac{3}{2}}^{(1),i}] \, \cdot \, [
\widetilde{\Phi}_{\frac{3}{2}}^{(1),j}] & = & [I^{ij}] + \delta^{ij} \, 
[\Phi_0^{(1)} \, \Phi_0^{(1)}] 
+[\Phi_0^{(1)} \, \Phi_{1}^{(1),ij}] +  [\Phi_{\frac{1}{2}}^{(1),i} \, 
\Phi_{\frac{1}{2}}^{(1),j}]+ \varepsilon^{ijkl} \, [\Phi_0^{(1)} \, 
\Phi_1^{(1),kl}]
\nonu \\
& + & \varepsilon^{ijkl} \, [\Phi_{\frac{1}{2}}^{(1),k} \, 
\Phi_{\frac{1}{2}}^{(1),l}]
+\delta^{ij} \, [\Phi_{\frac{1}{2}}^{(1),k} \, \widetilde{\Phi}_{\frac{3}{2}}^{(1),k}]
+\delta^{ij} \, [\Phi_0^{(1)}\, \widetilde{\Phi}_2^{(1)}]
\nonu \\
&+& \delta^{ij} \, [\Phi_0^{(2)}]+
 [\Phi_1^{(2),ij}] +\varepsilon^{ijkl} \, [\Phi_1^{(2),kl}]
+ \delta^{ij} \, [\widetilde{\Phi}_2^{(2)}].
\nonu 
\eea

The OPEs between the higher spin-$\frac{5}{2}$ currents and the
higher spin-$3$ current  are given by
\bea
&&\widetilde{\Phi}_{\frac{3}{2}}^{(1),i}(z)\:\widetilde{\Phi}_{2}^{(1)}(w)\;=\;\frac{1}{(z-w)^{5}}\,c_{1}\,\Gamma^{i}(w)
\nonu\\
&&+\frac{1}{(z-w)^{4}}\Bigg[\,c_{2}\,G^{i}+c_{3}\,U\,\Gamma^{i}+c_{4}\,\partial\Gamma^{i}+c_{5}\,T^{ij}\,\Gamma^{j}
+\varepsilon^{ijkl}\Big(c_{6}\,T^{jk}\,\Gamma^{l}+c_{7}\,\Gamma^{j}\,\Gamma^{k}\,\Gamma^{l}\Big)\Bigg](w)
\nonu\\
&&+\frac{1}{(z-w)^{3}}\Bigg[\,  c_{8}\, {\bf \Phi_{\frac{1}{2}}^{(2),i}}+
c_{9}\, {\bf \Phi_{0}^{(1)}\,\Phi_{\frac{1}{2}}^{(1),i}}+c_{10}\,L\,\Gamma^{i}+c_{11}\,\partial G^{i}+c_{12}\,G^{i}\,U+c_{13}\,\partial U\,\Gamma^{i}+c_{14}\,U\,\partial\Gamma^{i}
\nonu\\
&&+c_{15}\,U\,U\,\Gamma^{i}+c_{16}\,G^{j}\,T^{ij}+c_{17}\,\partial^{2}\Gamma^{i}+c_{18}\,(G^{j}\,\Gamma^{i}\,\Gamma^{j}+T^{ij}\,U\,\Gamma^{j})+c_{19}\,\partial T^{ij}\,\Gamma^{j}+c_{20}\,T^{ij}\,\partial\Gamma^{j}
\nonu\\
&&+c_{21}\,(\widetilde{T}^{ij}\,\widetilde{T}^{ij}\,\Gamma^{i}+T^{ij}\,T^{jk}\,\Gamma^{k})+c_{22}\,T^{jk}\,\Gamma^{i}\,\Gamma^{j}\,\Gamma^{k}+c_{23}\,\Gamma^{i}\,\partial\Gamma^{j}\,\Gamma^{j}
+\varepsilon^{ijkl}\Bigg\{\,c_{24}\,G^{j}\,T^{kl}
+c_{25}\,G^{j}\,\Gamma^{k}\,\Gamma^{l}
\nonu\\
&&+c_{26}\,\partial T^{jk}\,\Gamma^{l}+c_{27}\,T^{jk}\,\partial\Gamma^{l}+c_{28}\,T^{ij}\,T^{kl}\,\Gamma^{i}
+c_{29}\,T^{jk}\,U\,\Gamma^{l}+c_{30}\,U\,\Gamma^{j}\,\Gamma^{k}\,\Gamma^{l}+c_{31}\,\partial(\Gamma^{j}\,\Gamma^{k}\,\Gamma^{l})\Bigg\}\Bigg](w)
\nonu\\
&&+\frac{1}{(z-w)^{2}}\Bigg[\,  c_{32}\, {\bf 
\widetilde{\Phi}_{\frac{3}{2}}^{(2),i}}+c_{33}\, {\bf 
\partial\widetilde{\Phi}_{\frac{1}{2}}^{(2),i}}+c_{34}\,{\bf 
\Phi_{0}^{(1)}\,\widetilde{\Phi}_{\frac{3}{2}}^{(1),i}}+
c_{35}\, {\bf \Phi_{\frac{1}{2}}^{(1),j}\,\Phi_{1}^{(1),ij}}+
c_{36}\, {\bf \partial\Phi_{0}^{(1)}\,\Phi_{\frac{1}{2}}^{(1),i}}
\nonu\\
&& +c_{37}\,{\bf \Phi_{0}^{(1)}\,\partial\Phi_{\frac{1}{2}}^{(1),i}}
+c_{38}\,L\,G^{i}+c_{39}\,\partial L\,\Gamma^{i}+c_{40}\,L\,\partial\Gamma^{i}+c_{41}\,L\,U\,\Gamma^{i}+c_{42}\,L\,T^{ij}\,\Gamma^{j}+c_{43}\,\partial^{2}G^{i}
\nonu\\
&&+c_{44}\,G^{i}\,G^{j}\,\Gamma^{j}+c_{45}\,\partial G^{i}\,U+c_{46}\,G^{i}\,\partial U+c_{47}\,\partial G^{j}\,T^{ij}+c_{48}\,G^{j}\,\partial T^{ij}+c_{49}\,G^{i}\,\partial\Gamma^{i}\,\Gamma^{i}
\nonu\\
&&+c_{50}\,G^{i}\,T^{ij}\,T^{ij}+c_{51}\,\partial G^{j}\,\Gamma^{i}\,\Gamma^{j}+c_{52}\,G^{j}\,\partial\Gamma^{i}\,\Gamma^{j}+c_{53}\,G^{j}\,\Gamma^{i}\,\partial\Gamma^{j}+c_{54}\,G^{i}\,\partial\Gamma^{j}\,\Gamma^{j}+c_{55}\,G^{i}\,\widetilde{T}^{ij}\,\widetilde{T}^{ij}
\nonu\\
&&+c_{56}\,G^{i}\,T^{ij}\,\Gamma^{i}\,\Gamma^{j}+c_{57}\,G^{j}\,T^{ik}\,T^{kj}+c_{58}\,G^{j}\,T^{ij}\,U+c_{59}\,G^{i}\,U\,U+c_{60}\,\partial^{2}T^{ij}\,\Gamma^{j}+c_{61}\,\partial T^{ij}\,\partial\Gamma^{j}
\nonu\\
&&+c_{62}\,T^{ij}\,\partial^{2}\Gamma^{j}+c_{63}\,\partial T^{ij}\,T^{ij}\,\Gamma^{i}+c_{64}\,T^{ij}\,T^{ij}\,\partial\Gamma^{i}+c_{65}\,\partial\widetilde{T}^{ij}\,\widetilde{T}^{ij}\,\Gamma^{i}+c_{66}\,\widetilde{T}^{ij}\,\widetilde{T}^{ij}\,\partial\Gamma^{i}
\nonu\\
&&+c_{67}\,T^{ij}\,T^{ij}\,U\,\Gamma^{i}
+c_{68}\,\partial T^{ij}\,U\,\Gamma^{j}+c_{69}\,T^{ij}\,\partial U\,\Gamma^{j}+c_{70}\,T^{ij}\,U\,\partial\Gamma^{j}+c_{71}\,T^{ij}\,\partial\Gamma^{i}\,\Gamma^{i}\,\Gamma^{j}
\nonu\\
&&+c_{72}\,\partial(T^{jk}\,\Gamma^{i}\,\Gamma^{j}\,\Gamma^{k})
+c_{73}\,\partial^{2}U\,\Gamma^{i}+c_{74}\,\partial U\,\partial\Gamma^{i}+c_{75}\,U\,\partial^{2}\Gamma^{i}+c_{76}\,\partial U\,U\,\Gamma^{i}+c_{77}\,U\,U\,\partial\Gamma^{i}
\nonu\\
&&+c_{78}\,U\,U\,U\,\Gamma^{i}+c_{79}\,U\,\Gamma^{i}\,\partial\Gamma^{j}\,\Gamma^{j}+c_{80}\,\Gamma^{i}\,\partial^{2}\Gamma^{j}\,\Gamma^{j}+c_{81}\,\partial^{3}\Gamma^{i}
\nonu\\
&&+(1-\delta^{ij})\Bigg(c_{82}\,\Big[\frac{1}{2}(G^{i}\,T^{jk}\,\Gamma^{j}\,\Gamma^{k}-G^{i}\,T^{ij}\,\Gamma^{i}\,\Gamma^{j})+G^{j}\,T^{ik}\,\Gamma^{j}\,\Gamma^{k}\Big]+c_{83}\,\partial T^{ik}\,T^{kj}\,\Gamma^{j}
\nonu\\
&&+c_{84}\,T^{ik}\,\partial T^{kj}\,\Gamma^{j}+c_{85}\,T^{ik}\,T^{kj}\,\partial\Gamma^{j}+c_{86}\,T^{ik}\,T^{kj}\,U\,\Gamma^{j}+c_{87}\,T^{ik}\,\partial\Gamma^{j}\,\Gamma^{j}\,\Gamma^{k}
\nonu\\
&&+c_{88}\,(T^{jk}\,\partial\Gamma^{i}\,\Gamma^{j}\,\Gamma^{k}-T^{ij}\,\partial\Gamma^{i}\,\Gamma^{i}\,\Gamma^{j})+c_{89}\,\partial\Gamma^{i}\,\partial\Gamma^{j}\,\Gamma^{j}\Bigg)+\varepsilon^{abcd}c_{90}\,T^{iq}\,T^{ab}\,T^{cd}\,\Gamma^{q}
\nonu\\
&&+\varepsilon^{ijkl}\Bigg\{\,c_{91}\,L\,\Gamma^{j}\,\Gamma^{k}\,\Gamma^{l}+c_{92}\,L\,T^{jk}\,\Gamma^{l}+c_{93}\,G^{j}\,G^{k}\,\Gamma^{l}+c_{94}\,\partial G^{j}\,T^{kl}+c_{95}\,G^{j}\,\partial T^{kl}+c_{96}\,G^{i}\,T^{ij}\,T^{kl}
\nonu\\
&&+c_{97}\,G^{j}\,\partial(\Gamma^{k}\,\Gamma^{l})+c_{98}\,\partial G^{j}\,\Gamma^{k}\,\Gamma^{l}+c_{99}\,G^{j}\,T^{kl}\,U+c_{100}\,G^{j}\,U\,\Gamma^{k}\,\Gamma^{l}+c_{101}\,\partial^{2}T^{jk}\,\Gamma^{l}
\nonu\\
&&+c_{102}\,(\partial^{2}T^{jl}\,\Gamma^{k}+\partial^{2}T^{kl}\,\Gamma^{j})+c_{103}\,\partial T^{jk}\,\partial\Gamma^{l}+c_{104}\,T^{jk}\,\partial^{2}\Gamma^{l}+c_{105}\,(\partial T^{ij}\,T^{ik}\,\Gamma^{l}-T^{ij}\,\partial T^{ik}\,\Gamma^{l})
\nonu\\
&&+c_{106}\,\partial T^{ij}\,T^{kl}\,\Gamma^{i}+c_{107}\,T^{ij}\,\partial T^{kl}\,\Gamma^{i}+c_{108}\,T^{ij}\,T^{kl}\,\partial\Gamma^{i}+c_{109}\,(\partial T^{jk}\,T^{jl}\,\Gamma^{j}-T^{jk}\,\partial T^{jl}\,\Gamma^{j})
\nonu\\
&&+c_{110}\,\Big[T^{ij}\,\Gamma^{i}\,\partial(\Gamma^{k}\,\Gamma^{l})-\partial T^{ij}\,\Gamma^{i}\,\Gamma^{k}\,\Gamma^{l}\Big]+c_{111}\,T^{jk}\,\partial\Gamma^{j}\,\Gamma^{j}\,\Gamma^{l}+c_{112}\,\partial T^{jk}\,U\,\Gamma^{l}+c_{113}\,T^{jk}\,\partial U\,\Gamma^{l}
\nonu\\
&&+c_{114}\,T^{jk}\,U\,\partial\Gamma^{l}+c_{115}\,T^{jk}\,U\,U\,\Gamma^{l}+c_{116}\,\partial U\,\Gamma^{j}\,\Gamma^{k}\,\Gamma^{l}+c_{117}\,U\,\partial(\Gamma^{j}\,\Gamma^{k}\,\Gamma^{l})
\nonu\\
&&+c_{118}\,\partial^{2}\Gamma^{j}\,\Gamma^{k}\,\Gamma^{l}+c_{119}\,\Gamma^{j}\,\partial\Gamma^{k}\,\partial\Gamma^{l}\Bigg\}\Bigg](w)
\nonu\\
&&+\frac{1}{(z-w)}\Bigg[\,\frac{3}{7}\,\partial \Big( \mbox{pole-2}\Big)+
 c_{120}\, {\bf \partial^{2}\Phi_{\frac{1}{2}}^{(2),i}}+
c_{121}\, {\bf \partial\Phi_{0}^{(1)}\,\widetilde{\Phi}_{\frac{3}{2}}^{(1),i}}+
c_{122}\, {\bf \Phi_{0}^{(1)}\,\partial\widetilde{\Phi}_{\frac{3}{2}}^{(1),i}}+
c_{123}\, {\bf \partial^{2}(\Phi_{0}^{(1)}\,\Phi_{\frac{1}{2}}^{(1),i})}
\nonu\\
&& +c_{124}\,L\, {\bf \Phi_{\frac{1}{2}}^{(2),i}}
+c_{125}\,L\, {\bf \Phi_{0}^{(1)}\,\Phi_{\frac{1}{2}}^{(1),i}}
+c_{126}\,L\,\partial G^{i}+c_{127}\,\partial L\,G^{i}+c_{128}\,L\,G^{i}\,U+c_{129}\,L\,G^{j}\,T^{ij}
\nonu\\
&&+c_{130}\,(L\,G^{j}\,\Gamma^{i}\,\Gamma^{j}
+L\,T^{ij}\,U\,\Gamma^{j})+c_{131}\,\partial L\,T^{ij}\,\Gamma^{j}+c_{132}\,L\,\partial T^{ij}\,\Gamma^{j}+c_{133}\,L\,T^{ij}\,\partial\Gamma^{j}
\nonu\\
&&+c_{134}\,L\,\widetilde{T}^{ij}\,\widetilde{T}^{ij}\,\Gamma^{i}
+c_{135}\,L\,T^{jk}\,\Gamma^{i}\,\Gamma^{j}\,\Gamma^{k}+c_{136}\,L\,U\,U\,\Gamma^{i}+c_{137}\,L\,\partial U\,\Gamma^{i}+c_{138}\,L\,U\,\partial\Gamma^{i}
\nonu\\
&&+c_{139}\,\partial L\,U\,\Gamma^{i}+c_{140}\,L\,\Gamma^{i}\,\partial\Gamma^{j}\,\Gamma^{j}+c_{141}\,L\,\partial^{2}\Gamma^{i}+c_{142}\,\partial L\,\partial\Gamma^{i}+c_{143}\,\partial^{3}G^{i}+c_{144}\,\partial^{2}G^{j}\,\Gamma^{i}\,\Gamma^{j}
\nonu\\
&&+c_{145}\,\partial G^{i}\,\partial\Gamma^{i}\,\Gamma^{i}+c_{146}\,G^{i}\,\partial^{2}\Gamma^{i}\,\Gamma^{i}+c_{147}\,\partial G^{i}\,\widetilde{T}^{ij}\,\widetilde{T}^{ij}+c_{148}\,G^{i}\,\partial\widetilde{T}^{ij}\,\widetilde{T}^{ij}+c_{149}\,\partial^{2}G^{j}\,T^{ij}
\nonu\\
&&+c_{150}\,\partial G^{j}\,\partial T^{ij}+c_{151}\,G^{j}\,\partial^{2}T^{ij}+c_{152}\,\partial^{2}G^{i}\,U+c_{153}\,\partial G^{i}\,\partial U+c_{154}\,G^{i}\,\partial^{2}U+c_{155}\,\partial G^{i}\,U\,U
\nonu\\
&&+c_{156}\,G^{i}\,\partial U\,U+c_{157}\,\partial^{3}T^{ij}\,\Gamma^{j}+c_{158}\,\partial^{2}T^{ij}\,\partial\Gamma^{j}+c_{159}\,\partial T^{ij}\,\partial^{2}\Gamma^{j}+c_{160}\,T^{ij}\,\partial^{3}\Gamma^{j}
\nonu\\
&&+c_{161}\,\partial\widetilde{T}^{ij}\,\partial\widetilde{T}^{ij}\,\Gamma^{i}
+c_{162}\,\partial\widetilde{T}^{ij}\,\widetilde{T}^{ij}\,\partial\Gamma^{i}+c_{163}\,\widetilde{T}^{ij}\,\widetilde{T}^{ij}\,\partial^{2}\Gamma^{i}+c_{164}\,\Big[\partial^{2}T^{ij}\,U\,\Gamma^{i}+2(T^{ij}\,\partial U\,\partial\Gamma^{j})\Big]
\nonu\\
&&+c_{165}\,T^{ij}\,\partial^{2}\Gamma^{j}
+c_{166}\,T^{ij}\,U\,\partial^{2}\Gamma^{j}+c_{167}\,\partial T^{ij}\,\partial U\,\Gamma^{j}+c_{168}\,\partial T^{ij}\,U\,\Gamma^{j}+c_{169}\,\partial^{2}(T^{jk}\,\Gamma^{i}\,\Gamma^{j}\,\Gamma^{k})
\nonu\\
&&+c_{170}\,\partial^{3}U\,\Gamma^{i}
+c_{171}\,\partial^{2}U\,\partial\Gamma^{i}+c_{172}\,\partial U\,\partial^{2}\Gamma^{i}+c_{173}\,U\,\partial^{3}\Gamma^{i}+c_{174}\,\partial(\partial U\,U\,\Gamma^{i})+c_{175}\,\Big[2\, \partial U\, \partial U\,\Gamma^{i}
\nonu\\
&&+U\,U\,\partial^{2}\Gamma^{i}\Big]
+c_{176}\,\partial^{2}\Gamma^{i}\,\partial\Gamma^{i}\,\Gamma^{i}+c_{177}\,\Gamma^{i}\,\partial^{3}\Gamma^{j}\,\Gamma^{j}
+c_{178}\,\partial^{4}\Gamma^{i}
+(1-\delta^{ij})\Bigg(\,c_{179}\,L\,T^{ik}\,T^{kj}\,\Gamma^{j}
\nonu\\
&&+c_{180}\,(\partial G^{j}\,T^{ik}\,T^{kj}+G^{j}\,T^{ik}\,\partial T^{kj})+c_{181}\,G^{j}\,\partial T^{ik}\,T^{kj}
+c_{182}\,\partial G^{j}\,\Gamma^{i}\,\Gamma^{j}+c_{183}\,\partial G^{j}\,\Gamma^{i}\,\partial\Gamma^{j}
\nonu\\
&&+c_{184}\,G^{j}\,\partial^{2}\Gamma^{i}\,\Gamma^{j}+c_{185}\,G^{j}\,\partial\Gamma^{i}\,\partial\Gamma^{j}+c_{186}\,G^{j}\,\Gamma^{i}\,\partial^{2}\Gamma^{j}
+c_{187}\,\Big[\partial^{2}T^{ik}\,T^{kj}\,\Gamma^{j}+2(T^{ik}\,\partial T^{kj}\,\partial\Gamma^{j})\Big]
\nonu\\
&&+c_{188}\,T^{ik}\,\partial^{2}T^{kj}\,\Gamma^{j}+c_{189}\,T^{ik}\,T^{kj}\,\partial^{2}\Gamma^{j}
+c_{190}\,(\partial T^{ik}\,\partial T^{kj}\,\Gamma^{j}+\partial^{2}\widetilde{T}^{ij}\,\widetilde{T}^{ij}\,\Gamma^{i})+c_{191}\,\partial T^{ik}\,T^{kj}\,\partial\Gamma^{j}
\nonu\\
&&+c_{192}\,\partial\Gamma^{i}\,\partial^{2}\Gamma^{j}\,\Gamma^{j}+c_{193}\,\partial^{2}\Gamma^{i}\,\partial\Gamma^{j}\,\Gamma^{j}
+c_{194}\,\Gamma^{i}\,\partial^{2}\Gamma^{j}\,\partial\Gamma^{j}\Bigg)+\varepsilon^{ijkl}\Bigg\{\,c_{195}\,L\,G^{j}\,T^{kl}+c_{196}\,L\,G^{j}\,\Gamma^{k}\,\Gamma^{l}
\nonu\\
&&+c_{197}\,L\,T^{ij}\,T^{kl}\,\Gamma^{i}+c_{198}\,\partial L\,T^{jk}\,\Gamma^{l}+c_{199}\,L\,\partial T^{jk}\,\Gamma^{l}+c_{200}\,L\,T^{jk}\,\partial\Gamma^{l}+c_{201}\,L\,T^{jk}\,U\,\Gamma^{l}
\nonu\\
&&+c_{202}\,L\,U\,\Gamma^{j}\,\Gamma^{k}\,\Gamma^{l}+c_{203}\,\partial L\,\Gamma^{j}\,\Gamma^{k}\,\Gamma^{l}+c_{204}\,L\,\partial(\Gamma^{j}\,\Gamma^{k}\,\Gamma^{l})+c_{205}\,G^{j}\,G^{k}\,\partial\Gamma^{l}+c_{206}\,\partial(G^{j}\,G^{k})\,\Gamma^{l}
\nonu\\
&&+c_{207}\,\partial^{2}(G^{j}\,T^{kl})+c_{208}\,\partial G^{j}\,T^{kl}\,U+c_{209}\,G^{j}\,\partial(T^{kl}\,\Gamma^{l})+c_{210}\,\partial^{2}G^{j}\,\Gamma^{k}\,\Gamma^{l}
\nonu\\
&&+c_{211}\,(G^{j}\,\partial^{2}\Gamma^{k}\,\Gamma^{l}+G^{j}\,\Gamma^{k}\,\partial^{2}\Gamma^{l})+c_{212}\,G^{j}\,\partial\Gamma^{k}\,\partial\Gamma^{l}+c_{213}\,\partial G^{j}\,\partial(\Gamma^{k}\,\Gamma^{l})+c_{214}\,\partial^{3}T^{jk}\,\Gamma^{l}
\nonu\\
&&+c_{215}\,\partial^{2}T^{jk}\,\partial\Gamma^{l}+c_{216}\,\partial T^{jk}\,\partial^{2}\Gamma^{l}+c_{217}\,T^{jk}\,\partial^{3}\Gamma^{l}+c_{218}\,\partial^{2}T^{ij}\,T^{kl}\,\Gamma^{i}
\nonu\\
&&+c_{219}\,(T^{ij}\,\partial^{2}T^{kl}\,\Gamma^{i}+2\,\partial T^{ij}\,T^{kl}\,\partial\Gamma^{i})+c_{220}\,T^{ij}\,T^{kl}\,\partial^{2}\Gamma^{i}+c_{221}\,\partial T^{ij}\,\partial T^{kl}\,\Gamma^{i}+c_{222}\,T^{ij}\,\partial T^{kl}\,\partial\Gamma^{i}
\nonu\\
&&+c_{223}\,\partial^{2}T^{jk}\,T^{jl}\,\Gamma^{j}+c_{224}\,\partial T^{jk}\,T^{jl}\,\Gamma^{j}+c_{225}\,\partial^{2}T^{jk}\,U\,\Gamma^{l}+c_{226}\,T^{jk}\,\partial^{2}T^{jl}\,\Gamma^{j}+c_{227}\,T^{jk}\,T^{jl}\,\partial^{2}\Gamma^{j}
\nonu\\
&&+c_{228}\,\partial T^{jk}\,\partial U\,\Gamma^{l}+c_{229}\,\partial T^{jk}\,U\,\partial\Gamma^{l}+c_{230}\,T^{jk}\,\partial U\,\partial\Gamma^{l}+c_{231}\,\partial^{2}(U\,\Gamma^{j}\,\Gamma^{k}\,\Gamma^{l})+c_{232}\,\partial\Gamma^{j}\,\partial\Gamma^{k}\,\partial\Gamma^{l}
\nonu\\
&&+c_{233}\,(\partial\Gamma^{j}\,\Gamma^{k}\,\partial^{2}\Gamma^{l}+\partial\Gamma^{j}\,\partial^{2}\Gamma^{k}\,\Gamma^{l})+c_{234}\,\partial^{3}\Gamma^{j}\,\Gamma^{k}\,\Gamma^{l}\Bigg\}\Bigg](w)+\cdots,
\nonu
\eea
where the coefficients are
\bea
c_{1}&=&-\frac{32i\,k\,N(k-N)}{(2+k+N)^{2}(5+4k+4N+3k\,N)},\nonu\\
c_ {2} & = & - \frac {1}{3 (2 + k + N)^{2} (5 + 4 k + 4 N + 3 k\, N)}8 (45 + 216 k + 220 k^{2} + 68 k^{3} + 216 N + 
       613 k\, N 
       \nonu\\&+& 400 k^{2} N + 60 k^{3} N + 220 N^{2} + 400 k\, 
      N^{2} + 150 k^{2} N^{2} + 68 N^{3} + 60 k\, 
      N^{3}), \nonu\\
c_ {3} & = &\frac {1} {3 (2 + k + N)^{3} (5 + 4 k + 4 N + 3 k\, N)}16 i (45 + 216 k + 220 k^{2} + 68 k^{3} + 216 N + 
      478 k\, N 
      \nonu\\&+& 292 k^{2} N + 60 k^{3} N + 220 N^{2} + 292 k\, 
     N^{2} + 69 k^{2} N^{2} + 68 N^{3} + 60 k\, 
     N^{3}), \nonu\\
c_ {4} & = & - \frac {48 i (k - N) (30 + 38 k + 13 k^{2} + 38 N + 
       44 k\, N + 12 k^{2} N + 13 N^{2} + 12 k\, 
      N^{2})} {(2 + k + N)^{3} (5 + 4 k + 4 N + 3 k\, N)}, \nonu\\
c_ {5} & = &\frac {16 (k - N) (30 + 38 k + 13 k^{2} + 38 N + 44 k\, 
     N + 12 k^{2} N + 13 N^{2} + 12 k\, 
     N^{2})} {(2 + k + N)^{3} (5 + 4 k + 4 N + 3 k\, N)}, \nonu\\
c_ {6} & = & - \frac {1}{3 (2 + k + N)^{3} (5 + 4 k + 4 N + 3 k\, N)}8 (45 + 126 k + 106 k^{2} + 29 k^{3} + 126 N + 
       238 k\, N 
       \nonu\\&+& 133 k^{2} N + 24 k^{3} N + 106 N^{2} + 133 k\, 
      N^{2} + 33 k^{2} N^{2} + 29 N^{3} + 24 k\, 
      N^{3}), \nonu\\
c_ {7} & = &\frac{1}{9 (2 + k + N)^{4} (5 + 4 k + 4 N + 3 k\, N)}16 i (90 + 162 k + 98 k^{2} + 19 k^{3} + 162 N + 
      236 k\, N 
      \nonu\\&+& 107 k^{2} N + 12 k^{3} N + 98 N^{2} + 107 k\, 
     N^{2} + 30 k^{2} N^{2} + 19 N^{3} + 12 k\, 
     N^{3}), \nonu\\
c_ {8} & = &\frac {2 (k - N) (40 + 23 k + 23 N + 6 k\, 
     N)} {3 (2 + k + N) (5 + 4 k + 4 N + 3 k\, N)}, \nonu\\
c_ {9} & = & - \frac{1}{3 (2 + N) (2 + k + N)^{3} (5 + 4 k + 4 N + 
       3 k\, N)}2 (k - N) (40 + 23 k + 23 N + 6 k\, 
      N) (60 
      \nonu\\&+& 77 k + 22 k^{2} + 121 N + 115 k\, 
      N + 20 k^{2} N + 79 N^{2} + 42 k\, 
      N^{2} + 16 N^{3}), \nonu\\
c_ {10} & = & - \frac {64 i (k - N) (5 + 4 k + 
       4 N)} {3 (2 + k + N)^{2} (5 + 4 k + 4 N + 3 k\, N)}, \nonu\\
c_ {11} & = & - \frac {1}{9 (2 + N) (2 + k + N)^{3} (5 + 4 k + 4 N + 3 k\, N)}2 (-2160 - 408 k + 3692 k^{2} + 3383 k^{3} 
\nonu\\&+&  866 k^{4} - 6288 N - 460 k\, 
      N + 6179 k^{2} N + 3331 k^{3} N + 376 k^{4} N - 3592 N^{2} + 
       6401 k\, 
      N^{2} 
      \nonu\\&+&8407 k^{2} N^{2} + 2328 k^{3} N^{2} + 120 k^{4} N^{2} + 
       2685 N^{3} + 9065 k\, 
      N^{3} + 5308 k^{2} N^{3} + 708 k^{3} N^{3} 
      \nonu\\&+& 2883 N^{4} + 
       3588 k\, N^{4} + 996 k^{2} N^{4} + 640 N^{5} + 336 k\, 
      N^{5}), \nonu\\
c_ {12} & = & - \frac {8 (k - N) (-10 + k + N + 12 k\, 
      N)} {3 (2 + k + N)^{2} (5 + 4 k + 4 N + 3 k\, N)}, \nonu\\
c_ {13} & = &\frac {1}{9 (2 + N) (2 + k + N)^{4} (5 + 4 k + 4 N + 3 k\, N)}4 i (-2160 + 1752 k + 6500 k^{2} + 4247 k^{3}
\nonu\\&+&  866 k^{4} - 4128 N + 5516 k\, 
     N + 10535 k^{2} N + 4123 k^{3} N + 376 k^{4} N + 296 N^{2} + 
      11801 k\, 
     N^{2} 
     \nonu\\&+&10171 k^{2} N^{2} + 2292 k^{3} N^{2} + 120 k^{4} N^{2} + 
      4953 N^{3} + 10901 k\, 
     N^{3} + 5236 k^{2} N^{3} + 600 k^{3} N^{3} 
     \nonu\\&+& 3315 N^{4} + 
      3768 k\, N^{4} + 888 k^{2} N^{4} + 640 N^{5} + 336 k\, 
     N^{5}), \nonu\\
c_ {14} & = &\frac{1}{9 (2 + k + N)^{4} (5 + 4 k + 4 N + 3 k\, N)}8 i (540 + 42 k - 868 k^{2} - 565 k^{3} - 
      94 k^{4} + 2442 N 
      \nonu\\&+& 2132 k\, 
     N + 10 k^{2} N - 75 k^{3} N + 60 k^{4} N + 3272 N^{2} + 2917 k\, 
     N^{2} + 380 k^{2} N^{2} - 24 k^{3} N^{2} 
     \nonu\\&+& 1742 N^{3} + 1305 k\, 
     N^{3} + 120 k^{2} N^{3} + 320 N^{4} + 168 k\, 
     N^{4}), \nonu\\
c_ {15} & = & - \frac {80 i (k - N)} {(2 + k + N)^{2} (5 + 4 k + 
       4 N + 3 k\, N)}, \nonu\\
c_ {16} & = &\frac{1}{3 (2 + N) (2 + k + N)^{3} (5 + 4 k + 4 N + 3 k\, N)}2 i (-480 - 304 k + 484 k^{2} + 531 k^{3} + 
      138 k^{4} 
      \nonu\\&-& 2144 N - 2312 k\, 
     N - 441 k^{2} N + 185 k^{3} N + 36 k^{4} N - 2828 N^{2} - 
      2811 k\, 
     N^{2} - 745 k^{2} N^{2} 
     \nonu\\&-&30 k^{3} N^{2} - 1455 N^{3} - 1097 k\, 
     N^{3} - 180 k^{2} N^{3} - 257 N^{4} - 114 k\, 
     N^{4}), \nonu\\
c_ {17} & = & - \frac {16 i (k - N) (130 + 217 k + 85 k^{2} + 217 N + 
       248 k\, N + 57 k^{2} N + 85 N^{2} + 57 k\, 
      N^{2})} {3 (2 + k + N)^{3} (5 + 4 k + 4 N + 3 k\, N)}, \nonu\\
c_ {18} & = &\frac {1} {3 (2 + N) (2 + k + N)^{4} (5 + 4 k + 4 N + 3 k\, N)}4 (-480 - 64 k + 796 k^{2} + 627 k^{3} + 
      138 k^{4} 
      \nonu\\&-& 1904 N - 1568 k\, 
     N + 147 k^{2} N + 305 k^{3} N + 36 k^{4} N - 2396 N^{2} - 
      2067 k\, 
     N^{2} - 385 k^{2} N^{2} 
      \nonu\\&+& 6 k^{3} N^{2} - 1203 N^{3} - 809 k\, 
     N^{3} - 108 k^{2} N^{3} - 209 N^{4} - 78 k\, 
     N^{4}),\nonu\\ 
c_ {19} & = &\frac {16 (k - N) (-20 + 26 k + 21 k^{2} + 26 N + 66 k\, 
     N + 18 k^{2} N + 21 N^{2} + 18 k\, 
     N^{2})} {3 (2 + k + N)^{3} (5 + 4 k + 4 N + 3 k\, N)}, \nonu\\
c_ {20} & = & - \frac {16 (k - N) (-10 - 2 k + 3 k^{2} - 2 N + 
       3 N^{2})} {3 (2 + k + N)^{3} (5 + 4 k + 4 N + 3 k\, N)}, \nonu\\
c_ {21} & = & - \frac{1}{3 (2 + N) (2 + k + N)^{4} (5 + 4 k + 4 N + 3 k\, N)}4 i (k - N) (560 + 1104 k + 691 k^{2} + 
       138 k^{3} 
       \nonu\\&+& 984 N + 1372 k\, 
      N + 523 k^{2} N + 36 k^{3} N + 553 N^{2} + 448 k\, 
      N^{2} + 66 k^{2} N^{2} + 99 N^{3} + 18 k\, 
      N^{3}), \nonu\\
c_ {22} & = & - \frac {4 (k - N) (40 + 23 k + 23 N + 6 k\, 
      N) (32 + 55 k + 18 k^{2} + 41 N + 35 k\, 
      N + 11 N^{2})} {3 (2 + N) (2 + k + N)^{5} (5 + 4 k + 4 N + 
       3 k\, N)}, \nonu\\
c_ {23} & = & - \frac {1}{3 (2 + N) (2 + k + N)^{5} (5 + 4 k + 4 N + 3 k\, N)}16 i (k - N) (320 + 1096 k + 875 k^{2} + 
       202 k^{3} 
       \nonu\\&+& 856 N + 1880 k\, 
      N + 975 k^{2} N + 116 k^{3} N + 733 N^{2} + 1064 k\, 
      N^{2} + 342 k^{2} N^{2} + 24 k^{3} N^{2} 
       \nonu\\&+& 255 N^{3} + 246 k\, 
      N^{3} + 48 k^{2} N^{3} + 32 N^{4} + 24 k\, 
      N^{4}), \nonu\\
c_ {24} & = &\frac { i (k - N) (40 + 23 k + 23 N + 6 k\, 
     N) (16 + 21 k + 6 k^{2} + 19 N + 13 k\, 
     N + 5 N^{2})} {3 (2 + N) (2 + k + N)^{3} (5 + 4 k + 4 N + 3 k\, 
     N)},\nonu\\
 c_ {25} & = &\frac{1}{3 (2 + N) (2 + k + N)^{4} (5 + 4 k + 4 N + 3 k\, N)}2 (k - N) (80 + 624 k + 571 k^{2} + 
      138 k^{3} 
      \nonu\\&+& 264 N + 892 k\, 
     N + 463 k^{2} N + 36 k^{3} N + 193 N^{2} + 328 k\, 
     N^{2} + 66 k^{2} N^{2} + 39 N^{3} + 18 k\, 
     N^{3}), \nonu\\
c_ {26} & = &\frac {1}{9 (2 + N) (2 + k + N)^{4} (5 + 4 k + 4 N + 3 k\, N)}4 (360 + 468 k + 392 k^{2} + 185 k^{3} + 
      26 k^{4} 
      \nonu\\&+& 648 N + 1082 k\, 
     N + 1721 k^{2} N + 1096 k^{3} N + 208 k^{4} N - 574 N^{2} - 
      721 k\, N^{2} + 637 k^{2} N^{2} 
       \nonu\\&+& 519 k^{3} N^{2} + 
      48 k^{4} N^{2} - 1689 N^{3} - 1876 k\, 
     N^{3} - 368 k^{2} N^{3} + 24 k^{3} N^{3} - 1035 N^{4} - 777 k\, 
     N^{4} 
      \nonu\\&-& 120 k^{2} N^{4} - 194 N^{5} - 60 k\, 
     N^{5}), \nonu\\
c_ {27} & = & - \frac {1}{9 (2 + k + N)^{4} (5 + 4 k + 4 N + 3 k\, N)}4 (540 + 402 k - 328 k^{2} - 349 k^{3} - 
       76 k^{4} + 2802 N 
       \nonu\\&+& 3644 k\, 
      N + 1522 k^{2} N + 321 k^{3} N + 60 k^{4} N + 3812 N^{2} + 
       4429 k\, 
      N^{2} + 1568 k^{2} N^{2} + 192 k^{3} N^{2}
       \nonu\\&+& 1958 N^{3} + 
       1701 k\, N^{3} + 336 k^{2} N^{3} + 338 N^{4} + 168 k\, 
      N^{4}), \nonu\\
c_ {28} & = & - \frac {1}{3 (2 + N) (2 + k + N)^{4} (5 + 4 k + 4 N + 3 k\, N)}2 i (-480 - 64 k + 796 k^{2} + 627 k^{3} + 
       138 k^{4} 
       \nonu\\&-& 1904 N - 1568 k\, 
      N + 147 k^{2} N + 305 k^{3} N + 36 k^{4} N - 2396 N^{2} - 
       2067 k\, 
      N^{2} - 385 k^{2} N^{2} 
       \nonu\\&+&6 k^{3} N^{2} - 1203 N^{3} - 809 k\, 
      N^{3} - 108 k^{2} N^{3} - 209 N^{4} - 78 k\, 
      N^{4}), \nonu\\
c_ {29} & = &\frac {1}{3 (2 + N) (2 + k + N)^{4} (5 + 4 k + 4 N + 3 k\, N)}2 (k - N) (1040 + 1584 k + 811 k^{2} + 
      138 k^{3} 
      \nonu\\&+& 1704 N + 1852 k\, 
     N + 583 k^{2} N + 36 k^{3} N + 913 N^{2} + 568 k\, 
     N^{2} + 66 k^{2} N^{2} + 159 N^{3} + 18 k\, 
     N^{3}), \nonu\\
c_ {30} & = & - \frac {4 i (k - N) (40 + 23 k + 23 N + 6 k\, 
      N) (32 + 55 k + 18 k^{2} + 41 N + 35 k\, 
      N + 11 N^{2})} {9 (2 + N) (2 + k + N)^{5} (5 + 4 k + 4 N + 
       3 k\, N)}, \nonu\\
c_ {31} & = & - \frac {1}{27 (2 + N) (2 + k + N)^{5} (5 + 4 k + 4 N + 3 k\, N)}8 i (-720 + 744 k + 3028 k^{2} + 2107 k^{3} 
\nonu\\&+& 
       430 k^{4} - 4416 N - 2252 k\, 
      N + 3955 k^{2} N + 3035 k^{3} N + 506 k^{4} N - 8480 N^{2} - 
       7355 k\, 
      N^{2}
       \nonu\\&+& 263 k^{2} N^{2} + 1056 k^{3} N^{2} + 96 k^{4} N^{2} - 
       7203 N^{3} - 6107 k\, 
      N^{3} - 1024 k^{2} N^{3} + 48 k^{3} N^{3} 
       \nonu\\&-& 2805 N^{4} - 
       1788 k\, N^{4} - 240 k^{2} N^{4} - 406 N^{5} - 120 k\, 
      N^{5}), \nonu\\
c_ {32} & = &\frac {7} {2}, \nonu\\
c_ {33} & = &\frac{1}{60 (2 + N) (2 + k + N)^{2} (5 + 4 k + 4 N + 3 k\, N)}(31500 + 69145 k + 47602 k^{2} + 10216 k^{3}
\nonu\\&+&
     85205 N + 164195 k\, 
    N + 95511 k^{2} N + 16010 k^{3} N + 86823 N^{2} + 140285 k\, 
    N^{2} + 62945 k^{2} N^{2} 
     \nonu\\&+& 6396 k^{3} N^{2} + 38748 N^{3} + 
     48373 k\, N^{3} + 13230 k^{2} N^{3} + 6232 N^{4} + 4944 k\, 
    N^{4}), \nonu\\
c_ {34} & = & - \frac {1}{2 (2 + N) (2 + k + N)^{2} (5 + 4 k + 4 N + 3 k\, N)}(2100 + 4055 k + 2606 k^{2} + 536 k^{3} + 
      6235 N 
      \nonu\\&+& 10669 k\, 
     N + 5913 k^{2} N + 982 k^{3} N + 6633 N^{2} + 9523 k\, 
     N^{2} + 4111 k^{2} N^{2} + 420 k^{3} N^{2} 
      \nonu\\&+& 3012 N^{3} + 
      3323 k\, N^{3} + 882 k^{2} N^{3} + 488 N^{4} + 336 k\, 
     N^{4}), \nonu\\
c_ {35} & = &\frac {7 (60 + 77 k + 22 k^{2} + 121 N + 115 k\, 
     N + 20 k^{2} N + 79 N^{2} + 42 k\, 
     N^{2} + 16 N^{3})} {2 (2 + N) (2 + k + N)^{2}}, \nonu\\
c_ {36} & = & - \frac {1}{12 (2 + N)^{2} (2 + k + N)^{4} (5 + 4 k + 4 N + 3 k\, 
      N)}(60 + 77 k + 22 k^{2} + 121 N + 115 k\, 
      N 
      \nonu\\&+& 20 k^{2} N + 79 N^{2} + 42 k\, 
      N^{2} + 16 N^{3}) (6300 + 13661 k + 9302 k^{2} + 1976 k^{3} + 
       17209 N 
        \nonu\\&+& 32755 k\, 
      N + 18825 k^{2} N + 3118 k^{3} N + 17667 N^{2} + 28225 k\, 
      N^{2} + 12505 k^{2} N^{2} + 1254 k^{3} N^{2} 
       \nonu\\&+& 7926 N^{3} + 
       9809 k\, N^{3} + 2646 k^{2} N^{3} + 1280 N^{4} + 1014 k\, 
      N^{4}),\nonu\\
 c_ {37} & = & - \frac{1}{4 (2 + N)^{2} (2 + k + N)^{4} (5 + 4 k + 4 N + 3 k\, 
      N)}3 (60 + 77 k + 22 k^{2} + 121 N + 
       115 k\, N 
       \nonu\\&+& 20 k^{2} N + 79 N^{2} + 42 k\, 
      N^{2} + 16 N^{3}) (700 + 1549 k + 1074 k^{2} + 232 k^{3} + 
       1881 N + 3655 k\, 
      N 
       \nonu\\&+& 2143 k^{2} N + 362 k^{3} N + 1907 N^{2} + 3105 k\, 
      N^{2} + 1405 k^{2} N^{2} + 144 k^{3} N^{2} + 848 N^{3} + 
       1065 k\, N^{3} 
        \nonu\\&+& 294 k^{2} N^{3} + 136 N^{4} + 108 k\, 
      N^{4}),\nonu\\
 c_ {38} & = &\frac {1}{3 (2 + N) (2 + k + N)^{2} (5 + 4 k + 4 N + 
      3 k\, N)}(-3120 - 8151 k - 6158 k^{2} - 
     1384 k^{3} 
      \nonu\\&-& 6561 N - 13775 k\, 
    N - 7797 k^{2} N - 1070 k^{3} N - 4091 N^{2} - 6591 k\, 
    N^{2} - 2233 k^{2} N^{2} - 620 N^{3} 
     \nonu\\&-& 721 k\, 
    N^{3} + 64 N^{4}), \nonu\\
c_ {39} & = &\frac {7 i (420 + 553 k + 158 k^{2} + 893 N + 881 k\, 
     N + 160 k^{2} N + 611 N^{2} + 336 k\, 
     N^{2} + 128 N^{3})} {3 (2 + N) (2 + k + N)^{3}}, \nonu\\
c_ {40} & = &\frac {1}{(2 + N) (2 + k + N)^{3} (5 + 4 k + 4 N + 3 k\, N)}2 i (700 + 1195 k + 790 k^{2} + 168 k^{3} + 
      2935 N 
      \nonu\\&+& 4843 k\, 
     N + 2873 k^{2} N + 518 k^{3} N + 3733 N^{2} + 5285 k\, 
     N^{2} + 2457 k^{2} N^{2} + 280 k^{3} N^{2} 
      \nonu\\&+& 1908 N^{3} + 
      2079 k\, N^{3} + 588 k^{2} N^{3} + 336 N^{4} + 224 k\, 
     N^{4}), \nonu\\
c_ {41} & = & - \frac{1}{3 (2 + N) (2 + k + N)^{3} (5 + 4 k + 4 N + 
       3 k\, N)}2 i (-2880 - 7839 k - 6062 k^{2} - 
       1384 k^{3} 
       \nonu\\&-& 6129 N - 13283 k\, 
      N - 7677 k^{2} N - 1070 k^{3} N - 3839 N^{2} - 6351 k\, 
      N^{2} - 2197 k^{2} N^{2} - 572 N^{3} 
       \nonu\\&-& 685 k\, 
      N^{3} + 64 N^{4}), \nonu\\
c_ {42} & = &\frac {1}{(2 + N) (2 + k + N)^{3} (5 + 4 k + 4 N + 
      3 k\, N)}2 (700 + 1295 k + 734 k^{2} + 136 k^{3} + 
      1435 N 
      \nonu\\&+& 2163 k\, 
     N + 909 k^{2} N + 110 k^{3} N + 1009 N^{2} + 1117 k\, 
     N^{2} + 257 k^{2} N^{2} + 260 N^{3} + 163 k\, 
     N^{3} 
      \nonu\\&+& 16 N^{4}), \nonu\\
c_ {43} & = &\frac {1}{36 (2 + N)^{2} (2 + k + N)^{4} (5 + 4 k + 4 N + 3 k\, 
     N)}(-118800 - 433920 k - 661969 k^{2} 
     \nonu\\&-& 
     520992 k^{3} - 205572 k^{4} - 31664 k^{5} - 337200 N - 
     967774 k\, 
    N - 1079260 k^{2} N - 598983 k^{3} N 
     \nonu\\&-&159116 k^{4} N - 
     13580 k^{5} N - 110821 N^{2} + 209528 k\, 
    N^{2} + 899351 k^{2} N^{2} + 830134 k^{3} N^{2}
     \nonu\\&+& 
     296788 k^{4} N^{2} + 37600 k^{5} N^{2} + 637388 N^{3} + 
     2386943 k\, 
    N^{3} + 3060532 k^{2} N^{3} + 1653529 k^{3} N^{3} 
     \nonu\\&+& 
     363190 k^{4} N^{3} + 23340 k^{5} N^{3} + 990857 N^{4} + 
     2653606 k\, 
    N^{4} + 2433814 k^{2} N^{4} + 879964 k^{3} N^{4} 
     \nonu\\&+& 
     100782 k^{4} N^{4} + 625324 N^{5} + 1253557 k\, 
    N^{5} + 782542 k^{2} N^{5} + 148368 k^{3} N^{5} + 184744 N^{6} 
     \nonu\\&+&
     258088 k\, N^{6} + 85410 k^{2} N^{6} + 20992 N^{7} + 16752 k\, 
    N^{7}), \nonu\\
c_ {44} & = & - \frac {7 i (20 + 21 k + 6 k^{2} + 25 N + 
       13 k\, N + 7 N^{2})} {(2 + N) (2 + k + N)^{3}}, \nonu\\
c_ {45} & = & - \frac{1}{(2 + N) (2 + k + N)^{3} (5 + 4 k + 4 N + 3 k\, N)}(-700 - 897 k - 258 k^{2} - 8 k^{3} - 1133 N 
\nonu\\&-& 
      599 k\, N + 665 k^{2} N + 278 k^{3} N - 319 N^{2} + 599 k\, 
     N^{2} + 1027 k^{2} N^{2} + 204 k^{3} N^{2} + 228 N^{3} 
      \nonu\\&+& 483 k\, 
     N^{3} + 294 k^{2} N^{3} + 88 N^{4} + 48 k\, 
     N^{4}), \nonu\\
c_ {46} & = & - \frac{1} {3 (2 + N) (2 + k + N)^{3} (5 + 4 k + 4 N + 3 k\, N)}(6300 + 9799 k + 4942 k^{2} + 856 k^{3} 
\nonu\\&+&
      14771 N + 18539 k\, 
     N + 6213 k^{2} N + 422 k^{3} N + 11673 N^{2} + 11189 k\, 
     N^{2} + 1745 k^{2} N^{2} 
      \nonu\\&-& 192 k^{3} N^{2} + 3540 N^{3} + 
      2419 k\, N^{3} + 328 N^{4} + 192 k\, 
     N^{4}), \nonu\\
c_ {47} & = &\frac {1}{12 (2 + N)^{2} (2 + k + N)^{4} (5 + 4 k + 4 N + 3 k\, 
     N)}i (18480 + 65584 k + 128425 k^{2} + 
      128920 k^{3} 
      \nonu\\&+& 59124 k^{4} + 9840 k^{5} + 155024 N + 506074 k\, 
     N + 693328 k^{2} N + 470271 k^{3} N + 147668 k^{4} N 
      \nonu\\&+& 
      16188 k^{5} N + 384313 N^{2} + 1077426 k\, 
     N^{2} + 1146203 k^{2} N^{2} + 561972 k^{3} N^{2} + 
      116776 k^{4} N^{2}
       \nonu\\&+& 6768 k^{5} N^{2} + 434986 N^{3} + 
      1008693 k\, 
     N^{3} + 815384 k^{2} N^{3} + 269699 k^{3} N^{3} + 
      28956 k^{4} N^{3} 
       \nonu\\&+& 250789 N^{4} + 457896 k\, 
     N^{4} + 257130 k^{2} N^{4} + 44382 k^{3} N^{4} + 71600 N^{5} + 
      94531 k\, N^{5} 
       \nonu\\&+& 28590 k^{2} N^{5} + 8024 N^{6} + 6396 k\, 
     N^{6}), \nonu\\
c_ {48} & = &\frac {1}{12 (2 + N)^{2} (2 + k + N)^{4} (5 + 4 k + 4 N + 3 k\, 
     N)}i (191280 + 648304 k + 821569 k^{2} 
     \nonu\\&+& 
      493348 k^{3} + 142188 k^{4} + 15888 k^{5} + 708944 N + 
      2099962 k\, 
     N + 2299648 k^{2} N 
      \nonu\\&+& 1165497 k^{3} N + 274004 k^{4} N + 
      23748 k^{5} N + 1083457 N^{2} + 2761638 k\, 
     N^{2} + 2534345 k^{2} N^{2} 
      \nonu\\&+& 1024170 k^{3} N^{2} + 
      175774 k^{4} N^{2} + 9036 k^{5} N^{2} + 874426 N^{3} + 
      1867899 k\, 
     N^{3} + 1362530 k^{2} N^{3} 
      \nonu\\&+& 393605 k^{3} N^{3} + 
      37254 k^{4} N^{3} + 392767 N^{4} + 672306 k\, 
     N^{4} + 351612 k^{2} N^{4} + 54912 k^{3} N^{4} 
      \nonu\\&+& 92870 N^{5} + 
      117529 k\, N^{5} + 33828 k^{2} N^{5} + 9008 N^{6} + 7134 k\, 
     N^{6}),\nonu\\
 c_ {49} & = &\frac {1} {(2 + N) (2 + k + N)^{4}} 3 (512 + 
    1016 k + 443 k^{2} + 42 k^{3} + 712 N + 792 k\, 
   N + 113 k^{2} N 
   \nonu\\&+& 237 N^{2} + 86 k\, N^{2} + 15 N^{3}),\nonu\\ 
c_ {50} & = &\frac {7 (32 + 55 k + 18 k^{2} + 41 N + 35 k\, 
     N + 11 N^{2})} {(2 + N) (2 + k + N)^{3}}, \nonu\\
c_ {51} & = &\frac {1}{6 (2 + N)^{2} (2 + k + N)^{5} (5 + 4 k + 4 N + 3 k\, N)}(56880 + 186064 k + 259513 k^{2} + 188344 k^{3} \nonu\\&+&  68724 k^{4} + 9840 k^{5} + 280304 N + 836170 k\, 
    N + 991528 k^{2} N + 577791 k^{3} N + 160436 k^{4} N 
     \nonu\\&+& 
     16188 k^{5} N + 546361 N^{2} + 1428162 k\, 
    N^{2} + 1397795 k^{2} N^{2} + 627708 k^{3} N^{2} + 
     121336 k^{4} N^{2} 
      \nonu\\&+& 6768 k^{5} N^{2} + 540706 N^{3} + 
     1190373 k\, 
    N^{3} + 911528 k^{2} N^{3} + 284843 k^{3} N^{3} + 
     29244 k^{4} N^{3} 
      \nonu\\&+& 287149 N^{4} + 505200 k\, 
    N^{4} + 273114 k^{2} N^{4} + 45246 k^{3} N^{4} + 77744 N^{5} + 
     100315 k\, N^{5} 
      \nonu\\&+& 29454 k^{2} N^{5} + 8408 N^{6} + 6684 k\, 
    N^{6}),\nonu\\
 c_ {52} & = &\frac{1}{6 (2 + N)^{2} (2 + k + N)^{5} (5 + 4 k + 4 N + 3 k\, N)}(172080 + 588784 k + 756481 k^{2} + 
     463300 k^{3}
     \nonu\\&+& 137196 k^{4} + 15888 k^{5} + 630224 N + 
     1898746 k\, 
    N + 2117920 k^{2} N + 1097625 k^{3} N 
     \nonu\\&+& 265268 k^{4} N + 
     23748 k^{5} N + 954049 N^{2} + 2488422 k\, 
    N^{2} + 2335769 k^{2} N^{2} + 967602 k^{3} N^{2}
     \nonu\\&+& 
     170782 k^{4} N^{2} + 9036 k^{5} N^{2} + 764410 N^{3} + 
     1678011 k\, 
    N^{3} + 1256474 k^{2} N^{3} + 372917 k^{3} N^{3} 
     \nonu\\&+& 
     36318 k^{4} N^{3} + 341455 N^{4} + 601578 k\, 
    N^{4} + 323964 k^{2} N^{4} + 52104 k^{3} N^{4} + 80366 N^{5}
     \nonu\\&+& 
     104329 k\, N^{5} + 31020 k^{2} N^{5} + 7760 N^{6} + 6198 k\, 
    N^{6}),\nonu\\
 c_ {53} & = &\frac {1} {6 (2 + N)^{2} (2 + k + N)^{5} (5 + 4 k + 4 N + 3 k\, N)}(149040 + 502192 k + 654433 k^{2} + 
     415684 k^{3} 
     \nonu\\&+& 129516 k^{4} + 15888 k^{5} + 587792 N + 
     1745434 k\, 
    N + 1962160 k^{2} N + 1042521 k^{3} N 
     \nonu\\&+& 259892 k^{4} N + 
     23748 k^{5} N + 957889 N^{2} + 2458566 k\, 
    N^{2} + 2307497 k^{2} N^{2} + 968562 k^{3} N^{2}
     \nonu\\&+& 
     173182 k^{4} N^{2} + 9036 k^{5} N^{2} + 822922 N^{3} + 
     1771995 k\, 
    N^{3} + 1314602 k^{2} N^{3} + 390725 k^{3} N^{3}
     \nonu\\&+& 
     37902 k^{4} N^{3} + 391519 N^{4} + 674442 k\, 
    N^{4} + 356508 k^{2} N^{4} + 56856 k^{3} N^{4} + 97406 N^{5}
     \nonu\\&+&
     123577 k\, N^{5} + 35772 k^{2} N^{5} + 9872 N^{6} + 7782 k\, 
    N^{6}),\nonu\\
 c_ {54} & = &\frac {1} {(2 + N) (2 + k + N)^{4}} 7 (192 + 368 k + 
    167 k^{2} + 18 k^{3} + 304 N + 336 k\, 
   N + 61 k^{2} N 
   \nonu\\&+& 137 N^{2} + 62 k\, N^{2} + 19 N^{3}), \nonu\\
c_ {55} & = &\frac {(80 + 131 k + 42 k^{2} + 101 N + 83 k\, 
    N + 27 N^{2})} {(2 + N) (2 + k + N)^{3}}, \nonu\\
c_ {56} & = & - \frac {28 i (32 + 55 k + 18 k^{2} + 41 N + 35 k\, 
      N + 11 N^{2})} {(2 + N) (2 + k + N)^{4}}, \nonu\\
c_ {57} & = & - \frac {2 (72 + 127 k + 42 k^{2} + 93 N + 81 k\, 
      N + 25 N^{2})} {(2 + N) (2 + k + N)^{3}}, \nonu\\
c_ {58} & = & - \frac {7 i (20 + 21 k + 6 k^{2} + 25 N + 13 k\, 
      N + 7 N^{2})} {(2 + N) (2 + k + N)^{3}}, \nonu\\
c_ {59} & = & - \frac {3 (80 + 131 k + 42 k^{2} + 101 N + 83 k\, 
      N + 27 N^{2})} {(2 + N) (2 + k + N)^{3}}, \nonu\\
c_ {60} & = &\frac{1}{3 (2 + N) (2 + k + N)^{4} (5 + 4 k + 4 N + 3 k\, N)}2 (4200 + 8808 k + 6989 k^{2} + 2330 k^{3} + 
      272 k^{4} 
      \nonu\\&+& 13872 N + 26832 k\, 
     N + 19308 k^{2} N + 5657 k^{3} N + 574 k^{4} N + 16285 N^{2} + 
      27751 k\, 
     N^{2} 
      \nonu\\&+& 16921 k^{2} N^{2} + 3860 k^{3} N^{2} + 282 k^{4} N^{2} + 
      8487 N^{3} + 12082 k\, 
     N^{3} + 5589 k^{2} N^{3} + 723 k^{3} N^{3}
      \nonu\\&+& 1860 N^{4} + 
      2041 k\, N^{4} + 537 k^{2} N^{4} + 116 N^{5} + 96 k\, 
     N^{5}), \nonu\\
c_ {61} & = &\frac {1}{(2 + N) (2 + k + N)^{4} (5 + 4 k + 4 N + 3 k\, N)}(15400 + 30942 k + 21057 k^{2} + 5194 k^{3}
\nonu\\&+& 
     280 k^{4} + 44518 N + 77792 k\, 
    N + 43537 k^{2} N + 7603 k^{3} N + 74 k^{4} N + 47143 N^{2} + 
     68586 k\, 
    N^{2} 
     \nonu\\&+& 28756 k^{2} N^{2} + 2605 k^{3} N^{2} - 96 k^{4} N^{2} + 
     21899 N^{3} + 24709 k\, 
    N^{3} + 6156 k^{2} N^{3} - 96 k^{3} N^{3} 
     \nonu\\&+& 3948 N^{4} + 3065 k\, 
    N^{4} + 96 k^{2} N^{4} + 112 N^{5} + 96 k\, 
    N^{5}), \nonu\\
c_ {62} & = & - \frac{1}{3 (2 + N) (2 + k + N)^{4} (5 + 4 k + 4 N + 3 k\, N)}(-25200 - 51460 k - 30817 k^{2} - 3250 k^{3}
 \nonu\\&+& 
      1080 k^{4} - 72020 N - 132578 k\, 
     N - 72435 k^{2} N - 9179 k^{3} N + 630 k^{4} N - 75501 N^{2} 
      \nonu\\&-& 
      118712 k\, 
     N^{2} - 53162 k^{2} N^{2} - 5325 k^{3} N^{2} - 144 k^{4} N^{2} - 
      33411 N^{3} - 40903 k\, 
     N^{3} 
      \nonu\\&-& 11808 k^{2} N^{3} - 144 k^{3} N^{3} - 4684 N^{4} - 
      3369 k\, N^{4} + 144 k^{2} N^{4} + 216 N^{5} + 144 k\, 
     N^{5}), \nonu\\
c_ {63} & = & - \frac {7 i (200 + 246 k + 55 k^{2} - 6 k^{3} + 
       414 N + 388 k\, N + 61 k^{2} N + 277 N^{2} + 148 k\, 
      N^{2} + 57 N^{3})} {(2 + N) (2 + k + N)^{4}}, \nonu\\
c_ {64} & = & - \frac {14 i (20 + 21 k + 6 k^{2} + 25 N + 13 k\, 
      N + 7 N^{2})} {(2 + N) (2 + k + N)^{3}}, \nonu\\
c_ {65} & = & - \frac{1}{3 (2 + N)^{2} (2 + k + N)^{5} (5 + 4 k + 4 N + 3 k\, 
      N)}i (16 + 21 k + 6 k^{2} + 19 N + 13 k\, 
      N + 5 N^{2})
      \nonu\\&& (6300 + 13661 k + 9302 k^{2} + 1976 k^{3} + 
       17209 N + 32755 k\, 
      N + 18825 k^{2} N + 3118 k^{3} N 
       \nonu\\&+& 17667 N^{2} + 28225 k\, 
      N^{2} + 12505 k^{2} N^{2} + 1254 k^{3} N^{2} + 7926 N^{3} + 
       9809 k\, N^{3} + 2646 k^{2} N^{3} 
        \nonu\\&+& 1280 N^{4} + 1014 k\, 
      N^{4}), \nonu\\
c_ {66} & = & - \frac {1}{6 (2 + N)^{2} (2 + k + N)^{5} (5 + 4 k + 4 N + 3 k\, 
      N)}i (100800 + 359196 k + 
       499409 k^{2} 
       \nonu\\&+& 336896 k^{3} + 109892 k^{4} + 13872 k^{5} + 
       386724 N + 1226848 k\, 
      N + 1486600 k^{2} N + 848643 k^{3} N
       \nonu\\&+& 223828 k^{4} N + 
       21228 k^{5} N + 615327 N^{2} + 1700210 k\, 
      N^{2} + 1735317 k^{2} N^{2} + 792108 k^{3} N^{2} 
       \nonu\\&+& 
       151500 k^{4} N^{2} + 8280 k^{5} N^{2} + 517366 N^{3} + 
       1203889 k\, 
      N^{3} + 977768 k^{2} N^{3} + 320275 k^{3} N^{3} 
       \nonu\\&+& 
       33624 k^{4} N^{3} + 241107 N^{4} + 447660 k\, 
      N^{4} + 259768 k^{2} N^{4} + 46434 k^{3} N^{4} + 58716 N^{5} 
       \nonu\\&+& 
       79061 k\, N^{5} + 24966 k^{2} N^{5} + 5816 N^{6} + 4632 k\, 
      N^{6}), \nonu\\
c_ {67} & = & - \frac {14 i (32 + 55 k + 18 k^{2} + 41 N + 
       35 k\, N + 11 N^{2})} {(2 + N) (2 + k + N)^{4}}, \nonu\\
c_ {68} & = &\frac {1} {6 (2 + N)^{2} (2 + k + N)^{5} (5 + 4 k + 4 N + 3 k\, N)}(241200 + 791440 k + 968209 k^{2} + 
     557092 k^{3}
      \nonu\\&+& 152172 k^{4} + 15888 k^{5} + 868400 N + 
     2490730 k\, 
    N + 2634424 k^{2} N + 1281753 k^{3} N 
     \nonu\\&+& 287444 k^{4} N + 
     23748 k^{5} N + 1286593 N^{2} + 3177078 k\, 
    N^{2} + 2818913 k^{2} N^{2} + 1096266 k^{3} N^{2} 
     \nonu\\&+& 
     180718 k^{4} N^{2} + 9036 k^{5} N^{2} + 1005682 N^{3} + 
     2084523 k\, 
    N^{3} + 1472858 k^{2} N^{3} + 410717 k^{3} N^{3} 
     \nonu\\&+& 
     37614 k^{4} N^{3} + 437719 N^{4} + 729570 k\, 
    N^{4} + 370476 k^{2} N^{4} + 55992 k^{3} N^{4} + 100478 N^{5} 
     \nonu\\&+&
     124705 k\, N^{5} + 34908 k^{2} N^{5} + 9488 N^{6} + 7494 k\, 
    N^{6}),\nonu\\
 c_ {69} & = &\frac {1}{6 (2 + N)^{2} (2 + k + N)^{5} (5 + 4 k + 4 N + 3 k\, N)}(10800 + 36400 k + 92713 k^{2} + 111256 k^{3} \nonu\\&+&   56052 k^{4} + 9840 k^{5} + 118160 N + 400858 k\, 
    N + 590800 k^{2} N + 429951 k^{3} N + 142292 k^{4} N 
     \nonu\\&+&
     16188 k^{5} N + 317497 N^{2} + 927282 k\, 
    N^{2} + 1031291 k^{2} N^{2} + 527988 k^{3} N^{2} + 
     113704 k^{4} N^{2} 
      \nonu\\&+&6768 k^{5} N^{2} + 374314 N^{3} + 
     900213 k\, 
    N^{3} + 752600 k^{2} N^{3} + 257123 k^{3} N^{3} + 
     28380 k^{4} N^{3} 
      \nonu\\&+& 221125 N^{4} + 416232 k\, 
    N^{4} + 240426 k^{2} N^{4} + 42654 k^{3} N^{4} + 64112 N^{5} + 
     86563 k\, N^{5} 
      \nonu\\&+& 26862 k^{2} N^{5} + 7256 N^{6} + 5820 k\, 
    N^{6}),\nonu\\
 c_ {70} & = &\frac{1}{6 (2 + N)^{2} (2 + k + N)^{5} (5 + 4 k + 4 N + 3 k\, N)}(202800 + 642640 k + 755065 k^{2} + 
     412408 k^{3} 
      \nonu\\&+& 104820 k^{4} + 9840 k^{5} + 748400 N + 
     2069002 k\, 
    N + 2103112 k^{2} N + 976959 k^{3} N 
     \nonu\\&+& 207476 k^{4} N + 
     16188 k^{5} N + 1138489 N^{2} + 2710146 k\, 
    N^{2} + 2314595 k^{2} N^{2} + 864780 k^{3} N^{2} 
     \nonu\\&+& 
     137272 k^{4} N^{2} + 6768 k^{5} N^{2} + 914818 N^{3} + 
     1831845 k\, 
    N^{3} + 1247864 k^{2} N^{3} + 335915 k^{3} N^{3}
     \nonu\\&+& 
     29964 k^{4} N^{3} + 409453 N^{4} + 662304 k\, 
    N^{4} + 324378 k^{2} N^{4} + 47406 k^{3} N^{4} + 96608 N^{5} 
     \nonu\\&+& 
     117403 k\, N^{5} + 31614 k^{2} N^{5} + 9368 N^{6} + 7404 k\, 
    N^{6}),\nonu\\
 c_ {71} & = & - \frac {28 (k - N) (32 + 55 k + 18 k^{2} + 41 N + 
       35 k\, N + 11 N^{2})} {(2 + N) (2 + k + N)^{5}}, \nonu\\
c_ {72} & = & - \frac{1}{6 (2 + N)^{2} (2 + k + N)^{6} (5 + 4 k + 4 N + 3 k\, 
      N)}(32 + 55 k + 18 k^{2} + 41 N + 35 k\, 
      N + 11 N^{2})
       \nonu\\&+& (6300 + 13661 k + 9302 k^{2} + 1976 k^{3} + 
       17209 N + 32755 k\, 
      N + 18825 k^{2} N + 3118 k^{3} N 
       \nonu\\&+& 17667 N^{2} + 28225 k\, 
      N^{2} + 12505 k^{2} N^{2} + 1254 k^{3} N^{2} + 7926 N^{3} + 
       9809 k\, N^{3} + 2646 k^{2} N^{3} 
        \nonu\\&+& 1280 N^{4} + 1014 k\, 
      N^{4}), \nonu\\
c_ {73} & = & - \frac{1}{18 (2 + N)^{2} (2 + k + N)^{5} (5 + 4 k + 4 N + 3 k\, 
      N)}i (-216720 - 612696 k - 
       723961 k^{2}
        \nonu\\&-& 461574 k^{3} - 158448 k^{4} - 22592 k^{5} - 
       689496 N - 1502626 k\, 
      N - 1192390 k^{2} N - 435120 k^{3} N 
       \nonu\\&-& 71060 k^{4} N - 
       2240 k^{5} N - 636889 N^{2} - 442342 k\, 
      N^{2} + 848078 k^{2} N^{2} + 1004041 k^{3} N^{2} 
       \nonu\\&+& 
       353929 k^{4} N^{2} + 41002 k^{5} N^{2} + 220058 N^{3} + 
       1974932 k\, 
      N^{3} + 3083887 k^{2} N^{3} + 1743718 k^{3} N^{3} 
       \nonu\\&+& 
       378229 k^{4} N^{3} + 23340 k^{5} N^{3} + 805394 N^{4} + 
       2512999 k\, 
      N^{4} + 2463757 k^{2} N^{4} + 903535 k^{3} N^{4}
       \nonu\\&+& 
       102078 k^{4} N^{4} + 581557 N^{5} + 1229644 k\, 
      N^{5} + 792991 k^{2} N^{5} + 150960 k^{3} N^{5} + 
       180460 N^{6} 
        \nonu\\&+& 256603 k\, 
      N^{6} + 86706 k^{2} N^{6} + 20992 N^{7} + 16752 k\, 
      N^{7}),\nonu\\
 c_ {74} & = & - \frac {1}{18 (2 + N)^{2} (2 + k + N)^{5} (5 + 4 k + 4 N + 3 k\, 
      N)}i (32400 - 235680 k - 
       410465 k^{2}
        \nonu\\&-& 176220 k^{3} - 3228 k^{4} + 7088 k^{5} + 
       1017600 N + 2493046 k\, 
      N + 3103276 k^{2} N + 2210811 k^{3} N 
       \nonu\\&+& 763820 k^{4} N + 
       95852 k^{5} N + 3584287 N^{2} + 9622510 k\, 
      N^{2} + 10651507 k^{2} N^{2} + 5828042 k^{3} N^{2}
       \nonu\\&+& 
       1468070 k^{4} N^{2} + 129308 k^{5} N^{2} + 5403382 N^{3} + 
       13297777 k\, 
      N^{3} + 12389402 k^{2} N^{3} 
       \nonu\\&+& 5268275 k^{3} N^{3} + 
       932858 k^{4} N^{3} + 46680 k^{5} N^{3} + 4312525 N^{4} + 
       9138818 k\, 
      N^{4} + 6771128 k^{2} N^{4} 
       \nonu\\&+& 2040080 k^{3} N^{4} + 
       195516 k^{4} N^{4} + 1908710 N^{5} + 3284759 k\, 
      N^{5} + 1743176 k^{2} N^{5} 
       \nonu\\&+& 284640 k^{3} N^{5} 
      +442784 N^{6} + 566486 k\, 
      N^{6} + 164772 k^{2} N^{6} + 41984 N^{7} + 33504 k\, 
      N^{7}),\nonu\\
 c_ {75} & = & - \frac {1}{18 (2 + N) (2 + k + N)^{5} (5 + 4 k + 4 N + 3 k\, N)}i (568080 + 1958124 k + 
       2529508 k^{2} 
        \nonu\\&+&1548333 k^{3} + 453306 k^{4} + 51128 k^{5} + 
       2026644 N + 6126382 k\, 
      N + 6840035 k^{2} N 
       \nonu\\&+& 3514587 k^{3} N + 824243 k^{4} N + 
       68842 k^{5} N + 2969122 N^{2} + 7776467 k\, 
      N^{2} + 7313697 k^{2} N^{2} 
       \nonu\\&+& 3000251 k^{3} N^{2} + 
       506785 k^{4} N^{2} + 23340 k^{5} N^{2} + 2296361 N^{3} + 
       5086737 k\, 
      N^{3} + 3843545 k^{2} N^{3}
       \nonu\\&+& 1140697 k^{3} N^{3} + 
       106830 k^{4} N^{3} + 990513 N^{4} + 1773221 k\, 
      N^{4} + 975277 k^{2} N^{4} + 160464 k^{3} N^{4}
       \nonu\\&+& 225236 N^{5} + 297907 k\, 
      N^{5} + 91458 k^{2} N^{5} + 20992 N^{6} + 16752 k\, 
      N^{6}),\nonu\\ 
c_ {76} & = &\frac{1}{(2 + N) (2 + k + N)^{4} (5 + 4 k + 4 N + 3 k\, N)}6 i (2100 + 4247 k + 2798 k^{2} + 584 k^{3} + 
      6043 N 
       \nonu\\&+& 10765 k\, 
     N + 6057 k^{2} N + 1006 k^{3} N + 6345 N^{2} + 9475 k\, 
     N^{2} + 4135 k^{2} N^{2} + 420 k^{3} N^{2} 
      \nonu\\&+& 2868 N^{3} + 
      3299 k\, N^{3} + 882 k^{2} N^{3} + 464 N^{4} + 336 k\, 
     N^{4}), \nonu\\
c_ {77} & = &\frac {1} {(2 + N) (2 + k + N)^{4} (5 + 4 k + 4 N + 3 k\, N)}2 i (2100 + 4183 k + 2734 k^{2} + 568 k^{3} + 
      6107 N 
       \nonu\\&+& 10733 k\, 
     N + 6009 k^{2} N + 998 k^{3} N + 6441 N^{2} + 9491 k\, 
     N^{2} + 4127 k^{2} N^{2} + 420 k^{3} N^{2} 
      \nonu\\&+& 2916 N^{3} + 
      3307 k\, N^{3} + 882 k^{2} N^{3} + 472 N^{4} + 336 k\, 
     N^{4}), \nonu\\
c_ {78} & = &\frac {14 i (32 + 55 k + 18 k^{2} + 41 N + 35 k\, 
     N + 11 N^{2})} {(2 + N) (2 + k + N)^{4}}, \nonu\\
c_ {79} & = & - \frac {28 i (32 + 55 k + 18 k^{2} + 41 N + 35 k\, 
      N + 11 N^{2})} {(2 + N) (2 + k + N)^{4}}, \nonu\\
c_ {80} & = & - \frac{1}{3 (2 + N)^{2} (2 + k + N)^{6} (5 + 4 k + 4 N + 3 k\, 
      N)}2 i (84000 + 340900 k + 528433 k^{2} 
       \nonu\\&+&
       388322 k^{3} + 135400 k^{4} + 18016 k^{5} + 381500 N + 
       1337660 k\, 
      N + 1779414 k^{2} N 
       \nonu\\&+& 1102066 k^{3} N + 312948 k^{4} N + 
       32008 k^{5} N + 721947 N^{2} + 2158908 k\, 
      N^{2} + 2402234 k^{2} N^{2} 
       \nonu\\&+& 1199265 k^{3} N^{2} + 
       255509 k^{4} N^{2} + 16834 k^{5} N^{2} + 738220 N^{3} + 
       1840266 k\, 
      N^{3} + 1644813 k^{2} N^{3} 
       \nonu\\&+& 614512 k^{3} N^{3} + 
       83757 k^{4} N^{3} + 2100 k^{5} N^{3} + 439850 N^{4} + 
       883811 k\, 
      N^{4} + 595331 k^{2} N^{4} 
       \nonu\\&+& 145427 k^{3} N^{4} + 
       8610 k^{4} N^{4} + 152571 N^{5} + 235876 k\, 
      N^{5} + 107497 k^{2} N^{5} + 12600 k^{3} N^{5} 
       \nonu\\&+& 28580 N^{6} + 
       31989 k\, N^{6} + 7770 k^{2} N^{6} + 2240 N^{7} + 1680 k\, 
      N^{7}), \nonu\\
c_ {81} & = & - \frac {1}{3 (2 + N) (2 + k + N)^{3} (5 + 4 k + 4 N + 3 k\, N)}i (35700 + 76441 k + 53602 k^{2} + 
       11944 k^{3} 
        \nonu\\&+& 100589 N + 189221 k\, 
      N + 112099 k^{2} N + 19498 k^{3} N + 103803 N^{2} + 164119 k\, 
      N^{2} 
       \nonu\\&+& 75199 k^{2} N^{2} + 7960 k^{3} N^{2} + 46284 N^{3} + 
       56553 k\, N^{3} + 15876 k^{2} N^{3} + 7384 N^{4} + 5648 k\, 
      N^{4}), \nonu\\
c_ {82} & = & - \frac {14 i (32 + 55 k + 18 k^{2} + 41 N + 35 k\, 
      N + 11 N^{2})} {(2 + N) (2 + k + N)^{4}}, \nonu\\
c_ {83} & = & - \frac {1}{6 (2 + N)^{2} (2 + k + N)^{5} (5 + 4 k + 4 N + 3 k\, 
      N)}i (91756 k + 230369 k^{2} + 210476 k^{3} 
       \nonu\\&+& 
       82780 k^{4} + 11856 k^{5} + 9044 N + 331256 k\, 
      N + 702868 k^{2} N + 541429 k^{3} N + 172644 k^{4} N 
       \nonu\\&+& 
       18708 k^{5} N + 33095 N^{2} + 490974 k\, 
      N^{2} + 845211 k^{2} N^{2} + 517606 k^{3} N^{2} + 
       119810 k^{4} N^{2} 
        \nonu\\&+& 7524 k^{5} N^{2} + 45186 N^{3} + 
       368559 k\, 
      N^{3} + 487334 k^{2} N^{3} + 213509 k^{3} N^{3} + 
       27186 k^{4} N^{3} 
        \nonu\\&+& 28837 N^{4} + 140594 k\, 
      N^{4} + 129490 k^{2} N^{4} + 31176 k^{3} N^{4} + 8590 N^{5} + 
       23539 k\, N^{5} 
        \nonu\\&+& 11748 k^{2} N^{5} + 960 N^{6} + 990 k\, 
      N^{6}), \nonu\\
c_ {84} & = & - \frac {1}{6 (2 + N)^{2} (2 + k + N)^{5} (5 + 4 k + 4 N + 3 k\, 
      N)}i (33600 + 184956 k + 
       334625 k^{2} 
        \nonu\\&+& 269696 k^{3} + 99908 k^{4} + 13872 k^{5} + 
       130884 N + 628240 k\, 
      N + 985600 k^{2} N + 673059 k^{3} N 
       \nonu\\&+& 202324 k^{4} N + 
       21228 k^{5} N + 214143 N^{2} + 871634 k\, 
      N^{2} + 1143117 k^{2} N^{2} + 624492 k^{3} N^{2} 
       \nonu\\&+& 
       136476 k^{4} N^{2} + 8280 k^{5} N^{2} + 186430 N^{3} + 
       618145 k\, 
      N^{3} + 639248 k^{2} N^{3} + 251083 k^{3} N^{3} 
       \nonu\\&+& 
       30240 k^{4} N^{3} + 89931 N^{4} + 227964 k\, 
      N^{4} + 166792 k^{2} N^{4} + 35994 k^{3} N^{4} + 22500 N^{5}
       \nonu\\&+& 
       38717 k\, N^{5} + 15246 k^{2} N^{5} + 2264 N^{6} + 1968 k\, 
      N^{6}), \nonu\\
c_ {85} & = & - \frac{1}{6 (2 + N)^{2} (2 + k + N)^{5} (5 + 4 k + 4 N + 3 k\, 
      N)}i (33600 + 167676 k + 
       281681 k^{2} 
        \nonu\\&+& 213080 k^{3} + 74612 k^{4} + 9840 k^{5} + 
       148164 N + 627760 k\, 
      N + 903472 k^{2} N + 574887 k^{3} N 
       \nonu\\&+& 162676 k^{4} N + 
       16188 k^{5} N + 267567 N^{2} + 946058 k\, 
      N^{2} + 1128753 k^{2} N^{2} + 571104 k^{3} N^{2} 
       \nonu\\&+&
       117144 k^{4} N^{2} + 6768 k^{5} N^{2} + 250750 N^{3} + 
       718285 k\, 
      N^{3} + 672596 k^{2} N^{3} + 243079 k^{3} N^{3} 
       \nonu\\&+& 
       27324 k^{4} N^{3} + 127623 N^{4} + 280920 k\, 
      N^{4} + 185932 k^{2} N^{4} + 36606 k^{3} N^{4} + 33264 N^{5} 
       \nonu\\&+& 
       50753 k\, N^{5} + 18162 k^{2} N^{5} + 3464 N^{6} + 2868 k\, 
      N^{6}),\nonu\\
 c_ {86} & = &\frac {14 i (32 + 55 k + 18 k^{2} + 41 N + 
      35 k\, N + 11 N^{2})} {(2 + N) (2 + k + N)^{4}}, \nonu\\
c_ {87} & = & - \frac {14 (k - N) (32 + 55 k + 18 k^{2} + 41 N + 
       35 k\, N + 11 N^{2})} {(2 + N) (2 + k + N)^{5}}, \nonu\\
c_ {88} & = & - \frac {7 (k - N) (32 + 55 k + 18 k^{2} + 41 N + 
       35 k\, N + 11 N^{2})} {(2 + N) (2 + k + N)^{5}}, \nonu\\
c_ {89} & = & - \frac{1} {3 (2 + N)^{2} (2 + k + N)^{6} (5 + 4 k + 4 N + 3 k\, 
      N)}2 i (67200 + 309580 k + 523849 k^{2}
       \nonu\\&+& 
       410408 k^{3} + 150388 k^{4} + 20848 k^{5} + 288500 N + 
       1146488 k\, 
      N + 1670100 k^{2} N 
       \nonu\\&+& 1107271 k^{3} N + 331620 k^{4} N + 
       35452 k^{5} N + 519039 N^{2} + 1754202 k\, 
      N^{2} + 2141525 k^{2} N^{2} 
       \nonu\\&+& 1147860 k^{3} N^{2} + 
       258368 k^{4} N^{2} + 17776 k^{5} N^{2} + 507082 N^{3} + 
       1420581 k\, 
      N^{3} + 1393740 k^{2} N^{3} 
       \nonu\\&+& 560035 k^{3} N^{3} + 
       80484 k^{4} N^{3} + 2064 k^{5} N^{3} + 289451 N^{4} + 
       646652 k\, 
      N^{4} + 477524 k^{2} N^{4} 
       \nonu\\&+& 125270 k^{3} N^{4} + 
       7656 k^{4} N^{4} + 96144 N^{5} + 162085 k\, 
      N^{5} + 80818 k^{2} N^{5} + 10080 k^{3} N^{5} 
       \nonu\\&+&17192 N^{6} + 
       20292 k\, N^{6} + 5448 k^{2} N^{6} + 1280 N^{7} + 960 k\, 
      N^{7}), \nonu\\
c_ {90} & = &\frac {7 (32 + 55 k + 18 k^{2} + 41 N + 
      35 k\, N + 11 N^{2})} {4(2 + N) (2 + k + N)^{4}}, \nonu\\
c_ {91} & = &\frac{1}{9 (2 + N) (2 + k + N)^{4} (5 + 4 k + 4 N + 
      3 k\, N)}2 i (3360 + 8463 k + 6574 k^{2} + 1544 k^{3} 
       \nonu\\&+& 
      6993 N + 13627 k\, 
     N + 7917 k^{2} N + 1150 k^{3} N + 4663 N^{2} + 6351 k\, 
     N^{2} + 2189 k^{2} N^{2} + 988 N^{3} 
      \nonu\\&+& 677 k\, 
     N^{3} + 16 N^{4}), \nonu\\
c_ {92} & = &\frac{1} {3 (2 + N) (2 + k + N)^{3} (5 + 4 k + 4 N + 
      3 k\, N)}2 (-960 - 2613 k - 2074 k^{2} - 488 k^{3} 
       \nonu\\&-& 
      2043 N - 4321 k\, 
     N - 2559 k^{2} N - 370 k^{3} N - 1333 N^{2} - 2037 k\, 
     N^{2} - 719 k^{2} N^{2} - 244 N^{3} 
      \nonu\\&-& 215 k\,N^{3} + 8 N^{4}), \nonu\\
c_ {93} & = &\frac {i (64 + 123 k + 42 k^{2} + 85 N + 79 k\, 
     N + 23 N^{2})} {(2 + N) (2 + k + N)^{3}}, \nonu\\
c_ {94} & = &\frac{1}{24 (2 + N)^{2} (2 + k + N)^{4} (5 + 4 k + 4 N + 3 k\, 
     N)}i (33600 + 163836 k + 274769 k^{2} 
      \nonu\\&+& 
      209048 k^{3} + 73844 k^{4} + 9840 k^{5} + 152004 N + 623920 k\, 
     N + 890224 k^{2} N + 567015 k^{3} N 
      \nonu\\&+& 161332 k^{4} N + 
      16188 k^{5} N + 278319 N^{2} + 951434 k\, 
     N^{2} + 1120689 k^{2} N^{2} + 565680 k^{3} N^{2}
      \nonu\\&+& 
      116376 k^{4} N^{2} + 6768 k^{5} N^{2} + 262654 N^{3} + 
      728461 k\, 
     N^{3} + 671588 k^{2} N^{3} + 241543 k^{3} N^{3} 
      \nonu\\&+& 
      27180 k^{4} N^{3} + 134151 N^{4} + 286920 k\, 
     N^{4} + 186508 k^{2} N^{4} + 36462 k^{3} N^{4} + 35040 N^{5} 
      \nonu\\&+& 
      52289 k\, N^{5} + 18306 k^{2} N^{5} + 3656 N^{6} + 3012 k\, 
     N^{6}),\nonu\\
 c_ {95} & = &\frac {1}{24 (2 + N)^{2} (2 + k + N)^{4} (5 + 4 k + 4 N + 3 k\, 
     N)}i (16 + 21 k + 6 k^{2} + 19 N + 13 k\, 
     N + 5 N^{2}) 
      \nonu\\&+&(6300 + 15341 k + 11486 k^{2} + 2648 k^{3} + 
      15529 N + 33595 k\, 
     N + 21597 k^{2} N + 3958 k^{3} N 
      \nonu\\&+&14643 N^{2} + 26545 k\, 
     N^{2} + 13345 k^{2} N^{2} + 1506 k^{3} N^{2} + 6162 N^{3} + 
      8465 k\, N^{3} + 2646 k^{2} N^{3} 
       \nonu\\&+& 944 N^{4} + 762 k\, 
     N^{4}),\nonu\\
 c_ {96} & = &\frac {7 (20 + 21 k + 6 k^{2} + 25 N + 13 k\, 
     N + 7 N^{2})} {2 (2 + N) (2 + k + N)^{3}}, \nonu\\
c_ {97} & = &\frac {1}{12 (2 + N)^{2} (2 + k + N)^{5} (5 + 4 k + 4 N + 3 k\, N)}(100800 + 356380 k + 508025 k^{2}  \nonu\\&+& 
     355940 k^{3} + 120940 k^{4} + 15888 k^{5} + 389540 N + 
     1215632 k\, 
    N + 1499200 k^{2} N + 883921 k^{3} N 
     \nonu\\&+& 241716 k^{4} N + 
     23748 k^{5} N + 617927 N^{2} + 1670742 k\, 
    N^{2} + 1727427 k^{2} N^{2} + 811606 k^{3} N^{2}
     \nonu\\&+& 
     160430 k^{4} N^{2} + 9036 k^{5} N^{2} + 515190 N^{3} + 
     1168971 k\, 
    N^{3} + 958778 k^{2} N^{3} + 322805 k^{3} N^{3} 
     \nonu\\&+& 
     34998 k^{4} N^{3} + 237589 N^{4} + 428714 k\, 
    N^{4} + 250534 k^{2} N^{4} + 46044 k^{3} N^{4} + 57250 N^{5} 
     \nonu\\&+&
     74515 k\, N^{5} + 23592 k^{2} N^{5} + 5616 N^{6} + 4266 k\, 
    N^{6}),\nonu\\
 c_ {98} & = &\frac {1}{12 (2 + N)^{2} (2 + k + N)^{5} (5 + 4 k + 4 N + 3 k\, 
     N)}(67200 + 247820 k + 354281 k^{2} 
      \nonu\\&+& 
     242120 k^{3} + 78900 k^{4} + 9840 k^{5} + 283060 N + 920088 k\, 
    N + 1140964 k^{2} N + 658055 k^{3} N 
     \nonu\\&+& 173060 k^{4} N + 
     16188 k^{5} N + 484927 N^{2} + 1361818 k\, 
    N^{2} + 1419189 k^{2} N^{2} + 654044 k^{3} N^{2}
     \nonu\\&+& 
     124816 k^{4} N^{2} + 6768 k^{5} N^{2} + 432314 N^{3} + 
     1014437 k\, 
    N^{3} + 840716 k^{2} N^{3} + 277995 k^{3} N^{3} 
     \nonu\\&+& 
     29100 k^{4} N^{3} + 210923 N^{4} + 391300 k\, 
    N^{4} + 232084 k^{2} N^{4} + 41910 k^{3} N^{4} + 53216 N^{5} 
     \nonu\\&+&
     70605 k\, N^{5} + 22938 k^{2} N^{5} + 5416 N^{6} + 4116 k\, 
    N^{6}),\nonu\\
c_ {99} & = & - \frac {i (64 + 123 k + 42 k^{2} + 85 N + 
       79 k\, N + 23 N^{2})} {2 (2 + N) (2 + k + N)^{3}}, \nonu\\
c_ {100} & = &\frac {7 (32 + 55 k + 18 k^{2} + 41 N + 35 k\, 
     N + 11 N^{2})} {(2 + N) (2 + k + N)^{4}}, \nonu\\
c_ {101} & = &\frac{1}{54 (2 + N)^{2} (2 + k + N)^{5} (5 + 4 k + 4 N + 3 k\, 
     N)}(63360 + 266868 k + 407983 k^{2} 
      \nonu\\&+& 
     286422 k^{3} + 93000 k^{4} + 11264 k^{5} + 380628 N + 
     1402234 k\, 
    N + 1911943 k^{2} N + 1202784 k^{3} N 
     \nonu\\&+& 348428 k^{4} N + 
     37352 k^{5} N + 930391 N^{2} + 3002479 k\, 
    N^{2} + 3577690 k^{2} N^{2} + 1936523 k^{3} N^{2} 
     \nonu\\&+& 
     466691 k^{4} N^{2} + 39182 k^{5} N^{2} + 1221358 N^{3} + 
     3402304 k\, 
    N^{3} + 3424703 k^{2} N^{3} + 1500095 k^{3} N^{3} 
     \nonu\\&+& 
     267542 k^{4} N^{3} + 13362 k^{5} N^{3} + 934372 N^{4} + 
     2179703 k\, 
    N^{4} + 1750181 k^{2} N^{4} + 554378 k^{3} N^{4}
     \nonu\\&+& 
     54981 k^{4} N^{4} + 415919 N^{5} + 771281 k\, 
    N^{5} + 443639 k^{2} N^{5} + 76956 k^{3} N^{5} + 99356 N^{6} 
     \nonu\\&+& 
     133397 k\, N^{6} + 42003 k^{2} N^{6} + 9800 N^{7} + 7800 k\, 
    N^{7}), \nonu\\
c_ {102} & = &\frac{1}{54 (2 + N)^{2} (2 + k + N)^{5} (5 + 4 k + 4 N + 3 k\, 
     N)}(63360 + 266868 k + 407983 k^{2} 
      \nonu\\&+& 
     286422 k^{3} + 93000 k^{4} + 11264 k^{5} + 380628 N + 
     1402234 k\, 
    N + 1911943 k^{2} N + 1202784 k^{3} N 
     \nonu\\&+& 348428 k^{4} N + 
     37352 k^{5} N + 930391 N^{2} + 3002479 k\, 
    N^{2} + 3577690 k^{2} N^{2} + 1936523 k^{3} N^{2} 
     \nonu\\&+& 
     466691 k^{4} N^{2} + 39182 k^{5} N^{2} + 1221358 N^{3} + 
     3402304 k\, 
    N^{3} + 3424703 k^{2} N^{3} + 1500095 k^{3} N^{3} 
     \nonu\\&+& 
     267542 k^{4} N^{3} + 13362 k^{5} N^{3} + 934372 N^{4} + 
     2179703 k\, 
    N^{4} + 1750181 k^{2} N^{4} + 554378 k^{3} N^{4} 
     \nonu\\&+&
     54981 k^{4} N^{4} + 415919 N^{5} + 771281 k\, 
    N^{5} + 443639 k^{2} N^{5} + 76956 k^{3} N^{5} + 99356 N^{6} 
     \nonu\\&+&
     133397 k\, N^{6} + 42003 k^{2} N^{6} + 9800 N^{7} + 7800 k\, 
    N^{7}),\nonu\\
 c_ {103} & = &\frac{1}{18 (2 + N)^{2} (2 + k + N)^{5} (5 + 4 k + 4 N + 3 k\, 
     N)}(228240 + 698232 k + 886145 k^{2} 
      \nonu\\&+& 
     573144 k^{3} + 184596 k^{4} + 23344 k^{5} + 1240152 N + 
     3737090 k\, 
    N + 4467662 k^{2} N 
     \nonu\\&+& 2606691 k^{3} N + 726532 k^{4} N + 
     76060 k^{5} N + 2803889 N^{2} + 7850816 k\, 
    N^{2} + 8382935 k^{2} N^{2} 
     \nonu\\&+& 4185928 k^{3} N^{2} + 
     945136 k^{4} N^{2} + 73744 k^{5} N^{2} + 3409754 N^{3} + 
     8506529 k\, 
    N^{3} + 7746916 k^{2} N^{3} 
     \nonu\\&+& 3107041 k^{3} N^{3} + 
     507514 k^{4} N^{3} + 22476 k^{5} N^{3} + 2408021 N^{4} 
     +  5135032 k\, N^{4} + 3754726 k^{2} N^{4} 
      \nonu\\&+& 1083796 k^{3} N^{4} + 
     96894 k^{4} N^{4} + 987928 N^{5} + 1705633 k\, 
    N^{5} + 900262 k^{2} N^{5} + 142320 k^{3} N^{5} 
     \nonu\\&+& 218056 N^{6} + 
     279148 k\, N^{6} + 81522 k^{2} N^{6} + 19984 N^{7} + 15888 k\, 
    N^{7}),\nonu\\
 c_ {104} & = &\frac {1}{36 (2 + N) (2 + k + N)^{5} (5 + 4 k + 4 N + 3 k\, N)}(435600 + 1491420 k + 1951816 k^{2}  \nonu\\&+& 
     1229643 k^{3} + 375366 k^{4} + 44648 k^{5} + 1644900 N + 
     4969342 k\, 
    N + 5638805 k^{2} N 
     \nonu\\&+& 2990805 k^{3} N + 734045 k^{4} N + 
     65062 k^{5} N + 2566318 N^{2} + 6728453 k\, 
    N^{2} + 6426855 k^{2} N^{2} 
     \nonu\\&+& 2720873 k^{3} N^{2} + 
     482467 k^{4} N^{2} + 24204 k^{5} N^{2} + 2113319 N^{3} + 
     4672647 k\,N^{3} + 3565175 k^{2} N^{3} 
     \nonu\\&+& 1085347 k^{3} N^{3} + 
     105966 k^{4} N^{3} + 964719 N^{4} + 1712615 k\, 
    N^{4} + 940987 k^{2} N^{4} + 157008 k^{3} N^{4} 
     \nonu\\&+& 230060 N^{5} + 
     299725 k\, N^{5} + 90594 k^{2} N^{5} + 22288 N^{6} + 17616 k\, 
    N^{6}), \nonu\\
c_ {105} & = & - \frac {7 i (48 + 97 k + 47 k^{2} + 6 k^{3} + 71 N + 
       85 k\, N + 17 k^{2} N + 28 N^{2} + 14 k\, 
      N^{2} + 3 N^{3})} {(2 + N) (2 + k + N)^{4}}, \nonu\\
c_ {106} & = & - \frac {1} {12 (2 + N)^{2} (2 + k + N)^{5} (5 + 4 k + 4 N + 3 k\, 
      N)}i (133680 + 445024 k + 575089 k^{2} 
       \nonu\\&+& 
       363472 k^{3} + 113028 k^{4} + 13872 k^{5} + 532064 N + 
       1566562 k\, 
      N + 1750660 k^{2} N + 927483 k^{3} N
       \nonu\\&+& 231332 k^{4} N + 
       21228 k^{5} N + 875497 N^{2} + 2234850 k\, 
      N^{2} + 2089499 k^{2} N^{2} + 876228 k^{3} N^{2} 
       \nonu\\&+& 
       157156 k^{4} N^{2} + 8280 k^{5} N^{2} + 759310 N^{3} + 
       1630209 k\, 
      N^{3} + 1206620 k^{2} N^{3} + 358763 k^{3} N^{3} 
       \nonu\\&+& 
       34968 k^{4} N^{3} + 364429 N^{4} + 627216 k\, 
      N^{4} + 331038 k^{2} N^{4} + 52842 k^{3} N^{4} + 91364 N^{5} 
       \nonu\\&+& 
       116011 k\, N^{5} + 33522 k^{2} N^{5} + 9320 N^{6} + 7368 k\, 
      N^{6}), \nonu\\
c_ {107} & = & - \frac {1}{12 (2 + N)^{2} (2 + k + N)^{5} (5 + 4 k + 4 N + 3 k\, 
      N)}i (49200 + 146560 k + 
       195145 k^{2} 
        \nonu\\&+& 141700 k^{3} + 53004 k^{4} + 7824 k^{5} + 
       249920 N + 709522 k\, 
      N + 818620 k^{2} N + 474849 k^{3} N 
       \nonu\\&+& 133220 k^{4} N + 
       13668 k^{5} N + 498529 N^{2} + 1267038 k\, 
      N^{2} + 1218677 k^{2} N^{2} + 545022 k^{3} N^{2} 
       \nonu\\&+& 106078 k^{4} N^{2} + 6012 k^{5} N^{2} + 502054 N^{3} + 
       1087371 k\, 
      N^{3} + 822698 k^{2} N^{3} + 256241 k^{3} N^{3}
       \nonu\\&+& 
       26454 k^{4} N^{3} + 270139 N^{4} + 470982 k\, 
      N^{4} + 252252 k^{2} N^{4} + 41664 k^{3} N^{4} + 73862 N^{5} 
       \nonu\\&+& 
       94957 k\, N^{5} + 27636 k^{2} N^{5} + 8048 N^{6} + 6414 k\, 
      N^{6}),\nonu\\
 c_ {108} & = & - \frac {1}{12 (2 + N)^{2} (2 + k + N)^{5} (5 + 4 k + 4 N + 3 k\, 
      N)}i (233520 + 790576 k + 
       992713 k^{2} 
        \nonu\\&+& 586216 k^{3} + 165012 k^{4} + 17904 k^{5} + 
       826256 N + 2450266 k\, 
      N + 2665552 k^{2} N 
       \nonu\\&+& 1332279 k^{3} N + 307604 k^{4} N + 
       26268 k^{5} N + 1200793 N^{2} + 3071394 k\, 
      N^{2} + 2809475 k^{2} N^{2} 
       \nonu\\&+&1123788 k^{3} N^{2} + 
       190576 k^{4} N^{2} + 9792 k^{5} N^{2} + 919402 N^{3} + 
       1976061 k\, 
      N^{3} + 1443584 k^{2} N^{3} 
       \nonu\\&+& 414875 k^{3} N^{3} + 
       39108 k^{4} N^{3} + 391645 N^{4} + 676368 k\, 
      N^{4} + 356418 k^{2} N^{4} + 55686 k^{3} N^{4} 
       \nonu\\&+& 87944 N^{5} + 
       112459 k\, N^{5} + 32838 k^{2} N^{5} + 8120 N^{6} + 6468 k\, 
      N^{6}), \nonu\\
c_ {109} & = &\frac {i (16 + 99 k + 42 k^{2} + 37 N + 
      67 k\, N + 11 N^{2})} {2 (2 + N) (2 + k + N)^{3}},\nonu\\ 
c_ {110} & = &\frac {7 (32 + 55 k + 18 k^{2} + 41 N + 35 k\, 
     N + 11 N^{2})} {(2 + N) (2 + k + N)^{4}}, \nonu\\
c_ {111} & = &\frac {14 (32 + 55 k + 18 k^{2} + 41 N + 35 k\, 
     N + 11 N^{2})} {(2 + N) (2 + k + N)^{4}}, \nonu\\
c_ {112} & = &\frac{1}{12 (2 + N)^{2} (2 + k + N)^{5} (5 + 4 k + 4 N + 3 k\, 
     N)}(67200 + 289292 k + 456161 k^{2} 
      \nonu\\&+& 
     337844 k^{3}+118572 k^{4} + 15888 k^{5} + 241588 N + 
     936360 k\, 
    N + 1295692 k^{2} N + 817841 k^{3} N 
     \nonu\\&+&233828 k^{4} N + 
     23748 k^{5} N + 366775 N^{2} + 1244998 k\, 
    N^{2} + 1455423 k^{2} N^{2} + 738122 k^{3} N^{2}
     \nonu\\&+& 
     153814 k^{4} N^{2} + 9036 k^{5} N^{2} + 298682 N^{3} + 
     853427 k\, 
    N^{3} + 793394 k^{2} N^{3} + 290001 k^{3} N^{3} 
     \nonu\\&+& 
     33366 k^{4} N^{3} + 136241 N^{4} + 308302 k\, 
    N^{4} + 203806 k^{2} N^{4} + 40884 k^{3} N^{4} + 32642 N^{5} 
     \nonu\\&+& 
     52551 k\, N^{5} + 18672 k^{2} N^{5} + 3184 N^{6} + 2874 k\, 
    N^{6}), \nonu\\
c_ {113} & = &\frac{1} {12 (2 + N)^{2} (2 + k + N)^{5} (5 + 4 k + 4 N + 3 k\, 
     N)}(33600 + 178172 k + 297809 k^{2}
      \nonu\\&+& 
     221336 k^{3} +76020 k^{4} + 9840 k^{5} + 137668 N + 638256 k\, 
    N + 928624 k^{2} N + 586727 k^{3} N
     \nonu\\&+& 164276 k^{4} N 
    + 16188 k^{5} N + 240943 N^{2} + 939658 k\, 
    N^{2} + 1141809 k^{2} N^{2} + 576944 k^{3} N^{2} 
     \nonu\\&+&  117688 k^{4} N^{2} 
     +6768 k^{5} N^{2} + 223742 N^{3} + 
     705677 k\,N^{3} + 674660 k^{2} N^{3} + 244167 k^{3} N^{3} 
      \nonu\\&+& 
     27372 k^{4} N^{3} 
    +113927 N^{4} + 274888 k\, 
    N^{4} + 185740 k^{2} N^{4} + 36654 k^{3} N^{4} + 29792 N^{5} 
     \nonu\\&+& 
     49665 k\, N^{5} + 18114 k^{2} N^{5} + 3112 N^{6} + 2820 k\, 
    N^{6}),\nonu\\
 c_ {114} & = &\frac {1}{12 (2 + N)^{2} (2 + k + N)^{5} (5 + 4 k + 4 N + 3 k\, 
     N)}(92652 k + 218297 k^{2} + 
     189416 k^{3} 
      \nonu\\&+& 71348 k^{4} + 9840 k^{5} + 8148 N + 340552 k\, 
    N + 683644 k^{2} N + 502215 k^{3} N + 154084 k^{4} N 
     \nonu\\&+& 
     16188 k^{5} N + 35871 N^{2} + 523130 k\, 
    N^{2} + 849069 k^{2} N^{2} + 495396 k^{3} N^{2} + 
     110496 k^{4} N^{2} 
      \nonu\\&+& 6768 k^{5} N^{2} + 53314 N^{3} + 408565 k\, 
    N^{3} + 505820 k^{2} N^{3} + 210211 k^{3} N^{3} + 
     25740 k^{4} N^{3} 
      \nonu\\&+& 35619 N^{4} + 162540 k\, 
    N^{4} + 139012 k^{2} N^{4} + 31494 k^{3} N^{4} + 10944 N^{5} + 
     28853 k\, N^{5} 
      \nonu\\&+&13194 k^{2} N^{5} + 1256 N^{6} + 1428 k\, 
    N^{6}),\nonu\\
 c_ {115} & = & - \frac {7 (32 + 55 k + 18 k^{2} + 41 N + 
       35 k\, N + 11 N^{2})} {(2 + N) (2 + k + N)^{4}}, \nonu\\
c_ {116} & = & - \frac {1}{18 (2 + N)^{2} (2 + k + N)^{6} (5 + 4 k + 4 N + 3 k\, 
      N)}i (32 + 55 k + 18 k^{2} + 41 N + 35 k\, 
      N 
       \nonu\\&+& 11 N^{2}) (6300 + 12821 k + 8210 k^{2} + 1640 k^{3} + 
       18049 N + 32335 k\, 
      N + 17439 k^{2} N 
       \nonu\\&+&2698 k^{3} N + 19179 N^{2} + 29065 k\, 
      N^{2} + 12085 k^{2} N^{2} + 1128 k^{3} N^{2} + 8808 N^{3} + 
       10481 k\, N^{3}
        \nonu\\&+& 2646 k^{2} N^{3} + 1448 N^{4} + 1140 k\, 
      N^{4}),\nonu\\
 c_ {117} & = & - \frac {1}{18 (2 + N)^{2} (2 + k + N)^{6} (5 + 4 k + 4 N + 3 k\, 
      N)}i (32 + 55 k + 18 k^{2} + 41 N + 
       35 k\, N 
        \nonu\\&+& 11 N^{2}) (6300 + 14501 k + 10394 k^{2} + 
       2312 k^{3} + 16369 N + 33175 k\, 
      N + 20211 k^{2} N 
       \nonu\\&+& 3538 k^{3} N + 16155 N^{2} + 27385 k\, 
      N^{2} + 12925 k^{2} N^{2} + 1380 k^{3} N^{2} + 7044 N^{3} + 
       9137 k\, N^{3} 
        \nonu\\&+&2646 k^{2} N^{3} + 1112 N^{4} + 888 k\, 
      N^{4}),\nonu\\
  c_ {118} & = & - \frac {1}{18 (2 + N)^{2} (2 + k + N)^{6} (5 + 4 k + 4 N + 3 k\, 
      N)}i (458640 + 1621272 k + 
       2214245 k^{2} 
        \nonu\\&+& 1461066 k^{3} + 466152 k^{4} + 57664 k^{5} + 
       2091192 N + 6691634 k\, 
      N + 8174966 k^{2} N 
       \nonu\\&+&4737738 k^{3} N + 1292068 k^{4} N + 
       131368 k^{5} N + 4081949 N^{2} + 11666606 k\, 
      N^{2} 
       \nonu\\&+& 12484052 k^{2} N^{2} + 6147553 k^{3} N^{2} + 
       1355449 k^{4} N^{2} + 102178 k^{5} N^{2} + 4413458 N^{3} 
        \nonu\\&+& 
       11043842 k\, 
      N^{3} + 10019815 k^{2} N^{3} + 3966976 k^{3} N^{3} + 
       635209 k^{4} N^{3} + 27156 k^{5} N^{3} 
        \nonu\\&+& 2846588 N^{4} + 
       6048535 k\, 
      N^{4} + 4405969 k^{2} N^{4} + 1260883 k^{3} N^{4} + 
       111258 k^{4} N^{4} 
        \nonu\\&+&  1091053 N^{5}+1874350 k\,N^{5} + 987655 k^{2} N^{5} + 155640 k^{3} N^{5} + 
       229084 N^{6} + 292003 k\, 
      N^{6} 
       \nonu\\&+&  85302 k^{2} N^{6}+ 20248 N^{7} + 16032 k\, 
      N^{7}),\nonu\\
 c_ {119} & = & - \frac {1}{9 (2 + N)^{2} (2 + k + N)^{6} (5 + 4 k + 4 N + 3 k\, 
      N)} i (285840 + 981192 k + 
       1314233 k^{2}
        \nonu\\&+& 854304 k^{3} + 268548 k^{4} + 32656 k^{5} + 
       1429512 N + 4560890 k\, 
      N + 5584178 k^{2} N 
       \nonu\\&+& 3254691 k^{3} N + 894892 k^{4} N + 
       92140 k^{5} N + 3032897 N^{2} + 8774840 k\, 
      N^{2} + 9519467 k^{2} N^{2} 
       \nonu\\&+& 4762402 k^{3} N^{2} + 
       1069972 k^{4} N^{2} + 82840 k^{5} N^{2} + 3523958 N^{3} + 
       8992037 k\, 
      N^{3} 
       \nonu\\&+& 8312068 k^{2} N^{3} + 3354871 k^{3} N^{3} + 
       548812 k^{4} N^{3} + 24168 k^{5} N^{3} + 2413325 N^{4} 
        \nonu\\&+& 
       5239330 k\,N^{4} + 3892342 k^{2} N^{4}+ 1135156 k^{3} N^{4} + 
       102132 k^{4} N^{4} + 970924 N^{5} 
        \nonu\\&+& 1701031 k\,N^{5} + 913336 k^{2} N^{5} + 146388 k^{3} N^{5}+211816 N^{6} + 273880 k\, 
      N^{6} + 81468 k^{2} N^{6} 
       \nonu\\&+& 19288 N^{7} + 15312 k\,N^{7}),\nonu\\
       c_ {120} & = & - \frac {(k - N) (76 + 41 k + 41 N + 6 k\, 
      N)} {21 (2 + k + N) (5 + 4 k + 4 N + 3 k\, N)}, \nonu\\
c_ {121} & = & - \frac {60 (k - N)} {7 (5 + 4 k + 4 N + 3 k\, N)}, \qquad
c_ {122}  = \frac {24 (k - N)} {7 (5 + 4 k + 4 N + 3 k\, N)}, \nonu\\
c_ {123} & = & - \frac {1}{21 (2 + N) (2 + k + N)^{3} (5 + 4 k + 4 N + 
       3 k\, N)}(-k + N) (76 + 41 k + 41 N + 6 k\, 
      N) 
      \nonu\\&&(60 + 77 k + 22 k^{2} + 121 N + 115 k\, 
      N + 20 k^{2} N + 79 N^{2} + 42 k\, 
      N^{2} + 16 N^{3}), \nonu\\
c_ {124} & = &\frac {4 (k - N)} {(5 + 4 k + 4 N + 3 k\, N)}, \nonu\\
c_ {125} & = &\frac {4 (-k + N) (60 + 77 k + 22 k^{2} + 121 N + 
      115 k\, N + 20 k^{2} N + 79 N^{2} + 42 k\, 
     N^{2} + 16 N^{3})} {(2 + N) (2 + k + N)^{2} (5 + 4 k + 4 N + 
      3 k\, N)}, \nonu\\
c_ {126} & = & - \frac {1}{7 (2 + N) (2 + k + N)^{2} (5 + 4 k + 4 N + 
       3 k\, N)}4 (-640 - 692 k - 227 k^{2} - 10 k^{3} - 
       1292 N 
       \nonu\\&-& 1188 k\,N - 439 k^{2} N - 68 k^{3} N - 825 N^{2} - 512 k\, 
      N^{2} - 126 k^{2} N^{2} - 127 N^{3} - 14 k\, 
      N^{3} + 16 N^{4}), \nonu\\
c_ {127} & = & - \frac {32 (15 + 12 k + 2 k^{2} + 12 N + 5 k\, 
      N + 2 N^{2})} {7 (2 + k + N) (5 + 4 k + 4 N + 3 k\, N)}, \nonu\\
c_ {128} & = &\frac {16 (k - N)} {(2 + k + N) (5 + 4 k + 4 N + 3 k\, 
     N)}, 
\nonu\\c_ {129} & = & - \frac {4 i (-k + N) (20 + 21 k + 6 k^{2} + 
       25 N + 13 k\, 
      N + 7 N^{2})} {(2 + N) (2 + k + N)^{2} (5 + 4 k + 4 N + 3 k\, 
      N)},\nonu\\
 c_ {130} & = & - \frac {8 (-k + N) (20 + 21 k + 6 k^{2} + 
       25 N + 13 k\, 
      N + 7 N^{2})} {(2 + N) (2 + k + N)^{3} (5 + 4 k + 4 N + 3 k\, 
      N)},\nonu\\
 c_ {131} & = &\frac {96 (k - N) (k + 
      N)} {7 (2 + k + N)^{2} (5 + 4 k + 4 N + 3 k\, N)}, \nonu\\
c_ {132} & = & - \frac {64 (k - N) (7 + 2 k + 
       2 N)} {7 (2 + k + N)^{2} (5 + 4 k + 4 N + 3 k\, N)}, \nonu\\
c_ {133} & = & - \frac {128 (k - N) (k + N)} {7 (2 + k + N)^{2} (5 + 
       4 k + 4 N + 3 k\, N)}, \nonu\\
c_ {134} & = &\frac {8 i (-k + N) (16 + 21 k + 6 k^{2} + 19 N + 
      13 k\, N + 5 N^{2})} {(2 + N) (2 + k + N)^{3} (5 + 4 k + 4 N + 
      3 k\, N)}, \nonu\\
c_ {135} & = &\frac {8 (-k + N) (32 + 55 k + 18 k^{2} + 41 N + 35 k\, 
     N + 11 N^{2})} {(2 + N) (2 + k + N)^{4} (5 + 4 k + 4 N + 3 k\, 
     N)},\nonu\\
 c_ {136} & = & - \frac {32 i (k - N)} {(2 + k + N)^{2} (5 + 
       4 k + 4 N + 3 k\, N)}, \nonu\\
c_ {137} & = &\frac {1} {7 (2 + N) (2 + k + N)^{3} (5 + 4 k + 4 N + 
      3 k\, N)}8 i (-720 - 796 k - 259 k^{2} - 10 k^{3} - 
      1436 N 
      \nonu\\&-& 1352 k\, 
     N - 479 k^{2} N - 68 k^{3} N - 909 N^{2} - 592 k\, 
     N^{2} - 138 k^{2} N^{2} - 143 N^{3} - 26 k\, 
     N^{3} + 16 N^{4}), \nonu\\
c_ {138} & = & - \frac {16 i (40 + 122 k + 119 k^{2} + 34 k^{3} - 
       18 N + 11 k\, N + 27 k^{2} N - 42 N^{2} - 29 k\, 
      N^{2} - 8 N^{3})} {7 (2 + k + N)^{3} (5 + 4 k + 4 N + 3 k\, N)},\nonu\\
 c_ {139} & = &\frac {32 i (25 + 20 k + 4 k^{2} + 20 N + 7 k\, 
     N + 4 N^{2})} {7 (2 + k + N)^{2} (5 + 4 k + 4 N + 3 k\, N)}, \nonu\\
c_ {140} & = &\frac {32 i (-k + N) (13 k + 6 k^{2} + 3 N + 9 k\, 
     N + N^{2})} {(2 + N) (2 + k + N)^{4} (5 + 4 k + 4 N + 3 k\, N)}, \nonu\\
c_ {141} & = & - \frac {64 i (k - N) (11 + 7 k + 
       7 N)} {7 (2 + k + N)^{2} (5 + 4 k + 4 N + 3 k\, N)}, \nonu\\
c_ {142} & = &\frac {32 i (k - N) (20 + 7 k + 
      7 N)} {7 (2 + k + N)^{2} (5 + 4 k + 4 N + 3 k\, N)}, \nonu\\
c_ {143} & = &\frac {1} {21 (2 + N) (2 + k + N)^{3} (5 + 4 k + 4 N + 3 k\, N)}(-1920 - 1936 k - 40 k^{2} + 683 k^{3} + 
     234 k^{4} 
     \nonu\\&-& 5936 N - 5328 k\, 
    N - 1677 k^{2} N - 349 k^{3} N - 96 k^{4} N - 5384 N^{2} - 
     2187 k\, 
    N^{2} + 295 k^{2} N^{2} 
    \nonu\\&+& 140 k^{3} N^{2} + 24 k^{4} N^{2} - 
     851 N^{3} + 1917 k\, 
    N^{3} + 1260 k^{2} N^{3} + 180 k^{3} N^{3} + 783 N^{4} + 1144 k\, 
    N^{4} 
    \nonu\\&+& 276 k^{2} N^{4} + 240 N^{5} + 96 k\, 
    N^{5}), \nonu\\
c_ {144} & = &\frac {1}{21 (2 + N) (2 + k + N)^{4} (5 + 4 k + 4 N + 3 k\, N)}2 (-480 - 2384 k - 2920 k^{2} - 1413 k^{3} 
\nonu\\&-& 246 k^{4} + 416 N - 2032 k\, 
     N - 2505 k^{2} N - 779 k^{3} N - 36 k^{4} N + 1784 N^{2} + 
      213 k\, N^{2} 
      \nonu\\&-& 497 k^{2} N^{2} - 78 k^{3} N^{2} + 1209 N^{3} + 
      323 k\, N^{3} - 36 k^{2} N^{3} + 239 N^{4} + 6 k\, 
     N^{4}), \nonu\\
c_ {145} & = & - \frac {128} {7 (2 + k + N)^{2}}, \qquad
c_ {146}  = \frac {96} {7 (2 + k + N)^{2}}, \nonu\\
c_ {147} & = & - \frac {32} {7 (2 + k + N)^{2}}, \qquad
c_ {148}  = \frac {48} {7 (2 + k + N)^{2}}, \nonu\\
c_ {149} & = & - \frac{1}{21 (2 + N) (2 + k + N)^{3} (5 + 4 k + 4 N + 3 k\, N)}i (3360 + 6368 k + 4384 k^{2} + 1509 k^{3}
\nonu\\&+& 246 k^{4} + 5008 N + 8680 k\, 
      N + 4533 k^{2} N + 899 k^{3} N + 36 k^{4} N + 1672 N^{2} + 
       3411 k\, 
      N^{2} 
      \nonu\\&+& 1289 k^{2} N^{2} + 114 k^{3} N^{2} - 381 N^{3} + 
       397 k\, N^{3} + 108 k^{2} N^{3} - 191 N^{4} + 30 k\, 
      N^{4}), \nonu\\
c_ {150} & = &\frac {1}{21 (2 + N) (2 + k + N)^{3} (5 + 4 k + 4 N + 3 k\, N)}2 i (8160 + 11488 k + 3968 k^{2} - 357 k^{3} 
\nonu\\&-& 
      246 k^{4} + 18608 N + 23864 k\, 
     N + 8283 k^{2} N + 541 k^{3} N - 36 k^{4} N + 15608 N^{2} + 
      17037 k\, 
     N^{2} 
     \nonu\\&+& 4759 k^{2} N^{2} + 318 k^{3} N^{2} + 5709 N^{3} + 
      4787 k\, N^{3} + 756 k^{2} N^{3} + 767 N^{4} + 402 k\, 
     N^{4}), \nonu\\
c_ {151} & = & - \frac {1} {21 (2 + N) (2 + k + N)^{3} (5 + 4 k + 4 N + 3 k\, N)}i (7200 + 12320 k + 7168 k^{2} + 1893 k^{3} 
\nonu\\&+&
       246 k^{4} + 12880 N + 19528 k\, 
      N + 8805 k^{2} N + 1379 k^{3} N + 36 k^{4} N + 7432 N^{2} + 
       10227 k\, 
      N^{2} 
      \nonu\\&+& 3305 k^{2} N^{2} + 258 k^{3} N^{2} + 1395 N^{3} + 
       2125 k\, N^{3} + 396 k^{2} N^{3} + N^{4} + 174 k\, 
      N^{4}), \nonu\\
c_ {152} & = & - \frac {4 (k - N) (106 + 65 k + 65 N + 24 k\, 
      N)} {21 (2 + k + N)^{2} (5 + 4 k + 4 N + 3 k\, N)}, \nonu\\
c_ {153} & = &\frac {8 (k - N) (74 + 79 k + 79 N + 84 k\, 
     N)} {21 (2 + k + N)^{2} (5 + 4 k + 4 N + 3 k\, N)}, \nonu\\
c_ {154} & = & - \frac {4 (k - N) (166 + 113 k + 113 N + 60 k\, 
      N)} {21 (2 + k + N)^{2} (5 + 4 k + 4 N + 3 k\, N)}, \nonu\\
c_ {155} & = &\frac {96} {7 (2 + k + N)^{2}}, \qquad
c_ {156}  =  - \frac {144} {7 (2 + k + N)^{2}}, \nonu\\
c_ {157} & = & - \frac {8 (k - N) (-137 - 74 k + 4 k^{2} - 74 N + 
       5 k\, N + 12 k^{2} N + 4 N^{2} + 12 k\, 
      N^{2})} {21 (2 + k + N)^{3} (5 + 4 k + 4 N + 3 k\, N)}, \nonu\\
c_ {158} & = & - \frac {32 (k - N) (29 + 55 k + 15 k^{2} + 55 N + 
       87 k\, N + 18 k^{2} N + 15 N^{2} + 18 k\, 
      N^{2})} {21 (2 + k + N)^{3} (5 + 4 k + 4 N + 3 k\, N)}, \nonu\\
c_ {159} & = &\frac {16 (k - N) (211 + 230 k + 54 k^{2} + 230 N + 
      195 k\, N + 27 k^{2} N + 54 N^{2} + 27 k\, 
     N^{2})} {21 (2 + k + N)^{3} (5 + 4 k + 4 N + 3 k\, N)}, \nonu\\
c_ {160} & = & - \frac {32 (k - N) (15 + 11 k + k^{2} + 11 N + 11 k\, 
      N + 3 k^{2} N + N^{2} + 3 k\, 
      N^{2})} {21 (2 + k + N)^{3} (5 + 4 k + 4 N + 3 k\, N)}, \nonu\\
c_ {161} & = & - \frac {4 i (-k + N) (76 + 41 k + 41 N + 6 k\, 
      N) (16 + 21 k + 6 k^{2} + 19 N + 13 k\, 
      N + 5 N^{2})} {21 (2 + N) (2 + k + N)^{4} (5 + 4 k + 4 N + 
       3 k\, N)}, \nonu\\
c_ {162} & = &\frac {1}{21 (2 + N) (2 + k + N)^{4} (5 + 4 k + 4 N + 3 k\, N)}8 i (-k + N) (-736 - 1628 k - 1125 k^{2}\nonu\\&-& 
      246 k^{3} - 1236 N - 1740 k\, 
     N - 665 k^{2} N - 36 k^{3} N - 655 N^{2} - 372 k\, 
     N^{2} - 6 k^{2} N^{2} - 109 N^{3} 
     \nonu\\&+& 42 k\, 
     N^{3}), \nonu\\
c_ {163} & = & - \frac {1}{21 (2 + N) (2 + k + N)^{4} (5 + 4 k + 4 N + 3 k\, N)}2 i (-k + N) (1936 + 3188 k + 1605 k^{2} 
\nonu\\&+& 
       246 k^{3} + 3396 N + 4200 k\, 
      N + 1265 k^{2} N + 36 k^{3} N + 1915 N^{2} + 1572 k\, 
      N^{2} + 186 k^{2} N^{2} 
      \nonu\\&+& 349 N^{3} + 138 k\, 
      N^{3}),\nonu\\ 
c_ {164} & = & - \frac {1}{21 (2 + N) (2 + k + N)^{4} (5 + 4 k + 4 N + 3 k\, N)}2 (1440 + 4112 k + 3928 k^{2} + 1605 k^{3}
\nonu\\&+&
       246 k^{4} + 1792 N + 5488 k\, 
      N + 4161 k^{2} N + 1019 k^{3} N + 36 k^{4} N + 88 N^{2} + 
       2235 k\, 
      N^{2} 
      \nonu\\&+& 1361 k^{2} N^{2} + 150 k^{3} N^{2} - 513 N^{3} + 
       397 k\, N^{3} + 180 k^{2} N^{3} - 143 N^{4} + 66 k\, 
      N^{4}),\nonu\\ 
c_ {165} & = &\frac {1}{21 (2 + N) (2 + k + N)^{4} (5 + 4 k + 4 N + 3 k\, N)}2 (1440 + 1072 k - 904 k^{2} - 1029 k^{3} - 
      246 k^{4} 
      \nonu\\&+& 4832 N + 4880 k\, 
     N + 807 k^{2} N - 299 k^{3} N - 36 k^{4} N + 5528 N^{2} + 
      5109 k\, 
     N^{2} + 1231 k^{2} N^{2} 
     \nonu\\&+& 66 k^{3} N^{2} + 2601 N^{3} + 
      1763 k\, N^{3} + 252 k^{2} N^{3} + 431 N^{4} + 150 k\, 
     N^{4}), \nonu\\
c_ {166} & = & - \frac {1}{21 (2 + N) (2 + k + N)^{4} (5 + 4 k + 4 N + 3 k\, N)}2 (4320 + 9296 k + 6952 k^{2} + 2181 k^{3} 
\nonu\\&+& 246 k^{4} +8416 N + 15856 k\, 
      N + 9129 k^{2} N + 1739 k^{3} N + 36 k^{4} N + 5704 N^{2} + 
       9579 k\, N^{2} 
      \nonu\\&+& 3953 k^{2} N^{2} +366 k^{3} N^{2} + 1575 N^{3} + 
       2557 k\, N^{3} + 612 k^{2} N^{3} + 145 N^{4} + 282 k\, 
      N^{4}),\nonu\\ 
c_ {167} & = &\frac {4 (-k + N) (76 + 41 k + 41 N + 6 k\, 
     N) (20 + 21 k + 6 k^{2} + 25 N + 13 k\, 
     N + 7 N^{2})} {21 (2 + N) (2 + k + N)^{4} (5 + 4 k + 4 N + 3 k\, 
     N)}, \nonu\\
c_ {168} & = &\frac {1}{21 (2 + N) (2 + k + N)^{4} (5 + 4 k + 4 N + 3 k\, N)}4 (7200 + 11440 k + 5144 k^{2} + 
      123 k^{3} - 246 k^{4} 
      \nonu\\&+&18080 N + 25616 k\, 
     N + 10743 k^{2} N + 1141 k^{3} N - 36 k^{4} N + 16760 N^{2} + 
      19797 k\, 
     N^{2} 
     \nonu\\&+& 6415 k^{2} N^{2} + 498 k^{3} N^{2} + 6777 N^{3} + 
      6083 k\, N^{3} + 1116 k^{2} N^{3} + 1007 N^{4} + 582 k\, 
     N^{4}), \nonu\\
c_ {169} & = & - \frac {2 (-k + N) (76 + 41 k + 41 N + 6 k\, 
      N) (32 + 55 k + 18 k^{2} + 41 N + 35 k\, 
      N + 11 N^{2})} {21 (2 + N) (2 + k + N)^{5} (5 + 4 k + 4 N + 
       3 k\, N)}, \nonu\\
c_ {170} & = & - \frac{1}{21 (2 + N) (2 + k + N)^{4} (5 + 4 k + 4 N + 3 k\, N)}2 i (-2240 - 1792 k + 560 k^{2} + 907 k^{3} 
\nonu\\&+& 
       234 k^{4} - 5952 N - 4728 k\, 
      N - 1089 k^{2} N - 261 k^{3} N - 96 k^{4} N - 4712 N^{2} - 
       1635 k\, 
      N^{2} 
      \nonu\\&+& 103 k^{2} N^{2} - 16 k^{3} N^{2} + 24 k^{4} N^{2} - 
       327 N^{3} + 2037 k\, 
      N^{3} + 948 k^{2} N^{3} + 108 k^{3} N^{3} + 895 N^{4} 
      \nonu\\&+& 
       1132 k\, N^{4} + 204 k^{2} N^{4} + 240 N^{5} + 96 k\, 
      N^{5}),\nonu\\ 
c_ {171} & = & - \frac {1}{7 (2 + N) (2 + k + N)^{4} (5 + 4 k + 4 N + 3 k\, N)}4 i (1360 + 3168 k + 2144 k^{2} + 435 k^{3} - 
       6 k^{4} 
       \nonu\\&+& 3848 N + 8248 k\, 
      N + 4713 k^{2} N + 628 k^{3} N - 66 k^{4} N + 4488 N^{2} + 
       8637 k\, 
      N^{2} + 4267 k^{2} N^{2} 
      \nonu\\&+& 571 k^{3} N^{2} + 12 k^{4} N^{2} + 
       2791 N^{3} + 4408 k\, 
      N^{3} + 1692 k^{2} N^{3} + 162 k^{3} N^{3} + 911 N^{4} + 
       971 k\, N^{4} 
       \nonu\\&+&210 k^{2} N^{4} + 120 N^{5} + 48 k\, 
      N^{5}), \nonu\\
c_ {172} & = & - \frac{1}{7 (2 + N) (2 + k + N)^{4} (5 + 4 k + 4 N + 3 k\, N)}2 i (-5760 - 12608 k - 9840 k^{2} - 
       3007 k^{3} 
       \nonu\\&-&258 k^{4} - 12448 N - 25352 k\, 
      N - 17787 k^{2} N - 4331 k^{3} N - 168 k^{4} N - 8728 N^{2} - 
       17385 k\, 
      N^{2} 
      \nonu\\&-& 11259 k^{2} N^{2} - 2020 k^{3} N^{2} + 24 k^{4} N^{2} - 
       1277 N^{3} - 3989 k\, 
      N^{3} - 2820 k^{2} N^{3} - 324 k^{3} N^{3} 
      \nonu\\&+& 829 N^{4} + 
       160 k\, N^{4} - 228 k^{2} N^{4} + 240 N^{5} + 96 k\, 
      N^{5}), \nonu\\
c_ {173} & = & - \frac {1}{21 (2 + k + N)^{4} (5 + 4 k + 4 N + 3 k\, N)}4 i (920 + 1136 k - 40 k^{2} - 441 k^{3} - 
       126 k^{4} 
       \nonu\\&+& 2656 N + 3656 k\, 
      N + 1346 k^{2} N + 109 k^{3} N + 12 k^{4} N + 2528 N^{2} + 
       3017 k\, 
      N^{2} + 972 k^{2} N^{2} 
      \nonu\\&+& 90 k^{3} N^{2} + 958 N^{3} + 821 k\, 
      N^{3} + 138 k^{2} N^{3} + 120 N^{4} + 48 k\, 
      N^{4}), \nonu\\
c_ {174} & = &\frac {16 i (k - N) (136 + 89 k + 89 N + 42 k\, 
     N)} {21 (2 + k + N)^{3} (5 + 4 k + 4 N + 3 k\, N)}, \nonu\\
c_ {175} & = & - \frac {8 i (k - N) (74 + 79 k + 79 N + 84 k\, 
      N)} {21 (2 + k + N)^{3} (5 + 4 k + 4 N + 3 k\, N)}, \qquad
c_ {176}  = \frac {32 i (k - N)} {(2 + k + N)^{3}}, \nonu\\
c_ {177} & = & - \frac {8 i (-k + N) (76 + 41 k + 41 N + 6 k\, 
      N) (13 k + 6 k^{2} + 3 N + 9 k\, 
      N + N^{2})} {21 (2 + N) (2 + k + N)^{5} (5 + 4 k + 4 N + 3 k\, 
      N)},\nonu\\
 c_ {178} & = &\frac {4 i (k - N) (332 + 506 k + 
      188 k^{2} + 506 N + 442 k\, N + 69 k^{2} N + 188 N^{2} + 69 k\, 
     N^{2})} {21 (2 + k + N)^{3} (5 + 4 k + 4 N + 3 k\, N)}, \nonu\\
c_ {179} & = &\frac {8 i (-k + N) (16 + 21 k + 6 k^{2} + 19 N + 
      13 k\, N + 5 N^{2})} {(2 + N) (2 + k + N)^{3} (5 + 4 k + 4 N + 
      3 k\, N)},\nonu\\
 c_ {180} & = & - \frac {32} {7 (2 + k + N)^{2}}, \qquad
c_ {181}  = \frac {80} {7 (2 + k + N)^{2}}, \nonu\\
c_ {182} & = &\frac {1}{21 (2 + N) (2 + k + N)^{4} (5 + 4 k + 4 N + 3 k\, N)}4 (480 - 656 k - 1912 k^{2} - 1221 k^{3} - 
      246 k^{4} 
      \nonu\\&+& 2624 N + 1424 k\, 
     N - 849 k^{2} N - 539 k^{3} N - 36 k^{4} N + 3656 N^{2} + 
      2661 k\, 
     N^{2} + 367 k^{2} N^{2} 
     \nonu\\&-& 6 k^{3} N^{2} + 1905 N^{3} + 1043 k\, 
     N^{3} + 108 k^{2} N^{3} + 335 N^{4} + 78 k\, 
     N^{4}), \nonu\\
c_ {183} & = &\frac {1}{21 (2 + N) (2 + k + N)^{4} (5 + 4 k + 4 N + 3 k\, N)}4 (4320 + 6256 k + 2120 k^{2} - 453 k^{3} - 
      246 k^{4} 
      \nonu\\&+& 11456 N + 15248 k\, 
     N + 5775 k^{2} N + 421 k^{3} N - 36 k^{4} N + 11144 N^{2} + 
      12453 k\, 
     N^{2} 
     \nonu\\&+& 3823 k^{2} N^{2} + 282 k^{3} N^{2} + 4689 N^{3} + 
      3923 k\, N^{3} + 684 k^{2} N^{3} + 719 N^{4} + 366 k\, 
     N^{4}), \nonu\\
c_ {184} & = &\frac {1}{21 (2 + N) (2 + k + N)^{4} (5 + 4 k + 4 N + 3 k\, N)}2 (1440 + 1072 k - 904 k^{2} - 1029 k^{3} - 
      246 k^{4} 
      \nonu\\&+& 4832 N + 4880 k\, 
     N + 807 k^{2} N - 299 k^{3} N - 36 k^{4} N + 5528 N^{2} + 
      5109 k\, 
     N^{2} + 1231 k^{2} N^{2}
     \nonu\\&+& 66 k^{3} N^{2} + 2601 N^{3} + 
      1763 k\, N^{3} + 252 k^{2} N^{3} + 431 N^{4} + 150 k\, 
     N^{4}), \nonu\\
c_ {185} & = & - \frac{1}{21 (2 + N) (2 + k + N)^{4} (5 + 4 k + 4 N + 3 k\, N)}4 (1440 + 4112 k + 3928 k^{2} + 1605 k^{3} 
\nonu\\&+& 
       246 k^{4} + 1792 N + 5488 k\, 
      N + 4161 k^{2} N + 1019 k^{3} N + 36 k^{4} N + 88 N^{2} + 
       2235 k\, 
      N^{2} 
      \nonu\\&+& 1361 k^{2} N^{2} + 150 k^{3} N^{2} - 513 N^{3} + 
       397 k\, N^{3} + 180 k^{2} N^{3} - 143 N^{4} + 66 k\, 
      N^{4}),\nonu\\ 
c_ {186} & = & - \frac{1}{21 (2 + N) (2 + k + N)^{4} (5 + 4 k + 4 N + 3 k\, N)}2 (4320 + 9296 k + 6952 k^{2} + 2181 k^{3} 
\nonu\\&+&
       246 k^{4} + 8416 N + 15856 k\, 
      N + 9129 k^{2} N + 1739 k^{3} N + 36 k^{4} N + 5704 N^{2} + 
       9579 k\, 
      N^{2} 
      \nonu\\&+&3953 k^{2} N^{2} + 366 k^{3} N^{2} + 1575 N^{3} + 
       2557 k\, N^{3} + 612 k^{2} N^{3} + 145 N^{4} + 282 k\, 
      N^{4}),\nonu\\ 
c_ {187} & = & - \frac {1}{21 (2 + N) (2 + k + N)^{4} (5 + 4 k + 4 N + 3 k\, N)}2 i (-k + N) (1456 + 2564 k + 1413 k^{2} 
\nonu\\&+& 
       246 k^{3} + 2532 N + 3216 k\, 
      N + 1025 k^{2} N + 36 k^{3} N + 1411 N^{2} + 1092 k\, 
      N^{2} + 114 k^{2} N^{2} 
      \nonu\\&+&253 N^{3} + 66 k\, 
      N^{3}),\nonu\\ 
c_ {188} & = &\frac {1}{21 (2 + N) (2 + k + N)^{4} (5 + 4 k + 4 N + 3 k\, N)}2 i (-k + N) (-976 - 1940 k - 1221 k^{2} 
\nonu\\&-& 
      246 k^{3} - 1668 N - 2232 k\, 
     N - 785 k^{2} N - 36 k^{3} N - 907 N^{2} - 612 k\, 
     N^{2} - 42 k^{2} N^{2} - 157 N^{3} 
     \nonu\\&+& 6 k\, 
     N^{3}), \nonu\\
c_ {189} & = & - \frac {1} {21 (2 + N) (2 + k + N)^{4} (5 + 4 k + 4 N + 3 k\, N)}2 i (-k + N) (1936 + 3188 k + 1605 k^{2} 
\nonu\\&+& 
       246 k^{3} + 3396 N + 4200 k\, 
      N + 1265 k^{2} N + 36 k^{3} N + 1915 N^{2} + 1572 k\, 
      N^{2} + 186 k^{2} N^{2} 
      \nonu\\&+& 349 N^{3} + 138 k\, 
      N^{3}),\nonu\\ 
c_ {190} & = & - \frac {4 i (-k + N) (76 + 41 k + 41 N + 6 k\, 
      N) (16 + 21 k + 6 k^{2} + 19 N + 13 k\, 
      N + 5 N^{2})} {21 (2 + N) (2 + k + N)^{4} (5 + 4 k + 4 N + 
       3 k\, N)}, \nonu\\
c_ {191} & = &\frac{1}{21 (2 + N) (2 + k + N)^{4} (5 + 4 k + 4 N + 3 k\, N)}4 i (-k + N) (-16 - 692 k - 837 k^{2} - 
      246 k^{3} 
      \nonu\\&+& 60 N - 264 k\, 
     N - 305 k^{2} N - 36 k^{3} N + 101 N^{2} + 348 k\, 
     N^{2} + 102 k^{2} N^{2} + 35 N^{3} + 150 k\, 
     N^{3}), \nonu\\
c_ {192} & = & - \frac{1}{21 (2 + N) (2 + k + N)^{5} (5 + 4 k + 4 N + 3 k\, N)}16 i (-k + N) (240 + 1420 k + 1241 k^{2} 
\nonu\\&+& 
       294 k^{3} + 780 N + 2204 k\, 
      N + 1107 k^{2} N + 96 k^{3} N + 667 N^{2} + 1040 k\, 
      N^{2} + 270 k^{2} N^{2} 
      \nonu\\&+& 18 k^{3} N^{2} + 215 N^{3} + 186 k\, 
      N^{3} + 36 k^{2} N^{3} + 24 N^{4} + 18 k\, 
      N^{4}),\nonu\\ 
c_ {193} & = & - \frac {1}{21 (2 + N) (2 + k + N)^{5} (5 + 4 k + 4 N + 3 k\, N)}8 i (-k + N) (480 + 1852 k + 1493 k^{2} 
\nonu\\&+& 
       342 k^{3} + 1332 N + 3068 k\, 
      N + 1521 k^{2} N + 156 k^{3} N + 1135 N^{2} + 1652 k\, 
      N^{2} + 486 k^{2} N^{2} 
      \nonu\\&+& 36 k^{3} N^{2} + 389 N^{3} + 366 k\, 
      N^{3} + 72 k^{2} N^{3} + 48 N^{4} + 36 k\, 
      N^{4}),\nonu\\ 
c_ {194} & = & - \frac {1}{21 (2 + N) (2 + k + N)^{5} (5 + 4 k + 4 N + 3 k\, N)}8 i (-k + N) (3360 + 7036 k + 4517 k^{2} 
\nonu\\&+& 
       918 k^{3} + 7956 N + 13436 k\, 
      N + 6489 k^{2} N + 876 k^{3} N + 6751 N^{2} + 8996 k\, 
      N^{2} + 3078 k^{2} N^{2} 
      \nonu\\&+& 252 k^{3} N^{2} + 2477 N^{3} + 
       2526 k\, N^{3} + 504 k^{2} N^{3} + 336 N^{4} + 252 k\, 
      N^{4}),\nonu\\ 
c_ {195} & = & - \frac {2 i (-k + N) (16 + 21 k + 6 k^{2} + 19 N + 
       13 k\, N + 5 N^{2})} {(2 + N) (2 + k + N)^{2} (5 + 4 k + 4 N + 
       3 k\, N)}, \nonu\\
c_ {196} & = & - \frac {4 (-k + N) (13 k + 6 k^{2} + 3 N + 9 k\, 
      N + N^{2})} {(2 + N) (2 + k + N)^{3} (5 + 4 k + 4 N + 3 k\, N)},\nonu\\
 c_ {197} & = &\frac {4 i (-k + N) (20 + 21 k + 6 k^{2} + 25 N + 
      13 k\, N + 7 N^{2})} {(2 + N) (2 + k + N)^{3} (5 + 4 k + 4 N + 
      3 k\, N)}, \nonu\\
c_ {198} & = & - \frac {16 (5 + 4 k + k^{2} + 4 N + k\, 
      N + N^{2})} {7 (2 + k + N)^{2} (5 + 4 k + 4 N + 3 k\, N)}, \nonu\\
c_ {199} & = & - \frac {1}{7 (2 + N) (2 + k + N)^{3} (5 + 4 k + 4 N + 
       3 k\, N)}8 (40 + 52 k - 11 k^{2} - 10 k^{3} + 72 N + 
       66 k\, N
       \nonu\\&-& 64 k^{2} N - 26 k^{3} N + 85 N^{2} + 77 k\, 
      N^{2} - 17 k^{2} N^{2} + 65 N^{3} + 39 k\, 
      N^{3} + 16 N^{4}), \nonu\\
c_ {200} & = &\frac {8 (120 + 226 k + 135 k^{2} + 26 k^{3} + 86 N + 
      155 k\, N + 59 k^{2} N - 26 N^{2} + 3 k\, 
     N^{2} - 16 N^{3})} {7 (2 + k + N)^{3} (5 + 4 k + 4 N + 3 k\, N)},\nonu\\
 c_ {201} & = & - \frac {4 (-k + N) (32 + 29 k + 6 k^{2} + 35 N + 
       17 k\, N + 9 N^{2})} {(2 + N) (2 + k + N)^{3} (5 + 4 k + 4 N + 
       3 k\, N)}, \nonu\\
c_ {202} & = &\frac {8 i (-k + N) (32 + 55 k + 18 k^{2} + 41 N + 
      35 k\, N + 11 N^{2})} {3 (2 + N) (2 + k + N)^{4} (5 + 4 k + 
      4 N + 3 k\, N)}, \nonu\\
c_ {203} & = & - \frac {32 i (k - N)^{2}} {21 (2 + k + N)^{3} (5 + 
       4 k + 4 N + 3 k\, N)}, \nonu\\
c_ {204} & = &\frac {1}{21 (2 + N) (2 + k + N)^{4} (5 + 4 k + 4 N + 
      3 k\, N)}16 i (-k + N) (140 + 169 k + 46 k^{2} + 293 N
      \nonu\\&+& 
      263 k\, N + 44 k^{2} N + 195 N^{2} + 98 k\, 
     N^{2} + 40 N^{3}),\nonu\\
c_ {205} & = & - \frac {48 i} {7 (2 + k + N)^{2}}, \qquad
c_ {206}  = \frac {8 i} {7 (2 + k + N)^{2}}, \nonu\\
c_ {207} & = &\frac {i (-k + N) (76 + 41 k + 41 N + 6 k\, 
     N) (16 + 21 k + 6 k^{2} + 19 N + 13 k\, 
     N + 5 N^{2})} {42 (2 + N) (2 + k + N)^{3} (5 + 4 k + 4 N + 3 k\, 
     N)},\nonu\\
c_ {208} & = & - \frac {32 i} {7 (2 + k + N)^{2}}, \qquad
c_ {209}  = \frac {24 i} {7 (2 + k + N)^{2}}, \nonu\\
c_ {210} & = &\frac {1} {21 (2 + N) (2 + k + N)^{4} (5 + 4 k + 4 N + 3 k\, N)}(-k + N) (240 + 1300 k + 1085 k^{2} + 
      246 k^{3} 
      \nonu\\&+& 660 N + 1832 k\, 
     N + 813 k^{2} N + 36 k^{3} N + 451 N^{2} + 668 k\, 
     N^{2} + 90 k^{2} N^{2} + 89 N^{3} + 42 k\, 
     N^{3}), \nonu\\
c_ {211} & = & - \frac{1}{21 (2 + N) (2 + k + N)^{4} (5 + 4 k + 4 N + 3 k\, N)}(-k + N) (720 - 52 k - 701 k^{2} - 
       246 k^{3} 
       \nonu\\&+& 1068 N + 136 k\, 
      N - 333 k^{2} N - 36 k^{3} N + 557 N^{2} + 292 k\, 
      N^{2} + 54 k^{2} N^{2} + 103 N^{3} + 102 k\, 
      N^{3}), \nonu\\
c_ {212} & = &\frac {1}{21 (2 + N) (2 + k + N)^{4} (5 + 4 k + 4 N + 3 k\, N)}2 (-k + N) (2640 + 4420 k + 2045 k^{2} 
\nonu\\&+& 
      246 k^{3} + 4980 N + 6752 k\, 
     N + 2013 k^{2} N + 36 k^{3} N + 2971 N^{2} + 3068 k\, 
     N^{2} + 450 k^{2} N^{2} 
     \nonu\\&+& 569 N^{3} + 402 k\, 
     N^{3}), \nonu\\
c_ {213} & = & - \frac{1}{21 (2 + N) (2 + k + N)^{4} (5 + 4 k + 4 N + 3 k\, N)}2 (-k + N) (240 - 676 k - 893 k^{2} - 
       246 k^{3} 
       \nonu\\&+& 204 N - 848 k\, 
      N - 573 k^{2} N - 36 k^{3} N + 53 N^{2} - 188 k\, 
      N^{2} - 18 k^{2} N^{2} + 7 N^{3} + 30 k\, 
      N^{3}), \nonu\\
c_ {214} & = &\frac {1}{21 (2 + N) (2 + k + N)^{4} (5 + 4 k + 4 N + 3 k\, N)}2 (80 + 264 k + 52 k^{2} - 77 k^{3} - 22 k^{4} + 
      304 N 
      \nonu\\&+&516 k\, 
     N - 493 k^{2} N - 576 k^{3} N - 122 k^{4} N + 944 N^{2} + 
      1337 k\, 
     N^{2} - 13 k^{2} N^{2} - 215 k^{3} N^{2} 
     \nonu\\&-& 12 k^{4} N^{2} + 
      1233 N^{3} + 1380 k\, 
     N^{3} + 280 k^{2} N^{3} + 6 k^{3} N^{3} + 639 N^{4} + 449 k\, 
     N^{4} + 54 k^{2} N^{4} 
     \nonu\\&+& 112 N^{5} + 24 k\, 
     N^{5}), \nonu\\
c_ {215} & = &\frac{1}{21 (2 + 
      N) (2 + k + N)^{4} (5 + 4 k + 4 N + 3 k\, N)}2 (-4080 - 9584 k - 8680 k^{2} - 3468 k^{3} 
      \nonu\\&-& 
      504 k^{4} - 8584 N - 18856 k\, 
     N - 15678 k^{2} N - 5471 k^{3} N - 642 k^{4} N - 5296 N^{2} - 
      11700 k\, 
     N^{2} 
     \nonu\\&-& 8768 k^{2} N^{2} - 2253 k^{3} N^{2} - 108 k^{4} N^{2} + 
      126 N^{3} - 1957 k\, 
     N^{3} - 1632 k^{2} N^{3} - 234 k^{3} N^{3} 
     \nonu\\&+& 1052 N^{4} + 
      183 k\, N^{4} - 90 k^{2} N^{4} + 240 N^{5}), \nonu\\
c_ {216} & = &\frac {1}{21 (2 + N) (2 + k + N)^{4} (5 + 4 k + 4 N + 3 k\, N)}2 (7440 + 12872 k + 6892 k^{2} + 999 k^{3} - 
      78 k^{4} 
      \nonu\\&+& 22672 N + 35476 k\, 
     N + 17535 k^{2} N + 3062 k^{3} N + 162 k^{4} N + 27400 N^{2} + 
      36957 k\, 
     N^{2} 
     \nonu\\&+&15227 k^{2} N^{2} + 2295 k^{3} N^{2} + 144 k^{4} N^{2} + 
      16461 N^{3} + 17974 k\, 
     N^{3} + 5400 k^{2} N^{3} + 486 k^{3} N^{3} 
     \nonu\\&+& 4903 N^{4} + 
      3879 k\, N^{4} + 630 k^{2} N^{4} + 576 N^{5} + 252 k\, 
     N^{5}), \nonu\\
c_ {217} & = & - \frac{1}{21 (2 + k + N)^{4} (5 + 4 k + 4 N + 3 k\, N)}2 (680 + 1984 k + 2016 k^{2} + 865 k^{3} + 
       134 k^{4} 
       \nonu\\&+& 464 N + 1800 k\, 
      N + 1622 k^{2} N + 443 k^{3} N + 12 k^{4} N - 552 N^{2} - 
       49 k\, N^{2} + 212 k^{2} N^{2} 
       \nonu\\&+& 30 k^{3} N^{2} - 534 N^{3} - 
       269 k\, N^{3} - 18 k^{2} N^{3} - 112 N^{4} - 24 k\, 
      N^{4}), \nonu\\
c_ {218} & = &\frac{1}{21 (2 + N) (2 + k + N)^{4} (5 + 4 k + 4 N + 3 k\, N)}i (1440 + 4112 k + 3928 k^{2} + 1605 k^{3} 
\nonu\\&+& 
      246 k^{4} + 1792 N + 5488 k\, 
     N + 4161 k^{2} N + 1019 k^{3} N + 36 k^{4} N + 88 N^{2} + 
      2235 k\, 
     N^{2} 
     \nonu\\&+& 1361 k^{2} N^{2} + 150 k^{3} N^{2} - 513 N^{3} + 397 k\, 
     N^{3} + 180 k^{2} N^{3} - 143 N^{4} + 66 k\, 
     N^{4}), \nonu\\
c_ {219} & = &\frac {1}{21 (2 + N) (2 + k + N)^{4} (5 + 4 k + 4 N + 3 k\, N)}i (2400 + 5840 k + 4936 k^{2} + 1797 k^{3}
\nonu\\&+& 
      246 k^{4} + 4000 N + 8944 k\, 
     N + 5817 k^{2} N + 1259 k^{3} N + 36 k^{4} N + 1960 N^{2} + 
      4683 k\, 
     N^{2} 
     \nonu\\&+& 2225 k^{2} N^{2} + 222 k^{3} N^{2} + 183 N^{3} + 
      1117 k\, N^{3} + 324 k^{2} N^{3} - 47 N^{4} + 138 k\, 
     N^{4}), \nonu\\
c_ {220} & = & - \frac{1}{21 (2 + N) (2 + k + N)^{4} (5 + 4 k + 4 N + 3 k\, N)}i (3360 + 4528 k + 1112 k^{2} - 645 k^{3} 
\nonu\\&-& 
       246 k^{4} + 9248 N + 11792 k\, 
      N + 4119 k^{2} N + 181 k^{3} N - 36 k^{4} N + 9272 N^{2} + 
       10005 k\, 
      N^{2}
      \nonu\\&+& 2959 k^{2} N^{2} + 210 k^{3} N^{2} + 3993 N^{3} + 
       3203 k\, N^{3} + 540 k^{2} N^{3} + 623 N^{4} + 294 k\, 
      N^{4}),\nonu\\ 
c_ {221} & = & - \frac {1}{21 (2 + N) (2 + k + N)^{4} (5 + 4 k + 4 N + 3 k\, N)}2 i (4800 + 7120 k + 2624 k^{2} - 357 k^{3}
\nonu\\&-& 
       246 k^{4} + 12560 N + 16976 k\, 
      N + 6603 k^{2} N + 541 k^{3} N - 36 k^{4} N + 12080 N^{2} + 
       13677 k\, 
      N^{2} 
      \nonu\\&+&4255 k^{2} N^{2} + 318 k^{3} N^{2} + 5037 N^{3} + 
       4283 k\, N^{3} + 756 k^{2} N^{3} + 767 N^{4} + 402 k\, 
      N^{4}),\nonu\\ 
c_ {222} & = & - \frac{1}{21 (2 + N) (2 + k + N)^{4} (5 + 4 k + 4 N + 3 k\, N)}2 i (480 - 656 k - 1912 k^{2} - 1221 k^{3}
\nonu\\&-& 
       246 k^{4} + 2624 N + 1424 k\, 
      N - 849 k^{2} N - 539 k^{3} N - 36 k^{4} N + 3656 N^{2} + 
       2661 k\, 
      N^{2} 
      \nonu\\&+& 367 k^{2} N^{2} - 6 k^{3} N^{2} + 1905 N^{3} + 1043 k\, 
      N^{3} + 108 k^{2} N^{3} + 335 N^{4} + 78 k\, 
      N^{4}),\nonu\\ 
c_ {223} & = &\frac {16 i} {7 (2 + k + N)^{2}}, \qquad
c_ {224}  =  - \frac {96 i} {7 (2 + k + N)^{2}}, \nonu\\
c_ {225} & = &\frac {1}{21 (2 + N) (2 + k + N)^{4} (5 + 4 k + 4 N + 3 k\, N)}(-k + N) (2192 + 3204 k + 1549 k^{2} \nonu\\&+& 
      246 k^{3}+ 3540 N + 3616 k\, 
     N + 997 k^{2} N + 36 k^{3} N + 1867 N^{2} + 1036 k\, 
     N^{2} + 66 k^{2} N^{2} 
     \nonu\\&+& 321 N^{3} + 18 k\, 
     N^{3}), \nonu\\
c_ {226} & = &\frac{1}{21 (2 + N) (2 + k + N)^{4} (5 + 4 k + 4 N + 3 k\, N)}(-k + N) (2672 + 3828 k + 1741 k^{2} \nonu\\&+& 
      246 k^{3}+ 4404 N + 4600 k\, 
     N + 1237 k^{2} N + 36 k^{3} N + 2371 N^{2} + 1516 k\, 
     N^{2} + 138 k^{2} N^{2} 
     \nonu\\&+& 417 N^{3} + 90 k\, 
     N^{3}), \nonu\\
c_ {227} & = &\frac{1}{21 (2 + N) (2 + k + N)^{4} (5 + 4 k + 4 N + 3 k\, N)}(-k + N) (4592 + 6324 k + 2509 k^{2} \nonu\\&+& 246 k^{3} + 7860 N + 8536 k\, 
     N + 2197 k^{2} N + 36 k^{3} N + 4387 N^{2} + 3436 k\, 
     N^{2} + 426 k^{2} N^{2} 
     \nonu\\&+& 801 N^{3}+ 378 k\, 
     N^{3}), \nonu\\
c_ {228} & = &\frac {1}{21 (2 + N) (2 + k + N)^{4} (5 + 4 k + 4 N + 3 k\, 
     N)}2 (-k + N) (76 + 41 k + 41 N + 6 k\, 
     N)
     \nonu\\&+& (32 + 29 k + 6 k^{2} + 35 N + 17 k\, 
     N + 9 N^{2}),\nonu\\
c_ {229} & = & - \frac {1}{21 (2 + N) (2 + k + N)^{4} (5 + 4 k + 4 N + 3 k\, N)}2 (-k + N) (-1712 - 2580 k - 
       1357 k^{2} 
       \nonu\\&-& 246 k^{3} - 2676 N - 2632 k\, 
      N - 757 k^{2} N - 36 k^{3} N - 1363 N^{2} - 556 k\, 
      N^{2} + 6 k^{2} N^{2} - 225 N^{3} 
      \nonu\\&+& 54 k\, 
      N^{3}),\nonu\\ 
c_ {230} & = & - \frac{1} {21 (2 + N) (2 + k + N)^{4} (5 + 4 k + 4 N + 3 k\, N)}2 (-k + N) (-272 - 708 k - 781 k^{2} - 
       246 k^{3} 
       \nonu\\&-& 84 N + 320 k\, 
      N - 37 k^{2} N - 36 k^{3} N + 149 N^{2} + 884 k\, 
      N^{2} + 222 k^{2} N^{2} + 63 N^{3} + 270 k\, 
      N^{3}), \nonu\\
c_ {231} & = & - \frac {2 i (-k + N) (76 + 41 k + 41 N + 6 k\, 
      N) (32 + 55 k + 18 k^{2} + 41 N + 35 k\, 
      N + 11 N^{2})} {63 (2 + N) (2 + k + N)^{5} (5 + 4 k + 4 N + 
       3 k\, N)}, \nonu\\
c_ {232} & = &\frac {1}{21 (2 + N) (2 + k + N)^{5} (5 + 4 k + 4 N + 3 k\, N)}8 i (960 + 3728 k + 4872 k^{2} + 2623 k^{3} + 
      498 k^{4} 
      \nonu\\&+& 1168 N + 6080 k\, 
     N + 7815 k^{2} N + 3731 k^{3} N + 576 k^{4} N - 680 N^{2} + 
      2481 k\, 
     N^{2} + 3627 k^{2} N^{2} 
     \nonu\\&+& 1384 k^{3} N^{2} + 120 k^{4} N^{2} - 
      1591 N^{3} - 163 k\, 
     N^{3} + 444 k^{2} N^{3} + 108 k^{3} N^{3} - 781 N^{4} - 76 k\, 
     N^{4} 
     \nonu\\&+&12 k^{2} N^{4} - 120 N^{5} + 48 k\, 
     N^{5}), \nonu\\
c_ {233} & = & - \frac {1}{7 (2 + N) (2 + k + N)^{5} (5 + 4 k + 4 N + 3 k\, N)}4 i (480 - 416 k - 1704 k^{2} - 1111 k^{3} - 
       210 k^{4} 
       \nonu\\&+& 2864 N + 1336 k\, 
      N - 2271 k^{2} N - 1535 k^{3} N - 216 k^{4} N + 5504 N^{2} + 
       4359 k\, 
      N^{2} - 207 k^{2} N^{2} 
      \nonu\\&-& 448 k^{3} N^{2} - 12 k^{4} N^{2} + 
       4687 N^{3} + 3583 k\, 
      N^{3} + 492 k^{2} N^{3} + 1825 N^{4} + 1012 k\, 
      N^{4} + 96 k^{2} N^{4} 
      \nonu\\&+& 264 N^{5} + 60 k\, 
      N^{5}), \nonu\\
c_ {234} & = & - \frac{1}{21 (2 + N) (2 + k + N)^{5} (5 + 4 k + 4 N + 3 k\, N)}4 i (-k + N) (1520 + 2760 k + 1615 k^{2} 
\nonu\\&+& 
       306 k^{3} + 3896 N + 5734 k\, 
      N + 2573 k^{2} N + 336 k^{3} N + 3655 N^{2} + 3920 k\, 
      N^{2} + 1096 k^{2} N^{2} 
      \nonu\\&+& 48 k^{3} N^{2} + 1477 N^{3} + 
       916 k\, N^{3} + 84 k^{2} N^{3} + 216 N^{4} + 24 k\, 
      N^{4}).\nonu
\eea

The fusion rule is
\bea
[\widetilde{\Phi}_{\frac{3}{2}}^{(1),i} ] \, \cdot \, 
[\widetilde{\Phi}_{2}^{(1)}] & = & [I^i]+
[\Phi_{0}^{(1)} \, 
\Phi_{\frac{1}{2}}^{(1),i}] +
[\Phi_{0}^{(1)} \, 
\widetilde{\Phi}_{\frac{3}{2}}^{(1),i}] +
[\Phi_{\frac{1}{2}}^{(1),1} \, 
\Phi_{1}^{(1),i j}]
+  [\Phi_{\frac{1}{2}}^{(2),i}]
+  [\widetilde{\Phi}_{\frac{3}{2}}^{(2),i}].
\nonu
\eea

\subsection{The OPE between the higher spin-$3$ current}

The OPEs between the higher spin-$3$ current   is given by
\bea
&&\widetilde{\Phi}_{2}^{(1)}(z)\:\widetilde{\Phi}_{2}^{(1)}(w)\;=\;\frac{1}{(z-w)^{6}}\,c_{1}
+\frac{1}{(z-w)^{4}}\Bigg[\, c_{2}\, {\bf \Phi_{0}^{(2)}}+
c_{3}\, {\bf \Phi_{0}^{(1)}\,\Phi_{0}^{(1)}}+c_{4}\,L+c_{5}\,G^{i}\,\Gamma^{i}
\nonu\\&&+c_{6}\,T^{ij}\,T^{ij}+c_{7}\,T^{ij}\,\Gamma^{i}\,\Gamma^{j}+c_{8}\,U\,U
+c_{9}\,\partial U+c_{10}\,\partial\Gamma^{i}\,\Gamma^{i}+\varepsilon^{ijkl}\Big(c_{11}\,T^{ij}\,T^{kl}+c_{12}\,T^{ij}\,\Gamma^{k}\,\Gamma^{l}
\nonu\\&&+c_{13}\,\Gamma^{i}\,\Gamma^{j}\Gamma^{k}\,\Gamma^{l}\Big)\Bigg](w)+\frac{1}{(z-w)^{3}}\Bigg[\,\frac{1}{2}\,\partial (\mbox{ pole-4}) \Bigg]
\nonu\\&&+\frac{1}{(z-w)^{2}}\Bigg[ 
 c_{14}\, {\bf \widetilde{\Phi}_{2}^{(2)}}+c_{15}\, {\bf 
\partial\Phi_{0}^{(2)}}+c_{16}\, {\bf 
\Phi_{0}^{(1)}\,\widetilde{\Phi}_{2}^{(1)}}+c_{17}\, {\bf 
\Phi_{\frac{1}{2}}^{(1),i}\,\widetilde{\Phi}_{\frac{3}{2}}^{(1),i}}+
c_{18}\, {\bf \partial\Phi_{\frac{1}{2}}^{(1),i}\,\Phi_{\frac{1}{2}}^{(1),i}}
\nonu\\&&  +c_{19}\, {\bf \partial^{2}\Phi_{0}^{(1)}\,\Phi_{0}^{(1)}}+
c_{20}\, {\bf \partial\Phi_{0}^{(1)}\,\partial\Phi_{0}^{(1)}}+
c_{21}\,L\, {\bf \Phi_{0}^{(2)}}+c_{22}\,L\, {\bf \Phi_{0}^{(1)}\,\Phi_{0}^{(1)}}+
c_{23}\,\partial^{2}L+c_{24}\,L\,L
\nonu\\&&+c_{25}\,\partial L\,U
+c_{26}\,L\,\partial U+c_{27}\,L\,G^{i}\,\Gamma^{i}+c_{28}\,L\,U\,U+c_{29}\,L\,T^{ij}\,T^{ij}+c_{30}\,L\,T^{ij}\,\Gamma^{i}\,\Gamma^{j}+c_{31}\,L\,\partial\Gamma^{i}\,\Gamma^{i}
\nonu\\&&+c_{32}\,\partial G^{i}\,G^{i}+c_{33}\,G^{i}\,G^{j}\,T^{ij}+c_{34}\,\partial(G^{i}\,T^{ij})\,\Gamma^{j}+c_{35}\,G^{i}\,T^{ij}\,\partial\Gamma^{j}+c_{36}\,\partial^{2}G^{i}\,\Gamma^{i}+c_{37}\,\partial G^{i}\,\partial\Gamma^{i}
\nonu\\&&+c_{38}\,G^{i}\,\partial^{2}\Gamma^{i}+c_{39}\,G^{i}\,G^{j}\,\Gamma^{i}\,\Gamma^{j}+c_{40}\,G^{i}\,T^{ij}\,U\,\Gamma^{j}+c_{41}\,\partial G^{i}\,U\,\Gamma^{i}+c_{42}\,G^{i}\,\partial U\,\Gamma^{i}
\nonu\\&&+c_{43}\,G^{i}\,U\,\partial\Gamma^{i}+c_{44}\,G^{i}\,\Gamma^{i}\,\partial\Gamma^{j}\,\Gamma^{j}+c_{45}\,\partial^{2}T^{ij}\,T^{ij}+c_{46}\,\partial T^{ij}\,\partial T^{ij}+c_{47}\,\partial^{2}T^{ij}\,\Gamma^{i}\,\Gamma^{j}
\nonu\\&&+c_{48}\,T^{ij}\,\partial^{2}\Gamma^{i}\,\Gamma^{j}+c_{49}\,\partial T^{ij}\,\partial(\Gamma^{i}\,\Gamma^{j})+c_{50}\,T^{ij}\,\partial\Gamma^{i}\,\partial\Gamma^{j}+c_{51}\,\partial T^{ij}\,T^{ij}\,U+c_{52}\,T^{ij}\,T^{ij}\,\partial U
\nonu\\&&+c_{53}\,T^{ij}\,T^{ij}\,U\,U+c_{54}\,T^{ij}\,T^{ij}\,\widetilde{T}^{ij}\,\widetilde{T}^{ij}+c_{55}\,T^{ij}\,T^{ij}\,\partial\Gamma^{i}\,\Gamma^{i}+c_{56}\,(\partial\widetilde{T}^{ij}\,\widetilde{T}^{ik}\,\Gamma^{k}\,\Gamma^{j}+\widetilde{T}^{ij}\,\widetilde{T}^{ik}\,\partial\Gamma^{k}\,\Gamma^{j})
\nonu\\&&+c_{57}\,(T^{ij}\,U\,\partial\Gamma^{i}\,\Gamma^{j}-\frac{1}{2}T^{ij}\,\partial U\,\Gamma^{i}\,\Gamma^{j})+c_{58}\,\partial^{3}U+c_{59}\,\partial^{2}U\,U+c_{60}\,\partial U\,U\,U+c_{61}\,\partial U\,\partial U
\nonu\\&&+c_{62}\,U\,U\,U\,U+c_{63}\,U\,U\,\partial\Gamma^{i}\,\Gamma^{i}+c_{64}\,U\,\partial^{2}\Gamma^{i}\,\Gamma^{i}+c_{65}\,\partial U\,\partial\Gamma^{i}\,\Gamma^{i}+c_{66}\,\partial^{2}\Gamma^{i}\,\partial\Gamma^{i}
\nonu\\&&+c_{67}\,\partial\Gamma^{i}\,\Gamma^{i}\,\partial\Gamma^{j}\,\Gamma^{j}
+c_{68}\,\partial^{3}\Gamma^{i}\,\Gamma^{i}+\varepsilon^{ijkl}\Bigg\{\,
 c_{69}\, {\bf \Phi_{1}^{(1),ij}\,\Phi_{1}^{(1),kl}}+c_{70}\,L\,T^{ij}\,T^{kl}+c_{71\,}L\,T^{ij}\,\Gamma^{k}\,\Gamma^{l}
\nonu\\&&+c_{72}\,L\,\Gamma^{i}\,\Gamma^{j}\,\Gamma^{k}\,\Gamma^{l}
+c_{73}\,(\partial G^{i}\,T^{jk}\,\Gamma^{l}+G^{i}\,T^{jk}\,\partial\Gamma^{l})+c_{74}\,G^{i}\,\partial T^{jk}\,\Gamma^{l}+c_{75}\,G^{a}\,T^{ij}\,T^{kl}\,\Gamma^{a}
\nonu\\&&+c_{76}\,\Big[\partial G^{i}\,\Gamma^{j}\,\Gamma^{k}\,\Gamma^{l}
-G^{i}\,\partial(\Gamma^{j}\,\Gamma^{k}\,\Gamma^{l})\Big]+c_{77}\,\partial^{2}T^{ij}\,T^{kl}+c_{78}\,\partial T^{ij}\,\partial T^{kl}+c_{79}\,\partial^{2}T^{ij}\,\Gamma^{k}\,\Gamma^{l}
\nonu\\&&+c_{80}\,T^{ij}\,\partial^{2}\Gamma^{k}\,\Gamma^{l}
+c_{81}\,T^{ij}\,\partial\Gamma^{k}\,\partial\Gamma^{l}+c_{82}\,\partial T^{ij}\,\partial(\Gamma^{k}\,\Gamma^{l})+c_{83}\,(\partial T^{ij}\,T^{kl}\,U-T^{ij}\,T^{kl}\,\partial U)
\nonu\\&&+c_{84}\,T^{ij}\,T^{kl}\,\partial\Gamma^{a}\,\Gamma^{a}
+c_{85}\,\partial\Gamma^{i}\,\partial(\Gamma^{j}\,\Gamma^{k}\,\Gamma^{l})+c_{86}\,\partial^{2}\Gamma^{i}\,\Gamma^{j}\,\Gamma^{k}\,\Gamma^{l}\Bigg\}
\nonu\\&&+(\varepsilon^{ijkl})^{2}\Big(\,c_{87}\,T^{ij}\,T^{ik}\,T^{jl}\,T^{kl}\Big)+c_{88}\,\Big[T^{ij}\,U\,\partial(\Gamma^{k}\,\Gamma^{k})-\partial T^{ij}\,U\,\Gamma^{k}\,\Gamma^{k}\Big](1-\delta^{ik}-\delta^{jk})\Bigg](w)
\nonu\\&&+\frac{1}{(z-w)}\Bigg[\,\frac{1}{2}\,\partial\Big(\mbox{ pole-2}\Big)
+ c_{89}\,{\bf \partial^{3}\Phi_{0}^{(2)}}+
c_{90}\, {\bf \partial^{3}(\Phi_{0}^{(2)}\,\Phi_{0}^{(2)})}
+c_{91}\,\partial^{3}L\,+c_{92}\,\partial^{3}(G^{i}\,\Gamma^{i})
\nonu\\&&+c_{93}\,\partial^{3}(T^{ij}\,T^{ij})
+c_{94}\,\partial^{3}(T^{ij}\,\Gamma^{i}\,\Gamma^{j})+c_{95}\,\partial^{4}U+c_{96}\,\partial^{3}(U\,U)+c_{97}\,\partial^{4}\Gamma^{i}\,\Gamma^{i}+c_{98}\,\partial^{3}\Gamma^{i}\,\partial\Gamma^{i}
\nonu\\&&+\varepsilon^{ijkl}\Bigg\{\,c_{99}\,\partial^{3}(T^{ij}\,T^{kl})
+c_{100}\,\partial^{3}(T^{ij}\,\Gamma^{k}\,\Gamma^{l})+c_{101}\,\partial^{3}(\Gamma^{i}\,\Gamma^{j}\,\Gamma^{k}\,\Gamma^{l})\Bigg\}\Bigg](w)+\cdots,
\nonu
\eea
where
the coefficients are given by
\bea
c_ {1} & = &\frac {1}{3 (2 + k + N)^{3} (5 + 4 k + 4 N + 3 k\, N)}64 k\, 
   N (225 + 405 k + 218 k^{2} + 34 k^{3} + 405 N 
   \nonu\\&+& 644 k\, 
     N + 281 k^{2} N + 30 k^{3} N + 218 N^{2} + 281 k\, 
     N^{2} + 75 k^{2} N^{2} + 34 N^{3} + 30 k\, 
     N^{3}), \nonu\\
c_ {2} & = & - \frac {8 (k - N) (-11 - 4 k - 4 N + 3 k\, 
      N)} {(2 + k + N) (5 + 4 k + 4 N + 3 k\, N)}, \nonu\\
c_ {3} & = &\frac{1} {(2 + N) (2 + k + N)^{3} (5 + 4 k + 4 N + 3 k\, N)^{2}}4 (k - N) (-3300 - 8203 k - 7482 k^{2} 
 \nonu\\&-&
      2928 k^{3} - 416 k^{4} - 10367 N - 22061 k\, 
     N - 16518 k^{2} N - 4876 k^{3} N - 448 k^{4} N - 12601 N^{2} 
      \nonu\\&-& 
      21336 k\, 
     N^{2} - 11130 k^{2} N^{2} - 1331 k^{3} N^{2} + 150 k^{4} N^{2} - 
      7392 N^{3} - 8654 k\, 
     N^{3} - 1491 k^{2} N^{3}
      \nonu\\&+& 987 k^{3} N^{3} + 180 k^{4} N^{3} - 
      2080 N^{4} - 1120 k\, 
     N^{4} + 759 k^{2} N^{4} + 378 k^{3} N^{4} - 224 N^{5} + 48 k\, 
     N^{5}
      \nonu\\&+& 144 k^{2} N^{5}), \nonu\\
c_ {4} & = &\frac{1}{(2 + k + N)^{3} (5 + 4 k + 4 N + 
       3 k\, N)^{2}}16 (600 + 3310 k + 5749 k^{2} + 4458 k^{3} + 
      1616 k^{4} 
       \nonu\\&+& 224 k^{5} + 2210 N + 10823 k\, 
     N + 16343 k^{2} N + 10630 k^{3} N + 3052 k^{4} N + 304 k^{5} N + 
      3204 N^{2} 
       \nonu\\&+& 13751 k\, 
     N^{2} + 17594 k^{2} N^{2} + 9128 k^{3} N^{2} + 
      1845 k^{4} N^{2} + 90 k^{5} N^{2} + 2336 N^{3} + 8584 k\, 
     N^{3} 
      \nonu\\&+& 8940 k^{2} N^{3} + 3415 k^{3} N^{3} + 387 k^{4} N^{3} + 
      864 N^{4} + 2624 k\, 
     N^{4} + 2052 k^{2} N^{4} + 459 k^{3} N^{4} 
      \nonu\\&+& 128 N^{5} + 304 k\, 
     N^{5} + 144 k^{2} N^{5}), \nonu\\
c_ {5} & = & - \frac {64 i (k - N) (7 + 5 k + 5 N + 3 k\, 
      N)} {(2 + k + N)^{2} (5 + 4 k + 4 N + 3 k\, N)}, \nonu\\
c_ {6} & = &\frac{1}{(2 + N) (2 + k + N)^{3} (5 + 4 k + 4 N + 3 k\, N)}2 (-240 - 772 k - 719 k^{2} - 246 k^{3} - 
      24 k^{4} 
       \nonu\\&-& 452 N - 1280 k\, 
     N - 825 k^{2} N - 109 k^{3} N + 18 k^{4} N - 329 N^{2} - 894 k\, 
     N^{2} - 388 k^{2} N^{2} 
      \nonu\\&-& 15 k^{3} N^{2} - 123 N^{3} - 347 k\, 
     N^{3} - 90 k^{2} N^{3} - 20 N^{4} - 57 k\, 
     N^{4}), \nonu\\
c_ {7} & = & - \frac {1}{(2 + N) (2 + k + N)^{4} (5 + 4 k + 4 N + 3 k\, N)}8 i (-240 - 652 k - 563 k^{2} - 198 k^{3} - 
       24 k^{4} 
        \nonu\\&-& 332 N - 908 k\, 
      N - 531 k^{2} N - 49 k^{3} N + 18 k^{4} N - 113 N^{2} - 522 k\, 
      N^{2} - 208 k^{2} N^{2} + 3 k^{3} N^{2} 
       \nonu\\&+& 3 N^{3} - 203 k\, 
      N^{3} - 54 k^{2} N^{3} + 4 N^{4} - 39 k\, 
      N^{4}), \nonu\\
c_ {8} & = &\frac {1} {(2 + k + N)^{4} (5 + 4 k + 4 N + 3 k\, N)}16 (120 + 566 k + 697 k^{2} + 334 k^{3} + 
      56 k^{4} + 346 N
       \nonu\\&+& 1243 k\, 
     N + 1125 k^{2} N + 355 k^{3} N + 30 k^{4} N + 364 N^{2} + 
      1005 k\, 
     N^{2} + 605 k^{2} N^{2} + 93 k^{3} N^{2} 
      \nonu\\&+& 176 N^{3} + 368 k\, 
     N^{3} + 117 k^{2} N^{3} + 32 N^{4} + 48 k\, 
     N^{4}), \nonu\\
c_ {9} & = & - \frac {256 (k - N) (7 + 5 k + 5 N + 3 k\, 
      N)} {(2 + k + N)^{2} (5 + 4 k + 4 N + 3 k\, N)}, \nonu\\
c_ {10} & = & - \frac{1} {(2 + N) (2 + k + N)^{4} (5 + 4 k + 4 N + 3 k\, N)}8 (1200 + 3860 k + 4009 k^{2} + 1786 k^{3} + 
       296 k^{4} 
        \nonu\\&+& 2260 N + 6992 k\, 
      N + 5669 k^{2} N + 1767 k^{3} N + 178 k^{4} N + 1839 N^{2} + 
       5660 k\, 
      N^{2} + 3278 k^{2} N^{2} 
       \nonu\\&+& 659 k^{3} N^{2} + 60 k^{4} N^{2} + 
       1045 N^{3} + 2903 k\, 
      N^{3} + 1012 k^{2} N^{3} + 78 k^{3} N^{3} + 396 N^{4} + 847 k\, 
      N^{4} 
      \nonu\\&+&  126 k^{2} N^{4} + 64 N^{5} + 96 k\, 
      N^{5}), \nonu\\
c_ {11} & = &\frac {1}{(2 + N) (2 + k + N)^{3} (5 + 4 k + 4 N + 3 k\, N)}(k - N) (240 + 169 k - 22 k^{2} - 24 k^{3} + 
      399 N
       \nonu\\&+& 329 k\, 
     N + 99 k^{2} N + 18 k^{3} N + 229 N^{2} + 209 k\, 
     N^{2} + 63 k^{2} N^{2} + 44 N^{3} + 39 k\, 
     N^{3}), \nonu\\
c_ {12} & = & - \frac {4 i (k - N) (-11 - 4 k - 4 N + 3 k\, 
      N) (8 + 17 k + 6 k^{2} + 11 N + 11 k\, 
      N + 3 N^{2})} {(2 + N) (2 + k + N)^{4} (5 + 4 k + 4 N + 3 k\, 
      N)},\nonu\\
 c_ {13} & = & - \frac {4 (k - N) (-11 - 4 k - 4 N + 3 k\, 
      N) (32 + 55 k + 18 k^{2} + 41 N + 35 k\, 
      N + 11 N^{2})} {3 (2 + N) (2 + k + N)^{5} (5 + 4 k + 4 N + 
       3 k\, N)},\nonu\\
c_ {14} & = & 4, \nonu\\
c_ {15} & = &\frac{1}{(2 + N) (2 + k + N)^{2} (59 + 37 k + 37 N + 15 k\, N)}2 (3540 + 6967 k + 4333 k^{2} + 856 k^{3} 
 \nonu\\&+& 
      9155 N + 15113 k\, 
     N + 7503 k^{2} N + 1073 k^{3} N + 8850 N^{2} + 11465 k\, 
     N^{2} + 4022 k^{2} N^{2} 
     \nonu\\&+&  291 k^{3} N^{2} + 3732 N^{3} + 
      3346 k\, N^{3} + 630 k^{2} N^{3} + 571 N^{4} + 249 k\, 
     N^{4}), \nonu\\
c_ {16} & = & - \frac{1}{(2 + N) (2 + k + N)^{2} (5 + 4 k + 4 N + 3 k\, N)}4 (300 + 529 k + 322 k^{2} + 64 k^{3} + 
       941 N 
        \nonu\\&+& 1499 k\, 
      N + 807 k^{2} N + 134 k^{3} N + 1023 N^{2} + 1373 k\, 
      N^{2} + 581 k^{2} N^{2} + 60 k^{3} N^{2} + 468 N^{3} 
      \nonu\\&+&  481 k\, 
      N^{3} + 126 k^{2} N^{3} + 76 N^{4} + 48 k\, 
      N^{4}), \nonu\\
c_ {17} & = & - \frac {4 (60 + 77 k + 22 k^{2} + 121 N + 115 k\, 
      N + 20 k^{2} N + 79 N^{2} + 42 k\, 
      N^{2} + 16 N^{3})} {(2 + N) (2 + k + N)^{2}}, \nonu\\
c_ {18} & = &\frac {4 (k - N) (60 + 77 k + 22 k^{2} + 121 N + 115 k\, 
     N + 20 k^{2} N + 79 N^{2} + 42 k\, 
     N^{2} + 16 N^{3})} {3 (2 + N) (2 + k + N)^{3}}, \nonu\\
c_ {19} & = & - \frac{1}{3 (2 + N)^{2} (2 + k + N)^{4} (5 + 
        4 k + 4 N + 3 k\, N)^{2}}2 (270000 + 1234200 k + 2248423 k^{2} 
         \nonu\\&+& 
       2109002 k^{3} + 1078176 k^{4} + 285344 k^{5} + 30592 k^{6} 
    + 1411800 N + 6109534 k\,N    
    \nonu\\&+&   10414874 k^{2} N + 9033455 k^{3} N + 4209758 k^{4} N + 
       997264 k^{5} N + 93472 k^{6} N + 3243943 N^{2}
          \nonu\\&+&   13158782 k\, 
      N^{2} + 20720450 k^{2} N^{2} + 16331581 k^{3} N^{2} + 
       6767761 k^{4} N^{2} + 1382326 k^{5} N^{2} 
          \nonu\\&+&  
       106672 k^{6} N^{2} + 4283022 N^{3} + 16093703 k\, 
      N^{3} + 23038371 k^{2} N^{3} + 16132888 k^{3} N^{3} 
         \nonu\\&+&   
       5744440 k^{4} N^{3} + 955517 k^{5} N^{3} + 54546 k^{6} N^{3} + 
       3548664 N^{4} + 12169899 k\, 
      N^{4} 
         \nonu\\&+&   15503581 k^{2} N^{4} + 9335680 k^{3} N^{4} + 
       2703938 k^{4} N^{4} + 329289 k^{5} N^{4} + 10620 k^{6} N^{4}
          \nonu\\&+&  
       1882855 N^{5} + 5780218 k\, 
      N^{5} + 6354592 k^{2} N^{5} + 3120773 k^{3} N^{5} + 
       662631 k^{4} N^{5} + 44982 k^{5} N^{5} 
          \nonu\\&+&   621816 N^{6} + 
       1664194 k\, 
      N^{6} + 1504244 k^{2} N^{6} + 544839 k^{3} N^{6} + 
       64944 k^{4} N^{6} + 116256 N^{7} 
          \nonu\\&+&   261080 k\, 
      N^{7} + 177927 k^{2} N^{7} + 36666 k^{3} N^{7} + 9376 N^{8} + 
       16656 k\, 
      N^{8} + 7056 k^{2} N^{8}), \nonu\\
c_ {20} & = & - \frac{1}{(2 + N)^{2} (2 + k + N)^{4}}2 (60 + 77 k + 22 k^{2} + 121 N + 115 k\, 
      N + 20 k^{2} N + 79 N^{2} + 42 k\, 
      N^{2} 
       \nonu\\&+& 16 N^{3}) (60 + 73 k + 20 k^{2} + 125 N + 113 k\, 
      N + 19 k^{2} N + 83 N^{2} + 42 k\, 
      N^{2} + 17 N^{3}), \nonu\\
c_ {21} & = & - \frac {32 (k - N) (-14 - k - N + 12 k\, 
      N)} {(5 + 4 k + 4 N + 3 k\, N) (59 + 37 k + 37 N + 15 k\, N)}, \nonu\\
c_ {22} & = &\frac{1}{(2 + N) (2 + k + N)^{2} (5 + 4 k + 4 N + 3 k\, N)^{2}}48 (k - N) (100 + 187 k + 118 k^{2} + 24 k^{3} 
 \nonu\\&+&
      303 N + 505 k\, 
     N + 277 k^{2} N + 46 k^{3} N + 325 N^{2} + 455 k\, 
     N^{2} + 195 k^{2} N^{2} + 20 k^{3} N^{2} + 148 N^{3} 
      \nonu\\&+& 159 k\, 
     N^{3} + 42 k^{2} N^{3} + 24 N^{4} + 16 k\, 
     N^{4}), \nonu\\
c_ {23} & = &\frac{1} {3 (2 + N) (2 + k + N)^{4}}4 (-1944 - 3798 k - 1217 k^{2} + 562 k^{3} + 
      232 k^{4} - 1530 N + 1603 k\, 
     N 
      \nonu\\&+& 5764 k^{2} N + 2867 k^{3} N + 314 k^{4} N + 2080 N^{2} + 
      7792 k\, 
     N^{2} + 6149 k^{2} N^{2} + 1137 k^{3} N^{2} 
      \nonu\\&+& 2496 N^{3} + 
      4174 k\, N^{3} + 1352 k^{2} N^{3} + 816 N^{4} + 591 k\, 
     N^{4} + 80 N^{5}), \nonu\\
c_ {24} & = & - \frac {1}{(2 + N) (2 + k + N)^{2} (5 + 4 k + 4 N + 3 k\, N)^{2}}16 (-800 - 3055 k - 3842 k^{2} - 1968 k^{3} 
 \nonu\\&-& 
       352 k^{4} - 1905 N - 7099 k\, 
      N - 8182 k^{2} N - 3652 k^{3} N - 528 k^{4} N - 1347 N^{2} - 
       5482 k\, 
      N^{2} 
       \nonu\\&-& 5790 k^{2} N^{2} - 2107 k^{3} N^{2} - 194 k^{4} N^{2} - 
       48 N^{3} - 1374 k\, 
      N^{3} - 1445 k^{2} N^{3} - 367 k^{3} N^{3} 
       \nonu\\&+& 256 N^{4} + 80 k\, 
      N^{4} - 55 k^{2} N^{4} + 64 N^{5} + 40 k\, 
      N^{5}),\nonu\\
 c_ {25} & = &\frac {1}{(2 + N) (2 + k + N)^{3} (5 + 4 k + 4 N + 3 k\, N)}32 (200 + 455 k + 326 k^{2} + 72 k^{3} + 575 N 
  \nonu\\&+& 
      1148 k\, N + 691 k^{2} N + 118 k^{3} N + 617 N^{2} + 1036 k\, 
     N^{2} + 480 k^{2} N^{2} + 50 k^{3} N^{2} + 288 N^{3} 
      \nonu\\&+&375 k\, 
     N^{3} + 105 k^{2} N^{3} + 48 N^{4} + 40 k\, 
     N^{4}), \nonu\\
c_ {26} & = &\frac{1}{(2 + k + N)^{3} (5 + 4 k + 4 N + 3 k\, N)}64 (50 + 77 k + 38 k^{2} + 6 k^{3} + 118 N + 
      157 k\, N 
       \nonu\\&+& 69 k^{2} N + 10 k^{3} N + 84 N^{2} + 80 k\, 
     N^{2} + 21 k^{2} N^{2} + 18 N^{3} + 8 k\, 
     N^{3}), \nonu\\
c_ {27} & = & - \frac {32 i (k - N)} {(2 + k + N) (5 + 4 k + 4 N + 
       3 k\, N)}, \nonu\\
c_ {28} & = & - \frac{1}{(2 + N) (2 + k + N)^{3} (5 + 4 k + 4 N + 
       3 k\, N)}32 (-160 - 443 k - 346 k^{2} - 80 k^{3} - 
       293 N 
        \nonu\\&-& 695 k\, 
      N - 415 k^{2} N - 58 k^{3} N - 143 N^{2} - 305 k\, 
      N^{2} - 115 k^{2} N^{2} - 27 k\, 
      N^{3} + 8 N^{4}), \nonu\\
c_ {29} & = &\frac{1}{(2 + N) (2 + k + N)^{3} (5 + 4 k + 4 N + 
      3 k\, N)}8 (-160 - 443 k - 372 k^{2} - 105 k^{3} - 
      6 k^{4} 
       \nonu\\&-&293 N - 663 k\, 
     N - 416 k^{2} N - 67 k^{3} N - 149 N^{2} - 274 k\, 
     N^{2} - 106 k^{2} N^{2} - 5 N^{3} - 20 k\, 
     N^{3} 
      \nonu\\&+&7 N^{4}), \nonu\\
c_ {30} & = & - \frac{1}{(2 + N) (2 + k + N)^{4} (5 + 4 k + 4 N + 
       3 k\, N)}32 i (-160 - 443 k - 372 k^{2} - 105 k^{3} - 
       6 k^{4} 
        \nonu\\&-& 293 N - 663 k\, 
      N - 416 k^{2} N - 67 k^{3} N - 149 N^{2} - 274 k\, 
      N^{2} - 106 k^{2} N^{2} - 5 N^{3} - 20 k\, 
      N^{3} 
       \nonu\\&+& 7 N^{4}), \nonu\\
c_ {31} & = &\frac{1}{(2 + N) (2 + k + N)^{4} (5 + 4 k + 4 N + 
      3 k\, N)}32 (-720 - 2231 k - 2167 k^{2} - 805 k^{3} - 
      98 k^{4} 
       \nonu\\&-& 1441 N - 3827 k\, 
     N - 2981 k^{2} N - 792 k^{3} N - 58 k^{4} N - 870 N^{2} - 
      2066 k\, 
     N^{2} - 1196 k^{2} N^{2} 
      \nonu\\&-& 167 k^{3} N^{2} - 100 N^{3} - 359 k\, 
     N^{3} - 130 k^{2} N^{3} + 45 N^{4} - 13 k\, 
     N^{4} + 8 N^{5}), \nonu\\
c_ {32} & = &\frac {1} {(2 + N) (2 + k + N)^{3}} 4 (224 + 440 k + 
    191 k^{2} + 18 k^{3} + 312 N + 344 k\, 
   N + 49 k^{2} N + 105 N^{2} 
    \nonu\\&+& 38 k\, N^{2} + 7 N^{3}), \nonu\\
c_ {33} & = &\frac {8 i (32 + 55 k + 18 k^{2} + 41 N + 35 k\, 
     N + 11 N^{2})} {(2 + N) (2 + k + N)^{3}}, \nonu\\
c_ {34} & = &\frac {8 (80 + 92 k + 11 k^{2} - 6 k^{3} + 172 N + 
      158 k\, N + 21 k^{2} N + 119 N^{2} + 64 k\, 
     N^{2} + 25 N^{3})} {(2 + N) (2 + k + N)^{4}}, \nonu\\
c_ {35} & = &\frac {8 (80 + 156 k + 121 k^{2} + 30 k^{3} + 108 N + 
      130 k\, N + 55 k^{2} N + 37 N^{2} + 16 k\, 
     N^{2} + 3 N^{3})} {(2 + N) (2 + k + N)^{4}}, \nonu\\
c_ {36} & = & - \frac {1}{3 (2 + N) (2 + k + N)^{4} (5 + 4 k + 4 N + 3 k\, N)}2 i (10200 + 22646 k + 16515 k^{2} + 
       4158 k^{3} 
        \nonu\\&+& 184 k^{4} + 31534 N + 63736 k\, 
      N + 42753 k^{2} N + 10517 k^{3} N + 806 k^{4} N + 37025 N^{2} 
       \nonu\\&+& 
       65040 k\, 
      N^{2} + 36966 k^{2} N^{2} + 7381 k^{3} N^{2} + 
       492 k^{4} N^{2} + 20073 N^{3} + 28753 k\, 
      N^{3} + 12150 k^{2} N^{3} 
       \nonu\\&+& 1374 k^{3} N^{3} + 4796 N^{4} + 
       5051 k\, N^{4} + 1146 k^{2} N^{4} + 376 N^{5} + 264 k\, 
      N^{5}), \nonu\\
c_ {37} & = & - \frac {1}{3 (2 + N) (2 + k + N)^{3} (5 + 4 k + 4 N + 3 k\, N)}4 i (7500 + 14357 k + 8954 k^{2} + 
       1832 k^{3} 
        \nonu\\&+& 18193 N + 29851 k\, 
      N + 14859 k^{2} N + 2146 k^{3} N + 15915 N^{2} + 21781 k\, 
      N^{2} + 7945 k^{2} N^{2}
       \nonu\\&+& 588 k^{3} N^{2} + 5892 N^{3} + 
       6401 k\, N^{3} + 1386 k^{2} N^{3} + 776 N^{4} + 600 k\, 
      N^{4}), \nonu\\
c_ {38} & = & - \frac {1}{(2 + N) (2 + k + N)^{4} (5 + 4 k + 4 N + 3 k\, N)}6 i (2200 + 5718 k + 5171 k^{2} + 1918 k^{3} 
 \nonu\\&+&
       248 k^{4} + 6862 N + 16072 k\, 
      N + 12889 k^{2} N + 4133 k^{3} N + 454 k^{4} N + 8433 N^{2} + 
       17128 k\, 
      N^{2} 
       \nonu\\&+& 11438 k^{2} N^{2} + 2813 k^{3} N^{2} + 
       204 k^{4} N^{2} + 5017 N^{3} + 8361 k\, 
      N^{3} + 4158 k^{2} N^{3} + 582 k^{3} N^{3} 
       \nonu\\&+& 1420 N^{4} + 
       1771 k\, N^{4} + 498 k^{2} N^{4} + 152 N^{5} + 120 k\, 
      N^{5}), \nonu\\
c_ {39} & = &\frac {16 (32 + 55 k + 18 k^{2} + 41 N + 35 k\, 
     N + 11 N^{2})} {(2 + N) (2 + k + N)^{4}}, \nonu\\
c_ {40} & = &\frac {32 (32 + 55 k + 18 k^{2} + 41 N + 35 k\, 
     N + 11 N^{2})} {(2 + N) (2 + k + N)^{4}}, \nonu\\
c_ {41} & = & - \frac {8 i (224 + 440 k + 191 k^{2} + 18 k^{3} + 
       312 N + 344 k\, N + 49 k^{2} N + 105 N^{2} + 38 k\, 
      N^{2} + 7 N^{3})} {(2 + N) (2 + k + N)^{4}}, \nonu\\
c_ {42} & = &\frac {24 i (64 + 136 k + 61 k^{2} + 6 k^{3} + 88 N + 
      104 k\, N + 15 k^{2} N + 27 N^{2} + 10 k\, 
     N^{2} + N^{3})} {(2 + N) (2 + k + N)^{4}}, \nonu\\
c_ {43} & = &\frac {24 i (-32 - 52 k - 3 k^{2} + 6 k^{3} - 28 N - 
      4 k\, N + 19 k^{2} N + 7 N^{2} + 18 k\, 
     N^{2} + 5 N^{3})} {(2 + N) (2 + k + N)^{4}}, \nonu\\
c_ {44} & = &\frac {32 i (k - N) (32 + 55 k + 18 k^{2} + 41 N + 
      35 k\, N + 11 N^{2})} {(2 + N) (2 + k + N)^{5}}, \nonu\\
c_ {45} & = & - \frac {1}{3 (2 + N)^{2} (2 + k + N)^{4}}(720 + 2136 k + 3431 k^{2} + 1810 k^{3} + 
      288 k^{4} + 4776 N + 12410 k\, 
     N 
      \nonu\\&+& 11728 k^{2} N + 3721 k^{3} N + 306 k^{4} N + 8435 N^{2} + 
      16432 k\, 
     N^{2} + 9814 k^{2} N^{2} + 1543 k^{3} N^{2}
      \nonu\\&+& 6138 N^{3} + 
      8017 k\, N^{3} + 2405 k^{2} N^{3} + 1980 N^{4} + 1307 k\, 
     N^{4} + 235 N^{5}), \nonu\\
c_ {46} & = & - \frac{1}{3 (2 + N)^{2} (2 + k + N)^{4}}(6864 + 17424 k + 13823 k^{2} + 4438 k^{3} + 
      504 k^{4} + 18288 N 
       \nonu\\&+& 37358 k\, 
     N + 23848 k^{2} N + 5791 k^{3} N + 414 k^{4} N + 18983 N^{2} + 
      30328 k\, 
     N^{2} + 14176 k^{2} N^{2} 
      \nonu\\&+& 1921 k^{3} N^{2} + 9750 N^{3} + 
      11035 k\, N^{3} + 2855 k^{2} N^{3} + 2490 N^{4} + 1505 k\, 
     N^{4} + 253 N^{5}), \nonu\\
c_ {47} & = &\frac {1} {3 (2 + N)^{2} (2 + k + N)^{5}}2 i (3792 + 8784 k + 7343 k^{2} + 2506 k^{3} + 
      288 k^{4} + 11376 N 
       \nonu\\&+& 22718 k\, 
     N + 15652 k^{2} N + 4057 k^{3} N + 306 k^{4} N + 13415 N^{2} + 
      21532 k\, 
     N^{2} + 10762 k^{2} N^{2} 
      \nonu\\&+& 1537 k^{3} N^{2} + 7698 N^{3} + 
      8785 k\, N^{3} + 2387 k^{2} N^{3} + 2136 N^{4} + 1289 k\, 
     N^{4} + 229 N^{5}), \nonu\\
c_ {48} & = &\frac {1}{3 (2 + N)^{2} (2 + k + N)^{5}}4 i (720 + 384 k + 959 k^{2} + 1066 k^{3} + 
      288 k^{4} + 5184 N + 8390 k\, 
     N 
      \nonu\\&+& 7084 k^{2} N + 2761 k^{3} N + 306 k^{4} N + 9935 N^{2} + 
      14020 k\, 
     N^{2} + 7258 k^{2} N^{2} + 1249 k^{3} N^{2}
      \nonu\\&+& 7938 N^{3} + 
      7897 k\, N^{3} + 1979 k^{2} N^{3} + 2832 N^{4} + 1457 k\, 
     N^{4} + 373 N^{5}), \nonu\\
c_ {49} & = &\frac {1}{3 (2 + N)^{2} (2 + k + N)^{5}}4 i (6480 + 15552 k + 12635 k^{2} + 4246 k^{3} + 
      504 k^{4} + 18432 N 
       \nonu\\&+& 35462 k\, 
     N + 22282 k^{2} N + 5419 k^{3} N + 414 k^{4} N + 20627 N^{2} + 
      30304 k\, 
     N^{2} + 13318 k^{2} N^{2} 
      \nonu\\&+&1783 k^{3} N^{2} + 11388 N^{3} + 
      11467 k\, N^{3} + 2669 k^{2} N^{3} + 3096 N^{4} + 1607 k\, 
     N^{4} + 331 N^{5}), \nonu\\
c_ {50} & = &\frac {1} {3 (2 + N)^{2} (2 + k + N)^{5}}4 i (6864 + 15888 k + 12047 k^{2} + 3466 k^{3} + 
      288 k^{4} + 19632 N
       \nonu\\&+& 36446 k\, 
     N + 21748 k^{2} N + 4729 k^{3} N + 306 k^{4} N + 21863 N^{2} + 
      31036 k\, 
     N^{2} + 13210 k^{2} N^{2} 
      \nonu\\&+& 1633 k^{3} N^{2} + 11874 N^{3} + 
      11569 k\, N^{3} + 2675 k^{2} N^{3} + 3144 N^{4} + 1577 k\, 
     N^{4} + 325 N^{5}), \nonu\\
c_ {51} & = & - \frac {8 (80 + 92 k + 11 k^{2} - 6 k^{3} + 172 N + 
       158 k\, N + 21 k^{2} N + 119 N^{2} + 64 k\, 
      N^{2} + 25 N^{3})} {(2 + N) (2 + k + N)^{4}}, \nonu\\
c_ {52} & = & - \frac {8 (20 + 21 k + 6 k^{2} + 25 N + 13 k\, 
      N + 7 N^{2})} {(2 + N) (2 + k + N)^{3}}, \nonu\\
c_ {53} & = & - \frac {8 (32 + 55 k + 18 k^{2} + 41 N + 35 k\, 
      N + 11 N^{2})} {(2 + N) (2 + k + N)^{4}}, \nonu\\
c_ {54} & = &\frac {4 (32 + 55 k + 18 k^{2} + 41 N + 35 k\, 
     N + 11 N^{2})} {(2 + N) (2 + k + N)^{4}}, \nonu\\
c_ {55} & = &\frac {16 (32 + 55 k + 18 k^{2} + 41 N + 35 k\, 
     N + 11 N^{2})} {(2 + N) (2 + k + N)^{4}}, \nonu\\
c_ {56} & = & - \frac {16 (32 + 55 k + 18 k^{2} + 41 N + 35 k\, 
      N + 11 N^{2})} {(2 + N) (2 + k + N)^{4}}, \nonu\\
c_ {57} & = & - \frac {32 i (k - N) (32 + 55 k + 18 k^{2} + 41 N + 
       35 k\, N + 11 N^{2})} {(2 + N) (2 + k + N)^{5}}, \nonu\\
c_ {58} & = & - \frac {1}{3 (2 + N) (2 + k + N)^{3} (5 + 4 k + 4 N + 3 k\, N)}8 (5100 + 11207 k + 7790 k^{2} + 1688 k^{3} 
 \nonu\\&+&
       14083 N + 27175 k\, 
      N + 16173 k^{2} N + 2806 k^{3} N + 14553 N^{2} + 23353 k\, 
      N^{2} + 10789 k^{2} N^{2} 
       \nonu\\&+& 1152 k^{3} N^{2} + 6564 N^{3} + 
       8003 k\, N^{3} + 2268 k^{2} N^{3} + 1064 N^{4} + 792 k\, 
      N^{4}), \nonu\\
c_ {59} & = &\frac {1}{3 (2 + N) (2 + k + N)^{5}}8 (744 + 1890 k + 2347 k^{2} + 1258 k^{3} + 
      232 k^{4} + 3342 N + 7963 k\, 
     N 
      \nonu\\&+& 7480 k^{2} N + 2771 k^{3} N + 314 k^{4} N + 4924 N^{2} + 
      9160 k\, 
     N^{2} + 5741 k^{2} N^{2} + 1077 k^{3} N^{2} 
      \nonu\\&+& 3084 N^{3} + 
      3886 k\, N^{3} + 1232 k^{2} N^{3} + 840 N^{4} + 531 k\, 
     N^{4} + 80 N^{5}), \nonu\\
c_ {60} & = &\frac {64 (60 + 77 k + 22 k^{2} + 121 N + 115 k\, 
     N + 20 k^{2} N + 79 N^{2} + 42 k\, 
     N^{2} + 16 N^{3})} {(2 + N) (2 + k + N)^{4}}, \nonu\\
c_ {61} & = &\frac{1}{3 (2 + N) (2 + k + N)^{5}}8 (2664 + 8298 k + 8071 k^{2} + 3190 k^{3} + 
      448 k^{4} + 7110 N + 17443 k\, 
     N 
      \nonu\\&+& 13072 k^{2} N + 3695 k^{3} N + 314 k^{4} N + 7192 N^{2} + 
      13660 k\, 
     N^{2} + 7337 k^{2} N^{2} + 1173 k^{3} N^{2}
      \nonu\\&+& 3540 N^{3} + 
      4786 k\, N^{3} + 1424 k^{2} N^{3} + 852 N^{4} + 627 k\, 
     N^{4} + 80 N^{5}), \nonu\\
c_ {62} & = &\frac {16 (32 + 55 k + 18 k^{2} + 41 N + 35 k\, 
     N + 11 N^{2})} {(2 + N) (2 + k + N)^{4}}, \nonu\\
c_ {63} & = & - \frac {48 (32 + 55 k + 18 k^{2} + 41 N + 35 k\, 
      N + 11 N^{2})} {(2 + N) (2 + k + N)^{4}}, \nonu\\
c_ {64} & = &\frac {64 (k - N) (17 + 10 k + 10 N + 3 k\, 
     N)} {3 (2 + k + N)^{3} (5 + 4 k + 4 N + 3 k\, N)}, \nonu\\
c_ {65} & = & - \frac {128 (k - N) (-1 + k + N + 3 k\, 
      N)} {3 (2 + k + N)^{3} (5 + 4 k + 4 N + 3 k\, N)}, \nonu\\
c_ {66} & = & - \frac {1}{3 (2 + N)^{2} (2 + k + N)^{5}}2 (35856 + 99720 k + 88801 k^{2} + 
       31798 k^{3} + 3952 k^{4} + 103176 N 
        \nonu\\&+& 236242 k\, 
      N + 167694 k^{2} N + 44887 k^{3} N + 3718 k^{4} N + 
       116785 N^{2} + 215754 k\, 
      N^{2} 
       \nonu\\&+& 115498 k^{2} N^{2} + 20095 k^{3} N^{2} + 
       628 k^{4} N^{2} + 66578 N^{3} + 94383 k\, 
      N^{3} + 33633 k^{2} N^{3} 
       \nonu\\&+& 2598 k^{3} N^{3} + 19992 N^{4} + 
       19529 k\, N^{4} + 3352 k^{2} N^{4} + 2945 N^{5} + 1506 k\, 
      N^{5} + 160 N^{6}), \nonu\\
c_ {67} & = &\frac {32 (2 + 2 k + k^{2} + 2 N + N^{2}) (32 + 55 k + 
      18 k^{2} + 41 N + 35 k\, 
     N + 11 N^{2})} {(2 + N) (2 + k + N)^{6}}, \nonu\\
c_ {68} & = & - \frac{1}{3 (2 + N)^{2} (2 + k + N)^{5}}2 (16912 + 42312 k + 39553 k^{2} + 
       15702 k^{3} + 2224 k^{4} + 55624 N 
        \nonu\\&+& 120178 k\, 
      N + 93150 k^{2} N + 28743 k^{3} N + 2854 k^{4} N + 
       73073 N^{2} + 128394 k\, 
      N^{2} + 75274 k^{2} N^{2} 
       \nonu\\&+& 15087 k^{3} N^{2} + 
       628 k^{4} N^{2} + 48226 N^{3} + 63231 k\, 
      N^{3} + 24081 k^{2} N^{3} + 2118 k^{3} N^{3} + 16568 N^{4} 
       \nonu\\&+& 
       13913 k\, N^{4} + 2392 k^{2} N^{4} + 2737 N^{5} + 1026 k\, 
      N^{5} + 160 N^{6}), \nonu\\
c_ {69} & = & - \frac {(60 + 77 k + 22 k^{2} + 121 N + 115 k\, 
     N + 20 k^{2} N + 79 N^{2} + 42 k\, 
     N^{2} + 16 N^{3})} {2 (2 + N) (2 + k + N)^{2}}, \nonu\\
c_ {70} & = & - \frac {4 (k - N) (13 k + 6 k^{2} + 3 N + 9 k\, 
      N + N^{2})} {(2 + N) (2 + k + N)^{2} (5 + 4 k + 4 N + 3 k\, N)},\nonu\\
 c_ {71} & = &\frac {16 i (k - N) (8 + 17 k + 6 k^{2} + 11 N + 11 k\, 
     N + 3 N^{2})} {(2 + N) (2 + k + N)^{3} (5 + 4 k + 4 N + 3 k\, 
     N)},\nonu\\
 c_ {72} & = &\frac {16 (k - N) (32 + 55 k + 18 k^{2} + 
      41 N + 35 k\, 
     N + 11 N^{2})} {3 (2 + N) (2 + k + N)^{4} (5 + 4 k + 4 N + 3 k\, 
     N)},\nonu\\
 c_ {73} & = &\frac {8 (48 + 97 k + 47 k^{2} + 6 k^{3} + 
      71 N + 85 k\, N + 17 k^{2} N + 28 N^{2} + 14 k\, 
     N^{2} + 3 N^{3})} {(2 + N) (2 + k + N)^{4}}, \nonu\\
c_ {74} & = & - \frac {8 (64 + 139 k + 76 k^{2} + 12 k^{3} + 101 N + 
       135 k\, N + 34 k^{2} N + 45 N^{2} + 28 k\, 
      N^{2} + 6 N^{3})} {(2 + N) (2 + k + N)^{4}}, \nonu\\
c_ {75} & = & - \frac {4 i (32 + 55 k + 18 k^{2} + 41 N + 35 k\, 
      N + 11 N^{2})} {(2 + N) (2 + k + N)^{4}}, \nonu\\
c_ {76} & = & - \frac {8 i (32 + 55 k + 18 k^{2} + 41 N + 35 k\, 
      N + 11 N^{2})} {3 (2 + N) (2 + k + N)^{4}}, \nonu\\
c_ {77} & = &\frac{1}{2 (2 + N)^{2} (2 + k + N)^{4}}(1280 + 1748 k + 347 k^{2} - 294 k^{3} - 
     96 k^{4} + 3116 N + 2960 k\, 
    N 
     \nonu\\&-&264 k^{2} N - 667 k^{3} N - 102 k^{4} N + 2933 N^{2} + 
     1612 k\, 
    N^{2} - 700 k^{2} N^{2} - 305 k^{3} N^{2} + 1330 N^{3} 
     \nonu\\&+& 275 k\, 
    N^{3} - 241 k^{2} N^{3} + 292 N^{4} - k\, 
    N^{4} + 25 N^{5}), \nonu\\
c_ {78} & = &\frac{1} {2 (2 + N)^{2} (2 + k + N)^{4}}(1280 + 868 k - 1005 k^{2} - 882 k^{3} - 
     168 k^{4} + 3516 N + 1768 k\, 
    N 
     \nonu\\&-&2048 k^{2} N - 1213 k^{3} N - 138 k^{4} N + 3653 N^{2} + 
     1032 k\, 
    N^{2} - 1506 k^{2} N^{2} - 431 k^{3} N^{2} 
     \nonu\\&+& 1762 N^{3} +113 k\,N^{3} - 367 k^{2} N^{3} + 390 N^{4} - 31 k\, 
    N^{4} + 31 N^{5}), \nonu\\
c_ {79} & = &\frac {1}{3 (2 + N)^{2} (2 + k + N)^{5} (5 + 4 k + 4 N + 3 k\, 
     N)}i (2400 + 11084 k + 18711 k^{2} + 15702 k^{3} 
      \nonu\\&+& 
      6760 k^{4} + 1152 k^{5} + 18676 N + 66620 k\, 
     N + 90042 k^{2} N + 58178 k^{3} N + 18196 k^{4} N 
      \nonu\\&+& 
      2088 k^{5} N + 41221 N^{2} + 121988 k\, 
     N^{2} + 135606 k^{2} N^{2} + 67979 k^{3} N^{2} + 
      14803 k^{4} N^{2} 
       \nonu\\&+&918 k^{5} N^{2} + 41044 N^{3} + 99658 k\, 
     N^{3} + 87861 k^{2} N^{3} + 31406 k^{3} N^{3} + 
      3711 k^{4} N^{3} + 20582 N^{4} 
       \nonu\\&+& 38989 k\, 
     N^{4} + 24777 k^{2} N^{4} + 4899 k^{3} N^{4} + 5039 N^{5} + 
      6634 k\, N^{5} + 2319 k^{2} N^{5} + 476 N^{6}
       \nonu\\&+& 321 k\, 
     N^{6}), \nonu\\
c_ {80} & = &\frac {1}{3 (2 + N)^{2} (2 + k + N)^{5} (5 + 4 k + 4 N + 3 k\, 
     N)}2 i (12000 + 42796 k + 56367 k^{2} + 
      34950 k^{3} 
       \nonu\\&+& 10280 k^{4} + 1152 k^{5} + 48404 N + 155356 k\, 
     N + 181326 k^{2} N + 97186 k^{3} N + 23780 k^{4} N 
      \nonu\\&+& 
      2088 k^{5} N + 81197 N^{2} + 227644 k\, 
     N^{2} + 225930 k^{2} N^{2} + 98023 k^{3} N^{2} + 
      17675 k^{4} N^{2} 
       \nonu\\&+& 918 k^{5} N^{2} + 71672 N^{3} + 168962 k\, 
     N^{3} + 134205 k^{2} N^{3} + 42058 k^{3} N^{3} + 
      4191 k^{4} N^{3} 
       \nonu\\&+&+ 34810 N^{4} 
     +65225 k\, 
     N^{4} + 37209 k^{2} N^{4} + 6387 k^{3} N^{4} + 8767 N^{5} + 
      11846 k\, N^{5} + 3711 k^{2} N^{5} 
       \nonu\\&+& 892 N^{6}+ 705 k\, 
     N^{6}),\nonu\\
 c_ {81} & = &\frac{1}{3 (2 + N)^{2} (2 + k + N)^{5} (5 + 4 k + 4 N + 3 k\, 
     N)}2 i (12000 + 38956 k + 46695 k^{2} + 
      27510 k^{3} 
       \nonu\\&+& 8552 k^{4} + 1152 k^{5} + 52244 N + 155116 k\, 
     N + 168210 k^{2} N + 86146 k^{3} N + 21620 k^{4} N
      \nonu\\&+& 
      2088 k^{5} N + 91109 N^{2} + 238948 k\, 
     N^{2} + 223470 k^{2} N^{2} + 93139 k^{3} N^{2} + 
      17027 k^{4} N^{2} 
       \nonu\\&+& 918 k^{5} N^{2} + 80924 N^{3} + 180506 k\, 
     N^{3} + 136773 k^{2} N^{3} + 41446 k^{3} N^{3} + 
      4191 k^{4} N^{3} + 38494 N^{4} 
       \nonu\\&+& 69173 k\, 
     N^{4} + 38073 k^{2} N^{4} + 6387 k^{3} N^{4} + 9295 N^{5} + 
      12242 k\, N^{5} + 3711 k^{2} N^{5} + 892 N^{6}
       \nonu\\&+& 705 k\, 
     N^{6}), \nonu\\
c_ {82} & = &\frac{1}{(2 + N)^{2} (2 + k + N)^{5}}2 i (480 + 1860 k + 2317 k^{2} + 
      1082 k^{3} + 168 k^{4} + 1404 N + 4468 k\, 
     N 
      \nonu\\&+& 4360 k^{2} N + 1461 k^{3} N + 138 k^{4} N + 1687 N^{2} + 
      4148 k\, 
     N^{2} + 2818 k^{2} N^{2} + 505 k^{3} N^{2} 
      \nonu\\&+& 1042 N^{3} + 
      1735 k\, N^{3} + 609 k^{2} N^{3} + 322 N^{4} + 269 k\, 
     N^{4} + 39 N^{5}), \nonu\\
c_ {83} & = &\frac {4 (64 + 136 k + 61 k^{2} + 6 k^{3} + 88 N + 
      104 k\, N + 15 k^{2} N + 27 N^{2} + 10 k\, 
     N^{2} + N^{3})} {(2 + N) (2 + k + N)^{4}}, \nonu\\
c_ {84} & = & - \frac {16 (k - N) (32 + 55 k + 18 k^{2} + 41 N + 
       35 k\, N + 11 N^{2})} {3 (2 + N) (2 + k + N)^{5}}, \nonu\\
c_ {85} & = &\frac{1}{(2 + N)^{2} (2 + k + N)^{6}}4 (32 + 55 k + 18 k^{2} + 41 N + 35 k\, 
     N + 11 N^{2}) (20 + 27 k + 8 k^{2} 
      \nonu\\&+& 39 N + 39 k\, 
     N + 7 k^{2} N + 25 N^{2} + 14 k\, 
     N^{2} + 5 N^{3}), \nonu\\
c_ {86} & = &\frac {1}{3 (2 + N)^{2} (2 + k + N)^{6}}4 (32 + 55 k + 18 k^{2} + 41 N + 35 k\, 
     N + 11 N^{2}) (60 + 65 k + 16 k^{2}
      \nonu\\&+& 133 N + 109 k\, 
     N + 17 k^{2} N + 91 N^{2} + 42 k\, 
     N^{2} + 19 N^{3}), \nonu\\
c_ {87} & = & - \frac {4 (32 + 55 k + 18 k^{2} + 41 N + 35 k\, 
      N + 11 N^{2})} {(2 + N) (2 + k + N)^{4}},\nonu\\
c_ {88} & = & - \frac {8 i (32 + 55 k + 18 k^{2} + 41 N + 35 k\, 
      N + 11 N^{2})} {(2 + N) (2 + k + N)^{4}}, \nonu\\
c_ {89} & = &\frac {(k - N) (-11 - 4 k - 4 N + 3 k\, 
     N)} {3 (2 + k + N) (5 + 4 k + 4 N + 3 k\, N)}, \nonu\\
c_ {90} & = &\frac {1}{6 (2 +N) (2 + k + N)^{3} (5 + 4 k + 4 N + 3 k\, N)^{2}}(-k + N) (-3300 - 8203 k - 7482 k^{2}
 \nonu\\&-& 
      2928 k^{3} - 416 k^{4} - 10367 N - 22061 k\, 
     N - 16518 k^{2} N - 4876 k^{3} N - 448 k^{4} N - 12601 N^{2} 
      \nonu\\&-&
      21336 k\, 
     N^{2} - 11130 k^{2} N^{2} - 1331 k^{3} N^{2} + 150 k^{4} N^{2} - 
      7392 N^{3} - 8654 k\, 
     N^{3} - 1491 k^{2} N^{3}
      \nonu\\&+& 987 k^{3} N^{3} + 180 k^{4} N^{3} - 
      2080 N^{4} - 1120 k\, 
     N^{4} + 759 k^{2} N^{4} + 378 k^{3} N^{4} - 224 N^{5} + 48 k\, 
     N^{5} 
      \nonu\\&+& 144 k^{2} N^{5}), \nonu\\
c_ {91} & = & - \frac {1} {3 (2 + k + N)^{3} (5 + 4 k + 4 N + 3 k\, 
       N)^{2}}2 (600 + 3310 k + 5749 k^{2} + 4458 k^{3} + 
       1616 k^{4} 
        \nonu\\&+&224 k^{5} + 2210 N + 10823 k\, 
      N + 16343 k^{2} N + 10630 k^{3} N + 3052 k^{4} N + 
       304 k^{5} N + 3204 N^{2} 
        \nonu\\&+& 13751 k\, 
      N^{2} + 17594 k^{2} N^{2} + 9128 k^{3} N^{2} + 
       1845 k^{4} N^{2} + 90 k^{5} N^{2} + 2336 N^{3} + 8584 k\, 
      N^{3} 
       \nonu\\&+& 8940 k^{2} N^{3} + 3415 k^{3} N^{3} + 387 k^{4} N^{3} + 
       864 N^{4} + 2624 k\, 
      N^{4} + 2052 k^{2} N^{4} + 459 k^{3} N^{4} 
       \nonu\\&+&128 N^{5} + 
       304 k\, N^{5} + 
       144 k^{2} N^{5}), \nonu\\
c_ {92} & = &\frac {8 i (k - N) (7 + 5 k + 5 N + 3 k\, 
     N)} {3 (2 + k + N)^{2} (5 + 4 k + 4 N + 3 k\, N)}, \nonu\\
c_ {93} & = &\frac{1}{12 (2 + N) (2 + k + N)^{3} (5 + 4 k + 4 N + 3 k\, N)}(240 + 772 k + 719 k^{2} + 246 k^{3} + 24 k^{4} 
 \nonu\\&+& 
     452 N + 1280 k\, 
    N + 825 k^{2} N + 109 k^{3} N - 18 k^{4} N + 329 N^{2} + 894 k\, 
    N^{2} + 388 k^{2} N^{2} 
     \nonu\\&+& 15 k^{3} N^{2} + 123 N^{3} + 347 k\, 
    N^{3} + 90 k^{2} N^{3} + 20 N^{4} + 57 k\, 
    N^{4}), \nonu\\
c_ {94} & = & - \frac{1}{3 (2 + N) (2 + k + N)^{4} (5 + 4 k + 4 N + 3 k\, N)}2 i (240 + 652 k + 563 k^{2} + 198 k^{3} + 
       24 k^{4} 
        \nonu\\&+& 332 N + 908 k\, 
      N + 531 k^{2} N + 49 k^{3} N - 18 k^{4} N + 113 N^{2} + 522 k\, 
      N^{2} + 208 k^{2} N^{2} - 3 k^{3} N^{2} 
       \nonu\\&-& 3 N^{3} + 203 k\, 
      N^{3} + 54 k^{2} N^{3} - 4 N^{4} + 39 k\, 
      N^{4}), \nonu\\
c_ {95} & = &\frac {32 (k - N) (7 + 5 k + 5 N + 3 k\, 
     N)} {3 (2 + k + N)^{2} (5 + 4 k + 4 N + 3 k\, N)}, \nonu\\
c_ {96} & = & - \frac{1}{3 (2 + k + N)^{4} (5 + 4 k + 4 N + 3 k\, N)}2 (120 + 566 k + 697 k^{2} + 334 k^{3} + 
       56 k^{4} + 346 N 
        \nonu\\&+& 1243 k\, 
      N + 1125 k^{2} N + 355 k^{3} N + 30 k^{4} N + 364 N^{2} + 
       1005 k\, 
      N^{2} + 605 k^{2} N^{2} + 93 k^{3} N^{2} 
       \nonu\\&+& 176 N^{3} + 368 k\, 
      N^{3} + 117 k^{2} N^{3} + 32 N^{4} + 48 k\, 
      N^{4}), \nonu\\
c_ {97} & = &\frac {1} {3 (2 + N) (2 + k + N)^{4} (5 + 4 k + 4 N + 3 k\, N)}(1200 + 3860 k + 4009 k^{2} + 1786 k^{3} + 
     296 k^{4} 
      \nonu\\&+& 2260 N + 6992 k\, 
    N + 5669 k^{2} N + 1767 k^{3} N + 178 k^{4} N + 1839 N^{2} + 
     5660 k\, 
    N^{2} + 3278 k^{2} N^{2}
     \nonu\\&+& 659 k^{3} N^{2} + 60 k^{4} N^{2} + 
     1045 N^{3} + 2903 k\, 
    N^{3} + 1012 k^{2} N^{3} + 78 k^{3} N^{3} + 396 N^{4} + 847 k\, 
    N^{4} 
     \nonu\\&+&126 k^{2} N^{4} + 64 N^{5} + 96 k\, 
    N^{5}), \nonu\\
c_ {98} & = &\frac {1}{3 (2 + N) (2 + k + N)^{4} (5 + 4 k + 4 N + 3 k\, N)}2 (1200 + 3860 k + 4009 k^{2} + 1786 k^{3} + 
      296 k^{4} 
       \nonu\\&+& 2260 N + 6992 k\, 
     N + 5669 k^{2} N + 1767 k^{3} N + 178 k^{4} N + 1839 N^{2} + 
      5660 k\, 
     N^{2} + 3278 k^{2} N^{2}
      \nonu\\&+& 659 k^{3} N^{2} + 60 k^{4} N^{2} + 
      1045 N^{3} + 2903 k\, 
     N^{3} + 1012 k^{2} N^{3} + 78 k^{3} N^{3} + 396 N^{4} + 847 k\, 
     N^{4} 
      \nonu\\&+& 126 k^{2} N^{4} + 64 N^{5} + 96 k\, 
     N^{5}), \nonu\\
c_ {99} & = &\frac {1}{24 (2 + N) (2 + k + N)^{3} (5 + 4 k + 4 N + 3 k\, N)}(-k + N) (240 + 169 k - 22 k^{2} - 24 k^{3} 
 \nonu\\&+& 
      399 N + 329 k\, 
     N + 99 k^{2} N + 18 k^{3} N + 229 N^{2} + 209 k\, 
     N^{2} + 63 k^{2} N^{2} + 44 N^{3} + 39 k\, 
     N^{3}), \nonu\\
c_ {100} & = & - \frac {i (-k + N) (-11 - 4 k - 4 N + 3 k\, 
      N) (8 + 17 k + 6 k^{2} + 11 N + 11 k\, 
      N + 3 N^{2})} {6 (2 + N) (2 + k + N)^{4} (5 + 4 k + 4 N + 3 k\, 
      N)},\nonu\\
c_ {101} & = & - \frac {(-k + N) (-11 - 4 k - 4 N + 3 k\, 
      N) (32 + 55 k + 18 k^{2} + 41 N + 35 k\, 
      N + 11 N^{2})} {18 (2 + N) (2 + k + N)^{5} (5 + 4 k + 4 N + 
       3 k\, N)}.
\nonu
\eea
The fusion rule is  
\bea
[\widetilde{\Phi}_2^{(1)}] \, \cdot \, [\widetilde{\Phi}_2^{(1)}] & = &
[I] + [\Phi_0^{(1)} \, \Phi_0^{(1)}] +[\Phi_0^{(1)} \, \widetilde{\Phi}_2^{(1)}]
+[\Phi_{\frac{1}{2}}^{(1),i} \, \widetilde{\Phi}_{\frac{3}{2}}^{(1),i}]
+\varepsilon^{ijkl} \, [\Phi_0^{(1),ij} \, \Phi_0^{(1),kl}]
\nonu \\
& + & [\Phi_0^{(2)}] + [\widetilde{\Phi}_2^{(2)}].
\nonu
\eea

\section{The structure constants appearing in the OPE in 
the ${\cal N}=4$ superspace of section $7$ and the $16$ fundamental OPEs}

The structure constants appearing in (\ref{finalPhiPhi}) 
of section $7$ can be 
summarized by
\bea
c_ {1} &=& \frac {4 k (k - N) N} {(2 + k + N)^{2}}, \qquad
c_ {2} = \frac {8 kN} {(2 + k + N)^{2}}, \qquad
c_ {3} = \frac {16 kN} {(2 + k + N)^{2}}, \qquad
c_ {4} = \frac {2 kN} {(2 + k + N)}, \nonu\\
c_ {5} &=& -\frac {2 (k - N)} {(2 + k + N)^{2}}, \qquad
c_ {6} = \frac {(k + N)} {(2 + k + N)}, \qquad
c_ {7} = -\frac {(k + N)} {(2 + k + N)^{2}}, \nonu\\
c_ {8} &=& \frac {(k - N)} {(2 + k + N)}, \qquad
c_ {9} = \frac {3 (k - N)} {(2 + k + N)}, \qquad
c_ {10} = \frac {(12 k + 3 k^{2} + 12 N + 10 k\, 
    N + 3 N^{2})} {(2 + k + N)^{2}}, \nonu\\
c_ {11} &=& \frac {2 (k + N)} {(2 + k + N)^{2}}, \qquad
c_ {12} = \frac {2 (-k + N)} {(2 + k + N)^{2}}, \qquad
c_ {13} = \frac {4 (-k + N)} {(2 + k + N)^{2}}, \qquad
c_ {14} = 2,  \nonu\\
c_ {15} &=& -\frac {(60 + 77 k + 22 k^{2} + 121 N + 115 k\, 
    N + 20 k^{2} N + 79 N^{2} + 42 k\, 
    N^{2} + 16 N^{3})} {(2 + N) (2 + k + N)^{2}}, \nonu\\
c_ {16} & = &\frac {2 (10 + 9 k + 2 k^{2} + 14 N + 7 k\, 
     N + 4 N^{2})} {(2 + k + N)^{2}}, \qquad
c_ {17}  =  - \frac {2} {(2 + k + N)}, \nonu\\
c_ {18} & = &\frac {(32 + 55 k + 18 k^{2} + 41 N + 35 k\, 
    N + 11 N^{2})} {3 (2 + N) (2 + k + N)^{4}}, \qquad
c_ {19}  = \frac {(13 k + 6 k^{2} + 3 N + 9 k\, 
    N + N^{2})} {4 (2 + N) (2 + k + N)^{2}}, \nonu\\
c_ {20} & = &\frac {(20 + 21 k + 6 k^{2} + 25 N + 13 k\, 
    N + 7 N^{2})} {2 (2 + N) (2 + k + N)^{2}}, \nonu\\
c_ {21} & = & - \frac {2 (8 + 17 k + 6 k^{2} + 11 N + 11 k\, 
      N + 3 N^{2})} {(2 + N) (2 + k + N)^{3}}, \nonu\\
c_ {22} & = & - \frac {2 (20 + 21 k + 6 k^{2} + 25 N + 13 k\, 
      N + 7 N^{2})} {(2 + N) (2 + k + N)^{3}}, \nonu\\
c_ {23} & = & - \frac {2 (16 + 26 k + 13 k^{2} + 2 k^{3} + 6 N - 
       11 k\, N - 7 k^{2} N - 10 N^{2} - 15 k\, 
      N^{2} - 4 N^{3})} {(2 + k + N)^{3}}, \nonu\\
c_ {24} & = &\frac {2 (100 + 95 k + 24 k^{2} + 155 N + 83 k\, 
     N + 3 k^{2} N + 69 N^{2} + 14 k\, 
     N^{2} + 9 N^{3})} {(2 + N) (2 + k + N)^{3}}, \nonu\\
c_ {25} & = &\frac {4 (10 + 9 k + 2 k^{2} + 14 N + 7 k\, 
     N + 4 N^{2})} {(2 + k + N)^{3}}, \qquad
c_ {26}  = \frac {1} {6}, \qquad
c_ {27}  = \frac {(k - N)} {6 (2 + k + N)}, \nonu\\
c_ {28} & = & - \frac {1} {(2 + k + N)}, \qquad
c_ {29}  =  - \frac {4} {3 (2 + k + N)^{2}},\qquad
c_ {30}  = \frac {1} {(2 + k + N)}, \nonu\\
c_ {31} & = &\frac {(k - N)} {(2 + k + N)},\qquad
c_ {32}  =  1, \qquad
c_ {33}  =  - \frac {(k - N)} {(2 + k + N)^{2}}, \qquad
c_ {34}  =  - \frac {1} {(2 + k + N)}, \nonu\\
c_ {35} & = & - \frac {2} {(2 + k + N)^{2}}, \qquad
c_ {36}  = \frac {1} {2}, \nonu\\
c_ {37} & = & - \frac {(60 + 77 k + 22 k^{2} + 121 N + 115 k\, 
     N + 20 k^{2} N + 79 N^{2} + 42 k\, 
     N^{2} + 16 N^{3})} {2 (2 + N) (2 + k + N)^{2}}, \nonu\\
c_ {38} & = &\frac {(16 + 21 k + 6 k^{2} + 19 N + 13 k\, 
    N + 5 N^{2})} {(2 + N) (2 + k + N)^{3}}, \nonu\\
c_ {39} & = & - \frac {(20 + 37 k + 14 k^{2} + 9 N + 21 k\, 
     N + 4 k^{2} N - 9 N^{2} - 4 N^{3})} {(2 + N) (2 + k + N)^{3}}, \nonu\\
c_ {40} & = &\frac {(34 k + 29 k^{2} + 6 k^{3} + 14 N + 37 k\, 
    N + 13 k^{2} N + 6 N^{2} + 5 k\, N^{2})} {(2 + k + N)^{3}}, \nonu\\
c_ {41} & = & - \frac {(80 + 88 k + 11 k^{2} - 6 k^{3} + 136 N + 
      110 k\, N + 9 k^{2} N + 75 N^{2} + 36 k\, 
     N^{2} + 13 N^{3})} {2 (2 + N) (2 + k + N)^{3}}, \nonu\\
c_ {42} & = & - \frac {(-72 k - 45 k^{2} - 6 k^{3} + 8 N - 40 k\, 
     N - 7 k^{2} N + 21 N^{2} + 6 k\, 
     N^{2} + 7 N^{3})} {(2 + N) (2 + k + N)^{4}}, \nonu\\
c_ {43} & = &\frac {(16 + 21 k + 6 k^{2} + 19 N + 13 k\, 
    N + 5 N^{2})} {2 (2 + N) (2 + k + N)^{2}}, \nonu\\
c_ {44} & = & - \frac {(k - N) (32 + 29 k + 6 k^{2} + 35 N + 17 k\, 
      N + 9 N^{2})} {2 (2 + N) (2 + k + N)^{3}}, \nonu\\
c_ {45} & = &\frac {(10 + 17 k + 6 k^{2} + 6 N + 7 k\, 
    N)} {(2 + k + N)^{2}}, \nonu\\
c_ {46} & = & - \frac {(20 + 31 k + 10 k^{2} + 35 N + 41 k\, 
     N + 8 k^{2} N + 21 N^{2} + 14 k\, 
     N^{2} + 4 N^{3})} {(2 + N) (2 + k + N)^{3}}, \nonu\\
c_ {47} & = &\frac {(20 + 21 k + 6 k^{2} + 25 N + 13 k\, 
    N + 7 N^{2})} {2 (2 + N) (2 + k + N)^{2}}, \nonu\\
c_ {48} & = & - \frac {(32 + 29 k + 6 k^{2} + 35 N + 17 k\, 
     N + 9 N^{2})} {(2 + N) (2 + k + N)^{3}}, \nonu\\
c_ {49} & = & - \frac {(13 k + 6 k^{2} + 3 N + 9 k\, 
     N + N^{2})} {2 (2 + N) (2 + k + N)^{2}}, \qquad
c_ {50}  =  - \frac {(13 k + 6 k^{2} + 3 N + 9 k\, 
     N + N^{2})} {2 (2 + N) (2 + k + N)^{3}}, \nonu\\
c_ {51} & = &\frac {(20 + 21 k + 6 k^{2} + 25 N + 13 k\, 
    N + 7 N^{2})} {2 (2 + N) (2 + k + N)^{3}}, \nonu\\
c_ {52} & = & - \frac {(32 + 55 k + 18 k^{2} + 41 N + 35 k\, 
     N + 11 N^{2})} {(2 + N) (2 + k + N)^{4}}, \nonu\\
c_ {53} & = &\frac {(4 + k + N) (13 k + 6 k^{2} + 3 N + 9 k\, 
     N + N^{2})} {4 (2 + N) (2 + k + N)^{3}}, \nonu\\
c_ {54} & = & - \frac {(32 + 55 k + 18 k^{2} + 41 N + 35 k\, 
     N + 11 N^{2})} {3 (2 + N) (2 + k + N)^{4}}, \nonu\\
c_ {55} & = & - \frac {(80 + 92 k + 11 k^{2} - 6 k^{3} + 172 N + 
      158 k\, N + 21 k^{2} N + 119 N^{2} + 64 k\, 
     N^{2} + 25 N^{3})} {6 (2 + N) (2 + k + N)^{4}}, \nonu\\
c_ {56} & = &\frac {(16 + 29 k + 10 k^{2} + 27 N + 25 k\, 
    N + 2 k^{2} N + 13 N^{2} + 4 k\, 
    N^{2} + 2 N^{3})} {(2 + N) (2 + k + N)^{3}}, \qquad
c_ {57}  = 2, \nonu\\
c_ {58} & = & - \frac {2 (60 + 77 k + 22 k^{2} + 121 N + 115 k\, 
      N + 20 k^{2} N + 79 N^{2} + 42 k\, 
      N^{2} + 16 N^{3})} {(2 + N) (2 + k + N)^{2}}, \nonu\\
c_ {59} & = & - \frac {8} {(2 + k + N)}, \qquad
c_ {60}  = \frac {2 (10 + 9 k + 2 k^{2} + 14 N + 7 k\, 
     N + 4 N^{2})} {(2 + k + N)^{2}}, \nonu\\
c_ {61} & = &\frac {(20 + 21 k + 6 k^{2} + 25 N + 13 k\, 
    N + 7 N^{2})} {2 (2 + N) (2 + k + N)^{2}}, \nonu\\
c_ {62} & = &\frac {8 (10 + 9 k + 2 k^{2} + 14 N + 7 k\, 
     N + 4 N^{2})} {(2 + k + N)^{3}}, \nonu\\
c_ {63} & = & - \frac {2 (32 + 42 k + 17 k^{2} + 2 k^{3} + 22 N + 
       5 k\, N - 3 k^{2} N - 6 N^{2} - 11 k\, 
      N^{2} - 4 N^{3})} {(2 + k + N)^{3}}, \nonu\\
c_ {64} & = &\frac {2 (100 + 99 k + 26 k^{2} + 151 N + 85 k\, 
     N + 4 k^{2} N + 65 N^{2} + 14 k\, 
     N^{2} + 8 N^{3})} {(2 + N) (2 + k + N)^{3}}, \nonu\\
c_ {65} & = & - \frac {2 (16 + 21 k + 6 k^{2} + 19 N + 13 k\, 
      N + 5 N^{2})} {(2 + N) (2 + k + N)^{3}}, \qquad
c_ {66}  = \frac {8} {(2 + k + N)^{2}}, \nonu\\
c_ {67} & = & - \frac {16} {(2 + k + N)^{2}}, \qquad
c_ {68}  = \frac {(32 + 55 k + 18 k^{2} + 41 N + 35 k\, 
    N + 11 N^{2})} {3 (2 + N) (2 + k + N)^{4}}, \nonu\\
c_ {69} & = &\frac {(13 k + 6 k^{2} + 3 N + 9 k\, 
    N + N^{2})} {4 (2 + N) (2 + k + N)^{2}},\qquad
c_ {70}  =  - \frac {(20 + 21 k + 6 k^{2} + 25 N + 13 k\, 
     N + 7 N^{2})} {(2 + N) (2 + k + N)^{3}}.
\nonu
\label{finalOPEcoeff}
\eea
Of course, if one uses the description in 
the basis of \cite{BCG} or of \cite{Ahn1504}, then 
the corresponding OPE, which is equivalent to (\ref{finalPhiPhi}),
can be obtained similarly.  

It is useful to rewrite the previous $16$ OPEs in Appendix $H.1$
in order to see how the single OPE (\ref{finalPhiPhi})
arises in the component result.
By collecting the spin-$\frac{3}{2}$ currents
with the derivative of spin-$\frac{1}{2}$ currents (and similarly 
the spin-$2$ current and the derivative of spin-$1$ current), 
one has the following $16$ OPEs 
\bea
\Phi_{0}^{(1)}(z)\;\Phi_{0}^{(1)}(w)&=&\frac{1}{(z-w)^{2}}\,c_{4}+\cdots,
\nonu \\
\Phi_{\frac{1}{2}}^{(1),i}(z)\:\Phi_{0}^{(1)}(w)&=&\frac{1}{(z-w)}\Bigg[-6\,c_{26}\,(G^{i}-2\alpha i\,\partial\Gamma^{i})-6\,c_{27}\,i\,\partial\Gamma^{i}
\nonu\\&-&c_{28}\,(2i\,U\,\Gamma^{i}-\varepsilon^{ijkl}\,T^{jk}\,\Gamma^{l})-c_{29}\,\varepsilon^{ijkl}\,i\,\Gamma^{j}\,\Gamma^{k}\,\Gamma^{l}
\Bigg](w)+\cdots,
\nonu \\
\Phi_{1}^{(1),ij}(z)\:\Phi_{0}^{(1)}(w)&=&\frac{1}{(z-w)^{2}}\Bigg[-2\,c_{5}\,\Gamma^{i}\,\Gamma^{j}+\varepsilon^{ijkl}\Big(\,c_{6}\,i\,T^{kl}-2\,c_{7}\,\Gamma^{k}\,\Gamma^{l}\Big)+2\,c_{8}\,i\,T^{ij}
\Bigg](w)
\nonu\\
&+&\frac{1}{(z-w)}\Bigg[2\,c_{30}\,(\,i\,G^{i}\,\Gamma^{j}-i\,G^{j}\,\Gamma^{i}+2\alpha\,\partial\Gamma^{i}\,\Gamma^{j}-2\alpha\,\partial\Gamma^{j}\,\Gamma^{i})+2\,c_{31}\,i\,\partial T^{ij}
\nonu\\
&+&2\,c_{32}\,\varepsilon^{ijkl}\,i\,\partial T^{kl}-2\,c_{33}\,\partial(\Gamma^{i}\,\Gamma^{j})-\varepsilon^{ijkl}\,c_{34}\,\partial(\Gamma^{k}\,\Gamma^{l})
\nonu\\
&-&2\,c_{35}\,\Big(U\,\Gamma^{i}\,\Gamma^{j}+\,i\,\varepsilon^{ijkl}(T^{ki}\,\Gamma^{il}+T^{kj}\,\Gamma^{jl})\Big)
\Bigg](w)+\cdots,
\nonu \\
\Phi_{\frac{3}{2}}^{(1),i}(z)\:\Phi_{0}^{(1)}(w)&=&\frac{1}{(z-w)^{3}}\,c_{2}\,i\,\Gamma^{i}(w)
+\frac{1}{(z-w)^{2}}\Bigg[\,c_{9}\,(G^{i}-2\alpha i\,\partial\Gamma^{i})+c_{10}\,i\,\partial\Gamma^{j}-2\,c_{11}\,T^{ij}\,\Gamma^{j}
\nonu\\
&+&\,c_{12}\,(i\,U\,\Gamma^{i}-\frac{1}{2}\,\varepsilon^{ijkl}\,T^{jk}\,\Gamma^{l})+c_{13}\,i\,U\,\Gamma^{i}
\Bigg](w)
\nonu\\
&+&\frac{1}{(z-w)}\Bigg[
c_{36}\,{\bf \Phi_{\frac{1}{2}}^{(2),i}}+c_{37}\, 
{\bf \Phi_{0}^{(1)}\,\Phi_{\frac{1}{2}}^{(1),i}} \nonu \\
& + & 
c_{38}\,(i\,U\,U\,\Gamma^{i}-i\,\widetilde{T}^{ij}\,
\widetilde{T}^{ij}\,\Gamma^{i}-i\,T^{ij}\,T^{jk}\,\Gamma^{k}
+i\,T^{ij}\,T^{ji}\,\Gamma^{i})
\nonu \\
& + & 
c_{39}\,i\,\partial U\,\Gamma^{i}+c_{40}\,i\,\partial^{2}\Gamma^{i}+c_{41}\,(i\,U\,\partial\Gamma^{i}-\frac{1}{2}\varepsilon^{ijkl}\,T^{jk}\,\partial\Gamma^{l})
\nonu\\
&-&c_{42}\,i\,\Gamma^{i}\,\partial\Gamma^{j}\,\Gamma^{j}+c_{43}\,
(U\,G^{i}-2i\alpha\,U \, \partial\Gamma^{i}+\frac{1}{2}\varepsilon^{ijkl}\,i\,T^{jk}\,G^{l}+\varepsilon^{ijkl}\, \alpha\,T^{jk}\,\partial\Gamma^{l})
\nonu\\
&+&c_{44}\,i\,U\,\partial\Gamma^{i}+c_{45}\,(\partial G^{i}-2i\alpha\,\partial^{2}\Gamma^{i})+c_{46}\,(i\,\partial U\,\Gamma^{i}-\frac{1}{2}\varepsilon^{ijkl}\,\partial T^{jk}\,\Gamma^{l})
\nonu\\
&+&c_{47}\,(i\,T^{ij}\,G^{j}-2i\alpha\,T^{ij}\,\partial\Gamma^{j})+
c_{48}\,(i\,U\,U\,\Gamma^{i}-\frac{1}{2}\varepsilon^{ijkl}\, T^{jk}\,U \, 
\Gamma^{l})
\nonu\\
&+&c_{49}\,U (G^{i}-2i\alpha\,  \partial\Gamma^{i})-
c_{50}\,\varepsilon^{ijkl}(\Gamma^{j}\,\Gamma^{k}\,G^{l}+
2i\alpha\,\Gamma^{j}\,\Gamma^{k}\,\partial\Gamma^{l})
+2\,c_{51}\,(\Gamma^{i}\,\Gamma^{j}\,G^{j}
\nonu\\
&+&2i\alpha\,\Gamma^{i}\,\partial\Gamma^{j}\,\Gamma^{j}+T^{ij}\,U\,\Gamma^{j}-\frac{1}{2}\varepsilon^{ijkl}\,T^{ij}\,T^{kl}\,\Gamma^{i})+c_{52}\,\varepsilon^{ijkl}\,T^{jk}\,\Gamma^{i}\,\Gamma^{j}\,\Gamma^{k}
\nonu\\
&-&2\,c_{53}\,T^{ij}\,\partial\Gamma^{j}+c_{54}\,\varepsilon^{ijkl}\,i\,U\,\Gamma^{j}\,\Gamma^{k}\,\Gamma^{l}+c_{55}\,\varepsilon^{ijkl}\,i\,\partial(\Gamma^{j}\,\Gamma^{k}\,\Gamma^{l})-c_{56}\,\partial T^{ij}\,\Gamma^{j}
\Bigg](w)
\nonu \\
& + & \cdots,
\nonu \\
\Phi_{2}^{(1)}(z)\:\Phi_{0}^{(1)}(w)&=&
\frac{1}{(z-w)^{4}}\,c_{1}+\frac{1}{(z-w)^{3}}\,c_{3}\,U(w)
\nonu\\
&+&\frac{1}{(z-w)^{2}}\Bigg[\,c_{14}\, {\bf \Phi_{0}^{(2)}}+
c_{15}\,{\bf \Phi_{0}^{(1)}\,\Phi_{0}^{(1)}}\nonu \\
& + & 
c_{16}\, 2 (L+\alpha\,\partial U) +  
c_{17}\,(i\,G^{i}\,\Gamma^{i}+2\alpha\, \partial \Gamma^{i}\, \Gamma^{i})
\nonu\\
&+&\varepsilon^{ijkl}\Big(\,c_{18}\,\Gamma^{i}\,\Gamma^{j}\,\Gamma^{k}\,\Gamma^{l}
-c_{19}\,T^{ij}\,T^{kl}\Big)
-c_{20}\,T^{ij}\,T^{ij}-c_{21}\,\frac{i}{2}\,\varepsilon^{ijkl}\,\,T^{ij}\,\Gamma^{k}\,\Gamma^{l}
\nonu\\
&-&c_{22}\,i\,T^{ij}\,\Gamma^{i}\,\Gamma^{j}+c_{23}\,\partial U-c_{24}\,\partial\Gamma^{i}\,\Gamma^{i}+c_{25}\,U\,U\Bigg](w)
\nonu\\
&+&\frac{1}{(z-w)}\Bigg[c_{57}\, {\bf \partial\Phi_{0}^{(2)}}+
c_{58}\, {\bf \partial\Phi_{0}^{(1)}\,\Phi_{0}^{(1)}}+
c_{59}\,i\,G^{i}\,\partial\Gamma^{i}
\nonu \\
& + & c_{60}\,2 \partial( L+\alpha\,\partial U)-c_{61}\,\partial(T^{ij}\,T^{ij})
\nonu\\
&+&c_{62}\,\partial U\,U+c_{63}\,\partial^{2}U-c_{64}\,\partial^{2}\Gamma^{i}\,\Gamma^{i}-\frac{i}{2}\,\varepsilon^{ijkl}\Big(c_{65}\,\partial(T^{ij}\,\Gamma^{k}\,\Gamma^{l})+c_{66}\,\Gamma^{i}\,\Gamma^{j}\,\partial T^{kl}\Big)
\nonu\\
&-&c_{67}\,U\,\partial\Gamma^{i}\,\Gamma^{i}
+\varepsilon^{ijkl}\Big(\,c_{68}\,\partial(\Gamma^{i}\,\Gamma^{j}\,\Gamma^{k}\,\Gamma^{l})-c_{69}\,\partial(T^{ij}\,T^{kl})\Big)\nonu \\
& - & c_{70}\, 2i\,
\partial(T^{ij}\,\Gamma^{i}\,\Gamma^{j})\Bigg](w)
+  \cdots,
\label{finalope}
\eea
where the seventy coefficients $c_1$-$c_{70}$ are given in 
(\ref{finalOPEcoeff}).
Here we intentionally put the 
same structure constants as the ones in section $7$. 
These are the fundamental OPEs which provide the single ${\cal N}=4$ OPE
in (\ref{finalPhiPhi}) because the seventy structure constants 
fix those in (\ref{finalPhiPhi}). As explained in Appendix $F$, 
via the replacement (\ref{compsuper}), one 
obtains the final OPE. 
As described in the introduction, this is the power of ${\cal N}=4$ 
supersymmetry. The other structures in the remaining $120$ OPEs 
can be determined automatically 
by ``supersymmetrizing'' the composite fields appearing in 
(\ref{finalope}). 

For convenience, we present the $X$-dependent coefficients 
(before we are using the value (\ref{X})) as follows:
\bea
c_ {15} & = & - \frac {1} {(k - N) (1 + N)} 2 (-8 - 8 k - 16 N - 
     8 kN - 8 N^{2} + 32 X + 72 kX + 60 k^{2} X + 22 k^{3} X 
     \nonu\\&+& 
     3 k^{4} X + 72 NX + 136 kNX + 90 k^{2} NX + 24 k^{3} NX + 
     2 k^{4} NX + 60 N^{2} X + 90 kN^{2} X 
     \nonu\\&+&  42 k^{2} N^{2} X + 
     6 k^{3} N^{2} X + 22 N^{3} X + 24 kN^{3} X + 6 k^{2} N^{3} X + 
     3 N^{4} X + 2 kN^{4} X),\nonu\\
c_ {16} & = & - \frac {1} {3 (1 + N) (-k + N) (2 + k + N)} 4 (-16 - 
     20 k - 6 k^{2} - 36 N - 24 kN - 6 k^{2} N - 26 N^{2} 
     \nonu\\&-& 
     4 kN^{2} - 6 N^{3} + 64 X + 160 kX + 160 k^{2} X + 80 k^{3} X + 
     20 k^{4} X + 2 k^{5} X + 160 NX 
     \nonu\\&+&  336 kNX + 272 k^{2} NX + 
     104 k^{3} NX + 18 k^{4} NX + k^{5} NX + 160 N^{2} X + 
     272 kN^{2} X 
     \nonu\\&+&  168 k^{2} N^{2} X + 44 k^{3} N^{2} X + 
     4 k^{4} N^{2} X + 80 N^{3} X + 104 kN^{3} X + 44 k^{2} N^{3} X + 
     6 k^{3} N^{3} X
     \nonu\\&+&  20 N^{4} X + 18 kN^{4} X + 4 k^{2} N^{4} X + 
     2 N^{5} X + kN^{5} X), \nonu\\
c_ {18} & = &\frac {2 (2 + k + N) X} {3 (1 + N)}, \nonu\\
c_ {19} & = &\frac {1} {6 (1 + N) (2 + k + N)} (-4 - 4 N + 16 X + 
    32 kX + 24 k^{2} X + 8 k^{3} X + k^{4} X + 32 NX 
    \nonu\\&+& 48 kNX + 
    24 k^{2} NX + 4 k^{3} NX + 24 N^{2} X + 24 kN^{2} X + 
    6 k^{2} N^{2} X + 8 N^{3} X + 4 kN^{3} X 
    \nonu\\&+&  N^{4} X), \nonu\\
c_ {20} & = & - \frac {1} {3 (1 + N) (-k + N) (2 + k + N)} (-16 - 
     16 k - 32 N - 16 kN - 16 N^{2} + 64 X + 144 kX 
     \nonu\\&+& 128 k^{2} X + 
     56 k^{3} X + 12 k^{4} X + k^{5} X + 144 NX + 256 kNX + 
     168 k^{2} NX + 48 k^{3} NX 
     \nonu\\&+&  5 k^{4} NX + 128 N^{2} X + 
     168 kN^{2} X + 72 k^{2} N^{2} X + 10 k^{3} N^{2} X + 
     56 N^{3} X + 48 kN^{3} X 
     \nonu\\&+&  10 k^{2} N^{3} X + 12 N^{4} X + 
     5 kN^{4} X + N^{5} X), \nonu\\
c_ {21} & = & - \frac {1} {3 (1 + N) (2 + k + N)^{2}} 4 (-1 - N + 
     16 X + 32 kX + 24 k^{2} X + 8 k^{3} X + k^{4} X + 32 NX 
     \nonu\\&+& 
     48 kNX + 24 k^{2} NX + 4 k^{3} NX + 24 N^{2} X + 24 kN^{2} X + 
     6 k^{2} N^{2} X + 8 N^{3} X + 4 kN^{3} X 
     \nonu\\&+&  N^{4} X), \nonu\\
c_ {22} & = & - \frac {1} {3 (k - N) (1 + 
       N) (2 + k + N)^{2}} 4 (-16 - 16 k - 32 N - 16 kN - 16 N^{2} + 
     64 X + 144 kX 
     \nonu\\&+& 128 k^{2} X + 56 k^{3} X + 12 k^{4} X + 
     k^{5} X + 144 NX + 256 kNX + 168 k^{2} NX + 48 k^{3} NX 
     \nonu\\&+&  
     5 k^{4} NX + 128 N^{2} X + 168 kN^{2} X + 72 k^{2} N^{2} X + 
     10 k^{3} N^{2} X + 56 N^{3} X + 48 kN^{3} X 
     \nonu\\&+&  10 k^{2} N^{3} X + 
     12 N^{4} X + 5 kN^{4} X + N^{5} X), \nonu\\
c_ {23} & = & - \frac {1} {3 (1 + N) (2 + k + N)^{2}} 4 (-4 - 14 k - 
     6 k^{2} - 18 N - 36 kN - 6 k^{2} N - 20 N^{2} - 22 kN^{2} 
     \nonu\\&-& 
     6 N^{3} + 64 X + 160 kX + 160 k^{2} X + 80 k^{3} X + 
     20 k^{4} X + 2 k^{5} X + 160 NX + 336 kNX 
     \nonu\\&+&  272 k^{2} NX + 
     104 k^{3} NX + 18 k^{4} NX + k^{5} NX + 160 N^{2} X + 
     272 kN^{2} X + 168 k^{2} N^{2} X 
     \nonu\\&+&  44 k^{3} N^{2} X + 
     4 k^{4} N^{2} X + 80 N^{3} X + 104 kN^{3} X + 44 k^{2} N^{3} X + 
     6 k^{3} N^{3} X + 20 N^{4} X 
     \nonu\\&+&  18 kN^{4} X + 4 k^{2} N^{4} X + 
     2 N^{5} X + kN^{5} X), \nonu\\
c_ {24} & = &\frac {1} {3 (k - N) (1 + N) (2 + k + N)^{2}}2 (-160 - 
    176 k - 27 k^{2} - 336 N - 186 kN - 27 k^{2} N 
    \nonu\\&-& 203 N^{2} - 
    10 kN^{2} - 27 N^{3} + 640 X + 1504 kX + 1408 k^{2} X + 
    656 k^{3} X + 152 k^{4} X + 14 k^{5} X 
    \nonu\\&+&  1504 NX + 2880 kNX + 
    2096 k^{2} NX + 704 k^{3} NX + 102 k^{4} NX + 4 k^{5} NX + 
    1408 N^{2} X 
    \nonu\\&+&  2096 kN^{2} X + 1104 k^{2} N^{2} X + 
    236 k^{3} N^{2} X + 16 k^{4} N^{2} X + 656 N^{3} X + 
    704 kN^{3} X 
    \nonu\\&+&  236 k^{2} N^{3} X + 24 k^{3} N^{3} X + 
    152 N^{4} X + 102 kN^{4} X + 16 k^{2} N^{4} X + 14 N^{5} X + 
    4 kN^{5} X), \nonu\\
c_ {25} & = &\frac {1} {3 (k - N) (1 + N) (2 + k + N)^{2}} 8 (-16 - 
    20 k - 6 k^{2} - 36 N - 24 kN - 6 k^{2} N - 26 N^{2} 
    \nonu\\&-& 4 kN^{2} - 
    6 N^{3} + 64 X + 160 kX + 160 k^{2} X + 80 k^{3} X + 20 k^{4} X + 
    2 k^{5} X + 160 NX 
    \nonu\\&+& 336 kNX + 272 k^{2} NX + 104 k^{3} NX + 
    18 k^{4} NX + k^{5} NX + 160 N^{2} X + 272 kN^{2} X 
    \nonu\\&+&  
    168 k^{2} N^{2} X + 44 k^{3} N^{2} X + 4 k^{4} N^{2} X + 
    80 N^{3} X + 104 kN^{3} X + 44 k^{2} N^{3} X + 6 k^{3} N^{3} X 
    \nonu\\&+& 
    20 N^{4} X + 18 kN^{4} X + 4 k^{2} N^{4} X + 2 N^{5} X + 
    kN^{5} X), \nonu\\
c_ {37} & = & - \frac {1} {(k - N) (1 + N)} (-8 - 8 k - 16 N - 8 kN - 
     8 N^{2} + 32 X + 72 kX + 60 k^{2} X + 22 k^{3} X 
     \nonu\\&+& 3 k^{4} X + 
     72 NX + 136 kNX + 90 k^{2} NX + 24 k^{3} NX + 2 k^{4} NX + 
     60 N^{2} X + 90 kN^{2} X 
     \nonu\\&+&  42 k^{2} N^{2} X + 6 k^{3} N^{2} X + 
     22 N^{3} X + 24 kN^{3} X + 6 k^{2} N^{3} X + 3 N^{4} X + 
     2 kN^{4} X), \nonu\\
c_ {38} & = &\frac {1} {3 (1 + N) (2 + k + N)^{2}} 2 (2 + 2 N + 
    16 X + 32 kX + 24 k^{2} X + 8 k^{3} X + k^{4} X + 32 NX 
    \nonu\\&+& 
    48 kNX + 24 k^{2} NX + 4 k^{3} NX + 24 N^{2} X + 24 kN^{2} X + 
    6 k^{2} N^{2} X + 8 N^{3} X + 4 kN^{3} X 
    \nonu\\&+&  N^{4} X), \nonu\\
c_ {39} & = & - \frac {1} {3 (k - N) (1 + 
       N) (2 + k + N)^{2}} 2 (-16 - 16 k + 6 k^{2} - 32 N - 28 kN + 
     6 k^{2} N - 10 N^{2} 
     \nonu\\&-& 12 kN^{2} + 6 N^{3} + 64 X + 144 kX + 
     128 k^{2} X + 56 k^{3} X + 12 k^{4} X + k^{5} X + 144 NX + 
     256 kNX 
     \nonu\\&+&  168 k^{2} NX + 48 k^{3} NX + 5 k^{4} NX + 
     128 N^{2} X + 168 kN^{2} X + 72 k^{2} N^{2} X + 
     10 k^{3} N^{2} X 
     \nonu\\&+&  56 N^{3} X + 48 kN^{3} X + 10 k^{2} N^{3} X + 
     12 N^{4} X + 5 kN^{4} X + N^{5} X), \nonu\\
c_ {40} & = &\frac {1} {3 (1 + N) (2 + k + N)^{2}} 2 (-16 - 2 k - 
    18 N - 2 N^{2} + 2 kN^{2} + 64 X + 160 kX + 160 k^{2} X 
    \nonu\\&+& 
    80 k^{3} X + 20 k^{4} X + 2 k^{5} X + 160 NX + 336 kNX + 
    272 k^{2} NX + 104 k^{3} NX + 18 k^{4} NX 
    \nonu\\&+&  k^{5} NX + 
    160 N^{2} X + 272 kN^{2} X + 168 k^{2} N^{2} X + 
    44 k^{3} N^{2} X + 4 k^{4} N^{2} X + 80 N^{3} X 
    \nonu\\&+&  104 kN^{3} X + 
    44 k^{2} N^{3} X + 6 k^{3} N^{3} X + 20 N^{4} X + 18 kN^{4} X + 
    4 k^{2} N^{4} X + 2 N^{5} X + kN^{5} X), \nonu\\
c_ {41} & = &\frac {1} {3 (k - N) (1 + N) (2 + k + N)^{2}} (64 + 
    80 k + 14 k^{2} + 144 N + 116 kN + 14 k^{2} N + 94 N^{2} 
    \nonu\\&+& 
    36 kN^{2} + 14 N^{3} - 256 X - 640 kX - 624 k^{2} X - 
    288 k^{3} X - 56 k^{4} X + k^{6} X - 640 NX 
    \nonu\\&-&  1376 kNX - 
    1120 k^{2} NX - 416 k^{3} NX - 64 k^{4} NX - 2 k^{5} NX - 
    624 N^{2} X - 1120 kN^{2} X 
    \nonu\\&-&  720 k^{2} N^{2} X - 
    192 k^{3} N^{2} X - 17 k^{4} N^{2} X - 288 N^{3} X - 
    416 kN^{3} X - 192 k^{2} N^{3} X - 28 k^{3} N^{3} X 
    \nonu\\&-&  
    56 N^{4} X - 64 kN^{4} X - 17 k^{2} N^{4} X - 2 kN^{5} X + 
    N^{6} X), \nonu\\
c_ {42} & = &\frac {1} {3 (1 + N) (2 + k + N)^{2}} 2 (-16 - 16 N + 
    64 X + 104 kX + 60 k^{2} X + 14 k^{3} X + k^{4} X 
    \nonu\\&+&  104 NX 
   +120 kNX + 42 k^{2} NX + 4 k^{3} NX + 60 N^{2} X + 42 kN^{2} X + 
    6 k^{2} N^{2} X + 14 N^{3} X 
    \nonu\\&+&  4 kN^{3} X + N^{4} X), \nonu\\
c_ {43} & = &\frac {1} {3 (1 + N) (2 + k + N)} (2 + 2 N + 16 X + 
    32 kX + 24 k^{2} X + 8 k^{3} X + k^{4} X + 32 NX 
    \nonu\\&+& 48 kNX + 
    24 k^{2} NX + 4 k^{3} NX + 24 N^{2} X + 24 kN^{2} X + 
    6 k^{2} N^{2} X + 8 N^{3} X + 4 kN^{3} X 
    \nonu\\&+&  N^{4} X), \nonu\\
c_ {44} & = & - \frac {1} {3 (1 + N) (2 + k + N)^{2}} (k - N) (8 + 
     8 N + 16 X + 32 kX + 24 k^{2} X + 8 k^{3} X + k^{4} X 
     \nonu\\&+&  32 NX + 
     48 kNX + 24 k^{2} NX + 4 k^{3} NX + 24 N^{2} X + 24 kN^{2} X + 
     6 k^{2} N^{2} X + 8 N^{3} X 
     \nonu\\&+&  4 kN^{3} X + N^{4} X), \nonu\\
c_ {45} & = & - \frac {1} {3 (1 + N) (-k + N) (2 + k + N)} 2 (-16 - 
     20 k - 36 N - 36 kN - 20 N^{2} - 16 kN^{2} + 64 X 
     \nonu\\&+& 160 kX + 
     160 k^{2} X + 80 k^{3} X + 20 k^{4} X + 2 k^{5} X + 160 NX + 
     336 kNX + 272 k^{2} NX 
     \nonu\\&+&  104 k^{3} NX + 18 k^{4} NX + k^{5} NX + 
     160 N^{2} X + 272 kN^{2} X + 168 k^{2} N^{2} X + 
     44 k^{3} N^{2} X 
     \nonu\\&+&  4 k^{4} N^{2} X + 80 N^{3} X + 104 kN^{3} X + 
     44 k^{2} N^{3} X + 6 k^{3} N^{3} X + 20 N^{4} X + 18 kN^{4} X + 
     4 k^{2} N^{4} X 
     \nonu\\&+&  2 N^{5} X + kN^{5} X), \nonu\\
c_ {46} & = & - \frac {1} {3 (k - N) (1 + 
       N) (2 + k + N)^{2}} 2 (-16 - 24 k - 6 k^{2} - 40 N - 44 kN - 
     6 k^{2} N - 30 N^{2} 
     \nonu\\&-&20 kN^{2} - 6 N^{3} + 64 X + 176 kX + 
     192 k^{2} X + 104 k^{3} X + 28 k^{4} X + 3 k^{5} X + 176 NX 
     \nonu\\&+&  
     416 kNX + 376 k^{2} NX + 160 k^{3} NX + 31 k^{4} NX + 
     2 k^{5} NX + 192 N^{2} X + 376 kN^{2} X 
     \nonu\\&+&  264 k^{2} N^{2} X + 
     78 k^{3} N^{2} X + 8 k^{4} N^{2} X + 104 N^{3} X + 
     160 kN^{3} X + 78 k^{2} N^{3} X + 12 k^{3} N^{3} X 
     \nonu\\&+&  
     28 N^{4} X + 31 kN^{4} X + 8 k^{2} N^{4} X + 3 N^{5} X + 
     2 kN^{5} X), \nonu\\
c_ {47} & = & - \frac {1} {3 (1 + N) (-k + N) (2 + k + N)} (-16 - 
     16 k - 32 N - 16 kN - 16 N^{2} + 64 X + 144 kX 
     \nonu\\&+& 128 k^{2} X + 
     56 k^{3} X + 12 k^{4} X + k^{5} X + 144 NX + 256 kNX + 
     168 k^{2} NX + 48 k^{3} NX 
     \nonu\\&+&  5 k^{4} NX + 128 N^{2} X + 
     168 kN^{2} X + 72 k^{2} N^{2} X + 10 k^{3} N^{2} X + 
     56 N^{3} X + 48 kN^{3} X 
     \nonu\\&+&  10 k^{2} N^{3} X + 12 N^{4} X + 
     5 kN^{4} X + N^{5} X), \nonu\\
c_ {48} & = & - \frac {1} {3 (1 + N) (2 + k + N)^{2}} 2 (8 + 8 N + 
     16 X + 32 kX + 24 k^{2} X + 8 k^{3} X + k^{4} X + 32 NX 
     \nonu\\&+& 
     48 kNX + 24 k^{2} NX + 4 k^{3} NX + 24 N^{2} X + 24 kN^{2} X + 
     6 k^{2} N^{2} X + 8 N^{3} X + 4 kN^{3} X 
     \nonu\\&+&  N^{4} X), \nonu\\
c_ {49} & = & - \frac {1} {3 (1 + N) (2 + k + N)} (-4 - 4 N + 16 X + 
     32 kX + 24 k^{2} X + 8 k^{3} X + k^{4} X + 32 NX 
     \nonu\\&+& 48 kNX + 
     24 k^{2} NX + 4 k^{3} NX + 24 N^{2} X + 24 kN^{2} X + 
     6 k^{2} N^{2} X + 8 N^{3} X + 4 kN^{3} X 
     \nonu\\&+&  N^{4} X), \nonu\\
c_ {50} & = & - \frac {1} {3 (1 + N) (2 + k + N)^{2}} (-4 - 4 N + 
     16 X + 32 kX + 24 k^{2} X + 8 k^{3} X + k^{4} X + 32 NX 
     \nonu\\&+& 
     48 kNX + 24 k^{2} NX + 4 k^{3} NX + 24 N^{2} X + 24 kN^{2} X + 
     6 k^{2} N^{2} X + 8 N^{3} X + 4 kN^{3} X 
     \nonu\\&+&  N^{4} X), \nonu\\
c_ {51} & = &\frac {1} {3 (k - N) (1 + N) (2 + k + N)^{2}} (-16 - 
    16 k - 32 N - 16 kN - 16 N^{2} + 64 X + 144 kX
    \nonu\\&+& 128 k^{2} X + 
    56 k^{3} X + 12 k^{4} X + k^{5} X + 144 NX + 256 kNX + 
    168 k^{2} NX + 48 k^{3} NX 
    \nonu\\&+&  5 k^{4} NX + 128 N^{2} X + 
    168 kN^{2} X + 72 k^{2} N^{2} X + 10 k^{3} N^{2} X + 56 N^{3} X + 
    48 kN^{3} X
    \nonu\\&+&  10 k^{2} N^{3} X + 12 N^{4} X + 5 kN^{4} X + 
    N^{5} X),\nonu\\
c_ {52} & = & - \frac {2 (2 + k + N) X} {(1 + N)}, \nonu\\
c_ {53} & = &\frac {1} {6 (1 + N) (2 + k + N)^{2}} (4 + k + N) (-4 - 
    4 N + 16 X + 32 kX + 24 k^{2} X + 8 k^{3} X + k^{4} X 
    \nonu\\&+& 32 NX + 
    48 kNX + 24 k^{2} NX + 4 k^{3} NX + 24 N^{2} X + 24 kN^{2} X + 
    6 k^{2} N^{2} X + 8 N^{3} X 
    \nonu\\&+&  4 kN^{3} X + N^{4} X), \nonu\\
c_ {54} & = & - \frac {2 (2 + k + N) X} {3 (1 + N)}, \nonu\\
c_ {55} & = &\frac {1} {9 (k - N) (1 + N) (2 + k + N)^{2}} (32 + 
    32 k + 64 N + 32 kN + 32 N^{2} - 128 X - 288 kX 
    \nonu\\&-& 232 k^{2} X - 
    76 k^{3} X - 6 k^{4} X + k^{5} X - 288 NX - 560 kNX - 
    372 k^{2} NX - 96 k^{3} NX 
    \nonu\\&-&  7 k^{4} NX - 232 N^{2} X - 
    372 kN^{2} X - 180 k^{2} N^{2} X - 26 k^{3} N^{2} X - 
    76 N^{3} X - 96 kN^{3} X 
    \nonu\\&-&  26 k^{2} N^{3} X - 6 N^{4} X - 
    7 kN^{4} X + N^{5} X), \nonu\\
c_ {56} & = &\frac {1} {3 (1 + N) (2 + k + N)^{2}} 2 (2 + 3 k + 5 N + 
    3 kN + 3 N^{2} + 16 X + 32 kX + 24 k^{2} X + 8 k^{3} X 
    \nonu\\&+& 
    k^{4} X + 32 NX + 48 kNX + 24 k^{2} NX + 4 k^{3} NX + 
    24 N^{2} X + 24 kN^{2} X + 6 k^{2} N^{2} X + 8 N^{3} X 
    \nonu\\&+&  
    4 kN^{3} X + N^{4} X), \nonu\\
c_ {58} & = & - \frac {1} {(k - N) (1 + N)} 4 (-8 - 8 k - 16 N - 
     8 kN - 8 N^{2} + 32 X + 72 kX + 60 k^{2} X + 22 k^{3} X 
     \nonu\\&+& 
     3 k^{4} X + 72 NX + 136 kNX + 90 k^{2} NX + 24 k^{3} NX + 
     2 k^{4} NX + 60 N^{2} X + 90 kN^{2} X 
     \nonu\\&+&  42 k^{2} N^{2} X + 
     6 k^{3} N^{2} X + 22 N^{3} X + 24 kN^{3} X + 6 k^{2} N^{3} X + 
     3 N^{4} X + 2 kN^{4} X),\nonu\\
 c_ {60} & = & - \frac {1} {3 (1 + N) (-k + N) (2 + k + N)} 4 (-16 - 
     20 k - 6 k^{2} - 36 N - 24 kN - 6 k^{2} N - 26 N^{2} 
     \nonu\\&-& 
     4 kN^{2} - 6 N^{3} + 64 X + 160 kX + 160 k^{2} X + 80 k^{3} X + 
     20 k^{4} X + 2 k^{5} X + 160 NX 
      \nonu\\&+& 336 kNX + 272 k^{2} NX + 
     104 k^{3} NX + 18 k^{4} NX + k^{5} NX + 160 N^{2} X + 
     272 kN^{2} X 
      \nonu\\&+& 168 k^{2} N^{2} X+ 44 k^{3} N^{2} X + 
     4 k^{4} N^{2} X + 80 N^{3} X + 104 kN^{3} X + 44 k^{2} N^{3} X + 
     6 k^{3} N^{3} X
      \nonu\\&+& 20 N^{4} X + 18 kN^{4} X + 4 k^{2} N^{4} X + 
     2 N^{5} X + kN^{5} X), \nonu\\
c_ {61} & = & - \frac {1} {3 (1 + N) (-k + N) (2 + k + N)} (-16 - 
     16 k - 32 N - 16 kN - 16 N^{2} + 64 X + 144 kX 
     \nonu\\&+& 128 k^{2} X + 
     56 k^{3} X + 12 k^{4} X + k^{5} X + 144 NX + 256 kNX + 
     168 k^{2} NX + 48 k^{3} NX 
     \nonu\\&+&  5 k^{4} NX + 128 N^{2} X + 
     168 kN^{2} X + 72 k^{2} N^{2} X + 10 k^{3} N^{2} X + 
     56 N^{3} X + 48 kN^{3} X 
     \nonu\\&+&  10 k^{2} N^{3} X + 12 N^{4} X + 
     5 kN^{4} X + N^{5} X), \nonu\\
c_ {62} & = &\frac {1} {3 (k - N) (1 + N) (2 + k + N)^{2}} 16 (-16 - 
    20 k - 6 k^{2} - 36 N - 24 kN - 6 k^{2} N - 26 N^{2} 
    \nonu\\&-& 4 kN^{2} - 
    6 N^{3} + 64 X + 160 kX + 160 k^{2} X + 80 k^{3} X + 20 k^{4} X + 
    2 k^{5} X + 160 NX  
    \nonu\\&+& 336 kNX +272 k^{2} NX + 104 k^{3} NX + 
    18 k^{4} NX + k^{5} NX + 160 N^{2} X + 272 kN^{2} X 
     \nonu\\&+& 
    168 k^{2} N^{2} X + 44 k^{3} N^{2} X + 4 k^{4} N^{2} X + 
    80 N^{3} X + 104 kN^{3} X + 44 k^{2} N^{3} X + 6 k^{3} N^{3} X 
     \nonu\\&+& 
    20 N^{4} X + 18 kN^{4} X + 4 k^{2} N^{4} X + 2 N^{5} X + 
    kN^{5} X), \nonu\\
c_ {63} & = & - \frac {1} {3 (1 + N) (2 + k + N)^{2}} 4 (8 - 8 k - 
     6 k^{2} - 24 kN - 6 k^{2} N - 14 N^{2} - 16 kN^{2} - 6 N^{3} + 
     64 X 
     \nonu\\&+& 160 kX + 160 k^{2} X + 80 k^{3} X + 20 k^{4} X + 
     2 k^{5} X + 160 NX + 336 kNX + 272 k^{2} NX 
     \nonu\\&+&  104 k^{3} NX + 
     18 k^{4} NX + k^{5} NX + 160 N^{2} X + 272 kN^{2} X + 
     168 k^{2} N^{2} X + 44 k^{3} N^{2} X 
      \nonu\\&+& 4 k^{4} N^{2} X + 
     80 N^{3} X + 104 kN^{3} X + 44 k^{2} N^{3} X + 6 k^{3} N^{3} X + 
     20 N^{4} X + 18 kN^{4} X + 4 k^{2} N^{4} X 
      \nonu\\&+& 2 N^{5} X + 
     kN^{5} X), \nonu\\
c_ {64} & = &\frac {1} {3 (k - N) (1 + N) (2 + k + N)^{2}} 4 (-80 - 
    88 k - 12 k^{2} - 168 N - 96 kN - 12 k^{2} N - 100 N^{2} 
    \nonu\\&-& 
    8 kN^{2} - 12 N^{3} + 320 X + 752 kX + 704 k^{2} X + 
    328 k^{3} X + 76 k^{4} X + 7 k^{5} X + 752 NX 
    \nonu\\&+&  1440 kNX + 
    1048 k^{2} NX + 352 k^{3} NX + 51 k^{4} NX + 2 k^{5} NX + 
    704 N^{2} X + 1048 kN^{2} X 
    \nonu\\&+&  552 k^{2} N^{2} X + 
    118 k^{3} N^{2} X + 8 k^{4} N^{2} X + 328 N^{3} X + 
    352 kN^{3} X + 118 k^{2} N^{3} X + 12 k^{3} N^{3} X 
     \nonu\\&+& 
    76 N^{4} X + 51 kN^{4} X + 8 k^{2} N^{4} X + 7 N^{5} X + 
    2 kN^{5} X), \nonu\\
c_ {65} & = & - \frac {1} {3 (1 + N) (2 + k + N)^{2}} 4 (2 + 2 N + 
     16 X + 32 kX + 24 k^{2} X + 8 k^{3} X + k^{4} X + 32 NX 
     \nonu\\&+& 
     48 kNX + 24 k^{2} NX + 4 k^{3} NX + 24 N^{2} X + 24 kN^{2} X + 
     6 k^{2} N^{2} X + 8 N^{3} X + 4 kN^{3} X 
     \nonu\\&+&  N^{4} X),\nonu\\ 
c_ {68} & = &\frac {2 (2 + k + N) X} {3 (1 + N)}, \nonu\\
c_ {69} & = &\frac {1} {6 (1 + N) (2 + k + N)} (-4 - 4 N + 16 X + 
    32 kX + 24 k^{2} X + 8 k^{3} X + k^{4} X + 32 NX 
    \nonu\\&+& 48 kNX + 
    24 k^{2} NX + 4 k^{3} NX + 24 N^{2} X + 24 kN^{2} X + 
    6 k^{2} N^{2} X + 8 N^{3} X + 4 kN^{3} X 
    \nonu\\&+&  N^{4} X), \nonu\\
c_ {70} & = & - \frac {1} {3 (k - N) (1 + 
       N) (2 + k + N)^{2}} 2 (-16 - 16 k - 32 N - 16 kN - 16 N^{2} + 
     64 X + 144 kX 
     \nonu\\&+&
      128 k^{2} X + 56 k^{3} X + 12 k^{4} X + 
     k^{5} X + 144 NX + 256 kNX + 168 k^{2} NX + 48 k^{3} NX 
     \nonu\\&+&  
     5 k^{4} NX + 128 N^{2} X + 168 kN^{2} X + 72 k^{2} N^{2} X + 
     10 k^{3} N^{2} X + 56 N^{3} X + 48 kN^{3} X 
     \nonu\\&+&  10 k^{2} N^{3} X + 
     12 N^{4} X + 5 kN^{4} X + N^{5} X).
\nonu
\eea
The remaining coefficients we do not present above are the same as the ones 
in (\ref{finalOPEcoeff}). 

\section{Some relations between the next ${\cal N}=4$ higher spin currents
in different bases   }

The exact relations between the higher spin currents in \cite{BCG}
and those in \cite{Ahn1504} can be obtained.
For example, the lowest higher spin-$2$ current living in 
the next ${\cal N}=4$ higher spin multiplet can be written as 
\footnote{
The notation for the higher spin currents with typewriter fonts  
in this Appendix 
is the same as the ones with boldface fonts in \cite{Ahn1504}.} 
\bea
V_{0}^{(2)}(z)&=&
c_{1}\,{\tt P}^{(2)}(z)+
c_{2}\,{\tt W}^{(2)}(z)+
c_{3}\,{\tt T}^{(2)}(z)+
c_{4}\,{\tt T}^{(1)}\,{\tt T}^{(1)}(z)+
c_{5}\,L(z)+c_{6}\,A_{3}\,A_{3}(z)
\nonu \\
& + & 
c_{7}\,\partial A_{3}(z)+c_{8}\,A_{3}\,F_{11}\,F_{22}(z)
+c_{9}\,A_{3}\,F_{12}\,F_{21}(z)+
c_{10}\,A_{-}\,F_{11}\,F_{12}(z)
\nonu \\
& + & c_{11}\,A_{+}\,A_{-}(z)+
c_{12}\,B_{3}\,B_{3}(z)+
c_{13}\,\partial B_{3}(z)+
c_{14}\,B_{3}\,F_{11}\,F_{22}(z)
\nonu\\
&+& c_{15}\,B_{3}\,F_{12}\,F_{21}(z)+
c_{16}\,B_{-}\,F_{12}\,F_{22}(z)+
c_{17}\,B_{+}\,B_{-}(z)+
c_{18}\,B_{+}\,F_{11}\,F_{21}(z)
\nonu \\
& + & 
c_{19}\,F_{11}\,G_{22}(z)
+ c_{20}\,F_{11}\,F_{12}\,F_{21}\,F_{22}(z)
+c_{21}\,\partial F_{11}\,F_{22}(z)+
c_{22}\,F_{11}\,\partial F_{22}(z)
\nonu \\
& + &
c_{23}\,F_{12}\,G_{21}(z)+c_{24}\,\partial F_{12}\,F_{21}(z)
+ c_{25}\,F_{12}\,\partial F_{21}(z)+
c_{26}\,F_{21}\,G_{12}(z)
\nonu \\
& + & c_{27}\,F_{22}\,G_{11}(z)+
c_{28}\,U\,A_{3}(z)+
c_{29}\,U\,B_{3}(z)+
c_{30}\,U\,U(z)
\nonu\\
&+& c_{31}\,U\,F_{11}\,F_{22}(z)
+c_{32}\,U\,F_{12}\,F_{21}(z)+
c_{33}\,A_{3}\,B_{3}(z),
\nonu
\eea
where 
the 
coefficients are given by
\bea
c_ {1} & = & 1,
\qquad
c_ {2}  =  - 4, 
\qquad
c_ {3}  =  - \frac {4 (k - N)} {3 (2 + k + N)}, 
\nonu\\
c_ {4} & = &\frac {(60 + 77 k + 22 k^{2} + 121 N + 115 k\, 
    N + 20 k^{2} N + 79 N^{2} + 42 k\, 
    N^{2} + 16 N^{3})} {2 (2 + N) (2 + k + N)^{2}}, 
\nonu\\
c_ {5} & = & - \frac {2 (3 + 2 k + N) (10 + 5 k + 
       8 N)} {3 (2 + k + N)^{2}}, 
\qquad
c_ {6}  =  - \frac {2 (15 + 10 k + 11 N)} {3 (2 + k + N)^{2}}, 
\nonu\\
c_ {7} & = & - \frac {2 i (15 + 10 k + 11 N)} {3 (2 + k + N)^{2}}, 
\qquad
c_ {8}  = \frac {4 i (9 + 2 k + 7 N)} {3 (2 + k + N)^{3}}, 
\qquad
c_ {9}  = \frac {4 i (3 + k + 2 N)} {(2 + k + N)^{3}}, 
\nonu\\
c_ {10} & = &\frac {4 i (18 + 5 k + 13 N)} {3 (2 + k + N)^{3}}, 
\qquad
c_ {11}  =  - \frac {2 (15 + 10 k + 11 N)} {3 (2 + k + N)^{2}}, 
\nonu\\
c_ {12} & = & - \frac {2 (30 + 47 k + 18 k^{2} + 34 N + 31 k\, 
      N + 8 N^{2})} {3 (2 + N) (2 + k + N)^{2}}, 
\nonu\\
c_ {13} & = & - \frac {2 i (30 + 47 k + 18 k^{2} + 34 N + 31 k\, 
      N + 8 N^{2})} {3 (2 + N) (2 + k + N)^{2}}, 
\nonu\\
c_ {14} & = & - \frac {4 i (42 + 55 k + 18 k^{2} + 56 N + 35 k\, 
      N + 16 N^{2})} {3 (2 + N) (2 + k + N)^{3}}, 
\nonu\\
c_ {15} & = &\frac {4 i (14 + 19 k + 6 k^{2} + 18 N + 12 k\, 
     N + 5 N^{2})} {(2 + N) (2 + k + N)^{3}}, 
\nonu\\
c_ {16} & = & - \frac {4 i (24 + 43 k + 18 k^{2} + 41 N + 29 k\, 
      N + 13 N^{2})} {3 (2 + N) (2 + k + N)^{3}}, 
\nonu\\
c_ {17} & = & - \frac {2 (30 + 47 k + 18 k^{2} + 34 N + 31 k\, 
      N + 8 N^{2})} {3 (2 + N) (2 + k + N)^{2}}, 
\nonu\\
c_ {18} & = & - \frac {4 i (20 + 23 k + 6 k^{2} + 23 N + 14 k\, 
      N + 6 N^{2})} {(2 + N) (2 + k + N)^{3}}, 
\nonu\\
c_ {19} & = &\frac {2 (3 + 5 k + 4 N)} {3 (2 + k + N)^{2}}, 
\qquad
c_ {20}  =  - \frac {4 (32 + 55 k + 18 k^{2} + 41 N + 35 k\, 
      N + 11 N^{2})} {(2 + N) (2 + k + N)^{4}}, 
\nonu\\
c_ {21} & = & - \frac {(156 + 225 k + 94 k^{2} + 309 N + 219 k\, 
     N + 20 k^{2} N + 143 N^{2} + 42 k\, 
     N^{2} + 16 N^{3})} {3 (2 + N) (2 + k + N)^{3}}, 
\nonu\\
c_ {22} & = &\frac {(444 + 369 k + 94 k^{2} + 597 N + 291 k\, 
    N + 20 k^{2} N + 215 N^{2} + 42 k\, 
    N^{2} + 16 N^{3})} {3 (2 + N) (2 + k + N)^{3}}, 
\nonu\\
c_ {23} & = & - \frac {4} {(2 + k + N)^{2}}, 
\nonu\\
c_ {24} & = & - \frac {(300 + 345 k + 94 k^{2} + 405 N + 279 k\, 
     N + 20 k^{2} N + 155 N^{2} + 42 k\, 
     N^{2} + 16 N^{3})} {3 (2 + N) (2 + k + N)^{3}}, 
\nonu\\
c_ {25} & = &\frac {(300 + 281 k + 94 k^{2} + 469 N + 247 k\, 
    N + 20 k^{2} N + 187 N^{2} + 42 k\, 
    N^{2} + 16 N^{3})} {3 (2 + N) (2 + k + N)^{3}}, 
\nonu\\
c_ {26} & = &\frac {4} {(2 + k + N)}, 
\qquad
c_ {27}  = \frac {2 (3 + 4 k + 5 N)} {3 (2 + k + N)^{2}}, 
\qquad
c_ {28}  =  - \frac {4 i\, N} {(2 + k + N)^{2}}, 
\nonu\\
c_ {29} & = & - \frac {4 i\, k} {(2 + k + N)^{2}}, 
\qquad
c_ {30}  =  - \frac {2 (3 + 2 k + N) (10 + 5 k + 
       8 N)} {3 (2 + k + N)^{3}}, 
\nonu\\
c_ {31} & = &\frac {16 (k - N)} {3 (2 + k + N)^{3}}, 
\qquad
c_ {32}  =  - \frac {12} {(2 + k + N)^{2}}, 
\qquad
c_ {33}  = \frac {4 (k - N)} {3 (2 + k + N)^{2}}.
\nonu
\eea

For the higher spin-$\frac{5}{2}$ current, one has the following 
relation  
\bea
V_{\frac{1}{2}}^{(2),0}(z)&=&c_{1}\,{\tt P}{}_{-}^{(\frac{5}{2})}(z)
+c_{2}\,{\tt P}{}_{+}^{(\frac{5}{2})}(z)
+c_{3}\,{\tt W}{}_{-}^{(\frac{5}{2})}(z)
+c_{4}\,{\tt W}{}_{+}^{(\frac{5}{2})}(z)
+c_{5}\,A_{3}\,G_{12}(z)+c_{6}\,A_{3}\,G_{21}(z)
\nonu \\
& + & c_{7}\,A_{3}\,A_{3}\,F_{12}(z)
+ c_{8}\,A_{3}\,A_{3}\,F_{21}(z)
+c_{9}\,A_{3}\,B_{3}\,F_{12}(z)
+c_{10}\,A_{3}\,B_{3}\,F_{21}(z)
\nonu \\
& + & c_{11}\,A_{3}\,B_{-}\,F_{22}(z)
+c_{12}\,A_{3}\,B_{+}\,F_{11}(z)
+c_{13}\,A_{3}\,F_{11}\,F_{12}\,F_{22}(z)
+c_{14}\,A_{3}\,F_{11}\,F_{21}\,F_{22}(z)
\nonu \\
& + & c_{15}\,A_{3}\,\partial F_{21}(z)
+c_{16}\,A_{-}\,G_{11}(z)+c_{17}\,A_{-}\,B_{3}\,F_{11}(z)
+c_{18}\,A_{-}\,B_{-}\,F_{12}(z)
\nonu \\
& + & c_{19}\,A_{-}\,F_{11}\,F_{12}\,F_{21}(z)+
c_{20}\,A_{-}\,\partial F_{11}(z)
+c_{21}\,A_{+}\,G_{22}(z)
+c_{22}\,A_{+}\,A_{-}\,F_{12}(z)
\nonu\\&+&c_{23}\,A_{+}\,A_{-}\,F_{21}(z)
+c_{24}\,A_{+}\,B_{3}\,F_{22}(z)
+c_{25}\,A_{+}\,B_{+}\,F_{21}(z)
+c_{26}\,A_{+}\,F_{12}\,F_{21}\,F_{22}(z)
\nonu \\
& + & c_{27}\,B_{3}\,G_{12}(z)
+ c_{28}\,B_{3}\,G_{21}(z)
+c_{29}\,B_{3}\,B_{3}\,F_{12}(z)
+c_{30}\,B_{3}\,B_{3}\,F_{21}(z)
\nonu \\
& + & c_{31}\,B_{3}\,F_{11}\,F_{12}\,F_{22}(z)
+c_{32}\,B_{3}\,F_{11}\,F_{21}\,F_{22}(z)
+c_{33}\,B_{3}\,\partial F_{21}(z)
+c_{34}\,B_{-}\,G_{22}(z)
\nonu \\
& + & c_{35}\,B_{-}\,F_{12}\,F_{21}\,F_{22}(z)
+c_{36}\,B_{-}\,\partial F_{22}(z)+c_{37}\,B_{+}\,G_{11}(z)
+c_{38}\,B_{+}\,B_{-}\,F_{12}(z)
\nonu \\
& + & c_{39}\,B_{+}\,B_{-}\,F_{21}(z)
+c_{40}\,B_{+}\,F_{11}\,F_{12}\,F_{21}(z)
+c_{41}\,F_{11}\,F_{12}\,G_{22}(z)
+c_{42}\,F_{11}\,F_{12}\,\partial F_{22}(z)
\nonu \\
& + & c_{43}\,F_{11}\,F_{21}\,G_{22}(z)
+c_{44}\,F_{11}\,F_{21}\,\partial F_{22}(z)
+c_{45}\,F_{11}\,F_{22}\,G_{12}(z)
+c_{46}\,F_{11}\,F_{22}\,G_{21}(z)
\nonu\\
&+&c_{47}\,F_{11}\,\partial F_{12}\,F_{22}(z)
+c_{48}\,F_{11}\,\partial F_{21}\,F_{22}(z)
+c_{49}\,F_{12}\,{\tt P}{}^{(2)}(z)
+c_{50}\,F_{12}\,F_{21}\,G_{12}(z)
\nonu \\
& + & c_{51}\,F_{12}\,F_{21}\,G_{21}(z)
+c_{52}\,F_{12}\,F_{22}\,G_{11}(z)
+c_{53}\,F_{12}\,\partial F_{12}\,F_{21}(z)
+c_{54}\,F_{21}\,{\tt P}^{(2)}(z)
\nonu \\
& + & c_{55}\,F_{21}\,F_{22}\,G_{11}(z)
+c_{56}\,F_{21}\,\partial F_{21}\,F_{12}(z)
+c_{57}\,G_{12}\,{\tt T}^{(1)}(z)
+c_{58}\,G_{21}\,{\tt T}^{(1)}(z)
\nonu \\
& + & c_{59}\,{\tt P}^{(2)}\,F_{12}(z)
+c_{60}\,{\tt P}^{(2)}\,F_{21}(z)
+c_{61}\,{\tt T}^{(1)}\,{\tt T}_{-}^{(\frac{3}{2})}(z)
+c_{62}\,{\tt T}^{(1)}\,{\tt T}_{+}^{(\frac{3}{2})}(z)
\nonu\\
&+&c_{63}\,U\,G_{12}(z)
+c_{64}\,U\,G_{21}(z)+c_{65}\,U\,A_{3}\,F_{12}(z)
+c_{66}\,U\,A_{3}\,F_{21}(z)
\nonu \\
& + & c_{67}\,U\,A_{-}\,F_{11}(z)
+c_{68}\,U\,A_{+}\,F_{22}(z)
+c_{69}\,U\,B_{3}\,F_{12}(z)
+c_{70}\,U\,B_{3}\,F_{21}(z)
\nonu \\
& + & c_{71}\,U\,B_{-}\,F_{22}(z)
+c_{72}\,U\,B_{+}\,F_{11}(z)
+c_{73}\,U\,F_{11}\,F_{12}\,F_{22}(z)
+c_{74}\,U\,F_{11}\,F_{21}\,F_{22}(z)
\nonu \\
& + & c_{75}\,U\,U\,F_{12}(z)+c_{76}\,U\,U\,F_{21}(z)
+c_{77}\,U\,\partial F_{21}(z)
+c_{78}\,\partial A_{3}\,F_{21}(z)
\nonu \\
& + & c_{79}\,\partial A_{-}\,F_{11}(z)
+c_{80}\,\partial A_{+}\,F_{22}(z)
+c_{81}\,\partial B_{3}\,F_{21}(z)
+c_{82}\,\partial B_{-}\,F_{22}(z)
\nonu \\
& + & c_{83}\,\partial B_{+}\,F_{11}(z)
+c_{84}\,\partial F_{11}\,F_{12}\,F_{22}(z)
+c_{85}\,\partial F_{11}\,F_{21}\,F_{22}(z)
+c_{86}\,\partial U\,F_{21}(z)
\nonu \\
& + & c_{87}\,\partial G_{12}(z)
+c_{88}\,\partial G_{21}(z)
+c_{89}\,\partial{\tt T}_{-}^{(\frac{3}{2})}(z)+
c_{90}\,\partial^{2}F_{21}(z),
\nonu
\eea
where the coefficients are given by
\bea
c_ {1} & = & - 2 i\sqrt {2}, \qquad
c_ {2}  =  - 2 i\sqrt {2}, \qquad
c_ {3}  =  2 i\sqrt {2}, \qquad
c_ {4}  =  - 2 i\sqrt {2}, \qquad
c_ {5}  =  - \frac {\sqrt {2} (1 + N)} {(2 + k + N)^{2}}, \nonu \\
c_ {6} & = & - \frac {\sqrt {2} (1 + N)} {(2 + k + N)^{2}}, \qquad
c_ {7}  =  - \frac {2 i\sqrt {2} (1 + N)} {(2 + k + 
       N)^{3}}, \qquad
c_ {8}  = \frac {2 i\sqrt {2} (1 + N)} {(2 + k + N)^{3}}, \nonu \\
c_ {9} & = & - \frac {2 i\sqrt {2} (16 + 21 k + 6 k^{2} + 19 N + 
       13 k\, N + 5 N^{2})} {(2 + N) (2 + k + N)^{3}}, \nonu \\
c_ {10} & = &\frac {2 i\sqrt {2} (16 + 21 k + 6 k^{2} + 19 N + 13 k\, 
     N + 5 N^{2})} {(2 + N) (2 + k + N)^{3}}, \nonu \\
c_ {11} & = & - \frac {2 i\sqrt {2} (16 + 21 k + 6 k^{2} + 19 N + 
       13 k\, N + 5 N^{2})} {(2 + N) (2 + k + N)^{3}}, \nonu \\
c_ {12} & = &\frac {2 i\sqrt {2} (16 + 21 k + 6 k^{2} + 19 N + 13 k\, 
     N + 5 N^{2})} {(2 + N) (2 + k + N)^{3}}, \nonu \\
c_ {13} & = & - \frac {2\sqrt {2} (32 + 55 k + 18 k^{2} + 41 N + 
       35 k\, N + 11 N^{2})} {(2 + N) (2 + k + N)^{4}}, \nonu \\
c_ {14} & = &\frac {2\sqrt {2} (32 + 55 k + 18 k^{2} + 41 N + 35 k\, 
     N + 11 N^{2})} {(2 + N) (2 + k + N)^{4}}, \nonu \\
c_ {15} & = &\frac{1}{3 (2 + N) (2 + k + N)^{4}}\sqrt {2} (60 + 169 k + 154 k^{2} + 40 k^{3} + 
      149 N + 299 k\, 
     N + 174 k^{2} N 
     \nonu \\&+& 20 k^{3} N + 159 N^{2} + 212 k\, 
     N^{2} + 62 k^{2} N^{2} + 84 N^{3} + 58 k\, 
     N^{3} + 16 N^{4}), \nonu \\
c_ {16} & = &\frac {\sqrt {2} (1 + N)} {(2 + k + N)^{2}}, \qquad
c_ {17}  =  - \frac {2 i\sqrt {2} (16 + 21 k + 6 k^{2} + 19 N + 
       13 k\, N + 5 N^{2})} {(2 + N) (2 + k + N)^{3}}, \nonu \\
c_ {18} & = & - \frac {2 i\sqrt {2} (16 + 21 k + 6 k^{2} + 19 N + 
       13 k\, N + 5 N^{2})} {(2 + N) (2 + k + N)^{3}}, \nonu \\
c_ {19} & = & - \frac {2\sqrt {2} (32 + 55 k + 18 k^{2} + 41 N + 
       35 k\, N + 11 N^{2})} {(2 + N) (2 + k + N)^{4}}, \nonu \\
c_ {20} & = &\frac {1}{3 (2 + N) (2 + k + N)^{4}}\sqrt {2} (-60 - 169 k - 154 k^{2} - 40 k^{3} + 
      31 N - 68 k\, 
     N - 108 k^{2} N 
     \nonu \\&-& 20 k^{3} N + 204 N^{2} + 133 k\, 
     N^{2} - 2 k^{2} N^{2} + 153 N^{3} + 68 k\, 
     N^{3} + 32 N^{4}), \nonu \\
c_ {21} & = &\frac {\sqrt {2} (1 + N)} {(2 + k + N)^{2}}, \qquad
c_ {22} =  - \frac {2 i\sqrt {2} (1 + N)} {(2 + k + N)^{3}}, \qquad
c_ {23}  = \frac {2 i\sqrt {2} (1 + N)} {(2 + k + N)^{3}}, \nonu \\
c_ {24} & = &\frac {2 i\sqrt {2} (16 + 21 k + 6 k^{2} + 19 N + 13 k\, 
     N + 5 N^{2})} {(2 + N) (2 + k + N)^{3}}, \nonu \\
c_ {25} & = &\frac {2 i\sqrt {2} (16 + 21 k + 6 k^{2} + 19 N + 13 k\, 
     N + 5 N^{2})} {(2 + N) (2 + k + N)^{3}}, \nonu \\
c_ {26} & = &\frac {2\sqrt {2} (32 + 55 k + 18 k^{2} + 41 N + 35 k\, 
     N + 11 N^{2})} {(2 + N) (2 + k + N)^{4}}, \nonu \\
c_ {27} & = & - \frac {\sqrt {2} (18 + 21 k + 6 k^{2} + 22 N + 13 k\, 
      N + 6 N^{2})} {(2 + N) (2 + k + N)^{2}}, \nonu \\
c_ {28} & = & - \frac {\sqrt {2} (18 + 21 k + 6 k^{2} + 22 N + 13 k\, 
      N + 6 N^{2})} {(2 + N) (2 + k + N)^{2}}, \nonu \\
c_ {29} & = &\frac {2 i\sqrt {2} (18 + 21 k + 6 k^{2} + 22 N + 13 k\, 
     N + 6 N^{2})} {(2 + N) (2 + k + N)^{3}}, \nonu \\
c_ {30} & = & - \frac {2 i\sqrt {2} (18 + 21 k + 6 k^{2} + 22 N + 
       13 k\, N + 6 N^{2})} {(2 + N) (2 + k + N)^{3}}, \nonu \\
c_ {31} & = &\frac {2\sqrt {2} (32 + 55 k + 18 k^{2} + 41 N + 35 k\, 
     N + 11 N^{2})} {(2 + N) (2 + k + N)^{4}}, \nonu \\
c_ {32} & = & - \frac {2\sqrt {2} (32 + 55 k + 18 k^{2} + 41 N + 
       35 k\, N + 11 N^{2})} {(2 + N) (2 + k + N)^{4}}, \nonu \\
c_ {33} & = & - \frac{1}{3 (2 + N) (2 + k + N)^{4}}\sqrt {2} (60 + 169 k + 154 k^{2} + 40 k^{3} + 
       149 N + 299 k\, 
      N + 174 k^{2} N 
      \nonu \\&+& 20 k^{3} N + 159 N^{2} + 212 k\, 
      N^{2} + 62 k^{2} N^{2} + 84 N^{3} + 58 k\, 
      N^{3} + 16 N^{4}), \nonu \\
c_ {34} & = &\frac {\sqrt {2} (18 + 21 k + 6 k^{2} + 22 N + 13 k\, 
     N + 6 N^{2})} {(2 + N) (2 + k + N)^{2}}, \nonu \\
c_ {35} & = & - \frac {2\sqrt {2} (32 + 55 k + 18 k^{2} + 41 N + 
       35 k\, N + 11 N^{2})} {(2 + N) (2 + k + N)^{4}}, \nonu \\
c_ {36} & = &\frac{1}{3 (2 + N) (2 + k + N)^{4}}\sqrt {2} (60 - 11 k - 77 k^{2} - 26 k^{3} + 
      149 N - 64 k\, 
     N - 171 k^{2} N 
     \nonu \\&-& 40 k^{3} N + 159 N^{2} - 25 k\, 
     N^{2} - 64 k^{2} N^{2} + 84 N^{3} + 10 k\, 
     N^{3} + 16 N^{4}), \nonu \\
c_ {37} & = &\frac {\sqrt {2} (18 + 21 k + 6 k^{2} + 22 N + 13 k\, 
     N + 6 N^{2})} {(2 + N) (2 + k + N)^{2}}, \nonu \\
c_ {38} & = &\frac {2 i\sqrt {2} (18 + 21 k + 6 k^{2} + 22 N + 13 k\, 
     N + 6 N^{2})} {(2 + N) (2 + k + N)^{3}}, \nonu \\
c_ {39} & = & - \frac {2 i\sqrt {2} (18 + 21 k + 6 k^{2} + 22 N + 
       13 k\, N + 6 N^{2})} {(2 + N) (2 + k + N)^{3}}, \nonu \\
c_ {40} & = &\frac {2\sqrt {2} (32 + 55 k + 18 k^{2} + 41 N + 35 k\, 
     N + 11 N^{2})} {(2 + N) (2 + k + N)^{4}}, \nonu \\
c_ {41} & = & - \frac {2 i\sqrt {2} (5 + 2 k + 3 N)} {(2 + k + 
       N)^{3}}, \qquad
c_ {42}  =  - \frac {8 i\sqrt {2} (5 + 2 k + 3 N)} {(2 + k + 
       N)^{4}}, \nonu \\
c_ {43} & = &\frac {2 i\sqrt {2} (10 + 17 k + 6 k^{2} + 14 N + 11 k\, 
     N + 4 N^{2})} {(2 + N) (2 + k + N)^{3}}, \nonu \\
c_ {44} & = & - \frac {2 i\sqrt {2} (60 + 155 k + 58 k^{2} + 139 N + 
       169 k\, N + 20 k^{2} N + 85 N^{2} + 42 k\, 
      N^{2} + 16 N^{3})} {3 (2 + N) (2 + k + N)^{4}}, \nonu \\
c_ {45} & = & - \frac {i\sqrt {2} (13 k + 6 k^{2} + 3 N + 9 k\, 
      N + N^{2})} {(2 + N) (2 + k + N)^{3}}, \nonu \\
c_ {46} & = & - \frac {i\sqrt {2} (13 k + 6 k^{2} + 3 N + 9 k\, 
      N + N^{2})} {(2 + N) (2 + k + N)^{3}}, \nonu \\
c_ {47} & = & - \frac {4 i\sqrt {2} (20 + 21 k + 6 k^{2} + 25 N + 
       13 k\, N + 7 N^{2})} {(2 + N) (2 + k + N)^{4}}, \nonu \\
c_ {48} & = & - \frac {2 i\sqrt {2} (60 + 77 k + 22 k^{2} + 121 N + 
       115 k\, N + 20 k^{2} N + 79 N^{2} + 42 k\, 
      N^{2} + 16 N^{3})} {3 (2 + N) (2 + k + N)^{4}}, \nonu \\
c_ {49} & = & - \frac {i (3 + 2 k + N) (10 + 5 k + 
       8 N)} {3\sqrt {2} (2 + k + N)^{2}}, \nonu \\
c_ {50} & = &\frac {i\sqrt {2} (20 + 21 k + 6 k^{2} + 25 N + 13 k\, 
     N + 7 N^{2})} {(2 + N) (2 + k + N)^{3}}, \nonu \\
c_ {51} & = &\frac {i\sqrt {2} (20 + 21 k + 6 k^{2} + 25 N + 13 k\, 
     N + 7 N^{2})} {(2 + N) (2 + k + N)^{3}}, \nonu \\
c_ {52} & = &\frac {2 i\sqrt {2} (10 + 17 k + 6 k^{2} + 14 N + 11 k\, 
     N + 4 N^{2})} {(2 + N) (2 + k + N)^{3}}, \nonu \\
c_ {53} & = &\frac {4 i\sqrt {2} (13 k + 6 k^{2} + 3 N + 9 k\, 
     N + N^{2})} {(2 + N) (2 + k + N)^{4}}, \nonu \\
c_ {54} & = & - \frac {i (60 + 77 k + 22 k^{2} + 121 N + 115 k\, 
      N + 20 k^{2} N + 79 N^{2} + 42 k\, 
      N^{2} + 16 N^{3})} {2\sqrt {2} (2 + 
        N) (2 + k + N)^{2}}, \nonu \\
c_ {55} & = & - \frac {2 i\sqrt {2} (5 + 2 k + 3 N)} {(2 + k + 
       N)^{3}}, \qquad
c_ {56}  =  - \frac {4 i\sqrt {2} (13 k + 6 k^{2} + 3 N + 9 k\, 
      N + N^{2})} {(2 + N) (2 + k + N)^{4}}, \nonu \\
c_ {57} & = &\frac {i (60 + 77 k + 22 k^{2} + 121 N + 115 k\, 
     N + 20 k^{2} N + 79 N^{2} + 42 k\, 
     N^{2} + 16 N^{3})} {\sqrt {2} (2 + 
       N) (2 + k + N)^{2}}, \nonu \\
c_ {58} & = &\frac {i (60 + 77 k + 22 k^{2} + 121 N + 115 k\, 
     N + 20 k^{2} N + 79 N^{2} + 42 k\, 
     N^{2} + 16 N^{3})} {\sqrt {2} (2 + 
       N) (2 + k + N)^{2}}, \nonu \\
c_ {59} & = &\frac {i (3 + 2 k + N) (10 + 5 k + 
      8 N)} {3\sqrt {2} (2 + k + N)^{2}}, \nonu \\
c_ {60} & = &\frac {i (60 + 77 k + 22 k^{2} + 121 N + 115 k\, 
     N + 20 k^{2} N + 79 N^{2} + 42 k\, 
     N^{2} + 16 N^{3})} {2\sqrt {2} (2 + 
       N) (2 + k + N)^{2}}, \nonu \\
c_ {61} & = & - \frac {i\sqrt {2} (60 + 77 k + 22 k^{2} + 121 N + 
       115 k\, N + 20 k^{2} N + 79 N^{2} + 42 k\, 
      N^{2} + 16 N^{3})} {(2 + N) (2 + k + N)^{2}}, \nonu \\
c_ {62} & = &\frac {i\sqrt {2} (60 + 77 k + 22 k^{2} + 121 N + 
      115 k\, N + 20 k^{2} N + 79 N^{2} + 42 k\, 
     N^{2} + 16 N^{3})} {(2 + N) (2 + k + N)^{2}}, \nonu \\
c_ {63} & = & - \frac {2 i\sqrt {2}} {(2 + k + N)}, \qquad
c_ {64}  = \frac {2 i\sqrt {2}} {(2 + k + N)}, \qquad
c_ {65}  = \frac {2\sqrt {2} (3 + 2 k + N)} {(2 + k + 
      N)^{3}}, \nonu \\
c_ {66} & = &\frac {2\sqrt {2} (3 + 2 k + N)} {(2 + k + 
      N)^{3}}, \qquad
c_ {67}  =  - \frac {2\sqrt {2} (3 + 2 k + N)} {(2 + k + 
       N)^{3}}, \qquad
c_ {68}  =  - \frac {2\sqrt {2} (3 + 2 k + N)} {(2 + k + 
       N)^{3}}, \nonu \\
c_ {69} & = & - \frac {2\sqrt {2} (26 + 25 k + 6 k^{2} + 30 N + 
       15 k\, N + 8 N^{2})} {(2 + N) (2 + k + N)^{3}}, \nonu \\
c_ {70} & = & - \frac {2\sqrt {2} (26 + 25 k + 6 k^{2} + 30 N + 
       15 k\, N + 8 N^{2})} {(2 + N) (2 + k + N)^{3}}, \nonu \\
c_ {71} & = &\frac {2\sqrt {2} (26 + 25 k + 6 k^{2} + 30 N + 15 k\, 
     N + 8 N^{2})} {(2 + N) (2 + k + N)^{3}}, \nonu \\
c_ {72} & = &\frac {2\sqrt {2} (26 + 25 k + 6 k^{2} + 30 N + 15 k\, 
     N + 8 N^{2})} {(2 + N) (2 + k + N)^{3}}, \nonu \\
c_ {73} & = &\frac {2 i\sqrt {2} (32 + 55 k + 18 k^{2} + 41 N + 
      35 k\, N + 11 N^{2})} {(2 + N) (2 + k + N)^{4}}, \nonu \\
c_ {74} & = &\frac {2 i\sqrt {2} (32 + 55 k + 18 k^{2} + 41 N + 
      35 k\, N + 11 N^{2})} {(2 + N) (2 + k + N)^{4}}, \nonu \\
c_ {75} & = & - \frac {4 i\sqrt {2}} {(2 + k + N)^{2}}, \qquad
c_ {76}  = \frac {4 i\sqrt {2}} {(2 + k + N)^{2}}, \nonu \\
c_ {77} & = &\frac {1}{3 (2 + N) (2 + k + N)^{4}}i\sqrt {2} (60 + 169 k + 154 k^{2} + 40 k^{3} + 
      149 N + 299 k\, 
     N + 174 k^{2} N 
     \nonu \\&+& 20 k^{3} N + 159 N^{2} + 212 k\, 
     N^{2} + 62 k^{2} N^{2} + 84 N^{3} + 58 k\, 
     N^{3} + 16 N^{4}), \nonu \\
c_ {78} & = &\frac {1}{3 (2 + N) (2 + k + N)^{4}}\sqrt {2} (12 + 145 k + 154 k^{2} + 40 k^{3} + 
      53 N + 263 k\, 
     N + 174 k^{2} N 
     \nonu \\&+& 20 k^{3} N + 99 N^{2} + 200 k\, 
     N^{2} + 62 k^{2} N^{2} + 72 N^{3} + 58 k\, 
     N^{3} + 16 N^{4}), \nonu \\
c_ {79} & = &\frac {1}{3 (2 + N) (2 + k + N)^{4}}\sqrt {2} (-36 - 157 k - 154 k^{2} - 40 k^{3} + 
      79 N - 50 k\, 
     N - 108 k^{2} N 
     \nonu \\&-& 20 k^{3} N + 234 N^{2} + 139 k\, 
     N^{2} - 2 k^{2} N^{2} + 159 N^{3} + 68 k\, 
     N^{3} + 32 N^{4}), \nonu \\
c_ {80} & = &\frac {2\sqrt {2} (1 + N)} {(2 + k + N)^{3}}, \nonu \\
c_ {81} & = & - \frac {1}{3 (2 + N) (2 + k + N)^{4}}\sqrt {2} (-372 - 551 k - 242 k^{2} - 
       32 k^{3} - 595 N - 529 k\, 
      N - 54 k^{2} N 
      \nonu \\&+& 20 k^{3} N - 249 N^{2} - 16 k\, 
      N^{2} + 62 k^{2} N^{2} + 12 N^{3} + 58 k\, 
      N^{3} + 16 N^{4}), \nonu \\
c_ {82} & = &\frac {1}{3 (2 + N) (2 + k + N)^{4}}\sqrt {2} (-156 - 371 k - 275 k^{2} - 62 k^{3} - 
      223 N - 478 k\, 
     N - 285 k^{2} N 
     \nonu \\&-& 40 k^{3} N - 45 N^{2} - 139 k\, 
     N^{2} - 64 k^{2} N^{2} + 48 N^{3} + 10 k\, 
     N^{3} + 16 N^{4}), \nonu \\
c_ {83} & = & - \frac {2\sqrt {2} (18 + 21 k + 6 k^{2} + 22 N + 
       13 k\, N + 6 N^{2})} {(2 + N) (2 + k + N)^{3}}, \nonu \\
c_ {84} & = & - \frac {8 i\sqrt {2} (10 + 17 k + 6 k^{2} + 14 N + 
       11 k\, N + 4 N^{2})} {(2 + N) (2 + k + N)^{4}}, \nonu \\
c_ {85} & = & - \frac {2 i\sqrt {2} (60 - k - 14 k^{2} + 103 N + 
       61 k\, N + 20 k^{2} N + 73 N^{2} + 42 k\, 
      N^{2} + 16 N^{3})} {3 (2 + N) (2 + k + N)^{4}}, \nonu \\
c_ {86} & = &\frac{1} {3 (2 + N) (2 + k + N)^{4}}i\sqrt {2} (60 + 169 k + 154 k^{2} + 40 k^{3} + 
      149 N + 299 k\, 
     N + 174 k^{2} N 
     \nonu \\&+& 20 k^{3} N + 159 N^{2} + 212 k\, 
     N^{2} + 62 k^{2} N^{2} + 84 N^{3} + 58 k\, 
     N^{3} + 16 N^{4}), \nonu \\
c_ {87} & = &\frac {i (300 + 371 k + 106 k^{2} + 559 N + 499 k\, 
     N + 80 k^{2} N + 337 N^{2} + 168 k\, 
     N^{2} + 64 N^{3})} {3\sqrt {2} (2 + 
       N) (2 + k + N)^{2}}, \nonu \\
c_ {88} & = &\frac {1}{3\sqrt {2} (2 + 
       N) (2 + k + N)^{3}}i (60 + 169 k + 154 k^{2} + 40 k^{3} + 149 N + 
      299 k\, N + 174 k^{2} N 
      \nonu \\&+& 20 k^{3} N + 159 N^{2} + 212 k\, 
     N^{2} + 62 k^{2} N^{2} + 84 N^{3} + 58 k\, 
     N^{3} + 16 N^{4}), \nonu \\
c_ {89} & = & - \frac {i\sqrt {2} (300 + 371 k + 106 k^{2} + 559 N + 
       499 k\, N + 80 k^{2} N + 337 N^{2} + 168 k\, 
      N^{2} + 64 N^{3})} {3 (2 + N) (2 + k + N)^{2}}, \nonu \\
c_ {90} & = &\frac {1} {(2 + N) (2 + k + N)^{4}} i\sqrt {2} (-k + 
    N) (60 + 77 k + 22 k^{2} + 121 N + 115 k\, 
   N + 20 k^{2} N 
   \nonu \\&+& 79 N^{2} + 42 k\, N^{2} + 16 N^{3}).
\nonu
\eea

For the other higher spin-$\frac{5}{2}$ current, one has the following 
relation  
\bea
V_{\frac{1}{2}}^{(2),1}(z) &=& 
c_ {1}\, {\tt Q}^{(\frac{5}{2})}(z)+ c_ {2}\,{\tt R}^{(\frac{5}{2})}(z) 
+ c_ {3}\, {\tt U}^{(\frac{5}{2})}(z) + c_ {4}\, {\tt V}^{(\frac{5}{2})}(z)
+ c_ {5}\, A_ {3}\, G_ {11}(z) + c_ {6}\, A_ {3}\, G_ {22}(z) 
\nonu \\
& + & c_ {7}\, A_ {3}\, A_ {3}\, F_ {11}(z) 
+ c_ {8}\, A_ {3}\, A_ {3}\, F_ {22}(z) 
+ c_ {9}\, A_ {3}\, B_ {3}\, F_ {11}(z)
+ c_ {10}\, A_ {3}\, B_ {3}\, F_ {22}(z) 
\nonu \\
& + & c_ {11}\, A_ {3}\, B_ {-}\, F_ {12}(z) 
+ c_ {12}\, A_ {3}\, B_{+}\, F_ {21}(z) 
+ c_ {13}\, A_ {3}\, F_ {11}\, F_ {12}\, F_ {21}(z) 
+ c_ {14}\, A_ {3}\, F_ {22}\, F_ {21}\, F_ {12}(z) 
\nonu \\
& + &  c_ {15}\, A_ {3}\, \partial F_ {11}(z)
+  c_ {16}\, A_ {3}\, \partial F_ {22}(z) 
+ c_ {17}\, A_ {-}\, G_ {12}(z) 
+  c_ {18}\, A_ {-}\, A_ {+}\, 
F_ {22}(z) 
\nonu \\
& + &  c_ {19}\, A_ {-}\, B_ {3}\, 
F_ {12}(z) + 
c_ {20}\, A_ {-}\, B_{+}\, 
F_ {11}(z) + 
c_ {21}\, A_ {-}\, F_ {22}\, F_ {12}\, 
F_ {11}(z) + 
c_ {22}\, A_ {-}\, \partial F_ {12}(z) 
\nonu\\&+&  
 c_ {23}\, A_ {+}\, G_ {21}(z) + c_ {24}\, A_ {+}\, A_ {-}\, F_ {11}(z) 
+ c_ {25}\, A_ {+}\, B_ {3}\, F_ {21}(z) 
+ c_ {26}\, A_ {+}\, B_ {-}\, F_ {22}(z) 
\nonu \\
& + & c_ {27}\, A_ {+}\, F_ {11}\, F_ {21}\, F_ {22}(z) 
+  c_ {28}\, A_ {+}\, \partial F_ {21}(z)+ 
 c_ {29}\, B_ {3}\, G_ {11}(z) + c_ {30}\, B_ {3}\, G_ {22}(z)
\nonu \\ 
& + & c_ {31}\, B_ {3}\, B_ {3}\, F_ {11}(z) 
+ c_ {32}\, B_ {3}\, B_ {3}\, F_ {22}(z) 
+  c_ {33}\, B_ {3}\, F_ {11}\, F_ {12}\, F_ {21}(z) 
+ c_ {34}\, B_ {3}\, F_ {22}\, F_ {21}\,F_ {12}(z) 
\nonu \\
& + & c_ {35}\, B_ {3}\,\partial F_ {11}(z) 
+ c_ {36}\, B_ {3}\, \partial F_ {22}(z) 
+ c_ {37}\, B_ {-}\, G_ {12}(z) 
+ c_ {38}\, B_ {-}\, B_{+}\, F_ {22}(z) 
\nonu \\
& + &  c_ {39}\, B_ {-}\, F_ {11}\, F_ {12}\, F_ {22}(z) 
+ c_ {40}\, B_ {-}\, \partial F_ {12}(z)
 +  c_ {41}\, B_{+}\, G_ {21}(z) + c_ {42}\, B_{+}\, B_ {-}\, F_ {11}(z) 
\nonu\\&+& c_ {43}\, B_{+}\, F_ {22}\, F_ {21}\, F_ {11}(z) 
+ c_ {44}\, B_{+}\, \partial F_ {21}(z) + 
 c_ {45}\, F_ {11}\, F_ {12}\, 
G_ {21}(z) + c_ {46}\, F_ {11}\, F_ {12}\,\partial F_ {21}(z)
\nonu\\&+& c_ {47}\, F_ {11}\, F_ {21}\, G_ {12}(z) 
+ c_ {48}\, F_ {11}\, F_ {22}\, G_ {11}(z) 
+ c_ {49}\, F_ {11}\, \partial F_ {11}\, F_ {22}(z) 
+ c_ {50}\, F_ {11}\, \partial F_ {12}\, F_ {21}(z) 
\nonu \\
& + & c_ {51}\, F_ {12}\, F_ {21}\, G_ {11}(z) 
+ c_ {52}\, F_ {21}\, F_ {12}\, G_ {22}(z) 
+ c_ {53}\, F_ {22}\,{\tt P}^{(2)}(z) + 
c_ {54}\, F_ {22}\, F_ {11}\, G_ {22}(z) 
\nonu \\
& + &  c_ {55}\, F_ {22}\, F_ {12}\, G_ {21}(z) 
+ c_ {56}\, F_ {22}\, F_ {21}\, G_ {12}(z) 
+ c_ {57}\, F_ {22}\, G_ {11}\, F_ {22}(z) 
+ c_ {58}\, F_ {22}\, \partial F_ {21}\, F_ {12}(z) 
\nonu \\
& + &  c_ {59}\, F_ {22}\, \partial F_ {22}\, F_ {11}(z) 
+ c_ {60}\, G_ {11}\, {\tt T}^{(1)}(z) + 
c_ {61}\, G_ {22}\,{\tt T}^{(1)}(z) 
+ c_ {62}\, {\tt P}^{(2)}\, F_ {22}(z) 
\nonu \\
& 
+& c_ {63}\, {\tt T}^{(1)}\, {\tt U}^{(\frac{3}{2})}(z) + 
c_ {64}\, {\tt T}^{(1)}\, {\tt V}^{(\frac{3}{2})}(z) + 
c_ {65}\, U\, G_ {11}(z) 
+ c_ {66}\, U\, G_ {22}(z) 
\nonu \\
& + &  c_ {67}\, U\, A_ {3}\, F_ {11}(z) 
+ c_ {68}\, U\, A_ {3}\, F_ {22}(z) + 
c_ {69}\, U\, A_ {-}\, F_ {12}(z) 
+ c_ {70}\, U\, A_ {+}\, F_ {21}(z) 
\nonu \\
& + &  c_ {71}\, U\, B_ {3}\, F_ {11}(z) 
+ c_ {72}\, U\, B_ {3}\, F_ {22}(z) 
+c_ {73}\, U\, B_ {-}\, F_ {12}(z)
+ c_ {74}\, U\, B_{+}\, F_ {21}(z) 
\nonu \\
& + &  c_ {75}\, U\, F_ {11}\, F_ {12}\, F_ {21}(z) 
+ c_ {76}\, U\, F_ {22}\, F_ {21}\, F_ {12}(z) 
+ c_ {77}\, U\, U\, F_ {11}(z) 
+c_ {78}\, U\, U\, F_ {22}(z) 
\nonu \\
& + & c_ {79}\, U\, \partial F_ {11}(z) + 
 c_ {80}\, U\,\partial F_ {22}(z) + 
 c_ {81}\, \partial A_ {3}\, F_ {11}(z) 
+ c_ {82}\, \partial A_ {3}\, F_ {22}(z) 
\nonu \\
& + &  c_ {83}\,\partial A_ {-}\, F_ {12}(z) 
+ c_ {84}\, \partial A_ {+}\, F_ {21}(z) 
+ c_ {85}\, \partial B_ {3}\, F_ {11}(z) 
+ c_ {86}\, \partial B_ {-}\, F_ {12}(z) 
\nonu \\
& + & c_ {87}\, \partial B_{+}\, F_ {21}(z) 
+ c_ {88}\,\partial F_ {11}\, F_ {12}\, F_ {21}(z) 
+ c_ {89}\, \partial F_ {22}\, F_ {21}\, F_ {12}(z) 
+ c_ {90}\,\partial U\, F_ {11}(z) 
\nonu \\
& + & c_ {91}\, \partial U\, F_ {22}(z) 
+ c_ {92}\, \partial G_ {11}(z) + 
 c_ {93}\, \partial G_ {22}(z)
 + c_ {94}\, \partial  {\tt U}^{(\frac{3}{2})}(z)
 \nonu\\
&+&  c_ {95}\, \partial {\tt V}^{(\frac{3}{2})}(z) + 
c_ {96}\,\partial^{2} 
F_ {22}(z),
 \nonu
\eea
where the coefficients are
given by
\bea
c_ {1} & = & -2\sqrt {2},\qquad
c_ {2}  = 2\sqrt {2}, \qquad
c_ {3}  =  - 2\sqrt {2}, \qquad
c_ {4}  =  - 2\sqrt {2}, \qquad
c_ {5}  =  - \frac {i\sqrt {2} (1 + N)} {(2 + k + N)^{2}}, \nonu\\
c_ {6} & = &\frac {i\sqrt {2} (1 + N)} {(2 + k + N)^{2}}, \qquad
c_ {7}  = \frac {2\sqrt {2} (1 + N)} {(2 + k + N)^{3}}, \qquad
c_ {8}  = \frac {2\sqrt {2} (1 + N)} {(2 + k + N)^{3}}, \nonu\\
c_ {9} & = & - \frac {2\sqrt {2} (16 + 21 k + 6 k^{2} + 19 N + 13 k\, 
      N + 5 N^{2})} {(2 + N) (2 + k + N)^{3}}, \nonu\\
c_ {10} & = & - \frac {2\sqrt {2} (16 + 21 k + 6 k^{2} + 19 N + 
       13 k\, N + 5 N^{2})} {(2 + N) (2 + k + N)^{3}}, \nonu\\
c_ {11} & = & - \frac {2\sqrt {2} (16 + 21 k + 6 k^{2} + 19 N + 
       13 k\, N + 5 N^{2})} {(2 + N) (2 + k + N)^{3}}, \nonu\\
c_ {12} & = & - \frac {2\sqrt {2} (16 + 21 k + 6 k^{2} + 19 N + 
       13 k\, N + 5 N^{2})} {(2 + N) (2 + k + N)^{3}}, \nonu\\
c_ {13} & = &\frac {2 i\sqrt {2} (32 + 55 k + 18 k^{2} + 41 N + 
      35 k\, N + 11 N^{2})} {(2 + N) (2 + k + N)^{4}}, \nonu\\
c_ {14} & = & - \frac {2 i\sqrt {2} (32 + 55 k + 18 k^{2} + 41 N + 
       35 k\, N + 11 N^{2})} {(2 + N) (2 + k + N)^{4}}, \nonu\\
c_ {15} & = &\frac {2 i\sqrt {2} (3 + 2 k + N) (10 + 5 k + 
      8 N)} {3 (2 + k + N)^{3}}, \nonu\\
c_ {16} & = &\frac {2 i\sqrt {2} (-30 - 32 k + 15 k^{2} + 10 k^{3} - 
      28 N - 23 k\, N + 11 k^{2} N - 22 N^{2} - 13 k\, 
     N^{2} - 8 N^{3})} {(6 + 7 k - N) (2 + k + N)^{3}}, \nonu\\
c_ {17} & = &\frac {i\sqrt {2} (1 + N)} {(2 + k + N)^{2}}, \qquad
c_ {18}  = \frac {2\sqrt {2} (1 + N)} {(2 + k + N)^{3}}, \nonu\\
c_ {19} & = & - \frac {2\sqrt {2} (16 + 21 k + 6 k^{2} + 19 N + 
       13 k\, N + 5 N^{2})} {(2 + N) (2 + k + N)^{3}}, \nonu\\
c_ {20} & = &\frac {2\sqrt {2} (16 + 21 k + 6 k^{2} + 19 N + 13 k\, 
     N + 5 N^{2})} {(2 + N) (2 + k + N)^{3}}, \nonu\\
c_ {21} & = & - \frac {2 i\sqrt {2} (32 + 55 k + 18 k^{2} + 41 N + 
       35 k\, N + 11 N^{2})} {(2 + N) (2 + k + N)^{4}}, \nonu\\
c_ {22} & = &\frac {2 i\sqrt {2} (-30 - 32 k + 15 k^{2} + 10 k^{3} - 
      28 N - 23 k\, N + 11 k^{2} N - 22 N^{2} - 13 k\, 
     N^{2} - 8 N^{3})} {(6 + 7 k - N) (2 + k + N)^{3}}, \nonu\\
c_ {23} & = & - \frac {i\sqrt {2} (1 + N)} {(2 + k + N)^{2}}, \qquad
c_ {24}  = \frac {2\sqrt {2} (1 + N)} {(2 + k + N)^{3}}, \nonu\\
c_ {25} & = & - \frac {2\sqrt {2} (16 + 21 k + 6 k^{2} + 19 N + 
       13 k\, N + 5 N^{2})} {(2 + N) (2 + k + N)^{3}}, \nonu\\
c_ {26} & = &\frac {2\sqrt {2} (16 + 21 k + 6 k^{2} + 19 N + 13 k\, 
     N + 5 N^{2})} {(2 + N) (2 + k + N)^{3}}, \nonu\\
c_ {27} & = &\frac {2 i\sqrt {2} (32 + 55 k + 18 k^{2} + 41 N + 
      35 k\, N + 11 N^{2})} {(2 + N) (2 + k + N)^{4}}, \nonu\\
c_ {28} & = &\frac {2 i\sqrt {2} (3 + 2 k + N) (10 + 5 k + 
      8 N)} {3 (2 + k + N)^{3}}, \nonu\\
c_ {29} & = &\frac {i\sqrt {2} (18 + 21 k + 6 k^{2} + 22 N + 13 k\, 
     N + 6 N^{2})} {(2 + N) (2 + k + N)^{2}}, \nonu\\
c_ {30} & = & - \frac {i\sqrt {2} (18 + 21 k + 6 k^{2} + 22 N + 
       13 k\, N + 6 N^{2})} {(2 + N) (2 + k + N)^{2}}, \nonu\\
c_ {31} & = & - \frac {2\sqrt {2} (18 + 21 k + 6 k^{2} + 22 N + 
       13 k\, N + 6 N^{2})} {(2 + N) (2 + k + N)^{3}}, \nonu\\
c_ {32} & = & - \frac {2\sqrt {2} (18 + 21 k + 6 k^{2} + 22 N + 
       13 k\, N + 6 N^{2})} {(2 + N) (2 + k + N)^{3}}, \nonu\\
c_ {33} & = &\frac {2 i\sqrt {2} (32 + 55 k + 18 k^{2} + 41 N + 
      35 k\, N + 11 N^{2})} {(2 + N) (2 + k + N)^{4}}, \nonu\\
c_ {34} & = & - \frac {2 i\sqrt {2} (32 + 55 k + 18 k^{2} + 41 N + 
       35 k\, N + 11 N^{2})} {(2 + N) (2 + k + N)^{4}}, \nonu\\
c_ {35} & = &\frac {2 i\sqrt {2} (3 + 2 k + N) (10 + 5 k + 
      8 N)} {3 (2 + k + N)^{3}}, \nonu\\
c_ {36} & = & - \frac {2 i\sqrt {2} (30 + 92 k + 101 k^{2} + 
       30 k^{3} - 32 N + 7 k\, N + 33 k^{2} N - 78 N^{2} - 39 k\, 
      N^{2} - 24 N^{3})} {(6 + 7 k - N) (2 + k + N)^{3}},\nonu\\ 
c_ {37} & = &\frac {i\sqrt {2} (18 + 21 k + 6 k^{2} + 22 N + 13 k\, 
     N + 6 N^{2})} {(2 + N) (2 + k + N)^{2}}, \nonu\\
c_ {38} & = & - \frac {2\sqrt {2} (18 + 21 k + 6 k^{2} + 22 N + 
       13 k\, N + 6 N^{2})} {(2 + N) (2 + k + N)^{3}}, \nonu\\
c_ {39} & = & - \frac {2 i\sqrt {2} (32 + 55 k + 18 k^{2} + 41 N + 
       35 k\, N + 11 N^{2})} {(2 + N) (2 + k + N)^{4}}, \nonu\\
c_ {40} & = &\frac {2 i\sqrt {2} (3 + 2 k + N) (10 + 5 k + 
      8 N)} {3 (2 + k + N)^{3}}, \nonu\\
c_ {41} & = & - \frac {i\sqrt {2} (18 + 21 k + 6 k^{2} + 22 N + 
       13 k\, N + 6 N^{2})} {(2 + N) (2 + k + N)^{2}}, \nonu\\
c_ {42} & = & - \frac {2\sqrt {2} (18 + 21 k + 6 k^{2} + 22 N + 
       13 k\, N + 6 N^{2})} {(2 + N) (2 + k + N)^{3}}, \nonu\\
c_ {43} & = &\frac {2 i\sqrt {2} (32 + 55 k + 18 k^{2} + 41 N + 
      35 k\, N + 11 N^{2})} {(2 + N) (2 + k + N)^{4}}, \nonu\\
c_ {44} & = &\frac {2 i\sqrt {2} (-30 - 32 k + 15 k^{2} + 10 k^{3} + 
      32 N + 93 k\, N + 51 k^{2} N + 78 N^{2} + 71 k\, 
     N^{2} + 24 N^{3})} {(6 + 7 k - N) (2 + k + N)^{3}}, \nonu\\
c_ {45} & = & - \frac {2\sqrt {2} (5 + 2 k + 3 N)} {(2 + k + N)^{3}}, \qquad
c_ {46}  = \frac {8\sqrt {2} (15 + 29 k + 10 k^{2} + 25 N + 21 k\, 
     N + 8 N^{2})} {3 (2 + k + N)^{4}}, \nonu\\
c_ {47} & = & - \frac {2\sqrt {2} (10 + 17 k + 6 k^{2} + 14 N + 
       11 k\, N + 4 N^{2})} {(2 + N) (2 + k + N)^{3}}, \nonu\\
c_ {48} & = & - \frac {\sqrt {2} (20 + 21 k + 6 k^{2} + 25 N + 13 k\, 
      N + 7 N^{2})} {(2 + N) (2 + k + N)^{3}}, \nonu\\
c_ {49} & = & - \frac {4\sqrt {2} (13 k + 6 k^{2} + 3 N + 9 k\, 
      N + N^{2})} {(2 + N) (2 + k + N)^{4}}, \nonu\\
c_ {50} & = &\frac {8\sqrt {2} (30 + 19 k + 2 k^{2} + 56 N + 44 k\, 
     N + 10 k^{2} N + 38 N^{2} + 21 k\, 
     N^{2} + 8 N^{3})} {3 (2 + N) (2 + k + N)^{4}}, \nonu\\
c_ {51} & = &\frac {\sqrt {2} (13 k + 6 k^{2} + 3 N + 9 k\, 
     N + N^{2})} {(2 + N) (2 + k + N)^{3}}, \qquad
c_ {52}  = \frac {\sqrt {2} (13 k + 6 k^{2} + 3 N + 9 k\, 
     N + N^{2})} {(2 + N) (2 + k + N)^{3}}, \nonu\\
c_ {53} & = &\frac {2\sqrt {2} (15 + 29 k + 10 k^{2} + 25 N + 21 k\, 
     N + 8 N^{2})} {(6 + 7 k - N) (2 + k + N)}, \nonu\\
c_ {54} & = & - \frac {\sqrt {2} (20 + 21 k + 6 k^{2} + 25 N + 13 k\, 
      N + 7 N^{2})} {(2 + N) (2 + k + N)^{3}}, \nonu\\
c_ {55} & = & - \frac {2\sqrt {2} (10 + 17 k + 6 k^{2} + 14 N + 
       11 k\, N + 4 N^{2})} {(2 + N) (2 + k + N)^{3}}, \qquad
c_ {56}  =  - \frac {2\sqrt {2} (5 + 2 k + 3 N)} {(2 + k + N)^{3}}, \nonu\\
c_ {57} & = & - \frac {1}{(6 + 7 k - N) (2 + N) (2 + k + N)^{3}}2\sqrt {2} (-156 - 320 k - 164 k^{2} - 
       24 k^{3} - 262 N 
       \nonu\\&-& 316 k\, 
      N - 3 k^{2} N + 30 k^{3} N - 216 N^{2} - 129 k\, 
      N^{2} + 33 k^{2} N^{2} - 114 N^{3} - 39 k\, 
      N^{3} - 24 N^{4}), \nonu\\
c_ {58} & = & - \frac {8\sqrt {2} (13 k + 6 k^{2} + 3 N + 9 k\, 
      N + N^{2})} {(2 + N) (2 + k + N)^{4}}, \qquad
c_ {59}  =  - \frac {4\sqrt {2} (13 k + 6 k^{2} + 3 N + 9 k\, 
      N + N^{2})} {(2 + N) (2 + k + N)^{4}}, \nonu\\
c_ {60} & = &\frac {(60 + 77 k + 22 k^{2} + 121 N + 115 k\, 
    N + 20 k^{2} N + 79 N^{2} + 42 k\, 
    N^{2} + 16 N^{3})} {\sqrt {2} (2 + N) (2 + k + N)^{2}}, \nonu\\
c_ {61} & = & - \frac {(60 + 77 k + 22 k^{2} + 121 N + 115 k\, 
     N + 20 k^{2} N + 79 N^{2} + 42 k\, 
     N^{2} + 16 N^{3})} {\sqrt {2} (2 + N) (2 + k + N)^{2}}, \nonu\\
c_ {62} & = & - \frac {2\sqrt {2} (15 + 29 k + 10 k^{2} + 25 N + 
       21 k\, N + 8 N^{2})} {(6 + 7 k - N) (2 + k + N)}, \nonu\\
c_ {63} & = &\frac {\sqrt {2} (60 + 77 k + 22 k^{2} + 121 N + 115 k\, 
     N + 20 k^{2} N + 79 N^{2} + 42 k\, 
     N^{2} + 16 N^{3})} {(2 + N) (2 + k + N)^{2}}, \nonu\\
c_ {64} & = &\frac {\sqrt {2} (60 + 77 k + 22 k^{2} + 121 N + 115 k\, 
     N + 20 k^{2} N + 79 N^{2} + 42 k\, 
     N^{2} + 16 N^{3})} {(2 + N) (2 + k + N)^{2}}, \nonu\\
c_ {65} & = &\frac {2\sqrt {2}} {(2 + k + N)}, \qquad
c_ {66}  = \frac {2\sqrt {2}} {(2 + k + N)}, \qquad
c_ {67}  = \frac {2 i\sqrt {2} (3 + 2 k + N)} {(2 + k + N)^{3}}, \nonu\\
c_ {68} & = & - \frac {2 i\sqrt {2} (3 + 2 k + N)} {(2 + k + N)^{3}}, \qquad
c_ {69}  =  - \frac {2 i\sqrt {2} (3 + 2 k + N)} {(2 + k + N)^{3}}, \qquad
c_ {70}  = \frac {2 i\sqrt {2} (3 + 2 k + N)} {(2 + k + N)^{3}}, \nonu\\
c_ {71} & = &\frac {2 i\sqrt {2} (26 + 25 k + 6 k^{2} + 30 N + 15 k\, 
     N + 8 N^{2})} {(2 + N) (2 + k + N)^{3}}, \nonu\\
c_ {72} & = & - \frac {2 i\sqrt {2} (26 + 25 k + 6 k^{2} + 30 N + 
       15 k\, N + 8 N^{2})} {(2 + N) (2 + k + N)^{3}}, \nonu\\
c_ {73} & = &\frac {2 i\sqrt {2} (26 + 25 k + 6 k^{2} + 30 N + 15 k\, 
     N + 8 N^{2})} {(2 + N) (2 + k + N)^{3}}, \nonu\\
c_ {74} & = & - \frac {2 i\sqrt {2} (26 + 25 k + 6 k^{2} + 30 N + 
       15 k\, N + 8 N^{2})} {(2 + N) (2 + k + N)^{3}}, \nonu\\
c_ {75} & = &\frac {2\sqrt {2} (32 + 55 k + 18 k^{2} + 41 N + 35 k\, 
     N + 11 N^{2})} {(2 + N) (2 + k + N)^{4}}, \nonu\\
c_ {76} & = &\frac {2\sqrt {2} (32 + 55 k + 18 k^{2} + 41 N + 35 k\, 
     N + 11 N^{2})} {(2 + N) (2 + k + N)^{4}}, \nonu\\
c_ {77} & = &\frac {4\sqrt {2}} {(2 + k + N)^{2}}, \qquad
c_ {78}  = \frac {4\sqrt {2}} {(2 + k + N)^{2}}, \qquad
c_ {79}  =\frac {2\sqrt {2} (3 + 2 k + N) (10 + 5 k + 
      8 N)} {3 (2 + k + N)^{3}}, \nonu\\
c_ {80} & = & - \frac {2\sqrt {2} (-30 - 32 k + 15 k^{2} + 10 k^{3} + 
       32 N + 93 k\, N + 51 k^{2} N + 78 N^{2} + 71 k\, 
      N^{2} + 24 N^{3})} {(6 + 7 k - N) (2 + k + N)^{3}}, \nonu\\
c_ {81} & = &\frac {2 i\sqrt {2} (3 + 2 k + N) (10 + 5 k + 
      8 N)} {3 (2 + k + N)^{3}}, \nonu\\
c_ {82} & = &\frac{1} {(6 + 7 k - N) (2 + N) (2 + k + N)^{3}}4 i\sqrt {2} (-108 - 192 k - 67 k^{2} - 2 k^{3} - 
      174 N - 197 k\, 
     N 
     \nonu\\&+& 17 k^{2} N + 20 k^{3} N - 144 N^{2} - 89 k\, 
     N^{2} + 22 k^{2} N^{2} - 76 N^{3} - 26 k\, 
     N^{3} - 16 N^{4}), \nonu\\
c_ {83} & = &\frac {2 i\sqrt {2} (-24 - 25 k + 15 k^{2} + 10 k^{3} - 
      23 N - 16 k\, N + 11 k^{2} N - 23 N^{2} - 13 k\, 
     N^{2} - 8 N^{3})} {(6 + 7 k - N) (2 + k + N)^{3}}, \nonu\\
c_ {84} & = &\frac {2 i\sqrt {2} (27 + 35 k + 10 k^{2} + 31 N + 
      21 k\, N + 8 N^{2})} {3 (2 + k + N)^{3}}, \nonu\\
c_ {85} & = &\frac {2 i\sqrt {2} (-48 - 56 k - 16 k^{2} - 34 N - k\, 
     N + 10 k^{2} N + 14 N^{2} + 21 k\, 
     N^{2} + 8 N^{3})} {3 (2 + N) (2 + k + N)^{3}}, \nonu\\
c_ {86} & = &\frac {2 i\sqrt {2} (6 + 7 k + 2 k^{2} + 32 N + 38 k\, 
     N + 10 k^{2} N + 32 N^{2} + 21 k\, 
     N^{2} + 8 N^{3})} {3 (2 + N) (2 + k + N)^{3}}, \nonu\\
c_ {87} & = &\frac {1}{(6 + 7 k - N) (2 + N) (2 + k + N)^{3}}2 i\sqrt {2} (48 + 188 k + 213 k^{2} + 62 k^{3} + 
      148 N + 365 k\, N 
      \nonu\\&+& 202 k^{2} N + 10 k^{3} N + 202 N^{2} + 264 k\, 
     N^{2} + 51 k^{2} N^{2} + 120 N^{3} + 71 k\, 
     N^{3} + 24 N^{4}), \nonu\\
c_ {88} & = &\frac {4\sqrt {2} (60 + 77 k + 22 k^{2} + 121 N + 
      115 k\, N + 20 k^{2} N + 79 N^{2} + 42 k\, 
     N^{2} + 16 N^{3})} {3 (2 + N) (2 + k + N)^{4}}, \nonu\\
c_ {89} & = & - \frac {4\sqrt {2} (13 k + 6 k^{2} + 3 N + 9 k\, 
      N + N^{2})} {(2 + N) (2 + k + N)^{4}}, \qquad
c_ {90}  = \frac {2\sqrt {2} (3 + 2 k + N) (10 + 5 k + 
      8 N)} {3 (2 + k + N)^{3}}, \nonu\\
c_ {91} & = &\frac {1}{(6 + 7 k - N) (2 + N) (2 + k + N)^{3}}4\sqrt {2} (-48 - 128 k - 97 k^{2} - 22 k^{3} - 
      148 N - 235 k\,N 
     \nonu\\&-&60 k^{2} N + 10 k^{3} N - 202 N^{2} - 182 k\, 
     N^{2} - 9 k^{2} N^{2} - 120 N^{3} - 55 k\, 
     N^{3} - 24 N^{4}), \nonu\\
c_ {92} & = & - \frac {(60 + 91 k + 26 k^{2} + 167 N + 191 k\, 
     N + 40 k^{2} N + 137 N^{2} + 84 k\, 
     N^{2} + 32 N^{3})} {3\sqrt {2} (2 + N) (2 + k + N)^{2}}, \nonu\\
c_ {93} & = & - \frac {1}{3\sqrt {2} (6 + 7 k - N) (2 + 
        N) (2 + k + N)^{2}}(720 + 2502 k + 2657 k^{2} + 742 k^{3} + 
      1962 N 
      \nonu\\&+&5360 k\, 
     N + 3819 k^{2} N + 560 k^{3} N + 1667 N^{2} + 3240 k\, 
     N^{2} + 1216 k^{2} N^{2} + 479 N^{3}
     \nonu\\&+& 532 k\, 
     N^{3} + 32 N^{4}), \nonu\\
c_ {94} & = & - \frac {\sqrt {2} (60 + 77 k + 22 k^{2} + 121 N + 
       115 k\, N + 20 k^{2} N + 79 N^{2} + 42 k\, 
      N^{2} + 16 N^{3})} {(2 + N) (2 + k + N)^{2}}, \nonu\\
c_ {95} & = &\frac{1}{3 (6 + 7 k - N) (2 + N) (2 + k + N)^{2}}\sqrt {2} (1080 + 2886 k + 2477 k^{2} + 
      622 k^{3} + 1758 N 
      \nonu\\&+&4436 k\, 
     N + 3117 k^{2} N + 500 k^{3} N + 539 N^{2} + 1830 k\, 
     N^{2} + 910 k^{2} N^{2} - 277 N^{3} + 106 k\, 
     N^{3} 
     \nonu\\&-&- 112 N^{4}), \nonu\\
c_ {96} & = &\frac {4\sqrt {2} (k - 3 N) (15 + 29 k + 10 k^{2} + 
      25 N + 21 k\, N + 8 N^{2})} {(6 + 7 k - N) (2 + k + N)^{3}}.   
\nonu
\eea
Similarly, 
for the other higher spin-$\frac{5}{2}$ current, the following 
relation holds  
\bea
V_{\frac{1}{2}}^{(2),2}(z) &=& 
c_ {1}\, {\tt Q}^{(\frac{5}{2})}(z)+ 
c_ {2}\, {\tt R}^{(\frac{5}{2})}(z)+ 
c_ {3}\, {\tt U}^{(\frac{5}{2})}(z) 
+ c_ {4}\, {\tt V}^{(\frac{5}{2})}(z) +
  c_ {5}\, A_ {3}\, G_ {11}(z) + 
c_ {6}\, A_ {3}\, G_ {22}(z) 
\nonu \\
& + & c_ {7}\, A_ {3}\, A_ {3}\, F_ {11}(z) 
+ c_ {8}\, A_ {3}\, A_ {3}\, F_ {22}(z) + 
c_ {9}\, A_ {3}\, B_ {3}\, F_ {11}(z) 
+ c_ {10}\, A_ {3}\, B_ {3}\, F_ {22}(z)
\nonu \\ 
& + &  c_ {11}\, A_ {3}\, B_ {-}\, F_ {12}(z) 
+ c_ {12}\, A_ {3}\, B_{+}\, F_ {21}(z) 
+ c_ {13}\, A_ {3}\, F_ {11}\, F_ {12}\, F_ {21}(z) 
+ c_ {14}\, A_ {3}\, F_ {22}\, F_ {21}\, F_ {12}(z) 
\nonu \\
&
+& c_ {15}\, A_ {3}\, \partial F_ {11}(z) +  
c_ {16}\, A_ {3}\, \partial F_ {22}(z) 
+ c_ {17}\, A_ {-}\, G_ {12}(z) 
+ c_ {18}\, A_ {-}\, A_ {+}\, F_ {22}(z) 
\nonu \\
& + & c_ {19}\, A_ {-}\, B_ {3}\, F_ {12}(z) 
+ c_ {20}\, A_ {-}\, B_{+}\, F_ {11}(z) 
+ c_ {21}\, A_ {-}\, F_ {22}\, F_ {12}\, F_ {11}(z) 
+ c_ {22}\, A_ {-}\, \partial F_ {12}(z) 
\nonu\\
&+
& c_ {23}\, A_ {+}\, G_ {21}(z) + 
c_ {24}\, A_ {+}\, A_ {-}\, F_ {11}(z) 
+ c_ {25}\, A_ {+}\, B_ {3}\, F_ {21}(z) 
+ c_ {26}\, A_ {+}\, B_ {-}\, F_ {22}(z) 
\nonu \\
& + & c_ {27}\, A_ {+}\, F_ {11}\, F_ {21}\, F_ {22}(z) 
+ c_ {28}\, A_ {+}\, \partial F_ {21}(z) 
+  c_ {29}\, B_ {3}\, G_ {11}(z) + 
c_ {30}\, B_ {3}\, G_ {22}(z) 
\nonu \\
& + & c_ {31}\, B_ {3}\, B_ {3}\, F_ {11}(z) 
+ c_ {32}\, B_ {3}\, B_ {3}\, F_ {22}(z) 
+ c_ {33}\, B_ {3}\, F_ {11}\, F_ {12}\, F_ {21}(z) 
+ c_ {34}\, B_ {3}\, F_ {22}\, F_ {21}\, F_ {12}(z) 
\nonu \\
& + &  c_ {35}\, B_ {3}\, \partial F_ {11}(z) 
+ c_ {36}\, B_ {3}\, \partial F_ {22}(z) + 
c_ {37}\, B_ {-}\, G_ {12}(z) 
+ c_ {38}\, B_ {-}\, B_{+}\, F_ {22}(z) 
\nonu \\
& + & c_ {39}\, B_ {-}\, F_ {11}\, F_ {12}\, F_ {22}(z) 
+ c_ {40}\, B_ {-}\, \partial F_ {12}(z) 
+ c_ {41}\, B_{+}\, G_ {21}(z) + c_ {42}\, B_{+}\, B_ {-}\, F_ {11}(z) 
\nonu\\&+& c_ {43}\, B_{+}\, F_ {22}\, F_ {21}\, F_ {11}(z) 
+ c_ {44}\, B_{+}\, \partial F_ {21}(z) 
+ c_ {45}\, F_ {11}\, F_ {12}\, G_ {21}(z) 
+ c_ {46}\, F_ {11}\, F_ {12}\, \partial F_ {21}(z)
\nonu \\
& +& c_ {47}\, F_ {11}\, F_ {21}\, G_ {12}(z) 
+ c_ {48}\, F_ {11}\, F_ {22}\, G_ {11}(z) 
+ c_ {49}\, F_ {11}\, \partial F_ {11}\, F_ {22}(z) 
+ c_ {50}\, F_ {11}\,\partial F_ {12}\, F_ {21}(z) 
\nonu \\
& + & c_ {51}\, F_ {12}\, F_ {21}\, G_ {11}(z) 
+ c_ {52}\, F_ {21}\, F_ {12}\, G_ {22}(z) 
+ c_ {53}\, F_ {22}\, {\tt P}^{(2)}(z) + 
c_ {54}\, F_ {22}\, F_ {11}\, G_ {22}(z) 
\nonu \\
& + &  c_ {55}\, F_ {22}\, F_ {12}\, G_ {21}(z) 
+ c_ {56}\, F_ {22}\, F_ {21}\, G_ {12}(z) 
+ c_ {57}\, F_ {22}\, G_ {11}\, F_ {22}(z) 
+ c_ {58}\, F_ {22}\, \partial F_ {21}\, F_ {12}(z) 
\nonu \\
& + & c_ {59}\, F_ {22}\, \partial F_ {22}\, F_ {11}(z) 
+ c_ {60}\, G_ {11}\, {\tt T}^{(1)}(z) + 
c_ {61}\, G_ {22}\, {\tt T}^{(1)}(z)
+  c_ {62}\, {\tt P}^{(2)}\, F_ {22}(z)
\nonu \\ 
& + & 
c_ {63}\, {\tt T}^{(1)}\, {\tt U}^{(\frac{3}{2})}(z) + 
c_ {64}\, {\tt T}^{(1)} \, {\tt V}^{(\frac{3}{2})}(z) + 
c_ {65}\, U\, G_ {11}(z) 
+ c_ {66}\, U\, G_ {22}(z) 
\nonu \\
& + & c_ {67}\, U\, A_ {3}\, F_ {11}(z) 
+ c_ {68}\, U\, A_ {3}\, F_ {22}(z) + 
c_ {69}\, U\, A_ {-}\, F_ {12}(z) 
+ c_ {70}\, U\, A_ {+}\, F_ {21}(z) 
\nonu \\
& + &  c_ {71}\, U\, B_ {3}\, F_ {11}(z) 
+ c_ {72}\, U\, B_ {3}\, F_ {22}(z) 
+ c_ {73}\, U\, B_ {-}\, F_ {12}(z) 
+ c_ {74}\, U\, B_{+}\, F_ {21}(z) 
\nonu \\
& + & c_ {75}\, U\, F_ {11}\, F_ {12}\, F_ {21}(z) 
+ c_ {76}\, U\, F_ {22}\, F_ {21}\, F_ {12}(z) 
+ c_ {77}\, U\, U\, F_ {11}(z) 
+ c_ {78}\, U\, U\, F_ {22}(z) 
\nonu \\
& + & c_ {79}\, U\, \partial F_ {11}(z) 
+ c_ {80}\, U\, \partial F_ {22}(z) 
+ c_ {81}\, \partial A_ {3}\, F_ {11}(z) 
+ c_ {82}\, \partial A_ {3}\, F_ {22}(z) 
\nonu \\
&+& c_ {83}\, \partial A_ {-}\, F_ {12}(z) 
+ c_ {84}\, \partial A_ {+}\, F_ {21}(z) 
+ c_ {85}\, \partial B_ {3}\, F_ {11}(z) 
+ c_ {86}\, \partial B_ {-}\, F_ {12}(z) 
\nonu \\
& + & c_ {87}\, \partial B_{+}\, F_ {21}(z) 
+ c_ {88}\, \partial F_ {11}\, F_ {12}\, F_ {21}(z) 
+ c_ {89}\, \partial F_ {22}\, F_ {21}\, F_ {12}(z) 
+ c_ {90}\,\partial U\, F_ {11}(z) 
\nonu \\
& + & c_ {91 }\, \partial U\, F_ {22}(z) 
+ c_ {92}\, \partial G_ {11}(z)
+ c_ {93}\,\partial G_ {22}(z) + 
c_ {94}\, \partial {\tt U}^{(\frac{3}{2})}(z)
\nonu\\&+&  c_ {95}\, \partial {\tt V}^{(\frac{3}{2})}(z)+ 
c_ {96}\, \partial^{2} 
F_ {22}(z),
\nonu
\eea
where the coefficients are given by
\bea
c_ {1} & = & - 2 i\sqrt {2}, \qquad
c_ {2}  =  - 2 i\sqrt {2}, \qquad
c_ {3}  =  - 2 i\sqrt {2}, \qquad
c_ {4}  =  2 i\sqrt {2},\qquad
c_ {5}  = \frac {\sqrt {2} (1 + N)} {(2 + k + N)^{2}}, \nonu\\
c_ {6} & = &\frac {\sqrt {2} (1 + N)} {(2 + k + N)^{2}}, \qquad
c_ {7}  = \frac {2 i\sqrt {2} (1 + N)} {(2 + k + N)^{3}}, \qquad
c_ {8}  =  - \frac {2 i\sqrt {2} (1 + N)} {(2 + k + N)^{3}}, \nonu\\
c_ {9} & = & - \frac {2 i\sqrt {2} (16 + 21 k + 6 k^{2} + 19 N + 
       13 k\, N + 5 N^{2})} {(2 + N) (2 + k + N)^{3}}, \nonu\\
c_ {10} & = &\frac {2 i\sqrt {2} (16 + 21 k + 6 k^{2} + 19 N + 13 k\, 
     N + 5 N^{2})} {(2 + N) (2 + k + N)^{3}}, \nonu\\
c_ {11} & = & - \frac {2 i\sqrt {2} (16 + 21 k + 6 k^{2} + 19 N + 
       13 k\, N + 5 N^{2})} {(2 + N) (2 + k + N)^{3}}, \nonu\\
c_ {12} & = &\frac {2 i\sqrt {2} (16 + 21 k + 6 k^{2} + 19 N + 13 k\, 
     N + 5 N^{2})} {(2 + N) (2 + k + N)^{3}}, \nonu\\
c_ {13} & = & - \frac {2\sqrt {2} (32 + 55 k + 18 k^{2} + 41 N + 
       35 k\, N + 11 N^{2})} {(2 + N) (2 + k + N)^{4}}, \nonu\\
c_ {14} & = & - \frac {2\sqrt {2} (32 + 55 k + 18 k^{2} + 41 N + 
       35 k\, N + 11 N^{2})} {(2 + N) (2 + k + N)^{4}}, \nonu\\
c_ {15} & = & - \frac {2\sqrt {2} (3 + 2 k + N) (10 + 5 k + 
       8 N)} {3 (2 + k + N)^{3}}, \nonu\\
c_ {16} & = &\frac {2\sqrt {2} (-30 - 32 k + 15 k^{2} + 10 k^{3} - 
      28 N - 23 k\, N + 11 k^{2} N - 22 N^{2} - 13 k\, 
     N^{2} - 8 N^{3})} {(6 + 7 k - N) (2 + k + N)^{3}}, \nonu\\
c_ {17} & = &\frac {\sqrt {2} (1 + N)} {(2 + k + N)^{2}}, \qquad
c_ {18} =  - \frac {2 i\sqrt {2} (1 + N)} {(2 + k + N)^{3}}, \nonu\\
c_ {19} & = &\frac {2 i\sqrt {2} (16 + 21 k + 6 k^{2} + 19 N + 13 k\, 
     N + 5 N^{2})} {(2 + N) (2 + k + N)^{3}}, \nonu\\
c_ {20} & = & - \frac {2 i\sqrt {2} (16 + 21 k + 6 k^{2} + 19 N + 
       13 k\, N + 5 N^{2})} {(2 + N) (2 + k + N)^{3}}, \nonu\\
c_ {21} & = & - \frac {2\sqrt {2} (32 + 55 k + 18 k^{2} + 41 N + 
       35 k\, N + 11 N^{2})} {(2 + N) (2 + k + N)^{4}}, \nonu\\
c_ {22} & = &\frac {2\sqrt {2} (-30 - 32 k + 15 k^{2} + 10 k^{3} - 
      28 N - 23 k\, N + 11 k^{2} N - 22 N^{2} - 13 k\, 
     N^{2} - 8 N^{3})} {(6 + 7 k - N) (2 + k + N)^{3}}, \nonu\\
c_ {23} & = &\frac {\sqrt {2} (1 + N)} {(2 + k + N)^{2}}, \qquad
c_ {24}  = \frac {2 i\sqrt {2} (1 + N)} {(2 + k + N)^{3}}, \nonu\\
c_ {25} & = & - \frac {2 i\sqrt {2} (16 + 21 k + 6 k^{2} + 19 N + 
       13 k\, N + 5 N^{2})} {(2 + N) (2 + k + N)^{3}}, \nonu\\
c_ {26} & = &\frac {2 i\sqrt {2} (16 + 21 k + 6 k^{2} + 19 N + 13 k\, 
     N + 5 N^{2})} {(2 + N) (2 + k + N)^{3}}, \nonu\\
c_ {27} & = & - \frac {2\sqrt {2} (32 + 55 k + 18 k^{2} + 41 N + 
       35 k\, N + 11 N^{2})} {(2 + N) (2 + k + N)^{4}}, \nonu\\
c_ {28} & = & - \frac {2\sqrt {2} (3 + 2 k + N) (10 + 5 k + 
       8 N)} {3 (2 + k + N)^{3}}, \nonu\\
c_ {29} & = & - \frac {\sqrt {2} (18 + 21 k + 6 k^{2} + 22 N + 13 k\, 
      N + 6 N^{2})} {(2 + N) (2 + k + N)^{2}}, \nonu\\
c_ {30} & = & - \frac {\sqrt {2} (18 + 21 k + 6 k^{2} + 22 N + 13 k\, 
      N + 6 N^{2})} {(2 + N) (2 + k + N)^{2}}, \nonu\\
c_ {31} & = & - \frac {2 i\sqrt {2} (18 + 21 k + 6 k^{2} + 22 N + 
       13 k\, N + 6 N^{2})} {(2 + N) (2 + k + N)^{3}}, \nonu\\
c_ {32} & = &\frac {2 i\sqrt {2} (18 + 21 k + 6 k^{2} + 22 N + 13 k\, 
     N + 6 N^{2})} {(2 + N) (2 + k + N)^{3}}, \nonu\\
c_ {33} & = & - \frac {2\sqrt {2} (32 + 55 k + 18 k^{2} + 41 N + 
       35 k\, N + 11 N^{2})} {(2 + N) (2 + k + N)^{4}}, \nonu\\
c_ {34} & = & - \frac {2\sqrt {2} (32 + 55 k + 18 k^{2} + 41 N + 
       35 k\, N + 11 N^{2})} {(2 + N) (2 + k + N)^{4}}, \nonu\\
c_ {35} & = & - \frac {2\sqrt {2} (3 + 2 k + N) (10 + 5 k + 
       8 N)} {3 (2 + k + N)^{3}},  \nonu\\
c_ {36} & = & - \frac {2\sqrt {2} (30 + 92 k + 101 k^{2} + 30 k^{3} - 
       32 N + 7 k\, N + 33 k^{2} N - 78 N^{2} - 39 k\, 
      N^{2} - 24 N^{3})} {(6 + 7 k - N) (2 + k + N)^{3}}, \nonu\\
c_ {37} & = & - \frac {\sqrt {2} (18 + 21 k + 6 k^{2} + 22 N + 13 k\, 
      N + 6 N^{2})} {(2 + N) (2 + k + N)^{2}}, \nonu\\
c_ {38} & = &\frac {2 i\sqrt {2} (18 + 21 k + 6 k^{2} + 22 N + 13 k\, 
     N + 6 N^{2})} {(2 + N) (2 + k + N)^{3}}, \nonu\\
c_ {39} & = &\frac {2\sqrt {2} (32 + 55 k + 18 k^{2} + 41 N + 35 k\, 
     N + 11 N^{2})} {(2 + N) (2 + k + N)^{4}}, \nonu\\
c_ {40} & = & - \frac {2\sqrt {2} (3 + 2 k + N) (10 + 5 k + 
       8 N)} {3 (2 + k + N)^{3}}, \nonu\\
c_ {41} & = & - \frac {\sqrt {2} (18 + 21 k + 6 k^{2} + 22 N + 13 k\, 
      N + 6 N^{2})} {(2 + N) (2 + k + N)^{2}}, \nonu\\
c_ {42} & = & - \frac {2 i\sqrt {2} (18 + 21 k + 6 k^{2} + 22 N + 
       13 k\, N + 6 N^{2})} {(2 + N) (2 + k + N)^{3}}, \nonu\\
c_ {43} & = &\frac {2\sqrt {2} (32 + 55 k + 18 k^{2} + 41 N + 35 k\, 
     N + 11 N^{2})} {(2 + N) (2 + k + N)^{4}}, \nonu\\
c_ {44} & = &\frac {2\sqrt {2} (-30 - 32 k + 15 k^{2} + 10 k^{3} + 
      32 N + 93 k\, N + 51 k^{2} N + 78 N^{2} + 71 k\, 
     N^{2} + 24 N^{3})} {(6 + 7 k - N) (2 + k + N)^{3}}, \nonu\\
c_ {45} & = & - \frac {2 i\sqrt {2} (5 + 2 k + 3 N)} {(2 + k + 
       N)^{3}}, \qquad
c_ {46}  = \frac {8 i\sqrt {2} (15 + 29 k + 10 k^{2} + 25 N + 
      21 k\, N + 8 N^{2})} {3 (2 + k + N)^{4}}, \nonu\\
c_ {47} & = & - \frac {2 i\sqrt {2} (10 + 17 k + 6 k^{2} + 14 N + 
       11 k\, N + 4 N^{2})} {(2 + N) (2 + k + N)^{3}}, \nonu\\
c_ {48} & = & - \frac {i\sqrt {2} (20 + 21 k + 6 k^{2} + 25 N + 
       13 k\, N + 7 N^{2})} {(2 + N) (2 + k + N)^{3}}, \nonu\\
c_ {49} & = & - \frac {4 i\sqrt {2} (13 k + 6 k^{2} + 3 N + 9 k\, 
      N + N^{2})} {(2 + N) (2 + k + N)^{4}}, \nonu\\
c_ {50} & = &\frac {8 i\sqrt {2} (30 + 19 k + 2 k^{2} + 56 N + 44 k\, 
     N + 10 k^{2} N + 38 N^{2} + 21 k\, 
     N^{2} + 8 N^{3})} {3 (2 + N) (2 + k + N)^{4}}, \nonu\\
c_ {51} & = &\frac {i\sqrt {2} (13 k + 6 k^{2} + 3 N + 9 k\, 
     N + N^{2})} {(2 + N) (2 + k + N)^{3}}, \nonu\\
c_ {52} & = & - \frac {i\sqrt {2} (13 k + 6 k^{2} + 3 N + 9 k\, 
      N + N^{2})} {(2 + N) (2 + k + N)^{3}}, \nonu\\
c_ {53} & = & - \frac {2 i\sqrt {2} (15 + 29 k + 10 k^{2} + 25 N + 
       21 k\, N + 8 N^{2})} {(6 + 7 k - N) (2 + k + N)}, \nonu\\
c_ {54} & = &\frac {i\sqrt {2} (20 + 21 k + 6 k^{2} + 25 N + 13 k\, 
     N + 7 N^{2})} {(2 + N) (2 + k + N)^{3}}, \nonu\\
c_ {55} & = &\frac {2 i\sqrt {2} (10 + 17 k + 6 k^{2} + 14 N + 11 k\, 
     N + 4 N^{2})} {(2 + N) (2 + k + N)^{3}}, \qquad
c_ {56}  = \frac {2 i\sqrt {2} (5 + 2 k + 3 N)} {(2 + k + N)^{3}}, \nonu\\
c_ {57} & = &\frac {1} {(6 + 7 k - N) (2 + N) (2 + k + N)^{3}}2 i\sqrt {2} (-156 - 320 k - 164 k^{2} - 
      24 k^{3} - 262 N 
      \nonu\\&-& 316 k\, 
     N - 3 k^{2} N + 30 k^{3} N - 216 N^{2} - 129 k\, 
     N^{2} + 33 k^{2} N^{2} - 114 N^{3} - 39 k\, 
     N^{3} - 24 N^{4}), \nonu\\
c_ {58} & = &\frac {8 i\sqrt {2} (13 k + 6 k^{2} + 3 N + 9 k\, 
     N + N^{2})} {(2 + N) (2 + k + N)^{4}}, \qquad
c_ {59}  = \frac {4 i\sqrt {2} (13 k + 6 k^{2} + 3 N + 9 k\, 
     N + N^{2})} {(2 + N) (2 + k + N)^{4}}, \nonu\\
c_ {60} & = &\frac {i (60 + 77 k + 22 k^{2} + 121 N + 115 k\, 
     N + 20 k^{2} N + 79 N^{2} + 42 k\, 
     N^{2} + 16 N^{3})} {\sqrt {2} (2 + N) (2 + k + N)^{2}}, \nonu\\
c_ {61} & = &\frac {i (60 + 77 k + 22 k^{2} + 121 N + 115 k\, 
     N + 20 k^{2} N + 79 N^{2} + 42 k\, 
     N^{2} + 16 N^{3})} {\sqrt {2} (2 + N) (2 + k + N)^{2}}, \nonu\\
c_ {62} & = &\frac {2 i\sqrt {2} (15 + 29 k + 10 k^{2} + 25 N + 
      21 k\, N + 8 N^{2})} {(6 + 7 k - N) (2 + k + N)}, \nonu\\
c_ {63} & = &\frac {i\sqrt {2} (60 + 77 k + 22 k^{2} + 121 N + 
      115 k\, N + 20 k^{2} N + 79 N^{2} + 42 k\, 
     N^{2} + 16 N^{3})} {(2 + N) (2 + k + N)^{2}}, \nonu\\
c_ {64} & = & - \frac {i\sqrt {2} (60 + 77 k + 22 k^{2} + 121 N + 
       115 k\, N + 20 k^{2} N + 79 N^{2} + 42 k\, 
      N^{2} + 16 N^{3})} {(2 + N) (2 + k + N)^{2}}, \nonu\\
c_ {65} & = &\frac {2 i\sqrt {2}} {(2 + k + N)}, \qquad
c_ {66}  =  - \frac {2 i\sqrt {2}} {(2 + k + N)}, \qquad
c_ {67}  =  - \frac {2\sqrt {2} (3 + 2 k + N)} {(2 + k + N)^{3}}, \nonu\\
c_ {68} & = & - \frac {2\sqrt {2} (3 + 2 k + N)} {(2 + k + N)^{3}}, \qquad
c_ {69}  =  - \frac {2\sqrt {2} (3 + 2 k + N)} {(2 + k + N)^{3}}, \qquad
c_ {70}  =  - \frac {2\sqrt {2} (3 + 2 k + N)} {(2 + k + N)^{3}},\nonu\\
c_ {71} & = & - \frac {2\sqrt {2} (26 + 25 k + 6 k^{2} + 30 N + 
       15 k\, N + 8 N^{2})} {(2 + N) (2 + k + N)^{3}}, \nonu\\
c_ {72} & = & - \frac {2\sqrt {2} (26 + 25 k + 6 k^{2} + 30 N + 
       15 k\, N + 8 N^{2})} {(2 + N) (2 + k + N)^{3}}, \nonu\\
c_ {73} & = & - \frac {2\sqrt {2} (26 + 25 k + 6 k^{2} + 30 N + 
       15 k\, N + 8 N^{2})} {(2 + N) (2 + k + N)^{3}}, \nonu\\
c_ {74} & = & - \frac {2\sqrt {2} (26 + 25 k + 6 k^{2} + 30 N + 
       15 k\, N + 8 N^{2})} {(2 + N) (2 + k + N)^{3}}, \nonu\\
c_ {75} & = &\frac {2 i\sqrt {2} (32 + 55 k + 18 k^{2} + 41 N + 
      35 k\, N + 11 N^{2})} {(2 + N) (2 + k + N)^{4}}, \nonu\\
c_ {76} & = & - \frac {2 i\sqrt {2} (32 + 55 k + 18 k^{2} + 41 N + 
       35 k\, N + 11 N^{2})} {(2 + N) (2 + k + N)^{4}}, \nonu\\
c_ {77} & = &\frac {4 i\sqrt {2}} {(2 + k + N)^{2}}, \qquad
c_ {78}  =  - \frac {4 i\sqrt {2}} {(2 + k + N)^{2}}, \qquad
c_ {79}  = \frac {2 i\sqrt {2} (3 + 2 k + N) (10 + 5 k + 
      8 N)} {3 (2 + k + N)^{3}}, \nonu\\
c_ {80} & = &\frac {2 i\sqrt {2} (-30 - 32 k + 15 k^{2} + 10 k^{3} + 
      32 N + 93 k\, N + 51 k^{2} N + 78 N^{2} + 71 k\, 
     N^{2} + 24 N^{3})} {(6 + 7 k - N) (2 + k + N)^{3}}, \nonu\\
c_ {81} & = & - \frac {2\sqrt {2} (3 + 2 k + N) (10 + 5 k + 
       8 N)} {3 (2 + k + N)^{3}}, \nonu\\
c_ {82} & = &\frac {1}{(6 + 7 k - N) (2 + N) (2 + k + N)^{3}}4\sqrt {2} (-108 - 192 k - 67 k^{2} - 2 k^{3} - 
      174 N - 197 k\, 
     N 
     \nonu\\&+& 17 k^{2} N + 20 k^{3} N - 144 N^{2} - 89 k\, 
     N^{2} + 22 k^{2} N^{2} - 76 N^{3} - 26 k\, 
     N^{3} - 16 N^{4}), \nonu\\
c_ {83} & = &\frac {2\sqrt {2} (-24 - 25 k + 15 k^{2} + 10 k^{3} - 
      23 N - 16 k\, N + 11 k^{2} N - 23 N^{2} - 13 k\, 
     N^{2} - 8 N^{3})} {(6 + 7 k - N) (2 + k + N)^{3}}, \nonu\\
c_ {84} & = & - \frac {2\sqrt {2} (27 + 35 k + 10 k^{2} + 31 N + 
       21 k\, N + 8 N^{2})} {3 (2 + k + N)^{3}}, \nonu\\
c_ {85} & = & - \frac {2\sqrt {2} (-48 - 56 k - 16 k^{2} - 34 N - k\, 
      N + 10 k^{2} N + 14 N^{2} + 21 k\, 
      N^{2} + 8 N^{3})} {3 (2 + N) (2 + k + N)^{3}}, \nonu\\
c_ {86} & = & - \frac {2\sqrt {2} (6 + 7 k + 2 k^{2} + 32 N + 38 k\, 
      N + 10 k^{2} N + 32 N^{2} + 21 k\, 
      N^{2} + 8 N^{3})} {3 (2 + N) (2 + k + N)^{3}}, \nonu\\
c_ {87} & = &\frac{1}{(6 + 7 k - N) (2 + N) (2 + k + N)^{3}}2\sqrt {2} (48 + 188 k + 213 k^{2} + 62 k^{3} + 
      148 N + 365 k\, 
     N \nonu\\&+& 202 k^{2} N + 10 k^{3} N 
      +202 N^{2} + 264 k\, 
     N^{2} + 51 k^{2} N^{2} + 120 N^{3} + 71 k\, 
     N^{3} + 24 N^{4}), \nonu\\
c_ {88} & = &\frac {4 i\sqrt {2} (60 + 77 k + 22 k^{2} + 121 N + 
      115 k\, N + 20 k^{2} N + 79 N^{2} + 42 k\, 
     N^{2} + 16 N^{3})} {3 (2 + N) (2 + k + N)^{4}}, \nonu\\
c_ {89} & = &\frac {4 i\sqrt {2} (13 k + 6 k^{2} + 3 N + 9 k\, 
     N + N^{2})} {(2 + N) (2 + k + N)^{4}}, \qquad
c_ {90}  = \frac {2 i\sqrt {2} (3 + 2 k + N) (10 + 5 k + 
      8 N)} {3 (2 + k + N)^{3}}, \nonu\\
c_ {91} & = & - \frac{1}{(6 + 7 k - N) (2 + N) (2 + k + N)^{3}}4 i\sqrt {2} (-48 - 128 k - 97 k^{2} - 
       22 k^{3} - 148 N 
       \nonu\\&-& 235 k\, 
      N - 60 k^{2} N + 10 k^{3} N - 202 N^{2} - 182 k\, 
      N^{2} - 9 k^{2} N^{2} - 120 N^{3} - 55 k\, 
      N^{3} - 24 N^{4}), \nonu\\
c_ {92} & = & - \frac {i (60 + 91 k + 26 k^{2} + 167 N + 191 k\, 
      N + 40 k^{2} N + 137 N^{2} + 84 k\, 
      N^{2} + 32 N^{3})} {3\sqrt {2} (2 + N) (2 + k + N)^{2}}, \nonu\\
c_ {93} & = &\frac {1} {3\sqrt {2} (6 + 7 k - N) (2 + 
       N) (2 + k + N)^{2}}i (720 + 2502 k + 2657 k^{2} + 742 k^{3} + 1962 N 
      \nonu\\&+& 5360 k\, 
     N + 3819 k^{2} N + 560 k^{3} N + 1667 N^{2} + 3240 k\, 
     N^{2} + 1216 k^{2} N^{2} + 479 N^{3} 
     \nonu\\&+& 532 k\, 
     N^{3} + 32 N^{4}), \nonu\\
c_ {94} & = & - \frac {i\sqrt {2} (60 + 77 k + 22 k^{2} + 121 N + 
       115 k\, N + 20 k^{2} N + 79 N^{2} + 42 k\, 
      N^{2} + 16 N^{3})} {(2 + N) (2 + k + N)^{2}}, \nonu\\
c_ {95} & = & - \frac {1} {3 (6 + 7 k - N) (2 + N) (2 + k + N)^{2}}i\sqrt {2} (1080 + 2886 k + 2477 k^{2} + 
       622 k^{3} + 1758 N 
       \nonu\\&+&4436 k\, 
      N + 3117 k^{2} N + 500 k^{3} N + 539 N^{2} + 1830 k\, 
      N^{2} + 910 k^{2} N^{2} - 277 N^{3} + 106 k\, 
      N^{3} 
      \nonu\\&-& 112 N^{4}), \nonu\\
c_ {96} & = & - \frac {4 i\sqrt {2} (k - 3 N) (15 + 29 k + 10 k^{2} + 
       25 N + 21 k\, N + 8 N^{2})} {(6 + 7 k - N) (2 + k + N)^{3}}.    
\nonu
\eea

Finally, 
for the remaining higher spin-$\frac{5}{2}$ current, one has the following 
relation  
\bea
V_{\frac{1}{2}}^{(2),3}(z) &=& 
c_ {1}\, {\tt P}_{-}^{(\frac{5}{2})}(z) + 
c_ {2}\, {\tt P}_{+}^{(\frac{5}{2})}(z) 
+ c_ {3}\, {\tt W}_{-}^{(\frac{5}{2})}(z) 
+  c_ {4}\, {\tt W}_{+}^{(\frac{5}{2})}(z)  
+ c_ {5}\, A_ {3}\, G_ {12}(z) + 
c_ {6}\, A_ {3}\, G_ {21}(z) 
\nonu \\
& + &  c_ {7}\, A_ {3}\, A_ {3}\, F_ {12}(z) 
+  c_ {8}\, A_ {3}\, A_ {3}\, F_ {21}(z) 
+ c_ {9}\, A_ {3}\, B_ {3}\, F_ {12}(z) 
+ c_ {10}\, A_ {3}\, B_ {3}\, F_ {21}(z) 
\nonu \\
& + &  c_ {11}\, A_ {3}\, B_ {-}\, F_ {22}(z) 
+ c_ {12}\, A_ {3}\, B_ {+}\, F_ {11}(z) 
+  c_ {13}\, A_ {3}\, F_ {11}\, F_ {12}\, F_ {22}(z) 
+ c_ {14}\, A_ {3}\, F_ {11}\, F_ {21}\, F_ {22}(z) 
\nonu \\
& + &  c_ {15}\, A_ {3}\, \partial F_ {21}(z) + 
 c_ {16}\, A_ {-}\, G_ {11}(z) + 
c_ {17}\, A_ {-}\, B_ {3}\, F_ {11}(z) 
+ c_ {18}\, A_ {-}\, B_ {-}\, F_ {12}(z) 
\nonu \\
& + & c_ {19}\, A_ {-}\, F_ {11}\, F_ {12}\, F_ {21}(z)
+ c_ {20}\, A_ {-}\,\partial F_ {11}(z) 
+  c_ {21}\, A_ {+}\, G_ {22}(z) + 
c_ {22}\, A_ {+}\, A_ {-}\, F_ {12}(z) 
\nonu\\&+&  
c_ {23}\, A_ {+}\, A_ {-}\, F_ {21}(z) 
+ c_ {24}\, A_ {+}\, B_ {3}\, F_ {22}(z) + 
c_ {25}\, A_ {+}\, B_ {+}\, F_ {21}(z) 
+ c_ {26}\, A_ {+}\, F_ {12}\, F_ {21}\, F_ {22}(z) 
\nonu \\
& + &  c_ {27}\, B_ {3}\, G_ {12}(z) 
+  c_ {28}\, B_ {3}\, G_ {21}(z) 
+ c_ {29}\, B_ {3}\, B_ {3}\, F_ {12}(z) 
+ c_ {30}\, B_ {3}\, B_ {3}\, F_ {21}(z) 
\nonu \\
& + & c_ {31}\, B_ {3}\, F_ {11}\, F_ {12}\, F_ {22}(z) 
+ c_ {32}\, B_ {3}\, F_ {11}\, F_ {21}\, F_ {22}(z) 
+  c_ {33 }\, B_ {3}\, \partial F_ {21}(z) 
+  c_ {34}\, B_ {-}\, G_ {22}(z) 
\nonu \\
& + &  c_ {35}\, B_ {-}\, F_ {12}\, F_ {21}\, F_ {22}(z) 
+ c_ {36}\, B_ {-}\, \partial F_ {22}(z)
+  c_ {37}\, B_ {+}\, G_ {11}(z) 
+  c_ {38}\, B_ {+}\, B_ {-}\, F_ {12}(z) 
\nonu \\
&
+& c_ {39}\, B_ {+}\, B_ {-}\, F_ {21}(z) 
+ c_ {40}\, B_ {+}\, F_ {11}\, F_ {12}\, F_ {21}(z) 
+ c_ {41}\, F_ {11}\, F_ {12}\, G_ {22}(z) 
+ c_ {42}\, F_ {11}\, F_ {12}\, \partial F_ {22}(z) 
\nonu\\&+&   
c_ {43 }\, F_ {11}\, F_ {21}\, G_ {22}(z) 
+ c_ {44}\, F_ {11}\, F_ {21}\, \partial F_ {22}(z) 
+  c_ {45}\, F_ {11}\, F_ {22}\, G_ {12}(z) 
+ c_ {46}\, F_ {11}\, F_ {22}\, G_ {21}(z) 
\nonu \\
& + & c_ {47}\, F_ {11}\, \partial F_ {12}\, F_ {22}(z) 
+  c_ {48}\, F_ {11}\, \partial F_ {21}\, F_ {22}(z) 
+ c_ {49}\, F_ {12}\,  {\tt P}^{(2)}(z) + 
c_ {50}\, F_ {12}\, F_ {21}\, G_ {12}(z) 
\nonu \\
& + &  c_ {51}\, F_ {12}\, F_ {21}\, G_ {21}(z) 
+ c_ {52}\, F_ {12}\, F_ {22}\, G_ {11}(z) 
+ c_ {53}\, F_ {12}\, \partial F_ {12}\, F_ {21}(z) 
+ c_ {54}\, F_ {21}\,  {\tt P}^{(2)}(z) 
\nonu \\
& + &  c_ {55}\, F_ {21}\, F_ {22}\,G_ {11}(z) 
+ c_ {56}\, F_ {21}\, \partial F_ {21}\, F_ {12}(z) 
+ c_ {57}\, G_ {12}\, {\tt T}^{(1)}(z) 
+  c_ {58}\, G_ {21}\,{\tt T}^{(1)}(z) 
\nonu \\
& + &  c_ {59}\, {\tt P}^{(2)}\, F_ {12}(z) 
+ c_ {60}\, {\tt P}^{(2)}\, F_ {21}(z) 
+ c_ {61}\, {\tt T}^{(1)}\, {\tt T}^{(\frac{3}{2})}_{-}(z)
+ c_ {62}\, {\tt T}^{(1)}\, {\tt T}^{(\frac{3}{2})}_{+}(z) 
\nonu \\
& + &  c_ {63}\, U\, G_ {12}(z) 
+  c_ {64}\, U\, G_ {21}(z) 
+ c_ {65}\, U\, A_ {3}\, F_ {12}(z) 
+ c_ {66}\, U\, A_ {3}\, F_ {21}(z) 
\nonu \\
& + & c_ {67}\, U\, A_ {-}\, F_ {11}(z) 
+ c_ {68}\, U\, A_ {+}\, F_ {22}(z) + 
c_ {69}\, U\, B_ {3}\, F_ {12}(z) 
+   c_ {70}\, U\, B_ {3}\, F_ {21}(z) 
\nonu \\
& + & c_ {71}\, U\, B_ {-}\, F_ {22}(z) 
+ c_ {72}\, U\, B_ {+}\, F_ {11}(z) 
+ c_ {73}\, U\, F_ {11}\, F_ {12}\, F_ {22}(z) 
+ c_ {74}\, U\, F_ {11}\, F_ {21}\, F_ {22}(z) 
\nonu\\&+&  c_ {75}\, U\, U\, F_ {12}(z) + 
c_ {76}\, U\, U\, F_ {21}(z) 
+ c_ {77}\, U\, \partial F_ {21}(z) 
+  c_ {78}\, \partial A_ {3}\, F_ {21}(z)
\nonu \\ 
& + & c_ {79}\, \partial A_ {-}\, F_ {11}(z) 
+ c_ {80}\, \partial A_ {+}\, F_ {22}(z) 
+  c_ {81}\, \partial B_ {3}\, F_ {21}(z) 
+ c_ {82}\, \partial B_ {-}\, F_ {22}(z) 
\nonu \\
& + &  c_ {83}\, \partial B_ {+}\, F_ {11}(z) 
+ c_ {84}\, \partial F_ {11}\, F_ {12}\, F_ {22}(z) 
+ c_ {85}\, \partial F_ {11}\, F_ {21}\, F_ {22}(z) 
+   c_ {86}\, \partial U\, F_ {21}(z) 
\nonu \\
& + &  c_ {87}\, \partial G_ {12}(z)
+ c_ {88}\, \partial G_ {21}(z)  
+ c_ {89}\, \partial {\tt T}^{(\frac{3}{2})}_{-}(z) 
+  c_ {90}\, \partial^{2} F_ {21}(z),
\nonu
\eea
where the coefficients are given by
\bea
c_ {1} & = & - 2\sqrt {2},\qquad
c_ {2}  =  2\sqrt {2}, \qquad
c_ {3}  =  2\sqrt {2}, \qquad
c_ {4}  =  2\sqrt {2}, \qquad
c_ {5}  = \frac {i\sqrt {2} (1 + N)} {(2 + k + N)^{2}}, \nonu\\
c_ {6} & = & - \frac {i\sqrt {2} (1 + N)} {(2 + k + N)^{2}}, \qquad
c_ {7}  =  - \frac {2\sqrt {2} (1 + N)} {(2 + k + N)^{3}}, \qquad
c_ {8}  =  - \frac {2\sqrt {2} (1 + N)} {(2 + k + N)^{3}}, \nonu\\
c_ {9} & = & - \frac {2\sqrt {2} (16 + 21 k + 6 k^{2} + 19 N + 13 k\, 
      N + 5 N^{2})} {(2 + N) (2 + k + N)^{3}}, \nonu\\
c_ {10} & = & - \frac {2\sqrt {2} (16 + 21 k + 6 k^{2} + 19 N + 
       13 k\, N + 5 N^{2})} {(2 + N) (2 + k + N)^{3}}, \nonu\\
c_ {11} & = &\frac {2\sqrt {2} (16 + 21 k + 6 k^{2} + 19 N + 13 k\, 
     N + 5 N^{2})} {(2 + N) (2 + k + N)^{3}}, \nonu\\
c_ {12} & = &\frac {2\sqrt {2} (16 + 21 k + 6 k^{2} + 19 N + 13 k\, 
     N + 5 N^{2})} {(2 + N) (2 + k + N)^{3}}, \nonu\\
c_ {13} & = &\frac {2 i\sqrt {2} (32 + 55 k + 18 k^{2} + 41 N + 
      35 k\, N + 11 N^{2})} {(2 + N) (2 + k + N)^{4}}, \nonu\\
c_ {14} & = &\frac {2 i\sqrt {2} (32 + 55 k + 18 k^{2} + 41 N + 
      35 k\, N + 11 N^{2})} {(2 + N) (2 + k + N)^{4}}, \nonu\\
c_ {15} & = &\frac {1}{3 (2 + N) (2 + k + N)^{4}}i\sqrt {2} (60 + 169 k + 154 k^{2} + 40 k^{3} + 
      149 N + 299 k\, 
     N + 174 k^{2} N 
     \nonu\\&+& 20 k^{3} N + 159 N^{2} + 212 k\, 
     N^{2} + 62 k^{2} N^{2} + 84 N^{3} + 58 k\, 
     N^{3} + 16 N^{4}), \nonu\\
c_ {16} & = &\frac {i\sqrt {2} (1 + N)} {(2 + k + N)^{2}}, \qquad
c_ {17}  = \frac {2\sqrt {2} (16 + 21 k + 6 k^{2} + 19 N + 13 k\, 
     N + 5 N^{2})} {(2 + N) (2 + k + N)^{3}}, \nonu\\
c_ {18} & = &\frac {2\sqrt {2} (16 + 21 k + 6 k^{2} + 19 N + 13 k\, 
     N + 5 N^{2})} {(2 + N) (2 + k + N)^{3}}, \nonu\\
c_ {19} & = & - \frac {2 i\sqrt {2} (32 + 55 k + 18 k^{2} + 41 N + 
       35 k\, N + 11 N^{2})} {(2 + N) (2 + k + N)^{4}}, \nonu\\
c_ {20} & = &\frac {1}{3 (2 + N) (2 + k + N)^{4}}i\sqrt {2} (-60 - 169 k - 154 k^{2} - 40 k^{3} + 
      31 N - 68 k\, 
     N - 108 k^{2} N 
     \nonu\\&-& 20 k^{3} N + 204 N^{2} + 133 k\, 
     N^{2} - 2 k^{2} N^{2} + 153 N^{3} + 68 k\, 
     N^{3} + 32 N^{4}), \nonu\\
c_ {21} & = & - \frac {i\sqrt {2} (1 + N)} {(2 + k + N)^{2}}, \qquad
c_ {22}  =  - \frac {2\sqrt {2} (1 + N)} {(2 + k + N)^{3}}, \qquad
c_ {23}  =  - \frac {2\sqrt {2} (1 + N)} {(2 + k + N)^{3}}, \nonu\\
c_ {24} & = &\frac {2\sqrt {2} (16 + 21 k + 6 k^{2} + 19 N + 13 k\, 
     N + 5 N^{2})} {(2 + N) (2 + k + N)^{3}}, \nonu\\
c_ {25} & = &\frac {2\sqrt {2} (16 + 21 k + 6 k^{2} + 19 N + 13 k\, 
     N + 5 N^{2})} {(2 + N) (2 + k + N)^{3}}, \nonu\\
c_ {26} & = & - \frac {2 i\sqrt {2} (32 + 55 k + 18 k^{2} + 41 N + 
       35 k\, N + 11 N^{2})} {(2 + N) (2 + k + N)^{4}}, \nonu\\
c_ {27} & = &\frac {i\sqrt {2} (18 + 21 k + 6 k^{2} + 22 N + 13 k\, 
     N + 6 N^{2})} {(2 + N) (2 + k + N)^{2}}, \nonu\\
c_ {28} & = & - \frac {i\sqrt {2} (18 + 21 k + 6 k^{2} + 22 N + 
       13 k\, N + 6 N^{2})} {(2 + N) (2 + k + N)^{2}}, \nonu\\
c_ {29} & = &\frac {2\sqrt {2} (18 + 21 k + 6 k^{2} + 22 N + 13 k\, 
     N + 6 N^{2})} {(2 + N) (2 + k + N)^{3}}, \nonu\\
c_ {30} & = &\frac {2\sqrt {2} (18 + 21 k + 6 k^{2} + 22 N + 13 k\, 
     N + 6 N^{2})} {(2 + N) (2 + k + N)^{3}}, \nonu\\
c_ {31} & = & - \frac {2 i\sqrt {2} (32 + 55 k + 18 k^{2} + 41 N + 
       35 k\, N + 11 N^{2})} {(2 + N) (2 + k + N)^{4}}, \nonu\\
c_ {32} & = & - \frac {2 i\sqrt {2} (32 + 55 k + 18 k^{2} + 41 N + 
       35 k\, N + 11 N^{2})} {(2 + N) (2 + k + N)^{4}}, \nonu\\
c_ {33} & = & - \frac {1}{3 (2 + N) (2 + k + N)^{4}}i\sqrt {2} (60 + 169 k + 154 k^{2} + 
       40 k^{3} + 149 N + 299 k\, 
      N + 174 k^{2} N 
      \nonu\\&+& 20 k^{3} N + 159 N^{2} + 212 k\, 
      N^{2} + 62 k^{2} N^{2} + 84 N^{3} + 58 k\, 
      N^{3} + 16 N^{4}), \nonu\\
c_ {34} & = &\frac {i\sqrt {2} (18 + 21 k + 6 k^{2} + 22 N + 13 k\, 
     N + 6 N^{2})} {(2 + N) (2 + k + N)^{2}}, \nonu\\
c_ {35} & = & - \frac {2 i\sqrt {2} (32 + 55 k + 18 k^{2} + 41 N + 
       35 k\, N + 11 N^{2})} {(2 + N) (2 + k + N)^{4}}, \nonu\\
c_ {36} & = &\frac {1}{3 (2 + N) (2 + k + N)^{4}}i\sqrt {2} (60 - 11 k - 77 k^{2} - 26 k^{3} + 
      149 N - 64 k\, 
     N - 171 k^{2} N 
     \nonu\\&-& 40 k^{3} N + 159 N^{2} - 25 k\, 
     N^{2} - 64 k^{2} N^{2} + 84 N^{3} + 10 k\, 
     N^{3} + 16 N^{4}), \nonu\\
c_ {37} & = & - \frac {i\sqrt {2} (18 + 21 k + 6 k^{2} + 22 N + 
       13 k\, N + 6 N^{2})} {(2 + N) (2 + k + N)^{2}}, \nonu\\
c_ {38} & = &\frac {2\sqrt {2} (18 + 21 k + 6 k^{2} + 22 N + 13 k\, 
     N + 6 N^{2})} {(2 + N) (2 + k + N)^{3}}, \nonu\\
c_ {39} & = &\frac {2\sqrt {2} (18 + 21 k + 6 k^{2} + 22 N + 13 k\, 
     N + 6 N^{2})} {(2 + N) (2 + k + N)^{3}}, \nonu\\
c_ {40} & = & - \frac {2 i\sqrt {2} (32 + 55 k + 18 k^{2} + 41 N + 
       35 k\, N + 11 N^{2})} {(2 + N) (2 + k + N)^{4}}, \nonu\\
c_ {41} & = & - \frac {2\sqrt {2} (5 + 2 k + 3 N)} {(2 + k + N)^{3}}, \qquad
c_ {42} =  - \frac {8\sqrt {2} (5 + 2 k + 3 N)} {(2 + k + N)^{4}}, \nonu\\
c_ {43} & = & - \frac {2\sqrt {2} (10 + 17 k + 6 k^{2} + 14 N + 
       11 k\, N + 4 N^{2})} {(2 + N) (2 + k + N)^{3}}, \nonu\\
c_ {44} & = &\frac {2\sqrt {2} (60 + 155 k + 58 k^{2} + 139 N + 
      169 k\, N + 20 k^{2} N + 85 N^{2} + 42 k\, 
     N^{2} + 16 N^{3})} {3 (2 + N) (2 + k + N)^{4}}, \nonu\\
c_ {45} & = & - \frac {\sqrt {2} (13 k + 6 k^{2} + 3 N + 9 k\, 
      N + N^{2})} {(2 + N) (2 + k + N)^{3}}, \qquad
c_ {46}  = \frac {\sqrt {2} (13 k + 6 k^{2} + 3 N + 9 k\, 
     N + N^{2})} {(2 + N) (2 + k + N)^{3}}, \nonu\\
c_ {47} & = & - \frac {4\sqrt {2} (20 + 21 k + 6 k^{2} + 25 N + 
       13 k\, N + 7 N^{2})} {(2 + N) (2 + k + N)^{4}}, \nonu\\
c_ {48} & = &\frac {2\sqrt {2} (60 + 77 k + 22 k^{2} + 121 N + 
      115 k\, N + 20 k^{2} N + 79 N^{2} + 42 k\, 
     N^{2} + 16 N^{3})} {3 (2 + N) (2 + k + N)^{4}}, \nonu\\
c_ {49} & = & - \frac {(3 + 2 k + N) (10 + 5 k + 
       8 N)} {(3\sqrt {2} (2 + k + N)^{2})}, \qquad
c_ {50}  = \frac {\sqrt {2} (20 + 21 k + 6 k^{2} + 25 N + 13 k\, 
     N + 7 N^{2})} {(2 + N) (2 + k + N)^{3}}, \nonu\\
c_ {51} & = & - \frac {\sqrt {2} (20 + 21 k + 6 k^{2} + 25 N + 13 k\, 
      N + 7 N^{2})} {(2 + N) (2 + k + N)^{3}}, \nonu\\
c_ {52} & = &\frac {2\sqrt {2} (10 + 17 k + 6 k^{2} + 14 N + 11 k\, 
     N + 4 N^{2})} {(2 + N) (2 + k + N)^{3}}, \nonu\\
c_ {53} & = &\frac {4\sqrt {2} (13 k + 6 k^{2} + 3 N + 9 k\, 
     N + N^{2})} {(2 + N) (2 + k + N)^{4}}, \nonu\\
c_ {54} & = &\frac {(60 + 77 k + 22 k^{2} + 121 N + 115 k\, 
    N + 20 k^{2} N + 79 N^{2} + 42 k\, 
    N^{2} + 16 N^{3})} {2\sqrt {2} (2 + N) (2 + k + N)^{2}}, \nonu\\
c_ {55} & = &\frac {2\sqrt {2} (5 + 2 k + 3 N)} {(2 + k + N)^{3}}, \qquad
c_ {56}  = \frac {4\sqrt {2} (13 k + 6 k^{2} + 3 N + 9 k\, 
     N + N^{2})} {(2 + N) (2 + k + N)^{4}}, \nonu\\
c_ {57} & = &\frac {(60 + 77 k + 22 k^{2} + 121 N + 115 k\, 
    N + 20 k^{2} N + 79 N^{2} + 42 k\, 
    N^{2} + 16 N^{3})} {\sqrt {2} (2 + N) (2 + k + N)^{2}}, \nonu\\
c_ {58} & = & - \frac {(60 + 77 k + 22 k^{2} + 121 N + 115 k\, 
     N + 20 k^{2} N + 79 N^{2} + 42 k\, 
     N^{2} + 16 N^{3})} {\sqrt {2} (2 + N) (2 + k + N)^{2}}, \nonu\\
c_ {59} & = &\frac {(3 + 2 k + N) (10 + 5 k + 
      8 N)} {3\sqrt {2} (2 + k + N)^{2}}, \nonu\\
c_ {60} & = & - \frac {(60 + 77 k + 22 k^{2} + 121 N + 115 k\, 
     N + 20 k^{2} N + 79 N^{2} + 42 k\, 
     N^{2} + 16 N^{3})} {2\sqrt {2} (2 + N) (2 + k + N)^{2}}, \nonu\\
c_ {61} & = & - \frac {\sqrt {2} (60 + 77 k + 22 k^{2} + 121 N + 
       115 k\, N + 20 k^{2} N + 79 N^{2} + 42 k\, 
      N^{2} + 16 N^{3})} {(2 + N) (2 + k + N)^{2}}, \nonu\\
c_ {62} & = & - \frac {\sqrt {2} (60 + 77 k + 22 k^{2} + 121 N + 
       115 k\, N + 20 k^{2} N + 79 N^{2} + 42 k\, 
      N^{2} + 16 N^{3})} {(2 + N) (2 + k + N)^{2}}, \nonu\\
c_ {63} & = & - \frac {2\sqrt {2}} {(2 + k + N)}, \qquad
c_ {64}  =  - \frac {2\sqrt {2}} {(2 + k + N)}, \qquad
c_ {65}  =  - \frac {2 i\sqrt {2} (3 + 2 k + N)} {(2 + k + N)^{3}}, \nonu\\
c_ {66} & = &\frac {2 i\sqrt {2} (3 + 2 k + N)} {(2 + k + N)^{3}}, \qquad
c_ {67}  =  - \frac {2 i\sqrt {2} (3 + 2 k + N)} {(2 + k + N)^{3}}, \qquad
c_ {68}  = \frac {2 i\sqrt {2} (3 + 2 k + N)} {(2 + k + N)^{3}}, \nonu\\
c_ {69} & = &\frac {2 i\sqrt {2} (26 + 25 k + 6 k^{2} + 30 N + 15 k\, 
     N + 8 N^{2})} {(2 + N) (2 + k + N)^{3}}, \nonu\\
c_ {70} & = & - \frac {2 i\sqrt {2} (26 + 25 k + 6 k^{2} + 30 N + 
       15 k\, N + 8 N^{2})} {(2 + N) (2 + k + N)^{3}}, \nonu\\
c_ {71} & = &\frac {2 i\sqrt {2} (26 + 25 k + 6 k^{2} + 30 N + 15 k\, 
     N + 8 N^{2})} {(2 + N) (2 + k + N)^{3}}, \nonu\\
c_ {72} & = & - \frac {2 i\sqrt {2} (26 + 25 k + 6 k^{2} + 30 N + 
       15 k\, N + 8 N^{2})} {(2 + N) (2 + k + N)^{3}}, \nonu\\
c_ {73} & = &\frac {2\sqrt {2} (32 + 55 k + 18 k^{2} + 41 N + 35 k\, 
     N + 11 N^{2})} {(2 + N) (2 + k + N)^{4}}, \nonu\\
c_ {74} & = & - \frac {2\sqrt {2} (32 + 55 k + 18 k^{2} + 41 N + 
       35 k\, N + 11 N^{2})} {(2 + N) (2 + k + N)^{4}}, \nonu\\
c_ {75} & = & - \frac {4\sqrt {2}} {(2 + k + N)^{2}}, \qquad
c_ {76}  =  - \frac {4\sqrt {2}} {(2 + k + N)^{2}}, \nonu\\
c_ {77} & = & - \frac{1}{3 (2 + N) (2 + k + N)^{4}}\sqrt {2} (60 + 169 k + 154 k^{2} + 40 k^{3} + 
       149 N + 299 k\, 
      N + 174 k^{2} N 
      \nonu\\&+& 20 k^{3} N + 159 N^{2} + 212 k\, 
      N^{2} + 62 k^{2} N^{2} + 84 N^{3} + 58 k\, 
      N^{3} + 16 N^{4}), \nonu\\
c_ {78} & = &\frac{1} {3 (2 + N) (2 + k + N)^{4}}i\sqrt {2} (12 + 145 k + 154 k^{2} + 40 k^{3} + 
      53 N + 263 k\, 
     N + 174 k^{2} N 
     \nonu\\&+& 20 k^{3} N + 99 N^{2} + 200 k\, 
     N^{2} + 62 k^{2} N^{2} + 72 N^{3} + 58 k\, 
     N^{3} + 16 N^{4}), \nonu\\
c_ {79} & = &\frac {1}{3 (2 + N) (2 + k + N)^{4}}i\sqrt {2} (-36 - 157 k - 154 k^{2} - 40 k^{3} + 
      79 N - 50 k\, 
     N - 108 k^{2} N 
     \nonu\\&-& 20 k^{3} N + 234 N^{2} + 139 k\, 
     N^{2} - 2 k^{2} N^{2} + 159 N^{3} + 68 k\, 
     N^{3} + 32 N^{4}), \nonu\\
c_ {80} & = & - \frac {2 i\sqrt {2} (1 + N)} {(2 + k + N)^{3}}, \nonu\\
c_ {81} & = & - \frac {1}{3 (2 + N) (2 + k + N)^{4}}i\sqrt {2} (-372 - 551 k - 242 k^{2} - 
       32 k^{3} - 595 N - 529 k\, 
      N - 54 k^{2} N 
      \nonu\\&+& 20 k^{3} N - 249 N^{2} - 16 k\, 
      N^{2} + 62 k^{2} N^{2} + 12 N^{3} + 58 k\, 
      N^{3} + 16 N^{4}), \nonu\\
c_ {82} & = &\frac{1} {3 (2 + N) (2 + k + N)^{4}}i\sqrt {2} (-156 - 371 k - 275 k^{2} - 62 k^{3} - 
      223 N - 478 k\, 
     N - 285 k^{2} N 
     \nonu\\&-& 40 k^{3} N - 45 N^{2} - 139 k\, 
     N^{2} - 64 k^{2} N^{2} + 48 N^{3} + 10 k\, 
     N^{3} + 16 N^{4}), \nonu\\
c_ {83} & = &\frac {2 i\sqrt {2} (18 + 21 k + 6 k^{2} + 22 N + 13 k\, 
     N + 6 N^{2})} {(2 + N) (2 + k + N)^{3}}, \nonu\\
c_ {84} & = & - \frac {8\sqrt {2} (10 + 17 k + 6 k^{2} + 14 N + 
       11 k\, N + 4 N^{2})} {(2 + N) (2 + k + N)^{4}}, \nonu\\
c_ {85} & = &\frac {2\sqrt {2} (60 - k - 14 k^{2} + 103 N + 61 k\, 
     N + 20 k^{2} N + 73 N^{2} + 42 k\, 
     N^{2} + 16 N^{3})} {3 (2 + N) (2 + k + N)^{4}}, \nonu\\
c_ {86} & = & - \frac{1}{(3 (2 + N) (2 + k + N)^{4})}\sqrt {2} (60 + 169 k + 154 k^{2} + 40 k^{3} + 
       149 N + 299 k\, 
      N + 174 k^{2} N 
      \nonu\\&+& 20 k^{3} N + 159 N^{2} + 212 k\, 
      N^{2} + 62 k^{2} N^{2} + 84 N^{3} + 58 k\, 
      N^{3} + 16 N^{4}), \nonu\\
c_ {87} & = &\frac {(300 + 371 k + 106 k^{2} + 559 N + 499 k\, 
    N + 80 k^{2} N + 337 N^{2} + 168 k\, 
    N^{2} + 64 N^{3})} {3\sqrt {2} (2 + N) (2 + k + N)^{2}}, \nonu\\
c_ {88} & = & - \frac {1}{3\sqrt {2} (2 + N) (2 + k + N)^{3}}(60 + 169 k + 154 k^{2} + 40 k^{3} + 149 N + 
      299 k\, N + 174 k^{2} N 
      \nonu\\&+& 20 k^{3} N + 159 N^{2} + 212 k\, 
     N^{2} + 62 k^{2} N^{2} + 84 N^{3} + 58 k\, 
     N^{3} + 16 N^{4}), \nonu\\
c_ {89} & = & - \frac {\sqrt {2} (300 + 371 k + 106 k^{2} + 559 N + 
       499 k\, N + 80 k^{2} N + 337 N^{2} + 168 k\, 
      N^{2} + 64 N^{3})} {3 (2 + N) (2 + k + N)^{2}}, \nonu\\
c_ {90} & = & - \frac{1}{(2 + N) (2 + k + N)^{4}}\sqrt {2} (-k + N) (60 + 77 k + 22 k^{2} + 
       121 N + 115 k\, N + 20 k^{2} N 
       \nonu\\&+& 79 N^{2} + 42 k\, 
      N^{2} + 16 N^{3}).
\nonu
\eea
Similarly, the exact relations between 
the remaining higher spin-$3, \frac{7}{2}, 4$ currents, which 
are not presented in this paper, 
can be obtained.



\begin{thebibliography}{99}

\bibitem{GG1305} 
  M.~R.~Gaberdiel and R.~Gopakumar,
  ``Large N=4 Holography,''
  JHEP {\bf 1309}, 036 (2013)
  [arXiv:1305.4181 [hep-th]].

\bibitem{GG1406} 
  M.~R.~Gaberdiel and R.~Gopakumar,
  ``Higher Spins $\&$ Strings,''
  JHEP {\bf 1411}, 044 (2014)
  [arXiv:1406.6103 [hep-th]].

\bibitem{GG1501} 
  M.~R.~Gaberdiel and R.~Gopakumar,
  ``Stringy Symmetries and the Higher Spin Square,''
  J.\ Phys.\ A {\bf 48}, no. 18, 185402 (2015)
  [arXiv:1501.07236 [hep-th]].

\bibitem{BGP} 
  M.~Baggio, M.~R.~Gaberdiel and C.~Peng,
  ``Higher spins in the symmetric orbifold of K3,''
  Phys.\ Rev.\ D {\bf 92}, no. 2, 026007 (2015)
  [arXiv:1504.00926 [hep-th]].

\bibitem{GPZ} 
  M.~R.~Gaberdiel, C.~Peng and I.~G.~Zadeh,
  ``Higgsing the stringy higher spin symmetry,''
  arXiv:1506.02045 [hep-th].

\bibitem{AK1506} 
  C.~Ahn and H.~Kim,
  ``Three Point Functions in the Large N=4 Holography,''
  arXiv:1506.00357 [hep-th].

\bibitem{GG1011} 
  M.~R.~Gaberdiel and R.~Gopakumar,
  ``An $AdS_3$ Dual for Minimal Model CFTs,''  
Phys.\ Rev.\ D {\bf 83}, 066007 (2011)  
[arXiv:1011.2986 [hep-th]].  

\bibitem{GG1205} 
  M.~R.~Gaberdiel and R.~Gopakumar,
  ``Triality in Minimal Model Holography,''
  JHEP {\bf 1207}, 127 (2012)
  [arXiv:1205.2472 [hep-th]].

\bibitem{AK1308} 
  C.~Ahn and H.~Kim,
  ``Spin-5 Casimir operator its three-point functions with two scalars,''
  JHEP {\bf 1401}, 012 (2014)
  [JHEP {\bf 1401}, 174 (2014)]
  [arXiv:1308.1726 [hep-th]].

\bibitem{Ahn1111} 
  C.~Ahn,
  ``The Coset Spin-4 Casimir Operator and Its Three-Point Functions with Scalars,''
  JHEP {\bf 1202}, 027 (2012)
  [arXiv:1111.0091 [hep-th]].

\bibitem{Ahn1504} 
  C.~Ahn,
  ``Higher Spin Currents in Wolf Space: Part III,''
  Class.\ Quant.\ Grav.\  {\bf 32}, no. 18, 185001 (2015)
  [arXiv:1504.00070 [hep-th]].

\bibitem{Thielemans} 
  K.~Thielemans,
  ``A Mathematica package for computing operator product expansions,''
  Int.\ J.\ Mod.\ Phys.\ C {\bf 2}, 787 (1991).

\bibitem{BS} 
  P.~Bouwknegt and K.~Schoutens,
  ``W symmetry in conformal field theory,''
  Phys.\ Rept.\  {\bf 223}, 183 (1993)
  [hep-th/9210010].

\bibitem{KT} 
  S.~Krivonos and K.~Thielemans,
  ``A Mathematica package for computing N=2 superfield operator product expansions,''
  Class.\ Quant.\ Grav.\  {\bf 13}, 2899 (1996)
  [hep-th/9512029].

\bibitem{RASS} 
  M.~Rocek, C.~Ahn, K.~Schoutens and A.~Sevrin,
  ``Superspace WZW models and black holes,''
  hep-th/9110035.

\bibitem{BO} 
  F.~Bastianelli and N.~Ohta,
  ``The Large N=4 superconformal algebra and its BRST operator,''
  Phys.\ Rev.\ D {\bf 50}, 4051 (1994)
  [hep-th/9402118].

\bibitem{Schoutensnpb} 
  K.~Schoutens,
  ``O(n) Extended Superconformal Field Theory in Superspace,''
  Nucl.\ Phys.\ B {\bf 295}, 634 (1988).

\bibitem{Ahn1992} 
  C.~Ahn,
  ``Extended conformal symmetry in two-dimensional quantum field theory,''
  UMI-93-10057.

\bibitem{Ademolloetal} 
  M.~Ademollo {\it et al.},
  ``Supersymmetric Strings and Color Confinement,''
  Phys.\ Lett.\ B {\bf 62}, 105 (1976).

\bibitem{Schoutensplb} 
  K.~Schoutens,
  ``A Nonlinear Representation of the $d=2$ SO(4) Extended Superconformal Algebra,''
  Phys.\ Lett.\ B {\bf 194}, 75 (1987).

\bibitem{IKL1} 
  E.~A.~Ivanov, S.~O.~Krivonos and V.~M.~Leviant,
  ``N=3 And N=4 Superconformal Wznw Sigma Models In Superspace. 1. General Formalism And N=3 Case,''
  Int.\ J.\ Mod.\ Phys.\ A {\bf 6}, 2147 (1991).

\bibitem{IKL2} 
  E.~A.~Ivanov, S.~O.~Krivonos and V.~M.~Leviant,
  ``N=3 and N=4 superconformal WZNW sigma models in superspace. 2: The N=4 case,''
  Int.\ J.\ Mod.\ Phys.\ A {\bf 7}, 287 (1992).

\bibitem{BCG} 
  M.~Beccaria, C.~Candu and M.~R.~Gaberdiel,
  ``The large N = 4 superconformal $W_{\infty}$ algebra,''
  JHEP {\bf 1406}, 117 (2014)
  [arXiv:1404.1694 [hep-th]].

\bibitem{STV} 
  A.~Sevrin, W.~Troost and A.~Van Proeyen,
  ``Superconformal Algebras in Two-Dimensions with N=4,''
  Phys.\ Lett.\ B {\bf 208}, 447 (1988).

\bibitem{Saulina} 
  N.~Saulina,
  ``Geometric interpretation of the large N=4 index,''
  Nucl.\ Phys.\ B {\bf 706}, 491 (2005)
  [hep-th/0409175].

\bibitem{ST} 
  A.~Sevrin and G.~Theodoridis,
  ``N=4 Superconformal Coset Theories,''
  Nucl.\ Phys.\ B {\bf 332}, 380 (1990).

\bibitem{GS} 
  P.~Goddard and A.~Schwimmer,
  ``Factoring Out Free Fermions and Superconformal Algebras,''
  Phys.\ Lett.\ B {\bf 214}, 209 (1988).

\bibitem{CK} 
  D.~Chang and A.~Kumar,
  ``Representations of $N=3$ Superconformal Algebra,''
  Phys.\ Lett.\ B {\bf 193}, 181 (1987).

\bibitem{OS} 
  N.~Ohta and T.~Shimizu,
  ``Universal string and small N=4 superstring,''
  Phys.\ Lett.\ B {\bf 355}, 127 (1995)
  [hep-th/9504099].

\bibitem{ASS} 
  C.~Ahn, K.~Schoutens and A.~Sevrin,
  ``The full structure of the super W(3) algebra,''
  Int.\ J.\ Mod.\ Phys.\ A {\bf 6}, 3467 (1991).

\bibitem{Ahn1208} 
  C.~Ahn,
  ``The Operator Product Expansion of the Lowest Higher Spin Current at Finite N,''
  JHEP {\bf 1301}, 041 (2013)
  [arXiv:1208.0058 [hep-th]].

\bibitem{Ahn1206} 
  C.~Ahn,
  ``The Large N 't Hooft Limit of Kazama-Suzuki Model,''
  JHEP {\bf 1208}, 047 (2012)
  [arXiv:1206.0054 [hep-th]].

\bibitem{AP} 
  C.~Ahn and J.~Paeng,
  ``Higher Spin Currents in Orthogonal Wolf Space,''
  Class.\ Quant.\ Grav.\  {\bf 32}, no. 4, 045011 (2015)
  [arXiv:1410.0080 [hep-th]].

\bibitem{FG} 
  K.~Ferreira and M.~R.~Gaberdiel,
  ``The so-Kazama-Suzuki Models at Large Level,''
  JHEP {\bf 1504}, 017 (2015)
  [arXiv:1412.7213 [hep-th]].

\bibitem{AP1310} 
  C.~Ahn and J.~Paeng,
  ``Higher Spin Currents in the Holographic $\mathcal{N} = 1$ Coset Minimal Model,''
  JHEP {\bf 1401}, 007 (2014)
  [arXiv:1310.6185 [hep-th]].

\bibitem{AP1301} 
  C.~Ahn and J.~Paeng,
  ``The OPEs of Spin-4 Casimir Currents in the Holographic $SO(N)$ Coset Minimal Models,''
  Class.\ Quant.\ Grav.\  {\bf 30}, 175004 (2013)
  [arXiv:1301.0208 [hep-th]].

\bibitem{Ahn1202} 
  C.~Ahn,
  ``The Primary Spin-4 Casimir Operators in the Holographic SO(N) Coset Minimal Models,''
  JHEP {\bf 1205}, 040 (2012)
  [arXiv:1202.0074 [hep-th]].

\bibitem{Ahn1106} 
  C.~Ahn,
  ``The Large N 't Hooft Limit of Coset Minimal Models,''
  JHEP {\bf 1110}, 125 (2011)
  [arXiv:1106.0351 [hep-th]].

\bibitem{AK1411} 
  C.~Ahn and H.~Kim,
  ``Higher Spin Currents in Wolf Space for Generic N,''
  JHEP {\bf 1412}, 109 (2014)
  [arXiv:1411.0356 [hep-th]].

\bibitem{Ahn1408} 
  C.~Ahn,
  ``Higher Spin Currents in Wolf Space: Part II,''
  Class.\ Quant.\ Grav.\  {\bf 32}, no. 1, 015023 (2015)
  [arXiv:1408.0655 [hep-th]].

\bibitem{Ahn1311} 
  C.~Ahn,
  ``Higher Spin Currents in Wolf Space. Part I,''
  JHEP {\bf 1403}, 091 (2014)
  [arXiv:1311.6205 [hep-th]].

\bibitem{CH1506} 
  T.~Creutzig and Y.~Hikida,
  ``Higgs phenomenon for higher spin fields on $AdS_3$,''
  arXiv:1506.04465 [hep-th].

\bibitem{HR1503} 
  Y.~Hikida and P.~B.~R©ªnne,
  ``Marginal deformations and the Higgs phenomenon in higher spin AdS$_{3}$ 
holography,''
  JHEP {\bf 1507}, 125 (2015)
  [arXiv:1503.03870 [hep-th]].

\bibitem{CHR1406} 
  T.~Creutzig, Y.~Hikida and P.~B.~Ronne,
  ``Higher spin AdS$_{3}$ holography with extended supersymmetry,''
  JHEP {\bf 1410}, 163 (2014)
  [arXiv:1406.1521 [hep-th]].

\bibitem{Ahn1305} 
  C.~Ahn,
  ``Higher Spin Currents with Arbitrary N in the ${ \cal N} = 1$ 
Stringy Coset Minimal Model,''
  JHEP {\bf 1307}, 141 (2013)
  [arXiv:1305.5892 [hep-th]].

\bibitem{Ahn1211} 
  C.~Ahn,
  ``The Higher Spin Currents in the N=1 Stringy Coset Minimal Model,''
  JHEP {\bf 1304}, 033 (2013)
  [arXiv:1211.2589 [hep-th]].

\bibitem{STVS} 
  A.~Sevrin, W.~Troost, A.~Van Proeyen and P.~Spindel,
  ``EXTENDED SUPERSYMMETRIC sigma MODELS ON GROUP MANIFOLDS. 2. CURRENT ALGEBRAS,''
  Nucl.\ Phys.\ B {\bf 311}, 465 (1988).

\end{thebibliography}
\end{document}